\documentclass{article}[10pt,letterpaper]
\usepackage{fancyhdr}
\usepackage{supertabular}
\renewcommand{\sectionmark}[1]{\markboth{#1}{}}
\usepackage{color}
\usepackage{amsmath}
\usepackage{pbox}
\usepackage{amssymb}
\usepackage{amsbsy}
\usepackage{scalefnt}
\usepackage{ifpdf}
\usepackage{graphicx}
\usepackage{multicol}
\usepackage[colorlinks=true, linkcolor=blue, linktocpage=true,urlcolor=blue]{hyperref}
\usepackage{longtable}
\usepackage{needspace}
\usepackage{eso-pic}
\usepackage{mathptmx}
\usepackage{multirow}
\usepackage{lscape}
\usepackage{textcomp}
\usepackage{changepage}
\usepackage{xfrac}
\usepackage{txfonts}
\usepackage{pdfcomment}
\usepackage[absolute]{textpos}
\ifpdf
\fi
\makeatletter
\renewcommand{\l@section}{\@dottedtocline{2}{1em}{0em}}
\renewcommand{\l@subsection}{\@dottedtocline{3}{2.5cm}{0em}}
\renewcommand{\hrulefill}{\leavevmode \leaders \hrule \@height 1pt \hfill \kern\z@}
\makeatother
\renewcommand{\underline}[1]{\begin{tabular}{@{\extracolsep{\fill}}c@{\extracolsep{\fill}}}#1\\[-0.2cm]\hrulefill\end{tabular}}
\renewcommand{\footrulewidth}{1pt}
\renewcommand{\headrulewidth}{1pt}
\newcounter{chappage}
\AddToShipoutPicture{\stepcounter{chappage}}
\fancypagestyle{plain}{\lhead{}\rhead{}}
\fancypagestyle{single}{\lhead{\leftmark \ \\}\rhead{\leftmark \ \\}}
\fancypagestyle{bob}{\lhead{\leftmark -\thechappage \ \\}\rhead{\leftmark -\thechappage \ \\}}
\begin{document}
\fontsize{9}{10}\selectfont
\fontdimen2\font=1.3\fontdimen2\font
\thispagestyle{empty}
\setcounter{page}{1}
\begin{center}

\vspace{0.5cm}
{ \huge Energy Levels of \ensuremath{^{\textnormal{19}}}Ne*}\\
\vspace{1.0cm}
{ \normalsize K. Setoodehnia\ensuremath{^{\textnormal{1}}}, J. H. Kelley\ensuremath{^{\textnormal{1,2}}} and C. G. Sheu\ensuremath{^{\textnormal{1}}}}\\
\vspace{0.2in}
{ \small \it \ensuremath{^{\textnormal{1}}}Triangle Universities Nuclear Laboratory, Duke University,\\
  Durham North Carolina 27708, USA.\\
  \ensuremath{^{\textnormal{2}}}Department of Physics, North Carolina State University\\
  Raleigh, North Carolina 27607, USA}\\
\vspace{0.2in}
\end{center}

\setlength{\parindent}{-0.5cm}
\addtolength{\leftskip}{2cm}
\addtolength{\rightskip}{2cm}
{\bf Abstract: }
In this document, experimental nuclear structure data are evaluated for \ensuremath{^{\textnormal{19}}}Ne. \ensuremath{^{\textnormal{19}}}Ne was first identified by (\href{https://www.nndc.bnl.gov/nsr/nsrlink.jsp?1939Fo01,B}{1939Fo01}), see (\href{https://www.nndc.bnl.gov/nsr/nsrlink.jsp?2012Th01,B}{2012Th01}). The details of each reaction and decay experiment populating \ensuremath{^{\textnormal{19}}}Ne levels are compiled and evaluated. The combined results provide a set of adopted values that include level energies, spins and parities, level half-lives, \ensuremath{\gamma}-ray energies, decay types and branching ratios, and other nuclear properties. This work supersedes the earlier work by Ron Tilley (\href{https://www.nndc.bnl.gov/nsr/nsrlink.jsp?1995Ti07,B}{1995Ti07}) published in Nuclear Physics A \textbf{595} (1995) 1. The earlier evaluations were published by Fay Ajzenberg-Selove in (\href{https://www.nndc.bnl.gov/nsr/nsrlink.jsp?1959Aj76,B}{1959Aj76}, \href{https://www.nndc.bnl.gov/nsr/nsrlink.jsp?1972Aj02,B}{1972Aj02}, \href{https://www.nndc.bnl.gov/nsr/nsrlink.jsp?1978Aj03,B}{1978Aj03}, \href{https://www.nndc.bnl.gov/nsr/nsrlink.jsp?1983Aj01,B}{1983Aj01}, and \href{https://www.nndc.bnl.gov/nsr/nsrlink.jsp?1987Aj02,B}{1987Aj02}).\\

{\bf Cutoff Date: }
Literature available up to June 30, 2025 has been considered; the primary bibliographic source, the NSR database (\href{https://www.nndc.bnl.gov/nsr/nsrlink.jsp?2011Pr03,B}{2011Pr03}) available at Brookhaven National Laboratory web page: www.nndc.bnl.gov/nsr/.\\

{\bf General Policies and Organization of Material: }
See the April 2025 issue of the {\it Nuclear Data Sheets} or \\https://www.nndc.bnl.gov/nds/docs/NDSPolicies.pdf. \\

{\bf Acknowledgements: }
The authors expresses her gratitude to personnel at the National Nuclear Data Center (NNDC) at Brookhaven National Laboratory for facilitating this work.\\

\vfill

* This work is supported by the Office of Nuclear Physics, Office of Science, U.S. Department of Energy under contracts: DE-FG02-97ER41042 {\textminus} North Carolina State University and DE-FG02-97ER41033 {\textminus} Duke University\\

\setlength{\parindent}{+0.5cm}
\addtolength{\leftskip}{-2cm}
\addtolength{\rightskip}{-2cm}
\newpage
\pagestyle{plain}
\setlength{\columnseprule}{1pt}
\setlength{\columnsep}{1cm}
\begin{center}
\underline{\normalsize Index for A=19}
\end{center}
\hspace{.3cm}\raggedright\underline{Nuclide}\hspace{1cm}\underline{Data Type\mbox{\hspace{2.3cm}}}\hspace{2cm}\underline{Page}\hspace{1cm}
\raggedright\underline{Nuclide}\hspace{1cm}\underline{Data Type\mbox{\hspace{2.3cm}}}\hspace{2cm}\underline{Page}
\begin{adjustwidth}{}{0.05\textwidth}
\begin{multicols}{2}
\setcounter{tocdepth}{3}
\renewcommand{\contentsname}{\protect\vspace{-0.8cm}}
\tableofcontents
\end{multicols}
\end{adjustwidth}
\clearpage
\thispagestyle{empty}
\mbox{}
\clearpage
\clearpage
\pagestyle{bob}
\begin{center}
\section[\ensuremath{^{19}_{10}}Ne\ensuremath{_{9}^{~}}]{ }
\vspace{-30pt}
\setcounter{chappage}{1}
\subsection[\hspace{-0.2cm}Adopted Levels, Gammas]{ }
\vspace{-20pt}
\vspace{0.3cm}
\hypertarget{NE0}{{\bf \small \underline{Adopted \hyperlink{19NE_LEVEL}{Levels}, \hyperlink{19NE_GAMMA}{Gammas}}}}\\
\vspace{4pt}
\vspace{8pt}
\parbox[b][0.3cm]{17.7cm}{\addtolength{\parindent}{-0.2in}Q(\ensuremath{\beta^-})=$-$11177 {\it 11}; S(n)=11636.9 {\it 4}; S(p)=6410.0 {\it 5}; Q(\ensuremath{\alpha})=$-$3528.5 {\it 5}\hspace{0.2in}\href{https://www.nndc.bnl.gov/nsr/nsrlink.jsp?2021Wa16,B}{2021Wa16}}\\
\parbox[b][0.3cm]{17.7cm}{\addtolength{\parindent}{-0.2in}S\ensuremath{_{\textnormal{2n}}}=30891.0 keV \textit{4}, S\ensuremath{_{\textnormal{2p}}}=12017.12 keV \textit{16} (\href{https://www.nndc.bnl.gov/nsr/nsrlink.jsp?2021Wa16,B}{2021Wa16}).}\\

\vspace{0.385cm}
\parbox[b][0.3cm]{17.7cm}{\addtolength{\parindent}{-0.2in}\textit{Experimental (see also the individual datasets)}:}\\
\parbox[b][0.3cm]{17.7cm}{\addtolength{\parindent}{-0.2in}\textit{Measured mass, charge radius, and matter radius}: \href{https://www.nndc.bnl.gov/nsr/nsrlink.jsp?2004Bl20,B}{2004Bl20}, \href{https://www.nndc.bnl.gov/nsr/nsrlink.jsp?2008Ge07,B}{2008Ge07}, \href{https://www.nndc.bnl.gov/nsr/nsrlink.jsp?2011Ma48,B}{2011Ma48}.}\\
\parbox[b][0.3cm]{17.7cm}{\addtolength{\parindent}{-0.2in}\textit{Measured NMR, deduced \ensuremath{\mu}, \ensuremath{^{19}}Ne(\ensuremath{\beta}\ensuremath{^{\textnormal{+}}})}: \href{https://www.nndc.bnl.gov/nsr/nsrlink.jsp?1954Ki53,B}{1954Ki53} (compilation), \href{https://www.nndc.bnl.gov/nsr/nsrlink.jsp?1960Wa04,B}{1960Wa04}, \href{https://www.nndc.bnl.gov/nsr/nsrlink.jsp?1962Ea02,B}{1962Ea02}, \href{https://www.nndc.bnl.gov/nsr/nsrlink.jsp?1964Va23,B}{1964Va23}, \href{https://www.nndc.bnl.gov/nsr/nsrlink.jsp?1969Ca14,B}{1969Ca14}, \href{https://www.nndc.bnl.gov/nsr/nsrlink.jsp?1974Ca17,B}{1974Ca17},}\\
\parbox[b][0.3cm]{17.7cm}{\href{https://www.nndc.bnl.gov/nsr/nsrlink.jsp?1974Ma31,B}{1974Ma31}, \href{https://www.nndc.bnl.gov/nsr/nsrlink.jsp?1975Ca28,B}{1975Ca28}, \href{https://www.nndc.bnl.gov/nsr/nsrlink.jsp?1975Fr15,B}{1975Fr15}, \href{https://www.nndc.bnl.gov/nsr/nsrlink.jsp?1975FrZY,B}{1975FrZY}, \href{https://www.nndc.bnl.gov/nsr/nsrlink.jsp?1975MaXA,B}{1975MaXA}, \href{https://www.nndc.bnl.gov/nsr/nsrlink.jsp?1975MaXV,B}{1975MaXV}, \href{https://www.nndc.bnl.gov/nsr/nsrlink.jsp?1975VaZR,B}{1975VaZR}, \href{https://www.nndc.bnl.gov/nsr/nsrlink.jsp?1976Al07,B}{1976Al07}, \href{https://www.nndc.bnl.gov/nsr/nsrlink.jsp?1980MaZR,B}{1980MaZR}, \href{https://www.nndc.bnl.gov/nsr/nsrlink.jsp?1981Ad05,B}{1981Ad05}, \href{https://www.nndc.bnl.gov/nsr/nsrlink.jsp?1982Ma39,B}{1982Ma39},}\\
\parbox[b][0.3cm]{17.7cm}{\href{https://www.nndc.bnl.gov/nsr/nsrlink.jsp?1983Ad03,B}{1983Ad03}, \href{https://www.nndc.bnl.gov/nsr/nsrlink.jsp?1983HaZD,B}{1983HaZD}, \href{https://www.nndc.bnl.gov/nsr/nsrlink.jsp?1983MaZA,B}{1983MaZA}, \href{https://www.nndc.bnl.gov/nsr/nsrlink.jsp?1983Sc32,B}{1983Sc32}, \href{https://www.nndc.bnl.gov/nsr/nsrlink.jsp?1983ScZM,B}{1983ScZM}, \href{https://www.nndc.bnl.gov/nsr/nsrlink.jsp?1983ScZQ,B}{1983ScZQ}, \href{https://www.nndc.bnl.gov/nsr/nsrlink.jsp?1984Ha01,B}{1984Ha01}, \href{https://www.nndc.bnl.gov/nsr/nsrlink.jsp?1985PiZY,B}{1985PiZY}, \href{https://www.nndc.bnl.gov/nsr/nsrlink.jsp?1987SeZL,B}{1987SeZL}, \href{https://www.nndc.bnl.gov/nsr/nsrlink.jsp?1987SeZR,B}{1987SeZR}, \href{https://www.nndc.bnl.gov/nsr/nsrlink.jsp?1988Se11,B}{1988Se11},}\\
\parbox[b][0.3cm]{17.7cm}{\href{https://www.nndc.bnl.gov/nsr/nsrlink.jsp?1992Ge08,B}{1992Ge08}, \href{https://www.nndc.bnl.gov/nsr/nsrlink.jsp?1993Sa32,B}{1993Sa32}, \href{https://www.nndc.bnl.gov/nsr/nsrlink.jsp?2009Na06,B}{2009Na06} (Gamow-Teller to Fermi mixing ratio for \ensuremath{^{\textnormal{19}}}Ne(\ensuremath{\beta}) decay), \href{https://www.nndc.bnl.gov/nsr/nsrlink.jsp?2011TrZX,B}{2011TrZX}, \href{https://www.nndc.bnl.gov/nsr/nsrlink.jsp?2012Tr09,B}{2012Tr09}, \href{https://www.nndc.bnl.gov/nsr/nsrlink.jsp?2012Uj01,B}{2012Uj01},}\\
\parbox[b][0.3cm]{17.7cm}{\href{https://www.nndc.bnl.gov/nsr/nsrlink.jsp?2018Fa02,B}{2018Fa02}, \href{https://www.nndc.bnl.gov/nsr/nsrlink.jsp?2019Re07,B}{2019Re07}.}\\
\parbox[b][0.3cm]{17.7cm}{\addtolength{\parindent}{-0.2in}\textit{Measured nuclear moment}: \href{https://www.nndc.bnl.gov/nsr/nsrlink.jsp?1962Do03,B}{1962Do03}, \href{https://www.nndc.bnl.gov/nsr/nsrlink.jsp?1963Co22,B}{1963Co22}, \href{https://www.nndc.bnl.gov/nsr/nsrlink.jsp?1963Do15,B}{1963Do15}.}\\
\parbox[b][0.3cm]{17.7cm}{\addtolength{\parindent}{-0.2in}{\ensuremath{\mu}, Quadrupole moment}: \href{https://www.nndc.bnl.gov/nsr/nsrlink.jsp?1982Ma39,B}{1982Ma39}: \ensuremath{\mu}={\textminus}1.88542 \ensuremath{\mu}\ensuremath{_{\textnormal{N}}} \textit{8}, \href{https://www.nndc.bnl.gov/nsr/nsrlink.jsp?2005Ge06,B}{2005Ge06}: \ensuremath{\mu}={\textminus}1.8846 \ensuremath{\mu}\ensuremath{_{\textnormal{N}}} \textit{8}.}\\
\parbox[b][0.3cm]{17.7cm}{\addtolength{\parindent}{-0.2in}\textit{Discussion on superallowed \ensuremath{\beta} decays, \ensuremath{^{19}}Ne(\ensuremath{\beta}\ensuremath{^{\textnormal{+}}},EC)}: \href{https://www.nndc.bnl.gov/nsr/nsrlink.jsp?2015Gr05,B}{2015Gr05}.}\\
\parbox[b][0.3cm]{17.7cm}{\addtolength{\parindent}{-0.2in}\textit{Other reactions that populated \ensuremath{^{19}}Ne states}:}\\
\parbox[b][0.3cm]{17.7cm}{\addtolength{\parindent}{-0.2in}R. Lewis, J. A. Caggiano, D. W. Visser, P. D. Parker, A. A. Chen, W. B. Handler,}\\
\parbox[b][0.3cm]{17.7cm}{\textit{Structure of \ensuremath{^{19}}Ne from the \ensuremath{^{\textnormal{12}}}C(\ensuremath{^{\textnormal{10}}}B,t) reaction}. Fall Meeting of the Division of Nuclear Physics, East Lansing, MI, Bulletin of the}\\
\parbox[b][0.3cm]{17.7cm}{American Physical Society 47 (2002) 68: \ensuremath{^{\textnormal{12}}}C(\ensuremath{^{\textnormal{10}}}B,t) E=29 and 35 MeV; studied \ensuremath{^{\textnormal{19}}}Ne* states with E\ensuremath{_{\textnormal{x}}}=0-10 MeV using the Yale}\\
\parbox[b][0.3cm]{17.7cm}{Enge split-pole spectrograph placed at \ensuremath{\theta}\ensuremath{_{\textnormal{lab}}}=5\ensuremath{^\circ}, 10\ensuremath{^\circ} and 15\ensuremath{^\circ}. The results are not published.}\\
\parbox[b][0.3cm]{17.7cm}{\addtolength{\parindent}{-0.2in}\href{https://www.nndc.bnl.gov/nsr/nsrlink.jsp?2016Be32,B}{2016Be32}: \ensuremath{^{\textnormal{2}}}H(\ensuremath{^{\textnormal{19}}}Ne,\ensuremath{^{\textnormal{20}}}Na*\ensuremath{\rightarrow}p+\ensuremath{^{\textnormal{19}}}Ne*(0, 238)) E=86 MeV; measured decay products from the populated, low-lying proton}\\
\parbox[b][0.3cm]{17.7cm}{resonances in \ensuremath{^{\textnormal{20}}}Na*, which proton decayed to the \ensuremath{^{\textnormal{19}}}Ne\ensuremath{_{\textnormal{g.s.}}} and \ensuremath{^{\textnormal{19}}}Ne*(238) states.}\\
\vspace{0.385cm}
\parbox[b][0.3cm]{17.7cm}{\addtolength{\parindent}{-0.2in}\textit{Theory (see also individual datasets)}:}\\
\parbox[b][0.3cm]{17.7cm}{\addtolength{\parindent}{-0.2in}\textit{Reaction cross sections}: \href{https://www.nndc.bnl.gov/nsr/nsrlink.jsp?1991Re10,B}{1991Re10}: \ensuremath{^{\textnormal{20}}}Ne(n,2n) E=0-6 MeV; evaluated reaction cross section by modeling of the reaction}\\
\parbox[b][0.3cm]{17.7cm}{mechanisms involved.}\\
\parbox[b][0.3cm]{17.7cm}{\addtolength{\parindent}{-0.2in}\textit{\ensuremath{\alpha} Cluster structure and Coulomb displacement energies}: \href{https://www.nndc.bnl.gov/nsr/nsrlink.jsp?1972Ne22,B}{1972Ne22}, \href{https://www.nndc.bnl.gov/nsr/nsrlink.jsp?1977Bu05,B}{1977Bu05} (\ensuremath{\alpha}-cluster model), \href{https://www.nndc.bnl.gov/nsr/nsrlink.jsp?1979Sa41,B}{1979Sa41} (semi-microscopic}\\
\parbox[b][0.3cm]{17.7cm}{model), \href{https://www.nndc.bnl.gov/nsr/nsrlink.jsp?1979Sa43,B}{1979Sa43} (semi-microscopic model), \href{https://www.nndc.bnl.gov/nsr/nsrlink.jsp?2008Ne13,B}{2008Ne13} (fermionic molecular dynamics model), \href{https://www.nndc.bnl.gov/nsr/nsrlink.jsp?2008NeZX,B}{2008NeZX}.}\\
\parbox[b][0.3cm]{17.7cm}{\addtolength{\parindent}{-0.2in}\textit{Giant multipole resonances}: \href{https://www.nndc.bnl.gov/nsr/nsrlink.jsp?1972Le06,B}{1972Le06}, \href{https://www.nndc.bnl.gov/nsr/nsrlink.jsp?1977Sc08,B}{1977Sc08}, \href{https://www.nndc.bnl.gov/nsr/nsrlink.jsp?1978Sc19,B}{1978Sc19}.}\\
\parbox[b][0.3cm]{17.7cm}{\addtolength{\parindent}{-0.2in}\textit{Calculated \ensuremath{\mu}, quadrupole moment, and electromagnetic moment}: \href{https://www.nndc.bnl.gov/nsr/nsrlink.jsp?1968Pe16,B}{1968Pe16}, \href{https://www.nndc.bnl.gov/nsr/nsrlink.jsp?1971Ar25,B}{1971Ar25}, \href{https://www.nndc.bnl.gov/nsr/nsrlink.jsp?1973MeZG,B}{1973MeZG}, \href{https://www.nndc.bnl.gov/nsr/nsrlink.jsp?1976Pa03,B}{1976Pa03}, \href{https://www.nndc.bnl.gov/nsr/nsrlink.jsp?1977Bu05,B}{1977Bu05} (see}\\
\parbox[b][0.3cm]{17.7cm}{Table 9 and references therein) \href{https://www.nndc.bnl.gov/nsr/nsrlink.jsp?1978Le03,B}{1978Le03}, \href{https://www.nndc.bnl.gov/nsr/nsrlink.jsp?1978Ma54,B}{1978Ma54}, \href{https://www.nndc.bnl.gov/nsr/nsrlink.jsp?1979Sa43,B}{1979Sa43}.}\\
\parbox[b][0.3cm]{17.7cm}{\addtolength{\parindent}{-0.2in}\textit{Calculated quadrupole deformation}: \href{https://www.nndc.bnl.gov/nsr/nsrlink.jsp?2022Su17,B}{2022Su17}.}\\
\parbox[b][0.3cm]{17.7cm}{\addtolength{\parindent}{-0.2in}\textit{Mass excess}: \href{https://www.nndc.bnl.gov/nsr/nsrlink.jsp?1986RoZQ,B}{1986RoZQ}.}\\
\parbox[b][0.3cm]{17.7cm}{\addtolength{\parindent}{-0.2in}\textit{Matter radius}: \href{https://www.nndc.bnl.gov/nsr/nsrlink.jsp?2001Oz04,B}{2001Oz04}, \href{https://www.nndc.bnl.gov/nsr/nsrlink.jsp?2008Ge07,B}{2008Ge07}, \href{https://www.nndc.bnl.gov/nsr/nsrlink.jsp?2019Oh03,B}{2019Oh03}, \href{https://www.nndc.bnl.gov/nsr/nsrlink.jsp?2024Zh35,B}{2024Zh35}.}\\
\parbox[b][0.3cm]{17.7cm}{\addtolength{\parindent}{-0.2in}\textit{Mirror nuclei}: \href{https://www.nndc.bnl.gov/nsr/nsrlink.jsp?1969Mu09,B}{1969Mu09}, \href{https://www.nndc.bnl.gov/nsr/nsrlink.jsp?1970St04,B}{1970St04}, \href{https://www.nndc.bnl.gov/nsr/nsrlink.jsp?1972Ga14,B}{1972Ga14}, \href{https://www.nndc.bnl.gov/nsr/nsrlink.jsp?1974Ts03,B}{1974Ts03}, \href{https://www.nndc.bnl.gov/nsr/nsrlink.jsp?1992Wa22,B}{1992Wa22}, \href{https://www.nndc.bnl.gov/nsr/nsrlink.jsp?2000Fo01,B}{2000Fo01}, \href{https://www.nndc.bnl.gov/nsr/nsrlink.jsp?2003Fo15,B}{2003Fo15}, \href{https://www.nndc.bnl.gov/nsr/nsrlink.jsp?2007Ti02,B}{2007Ti02}, \href{https://www.nndc.bnl.gov/nsr/nsrlink.jsp?2010Fo07,B}{2010Fo07}, \href{https://www.nndc.bnl.gov/nsr/nsrlink.jsp?2018Fo04,B}{2018Fo04},}\\
\parbox[b][0.3cm]{17.7cm}{\href{https://www.nndc.bnl.gov/nsr/nsrlink.jsp?2021Am03,B}{2021Am03}, \href{https://www.nndc.bnl.gov/nsr/nsrlink.jsp?2021Ma33,B}{2021Ma33}, \href{https://www.nndc.bnl.gov/nsr/nsrlink.jsp?2022Zo01,B}{2022Zo01}, \href{https://www.nndc.bnl.gov/nsr/nsrlink.jsp?2023Li03,B}{2023Li03}.}\\
\parbox[b][0.3cm]{17.7cm}{\addtolength{\parindent}{-0.2in}{Shell model}: J. P. Elliott and B. H. Flowers, Proc. R. Soc. Lond. A 229 (1955) 536, \href{https://www.nndc.bnl.gov/nsr/nsrlink.jsp?1955Re53,B}{1955Re53}, \href{https://www.nndc.bnl.gov/nsr/nsrlink.jsp?1967En01,B}{1967En01}, \href{https://www.nndc.bnl.gov/nsr/nsrlink.jsp?1970El23,B}{1970El23}, \href{https://www.nndc.bnl.gov/nsr/nsrlink.jsp?1971Ar25,B}{1971Ar25},}\\
\parbox[b][0.3cm]{17.7cm}{\href{https://www.nndc.bnl.gov/nsr/nsrlink.jsp?1973Mc06,B}{1973Mc06}, \href{https://www.nndc.bnl.gov/nsr/nsrlink.jsp?1977An12,B}{1977An12}, \href{https://www.nndc.bnl.gov/nsr/nsrlink.jsp?1982Ki02,B}{1982Ki02}, \href{https://www.nndc.bnl.gov/nsr/nsrlink.jsp?1987Po01,B}{1987Po01}, \href{https://www.nndc.bnl.gov/nsr/nsrlink.jsp?1997Bo47,B}{1997Bo47}, \href{https://www.nndc.bnl.gov/nsr/nsrlink.jsp?2016Ja03,B}{2016Ja03}, \href{https://www.nndc.bnl.gov/nsr/nsrlink.jsp?2016St12,B}{2016St12}, \href{https://www.nndc.bnl.gov/nsr/nsrlink.jsp?2017Sa48,B}{2017Sa48}, \href{https://www.nndc.bnl.gov/nsr/nsrlink.jsp?2018Ka12,B}{2018Ka12}, \href{https://www.nndc.bnl.gov/nsr/nsrlink.jsp?2018Mi22,B}{2018Mi22}, \href{https://www.nndc.bnl.gov/nsr/nsrlink.jsp?2019BaZS,B}{2019BaZS},}\\
\parbox[b][0.3cm]{17.7cm}{\href{https://www.nndc.bnl.gov/nsr/nsrlink.jsp?2019Sa38,B}{2019Sa38}, \href{https://www.nndc.bnl.gov/nsr/nsrlink.jsp?2020Ma25,B}{2020Ma25}, \href{https://www.nndc.bnl.gov/nsr/nsrlink.jsp?2021Sa49,B}{2021Sa49}.}\\
\parbox[b][0.3cm]{17.7cm}{\addtolength{\parindent}{-0.2in}\textit{Other theoretical analyses}: \href{https://www.nndc.bnl.gov/nsr/nsrlink.jsp?1972En03,B}{1972En03}, \href{https://www.nndc.bnl.gov/nsr/nsrlink.jsp?1973Pe09,B}{1973Pe09}, \href{https://www.nndc.bnl.gov/nsr/nsrlink.jsp?1976Iw03,B}{1976Iw03}, \href{https://www.nndc.bnl.gov/nsr/nsrlink.jsp?1977Sh13,B}{1977Sh13}, \href{https://www.nndc.bnl.gov/nsr/nsrlink.jsp?1979Ma27,B}{1979Ma27}, \href{https://www.nndc.bnl.gov/nsr/nsrlink.jsp?1985Al21,B}{1985Al21}, \href{https://www.nndc.bnl.gov/nsr/nsrlink.jsp?1995Ho13,B}{1995Ho13}, \href{https://www.nndc.bnl.gov/nsr/nsrlink.jsp?1996Go38,B}{1996Go38}, \href{https://www.nndc.bnl.gov/nsr/nsrlink.jsp?1997Po23,B}{1997Po23},}\\
\parbox[b][0.3cm]{17.7cm}{\href{https://www.nndc.bnl.gov/nsr/nsrlink.jsp?2001Ga46,B}{2001Ga46}, \href{https://www.nndc.bnl.gov/nsr/nsrlink.jsp?2001Na02,B}{2001Na02}, \href{https://www.nndc.bnl.gov/nsr/nsrlink.jsp?2001Oz04,B}{2001Oz04}, \href{https://www.nndc.bnl.gov/nsr/nsrlink.jsp?2004Ge02,B}{2004Ge02}, \href{https://www.nndc.bnl.gov/nsr/nsrlink.jsp?2004La24,B}{2004La24}, \href{https://www.nndc.bnl.gov/nsr/nsrlink.jsp?2004Sa58,B}{2004Sa58}, \href{https://www.nndc.bnl.gov/nsr/nsrlink.jsp?2005Ch71,B}{2005Ch71}, \href{https://www.nndc.bnl.gov/nsr/nsrlink.jsp?2005Ni24,B}{2005Ni24}, \href{https://www.nndc.bnl.gov/nsr/nsrlink.jsp?2014Ch39,B}{2014Ch39}, \href{https://www.nndc.bnl.gov/nsr/nsrlink.jsp?2017Ah08,B}{2017Ah08}, \href{https://www.nndc.bnl.gov/nsr/nsrlink.jsp?2019Ra09,B}{2019Ra09},}\\
\parbox[b][0.3cm]{17.7cm}{\href{https://www.nndc.bnl.gov/nsr/nsrlink.jsp?2020An13,B}{2020An13}, \href{https://www.nndc.bnl.gov/nsr/nsrlink.jsp?2020Ma17,B}{2020Ma17}, \href{https://www.nndc.bnl.gov/nsr/nsrlink.jsp?2021He03,B}{2021He03}, \href{https://www.nndc.bnl.gov/nsr/nsrlink.jsp?2022Gu11,B}{2022Gu11}, \href{https://www.nndc.bnl.gov/nsr/nsrlink.jsp?2022St03,B}{2022St03}, \href{https://www.nndc.bnl.gov/nsr/nsrlink.jsp?2023Al14,B}{2023Al14}, \href{https://www.nndc.bnl.gov/nsr/nsrlink.jsp?2023Di08,B}{2023Di08}, \href{https://www.nndc.bnl.gov/nsr/nsrlink.jsp?2023Fo05,B}{2023Fo05}, \href{https://www.nndc.bnl.gov/nsr/nsrlink.jsp?2023Sa22,B}{2023Sa22}, \href{https://www.nndc.bnl.gov/nsr/nsrlink.jsp?2024Xu14,B}{2024Xu14}.}\\
\parbox[b][0.3cm]{17.7cm}{\addtolength{\parindent}{-0.2in}\textit{Superallowed \ensuremath{\beta} decays, \ensuremath{^{19}}Ne(\ensuremath{\beta}\ensuremath{^{\textnormal{+}}}, EC)}: \href{https://www.nndc.bnl.gov/nsr/nsrlink.jsp?1952Ko35,B}{1952Ko35}, \href{https://www.nndc.bnl.gov/nsr/nsrlink.jsp?1952Tr01,B}{1952Tr01}, A. Winther and O. Kofoed-Hansen, Kgl. Danske Videnskab.}\\
\parbox[b][0.3cm]{17.7cm}{Selskab, Mat.-fys. Medd. 27, No. 14 (1953), \href{https://www.nndc.bnl.gov/nsr/nsrlink.jsp?1963Ba21,B}{1963Ba21}, \href{https://www.nndc.bnl.gov/nsr/nsrlink.jsp?1963Ba72,B}{1963Ba72}, \href{https://www.nndc.bnl.gov/nsr/nsrlink.jsp?1965Ha31,B}{1965Ha31}, \href{https://www.nndc.bnl.gov/nsr/nsrlink.jsp?1970Br24,B}{1970Br24}, \href{https://www.nndc.bnl.gov/nsr/nsrlink.jsp?1970Ko41,B}{1970Ko41}, \href{https://www.nndc.bnl.gov/nsr/nsrlink.jsp?1970Mc23,B}{1970Mc23}, \href{https://www.nndc.bnl.gov/nsr/nsrlink.jsp?1970Va29,B}{1970Va29},}\\
\parbox[b][0.3cm]{17.7cm}{\href{https://www.nndc.bnl.gov/nsr/nsrlink.jsp?1973Su04,B}{1973Su04}, \href{https://www.nndc.bnl.gov/nsr/nsrlink.jsp?1973Wi04,B}{1973Wi04}, \href{https://www.nndc.bnl.gov/nsr/nsrlink.jsp?1973Wi11,B}{1973Wi11}, \href{https://www.nndc.bnl.gov/nsr/nsrlink.jsp?1975Ha29,B}{1975Ha29}, \href{https://www.nndc.bnl.gov/nsr/nsrlink.jsp?1976Ba19,B}{1976Ba19}, \href{https://www.nndc.bnl.gov/nsr/nsrlink.jsp?1977Kl09,B}{1977Kl09}, \href{https://www.nndc.bnl.gov/nsr/nsrlink.jsp?1978Ig03,B}{1978Ig03}, \href{https://www.nndc.bnl.gov/nsr/nsrlink.jsp?1979Lo14,B}{1979Lo14}, \href{https://www.nndc.bnl.gov/nsr/nsrlink.jsp?1980An31,B}{1980An31}, \href{https://www.nndc.bnl.gov/nsr/nsrlink.jsp?1983Ca03,B}{1983Ca03}, \href{https://www.nndc.bnl.gov/nsr/nsrlink.jsp?1983Vo05,B}{1983Vo05},}\\
\parbox[b][0.3cm]{17.7cm}{\href{https://www.nndc.bnl.gov/nsr/nsrlink.jsp?1985Gi09,B}{1985Gi09}, \href{https://www.nndc.bnl.gov/nsr/nsrlink.jsp?1988HaZM,B}{1988HaZM}, \href{https://www.nndc.bnl.gov/nsr/nsrlink.jsp?1989Sa55,B}{1989Sa55}, \href{https://www.nndc.bnl.gov/nsr/nsrlink.jsp?1992Ca12,B}{1992Ca12}, \href{https://www.nndc.bnl.gov/nsr/nsrlink.jsp?1992He12,B}{1992He12}, \href{https://www.nndc.bnl.gov/nsr/nsrlink.jsp?1992Se08,B}{1992Se08}, \href{https://www.nndc.bnl.gov/nsr/nsrlink.jsp?1995Go34,B}{1995Go34}, \href{https://www.nndc.bnl.gov/nsr/nsrlink.jsp?1997Kl06,B}{1997Kl06}, \href{https://www.nndc.bnl.gov/nsr/nsrlink.jsp?2008Pa02,B}{2008Pa02}, \href{https://www.nndc.bnl.gov/nsr/nsrlink.jsp?2008Pe13,B}{2008Pe13}, \href{https://www.nndc.bnl.gov/nsr/nsrlink.jsp?2008Se10,B}{2008Se10},}\\
\parbox[b][0.3cm]{17.7cm}{\href{https://www.nndc.bnl.gov/nsr/nsrlink.jsp?2012Sa50,B}{2012Sa50}, \href{https://www.nndc.bnl.gov/nsr/nsrlink.jsp?2015To02,B}{2015To02}, \href{https://www.nndc.bnl.gov/nsr/nsrlink.jsp?2019Gy02,B}{2019Gy02}, \href{https://www.nndc.bnl.gov/nsr/nsrlink.jsp?2020Oh01,B}{2020Oh01}, \href{https://www.nndc.bnl.gov/nsr/nsrlink.jsp?2022Gl04,B}{2022Gl04}, \href{https://www.nndc.bnl.gov/nsr/nsrlink.jsp?2022Ko18,B}{2022Ko18}, \href{https://www.nndc.bnl.gov/nsr/nsrlink.jsp?2022Va06,B}{2022Va06}, \href{https://www.nndc.bnl.gov/nsr/nsrlink.jsp?2023Li31,B}{2023Li31}, \href{https://www.nndc.bnl.gov/nsr/nsrlink.jsp?2023Se01,B}{2023Se01}, \href{https://www.nndc.bnl.gov/nsr/nsrlink.jsp?2023Xu10,B}{2023Xu10}, \href{https://www.nndc.bnl.gov/nsr/nsrlink.jsp?2024Fa01,B}{2024Fa01}.}\\
\vspace{0.385cm}
\parbox[b][0.3cm]{17.7cm}{\addtolength{\parindent}{-0.2in}\textit{Previous \ensuremath{^{19}}Ne evaluations}: \href{https://www.nndc.bnl.gov/nsr/nsrlink.jsp?1959Aj76,B}{1959Aj76}, \href{https://www.nndc.bnl.gov/nsr/nsrlink.jsp?1972Aj02,B}{1972Aj02}, \href{https://www.nndc.bnl.gov/nsr/nsrlink.jsp?1978Aj03,B}{1978Aj03}, \href{https://www.nndc.bnl.gov/nsr/nsrlink.jsp?1983Aj01,B}{1983Aj01}, \href{https://www.nndc.bnl.gov/nsr/nsrlink.jsp?1987Aj02,B}{1987Aj02}, \href{https://www.nndc.bnl.gov/nsr/nsrlink.jsp?1995Ti07,B}{1995Ti07}.}\\
\vspace{12pt}
\clearpage
\vspace{0.3cm}
\vspace*{-0.5cm}
{\bf \small \underline{Adopted \hyperlink{19NE_LEVEL}{Levels}, \hyperlink{19NE_GAMMA}{Gammas} (continued)}}\\
\vspace{0.3cm}
\hypertarget{19NE_LEVEL}{\underline{$^{19}$Ne Levels}}\\
\vspace{0.34cm}
\parbox[b][0.3cm]{17.7cm}{\addtolength{\parindent}{-0.254cm}\textit{Notes}:}\\
\parbox[b][0.3cm]{17.7cm}{\addtolength{\parindent}{-0.254cm}(1) For K\ensuremath{^{\ensuremath{\pi}}}=1/2\ensuremath{^{\textnormal{+}}} band, see (\href{https://www.nndc.bnl.gov/nsr/nsrlink.jsp?1970Ga18,B}{1970Ga18}, \href{https://www.nndc.bnl.gov/nsr/nsrlink.jsp?1971Bi06,B}{1971Bi06}, \href{https://www.nndc.bnl.gov/nsr/nsrlink.jsp?1972Ga08,B}{1972Ga08}, \href{https://www.nndc.bnl.gov/nsr/nsrlink.jsp?1972Pa29,B}{1972Pa29}, \href{https://www.nndc.bnl.gov/nsr/nsrlink.jsp?1974Ga11,B}{1974Ga11}, \href{https://www.nndc.bnl.gov/nsr/nsrlink.jsp?1976Ha06,B}{1976Ha06}, \href{https://www.nndc.bnl.gov/nsr/nsrlink.jsp?1981Ov01,B}{1981Ov01}, \href{https://www.nndc.bnl.gov/nsr/nsrlink.jsp?1983Cu02,B}{1983Cu02}, \href{https://www.nndc.bnl.gov/nsr/nsrlink.jsp?1988Kr11,B}{1988Kr11},}\\
\parbox[b][0.3cm]{17.7cm}{\href{https://www.nndc.bnl.gov/nsr/nsrlink.jsp?2000Du09,B}{2000Du09}, \href{https://www.nndc.bnl.gov/nsr/nsrlink.jsp?2017To14,B}{2017To14}). For K\ensuremath{^{\ensuremath{\pi}}}=1/2\ensuremath{^{-}} band, see (\href{https://www.nndc.bnl.gov/nsr/nsrlink.jsp?1970Ga18,B}{1970Ga18}, \href{https://www.nndc.bnl.gov/nsr/nsrlink.jsp?1971Bi06,B}{1971Bi06}, \href{https://www.nndc.bnl.gov/nsr/nsrlink.jsp?1978Pi06,B}{1978Pi06}, \href{https://www.nndc.bnl.gov/nsr/nsrlink.jsp?2014Ot03,B}{2014Ot03}). For K\ensuremath{^{\ensuremath{\pi}}}=3/2\ensuremath{^{\textnormal{+}}} band, see (\href{https://www.nndc.bnl.gov/nsr/nsrlink.jsp?1978Fo26,B}{1978Fo26},}\\
\parbox[b][0.3cm]{17.7cm}{\href{https://www.nndc.bnl.gov/nsr/nsrlink.jsp?1995Wi26,B}{1995Wi26}). For K\ensuremath{^{\ensuremath{\pi}}}=3/2\ensuremath{^{-}}, see (\href{https://www.nndc.bnl.gov/nsr/nsrlink.jsp?1995Wi26,B}{1995Wi26}).}\\
\parbox[b][0.3cm]{17.7cm}{\addtolength{\parindent}{-0.254cm}(2) Unless otherwise noted, the excitation energies from (\href{https://www.nndc.bnl.gov/nsr/nsrlink.jsp?2017To14,B}{2017To14}) were not considered for the \ensuremath{^{\textnormal{19}}}Ne Adopted Levels due to their}\\
\parbox[b][0.3cm]{17.7cm}{large systematic uncertainties of \textit{50} keV.}\\
\parbox[b][0.3cm]{17.7cm}{\addtolength{\parindent}{-0.254cm}(3) When E\ensuremath{_{\textnormal{c.m.}}}(\ensuremath{^{\textnormal{15}}}O+\ensuremath{\alpha}) or E\ensuremath{_{\textnormal{c.m.}}}(\ensuremath{^{\textnormal{18}}}F+p) are directly measured, we have used those values together with the S\ensuremath{_{\ensuremath{\alpha}}}(\ensuremath{^{\textnormal{19}}}Ne) and the}\\
\parbox[b][0.3cm]{17.7cm}{S\ensuremath{_{\textnormal{p}}}(\ensuremath{^{\textnormal{19}}}Ne), respectively, to deduce the corresponding excitation energies given in comments. The separation energies are taken from}\\
\parbox[b][0.3cm]{17.7cm}{(\href{https://www.nndc.bnl.gov/nsr/nsrlink.jsp?2021Wa16,B}{2021Wa16}).}\\
\parbox[b][0.3cm]{17.7cm}{\addtolength{\parindent}{-0.254cm}(4) The results of (\href{https://www.nndc.bnl.gov/nsr/nsrlink.jsp?2025PhZZ,B}{2025PhZZ}: PhD Thesis) are unpublished. Therefore, we only used selected information that helped guiding our}\\
\parbox[b][0.3cm]{17.7cm}{judgments.}\\
\parbox[b][0.3cm]{17.7cm}{\addtolength{\parindent}{-0.254cm}(5) In the comments below, \ensuremath{\tau} is the mean lifetime.}\\
\parbox[b][0.3cm]{17.7cm}{\addtolength{\parindent}{-0.254cm}(6) Throughout this document and unless otherwise noted, \ensuremath{\Gamma}\ensuremath{_{\ensuremath{\alpha}}}/\ensuremath{\Gamma}, \ensuremath{\Gamma}\ensuremath{_{\textnormal{p}}}/\ensuremath{\Gamma}, \ensuremath{\Gamma}\ensuremath{_{\textnormal{p}}} and \ensuremath{\Gamma}\ensuremath{_{\ensuremath{\alpha}}} refer to the \ensuremath{\Gamma}\ensuremath{_{\ensuremath{\alpha}_{\textnormal{0}}}}/\ensuremath{\Gamma}, \ensuremath{\Gamma}\ensuremath{_{\textnormal{p}_{\textnormal{0}}}}/\ensuremath{\Gamma}, \ensuremath{\Gamma}\ensuremath{_{\ensuremath{\alpha}_{\textnormal{0}}}} and}\\
\parbox[b][0.3cm]{17.7cm}{\ensuremath{\Gamma}\ensuremath{_{\textnormal{p}_{\textnormal{0}}}}.}\\
\vspace{0.34cm}
\begin{longtable}[c]{llllll}
\multicolumn{6}{c}{\underline{Cross Reference (XREF) Flags}}\\
 \\
\hyperlink{MG1}{\texttt{A }}& \ensuremath{^{\textnormal{20}}}Mg \ensuremath{\beta}\ensuremath{^{\textnormal{+}}}p decay & \hyperlink{NE15}{\texttt{O }}& \ensuremath{^{\textnormal{9}}}Be(\ensuremath{^{\textnormal{20}}}Ne,\ensuremath{^{\textnormal{19}}}Ne) & \hyperlink{NE28}{\texttt{AB }}& \ensuremath{^{\textnormal{16}}}O(\ensuremath{^{\textnormal{12}}}C,\ensuremath{^{\textnormal{9}}}Be)\\
\hyperlink{NE2}{\texttt{B }}& \ensuremath{^{\textnormal{1}}}H(\ensuremath{^{\textnormal{18}}}F,p),(\ensuremath{^{\textnormal{18}}}F,\ensuremath{\alpha}):res & \hyperlink{NE16}{\texttt{P }}& \ensuremath{^{\textnormal{10}}}B(\ensuremath{^{\textnormal{10}}}B,\ensuremath{^{\textnormal{19}}}Ne),(\ensuremath{^{\textnormal{11}}}B,\ensuremath{^{\textnormal{19}}}Ne) & \hyperlink{NE29}{\texttt{AC }}& \ensuremath{^{\textnormal{17}}}O(\ensuremath{^{\textnormal{3}}}He,n),(\ensuremath{^{\textnormal{3}}}He,n\ensuremath{\gamma})\\
\hyperlink{NE3}{\texttt{C }}& \ensuremath{^{\textnormal{1}}}H(\ensuremath{^{\textnormal{18}}}F,\ensuremath{^{\textnormal{19}}}Ne) & \hyperlink{NE17}{\texttt{Q }}& \ensuremath{^{\textnormal{10}}}B(\ensuremath{^{\textnormal{14}}}N,\ensuremath{\alpha}n) & \hyperlink{NE30}{\texttt{AD }}& \ensuremath{^{\textnormal{18}}}O(p,\ensuremath{\pi}\ensuremath{^{-}})\\
\hyperlink{NE4}{\texttt{D }}& \ensuremath{^{\textnormal{1}}}H(\ensuremath{^{\textnormal{19}}}F,\ensuremath{^{\textnormal{19}}}Ne) & \hyperlink{NE18}{\texttt{R }}& \ensuremath{^{\textnormal{12}}}C(\ensuremath{^{\textnormal{12}}}C,n\ensuremath{\alpha}) & \hyperlink{NE31}{\texttt{AE }}& \ensuremath{^{\textnormal{19}}}F(p,n),(p,n\ensuremath{\gamma}),(d,2n\ensuremath{\gamma})\\
\end{longtable}
\vspace{-0.5cm}
\begin{longtable}{ccccccc@{\extracolsep{\fill}}c}
\multicolumn{2}{c}{E(level)$^{}$}&J$^{\pi}$$^{}$&\multicolumn{2}{c}{T\ensuremath{_{\textnormal{1/2}}} or \ensuremath{\Gamma}$^{}$}&XREF&Comments&\\[-.2cm]
\multicolumn{2}{c}{\hrulefill}&\hrulefill&\multicolumn{2}{c}{\hrulefill}&\hrulefill&\hrulefill&
\endfirsthead
\multicolumn{1}{r@{}}{0}&\multicolumn{1}{@{}l}{\ensuremath{^{{\hyperlink{NE0LEVEL0}{a}}}}}&\multicolumn{1}{l}{1/2\ensuremath{^{+}}}&\multicolumn{1}{r@{}}{17}&\multicolumn{1}{@{.}l}{2562 s {\it 30}}&\multicolumn{1}{l}{\texttt{\hyperlink{MG1}{A}\ \hyperlink{NE3}{C}\hyperlink{NE4}{D}\ \hyperlink{NE6}{F}\hyperlink{NE7}{G}\ \hyperlink{NE9}{I}\ \hyperlink{NE11}{K}\ \ \hyperlink{NE14}{N}\hyperlink{NE15}{O}\hyperlink{NE16}{P}\hyperlink{NE17}{Q}\hyperlink{NE18}{R}\hyperlink{NE19}{S}\hyperlink{NE20}{T}\ \ \ \hyperlink{NE24}{X}\hyperlink{NE25}{Y}\hyperlink{NE26}{Z}} }&\parbox[t][0.3cm]{8.453941cm}{\raggedright XREF: Others: \hyperlink{NE28}{\texttt{AB}}, \hyperlink{NE29}{\texttt{AC}}, \hyperlink{NE30}{\texttt{AD}}, \hyperlink{NE31}{\texttt{AE}}, \hyperlink{NE32}{\texttt{AF}}, \hyperlink{NE33}{\texttt{AG}}, \hyperlink{NE34}{\texttt{AH}}, \hyperlink{NE35}{\texttt{AI}}, \hyperlink{NE36}{\texttt{AJ}}, \hyperlink{NE38}{\texttt{AL}},\vspace{0.1cm}}&\\
&&&&&&\parbox[t][0.3cm]{8.453941cm}{\raggedright {\ }{\ }{\ }\hyperlink{NE39}{\texttt{AM}}, \hyperlink{NE40}{\texttt{AN}}, \hyperlink{NE41}{\texttt{AO}}\vspace{0.1cm}}&\\
&&&&&&\parbox[t][0.3cm]{8.453941cm}{\raggedright \%\ensuremath{\varepsilon}+\%\ensuremath{\beta}\ensuremath{^{\textnormal{+}}}=100 (\href{https://www.nndc.bnl.gov/nsr/nsrlink.jsp?2013Uj01,B}{2013Uj01})\vspace{0.1cm}}&\\
&&&&&&\parbox[t][0.3cm]{8.453941cm}{\raggedright T=1/2 (\href{https://www.nndc.bnl.gov/nsr/nsrlink.jsp?1969Ha38,B}{1969Ha38},\href{https://www.nndc.bnl.gov/nsr/nsrlink.jsp?1984Ra22,B}{1984Ra22},\href{https://www.nndc.bnl.gov/nsr/nsrlink.jsp?2012Tr06,B}{2012Tr06})\vspace{0.1cm}}&\\
&&&&&&\parbox[t][0.3cm]{8.453941cm}{\raggedright \ensuremath{\mu}={\textminus}1.88515 \textit{7} (\href{https://www.nndc.bnl.gov/nsr/nsrlink.jsp?2019StZV,B}{2019StZV})\vspace{0.1cm}}&\\
&&&&&&\parbox[t][0.3cm]{8.453941cm}{\raggedright g=0.742 (\href{https://www.nndc.bnl.gov/nsr/nsrlink.jsp?1969Bl02,B}{1969Bl02})\vspace{0.1cm}}&\\
\end{longtable}
\begin{textblock}{29}(0,27.3)
Continued on next page (footnotes at end of table)
\end{textblock}
\clearpage
\begin{longtable}{ccccccc@{\extracolsep{\fill}}c}
\\[-.4cm]
\multicolumn{8}{c}{{\bf \small \underline{Adopted \hyperlink{19NE_LEVEL}{Levels}, \hyperlink{19NE_GAMMA}{Gammas} (continued)}}}\\
\multicolumn{8}{c}{~}\\
\multicolumn{8}{c}{\underline{\ensuremath{^{19}}Ne Levels (continued)}}\\
\multicolumn{8}{c}{~}\\
\multicolumn{2}{c}{E(level)$^{}$}&J$^{\pi}$$^{}$&\multicolumn{2}{c}{T\ensuremath{_{\textnormal{1/2}}} or \ensuremath{\Gamma}$^{}$}&XREF&Comments&\\[-.2cm]
\multicolumn{2}{c}{\hrulefill}&\hrulefill&\multicolumn{2}{c}{\hrulefill}&\hrulefill&\hrulefill&
\endhead
&&&&&&\parbox[t][0.3cm]{12.363561cm}{\raggedright {\ }{\ }{\ }Thesis, Princeton University, (1984), unpublished) as cited in (\href{https://www.nndc.bnl.gov/nsr/nsrlink.jsp?2008Se10,B}{2008Se10}); 17.254 s \textit{5}\vspace{0.1cm}}&\\
&&&&&&\parbox[t][0.3cm]{12.363561cm}{\raggedright {\ }{\ }{\ }(\href{https://www.nndc.bnl.gov/nsr/nsrlink.jsp?2013Uj01,B}{2013Uj01}); and 17.2569 s \textit{19} (stat.) \textit{9} (sys.) (\href{https://www.nndc.bnl.gov/nsr/nsrlink.jsp?2017Fo24,B}{2017Fo24}).\vspace{0.1cm}}&\\
&&&&&&\parbox[t][0.3cm]{12.363561cm}{\raggedright T\ensuremath{_{1/2}}: See also the excluded values of 20 s (\href{https://www.nndc.bnl.gov/nsr/nsrlink.jsp?1939Fo01,B}{1939Fo01}); 20.3 s \textit{5} (\href{https://www.nndc.bnl.gov/nsr/nsrlink.jsp?1939Wh02,B}{1939Wh02}); 18.2 s \textit{6}\vspace{0.1cm}}&\\
&&&&&&\parbox[t][0.3cm]{12.363561cm}{\raggedright {\ }{\ }{\ }(\href{https://www.nndc.bnl.gov/nsr/nsrlink.jsp?1949Sh25,B}{1949Sh25}); 18.6 s \textit{4} (\href{https://www.nndc.bnl.gov/nsr/nsrlink.jsp?1951Bl75,B}{1951Bl75}); 18.5 s \textit{5} (\href{https://www.nndc.bnl.gov/nsr/nsrlink.jsp?1952Sc15,B}{1952Sc15}); 19 s \textit{1} (\href{https://www.nndc.bnl.gov/nsr/nsrlink.jsp?1954Na29,B}{1954Na29}); 18.3 s \textit{5}\vspace{0.1cm}}&\\
&&&&&&\parbox[t][0.3cm]{12.363561cm}{\raggedright {\ }{\ }{\ }(\href{https://www.nndc.bnl.gov/nsr/nsrlink.jsp?1957Al29,B}{1957Al29}); 17.7 s \textit{1} (\href{https://www.nndc.bnl.gov/nsr/nsrlink.jsp?1957Pe12,B}{1957Pe12}); 19.5 s \textit{10} (\href{https://www.nndc.bnl.gov/nsr/nsrlink.jsp?1958We25,B}{1958We25}, \href{https://www.nndc.bnl.gov/nsr/nsrlink.jsp?1960Wa04,B}{1960Wa04}); 16.72 s \textit{5}\vspace{0.1cm}}&\\
&&&&&&\parbox[t][0.3cm]{12.363561cm}{\raggedright {\ }{\ }{\ }(\href{https://www.nndc.bnl.gov/nsr/nsrlink.jsp?1960Ja12,B}{1960Ja12}); 16.5 s \textit{10} (\href{https://www.nndc.bnl.gov/nsr/nsrlink.jsp?1964Va23,B}{1964Va23}); 17.3 s \textit{2} (J. Goss, J. Harris, N. Khelil, F. Riggle, and\vspace{0.1cm}}&\\
\end{longtable}
\begin{textblock}{29}(0,27.3)
Continued on next page (footnotes at end of table)
\end{textblock}
\clearpage
\begin{longtable}{ccccccc@{\extracolsep{\fill}}c}
\\[-.4cm]
\multicolumn{8}{c}{{\bf \small \underline{Adopted \hyperlink{19NE_LEVEL}{Levels}, \hyperlink{19NE_GAMMA}{Gammas} (continued)}}}\\
\multicolumn{8}{c}{~}\\
\multicolumn{8}{c}{\underline{\ensuremath{^{19}}Ne Levels (continued)}}\\
\multicolumn{8}{c}{~}\\
\multicolumn{2}{c}{E(level)$^{}$}&J$^{\pi}$$^{}$&\multicolumn{2}{c}{T\ensuremath{_{\textnormal{1/2}}} or \ensuremath{\Gamma}$^{}$}&XREF&Comments&\\[-.2cm]
\multicolumn{2}{c}{\hrulefill}&\hrulefill&\multicolumn{2}{c}{\hrulefill}&\hrulefill&\hrulefill&
\endhead
&&&&&&\parbox[t][0.3cm]{8.578061cm}{\raggedright See also r\ensuremath{_{\textnormal{ch}}}=3.00 fm and r\ensuremath{_{\textnormal{m}}}=2.84 fm, which are calculated\vspace{0.1cm}}&\\
&&&&&&\parbox[t][0.3cm]{8.578061cm}{\raggedright {\ }{\ }{\ }by (\href{https://www.nndc.bnl.gov/nsr/nsrlink.jsp?2008Ge07,B}{2008Ge07}) using fermionic molecular dynamics from the\vspace{0.1cm}}&\\
&&&&&&\parbox[t][0.3cm]{8.578061cm}{\raggedright {\ }{\ }{\ }point proton radius and by taking the charge radius of\vspace{0.1cm}}&\\
&&&&&&\parbox[t][0.3cm]{8.578061cm}{\raggedright {\ }{\ }{\ }protons and neutrons into account.\vspace{0.1cm}}&\\
&&&&&&\parbox[t][0.3cm]{8.578061cm}{\raggedright Decay mode: \ensuremath{^{\textnormal{19}}}Ne\ensuremath{_{\textnormal{g.s.}}} decays almost exclusively to the\vspace{0.1cm}}&\\
&&&&&&\parbox[t][0.3cm]{8.578061cm}{\raggedright {\ }{\ }{\ }\ensuremath{^{\textnormal{19}}}F\ensuremath{_{\textnormal{g.s.}}} with BR=99.9878\% \textit{7} (\href{https://www.nndc.bnl.gov/nsr/nsrlink.jsp?2019Re07,B}{2019Re07}). See also\vspace{0.1cm}}&\\
&&&&&&\parbox[t][0.3cm]{8.578061cm}{\raggedright {\ }{\ }{\ }BR=99.99\% assumed by (\href{https://www.nndc.bnl.gov/nsr/nsrlink.jsp?1981Ad05,B}{1981Ad05}, \href{https://www.nndc.bnl.gov/nsr/nsrlink.jsp?1983Ad03,B}{1983Ad03}); and\vspace{0.1cm}}&\\
&&&&&&\parbox[t][0.3cm]{8.578061cm}{\raggedright {\ }{\ }{\ }BR=99.9858\% \textit{20} from the compilation of (\href{https://www.nndc.bnl.gov/nsr/nsrlink.jsp?2008Se10,B}{2008Se10}). This\vspace{0.1cm}}&\\
\end{longtable}
\begin{textblock}{29}(0,27.3)
Continued on next page (footnotes at end of table)
\end{textblock}
\clearpage
\begin{longtable}{ccccccc@{\extracolsep{\fill}}c}
\\[-.4cm]
\multicolumn{8}{c}{{\bf \small \underline{Adopted \hyperlink{19NE_LEVEL}{Levels}, \hyperlink{19NE_GAMMA}{Gammas} (continued)}}}\\
\multicolumn{8}{c}{~}\\
\multicolumn{8}{c}{\underline{\ensuremath{^{19}}Ne Levels (continued)}}\\
\multicolumn{8}{c}{~}\\
\multicolumn{2}{c}{E(level)$^{}$}&J$^{\pi}$$^{}$&\multicolumn{2}{c}{T\ensuremath{_{\textnormal{1/2}}} or \ensuremath{\Gamma}$^{}$}&XREF&Comments&\\[-.2cm]
\multicolumn{2}{c}{\hrulefill}&\hrulefill&\multicolumn{2}{c}{\hrulefill}&\hrulefill&\hrulefill&
\endhead
&&&&&&\parbox[t][0.3cm]{8.203561cm}{\raggedright {\ }{\ }{\ }From comparison of the measured \ensuremath{\gamma}-ray decay scheme of\vspace{0.1cm}}&\\
&&&&&&\parbox[t][0.3cm]{8.203561cm}{\raggedright {\ }{\ }{\ }the \ensuremath{^{\textnormal{19}}}Ne*(238) level with that of the mirror state.\vspace{0.1cm}}&\\
&&&&&&\parbox[t][0.3cm]{8.203561cm}{\raggedright \ensuremath{\mu}: Adopted by (\href{https://www.nndc.bnl.gov/nsr/nsrlink.jsp?2020StZV,B}{2020StZV}) from \ensuremath{\mu}={\textminus}0.740 \ensuremath{\mu}\ensuremath{_{\textnormal{N}}}\vspace{0.1cm}}&\\
&&&&&&\parbox[t][0.3cm]{8.203561cm}{\raggedright {\ }{\ }{\ }(\href{https://www.nndc.bnl.gov/nsr/nsrlink.jsp?1969Bl02,B}{1969Bl02}), which is deduced using the Time Dependent\vspace{0.1cm}}&\\
&&&&&&\parbox[t][0.3cm]{8.203561cm}{\raggedright {\ }{\ }{\ }Perturbed Angular Distribution method. The adopted value\vspace{0.1cm}}&\\
&&&&&&\parbox[t][0.3cm]{8.203561cm}{\raggedright {\ }{\ }{\ }is in units of \ensuremath{\mu}\ensuremath{_{\textnormal{N}}}.\vspace{0.1cm}}&\\
&&&&&&\parbox[t][0.3cm]{8.203561cm}{\raggedright g: From a Larmor frequency ratio of\vspace{0.1cm}}&\\
&&&&&&\parbox[t][0.3cm]{8.203561cm}{\raggedright {\ }{\ }{\ }\ensuremath{\omega}B(Ne)/\ensuremath{\omega}B(F)={\textminus}0.2054 \textit{15} (\href{https://www.nndc.bnl.gov/nsr/nsrlink.jsp?1969Bl02,B}{1969Bl02}), which is\vspace{0.1cm}}&\\
\end{longtable}
\begin{textblock}{29}(0,27.3)
Continued on next page (footnotes at end of table)
\end{textblock}
\clearpage
\begin{longtable}{ccccccc@{\extracolsep{\fill}}c}
\\[-.4cm]
\multicolumn{8}{c}{{\bf \small \underline{Adopted \hyperlink{19NE_LEVEL}{Levels}, \hyperlink{19NE_GAMMA}{Gammas} (continued)}}}\\
\multicolumn{8}{c}{~}\\
\multicolumn{8}{c}{\underline{\ensuremath{^{19}}Ne Levels (continued)}}\\
\multicolumn{8}{c}{~}\\
\multicolumn{2}{c}{E(level)$^{}$}&J$^{\pi}$$^{}$&\multicolumn{2}{c}{T\ensuremath{_{\textnormal{1/2}}} or \ensuremath{\Gamma}$^{}$}&XREF&Comments&\\[-.2cm]
\multicolumn{2}{c}{\hrulefill}&\hrulefill&\multicolumn{2}{c}{\hrulefill}&\hrulefill&\hrulefill&
\endhead
&&&&&&\parbox[t][0.3cm]{7.523041cm}{\raggedright \%IT=100\vspace{0.1cm}}&\\
&&&&&&\parbox[t][0.3cm]{7.523041cm}{\raggedright XREF: ab(1.54E3)AH(?)aj(1524)\vspace{0.1cm}}&\\
&&&&&&\parbox[t][0.3cm]{7.523041cm}{\raggedright E(level): Weighted average of 1504.0 keV \textit{30}\vspace{0.1cm}}&\\
&&&&&&\parbox[t][0.3cm]{7.523041cm}{\raggedright {\ }{\ }{\ }(\href{https://www.nndc.bnl.gov/nsr/nsrlink.jsp?1972Ha03,B}{1972Ha03}); 1507.8 keV \textit{6} (\href{https://www.nndc.bnl.gov/nsr/nsrlink.jsp?1977Le03,B}{1977Le03}); 1507.51\vspace{0.1cm}}&\\
&&&&&&\parbox[t][0.3cm]{7.523041cm}{\raggedright {\ }{\ }{\ }keV \textit{35} (\href{https://www.nndc.bnl.gov/nsr/nsrlink.jsp?2005Ta28,B}{2005Ta28}); and 1507.43 keV \textit{17} from the\vspace{0.1cm}}&\\
&&&&&&\parbox[t][0.3cm]{7.523041cm}{\raggedright {\ }{\ }{\ }least-squares fit to E\ensuremath{_{\ensuremath{\gamma}}} values measured by\vspace{0.1cm}}&\\
&&&&&&\parbox[t][0.3cm]{7.523041cm}{\raggedright {\ }{\ }{\ }(\href{https://www.nndc.bnl.gov/nsr/nsrlink.jsp?1970Gi09,B}{1970Gi09}, \href{https://www.nndc.bnl.gov/nsr/nsrlink.jsp?1971It02,B}{1971It02}, \href{https://www.nndc.bnl.gov/nsr/nsrlink.jsp?2019Gl02,B}{2019Gl02}, \href{https://www.nndc.bnl.gov/nsr/nsrlink.jsp?2020Ha31,B}{2020Ha31}).\vspace{0.1cm}}&\\
&&&&&&\parbox[t][0.3cm]{7.523041cm}{\raggedright T\ensuremath{_{1/2}}: From \ensuremath{\tau}=1.8 ps \textit{3}: Weighted average (with\vspace{0.1cm}}&\\
\end{longtable}
\begin{textblock}{29}(0,27.3)
Continued on next page (footnotes at end of table)
\end{textblock}
\clearpage
\begin{longtable}{ccccccc@{\extracolsep{\fill}}c}
\\[-.4cm]
\multicolumn{8}{c}{{\bf \small \underline{Adopted \hyperlink{19NE_LEVEL}{Levels}, \hyperlink{19NE_GAMMA}{Gammas} (continued)}}}\\
\multicolumn{8}{c}{~}\\
\multicolumn{8}{c}{\underline{\ensuremath{^{19}}Ne Levels (continued)}}\\
\multicolumn{8}{c}{~}\\
\multicolumn{2}{c}{E(level)$^{}$}&J$^{\pi}$$^{}$&\multicolumn{2}{c}{T\ensuremath{_{\textnormal{1/2}}} or \ensuremath{\Gamma}$^{}$}&XREF&Comments&\\[-.2cm]
\multicolumn{2}{c}{\hrulefill}&\hrulefill&\multicolumn{2}{c}{\hrulefill}&\hrulefill&\hrulefill&
\endhead
&&&&&&\parbox[t][0.3cm]{8.261341cm}{\raggedright {\ }{\ }{\ }(\href{https://www.nndc.bnl.gov/nsr/nsrlink.jsp?2020Ha31,B}{2020Ha31}): From comparison of the measured \ensuremath{\gamma}-ray\vspace{0.1cm}}&\\
&&&&&&\parbox[t][0.3cm]{8.261341cm}{\raggedright {\ }{\ }{\ }decay scheme of the \ensuremath{^{\textnormal{19}}}Ne*(1536) state to that of the\vspace{0.1cm}}&\\
&&&&&&\parbox[t][0.3cm]{8.261341cm}{\raggedright {\ }{\ }{\ }mirror level. This finding supports the J\ensuremath{^{\ensuremath{\pi}}}=3/2\ensuremath{^{\textnormal{+}}}\vspace{0.1cm}}&\\
&&&&&&\parbox[t][0.3cm]{8.261341cm}{\raggedright {\ }{\ }{\ }assignment for \ensuremath{^{\textnormal{19}}}Ne*(1536).\vspace{0.1cm}}&\\
&&&&&&\parbox[t][0.3cm]{8.261341cm}{\raggedright B(GT)=0.045 \textit{15} (\href{https://www.nndc.bnl.gov/nsr/nsrlink.jsp?1984Ra22,B}{1984Ra22}).\vspace{0.1cm}}&\\
\multicolumn{1}{r@{}}{1615}&\multicolumn{1}{@{.}l}{28\ensuremath{^{{\hyperlink{NE0LEVEL1}{b}}}} {\it 22}}&\multicolumn{1}{l}{3/2\ensuremath{^{-}}}&\multicolumn{1}{r@{}}{64}&\multicolumn{1}{@{ }l}{fs {\it 13}}&\multicolumn{1}{l}{\texttt{\hyperlink{MG1}{A}\ \ \ \ \ \ \ \ \ \hyperlink{NE11}{K}\ \ \ \ \ \ \ \ \ \ \ \ \hyperlink{NE24}{X}\hyperlink{NE25}{Y}\ } }&\parbox[t][0.3cm]{8.261341cm}{\raggedright XREF: Others: \hyperlink{NE29}{\texttt{AC}}, \hyperlink{NE31}{\texttt{AE}}, \hyperlink{NE33}{\texttt{AG}}, \hyperlink{NE34}{\texttt{AH}}, \hyperlink{NE36}{\texttt{AJ}}, \hyperlink{NE38}{\texttt{AL}}, \hyperlink{NE39}{\texttt{AM}}\vspace{0.1cm}}&\\
&&&&&&\parbox[t][0.3cm]{8.261341cm}{\raggedright \%IT=100\vspace{0.1cm}}&\\
&&&&&&\parbox[t][0.3cm]{8.261341cm}{\raggedright E(level): Weighted average (with external errors) of 1620\vspace{0.1cm}}&\\
\end{longtable}
\begin{textblock}{29}(0,27.3)
Continued on next page (footnotes at end of table)
\end{textblock}
\clearpage
\begin{longtable}{ccccccc@{\extracolsep{\fill}}c}
\\[-.4cm]
\multicolumn{8}{c}{{\bf \small \underline{Adopted \hyperlink{19NE_LEVEL}{Levels}, \hyperlink{19NE_GAMMA}{Gammas} (continued)}}}\\
\multicolumn{8}{c}{~}\\
\multicolumn{8}{c}{\underline{\ensuremath{^{19}}Ne Levels (continued)}}\\
\multicolumn{8}{c}{~}\\
\multicolumn{2}{c}{E(level)$^{}$}&J$^{\pi}$$^{}$&\multicolumn{2}{c}{T\ensuremath{_{\textnormal{1/2}}} or \ensuremath{\Gamma}$^{}$}&XREF&Comments&\\[-.2cm]
\multicolumn{2}{c}{\hrulefill}&\hrulefill&\multicolumn{2}{c}{\hrulefill}&\hrulefill&\hrulefill&
\endhead
&&&&&&\parbox[t][0.3cm]{8.578061cm}{\raggedright {\ }{\ }{\ }(\href{https://www.nndc.bnl.gov/nsr/nsrlink.jsp?1981Ov01,B}{1981Ov01}: \ensuremath{^{\textnormal{16}}}O(\ensuremath{\alpha},n)) with L=(4), see text, where DWBA\vspace{0.1cm}}&\\
&&&&&&\parbox[t][0.3cm]{8.578061cm}{\raggedright {\ }{\ }{\ }curves were reported to be J-dependent; and (\href{https://www.nndc.bnl.gov/nsr/nsrlink.jsp?1972Ga08,B}{1972Ga08}:\vspace{0.1cm}}&\\
&&&&&&\parbox[t][0.3cm]{8.578061cm}{\raggedright {\ }{\ }{\ }\ensuremath{^{\textnormal{16}}}O(\ensuremath{^{\textnormal{6}}}Li,t)): J\ensuremath{^{\ensuremath{\pi}}}=9/2\ensuremath{^{\textnormal{+}}}; (2) the exact finite-range DWBA and\vspace{0.1cm}}&\\
&&&&&&\parbox[t][0.3cm]{8.578061cm}{\raggedright {\ }{\ }{\ }spectroscopic factor analyses of (\href{https://www.nndc.bnl.gov/nsr/nsrlink.jsp?1981Go11,B}{1981Go11}, \href{https://www.nndc.bnl.gov/nsr/nsrlink.jsp?1988Kr11,B}{1988Kr11}:\vspace{0.1cm}}&\\
&&&&&&\parbox[t][0.3cm]{8.578061cm}{\raggedright {\ }{\ }{\ }\ensuremath{^{\textnormal{16}}}O(\ensuremath{^{\textnormal{12}}}C,\ensuremath{^{\textnormal{9}}}Be)): J\ensuremath{^{\ensuremath{\pi}}}=9/2\ensuremath{^{\textnormal{+}}}, see Figs. 5b and 15, respectively;\vspace{0.1cm}}&\\
&&&&&&\parbox[t][0.3cm]{8.578061cm}{\raggedright {\ }{\ }{\ }and (3) finite-range DWBA analysis of (\href{https://www.nndc.bnl.gov/nsr/nsrlink.jsp?2017Ba42,B}{2017Ba42}:\vspace{0.1cm}}&\\
&&&&&&\parbox[t][0.3cm]{8.578061cm}{\raggedright {\ }{\ }{\ }\ensuremath{^{\textnormal{20}}}Ne(p,d)): J\ensuremath{^{\ensuremath{\pi}}}=(7/2\ensuremath{^{\textnormal{+}}}, 9/2\ensuremath{^{\textnormal{+}}}) with L=4 using TWOFNR18.\vspace{0.1cm}}&\\
&&&&&&\parbox[t][0.3cm]{8.578061cm}{\raggedright This level is the mirror state for \ensuremath{^{\textnormal{19}}}F*(2779, 9/2\ensuremath{^{\textnormal{+}}})\vspace{0.1cm}}&\\
\end{longtable}
\begin{textblock}{29}(0,27.3)
Continued on next page (footnotes at end of table)
\end{textblock}
\clearpage
\begin{longtable}{ccccccc@{\extracolsep{\fill}}c}
\\[-.4cm]
\multicolumn{8}{c}{{\bf \small \underline{Adopted \hyperlink{19NE_LEVEL}{Levels}, \hyperlink{19NE_GAMMA}{Gammas} (continued)}}}\\
\multicolumn{8}{c}{~}\\
\multicolumn{8}{c}{\underline{\ensuremath{^{19}}Ne Levels (continued)}}\\
\multicolumn{8}{c}{~}\\
\multicolumn{2}{c}{E(level)$^{}$}&J$^{\pi}$$^{}$&\multicolumn{2}{c}{T\ensuremath{_{\textnormal{1/2}}} or \ensuremath{\Gamma}$^{}$}&XREF&Comments&\\[-.2cm]
\multicolumn{2}{c}{\hrulefill}&\hrulefill&\multicolumn{2}{c}{\hrulefill}&\hrulefill&\hrulefill&
\endhead
&&&&&&\parbox[t][0.3cm]{12.363561cm}{\raggedright {\ }{\ }{\ }from a private communication of T. Motobayashi with (\href{https://www.nndc.bnl.gov/nsr/nsrlink.jsp?2009Ta09,B}{2009Ta09}).\vspace{0.1cm}}&\\
&&&&&&\parbox[t][0.3cm]{12.363561cm}{\raggedright \ensuremath{\Gamma}\ensuremath{_{\ensuremath{\gamma}}}=45 meV \textit{+200{\textminus}33} (\href{https://www.nndc.bnl.gov/nsr/nsrlink.jsp?2000Ha26,B}{2000Ha26}). This result is from a combined analysis of Coulomb\vspace{0.1cm}}&\\
&&&&&&\parbox[t][0.3cm]{12.363561cm}{\raggedright {\ }{\ }{\ }excitation (\href{https://www.nndc.bnl.gov/nsr/nsrlink.jsp?2000Ha26,B}{2000Ha26}: \ensuremath{^{\textnormal{197}}}Au(\ensuremath{^{\textnormal{19}}}Ne,\ensuremath{^{\textnormal{19}}}Ne\ensuremath{'})) and DSAM (\href{https://www.nndc.bnl.gov/nsr/nsrlink.jsp?1973Da31,B}{1973Da31}: \ensuremath{^{\textnormal{17}}}O(\ensuremath{^{\textnormal{3}}}He,n\ensuremath{\gamma}))\vspace{0.1cm}}&\\
&&&&&&\parbox[t][0.3cm]{12.363561cm}{\raggedright {\ }{\ }{\ }assuming pure M1 transition and is recommended by (\href{https://www.nndc.bnl.gov/nsr/nsrlink.jsp?2003Da13,B}{2003Da13}). Considering that the\vspace{0.1cm}}&\\
&&&&&&\parbox[t][0.3cm]{12.363561cm}{\raggedright {\ }{\ }{\ }total width of this state is dominated by \ensuremath{\Gamma}\ensuremath{_{\ensuremath{\gamma}}}, this value is consistent with the adopted\vspace{0.1cm}}&\\
&&&&&&\parbox[t][0.3cm]{12.363561cm}{\raggedright {\ }{\ }{\ }half-life of this state within 1\ensuremath{\sigma}; however, the reported \ensuremath{\Gamma}\ensuremath{_{\ensuremath{\gamma}}} is less precise.\vspace{0.1cm}}&\\
&&&&&&\parbox[t][0.3cm]{12.363561cm}{\raggedright \ensuremath{\Gamma}\ensuremath{_{\ensuremath{\gamma}}}: See also (1) \ensuremath{\Gamma}\ensuremath{_{\ensuremath{\gamma}}}=73 meV \textit{41} (\href{https://www.nndc.bnl.gov/nsr/nsrlink.jsp?1990Ma05,B}{1990Ma05}: See Table 1) deduced from the\vspace{0.1cm}}&\\
\end{longtable}
\begin{textblock}{29}(0,27.3)
Continued on next page (footnotes at end of table)
\end{textblock}
\clearpage
\begin{longtable}{ccccccc@{\extracolsep{\fill}}c}
\\[-.4cm]
\multicolumn{8}{c}{{\bf \small \underline{Adopted \hyperlink{19NE_LEVEL}{Levels}, \hyperlink{19NE_GAMMA}{Gammas} (continued)}}}\\
\multicolumn{8}{c}{~}\\
\multicolumn{8}{c}{\underline{\ensuremath{^{19}}Ne Levels (continued)}}\\
\multicolumn{8}{c}{~}\\
\multicolumn{2}{c}{E(level)$^{}$}&J$^{\pi}$$^{}$&\multicolumn{2}{c}{T\ensuremath{_{\textnormal{1/2}}} or \ensuremath{\Gamma}$^{}$}&XREF&Comments&\\[-.2cm]
\multicolumn{2}{c}{\hrulefill}&\hrulefill&\multicolumn{2}{c}{\hrulefill}&\hrulefill&\hrulefill&
\endhead
\multicolumn{1}{r@{}}{4142}&\multicolumn{1}{@{.}l}{8\ensuremath{^{{\hyperlink{NE0LEVEL1}{b}}}} {\it 6}}&\multicolumn{1}{l}{(7/2\ensuremath{^{-}})\ensuremath{^{{\hyperlink{NE0LEVEL6}{g}}}}}&\multicolumn{1}{r@{}}{11}&\multicolumn{1}{@{.}l}{7 fs {\it +16\textminus22}}&\multicolumn{1}{l}{\texttt{\ \ \ \ \ \ \hyperlink{NE7}{G}\ \hyperlink{NE9}{I}\hyperlink{NE10}{J}\hyperlink{NE11}{K}\hyperlink{NE12}{L}\hyperlink{NE13}{M}\hyperlink{NE14}{N}\ \ \ \ \ \ \ \hyperlink{NE22}{V}\ \ \hyperlink{NE25}{Y}\ } }&\parbox[t][0.3cm]{7.437441cm}{\raggedright XREF: Others: \hyperlink{NE29}{\texttt{AC}}, \hyperlink{NE32}{\texttt{AF}}, \hyperlink{NE33}{\texttt{AG}}, \hyperlink{NE36}{\texttt{AJ}}, \hyperlink{NE38}{\texttt{AL}}\vspace{0.1cm}}&\\
&&&&&&\parbox[t][0.3cm]{7.437441cm}{\raggedright \ensuremath{\Gamma}\ensuremath{\alpha}/\ensuremath{\Gamma}\ensuremath{<}0.01 (\href{https://www.nndc.bnl.gov/nsr/nsrlink.jsp?2002La29,B}{2002La29})\vspace{0.1cm}}&\\
&&&&&&\parbox[t][0.3cm]{7.437441cm}{\raggedright E(level): Weighted average of 4152 keV \textit{15}\vspace{0.1cm}}&\\
&&&&&&\parbox[t][0.3cm]{7.437441cm}{\raggedright {\ }{\ }{\ }(\href{https://www.nndc.bnl.gov/nsr/nsrlink.jsp?1967Gr04,B}{1967Gr04}); 4.13 MeV \textit{3} (\href{https://www.nndc.bnl.gov/nsr/nsrlink.jsp?1968Gu07,B}{1968Gu07}); 4142 keV \textit{10}\vspace{0.1cm}}&\\
&&&&&&\parbox[t][0.3cm]{7.437441cm}{\raggedright {\ }{\ }{\ }(\href{https://www.nndc.bnl.gov/nsr/nsrlink.jsp?1970Ga18,B}{1970Ga18}); 4.15 MeV \textit{4} (\href{https://www.nndc.bnl.gov/nsr/nsrlink.jsp?1972Pa29,B}{1972Pa29}); 4140 keV \textit{4}\vspace{0.1cm}}&\\
&&&&&&\parbox[t][0.3cm]{7.437441cm}{\raggedright {\ }{\ }{\ }(\href{https://www.nndc.bnl.gov/nsr/nsrlink.jsp?1973Da31,B}{1973Da31}); 4153 keV \textit{4} (stat.) \textit{3} (sys.)\vspace{0.1cm}}&\\
\end{longtable}
\begin{textblock}{29}(0,27.3)
Continued on next page (footnotes at end of table)
\end{textblock}
\clearpage
\begin{longtable}{ccccccc@{\extracolsep{\fill}}c}
\\[-.4cm]
\multicolumn{8}{c}{{\bf \small \underline{Adopted \hyperlink{19NE_LEVEL}{Levels}, \hyperlink{19NE_GAMMA}{Gammas} (continued)}}}\\
\multicolumn{8}{c}{~}\\
\multicolumn{8}{c}{\underline{\ensuremath{^{19}}Ne Levels (continued)}}\\
\multicolumn{8}{c}{~}\\
\multicolumn{2}{c}{E(level)$^{}$}&J$^{\pi}$$^{}$&\multicolumn{2}{c}{T\ensuremath{_{\textnormal{1/2}}} or \ensuremath{\Gamma}$^{}$}&XREF&Comments&\\[-.2cm]
\multicolumn{2}{c}{\hrulefill}&\hrulefill&\multicolumn{2}{c}{\hrulefill}&\hrulefill&\hrulefill&
\endhead
&&&&&&\parbox[t][0.3cm]{8.156481cm}{\raggedright {\ }{\ }{\ }ray depopulating the 4199.8-keV state. A transition from\vspace{0.1cm}}&\\
&&&&&&\parbox[t][0.3cm]{8.156481cm}{\raggedright {\ }{\ }{\ }the 4200-keV state to the 238-keV state was not observed\vspace{0.1cm}}&\\
&&&&&&\parbox[t][0.3cm]{8.156481cm}{\raggedright {\ }{\ }{\ }by (\href{https://www.nndc.bnl.gov/nsr/nsrlink.jsp?2019Ha14,B}{2019Ha14}, \href{https://www.nndc.bnl.gov/nsr/nsrlink.jsp?2020Ha31,B}{2020Ha31}) even though their sensitivity\vspace{0.1cm}}&\\
&&&&&&\parbox[t][0.3cm]{8.156481cm}{\raggedright {\ }{\ }{\ }was such that they would observe the\vspace{0.1cm}}&\\
&&&&&&\parbox[t][0.3cm]{8.156481cm}{\raggedright {\ }{\ }{\ }\ensuremath{^{\textnormal{19}}}Ne*(4200)\ensuremath{\rightarrow}\ensuremath{^{\textnormal{19}}}Ne*(238) transition if it were to exist.\vspace{0.1cm}}&\\
&&&&&&\parbox[t][0.3cm]{8.156481cm}{\raggedright J\ensuremath{^{\pi}}: Based on the comparison of the \ensuremath{\gamma}-ray decay scheme\vspace{0.1cm}}&\\
&&&&&&\parbox[t][0.3cm]{8.156481cm}{\raggedright {\ }{\ }{\ }measured by (\href{https://www.nndc.bnl.gov/nsr/nsrlink.jsp?2019Ha14,B}{2019Ha14}, \href{https://www.nndc.bnl.gov/nsr/nsrlink.jsp?2020Ha31,B}{2020Ha31}) with those of\vspace{0.1cm}}&\\
&&&&&&\parbox[t][0.3cm]{8.156481cm}{\raggedright {\ }{\ }{\ }\ensuremath{^{\textnormal{19}}}F*(3999, 4032), we assigned J\ensuremath{^{\ensuremath{\pi}}}=(9/2\ensuremath{^{-}}) to this state\vspace{0.1cm}}&\\
\end{longtable}
\begin{textblock}{29}(0,27.3)
Continued on next page (footnotes at end of table)
\end{textblock}
\clearpage
\begin{longtable}{ccccccc@{\extracolsep{\fill}}c}
\\[-.4cm]
\multicolumn{8}{c}{{\bf \small \underline{Adopted \hyperlink{19NE_LEVEL}{Levels}, \hyperlink{19NE_GAMMA}{Gammas} (continued)}}}\\
\multicolumn{8}{c}{~}\\
\multicolumn{8}{c}{\underline{\ensuremath{^{19}}Ne Levels (continued)}}\\
\multicolumn{8}{c}{~}\\
\multicolumn{2}{c}{E(level)$^{}$}&J$^{\pi}$$^{}$&\multicolumn{2}{c}{T\ensuremath{_{\textnormal{1/2}}} or \ensuremath{\Gamma}$^{}$}&XREF&Comments&\\[-.2cm]
\multicolumn{2}{c}{\hrulefill}&\hrulefill&\multicolumn{2}{c}{\hrulefill}&\hrulefill&\hrulefill&
\endhead
&&&&&&\parbox[t][0.3cm]{8.062321cm}{\raggedright {\ }{\ }{\ }study.\vspace{0.1cm}}&\\
&&&&&&\parbox[t][0.3cm]{8.062321cm}{\raggedright See \ensuremath{\Gamma}\ensuremath{_{\ensuremath{\gamma}}}=458 meV \textit{92} (\href{https://www.nndc.bnl.gov/nsr/nsrlink.jsp?2003Da03,B}{2003Da03}): From shell model\vspace{0.1cm}}&\\
&&&&&&\parbox[t][0.3cm]{8.062321cm}{\raggedright {\ }{\ }{\ }calculation by B. A. Brown (priv. comm. with\vspace{0.1cm}}&\\
&&&&&&\parbox[t][0.3cm]{8.062321cm}{\raggedright {\ }{\ }{\ }\href{https://www.nndc.bnl.gov/nsr/nsrlink.jsp?2003Da03,B}{2003Da03}, \href{https://www.nndc.bnl.gov/nsr/nsrlink.jsp?2003Da13,B}{2003Da13}), where a 1\ensuremath{\sigma} uncertainty of 20\%\vspace{0.1cm}}&\\
&&&&&&\parbox[t][0.3cm]{8.062321cm}{\raggedright {\ }{\ }{\ }was assigned by those authors.\vspace{0.1cm}}&\\
&&&&&&\parbox[t][0.3cm]{8.062321cm}{\raggedright \ensuremath{\Gamma}\ensuremath{_{\ensuremath{\alpha}}}\ensuremath{<}2.3 meV (\href{https://www.nndc.bnl.gov/nsr/nsrlink.jsp?2003Da03,B}{2003Da03}, \href{https://www.nndc.bnl.gov/nsr/nsrlink.jsp?2003Da13,B}{2003Da13}): Deduced using a\vspace{0.1cm}}&\\
&&&&&&\parbox[t][0.3cm]{8.062321cm}{\raggedright {\ }{\ }{\ }1.28\ensuremath{\sigma} upper limit on the \ensuremath{\Gamma}\ensuremath{_{\ensuremath{\gamma}}} (\href{https://www.nndc.bnl.gov/nsr/nsrlink.jsp?2003Da13,B}{2003Da13}) and is\vspace{0.1cm}}&\\
&&&&&&\parbox[t][0.3cm]{8.062321cm}{\raggedright {\ }{\ }{\ }reported at 90\% C.L. See also \ensuremath{\Gamma}\ensuremath{_{\ensuremath{\alpha}}}\ensuremath{\leq}5.6 meV at 99.73\%\vspace{0.1cm}}&\\
\end{longtable}
\begin{textblock}{29}(0,27.3)
Continued on next page (footnotes at end of table)
\end{textblock}
\clearpage
\begin{longtable}{ccccccc@{\extracolsep{\fill}}c}
\\[-.4cm]
\multicolumn{8}{c}{{\bf \small \underline{Adopted \hyperlink{19NE_LEVEL}{Levels}, \hyperlink{19NE_GAMMA}{Gammas} (continued)}}}\\
\multicolumn{8}{c}{~}\\
\multicolumn{8}{c}{\underline{\ensuremath{^{19}}Ne Levels (continued)}}\\
\multicolumn{8}{c}{~}\\
\multicolumn{2}{c}{E(level)$^{}$}&J$^{\pi}$$^{}$&\multicolumn{2}{c}{T\ensuremath{_{\textnormal{1/2}}} or \ensuremath{\Gamma}$^{}$}&XREF&Comments&\\[-.2cm]
\multicolumn{2}{c}{\hrulefill}&\hrulefill&\multicolumn{2}{c}{\hrulefill}&\hrulefill&\hrulefill&
\endhead
&&&&&&\parbox[t][0.3cm]{8.578061cm}{\raggedright \ensuremath{\Gamma}\ensuremath{_{\ensuremath{\alpha}}}: Other values: 3.4 meV (\href{https://www.nndc.bnl.gov/nsr/nsrlink.jsp?1995Wi26,B}{1995Wi26}) calculated from\vspace{0.1cm}}&\\
&&&&&&\parbox[t][0.3cm]{8.578061cm}{\raggedright {\ }{\ }{\ }potential model; 3.2 meV (\href{https://www.nndc.bnl.gov/nsr/nsrlink.jsp?1995Wi26,B}{1995Wi26}) deduced from Wigner\vspace{0.1cm}}&\\
&&&&&&\parbox[t][0.3cm]{8.578061cm}{\raggedright {\ }{\ }{\ }limit with a channel radius R=6.0 fm; 2.9 meV \textit{+17{\textminus}14}\vspace{0.1cm}}&\\
&&&&&&\parbox[t][0.3cm]{8.578061cm}{\raggedright {\ }{\ }{\ }evaluated by (\href{https://www.nndc.bnl.gov/nsr/nsrlink.jsp?1997De14,B}{1997De14}); 5 \ensuremath{\mu}eV (\href{https://www.nndc.bnl.gov/nsr/nsrlink.jsp?1987MaZQ,B}{1987MaZQ}) from the\vspace{0.1cm}}&\\
&&&&&&\parbox[t][0.3cm]{8.578061cm}{\raggedright {\ }{\ }{\ }analog state; and \ensuremath{<}3.8 meV (\href{https://www.nndc.bnl.gov/nsr/nsrlink.jsp?1987Ma31,B}{1987Ma31}) deduced for the\vspace{0.1cm}}&\\
&&&&&&\parbox[t][0.3cm]{8.578061cm}{\raggedright {\ }{\ }{\ }analog state from \ensuremath{^{\textnormal{15}}}N(\ensuremath{\alpha},\ensuremath{\gamma})\ensuremath{^{\textnormal{19}}}F, see Table 1 in (\href{https://www.nndc.bnl.gov/nsr/nsrlink.jsp?1990Ma05,B}{1990Ma05}).\vspace{0.1cm}}&\\
&&&&&&\parbox[t][0.3cm]{8.578061cm}{\raggedright \ensuremath{\Gamma}\ensuremath{_{\ensuremath{\alpha}}}/\ensuremath{\Gamma}=0.16 \textit{4} (stat.) (\href{https://www.nndc.bnl.gov/nsr/nsrlink.jsp?2003Da03,B}{2003Da03}, \href{https://www.nndc.bnl.gov/nsr/nsrlink.jsp?2003Da13,B}{2003Da13}). Other values:\vspace{0.1cm}}&\\
\end{longtable}
\begin{textblock}{29}(0,27.3)
Continued on next page (footnotes at end of table)
\end{textblock}
\clearpage
\begin{longtable}{ccccccc@{\extracolsep{\fill}}c}
\\[-.4cm]
\multicolumn{8}{c}{{\bf \small \underline{Adopted \hyperlink{19NE_LEVEL}{Levels}, \hyperlink{19NE_GAMMA}{Gammas} (continued)}}}\\
\multicolumn{8}{c}{~}\\
\multicolumn{8}{c}{\underline{\ensuremath{^{19}}Ne Levels (continued)}}\\
\multicolumn{8}{c}{~}\\
\multicolumn{2}{c}{E(level)$^{}$}&J$^{\pi}$$^{}$&\multicolumn{2}{c}{T\ensuremath{_{\textnormal{1/2}}} or \ensuremath{\Gamma}$^{}$}&XREF&Comments&\\[-.2cm]
\multicolumn{2}{c}{\hrulefill}&\hrulefill&\multicolumn{2}{c}{\hrulefill}&\hrulefill&\hrulefill&
\endhead
&&&&&&\parbox[t][0.3cm]{8.419701cm}{\raggedright \ensuremath{\Gamma}\ensuremath{_{\ensuremath{\alpha}}}=33 meV \textit{+24{\textminus}10}: Weighted average of 43 meV \textit{24}\vspace{0.1cm}}&\\
&&&&&&\parbox[t][0.3cm]{8.419701cm}{\raggedright {\ }{\ }{\ }(\href{https://www.nndc.bnl.gov/nsr/nsrlink.jsp?2003Da03,B}{2003Da03}, \href{https://www.nndc.bnl.gov/nsr/nsrlink.jsp?2003Da13,B}{2003Da13}) and 24 meV \textit{+33{\textminus}10} (\href{https://www.nndc.bnl.gov/nsr/nsrlink.jsp?2009Ta09,B}{2009Ta09}).\vspace{0.1cm}}&\\
&&&&&&\parbox[t][0.3cm]{8.419701cm}{\raggedright \ensuremath{\Gamma}\ensuremath{_{\ensuremath{\alpha}}}: Other values: \ensuremath{\Gamma}\ensuremath{_{\ensuremath{\alpha}}}=32 \ensuremath{\mu}eV \textit{7} (\href{https://www.nndc.bnl.gov/nsr/nsrlink.jsp?1987MaZQ,B}{1987MaZQ}) deduced\vspace{0.1cm}}&\\
&&&&&&\parbox[t][0.3cm]{8.419701cm}{\raggedright {\ }{\ }{\ }from the analog state; \ensuremath{\Gamma}\ensuremath{_{\ensuremath{\alpha}}}=88 meV \textit{18} (\href{https://www.nndc.bnl.gov/nsr/nsrlink.jsp?1987Ma31,B}{1987Ma31}) from\vspace{0.1cm}}&\\
&&&&&&\parbox[t][0.3cm]{8.419701cm}{\raggedright {\ }{\ }{\ }\ensuremath{^{\textnormal{15}}}N(\ensuremath{\alpha},\ensuremath{\gamma})\ensuremath{^{\textnormal{19}}}F mirror analysis, see Table 1 in (\href{https://www.nndc.bnl.gov/nsr/nsrlink.jsp?1990Ma05,B}{1990Ma05});\vspace{0.1cm}}&\\
&&&&&&\parbox[t][0.3cm]{8.419701cm}{\raggedright {\ }{\ }{\ }83 meV (\href{https://www.nndc.bnl.gov/nsr/nsrlink.jsp?1995Wi26,B}{1995Wi26}) calculated from potential model;\vspace{0.1cm}}&\\
&&&&&&\parbox[t][0.3cm]{8.419701cm}{\raggedright {\ }{\ }{\ }\ensuremath{\Gamma}\ensuremath{_{\ensuremath{\alpha}}}=81.9 meV (\href{https://www.nndc.bnl.gov/nsr/nsrlink.jsp?1995Wi26,B}{1995Wi26}) deduced from Wigner limit\vspace{0.1cm}}&\\
&&&&&&\parbox[t][0.3cm]{8.419701cm}{\raggedright {\ }{\ }{\ }with a channel radius R=6.0 fm; \ensuremath{\Gamma}\ensuremath{_{\ensuremath{\alpha}}}=33 meV \textit{18} evaluated\vspace{0.1cm}}&\\
\end{longtable}
\begin{textblock}{29}(0,27.3)
Continued on next page (footnotes at end of table)
\end{textblock}
\clearpage
\begin{longtable}{ccccc@{\extracolsep{\fill}}c}
\\[-.4cm]
\multicolumn{6}{c}{{\bf \small \underline{Adopted \hyperlink{19NE_LEVEL}{Levels}, \hyperlink{19NE_GAMMA}{Gammas} (continued)}}}\\
\multicolumn{6}{c}{~}\\
\multicolumn{6}{c}{\underline{\ensuremath{^{19}}Ne Levels (continued)}}\\
\multicolumn{6}{c}{~}\\
\multicolumn{2}{c}{E(level)$^{}$}&J$^{\pi}$$^{}$&XREF&Comments&\\[-.2cm]
\multicolumn{2}{c}{\hrulefill}&\hrulefill&\hrulefill&\hrulefill&
\endhead
&&&&\parbox[t][0.3cm]{10.465361cm}{\raggedright {\ }{\ }{\ }populating the \ensuremath{^{\textnormal{19}}}F* and \ensuremath{^{\textnormal{19}}}Ne* analog states, respectively. Furthermore,\vspace{0.1cm}}&\\
&&&&\parbox[t][0.3cm]{10.465361cm}{\raggedright {\ }{\ }{\ }(\href{https://www.nndc.bnl.gov/nsr/nsrlink.jsp?1986Ke04,B}{1986Ke04}: \ensuremath{^{\textnormal{18}}}O(pol. p, \ensuremath{\pi}\ensuremath{^{-}})) reported that the pattern of the measured\vspace{0.1cm}}&\\
&&&&\parbox[t][0.3cm]{10.465361cm}{\raggedright {\ }{\ }{\ }analyzing power for this state supports J\ensuremath{^{\ensuremath{\pi}}}=13/2\ensuremath{^{\textnormal{+}}}.\vspace{0.1cm}}&\\
&&&&\parbox[t][0.3cm]{10.465361cm}{\raggedright The J\ensuremath{^{\ensuremath{\pi}}}=13/2\ensuremath{^{\textnormal{+}}} assignment is supported by the comparison of the \ensuremath{\gamma}-ray\vspace{0.1cm}}&\\
&&&&\parbox[t][0.3cm]{10.465361cm}{\raggedright {\ }{\ }{\ }decay scheme of this state with that of the mirror state, \ensuremath{^{\textnormal{19}}}F*(4648),\vspace{0.1cm}}&\\
&&&&\parbox[t][0.3cm]{10.465361cm}{\raggedright {\ }{\ }{\ }proposed by (\href{https://www.nndc.bnl.gov/nsr/nsrlink.jsp?1973Da31,B}{1973Da31}, \href{https://www.nndc.bnl.gov/nsr/nsrlink.jsp?2009Ta09,B}{2009Ta09}, \href{https://www.nndc.bnl.gov/nsr/nsrlink.jsp?2020Ha31,B}{2020Ha31}). (\href{https://www.nndc.bnl.gov/nsr/nsrlink.jsp?1973Da31,B}{1973Da31}: \ensuremath{^{\textnormal{17}}}O(\ensuremath{^{\textnormal{3}}}He,n\ensuremath{\gamma}))\vspace{0.1cm}}&\\
&&&&\parbox[t][0.3cm]{10.465361cm}{\raggedright {\ }{\ }{\ }reported that if J=(9/2, 11/2), the M1 or E1 transition strengths would be\vspace{0.1cm}}&\\
&&&&\parbox[t][0.3cm]{10.465361cm}{\raggedright {\ }{\ }{\ }too weak to be observed.\vspace{0.1cm}}&\\
\end{longtable}
\begin{textblock}{29}(0,27.3)
Continued on next page (footnotes at end of table)
\end{textblock}
\clearpage
\begin{longtable}{ccccccc@{\extracolsep{\fill}}c}
\\[-.4cm]
\multicolumn{8}{c}{{\bf \small \underline{Adopted \hyperlink{19NE_LEVEL}{Levels}, \hyperlink{19NE_GAMMA}{Gammas} (continued)}}}\\
\multicolumn{8}{c}{~}\\
\multicolumn{8}{c}{\underline{\ensuremath{^{19}}Ne Levels (continued)}}\\
\multicolumn{8}{c}{~}\\
\multicolumn{2}{c}{E(level)$^{}$}&J$^{\pi}$$^{}$&\multicolumn{2}{c}{T\ensuremath{_{\textnormal{1/2}}} or \ensuremath{\Gamma}$^{}$}&XREF&Comments&\\[-.2cm]
\multicolumn{2}{c}{\hrulefill}&\hrulefill&\multicolumn{2}{c}{\hrulefill}&\hrulefill&\hrulefill&
\endhead
&&&&&&\parbox[t][0.3cm]{8.383321cm}{\raggedright XREF: y(4.74E3)\vspace{0.1cm}}&\\
&&&&&&\parbox[t][0.3cm]{8.383321cm}{\raggedright E(level): From (\href{https://www.nndc.bnl.gov/nsr/nsrlink.jsp?1970Ga18,B}{1970Ga18}: \ensuremath{^{\textnormal{20}}}Ne(\ensuremath{^{\textnormal{3}}}He,\ensuremath{\alpha})), where the state is\vspace{0.1cm}}&\\
&&&&&&\parbox[t][0.3cm]{8.383321cm}{\raggedright {\ }{\ }{\ }populated very weakly. See also (\href{https://www.nndc.bnl.gov/nsr/nsrlink.jsp?2023Ma57,B}{2023Ma57}), who\vspace{0.1cm}}&\\
&&&&&&\parbox[t][0.3cm]{8.383321cm}{\raggedright {\ }{\ }{\ }populated a state at E\ensuremath{_{\textnormal{x}}}=4.74 MeV \textit{10} (sys.) but chose to\vspace{0.1cm}}&\\
&&&&&&\parbox[t][0.3cm]{8.383321cm}{\raggedright {\ }{\ }{\ }associate their state with E\ensuremath{_{\textnormal{x}}}=4712 keV \textit{10} of (\href{https://www.nndc.bnl.gov/nsr/nsrlink.jsp?2009Ta09,B}{2009Ta09}).\vspace{0.1cm}}&\\
&&&&&&\parbox[t][0.3cm]{8.383321cm}{\raggedright {\ }{\ }{\ }We considered this state questionable as it was disputed by\vspace{0.1cm}}&\\
&&&&&&\parbox[t][0.3cm]{8.383321cm}{\raggedright {\ }{\ }{\ }(\href{https://www.nndc.bnl.gov/nsr/nsrlink.jsp?1973Da31,B}{1973Da31}), who suggested that it is erroneous. See the\vspace{0.1cm}}&\\
&&&&&&\parbox[t][0.3cm]{8.383321cm}{\raggedright {\ }{\ }{\ }discussion in the \ensuremath{^{\textnormal{17}}}O(\ensuremath{^{\textnormal{3}}}He,n) dataset.\vspace{0.1cm}}&\\
\end{longtable}
\begin{textblock}{29}(0,27.3)
Continued on next page (footnotes at end of table)
\end{textblock}
\clearpage
\begin{longtable}{ccccccc@{\extracolsep{\fill}}c}
\\[-.4cm]
\multicolumn{8}{c}{{\bf \small \underline{Adopted \hyperlink{19NE_LEVEL}{Levels}, \hyperlink{19NE_GAMMA}{Gammas} (continued)}}}\\
\multicolumn{8}{c}{~}\\
\multicolumn{8}{c}{\underline{\ensuremath{^{19}}Ne Levels (continued)}}\\
\multicolumn{8}{c}{~}\\
\multicolumn{2}{c}{E(level)$^{}$}&J$^{\pi}$$^{}$&\multicolumn{2}{c}{T\ensuremath{_{\textnormal{1/2}}} or \ensuremath{\Gamma}$^{}$}&XREF&Comments&\\[-.2cm]
\multicolumn{2}{c}{\hrulefill}&\hrulefill&\multicolumn{2}{c}{\hrulefill}&\hrulefill&\hrulefill&
\endhead
&&&&&&\parbox[t][0.3cm]{8.233521cm}{\raggedright {\ }{\ }{\ }\ensuremath{\Gamma}\ensuremath{_{\ensuremath{\gamma}}}=196 meV \textit{39} deduced using shell model calculations\vspace{0.1cm}}&\\
&&&&&&\parbox[t][0.3cm]{8.233521cm}{\raggedright {\ }{\ }{\ }obtained by (\href{https://www.nndc.bnl.gov/nsr/nsrlink.jsp?2003Da03,B}{2003Da03}) via a priv. comm. with B. A.\vspace{0.1cm}}&\\
&&&&&&\parbox[t][0.3cm]{8.233521cm}{\raggedright {\ }{\ }{\ }Brown. (\href{https://www.nndc.bnl.gov/nsr/nsrlink.jsp?2003Da03,B}{2003Da03}) assigned a 1\ensuremath{\sigma} uncertainty of 20\% to\vspace{0.1cm}}&\\
&&&&&&\parbox[t][0.3cm]{8.233521cm}{\raggedright {\ }{\ }{\ }\ensuremath{\Gamma}\ensuremath{_{\ensuremath{\gamma}}}=196 meV.\vspace{0.1cm}}&\\
&&&&&&\parbox[t][0.3cm]{8.233521cm}{\raggedright Decay mode: Predominantly \ensuremath{\alpha} (\href{https://www.nndc.bnl.gov/nsr/nsrlink.jsp?2003Da03,B}{2003Da03}, \href{https://www.nndc.bnl.gov/nsr/nsrlink.jsp?2003Da13,B}{2003Da13},\vspace{0.1cm}}&\\
&&&&&&\parbox[t][0.3cm]{8.233521cm}{\raggedright {\ }{\ }{\ }\href{https://www.nndc.bnl.gov/nsr/nsrlink.jsp?2011Ad05,B}{2011Ad05}, \href{https://www.nndc.bnl.gov/nsr/nsrlink.jsp?2011Ad24,B}{2011Ad24}).\vspace{0.1cm}}&\\
\multicolumn{1}{r@{}}{5350}&\multicolumn{1}{@{ }l}{{\it 10}}&\multicolumn{1}{l}{1/2\ensuremath{^{+}}}&\multicolumn{1}{r@{}}{6}&\multicolumn{1}{@{.}l}{6 keV {\it 34}}&\multicolumn{1}{l}{\texttt{\ \hyperlink{NE2}{B}\ \ \ \ \ \hyperlink{NE8}{H}\hyperlink{NE9}{I}\ \ \ \hyperlink{NE13}{M}\ \ \ \ \ \ \ \ \ \ \ \hyperlink{NE25}{Y}\ } }&\parbox[t][0.3cm]{8.233521cm}{\raggedright XREF: Others: \hyperlink{NE31}{\texttt{AE}}, \hyperlink{NE32}{\texttt{AF}}, \hyperlink{NE38}{\texttt{AL}}\vspace{0.1cm}}&\\
&&&&&&\parbox[t][0.3cm]{8.233521cm}{\raggedright \%\ensuremath{\alpha}=100 (\href{https://www.nndc.bnl.gov/nsr/nsrlink.jsp?1990Ma05,B}{1990Ma05},\href{https://www.nndc.bnl.gov/nsr/nsrlink.jsp?2009Ta09,B}{2009Ta09})\vspace{0.1cm}}&\\
\end{longtable}
\begin{textblock}{29}(0,27.3)
Continued on next page (footnotes at end of table)
\end{textblock}
\clearpage
\begin{longtable}{ccccccc@{\extracolsep{\fill}}c}
\\[-.4cm]
\multicolumn{8}{c}{{\bf \small \underline{Adopted \hyperlink{19NE_LEVEL}{Levels}, \hyperlink{19NE_GAMMA}{Gammas} (continued)}}}\\
\multicolumn{8}{c}{~}\\
\multicolumn{8}{c}{\underline{\ensuremath{^{19}}Ne Levels (continued)}}\\
\multicolumn{8}{c}{~}\\
\multicolumn{2}{c}{E(level)$^{}$}&J$^{\pi}$$^{}$&\multicolumn{2}{c}{T\ensuremath{_{\textnormal{1/2}}} or \ensuremath{\Gamma}$^{}$}&XREF&Comments&\\[-.2cm]
\multicolumn{2}{c}{\hrulefill}&\hrulefill&\multicolumn{2}{c}{\hrulefill}&\hrulefill&\hrulefill&
\endhead
&&&&&&\parbox[t][0.3cm]{7.906101cm}{\raggedright {\ }{\ }{\ }9/2\ensuremath{^{\textnormal{+}}}) and L=4. Those authors selected J\ensuremath{^{\ensuremath{\pi}}}=(7/2\ensuremath{^{\textnormal{+}}}) based\vspace{0.1cm}}&\\
&&&&&&\parbox[t][0.3cm]{7.906101cm}{\raggedright {\ }{\ }{\ }on mirror analysis confirmed by (\href{https://www.nndc.bnl.gov/nsr/nsrlink.jsp?1971Bi06,B}{1971Bi06}).\vspace{0.1cm}}&\\
&&&&&&\parbox[t][0.3cm]{7.906101cm}{\raggedright \ensuremath{\Gamma}\ensuremath{_{\ensuremath{\alpha}}}/\ensuremath{\Gamma}: No evidence for the \ensuremath{\gamma} decay of this state was\vspace{0.1cm}}&\\
&&&&&&\parbox[t][0.3cm]{7.906101cm}{\raggedright {\ }{\ }{\ }observed by (\href{https://www.nndc.bnl.gov/nsr/nsrlink.jsp?2003Da13,B}{2003Da13}). Therefore, they reported\vspace{0.1cm}}&\\
&&&&&&\parbox[t][0.3cm]{7.906101cm}{\raggedright {\ }{\ }{\ }\ensuremath{\Gamma}\ensuremath{_{\ensuremath{\alpha}}}/\ensuremath{\Gamma}=100\% for this state. This was confirmed by\vspace{0.1cm}}&\\
&&&&&&\parbox[t][0.3cm]{7.906101cm}{\raggedright {\ }{\ }{\ }(\href{https://www.nndc.bnl.gov/nsr/nsrlink.jsp?2023Ma57,B}{2023Ma57}). However, (\href{https://www.nndc.bnl.gov/nsr/nsrlink.jsp?2025PhZZ,B}{2025PhZZ}) reports\vspace{0.1cm}}&\\
&&&&&&\parbox[t][0.3cm]{7.906101cm}{\raggedright {\ }{\ }{\ }\ensuremath{\Gamma}\ensuremath{_{\ensuremath{\alpha}}}/\ensuremath{\Gamma}=0.71 \textit{11} (unpublished).\vspace{0.1cm}}&\\
&&&&&&\parbox[t][0.3cm]{7.906101cm}{\raggedright Decay mode: \ensuremath{\alpha} (\href{https://www.nndc.bnl.gov/nsr/nsrlink.jsp?2003Da03,B}{2003Da03}, \href{https://www.nndc.bnl.gov/nsr/nsrlink.jsp?2003Da13,B}{2003Da13}).\vspace{0.1cm}}&\\
\end{longtable}
\begin{textblock}{29}(0,27.3)
Continued on next page (footnotes at end of table)
\end{textblock}
\clearpage
\begin{longtable}{ccccccc@{\extracolsep{\fill}}c}
\\[-.4cm]
\multicolumn{8}{c}{{\bf \small \underline{Adopted \hyperlink{19NE_LEVEL}{Levels}, \hyperlink{19NE_GAMMA}{Gammas} (continued)}}}\\
\multicolumn{8}{c}{~}\\
\multicolumn{8}{c}{\underline{\ensuremath{^{19}}Ne Levels (continued)}}\\
\multicolumn{8}{c}{~}\\
\multicolumn{2}{c}{E(level)$^{}$}&J$^{\pi}$$^{}$&\multicolumn{2}{c}{T\ensuremath{_{\textnormal{1/2}}} or \ensuremath{\Gamma}$^{}$}&XREF&Comments&\\[-.2cm]
\multicolumn{2}{c}{\hrulefill}&\hrulefill&\multicolumn{2}{c}{\hrulefill}&\hrulefill&\hrulefill&
\endhead
&&&&&&\parbox[t][0.3cm]{7.62576cm}{\raggedright \ensuremath{\Gamma}\ensuremath{\alpha}/\ensuremath{\Gamma}=1.0 (\href{https://www.nndc.bnl.gov/nsr/nsrlink.jsp?2003Da13,B}{2003Da13})\vspace{0.1cm}}&\\
&&&&&&\parbox[t][0.3cm]{7.62576cm}{\raggedright E(level): From 5830 keV \textit{6} (stat.) \textit{10} (sys.): Weighted\vspace{0.1cm}}&\\
&&&&&&\parbox[t][0.3cm]{7.62576cm}{\raggedright {\ }{\ }{\ }average of 5831 keV \textit{10} (\href{https://www.nndc.bnl.gov/nsr/nsrlink.jsp?1970Ga18,B}{1970Ga18}); 5837 keV \textit{20}\vspace{0.1cm}}&\\
&&&&&&\parbox[t][0.3cm]{7.62576cm}{\raggedright {\ }{\ }{\ }(\href{https://www.nndc.bnl.gov/nsr/nsrlink.jsp?1972Ha03,B}{1972Ha03}); 5836 keV \textit{17} (stat.) \textit{10} (sys.) from\vspace{0.1cm}}&\\
&&&&&&\parbox[t][0.3cm]{7.62576cm}{\raggedright {\ }{\ }{\ }E\ensuremath{_{\textnormal{c.m.}}}(\ensuremath{^{\textnormal{18}}}F+p) given in (\href{https://www.nndc.bnl.gov/nsr/nsrlink.jsp?2015Ch41,B}{2015Ch41}); and 5824 keV \textit{10}\vspace{0.1cm}}&\\
&&&&&&\parbox[t][0.3cm]{7.62576cm}{\raggedright {\ }{\ }{\ }(\href{https://www.nndc.bnl.gov/nsr/nsrlink.jsp?2019Ka15,B}{2019Ka15}).\vspace{0.1cm}}&\\
&&&&&&\parbox[t][0.3cm]{7.62576cm}{\raggedright J\ensuremath{^{\pi}}: From (\href{https://www.nndc.bnl.gov/nsr/nsrlink.jsp?2019Ka15,B}{2019Ka15}): \ensuremath{^{\textnormal{19}}}F(\ensuremath{^{\textnormal{3}}}He,t), \ensuremath{\Delta}L=0, which leads\vspace{0.1cm}}&\\
&&&&&&\parbox[t][0.3cm]{7.62576cm}{\raggedright {\ }{\ }{\ }to J\ensuremath{^{\ensuremath{\pi}}}=1/2\ensuremath{^{\textnormal{+}}}, 3/2\ensuremath{^{\textnormal{+}}}. (\href{https://www.nndc.bnl.gov/nsr/nsrlink.jsp?2019Ka15,B}{2019Ka15}) suggested J\ensuremath{^{\ensuremath{\pi}}}=1/2\ensuremath{^{\textnormal{+}}}\vspace{0.1cm}}&\\
\end{longtable}
\begin{textblock}{29}(0,27.3)
Continued on next page (footnotes at end of table)
\end{textblock}
\clearpage
\begin{longtable}{ccccc@{\extracolsep{\fill}}c}
\\[-.4cm]
\multicolumn{6}{c}{{\bf \small \underline{Adopted \hyperlink{19NE_LEVEL}{Levels}, \hyperlink{19NE_GAMMA}{Gammas} (continued)}}}\\
\multicolumn{6}{c}{~}\\
\multicolumn{6}{c}{\underline{\ensuremath{^{19}}Ne Levels (continued)}}\\
\multicolumn{6}{c}{~}\\
\multicolumn{2}{c}{E(level)$^{}$}&J$^{\pi}$$^{}$&XREF&Comments&\\[-.2cm]
\multicolumn{2}{c}{\hrulefill}&\hrulefill&\hrulefill&\hrulefill&
\endhead
&&&&\parbox[t][0.3cm]{9.493801cm}{\raggedright {\ }{\ }{\ }J\ensuremath{^{\ensuremath{\pi}}}=(3/2\ensuremath{^{\textnormal{+}}}, 5/2\ensuremath{^{-}}). This result was also recommended by the\vspace{0.1cm}}&\\
&&&&\parbox[t][0.3cm]{9.493801cm}{\raggedright {\ }{\ }{\ }reanalysis of the same data by (\href{https://www.nndc.bnl.gov/nsr/nsrlink.jsp?2015Ch41,B}{2015Ch41}). Furthermore, from the\vspace{0.1cm}}&\\
&&&&\parbox[t][0.3cm]{9.493801cm}{\raggedright {\ }{\ }{\ }analysis of the \ensuremath{\alpha} angular distribution of \ensuremath{^{\textnormal{19}}}F(\ensuremath{^{\textnormal{3}}}He,t)\ensuremath{^{\textnormal{19}}}Ne*(\ensuremath{\alpha}) with\vspace{0.1cm}}&\\
&&&&\parbox[t][0.3cm]{9.493801cm}{\raggedright {\ }{\ }{\ }t-\ensuremath{\alpha} coincidence measurement by (\href{https://www.nndc.bnl.gov/nsr/nsrlink.jsp?2025PhZZ,B}{2025PhZZ}), angular momentum\vspace{0.1cm}}&\\
&&&&\parbox[t][0.3cm]{9.493801cm}{\raggedright {\ }{\ }{\ }transfers of L=1,2 were determined, which led to J\ensuremath{^{\ensuremath{\pi}}}=1/2\ensuremath{^{\textnormal{+}}}, 3/2,\vspace{0.1cm}}&\\
&&&&\parbox[t][0.3cm]{9.493801cm}{\raggedright {\ }{\ }{\ }5/2\ensuremath{^{-}}. Those authors preferred L=1 and J\ensuremath{^{\ensuremath{\pi}}}=1/2\ensuremath{^{\textnormal{+}}}, 3/2\ensuremath{^{\textnormal{+}}}. We note that\vspace{0.1cm}}&\\
&&&&\parbox[t][0.3cm]{9.493801cm}{\raggedright {\ }{\ }{\ }both L-values equally describe the data of (\href{https://www.nndc.bnl.gov/nsr/nsrlink.jsp?2025PhZZ,B}{2025PhZZ}).\vspace{0.1cm}}&\\
&&&&\parbox[t][0.3cm]{9.493801cm}{\raggedright See \ensuremath{\Gamma}\ensuremath{_{\ensuremath{\alpha}}}/\ensuremath{\Gamma}=0.79 \textit{7} (\href{https://www.nndc.bnl.gov/nsr/nsrlink.jsp?2025PhZZ,B}{2025PhZZ}, unpublished).\vspace{0.1cm}}&\\
\end{longtable}
\begin{textblock}{29}(0,27.3)
Continued on next page (footnotes at end of table)
\end{textblock}
\clearpage
\begin{longtable}{ccccccc@{\extracolsep{\fill}}c}
\\[-.4cm]
\multicolumn{8}{c}{{\bf \small \underline{Adopted \hyperlink{19NE_LEVEL}{Levels}, \hyperlink{19NE_GAMMA}{Gammas} (continued)}}}\\
\multicolumn{8}{c}{~}\\
\multicolumn{8}{c}{\underline{\ensuremath{^{19}}Ne Levels (continued)}}\\
\multicolumn{8}{c}{~}\\
\multicolumn{2}{c}{E(level)$^{}$}&J$^{\pi}$$^{}$&\multicolumn{2}{c}{T\ensuremath{_{\textnormal{1/2}}} or \ensuremath{\Gamma}$^{}$}&XREF&Comments&\\[-.2cm]
\multicolumn{2}{c}{\hrulefill}&\hrulefill&\multicolumn{2}{c}{\hrulefill}&\hrulefill&\hrulefill&
\endhead
&&&&&&\parbox[t][0.3cm]{7.446cm}{\raggedright {\ }{\ }{\ }J\ensuremath{^{\ensuremath{\pi}}}=3/2\ensuremath{^{\textnormal{+}}} assignment for \ensuremath{^{\textnormal{19}}}Ne*(6132.3). (\href{https://www.nndc.bnl.gov/nsr/nsrlink.jsp?2021Ka51,B}{2021Ka51})\vspace{0.1cm}}&\\
&&&&&&\parbox[t][0.3cm]{7.446cm}{\raggedright {\ }{\ }{\ }supports the J\ensuremath{^{\ensuremath{\pi}}}=1/2\ensuremath{^{\textnormal{+}}} assignment, while (\href{https://www.nndc.bnl.gov/nsr/nsrlink.jsp?2021Ri04,B}{2021Ri04})\vspace{0.1cm}}&\\
&&&&&&\parbox[t][0.3cm]{7.446cm}{\raggedright {\ }{\ }{\ }reports that literature results support J\ensuremath{^{\ensuremath{\pi}}}=3/2\ensuremath{^{\textnormal{+}}}. So\vspace{0.1cm}}&\\
&&&&&&\parbox[t][0.3cm]{7.446cm}{\raggedright {\ }{\ }{\ }we accepted both and made them tentative.\vspace{0.1cm}}&\\
&&&&&&\parbox[t][0.3cm]{7.446cm}{\raggedright \ensuremath{\Gamma}\ensuremath{_{\ensuremath{\alpha}}}/\ensuremath{\Gamma}=0.86 \textit{5} (\href{https://www.nndc.bnl.gov/nsr/nsrlink.jsp?2025PhZZ,B}{2025PhZZ}, unpublished).\vspace{0.1cm}}&\\
\multicolumn{1}{r@{}}{6197}&\multicolumn{1}{@{ }l}{{\it 51}}&\multicolumn{1}{l}{(1/2)}&\multicolumn{1}{r@{}}{16}&\multicolumn{1}{@{ }l}{keV {\it 7}}&\multicolumn{1}{l}{\texttt{\ \ \ \ \ \ \ \ \ \ \ \ \hyperlink{NE13}{M}\ \ \ \ \ \ \ \ \ \ \hyperlink{NE24}{x}\ \ } }&\parbox[t][0.3cm]{7.446cm}{\raggedright XREF: Others: \hyperlink{NE31}{\texttt{AE}}\vspace{0.1cm}}&\\
&&&&&&\parbox[t][0.3cm]{7.446cm}{\raggedright \ensuremath{\Gamma}\ensuremath{\alpha}=16 keV \textit{5} (\href{https://www.nndc.bnl.gov/nsr/nsrlink.jsp?2017To14,B}{2017To14})\vspace{0.1cm}}&\\
&&&&&&\parbox[t][0.3cm]{7.446cm}{\raggedright XREF: x(6.2E3)ae(6.2E3)\vspace{0.1cm}}&\\
\end{longtable}%
\parbox[b][0.3cm]{17.7cm}{\makebox[1ex]{\ensuremath{^{\hypertarget{NE0LEVEL0}{a}}}} Seq.(A): K\ensuremath{^{\ensuremath{\pi}}}=1/2\ensuremath{^{+}} g.s. band (\href{https://www.nndc.bnl.gov/nsr/nsrlink.jsp?1971Bi06,B}{1971Bi06}).}\\
\parbox[b][0.3cm]{17.7cm}{\makebox[1ex]{\ensuremath{^{\hypertarget{NE0LEVEL1}{b}}}} Seq.(B): K\ensuremath{^{\ensuremath{\pi}}}=1/2\ensuremath{^{-}} band (\href{https://www.nndc.bnl.gov/nsr/nsrlink.jsp?1970Ga18,B}{1970Ga18},\href{https://www.nndc.bnl.gov/nsr/nsrlink.jsp?1971Bi06,B}{1971Bi06}).}\\
\parbox[b][0.3cm]{17.7cm}{\makebox[1ex]{\ensuremath{^{\hypertarget{NE0LEVEL2}{c}}}} Seq.(C): K\ensuremath{^{\ensuremath{\pi}}}=3/2\ensuremath{^{+}} band (\href{https://www.nndc.bnl.gov/nsr/nsrlink.jsp?1995Wi26,B}{1995Wi26}).}\\
\parbox[b][0.3cm]{17.7cm}{\makebox[1ex]{\ensuremath{^{\hypertarget{NE0LEVEL3}{d}}}} Seq.(D): K\ensuremath{^{\ensuremath{\pi}}}=3/2\ensuremath{^{-}} band (\href{https://www.nndc.bnl.gov/nsr/nsrlink.jsp?1995Wi26,B}{1995Wi26}).}\\
\parbox[b][0.3cm]{17.7cm}{\makebox[1ex]{\ensuremath{^{\hypertarget{NE0LEVEL4}{e}}}} Level energies are deduced using the listed E\ensuremath{_{\textnormal{lab}}}(\ensuremath{^{\textnormal{3}}}He) and the \ensuremath{^{\textnormal{3}}}He, \ensuremath{^{\textnormal{16}}}O and \ensuremath{^{\textnormal{19}}}Ne masses from (\href{https://www.nndc.bnl.gov/nsr/nsrlink.jsp?2021Wa16,B}{2021Wa16}: AME-2020).}\\
\parbox[b][0.3cm]{17.7cm}{{\ }{\ }E\ensuremath{_{\textnormal{x}}}=S\ensuremath{_{^{\textnormal{3}}\textnormal{He}}}+E\ensuremath{_{\textnormal{c.m.}}} (relativistic).}\\
\parbox[b][0.3cm]{17.7cm}{\makebox[1ex]{\ensuremath{^{\hypertarget{NE0LEVEL5}{f}}}} (\href{https://www.nndc.bnl.gov/nsr/nsrlink.jsp?2008Oh03,B}{2008Oh03}) reports that this state should be regarded as a member of an N=8 higher nodal (vibrational state with \ensuremath{^{\textnormal{3}}}He cluster)}\\
\parbox[b][0.3cm]{17.7cm}{{\ }{\ }band, and that this state may be considered to be fragmented from the higher nodal L=2 state.}\\
\parbox[b][0.3cm]{17.7cm}{\makebox[1ex]{\ensuremath{^{\hypertarget{NE0LEVEL6}{g}}}} Conflicting results exist in the literature for the J\ensuremath{^{\ensuremath{\pi}}} assignments of these states: (i) J\ensuremath{^{\ensuremath{\pi}}}=(7/2\ensuremath{^{-}}) for \ensuremath{^{\textnormal{19}}}Ne*(4143) and J\ensuremath{^{\ensuremath{\pi}}}=(9/2\ensuremath{^{-}}) for}\\
\parbox[b][0.3cm]{17.7cm}{{\ }{\ }\ensuremath{^{\textnormal{19}}}Ne*(4200) from: (1) (\href{https://www.nndc.bnl.gov/nsr/nsrlink.jsp?1970Ga18,B}{1970Ga18}: \ensuremath{^{\textnormal{20}}}Ne(\ensuremath{^{\textnormal{3}}}He,\ensuremath{\alpha})) based on comparison of rotational bands in \ensuremath{^{\textnormal{19}}}F* and \ensuremath{^{\textnormal{19}}}Ne* mirror states. (2)}\\
\parbox[b][0.3cm]{17.7cm}{{\ }{\ }(\href{https://www.nndc.bnl.gov/nsr/nsrlink.jsp?2009Ta09,B}{2009Ta09}: \ensuremath{^{\textnormal{19}}}F(\ensuremath{^{\textnormal{3}}}He,t)\ensuremath{^{\textnormal{19}}}Ne*(\ensuremath{\alpha})) and the evaluation of (\href{https://www.nndc.bnl.gov/nsr/nsrlink.jsp?2011Da24,B}{2011Da24}) and (\href{https://www.nndc.bnl.gov/nsr/nsrlink.jsp?2020Ha31,B}{2020Ha31}: \ensuremath{^{\textnormal{19}}}F(\ensuremath{^{\textnormal{3}}}He,t\ensuremath{\gamma})) based on comparisons of the}\\
\parbox[b][0.3cm]{17.7cm}{{\ }{\ }\ensuremath{\gamma}-ray decay schemes (\href{https://www.nndc.bnl.gov/nsr/nsrlink.jsp?2009Ta09,B}{2009Ta09}, \href{https://www.nndc.bnl.gov/nsr/nsrlink.jsp?2020Ha31,B}{2020Ha31}) and the calculated (by \href{https://www.nndc.bnl.gov/nsr/nsrlink.jsp?2011Da24,B}{2011Da24}) M1 and E2 reduced transition strengths of these}\\
\begin{textblock}{29}(0,27.3)
Continued on next page (footnotes at end of table)
\end{textblock}
\clearpage
\vspace*{-0.5cm}
{\bf \small \underline{Adopted \hyperlink{19NE_LEVEL}{Levels}, \hyperlink{19NE_GAMMA}{Gammas} (continued)}}\\
\vspace{0.3cm}
\underline{$^{19}$Ne Levels (continued)}\\
\vspace{0.3cm}
\parbox[b][0.3cm]{17.7cm}{{\ }{\ }two \ensuremath{^{\textnormal{19}}}Ne* states with those of the \ensuremath{^{\textnormal{19}}}F*(3999, 4033) mirror levels, and from a comparison of the calculated single-particle}\\
\parbox[b][0.3cm]{17.7cm}{{\ }{\ }\ensuremath{\alpha}-widths and the spectroscopic factors for the two \ensuremath{^{\textnormal{19}}}Ne* states from (\href{https://www.nndc.bnl.gov/nsr/nsrlink.jsp?2009Ta09,B}{2009Ta09}). (ii) J\ensuremath{^{\ensuremath{\pi}}}=(9/2\ensuremath{^{-}}) for \ensuremath{^{\textnormal{19}}}Ne*(4143) and J\ensuremath{^{\ensuremath{\pi}}}=(7/2\ensuremath{^{-}})}\\
\parbox[b][0.3cm]{17.7cm}{{\ }{\ }for \ensuremath{^{\textnormal{19}}}Ne*(4200) from (1) (\href{https://www.nndc.bnl.gov/nsr/nsrlink.jsp?1970Ga18,B}{1970Ga18}) based on zero-range DWBA analysis of \ensuremath{^{\textnormal{20}}}Ne(\ensuremath{^{\textnormal{3}}}He,\ensuremath{\alpha}) using the JULIE code (L not}\\
\parbox[b][0.3cm]{17.7cm}{{\ }{\ }reported). (2) (\href{https://www.nndc.bnl.gov/nsr/nsrlink.jsp?1971Bi06,B}{1971Bi06}: \ensuremath{^{\textnormal{16}}}O(\ensuremath{^{\textnormal{6}}}Li,\ensuremath{^{\textnormal{3}}}He) and \ensuremath{^{\textnormal{16}}}O(\ensuremath{^{\textnormal{6}}}Li,t)) and (\href{https://www.nndc.bnl.gov/nsr/nsrlink.jsp?1973Da31,B}{1973Da31}: \ensuremath{^{\textnormal{17}}}O(\ensuremath{^{\textnormal{3}}}He,n\ensuremath{\gamma})) based on mirror assignments. These were}\\
\parbox[b][0.3cm]{17.7cm}{{\ }{\ }adopted by the last evaluation of A=19 (\href{https://www.nndc.bnl.gov/nsr/nsrlink.jsp?1995Ti07,B}{1995Ti07}). (3) The evaluation of (\href{https://www.nndc.bnl.gov/nsr/nsrlink.jsp?2011Da24,B}{2011Da24}) based on comparison of the \ensuremath{\gamma} decay}\\
\parbox[b][0.3cm]{17.7cm}{{\ }{\ }branching ratios for these two states with those of their respective mirror levels in \ensuremath{^{\textnormal{19}}}F. (4) (\href{https://www.nndc.bnl.gov/nsr/nsrlink.jsp?2015Pa46,B}{2015Pa46}: \ensuremath{^{\textnormal{19}}}F(\ensuremath{^{\textnormal{3}}}He,t)) based on}\\
\parbox[b][0.3cm]{17.7cm}{{\ }{\ }finite-range coupled-channels analysis via FRESCO. In that study, J\ensuremath{^{\ensuremath{\pi}}}=7/2\ensuremath{^{\textnormal{+}}} for the \ensuremath{^{\textnormal{19}}}Ne*(4142) state fit the data as well. While,}\\
\parbox[b][0.3cm]{17.7cm}{{\ }{\ }J\ensuremath{^{\ensuremath{\pi}}}=7/2\ensuremath{^{-}} was a poorer fit to the data particularly at forward angles. (\href{https://www.nndc.bnl.gov/nsr/nsrlink.jsp?2019Ha14,B}{2019Ha14}, \href{https://www.nndc.bnl.gov/nsr/nsrlink.jsp?2020Ha31,B}{2020Ha31}) searched for (using \ensuremath{\gamma}-\ensuremath{\gamma} coincidence)}\\
\parbox[b][0.3cm]{17.7cm}{{\ }{\ }the weak \ensuremath{\gamma}-ray transition from the \ensuremath{^{\textnormal{19}}}Ne*(4200)\ensuremath{\rightarrow}\ensuremath{^{\textnormal{19}}}Ne*(238) decay, which was previously reported by (\href{https://www.nndc.bnl.gov/nsr/nsrlink.jsp?1973Da31,B}{1973Da31}) via an n-\ensuremath{\gamma}}\\
\parbox[b][0.3cm]{17.7cm}{{\ }{\ }coincidence, and found no evidence for it. From this, we made the assignments presented here.}\\
\parbox[b][0.3cm]{17.7cm}{\makebox[1ex]{\ensuremath{^{\hypertarget{NE0LEVEL7}{h}}}} From (\href{https://www.nndc.bnl.gov/nsr/nsrlink.jsp?1972Ha03,B}{1972Ha03}).}\\
\parbox[b][0.3cm]{17.7cm}{\makebox[1ex]{\ensuremath{^{\hypertarget{NE0LEVEL8}{i}}}} From (\href{https://www.nndc.bnl.gov/nsr/nsrlink.jsp?1979Ma26,B}{1979Ma26}).}\\
\parbox[b][0.3cm]{17.7cm}{\makebox[1ex]{\ensuremath{^{\hypertarget{NE0LEVEL9}{j}}}} From (\href{https://www.nndc.bnl.gov/nsr/nsrlink.jsp?1983Wa05,B}{1983Wa05}).}\\
\vspace{0.5cm}
\clearpage
\vspace{0.3cm}
\begin{landscape}
\vspace*{-0.5cm}
{\bf \small \underline{Adopted \hyperlink{19NE_LEVEL}{Levels}, \hyperlink{19NE_GAMMA}{Gammas} (continued)}}\\
\vspace{0.3cm}
\hypertarget{19NE_GAMMA}{\underline{$\gamma$($^{19}$Ne)}}\\

\parbox[b][0.3cm]{22.5cm}{\makebox[1ex]{\ensuremath{^{\hypertarget{NE0GAMMA0}{a}}}} This previously known \ensuremath{\gamma} ray was observed by (\href{https://www.nndc.bnl.gov/nsr/nsrlink.jsp?2017Wr02,B}{2017Wr02}) as part of the \ensuremath{^{\textnormal{20}}}Mg\ensuremath{_{\textnormal{g.s.}}}\ensuremath{\rightarrow}\ensuremath{^{\textnormal{20}}}Na*\ensuremath{\rightarrow}p+\ensuremath{^{\textnormal{19}}}Ne*\ensuremath{\rightarrow}p+\ensuremath{\gamma}+\ensuremath{^{\textnormal{19}}}Ne\ensuremath{_{\textnormal{g.s.}}} decay for the first time.}\\
\parbox[b][0.3cm]{22.5cm}{\makebox[1ex]{\ensuremath{^{\hypertarget{NE0GAMMA1}{b}}}} This previously known \ensuremath{\gamma} ray was observed by (\href{https://www.nndc.bnl.gov/nsr/nsrlink.jsp?2019Gl02,B}{2019Gl02}) as part of the \ensuremath{\beta}\ensuremath{^{\textnormal{+}}}-delayed p\ensuremath{\gamma} decay of \ensuremath{^{\textnormal{20}}}Mg\ensuremath{_{\textnormal{g.s.}}} for the first time.}\\
\parbox[b][0.3cm]{22.5cm}{\makebox[1ex]{\ensuremath{^{\hypertarget{NE0GAMMA2}{c}}}} From (\href{https://www.nndc.bnl.gov/nsr/nsrlink.jsp?2019Ha08,B}{2019Ha08}, \href{https://www.nndc.bnl.gov/nsr/nsrlink.jsp?2020Ha31,B}{2020Ha31}): \ensuremath{^{\textnormal{19}}}F(\ensuremath{^{\textnormal{3}}}He,t\ensuremath{\gamma}).}\\
\parbox[b][0.3cm]{22.5cm}{\makebox[1ex]{\ensuremath{^{\hypertarget{NE0GAMMA3}{d}}}} From (\href{https://www.nndc.bnl.gov/nsr/nsrlink.jsp?2019Ha14,B}{2019Ha14}, \href{https://www.nndc.bnl.gov/nsr/nsrlink.jsp?2020Ha31,B}{2020Ha31}): \ensuremath{^{\textnormal{19}}}F(\ensuremath{^{\textnormal{3}}}He,t\ensuremath{\gamma}).}\\
\parbox[b][0.3cm]{22.5cm}{\makebox[1ex]{\ensuremath{^{\hypertarget{NE0GAMMA4}{e}}}} From (\href{https://www.nndc.bnl.gov/nsr/nsrlink.jsp?2020Ha31,B}{2020Ha31}): \ensuremath{^{\textnormal{19}}}F(\ensuremath{^{\textnormal{3}}}He,t\ensuremath{\gamma}).}\\
\parbox[b][0.3cm]{22.5cm}{\makebox[1ex]{\ensuremath{^{\hypertarget{NE0GAMMA5}{f}}}} From (\href{https://www.nndc.bnl.gov/nsr/nsrlink.jsp?2019Gl02,B}{2019Gl02}: \ensuremath{^{\textnormal{20}}}Mg(\ensuremath{\beta}\ensuremath{^{\textnormal{+}}}p\ensuremath{\gamma})) and (\href{https://www.nndc.bnl.gov/nsr/nsrlink.jsp?2020Ha31,B}{2020Ha31}: \ensuremath{^{\textnormal{19}}}F(\ensuremath{^{\textnormal{3}}}He,t\ensuremath{\gamma})).}\\
\parbox[b][0.3cm]{22.5cm}{\makebox[1ex]{\ensuremath{^{\hypertarget{NE0GAMMA6}{g}}}} From (\href{https://www.nndc.bnl.gov/nsr/nsrlink.jsp?2020Ha31,B}{2020Ha31}): \ensuremath{^{\textnormal{19}}}F(\ensuremath{^{\textnormal{3}}}He,t\ensuremath{\gamma}).}\\
\parbox[b][0.3cm]{22.5cm}{\makebox[1ex]{\ensuremath{^{\hypertarget{NE0GAMMA7}{h}}}} The intensities given in the table are normalized to the strongest branch, for which I\ensuremath{_{\ensuremath{\gamma}}} is set to be 100\%.}\\
\parbox[b][0.3cm]{22.5cm}{\makebox[1ex]{\ensuremath{^{\hypertarget{NE0GAMMA8}{i}}}} Total theoretical internal conversion coefficients, calculated using the BrIcc code (\href{https://www.nndc.bnl.gov/nsr/nsrlink.jsp?2008Ki07,B}{2008Ki07}) with ``Frozen Orbitals'' approximation based on \ensuremath{\gamma}-ray energies, assigned}\\
\parbox[b][0.3cm]{22.5cm}{{\ }{\ }multipolarities, and mixing ratios, unless otherwise specified.}\\
\parbox[b][0.3cm]{22.5cm}{\makebox[1ex]{\ensuremath{^{\hypertarget{NE0GAMMA9}{j}}}} Placement of transition in the level scheme is uncertain.}\\
\vspace{0.5cm}
\end{landscape}\clearpage
\clearpage
\begin{figure}[h]
\begin{center}
\includegraphics{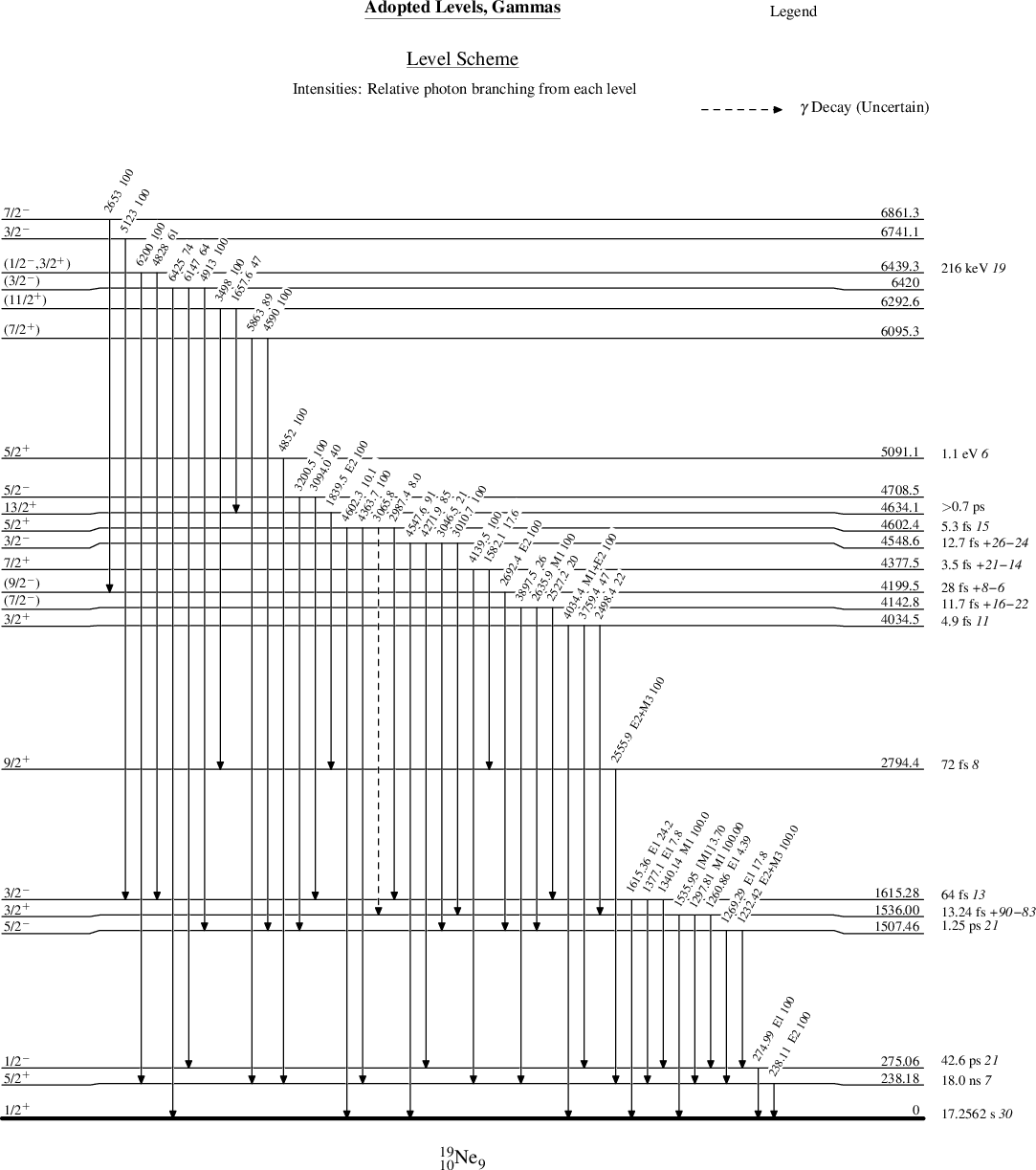}\\
\end{center}
\end{figure}
\clearpage
\clearpage
\begin{figure}[h]
\begin{center}
\includegraphics{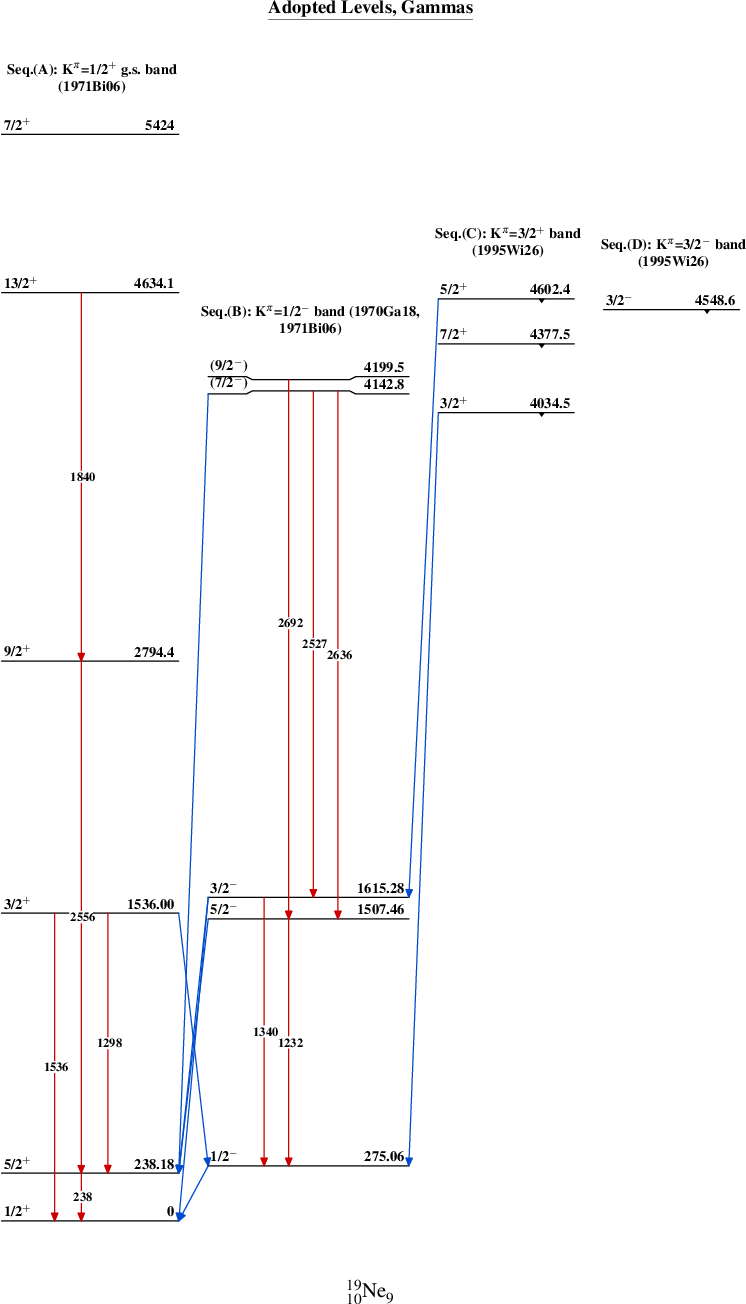}\\
\end{center}
\end{figure}
\clearpage
\subsection[\hspace{-0.2cm}\ensuremath{^{\textnormal{20}}}Mg \ensuremath{\beta}\ensuremath{^{\textnormal{+}}}p decay]{ }
\vspace{-27pt}
\vspace{0.3cm}
\hypertarget{MG1}{{\bf \small \underline{\ensuremath{^{\textnormal{20}}}Mg \ensuremath{\beta}\ensuremath{^{\textnormal{+}}}p decay\hspace{0.2in}\href{https://www.nndc.bnl.gov/nsr/nsrlink.jsp?1995Pi03,B}{1995Pi03},\href{https://www.nndc.bnl.gov/nsr/nsrlink.jsp?2016Li45,B}{2016Li45},\href{https://www.nndc.bnl.gov/nsr/nsrlink.jsp?2019Gl02,B}{2019Gl02}}}}\\
\vspace{4pt}
\vspace{8pt}
\parbox[b][0.3cm]{17.7cm}{\addtolength{\parindent}{-0.2in}Parent: $^{20}$Mg: E=0; J$^{\pi}$=0\ensuremath{^{+}}; T$_{1/2}$=90.4 ms {\it 7}; Q(\ensuremath{\beta}\ensuremath{^{\textnormal{+}}}p)=8436.7 {\it 19}; \%\ensuremath{\beta}\ensuremath{^{\textnormal{+}}}p decay=30.3 {\it 12}

}\\
\parbox[b][0.3cm]{17.7cm}{\addtolength{\parindent}{-0.2in}\ensuremath{^{20}}Mg-T=2 (\href{https://www.nndc.bnl.gov/nsr/nsrlink.jsp?1979Mo02,B}{1979Mo02}, \href{https://www.nndc.bnl.gov/nsr/nsrlink.jsp?1981Ay01,B}{1981Ay01}, \href{https://www.nndc.bnl.gov/nsr/nsrlink.jsp?2015Gl03,B}{2015Gl03}).}\\
\parbox[b][0.3cm]{17.7cm}{\addtolength{\parindent}{-0.2in}\ensuremath{^{20}}Mg-E,J$^{\pi}$: From the Adopted Levels of \ensuremath{^{\textnormal{20}}}Mg in the ENSDF database.}\\
\parbox[b][0.3cm]{17.7cm}{\addtolength{\parindent}{-0.2in}\ensuremath{^{20}}Mg-T$_{1/2}$: Weighted average (with external errors) of 95 ms \textit{+80{\textminus}50} (\href{https://www.nndc.bnl.gov/nsr/nsrlink.jsp?1979Mo02,B}{1979Mo02}, \href{https://www.nndc.bnl.gov/nsr/nsrlink.jsp?1981Ay01,B}{1981Ay01}); 82 ms \textit{4} (\href{https://www.nndc.bnl.gov/nsr/nsrlink.jsp?1992Go10,B}{1992Go10}); 114 ms \textit{17}}\\
\parbox[b][0.3cm]{17.7cm}{(\href{https://www.nndc.bnl.gov/nsr/nsrlink.jsp?1992Ku07,B}{1992Ku07}); 95 ms \textit{3} (\href{https://www.nndc.bnl.gov/nsr/nsrlink.jsp?1995Pi03,B}{1995Pi03}); 91.4 ms \textit{10} (\href{https://www.nndc.bnl.gov/nsr/nsrlink.jsp?2016Lu13,B}{2016Lu13}); and 90.0 ms \textit{6} (\href{https://www.nndc.bnl.gov/nsr/nsrlink.jsp?2017Su05,B}{2017Su05}).}\\
\parbox[b][0.3cm]{17.7cm}{\addtolength{\parindent}{-0.2in}\ensuremath{^{20}}Mg-T$_{1/2}$: See also \ensuremath{\sim}90 ms (\href{https://www.nndc.bnl.gov/nsr/nsrlink.jsp?2012Wa15,B}{2012Wa15}); and 93 ms \textit{5} (\href{https://www.nndc.bnl.gov/nsr/nsrlink.jsp?2017Wr02,B}{2017Wr02}) which is reported from (\href{https://www.nndc.bnl.gov/nsr/nsrlink.jsp?2017Au03,B}{2017Au03}).}\\
\parbox[b][0.3cm]{17.7cm}{\addtolength{\parindent}{-0.2in}\ensuremath{^{20}}Mg-Q(\ensuremath{\beta}\ensuremath{^{\textnormal{+}}}p): From (\href{https://www.nndc.bnl.gov/nsr/nsrlink.jsp?1995Pi03,B}{1995Pi03}).}\\
\parbox[b][0.3cm]{17.7cm}{\addtolength{\parindent}{-0.2in}\ensuremath{^{20}}Mg-\%\ensuremath{\beta}\ensuremath{^{\textnormal{+}}}\ensuremath{\gamma} decay: 70.0\% \textit{11} for \ensuremath{^{\textnormal{20}}}Mg(\ensuremath{\beta}\ensuremath{^{\textnormal{+}}})\ensuremath{^{\textnormal{20}}}Na*(984)\ensuremath{\rightarrow}\ensuremath{^{\textnormal{20}}}Na\ensuremath{_{\textnormal{g.s.}}}+\ensuremath{\gamma}: Weighted average of 69.7\% \textit{12} (\href{https://www.nndc.bnl.gov/nsr/nsrlink.jsp?1995Pi03,B}{1995Pi03}); 72.0\% \textit{25}}\\
\parbox[b][0.3cm]{17.7cm}{(\href{https://www.nndc.bnl.gov/nsr/nsrlink.jsp?2016Lu13,B}{2016Lu13}); and 66.9\% \textit{46} (\href{https://www.nndc.bnl.gov/nsr/nsrlink.jsp?2017Su05,B}{2017Su05}).}\\
\parbox[b][0.3cm]{17.7cm}{\addtolength{\parindent}{-0.2in}\ensuremath{^{20}}Mg-\%\ensuremath{\beta}\ensuremath{^{\textnormal{+}}}\ensuremath{\gamma} decay: See also I\ensuremath{_{\ensuremath{\beta}\ensuremath{\gamma}}}=74\% \textit{7} (\href{https://www.nndc.bnl.gov/nsr/nsrlink.jsp?1992Go10,B}{1992Go10}) deduced from the strength of the secondary \ensuremath{\alpha} decay of \ensuremath{^{\textnormal{20}}}Na*; I\ensuremath{_{\ensuremath{\beta}\ensuremath{\gamma}}}=85\%}\\
\parbox[b][0.3cm]{17.7cm}{(\href{https://www.nndc.bnl.gov/nsr/nsrlink.jsp?1992Ku07,B}{1992Ku07}); and I\ensuremath{_{\ensuremath{\beta}\ensuremath{\gamma}}}=70.5\% \textit{14} (\href{https://www.nndc.bnl.gov/nsr/nsrlink.jsp?1995Pi03,B}{1995Pi03}) determined from the \ensuremath{\gamma} data for the decay from \ensuremath{^{\textnormal{20}}}Mg\ensuremath{_{\textnormal{g.s.}}} to the \ensuremath{^{\textnormal{20}}}Na*(984) bound}\\
\parbox[b][0.3cm]{17.7cm}{level. This value is reported by those authors to be biased because of the energy dependence for the \ensuremath{\beta} detection efficiency of the}\\
\parbox[b][0.3cm]{17.7cm}{detector used for \ensuremath{^{\textnormal{20}}}Mg implantation.}\\
\parbox[b][0.3cm]{17.7cm}{\addtolength{\parindent}{-0.2in}\ensuremath{^{20}}Mg-\%\ensuremath{\beta}\ensuremath{^{\textnormal{+}}}\ensuremath{\gamma} decay: Evaluator highlights a misprint in (\href{https://www.nndc.bnl.gov/nsr/nsrlink.jsp?2019Gl02,B}{2019Gl02}): They reported I\ensuremath{_{\ensuremath{\beta}\ensuremath{\gamma}}}=72.5\% \textit{25} from (\href{https://www.nndc.bnl.gov/nsr/nsrlink.jsp?2016Lu13,B}{2016Lu13}) for the decay of}\\
\parbox[b][0.3cm]{17.7cm}{\ensuremath{^{\textnormal{20}}}Mg\ensuremath{_{\textnormal{g.s.}}} to the \ensuremath{^{\textnormal{20}}}Na*(984) level. The correct value is 72.0\% \textit{25}.}\\
\parbox[b][0.3cm]{17.7cm}{\addtolength{\parindent}{-0.2in}\ensuremath{^{20}}Mg-\%\ensuremath{\beta}\ensuremath{^{\textnormal{+}}}p decay: \%\ensuremath{\beta}\ensuremath{^{\textnormal{+}}}p=30.3\% \textit{12} (\href{https://www.nndc.bnl.gov/nsr/nsrlink.jsp?1995Pi03,B}{1995Pi03}) from analysis of the full \ensuremath{\beta}\ensuremath{^{\textnormal{+}}}+p decay scheme.}\\
\parbox[b][0.3cm]{17.7cm}{\addtolength{\parindent}{-0.2in}\ensuremath{^{20}}Mg-\%\ensuremath{\beta}\ensuremath{^{\textnormal{+}}}p decay: See also (1) I\ensuremath{_{\ensuremath{\beta}\textnormal{p}}}=27.7\% \textit{8} (stat.) \textit{29} (sys.) (\href{https://www.nndc.bnl.gov/nsr/nsrlink.jsp?2016Lu13,B}{2016Lu13}): They did not observe the previously known proton}\\
\parbox[b][0.3cm]{17.7cm}{branches of \ensuremath{^{\textnormal{20}}}Na*(3075)\ensuremath{\rightarrow}\ensuremath{^{\textnormal{19}}}Ne\ensuremath{_{\textnormal{g.s.}}} (\href{https://www.nndc.bnl.gov/nsr/nsrlink.jsp?2012Wa15,B}{2012Wa15}) and \ensuremath{^{\textnormal{20}}}Na*(3860)\ensuremath{\rightarrow}\ensuremath{^{\textnormal{19}}}Ne*(238+275) due to the large contamination from \ensuremath{^{\textnormal{20}}}Na in}\\
\parbox[b][0.3cm]{17.7cm}{their \ensuremath{^{\textnormal{20}}}Mg beam. Those authors therefore excluded the two unobserved branches mentioned above and renormalized the \%\ensuremath{\beta}\ensuremath{^{\textnormal{+}}}p}\\
\parbox[b][0.3cm]{17.7cm}{ratio. The systematic uncertainty is from a 10.4\% relative systematic uncertainty discussed in the text. (2) I\ensuremath{_{\ensuremath{\beta}\textnormal{p}}}=26.9\% \textit{32} obtained}\\
\parbox[b][0.3cm]{17.7cm}{by (\href{https://www.nndc.bnl.gov/nsr/nsrlink.jsp?2016Lu13,B}{2016Lu13}) from the data of (\href{https://www.nndc.bnl.gov/nsr/nsrlink.jsp?1995Pi03,B}{1995Pi03}) when excluding the aforementioned proton branches.}\\
\vspace{0.385cm}
\parbox[b][0.3cm]{17.7cm}{\addtolength{\parindent}{-0.2in}\href{https://www.nndc.bnl.gov/nsr/nsrlink.jsp?1963Ka36,B}{1963Ka36}: \ensuremath{^{\textnormal{20}}}Mg(EC+\ensuremath{\beta}\ensuremath{^{\textnormal{+}}})\ensuremath{^{\textnormal{20}}}Na*(p); measured decay products, E\ensuremath{_{\textnormal{p}}}, I\ensuremath{_{\textnormal{p}}}; deduced T\ensuremath{_{\textnormal{1/2}}}(\ensuremath{^{\textnormal{20}}}Mg), delayed p-emission.}\\
\parbox[b][0.3cm]{17.7cm}{\addtolength{\parindent}{-0.2in}\href{https://www.nndc.bnl.gov/nsr/nsrlink.jsp?1979Mo02,B}{1979Mo02}, \href{https://www.nndc.bnl.gov/nsr/nsrlink.jsp?1981Ay01,B}{1981Ay01}: \ensuremath{^{\textnormal{20}}}Ne(\ensuremath{^{\textnormal{3}}}He,3n)\ensuremath{^{\textnormal{20}}}Mg(\ensuremath{\beta}\ensuremath{^{\textnormal{+}}})\ensuremath{^{\textnormal{20}}}Na*(p) E=70 MeV; thermalized the \ensuremath{^{\textnormal{20}}}Mg recoils in a spark chamber filled with}\\
\parbox[b][0.3cm]{17.7cm}{He-Ne mixture and transported (using a He jet) \ensuremath{^{\textnormal{20}}}Mg recoils to an ion source at 1300\ensuremath{^\circ} C, which ionized the recoils. These ions}\\
\parbox[b][0.3cm]{17.7cm}{were momentum analyzed by the RAMA mass analyzer and implanted (for 280 ms) into an aluminized polyethylene foil located on}\\
\parbox[b][0.3cm]{17.7cm}{the focal plane, where a large \ensuremath{\Delta}E-E telescope measured the \ensuremath{\beta}-delayed protons. The telescope consisted of Si surface barrier}\\
\parbox[b][0.3cm]{17.7cm}{detectors with a resolution of 55 keV (FWHM). Measured half-life of \ensuremath{^{\textnormal{20}}}Mg and two \ensuremath{\beta}-delayed proton groups at E\ensuremath{_{\textnormal{p}}}=3.95 MeV \textit{6}}\\
\parbox[b][0.3cm]{17.7cm}{and E\ensuremath{_{\textnormal{p}}}=4.16 MeV \textit{5}. These were attributed to the isospin forbidden proton decay of the lowest T=2, J\ensuremath{^{\ensuremath{\pi}}}=0\ensuremath{^{\textnormal{+}}} state at E\ensuremath{_{\textnormal{x}}}=6534 keV}\\
\parbox[b][0.3cm]{17.7cm}{in \ensuremath{^{\textnormal{20}}}Na (which is fed by the superallowed Fermi \ensuremath{\beta}\ensuremath{^{\textnormal{+}}} decay of T=2, J\ensuremath{^{\ensuremath{\pi}}}=0\ensuremath{^{\textnormal{+}}} \ensuremath{^{\textnormal{20}}}Mg\ensuremath{_{\textnormal{g.s.}}}) to \ensuremath{^{\textnormal{19}}}Ne*(0, 238) levels, respectively.}\\
\parbox[b][0.3cm]{17.7cm}{\addtolength{\parindent}{-0.2in}\href{https://www.nndc.bnl.gov/nsr/nsrlink.jsp?1992Go10,B}{1992Go10}: \ensuremath{^{\textnormal{20}}}Mg(EC+\ensuremath{\beta}\ensuremath{^{\textnormal{+}}})\ensuremath{^{\textnormal{20}}}Na*(p); a \ensuremath{^{\textnormal{20}}}Mg beam produced from the projectile fragmentation of \ensuremath{^{\textnormal{36}}}Ar on a Ni target at 80}\\
\parbox[b][0.3cm]{17.7cm}{MeV/nucleon was purified using the A1200 fragment separator and the recoil particle mass separator (RPMS) and was implanted}\\
\parbox[b][0.3cm]{17.7cm}{into the second detector of a Si \ensuremath{\Delta}E-\ensuremath{\Delta}E-\ensuremath{\Delta}E-E telescope at the RPMS$'$ focal plane. Measured the energies and branching ratios of}\\
\parbox[b][0.3cm]{17.7cm}{the protons from the decay of \ensuremath{^{\textnormal{20}}}Mg(\ensuremath{\beta}\ensuremath{^{\textnormal{+}}})\ensuremath{^{\textnormal{20}}}Na*(p)\ensuremath{^{\textnormal{19}}}Ne in coincidence with the implanted beam particles. Observed strong proton}\\
\parbox[b][0.3cm]{17.7cm}{groups at 807 and 1670 keV from \ensuremath{^{\textnormal{20}}}Na*(p) decay and an \ensuremath{\alpha} group from the \ensuremath{\alpha}-decay of the \ensuremath{^{\textnormal{20}}}Na*(7145) level. Deduced}\\
\parbox[b][0.3cm]{17.7cm}{T\ensuremath{_{\textnormal{1/2}}}(\ensuremath{^{\textnormal{20}}}Mg) and log \textit{ft} for the observed transitions. Discussed the \ensuremath{^{\textnormal{19}}}Ne(p,\ensuremath{\gamma}) reaction rate.}\\
\parbox[b][0.3cm]{17.7cm}{\addtolength{\parindent}{-0.2in}\href{https://www.nndc.bnl.gov/nsr/nsrlink.jsp?1992PiZT,B}{1992PiZT}: \ensuremath{^{\textnormal{20}}}Mg(EC+\ensuremath{\beta}\ensuremath{^{\textnormal{+}}})\ensuremath{^{\textnormal{20}}}Na*(p); produced \ensuremath{^{\textnormal{20}}}Mg beam from fragmentation of a \ensuremath{^{\textnormal{24}}}Mg beam at E=95 MeV/nucleon on a Ni}\\
\parbox[b][0.3cm]{17.7cm}{target; measured \ensuremath{\beta}-delayed particle spectra, E\ensuremath{_{\ensuremath{\beta}}}, I\ensuremath{_{\ensuremath{\beta}}}, \ensuremath{\gamma}\ensuremath{\gamma} coincidence; deduced \ensuremath{^{\textnormal{20}}}Na levels, Fermi, and Gamow-Teller transition}\\
\parbox[b][0.3cm]{17.7cm}{strengths; deduced r-process implications.}\\
\parbox[b][0.3cm]{17.7cm}{\addtolength{\parindent}{-0.2in}\href{https://www.nndc.bnl.gov/nsr/nsrlink.jsp?1992KuZO,B}{1992KuZO}, \href{https://www.nndc.bnl.gov/nsr/nsrlink.jsp?1992KuZQ,B}{1992KuZQ}, \href{https://www.nndc.bnl.gov/nsr/nsrlink.jsp?1992Ku07,B}{1992Ku07}, \href{https://www.nndc.bnl.gov/nsr/nsrlink.jsp?1992Ku24,B}{1992Ku24}: \ensuremath{^{\textnormal{20}}}Mg(\ensuremath{\beta}\ensuremath{^{\textnormal{+}}})\ensuremath{^{\textnormal{20}}}Na*(p) E=100 MeV/nucleon. The \ensuremath{^{\textnormal{20}}}Mg beam was produced from}\\
\parbox[b][0.3cm]{17.7cm}{projectile fragmentation of a \ensuremath{^{\textnormal{24}}}Mg beam on a \ensuremath{^{\textnormal{9}}}Be target using the RIPS facility in RIKEN and stopped in the third position}\\
\parbox[b][0.3cm]{17.7cm}{sensitive Si detector in an array consisting of a stack of 5 such detectors at \ensuremath{\theta}\ensuremath{_{\textnormal{lab}}}=0\ensuremath{^\circ} to measure the delayed protons. This array was}\\
\parbox[b][0.3cm]{17.7cm}{surrounded by 3 \ensuremath{\Delta}E-E plastic scintillator telescopes at \ensuremath{\theta}\ensuremath{_{\textnormal{lab}}}=+90\ensuremath{^\circ} to measure \ensuremath{\beta}\ensuremath{^{\textnormal{+}}} particles; and 2 HPGe detectors at \ensuremath{\theta}\ensuremath{_{\textnormal{lab}}}={\textminus}90\ensuremath{^\circ} to}\\
\parbox[b][0.3cm]{17.7cm}{measure the \ensuremath{\gamma} rays. Measured a few proton groups at E\ensuremath{_{\textnormal{c.m.}}}=847, 1669, and 1891 keV, as well as two probably proton (or \ensuremath{\alpha})}\\
\parbox[b][0.3cm]{17.7cm}{groups at 3990 and 4239 keV (\href{https://www.nndc.bnl.gov/nsr/nsrlink.jsp?1992Ku07,B}{1992Ku07}) (see also E\ensuremath{_{\textnormal{c.m.}}}=857 keV and 1740 keV: Preliminary results in \href{https://www.nndc.bnl.gov/nsr/nsrlink.jsp?1992Ku24,B}{1992Ku24}) from}\\
\parbox[b][0.3cm]{17.7cm}{\ensuremath{\beta}\ensuremath{^{\textnormal{+}}}-delayed proton decay to \ensuremath{^{\textnormal{19}}}Ne\ensuremath{_{\textnormal{g.s.}}}. No \ensuremath{\gamma} rays from \ensuremath{^{\textnormal{20}}}Na decay were measured. No protons were observed from the decay of}\\
\parbox[b][0.3cm]{17.7cm}{\ensuremath{^{\textnormal{20}}}Na*(2645) state. These authors deduced an upper limit branching ratio of I\ensuremath{_{\ensuremath{\beta}}}\ensuremath{<}1\% for the \ensuremath{\beta}\ensuremath{^{\textnormal{+}}} decay of \ensuremath{^{\textnormal{20}}}Mg to the \ensuremath{^{\textnormal{20}}}Na*(2645)}\\
\parbox[b][0.3cm]{17.7cm}{state (\href{https://www.nndc.bnl.gov/nsr/nsrlink.jsp?1992Ku07,B}{1992Ku07}) (see I\ensuremath{_{\ensuremath{\beta}}}\ensuremath{<}2\%: Preliminary result of \href{https://www.nndc.bnl.gov/nsr/nsrlink.jsp?1992Ku24,B}{1992Ku24}). Discussed the J\ensuremath{^{\ensuremath{\pi}}} assignment of the \ensuremath{^{\textnormal{20}}}Na*(2645) level.}\\
\parbox[b][0.3cm]{17.7cm}{\addtolength{\parindent}{-0.2in}\href{https://www.nndc.bnl.gov/nsr/nsrlink.jsp?1993PiZZ,B}{1993PiZZ}: \ensuremath{^{\textnormal{20}}}Mg(EC+\ensuremath{\beta}\ensuremath{^{\textnormal{+}}})\ensuremath{^{\textnormal{20}}}Na*(p); measured \ensuremath{\beta}-delayed particle, and \ensuremath{\gamma}-spectra; deduced log \textit{ft} and discussed consequences on the}\\
\parbox[b][0.3cm]{17.7cm}{\ensuremath{^{\textnormal{19}}}Ne(p,\ensuremath{\gamma}) reaction; deduced \ensuremath{^{\textnormal{20}}}Na levels, J, \ensuremath{\pi}, \ensuremath{\beta}-branching ratios.}\\
\parbox[b][0.3cm]{17.7cm}{\addtolength{\parindent}{-0.2in}\href{https://www.nndc.bnl.gov/nsr/nsrlink.jsp?1995Pi03,B}{1995Pi03}: \ensuremath{^{\textnormal{20}}}Mg(\ensuremath{\beta}\ensuremath{^{\textnormal{+}}})\ensuremath{^{\textnormal{20}}}Na*(p\ensuremath{\gamma})\ensuremath{^{\textnormal{19}}}Ne*(\ensuremath{\gamma}); implanted a \ensuremath{^{\textnormal{20}}}Mg beam, produced from projectile fragmentation of 95 MeV/nucleon}\\
\parbox[b][0.3cm]{17.7cm}{\ensuremath{^{\textnormal{24}}}Mg beam on a \ensuremath{^{\textnormal{nat}}}Ni target using the LISE3 facility, into a position sensitive Si detector tilted at \ensuremath{\theta}\ensuremath{_{\textnormal{lab}}}=45\ensuremath{^\circ} and placed}\\
\parbox[b][0.3cm]{17.7cm}{downstream of an annular Si detector (used for \ensuremath{\beta} energy loss measurements). These two detectors were sandwiched between two}\\
\clearpage
\vspace{0.3cm}
{\bf \small \underline{\ensuremath{^{\textnormal{20}}}Mg \ensuremath{\beta}\ensuremath{^{\textnormal{+}}}p decay\hspace{0.2in}\href{https://www.nndc.bnl.gov/nsr/nsrlink.jsp?1995Pi03,B}{1995Pi03},\href{https://www.nndc.bnl.gov/nsr/nsrlink.jsp?2016Li45,B}{2016Li45},\href{https://www.nndc.bnl.gov/nsr/nsrlink.jsp?2019Gl02,B}{2019Gl02} (continued)}}\\
\vspace{0.3cm}
\parbox[b][0.3cm]{17.7cm}{large position sensitive Si detectors (used to measure \ensuremath{\beta} rays and as veto detectors for low energy protons from \ensuremath{\beta}p decays), which}\\
\parbox[b][0.3cm]{17.7cm}{were surrounded by 3 HPGe detectors. Identified \ensuremath{^{\textnormal{20}}}Mg nuclei via \ensuremath{\Delta}E-E and ToF; measured \ensuremath{\beta}-delayed p spectra; \ensuremath{\gamma}-ray spectra}\\
\parbox[b][0.3cm]{17.7cm}{and p-\ensuremath{\gamma} coincidence events; and energies and positions of \ensuremath{\beta}-delayed protons. Deduced \ensuremath{\beta}p branching ratios for decays to \ensuremath{^{\textnormal{19}}}Ne*(0,}\\
\parbox[b][0.3cm]{17.7cm}{238, 275, 1508, 1536) levels. Comparison with shell model calculations are provided.}\\
\parbox[b][0.3cm]{17.7cm}{\addtolength{\parindent}{-0.2in}\href{https://www.nndc.bnl.gov/nsr/nsrlink.jsp?2012Wa15,B}{2012Wa15}: \ensuremath{^{\textnormal{20}}}Mg(\ensuremath{\beta}\ensuremath{^{\textnormal{+}}})\ensuremath{^{\textnormal{20}}}Na*(p)\ensuremath{^{\textnormal{19}}}Ne E\ensuremath{\sim}380 MeV; momentum analyzed \ensuremath{^{\textnormal{20}}}Mg ions produced from \ensuremath{^{\textnormal{3}}}He(\ensuremath{^{\textnormal{20}}}Ne,3n) using MARS}\\
\parbox[b][0.3cm]{17.7cm}{recoil separator; implanted \ensuremath{^{\textnormal{20}}}Mg ions (200 ms beam on, 200 ms beam off) into a position sensitive Si detector sandwiched between}\\
\parbox[b][0.3cm]{17.7cm}{two similar but thicker detectors. The assembly was at \ensuremath{\theta}\ensuremath{_{\textnormal{lab}}}=45\ensuremath{^\circ}. Measured \ensuremath{\beta}-delayed protons from the decay of implanted \ensuremath{^{\textnormal{20}}}Mg}\\
\parbox[b][0.3cm]{17.7cm}{ions. The authors focused on the \ensuremath{\beta}-delayed decay of \ensuremath{^{\textnormal{20}}}Na*(2647) to \ensuremath{^{\textnormal{19}}}Ne\ensuremath{_{\textnormal{g.s.}}}. Deduced I\ensuremath{_{\ensuremath{\beta}}}\ensuremath{<}0.02 (90\% C.L.) for this decay,}\\
\parbox[b][0.3cm]{17.7cm}{making the \ensuremath{^{\textnormal{20}}}Na*(2647) a J\ensuremath{^{\ensuremath{\pi}}}=3\ensuremath{^{\textnormal{+}}} state. Observed a new decay branch at E\ensuremath{_{\textnormal{decay}}}=885 keV \textit{15} from the \ensuremath{^{\textnormal{20}}}Na*(3075 keV, 0\ensuremath{^{\textnormal{+}}}) state.}\\
\parbox[b][0.3cm]{17.7cm}{Deduced I\ensuremath{_{\ensuremath{\beta}}}=0.5\% \textit{1} for this branch.}\\
\parbox[b][0.3cm]{17.7cm}{\addtolength{\parindent}{-0.2in}\href{https://www.nndc.bnl.gov/nsr/nsrlink.jsp?2016Li45,B}{2016Li45}, \href{https://www.nndc.bnl.gov/nsr/nsrlink.jsp?2017Su05,B}{2017Su05}: \ensuremath{^{\textnormal{20}}}Mg(\ensuremath{\beta}\ensuremath{^{\textnormal{+}}})\ensuremath{^{\textnormal{20}}}Na*(p\ensuremath{\gamma})\ensuremath{^{\textnormal{19}}}Ne*(\ensuremath{\gamma}); implanted \ensuremath{^{\textnormal{20}}}Mg beam (produced by projectile fragmentation of a \ensuremath{^{\textnormal{28}}}Si beam}\\
\parbox[b][0.3cm]{17.7cm}{with 75.8 MeV/nucleon on a \ensuremath{^{\textnormal{9}}}Be target) into a position sensitive Si detector at \ensuremath{\theta}\ensuremath{_{\textnormal{lab}}}=45\ensuremath{^\circ}, which was surrounded by a cube made}\\
\parbox[b][0.3cm]{17.7cm}{of position sensitive Si \ensuremath{\Delta}E-E telescopes. This assembly was surrounded by 5 clover HPGe detectors. Three \ensuremath{\Delta}E Si detectors}\\
\parbox[b][0.3cm]{17.7cm}{upstream of this detection setup measured the energy loss. Measured decay products, \ensuremath{\gamma}-p and \ensuremath{\beta}-\ensuremath{\gamma} coincidence events. Found a new}\\
\parbox[b][0.3cm]{17.7cm}{proton branch with an energy of 2256 keV. Deduced \ensuremath{^{\textnormal{20}}}Mg decay scheme and half-life; deduced \ensuremath{^{\textnormal{20}}}Na levels and decay branching}\\
\parbox[b][0.3cm]{17.7cm}{ratios. The statistics of the (\href{https://www.nndc.bnl.gov/nsr/nsrlink.jsp?2017Su05,B}{2017Su05}) is low. The sensitivity of detecting protons was increased by \ensuremath{\beta}-p coincident events. The}\\
\parbox[b][0.3cm]{17.7cm}{p-\ensuremath{\gamma} coincidence analysis was used to identify p-branches.}\\
\parbox[b][0.3cm]{17.7cm}{\addtolength{\parindent}{-0.2in}\href{https://www.nndc.bnl.gov/nsr/nsrlink.jsp?2016Lu13,B}{2016Lu13}: \ensuremath{^{\textnormal{20}}}Mg(\ensuremath{\beta}\ensuremath{^{\textnormal{+}}})\ensuremath{^{\textnormal{20}}}Na*(p\ensuremath{\gamma})\ensuremath{^{\textnormal{19}}}Ne*(\ensuremath{\gamma}) E=30 keV; momentum analyzed the \ensuremath{^{\textnormal{20}}}Mg beam using the ISOLDE High Resolution}\\
\parbox[b][0.3cm]{17.7cm}{Separator; implanted these ions into a thin carbon foil at the center of the detection setup. The foil was surrounded by a 5-sides}\\
\parbox[b][0.3cm]{17.7cm}{cube, which consisted of 4 position sensitive Si \ensuremath{\Delta}E-E telescopes and a position sensitive Si detector at the bottom. This assembly}\\
\parbox[b][0.3cm]{17.7cm}{was surrounded by 4 clover Ge detectors. Measured p-\ensuremath{\gamma} coincidence events; and T\ensuremath{_{\textnormal{1/2}}}(\ensuremath{^{\textnormal{20}}}Mg); measured proton and \ensuremath{\gamma} spectra;}\\
\parbox[b][0.3cm]{17.7cm}{deduced 27 delayed proton branches, including seven new ones. Deduced decay scheme; observed \ensuremath{\beta}-decay feeding to two \ensuremath{^{\textnormal{20}}}Na*}\\
\parbox[b][0.3cm]{17.7cm}{states above the \ensuremath{^{\textnormal{20}}}Na*(IAS). This study could eliminate the \ensuremath{\alpha} particles from the \ensuremath{^{\textnormal{20}}}Na* decay using \ensuremath{\Delta}E-E software gates. The}\\
\parbox[b][0.3cm]{17.7cm}{identification of proton branches was carried out using the ratio of the efficiency corrected number of \ensuremath{\gamma} rays and the integrated}\\
\parbox[b][0.3cm]{17.7cm}{number of coincident protons.}\\
\parbox[b][0.3cm]{17.7cm}{\addtolength{\parindent}{-0.2in}\href{https://www.nndc.bnl.gov/nsr/nsrlink.jsp?2017Wr02,B}{2017Wr02}, \href{https://www.nndc.bnl.gov/nsr/nsrlink.jsp?2019Gl02,B}{2019Gl02}: \ensuremath{^{\textnormal{20}}}Mg(\ensuremath{\beta}\ensuremath{^{\textnormal{+}}})\ensuremath{^{\textnormal{20}}}Na*(p\ensuremath{\gamma})\ensuremath{^{\textnormal{19}}}Ne*(\ensuremath{\gamma}); produced a \ensuremath{^{\textnormal{20}}}Mg beam from projectile fragmentation of a 170-MeV/nucleon}\\
\parbox[b][0.3cm]{17.7cm}{\ensuremath{^{\textnormal{24}}}Mg beam on a \ensuremath{^{\textnormal{9}}}Be target and momentum analyzed the beam by the A1900 fragment separator. Implanted this beam into a plastic}\\
\parbox[b][0.3cm]{17.7cm}{scintillator, which measured the \ensuremath{\beta} rays, decay products and heavy implanted ions. The scintillator was surrounded by the SeGA}\\
\parbox[b][0.3cm]{17.7cm}{array, which measured the \ensuremath{\beta}-delayed \ensuremath{\gamma} rays in coincidence with \ensuremath{\beta}-particles. Measured for the first time the known \ensuremath{\gamma} transitions at}\\
\parbox[b][0.3cm]{17.7cm}{1261, 1269, 1340, 4033 keV from \ensuremath{\beta}-delayed proton decays to the \ensuremath{^{\textnormal{19}}}Ne*(1508, 1536, 1616, 4033) states, respectively. Deduced the}\\
\parbox[b][0.3cm]{17.7cm}{energies of the delayed protons by analyzing the Doppler broadening of the \ensuremath{\gamma}-ray energies using a Monte Carlo simulation.}\\
\parbox[b][0.3cm]{17.7cm}{Deduced the lifetime of the \ensuremath{^{\textnormal{19}}}Ne*(1507.5) state and \ensuremath{^{\textnormal{20}}}Mg decay scheme assuming an isotropic distribution of \ensuremath{\gamma} rays with respect}\\
\parbox[b][0.3cm]{17.7cm}{to proton distribution. Deduced I\ensuremath{_{\ensuremath{\beta}\textnormal{p}}} for the transition to the \ensuremath{^{\textnormal{19}}}Ne*(4033) state.}\\
\vspace{0.385cm}
\parbox[b][0.3cm]{17.7cm}{\addtolength{\parindent}{-0.2in}\textit{Theory}:}\\
\parbox[b][0.3cm]{17.7cm}{\addtolength{\parindent}{-0.2in}\href{https://www.nndc.bnl.gov/nsr/nsrlink.jsp?2017Me01,B}{2017Me01}: \ensuremath{^{\textnormal{20}}}Mg(\ensuremath{\beta}\ensuremath{^{\textnormal{+}}})\ensuremath{^{\textnormal{20}}}Na*(p); used available data from (\href{https://www.nndc.bnl.gov/nsr/nsrlink.jsp?1995Pi03,B}{1995Pi03}, \href{https://www.nndc.bnl.gov/nsr/nsrlink.jsp?2012Wa15,B}{2012Wa15}) to calculate \ensuremath{\beta}-summing for \ensuremath{\beta}-delayed proton}\\
\parbox[b][0.3cm]{17.7cm}{emitting nuclei (such as \ensuremath{^{\textnormal{20}}}Mg) that are detected via implantation in DSSSD detectors. Found a \ensuremath{\beta}-summing correction on the order}\\
\parbox[b][0.3cm]{17.7cm}{of 18 keV \textit{6} for a mean implantation depth of \ensuremath{\sim}27 \ensuremath{\mu}m.}\\
\vspace{12pt}
\underline{$^{19}$Ne Levels}\\
\begin{longtable}{cccccc@{\extracolsep{\fill}}c}
\multicolumn{2}{c}{E(level)$^{{\hyperlink{NE1LEVEL0}{a}}}$}&J$^{\pi}$$^{{\hyperlink{NE1LEVEL1}{b}}}$&\multicolumn{2}{c}{T$_{1/2}$$^{}$}&Comments&\\[-.2cm]
\multicolumn{2}{c}{\hrulefill}&\hrulefill&\multicolumn{2}{c}{\hrulefill}&\hrulefill&
\endfirsthead
\multicolumn{1}{r@{}}{0}&\multicolumn{1}{@{}l}{}&\multicolumn{1}{l}{1/2\ensuremath{^{+}}}&&&&\\
\multicolumn{1}{r@{}}{238}&\multicolumn{1}{@{.}l}{03 {\it 9}}&\multicolumn{1}{l}{5/2\ensuremath{^{+}}}&&&&\\
\multicolumn{1}{r@{}}{274}&\multicolumn{1}{@{.}l}{96 {\it 10}}&\multicolumn{1}{l}{1/2\ensuremath{^{-}}}&&&&\\
\multicolumn{1}{r@{}}{1507}&\multicolumn{1}{@{.}l}{52 {\it 18}}&\multicolumn{1}{l}{5/2\ensuremath{^{-}}}&\multicolumn{1}{r@{}}{3}&\multicolumn{1}{@{.}l}{0 ps {\it +9\textminus8}}&\parbox[t][0.3cm]{12.71648cm}{\raggedright T\ensuremath{_{1/2}}: From \ensuremath{\tau}=4.3 ps \textit{+13{\textminus}11} (\href{https://www.nndc.bnl.gov/nsr/nsrlink.jsp?2019Gl02,B}{2019Gl02}).\vspace{0.1cm}}&\\
\multicolumn{1}{r@{}}{1535}&\multicolumn{1}{@{.}l}{96 {\it 15}}&\multicolumn{1}{l}{3/2\ensuremath{^{+}}}&&&&\\
\multicolumn{1}{r@{}}{1615}&\multicolumn{1}{@{.}l}{24 {\it 17}}&\multicolumn{1}{l}{3/2\ensuremath{^{-}}}&&&&\\
\multicolumn{1}{r@{}}{4034}&\multicolumn{1}{@{.}l}{7 {\it 16}}&\multicolumn{1}{l}{3/2\ensuremath{^{+}}}&&&\parbox[t][0.3cm]{12.71648cm}{\raggedright Decay mode: Predominantly \ensuremath{\gamma}, and \ensuremath{\alpha} (\href{https://www.nndc.bnl.gov/nsr/nsrlink.jsp?2017Wr02,B}{2017Wr02}).\vspace{0.1cm}}&\\
\end{longtable}
\parbox[b][0.3cm]{17.7cm}{\makebox[1ex]{\ensuremath{^{\hypertarget{NE1LEVEL0}{a}}}} From a least-squares fit to the \ensuremath{\gamma}-ray energies.}\\
\parbox[b][0.3cm]{17.7cm}{\makebox[1ex]{\ensuremath{^{\hypertarget{NE1LEVEL1}{b}}}} From the \ensuremath{^{\textnormal{19}}}Ne Adopted Levels unless otherwise noted.}\\
\vspace{0.5cm}
\clearpage
\vspace{0.3cm}
\vspace*{-0.5cm}
{\bf \small \underline{\ensuremath{^{\textnormal{20}}}Mg \ensuremath{\beta}\ensuremath{^{\textnormal{+}}}p decay\hspace{0.2in}\href{https://www.nndc.bnl.gov/nsr/nsrlink.jsp?1995Pi03,B}{1995Pi03},\href{https://www.nndc.bnl.gov/nsr/nsrlink.jsp?2016Li45,B}{2016Li45},\href{https://www.nndc.bnl.gov/nsr/nsrlink.jsp?2019Gl02,B}{2019Gl02} (continued)}}\\
\vspace{0.3cm}
\underline{$\gamma$($^{19}$Ne)}\\
\begin{longtable}{ccccccccc@{}cc@{\extracolsep{\fill}}c}
\multicolumn{2}{c}{E\ensuremath{_{\gamma}}\ensuremath{^{\hyperlink{MG1GAMMA0}{a}}}}&\multicolumn{2}{c}{I\ensuremath{_{\gamma}}\ensuremath{^{\hyperlink{MG1GAMMA0}{a}\hyperlink{MG1GAMMA3}{d}}}}&\multicolumn{2}{c}{E\ensuremath{_{i}}(level)}&J\ensuremath{^{\pi}_{i}}&\multicolumn{2}{c}{E\ensuremath{_{f}}}&J\ensuremath{^{\pi}_{f}}&Comments&\\[-.2cm]
\multicolumn{2}{c}{\hrulefill}&\multicolumn{2}{c}{\hrulefill}&\multicolumn{2}{c}{\hrulefill}&\hrulefill&\multicolumn{2}{c}{\hrulefill}&\hrulefill&\hrulefill&
\endfirsthead
\multicolumn{1}{r@{}}{238}&\multicolumn{1}{@{.}l}{04 {\it 10}}&\multicolumn{1}{r@{}}{3}&\multicolumn{1}{@{.}l}{80 {\it 11}}&\multicolumn{1}{r@{}}{238}&\multicolumn{1}{@{.}l}{03}&\multicolumn{1}{l}{5/2\ensuremath{^{+}}}&\multicolumn{1}{r@{}}{0}&\multicolumn{1}{@{}l}{}&\multicolumn{1}{@{}l}{1/2\ensuremath{^{+}}}&\parbox[t][0.3cm]{9.116961cm}{\raggedright E\ensuremath{_{\gamma}}: See also 238 keV (\href{https://www.nndc.bnl.gov/nsr/nsrlink.jsp?1995Pi03,B}{1995Pi03}, \href{https://www.nndc.bnl.gov/nsr/nsrlink.jsp?2016Lu13,B}{2016Lu13}, \href{https://www.nndc.bnl.gov/nsr/nsrlink.jsp?2017Su05,B}{2017Su05},\vspace{0.1cm}}&\\
&&&&&&&&&&\parbox[t][0.3cm]{9.116961cm}{\raggedright {\ }{\ }{\ }\href{https://www.nndc.bnl.gov/nsr/nsrlink.jsp?2017Wr02,B}{2017Wr02}, \href{https://www.nndc.bnl.gov/nsr/nsrlink.jsp?2018Gl01,B}{2018Gl01}).\vspace{0.1cm}}&\\
&&&&&&&&&&\parbox[t][0.3cm]{9.116961cm}{\raggedright I\ensuremath{_{\gamma}}: From 3.80\ensuremath{\times}10\ensuremath{^{\textnormal{$-$2}}} \textit{7} (stat.) \textit{8} (sys.) corresponding to a 100\%\vspace{0.1cm}}&\\
&&&&&&&&&&\parbox[t][0.3cm]{9.116961cm}{\raggedright {\ }{\ }{\ }\ensuremath{\gamma}-ray branching ratio (\href{https://www.nndc.bnl.gov/nsr/nsrlink.jsp?2019Gl02,B}{2019Gl02}).\vspace{0.1cm}}&\\
\multicolumn{1}{r@{}}{274}&\multicolumn{1}{@{.}l}{97 {\it 9}}&\multicolumn{1}{r@{}}{3}&\multicolumn{1}{@{.}l}{59 {\it 10}}&\multicolumn{1}{r@{}}{274}&\multicolumn{1}{@{.}l}{96}&\multicolumn{1}{l}{1/2\ensuremath{^{-}}}&\multicolumn{1}{r@{}}{0}&\multicolumn{1}{@{}l}{}&\multicolumn{1}{@{}l}{1/2\ensuremath{^{+}}}&\parbox[t][0.3cm]{9.116961cm}{\raggedright E\ensuremath{_{\gamma}}: See also 275 keV (\href{https://www.nndc.bnl.gov/nsr/nsrlink.jsp?1995Pi03,B}{1995Pi03}, \href{https://www.nndc.bnl.gov/nsr/nsrlink.jsp?2016Lu13,B}{2016Lu13}, \href{https://www.nndc.bnl.gov/nsr/nsrlink.jsp?2017Su05,B}{2017Su05},\vspace{0.1cm}}&\\
&&&&&&&&&&\parbox[t][0.3cm]{9.116961cm}{\raggedright {\ }{\ }{\ }\href{https://www.nndc.bnl.gov/nsr/nsrlink.jsp?2017Wr02,B}{2017Wr02}, \href{https://www.nndc.bnl.gov/nsr/nsrlink.jsp?2018Gl01,B}{2018Gl01}).\vspace{0.1cm}}&\\
&&&&&&&&&&\parbox[t][0.3cm]{9.116961cm}{\raggedright I\ensuremath{_{\gamma}}: From 3.59\ensuremath{\times}10\ensuremath{^{\textnormal{$-$2}}} \textit{6} (stat.) \textit{8} (sys.) corresponding to a 100\%\vspace{0.1cm}}&\\
&&&&&&&&&&\parbox[t][0.3cm]{9.116961cm}{\raggedright {\ }{\ }{\ }\ensuremath{\gamma}-ray branching ratio (\href{https://www.nndc.bnl.gov/nsr/nsrlink.jsp?2019Gl02,B}{2019Gl02}).\vspace{0.1cm}}&\\
\multicolumn{1}{r@{}}{1232}&\multicolumn{1}{@{.}l}{49 {\it 22}}&\multicolumn{1}{r@{}}{0}&\multicolumn{1}{@{.}l}{236 {\it 6}}&\multicolumn{1}{r@{}}{1507}&\multicolumn{1}{@{.}l}{52}&\multicolumn{1}{l}{5/2\ensuremath{^{-}}}&\multicolumn{1}{r@{}}{274}&\multicolumn{1}{@{.}l}{96 }&\multicolumn{1}{@{}l}{1/2\ensuremath{^{-}}}&\parbox[t][0.3cm]{9.116961cm}{\raggedright E\ensuremath{_{\gamma}}: See also 1232 keV (\href{https://www.nndc.bnl.gov/nsr/nsrlink.jsp?1995Pi03,B}{1995Pi03}, \href{https://www.nndc.bnl.gov/nsr/nsrlink.jsp?2017Wr02,B}{2017Wr02}, \href{https://www.nndc.bnl.gov/nsr/nsrlink.jsp?2018Gl01,B}{2018Gl01}); and\vspace{0.1cm}}&\\
&&&&&&&&&&\parbox[t][0.3cm]{9.116961cm}{\raggedright {\ }{\ }{\ }1233 keV (\href{https://www.nndc.bnl.gov/nsr/nsrlink.jsp?2017Su05,B}{2017Su05}).\vspace{0.1cm}}&\\
&&&&&&&&&&\parbox[t][0.3cm]{9.116961cm}{\raggedright I\ensuremath{_{\gamma}}: From 2.36\ensuremath{\times}10\ensuremath{^{\textnormal{$-$3}}} \textit{4} (stat.) \textit{5} (sys.) corresponding to a 84.9\% \textit{4}\vspace{0.1cm}}&\\
&&&&&&&&&&\parbox[t][0.3cm]{9.116961cm}{\raggedright {\ }{\ }{\ }\ensuremath{\gamma}-ray branching ratio (\href{https://www.nndc.bnl.gov/nsr/nsrlink.jsp?2019Gl02,B}{2019Gl02}).\vspace{0.1cm}}&\\
\multicolumn{1}{r@{}}{1260}&\multicolumn{1}{@{.}l}{87\ensuremath{^{\hyperlink{MG1GAMMA1}{b}}} {\it 24}}&\multicolumn{1}{r@{}}{0}&\multicolumn{1}{@{.}l}{0675 {\it 21}}&\multicolumn{1}{r@{}}{1535}&\multicolumn{1}{@{.}l}{96}&\multicolumn{1}{l}{3/2\ensuremath{^{+}}}&\multicolumn{1}{r@{}}{274}&\multicolumn{1}{@{.}l}{96 }&\multicolumn{1}{@{}l}{1/2\ensuremath{^{-}}}&\parbox[t][0.3cm]{9.116961cm}{\raggedright E\ensuremath{_{\gamma}}: See also 1261 keV (\href{https://www.nndc.bnl.gov/nsr/nsrlink.jsp?2017Wr02,B}{2017Wr02}).\vspace{0.1cm}}&\\
&&&&&&&&&&\parbox[t][0.3cm]{9.116961cm}{\raggedright I\ensuremath{_{\gamma}}: From 6.75\ensuremath{\times}10\ensuremath{^{\textnormal{$-$4}}} \textit{15} (stat.) \textit{15} (sys.) corresponding to a 4.05\%\vspace{0.1cm}}&\\
&&&&&&&&&&\parbox[t][0.3cm]{9.116961cm}{\raggedright {\ }{\ }{\ }\textit{16} \ensuremath{\gamma}-ray branching ratio (\href{https://www.nndc.bnl.gov/nsr/nsrlink.jsp?2019Gl02,B}{2019Gl02}).\vspace{0.1cm}}&\\
\multicolumn{1}{r@{}}{1269}&\multicolumn{1}{@{.}l}{47\ensuremath{^{\hyperlink{MG1GAMMA1}{b}}} {\it 24}}&\multicolumn{1}{r@{}}{0}&\multicolumn{1}{@{.}l}{0418 {\it 15}}&\multicolumn{1}{r@{}}{1507}&\multicolumn{1}{@{.}l}{52}&\multicolumn{1}{l}{5/2\ensuremath{^{-}}}&\multicolumn{1}{r@{}}{238}&\multicolumn{1}{@{.}l}{03 }&\multicolumn{1}{@{}l}{5/2\ensuremath{^{+}}}&\parbox[t][0.3cm]{9.116961cm}{\raggedright E\ensuremath{_{\gamma}}: See also 1269 keV (\href{https://www.nndc.bnl.gov/nsr/nsrlink.jsp?2017Wr02,B}{2017Wr02}).\vspace{0.1cm}}&\\
&&&&&&&&&&\parbox[t][0.3cm]{9.116961cm}{\raggedright I\ensuremath{_{\gamma}}: From 4.18\ensuremath{\times}10\ensuremath{^{\textnormal{$-$4}}} \textit{12} (stat.) \textit{9} (sys.) corresponding to a 15.1\% \textit{4}\vspace{0.1cm}}&\\
&&&&&&&&&&\parbox[t][0.3cm]{9.116961cm}{\raggedright {\ }{\ }{\ }\ensuremath{\gamma}-ray branching ratio (\href{https://www.nndc.bnl.gov/nsr/nsrlink.jsp?2019Gl02,B}{2019Gl02}).\vspace{0.1cm}}&\\
\multicolumn{1}{r@{}}{1297}&\multicolumn{1}{@{.}l}{94 {\it 22}}&\multicolumn{1}{r@{}}{1}&\multicolumn{1}{@{.}l}{539 {\it 43}}&\multicolumn{1}{r@{}}{1535}&\multicolumn{1}{@{.}l}{96}&\multicolumn{1}{l}{3/2\ensuremath{^{+}}}&\multicolumn{1}{r@{}}{238}&\multicolumn{1}{@{.}l}{03 }&\multicolumn{1}{@{}l}{5/2\ensuremath{^{+}}}&\parbox[t][0.3cm]{9.116961cm}{\raggedright E\ensuremath{_{\gamma}}: See also 1298 keV (\href{https://www.nndc.bnl.gov/nsr/nsrlink.jsp?1995Pi03,B}{1995Pi03}, \href{https://www.nndc.bnl.gov/nsr/nsrlink.jsp?2016Lu13,B}{2016Lu13}, \href{https://www.nndc.bnl.gov/nsr/nsrlink.jsp?2017Su05,B}{2017Su05},\vspace{0.1cm}}&\\
&&&&&&&&&&\parbox[t][0.3cm]{9.116961cm}{\raggedright {\ }{\ }{\ }\href{https://www.nndc.bnl.gov/nsr/nsrlink.jsp?2017Wr02,B}{2017Wr02}, \href{https://www.nndc.bnl.gov/nsr/nsrlink.jsp?2018Gl01,B}{2018Gl01}).\vspace{0.1cm}}&\\
&&&&&&&&&&\parbox[t][0.3cm]{9.116961cm}{\raggedright I\ensuremath{_{\gamma}}: From 1.539\ensuremath{\times}10\ensuremath{^{\textnormal{$-$2}}} \textit{27} (stat.) \textit{33} (sys.) corresponding to a\vspace{0.1cm}}&\\
&&&&&&&&&&\parbox[t][0.3cm]{9.116961cm}{\raggedright {\ }{\ }{\ }92.53\% \textit{35} \ensuremath{\gamma}-ray branching ratio (\href{https://www.nndc.bnl.gov/nsr/nsrlink.jsp?2019Gl02,B}{2019Gl02}).\vspace{0.1cm}}&\\
\multicolumn{1}{r@{}}{1340}&\multicolumn{1}{@{.}l}{27\ensuremath{^{\hyperlink{MG1GAMMA1}{b}}} {\it 25}}&\multicolumn{1}{r@{}}{0}&\multicolumn{1}{@{.}l}{157 {\it 4}}&\multicolumn{1}{r@{}}{1615}&\multicolumn{1}{@{.}l}{24}&\multicolumn{1}{l}{3/2\ensuremath{^{-}}}&\multicolumn{1}{r@{}}{274}&\multicolumn{1}{@{.}l}{96 }&\multicolumn{1}{@{}l}{1/2\ensuremath{^{-}}}&\parbox[t][0.3cm]{9.116961cm}{\raggedright E\ensuremath{_{\gamma}}: See also 1340 keV (\href{https://www.nndc.bnl.gov/nsr/nsrlink.jsp?2017Wr02,B}{2017Wr02}).\vspace{0.1cm}}&\\
&&&&&&&&&&\parbox[t][0.3cm]{9.116961cm}{\raggedright I\ensuremath{_{\gamma}}: From 1.57\ensuremath{\times}10\ensuremath{^{\textnormal{$-$3}}} \textit{3} (stat.) \textit{3} (sys.) corresponding to a 74.0\% \textit{17}\vspace{0.1cm}}&\\
&&&&&&&&&&\parbox[t][0.3cm]{9.116961cm}{\raggedright {\ }{\ }{\ }\ensuremath{\gamma}-ray branching ratio (\href{https://www.nndc.bnl.gov/nsr/nsrlink.jsp?2019Gl02,B}{2019Gl02}).\vspace{0.1cm}}&\\
\multicolumn{1}{r@{}}{1377}&\multicolumn{1}{@{.}l}{1\ensuremath{^{\hyperlink{MG1GAMMA2}{c}}} {\it 3}}&\multicolumn{1}{r@{}}{0}&\multicolumn{1}{@{.}l}{0182 {\it 41}}&\multicolumn{1}{r@{}}{1615}&\multicolumn{1}{@{.}l}{24}&\multicolumn{1}{l}{3/2\ensuremath{^{-}}}&\multicolumn{1}{r@{}}{238}&\multicolumn{1}{@{.}l}{03 }&\multicolumn{1}{@{}l}{5/2\ensuremath{^{+}}}&\parbox[t][0.3cm]{9.116961cm}{\raggedright I\ensuremath{_{\gamma}}: From 1.82\ensuremath{\times}10\ensuremath{^{\textnormal{$-$4}}} \textit{41} (stat.) \textit{4} (sys.) corresponding to a 8.6\% \textit{18}\vspace{0.1cm}}&\\
&&&&&&&&&&\parbox[t][0.3cm]{9.116961cm}{\raggedright {\ }{\ }{\ }\ensuremath{\gamma}-ray branching ratio (\href{https://www.nndc.bnl.gov/nsr/nsrlink.jsp?2019Gl02,B}{2019Gl02}).\vspace{0.1cm}}&\\
\multicolumn{1}{r@{}}{1535}&\multicolumn{1}{@{.}l}{90 {\it 24}}&\multicolumn{1}{r@{}}{0}&\multicolumn{1}{@{.}l}{0568 {\it 47}}&\multicolumn{1}{r@{}}{1535}&\multicolumn{1}{@{.}l}{96}&\multicolumn{1}{l}{3/2\ensuremath{^{+}}}&\multicolumn{1}{r@{}}{0}&\multicolumn{1}{@{}l}{}&\multicolumn{1}{@{}l}{1/2\ensuremath{^{+}}}&\parbox[t][0.3cm]{9.116961cm}{\raggedright E\ensuremath{_{\gamma}}: This \ensuremath{\gamma}-ray branch was observed for the first time in\vspace{0.1cm}}&\\
&&&&&&&&&&\parbox[t][0.3cm]{9.116961cm}{\raggedright {\ }{\ }{\ }(\href{https://www.nndc.bnl.gov/nsr/nsrlink.jsp?2019Gl02,B}{2019Gl02}).\vspace{0.1cm}}&\\
&&&&&&&&&&\parbox[t][0.3cm]{9.116961cm}{\raggedright I\ensuremath{_{\gamma}}: From 5.68\ensuremath{\times}10\ensuremath{^{\textnormal{$-$4}}} \textit{44} (stat.) \textit{17} (sys.) corresponding to a 3.42\%\vspace{0.1cm}}&\\
&&&&&&&&&&\parbox[t][0.3cm]{9.116961cm}{\raggedright {\ }{\ }{\ }\textit{29} \ensuremath{\gamma}-ray branching ratio (\href{https://www.nndc.bnl.gov/nsr/nsrlink.jsp?2019Gl02,B}{2019Gl02}).\vspace{0.1cm}}&\\
\multicolumn{1}{r@{}}{1615}&\multicolumn{1}{@{.}l}{16\ensuremath{^{\hyperlink{MG1GAMMA2}{c}}} {\it 30}}&\multicolumn{1}{r@{}}{0}&\multicolumn{1}{@{.}l}{0368 {\it 20}}&\multicolumn{1}{r@{}}{1615}&\multicolumn{1}{@{.}l}{24}&\multicolumn{1}{l}{3/2\ensuremath{^{-}}}&\multicolumn{1}{r@{}}{0}&\multicolumn{1}{@{}l}{}&\multicolumn{1}{@{}l}{1/2\ensuremath{^{+}}}&\parbox[t][0.3cm]{9.116961cm}{\raggedright I\ensuremath{_{\gamma}}: From 3.68\ensuremath{\times}10\ensuremath{^{\textnormal{$-$4}}} \textit{18} (stat) \textit{8} (sys.) corresponding to a 17.4\% \textit{9}\vspace{0.1cm}}&\\
&&&&&&&&&&\parbox[t][0.3cm]{9.116961cm}{\raggedright {\ }{\ }{\ }\ensuremath{\gamma}-ray branching ratio (\href{https://www.nndc.bnl.gov/nsr/nsrlink.jsp?2019Gl02,B}{2019Gl02}).\vspace{0.1cm}}&\\
\multicolumn{1}{r@{}}{4034}&\multicolumn{1}{@{.}l}{2 {\it 16}}&\multicolumn{1}{r@{}}{0}&\multicolumn{1}{@{.}l}{0119 {\it 17}}&\multicolumn{1}{r@{}}{4034}&\multicolumn{1}{@{.}l}{7}&\multicolumn{1}{l}{3/2\ensuremath{^{+}}}&\multicolumn{1}{r@{}}{0}&\multicolumn{1}{@{}l}{}&\multicolumn{1}{@{}l}{1/2\ensuremath{^{+}}}&\parbox[t][0.3cm]{9.116961cm}{\raggedright E\ensuremath{_{\gamma}}: See also 4033.4 keV \textit{17} (\href{https://www.nndc.bnl.gov/nsr/nsrlink.jsp?2017Wr02,B}{2017Wr02}, \href{https://www.nndc.bnl.gov/nsr/nsrlink.jsp?2018Gl01,B}{2018Gl01}).\vspace{0.1cm}}&\\
&&&&&&&&&&\parbox[t][0.3cm]{9.116961cm}{\raggedright I\ensuremath{_{\gamma}}: From 1.19\ensuremath{\times}10\ensuremath{^{\textnormal{$-$4}}} \textit{12} (stat.) \textit{12} (sys.) (\href{https://www.nndc.bnl.gov/nsr/nsrlink.jsp?2019Gl02,B}{2019Gl02}). This\vspace{0.1cm}}&\\
&&&&&&&&&&\parbox[t][0.3cm]{9.116961cm}{\raggedright {\ }{\ }{\ }corresponds to a 80\% \textit{15} \ensuremath{\gamma}-ray branching ratio suggested by\vspace{0.1cm}}&\\
&&&&&&&&&&\parbox[t][0.3cm]{9.116961cm}{\raggedright {\ }{\ }{\ }(\href{https://www.nndc.bnl.gov/nsr/nsrlink.jsp?1995Ti07,B}{1995Ti07}, \href{https://www.nndc.bnl.gov/nsr/nsrlink.jsp?2019Gl02,B}{2019Gl02}). See also I\ensuremath{_{\ensuremath{\beta}\textnormal{p}\ensuremath{\gamma}}}=1.25\ensuremath{\times}10\ensuremath{^{\textnormal{$-$4}}} \textit{20} per decay\vspace{0.1cm}}&\\
&&&&&&&&&&\parbox[t][0.3cm]{9.116961cm}{\raggedright {\ }{\ }{\ }of one \ensuremath{^{\textnormal{20}}}Mg parent nucleus (\href{https://www.nndc.bnl.gov/nsr/nsrlink.jsp?2017Wr02,B}{2017Wr02}), which is likely to be\vspace{0.1cm}}&\\
&&&&&&&&&&\parbox[t][0.3cm]{9.116961cm}{\raggedright {\ }{\ }{\ }the preceding result.\vspace{0.1cm}}&\\
\end{longtable}
\parbox[b][0.3cm]{17.7cm}{\makebox[1ex]{\ensuremath{^{\hypertarget{MG1GAMMA0}{a}}}} From (\href{https://www.nndc.bnl.gov/nsr/nsrlink.jsp?2019Gl02,B}{2019Gl02}) unless otherwise noted. We note that the I\ensuremath{_{\ensuremath{\gamma}}} values given in the comments are the total intensities per one}\\
\parbox[b][0.3cm]{17.7cm}{{\ }{\ }\ensuremath{^{\textnormal{20}}}Mg decay as reported by (\href{https://www.nndc.bnl.gov/nsr/nsrlink.jsp?2019Gl02,B}{2019Gl02}).}\\
\parbox[b][0.3cm]{17.7cm}{\makebox[1ex]{\ensuremath{^{\hypertarget{MG1GAMMA1}{b}}}} This transition was observed for the first time as part of the \ensuremath{\beta} decay of \ensuremath{^{\textnormal{20}}}Mg in (\href{https://www.nndc.bnl.gov/nsr/nsrlink.jsp?2017Wr02,B}{2017Wr02}).}\\
\parbox[b][0.3cm]{17.7cm}{\makebox[1ex]{\ensuremath{^{\hypertarget{MG1GAMMA2}{c}}}} This transition was observed for the first time as part of the \ensuremath{^{\textnormal{20}}}Mg(\ensuremath{\beta}p\ensuremath{\gamma}) decay in (\href{https://www.nndc.bnl.gov/nsr/nsrlink.jsp?2019Gl02,B}{2019Gl02}).}\\
\parbox[b][0.3cm]{17.7cm}{\makebox[1ex]{\ensuremath{^{\hypertarget{MG1GAMMA3}{d}}}} Absolute intensity per 100 decays.}\\
\vspace{0.5cm}
\clearpage
\vspace{0.3cm}
\vspace*{-0.5cm}
{\bf \small \underline{\ensuremath{^{\textnormal{20}}}Mg \ensuremath{\beta}\ensuremath{^{\textnormal{+}}}p decay\hspace{0.2in}\href{https://www.nndc.bnl.gov/nsr/nsrlink.jsp?1995Pi03,B}{1995Pi03},\href{https://www.nndc.bnl.gov/nsr/nsrlink.jsp?2016Li45,B}{2016Li45},\href{https://www.nndc.bnl.gov/nsr/nsrlink.jsp?2019Gl02,B}{2019Gl02} (continued)}}\\
\vspace{0.3cm}
\underline{Delayed Protons ($^{19}$Ne)}\\
\vspace{0.34cm}
\parbox[b][0.3cm]{17.7cm}{\addtolength{\parindent}{-0.254cm}\textit{Notes}:}\\
\parbox[b][0.3cm]{17.7cm}{\addtolength{\parindent}{-0.254cm}(1) (\href{https://www.nndc.bnl.gov/nsr/nsrlink.jsp?1995Pi03,B}{1995Pi03}) deduced a branching ratio of 6.95\% \textit{80} for the \ensuremath{\beta}-delayed proton decay to the \ensuremath{^{\textnormal{19}}}Ne*(238, 275, 1507, 1536) levels.}\\
\parbox[b][0.3cm]{17.7cm}{From this and the total branching ratio of 30.3\% \textit{12} for the \ensuremath{\beta}-delayed proton decay of \ensuremath{^{\textnormal{20}}}Mg (\href{https://www.nndc.bnl.gov/nsr/nsrlink.jsp?1995Pi03,B}{1995Pi03}), they deduced a branching}\\
\parbox[b][0.3cm]{17.7cm}{ratio of 23.35\% \textit{150} for the \ensuremath{\beta}-delayed proton decay to \ensuremath{^{\textnormal{19}}}Ne\ensuremath{_{\textnormal{g.s.}}} (see Fig. 7). This value is consistent with \ensuremath{\sum}\ensuremath{>}22.9\% (including the}\\
\parbox[b][0.3cm]{17.7cm}{cumulative 3\% feeding to the unresolved \ensuremath{^{\textnormal{20}}}Na*(4800, 5600) states from that study), which is given at the bottom of Table 2 of}\\
\parbox[b][0.3cm]{17.7cm}{(\href{https://www.nndc.bnl.gov/nsr/nsrlink.jsp?1995Pi03,B}{1995Pi03}). (\href{https://www.nndc.bnl.gov/nsr/nsrlink.jsp?2019Gl02,B}{2019Gl02}) measured the branching ratio to the \ensuremath{^{\textnormal{19}}}Ne\ensuremath{_{\textnormal{g.s.}}} to be I\ensuremath{_{\ensuremath{\beta}\textnormal{p}}}=22.8\% \textit{12}. (\href{https://www.nndc.bnl.gov/nsr/nsrlink.jsp?2019Gl02,B}{2019Gl02}) did not report the proton}\\
\parbox[b][0.3cm]{17.7cm}{energies for \ensuremath{\beta}\ensuremath{^{\textnormal{+}}}p discrete transitions to the bound states of \ensuremath{^{\textnormal{19}}}Ne, which they showed by symbol x representing a broad range of}\\
\parbox[b][0.3cm]{17.7cm}{proton energies from the decay of unresolved states in \ensuremath{^{\textnormal{20}}}Na, which were represented by symbol U.}\\
\parbox[b][0.3cm]{17.7cm}{\addtolength{\parindent}{-0.254cm}(2) (\href{https://www.nndc.bnl.gov/nsr/nsrlink.jsp?2019Gl02,B}{2019Gl02}) determined I\ensuremath{_{\ensuremath{\beta}\textnormal{p}}}=2.21\% \textit{14} for feeding the \ensuremath{^{\textnormal{19}}}Ne*(238) level; I\ensuremath{_{\ensuremath{\beta}\textnormal{p}}}=3.13\% \textit{15} for feeding the \ensuremath{^{\textnormal{19}}}Ne*(275) level;}\\
\parbox[b][0.3cm]{17.7cm}{I\ensuremath{_{\ensuremath{\beta}\textnormal{p}}}=0.278\% \textit{7} for feeding the \ensuremath{^{\textnormal{19}}}Ne*(1507.5) level; and I\ensuremath{_{\ensuremath{\beta}\textnormal{p}}}=1.663\% \textit{45} for feeding the \ensuremath{^{\textnormal{19}}}Ne*(1535.95) level. They reported that}\\
\parbox[b][0.3cm]{17.7cm}{using the I\ensuremath{_{\ensuremath{\beta}\textnormal{p}}} deduced by (\href{https://www.nndc.bnl.gov/nsr/nsrlink.jsp?2016Lu13,B}{2016Lu13}) did not fit their data and that those of (\href{https://www.nndc.bnl.gov/nsr/nsrlink.jsp?1995Pi03,B}{1995Pi03}) resulted in more reasonable reduced \ensuremath{\chi}\ensuremath{^{\textnormal{2}}}}\\
\parbox[b][0.3cm]{17.7cm}{for this branch.}\\
\parbox[b][0.3cm]{17.7cm}{\addtolength{\parindent}{-0.254cm}(3) Based on the proton peak labeling convention used by (\href{https://www.nndc.bnl.gov/nsr/nsrlink.jsp?2016Lu13,B}{2016Lu13}), the p\ensuremath{_{\textnormal{I}}} and p\ensuremath{_{\textnormal{II}}} transitions (see \href{https://www.nndc.bnl.gov/nsr/nsrlink.jsp?2016Lu13,B}{2016Lu13}) interfere}\\
\parbox[b][0.3cm]{17.7cm}{destructively, so do p\ensuremath{_{\textnormal{4}}} and p\ensuremath{_{\textnormal{7}}}, and possibly p\ensuremath{_{\textnormal{IV}}} and p\ensuremath{_{\textnormal{III}}}, as well as p\ensuremath{_{\textnormal{III}}} and p\ensuremath{_{\textnormal{9}}}.}\\
\parbox[b][0.3cm]{17.7cm}{\addtolength{\parindent}{-0.254cm}(4) (\href{https://www.nndc.bnl.gov/nsr/nsrlink.jsp?2016Lu13,B}{2016Lu13}) obtained BR\ensuremath{_{\ensuremath{\beta}\ensuremath{\gamma}}}=82.4\% \textit{13} for the emission of the \ensuremath{\beta}-delayed 1634-keV \ensuremath{\gamma} ray in \ensuremath{^{\textnormal{20}}}Na.}\\
\parbox[b][0.3cm]{17.7cm}{\addtolength{\parindent}{-0.254cm}(5) (\href{https://www.nndc.bnl.gov/nsr/nsrlink.jsp?2016Lu13,B}{2016Lu13}) determined the \ensuremath{\beta} decay strength going to the T=1 \ensuremath{^{\textnormal{20}}}Na* states above the \ensuremath{^{\textnormal{20}}}Na*(IAS) level to be I\ensuremath{_{\ensuremath{\beta}}}=0.67\% \textit{9}.}\\
\parbox[b][0.3cm]{17.7cm}{\addtolength{\parindent}{-0.254cm}(6) (\href{https://www.nndc.bnl.gov/nsr/nsrlink.jsp?2016Lu13,B}{2016Lu13}) observed no evidence for feeding of the 2647 keV \textit{3} (\href{https://www.nndc.bnl.gov/nsr/nsrlink.jsp?2012Wa15,B}{2012Wa15}) level in \ensuremath{^{\textnormal{20}}}Na. See also (\href{https://www.nndc.bnl.gov/nsr/nsrlink.jsp?2018Gl01,B}{2018Gl01}).}\\
\parbox[b][0.3cm]{17.7cm}{\addtolength{\parindent}{-0.254cm}(7) (\href{https://www.nndc.bnl.gov/nsr/nsrlink.jsp?2017Su05,B}{2017Su05}) found no discernible proton peak around the expected center-of-mass decay energy of 455 keV and placed an upper}\\
\parbox[b][0.3cm]{17.7cm}{limit of 0.24\% \textit{3} for its branching ratio.}\\
\vspace{0.34cm}

\begin{textblock}{29}(0,27.3)
Continued on next page (footnotes at end of table)
\end{textblock}
\clearpage
%
\begin{textblock}{29}(0,27.3)
Continued on next page (footnotes at end of table)
\end{textblock}
\clearpage
%
\begin{textblock}{29}(0,27.3)
Continued on next page (footnotes at end of table)
\end{textblock}
\clearpage
%
\parbox[b][0.3cm]{17.7cm}{\makebox[1ex]{\ensuremath{^{\hypertarget{MG1DELAY0}{a}}}} Deduced from Q\ensuremath{_{\textnormal{EC}}}(\ensuremath{^{\textnormal{20}}}Mg) and S\ensuremath{_{\textnormal{p}}}(\ensuremath{^{\textnormal{20}}}Na) from (\href{https://www.nndc.bnl.gov/nsr/nsrlink.jsp?2021Wa16,B}{2021Wa16}), the \ensuremath{^{\textnormal{20}}}Na level-energies which are evaluated in this dataset (based}\\
\parbox[b][0.3cm]{17.7cm}{{\ }{\ }on the articles of interest to this dataset only), and the \ensuremath{^{\textnormal{19}}}Ne level-energies from the Adopted Levels. In the articles provided in}\\
\parbox[b][0.3cm]{17.7cm}{{\ }{\ }this dataset, E\ensuremath{_{\textnormal{p}}} refers to the directly measured total decay energy in the center-of-mass frame. These E\ensuremath{_{\textnormal{p}}}(c.m.) values are provided}\\
\parbox[b][0.3cm]{17.7cm}{{\ }{\ }for each transition in the comments section.}\\
\parbox[b][0.3cm]{17.7cm}{\makebox[1ex]{\ensuremath{^{\hypertarget{MG1DELAY1}{b}}}} (\href{https://www.nndc.bnl.gov/nsr/nsrlink.jsp?1995Pi03,B}{1995Pi03}) reported unresolved transitions from the decay of those levels in \ensuremath{^{\textnormal{20}}}Na with excitation energies in the range of}\\
\parbox[b][0.3cm]{17.7cm}{{\ }{\ }E\ensuremath{_{\textnormal{x}}}=4.9-5.8 MeV (see Table 2). That study assigned a branching ratio of 0.10\% \textit{1} to those unresolved branches. (\href{https://www.nndc.bnl.gov/nsr/nsrlink.jsp?2012Wa15,B}{2012Wa15}) also}\\
\parbox[b][0.3cm]{17.7cm}{{\ }{\ }reported an unresolved proton group with a total c.m. decay energy of \ensuremath{\approx}1050 keV \textit{16} from the \ensuremath{^{\textnormal{20}}}Mg \ensuremath{\beta}-delayed proton decay to}\\
\parbox[b][0.3cm]{17.7cm}{{\ }{\ }the \ensuremath{^{\textnormal{19}}}Ne*(1508+1536) states through a \ensuremath{^{\textnormal{20}}}Na* level whose excitation energy was reported by (\href{https://www.nndc.bnl.gov/nsr/nsrlink.jsp?2012Wa15,B}{2012Wa15}) to be E\ensuremath{_{\textnormal{x}}}\ensuremath{\approx}4780 keV.}\\
\parbox[b][0.3cm]{17.7cm}{{\ }{\ }(\href{https://www.nndc.bnl.gov/nsr/nsrlink.jsp?2016Lu13,B}{2016Lu13}) resolved those unresolved transitions reported by (\href{https://www.nndc.bnl.gov/nsr/nsrlink.jsp?1995Pi03,B}{1995Pi03}) (see Table 5 in \href{https://www.nndc.bnl.gov/nsr/nsrlink.jsp?2016Lu13,B}{2016Lu13}). Some of the above}\\
\parbox[b][0.3cm]{17.7cm}{{\ }{\ }mentioned transitions were also measured by (\href{https://www.nndc.bnl.gov/nsr/nsrlink.jsp?2017Su05,B}{2017Su05}). However, (\href{https://www.nndc.bnl.gov/nsr/nsrlink.jsp?2016Lu13,B}{2016Lu13}, \href{https://www.nndc.bnl.gov/nsr/nsrlink.jsp?2017Su05,B}{2017Su05}) did not report a decay branch to the}\\
\parbox[b][0.3cm]{17.7cm}{{\ }{\ }\ensuremath{^{\textnormal{19}}}Ne*(1508) level from a \ensuremath{^{\textnormal{20}}}Na* level with E\ensuremath{_{\textnormal{x}}}=4.7-4.8 MeV. Therefore, we have omitted the E\ensuremath{_{\textnormal{p}}}\ensuremath{\approx}1046 keV transition to the}\\
\parbox[b][0.3cm]{17.7cm}{{\ }{\ }\ensuremath{^{\textnormal{19}}}Ne*(1507.56) level, which was adopted by (\href{https://www.nndc.bnl.gov/nsr/nsrlink.jsp?1995Ti07,B}{1995Ti07}) from (\href{https://www.nndc.bnl.gov/nsr/nsrlink.jsp?1995Pi03,B}{1995Pi03}).}\\
\parbox[b][0.3cm]{17.7cm}{\makebox[1ex]{\ensuremath{^{\hypertarget{MG1DELAY2}{c}}}} (\href{https://www.nndc.bnl.gov/nsr/nsrlink.jsp?2017Su05,B}{2017Su05}) attributed the proton group to the decay branches from \ensuremath{^{\textnormal{20}}}Na*(3863)\ensuremath{\rightarrow}\ensuremath{^{\textnormal{19}}}Ne\ensuremath{_{\textnormal{g.s.}}} and \ensuremath{^{\textnormal{20}}}Na*(4130)\ensuremath{\rightarrow}\ensuremath{^{\textnormal{19}}}Ne*(275),}\\
\parbox[b][0.3cm]{17.7cm}{{\ }{\ }which were distinguished with \ensuremath{\gamma}-ray coincidence events and would otherwise be indistinguishable. The \ensuremath{^{\textnormal{20}}}Na*(3683, 4130)}\\
\parbox[b][0.3cm]{17.7cm}{{\ }{\ }level-energies are quoted from that study.}\\
\parbox[b][0.3cm]{17.7cm}{\makebox[1ex]{\ensuremath{^{\hypertarget{MG1DELAY3}{d}}}} Transition was used as a calibration point for protons in (\href{https://www.nndc.bnl.gov/nsr/nsrlink.jsp?2012Wa15,B}{2012Wa15}).}\\
\parbox[b][0.3cm]{17.7cm}{\makebox[1ex]{\ensuremath{^{\hypertarget{MG1DELAY4}{e}}}} (\href{https://www.nndc.bnl.gov/nsr/nsrlink.jsp?2017Su05,B}{2017Su05}) attributed the proton group to the decay branches from \ensuremath{^{\textnormal{20}}}Na*(4130)\ensuremath{\rightarrow}\ensuremath{^{\textnormal{19}}}Ne\ensuremath{_{\textnormal{g.s.}}}, \ensuremath{^{\textnormal{20}}}Na*(4362)\ensuremath{\rightarrow}\ensuremath{^{\textnormal{19}}}Ne*(275)(?), and}\\
\parbox[b][0.3cm]{17.7cm}{{\ }{\ }\ensuremath{^{\textnormal{20}}}Na*(5595)\ensuremath{\rightarrow}\ensuremath{^{\textnormal{19}}}Ne*(1508), which were distinguished using \ensuremath{\gamma}-ray coincidence events and would otherwise be indistinguishable.}\\
\parbox[b][0.3cm]{17.7cm}{{\ }{\ }The \ensuremath{^{\textnormal{20}}}Na*(4130, 4362, 5595, 1508) level-energies are quoted from that study. They observed one event in the proton-coincident}\\
\begin{textblock}{29}(0,27.3)
Continued on next page (footnotes at end of table)
\end{textblock}
\clearpage
\vspace*{-0.5cm}
{\bf \small \underline{\ensuremath{^{\textnormal{20}}}Mg \ensuremath{\beta}\ensuremath{^{\textnormal{+}}}p decay\hspace{0.2in}\href{https://www.nndc.bnl.gov/nsr/nsrlink.jsp?1995Pi03,B}{1995Pi03},\href{https://www.nndc.bnl.gov/nsr/nsrlink.jsp?2016Li45,B}{2016Li45},\href{https://www.nndc.bnl.gov/nsr/nsrlink.jsp?2019Gl02,B}{2019Gl02} (continued)}}\\
\vspace{0.3cm}
\underline{Delayed Protons ($^{19}$Ne) (continued)}\\
\vspace{0.3cm}
\parbox[b][0.3cm]{17.7cm}{{\ }{\ }\ensuremath{\gamma}-ray spectrum for the \ensuremath{^{\textnormal{20}}}Na*(4362)\ensuremath{\rightarrow}\ensuremath{^{\textnormal{19}}}Ne*(275); \ensuremath{^{\textnormal{20}}}Na*(5142)\ensuremath{\rightarrow}\ensuremath{^{\textnormal{19}}}Ne*(1536); and \ensuremath{^{\textnormal{20}}}Na*(5982)\ensuremath{\rightarrow}\ensuremath{^{\textnormal{19}}}Ne*(1536) branches (see}\\
\parbox[b][0.3cm]{17.7cm}{{\ }{\ }p\ensuremath{_{\textnormal{5}}}, p\ensuremath{_{\textnormal{3}}}, and p\ensuremath{_{\textnormal{x}}} groups in Table III, respectively, where the E\ensuremath{_{\textnormal{x}}}(\ensuremath{^{\textnormal{20}}}Na) are from that study). The above mentioned \ensuremath{^{\textnormal{20}}}Na states are}\\
\parbox[b][0.3cm]{17.7cm}{{\ }{\ }not reported in Table IV of (\href{https://www.nndc.bnl.gov/nsr/nsrlink.jsp?2017Su05,B}{2017Su05}). So we assumed that these questionable branches were not clearly identified, and thus we}\\
\parbox[b][0.3cm]{17.7cm}{{\ }{\ }ignored them.}\\
\parbox[b][0.3cm]{17.7cm}{\makebox[1ex]{\ensuremath{^{\hypertarget{MG1DELAY5}{f}}}} (\href{https://www.nndc.bnl.gov/nsr/nsrlink.jsp?2017Su05,B}{2017Su05}) attributed the proton group to the decay branches from \ensuremath{^{\textnormal{20}}}Na*(6523)\ensuremath{\rightarrow}\ensuremath{^{\textnormal{19}}}Ne*(238) and \ensuremath{^{\textnormal{20}}}Na*(6523)\ensuremath{\rightarrow}\ensuremath{^{\textnormal{19}}}Ne*(275)}\\
\parbox[b][0.3cm]{17.7cm}{{\ }{\ }(with \ensuremath{^{\textnormal{20}}}Na* excitation energies quoted from \href{https://www.nndc.bnl.gov/nsr/nsrlink.jsp?2017Su05,B}{2017Su05}), which were distinguished with \ensuremath{\gamma}-ray coincidence events and would}\\
\parbox[b][0.3cm]{17.7cm}{{\ }{\ }otherwise be indistinguishable.}\\
\parbox[b][0.3cm]{17.7cm}{\makebox[1ex]{\ensuremath{^{\hypertarget{MG1DELAY6}{g}}}} For E\ensuremath{_{\textnormal{p}}}(lab)=4329 and 4741 keV transitions, no conclusive results were obtained by (\href{https://www.nndc.bnl.gov/nsr/nsrlink.jsp?2016Lu13,B}{2016Lu13}) from \ensuremath{\gamma}-coincidence events}\\
\parbox[b][0.3cm]{17.7cm}{{\ }{\ }regarding which \ensuremath{^{\textnormal{19}}}Ne level is fed by the decay of the \ensuremath{^{\textnormal{20}}}Na*(6749, 7183) states, respectively, and how the decay strength is}\\
\parbox[b][0.3cm]{17.7cm}{{\ }{\ }distributed among the various possible branches. Therefore, (\href{https://www.nndc.bnl.gov/nsr/nsrlink.jsp?2016Lu13,B}{2016Lu13}) chose the simple interpretation that those transitions only}\\
\parbox[b][0.3cm]{17.7cm}{{\ }{\ }contain feeding to the \ensuremath{^{\textnormal{19}}}Ne\ensuremath{_{\textnormal{g.s.}}}.}\\
\parbox[b][0.3cm]{17.7cm}{\makebox[1ex]{\ensuremath{^{\hypertarget{MG1DELAY7}{h}}}} Transition is observed for the first time in (\href{https://www.nndc.bnl.gov/nsr/nsrlink.jsp?2016Lu13,B}{2016Lu13}).}\\
\parbox[b][0.3cm]{17.7cm}{\makebox[1ex]{\ensuremath{^{\hypertarget{MG1DELAY8}{i}}}} (\href{https://www.nndc.bnl.gov/nsr/nsrlink.jsp?2017Su05,B}{2017Su05}) attributed the proton group to the branches from \ensuremath{^{\textnormal{20}}}Na*(4721)\ensuremath{\rightarrow}\ensuremath{^{\textnormal{19}}}Ne*(275) and \ensuremath{^{\textnormal{20}}}Na*(5982)\ensuremath{\rightarrow}\ensuremath{^{\textnormal{19}}}Ne*(1536)(?). The}\\
\parbox[b][0.3cm]{17.7cm}{{\ }{\ }latter transition is ignored due to low statistics (see previous comments). The \ensuremath{^{\textnormal{20}}}Na*(4721, 5982) level-energies are from}\\
\parbox[b][0.3cm]{17.7cm}{{\ }{\ }(\href{https://www.nndc.bnl.gov/nsr/nsrlink.jsp?2017Su05,B}{2017Su05}).}\\
\parbox[b][0.3cm]{17.7cm}{\makebox[1ex]{\ensuremath{^{\hypertarget{MG1DELAY9}{j}}}} Transition is observed for the first time in (\href{https://www.nndc.bnl.gov/nsr/nsrlink.jsp?2017Su05,B}{2017Su05}).}\\
\parbox[b][0.3cm]{17.7cm}{\makebox[1ex]{\ensuremath{^{\hypertarget{MG1DELAY10}{k}}}} (\href{https://www.nndc.bnl.gov/nsr/nsrlink.jsp?1992Go10,B}{1992Go10}) estimated the branching ratio of the unaccounted, weak transitions, which remained unobserved in their study, to be}\\
\parbox[b][0.3cm]{17.7cm}{{\ }{\ }7\%.}\\
\parbox[b][0.3cm]{17.7cm}{\makebox[1ex]{\ensuremath{^{\hypertarget{MG1DELAY11}{l}}}} Each branching ratio from (\href{https://www.nndc.bnl.gov/nsr/nsrlink.jsp?1995Pi03,B}{1995Pi03}) has an uncertainty of 12\% reported by those authors (see the caption of Table 2).}\\
\parbox[b][0.3cm]{17.7cm}{\makebox[1ex]{\ensuremath{^{\hypertarget{MG1DELAY12}{m}}}} A recommended, additional 10\% relative systematic uncertainty (uncertainty/value=10\%) is added in quadrature to the statistical}\\
\parbox[b][0.3cm]{17.7cm}{{\ }{\ }uncertainties in branching ratios reported by (\href{https://www.nndc.bnl.gov/nsr/nsrlink.jsp?2016Lu13,B}{2016Lu13}).}\\
\parbox[b][0.3cm]{17.7cm}{\makebox[1ex]{\ensuremath{^{\hypertarget{MG1DELAY13}{n}}}} (\href{https://www.nndc.bnl.gov/nsr/nsrlink.jsp?1995Pi03,B}{1995Pi03}) reported two \ensuremath{\beta}-delayed proton decay branches from the \ensuremath{^{\textnormal{20}}}Na* state at E\ensuremath{_{\textnormal{x}}}=6920 keV \textit{100} to the \ensuremath{^{\textnormal{19}}}Ne*(238) and}\\
\parbox[b][0.3cm]{17.7cm}{{\ }{\ }\ensuremath{^{\textnormal{19}}}Ne*(1536) states with I\ensuremath{_{\ensuremath{\beta}\textnormal{p}}}=0.01\% and 0.02\%, respectively; and another \ensuremath{\beta}-delayed proton decay branch from the \ensuremath{^{\textnormal{20}}}Na*(7440}\\
\parbox[b][0.3cm]{17.7cm}{{\ }{\ }keV \textit{100}) state to the \ensuremath{^{\textnormal{19}}}Ne*(1536) state with I\ensuremath{_{\ensuremath{\beta}\textnormal{p}}}\ensuremath{\geq}0.01. The aforementioned branching ratios were deduced from \ensuremath{\gamma}-p coincidence}\\
\parbox[b][0.3cm]{17.7cm}{{\ }{\ }events (see Table 2 in \href{https://www.nndc.bnl.gov/nsr/nsrlink.jsp?1995Pi03,B}{1995Pi03}). Figure 8d of (\href{https://www.nndc.bnl.gov/nsr/nsrlink.jsp?1995Pi03,B}{1995Pi03}) does not show a clear evidence for the proton groups mentioned}\\
\parbox[b][0.3cm]{17.7cm}{{\ }{\ }above. (\href{https://www.nndc.bnl.gov/nsr/nsrlink.jsp?2016Lu13,B}{2016Lu13}) did not observe any evidence for those branches or those \ensuremath{^{\textnormal{20}}}Na* states.}\\
\parbox[b][0.3cm]{17.7cm}{\makebox[1ex]{\ensuremath{^{\hypertarget{MG1DELAY14}{o}}}} The branching ratio was measured for the first time by (\href{https://www.nndc.bnl.gov/nsr/nsrlink.jsp?2016Lu13,B}{2016Lu13}).}\\
\parbox[b][0.3cm]{17.7cm}{\makebox[1ex]{\ensuremath{^{\hypertarget{MG1DELAY15}{p}}}} The branching ratio reported by (\href{https://www.nndc.bnl.gov/nsr/nsrlink.jsp?2016Lu13,B}{2016Lu13}) is an unweighted average between the branching ratios deduced from intensities}\\
\parbox[b][0.3cm]{17.7cm}{{\ }{\ }measured in different detectors. An unweighted average was used due to inconsistent intensities measured by different detectors.}\\
\parbox[b][0.3cm]{17.7cm}{{\ }{\ }The uncertainty is estimated by (\href{https://www.nndc.bnl.gov/nsr/nsrlink.jsp?2016Lu13,B}{2016Lu13}).}\\
\parbox[b][0.3cm]{17.7cm}{\makebox[1ex]{\ensuremath{^{\hypertarget{MG1DELAY16}{q}}}} The excitation energies of the intermediate \ensuremath{^{\textnormal{20}}}Na* states are not taken from the \ensuremath{^{\textnormal{20}}}Na Adopted Levels in ENSDF database.}\\
\parbox[b][0.3cm]{17.7cm}{{\ }{\ }Instead, they are reevaluated based on the articles of interest to this dataset only. See the comments on each individual transition}\\
\parbox[b][0.3cm]{17.7cm}{{\ }{\ }for the details.}\\
\parbox[b][0.3cm]{17.7cm}{\makebox[1ex]{\ensuremath{^{\hypertarget{MG1DELAY17}{r}}}} The \ensuremath{^{\textnormal{20}}}Na*(4760, 5603) levels were reported to be a group of unresolved states in (\href{https://www.nndc.bnl.gov/nsr/nsrlink.jsp?1995Pi03,B}{1995Pi03}).}\\
\parbox[b][0.3cm]{17.7cm}{\makebox[1ex]{\ensuremath{^{\hypertarget{MG1DELAY18}{s}}}} The \ensuremath{^{\textnormal{20}}}Na* intermediate level was observed for the first time in (\href{https://www.nndc.bnl.gov/nsr/nsrlink.jsp?2016Lu13,B}{2016Lu13}).}\\
\parbox[b][0.3cm]{17.7cm}{\makebox[1ex]{\ensuremath{^{\hypertarget{MG1DELAY19}{t}}}} (\href{https://www.nndc.bnl.gov/nsr/nsrlink.jsp?2017Su05,B}{2017Su05}) listed several proton groups (and intensities) that are attributed to multiple initial \ensuremath{^{\textnormal{20}}}Na state\ensuremath{\rightarrow}final \ensuremath{^{\textnormal{19}}}Ne state}\\
\parbox[b][0.3cm]{17.7cm}{{\ }{\ }combinations. Their Table III clarifies these unresolved placements. The text does not describe how the proton group intensities}\\
\parbox[b][0.3cm]{17.7cm}{{\ }{\ }are unfolded to obtain the I(\ensuremath{\varepsilon}+\ensuremath{\beta}\ensuremath{^{\textnormal{+}}}) feedings.}\\
\parbox[b][0.3cm]{17.7cm}{\makebox[1ex]{\ensuremath{^{\hypertarget{MG1DELAY20}{u}}}} Absolute intensity per 100 decays.}\\
\parbox[b][0.3cm]{17.7cm}{\makebox[1ex]{\ensuremath{^{\hypertarget{MG1DELAY21}{v}}}} Placement of transition in the level scheme is uncertain.}\\
\vspace{0.5cm}
\clearpage
\begin{figure}[h]
\begin{center}
\includegraphics{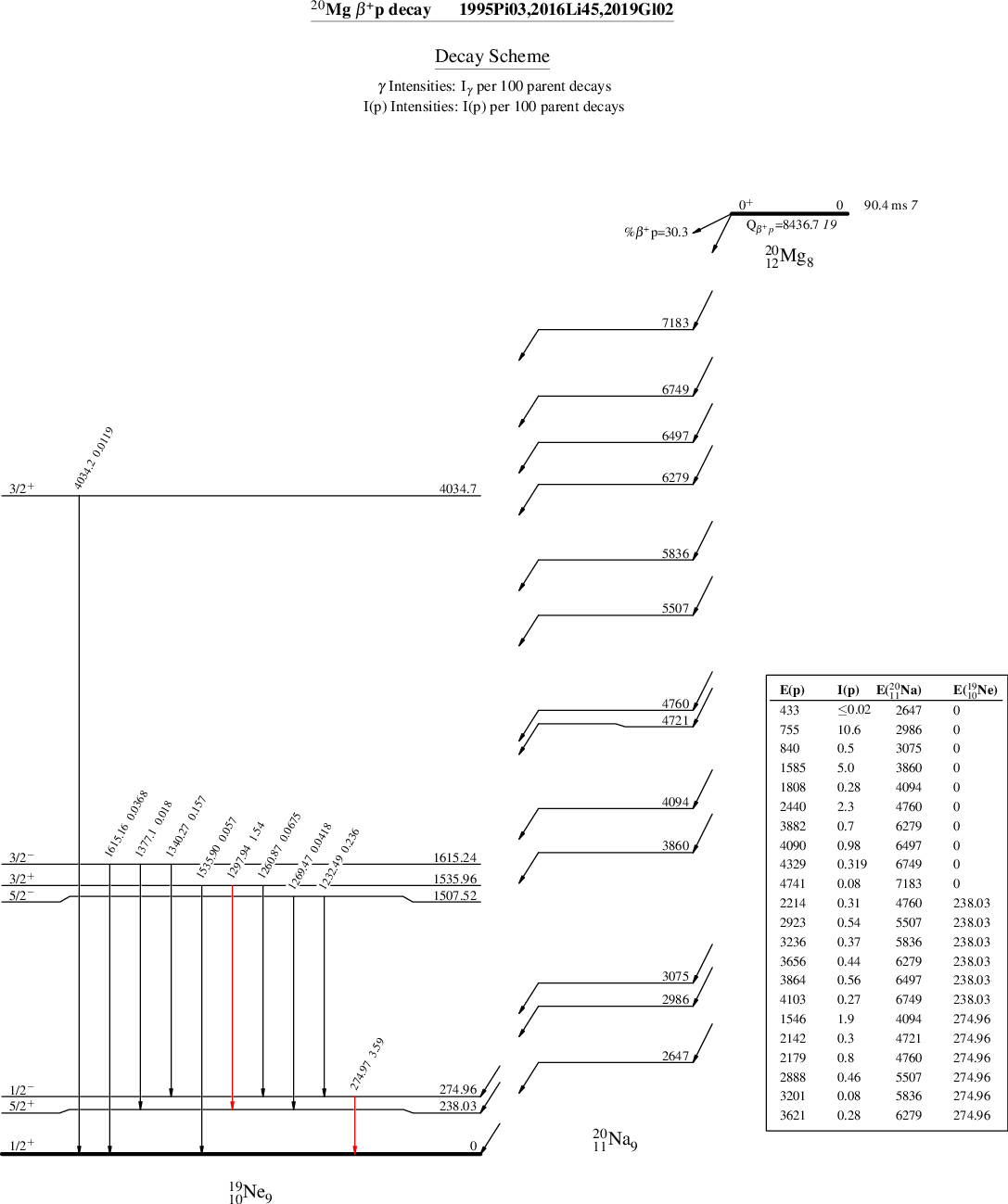}\\
\end{center}
\end{figure}
\clearpage
\begin{figure}[h]
\begin{center}
\includegraphics{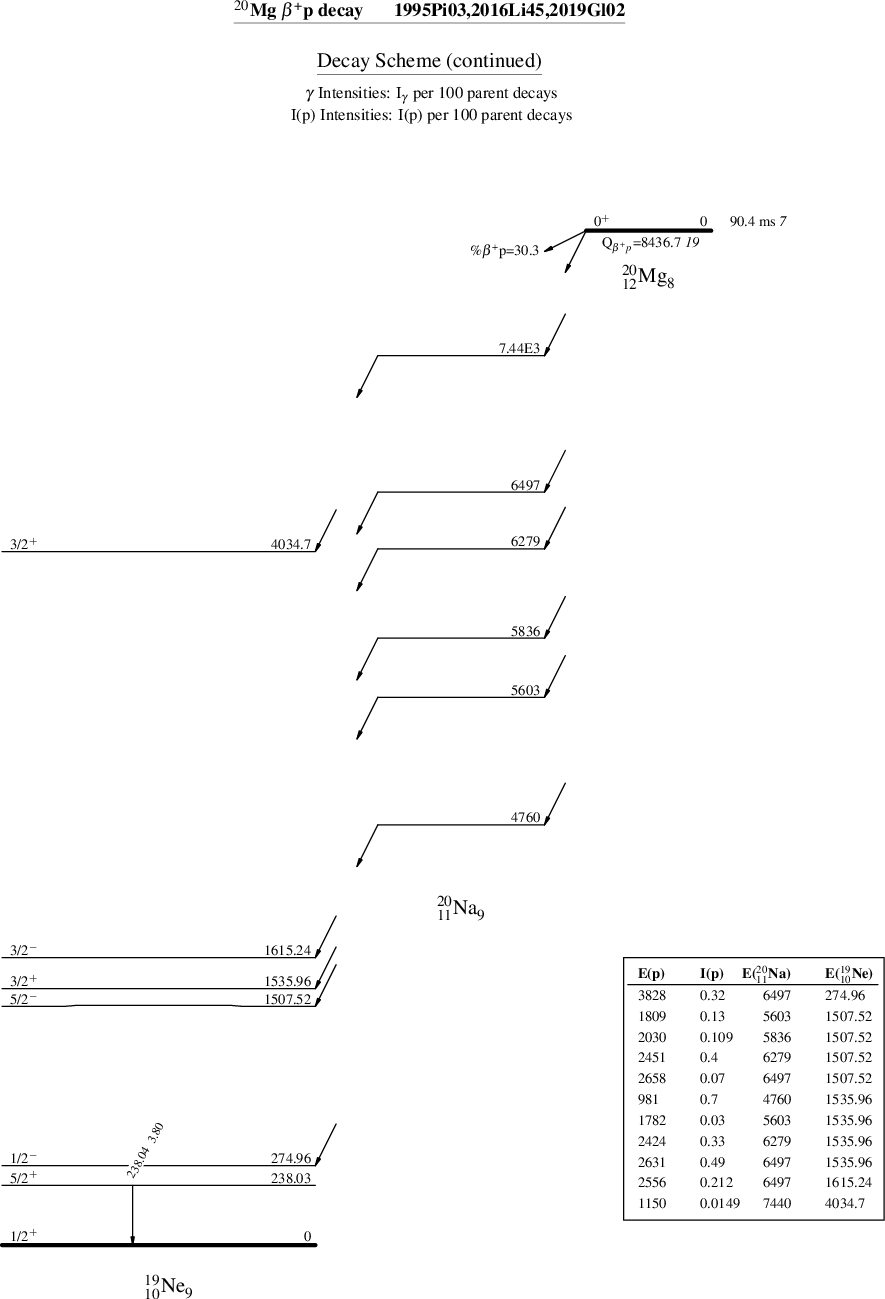}\\
\end{center}
\end{figure}
\clearpage
\subsection[\hspace{-0.2cm}\ensuremath{^{\textnormal{1}}}H(\ensuremath{^{\textnormal{18}}}F,p),(\ensuremath{^{\textnormal{18}}}F,\ensuremath{\alpha}):res]{ }
\vspace{-27pt}
\vspace{0.3cm}
\hypertarget{NE2}{{\bf \small \underline{\ensuremath{^{\textnormal{1}}}H(\ensuremath{^{\textnormal{18}}}F,p),(\ensuremath{^{\textnormal{18}}}F,\ensuremath{\alpha}):res\hspace{0.2in}\href{https://www.nndc.bnl.gov/nsr/nsrlink.jsp?2009Mu07,B}{2009Mu07},\href{https://www.nndc.bnl.gov/nsr/nsrlink.jsp?2012Mo03,B}{2012Mo03}}}}\\
\vspace{4pt}
\vspace{8pt}
\parbox[b][0.3cm]{17.7cm}{\addtolength{\parindent}{-0.2in}\ensuremath{^{\textnormal{18}}}F(p,p) and \ensuremath{^{\textnormal{18}}}F(p,\ensuremath{\alpha}) resonant reactions in inverse kinematics.}\\
\parbox[b][0.3cm]{17.7cm}{\addtolength{\parindent}{-0.2in}J\ensuremath{^{\ensuremath{\pi}}}(\ensuremath{^{\textnormal{18}}}F\ensuremath{_{\textnormal{g.s.}}})=1\ensuremath{^{\textnormal{+}}} and J\ensuremath{^{\ensuremath{\pi}}}(p)=1/2\ensuremath{^{\textnormal{+}}}.}\\
\parbox[b][0.3cm]{17.7cm}{\addtolength{\parindent}{-0.2in}\href{https://www.nndc.bnl.gov/nsr/nsrlink.jsp?1995Co23,B}{1995Co23}: \ensuremath{^{\textnormal{1}}}H(\ensuremath{^{\textnormal{18}}}F,\ensuremath{\alpha}) and \ensuremath{^{\textnormal{1}}}H(\ensuremath{^{\textnormal{18}}}F, p) E=14 MeV; measured the \ensuremath{^{\textnormal{1}}}H(\ensuremath{^{\textnormal{18}}}F,\ensuremath{\alpha}) excitation function in the E\ensuremath{_{\textnormal{c.m.}}}=550-740 MeV range;}\\
\parbox[b][0.3cm]{17.7cm}{measured energy and time-of-flight of the elastically scattered protons and the \ensuremath{\alpha}-particles from the reaction using the LEDA array}\\
\parbox[b][0.3cm]{17.7cm}{(position sensitive annular Si detectors) covering \ensuremath{\theta}\ensuremath{_{\textnormal{lab}}}=12\ensuremath{^\circ}{\textminus}28\ensuremath{^\circ}. Each segment of the array (8 in total) covered \ensuremath{\phi}\ensuremath{_{\textnormal{lab}}}=41\ensuremath{^\circ}. The}\\
\parbox[b][0.3cm]{17.7cm}{excitation functions showed evidence of a resonance at E\ensuremath{_{\textnormal{c.m.}}}=638 keV \textit{15}, whose width, resonance strength, and J\ensuremath{^{\ensuremath{\pi}}} assignment}\\
\parbox[b][0.3cm]{17.7cm}{were deduced. This resonance energy is in agreement with the 657-keV resonance with \ensuremath{\Gamma}\ensuremath{\approx}40 keV predicted by M. Wiescher and}\\
\parbox[b][0.3cm]{17.7cm}{K.-U., Kettner, Astrophys. J. 263 (1982) 891.}\\
\parbox[b][0.3cm]{17.7cm}{\addtolength{\parindent}{-0.2in}\href{https://www.nndc.bnl.gov/nsr/nsrlink.jsp?1995Re11,B}{1995Re11}: \ensuremath{^{\textnormal{1}}}H(\ensuremath{^{\textnormal{18}}}F,\ensuremath{^{\textnormal{15}}}O) E=11.7-15.1 MeV; momentum analyzed the heavy reaction products (\ensuremath{^{\textnormal{15}}}O) via a N\ensuremath{_{\textnormal{2}}} gas-filled Enge}\\
\parbox[b][0.3cm]{17.7cm}{spectrograph utilized for Z identification placed at \ensuremath{\theta}\ensuremath{_{\textnormal{c.m.}}}=110\ensuremath{^\circ}. The position and time-of-flight of the \ensuremath{^{\textnormal{15}}}O ions were measured}\\
\parbox[b][0.3cm]{17.7cm}{using a parallel grid avalanche counter at the focal plane. \ensuremath{\sigma}(E) of the \ensuremath{^{\textnormal{18}}}F(p,\ensuremath{\alpha}) reaction was measured at 3 energies (11.7, 13.4 and}\\
\parbox[b][0.3cm]{17.7cm}{15.1 MeV, see \href{https://www.nndc.bnl.gov/nsr/nsrlink.jsp?1996Re05,B}{1996Re05}). The excitation function showed evidence for an \textit{s}-wave \ensuremath{^{\textnormal{19}}}Ne resonance at 660 keV, whose proton and}\\
\parbox[b][0.3cm]{17.7cm}{\ensuremath{\alpha} partial widths were deduced.}\\
\parbox[b][0.3cm]{17.7cm}{\addtolength{\parindent}{-0.2in}\href{https://www.nndc.bnl.gov/nsr/nsrlink.jsp?1996Re05,B}{1996Re05}, \href{https://www.nndc.bnl.gov/nsr/nsrlink.jsp?1996Re28,B}{1996Re28}, \href{https://www.nndc.bnl.gov/nsr/nsrlink.jsp?1997Re05,B}{1997Re05}, \href{https://www.nndc.bnl.gov/nsr/nsrlink.jsp?1997Re17,B}{1997Re17}: \ensuremath{^{\textnormal{1}}}H(\ensuremath{^{\textnormal{18}}}F,\ensuremath{^{\textnormal{15}}}O) E=12.0, 12.4, 12.9, and 13.4 MeV; measured the excitation function for}\\
\parbox[b][0.3cm]{17.7cm}{\ensuremath{^{\textnormal{18}}}F(p,\ensuremath{\alpha}) using the same technique as that of (\href{https://www.nndc.bnl.gov/nsr/nsrlink.jsp?1995Re11,B}{1995Re11}) except that the parallel grid avalanche counter at the focal plane was}\\
\parbox[b][0.3cm]{17.7cm}{replaced with an array of 24 position sensitive Si detectors, which measured the position, time of arrival, and the residual energy of}\\
\parbox[b][0.3cm]{17.7cm}{the \ensuremath{^{\textnormal{15}}}O ions. Furthermore, the \ensuremath{\alpha}-particles were measured in coincidence with the \ensuremath{^{\textnormal{15}}}O ions at \ensuremath{\theta}\ensuremath{_{\textnormal{lab}}}=60\ensuremath{^\circ} using a Si surface barrier}\\
\parbox[b][0.3cm]{17.7cm}{detector. The spectrograph was placed at \ensuremath{\theta}\ensuremath{_{\textnormal{lab}}}=13\ensuremath{^\circ}. Considering the data of (\href{https://www.nndc.bnl.gov/nsr/nsrlink.jsp?1995Re11,B}{1995Re11}), the authors measured the excitation}\\
\parbox[b][0.3cm]{17.7cm}{function at 7 energies corresponding to E\ensuremath{_{\textnormal{c.m.}}}=616-795 keV and deduced a resonance at E\ensuremath{_{\textnormal{c.m.}}}=652 keV \textit{4}, \ensuremath{\omega}\ensuremath{\gamma}\ensuremath{_{\textnormal{(p,}\ensuremath{\alpha}\textnormal{)}}}=2.1 keV \textit{7},}\\
\parbox[b][0.3cm]{17.7cm}{\ensuremath{\Gamma}\ensuremath{_{\textnormal{p}}}=5.0 keV \textit{16}, \ensuremath{\Gamma}\ensuremath{_{\ensuremath{\alpha}}}=8.6 keV \textit{25}, and \ensuremath{\Gamma}=13.6 keV \textit{46} from a least-squares fit to the data for the \ensuremath{^{\textnormal{19}}}Ne*(7063) state. Deduced the}\\
\parbox[b][0.3cm]{17.7cm}{\ensuremath{^{\textnormal{18}}}F(p,\ensuremath{\alpha}) reaction rate and discussed the astrophysical implications.}\\
\parbox[b][0.3cm]{17.7cm}{\addtolength{\parindent}{-0.2in}\href{https://www.nndc.bnl.gov/nsr/nsrlink.jsp?1997Gr23,B}{1997Gr23}: \ensuremath{^{\textnormal{1}}}H(\ensuremath{^{\textnormal{18}}}F,\ensuremath{\alpha}), \ensuremath{^{\textnormal{1}}}H(\ensuremath{^{\textnormal{18}}}F,p) and \ensuremath{^{\textnormal{1}}}H(\ensuremath{^{\textnormal{18}}}F,\ensuremath{^{\textnormal{18}}}F) E=5 and 14 MeV; measured the excitation function of the \ensuremath{^{\textnormal{18}}}F(p,\ensuremath{\alpha}) and}\\
\parbox[b][0.3cm]{17.7cm}{\ensuremath{^{\textnormal{1}}}H(\ensuremath{^{\textnormal{18}}}F,p) reactions corresponding to E\ensuremath{_{\textnormal{c.m.}}}=265-535 and E\ensuremath{_{\textnormal{c.m.}}}=550-740 MeV. Measured energy, angle and time-of-flight of}\\
\parbox[b][0.3cm]{17.7cm}{protons, \ensuremath{\alpha}-particles, and \ensuremath{^{\textnormal{18}}}F ions (only for the low energy measurement) using the LEDA array covering \ensuremath{\theta}\ensuremath{_{\textnormal{lab}}}=12\ensuremath{^\circ}{\textminus}26\ensuremath{^\circ}. At the}\\
\parbox[b][0.3cm]{17.7cm}{low beam energy, elastically scattered protons were contaminated with the positrons from the \ensuremath{\beta}\ensuremath{^{\textnormal{+}}} decay of \ensuremath{^{\textnormal{18}}}F, and were thus not}\\
\parbox[b][0.3cm]{17.7cm}{used for analysis. The authors measured a resonance at E\ensuremath{_{\textnormal{c.m.}}}=324 keV \textit{7} and another at E\ensuremath{_{\textnormal{c.m.}}}=653 keV and deduced their}\\
\parbox[b][0.3cm]{17.7cm}{resonance strengths and total widths from the thick target yield curve method. The \ensuremath{^{\textnormal{18}}}F(p,\ensuremath{\alpha}) reaction rate was computed for}\\
\parbox[b][0.3cm]{17.7cm}{T=0.1-1 GK and the astrophysical implications were discussed.}\\
\parbox[b][0.3cm]{17.7cm}{\addtolength{\parindent}{-0.2in}\href{https://www.nndc.bnl.gov/nsr/nsrlink.jsp?1997Le13,B}{1997Le13}: \ensuremath{^{\textnormal{1}}}H(\ensuremath{^{\textnormal{18}}}F,\ensuremath{\alpha}), \ensuremath{^{\textnormal{1}}}H(\ensuremath{^{\textnormal{18}}}F,p) and \ensuremath{^{\textnormal{1}}}H(\ensuremath{^{\textnormal{18}}}F,\ensuremath{^{\textnormal{18}}}F) E=10 and 14 MeV; measured the \ensuremath{^{\textnormal{18}}}F(p,\ensuremath{\alpha}) excitation function at}\\
\parbox[b][0.3cm]{17.7cm}{E\ensuremath{_{\textnormal{c.m.}}}=250-500 and 550-740 MeV. Measured the position, energy and time-of-flight of the \ensuremath{\alpha}-particles, protons, and \ensuremath{^{\textnormal{18}}}F scattered}\\
\parbox[b][0.3cm]{17.7cm}{particles using the LEDA array covering \ensuremath{\theta}\ensuremath{_{\textnormal{lab}}}=12\ensuremath{^\circ}{\textminus}28\ensuremath{^\circ}. Results were not published but the authors defended the results reported in}\\
\parbox[b][0.3cm]{17.7cm}{(\href{https://www.nndc.bnl.gov/nsr/nsrlink.jsp?1995Co23,B}{1995Co23}).}\\
\parbox[b][0.3cm]{17.7cm}{\addtolength{\parindent}{-0.2in}\href{https://www.nndc.bnl.gov/nsr/nsrlink.jsp?2000Ba87,B}{2000Ba87}, \href{https://www.nndc.bnl.gov/nsr/nsrlink.jsp?2004Ba17,B}{2004Ba17}, \href{https://www.nndc.bnl.gov/nsr/nsrlink.jsp?2004Bb17,B}{2004Bb17}: \ensuremath{^{\textnormal{1}}}H(\ensuremath{^{\textnormal{18}}}F,p), \ensuremath{^{\textnormal{1}}}H(\ensuremath{^{\textnormal{18}}}F,\ensuremath{^{\textnormal{18}}}F) and \ensuremath{^{\textnormal{1}}}H(\ensuremath{^{\textnormal{18}}}F,\ensuremath{\alpha}) E=10-14 MeV; measured p-\ensuremath{^{\textnormal{18}}}F (recoils) coincidences as}\\
\parbox[b][0.3cm]{17.7cm}{well as \ensuremath{\alpha}-\ensuremath{^{\textnormal{15}}}O coincidences using the SIDAR array to measure protons and \ensuremath{\alpha}s, and an ionization chamber downstream of SIDAR}\\
\parbox[b][0.3cm]{17.7cm}{to measure the \ensuremath{^{\textnormal{18}}}F and \ensuremath{^{\textnormal{15}}}O recoils. The SIDAR array was tilted at \ensuremath{\theta}\ensuremath{_{\textnormal{lab}}}=41\ensuremath{^\circ} (see also 43\ensuremath{^\circ} \href{https://www.nndc.bnl.gov/nsr/nsrlink.jsp?2001Ba49,B}{2001Ba49}) and covered \ensuremath{\theta}\ensuremath{_{\textnormal{lab}}}=15\ensuremath{^\circ}{\textminus}43\ensuremath{^\circ}.}\\
\parbox[b][0.3cm]{17.7cm}{(\href{https://www.nndc.bnl.gov/nsr/nsrlink.jsp?2000Ba87,B}{2000Ba87}) analyzed only the data from the \ensuremath{^{\textnormal{18}}}F(p,p) channel. A resonance was measured at E\ensuremath{_{\textnormal{c.m.}}}=665.3 keV \textit{17} in the \ensuremath{^{\textnormal{18}}}F(p,p)}\\
\parbox[b][0.3cm]{17.7cm}{excitation function (\href{https://www.nndc.bnl.gov/nsr/nsrlink.jsp?2000Ba87,B}{2000Ba87}), which was fitted using Breit-Wigner formalism, as well as R-matrix via the MULTI computer code.}\\
\parbox[b][0.3cm]{17.7cm}{Deduced \ensuremath{\Gamma}=38.5 keV \textit{34}, \ensuremath{\Gamma}\ensuremath{_{\textnormal{p}}}=15.8 keV \textit{16}, and \ensuremath{\omega}\ensuremath{\gamma}\ensuremath{_{\textnormal{(p,}\ensuremath{\alpha}\textnormal{)}}}=6.2 keV \textit{6} for this resonance.}\\
\parbox[b][0.3cm]{17.7cm}{\addtolength{\parindent}{-0.2in}\href{https://www.nndc.bnl.gov/nsr/nsrlink.jsp?2001Ba49,B}{2001Ba49}, \href{https://www.nndc.bnl.gov/nsr/nsrlink.jsp?2001Ba59,B}{2001Ba59}, \href{https://www.nndc.bnl.gov/nsr/nsrlink.jsp?2011Be59,B}{2011Be59}: \ensuremath{^{\textnormal{1}}}H(\ensuremath{^{\textnormal{18}}}F,p), \ensuremath{^{\textnormal{1}}}H(\ensuremath{^{\textnormal{18}}}F,\ensuremath{^{\textnormal{18}}}F) and \ensuremath{^{\textnormal{1}}}H(\ensuremath{^{\textnormal{18}}}F,\ensuremath{\alpha}) E=10-14 MeV; the experiment was performed and}\\
\parbox[b][0.3cm]{17.7cm}{described by (\href{https://www.nndc.bnl.gov/nsr/nsrlink.jsp?2000Ba87,B}{2000Ba87}). For the \ensuremath{^{\textnormal{18}}}F(p,\ensuremath{\alpha}) measurement, the \ensuremath{\alpha}-\ensuremath{^{\textnormal{15}}}O coincidence events were measured at \ensuremath{\theta}\ensuremath{_{\textnormal{c.m.}}}\ensuremath{\approx}95\ensuremath{^\circ}{\textminus}125\ensuremath{^\circ}.}\\
\parbox[b][0.3cm]{17.7cm}{(\href{https://www.nndc.bnl.gov/nsr/nsrlink.jsp?2001Ba49,B}{2001Ba49}) analyzed the data from the \ensuremath{^{\textnormal{18}}}F(p,\ensuremath{\alpha}) channel of the experiment by (\href{https://www.nndc.bnl.gov/nsr/nsrlink.jsp?2000Ba87,B}{2000Ba87}) and fitted the \ensuremath{^{\textnormal{18}}}F(p,p) and \ensuremath{^{\textnormal{18}}}F(p,\ensuremath{\alpha})}\\
\parbox[b][0.3cm]{17.7cm}{excitation functions simultaneously using the Breit-Wigner formalism and R-matrix for the \ensuremath{^{\textnormal{18}}}F(p,p), and the Breit-Wigner}\\
\parbox[b][0.3cm]{17.7cm}{formalism for an isolated isotropic resonance for the \ensuremath{^{\textnormal{18}}}F(p,\ensuremath{\alpha}) channel. A resonance was measured at E\ensuremath{_{\textnormal{c.m.}}}=664.7 keV \textit{16}, for}\\
\parbox[b][0.3cm]{17.7cm}{which \ensuremath{\Gamma}=39.0 keV \textit{16}, \ensuremath{\Gamma}\ensuremath{_{\textnormal{p}}}/\ensuremath{\Gamma}=0.39 \textit{2}, and \ensuremath{\omega}\ensuremath{\gamma}\ensuremath{_{\textnormal{(p,}\ensuremath{\alpha}\textnormal{)}}}=6.2 keV \textit{3} were deduced. Determined the \ensuremath{^{\textnormal{18}}}F(p,\ensuremath{\alpha}) reaction rate at T\ensuremath{\leq}2 GK}\\
\parbox[b][0.3cm]{17.7cm}{and discussed the astrophysical implications.}\\
\parbox[b][0.3cm]{17.7cm}{\addtolength{\parindent}{-0.2in}\href{https://www.nndc.bnl.gov/nsr/nsrlink.jsp?2001Gr01,B}{2001Gr01}, \href{https://www.nndc.bnl.gov/nsr/nsrlink.jsp?2001Gr12,B}{2001Gr12}: \ensuremath{^{\textnormal{1}}}H(\ensuremath{^{\textnormal{18}}}F,p), \ensuremath{^{\textnormal{1}}}H(\ensuremath{^{\textnormal{18}}}F,\ensuremath{\alpha}) and \ensuremath{^{\textnormal{1}}}H(\ensuremath{^{\textnormal{18}}}F,\ensuremath{^{\textnormal{18}}}F) E=14 MeV; measured energy and time-of-flight of the protons, \ensuremath{\alpha}s,}\\
\parbox[b][0.3cm]{17.7cm}{and \ensuremath{^{\textnormal{18}}}F recoils using the position sensitive LEDA annular array covering \ensuremath{\theta}\ensuremath{_{\textnormal{lab}}}=12\ensuremath{^\circ}{\textminus}26\ensuremath{^\circ} with a 90\% coverage of the azimuthal}\\
\parbox[b][0.3cm]{17.7cm}{angle. This measurement has a much improved statistics over the ones in (\href{https://www.nndc.bnl.gov/nsr/nsrlink.jsp?1995Co23,B}{1995Co23}, \href{https://www.nndc.bnl.gov/nsr/nsrlink.jsp?1997Gr23,B}{1997Gr23}). Both excitation functions}\\
\parbox[b][0.3cm]{17.7cm}{displayed a resonance. These spectra were fitted using least-squares fits, which consisted of a Coulomb term, a Breit-Wigner}\\
\parbox[b][0.3cm]{17.7cm}{resonance, and an interference term. Consequently, E\ensuremath{_{\textnormal{c.m.}}}=657.5 keV \textit{7} (stat.) \textit{17} (sys.), \ensuremath{\Gamma}=34.2 keV \textit{14} (stat.) \textit{17} (sys.), and}\\
\parbox[b][0.3cm]{17.7cm}{\ensuremath{\omega}\ensuremath{\gamma}\ensuremath{_{\textnormal{(p,}\ensuremath{\alpha}\textnormal{)}}}=4.70 keV \textit{10} (stat.) \textit{15} (sys.) were deduced for this resonance.}\\
\parbox[b][0.3cm]{17.7cm}{\addtolength{\parindent}{-0.2in}\href{https://www.nndc.bnl.gov/nsr/nsrlink.jsp?2002Bb02,B}{2002Bb02}: \ensuremath{^{\textnormal{1}}}H(\ensuremath{^{\textnormal{18}}}F,\ensuremath{\alpha}) E\ensuremath{_{\textnormal{c.m.}}}=330 keV; measured the thick target yield curve to study a resonance at E\ensuremath{_{\textnormal{c.m.}}}=332 keV \textit{17}; measured}\\
\parbox[b][0.3cm]{17.7cm}{\ensuremath{\alpha}-\ensuremath{^{\textnormal{15}}}O coincidences using SIDAR array covering \ensuremath{\theta}\ensuremath{_{\textnormal{lab}}}=18\ensuremath{^\circ}{\textminus}48\ensuremath{^\circ}; measured the yield of the \ensuremath{^{\textnormal{1}}}H(\ensuremath{^{\textnormal{18}}}F,p) reaction on and off resonance}\\
\clearpage
\vspace{0.3cm}
{\bf \small \underline{\ensuremath{^{\textnormal{1}}}H(\ensuremath{^{\textnormal{18}}}F,p),(\ensuremath{^{\textnormal{18}}}F,\ensuremath{\alpha}):res\hspace{0.2in}\href{https://www.nndc.bnl.gov/nsr/nsrlink.jsp?2009Mu07,B}{2009Mu07},\href{https://www.nndc.bnl.gov/nsr/nsrlink.jsp?2012Mo03,B}{2012Mo03} (continued)}}\\
\vspace{0.3cm}
\parbox[b][0.3cm]{17.7cm}{using \ensuremath{^{\textnormal{18}}}F beams with E\ensuremath{_{\textnormal{lab}}}=6.6 and 7.5 MeV, respectively. Deduced cross sections on and off resonance, from which \ensuremath{\Gamma}\ensuremath{_{\textnormal{p}}}=2.22 eV}\\
\parbox[b][0.3cm]{17.7cm}{\textit{69} and \ensuremath{\omega}\ensuremath{\gamma}\ensuremath{_{\textnormal{(p,}\ensuremath{\alpha}\textnormal{)}}}=1.48 eV \textit{46} were determined. Deduced the \ensuremath{^{\textnormal{18}}}F(p,\ensuremath{\alpha}) reaction rate and discussed astrophysical implications.}\\
\parbox[b][0.3cm]{17.7cm}{\addtolength{\parindent}{-0.2in}\href{https://www.nndc.bnl.gov/nsr/nsrlink.jsp?2004Ba63,B}{2004Ba63}, \href{https://www.nndc.bnl.gov/nsr/nsrlink.jsp?2004Bb08,B}{2004Bb08}, \href{https://www.nndc.bnl.gov/nsr/nsrlink.jsp?2004Bb10,B}{2004Bb10}: \ensuremath{^{\textnormal{1}}}H(\ensuremath{^{\textnormal{18}}}F,p) E\ensuremath{_{\textnormal{lab}}}=24 MeV; measured thick target yield curve; measured positions and energies of}\\
\parbox[b][0.3cm]{17.7cm}{scattered protons using a position sensitive Si detector, which covered \ensuremath{\theta}\ensuremath{_{\textnormal{lab}}}=8\ensuremath{^\circ}{\textminus}16\ensuremath{^\circ}. Measured the \ensuremath{^{\textnormal{1}}}H(\ensuremath{^{\textnormal{18}}}F,p) excitation function at}\\
\parbox[b][0.3cm]{17.7cm}{E\ensuremath{_{\textnormal{c.m.}}}\ensuremath{\approx}0.3-1.3 MeV. Two resonances were measured: A previously known resonance at E\ensuremath{_{\textnormal{c.m.}}}=665 keV and a newly found}\\
\parbox[b][0.3cm]{17.7cm}{resonance at E\ensuremath{_{\textnormal{c.m.}}}=1009 keV \textit{14}, for which \ensuremath{\Gamma}\ensuremath{_{\textnormal{p}}}=27 keV \textit{4} and \ensuremath{\Gamma}\ensuremath{_{\ensuremath{\alpha}}}=71 keV \textit{11} were deduced using an R-matrix fit performed by}\\
\parbox[b][0.3cm]{17.7cm}{the MULTI computer code. Based on the measured excitation function and the \ensuremath{\Gamma}\ensuremath{_{\textnormal{p}}}/\ensuremath{\Gamma}\ensuremath{_{\ensuremath{\alpha}}}=0.19 from (\href{https://www.nndc.bnl.gov/nsr/nsrlink.jsp?1998Ut02,B}{1998Ut02}), these authors}\\
\parbox[b][0.3cm]{17.7cm}{estimated an upper limit of \ensuremath{\Gamma}\ensuremath{_{\textnormal{p}}}\ensuremath{<}2.5 keV (at 90\% C.L.) for the resonance at E\ensuremath{_{\textnormal{c.m.}}}=1.09 MeV. Deduced the \ensuremath{^{\textnormal{18}}}F(p,\ensuremath{\alpha}) and \ensuremath{^{\textnormal{18}}}F(p,\ensuremath{\gamma})}\\
\parbox[b][0.3cm]{17.7cm}{reaction rates and discussed astrophysical implications.}\\
\parbox[b][0.3cm]{17.7cm}{\addtolength{\parindent}{-0.2in}\href{https://www.nndc.bnl.gov/nsr/nsrlink.jsp?2005Ba82,B}{2005Ba82}, \href{https://www.nndc.bnl.gov/nsr/nsrlink.jsp?2005Bb05,B}{2005Bb05}: \ensuremath{^{\textnormal{1}}}H(\ensuremath{^{\textnormal{18}}}F,p) E\ensuremath{_{\textnormal{lab}}}=24 MeV corresponding to E\ensuremath{_{\textnormal{c.m.}}}\ensuremath{\sim}0.3-1.3 MeV; measured the scattered protons using a}\\
\parbox[b][0.3cm]{17.7cm}{position sensitive Si detector, which covered \ensuremath{\theta}\ensuremath{_{\textnormal{lab}}}=8\ensuremath{^\circ}{\textminus}16\ensuremath{^\circ}. Beam was stopped inside the thick target. Measured the \ensuremath{^{\textnormal{1}}}H(\ensuremath{^{\textnormal{18}}}F,p)}\\
\parbox[b][0.3cm]{17.7cm}{excitation function and measured a resonance at E\ensuremath{_{\textnormal{c.m.}}}=1.01 MeV. Deduced resonance energy, J\ensuremath{^{\ensuremath{\pi}}} assignment, \ensuremath{\Gamma}\ensuremath{_{\textnormal{p}}}, and \ensuremath{\Gamma}\ensuremath{_{\ensuremath{\alpha}}} for this}\\
\parbox[b][0.3cm]{17.7cm}{newly observed resonance using an R-matrix analysis via the MULTI computer code. Determined the \ensuremath{^{\textnormal{18}}}F(p,\ensuremath{\gamma}) and \ensuremath{^{\textnormal{18}}}F(p,\ensuremath{\alpha})}\\
\parbox[b][0.3cm]{17.7cm}{reaction rates.}\\
\parbox[b][0.3cm]{17.7cm}{\addtolength{\parindent}{-0.2in}\href{https://www.nndc.bnl.gov/nsr/nsrlink.jsp?2006Ch30,B}{2006Ch30}: \ensuremath{^{\textnormal{1}}}H(\ensuremath{^{\textnormal{18}}}F,\ensuremath{\alpha}) and \ensuremath{^{\textnormal{1}}}H(\ensuremath{^{\textnormal{18}}}F,\ensuremath{^{\textnormal{18}}}F) E\ensuremath{_{\textnormal{c.m.}}}\ensuremath{\sim}663-877 keV; measured excitation function off-resonance to determine the}\\
\parbox[b][0.3cm]{17.7cm}{interferences between 3 resonances at E\ensuremath{_{\textnormal{c.m.}}}=8, 38, and 665 keV with J\ensuremath{^{\ensuremath{\pi}}}=3/2\ensuremath{^{\textnormal{+}}}. Measured \ensuremath{\alpha}-\ensuremath{^{\textnormal{15}}}O coincidences using SIDAR array}\\
\parbox[b][0.3cm]{17.7cm}{tilted to \ensuremath{\theta}\ensuremath{_{\textnormal{lab}}}=43\ensuremath{^\circ} covering \ensuremath{\theta}\ensuremath{_{\textnormal{lab}}}=29\ensuremath{^\circ}{\textminus}73\ensuremath{^\circ} for \ensuremath{\alpha}-particles and the MINI Si detector array covering \ensuremath{\theta}\ensuremath{_{\textnormal{lab}}}=11.5\ensuremath{^\circ}{\textminus}22.5\ensuremath{^\circ} for the \ensuremath{^{\textnormal{15}}}O}\\
\parbox[b][0.3cm]{17.7cm}{recoils. Scattered \ensuremath{^{\textnormal{18}}}F ions and \ensuremath{^{\textnormal{18}}}O contaminants were measured using an ionization chamber downstream the Si detectors. Using}\\
\parbox[b][0.3cm]{17.7cm}{R-matrix, the authors found that a constructive interference with the 665-keV resonance would best fit the data, while the sign of}\\
\parbox[b][0.3cm]{17.7cm}{the interference for the 8 and 38 keV resonances could be {\textminus} or +. Set upper limits of \ensuremath{\Gamma}\ensuremath{_{\textnormal{p}}}\ensuremath{\leq}1.17 keV and \ensuremath{\Gamma}\ensuremath{_{\textnormal{p}}}\ensuremath{\leq}1.65 keV for the}\\
\parbox[b][0.3cm]{17.7cm}{higher energy resonances at E\ensuremath{_{\textnormal{c.m.}}}=827 keV and 842 keV, respectively.}\\
\parbox[b][0.3cm]{17.7cm}{\addtolength{\parindent}{-0.2in}\href{https://www.nndc.bnl.gov/nsr/nsrlink.jsp?2009De03,B}{2009De03}: \ensuremath{^{\textnormal{1}}}H(\ensuremath{^{\textnormal{18}}}F,\ensuremath{\alpha}) E=13.8 MeV; measured \ensuremath{\sigma}(\ensuremath{\theta}) by measuring \ensuremath{\alpha}-\ensuremath{^{\textnormal{15}}}O coincidence events using two position sensitive Si}\\
\parbox[b][0.3cm]{17.7cm}{detectors (LEDA) covering \ensuremath{\theta}\ensuremath{_{\textnormal{lab}}}=30.5\ensuremath{^\circ}{\textminus}56.8\ensuremath{^\circ} (for \ensuremath{\alpha}-particles) and \ensuremath{\theta}\ensuremath{_{\textnormal{lab}}}=8.9\ensuremath{^\circ}{\textminus}22.1\ensuremath{^\circ} (for \ensuremath{^{\textnormal{15}}}O recoils). Using an Al degrader,}\\
\parbox[b][0.3cm]{17.7cm}{measured \ensuremath{\sigma}(E) at E\ensuremath{_{\textnormal{beam}}}=13.1, 9.9, and 8.6 MeV corresponding to E\ensuremath{_{\textnormal{c.m.}}}=665 keV, off resonance, and the low energy tail of the}\\
\parbox[b][0.3cm]{17.7cm}{resonance, respectively. Deduced total cross section for the \ensuremath{^{\textnormal{18}}}F(p,\ensuremath{\alpha}) reaction down to E\ensuremath{_{\textnormal{c.m.}}}=400 keV. Performed R-matrix}\\
\parbox[b][0.3cm]{17.7cm}{calculations. Deduced the astrophysical S-factor for E\ensuremath{_{\textnormal{c.m.}}}\ensuremath{<}1 MeV. Discussed interferences between 3/2\ensuremath{^{\textnormal{+}}} resonances in \ensuremath{^{\textnormal{19}}}Ne.}\\
\parbox[b][0.3cm]{17.7cm}{\addtolength{\parindent}{-0.2in}\href{https://www.nndc.bnl.gov/nsr/nsrlink.jsp?2009Mu07,B}{2009Mu07}: \ensuremath{^{\textnormal{1}}}H(\ensuremath{^{\textnormal{18}}}F,\ensuremath{\alpha}) and \ensuremath{^{\textnormal{1}}}H(\ensuremath{^{\textnormal{18}}}F,p) E=1.750 MeV/nucleon; measured energy and TOF of the reaction protons and \ensuremath{\alpha}-particles}\\
\parbox[b][0.3cm]{17.7cm}{using the position sensitive annular Si TUDA array covering \ensuremath{\theta}\ensuremath{_{\textnormal{lab}}}=7\ensuremath{^\circ}{\textminus}16.6\ensuremath{^\circ}. Measured the \ensuremath{^{\textnormal{1}}}H(\ensuremath{^{\textnormal{18}}}F,p) and \ensuremath{^{\textnormal{1}}}H(\ensuremath{^{\textnormal{18}}}F,\ensuremath{\alpha}) excitation}\\
\parbox[b][0.3cm]{17.7cm}{functions at E\ensuremath{_{\textnormal{c.m.}}}=0.5-2 MeV. A simultaneous multi-channel R-matrix fit to both data sets was performed. A new resonance and a}\\
\parbox[b][0.3cm]{17.7cm}{candidate for a previously predicted broad J\ensuremath{^{\ensuremath{\pi}}}=1/2\ensuremath{^{\textnormal{+}}} resonance were observed at E\ensuremath{_{\textnormal{c.m.}}}=1347 keV \textit{5} and E\ensuremath{_{\textnormal{c.m.}}}=1573 keV \textit{8},}\\
\parbox[b][0.3cm]{17.7cm}{respectively. Their J\ensuremath{^{\ensuremath{\pi}}}, \ensuremath{\Gamma}\ensuremath{_{\textnormal{p}}} and \ensuremath{\Gamma}\ensuremath{_{\ensuremath{\alpha}}} were deduced.}\\
\parbox[b][0.3cm]{17.7cm}{\addtolength{\parindent}{-0.2in}\href{https://www.nndc.bnl.gov/nsr/nsrlink.jsp?2011Be11,B}{2011Be11}: \ensuremath{^{\textnormal{1}}}H(\ensuremath{^{\textnormal{18}}}F,\ensuremath{\alpha}) E\ensuremath{_{\textnormal{c.m.}}}=250, 330, 453, 673 keV; measured energy, position and TOF for \ensuremath{\alpha}-particles and \ensuremath{^{\textnormal{15}}}O recoils using the}\\
\parbox[b][0.3cm]{17.7cm}{position sensitive Si annular TUDA array covering \ensuremath{\theta}\ensuremath{_{\textnormal{lab}}}=4\ensuremath{^\circ}{\textminus}69\ensuremath{^\circ}. Measured I\ensuremath{_{\ensuremath{\alpha}}}, \ensuremath{\alpha}-\ensuremath{^{\textnormal{15}}}O coincidences, \ensuremath{\alpha} angular distributions, and the}\\
\parbox[b][0.3cm]{17.7cm}{\ensuremath{^{\textnormal{18}}}F(p,\ensuremath{\alpha}) reaction cross section. Deduced astrophysical S-factor using multi-channel R-matrix analysis performed via the DREAM}\\
\parbox[b][0.3cm]{17.7cm}{computer code and discussed implications for the \ensuremath{^{\textnormal{18}}}F(p,\ensuremath{\alpha}) reaction rate in the context of nova nucleosynthesis.}\\
\parbox[b][0.3cm]{17.7cm}{\addtolength{\parindent}{-0.2in}\href{https://www.nndc.bnl.gov/nsr/nsrlink.jsp?2012Mo03,B}{2012Mo03}: \ensuremath{^{\textnormal{1}}}H(\ensuremath{^{\textnormal{18}}}F,p) and \ensuremath{^{\textnormal{1}}}H(\ensuremath{^{\textnormal{18}}}F,\ensuremath{\alpha}) E=1.7 MeV/nucleon; measured energy and TOF of protons and \ensuremath{\alpha} particles from the reaction}\\
\parbox[b][0.3cm]{17.7cm}{using a position sensitive Si detector and an MCP detector upstream of the Si detector; analyzed the measured excitation functions}\\
\parbox[b][0.3cm]{17.7cm}{(at E\ensuremath{_{\textnormal{c.m.}}}=0.5-1.9 MeV) using a simultaneous multi-channel R-matrix analysis; deduced \ensuremath{^{\textnormal{19}}}Ne resonances and their partial widths.}\\
\parbox[b][0.3cm]{17.7cm}{J\ensuremath{^{\ensuremath{\pi}}} assignments were based on literature due to lack of angular distribution information. The authors deduced the astrophysical}\\
\parbox[b][0.3cm]{17.7cm}{S-factor at E\ensuremath{_{\textnormal{c.m.}}}\ensuremath{<}0.5 MeV and discussed the interferences between observed resonances. We highlight that there are inconsistencies}\\
\parbox[b][0.3cm]{17.7cm}{between the results of this study and those of (\href{https://www.nndc.bnl.gov/nsr/nsrlink.jsp?2009Mu07,B}{2009Mu07}).}\\
\vspace{0.385cm}
\parbox[b][0.3cm]{17.7cm}{\addtolength{\parindent}{-0.2in}\textit{Related Experiments on the Properties of \ensuremath{^{19}}F* Mirror Levels}:}\\
\parbox[b][0.3cm]{17.7cm}{\addtolength{\parindent}{-0.2in}\href{https://www.nndc.bnl.gov/nsr/nsrlink.jsp?1998Bu13,B}{1998Bu13}: \ensuremath{^{\textnormal{15}}}N(\ensuremath{\alpha},\ensuremath{\gamma}) E=2.6-3.93 MeV; Deduced \ensuremath{\Gamma}=28 keV \textit{1}, \ensuremath{\Gamma}\ensuremath{_{\ensuremath{\gamma}}}=0.39 eV \textit{6} and \ensuremath{\omega}\ensuremath{\gamma}\ensuremath{_{\textnormal{(}\ensuremath{\alpha}\textnormal{,}\ensuremath{\gamma}\textnormal{)}}}=0.77 eV \textit{11} for the \ensuremath{^{\textnormal{19}}}F*(7101) state}\\
\parbox[b][0.3cm]{17.7cm}{by measuring E\ensuremath{_{\ensuremath{\gamma}}} and I\ensuremath{_{\ensuremath{\gamma}}} of the \ensuremath{\gamma} decay of this state using the RHINOCEROS windowless gas target and a Compton suppressed}\\
\parbox[b][0.3cm]{17.7cm}{HPGe detector at \ensuremath{\theta}\ensuremath{_{\textnormal{lab}}}=90\ensuremath{^\circ}. This \ensuremath{^{\textnormal{19}}}F state was proposed to be the mirror state of the \ensuremath{^{\textnormal{19}}}Ne*(7.07 MeV) level. The measured \ensuremath{\Gamma}\ensuremath{_{\ensuremath{\gamma}}}}\\
\parbox[b][0.3cm]{17.7cm}{for the \ensuremath{^{\textnormal{19}}}F*(7101) state was used as the \ensuremath{\Gamma}\ensuremath{_{\ensuremath{\gamma}}} for the resonance corresponding to the \ensuremath{^{\textnormal{19}}}Ne*(7.07 MeV) mirror state. (\href{https://www.nndc.bnl.gov/nsr/nsrlink.jsp?1998Bu13,B}{1998Bu13})}\\
\parbox[b][0.3cm]{17.7cm}{also obtained \ensuremath{\Gamma}\ensuremath{_{\ensuremath{\alpha}}}\ensuremath{\approx}30 keV for this \ensuremath{^{\textnormal{19}}}Ne level assuming the same reduced alpha-width as that of the \ensuremath{^{\textnormal{19}}}F*(7101) state. Discussed}\\
\parbox[b][0.3cm]{17.7cm}{the astrophysical implications. Evaluator notes that (\href{https://www.nndc.bnl.gov/nsr/nsrlink.jsp?2000Fo01,B}{2000Fo01}) disputed the proposed \ensuremath{^{\textnormal{19}}}F*(7.10 MeV) state as the mirror to the}\\
\parbox[b][0.3cm]{17.7cm}{\ensuremath{^{\textnormal{19}}}Ne*(7.07 MeV) level.}\\
\parbox[b][0.3cm]{17.7cm}{\addtolength{\parindent}{-0.2in}\href{https://www.nndc.bnl.gov/nsr/nsrlink.jsp?2003DeZZ,B}{2003DeZZ}, \href{https://www.nndc.bnl.gov/nsr/nsrlink.jsp?2003De15,B}{2003De15}, \href{https://www.nndc.bnl.gov/nsr/nsrlink.jsp?2003De26,B}{2003De26}, \href{https://www.nndc.bnl.gov/nsr/nsrlink.jsp?2005De45,B}{2005De45}, \href{https://www.nndc.bnl.gov/nsr/nsrlink.jsp?2007De47,B}{2007De47}: \ensuremath{^{\textnormal{2}}}H(\ensuremath{^{\textnormal{18}}}F,p)\ensuremath{^{\textnormal{19}}}F*(\ensuremath{\alpha}) E=14 MeV; measured energies and emission angles of}\\
\parbox[b][0.3cm]{17.7cm}{the outgoing protons in coincidence with the \ensuremath{^{\textnormal{15}}}N decay products using two position sensitive Si arrays, LAMP and LEDA,}\\
\parbox[b][0.3cm]{17.7cm}{respectively. The LAMP array covered \ensuremath{\theta}\ensuremath{_{\textnormal{lab}}}=110\ensuremath{^\circ}{\textminus}157\ensuremath{^\circ}, while the LEDA array covered \ensuremath{\theta}\ensuremath{_{\textnormal{lab}}}=7\ensuremath{^\circ}{\textminus}18\ensuremath{^\circ}. Measured TOF for p-\ensuremath{^{\textnormal{15}}}N}\\
\clearpage
\vspace{0.3cm}
{\bf \small \underline{\ensuremath{^{\textnormal{1}}}H(\ensuremath{^{\textnormal{18}}}F,p),(\ensuremath{^{\textnormal{18}}}F,\ensuremath{\alpha}):res\hspace{0.2in}\href{https://www.nndc.bnl.gov/nsr/nsrlink.jsp?2009Mu07,B}{2009Mu07},\href{https://www.nndc.bnl.gov/nsr/nsrlink.jsp?2012Mo03,B}{2012Mo03} (continued)}}\\
\vspace{0.3cm}
\parbox[b][0.3cm]{17.7cm}{coincidence events. Deduced \ensuremath{^{\textnormal{19}}}F* levels, including two levels at 6497 and 6528 keV, which were proposed to be mirror states to}\\
\parbox[b][0.3cm]{17.7cm}{the astrophysically significant \ensuremath{^{\textnormal{19}}}Ne*(6420, 6449) states with the proposed J\ensuremath{^{\ensuremath{\pi}}}=3/2\ensuremath{^{\textnormal{+}}}. Performed DWBA calculations and obtained}\\
\parbox[b][0.3cm]{17.7cm}{L=0,2 (with strong dominance of L=0) for the aforementioned \ensuremath{^{\textnormal{19}}}F states. Deduced neutron spectroscopic factor of 0.17 when}\\
\parbox[b][0.3cm]{17.7cm}{considering compound reaction contribution, see (\href{https://www.nndc.bnl.gov/nsr/nsrlink.jsp?2005De45,B}{2005De45}, \href{https://www.nndc.bnl.gov/nsr/nsrlink.jsp?2007De47,B}{2007De47}) for the sum of these two levels for L=0. Determined}\\
\parbox[b][0.3cm]{17.7cm}{S\ensuremath{_{\textnormal{n}}}\ensuremath{<}0.33 (\href{https://www.nndc.bnl.gov/nsr/nsrlink.jsp?2007De47,B}{2007De47}) for an unobserved J\ensuremath{^{\ensuremath{\pi}}}=1/2\ensuremath{^{-}} level in \ensuremath{^{\textnormal{19}}}F at 6429 keV assuming L=1. This state was proposed to be the mirror}\\
\parbox[b][0.3cm]{17.7cm}{to the \ensuremath{^{\textnormal{19}}}Ne*(6437 keV, 1/2\ensuremath{^{-}}) state. Using the deduced neutron spectroscopic factors, the authors deduced the astrophysical S-factor}\\
\parbox[b][0.3cm]{17.7cm}{for the \ensuremath{^{\textnormal{18}}}F(p,\ensuremath{\alpha}) reaction at E\ensuremath{_{\textnormal{c.m.}}}\ensuremath{<}1 MeV using R-matrix. Deduced the \ensuremath{^{\textnormal{18}}}F(p,\ensuremath{\alpha}) reaction rate at T=0.01-1 GK. These authors}\\
\parbox[b][0.3cm]{17.7cm}{suggested that \ensuremath{^{\textnormal{19}}}F*(6497, 6528) could be the mirrors to the \ensuremath{^{\textnormal{19}}}Ne*(6449, 6420) states, respectively.}\\
\parbox[b][0.3cm]{17.7cm}{\addtolength{\parindent}{-0.2in}\href{https://www.nndc.bnl.gov/nsr/nsrlink.jsp?2005Ba06,B}{2005Ba06}: \ensuremath{^{\textnormal{15}}}N(\ensuremath{\alpha},\ensuremath{\alpha}\ensuremath{'}) E=1.9-4.2 MeV; analyzed \ensuremath{\sigma}(\ensuremath{\theta}); deduced \ensuremath{^{\textnormal{19}}}F energy-levels, widths, J, \ensuremath{\pi}; discussed \ensuremath{^{\textnormal{19}}}Ne-\ensuremath{^{\textnormal{19}}}F mirror levels;}\\
\parbox[b][0.3cm]{17.7cm}{and astrophysical implications for the \ensuremath{^{\textnormal{18}}}F(p,\ensuremath{\alpha}) and \ensuremath{^{\textnormal{18}}}F(p,\ensuremath{\gamma}) reaction rates.}\\
\parbox[b][0.3cm]{17.7cm}{\addtolength{\parindent}{-0.2in}\href{https://www.nndc.bnl.gov/nsr/nsrlink.jsp?2005Ko09,B}{2005Ko09}, \href{https://www.nndc.bnl.gov/nsr/nsrlink.jsp?2006Ko13,B}{2006Ko13}: \ensuremath{^{\textnormal{2}}}H(\ensuremath{^{\textnormal{18}}}F,p) E=108.49 MeV; measured neutron single-particle states in the \ensuremath{^{\textnormal{19}}}F mirror nucleus in the}\\
\parbox[b][0.3cm]{17.7cm}{excitation energy range corresponding to the \ensuremath{^{\textnormal{19}}}Ne region of interest to the \ensuremath{^{\textnormal{18}}}F(p,\ensuremath{\gamma}) and \ensuremath{^{\textnormal{18}}}F(p,\ensuremath{\alpha}) reaction rates at novae and}\\
\parbox[b][0.3cm]{17.7cm}{X-ray bursts temperatures. The results of this work on \ensuremath{^{\textnormal{19}}}F supersede those used by (\href{https://www.nndc.bnl.gov/nsr/nsrlink.jsp?2003Sh25,B}{2003Sh25}, \href{https://www.nndc.bnl.gov/nsr/nsrlink.jsp?2004Ba63,B}{2004Ba63}, \href{https://www.nndc.bnl.gov/nsr/nsrlink.jsp?2004Bb08,B}{2004Bb08}, \href{https://www.nndc.bnl.gov/nsr/nsrlink.jsp?2004Bb10,B}{2004Bb10}).}\\
\parbox[b][0.3cm]{17.7cm}{(\href{https://www.nndc.bnl.gov/nsr/nsrlink.jsp?2005Ko09,B}{2005Ko09}) tentatively assigned the \ensuremath{^{\textnormal{19}}}F*(6497 keV, 3/2\ensuremath{^{\textnormal{+}}}) level to be the mirror to the \ensuremath{^{\textnormal{19}}}Ne*(6418 keV, 3/2\ensuremath{^{\textnormal{+}}}) state; however,}\\
\parbox[b][0.3cm]{17.7cm}{they could not rule out the possibility that the aforementioned \ensuremath{^{\textnormal{19}}}F* state may be the mirror of the \ensuremath{^{\textnormal{19}}}Ne*(6448) state. These}\\
\parbox[b][0.3cm]{17.7cm}{authors also assigned the \ensuremath{^{\textnormal{19}}}F*(7262) level as a likely L=0 mirror level to the \ensuremath{^{\textnormal{19}}}Ne*(7076) state, which is consistent with the}\\
\parbox[b][0.3cm]{17.7cm}{mirror level assignment proposed by (\href{https://www.nndc.bnl.gov/nsr/nsrlink.jsp?2000Fo01,B}{2000Fo01}). The \ensuremath{^{\textnormal{18}}}F(p,\ensuremath{\alpha}) reaction rate was deduced at T=0.1-0.5 GK based on proton widths}\\
\parbox[b][0.3cm]{17.7cm}{that were determined by (\href{https://www.nndc.bnl.gov/nsr/nsrlink.jsp?2005Ko09,B}{2005Ko09}) from using neutron spectroscopic factors deduced in this work assuming S\ensuremath{_{\textnormal{n}}}=S\ensuremath{_{\textnormal{p}}}.}\\
\vspace{0.385cm}
\parbox[b][0.3cm]{17.7cm}{\addtolength{\parindent}{-0.2in}\textit{Theory}:}\\
\parbox[b][0.3cm]{17.7cm}{\addtolength{\parindent}{-0.2in}\href{https://www.nndc.bnl.gov/nsr/nsrlink.jsp?2000Fo01,B}{2000Fo01}: Calculated single-particle width \ensuremath{\Gamma}\ensuremath{_{\textnormal{sp}}}=28 keV for the \ensuremath{^{\textnormal{19}}}Ne*(7.07 MeV) level using a Woods{\textminus}Saxon+Coulomb potential}\\
\parbox[b][0.3cm]{17.7cm}{model assuming this state is constructed by \ensuremath{^{\textnormal{18}}}F+p, where the protons are in the 2\textit{s}\ensuremath{_{\textnormal{1/2}}} orbital. Through this calculation, they}\\
\parbox[b][0.3cm]{17.7cm}{determined the laboratory resonance energy corresponding to the 7.07-MeV state to be E\ensuremath{_{\textnormal{p}}}=655 keV. Using S\ensuremath{_{\textnormal{p}}}=\ensuremath{\Gamma}\ensuremath{_{\textnormal{p}}}/\ensuremath{\Gamma}\ensuremath{_{\textnormal{sp}}} and}\\
\parbox[b][0.3cm]{17.7cm}{\ensuremath{\Gamma}\ensuremath{_{\textnormal{p}}}=11 keV \textit{3} (an average obtained by \href{https://www.nndc.bnl.gov/nsr/nsrlink.jsp?2000Fo01,B}{2000Fo01} from \href{https://www.nndc.bnl.gov/nsr/nsrlink.jsp?1998Ut02,B}{1998Ut02}, \href{https://www.nndc.bnl.gov/nsr/nsrlink.jsp?1996Re05,B}{1996Re05}, \href{https://www.nndc.bnl.gov/nsr/nsrlink.jsp?1995Co23,B}{1995Co23}), (\href{https://www.nndc.bnl.gov/nsr/nsrlink.jsp?2000Fo01,B}{2000Fo01}) determined a theoretical}\\
\parbox[b][0.3cm]{17.7cm}{spectroscopic factor of S\ensuremath{_{\textnormal{p}}}=0.40 \textit{12} for the \ensuremath{^{\textnormal{19}}}Ne*(7.07 MeV) state assuming J\ensuremath{^{\ensuremath{\pi}}}=3/2\ensuremath{^{\textnormal{+}}} even though no evidence favored this}\\
\parbox[b][0.3cm]{17.7cm}{assignment over J\ensuremath{^{\ensuremath{\pi}}}=1/2\ensuremath{^{\textnormal{+}}}. (\href{https://www.nndc.bnl.gov/nsr/nsrlink.jsp?2000Fo01,B}{2000Fo01}) disputed the \ensuremath{^{\textnormal{19}}}F*(7.10 MeV) state as the mirror of the \ensuremath{^{\textnormal{19}}}Ne*(7.07 MeV) level proposed by}\\
\parbox[b][0.3cm]{17.7cm}{(\href{https://www.nndc.bnl.gov/nsr/nsrlink.jsp?1998Bu13,B}{1998Bu13}) and reported that it is impossible for these two states to be mirror levels. (\href{https://www.nndc.bnl.gov/nsr/nsrlink.jsp?2000Fo01,B}{2000Fo01}) deduced the mirror level to be a}\\
\parbox[b][0.3cm]{17.7cm}{state in \ensuremath{^{\textnormal{19}}}F* at E\ensuremath{_{\textnormal{x}}}=7.41 MeV \textit{10}. They attributed the large shift to the appreciable fraction of the 2\textit{s}\ensuremath{_{\textnormal{1/2}}} single-particle strength in}\\
\parbox[b][0.3cm]{17.7cm}{\ensuremath{^{\textnormal{19}}}Ne*(7.07 MeV).}\\
\parbox[b][0.3cm]{17.7cm}{\addtolength{\parindent}{-0.2in}\href{https://www.nndc.bnl.gov/nsr/nsrlink.jsp?2006Fo03,B}{2006Fo03}: Calculated single-particle alpha and proton widths for the 6.4-7.5 MeV excited levels in \ensuremath{^{\textnormal{19}}}Ne; deduced spectroscopic}\\
\parbox[b][0.3cm]{17.7cm}{factors for these levels. Based on the deduced proton spectroscopic factor for the \ensuremath{^{\textnormal{19}}}Ne*(7419) state (S\ensuremath{_{\textnormal{p}}}=8.2 \textit{12}, which was}\\
\parbox[b][0.3cm]{17.7cm}{significantly larger than the theoretical upper limit) using the J\ensuremath{^{\ensuremath{\pi}}}=7/2\ensuremath{^{\textnormal{+}}} assignment and the \ensuremath{\Gamma}\ensuremath{_{\textnormal{p}}}=27 keV \textit{4} deduced by (\href{https://www.nndc.bnl.gov/nsr/nsrlink.jsp?2004Ba63,B}{2004Ba63}),}\\
\parbox[b][0.3cm]{17.7cm}{(\href{https://www.nndc.bnl.gov/nsr/nsrlink.jsp?2006Fo03,B}{2006Fo03}) disputed either the J\ensuremath{^{\ensuremath{\pi}}} assignment, or the \ensuremath{\Gamma}\ensuremath{_{\textnormal{p}}}, or both and argued that one or both of these values could be incorrect.}\\
\parbox[b][0.3cm]{17.7cm}{\addtolength{\parindent}{-0.2in}\href{https://www.nndc.bnl.gov/nsr/nsrlink.jsp?2007Du09,B}{2007Du09}: Calculated the \ensuremath{^{\textnormal{18}}}F(p,\ensuremath{\alpha}) S-factor at E\ensuremath{_{\textnormal{c.m.}}}\ensuremath{\leq}1.5 MeV using a microscopic cluster mode with wave functions defined}\\
\parbox[b][0.3cm]{17.7cm}{using Generator Coordinate Method. Obtained the spectra of \textit{s}-wave resonances with J\ensuremath{^{\ensuremath{\pi}}}=1/2\ensuremath{^{\textnormal{+}}} and 3/2\ensuremath{^{\textnormal{+}}} and compared them with}\\
\parbox[b][0.3cm]{17.7cm}{experimental values. The authors suggested that the \ensuremath{^{\textnormal{18}}}F(p,\ensuremath{\alpha}) rate is dominated by an unobserved J\ensuremath{^{\ensuremath{\pi}}}=1/2\ensuremath{^{\textnormal{+}}} sub-threshold at E\ensuremath{_{\textnormal{x}}}\ensuremath{\sim}6}\\
\parbox[b][0.3cm]{17.7cm}{MeV and an unobserved J\ensuremath{^{\ensuremath{\pi}}}=1/2\ensuremath{^{\textnormal{+}}} broad resonance at E\ensuremath{_{\textnormal{c.m.}}}(\ensuremath{^{\textnormal{18}}}F+p)=1.49 MeV. They reported that both of these states should have}\\
\parbox[b][0.3cm]{17.7cm}{a strong single-particle structure.}\\
\parbox[b][0.3cm]{17.7cm}{\addtolength{\parindent}{-0.2in}\href{https://www.nndc.bnl.gov/nsr/nsrlink.jsp?2007Ne09,B}{2007Ne09}: Updated and expanded on the results of (\href{https://www.nndc.bnl.gov/nsr/nsrlink.jsp?2003Sh25,B}{2003Sh25}). Evaluated E\ensuremath{_{\textnormal{r}}}, J\ensuremath{^{\ensuremath{\pi}}}, \ensuremath{\Gamma}\ensuremath{_{\ensuremath{\gamma}}}, \ensuremath{\Gamma}\ensuremath{_{\textnormal{p}}}, \ensuremath{\Gamma}\ensuremath{_{\ensuremath{\alpha}}}, and \ensuremath{\theta}\ensuremath{_{\textnormal{p}}^{\textnormal{2}}} for \ensuremath{^{\textnormal{19}}}Ne levels with}\\
\parbox[b][0.3cm]{17.7cm}{E\ensuremath{_{\textnormal{x}}}=6.4-8.1 MeV, including unmeasured ones, based on all available (by then) experimental data and those of the analog states in}\\
\parbox[b][0.3cm]{17.7cm}{the mirror nucleus \ensuremath{^{\textnormal{19}}}F. Assumptions are made when properties are unknown.}\\
\parbox[b][0.3cm]{17.7cm}{\addtolength{\parindent}{-0.2in}\href{https://www.nndc.bnl.gov/nsr/nsrlink.jsp?2010Fo07,B}{2010Fo07}: Deduced \ensuremath{\Gamma}\ensuremath{_{\ensuremath{\alpha}}} for the \ensuremath{^{\textnormal{19}}}Ne*(4.03, 4.379, 4.6, 5.092, 5.351,7.42 MeV) levels using the experimental and theoretical}\\
\parbox[b][0.3cm]{17.7cm}{information available at the time in the literature. The \ensuremath{\alpha} spectroscopic factors (S\ensuremath{_{\ensuremath{\alpha}}}) were computed for the mirrors of the above}\\
\parbox[b][0.3cm]{17.7cm}{mentioned states in \ensuremath{^{\textnormal{19}}}F and used to compute the \ensuremath{\alpha} widths of the \ensuremath{^{\textnormal{19}}}Ne*(4.379, 4.6, 5.092, 5.351, 7.42 MeV) levels.}\\
\vspace{0.385cm}
\parbox[b][0.3cm]{17.7cm}{\addtolength{\parindent}{-0.2in}\textit{The \ensuremath{^{18}}F(p,\ensuremath{\alpha}) Astrophysical Reaction Rate}:}\\
\parbox[b][0.3cm]{17.7cm}{\addtolength{\parindent}{-0.2in}\textbf{Foreword:}}\\
\parbox[b][0.3cm]{17.7cm}{\addtolength{\parindent}{-0.2in}Most of the following studies are experimental and are presented in this or the other individual reaction datasets. The measured}\\
\parbox[b][0.3cm]{17.7cm}{resonances and the measured or deduced resonance properties from the following studies are important for the determination of the}\\
\parbox[b][0.3cm]{17.7cm}{\ensuremath{^{\textnormal{18}}}F(p,\ensuremath{\alpha})\ensuremath{^{\textnormal{15}}}O astrophysical reaction rate.}\\
\vspace{0.385cm}
\parbox[b][0.3cm]{17.7cm}{\addtolength{\parindent}{-0.2in}R. V. Wagoner, Astrophys. J. Suppl. Ser., 18 (1969) 247: Calculated the parameterized \ensuremath{^{\textnormal{18}}}F(p,\ensuremath{\alpha}) reaction rate as a function of}\\
\parbox[b][0.3cm]{17.7cm}{temperature in GK.}\\
\clearpage
\vspace{0.3cm}
{\bf \small \underline{\ensuremath{^{\textnormal{1}}}H(\ensuremath{^{\textnormal{18}}}F,p),(\ensuremath{^{\textnormal{18}}}F,\ensuremath{\alpha}):res\hspace{0.2in}\href{https://www.nndc.bnl.gov/nsr/nsrlink.jsp?2009Mu07,B}{2009Mu07},\href{https://www.nndc.bnl.gov/nsr/nsrlink.jsp?2012Mo03,B}{2012Mo03} (continued)}}\\
\vspace{0.3cm}
\parbox[b][0.3cm]{17.7cm}{\addtolength{\parindent}{-0.2in}\href{https://www.nndc.bnl.gov/nsr/nsrlink.jsp?1979Wo07,B}{1979Wo07}: Calculated statistical, parameterized reaction rates at T=0.05-10 GK.}\\
\parbox[b][0.3cm]{17.7cm}{\addtolength{\parindent}{-0.2in}M. Wiescher and K.-U., Kettner, Astrophys. J. 263 (1982) 891: Deduced the resonance properties (E\ensuremath{_{\textnormal{r}}}, J\ensuremath{^{\ensuremath{\pi}}}, \ensuremath{\Gamma}\ensuremath{_{\textnormal{p}}}, \ensuremath{\Gamma}\ensuremath{_{\ensuremath{\alpha}}}, and \ensuremath{\Gamma}\ensuremath{_{\ensuremath{\gamma}}}) for}\\
\parbox[b][0.3cm]{17.7cm}{the resonances associated with the \ensuremath{^{\textnormal{19}}}Ne states at E\ensuremath{_{\textnormal{x}}}=6437, 6500, 6540, 6742, 6790, and 6862 keV. Using these properties, they}\\
\parbox[b][0.3cm]{17.7cm}{deduced the \ensuremath{^{\textnormal{18}}}F(p,\ensuremath{\alpha}) reaction rate for T=0.04-0.6 GK. Compared this rate with the ones previously calculated and discussed the}\\
\parbox[b][0.3cm]{17.7cm}{astrophysical implications.}\\
\parbox[b][0.3cm]{17.7cm}{\addtolength{\parindent}{-0.2in}\href{https://www.nndc.bnl.gov/nsr/nsrlink.jsp?1992Ch50,B}{1992Ch50}: Reviewed the \ensuremath{^{\textnormal{18}}}F(p,\ensuremath{\gamma}) and \ensuremath{^{\textnormal{18}}}F(p,\ensuremath{\alpha}) reaction rates and recommended those of R. K. Wallace and S. E., Woosley,}\\
\parbox[b][0.3cm]{17.7cm}{Astrophys. J. Suppl. 45 (1981) 389 and (\href{https://www.nndc.bnl.gov/nsr/nsrlink.jsp?1990Ma05,B}{1990Ma05}).}\\
\parbox[b][0.3cm]{17.7cm}{\addtolength{\parindent}{-0.2in}\href{https://www.nndc.bnl.gov/nsr/nsrlink.jsp?1996Re05,B}{1996Re05}: Deduced the \ensuremath{^{\textnormal{18}}}F(p,\ensuremath{\alpha}) reaction rate at T=0.4-2 GK using the resonance properties they deduced for the 652-keV}\\
\parbox[b][0.3cm]{17.7cm}{resonance. For other resonances in \ensuremath{^{\textnormal{19}}}Ne in the E\ensuremath{_{\textnormal{c.m.}}}=330-658 keV region, they assumed proton widths of 1\% and 10\% of the}\\
\parbox[b][0.3cm]{17.7cm}{Wigner limits for the negative- and positive-parity states, respectively. The \ensuremath{\alpha}-widths were deduced from the \ensuremath{^{\textnormal{19}}}F* analog states.}\\
\parbox[b][0.3cm]{17.7cm}{\addtolength{\parindent}{-0.2in}\href{https://www.nndc.bnl.gov/nsr/nsrlink.jsp?1997Re02,B}{1997Re02}: Deduced the upper and lower limits for the ratio of the \ensuremath{^{\textnormal{18}}}F(p,\ensuremath{\alpha})/\ensuremath{^{\textnormal{18}}}F(p,\ensuremath{\gamma}) reaction rates at T=0.4-2 GK and discussed}\\
\parbox[b][0.3cm]{17.7cm}{the astrophysical implications.}\\
\parbox[b][0.3cm]{17.7cm}{\addtolength{\parindent}{-0.2in}\href{https://www.nndc.bnl.gov/nsr/nsrlink.jsp?1997Re05,B}{1997Re05}: Provided the first experimental limit for the ratio of the \ensuremath{^{\textnormal{18}}}F(p,\ensuremath{\alpha})/\ensuremath{^{\textnormal{18}}}F(p,\ensuremath{\gamma}) astrophysical reaction rates.}\\
\parbox[b][0.3cm]{17.7cm}{\addtolength{\parindent}{-0.2in}\href{https://www.nndc.bnl.gov/nsr/nsrlink.jsp?1997Gr23,B}{1997Gr23}: Deduced the \ensuremath{^{\textnormal{18}}}F(p,\ensuremath{\alpha}) reaction rate at T=0.1-1 GK based on two resonances at 653 keV and 324 keV and discussed the}\\
\parbox[b][0.3cm]{17.7cm}{astrophysical implications.}\\
\parbox[b][0.3cm]{17.7cm}{\addtolength{\parindent}{-0.2in}\href{https://www.nndc.bnl.gov/nsr/nsrlink.jsp?1998Ut02,B}{1998Ut02}: Deduced resonance properties, \ensuremath{\Gamma}\ensuremath{_{\textnormal{p}}}, \ensuremath{\Gamma}\ensuremath{_{\ensuremath{\alpha}}}, E\ensuremath{_{\textnormal{x}}}, E\ensuremath{_{\textnormal{c.m.}}}, \ensuremath{\Gamma}, J\ensuremath{^{\ensuremath{\pi}}}, and \ensuremath{\omega}\ensuremath{\gamma}\ensuremath{_{\textnormal{(p,}\ensuremath{\alpha}\textnormal{)}}} for resonances at E\ensuremath{_{\textnormal{c.m.}}}=0-1 MeV which}\\
\parbox[b][0.3cm]{17.7cm}{contribute to the \ensuremath{^{\textnormal{18}}}F(p,\ensuremath{\alpha}) reaction rate. Deduced the direct proton capture rate as well as resonant contributions to the (p,\ensuremath{\alpha}) total}\\
\parbox[b][0.3cm]{17.7cm}{reaction rate at T=0.1-1 GK (see the erratum at Phys. Rev. C 58 (1998) 1354). Provided REACLIB format.}\\
\parbox[b][0.3cm]{17.7cm}{\addtolength{\parindent}{-0.2in}\href{https://www.nndc.bnl.gov/nsr/nsrlink.jsp?1999He40,B}{1999He40}, \href{https://www.nndc.bnl.gov/nsr/nsrlink.jsp?2001Co14,B}{2001Co14}: Investigated nova nucleosynthesis using the \ensuremath{^{\textnormal{18}}}F(p,\ensuremath{\gamma}) and \ensuremath{^{\textnormal{18}}}F(p,\ensuremath{\alpha}) rates from (\href{https://www.nndc.bnl.gov/nsr/nsrlink.jsp?1997Gr23,B}{1997Gr23}, \href{https://www.nndc.bnl.gov/nsr/nsrlink.jsp?1998Ut02,B}{1998Ut02}) and}\\
\parbox[b][0.3cm]{17.7cm}{hydrodynamic models for CO and ONe novae with different white dwarf masses. Discussed implications for \ensuremath{\gamma}-ray emission from}\\
\parbox[b][0.3cm]{17.7cm}{\ensuremath{^{\textnormal{18}}}F in novae.}\\
\parbox[b][0.3cm]{17.7cm}{\addtolength{\parindent}{-0.2in}\href{https://www.nndc.bnl.gov/nsr/nsrlink.jsp?2000Co33,B}{2000Co33}: They deduced the \ensuremath{^{\textnormal{18}}}F(p,\ensuremath{\alpha}) reaction rate for T=0.03-3 GK based on the experimental information from literature and by}\\
\parbox[b][0.3cm]{17.7cm}{considering the tails of broad resonances. The S-factor was computed. Using the nucleosynthesis code SHIVA, they performed a}\\
\parbox[b][0.3cm]{17.7cm}{nova nucleosynthesis calculation using their updated reaction rate.}\\
\parbox[b][0.3cm]{17.7cm}{\addtolength{\parindent}{-0.2in}\href{https://www.nndc.bnl.gov/nsr/nsrlink.jsp?2001Ba49,B}{2001Ba49}: Deduced the contributions of individual resonances and the total \ensuremath{^{\textnormal{18}}}F(p,\ensuremath{\alpha}) reaction rate for T\ensuremath{\leq}2 GK and discussed the}\\
\parbox[b][0.3cm]{17.7cm}{astrophysical implications.}\\
\parbox[b][0.3cm]{17.7cm}{\addtolength{\parindent}{-0.2in}\href{https://www.nndc.bnl.gov/nsr/nsrlink.jsp?2002Il05,B}{2002Il05}: Recommended the reaction rate of (\href{https://www.nndc.bnl.gov/nsr/nsrlink.jsp?2000Co33,B}{2000Co33}), varied this rate by a factor of 30 up and down, and studied its effect on}\\
\parbox[b][0.3cm]{17.7cm}{nova nucleosynthesis.}\\
\parbox[b][0.3cm]{17.7cm}{\addtolength{\parindent}{-0.2in}\href{https://www.nndc.bnl.gov/nsr/nsrlink.jsp?2002Bb02,B}{2002Bb02}: Deduced resonance properties (E\ensuremath{_{\textnormal{r}}}, J\ensuremath{^{\ensuremath{\pi}}}, \ensuremath{\Gamma}\ensuremath{_{\textnormal{p}}}, \ensuremath{\Gamma}\ensuremath{_{\ensuremath{\alpha}}}, and \ensuremath{\theta}\ensuremath{_{\textnormal{p}}^{\textnormal{2}}}) for resonances at E\ensuremath{_{\textnormal{c.m.}}}=8, 26, 38, 287, 330, 665 keV.}\\
\parbox[b][0.3cm]{17.7cm}{Deduced the contributions of those individual resonances and the total \ensuremath{^{\textnormal{18}}}F(p,\ensuremath{\alpha}) reaction rate at T=0.1-0.5 GK. Performed a}\\
\parbox[b][0.3cm]{17.7cm}{multizone postprocessing nova nucleosynthesis calculations and discussed the results.}\\
\parbox[b][0.3cm]{17.7cm}{\addtolength{\parindent}{-0.2in}\href{https://www.nndc.bnl.gov/nsr/nsrlink.jsp?2003Sh25,B}{2003Sh25}: Collected all the published information on the \ensuremath{^{\textnormal{19}}}Ne excited states and deduced the \ensuremath{^{\textnormal{18}}}F(p,\ensuremath{\alpha}) reaction rate at T=0.03-3}\\
\parbox[b][0.3cm]{17.7cm}{GK using \ensuremath{\sim}30 levels of \ensuremath{^{\textnormal{19}}}Ne. The unknown properties of some of these levels were taken from \ensuremath{^{\textnormal{19}}}F mirror levels if possible. The}\\
\parbox[b][0.3cm]{17.7cm}{resonance properties (E\ensuremath{_{\textnormal{r}}}, J\ensuremath{^{\ensuremath{\pi}}}, \ensuremath{\Gamma}\ensuremath{_{\ensuremath{\gamma}}}, \ensuremath{\Gamma}\ensuremath{_{\textnormal{p}}}, \ensuremath{\Gamma}\ensuremath{_{\ensuremath{\alpha}}}, \ensuremath{\theta}\ensuremath{_{\textnormal{p}}^{\textnormal{2}}} and \ensuremath{\theta}\ensuremath{_{\ensuremath{\alpha}}^{\textnormal{2}}}) and their mirror levels in \ensuremath{^{\textnormal{19}}}F for the most important resonances are}\\
\parbox[b][0.3cm]{17.7cm}{evaluated by the authors, who also provided the reaction rate in the REACLIB format. Comparison with literature and astrophysical}\\
\parbox[b][0.3cm]{17.7cm}{implications are discussed.}\\
\parbox[b][0.3cm]{17.7cm}{\addtolength{\parindent}{-0.2in}\href{https://www.nndc.bnl.gov/nsr/nsrlink.jsp?2004Ba63,B}{2004Ba63}: Deduced upper limits on \ensuremath{\Gamma}\ensuremath{_{\textnormal{p}}} (at 90\% C.L.) for the resonances at E\ensuremath{_{\textnormal{c.m.}}}=827, 915, and 1089 keV. Tabulated the}\\
\parbox[b][0.3cm]{17.7cm}{properties for the resonances in the E\ensuremath{_{\textnormal{c.m.}}}=8-1122 keV region. Deduced the \ensuremath{^{\textnormal{18}}}F(p,\ensuremath{\alpha}) reaction rate at T=1-3 GK.}\\
\parbox[b][0.3cm]{17.7cm}{\addtolength{\parindent}{-0.2in}\href{https://www.nndc.bnl.gov/nsr/nsrlink.jsp?2005Ba82,B}{2005Ba82}, \href{https://www.nndc.bnl.gov/nsr/nsrlink.jsp?2005Bb05,B}{2005Bb05}: These authors deduced the total \ensuremath{^{\textnormal{18}}}F(p,\ensuremath{\alpha}) reaction rate and the contributions from individual resonances at}\\
\parbox[b][0.3cm]{17.7cm}{T=1-3 GK and discussed the astrophysical implications of a new resonance on the reaction rate.}\\
\parbox[b][0.3cm]{17.7cm}{\addtolength{\parindent}{-0.2in}\href{https://www.nndc.bnl.gov/nsr/nsrlink.jsp?2005Ko09,B}{2005Ko09}: Deduced the \ensuremath{^{\textnormal{18}}}F(p,\ensuremath{\alpha}) reaction rate at T=0.1-0.5 GK. The properties of \ensuremath{^{\textnormal{19}}}Ne resonances relevant for this temperature}\\
\parbox[b][0.3cm]{17.7cm}{range were deduced based on new mirror level assignments from the \ensuremath{^{\textnormal{2}}}H(\ensuremath{^{\textnormal{18}}}F,p) measurement by (\href{https://www.nndc.bnl.gov/nsr/nsrlink.jsp?2005Ko09,B}{2005Ko09}).}\\
\parbox[b][0.3cm]{17.7cm}{\addtolength{\parindent}{-0.2in}\href{https://www.nndc.bnl.gov/nsr/nsrlink.jsp?2006Ch30,B}{2006Ch30}: Deduced the \ensuremath{^{\textnormal{18}}}F(p,\ensuremath{\alpha}) astrophysical S-factor at E\ensuremath{_{\textnormal{c.m.}}}=5-1000 keV considering interferences between the J\ensuremath{^{\ensuremath{\pi}}}=3/2\ensuremath{^{\textnormal{+}}}}\\
\parbox[b][0.3cm]{17.7cm}{resonances at E\ensuremath{_{\textnormal{c.m.}}}=8, 38, and 665 keV. Performed a nova nucleosynthesis calculation and discussed the implications.}\\
\parbox[b][0.3cm]{17.7cm}{\addtolength{\parindent}{-0.2in}\href{https://www.nndc.bnl.gov/nsr/nsrlink.jsp?2009Mu07,B}{2009Mu07}: Determined the \ensuremath{^{\textnormal{18}}}F(p,\ensuremath{\alpha}) reaction rate at nova temperatures using the resonances observed in their experiment and with}\\
\parbox[b][0.3cm]{17.7cm}{addition of a few higher lying resonances. Their rate was consistent with those of (\href{https://www.nndc.bnl.gov/nsr/nsrlink.jsp?2003De15,B}{2003De15}, \href{https://www.nndc.bnl.gov/nsr/nsrlink.jsp?2005Ko09,B}{2005Ko09}).}\\
\parbox[b][0.3cm]{17.7cm}{\addtolength{\parindent}{-0.2in}\href{https://www.nndc.bnl.gov/nsr/nsrlink.jsp?2009Da07,B}{2009Da07}: Deduced the astrophysical S-factor for the \ensuremath{^{\textnormal{18}}}F(p,\ensuremath{\alpha}) reaction rate at E\ensuremath{_{\textnormal{c.m.}}}=0.1-1.6 MeV.}\\
\parbox[b][0.3cm]{17.7cm}{\addtolength{\parindent}{-0.2in}\href{https://www.nndc.bnl.gov/nsr/nsrlink.jsp?2010Il04,B}{2010Il04}, \href{https://www.nndc.bnl.gov/nsr/nsrlink.jsp?2010Il06,B}{2010Il06}: Re-evaluated the \ensuremath{^{\textnormal{18}}}F(p,\ensuremath{\alpha}) reaction rate and its uncertainty for T=0.01-10 GK using a Monte Carlo technique.}\\
\parbox[b][0.3cm]{17.7cm}{\addtolength{\parindent}{-0.2in}\href{https://www.nndc.bnl.gov/nsr/nsrlink.jsp?2011Ad05,B}{2011Ad05}: Determined the \ensuremath{^{\textnormal{18}}}F(p,\ensuremath{\alpha}) reaction rate for T=0.01-0.4 GK using the REACLIB format and by using the \ensuremath{\Gamma}\ensuremath{_{\textnormal{p}}} and \ensuremath{\Gamma}\ensuremath{_{\ensuremath{\alpha}}}}\\
\parbox[b][0.3cm]{17.7cm}{values which were estimated from the measured proton spectroscopic factors for the \ensuremath{^{\textnormal{19}}}Ne*(6420, 6449) levels corresponding to the}\\
\parbox[b][0.3cm]{17.7cm}{resonances at E\ensuremath{_{\textnormal{c.m.}}}=8 and 38 keV, respectively. These authors assumed a J\ensuremath{^{\ensuremath{\pi}}}=3/2\ensuremath{^{-}} for the 8-keV resonance. Discussed the effect of}\\
\parbox[b][0.3cm]{17.7cm}{the sub-threshold resonance corresponding to \ensuremath{^{\textnormal{19}}}Ne*(6289) and interferences between \ensuremath{^{\textnormal{19}}}Ne resonances and their effect on the}\\
\parbox[b][0.3cm]{17.7cm}{reaction rate. Performed nucleosynthesis calculations for novae.}\\
\clearpage
\vspace{0.3cm}
{\bf \small \underline{\ensuremath{^{\textnormal{1}}}H(\ensuremath{^{\textnormal{18}}}F,p),(\ensuremath{^{\textnormal{18}}}F,\ensuremath{\alpha}):res\hspace{0.2in}\href{https://www.nndc.bnl.gov/nsr/nsrlink.jsp?2009Mu07,B}{2009Mu07},\href{https://www.nndc.bnl.gov/nsr/nsrlink.jsp?2012Mo03,B}{2012Mo03} (continued)}}\\
\vspace{0.3cm}
\parbox[b][0.3cm]{17.7cm}{\addtolength{\parindent}{-0.2in}\href{https://www.nndc.bnl.gov/nsr/nsrlink.jsp?2011Be11,B}{2011Be11}: Measured the \ensuremath{^{\textnormal{18}}}F(p,\ensuremath{\alpha}) cross section at E\ensuremath{_{\textnormal{c.m.}}}=250, 330, 453, and 673 keV; deduced the \ensuremath{^{\textnormal{18}}}F(p,\ensuremath{\alpha}) S-factor at E\ensuremath{_{\textnormal{c.m.}}}\ensuremath{\leq}1}\\
\parbox[b][0.3cm]{17.7cm}{MeV using a multi-channel R-matrix analysis via the DREAM computer code; compared the results to the previous data and}\\
\parbox[b][0.3cm]{17.7cm}{R-matrix calculations. Discussed the resonance interferences and astrophysical implications for nova nucleosynthesis.}\\
\parbox[b][0.3cm]{17.7cm}{\addtolength{\parindent}{-0.2in}\href{https://www.nndc.bnl.gov/nsr/nsrlink.jsp?2012Mo03,B}{2012Mo03}: Deduced the astrophysical S-factor at E\ensuremath{_{\textnormal{c.m.}}}\ensuremath{<}0.5 MeV and discussed the interferences between the observed resonances.}\\
\parbox[b][0.3cm]{17.7cm}{\addtolength{\parindent}{-0.2in}\href{https://www.nndc.bnl.gov/nsr/nsrlink.jsp?2013La01,B}{2013La01}: On the contrary to a previous belief that there were two 3/2\ensuremath{^{\textnormal{+}}} states in \ensuremath{^{\textnormal{19}}}Ne near the proton threshold, these authors}\\
\parbox[b][0.3cm]{17.7cm}{deduced J\ensuremath{^{\ensuremath{\pi}}} values (via DWBA calculations) using \ensuremath{^{\textnormal{19}}}F(\ensuremath{^{\textnormal{3}}}He,t) where 3 states were observed at E\ensuremath{_{\textnormal{x}}}(\ensuremath{^{\textnormal{19}}}Ne)=6416, 6440, 6459 keV}\\
\parbox[b][0.3cm]{17.7cm}{near the proton threshold. The deduced J\ensuremath{^{\ensuremath{\pi}}} revealed that none of these state are consistent with 3/2\ensuremath{^{\textnormal{+}}} states. Deduced the \ensuremath{^{\textnormal{18}}}F(p,\ensuremath{\alpha})}\\
\parbox[b][0.3cm]{17.7cm}{reaction rate at nova temperatures with possible J permutations; performed a nova nucleosynthesis calculation; claimed that the}\\
\parbox[b][0.3cm]{17.7cm}{unknown proton width of the 48-keV resonance makes the reaction rate uncertain.}\\
\parbox[b][0.3cm]{17.7cm}{\addtolength{\parindent}{-0.2in}\href{https://www.nndc.bnl.gov/nsr/nsrlink.jsp?2015Ch41,B}{2015Ch41}: Populated resonances in \ensuremath{^{\textnormal{19}}}Ne at excitation energies 6255, 6459, 6536, 6754, 6966, and 7074 keV using Trojan Horse}\\
\parbox[b][0.3cm]{17.7cm}{Method (THM). Deduced the astrophysical S-factor for nova temperatures based on these resonances.}\\
\parbox[b][0.3cm]{17.7cm}{\addtolength{\parindent}{-0.2in}\href{https://www.nndc.bnl.gov/nsr/nsrlink.jsp?2015Ba51,B}{2015Ba51}, \href{https://www.nndc.bnl.gov/nsr/nsrlink.jsp?2015BaZQ,B}{2015BaZQ}: Deduced the \ensuremath{^{\textnormal{18}}}F(p,\ensuremath{\alpha}) reaction rate at nova temperatures and its S-factor at E\ensuremath{_{\textnormal{c.m.}}}\ensuremath{<}1 MeV using the}\\
\parbox[b][0.3cm]{17.7cm}{R-matrix code AZURE2 and resonance properties from literature. Discussed the astrophysical implications.}\\
\parbox[b][0.3cm]{17.7cm}{\addtolength{\parindent}{-0.2in}\href{https://www.nndc.bnl.gov/nsr/nsrlink.jsp?2015Pa46,B}{2015Pa46}: Confirmed the triplet of states observed around 6.4 MeV by (\href{https://www.nndc.bnl.gov/nsr/nsrlink.jsp?2013La01,B}{2013La01}). Proposed that the sub-threshold 6.29-MeV state}\\
\parbox[b][0.3cm]{17.7cm}{is either a doublet or a broad state, and suggested that the region around the 6.86-MeV state may have additional unidentified}\\
\parbox[b][0.3cm]{17.7cm}{levels.}\\
\parbox[b][0.3cm]{17.7cm}{\addtolength{\parindent}{-0.2in}\href{https://www.nndc.bnl.gov/nsr/nsrlink.jsp?2016Pi01,B}{2016Pi01}: Deduced the astrophysical S-factor for E\ensuremath{_{\textnormal{c.m.}}}\ensuremath{<}0.9 MeV and determined the \ensuremath{^{\textnormal{18}}}F(p,\ensuremath{\alpha}) reaction rate at T=0.05-1.15 GK}\\
\parbox[b][0.3cm]{17.7cm}{based on the \ensuremath{^{\textnormal{19}}}Ne states measured by (\href{https://www.nndc.bnl.gov/nsr/nsrlink.jsp?2015Ch41,B}{2015Ch41}).}\\
\parbox[b][0.3cm]{17.7cm}{\addtolength{\parindent}{-0.2in}\href{https://www.nndc.bnl.gov/nsr/nsrlink.jsp?2017Ba42,B}{2017Ba42}: Deduced the \ensuremath{^{\textnormal{18}}}F(p,\ensuremath{\alpha}) reaction rate in the REACLIB format; discussed the interferences between sub- and}\\
\parbox[b][0.3cm]{17.7cm}{near-threshold resonances and higher-lying broad \textit{s}-wave resonances; performed a hydrodynamic nucleosynthesis calculation and}\\
\parbox[b][0.3cm]{17.7cm}{discussed the astrophysical implications.}\\
\parbox[b][0.3cm]{17.7cm}{\addtolength{\parindent}{-0.2in}\href{https://www.nndc.bnl.gov/nsr/nsrlink.jsp?2017La12,B}{2017La12}, \href{https://www.nndc.bnl.gov/nsr/nsrlink.jsp?2019LaZX,B}{2019LaZX}: Using the resonance parameters from an R-matrix analysis of the THM data of (\href{https://www.nndc.bnl.gov/nsr/nsrlink.jsp?2015Ch41,B}{2015Ch41}, \href{https://www.nndc.bnl.gov/nsr/nsrlink.jsp?2016Pi01,B}{2016Pi01}), the}\\
\parbox[b][0.3cm]{17.7cm}{authors deduced the \ensuremath{^{\textnormal{18}}}F(p,\ensuremath{\alpha}) reaction rate at T=0.007-1 GK and discussed the changes to the S-factor from various interferences}\\
\parbox[b][0.3cm]{17.7cm}{assumed between the \ensuremath{^{\textnormal{19}}}Ne resonances. Comparison with the rates of (\href{https://www.nndc.bnl.gov/nsr/nsrlink.jsp?2010Cy01,B}{2010Cy01}, \href{https://www.nndc.bnl.gov/nsr/nsrlink.jsp?2015Ba51,B}{2015Ba51}) and the astrophysical implication of}\\
\parbox[b][0.3cm]{17.7cm}{the rate are discussed.}\\
\parbox[b][0.3cm]{17.7cm}{\addtolength{\parindent}{-0.2in}\href{https://www.nndc.bnl.gov/nsr/nsrlink.jsp?2019Ka15,B}{2019Ka15}: Deduced the astrophysical S-factor for the \ensuremath{^{\textnormal{18}}}F(p,\ensuremath{\alpha}) reaction at E\ensuremath{_{\textnormal{c.m.}}}\ensuremath{<}1 MeV using AZURE2 R-matrix code based on}\\
\parbox[b][0.3cm]{17.7cm}{the \ensuremath{^{\textnormal{19}}}Ne resonances with E\ensuremath{_{\textnormal{c.m.}}}={\textminus}278 to 1380 keV (the negative resonance energy indicates a sub-threshold resonance); discussed}\\
\parbox[b][0.3cm]{17.7cm}{the interferences of the J\ensuremath{^{\ensuremath{\pi}}}=1/2\ensuremath{^{\textnormal{+}}} and 3/2\ensuremath{^{\textnormal{+}}} resonances.}\\
\parbox[b][0.3cm]{17.7cm}{\addtolength{\parindent}{-0.2in}\href{https://www.nndc.bnl.gov/nsr/nsrlink.jsp?2019Ha08,B}{2019Ha08}: Deduced \ensuremath{\Gamma}\ensuremath{_{\textnormal{p}}}\ensuremath{\leq}3.9\ensuremath{\times}10\ensuremath{^{\textnormal{$-$29}}} keV and \ensuremath{\Gamma}\ensuremath{_{\ensuremath{\alpha}}}=1.2 keV; and \ensuremath{\Gamma}\ensuremath{_{\textnormal{p}}}\ensuremath{\leq}8.4\ensuremath{\times}10\ensuremath{^{\textnormal{$-$18}}} keV and \ensuremath{\Gamma}\ensuremath{_{\ensuremath{\alpha}}}=1.3 keV for the \ensuremath{^{\textnormal{19}}}Ne*(6423, 6441)}\\
\parbox[b][0.3cm]{17.7cm}{states (considering J\ensuremath{^{\ensuremath{\pi}}}=3/2\ensuremath{^{\textnormal{+}}} for both), which correspond to the resonance energies of 13 keV and 31 keV, respectively. Using these}\\
\parbox[b][0.3cm]{17.7cm}{together with other resonances in the center-of-mass energy range of {\textminus}124 to 1461 keV, they deduced the astrophysical S-factor at}\\
\parbox[b][0.3cm]{17.7cm}{E\ensuremath{_{\textnormal{c.m.}}}\ensuremath{<}1 MeV using R-matrix analysis via AZURE2 code (channel radius=5.2 fm). The resulting S-factor was used to calculate the}\\
\parbox[b][0.3cm]{17.7cm}{\ensuremath{^{\textnormal{18}}}F(p,\ensuremath{\alpha}) reaction rate at T=0.05-0.4 GK. They performed nucleosynthesis calculations and discussed the results.}\\
\parbox[b][0.3cm]{17.7cm}{\addtolength{\parindent}{-0.2in}\href{https://www.nndc.bnl.gov/nsr/nsrlink.jsp?2020Ha31,B}{2020Ha31}: Deduced \ensuremath{\Gamma}\ensuremath{_{\ensuremath{\alpha}}}, \ensuremath{\Gamma}\ensuremath{_{\textnormal{p}}}, \ensuremath{\Gamma}\ensuremath{_{\ensuremath{\gamma}}}, E\ensuremath{_{\textnormal{res}}}, \ensuremath{\theta}\ensuremath{_{\textnormal{p}}^{\textnormal{2}}}, and J\ensuremath{^{\ensuremath{\pi}}} for the \ensuremath{^{\textnormal{18}}}F(p,\ensuremath{\alpha}) resonances in the E\ensuremath{_{\textnormal{c.m.}}}={\textminus}124-1461 keV region. Deduced}\\
\parbox[b][0.3cm]{17.7cm}{the astrophysical S-factor at E\ensuremath{_{\textnormal{c.m.}}}\ensuremath{<}1 MeV using R-matrix formalism via AZURE2 to take into account the interferences. Deduced}\\
\parbox[b][0.3cm]{17.7cm}{the \ensuremath{^{\textnormal{18}}}F(p,\ensuremath{\alpha}) reaction rate at T=0.05-0.4 GK. Performed nucleosynthesis calculations and discussed the results. See (\href{https://www.nndc.bnl.gov/nsr/nsrlink.jsp?2021Ka51,B}{2021Ka51} for a}\\
\parbox[b][0.3cm]{17.7cm}{detailed assessment of the reaction rate deduced by \href{https://www.nndc.bnl.gov/nsr/nsrlink.jsp?2020Ha31,B}{2020Ha31}).}\\
\parbox[b][0.3cm]{17.7cm}{\addtolength{\parindent}{-0.2in}\href{https://www.nndc.bnl.gov/nsr/nsrlink.jsp?2021Ri04,B}{2021Ri04}: Deduced the \ensuremath{^{\textnormal{18}}}F(p,\ensuremath{\alpha}) reaction rate and its S-factor (for E\ensuremath{_{\textnormal{c.m.}}}\ensuremath{<}1.6 MeV) using R-matrix formalism via AZURE2 code}\\
\parbox[b][0.3cm]{17.7cm}{focusing on the impact of sub-threshold states at 6008 (a tentative state), 6132, and 6286 keV in \ensuremath{^{\textnormal{19}}}Ne. Discussed the implications}\\
\parbox[b][0.3cm]{17.7cm}{of the tentative state at 6008 keV.}\\
\parbox[b][0.3cm]{17.7cm}{\addtolength{\parindent}{-0.2in}\href{https://www.nndc.bnl.gov/nsr/nsrlink.jsp?2021Ka51,B}{2021Ka51}: Evaluated the \ensuremath{^{\textnormal{18}}}F(p,\ensuremath{\alpha}) reaction rate at T=0.1-10 GK based on various experimental results published after 2000.}\\
\parbox[b][0.3cm]{17.7cm}{Deduced resonance parameters (E\ensuremath{_{\textnormal{x}}}, E\ensuremath{_{\textnormal{c.m.}}}, J\ensuremath{^{\ensuremath{\pi}}}, ANC, \ensuremath{\Gamma}\ensuremath{_{\textnormal{p}}}, and \ensuremath{\Gamma}\ensuremath{_{\ensuremath{\alpha}}}) for the \ensuremath{^{\textnormal{19}}}Ne states that influence this rate at those}\\
\parbox[b][0.3cm]{17.7cm}{temperatures. Deduced the astrophysical S-factor at E\ensuremath{_{\textnormal{c.m.}}}\ensuremath{<}1 MeV and the reaction rate using R-matrix code AZURE2. The}\\
\parbox[b][0.3cm]{17.7cm}{sensitivity of the reaction rate to each significant parameter$'$s uncertainty was studied. Hydrodynamic simulations of nova explosions}\\
\parbox[b][0.3cm]{17.7cm}{and nucleosynthesis calculations were performed. Results, as well as the interferences between \ensuremath{^{\textnormal{19}}}Ne resonances are discussed.}\\
\parbox[b][0.3cm]{17.7cm}{\addtolength{\parindent}{-0.2in}\href{https://www.nndc.bnl.gov/nsr/nsrlink.jsp?2023Po03,B}{2023Po03}: Deduced the center-of-mass energies for the relevant resonances in the E\ensuremath{_{\textnormal{c.m.}}}={\textminus}277 keV to 1571 keV range; deduced J\ensuremath{^{\ensuremath{\pi}}}}\\
\parbox[b][0.3cm]{17.7cm}{values, ANC, \ensuremath{\Gamma}\ensuremath{_{\textnormal{p}}} and \ensuremath{\Gamma}\ensuremath{_{\ensuremath{\alpha}}} for these resonances; obtained the astrophysical S-factor for the \ensuremath{^{\textnormal{18}}}F(p,\ensuremath{\alpha}) reaction at E\ensuremath{_{\textnormal{c.m.}}}\ensuremath{\leq}1.7 MeV}\\
\parbox[b][0.3cm]{17.7cm}{using R-matrix analysis via AZURE2; determined the \ensuremath{^{\textnormal{18}}}F(p,\ensuremath{\alpha}) reaction rate at T\ensuremath{\leq}0.5 GK. Discussed the interferences between}\\
\parbox[b][0.3cm]{17.7cm}{resonances and their effects on the uncertainty of the reaction rate; discussed astrophysical implications.}\\
\parbox[b][0.3cm]{17.7cm}{\addtolength{\parindent}{-0.2in}D. Kahl, H. Yamaguchi and S. Hayakawa, Front. Phys. 11 (2023) 1189040: Reviewed the \ensuremath{^{\textnormal{18}}}F(p,\ensuremath{\alpha}) reaction rate and discussed the}\\
\parbox[b][0.3cm]{17.7cm}{resonance interferences.}\\
\vspace{12pt}
\clearpage
\vspace{0.3cm}
{\bf \small \underline{\ensuremath{^{\textnormal{1}}}H(\ensuremath{^{\textnormal{18}}}F,p),(\ensuremath{^{\textnormal{18}}}F,\ensuremath{\alpha}):res\hspace{0.2in}\href{https://www.nndc.bnl.gov/nsr/nsrlink.jsp?2009Mu07,B}{2009Mu07},\href{https://www.nndc.bnl.gov/nsr/nsrlink.jsp?2012Mo03,B}{2012Mo03} (continued)}}\\
\vspace{0.3cm}
\underline{$^{19}$Ne Levels}\\
\vspace{0.34cm}
\parbox[b][0.3cm]{17.7cm}{\addtolength{\parindent}{-0.254cm}\textit{Notes}:}\\
\parbox[b][0.3cm]{17.7cm}{\addtolength{\parindent}{-0.254cm}(1) The uncertainties in E\ensuremath{_{\textnormal{c.m.}}}, \ensuremath{\Gamma} and \ensuremath{\Gamma}\ensuremath{_{\textnormal{p}}}/\ensuremath{\Gamma} reported by (\href{https://www.nndc.bnl.gov/nsr/nsrlink.jsp?2000Ba87,B}{2000Ba87}, \href{https://www.nndc.bnl.gov/nsr/nsrlink.jsp?2001Ba49,B}{2001Ba49}) are quadrature sums of the statistical and}\\
\parbox[b][0.3cm]{17.7cm}{systematic uncertainties (see text).}\\
\parbox[b][0.3cm]{17.7cm}{\addtolength{\parindent}{-0.254cm}(2) (\href{https://www.nndc.bnl.gov/nsr/nsrlink.jsp?2005Ko09,B}{2005Ko09}) deduced neutron spectroscopic factors for \ensuremath{^{\textnormal{19}}}F* states and used them to determine proton partial widths for \ensuremath{^{\textnormal{19}}}Ne*}\\
\parbox[b][0.3cm]{17.7cm}{mirror states assuming S\ensuremath{_{\textnormal{p}}}=S\ensuremath{_{\textnormal{n}}}.}\\
\parbox[b][0.3cm]{17.7cm}{\addtolength{\parindent}{-0.254cm}(3) Evaluator notes that (\href{https://www.nndc.bnl.gov/nsr/nsrlink.jsp?2006Ch30,B}{2006Ch30}) found that each of the 8 sign combinations for interferences of the 3 resonances at}\\
\parbox[b][0.3cm]{17.7cm}{E\ensuremath{_{\textnormal{c.m.}}}(p+\ensuremath{^{\textnormal{18}}}F)=8, 38, and 665 keV gave a different \ensuremath{\chi}\ensuremath{^{\textnormal{2}}}. (\href{https://www.nndc.bnl.gov/nsr/nsrlink.jsp?2007Du09,B}{2007Du09}) found this conclusion surprising since a common change of all}\\
\parbox[b][0.3cm]{17.7cm}{signs should not modify the S-factor.}\\
\parbox[b][0.3cm]{17.7cm}{\addtolength{\parindent}{-0.254cm}(4) The widths deduced by (\href{https://www.nndc.bnl.gov/nsr/nsrlink.jsp?2006Fo03,B}{2006Fo03}) are obtained using the experimental widths of the \ensuremath{^{\textnormal{19}}}F* mirror levels and under the}\\
\parbox[b][0.3cm]{17.7cm}{assumption of mirror symmetry. For spectroscopic factors deduced by the theoretical work of (\href{https://www.nndc.bnl.gov/nsr/nsrlink.jsp?2006Fo03,B}{2006Fo03}), see Table I in that work.}\\
\parbox[b][0.3cm]{17.7cm}{These were deduced using an \ensuremath{\alpha}-particle well with R=3.45 fm and a=0.60 fm. Same assumptions were used for calculating \ensuremath{\Gamma}\ensuremath{_{\textnormal{sp,}\ensuremath{\alpha}}}.}\\
\parbox[b][0.3cm]{17.7cm}{The uncertainties in those S\ensuremath{_{\ensuremath{\alpha}}} and S\ensuremath{_{\textnormal{p}}} come only from experimental uncertainties in the measured widths. \ensuremath{\Gamma}\ensuremath{_{\textnormal{sp,p}}} values are proton}\\
\parbox[b][0.3cm]{17.7cm}{single-particle widths deduced by (\href{https://www.nndc.bnl.gov/nsr/nsrlink.jsp?2006Fo03,B}{2006Fo03}) using R=1.25(18)\ensuremath{^{\textnormal{1/3}}} and a=0.65 fm.}\\
\parbox[b][0.3cm]{17.7cm}{\addtolength{\parindent}{-0.254cm}(5) (\href{https://www.nndc.bnl.gov/nsr/nsrlink.jsp?2009Mu07,B}{2009Mu07}) deduced E\ensuremath{_{\textnormal{c.m.}}}(\ensuremath{^{\textnormal{18}}}F+p), J, \ensuremath{\Gamma}\ensuremath{_{\textnormal{p}}}, and \ensuremath{\Gamma}\ensuremath{_{\ensuremath{\alpha}}} from R-matrix analysis with the channel radius set to 5.0 fm. Evaluator}\\
\parbox[b][0.3cm]{17.7cm}{notes that the fit to the (p,\ensuremath{\alpha}) data of (\href{https://www.nndc.bnl.gov/nsr/nsrlink.jsp?2009Mu07,B}{2009Mu07}) in the E\ensuremath{_{\textnormal{c.m.}}}=0.8-1 MeV region is poorer than elsewhere. This is acknowledged}\\
\parbox[b][0.3cm]{17.7cm}{by the authors, and they mention that additional states or a need for revision of their results is required.}\\
\parbox[b][0.3cm]{17.7cm}{\addtolength{\parindent}{-0.254cm}(6) (\href{https://www.nndc.bnl.gov/nsr/nsrlink.jsp?2010Fo07,B}{2010Fo07}) computed \ensuremath{\Gamma}\ensuremath{_{\textnormal{sp}}} using r\ensuremath{_{\textnormal{0}}}=1.40 fm and a=0.60 fm for the geometry of the \ensuremath{\alpha} potential well, where R=r\ensuremath{_{\textnormal{0}}}(15)\ensuremath{^{\textnormal{1/3}}}.}\\
\parbox[b][0.3cm]{17.7cm}{These authors deduced S\ensuremath{_{\ensuremath{\alpha}}} (\ensuremath{\alpha} spectroscopic factors) from the mirror states in \ensuremath{^{\textnormal{19}}}F*.}\\
\parbox[b][0.3cm]{17.7cm}{\addtolength{\parindent}{-0.254cm}(7) (\href{https://www.nndc.bnl.gov/nsr/nsrlink.jsp?2012Mo03,B}{2012Mo03}) deduced E\ensuremath{_{\textnormal{c.m.}}}(\ensuremath{^{\textnormal{18}}}F+p), \ensuremath{\Gamma}\ensuremath{_{\textnormal{p}}}, \ensuremath{\Gamma}\ensuremath{_{\ensuremath{\alpha}}}, and \ensuremath{\Gamma}=\ensuremath{\Gamma}\ensuremath{_{\textnormal{p}}}+\ensuremath{\Gamma}\ensuremath{_{\ensuremath{\alpha}}} from R-matrix analysis with the channel radius set to 5 fm.}\\
\parbox[b][0.3cm]{17.7cm}{Due to the lack of angular distribution information in this study, their R-matrix analysis did not provide independent J\ensuremath{^{\ensuremath{\pi}}}}\\
\parbox[b][0.3cm]{17.7cm}{assignments. These authors recommended the J\ensuremath{^{\ensuremath{\pi}}} values taken from literature that resulted in their best R-matrix fit.}\\
\vspace{0.34cm}

\begin{textblock}{29}(0,27.3)
Continued on next page (footnotes at end of table)
\end{textblock}
\clearpage
%
\begin{textblock}{29}(0,27.3)
Continued on next page (footnotes at end of table)
\end{textblock}
\clearpage
%
\begin{textblock}{29}(0,27.3)
Continued on next page (footnotes at end of table)
\end{textblock}
\clearpage
%
\begin{textblock}{29}(0,27.3)
Continued on next page (footnotes at end of table)
\end{textblock}
\clearpage
\vspace*{-0.5cm}
{\bf \small \underline{\ensuremath{^{\textnormal{1}}}H(\ensuremath{^{\textnormal{18}}}F,p),(\ensuremath{^{\textnormal{18}}}F,\ensuremath{\alpha}):res\hspace{0.2in}\href{https://www.nndc.bnl.gov/nsr/nsrlink.jsp?2009Mu07,B}{2009Mu07},\href{https://www.nndc.bnl.gov/nsr/nsrlink.jsp?2012Mo03,B}{2012Mo03} (continued)}}\\
\vspace{0.3cm}
\underline{$^{19}$Ne Levels (continued)}\\
\vspace{0.3cm}
\parbox[b][0.3cm]{17.7cm}{\makebox[1ex]{\ensuremath{^{\hypertarget{NE2LEVEL0}{a}}}} This state is not directly measured using the \ensuremath{^{\textnormal{1}}}H(\ensuremath{^{\textnormal{18}}}F,p) or \ensuremath{^{\textnormal{1}}}H(\ensuremath{^{\textnormal{18}}}F,\ensuremath{\alpha}) reaction.}\\
\parbox[b][0.3cm]{17.7cm}{\makebox[1ex]{\ensuremath{^{\hypertarget{NE2LEVEL1}{b}}}} When possible, measured resonance energies in the center-of-mass have been used to deduce the corresponding E\ensuremath{_{\textnormal{x}}} using}\\
\parbox[b][0.3cm]{17.7cm}{{\ }{\ }S\ensuremath{_{\textnormal{p}}}(\ensuremath{^{\textnormal{19}}}Ne)=6410.0 keV \textit{5} (\href{https://www.nndc.bnl.gov/nsr/nsrlink.jsp?2021Wa16,B}{2021Wa16}). Note that this separation energy was considered to be 6411 keV by (\href{https://www.nndc.bnl.gov/nsr/nsrlink.jsp?1995Co23,B}{1995Co23}, \href{https://www.nndc.bnl.gov/nsr/nsrlink.jsp?1995Re11,B}{1995Re11},}\\
\parbox[b][0.3cm]{17.7cm}{{\ }{\ }\href{https://www.nndc.bnl.gov/nsr/nsrlink.jsp?1996Re05,B}{1996Re05}).}\\
\vspace{0.5cm}
\clearpage
\subsection[\hspace{-0.2cm}\ensuremath{^{\textnormal{1}}}H(\ensuremath{^{\textnormal{18}}}F,\ensuremath{^{\textnormal{19}}}Ne)]{ }
\vspace{-27pt}
\vspace{0.3cm}
\hypertarget{NE3}{{\bf \small \underline{\ensuremath{^{\textnormal{1}}}H(\ensuremath{^{\textnormal{18}}}F,\ensuremath{^{\textnormal{19}}}Ne)\hspace{0.2in}\href{https://www.nndc.bnl.gov/nsr/nsrlink.jsp?1997Re02,B}{1997Re02},\href{https://www.nndc.bnl.gov/nsr/nsrlink.jsp?2016Ak05,B}{2016Ak05}}}}\\
\vspace{4pt}
\vspace{8pt}
\parbox[b][0.3cm]{17.7cm}{\addtolength{\parindent}{-0.2in}\ensuremath{^{\textnormal{18}}}F(p,\ensuremath{\gamma})\ensuremath{^{\textnormal{19}}}Ne resonant reaction in inverse kinematics.}\\
\parbox[b][0.3cm]{17.7cm}{\addtolength{\parindent}{-0.2in}J\ensuremath{^{\ensuremath{\pi}}}(\ensuremath{^{\textnormal{18}}}F\ensuremath{_{\textnormal{g.s.}}})=1\ensuremath{^{\textnormal{+}}} and J\ensuremath{^{\ensuremath{\pi}}}(p)=1/2\ensuremath{^{\textnormal{+}}}.}\\
\parbox[b][0.3cm]{17.7cm}{\addtolength{\parindent}{-0.2in}\href{https://www.nndc.bnl.gov/nsr/nsrlink.jsp?1997Re02,B}{1997Re02}, \href{https://www.nndc.bnl.gov/nsr/nsrlink.jsp?1998Re24,B}{1998Re24}: \ensuremath{^{\textnormal{1}}}H(\ensuremath{^{\textnormal{18}}}F,\ensuremath{^{\textnormal{19}}}Ne) E\ensuremath{_{\textnormal{c.m.}}}=670 keV or E\ensuremath{_{\textnormal{lab}}}=13.4 MeV (at the center of target); attempted to measure the}\\
\parbox[b][0.3cm]{17.7cm}{excitation function of the \ensuremath{^{\textnormal{18}}}F(p,\ensuremath{\gamma}) reaction in inverse kinematics using the Fragment Mass Analyzer (FMA) at Argonne National}\\
\parbox[b][0.3cm]{17.7cm}{Laboratory together with a position sensitive \ensuremath{\Delta}E-E telescope at its focal plane. Measured 3 events, which were determined to be}\\
\parbox[b][0.3cm]{17.7cm}{\ensuremath{^{\textnormal{19}}}F. No \ensuremath{^{\textnormal{19}}}Ne events were detected. Deduced an upper limits of \ensuremath{\Gamma}\ensuremath{_{\ensuremath{\gamma}}}\ensuremath{\leq}3 eV and \ensuremath{\omega}\ensuremath{\gamma}\ensuremath{\leq}740 meV for the \ensuremath{^{\textnormal{19}}}Ne*(7.07 MeV, 3/2\ensuremath{^{\textnormal{+}}})}\\
\parbox[b][0.3cm]{17.7cm}{resonance. This \ensuremath{\Gamma}\ensuremath{_{\ensuremath{\gamma}}} corresponds to about 2\% of the single particle width for an E1 and 40\% for an M1 transition. Discussed the}\\
\parbox[b][0.3cm]{17.7cm}{astrophysical implications.}\\
\parbox[b][0.3cm]{17.7cm}{\addtolength{\parindent}{-0.2in}\href{https://www.nndc.bnl.gov/nsr/nsrlink.jsp?1997Re05,B}{1997Re05}: \ensuremath{^{\textnormal{1}}}H(\ensuremath{^{\textnormal{18}}}F,\ensuremath{^{\textnormal{19}}}Ne) E=11.7-15.1 MeV and E\ensuremath{_{\textnormal{c.m.}}}=670 keV; momentum analyzed and measured the \ensuremath{^{\textnormal{19}}}Ne recoils using a}\\
\parbox[b][0.3cm]{17.7cm}{position sensitive avalanche counter and an ionization chamber at the focal plane of the FMA spectrograph at Argonne National}\\
\parbox[b][0.3cm]{17.7cm}{Laboratory. This system suppressed 10\ensuremath{^{\textnormal{+12}}} beam particles per (p,\ensuremath{\gamma}) reaction product. Deduced an upper limit of \ensuremath{\sigma}=42 \ensuremath{\mu}b for the}\\
\parbox[b][0.3cm]{17.7cm}{\ensuremath{^{\textnormal{18}}}F(p,\ensuremath{\gamma}) reaction at E\ensuremath{_{\textnormal{c.m.}}}=670 keV. Obtained upper limits of \ensuremath{\omega}\ensuremath{\gamma}\ensuremath{_{\textnormal{(p,}\ensuremath{\gamma}\textnormal{)}}}\ensuremath{\leq}740 meV and \ensuremath{\Gamma}\ensuremath{_{\ensuremath{\gamma}}}\ensuremath{\leq}3 eV for the \ensuremath{^{\textnormal{19}}}Ne*(7063) state.}\\
\parbox[b][0.3cm]{17.7cm}{Discussed the astrophysical implications.}\\
\parbox[b][0.3cm]{17.7cm}{\addtolength{\parindent}{-0.2in}\href{https://www.nndc.bnl.gov/nsr/nsrlink.jsp?2013Ak03,B}{2013Ak03}, \href{https://www.nndc.bnl.gov/nsr/nsrlink.jsp?2016Ak05,B}{2016Ak05}: \ensuremath{^{\textnormal{1}}}H(\ensuremath{^{\textnormal{18}}}F,\ensuremath{^{\textnormal{19}}}Ne) E=12.9 MeV corresponding to E\ensuremath{_{\textnormal{c.m.}}}=665 keV; measured time-of-flight and energy losses of}\\
\parbox[b][0.3cm]{17.7cm}{the \ensuremath{^{\textnormal{19}}}Ne recoils associated with the population of the 665-keV resonance using the DRAGON recoil separator. The authors}\\
\parbox[b][0.3cm]{17.7cm}{deduced 2 events with a statistical significance of 2.0 counts \textit{+45{\textminus}17} (at 95\% C.L.) and 2.0 counts \textit{+18{\textminus}11} (at 68\% C.L.) using the}\\
\parbox[b][0.3cm]{17.7cm}{profile likelihood technique. No capture \ensuremath{\gamma} rays were measured in coincidence with these 2 recoils. The authors deduced}\\
\parbox[b][0.3cm]{17.7cm}{\ensuremath{\omega}\ensuremath{\gamma}\ensuremath{_{\textnormal{(p,}\ensuremath{\gamma}\textnormal{)}}}=26 meV \textit{+59{\textminus}22} and \ensuremath{\omega}\ensuremath{\gamma}\ensuremath{_{\textnormal{(p,}\ensuremath{\gamma}\textnormal{)}}}=26 meV \textit{+24{\textminus}14} at the 95\% and 68\% C.L., respectively; and \ensuremath{\Gamma}\ensuremath{_{\ensuremath{\gamma}}}=101 meV \textit{+226{\textminus}86} and}\\
\parbox[b][0.3cm]{17.7cm}{\ensuremath{\Gamma}\ensuremath{_{\ensuremath{\gamma}}}=101 meV \textit{+91{\textminus}55} at the 95\% and 68\% C.L., respectively, for this resonance (\href{https://www.nndc.bnl.gov/nsr/nsrlink.jsp?2016Ak05,B}{2016Ak05}: Results supersede those of}\\
\parbox[b][0.3cm]{17.7cm}{\href{https://www.nndc.bnl.gov/nsr/nsrlink.jsp?2013Ak03,B}{2013Ak03}). The final results are a factor of 10 smaller than what was previously assumed. Deduced the \ensuremath{^{\textnormal{18}}}F(p,\ensuremath{\gamma}) reaction rate and}\\
\parbox[b][0.3cm]{17.7cm}{discussed the astrophysical implications.}\\
\vspace{0.385cm}
\parbox[b][0.3cm]{17.7cm}{\addtolength{\parindent}{-0.2in}\textit{The \ensuremath{^{18}}F(p,\ensuremath{\gamma}) Astrophysical Reaction Rate}:}\\
\parbox[b][0.3cm]{17.7cm}{\addtolength{\parindent}{-0.2in}\textbf{Foreword:}}\\
\parbox[b][0.3cm]{17.7cm}{\addtolength{\parindent}{-0.2in}Most of the following studies are experimental and are presented in this or the other individual reaction datasets. The measured}\\
\parbox[b][0.3cm]{17.7cm}{resonances and the measured or deduced resonance properties from the following studies are important for the determination of the}\\
\parbox[b][0.3cm]{17.7cm}{\ensuremath{^{\textnormal{18}}}F(p,\ensuremath{\gamma})\ensuremath{^{\textnormal{19}}}Ne astrophysical reaction rate.}\\
\vspace{0.385cm}
\parbox[b][0.3cm]{17.7cm}{\addtolength{\parindent}{-0.2in}R. V. Wagoner, W. A., Fowler, and F. Hoyle, Astrophys. J., 148 (1967) 3, R. V. Wagoner, Astrophys. J. Suppl. Ser., 18 (1969)}\\
\parbox[b][0.3cm]{17.7cm}{247: Calculated the parameterized \ensuremath{^{\textnormal{18}}}F(p,\ensuremath{\gamma}) reaction rate as a function of temperature in GK.}\\
\parbox[b][0.3cm]{17.7cm}{\addtolength{\parindent}{-0.2in}\href{https://www.nndc.bnl.gov/nsr/nsrlink.jsp?1979Wo07,B}{1979Wo07}: Calculated statistical, parameterized reaction rates at T=0.05-10 GK.}\\
\parbox[b][0.3cm]{17.7cm}{\addtolength{\parindent}{-0.2in}M. Wiescher and K.-U., Kettner, Astrophys. J. 263 (1982) 891: Deduced the resonance properties (E\ensuremath{_{\textnormal{r}}}, J\ensuremath{^{\ensuremath{\pi}}}, \ensuremath{\Gamma}\ensuremath{_{\textnormal{p}}}, \ensuremath{\Gamma}\ensuremath{_{\ensuremath{\alpha}}}, and \ensuremath{\Gamma}\ensuremath{_{\ensuremath{\gamma}}}) for}\\
\parbox[b][0.3cm]{17.7cm}{the resonances associated with the \ensuremath{^{\textnormal{19}}}Ne states at E\ensuremath{_{\textnormal{x}}}=6437, 6500, 6540, 6742, 6790, and 6862 keV. Using these properties, they}\\
\parbox[b][0.3cm]{17.7cm}{deduced the \ensuremath{^{\textnormal{18}}}F(p,\ensuremath{\gamma}) reaction rate for T=0.04-0.6 GK. Deduced the ratio for \ensuremath{^{\textnormal{18}}}F(p,\ensuremath{\alpha})/\ensuremath{^{\textnormal{18}}}F(p,\ensuremath{\gamma}) rates. Compared the \ensuremath{^{\textnormal{18}}}F(p,\ensuremath{\gamma}) rate}\\
\parbox[b][0.3cm]{17.7cm}{with the ones previously calculated and discussed the astrophysical implications.}\\
\parbox[b][0.3cm]{17.7cm}{\addtolength{\parindent}{-0.2in}\href{https://www.nndc.bnl.gov/nsr/nsrlink.jsp?1992Ch50,B}{1992Ch50}: Reviewed the \ensuremath{^{\textnormal{18}}}F(p,\ensuremath{\gamma}) and \ensuremath{^{\textnormal{18}}}F(p,\ensuremath{\alpha}) reaction rates and recommended those of R. K. Wallace and S. E., Woosley,}\\
\parbox[b][0.3cm]{17.7cm}{Astrophys. J. Suppl. 45 (1981) 389 and (\href{https://www.nndc.bnl.gov/nsr/nsrlink.jsp?1990Ma05,B}{1990Ma05}).}\\
\parbox[b][0.3cm]{17.7cm}{\addtolength{\parindent}{-0.2in}\href{https://www.nndc.bnl.gov/nsr/nsrlink.jsp?1997Re02,B}{1997Re02}: Deduced the upper limit contribution of the \ensuremath{^{\textnormal{19}}}Ne*(7.07 MeV, 3/2\ensuremath{^{\textnormal{+}}}) resonance to the \ensuremath{^{\textnormal{18}}}F(p,\ensuremath{\gamma}) reaction rate at T=0.4-2}\\
\parbox[b][0.3cm]{17.7cm}{GK. Deduced the upper and lower limits for the ratio of the \ensuremath{^{\textnormal{18}}}F(p,\ensuremath{\alpha})/\ensuremath{^{\textnormal{18}}}F(p,\ensuremath{\gamma}) reaction rates at T=0.4-2 GK and discussed the}\\
\parbox[b][0.3cm]{17.7cm}{astrophysical implications.}\\
\parbox[b][0.3cm]{17.7cm}{\addtolength{\parindent}{-0.2in}\href{https://www.nndc.bnl.gov/nsr/nsrlink.jsp?1997Re05,B}{1997Re05}: Deduced resonance properties for \ensuremath{^{\textnormal{19}}}Ne relevant states from those of the mirror levels in \ensuremath{^{\textnormal{19}}}F. Deduced the \ensuremath{^{\textnormal{18}}}F(p,\ensuremath{\gamma})}\\
\parbox[b][0.3cm]{17.7cm}{reaction rate and the contributions of individual resonances to this rate for T=0.4-2 GK. Reported that the production of \ensuremath{^{\textnormal{19}}}Ne at}\\
\parbox[b][0.3cm]{17.7cm}{these temperatures is dominated by the \ensuremath{^{\textnormal{15}}}O(\ensuremath{\alpha},\ensuremath{\gamma}) reaction and that the \ensuremath{^{\textnormal{18}}}F(p,\ensuremath{\gamma}) rate plays a negligible role.}\\
\parbox[b][0.3cm]{17.7cm}{\addtolength{\parindent}{-0.2in}\href{https://www.nndc.bnl.gov/nsr/nsrlink.jsp?1998Ut02,B}{1998Ut02}: Deduced resonance properties, \ensuremath{\Gamma}\ensuremath{_{\textnormal{p}}}, \ensuremath{\Gamma}\ensuremath{_{\ensuremath{\gamma}}}, E\ensuremath{_{\textnormal{x}}}, E\ensuremath{_{\textnormal{c.m.}}}, \ensuremath{\Gamma}, J\ensuremath{^{\ensuremath{\pi}}}, and \ensuremath{\omega}\ensuremath{\gamma}\ensuremath{_{\textnormal{(p,}\ensuremath{\gamma}\textnormal{)}}} for resonances at E\ensuremath{_{\textnormal{c.m.}}}\ensuremath{\leq}1 MeV which}\\
\parbox[b][0.3cm]{17.7cm}{contribute to the \ensuremath{^{\textnormal{18}}}F(p,\ensuremath{\gamma}) reaction rate. Deduced the direct proton capture rate as well as resonant contributions to the (p,\ensuremath{\gamma}) total}\\
\parbox[b][0.3cm]{17.7cm}{reaction rate at T=0.1-1 GK (see the erratum at Phys. Rev. C 58 (1998) 1354). Provided the REACLIB format.}\\
\parbox[b][0.3cm]{17.7cm}{\addtolength{\parindent}{-0.2in}\href{https://www.nndc.bnl.gov/nsr/nsrlink.jsp?1999He40,B}{1999He40}, \href{https://www.nndc.bnl.gov/nsr/nsrlink.jsp?2001Co14,B}{2001Co14}: Investigated nova nucleosynthesis using the \ensuremath{^{\textnormal{18}}}F(p,\ensuremath{\gamma}) and \ensuremath{^{\textnormal{18}}}F(p,\ensuremath{\alpha}) rates from (\href{https://www.nndc.bnl.gov/nsr/nsrlink.jsp?1997Gr23,B}{1997Gr23}, \href{https://www.nndc.bnl.gov/nsr/nsrlink.jsp?1998Ut02,B}{1998Ut02}) and}\\
\parbox[b][0.3cm]{17.7cm}{hydrodynamic models for CO and ONe novae with different white dwarf masses. Discussed implications for \ensuremath{\gamma}-ray emission from}\\
\parbox[b][0.3cm]{17.7cm}{\ensuremath{^{\textnormal{18}}}F in novae.}\\
\parbox[b][0.3cm]{17.7cm}{\addtolength{\parindent}{-0.2in}\href{https://www.nndc.bnl.gov/nsr/nsrlink.jsp?2000Co33,B}{2000Co33}: Deduced the \ensuremath{^{\textnormal{18}}}F(p,\ensuremath{\gamma}) reaction rate for T=0.03-3 GK based on the experimental information from literature and by}\\
\parbox[b][0.3cm]{17.7cm}{considering the tails of broad resonances. The S-factor was computed. Using the nucleosynthesis code SHIVA, they performed a}\\
\parbox[b][0.3cm]{17.7cm}{nova nucleosynthesis calculation using their updated reaction rate.}\\
\parbox[b][0.3cm]{17.7cm}{\addtolength{\parindent}{-0.2in}\href{https://www.nndc.bnl.gov/nsr/nsrlink.jsp?2002Il05,B}{2002Il05}: Recommended the reaction rate of (\href{https://www.nndc.bnl.gov/nsr/nsrlink.jsp?2000Co33,B}{2000Co33}), varied this rate by a factor of 15 up and down, and studied its effect on}\\
\clearpage
\vspace{0.3cm}
{\bf \small \underline{\ensuremath{^{\textnormal{1}}}H(\ensuremath{^{\textnormal{18}}}F,\ensuremath{^{\textnormal{19}}}Ne)\hspace{0.2in}\href{https://www.nndc.bnl.gov/nsr/nsrlink.jsp?1997Re02,B}{1997Re02},\href{https://www.nndc.bnl.gov/nsr/nsrlink.jsp?2016Ak05,B}{2016Ak05} (continued)}}\\
\vspace{0.3cm}
\parbox[b][0.3cm]{17.7cm}{nova nucleosynthesis.}\\
\parbox[b][0.3cm]{17.7cm}{\addtolength{\parindent}{-0.2in}\href{https://www.nndc.bnl.gov/nsr/nsrlink.jsp?2003Li17,B}{2003Li17}: Calculated electron screening enhancement factors and discussed the implications for the astrophysical rp-process.}\\
\parbox[b][0.3cm]{17.7cm}{\addtolength{\parindent}{-0.2in}\href{https://www.nndc.bnl.gov/nsr/nsrlink.jsp?2003Sh25,B}{2003Sh25}: Collected all the published information on the \ensuremath{^{\textnormal{19}}}Ne excited states and deduced the \ensuremath{^{\textnormal{18}}}F(p,\ensuremath{\gamma}) reaction rate at T=0.03-3}\\
\parbox[b][0.3cm]{17.7cm}{GK using \ensuremath{\sim}30 levels of \ensuremath{^{\textnormal{19}}}Ne. The unknown properties of some of these levels were taken from \ensuremath{^{\textnormal{19}}}F mirror levels if possible. The}\\
\parbox[b][0.3cm]{17.7cm}{resonance properties (E\ensuremath{_{\textnormal{r}}}, J\ensuremath{^{\ensuremath{\pi}}}, \ensuremath{\Gamma}\ensuremath{_{\ensuremath{\gamma}}}, \ensuremath{\Gamma}\ensuremath{_{\textnormal{p}}}, \ensuremath{\Gamma}\ensuremath{_{\ensuremath{\alpha}}}, \ensuremath{\theta}\ensuremath{_{\textnormal{p}}^{\textnormal{2}}} and \ensuremath{\theta}\ensuremath{_{\ensuremath{\alpha}}^{\textnormal{2}}}) and their mirror levels in \ensuremath{^{\textnormal{19}}}F for the most important resonances are}\\
\parbox[b][0.3cm]{17.7cm}{evaluated by the authors. They also provided the reaction rate in the REACLIB format. Comparison with literature and}\\
\parbox[b][0.3cm]{17.7cm}{astrophysical implications are discussed.}\\
\parbox[b][0.3cm]{17.7cm}{\addtolength{\parindent}{-0.2in}\href{https://www.nndc.bnl.gov/nsr/nsrlink.jsp?2004Ba63,B}{2004Ba63}: Deduced upper limits on \ensuremath{\Gamma}\ensuremath{_{\textnormal{p}}} (at 90\% C.L.) for the resonances at E\ensuremath{_{\textnormal{c.m.}}}=827, 915, and 1089 keV. Tabulated the}\\
\parbox[b][0.3cm]{17.7cm}{properties for the resonances in the E\ensuremath{_{\textnormal{c.m.}}}=8-1122 keV region. Deduced the \ensuremath{^{\textnormal{18}}}F(p,\ensuremath{\gamma}) reaction rate at T=1-3 GK.}\\
\parbox[b][0.3cm]{17.7cm}{\addtolength{\parindent}{-0.2in}\href{https://www.nndc.bnl.gov/nsr/nsrlink.jsp?2005Ba82,B}{2005Ba82}, \href{https://www.nndc.bnl.gov/nsr/nsrlink.jsp?2005Bb05,B}{2005Bb05}: Measured a new resonance at E\ensuremath{_{\textnormal{c.m.}}}=1009 keV. Deduced the total \ensuremath{^{\textnormal{18}}}F(p,\ensuremath{\gamma}) reaction rate and the}\\
\parbox[b][0.3cm]{17.7cm}{contributions from individual resonances at T=1-3 GK and discussed the astrophysical implications of the new resonance on the}\\
\parbox[b][0.3cm]{17.7cm}{reaction rate.}\\
\parbox[b][0.3cm]{17.7cm}{\addtolength{\parindent}{-0.2in}\href{https://www.nndc.bnl.gov/nsr/nsrlink.jsp?2007Ne09,B}{2007Ne09}: Updated and expanded on the results of (\href{https://www.nndc.bnl.gov/nsr/nsrlink.jsp?2003Sh25,B}{2003Sh25}). Evaluated E\ensuremath{_{\textnormal{r}}}, J\ensuremath{^{\ensuremath{\pi}}}, \ensuremath{\Gamma}\ensuremath{_{\ensuremath{\gamma}}}, \ensuremath{\Gamma}\ensuremath{_{\textnormal{p}}}, \ensuremath{\Gamma}\ensuremath{_{\ensuremath{\alpha}}}, and \ensuremath{\theta}\ensuremath{_{\textnormal{p}}^{\textnormal{2}}} for the \ensuremath{^{\textnormal{19}}}Ne levels with}\\
\parbox[b][0.3cm]{17.7cm}{E\ensuremath{_{\textnormal{x}}}=6.4-8.1 MeV, including unmeasured ones, based on all available (at the time) experimental data on \ensuremath{^{\textnormal{19}}}Ne and those of the}\\
\parbox[b][0.3cm]{17.7cm}{analog states in the mirror nucleus \ensuremath{^{\textnormal{19}}}F. Assumptions are made when properties are unknown.}\\
\parbox[b][0.3cm]{17.7cm}{\addtolength{\parindent}{-0.2in}\href{https://www.nndc.bnl.gov/nsr/nsrlink.jsp?2010Il04,B}{2010Il04}, \href{https://www.nndc.bnl.gov/nsr/nsrlink.jsp?2010Il06,B}{2010Il06}: Re-evaluated the \ensuremath{^{\textnormal{18}}}F(p,\ensuremath{\gamma}) reaction rate and its uncertainty for T=0.01-10 GK using a Monte Carlo technique.}\\
\parbox[b][0.3cm]{17.7cm}{\addtolength{\parindent}{-0.2in}\href{https://www.nndc.bnl.gov/nsr/nsrlink.jsp?2013Ak03,B}{2013Ak03}, \href{https://www.nndc.bnl.gov/nsr/nsrlink.jsp?2016Ak05,B}{2016Ak05}: Using the directly determined (experimentally) \ensuremath{\omega}\ensuremath{\gamma}\ensuremath{_{\textnormal{(p,}\ensuremath{\gamma}\textnormal{)}}} and \ensuremath{\Gamma}\ensuremath{_{\ensuremath{\gamma}}} for the resonance at E\ensuremath{_{\textnormal{c.m.}}}=665 keV in}\\
\parbox[b][0.3cm]{17.7cm}{\ensuremath{^{\textnormal{19}}}Ne, these authors recalculated the \ensuremath{^{\textnormal{18}}}F(p,\ensuremath{\gamma}) reaction rate at T=0.1-0.4 GK. They used R-matrix to compute contributions of}\\
\parbox[b][0.3cm]{17.7cm}{individual resonances using resonance parameters from (\href{https://www.nndc.bnl.gov/nsr/nsrlink.jsp?2011Ad24,B}{2011Ad24}, \href{https://www.nndc.bnl.gov/nsr/nsrlink.jsp?2005Ba06,B}{2005Ba06}) and the AZURE2 code. Direct capture was also}\\
\parbox[b][0.3cm]{17.7cm}{included. The S-factor for the \ensuremath{^{\textnormal{18}}}F(p,\ensuremath{\gamma}) reaction rate was deduced for E\ensuremath{_{\textnormal{c.m.}}}=0.1-0.9 MeV. The resonance parameters (E\ensuremath{_{\textnormal{r}}}, J\ensuremath{^{\ensuremath{\pi}}}, \ensuremath{\Gamma}\ensuremath{_{\ensuremath{\alpha}}},}\\
\parbox[b][0.3cm]{17.7cm}{\ensuremath{\Gamma}\ensuremath{_{\textnormal{p}}}, and \ensuremath{\Gamma}\ensuremath{_{\ensuremath{\gamma}}}) are provided by (\href{https://www.nndc.bnl.gov/nsr/nsrlink.jsp?2016Ak05,B}{2016Ak05}). It was reported that the 665-keV resonance does not play a significant role on this rate}\\
\parbox[b][0.3cm]{17.7cm}{at temperatures associated with the ONe novae.}\\
\vspace{0.385cm}
\parbox[b][0.3cm]{17.7cm}{\addtolength{\parindent}{-0.2in}\textit{Other Related Astrophysical Articles}:}\\
\parbox[b][0.3cm]{17.7cm}{\addtolength{\parindent}{-0.2in}M. Hernanz, J. Gomez-Gomar, J. Jos\'{e}, New. Astron. Rev. 46 (2002) 559.}\\
\vspace{12pt}
\underline{$^{19}$Ne Levels}\\
\begin{longtable}{cccc@{\extracolsep{\fill}}c}
\multicolumn{2}{c}{E(level)$^{}$}&J$^{\pi}$$^{}$&Comments&\\[-.2cm]
\multicolumn{2}{c}{\hrulefill}&\hrulefill&\hrulefill&
\endfirsthead
\multicolumn{1}{r@{}}{0}&\multicolumn{1}{@{}l}{}&\multicolumn{1}{l}{1/2\ensuremath{^{+}}}&\parbox[t][0.3cm]{15.021081cm}{\raggedright E(level),J\ensuremath{^{\pi}}: From the Adopted Levels of \ensuremath{^{\textnormal{19}}}Ne.\vspace{0.1cm}}&\\
\multicolumn{1}{r@{}}{6861}&\multicolumn{1}{@{}l}{}&\multicolumn{1}{l}{7/2\ensuremath{^{-}}}&\parbox[t][0.3cm]{15.021081cm}{\raggedright E(level),J\ensuremath{^{\pi}}: Used in the R-matrix analysis of (\href{https://www.nndc.bnl.gov/nsr/nsrlink.jsp?2016Ak05,B}{2016Ak05}: See Table IV) based on the evaluation by\vspace{0.1cm}}&\\
&&&\parbox[t][0.3cm]{15.021081cm}{\raggedright {\ }{\ }{\ }(\href{https://www.nndc.bnl.gov/nsr/nsrlink.jsp?2007Ne09,B}{2007Ne09}).\vspace{0.1cm}}&\\
\multicolumn{1}{r@{}}{7068}&\multicolumn{1}{@{}l}{}&\multicolumn{1}{l}{[3/2\ensuremath{^{+}}]}&\parbox[t][0.3cm]{15.021081cm}{\raggedright \ensuremath{\Gamma}\ensuremath{_{\ensuremath{\gamma}}}=101\ensuremath{\times}10\ensuremath{^{\textnormal{$-$3}}} eV \textit{+91{\textminus}55} (\href{https://www.nndc.bnl.gov/nsr/nsrlink.jsp?2016Ak05,B}{2016Ak05})\vspace{0.1cm}}&\\
&&&\parbox[t][0.3cm]{15.021081cm}{\raggedright E(level): From E\ensuremath{_{\textnormal{c.m.}}}=658 keV, which is the unweighted average of E\ensuremath{_{\textnormal{c.m.}}}=652 keV (\href{https://www.nndc.bnl.gov/nsr/nsrlink.jsp?1997Re02,B}{1997Re02}, \href{https://www.nndc.bnl.gov/nsr/nsrlink.jsp?1997Re05,B}{1997Re05},\vspace{0.1cm}}&\\
&&&\parbox[t][0.3cm]{15.021081cm}{\raggedright {\ }{\ }{\ }\href{https://www.nndc.bnl.gov/nsr/nsrlink.jsp?1998Re24,B}{1998Re24}) and E\ensuremath{_{\textnormal{c.m.}}}=665 keV (\href{https://www.nndc.bnl.gov/nsr/nsrlink.jsp?2013Ak03,B}{2013Ak03}, \href{https://www.nndc.bnl.gov/nsr/nsrlink.jsp?2016Ak05,B}{2016Ak05}). S\ensuremath{_{\textnormal{p}}}(\ensuremath{^{\textnormal{19}}}Ne)=6410.0 keV \textit{5} (\href{https://www.nndc.bnl.gov/nsr/nsrlink.jsp?2021Wa16,B}{2021Wa16}).\vspace{0.1cm}}&\\
&&&\parbox[t][0.3cm]{15.021081cm}{\raggedright \ensuremath{\Gamma}\ensuremath{_{\ensuremath{\gamma}}}=0.101 eV \textit{+91{\textminus}55} (\href{https://www.nndc.bnl.gov/nsr/nsrlink.jsp?2016Ak05,B}{2016Ak05}) at 68\% C.L. and deduced from the measured resonance strength (see\vspace{0.1cm}}&\\
&&&\parbox[t][0.3cm]{15.021081cm}{\raggedright {\ }{\ }{\ }below) and \ensuremath{\Gamma}\ensuremath{_{\textnormal{p}}}=15.2 keV \textit{10} from (\href{https://www.nndc.bnl.gov/nsr/nsrlink.jsp?2001Ba49,B}{2001Ba49}).\vspace{0.1cm}}&\\
&&&\parbox[t][0.3cm]{15.021081cm}{\raggedright \ensuremath{\Gamma}\ensuremath{_{\ensuremath{\gamma}}}: See also \ensuremath{\Gamma}\ensuremath{_{\ensuremath{\gamma}}}=101 meV \textit{+226{\textminus}86} (\href{https://www.nndc.bnl.gov/nsr/nsrlink.jsp?2016Ak05,B}{2016Ak05}) at 95\% C.L.; \ensuremath{\Gamma}\ensuremath{_{\ensuremath{\gamma}}}=72 meV \textit{+172{\textminus}61} (\href{https://www.nndc.bnl.gov/nsr/nsrlink.jsp?2013Ak03,B}{2013Ak03}) (the\vspace{0.1cm}}&\\
&&&\parbox[t][0.3cm]{15.021081cm}{\raggedright {\ }{\ }{\ }results deduced by (\href{https://www.nndc.bnl.gov/nsr/nsrlink.jsp?2016Ak05,B}{2016Ak05}) supersede those of \href{https://www.nndc.bnl.gov/nsr/nsrlink.jsp?2013Ak03,B}{2013Ak03}); and \ensuremath{\Gamma}\ensuremath{_{\ensuremath{\gamma}}}\ensuremath{\leq}3 eV (\href{https://www.nndc.bnl.gov/nsr/nsrlink.jsp?1997Re02,B}{1997Re02}, \href{https://www.nndc.bnl.gov/nsr/nsrlink.jsp?1997Re05,B}{1997Re05},\vspace{0.1cm}}&\\
&&&\parbox[t][0.3cm]{15.021081cm}{\raggedright {\ }{\ }{\ }\href{https://www.nndc.bnl.gov/nsr/nsrlink.jsp?1998Re24,B}{1998Re24}). This last value was obtained from \ensuremath{\sigma}\ensuremath{\leq}42 \ensuremath{\mu}b (\href{https://www.nndc.bnl.gov/nsr/nsrlink.jsp?1998Re24,B}{1998Re24}) (for the \ensuremath{^{\textnormal{18}}}F(p,\ensuremath{\gamma}) reaction at\vspace{0.1cm}}&\\
&&&\parbox[t][0.3cm]{15.021081cm}{\raggedright {\ }{\ }{\ }E\ensuremath{_{\textnormal{c.m.}}}=670 keV), which was, in turn, deduced for the \ensuremath{^{\textnormal{19}}}Ne*(7.07 MeV, 3/2\ensuremath{^{\textnormal{+}}}) level by (\href{https://www.nndc.bnl.gov/nsr/nsrlink.jsp?1998Re24,B}{1998Re24}). An\vspace{0.1cm}}&\\
&&&\parbox[t][0.3cm]{15.021081cm}{\raggedright {\ }{\ }{\ }upper limit of \ensuremath{\Gamma}\ensuremath{_{\ensuremath{\gamma}}}\ensuremath{\leq}3 eV corresponds to about 2\% of the single particle width for an E1 and 40\% for an M1\vspace{0.1cm}}&\\
&&&\parbox[t][0.3cm]{15.021081cm}{\raggedright {\ }{\ }{\ }transition (\href{https://www.nndc.bnl.gov/nsr/nsrlink.jsp?1997Re02,B}{1997Re02}). Transitions with such strengths were observed in this mass region (\href{https://www.nndc.bnl.gov/nsr/nsrlink.jsp?1987Aj02,B}{1987Aj02}).\vspace{0.1cm}}&\\
&&&\parbox[t][0.3cm]{15.021081cm}{\raggedright J\ensuremath{^{\pi}}: Assumed by (\href{https://www.nndc.bnl.gov/nsr/nsrlink.jsp?1997Re02,B}{1997Re02}, \href{https://www.nndc.bnl.gov/nsr/nsrlink.jsp?1997Re05,B}{1997Re05}, \href{https://www.nndc.bnl.gov/nsr/nsrlink.jsp?1998Re24,B}{1998Re24}).\vspace{0.1cm}}&\\
&&&\parbox[t][0.3cm]{15.021081cm}{\raggedright \ensuremath{\omega}\ensuremath{\gamma}\ensuremath{_{\textnormal{(p,}\ensuremath{\gamma}\textnormal{)}}}=26 meV \textit{+59{\textminus}22} (\href{https://www.nndc.bnl.gov/nsr/nsrlink.jsp?2016Ak05,B}{2016Ak05}) at 95\% C.L. Other values: \ensuremath{\omega}\ensuremath{\gamma}\ensuremath{_{\textnormal{(p,}\ensuremath{\gamma}\textnormal{)}}}=26 meV \textit{+24{\textminus}14} (\href{https://www.nndc.bnl.gov/nsr/nsrlink.jsp?2016Ak05,B}{2016Ak05}) at\vspace{0.1cm}}&\\
&&&\parbox[t][0.3cm]{15.021081cm}{\raggedright {\ }{\ }{\ }68\% C.L.; \ensuremath{\omega}\ensuremath{\gamma}\ensuremath{_{\textnormal{(p,}\ensuremath{\gamma}\textnormal{)}}}=19 meV \textit{+45{\textminus}16} (\href{https://www.nndc.bnl.gov/nsr/nsrlink.jsp?2013Ak03,B}{2013Ak03}), where the results deduced by (\href{https://www.nndc.bnl.gov/nsr/nsrlink.jsp?2016Ak05,B}{2016Ak05}) supersede those\vspace{0.1cm}}&\\
&&&\parbox[t][0.3cm]{15.021081cm}{\raggedright {\ }{\ }{\ }of (\href{https://www.nndc.bnl.gov/nsr/nsrlink.jsp?2013Ak03,B}{2013Ak03}); and \ensuremath{\omega}\ensuremath{\gamma}\ensuremath{_{\textnormal{(p,}\ensuremath{\gamma}\textnormal{)}}}\ensuremath{\leq}740 meV (\href{https://www.nndc.bnl.gov/nsr/nsrlink.jsp?1997Re02,B}{1997Re02}). This last result was obtained using \ensuremath{\Gamma}\ensuremath{_{\textnormal{p}}} and \ensuremath{\Gamma} from\vspace{0.1cm}}&\\
&&&\parbox[t][0.3cm]{15.021081cm}{\raggedright {\ }{\ }{\ }(\href{https://www.nndc.bnl.gov/nsr/nsrlink.jsp?1996Re05,B}{1996Re05}: \ensuremath{^{\textnormal{18}}}F(p,\ensuremath{\alpha})). Note that the results of (\href{https://www.nndc.bnl.gov/nsr/nsrlink.jsp?2016Ak05,B}{2016Ak05}) are an order of magnitude smaller than what was\vspace{0.1cm}}&\\
&&&\parbox[t][0.3cm]{15.021081cm}{\raggedright {\ }{\ }{\ }obtained by (\href{https://www.nndc.bnl.gov/nsr/nsrlink.jsp?2007Ne09,B}{2007Ne09}: 1 eV, based on an evaluation by the authors) and a factor of 30 smaller than the\vspace{0.1cm}}&\\
&&&\parbox[t][0.3cm]{15.021081cm}{\raggedright {\ }{\ }{\ }upper limit determined by (\href{https://www.nndc.bnl.gov/nsr/nsrlink.jsp?1997Re02,B}{1997Re02}).\vspace{0.1cm}}&\\
&&&\parbox[t][0.3cm]{15.021081cm}{\raggedright (\href{https://www.nndc.bnl.gov/nsr/nsrlink.jsp?2016Ak05,B}{2016Ak05}) assumed that the \ensuremath{^{\textnormal{19}}}Ne*(7068) state decays to \ensuremath{^{\textnormal{19}}}Ne\ensuremath{_{\textnormal{g.s.}}} based on the decay scheme for the \ensuremath{^{\textnormal{19}}}F\vspace{0.1cm}}&\\
&&&\parbox[t][0.3cm]{15.021081cm}{\raggedright {\ }{\ }{\ }levels in this energy region (see Table I of that study). (\href{https://www.nndc.bnl.gov/nsr/nsrlink.jsp?2016Ak05,B}{2016Ak05}) also assumed E2 transitions from the\vspace{0.1cm}}&\\
&&&\parbox[t][0.3cm]{15.021081cm}{\raggedright {\ }{\ }{\ }\ensuremath{^{\textnormal{19}}}Ne*(7068) state to a nearby state with J\ensuremath{^{\ensuremath{\pi}}}=7/2\ensuremath{^{-}} and to another nearby state with J\ensuremath{^{\ensuremath{\pi}}}=11/2\ensuremath{^{-}}. However, no \ensuremath{\gamma}\vspace{0.1cm}}&\\
&&&\parbox[t][0.3cm]{15.021081cm}{\raggedright {\ }{\ }{\ }ray was observed in (\href{https://www.nndc.bnl.gov/nsr/nsrlink.jsp?2013Ak03,B}{2013Ak03}, \href{https://www.nndc.bnl.gov/nsr/nsrlink.jsp?2016Ak05,B}{2016Ak05}) from the decay of this state.\vspace{0.1cm}}&\\
\end{longtable}
\clearpage
\subsection[\hspace{-0.2cm}\ensuremath{^{\textnormal{1}}}H(\ensuremath{^{\textnormal{19}}}F,\ensuremath{^{\textnormal{19}}}Ne)]{ }
\vspace{-27pt}
\vspace{0.3cm}
\hypertarget{NE4}{{\bf \small \underline{\ensuremath{^{\textnormal{1}}}H(\ensuremath{^{\textnormal{19}}}F,\ensuremath{^{\textnormal{19}}}Ne)\hspace{0.2in}\href{https://www.nndc.bnl.gov/nsr/nsrlink.jsp?2014Br06,B}{2014Br06}}}}\\
\vspace{4pt}
\vspace{8pt}
\parbox[b][0.3cm]{17.7cm}{\addtolength{\parindent}{-0.2in}\ensuremath{^{\textnormal{19}}}F(p,n) charge exchange reaction in inverse kinematics, detecting the heavy residues.}\\
\parbox[b][0.3cm]{17.7cm}{\addtolength{\parindent}{-0.2in}J\ensuremath{^{\ensuremath{\pi}}}(\ensuremath{^{\textnormal{19}}}F\ensuremath{_{\textnormal{g.s.}}})=1/2\ensuremath{^{\textnormal{+}}} and J\ensuremath{^{\ensuremath{\pi}}}(p)=1/2\ensuremath{^{\textnormal{+}}}.}\\
\parbox[b][0.3cm]{17.7cm}{\addtolength{\parindent}{-0.2in}\href{https://www.nndc.bnl.gov/nsr/nsrlink.jsp?2014Br06,B}{2014Br06}: \ensuremath{^{\textnormal{1}}}H(\ensuremath{^{\textnormal{19}}}F,\ensuremath{^{\textnormal{19}}}Ne) E=10.5 MeV/nucleon; measured the half-life of \ensuremath{^{\textnormal{19}}}Ne using a \ensuremath{^{\textnormal{19}}}F beam impinging on a H\ensuremath{_{\textnormal{2}}} gas target at}\\
\parbox[b][0.3cm]{17.7cm}{the TRI\ensuremath{\mu}P separator facility at KVI. The \ensuremath{^{\textnormal{1}}}H(\ensuremath{^{\textnormal{19}}}F,\ensuremath{^{\textnormal{19}}}Ne) reaction was used to produce \ensuremath{^{\textnormal{19}}}Ne and to minimize the \ensuremath{^{\textnormal{15}}}O and \ensuremath{^{\textnormal{17}}}F}\\
\parbox[b][0.3cm]{17.7cm}{contaminants. The \ensuremath{^{\textnormal{19}}}Ne activity was implanted for 50 s at the depth of 25 \ensuremath{\mu}m in a 100 \ensuremath{\mu}m thick aluminum tape. The activity}\\
\parbox[b][0.3cm]{17.7cm}{was then transported to a counting area, where two HPGe clover detectors on either side of the tape measured the annihilation}\\
\parbox[b][0.3cm]{17.7cm}{photons in coincidence following \ensuremath{\beta}\ensuremath{^{\textnormal{+}}} decay for 70 s.}\\
\parbox[b][0.3cm]{17.7cm}{\addtolength{\parindent}{-0.2in}The authors discussed the uncertainty analysis; the blinded analysis procedure yielding T\ensuremath{_{\textnormal{1/2}}}=17.2832 s \textit{51} (stat.) \textit{58} (sys.);}\\
\parbox[b][0.3cm]{17.7cm}{treatment of time dependent backgrounds associated with \ensuremath{\beta}\ensuremath{^{\textnormal{+}}} emitting beam contaminants; and the effects of diffusion on the}\\
\parbox[b][0.3cm]{17.7cm}{half-life. The authors extracted the diffusion coefficient of 0.0100 \ensuremath{\mu}m\ensuremath{^{\textnormal{2}}}s\ensuremath{^{\textnormal{$-$1}}} \textit{52} (stat.), deduced T\ensuremath{_{\textnormal{1/2}}}=17.2826 s \textit{44} (stat.) \textit{64} (sys.)}\\
\parbox[b][0.3cm]{17.7cm}{from a post-blind analysis resulting in a recommended T\ensuremath{_{\textnormal{1/2}}}=17.2832 s \textit{51} (stat.) \textit{66} (sys.), and discussed the impact of this half-life}\\
\parbox[b][0.3cm]{17.7cm}{on the superallowed, mixed Fermi-Gamow-Teller decay of the \ensuremath{^{\textnormal{19}}}Ne\ensuremath{_{\textnormal{g.s.}}} with J\ensuremath{^{\ensuremath{\pi}}}=1/2\ensuremath{^{\textnormal{+}}} to the \ensuremath{^{\textnormal{19}}}F\ensuremath{_{\textnormal{g.s.}}} with J\ensuremath{^{\ensuremath{\pi}}}=1/2\ensuremath{^{\textnormal{+}}}. A discussion}\\
\parbox[b][0.3cm]{17.7cm}{on the significance of the superallowed \ensuremath{\beta} transitions, and physics beyond the standard model is provided.}\\
\vspace{12pt}
\underline{$^{19}$Ne Levels}\\
\vspace{0.34cm}
\parbox[b][0.3cm]{17.7cm}{\addtolength{\parindent}{-0.254cm}Evaluator highlights that the \ensuremath{^{\textnormal{19}}}Ne\ensuremath{_{\textnormal{g.s.}}} half-life measurement by (\href{https://www.nndc.bnl.gov/nsr/nsrlink.jsp?2014Br06,B}{2014Br06}) carries unaccounted systematic effects, and thus should}\\
\parbox[b][0.3cm]{17.7cm}{be excluded. This makes the other results deduced by (\href{https://www.nndc.bnl.gov/nsr/nsrlink.jsp?2014Br06,B}{2014Br06}) less reliable.}\\
\vspace{0.34cm}
\begin{longtable}{cccccc@{\extracolsep{\fill}}c}
\multicolumn{2}{c}{E(level)$^{}$}&J$^{\pi}$$^{}$&\multicolumn{2}{c}{T\ensuremath{_{\textnormal{1/2}}~_{\textnormal{1/2}}}$^{}$}&Comments&\\[-.2cm]
\multicolumn{2}{c}{\hrulefill}&\hrulefill&\multicolumn{2}{c}{\hrulefill}&\hrulefill&
\endfirsthead
\multicolumn{1}{r@{}}{0}&\multicolumn{1}{@{}l}{}&\multicolumn{1}{l}{1/2\ensuremath{^{+}}}&\multicolumn{1}{r@{}}{17}&\multicolumn{1}{@{.}l}{2832 s {\it 83}}&\parbox[t][0.3cm]{13.013941cm}{\raggedright E(level),J\ensuremath{^{\pi}}: From the Adopted Levels of \ensuremath{^{\textnormal{19}}}Ne.\vspace{0.1cm}}&\\
&&&&&\parbox[t][0.3cm]{13.013941cm}{\raggedright T\ensuremath{_{1/2}}: From T\ensuremath{_{\textnormal{1/2}}}=17.2832 s \textit{51} (stat.) \textit{66} (sys.) (\href{https://www.nndc.bnl.gov/nsr/nsrlink.jsp?2014Br06,B}{2014Br06}). This value is discrepant with other\vspace{0.1cm}}&\\
&&&&&\parbox[t][0.3cm]{13.013941cm}{\raggedright {\ }{\ }{\ }similarly precise recent measurements of T\ensuremath{_{\textnormal{1/2}}}=17.254 s \textit{5} (\href{https://www.nndc.bnl.gov/nsr/nsrlink.jsp?2013Uj01,B}{2013Uj01}: Using \ensuremath{\beta}-ray counter\vspace{0.1cm}}&\\
&&&&&\parbox[t][0.3cm]{13.013941cm}{\raggedright {\ }{\ }{\ }plus 2 EXOGAM HPGe clover detectors at GANIL), and T\ensuremath{_{\textnormal{1/2}}}=17.262 s \textit{7} (sys.) (\href{https://www.nndc.bnl.gov/nsr/nsrlink.jsp?2012Tr06,B}{2012Tr06}:\vspace{0.1cm}}&\\
&&&&&\parbox[t][0.3cm]{13.013941cm}{\raggedright {\ }{\ }{\ }Using the SCEPTAR electron-positron array at TRIUMF) by 2\ensuremath{\sigma} and 3\ensuremath{\sigma}, respectively.\vspace{0.1cm}}&\\
&&&&&\parbox[t][0.3cm]{13.013941cm}{\raggedright (\href{https://www.nndc.bnl.gov/nsr/nsrlink.jsp?2014Br06,B}{2014Br06}) found a weighted average of 17.2604 s \textit{34} (with reduced \ensuremath{\chi}\ensuremath{^{\textnormal{2}}}/\ensuremath{\nu}=6.3) by combining\vspace{0.1cm}}&\\
&&&&&\parbox[t][0.3cm]{13.013941cm}{\raggedright {\ }{\ }{\ }their value with the results of eight previous measurements from 1957 to 2013 (see Fig. 3 in\vspace{0.1cm}}&\\
&&&&&\parbox[t][0.3cm]{13.013941cm}{\raggedright {\ }{\ }{\ }that study). The large, deduced \ensuremath{\chi}\ensuremath{^{\textnormal{2}}}/\ensuremath{\nu} implies the presence of systematic effects that were\vspace{0.1cm}}&\\
&&&&&\parbox[t][0.3cm]{13.013941cm}{\raggedright {\ }{\ }{\ }unaccounted for. From the above mentioned weighted average, (\href{https://www.nndc.bnl.gov/nsr/nsrlink.jsp?2014Br06,B}{2014Br06}) deduced \textit{ft}=1719.8\vspace{0.1cm}}&\\
&&&&&\parbox[t][0.3cm]{13.013941cm}{\raggedright {\ }{\ }{\ }\textit{13} for the 1/2\ensuremath{^{\textnormal{+}}} to 1/2\ensuremath{^{\textnormal{+}}} \ensuremath{\beta} transition. This value combined with \ensuremath{\rho}=1.5995 \textit{45} extracted from\vspace{0.1cm}}&\\
&&&&&\parbox[t][0.3cm]{13.013941cm}{\raggedright {\ }{\ }{\ }the \ensuremath{\beta} asymmetry of A\ensuremath{_{\textnormal{0}}}={\textminus}0.0391 \textit{14} (\href{https://www.nndc.bnl.gov/nsr/nsrlink.jsp?1975Ca28,B}{1975Ca28}) resulted in the Cabibbo-Kobayashi-Maskawa\vspace{0.1cm}}&\\
&&&&&\parbox[t][0.3cm]{13.013941cm}{\raggedright {\ }{\ }{\ }matrix element V\ensuremath{_{\textnormal{ud}}}=0.9712 \textit{22}.\vspace{0.1cm}}&\\
&&&&&\parbox[t][0.3cm]{13.013941cm}{\raggedright (\href{https://www.nndc.bnl.gov/nsr/nsrlink.jsp?2014Br06,B}{2014Br06}) extracted the diffusion coefficient of 0.0100 \ensuremath{\mu}m\ensuremath{^{\textnormal{2}}}s\ensuremath{^{\textnormal{$-$1}}} \textit{52} (stat.) for \ensuremath{^{\textnormal{19}}}Ne.\vspace{0.1cm}}&\\
&&&&&\parbox[t][0.3cm]{13.013941cm}{\raggedright (\href{https://www.nndc.bnl.gov/nsr/nsrlink.jsp?2014Br06,B}{2014Br06}) deduced Fierz term, {\textminus}0.050\ensuremath{<}b\ensuremath{<}0.007 and {\textminus}0.006\ensuremath{<}C\ensuremath{_{\textnormal{T}}}/C\ensuremath{_{\textnormal{A}}}\ensuremath{<}0.034 (at 2\ensuremath{\sigma} C.L.), which\vspace{0.1cm}}&\\
&&&&&\parbox[t][0.3cm]{13.013941cm}{\raggedright {\ }{\ }{\ }is 10 times less precise that the results of (\href{https://www.nndc.bnl.gov/nsr/nsrlink.jsp?2014Wa04,B}{2014Wa04}).\vspace{0.1cm}}&\\
\end{longtable}
\clearpage
\subsection[\hspace{-0.2cm}\ensuremath{^{\textnormal{1}}}H(\ensuremath{^{\textnormal{19}}}Ne,p\ensuremath{'})]{ }
\vspace{-27pt}
\vspace{0.3cm}
\hypertarget{NE5}{{\bf \small \underline{\ensuremath{^{\textnormal{1}}}H(\ensuremath{^{\textnormal{19}}}Ne,p\ensuremath{'})\hspace{0.2in}\href{https://www.nndc.bnl.gov/nsr/nsrlink.jsp?2009Da07,B}{2009Da07}}}}\\
\vspace{4pt}
\vspace{8pt}
\parbox[b][0.3cm]{17.7cm}{\addtolength{\parindent}{-0.2in}\ensuremath{^{\textnormal{19}}}Ne(p,p\ensuremath{'}) inelastic scattering in inverse kinematics.}\\
\parbox[b][0.3cm]{17.7cm}{\addtolength{\parindent}{-0.2in}J\ensuremath{^{\ensuremath{\pi}}}(\ensuremath{^{\textnormal{19}}}Ne\ensuremath{_{\textnormal{g.s.}}})=1/2\ensuremath{^{\textnormal{+}}} and J\ensuremath{^{\ensuremath{\pi}}}(p)=1/2\ensuremath{^{\textnormal{+}}}.}\\
\parbox[b][0.3cm]{17.7cm}{\addtolength{\parindent}{-0.2in}\href{https://www.nndc.bnl.gov/nsr/nsrlink.jsp?2009Da07,B}{2009Da07}, \href{https://www.nndc.bnl.gov/nsr/nsrlink.jsp?2009De42,B}{2009De42}: \ensuremath{^{\textnormal{1}}}H(\ensuremath{^{\textnormal{19}}}Ne,p\ensuremath{'})\ensuremath{^{\textnormal{19}}}Ne*(p) E=9 MeV/nucleon; measured the secondary protons emitted from the decay of \ensuremath{^{\textnormal{19}}}Ne*}\\
\parbox[b][0.3cm]{17.7cm}{unbound states in coincidence with the scattered protons using an annular position sensitive Si \ensuremath{\Delta}E-E telescope covering}\\
\parbox[b][0.3cm]{17.7cm}{\ensuremath{\theta}\ensuremath{_{\textnormal{lab}}}=4.3\ensuremath{^\circ}{\textminus}21.6\ensuremath{^\circ}; measured inelastically scattered protons using a Si-Si(Li) \ensuremath{\Delta}E-E telescope cooled to {\textminus}25\ensuremath{^\circ}C at \ensuremath{\theta}\ensuremath{_{\textnormal{lab}}}=0\ensuremath{^\circ} downstream}\\
\parbox[b][0.3cm]{17.7cm}{the target; measured E\ensuremath{_{\textnormal{p}}}, I\ensuremath{_{\textnormal{p}}}, proton-proton angular correlations, decay protons$'$ angular distributions, and \ensuremath{\sigma}(\ensuremath{\theta}) for the inelastic}\\
\parbox[b][0.3cm]{17.7cm}{scattering. Deduced E\ensuremath{_{\textnormal{x}}}(\ensuremath{^{\textnormal{19}}}Ne*), \ensuremath{\Gamma}, and J\ensuremath{^{\ensuremath{\pi}}} assignments for E\ensuremath{_{\textnormal{x}}}(\ensuremath{^{\textnormal{19}}}Ne*)=6.9-8.4 MeV with a resolution of 30 keV (FWHM).}\\
\parbox[b][0.3cm]{17.7cm}{(\href{https://www.nndc.bnl.gov/nsr/nsrlink.jsp?2009Da07,B}{2009Da07}) deduced the astrophysical S-factor for the \ensuremath{^{\textnormal{18}}}F(p,\ensuremath{\alpha}) reaction and discussed the astrophysical implications.}\\
\parbox[b][0.3cm]{17.7cm}{\addtolength{\parindent}{-0.2in}\href{https://www.nndc.bnl.gov/nsr/nsrlink.jsp?2018Bo27,B}{2018Bo27}: \ensuremath{^{\textnormal{1}}}H(\ensuremath{^{\textnormal{19}}}Ne,p\ensuremath{'})\ensuremath{^{\textnormal{19}}}Ne*(p) and \ensuremath{^{\textnormal{1}}}H(\ensuremath{^{\textnormal{19}}}Ne,p\ensuremath{'})\ensuremath{^{\textnormal{19}}}Ne*(\ensuremath{\alpha}) E=10 MeV/nucleon; momentum analyzed and detected the scattered}\\
\parbox[b][0.3cm]{17.7cm}{protons using the VAMOS spectrometer and its focal plane system placed at \ensuremath{\theta}\ensuremath{_{\textnormal{lab}}}=0\ensuremath{^\circ}. Measured, in coincidence with the scattered}\\
\parbox[b][0.3cm]{17.7cm}{protons, the \ensuremath{\alpha}s and protons from the decay of \ensuremath{^{\textnormal{19}}}Ne* states using an annular position sensitive Si \ensuremath{\Delta}E-E telescope at \ensuremath{\theta}\ensuremath{_{\textnormal{lab}}}=0\ensuremath{^\circ}.}\\
\parbox[b][0.3cm]{17.7cm}{Deduced several \ensuremath{^{\textnormal{19}}}Ne* levels in the E\ensuremath{_{\textnormal{x}}}=4.5-8.5 MeV region. Results are only presented for the \ensuremath{^{\textnormal{19}}}Ne*(7076) state and are}\\
\parbox[b][0.3cm]{17.7cm}{preliminary.}\\
\vspace{12pt}
\underline{$^{19}$Ne Levels}\\
\begin{longtable}{cccccccc@{\extracolsep{\fill}}c}
\multicolumn{2}{c}{E(level)$^{{\hyperlink{NE5LEVEL1}{b}}}$}&J$^{\pi}$$^{{\hyperlink{NE5LEVEL0}{a}}{\hyperlink{NE5LEVEL3}{d}}}$&\multicolumn{2}{c}{\ensuremath{\Gamma}$^{{\hyperlink{NE5LEVEL0}{a}}}$}&\multicolumn{2}{c}{E\ensuremath{_{\textnormal{c.m.}}}(\ensuremath{^{\textnormal{18}}}F+p) (keV)$^{{\hyperlink{NE5LEVEL0}{a}}}$}&Comments&\\[-.2cm]
\multicolumn{2}{c}{\hrulefill}&\hrulefill&\multicolumn{2}{c}{\hrulefill}&\multicolumn{2}{c}{\hrulefill}&\hrulefill&
\endfirsthead
\multicolumn{1}{r@{}}{7079}&\multicolumn{1}{@{ }l}{{\it 5}}&\multicolumn{1}{l}{3/2\ensuremath{^{(+)}}}&\multicolumn{1}{r@{}}{32}&\multicolumn{1}{@{ }l}{keV {\it 8}}&\multicolumn{1}{r@{}}{669}&\multicolumn{1}{@{ }l}{{\it 5}}&\parbox[t][0.3cm]{8.52868cm}{\raggedright E(level): See also E\ensuremath{_{\textnormal{x}}}=7076 keV \textit{3}, which is the preliminary\vspace{0.1cm}}&\\
&&&&&&&\parbox[t][0.3cm]{8.52868cm}{\raggedright {\ }{\ }{\ }result of (\href{https://www.nndc.bnl.gov/nsr/nsrlink.jsp?2018Bo27,B}{2018Bo27}).\vspace{0.1cm}}&\\
&&&&&&&\parbox[t][0.3cm]{8.52868cm}{\raggedright \ensuremath{\Gamma}: See also \ensuremath{\Gamma}=35 keV \textit{4}, which is the preliminary result of\vspace{0.1cm}}&\\
&&&&&&&\parbox[t][0.3cm]{8.52868cm}{\raggedright {\ }{\ }{\ }(\href{https://www.nndc.bnl.gov/nsr/nsrlink.jsp?2018Bo27,B}{2018Bo27}). Those authors also deduced a preliminary\vspace{0.1cm}}&\\
&&&&&&&\parbox[t][0.3cm]{8.52868cm}{\raggedright {\ }{\ }{\ }branching ratio of \ensuremath{\Gamma}\ensuremath{_{\textnormal{p}}}/\ensuremath{\Gamma}\ensuremath{_{\ensuremath{\alpha}}}=0.64 \textit{5} for this state.\vspace{0.1cm}}&\\
&&&&&&&\parbox[t][0.3cm]{8.52868cm}{\raggedright J\ensuremath{^{\pi}}: (\href{https://www.nndc.bnl.gov/nsr/nsrlink.jsp?2009Da07,B}{2009Da07}) fitted the proton angular distribution\vspace{0.1cm}}&\\
&&&&&&&\parbox[t][0.3cm]{8.52868cm}{\raggedright {\ }{\ }{\ }corresponding to the population of this state with a quadratic\vspace{0.1cm}}&\\
&&&&&&&\parbox[t][0.3cm]{8.52868cm}{\raggedright {\ }{\ }{\ }polynomial (see Fig. 2), indicating J=3/2.\vspace{0.1cm}}&\\
\multicolumn{1}{r@{}}{7203}&\multicolumn{1}{@{ }l}{{\it 31}}&\multicolumn{1}{l}{3/2\ensuremath{^{(+)}}\ensuremath{^{{\hyperlink{NE5LEVEL4}{e}}}}}&\multicolumn{1}{r@{}}{35}&\multicolumn{1}{@{ }l}{keV {\it 12}}&\multicolumn{1}{r@{}}{793}&\multicolumn{1}{@{ }l}{{\it 31}}&&\\
\multicolumn{1}{r@{}}{7502}&\multicolumn{1}{@{ }l}{{\it 30}}&\multicolumn{1}{l}{5/2\ensuremath{^{(-)}}\ensuremath{^{{\hyperlink{NE5LEVEL4}{e}}}}}&\multicolumn{1}{r@{}}{17}&\multicolumn{1}{@{ }l}{keV {\it 7}}&\multicolumn{1}{r@{}}{1092}&\multicolumn{1}{@{ }l}{{\it 30}}&&\\
\multicolumn{1}{r@{}}{7616}&\multicolumn{1}{@{ }l}{{\it 5}}&\multicolumn{1}{l}{3/2\ensuremath{^{(+)}}}&\multicolumn{1}{r@{}}{21}&\multicolumn{1}{@{ }l}{keV {\it 10}}&\multicolumn{1}{r@{}}{1206}&\multicolumn{1}{@{ }l}{{\it 5}}&&\\
\multicolumn{1}{r@{}}{7862}&\multicolumn{1}{@{}l}{\ensuremath{^{{\hyperlink{NE5LEVEL2}{c}}}} {\it 39}}&\multicolumn{1}{l}{1/2\ensuremath{^{(+)}}}&\multicolumn{1}{r@{}}{292}&\multicolumn{1}{@{ }l}{keV {\it 107}}&\multicolumn{1}{r@{}}{1452}&\multicolumn{1}{@{ }l}{{\it 39}}&\parbox[t][0.3cm]{8.52868cm}{\raggedright E(level): (\href{https://www.nndc.bnl.gov/nsr/nsrlink.jsp?2009Da07,B}{2009Da07}) fitted this state with a Breit-Wigner\vspace{0.1cm}}&\\
&&&&&&&\parbox[t][0.3cm]{8.52868cm}{\raggedright {\ }{\ }{\ }function. Excellent agreement was found with the\vspace{0.1cm}}&\\
&&&&&&&\parbox[t][0.3cm]{8.52868cm}{\raggedright {\ }{\ }{\ }predictions of (\href{https://www.nndc.bnl.gov/nsr/nsrlink.jsp?2007Du09,B}{2007Du09}), which proposed the existence of\vspace{0.1cm}}&\\
&&&&&&&\parbox[t][0.3cm]{8.52868cm}{\raggedright {\ }{\ }{\ }a broad 1/2\ensuremath{^{\textnormal{+}}} state at E\ensuremath{_{\textnormal{x}}}=7901 keV with \ensuremath{\Gamma}=296 keV.\vspace{0.1cm}}&\\
&&&&&&&\parbox[t][0.3cm]{8.52868cm}{\raggedright \ensuremath{\Gamma}: (\href{https://www.nndc.bnl.gov/nsr/nsrlink.jsp?2009Da07,B}{2009Da07}) assumed \ensuremath{\Gamma}\ensuremath{_{\ensuremath{\alpha}}}=139 keV for this state based on\vspace{0.1cm}}&\\
&&&&&&&\parbox[t][0.3cm]{8.52868cm}{\raggedright {\ }{\ }{\ }the theoretical estimation by (\href{https://www.nndc.bnl.gov/nsr/nsrlink.jsp?2007Du09,B}{2007Du09}).\vspace{0.1cm}}&\\
&&&&&&&\parbox[t][0.3cm]{8.52868cm}{\raggedright J\ensuremath{^{\pi}}: The proton angular distribution corresponding to the\vspace{0.1cm}}&\\
&&&&&&&\parbox[t][0.3cm]{8.52868cm}{\raggedright {\ }{\ }{\ }population of this state is flat, indicating J=1/2 for an\vspace{0.1cm}}&\\
&&&&&&&\parbox[t][0.3cm]{8.52868cm}{\raggedright {\ }{\ }{\ }isotropic proton emission (\href{https://www.nndc.bnl.gov/nsr/nsrlink.jsp?2009Da07,B}{2009Da07}).\vspace{0.1cm}}&\\
\multicolumn{1}{r@{}}{7974}&\multicolumn{1}{@{ }l}{{\it 10}}&\multicolumn{1}{l}{(5/2\ensuremath{^{-}})\ensuremath{^{{\hyperlink{NE5LEVEL4}{e}}}}}&\multicolumn{1}{r@{}}{11}&\multicolumn{1}{@{ }l}{keV {\it 8}}&\multicolumn{1}{r@{}}{1564}&\multicolumn{1}{@{ }l}{{\it 10}}&&\\
\end{longtable}
\parbox[b][0.3cm]{17.7cm}{\makebox[1ex]{\ensuremath{^{\hypertarget{NE5LEVEL0}{a}}}} From (\href{https://www.nndc.bnl.gov/nsr/nsrlink.jsp?2009Da07,B}{2009Da07}).}\\
\parbox[b][0.3cm]{17.7cm}{\makebox[1ex]{\ensuremath{^{\hypertarget{NE5LEVEL1}{b}}}} From E\ensuremath{_{\textnormal{x}}}=S\ensuremath{_{\textnormal{p}}}+E\ensuremath{_{\textnormal{c.m.}}}(\ensuremath{^{\textnormal{18}}}F+p), where E\ensuremath{_{\textnormal{c.m.}}} is from (\href{https://www.nndc.bnl.gov/nsr/nsrlink.jsp?2009Da07,B}{2009Da07}) and S\ensuremath{_{\textnormal{p}}}=6410.0 keV \textit{5} is from (\href{https://www.nndc.bnl.gov/nsr/nsrlink.jsp?2021Wa16,B}{2021Wa16}).}\\
\parbox[b][0.3cm]{17.7cm}{\makebox[1ex]{\ensuremath{^{\hypertarget{NE5LEVEL2}{c}}}} This state was observed for the first time by (\href{https://www.nndc.bnl.gov/nsr/nsrlink.jsp?2009Da07,B}{2009Da07}).}\\
\parbox[b][0.3cm]{17.7cm}{\makebox[1ex]{\ensuremath{^{\hypertarget{NE5LEVEL3}{d}}}} From the analysis of the proton-proton angular correlation following the method outlined by (\href{https://www.nndc.bnl.gov/nsr/nsrlink.jsp?1973Pr08,B}{1973Pr08}, \href{https://www.nndc.bnl.gov/nsr/nsrlink.jsp?1976Ot02,B}{1976Ot02}). The angular}\\
\parbox[b][0.3cm]{17.7cm}{{\ }{\ }distributions are parity independent. So positive parities were assumed as favorable assignments for the states with J=1/2, 3/2, 7/2,}\\
\parbox[b][0.3cm]{17.7cm}{{\ }{\ }etc.; while negative parities were assigned to the states with J=5/2, 9/2, etc. based on the assumption that states with large proton}\\
\parbox[b][0.3cm]{17.7cm}{{\ }{\ }widths emit protons with the lowest possible angular momentum value to have the smallest centrifugal barrier.}\\
\parbox[b][0.3cm]{17.7cm}{\makebox[1ex]{\ensuremath{^{\hypertarget{NE5LEVEL4}{e}}}} Spin assignment was made for the first time by (\href{https://www.nndc.bnl.gov/nsr/nsrlink.jsp?2009Da07,B}{2009Da07}). The previously deduced assignments were based on mirror analysis.}\\
\vspace{0.5cm}
\clearpage
\subsection[\hspace{-0.2cm}\ensuremath{^{\textnormal{1}}}H(\ensuremath{^{\textnormal{20}}}Ne,X),(\ensuremath{^{\textnormal{22}}}Ne,X)]{ }
\vspace{-27pt}
\vspace{0.3cm}
\hypertarget{NE6}{{\bf \small \underline{\ensuremath{^{\textnormal{1}}}H(\ensuremath{^{\textnormal{20}}}Ne,X),(\ensuremath{^{\textnormal{22}}}Ne,X)\hspace{0.2in}\href{https://www.nndc.bnl.gov/nsr/nsrlink.jsp?1998We22,B}{1998We22}}}}\\
\vspace{4pt}
\vspace{8pt}
\parbox[b][0.3cm]{17.7cm}{\addtolength{\parindent}{-0.2in}J\ensuremath{^{\ensuremath{\pi}}}(\ensuremath{^{\textnormal{20}}}Ne\ensuremath{_{\textnormal{g.s.}}})=0\ensuremath{^{\textnormal{+}}}; J\ensuremath{^{\ensuremath{\pi}}}(\ensuremath{^{\textnormal{22}}}Ne\ensuremath{_{\textnormal{g.s.}}})=0\ensuremath{^{\textnormal{+}}} and J\ensuremath{^{\ensuremath{\pi}}}(p)=1/2\ensuremath{^{\textnormal{+}}}.}\\
\parbox[b][0.3cm]{17.7cm}{\addtolength{\parindent}{-0.2in}\href{https://www.nndc.bnl.gov/nsr/nsrlink.jsp?1998We22,B}{1998We22}: \ensuremath{^{\textnormal{1}}}H(\ensuremath{^{\textnormal{20}}}Ne,\ensuremath{^{\textnormal{19}}}Ne) E=414 MeV/nucleon, and \ensuremath{^{\textnormal{1}}}H(\ensuremath{^{\textnormal{22}}}Ne,\ensuremath{^{\textnormal{19}}}Ne) E=401 MeV/nucleon; measured \ensuremath{^{\textnormal{19}}}Ne production cross}\\
\parbox[b][0.3cm]{17.7cm}{section from projectile fragmentation using a liquid hydrogen target. Astrophysical implications are discussed.}\\
\vspace{12pt}
\underline{$^{19}$Ne Levels}\\
\begin{longtable}{cccc@{\extracolsep{\fill}}c}
\multicolumn{2}{c}{E(level)$^{{\hyperlink{NE6LEVEL0}{a}}}$}&J$^{\pi}$$^{{\hyperlink{NE6LEVEL0}{a}}}$&Comments&\\[-.2cm]
\multicolumn{2}{c}{\hrulefill}&\hrulefill&\hrulefill&
\endfirsthead
\multicolumn{1}{r@{}}{0}&\multicolumn{1}{@{}l}{}&\multicolumn{1}{l}{1/2\ensuremath{^{+}}}&\parbox[t][0.3cm]{15.02536cm}{\raggedright \ensuremath{\sigma}=32.5 mb for \ensuremath{^{\textnormal{1}}}H(\ensuremath{^{\textnormal{20}}}Ne,\ensuremath{^{\textnormal{19}}}Ne) at 414 MeV/nucleon using a liquid hydrogen target (\href{https://www.nndc.bnl.gov/nsr/nsrlink.jsp?1998We22,B}{1998We22}). The\vspace{0.1cm}}&\\
&&&\parbox[t][0.3cm]{15.02536cm}{\raggedright {\ }{\ }{\ }uncertainty in \ensuremath{\sigma} is 3-5\%. See also \ensuremath{\sigma}=28.0 mb (with an uncertainty of 3-5\%) for the same reaction at the\vspace{0.1cm}}&\\
&&&\parbox[t][0.3cm]{15.02536cm}{\raggedright {\ }{\ }{\ }same energy using a CH\ensuremath{_{\textnormal{2}}}-C target (\href{https://www.nndc.bnl.gov/nsr/nsrlink.jsp?1998We22,B}{1998We22}).\vspace{0.1cm}}&\\
&&&\parbox[t][0.3cm]{15.02536cm}{\raggedright \ensuremath{\sigma}=0.8 mb (with an uncertainty of 10-20\%) for \ensuremath{^{\textnormal{1}}}H(\ensuremath{^{\textnormal{22}}}Ne,\ensuremath{^{\textnormal{19}}}Ne) at 401 MeV/nucleon (\href{https://www.nndc.bnl.gov/nsr/nsrlink.jsp?1998We22,B}{1998We22}).\vspace{0.1cm}}&\\
\end{longtable}
\parbox[b][0.3cm]{17.7cm}{\makebox[1ex]{\ensuremath{^{\hypertarget{NE6LEVEL0}{a}}}} From the \ensuremath{^{\textnormal{19}}}Ne Adopted Levels.}\\
\vspace{0.5cm}
\clearpage
\subsection[\hspace{-0.2cm}\ensuremath{^{\textnormal{1}}}H(\ensuremath{^{\textnormal{21}}}Ne,t)]{ }
\vspace{-27pt}
\vspace{0.3cm}
\hypertarget{NE7}{{\bf \small \underline{\ensuremath{^{\textnormal{1}}}H(\ensuremath{^{\textnormal{21}}}Ne,t)\hspace{0.2in}\href{https://www.nndc.bnl.gov/nsr/nsrlink.jsp?1978Fo26,B}{1978Fo26},\href{https://www.nndc.bnl.gov/nsr/nsrlink.jsp?2003Da13,B}{2003Da13}}}}\\
\vspace{4pt}
\vspace{8pt}
\parbox[b][0.3cm]{17.7cm}{\addtolength{\parindent}{-0.2in}\ensuremath{^{\textnormal{21}}}Ne(p,t) two nucleon transfer reaction in normal and inverse kinematics.}\\
\parbox[b][0.3cm]{17.7cm}{\addtolength{\parindent}{-0.2in}J\ensuremath{^{\ensuremath{\pi}}}(\ensuremath{^{\textnormal{21}}}Ne\ensuremath{_{\textnormal{g.s.}}})=3/2\ensuremath{^{\textnormal{+}}} and J\ensuremath{^{\ensuremath{\pi}}}(p)=1/2\ensuremath{^{\textnormal{+}}}.}\\
\parbox[b][0.3cm]{17.7cm}{\addtolength{\parindent}{-0.2in}\textit{Measurements Performed in Normal Kinematics using a \ensuremath{^{21}}Ne Target:}}\\
\parbox[b][0.3cm]{17.7cm}{\addtolength{\parindent}{-0.2in}\href{https://www.nndc.bnl.gov/nsr/nsrlink.jsp?1969Ha38,B}{1969Ha38}: \ensuremath{^{\textnormal{21}}}Ne(p,t), \ensuremath{^{\textnormal{21}}}Ne(p,\ensuremath{^{\textnormal{3}}}He) E=45 MeV; measured \ensuremath{^{\textnormal{3}}}He and tritons using 2 sets of \ensuremath{\Delta}E-E Si telescopes that were placed on}\\
\parbox[b][0.3cm]{17.7cm}{opposite sides of the target covering \ensuremath{\theta}\ensuremath{_{\textnormal{lab}}}=11.7\ensuremath{^\circ}{\textminus}31.5\ensuremath{^\circ}. Measured \ensuremath{\sigma}(E\ensuremath{_{\textnormal{t}}},\ensuremath{\theta}) and \ensuremath{\sigma}(E\ensuremath{_{^{\textnormal{3}}\textnormal{He}}},\ensuremath{\theta}); performed DWBA calculations; deduced}\\
\parbox[b][0.3cm]{17.7cm}{\ensuremath{^{\textnormal{19}}}Ne level-energies, isobaric analog states, and Coulomb displacement energies.}\\
\parbox[b][0.3cm]{17.7cm}{\addtolength{\parindent}{-0.2in}\href{https://www.nndc.bnl.gov/nsr/nsrlink.jsp?1976Na18,B}{1976Na18}: \ensuremath{^{\textnormal{21}}}Ne(p,t), \ensuremath{^{\textnormal{21}}}Ne(p,\ensuremath{^{\textnormal{3}}}He) E=40 MeV; momentum analyzed tritons and \ensuremath{^{\textnormal{3}}}He charged-particles using a split-pole}\\
\parbox[b][0.3cm]{17.7cm}{spectrograph; measured \ensuremath{\sigma}(\ensuremath{\theta})\ensuremath{_{\textnormal{(p,}^{\textnormal{3}}\textnormal{He)}}}/\ensuremath{\sigma}(\ensuremath{\theta})\ensuremath{_{\textnormal{(p,t)}}} at \ensuremath{\theta}\ensuremath{_{\textnormal{lab}}}=6\ensuremath{^\circ}{\textminus}50\ensuremath{^\circ}; measured \ensuremath{^{\textnormal{21}}}Ne(p,p) at E=40 MeV and deduced \ensuremath{\sigma}\ensuremath{_{\textnormal{(p,p)}}}/\ensuremath{\sigma}\ensuremath{_{\textnormal{(p,t)}}} and}\\
\parbox[b][0.3cm]{17.7cm}{\ensuremath{\sigma}\ensuremath{_{\textnormal{(p,p)}}}/\ensuremath{\sigma}\ensuremath{_{\textnormal{(p,}^{\textnormal{3}}\textnormal{He)}}}; performed DWBA calculations for both reactions. The S=0, T=1 pickup strength reproduces the \ensuremath{^{\textnormal{3}}}He angular}\\
\parbox[b][0.3cm]{17.7cm}{distributions from \ensuremath{^{\textnormal{21}}}Ne(p,\ensuremath{^{\textnormal{3}}}He). The S=1,T=0 strength is quenched only for the ground-state transition. This result is not}\\
\parbox[b][0.3cm]{17.7cm}{explainable by spin-orbit inclusion in the DWBA calculations.}\\
\parbox[b][0.3cm]{17.7cm}{\addtolength{\parindent}{-0.2in}\href{https://www.nndc.bnl.gov/nsr/nsrlink.jsp?1978Fo26,B}{1978Fo26}: \ensuremath{^{\textnormal{21}}}Ne(p,t) E=40 MeV; measured \ensuremath{\sigma}(E\ensuremath{_{\textnormal{t}}},\ensuremath{\theta}) using a split-pole spectrograph and its focal plane detector positioned at}\\
\parbox[b][0.3cm]{17.7cm}{\ensuremath{\theta}\ensuremath{_{\textnormal{lab}}}=8\ensuremath{^\circ}{\textminus}50\ensuremath{^\circ}. Deduced \ensuremath{^{\textnormal{19}}}Ne levels; estimated the dominant wave function configuration for E\ensuremath{_{\textnormal{x}}}(\ensuremath{^{\textnormal{19}}}Ne)=4033 keV; performed}\\
\parbox[b][0.3cm]{17.7cm}{DWBA calculations with microscopic wave functions.}\\
\parbox[b][0.3cm]{17.7cm}{\addtolength{\parindent}{-0.2in}\href{https://www.nndc.bnl.gov/nsr/nsrlink.jsp?1979Fo06,B}{1979Fo06}: \ensuremath{^{\textnormal{21}}}Ne(p,t) E=40 MeV; measured \ensuremath{\sigma}(E\ensuremath{_{\textnormal{t}}},\ensuremath{\theta}); deduced a \ensuremath{^{\textnormal{19}}}Ne level at E\ensuremath{_{\textnormal{x}}}=5.09 MeV; deduced L and \ensuremath{\pi} for this state using}\\
\parbox[b][0.3cm]{17.7cm}{DWBA analysis.}\\
\vspace{0.385cm}
\parbox[b][0.3cm]{17.7cm}{\addtolength{\parindent}{-0.2in}\textit{Measurements Performed in Inverse Kinematics using a \ensuremath{^{21}}Ne Beam:}}\\
\parbox[b][0.3cm]{17.7cm}{\addtolength{\parindent}{-0.2in}\href{https://www.nndc.bnl.gov/nsr/nsrlink.jsp?2003Da03,B}{2003Da03}, \href{https://www.nndc.bnl.gov/nsr/nsrlink.jsp?2003Da13,B}{2003Da13}, \href{https://www.nndc.bnl.gov/nsr/nsrlink.jsp?2003DaZZ,B}{2003DaZZ}, \href{https://www.nndc.bnl.gov/nsr/nsrlink.jsp?2003Da25,B}{2003Da25}: \ensuremath{^{\textnormal{1}}}H(\ensuremath{^{\textnormal{21}}}Ne,t)\ensuremath{^{\textnormal{19}}}Ne*(\ensuremath{\alpha}) E=43 MeV/nucleon; measured TOF, energy loss, and total}\\
\parbox[b][0.3cm]{17.7cm}{energy of \ensuremath{^{\textnormal{19}}}Ne and \ensuremath{^{\textnormal{15}}}O from the \ensuremath{^{\textnormal{19}}}Ne*\ensuremath{\rightarrow}\ensuremath{^{\textnormal{15}}}O+\ensuremath{\alpha} decay; measured positions of backward emitted (in c.m. frame) tritons using the}\\
\parbox[b][0.3cm]{17.7cm}{focal plane detection system of the Big-Bite spectrometer, at KVI institute, placed at \ensuremath{\theta}\ensuremath{_{\textnormal{lab}}}=0\ensuremath{^\circ}. Measured \ensuremath{^{\textnormal{19}}}Ne-t coincidences}\\
\parbox[b][0.3cm]{17.7cm}{(representing \ensuremath{\gamma} decays of the \ensuremath{^{\textnormal{19}}}Ne* states) and \ensuremath{^{\textnormal{15}}}O-t coincidences (representing \ensuremath{\alpha} decays of the \ensuremath{^{\textnormal{19}}}Ne* states). Deduced \ensuremath{^{\textnormal{19}}}Ne}\\
\parbox[b][0.3cm]{17.7cm}{levels, \ensuremath{\alpha}-decay branching ratios, decay widths, and reduced \ensuremath{\alpha} widths. Energy resolution was 90 keV FWHM. Deduced the}\\
\parbox[b][0.3cm]{17.7cm}{\ensuremath{^{\textnormal{15}}}O(\ensuremath{\alpha},\ensuremath{\gamma}) astrophysical reaction rate and discussed its astrophysical implications.}\\
\vspace{0.385cm}
\parbox[b][0.3cm]{17.7cm}{\addtolength{\parindent}{-0.2in}\textit{Theory:}}\\
\parbox[b][0.3cm]{17.7cm}{\addtolength{\parindent}{-0.2in}\href{https://www.nndc.bnl.gov/nsr/nsrlink.jsp?2010Fo07,B}{2010Fo07}: Deduced \ensuremath{\Gamma}\ensuremath{_{\ensuremath{\alpha}}} for the \ensuremath{^{\textnormal{19}}}Ne*(4.03, 4.379, 4.6, 5.092, 5.351, 7.42 MeV) levels using the experimental and theoretical}\\
\parbox[b][0.3cm]{17.7cm}{information available at the time in the literature. The \ensuremath{\alpha} spectroscopic factors (S\ensuremath{_{\ensuremath{\alpha}}}) were computed for the mirrors of those states in}\\
\parbox[b][0.3cm]{17.7cm}{\ensuremath{^{\textnormal{19}}}F and were used to compute the \ensuremath{\alpha} widths of the \ensuremath{^{\textnormal{19}}}Ne*(4.379, 4.6, 5.092, 5.351, 7.42 MeV) levels.}\\
\parbox[b][0.3cm]{17.7cm}{\addtolength{\parindent}{-0.2in}\href{https://www.nndc.bnl.gov/nsr/nsrlink.jsp?2018Ge07,B}{2018Ge07} and its supplemented material: \ensuremath{^{\textnormal{1}}}H(\ensuremath{^{\textnormal{21}}}Ne,t), \ensuremath{^{\textnormal{1}}}H(\ensuremath{^{\textnormal{24,25,26}}}Mg, \ensuremath{^{\textnormal{6,7,8}}}Li) E=0.1-10 GeV/nucleon; analyzed production \ensuremath{\sigma}(E)}\\
\parbox[b][0.3cm]{17.7cm}{with benchmark parametrizations and compared with experimental data.}\\
\vspace{12pt}
\underline{$^{19}$Ne Levels}\\
\vspace{0.34cm}
\parbox[b][0.3cm]{17.7cm}{\addtolength{\parindent}{-0.254cm}\textit{Notes}:}\\
\parbox[b][0.3cm]{17.7cm}{\addtolength{\parindent}{-0.254cm}(1) (\href{https://www.nndc.bnl.gov/nsr/nsrlink.jsp?2003Da03,B}{2003Da03}, \href{https://www.nndc.bnl.gov/nsr/nsrlink.jsp?2003Da13,B}{2003Da13}) used \ensuremath{\Gamma}\ensuremath{_{\ensuremath{\gamma}}} from the literature and deduced \ensuremath{\Gamma}\ensuremath{_{\ensuremath{\alpha}}} using their measured values of B\ensuremath{_{\ensuremath{\alpha}}}\ensuremath{\equiv}\ensuremath{\Gamma}\ensuremath{_{\ensuremath{\alpha}}}/\ensuremath{\Gamma} and the}\\
\parbox[b][0.3cm]{17.7cm}{\ensuremath{\Gamma}\ensuremath{_{\ensuremath{\alpha}}}=B\ensuremath{_{\ensuremath{\alpha}}}/[\ensuremath{\Gamma}\ensuremath{_{\ensuremath{\gamma}}}(1{\textminus}B\ensuremath{_{\ensuremath{\alpha}}})] relation. The \ensuremath{\Gamma}\ensuremath{_{\ensuremath{\gamma}}} values used by those authors are given and explained in other datasets. We did not present}\\
\parbox[b][0.3cm]{17.7cm}{the deduced \ensuremath{\Gamma}\ensuremath{_{\ensuremath{\alpha}}} values. Instead, we only present the measured \ensuremath{\Gamma}\ensuremath{_{\ensuremath{\alpha}}}/\ensuremath{\Gamma} values.}\\
\parbox[b][0.3cm]{17.7cm}{\addtolength{\parindent}{-0.254cm}(2) \ensuremath{\theta}\ensuremath{^{\textnormal{2}}_{\ensuremath{\alpha}}} are the reduced \ensuremath{\alpha} widths deduced by (\href{https://www.nndc.bnl.gov/nsr/nsrlink.jsp?2003Da13,B}{2003Da13}), see Eq. (1) in that study.}\\
\vspace{0.34cm}
\begin{longtable}{ccccccc@{\extracolsep{\fill}}c}
\multicolumn{2}{c}{E(level)$^{}$}&J$^{\pi}$$^{{\hyperlink{NE7LEVEL5}{f}}}$&\multicolumn{2}{c}{\ensuremath{\Gamma}\ensuremath{\alpha}/\ensuremath{\Gamma}$^{{\hyperlink{NE7LEVEL7}{h}}}$}&L$^{}$&Comments&\\[-.2cm]
\multicolumn{2}{c}{\hrulefill}&\hrulefill&\multicolumn{2}{c}{\hrulefill}&\hrulefill&\hrulefill&
\endfirsthead
\multicolumn{1}{r@{}}{0}&\multicolumn{1}{@{}l}{}&\multicolumn{1}{l}{1/2\ensuremath{^{+}}}&&&\multicolumn{1}{l}{2}&\parbox[t][0.3cm]{11.5311cm}{\raggedright T=1/2 (\href{https://www.nndc.bnl.gov/nsr/nsrlink.jsp?1969Ha38,B}{1969Ha38})\vspace{0.1cm}}&\\
&&&&&&\parbox[t][0.3cm]{11.5311cm}{\raggedright E(level): From (\href{https://www.nndc.bnl.gov/nsr/nsrlink.jsp?1969Ha38,B}{1969Ha38}: See the top panel of Fig. 7) and (\href{https://www.nndc.bnl.gov/nsr/nsrlink.jsp?1976Na18,B}{1976Na18}).\vspace{0.1cm}}&\\
&&&&&&\parbox[t][0.3cm]{11.5311cm}{\raggedright J\ensuremath{^{\pi}},L: From DWBA calculations in (\href{https://www.nndc.bnl.gov/nsr/nsrlink.jsp?1969Ha38,B}{1969Ha38}).\vspace{0.1cm}}&\\
\multicolumn{1}{r@{}}{238}&\multicolumn{1}{@{}l}{}&\multicolumn{1}{l}{5/2\ensuremath{^{+}}}&&&\multicolumn{1}{l}{2}&\parbox[t][0.3cm]{11.5311cm}{\raggedright E(level): From (\href{https://www.nndc.bnl.gov/nsr/nsrlink.jsp?1978Fo26,B}{1978Fo26}). See also 0.24 MeV (\href{https://www.nndc.bnl.gov/nsr/nsrlink.jsp?1969Ha38,B}{1969Ha38}).\vspace{0.1cm}}&\\
&&&&&&\parbox[t][0.3cm]{11.5311cm}{\raggedright J\ensuremath{^{\pi}},L: From DWBA calculations of (\href{https://www.nndc.bnl.gov/nsr/nsrlink.jsp?1978Fo26,B}{1978Fo26}) using microscopic two-nucleon\vspace{0.1cm}}&\\
&&&&&&\parbox[t][0.3cm]{11.5311cm}{\raggedright {\ }{\ }{\ }transfer option of the DWUCK code.\vspace{0.1cm}}&\\
&&&&&&\parbox[t][0.3cm]{11.5311cm}{\raggedright (\href{https://www.nndc.bnl.gov/nsr/nsrlink.jsp?1979Fo06,B}{1979Fo06}) obtained a normalization coefficient of C\ensuremath{^{\textnormal{2}}}N=95 for this state, where\vspace{0.1cm}}&\\
&&&&&&\parbox[t][0.3cm]{11.5311cm}{\raggedright {\ }{\ }{\ }C\ensuremath{^{\textnormal{2}}}=2/3 is an isospin Clebsch-Gordan coefficient, which leads to N=142.\vspace{0.1cm}}&\\
\multicolumn{1}{r@{}}{2795}&\multicolumn{1}{@{}l}{\ensuremath{^{{\hyperlink{NE7LEVEL1}{b}}}}}&\multicolumn{1}{l}{9/2\ensuremath{^{+}}}&&&&\parbox[t][0.3cm]{11.5311cm}{\raggedright E(level): From Fig. 7 in (\href{https://www.nndc.bnl.gov/nsr/nsrlink.jsp?2003Da13,B}{2003Da13}). See also 2.78 MeV (\href{https://www.nndc.bnl.gov/nsr/nsrlink.jsp?1969Ha38,B}{1969Ha38}).\vspace{0.1cm}}&\\
\multicolumn{1}{r@{}}{4033}&\multicolumn{1}{@{}l}{\ensuremath{^{{\hyperlink{NE7LEVEL0}{a}}{\hyperlink{NE7LEVEL1}{b}}{\hyperlink{NE7LEVEL4}{e}}}}}&\multicolumn{1}{l}{3/2\ensuremath{^{+}}}&\multicolumn{1}{r@{}}{$<$4}&\multicolumn{1}{@{.}l}{3\ensuremath{\times10^{-4}}\ensuremath{^{{\hyperlink{NE7LEVEL6}{g}}{\hyperlink{NE7LEVEL8}{i}}}}}&\multicolumn{1}{l}{0}&\parbox[t][0.3cm]{11.5311cm}{\raggedright T=1/2 (\href{https://www.nndc.bnl.gov/nsr/nsrlink.jsp?1969Ha38,B}{1969Ha38},\href{https://www.nndc.bnl.gov/nsr/nsrlink.jsp?1978Fo26,B}{1978Fo26})\vspace{0.1cm}}&\\
\end{longtable}
\begin{textblock}{29}(0,27.3)
Continued on next page (footnotes at end of table)
\end{textblock}
\clearpage
\begin{longtable}{ccccccc@{\extracolsep{\fill}}c}
\\[-.4cm]
\multicolumn{8}{c}{{\bf \small \underline{\ensuremath{^{\textnormal{1}}}H(\ensuremath{^{\textnormal{21}}}Ne,t)\hspace{0.2in}\href{https://www.nndc.bnl.gov/nsr/nsrlink.jsp?1978Fo26,B}{1978Fo26},\href{https://www.nndc.bnl.gov/nsr/nsrlink.jsp?2003Da13,B}{2003Da13} (continued)}}}\\
\multicolumn{8}{c}{~}\\
\multicolumn{8}{c}{\underline{\ensuremath{^{19}}Ne Levels (continued)}}\\
\multicolumn{8}{c}{~}\\
\multicolumn{2}{c}{E(level)$^{}$}&J$^{\pi}$$^{{\hyperlink{NE7LEVEL5}{f}}}$&\multicolumn{2}{c}{\ensuremath{\Gamma}\ensuremath{\alpha}/\ensuremath{\Gamma}$^{{\hyperlink{NE7LEVEL7}{h}}}$}&L$^{}$&Comments&\\[-.2cm]
\multicolumn{2}{c}{\hrulefill}&\hrulefill&\multicolumn{2}{c}{\hrulefill}&\hrulefill&\hrulefill&
\endhead
&&&&&&\parbox[t][0.3cm]{11.12878cm}{\raggedright E(level): From (\href{https://www.nndc.bnl.gov/nsr/nsrlink.jsp?1978Fo26,B}{1978Fo26}, \href{https://www.nndc.bnl.gov/nsr/nsrlink.jsp?2003Da03,B}{2003Da03}, \href{https://www.nndc.bnl.gov/nsr/nsrlink.jsp?2003Da13,B}{2003Da13}, \href{https://www.nndc.bnl.gov/nsr/nsrlink.jsp?2003Da25,B}{2003Da25}). See also 4013 keV\vspace{0.1cm}}&\\
&&&&&&\parbox[t][0.3cm]{11.12878cm}{\raggedright {\ }{\ }{\ }(\href{https://www.nndc.bnl.gov/nsr/nsrlink.jsp?1969Ha38,B}{1969Ha38}: See the caption of Fig. 8).\vspace{0.1cm}}&\\
&&&&&&\parbox[t][0.3cm]{11.12878cm}{\raggedright J\ensuremath{^{\pi}},L: From DWBA calculations of (\href{https://www.nndc.bnl.gov/nsr/nsrlink.jsp?1978Fo26,B}{1978Fo26}) using microscopic two-nucleon\vspace{0.1cm}}&\\
&&&&&&\parbox[t][0.3cm]{11.12878cm}{\raggedright {\ }{\ }{\ }transfer option of the DWUCK code.\vspace{0.1cm}}&\\
&&&&&&\parbox[t][0.3cm]{11.12878cm}{\raggedright \ensuremath{\Gamma}\ensuremath{\alpha}/\ensuremath{\Gamma}: The upper limit is explained in (\href{https://www.nndc.bnl.gov/nsr/nsrlink.jsp?2011Da24,B}{2011Da24}), where it is\vspace{0.1cm}}&\\
&&&&&&\parbox[t][0.3cm]{11.12878cm}{\raggedright {\ }{\ }{\ }reported that a small number of excess counts above the background under this\vspace{0.1cm}}&\\
&&&&&&\parbox[t][0.3cm]{11.12878cm}{\raggedright {\ }{\ }{\ }peak was observed in (\href{https://www.nndc.bnl.gov/nsr/nsrlink.jsp?2003Da13,B}{2003Da13}) but were considered statistically insignificant,\vspace{0.1cm}}&\\
&&&&&&\parbox[t][0.3cm]{11.12878cm}{\raggedright {\ }{\ }{\ }which led to the upper limit reported by (\href{https://www.nndc.bnl.gov/nsr/nsrlink.jsp?2003Da13,B}{2003Da13}).\vspace{0.1cm}}&\\
&&&&&&\parbox[t][0.3cm]{11.12878cm}{\raggedright Decay mode: Predominantly \ensuremath{\gamma} (\href{https://www.nndc.bnl.gov/nsr/nsrlink.jsp?2003Da03,B}{2003Da03}, \href{https://www.nndc.bnl.gov/nsr/nsrlink.jsp?2003Da13,B}{2003Da13}).\vspace{0.1cm}}&\\
&&&&&&\parbox[t][0.3cm]{11.12878cm}{\raggedright (\href{https://www.nndc.bnl.gov/nsr/nsrlink.jsp?1979Fo06,B}{1979Fo06}) obtained a normalization coefficient of N=110 for this state.\vspace{0.1cm}}&\\
&&&&&&\parbox[t][0.3cm]{11.12878cm}{\raggedright (\href{https://www.nndc.bnl.gov/nsr/nsrlink.jsp?1978Fo26,B}{1978Fo26}): This state is consistent with a dominant 5p-2h configuration\vspace{0.1cm}}&\\
&&&&&&\parbox[t][0.3cm]{11.12878cm}{\raggedright {\ }{\ }{\ }((\textit{sd})\ensuremath{^{\textnormal{5}}}(1\textit{p})\ensuremath{^{\textnormal{$-$2}}}) with the \ensuremath{^{\textnormal{14}}}O\ensuremath{_{\textnormal{g.s.}}} core and an amplitude of 0.88. The particle\vspace{0.1cm}}&\\
&&&&&&\parbox[t][0.3cm]{11.12878cm}{\raggedright {\ }{\ }{\ }configuration is the same as that in the \ensuremath{^{\textnormal{21}}}Ne\ensuremath{_{\textnormal{g.s.}}} and the hole configuration is\vspace{0.1cm}}&\\
&&&&&&\parbox[t][0.3cm]{11.12878cm}{\raggedright {\ }{\ }{\ }that of the \ensuremath{^{\textnormal{14}}}O\ensuremath{_{\textnormal{g.s.}}}.\vspace{0.1cm}}&\\
\multicolumn{1}{r@{}}{4140}&\multicolumn{1}{@{}l}{\ensuremath{^{{\hyperlink{NE7LEVEL1}{b}}{\hyperlink{NE7LEVEL3}{d}}{\hyperlink{NE7LEVEL4}{e}}}}}&\multicolumn{1}{l}{(7/2\ensuremath{^{-}})}&&&&\parbox[t][0.3cm]{11.12878cm}{\raggedright E(level): From (\href{https://www.nndc.bnl.gov/nsr/nsrlink.jsp?2003Da03,B}{2003Da03}).\vspace{0.1cm}}&\\
&&&&&&\parbox[t][0.3cm]{11.12878cm}{\raggedright Decay mode: Predominantly \ensuremath{\gamma}. The \ensuremath{\alpha} decay is hindered by an L=4 centrifugal\vspace{0.1cm}}&\\
&&&&&&\parbox[t][0.3cm]{11.12878cm}{\raggedright {\ }{\ }{\ }barrier and a low decay energy (\href{https://www.nndc.bnl.gov/nsr/nsrlink.jsp?2003Da03,B}{2003Da03}).\vspace{0.1cm}}&\\
\multicolumn{1}{r@{}}{4197}&\multicolumn{1}{@{}l}{\ensuremath{^{{\hyperlink{NE7LEVEL1}{b}}{\hyperlink{NE7LEVEL3}{d}}{\hyperlink{NE7LEVEL4}{e}}}}}&\multicolumn{1}{l}{(9/2\ensuremath{^{-}})}&&&&\parbox[t][0.3cm]{11.12878cm}{\raggedright E(level): From (\href{https://www.nndc.bnl.gov/nsr/nsrlink.jsp?2003Da03,B}{2003Da03}).\vspace{0.1cm}}&\\
&&&&&&\parbox[t][0.3cm]{11.12878cm}{\raggedright Decay mode: Predominantly \ensuremath{\gamma}. The \ensuremath{\alpha} decay is hindered by an L=4 centrifugal\vspace{0.1cm}}&\\
&&&&&&\parbox[t][0.3cm]{11.12878cm}{\raggedright {\ }{\ }{\ }barrier and a low decay energy (\href{https://www.nndc.bnl.gov/nsr/nsrlink.jsp?2003Da03,B}{2003Da03}).\vspace{0.1cm}}&\\
\multicolumn{1}{r@{}}{4379}&\multicolumn{1}{@{}l}{\ensuremath{^{{\hyperlink{NE7LEVEL1}{b}}{\hyperlink{NE7LEVEL4}{e}}}}}&\multicolumn{1}{l}{7/2\ensuremath{^{+}}}&\multicolumn{1}{r@{}}{$<$3}&\multicolumn{1}{@{.}l}{9\ensuremath{\times10^{-3}}\ensuremath{^{{\hyperlink{NE7LEVEL6}{g}}{\hyperlink{NE7LEVEL8}{i}}}}}&&\parbox[t][0.3cm]{11.12878cm}{\raggedright T=1/2 (\href{https://www.nndc.bnl.gov/nsr/nsrlink.jsp?2003Da13,B}{2003Da13})\vspace{0.1cm}}&\\
&&&&&&\parbox[t][0.3cm]{11.12878cm}{\raggedright E(level): From (\href{https://www.nndc.bnl.gov/nsr/nsrlink.jsp?2003Da03,B}{2003Da03}, \href{https://www.nndc.bnl.gov/nsr/nsrlink.jsp?2003Da13,B}{2003Da13}, \href{https://www.nndc.bnl.gov/nsr/nsrlink.jsp?2003Da25,B}{2003Da25}).\vspace{0.1cm}}&\\
&&&&&&\parbox[t][0.3cm]{11.12878cm}{\raggedright \ensuremath{\Gamma}\ensuremath{\alpha}/\ensuremath{\Gamma}: (\href{https://www.nndc.bnl.gov/nsr/nsrlink.jsp?2011Da24,B}{2011Da24}) reported that this upper limit (at 90\% C.L., see\vspace{0.1cm}}&\\
&&&&&&\parbox[t][0.3cm]{11.12878cm}{\raggedright {\ }{\ }{\ }\href{https://www.nndc.bnl.gov/nsr/nsrlink.jsp?2010Fo07,B}{2010Fo07}) is based on the lack of observation by (\href{https://www.nndc.bnl.gov/nsr/nsrlink.jsp?2003Da13,B}{2003Da13}) of any excess\vspace{0.1cm}}&\\
&&&&&&\parbox[t][0.3cm]{11.12878cm}{\raggedright {\ }{\ }{\ }counts above the background for this state. (\href{https://www.nndc.bnl.gov/nsr/nsrlink.jsp?2010Fo07,B}{2010Fo07}) reports (via priv. comm.\vspace{0.1cm}}&\\
&&&&&&\parbox[t][0.3cm]{11.12878cm}{\raggedright {\ }{\ }{\ }with B. Davids) that the upper limit based on 68\% C.L. is \ensuremath{\Gamma}\ensuremath{_{\ensuremath{\alpha}}}/\ensuremath{\Gamma}\ensuremath{<}2.6\ensuremath{\times}10\ensuremath{^{\textnormal{$-$3}}}\vspace{0.1cm}}&\\
&&&&&&\parbox[t][0.3cm]{11.12878cm}{\raggedright {\ }{\ }{\ }(\href{https://www.nndc.bnl.gov/nsr/nsrlink.jsp?2010Fo07,B}{2010Fo07}: See Table III).\vspace{0.1cm}}&\\
&&&&&&\parbox[t][0.3cm]{11.12878cm}{\raggedright \ensuremath{\Gamma}\ensuremath{\alpha}/\ensuremath{\Gamma}: This result is a factor of 11 smaller than that of (\href{https://www.nndc.bnl.gov/nsr/nsrlink.jsp?1990Ma05,B}{1990Ma05}:\vspace{0.1cm}}&\\
&&&&&&\parbox[t][0.3cm]{11.12878cm}{\raggedright {\ }{\ }{\ }\ensuremath{^{\textnormal{19}}}F(\ensuremath{^{\textnormal{3}}}He,t)\ensuremath{^{\textnormal{19}}}Ne*(\ensuremath{\alpha})). (\href{https://www.nndc.bnl.gov/nsr/nsrlink.jsp?2003Da13,B}{2003Da13}) attributed this discrepancy to imperfect\vspace{0.1cm}}&\\
&&&&&&\parbox[t][0.3cm]{11.12878cm}{\raggedright {\ }{\ }{\ }background subtraction in (\href{https://www.nndc.bnl.gov/nsr/nsrlink.jsp?1990Ma05,B}{1990Ma05}).\vspace{0.1cm}}&\\
&&&&&&\parbox[t][0.3cm]{11.12878cm}{\raggedright \ensuremath{\theta}\ensuremath{^{\textnormal{2}}_{\ensuremath{\alpha}}}\ensuremath{<}0.095 (\href{https://www.nndc.bnl.gov/nsr/nsrlink.jsp?2003Da13,B}{2003Da13}).\vspace{0.1cm}}&\\
&&&&&&\parbox[t][0.3cm]{11.12878cm}{\raggedright Decay mode: Predominantly \ensuremath{\gamma} (\href{https://www.nndc.bnl.gov/nsr/nsrlink.jsp?2003Da03,B}{2003Da03}, \href{https://www.nndc.bnl.gov/nsr/nsrlink.jsp?2003Da13,B}{2003Da13}).\vspace{0.1cm}}&\\
\multicolumn{1}{r@{}}{4549}&\multicolumn{1}{@{}l}{\ensuremath{^{{\hyperlink{NE7LEVEL1}{b}}{\hyperlink{NE7LEVEL2}{c}}{\hyperlink{NE7LEVEL3}{d}}}}}&\multicolumn{1}{l}{3/2\ensuremath{^{-}}}&\multicolumn{1}{r@{}}{0}&\multicolumn{1}{@{.}l}{16 {\it 4}}&&\parbox[t][0.3cm]{11.12878cm}{\raggedright T=1/2 (\href{https://www.nndc.bnl.gov/nsr/nsrlink.jsp?2003Da13,B}{2003Da13})\vspace{0.1cm}}&\\
&&&&&&\parbox[t][0.3cm]{11.12878cm}{\raggedright E(level): From (\href{https://www.nndc.bnl.gov/nsr/nsrlink.jsp?2003Da03,B}{2003Da03}, \href{https://www.nndc.bnl.gov/nsr/nsrlink.jsp?2003Da13,B}{2003Da13}, \href{https://www.nndc.bnl.gov/nsr/nsrlink.jsp?2003Da25,B}{2003Da25}).\vspace{0.1cm}}&\\
&&&&&&\parbox[t][0.3cm]{11.12878cm}{\raggedright J\ensuremath{^{\pi}}: See also J\ensuremath{^{\ensuremath{\pi}}}=(1/2,3/2)\ensuremath{^{-}} mentioned by (\href{https://www.nndc.bnl.gov/nsr/nsrlink.jsp?2003Da03,B}{2003Da03}).\vspace{0.1cm}}&\\
&&&&&&\parbox[t][0.3cm]{11.12878cm}{\raggedright \ensuremath{\Gamma}\ensuremath{\alpha}/\ensuremath{\Gamma}: From (\href{https://www.nndc.bnl.gov/nsr/nsrlink.jsp?2003Da03,B}{2003Da03}, \href{https://www.nndc.bnl.gov/nsr/nsrlink.jsp?2003Da13,B}{2003Da13}).\vspace{0.1cm}}&\\
&&&&&&\parbox[t][0.3cm]{11.12878cm}{\raggedright \ensuremath{\theta}\ensuremath{^{\textnormal{2}}_{\ensuremath{\alpha}}}=0.0016 \textit{+15{\textminus}7} (\href{https://www.nndc.bnl.gov/nsr/nsrlink.jsp?2003Da13,B}{2003Da13}).\vspace{0.1cm}}&\\
&&&&&&\parbox[t][0.3cm]{11.12878cm}{\raggedright Decay mode: Predominantly \ensuremath{\alpha} (\href{https://www.nndc.bnl.gov/nsr/nsrlink.jsp?2003Da03,B}{2003Da03}, \href{https://www.nndc.bnl.gov/nsr/nsrlink.jsp?2003Da13,B}{2003Da13}).\vspace{0.1cm}}&\\
\multicolumn{1}{r@{}}{4600}&\multicolumn{1}{@{}l}{\ensuremath{^{{\hyperlink{NE7LEVEL1}{b}}{\hyperlink{NE7LEVEL2}{c}}{\hyperlink{NE7LEVEL3}{d}}}}}&\multicolumn{1}{l}{5/2\ensuremath{^{+}}}&\multicolumn{1}{r@{}}{0}&\multicolumn{1}{@{.}l}{32 {\it 4}}&&\parbox[t][0.3cm]{11.12878cm}{\raggedright T=1/2 (\href{https://www.nndc.bnl.gov/nsr/nsrlink.jsp?2003Da13,B}{2003Da13})\vspace{0.1cm}}&\\
&&&&&&\parbox[t][0.3cm]{11.12878cm}{\raggedright E(level): From (\href{https://www.nndc.bnl.gov/nsr/nsrlink.jsp?2003Da03,B}{2003Da03}, \href{https://www.nndc.bnl.gov/nsr/nsrlink.jsp?2003Da13,B}{2003Da13}, \href{https://www.nndc.bnl.gov/nsr/nsrlink.jsp?2003Da25,B}{2003Da25}).\vspace{0.1cm}}&\\
&&&&&&\parbox[t][0.3cm]{11.12878cm}{\raggedright \ensuremath{\Gamma}\ensuremath{\alpha}/\ensuremath{\Gamma}: From (\href{https://www.nndc.bnl.gov/nsr/nsrlink.jsp?2003Da03,B}{2003Da03}, \href{https://www.nndc.bnl.gov/nsr/nsrlink.jsp?2003Da13,B}{2003Da13}).\vspace{0.1cm}}&\\
&&&&&&\parbox[t][0.3cm]{11.12878cm}{\raggedright \ensuremath{\theta}\ensuremath{^{\textnormal{2}}_{\ensuremath{\alpha}}}=0.063 \textit{35} (\href{https://www.nndc.bnl.gov/nsr/nsrlink.jsp?2003Da13,B}{2003Da13}).\vspace{0.1cm}}&\\
&&&&&&\parbox[t][0.3cm]{11.12878cm}{\raggedright Decay mode: Predominantly \ensuremath{\alpha} (\href{https://www.nndc.bnl.gov/nsr/nsrlink.jsp?2003Da03,B}{2003Da03}, \href{https://www.nndc.bnl.gov/nsr/nsrlink.jsp?2003Da13,B}{2003Da13}).\vspace{0.1cm}}&\\
\multicolumn{1}{r@{}}{4712}&\multicolumn{1}{@{}l}{\ensuremath{^{{\hyperlink{NE7LEVEL1}{b}}{\hyperlink{NE7LEVEL2}{c}}}}}&\multicolumn{1}{l}{5/2\ensuremath{^{-}}}&\multicolumn{1}{r@{}}{0}&\multicolumn{1}{@{.}l}{85 {\it 4}}&&\parbox[t][0.3cm]{11.12878cm}{\raggedright T=1/2 (\href{https://www.nndc.bnl.gov/nsr/nsrlink.jsp?2003Da13,B}{2003Da13})\vspace{0.1cm}}&\\
&&&&&&\parbox[t][0.3cm]{11.12878cm}{\raggedright E(level): From (\href{https://www.nndc.bnl.gov/nsr/nsrlink.jsp?2003Da03,B}{2003Da03}, \href{https://www.nndc.bnl.gov/nsr/nsrlink.jsp?2003Da13,B}{2003Da13}, \href{https://www.nndc.bnl.gov/nsr/nsrlink.jsp?2003Da25,B}{2003Da25}).\vspace{0.1cm}}&\\
&&&&&&\parbox[t][0.3cm]{11.12878cm}{\raggedright \ensuremath{\Gamma}\ensuremath{\alpha}/\ensuremath{\Gamma}: From (\href{https://www.nndc.bnl.gov/nsr/nsrlink.jsp?2003Da03,B}{2003Da03}, \href{https://www.nndc.bnl.gov/nsr/nsrlink.jsp?2003Da13,B}{2003Da13}).\vspace{0.1cm}}&\\
&&&&&&\parbox[t][0.3cm]{11.12878cm}{\raggedright \ensuremath{\theta}\ensuremath{^{\textnormal{2}}_{\ensuremath{\alpha}}}=0.012 \textit{4} (\href{https://www.nndc.bnl.gov/nsr/nsrlink.jsp?2003Da13,B}{2003Da13}).\vspace{0.1cm}}&\\
&&&&&&\parbox[t][0.3cm]{11.12878cm}{\raggedright Decay mode: Predominantly \ensuremath{\alpha} (\href{https://www.nndc.bnl.gov/nsr/nsrlink.jsp?2003Da03,B}{2003Da03}, \href{https://www.nndc.bnl.gov/nsr/nsrlink.jsp?2003Da13,B}{2003Da13}).\vspace{0.1cm}}&\\
\multicolumn{1}{r@{}}{5092}&\multicolumn{1}{@{}l}{\ensuremath{^{{\hyperlink{NE7LEVEL0}{a}}{\hyperlink{NE7LEVEL1}{b}}{\hyperlink{NE7LEVEL2}{c}}}}}&\multicolumn{1}{l}{5/2\ensuremath{^{+}}}&\multicolumn{1}{r@{}}{0}&\multicolumn{1}{@{.}l}{90 {\it 6}}&\multicolumn{1}{l}{4}&\parbox[t][0.3cm]{11.12878cm}{\raggedright T=1/2 (\href{https://www.nndc.bnl.gov/nsr/nsrlink.jsp?2003Da13,B}{2003Da13})\vspace{0.1cm}}&\\
&&&&&&\parbox[t][0.3cm]{11.12878cm}{\raggedright E(level): From (\href{https://www.nndc.bnl.gov/nsr/nsrlink.jsp?2003Da03,B}{2003Da03}, \href{https://www.nndc.bnl.gov/nsr/nsrlink.jsp?2003Da13,B}{2003Da13}, \href{https://www.nndc.bnl.gov/nsr/nsrlink.jsp?2003Da25,B}{2003Da25}). See also 5.09 MeV\vspace{0.1cm}}&\\
&&&&&&\parbox[t][0.3cm]{11.12878cm}{\raggedright {\ }{\ }{\ }(\href{https://www.nndc.bnl.gov/nsr/nsrlink.jsp?1978Fo26,B}{1978Fo26}).\vspace{0.1cm}}&\\
&&&&&&\parbox[t][0.3cm]{11.12878cm}{\raggedright J\ensuremath{^{\pi}},L: From DWBA analysis of (\href{https://www.nndc.bnl.gov/nsr/nsrlink.jsp?1978Fo26,B}{1978Fo26}). DWBA curves with odd L gave a\vspace{0.1cm}}&\\
\end{longtable}
\begin{textblock}{29}(0,27.3)
Continued on next page (footnotes at end of table)
\end{textblock}
\clearpage
\begin{longtable}{ccccccc@{\extracolsep{\fill}}c}
\\[-.4cm]
\multicolumn{8}{c}{{\bf \small \underline{\ensuremath{^{\textnormal{1}}}H(\ensuremath{^{\textnormal{21}}}Ne,t)\hspace{0.2in}\href{https://www.nndc.bnl.gov/nsr/nsrlink.jsp?1978Fo26,B}{1978Fo26},\href{https://www.nndc.bnl.gov/nsr/nsrlink.jsp?2003Da13,B}{2003Da13} (continued)}}}\\
\multicolumn{8}{c}{~}\\
\multicolumn{8}{c}{\underline{\ensuremath{^{19}}Ne Levels (continued)}}\\
\multicolumn{8}{c}{~}\\
\multicolumn{2}{c}{E(level)$^{}$}&J$^{\pi}$$^{{\hyperlink{NE7LEVEL5}{f}}}$&\multicolumn{2}{c}{\ensuremath{\Gamma}\ensuremath{\alpha}/\ensuremath{\Gamma}$^{{\hyperlink{NE7LEVEL7}{h}}}$}&L$^{}$&Comments&\\[-.2cm]
\multicolumn{2}{c}{\hrulefill}&\hrulefill&\multicolumn{2}{c}{\hrulefill}&\hrulefill&\hrulefill&
\endhead
&&&&&&\parbox[t][0.3cm]{11.28286cm}{\raggedright {\ }{\ }{\ }very poor account of the triton angular distribution data, and were thus ruled out.\vspace{0.1cm}}&\\
&&&&&&\parbox[t][0.3cm]{11.28286cm}{\raggedright Based on a comparison with the potential mirror levels in \ensuremath{^{\textnormal{19}}}F, (\href{https://www.nndc.bnl.gov/nsr/nsrlink.jsp?1978Fo26,B}{1978Fo26})\vspace{0.1cm}}&\\
&&&&&&\parbox[t][0.3cm]{11.28286cm}{\raggedright {\ }{\ }{\ }reported that this state may be the J\ensuremath{^{\ensuremath{\pi}}}=5/2\ensuremath{^{\textnormal{+}}} member of the K\ensuremath{^{\ensuremath{\pi}}}=3/2\ensuremath{^{\textnormal{+}}}\vspace{0.1cm}}&\\
&&&&&&\parbox[t][0.3cm]{11.28286cm}{\raggedright {\ }{\ }{\ }core-excited rotational band, whose J\ensuremath{^{\ensuremath{\pi}}}=3/2\ensuremath{^{\textnormal{+}}} band head is the 4.03-MeV state in\vspace{0.1cm}}&\\
&&&&&&\parbox[t][0.3cm]{11.28286cm}{\raggedright {\ }{\ }{\ }\ensuremath{^{\textnormal{19}}}Ne (\href{https://www.nndc.bnl.gov/nsr/nsrlink.jsp?1978Fo26,B}{1978Fo26}).\vspace{0.1cm}}&\\
&&&&&&\parbox[t][0.3cm]{11.28286cm}{\raggedright \ensuremath{\Gamma}\ensuremath{\alpha}/\ensuremath{\Gamma}: From (\href{https://www.nndc.bnl.gov/nsr/nsrlink.jsp?2003Da03,B}{2003Da03}, \href{https://www.nndc.bnl.gov/nsr/nsrlink.jsp?2003Da13,B}{2003Da13}). Note that (\href{https://www.nndc.bnl.gov/nsr/nsrlink.jsp?2010Fo07,B}{2010Fo07}) cites this\vspace{0.1cm}}&\\
&&&&&&\parbox[t][0.3cm]{11.28286cm}{\raggedright {\ }{\ }{\ }value as \ensuremath{\Gamma}\ensuremath{_{\ensuremath{\alpha}}}/\ensuremath{\Gamma}=0.90 \textit{5} (see Table III in that study).\vspace{0.1cm}}&\\
&&&&&&\parbox[t][0.3cm]{11.28286cm}{\raggedright \ensuremath{\theta}\ensuremath{^{\textnormal{2}}_{\ensuremath{\alpha}}}=0.013 \textit{7} (\href{https://www.nndc.bnl.gov/nsr/nsrlink.jsp?2003Da13,B}{2003Da13}).\vspace{0.1cm}}&\\
&&&&&&\parbox[t][0.3cm]{11.28286cm}{\raggedright Decay mode: Predominantly \ensuremath{\alpha} (\href{https://www.nndc.bnl.gov/nsr/nsrlink.jsp?2003Da03,B}{2003Da03}, \href{https://www.nndc.bnl.gov/nsr/nsrlink.jsp?2003Da13,B}{2003Da13}).\vspace{0.1cm}}&\\
\multicolumn{1}{r@{}}{5424}&\multicolumn{1}{@{}l}{\ensuremath{^{{\hyperlink{NE7LEVEL2}{c}}{\hyperlink{NE7LEVEL10}{k}}}}}&\multicolumn{1}{l}{7/2\ensuremath{^{+}}}&\multicolumn{1}{r@{}}{1}&\multicolumn{1}{@{.}l}{0\ensuremath{^{{\hyperlink{NE7LEVEL9}{j}}}}}&&&\\
\multicolumn{1}{r@{}}{5539}&\multicolumn{1}{@{}l}{\ensuremath{^{{\hyperlink{NE7LEVEL2}{c}}{\hyperlink{NE7LEVEL10}{k}}}}}&&\multicolumn{1}{r@{}}{1}&\multicolumn{1}{@{.}l}{0\ensuremath{^{{\hyperlink{NE7LEVEL9}{j}}}}}&&&\\
\multicolumn{1}{r@{}}{5832}&\multicolumn{1}{@{}l}{\ensuremath{^{{\hyperlink{NE7LEVEL2}{c}}{\hyperlink{NE7LEVEL10}{k}}}}}&\multicolumn{1}{l}{(1/2\ensuremath{^{+}})}&\multicolumn{1}{r@{}}{1}&\multicolumn{1}{@{.}l}{0\ensuremath{^{{\hyperlink{NE7LEVEL9}{j}}}}}&&&\\
\multicolumn{1}{r@{}}{7620}&\multicolumn{1}{@{ }l}{{\it 25}}&\multicolumn{1}{l}{3/2\ensuremath{^{+}}}&&&\multicolumn{1}{l}{0}&\parbox[t][0.3cm]{11.28286cm}{\raggedright T=3/2 (\href{https://www.nndc.bnl.gov/nsr/nsrlink.jsp?1969Ha38,B}{1969Ha38})\vspace{0.1cm}}&\\
&&&&&&\parbox[t][0.3cm]{11.28286cm}{\raggedright E(level): From (\href{https://www.nndc.bnl.gov/nsr/nsrlink.jsp?1969Ha38,B}{1969Ha38}: See the top panel of Fig. 7).\vspace{0.1cm}}&\\
&&&&&&\parbox[t][0.3cm]{11.28286cm}{\raggedright J\ensuremath{^{\pi}},L: From DWBA calculations in (\href{https://www.nndc.bnl.gov/nsr/nsrlink.jsp?1969Ha38,B}{1969Ha38}).\vspace{0.1cm}}&\\
&&&&&&\parbox[t][0.3cm]{11.28286cm}{\raggedright (\href{https://www.nndc.bnl.gov/nsr/nsrlink.jsp?1969Ha38,B}{1969Ha38}) considered the \ensuremath{^{\textnormal{19}}}F*(7.66 MeV) as the mirror state and reported that\vspace{0.1cm}}&\\
&&&&&&\parbox[t][0.3cm]{11.28286cm}{\raggedright {\ }{\ }{\ }the \ensuremath{^{\textnormal{19}}}Ne*(7620) and \ensuremath{^{\textnormal{19}}}F*(7.66 MeV) mirror state are not the lowest-energy\vspace{0.1cm}}&\\
&&&&&&\parbox[t][0.3cm]{11.28286cm}{\raggedright {\ }{\ }{\ }T=3/2 levels in A=19 nuclei, but they are the analog states to the first excited\vspace{0.1cm}}&\\
&&&&&&\parbox[t][0.3cm]{11.28286cm}{\raggedright {\ }{\ }{\ }state of \ensuremath{^{\textnormal{19}}}O* at E\ensuremath{_{\textnormal{x}}}=95 keV.\vspace{0.1cm}}&\\
\end{longtable}
\parbox[b][0.3cm]{17.7cm}{\makebox[1ex]{\ensuremath{^{\hypertarget{NE7LEVEL0}{a}}}} Seq.(A): K\ensuremath{^{\ensuremath{\pi}}}=3/2\ensuremath{^{+}} band (\href{https://www.nndc.bnl.gov/nsr/nsrlink.jsp?1978Fo26,B}{1978Fo26}).}\\
\parbox[b][0.3cm]{17.7cm}{\makebox[1ex]{\ensuremath{^{\hypertarget{NE7LEVEL1}{b}}}} From the \ensuremath{^{\textnormal{19}}}Ne-\ensuremath{^{\textnormal{3}}}H coincidence events in (\href{https://www.nndc.bnl.gov/nsr/nsrlink.jsp?2003Da03,B}{2003Da03}, \href{https://www.nndc.bnl.gov/nsr/nsrlink.jsp?2003Da13,B}{2003Da13}) representing a \ensuremath{\gamma} decaying \ensuremath{^{\textnormal{19}}}Ne* state.}\\
\parbox[b][0.3cm]{17.7cm}{\makebox[1ex]{\ensuremath{^{\hypertarget{NE7LEVEL2}{c}}}} From the \ensuremath{^{\textnormal{15}}}O-\ensuremath{^{\textnormal{3}}}H coincidence events in (\href{https://www.nndc.bnl.gov/nsr/nsrlink.jsp?2003Da03,B}{2003Da03}, \href{https://www.nndc.bnl.gov/nsr/nsrlink.jsp?2003Da13,B}{2003Da13}) representing an \ensuremath{\alpha} decaying \ensuremath{^{\textnormal{19}}}Ne* state.}\\
\parbox[b][0.3cm]{17.7cm}{\makebox[1ex]{\ensuremath{^{\hypertarget{NE7LEVEL3}{d}}}} An unresolved doublet from (\href{https://www.nndc.bnl.gov/nsr/nsrlink.jsp?2003Da03,B}{2003Da03}, \href{https://www.nndc.bnl.gov/nsr/nsrlink.jsp?2003Da13,B}{2003Da13}).}\\
\parbox[b][0.3cm]{17.7cm}{\makebox[1ex]{\ensuremath{^{\hypertarget{NE7LEVEL4}{e}}}} No statistically significant evidence for \ensuremath{\alpha} decay from this state was observed by (\href{https://www.nndc.bnl.gov/nsr/nsrlink.jsp?2003Da03,B}{2003Da03}, \href{https://www.nndc.bnl.gov/nsr/nsrlink.jsp?2003Da13,B}{2003Da13}, \href{https://www.nndc.bnl.gov/nsr/nsrlink.jsp?2003DaZZ,B}{2003DaZZ}). The}\\
\parbox[b][0.3cm]{17.7cm}{{\ }{\ }\ensuremath{\alpha}-decay threshold is 3528.5 keV \textit{5} (\href{https://www.nndc.bnl.gov/nsr/nsrlink.jsp?2021Wa16,B}{2021Wa16}).}\\
\parbox[b][0.3cm]{17.7cm}{\makebox[1ex]{\ensuremath{^{\hypertarget{NE7LEVEL5}{f}}}} From the Adopted Levels of \ensuremath{^{\textnormal{19}}}Ne unless otherwise noted.}\\
\parbox[b][0.3cm]{17.7cm}{\makebox[1ex]{\ensuremath{^{\hypertarget{NE7LEVEL6}{g}}}} From (\href{https://www.nndc.bnl.gov/nsr/nsrlink.jsp?2003Da03,B}{2003Da03}, \href{https://www.nndc.bnl.gov/nsr/nsrlink.jsp?2003Da13,B}{2003Da13}). Values are reported at the 90\% C.L.}\\
\parbox[b][0.3cm]{17.7cm}{\makebox[1ex]{\ensuremath{^{\hypertarget{NE7LEVEL7}{h}}}} The 1\ensuremath{\sigma} uncertainties in the deduced branching ratios are statistical (\href{https://www.nndc.bnl.gov/nsr/nsrlink.jsp?2003Da03,B}{2003Da03}, \href{https://www.nndc.bnl.gov/nsr/nsrlink.jsp?2003Da13,B}{2003Da13}).}\\
\parbox[b][0.3cm]{17.7cm}{\makebox[1ex]{\ensuremath{^{\hypertarget{NE7LEVEL8}{i}}}} The upper limit branching ratio for this state was recommended by (\href{https://www.nndc.bnl.gov/nsr/nsrlink.jsp?2003Da03,B}{2003Da03}, \href{https://www.nndc.bnl.gov/nsr/nsrlink.jsp?2003Da13,B}{2003Da13}) from a Bayesian analysis, whose}\\
\parbox[b][0.3cm]{17.7cm}{{\ }{\ }result was more conservative than, but consistent with, that of a classical statistical analysis.}\\
\parbox[b][0.3cm]{17.7cm}{\makebox[1ex]{\ensuremath{^{\hypertarget{NE7LEVEL9}{j}}}} No evidence for the \ensuremath{\gamma} decay of this state was observed by (\href{https://www.nndc.bnl.gov/nsr/nsrlink.jsp?2003Da13,B}{2003Da13}). Therefore, \ensuremath{\Gamma}\ensuremath{_{\ensuremath{\alpha}}}/\ensuremath{\Gamma}=100\% was reported for this state by}\\
\parbox[b][0.3cm]{17.7cm}{{\ }{\ }(\href{https://www.nndc.bnl.gov/nsr/nsrlink.jsp?2003Da13,B}{2003Da13}).}\\
\parbox[b][0.3cm]{17.7cm}{\makebox[1ex]{\ensuremath{^{\hypertarget{NE7LEVEL10}{k}}}} From Fig. 8 in (\href{https://www.nndc.bnl.gov/nsr/nsrlink.jsp?2003Da13,B}{2003Da13}).}\\
\vspace{0.5cm}
\clearpage
\clearpage
\begin{figure}[h]
\begin{center}
\includegraphics{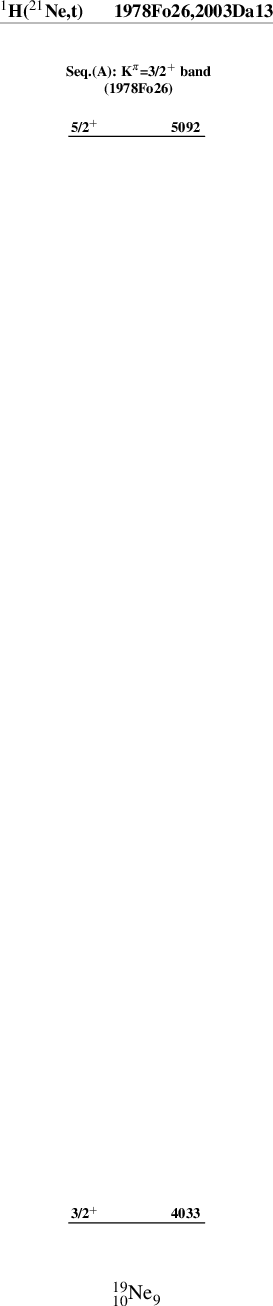}\\
\end{center}
\end{figure}
\clearpage
\subsection[\hspace{-0.2cm}\ensuremath{^{\textnormal{2}}}H(\ensuremath{^{\textnormal{18}}}F,\ensuremath{^{\textnormal{19}}}Ne)]{ }
\vspace{-27pt}
\vspace{0.3cm}
\hypertarget{NE8}{{\bf \small \underline{\ensuremath{^{\textnormal{2}}}H(\ensuremath{^{\textnormal{18}}}F,\ensuremath{^{\textnormal{19}}}Ne)\hspace{0.2in}\href{https://www.nndc.bnl.gov/nsr/nsrlink.jsp?2011Ad24,B}{2011Ad24},\href{https://www.nndc.bnl.gov/nsr/nsrlink.jsp?2015Ch41,B}{2015Ch41},\href{https://www.nndc.bnl.gov/nsr/nsrlink.jsp?2017La12,B}{2017La12}}}}\\
\vspace{4pt}
\vspace{8pt}
\parbox[b][0.3cm]{17.7cm}{\addtolength{\parindent}{-0.2in}One proton pickup reaction in inverse kinematics, detecting the heavy residues.}\\
\parbox[b][0.3cm]{17.7cm}{\addtolength{\parindent}{-0.2in}J\ensuremath{^{\ensuremath{\pi}}}(\ensuremath{^{\textnormal{2}}}H\ensuremath{_{\textnormal{g.s.}}})=1\ensuremath{^{\textnormal{+}}} and J\ensuremath{^{\ensuremath{\pi}}}(\ensuremath{^{\textnormal{18}}}F)=1\ensuremath{^{\textnormal{+}}}.}\\
\parbox[b][0.3cm]{17.7cm}{\addtolength{\parindent}{-0.2in}\href{https://www.nndc.bnl.gov/nsr/nsrlink.jsp?2011Ad05,B}{2011Ad05}, \href{https://www.nndc.bnl.gov/nsr/nsrlink.jsp?2011Ad24,B}{2011Ad24}, \href{https://www.nndc.bnl.gov/nsr/nsrlink.jsp?2015BaZQ,B}{2015BaZQ}: \ensuremath{^{\textnormal{2}}}H(\ensuremath{^{\textnormal{18}}}F,\ensuremath{^{\textnormal{19}}}Ne*\ensuremath{\rightarrow}\ensuremath{\alpha}+\ensuremath{^{\textnormal{15}}}O) and \ensuremath{^{\textnormal{2}}}H(\ensuremath{^{\textnormal{18}}}F,\ensuremath{^{\textnormal{19}}}F*\ensuremath{\rightarrow}\ensuremath{\alpha}+\ensuremath{^{\textnormal{15}}}N) E=150 MeV; measured positions and}\\
\parbox[b][0.3cm]{17.7cm}{energies of \ensuremath{\alpha}, \ensuremath{^{\textnormal{15}}}O, and \ensuremath{^{\textnormal{15}}}N particles from the in-flight decays of \ensuremath{^{\textnormal{19}}}Ne* and \ensuremath{^{\textnormal{19}}}F* reaction products using 6 \ensuremath{\Delta}E-E Si telescopes}\\
\parbox[b][0.3cm]{17.7cm}{with \ensuremath{\Delta}E detectors being position sensitive. Two of the telescopes covered \ensuremath{\theta}\ensuremath{_{\textnormal{lab}}}=\ensuremath{\pm}2.5\ensuremath{^\circ}{\textminus}8.5\ensuremath{^\circ} and were used to measure \ensuremath{^{\textnormal{15}}}N and}\\
\parbox[b][0.3cm]{17.7cm}{\ensuremath{^{\textnormal{15}}}O ions in coincidence with the \ensuremath{\alpha}-particles, which were measured by the 4 remaining telescopes covering \ensuremath{\theta}\ensuremath{_{\textnormal{lab}}}=\ensuremath{\pm}10.5\ensuremath{^\circ}{\textminus}16.5\ensuremath{^\circ}.}\\
\parbox[b][0.3cm]{17.7cm}{Reconstructed \ensuremath{^{\textnormal{19}}}Ne and \ensuremath{^{\textnormal{19}}}F excitation energies; deduced neutron angular distributions analyzed using DWBA; deduced proton}\\
\parbox[b][0.3cm]{17.7cm}{spectroscopic factors. (\href{https://www.nndc.bnl.gov/nsr/nsrlink.jsp?2011Ad05,B}{2011Ad05}) discussed 3 states near the proton threshold and deduced the \ensuremath{^{\textnormal{18}}}F(p,\ensuremath{\alpha}) reaction rate. (\href{https://www.nndc.bnl.gov/nsr/nsrlink.jsp?2011Ad24,B}{2011Ad24})}\\
\parbox[b][0.3cm]{17.7cm}{analyzed mirror levels from the measurement of the \ensuremath{^{\textnormal{2}}}H(\ensuremath{^{\textnormal{18}}}F,\ensuremath{\alpha}\ensuremath{^{\textnormal{15}}}N) reaction.}\\
\parbox[b][0.3cm]{17.7cm}{\addtolength{\parindent}{-0.2in}\href{https://www.nndc.bnl.gov/nsr/nsrlink.jsp?2012Ad05,B}{2012Ad05}: \ensuremath{^{\textnormal{2}}}H(\ensuremath{^{\textnormal{18}}}F,\ensuremath{^{\textnormal{19}}}Ne*\ensuremath{\rightarrow}p+\ensuremath{^{\textnormal{18}}}F) E=150 keV. Same data and experiment as (\href{https://www.nndc.bnl.gov/nsr/nsrlink.jsp?2011Ad05,B}{2011Ad05}, \href{https://www.nndc.bnl.gov/nsr/nsrlink.jsp?2011Ad24,B}{2011Ad24}). (\href{https://www.nndc.bnl.gov/nsr/nsrlink.jsp?2012Ad05,B}{2012Ad05}) analyzed the}\\
\parbox[b][0.3cm]{17.7cm}{\ensuremath{^{\textnormal{19}}}Ne*\ensuremath{\rightarrow}p+\ensuremath{^{\textnormal{18}}}F data for those \ensuremath{^{\textnormal{19}}}Ne* states in the E\ensuremath{_{\textnormal{x}}}=6.9{\textminus}8.4 MeV region, which decay via p+\ensuremath{^{\textnormal{18}}}F and \ensuremath{\alpha}+\ensuremath{^{\textnormal{15}}}O. Reconstructed the}\\
\parbox[b][0.3cm]{17.7cm}{\ensuremath{^{\textnormal{19}}}Ne excitation function from p+\ensuremath{^{\textnormal{18}}}F and \ensuremath{\alpha}+\ensuremath{^{\textnormal{15}}}O kinematics. Deduced \ensuremath{\Gamma}\ensuremath{_{\textnormal{p}}} and \ensuremath{\Gamma}\ensuremath{_{\textnormal{p}}}/\ensuremath{\Gamma}\ensuremath{_{\ensuremath{\alpha}}} for the \ensuremath{^{\textnormal{19}}}Ne*(7089, 7226, 7592, 7879,}\\
\parbox[b][0.3cm]{17.7cm}{8072) states, which are above the \ensuremath{\alpha} and proton threshold.}\\
\parbox[b][0.3cm]{17.7cm}{\addtolength{\parindent}{-0.2in}\href{https://www.nndc.bnl.gov/nsr/nsrlink.jsp?2013Gu34,B}{2013Gu34}, \href{https://www.nndc.bnl.gov/nsr/nsrlink.jsp?2015Ch41,B}{2015Ch41}: \ensuremath{^{\textnormal{2}}}H(\ensuremath{^{\textnormal{18}}}F,\ensuremath{^{\textnormal{19}}}Ne*\ensuremath{\rightarrow}\ensuremath{\alpha}+\ensuremath{^{\textnormal{15}}}O) E=47.9 MeV; measured the \ensuremath{^{\textnormal{1}}}H(\ensuremath{^{\textnormal{18}}}F,\ensuremath{\alpha}) excitation function at \ensuremath{\theta}\ensuremath{_{\textnormal{c.m.}}}=70\ensuremath{^\circ}{\textminus}120\ensuremath{^\circ} using}\\
\parbox[b][0.3cm]{17.7cm}{the indirect Trojan horse method (THM). Measured energy, position and TOF of the reaction products; measured \ensuremath{\alpha}-\ensuremath{^{\textnormal{15}}}O}\\
\parbox[b][0.3cm]{17.7cm}{coincidence events using a position sensitive Si \ensuremath{\Delta}E-E telescope covering \ensuremath{\theta}\ensuremath{_{\textnormal{lab}}}=11\ensuremath{^\circ}{\textminus}31\ensuremath{^\circ} to detect the \ensuremath{^{\textnormal{15}}}O recoils; used the silicon}\\
\parbox[b][0.3cm]{17.7cm}{array for the Trojan horse modular system (ASTRHO), which consisted of eight position sensitive Si detectors covering}\\
\parbox[b][0.3cm]{17.7cm}{\ensuremath{\theta}\ensuremath{_{\textnormal{lab}}}=2\ensuremath{^\circ}{\textminus}11\ensuremath{^\circ}, to detect the \ensuremath{\alpha}-particles. Reconstructed the ejectiles$'$ emission angle with an overall resolution of 0.2\ensuremath{^\circ}. Deduced \ensuremath{^{\textnormal{19}}}Ne}\\
\parbox[b][0.3cm]{17.7cm}{levels. Deduced the astrophysical S-factor for E\ensuremath{_{\textnormal{c.m.}}}\ensuremath{<}900 keV normalized to the results of the direct measurements of the}\\
\parbox[b][0.3cm]{17.7cm}{well-known resonance at 665 keV.}\\
\parbox[b][0.3cm]{17.7cm}{\addtolength{\parindent}{-0.2in}\href{https://www.nndc.bnl.gov/nsr/nsrlink.jsp?2016Pi01,B}{2016Pi01}: \ensuremath{^{\textnormal{2}}}H(\ensuremath{^{\textnormal{18}}}F,\ensuremath{^{\textnormal{19}}}Ne*\ensuremath{\rightarrow}\ensuremath{\alpha}+\ensuremath{^{\textnormal{15}}}O) E=52 MeV; the \ensuremath{^{\textnormal{18}}}F beam was separated and purified by the MARS spectrometer. Downstream}\\
\parbox[b][0.3cm]{17.7cm}{of the target, two position sensitive Si detectors covering \ensuremath{\theta}\ensuremath{_{\textnormal{lab}}}=3\ensuremath{^\circ}{\textminus}12\ensuremath{^\circ} measured energies and positions of the \ensuremath{^{\textnormal{15}}}O ions in}\\
\parbox[b][0.3cm]{17.7cm}{coincidence with the \ensuremath{\alpha}-particles, whose energies and positions were measured by the TExas Edinburgh Catania Silicon Array}\\
\parbox[b][0.3cm]{17.7cm}{(TECSA) placed downstream of the other two detectors. The TECSA array consisted of 8 position sensitive Si detectors covering}\\
\parbox[b][0.3cm]{17.7cm}{\ensuremath{\theta}\ensuremath{_{\textnormal{lab}}}=15\ensuremath{^\circ}{\textminus}40\ensuremath{^\circ}. Deduced the Q-value spectrum for the near proton threshold region; determined the trajectories of the undetected}\\
\parbox[b][0.3cm]{17.7cm}{neutrons using Trojan horse analysis. Extracted the d\ensuremath{\sigma}/d\ensuremath{\Omega} vs. E\ensuremath{_{\textnormal{c.m.}}} from THM. Individual states are not resolved. This spectrum}\\
\parbox[b][0.3cm]{17.7cm}{was fitted with 7 states with E\ensuremath{_{\textnormal{x}}}=5.8-7 MeV. Deduced astrophysical S-factor for the \ensuremath{^{\textnormal{18}}}F(p,\ensuremath{\alpha}) reaction rate at E\ensuremath{_{\textnormal{c.m.}}}\ensuremath{<}0.9 MeV and}\\
\parbox[b][0.3cm]{17.7cm}{obtained the reaction rate.}\\
\parbox[b][0.3cm]{17.7cm}{\addtolength{\parindent}{-0.2in}\href{https://www.nndc.bnl.gov/nsr/nsrlink.jsp?2017La12,B}{2017La12}, \href{https://www.nndc.bnl.gov/nsr/nsrlink.jsp?2019LaZX,B}{2019LaZX}: Performed an R-matrix analysis on the weighted average \ensuremath{^{\textnormal{18}}}F(p,\ensuremath{\alpha}) S-factor which was deduced from the}\\
\parbox[b][0.3cm]{17.7cm}{THM data of (\href{https://www.nndc.bnl.gov/nsr/nsrlink.jsp?2015Ch41,B}{2015Ch41}, \href{https://www.nndc.bnl.gov/nsr/nsrlink.jsp?2016Pi01,B}{2016Pi01}) at E\ensuremath{_{\textnormal{c.m.}}}\ensuremath{<}1 MeV. The R-matrix parameters were taken from (\href{https://www.nndc.bnl.gov/nsr/nsrlink.jsp?2015Ba51,B}{2015Ba51}). These parameters}\\
\parbox[b][0.3cm]{17.7cm}{were then varied until the THM S-factor data was best reproduced. Performed a check on the sensitivity of the S-factor to the}\\
\parbox[b][0.3cm]{17.7cm}{interferences between the \ensuremath{^{\textnormal{19}}}Ne resonances, which were used in the R-matrix analysis.}\\
\vspace{12pt}
\underline{$^{19}$Ne Levels}\\
\vspace{0.34cm}
\parbox[b][0.3cm]{17.7cm}{\addtolength{\parindent}{-0.254cm}\textit{Notes}:}\\
\parbox[b][0.3cm]{17.7cm}{\addtolength{\parindent}{-0.254cm}(1) The E\ensuremath{_{\textnormal{c.m.}}} values given here from (\href{https://www.nndc.bnl.gov/nsr/nsrlink.jsp?2012Ad05,B}{2012Ad05}, \href{https://www.nndc.bnl.gov/nsr/nsrlink.jsp?2015Ch41,B}{2015Ch41}, \href{https://www.nndc.bnl.gov/nsr/nsrlink.jsp?2016Pi01,B}{2016Pi01}, \href{https://www.nndc.bnl.gov/nsr/nsrlink.jsp?2017La12,B}{2017La12}, \href{https://www.nndc.bnl.gov/nsr/nsrlink.jsp?2019LaZX,B}{2019LaZX}) are the center-of-mass resonance}\\
\parbox[b][0.3cm]{17.7cm}{energies for \ensuremath{^{\textnormal{18}}}F+p populating \ensuremath{^{\textnormal{19}}}Ne* proton resonances, which are converted to excitation energies using S\ensuremath{_{\textnormal{p}}}(\ensuremath{^{\textnormal{19}}}Ne)=6410.0 keV \textit{5}}\\
\parbox[b][0.3cm]{17.7cm}{(\href{https://www.nndc.bnl.gov/nsr/nsrlink.jsp?2021Wa16,B}{2021Wa16}).}\\
\parbox[b][0.3cm]{17.7cm}{\addtolength{\parindent}{-0.254cm}(2) The uncertainties in the excitation energies, proton spectroscopic factors, and \ensuremath{\Gamma}\ensuremath{_{\textnormal{p}}} values reported by (\href{https://www.nndc.bnl.gov/nsr/nsrlink.jsp?2011Ad05,B}{2011Ad05}, \href{https://www.nndc.bnl.gov/nsr/nsrlink.jsp?2011Ad24,B}{2011Ad24}) are}\\
\parbox[b][0.3cm]{17.7cm}{statistical only. (\href{https://www.nndc.bnl.gov/nsr/nsrlink.jsp?2011Ad05,B}{2011Ad05}) estimated \ensuremath{\pm}10 keV, 40\%, and 30\% systematic uncertainties for E\ensuremath{_{\textnormal{x}}}, S\ensuremath{_{\textnormal{p}}}, and \ensuremath{\Gamma}\ensuremath{_{\textnormal{p}}}, respectively.}\\
\parbox[b][0.3cm]{17.7cm}{(\href{https://www.nndc.bnl.gov/nsr/nsrlink.jsp?2011Ad24,B}{2011Ad24}) estimated \ensuremath{\pm}10 keV, 30\%, and 20\% systematic uncertainties for E\ensuremath{_{\textnormal{x}}}, S\ensuremath{_{\textnormal{p}}}, and \ensuremath{\Gamma}\ensuremath{_{\textnormal{p}}}, respectively. These are included in}\\
\parbox[b][0.3cm]{17.7cm}{the uncertainties given here for each value from (\href{https://www.nndc.bnl.gov/nsr/nsrlink.jsp?2011Ad05,B}{2011Ad05}, \href{https://www.nndc.bnl.gov/nsr/nsrlink.jsp?2011Ad24,B}{2011Ad24}). Similarly, the uncertainties reported by (\href{https://www.nndc.bnl.gov/nsr/nsrlink.jsp?2012Ad05,B}{2012Ad05}) on the}\\
\parbox[b][0.3cm]{17.7cm}{proton widths are statistical only. These authors recommended 20\% systematic uncertainties in their proton widths, which are added}\\
\parbox[b][0.3cm]{17.7cm}{in quadrature to the values from (\href{https://www.nndc.bnl.gov/nsr/nsrlink.jsp?2012Ad05,B}{2012Ad05}).}\\
\parbox[b][0.3cm]{17.7cm}{\addtolength{\parindent}{-0.254cm}(3) \ensuremath{\Gamma}\ensuremath{_{\textnormal{p}}} values reported by (\href{https://www.nndc.bnl.gov/nsr/nsrlink.jsp?2011Ad05,B}{2011Ad05}, \href{https://www.nndc.bnl.gov/nsr/nsrlink.jsp?2011Ad24,B}{2011Ad24}) are deduced from \ensuremath{\Gamma}\ensuremath{_{\textnormal{p}}}=S\ensuremath{_{\textnormal{p}}}\ensuremath{\Gamma}\ensuremath{_{\textnormal{sp}}}, where \ensuremath{\Gamma}\ensuremath{_{\textnormal{sp}}} is the single-particle proton width}\\
\parbox[b][0.3cm]{17.7cm}{calculated by (\href{https://www.nndc.bnl.gov/nsr/nsrlink.jsp?2011Ad05,B}{2011Ad05}, \href{https://www.nndc.bnl.gov/nsr/nsrlink.jsp?2011Ad24,B}{2011Ad24}) using DWUCK4 assuming N=1.55 (\href{https://www.nndc.bnl.gov/nsr/nsrlink.jsp?2011Ad24,B}{2011Ad24}), and S\ensuremath{_{\textnormal{p}}} is the spectroscopic factor measured}\\
\parbox[b][0.3cm]{17.7cm}{by (\href{https://www.nndc.bnl.gov/nsr/nsrlink.jsp?2011Ad05,B}{2011Ad05}, \href{https://www.nndc.bnl.gov/nsr/nsrlink.jsp?2011Ad24,B}{2011Ad24}).}\\
\parbox[b][0.3cm]{17.7cm}{\addtolength{\parindent}{-0.254cm}(4) J\ensuremath{^{\ensuremath{\pi}}} values reported from (\href{https://www.nndc.bnl.gov/nsr/nsrlink.jsp?2011Ad05,B}{2011Ad05}, \href{https://www.nndc.bnl.gov/nsr/nsrlink.jsp?2011Ad24,B}{2011Ad24}) are obtained from DWBA analysis by those authors. The J\ensuremath{^{\ensuremath{\pi}}} values reported}\\
\parbox[b][0.3cm]{17.7cm}{from (\href{https://www.nndc.bnl.gov/nsr/nsrlink.jsp?2017La12,B}{2017La12}, \href{https://www.nndc.bnl.gov/nsr/nsrlink.jsp?2019LaZX,B}{2019LaZX}) are deduced using an R-matrix analysis of the weighted average astrophysical S-factors determined by}\\
\parbox[b][0.3cm]{17.7cm}{(\href{https://www.nndc.bnl.gov/nsr/nsrlink.jsp?2015Ch41,B}{2015Ch41}, \href{https://www.nndc.bnl.gov/nsr/nsrlink.jsp?2016Pi01,B}{2016Pi01}) for the \ensuremath{^{\textnormal{18}}}F(p,\ensuremath{\alpha}) reaction. Due to the lack of angular distribution data in (\href{https://www.nndc.bnl.gov/nsr/nsrlink.jsp?2015Ch41,B}{2015Ch41}), those authors could}\\
\parbox[b][0.3cm]{17.7cm}{not independently determine the J\ensuremath{^{\ensuremath{\pi}}} assignments for their observed \ensuremath{^{\textnormal{19}}}Ne states. For each populated \ensuremath{^{\textnormal{19}}}Ne excited state, a range of}\\
\parbox[b][0.3cm]{17.7cm}{J\ensuremath{^{\ensuremath{\pi}}} values was assumed by (\href{https://www.nndc.bnl.gov/nsr/nsrlink.jsp?2015Ch41,B}{2015Ch41}) from comparison of the J\ensuremath{^{\ensuremath{\pi}}} assignments available in the literature with those from the}\\
\clearpage
\vspace{0.3cm}
{\bf \small \underline{\ensuremath{^{\textnormal{2}}}H(\ensuremath{^{\textnormal{18}}}F,\ensuremath{^{\textnormal{19}}}Ne)\hspace{0.2in}\href{https://www.nndc.bnl.gov/nsr/nsrlink.jsp?2011Ad24,B}{2011Ad24},\href{https://www.nndc.bnl.gov/nsr/nsrlink.jsp?2015Ch41,B}{2015Ch41},\href{https://www.nndc.bnl.gov/nsr/nsrlink.jsp?2017La12,B}{2017La12} (continued)}}\\
\vspace{0.3cm}
\underline{$^{19}$Ne Levels (continued)}\\
\parbox[b][0.3cm]{17.7cm}{evaluation of (\href{https://www.nndc.bnl.gov/nsr/nsrlink.jsp?2007Ne09,B}{2007Ne09}). The resulting J\ensuremath{^{\ensuremath{\pi}}} assignments reported by (\href{https://www.nndc.bnl.gov/nsr/nsrlink.jsp?2015Ch41,B}{2015Ch41}) are all tentative. (\href{https://www.nndc.bnl.gov/nsr/nsrlink.jsp?2016Pi01,B}{2016Pi01}) also did not deduce}\\
\parbox[b][0.3cm]{17.7cm}{the J\ensuremath{^{\ensuremath{\pi}}} assignments independently. The values they report are based on discussions in (i) (\href{https://www.nndc.bnl.gov/nsr/nsrlink.jsp?2007Ne09,B}{2007Ne09}), which is an evaluation of all}\\
\parbox[b][0.3cm]{17.7cm}{experimental data on \ensuremath{^{\textnormal{19}}}Ne and \ensuremath{^{\textnormal{19}}}F mirror states; (ii) (\href{https://www.nndc.bnl.gov/nsr/nsrlink.jsp?2013La01,B}{2013La01}): \ensuremath{^{\textnormal{19}}}F(\ensuremath{^{\textnormal{3}}}He,t) using two-step finite-range DWBA; and (iii)}\\
\parbox[b][0.3cm]{17.7cm}{(\href{https://www.nndc.bnl.gov/nsr/nsrlink.jsp?2015Ch41,B}{2015Ch41}): See explanation given above.}\\
\parbox[b][0.3cm]{17.7cm}{\addtolength{\parindent}{-0.254cm}(5) \ensuremath{\Gamma}\ensuremath{_{\textnormal{p}}} and \ensuremath{\Gamma}\ensuremath{_{\ensuremath{\alpha}}} values reported by (\href{https://www.nndc.bnl.gov/nsr/nsrlink.jsp?2017La12,B}{2017La12}, \href{https://www.nndc.bnl.gov/nsr/nsrlink.jsp?2019LaZX,B}{2019LaZX}) are determined from an R-matrix analysis of the weighted average}\\
\parbox[b][0.3cm]{17.7cm}{astrophysical S-factors determined by (\href{https://www.nndc.bnl.gov/nsr/nsrlink.jsp?2015Ch41,B}{2015Ch41}, \href{https://www.nndc.bnl.gov/nsr/nsrlink.jsp?2016Pi01,B}{2016Pi01}) for the \ensuremath{^{\textnormal{18}}}F(p,\ensuremath{\alpha}) reaction.}\\
\parbox[b][0.3cm]{17.7cm}{\addtolength{\parindent}{-0.254cm}(6) Negative E\ensuremath{_{\textnormal{c.m.}}}(\ensuremath{^{\textnormal{18}}}F+p) values indicate resonances below the proton threshold.}\\
\vspace{0.34cm}

\begin{textblock}{29}(0,27.3)
Continued on next page (footnotes at end of table)
\end{textblock}
\clearpage
%
\begin{textblock}{29}(0,27.3)
Continued on next page (footnotes at end of table)
\end{textblock}
\clearpage
%
\begin{textblock}{29}(0,27.3)
Continued on next page (footnotes at end of table)
\end{textblock}
\clearpage
%
\parbox[b][0.3cm]{17.7cm}{\makebox[1ex]{\ensuremath{^{\hypertarget{NE8LEVEL0}{a}}}} Deduced from E\ensuremath{_{\textnormal{c.m.}}}(\ensuremath{^{\textnormal{18}}}F+p)+S\ensuremath{_{\textnormal{p}}}, where E\ensuremath{_{\textnormal{c.m.}}}(\ensuremath{^{\textnormal{18}}}F+p) is the center{\textminus}of{\textminus}mass \ensuremath{^{\textnormal{18}}}F+p resonance energy from (\href{https://www.nndc.bnl.gov/nsr/nsrlink.jsp?2015Ch41,B}{2015Ch41})}\\
\parbox[b][0.3cm]{17.7cm}{{\ }{\ }determined via the Trojan horse method and by the analysis of the \ensuremath{^{\textnormal{18}}}F(p,\ensuremath{\alpha}) astrophysical S-factor, and S\ensuremath{_{\textnormal{p}}}(\ensuremath{^{\textnormal{19}}}Ne)=6410.0 keV \textit{5}}\\
\parbox[b][0.3cm]{17.7cm}{{\ }{\ }(\href{https://www.nndc.bnl.gov/nsr/nsrlink.jsp?2021Wa16,B}{2021Wa16}). A 10-keV systematic uncertainty recommended by (\href{https://www.nndc.bnl.gov/nsr/nsrlink.jsp?2015Ch41,B}{2015Ch41}) is added in quadrature to the excitation energy}\\
\parbox[b][0.3cm]{17.7cm}{{\ }{\ }uncertainty given in (\href{https://www.nndc.bnl.gov/nsr/nsrlink.jsp?2015Ch41,B}{2015Ch41}) to reflect the fact that (\href{https://www.nndc.bnl.gov/nsr/nsrlink.jsp?2015Ch41,B}{2015Ch41}) fitted this state using a peak with a fixed Gaussian width,}\\
\parbox[b][0.3cm]{17.7cm}{{\ }{\ }which was considered to be 53 keV equal to their energy resolution. Also note that the E\ensuremath{_{\textnormal{x}}} listed here may differ by \ensuremath{\sim}1-2 keV}\\
\parbox[b][0.3cm]{17.7cm}{{\ }{\ }from that given in Figure 3 and Table I of (\href{https://www.nndc.bnl.gov/nsr/nsrlink.jsp?2015Ch41,B}{2015Ch41}) because we are using a revised S\ensuremath{_{\textnormal{p}}}(\ensuremath{^{\textnormal{19}}}Ne)=6410.0 keV \textit{5} from}\\
\parbox[b][0.3cm]{17.7cm}{{\ }{\ }(\href{https://www.nndc.bnl.gov/nsr/nsrlink.jsp?2021Wa16,B}{2021Wa16}). There are also discrepancies in excitation energies listed in Fig. 3 and Table I of (\href{https://www.nndc.bnl.gov/nsr/nsrlink.jsp?2015Ch41,B}{2015Ch41}). The E\ensuremath{_{\textnormal{x}}} value given}\\
\parbox[b][0.3cm]{17.7cm}{{\ }{\ }here are from Table I.}\\
\parbox[b][0.3cm]{17.7cm}{\makebox[1ex]{\ensuremath{^{\hypertarget{NE8LEVEL1}{b}}}} This state was first observed in (\href{https://www.nndc.bnl.gov/nsr/nsrlink.jsp?2015Ch41,B}{2015Ch41}).}\\
\parbox[b][0.3cm]{17.7cm}{\makebox[1ex]{\ensuremath{^{\hypertarget{NE8LEVEL2}{c}}}} A broad peak appeared in this energy region in the p+\ensuremath{^{\textnormal{18}}}F channel in (\href{https://www.nndc.bnl.gov/nsr/nsrlink.jsp?2012Ad05,B}{2012Ad05}). If this peak is assumed to be a single state, its}\\
\parbox[b][0.3cm]{17.7cm}{{\ }{\ }width would be \ensuremath{\sim}400 keV. (\href{https://www.nndc.bnl.gov/nsr/nsrlink.jsp?2012Ad05,B}{2012Ad05}) fitted two Gaussian peaks under this peak. As a result, this state is one of the unresolved}\\
\parbox[b][0.3cm]{17.7cm}{{\ }{\ }members observed.}\\
\parbox[b][0.3cm]{17.7cm}{\makebox[1ex]{\ensuremath{^{\hypertarget{NE8LEVEL3}{d}}}} From a zero-range DWBA analysis by (\href{https://www.nndc.bnl.gov/nsr/nsrlink.jsp?2011Ad24,B}{2011Ad24}) using DWUCK4 unless noted otherwise. A finite-range DWBA analysis was}\\
\begin{textblock}{29}(0,27.3)
Continued on next page (footnotes at end of table)
\end{textblock}
\clearpage
\vspace*{-0.5cm}
{\bf \small \underline{\ensuremath{^{\textnormal{2}}}H(\ensuremath{^{\textnormal{18}}}F,\ensuremath{^{\textnormal{19}}}Ne)\hspace{0.2in}\href{https://www.nndc.bnl.gov/nsr/nsrlink.jsp?2011Ad24,B}{2011Ad24},\href{https://www.nndc.bnl.gov/nsr/nsrlink.jsp?2015Ch41,B}{2015Ch41},\href{https://www.nndc.bnl.gov/nsr/nsrlink.jsp?2017La12,B}{2017La12} (continued)}}\\
\vspace{0.3cm}
\underline{$^{19}$Ne Levels (continued)}\\
\vspace{0.3cm}
\parbox[b][0.3cm]{17.7cm}{{\ }{\ }also performed by (\href{https://www.nndc.bnl.gov/nsr/nsrlink.jsp?2011Ad24,B}{2011Ad24}) using FRESCO only for proton bound states (E\ensuremath{_{\textnormal{x}}}\ensuremath{<}6410 keV) and the results were in agreement}\\
\parbox[b][0.3cm]{17.7cm}{{\ }{\ }with those using DWUCK4 (\href{https://www.nndc.bnl.gov/nsr/nsrlink.jsp?2011Ad24,B}{2011Ad24}).}\\
\parbox[b][0.3cm]{17.7cm}{\makebox[1ex]{\ensuremath{^{\hypertarget{NE8LEVEL4}{e}}}} From (\href{https://www.nndc.bnl.gov/nsr/nsrlink.jsp?2011Ad24,B}{2011Ad24}). To deduce the angular distributions of the undetected neutrons, (\href{https://www.nndc.bnl.gov/nsr/nsrlink.jsp?2011Ad24,B}{2011Ad24}) calculated angular correlations for}\\
\parbox[b][0.3cm]{17.7cm}{{\ }{\ }\ensuremath{\alpha}+\ensuremath{^{\textnormal{15}}}O using FRESCO. The obtained distributions were mostly isotropic, or with deviations less than 15-20\% from isotropy. Note}\\
\parbox[b][0.3cm]{17.7cm}{{\ }{\ }that the (2J+1)S\ensuremath{_{\textnormal{p}}} values deduced by (\href{https://www.nndc.bnl.gov/nsr/nsrlink.jsp?2011Ad24,B}{2011Ad24}) have an additional systematic uncertainty of 30\%, which are not reported in}\\
\parbox[b][0.3cm]{17.7cm}{{\ }{\ }(\href{https://www.nndc.bnl.gov/nsr/nsrlink.jsp?2011Ad24,B}{2011Ad24}), where only statistical uncertainties were listed.}\\
\vspace{0.5cm}
\clearpage
\subsection[\hspace{-0.2cm}\ensuremath{^{\textnormal{2}}}H(\ensuremath{^{\textnormal{18}}}Ne,\ensuremath{^{\textnormal{19}}}Ne),(\ensuremath{^{\textnormal{18}}}Ne,p)]{ }
\vspace{-27pt}
\vspace{0.3cm}
\hypertarget{NE9}{{\bf \small \underline{\ensuremath{^{\textnormal{2}}}H(\ensuremath{^{\textnormal{18}}}Ne,\ensuremath{^{\textnormal{19}}}Ne),(\ensuremath{^{\textnormal{18}}}Ne,p)\hspace{0.2in}\href{https://www.nndc.bnl.gov/nsr/nsrlink.jsp?2002La29,B}{2002La29},\href{https://www.nndc.bnl.gov/nsr/nsrlink.jsp?2002Os05,B}{2002Os05}}}}\\
\vspace{4pt}
\vspace{8pt}
\parbox[b][0.3cm]{17.7cm}{\addtolength{\parindent}{-0.2in}J\ensuremath{^{\ensuremath{\pi}}}(\ensuremath{^{\textnormal{2}}}H\ensuremath{_{\textnormal{g.s.}}})=1\ensuremath{^{\textnormal{+}}} and J\ensuremath{^{\ensuremath{\pi}}}(\ensuremath{^{\textnormal{18}}}Ne\ensuremath{_{\textnormal{g.s.}}})=0\ensuremath{^{\textnormal{+}}}.}\\
\parbox[b][0.3cm]{17.7cm}{\addtolength{\parindent}{-0.2in}\href{https://www.nndc.bnl.gov/nsr/nsrlink.jsp?1998LaZR,B}{1998LaZR}: \ensuremath{^{\textnormal{2}}}H(\ensuremath{^{\textnormal{18}}}Ne,p) E=45 MeV; measured proton spectra, p-\ensuremath{\alpha} coincidences following the \ensuremath{^{\textnormal{19}}}Ne* decay. Discussed implications}\\
\parbox[b][0.3cm]{17.7cm}{for astrophysical reaction rates.}\\
\parbox[b][0.3cm]{17.7cm}{\addtolength{\parindent}{-0.2in}\href{https://www.nndc.bnl.gov/nsr/nsrlink.jsp?1998Os02,B}{1998Os02}: \ensuremath{^{\textnormal{2}}}H(\ensuremath{^{\textnormal{18}}}Ne,p)\ensuremath{^{\textnormal{19}}}Ne*(\ensuremath{\alpha}) E=45 MeV; measured E and TOF for protons, as well as decay products using the position}\\
\parbox[b][0.3cm]{17.7cm}{sensitive LEDA Si array. Deduced proton energy spectrum. These preliminary data show very limited statistics but prove the}\\
\parbox[b][0.3cm]{17.7cm}{usefulness of this reaction.}\\
\parbox[b][0.3cm]{17.7cm}{\addtolength{\parindent}{-0.2in}\href{https://www.nndc.bnl.gov/nsr/nsrlink.jsp?2001Ch44,B}{2001Ch44}: \ensuremath{^{\textnormal{2}}}H(\ensuremath{^{\textnormal{18}}}Ne,p)\ensuremath{^{\textnormal{19}}}Ne*(\ensuremath{\alpha})\ensuremath{^{\textnormal{15}}}O E=45-54 MeV; measured E and TOF for protons, \ensuremath{\alpha}-particles and \ensuremath{^{\textnormal{15}}}O decay products;}\\
\parbox[b][0.3cm]{17.7cm}{measured p-\ensuremath{\alpha} coincidences using the position sensitive LEDA Si array covering almost full azimuthal angle and \ensuremath{\theta}\ensuremath{_{\textnormal{lab}}}=4\ensuremath{^\circ}{\textminus}27\ensuremath{^\circ} and}\\
\parbox[b][0.3cm]{17.7cm}{\ensuremath{\theta}\ensuremath{_{\textnormal{lab}}}=125\ensuremath{^\circ}{\textminus}155\ensuremath{^\circ}. Reconstructed \ensuremath{^{\textnormal{19}}}Ne decay Q-value spectrum. Discussed applicability of the results to the indirect measurement of}\\
\parbox[b][0.3cm]{17.7cm}{the \ensuremath{^{\textnormal{15}}}O(\ensuremath{\alpha},\ensuremath{\gamma}) astrophysical reaction rate.}\\
\parbox[b][0.3cm]{17.7cm}{\addtolength{\parindent}{-0.2in}\href{https://www.nndc.bnl.gov/nsr/nsrlink.jsp?1999LaZU,B}{1999LaZU}, \href{https://www.nndc.bnl.gov/nsr/nsrlink.jsp?2001La16,B}{2001La16}, \href{https://www.nndc.bnl.gov/nsr/nsrlink.jsp?2002La29,B}{2002La29}: \ensuremath{^{\textnormal{2}}}H(\ensuremath{^{\textnormal{18}}}Ne,p)\ensuremath{^{\textnormal{19}}}Ne*(\ensuremath{\alpha}) E=54.3 MeV; measured E and TOF for protons, \ensuremath{\alpha}-particles and \ensuremath{^{\textnormal{15}}}O decay}\\
\parbox[b][0.3cm]{17.7cm}{products; measured \ensuremath{^{\textnormal{15}}}O-\ensuremath{\alpha}-p triple coincidences using the position sensitive LEDA Si array covering \ensuremath{\theta}\ensuremath{_{\textnormal{lab}}}=120\ensuremath{^\circ}{\textminus}146\ensuremath{^\circ} (for protons),}\\
\parbox[b][0.3cm]{17.7cm}{\ensuremath{\theta}\ensuremath{_{\textnormal{lab}}}=14\ensuremath{^\circ}{\textminus}32\ensuremath{^\circ} (for \ensuremath{\alpha}s) and \ensuremath{\theta}\ensuremath{_{\textnormal{lab}}}=4\ensuremath{^\circ}{\textminus}10\ensuremath{^\circ} (for \ensuremath{^{\textnormal{15}}}O). Deduced \ensuremath{^{\textnormal{19}}}Ne excitation energy spectrum, and alpha branching ratios for the}\\
\parbox[b][0.3cm]{17.7cm}{\ensuremath{^{\textnormal{19}}}Ne*(4033, 4140, 4197, 4600, 5092, 5351/5424/5463, 6013/6092) states, where / means unresolved states. The results were}\\
\parbox[b][0.3cm]{17.7cm}{compared with those of (\href{https://www.nndc.bnl.gov/nsr/nsrlink.jsp?1990Ma05,B}{1990Ma05}). Performed DWBA calculations and deduced transferred angular momentum for a few states.}\\
\parbox[b][0.3cm]{17.7cm}{These latter results are only presented in A. Laird$'$s Ph.D. thesis.}\\
\parbox[b][0.3cm]{17.7cm}{\addtolength{\parindent}{-0.2in}\href{https://www.nndc.bnl.gov/nsr/nsrlink.jsp?2002Os05,B}{2002Os05}: \ensuremath{^{\textnormal{2}}}H(\ensuremath{^{\textnormal{18}}}Ne,p)\ensuremath{^{\textnormal{19}}}Ne*(\ensuremath{\alpha}) E=44.1 MeV; measured E and TOF for protons, \ensuremath{\alpha}-particles and \ensuremath{^{\textnormal{15}}}O decay products using}\\
\parbox[b][0.3cm]{17.7cm}{position sensitive LEDA Si array positioned at \ensuremath{\theta}\ensuremath{_{\textnormal{lab}}}=131\ensuremath{^\circ}{\textminus}154\ensuremath{^\circ}, \ensuremath{\theta}\ensuremath{_{\textnormal{lab}}}=5\ensuremath{^\circ}{\textminus}10\ensuremath{^\circ}, and \ensuremath{\theta}\ensuremath{_{\textnormal{lab}}}=15\ensuremath{^\circ}{\textminus}35\ensuremath{^\circ} to measure (in triple coincidence}\\
\parbox[b][0.3cm]{17.7cm}{mode) protons, \ensuremath{\alpha}-particles, and \ensuremath{^{\textnormal{15}}}O decay products, respectively. Deduced \ensuremath{^{\textnormal{19}}}Ne levels and \ensuremath{\alpha}-decay widths. Discussed}\\
\parbox[b][0.3cm]{17.7cm}{astrophysical implications.}\\
\vspace{12pt}
\underline{$^{19}$Ne Levels}\\
\begin{longtable}{ccccccc@{\extracolsep{\fill}}c}
\multicolumn{2}{c}{E(level)$^{}$}&J$^{\pi}$$^{{\hyperlink{NE9LEVEL0}{a}}}$&L$^{}$&\multicolumn{2}{c}{\ensuremath{\Gamma}\ensuremath{\alpha}/\ensuremath{\Gamma}$^{{\hyperlink{NE9LEVEL3}{d}}}$}&Comments&\\[-.2cm]
\multicolumn{2}{c}{\hrulefill}&\hrulefill&\hrulefill&\multicolumn{2}{c}{\hrulefill}&\hrulefill&
\endfirsthead
\multicolumn{1}{r@{}}{0}&\multicolumn{1}{@{}l}{\ensuremath{^{{\hyperlink{NE9LEVEL1}{b}}{\hyperlink{NE9LEVEL4}{e}}}}}&\multicolumn{1}{l}{1/2\ensuremath{^{+}}}&&&&&\\
\multicolumn{1}{r@{}}{238}&\multicolumn{1}{@{}l}{\ensuremath{^{{\hyperlink{NE9LEVEL4}{e}}}}}&\multicolumn{1}{l}{5/2\ensuremath{^{+}}}&&&&\parbox[t][0.3cm]{11.3599cm}{\raggedright J\ensuremath{^{\pi}}: Fig. 2 of (\href{https://www.nndc.bnl.gov/nsr/nsrlink.jsp?2002Os05,B}{2002Os05}) mistakenly reported a J\ensuremath{^{\ensuremath{\pi}}}=1/2\ensuremath{^{\textnormal{+}}} for this state.\vspace{0.1cm}}&\\
\multicolumn{1}{r@{}}{1536}&\multicolumn{1}{@{}l}{\ensuremath{^{{\hyperlink{NE9LEVEL4}{e}}}}}&\multicolumn{1}{l}{3/2\ensuremath{^{+}}}&&&&&\\
\multicolumn{1}{r@{}}{2794}&\multicolumn{1}{@{}l}{\ensuremath{^{{\hyperlink{NE9LEVEL4}{e}}}}}&\multicolumn{1}{l}{9/2\ensuremath{^{+}}}&&&&&\\
\multicolumn{1}{r@{}}{4033}&\multicolumn{1}{@{}l}{}&\multicolumn{1}{l}{3/2\ensuremath{^{+}}}&&\multicolumn{1}{r@{}}{$<$0}&\multicolumn{1}{@{.}l}{01\ensuremath{^{{\hyperlink{NE9LEVEL5}{f}}}}}&\parbox[t][0.3cm]{11.3599cm}{\raggedright E(level): From (\href{https://www.nndc.bnl.gov/nsr/nsrlink.jsp?2001La16,B}{2001La16}, \href{https://www.nndc.bnl.gov/nsr/nsrlink.jsp?2002La29,B}{2002La29}, \href{https://www.nndc.bnl.gov/nsr/nsrlink.jsp?2002Os05,B}{2002Os05}). This state was a member of an\vspace{0.1cm}}&\\
&&&&&&\parbox[t][0.3cm]{11.3599cm}{\raggedright {\ }{\ }{\ }unresolved triplet states together with \ensuremath{^{\textnormal{19}}}Ne*(4140, 4197) states in (\href{https://www.nndc.bnl.gov/nsr/nsrlink.jsp?2002La29,B}{2002La29}).\vspace{0.1cm}}&\\
&&&&&&\parbox[t][0.3cm]{11.3599cm}{\raggedright (d\ensuremath{\sigma}/d\ensuremath{\Omega})\ensuremath{_{\textnormal{lab}}}(\ensuremath{\theta}\ensuremath{_{\textnormal{lab}}}=138\ensuremath{^\circ})=0.5 mb/sr \textit{2} (\href{https://www.nndc.bnl.gov/nsr/nsrlink.jsp?2002Os05,B}{2002Os05}). No \ensuremath{\alpha}-decay event was observed\vspace{0.1cm}}&\\
&&&&&&\parbox[t][0.3cm]{11.3599cm}{\raggedright {\ }{\ }{\ }for this state in (\href{https://www.nndc.bnl.gov/nsr/nsrlink.jsp?2002Os05,B}{2002Os05}).\vspace{0.1cm}}&\\
&&&&&&\parbox[t][0.3cm]{11.3599cm}{\raggedright \ensuremath{\Gamma}\ensuremath{\alpha}/\ensuremath{\Gamma}: See also (\href{https://www.nndc.bnl.gov/nsr/nsrlink.jsp?2001La16,B}{2001La16}).\vspace{0.1cm}}&\\
\multicolumn{1}{r@{}}{4140}&\multicolumn{1}{@{}l}{}&\multicolumn{1}{l}{(7/2\ensuremath{^{-}})}&&\multicolumn{1}{r@{}}{$<$0}&\multicolumn{1}{@{.}l}{01\ensuremath{^{{\hyperlink{NE9LEVEL5}{f}}}}}&\parbox[t][0.3cm]{11.3599cm}{\raggedright E(level): From (\href{https://www.nndc.bnl.gov/nsr/nsrlink.jsp?2002La29,B}{2002La29}): A member of an unresolved triplet states together with\vspace{0.1cm}}&\\
&&&&&&\parbox[t][0.3cm]{11.3599cm}{\raggedright {\ }{\ }{\ }\ensuremath{^{\textnormal{19}}}Ne*(4033, 4197) states.\vspace{0.1cm}}&\\
\multicolumn{1}{r@{}}{4197}&\multicolumn{1}{@{}l}{\ensuremath{^{{\hyperlink{NE9LEVEL1}{b}}}}}&\multicolumn{1}{l}{(9/2\ensuremath{^{-}})}&&\multicolumn{1}{r@{}}{$<$0}&\multicolumn{1}{@{.}l}{01\ensuremath{^{{\hyperlink{NE9LEVEL5}{f}}}}}&\parbox[t][0.3cm]{11.3599cm}{\raggedright E(level): From (\href{https://www.nndc.bnl.gov/nsr/nsrlink.jsp?2002La29,B}{2002La29}): A member of an unresolved triplet states together with\vspace{0.1cm}}&\\
&&&&&&\parbox[t][0.3cm]{11.3599cm}{\raggedright {\ }{\ }{\ }\ensuremath{^{\textnormal{19}}}Ne*(4033, 4140) states.\vspace{0.1cm}}&\\
\multicolumn{1}{r@{}}{4549}&\multicolumn{1}{@{}l}{\ensuremath{^{{\hyperlink{NE9LEVEL1}{b}}{\hyperlink{NE9LEVEL2}{c}}}}}&\multicolumn{1}{l}{3/2\ensuremath{^{-}}}&&&&\parbox[t][0.3cm]{11.3599cm}{\raggedright E(level): From (\href{https://www.nndc.bnl.gov/nsr/nsrlink.jsp?2002La29,B}{2002La29}): A member of an unresolved quadruplet states together\vspace{0.1cm}}&\\
&&&&&&\parbox[t][0.3cm]{11.3599cm}{\raggedright {\ }{\ }{\ }with \ensuremath{^{\textnormal{19}}}Ne*(4600, 4635, 4712) states.\vspace{0.1cm}}&\\
\multicolumn{1}{r@{}}{4600}&\multicolumn{1}{@{}l}{\ensuremath{^{{\hyperlink{NE9LEVEL2}{c}}}}}&\multicolumn{1}{l}{5/2\ensuremath{^{+}}}&\multicolumn{1}{l}{2}&\multicolumn{1}{r@{}}{0}&\multicolumn{1}{@{.}l}{32 {\it 3}}&\parbox[t][0.3cm]{11.3599cm}{\raggedright E(level): From (\href{https://www.nndc.bnl.gov/nsr/nsrlink.jsp?2001La16,B}{2001La16}, \href{https://www.nndc.bnl.gov/nsr/nsrlink.jsp?2002Os05,B}{2002Os05}, \href{https://www.nndc.bnl.gov/nsr/nsrlink.jsp?2002La29,B}{2002La29}): A member of an unresolved\vspace{0.1cm}}&\\
&&&&&&\parbox[t][0.3cm]{11.3599cm}{\raggedright {\ }{\ }{\ }quadruplet states together with \ensuremath{^{\textnormal{19}}}Ne*(4549, 4635, 4712) states.\vspace{0.1cm}}&\\
&&&&&&\parbox[t][0.3cm]{11.3599cm}{\raggedright J\ensuremath{^{\pi}},L: From DWBA calculations presented in (A. M. Laird, Ph.D. Thesis,\vspace{0.1cm}}&\\
&&&&&&\parbox[t][0.3cm]{11.3599cm}{\raggedright {\ }{\ }{\ }University of Edinburgh, (2000), unpublished). These results are discussed\vspace{0.1cm}}&\\
&&&&&&\parbox[t][0.3cm]{11.3599cm}{\raggedright {\ }{\ }{\ }(without presentation of the data) in (\href{https://www.nndc.bnl.gov/nsr/nsrlink.jsp?2002La29,B}{2002La29}: See text).\vspace{0.1cm}}&\\
&&&&&&\parbox[t][0.3cm]{11.3599cm}{\raggedright \ensuremath{\Gamma}\ensuremath{\alpha}/\ensuremath{\Gamma}: Weighted average of \ensuremath{\Gamma}\ensuremath{_{\ensuremath{\alpha}}}/\ensuremath{\Gamma}=0.32 \textit{3} (\href{https://www.nndc.bnl.gov/nsr/nsrlink.jsp?2001La16,B}{2001La16}, \href{https://www.nndc.bnl.gov/nsr/nsrlink.jsp?2002La29,B}{2002La29}),\vspace{0.1cm}}&\\
&&&&&&\parbox[t][0.3cm]{11.3599cm}{\raggedright {\ }{\ }{\ }where the uncertainty is a quadratic sum of the statistical and systematic\vspace{0.1cm}}&\\
&&&&&&\parbox[t][0.3cm]{11.3599cm}{\raggedright {\ }{\ }{\ }uncertainties, and \ensuremath{\Gamma}\ensuremath{_{\ensuremath{\alpha}}}/\ensuremath{\Gamma}=0.28 \textit{13} (\href{https://www.nndc.bnl.gov/nsr/nsrlink.jsp?2002Os05,B}{2002Os05}). Other value: \ensuremath{\Gamma}\ensuremath{_{\ensuremath{\alpha}}}/\ensuremath{\Gamma}=0.28 \textit{13}\vspace{0.1cm}}&\\
&&&&&&\parbox[t][0.3cm]{11.3599cm}{\raggedright {\ }{\ }{\ }(\href{https://www.nndc.bnl.gov/nsr/nsrlink.jsp?1999LaZU,B}{1999LaZU}: Preliminary results).\vspace{0.1cm}}&\\
\multicolumn{1}{r@{}}{4635}&\multicolumn{1}{@{}l}{\ensuremath{^{{\hyperlink{NE9LEVEL2}{c}}}}}&\multicolumn{1}{l}{13/2\ensuremath{^{+}}}&&&&\parbox[t][0.3cm]{11.3599cm}{\raggedright E(level): From (\href{https://www.nndc.bnl.gov/nsr/nsrlink.jsp?2002La29,B}{2002La29}): A member of an unresolved quadruplet states together\vspace{0.1cm}}&\\
&&&&&&\parbox[t][0.3cm]{11.3599cm}{\raggedright {\ }{\ }{\ }with \ensuremath{^{\textnormal{19}}}Ne*(4549, 4600, 4712) states.\vspace{0.1cm}}&\\
\multicolumn{1}{r@{}}{4712}&\multicolumn{1}{@{}l}{\ensuremath{^{{\hyperlink{NE9LEVEL2}{c}}}}}&\multicolumn{1}{l}{5/2\ensuremath{^{-}}}&&&&\parbox[t][0.3cm]{11.3599cm}{\raggedright E(level): From (\href{https://www.nndc.bnl.gov/nsr/nsrlink.jsp?2002La29,B}{2002La29}): A member of an unresolved quadruplet states together\vspace{0.1cm}}&\\
&&&&&&\parbox[t][0.3cm]{11.3599cm}{\raggedright {\ }{\ }{\ }with \ensuremath{^{\textnormal{19}}}Ne*(4549, 4600, 4635) states.\vspace{0.1cm}}&\\
\multicolumn{1}{r@{}}{5092}&\multicolumn{1}{@{}l}{\ensuremath{^{{\hyperlink{NE9LEVEL1}{b}}}}}&\multicolumn{1}{l}{5/2\ensuremath{^{+}}}&&\multicolumn{1}{r@{}}{1}&\multicolumn{1}{@{.}l}{8 {\it 9}}&&\\
\end{longtable}
\begin{textblock}{29}(0,27.3)
Continued on next page (footnotes at end of table)
\end{textblock}
\clearpage
\begin{longtable}{ccccccc@{\extracolsep{\fill}}c}
\\[-.4cm]
\multicolumn{8}{c}{{\bf \small \underline{\ensuremath{^{\textnormal{2}}}H(\ensuremath{^{\textnormal{18}}}Ne,\ensuremath{^{\textnormal{19}}}Ne),(\ensuremath{^{\textnormal{18}}}Ne,p)\hspace{0.2in}\href{https://www.nndc.bnl.gov/nsr/nsrlink.jsp?2002La29,B}{2002La29},\href{https://www.nndc.bnl.gov/nsr/nsrlink.jsp?2002Os05,B}{2002Os05} (continued)}}}\\
\multicolumn{8}{c}{~}\\
\multicolumn{8}{c}{\underline{\ensuremath{^{19}}Ne Levels (continued)}}\\
\multicolumn{8}{c}{~}\\
\multicolumn{2}{c}{E(level)$^{}$}&J$^{\pi}$$^{{\hyperlink{NE9LEVEL0}{a}}}$&L$^{}$&\multicolumn{2}{c}{\ensuremath{\Gamma}\ensuremath{\alpha}/\ensuremath{\Gamma}$^{{\hyperlink{NE9LEVEL3}{d}}}$}&Comments&\\[-.2cm]
\multicolumn{2}{c}{\hrulefill}&\hrulefill&\hrulefill&\multicolumn{2}{c}{\hrulefill}&\hrulefill&
\endhead
\multicolumn{1}{r@{}}{5351}&\multicolumn{1}{@{}l}{\ensuremath{^{{\hyperlink{NE9LEVEL1}{b}}}}}&\multicolumn{1}{l}{1/2\ensuremath{^{+}}}&&\multicolumn{1}{r@{}}{1}&\multicolumn{1}{@{.}l}{3 {\it 3}}&\parbox[t][0.3cm]{11.4027cm}{\raggedright E(level): From (\href{https://www.nndc.bnl.gov/nsr/nsrlink.jsp?2002La29,B}{2002La29}): A member of an unresolved triplet states together with\vspace{0.1cm}}&\\
&&&&&&\parbox[t][0.3cm]{11.4027cm}{\raggedright {\ }{\ }{\ }\ensuremath{^{\textnormal{19}}}Ne*(5424, 5463) states.\vspace{0.1cm}}&\\
&&&&&&\parbox[t][0.3cm]{11.4027cm}{\raggedright \ensuremath{\Gamma}\ensuremath{\alpha}/\ensuremath{\Gamma}: The \ensuremath{\alpha}-branching ratio reported here is for the unresolved triplet\vspace{0.1cm}}&\\
&&&&&&\parbox[t][0.3cm]{11.4027cm}{\raggedright {\ }{\ }{\ }states at 5351/5424/5463 keV (\href{https://www.nndc.bnl.gov/nsr/nsrlink.jsp?2002La29,B}{2002La29}).\vspace{0.1cm}}&\\
\multicolumn{1}{r@{}}{5424}&\multicolumn{1}{@{}l}{}&\multicolumn{1}{l}{7/2\ensuremath{^{+}}}&&&&\parbox[t][0.3cm]{11.4027cm}{\raggedright E(level): From (\href{https://www.nndc.bnl.gov/nsr/nsrlink.jsp?2002Os05,B}{2002Os05}, \href{https://www.nndc.bnl.gov/nsr/nsrlink.jsp?2002La29,B}{2002La29}): A member of an unresolved triplet states\vspace{0.1cm}}&\\
&&&&&&\parbox[t][0.3cm]{11.4027cm}{\raggedright {\ }{\ }{\ }together with \ensuremath{^{\textnormal{19}}}Ne*(5351, 5463) states.\vspace{0.1cm}}&\\
\multicolumn{1}{r@{}}{5463}&\multicolumn{1}{@{}l}{}&&&&&\parbox[t][0.3cm]{11.4027cm}{\raggedright E(level): From (\href{https://www.nndc.bnl.gov/nsr/nsrlink.jsp?2002La29,B}{2002La29}): A member of an unresolved triplet states together with\vspace{0.1cm}}&\\
&&&&&&\parbox[t][0.3cm]{11.4027cm}{\raggedright {\ }{\ }{\ }\ensuremath{^{\textnormal{19}}}Ne*(5351, 5424) states.\vspace{0.1cm}}&\\
\multicolumn{1}{r@{}}{6013}&\multicolumn{1}{@{}l}{}&\multicolumn{1}{l}{(3/2\ensuremath{^{-}})}&&\multicolumn{1}{r@{}}{0}&\multicolumn{1}{@{.}l}{96 {\it 20}}&\parbox[t][0.3cm]{11.4027cm}{\raggedright E(level): From (\href{https://www.nndc.bnl.gov/nsr/nsrlink.jsp?2002Os05,B}{2002Os05}, \href{https://www.nndc.bnl.gov/nsr/nsrlink.jsp?2002La29,B}{2002La29}): A member of an unresolved doublet states\vspace{0.1cm}}&\\
&&&&&&\parbox[t][0.3cm]{11.4027cm}{\raggedright {\ }{\ }{\ }together with the \ensuremath{^{\textnormal{19}}}Ne*(6092) state.\vspace{0.1cm}}&\\
&&&&&&\parbox[t][0.3cm]{11.4027cm}{\raggedright \ensuremath{\Gamma}\ensuremath{\alpha}/\ensuremath{\Gamma}: The \ensuremath{\alpha}-branching ratio reported here is for the unresolved\vspace{0.1cm}}&\\
&&&&&&\parbox[t][0.3cm]{11.4027cm}{\raggedright {\ }{\ }{\ }doublet states at 6013 keV and 6092 keV (\href{https://www.nndc.bnl.gov/nsr/nsrlink.jsp?2002La29,B}{2002La29}).\vspace{0.1cm}}&\\
\multicolumn{1}{r@{}}{6092}&\multicolumn{1}{@{}l}{}&\multicolumn{1}{l}{1/2\ensuremath{^{+}}}&\multicolumn{1}{l}{0}&&&\parbox[t][0.3cm]{11.4027cm}{\raggedright E(level): From (\href{https://www.nndc.bnl.gov/nsr/nsrlink.jsp?2002La29,B}{2002La29}): A member of an unresolved doublet states together\vspace{0.1cm}}&\\
&&&&&&\parbox[t][0.3cm]{11.4027cm}{\raggedright {\ }{\ }{\ }with the \ensuremath{^{\textnormal{19}}}Ne*(6013) state.\vspace{0.1cm}}&\\
&&&&&&\parbox[t][0.3cm]{11.4027cm}{\raggedright J\ensuremath{^{\pi}},L: From DWBA calculations presented in (A. M. Laird, Ph.D. Thesis,\vspace{0.1cm}}&\\
&&&&&&\parbox[t][0.3cm]{11.4027cm}{\raggedright {\ }{\ }{\ }University of Edinburgh (2000), unpublished). These results are discussed\vspace{0.1cm}}&\\
&&&&&&\parbox[t][0.3cm]{11.4027cm}{\raggedright {\ }{\ }{\ }(without presentation of the data) in (\href{https://www.nndc.bnl.gov/nsr/nsrlink.jsp?2002La29,B}{2002La29}: See text).\vspace{0.1cm}}&\\
\end{longtable}
\parbox[b][0.3cm]{17.7cm}{\makebox[1ex]{\ensuremath{^{\hypertarget{NE9LEVEL0}{a}}}} From the Adopted Levels of \ensuremath{^{\textnormal{19}}}Ne unless otherwise mentioned.}\\
\parbox[b][0.3cm]{17.7cm}{\makebox[1ex]{\ensuremath{^{\hypertarget{NE9LEVEL1}{b}}}} This state was observed in (\href{https://www.nndc.bnl.gov/nsr/nsrlink.jsp?1998Os02,B}{1998Os02}) but the statistics were very limited (see Fig. 5).}\\
\parbox[b][0.3cm]{17.7cm}{\makebox[1ex]{\ensuremath{^{\hypertarget{NE9LEVEL2}{c}}}} No attempt was made by (\href{https://www.nndc.bnl.gov/nsr/nsrlink.jsp?2002La29,B}{2002La29}) to fit separate components in the observed, unresolved quadruplet peak.}\\
\parbox[b][0.3cm]{17.7cm}{\makebox[1ex]{\ensuremath{^{\hypertarget{NE9LEVEL3}{d}}}} From (\href{https://www.nndc.bnl.gov/nsr/nsrlink.jsp?2002La29,B}{2002La29}) unless otherwise mentioned.}\\
\parbox[b][0.3cm]{17.7cm}{\makebox[1ex]{\ensuremath{^{\hypertarget{NE9LEVEL4}{e}}}} From (\href{https://www.nndc.bnl.gov/nsr/nsrlink.jsp?2002Os05,B}{2002Os05}).}\\
\parbox[b][0.3cm]{17.7cm}{\makebox[1ex]{\ensuremath{^{\hypertarget{NE9LEVEL5}{f}}}} The upper limit is deduced from the fact that no \ensuremath{\alpha}-decay event was observed for this state by (\href{https://www.nndc.bnl.gov/nsr/nsrlink.jsp?2002La29,B}{2002La29}).}\\
\vspace{0.5cm}
\clearpage
\subsection[\hspace{-0.2cm}\ensuremath{^{\textnormal{2}}}H(\ensuremath{^{\textnormal{20}}}Ne,\ensuremath{^{\textnormal{19}}}Ne),(\ensuremath{^{\textnormal{20}}}Ne,t)]{ }
\vspace{-27pt}
\vspace{0.3cm}
\hypertarget{NE10}{{\bf \small \underline{\ensuremath{^{\textnormal{2}}}H(\ensuremath{^{\textnormal{20}}}Ne,\ensuremath{^{\textnormal{19}}}Ne),(\ensuremath{^{\textnormal{20}}}Ne,t)\hspace{0.2in}\href{https://www.nndc.bnl.gov/nsr/nsrlink.jsp?2003Re16,B}{2003Re16},\href{https://www.nndc.bnl.gov/nsr/nsrlink.jsp?2003Re25,B}{2003Re25}}}}\\
\vspace{4pt}
\vspace{8pt}
\parbox[b][0.3cm]{17.7cm}{\addtolength{\parindent}{-0.2in}\ensuremath{^{\textnormal{20}}}Ne(d,t) reaction in inverse kinematics.}\\
\parbox[b][0.3cm]{17.7cm}{\addtolength{\parindent}{-0.2in}J\ensuremath{^{\ensuremath{\pi}}}(\ensuremath{^{\textnormal{2}}}H\ensuremath{_{\textnormal{g.s.}}})=1\ensuremath{^{\textnormal{+}}} and J\ensuremath{^{\ensuremath{\pi}}}(\ensuremath{^{\textnormal{20}}}Ne\ensuremath{_{\textnormal{g.s.}}})=0\ensuremath{^{\textnormal{+}}}.}\\
\parbox[b][0.3cm]{17.7cm}{\addtolength{\parindent}{-0.2in}\href{https://www.nndc.bnl.gov/nsr/nsrlink.jsp?2003Re16,B}{2003Re16}, \href{https://www.nndc.bnl.gov/nsr/nsrlink.jsp?2003Re25,B}{2003Re25}: \ensuremath{^{\textnormal{2}}}H(\ensuremath{^{\textnormal{20}}}Ne,t) and \ensuremath{^{\textnormal{2}}}H(\ensuremath{^{\textnormal{20}}}Ne,\ensuremath{^{\textnormal{19}}}Ne*\ensuremath{\rightarrow}\ensuremath{\alpha}+\ensuremath{^{\textnormal{15}}}O) E not given; solid (CD\ensuremath{_{\textnormal{2}}})\ensuremath{_{\textnormal{n}}} target or a gas cell filled with D\ensuremath{_{\textnormal{2}}} gas}\\
\parbox[b][0.3cm]{17.7cm}{(pressure not given). Measured \ensuremath{^{\textnormal{2}}}H(\ensuremath{^{\textnormal{20}}}Ne,t)\ensuremath{^{\textnormal{19}}}Ne*, where the \ensuremath{^{\textnormal{15}}}O ions from the \ensuremath{^{\textnormal{19}}}Ne*\ensuremath{\rightarrow}\ensuremath{\alpha}+\ensuremath{^{\textnormal{15}}}O decay were measured in}\\
\parbox[b][0.3cm]{17.7cm}{coincidence with the tritons.}\\
\vspace{12pt}
\underline{$^{19}$Ne Levels}\\
\begin{longtable}{ccc@{\extracolsep{\fill}}c}
\multicolumn{2}{c}{E(level)$^{{\hyperlink{NE10LEVEL0}{a}}}$}&Comments&\\[-.2cm]
\multicolumn{2}{c}{\hrulefill}&\hrulefill&
\endfirsthead
\multicolumn{1}{r@{}}{4033}&\multicolumn{1}{@{}l}{}&\parbox[t][0.3cm]{16.016cm}{\raggedright Reaction cross section for populating this level was found to be of the order of 20 \ensuremath{\mu}b/sr.\vspace{0.1cm}}&\\
\multicolumn{1}{r@{}}{4140}&\multicolumn{1}{@{}l}{\ensuremath{^{{\hyperlink{NE10LEVEL1}{b}}}}}&&\\
\multicolumn{1}{r@{}}{4197}&\multicolumn{1}{@{}l}{\ensuremath{^{{\hyperlink{NE10LEVEL1}{b}}}}}&&\\
\end{longtable}
\parbox[b][0.3cm]{17.7cm}{\makebox[1ex]{\ensuremath{^{\hypertarget{NE10LEVEL0}{a}}}} These states are proton bound and \ensuremath{\alpha} unbound: S\ensuremath{_{\ensuremath{\alpha}}}(\ensuremath{^{\textnormal{19}}}Ne)=3528.5 keV \textit{5} (\href{https://www.nndc.bnl.gov/nsr/nsrlink.jsp?2021Wa16,B}{2021Wa16}).}\\
\parbox[b][0.3cm]{17.7cm}{\makebox[1ex]{\ensuremath{^{\hypertarget{NE10LEVEL1}{b}}}} Comparable yield to that of (\href{https://www.nndc.bnl.gov/nsr/nsrlink.jsp?2002Ku12,B}{2002Ku12}: \ensuremath{^{\textnormal{20}}}Ne(d,t) in normal kinematics) was observed for this state. Evaluator notes that}\\
\parbox[b][0.3cm]{17.7cm}{{\ }{\ }(\href{https://www.nndc.bnl.gov/nsr/nsrlink.jsp?2002Ku12,B}{2002Ku12}) does not present the results. Instead, see (K. Kumagai, M.Sc. Thesis, Tohoku University (1999), unpublished).}\\
\vspace{0.5cm}
\clearpage
\subsection[\hspace{-0.2cm}\ensuremath{^{\textnormal{3}}}He(\ensuremath{^{\textnormal{20}}}Ne,\ensuremath{\alpha}\ensuremath{\gamma})]{ }
\vspace{-27pt}
\vspace{0.3cm}
\hypertarget{NE11}{{\bf \small \underline{\ensuremath{^{\textnormal{3}}}He(\ensuremath{^{\textnormal{20}}}Ne,\ensuremath{\alpha}\ensuremath{\gamma})\hspace{0.2in}\href{https://www.nndc.bnl.gov/nsr/nsrlink.jsp?1970Bh02,B}{1970Bh02},\href{https://www.nndc.bnl.gov/nsr/nsrlink.jsp?2008My01,B}{2008My01}}}}\\
\vspace{4pt}
\vspace{8pt}
\parbox[b][0.3cm]{17.7cm}{\addtolength{\parindent}{-0.2in}\ensuremath{^{\textnormal{20}}}Ne(\ensuremath{^{\textnormal{3}}}He,\ensuremath{\alpha}\ensuremath{\gamma}) reaction in inverse kinematics.}\\
\parbox[b][0.3cm]{17.7cm}{\addtolength{\parindent}{-0.2in}J\ensuremath{^{\ensuremath{\pi}}}(\ensuremath{^{\textnormal{3}}}He\ensuremath{_{\textnormal{g.s.}}})=1/2\ensuremath{^{\textnormal{+}}} and J\ensuremath{^{\ensuremath{\pi}}}(\ensuremath{^{\textnormal{20}}}Ne\ensuremath{_{\textnormal{g.s.}}})=0\ensuremath{^{\textnormal{+}}}.}\\
\parbox[b][0.3cm]{17.7cm}{\addtolength{\parindent}{-0.2in}\href{https://www.nndc.bnl.gov/nsr/nsrlink.jsp?1970Bh02,B}{1970Bh02}: \ensuremath{^{\textnormal{3}}}He(\ensuremath{^{\textnormal{20}}}Ne,\ensuremath{\alpha}\ensuremath{\gamma}) E=31.18 MeV; measured E\ensuremath{_{\ensuremath{\gamma}}} from the de-excitation of \ensuremath{^{\textnormal{19}}}Ne*(238, 275) states using a Ge(Li) detector}\\
\parbox[b][0.3cm]{17.7cm}{with a resolution of 1.6 keV placed at \ensuremath{\theta}\ensuremath{_{\textnormal{lab}}}=0\ensuremath{^\circ}. Measured Doppler shift for the 275-keV \ensuremath{\gamma} ray. Deduced \ensuremath{^{\textnormal{19}}}Ne levels and lifetime}\\
\parbox[b][0.3cm]{17.7cm}{for the 275-keV state using the recoil distance method.}\\
\parbox[b][0.3cm]{17.7cm}{\addtolength{\parindent}{-0.2in}\href{https://www.nndc.bnl.gov/nsr/nsrlink.jsp?2006Ka50,B}{2006Ka50}: \ensuremath{^{\textnormal{3}}}He(\ensuremath{^{\textnormal{20}}}Ne,\ensuremath{\alpha}\ensuremath{\gamma}) E=34 MeV; stopped the \ensuremath{^{\textnormal{19}}}Ne recoils in a \ensuremath{^{\textnormal{3}}}He-implanted Au foil cooled to below room temperature.}\\
\parbox[b][0.3cm]{17.7cm}{Measured lifetime of the \ensuremath{^{\textnormal{19}}}Ne*(4034) level using Doppler-shift attenuation method. Measured \ensuremath{\alpha}-\ensuremath{\gamma} coincidences using a regular}\\
\parbox[b][0.3cm]{17.7cm}{\ensuremath{\Delta}E-E Si telescope (covering \ensuremath{\theta}\ensuremath{_{\textnormal{lab}}}\ensuremath{<}20\ensuremath{^\circ}) and a HPGe detector downstream the telescope, both at \ensuremath{\theta}\ensuremath{_{\textnormal{lab}}}=0\ensuremath{^\circ}. A second HPGe detector}\\
\parbox[b][0.3cm]{17.7cm}{was used to measure the unshifted \ensuremath{\gamma}-ray energies at \ensuremath{\theta}\ensuremath{_{\textnormal{lab}}}=90\ensuremath{^\circ}. Observed the \ensuremath{\gamma} ray from the direct transition to the ground state}\\
\parbox[b][0.3cm]{17.7cm}{from the 4034-keV level, and the decays from the \ensuremath{^{\textnormal{19}}}Ne*(238, 275, 1508) levels.}\\
\parbox[b][0.3cm]{17.7cm}{\addtolength{\parindent}{-0.2in}\href{https://www.nndc.bnl.gov/nsr/nsrlink.jsp?2008My01,B}{2008My01}: \ensuremath{^{\textnormal{3}}}He(\ensuremath{^{\textnormal{20}}}Ne,\ensuremath{\alpha}\ensuremath{\gamma}) E=34 MeV; stopped the \ensuremath{^{\textnormal{19}}}Ne recoils using a cold Au foil with a layer of implanted \ensuremath{^{\textnormal{3}}}He; measured}\\
\parbox[b][0.3cm]{17.7cm}{lifetimes of 6 levels above the \ensuremath{^{\textnormal{15}}}O+\ensuremath{\alpha} breakup threshold of \ensuremath{^{\textnormal{19}}}Ne using Doppler-shift attenuation method; measured \ensuremath{\alpha}-\ensuremath{\gamma}}\\
\parbox[b][0.3cm]{17.7cm}{coincidences using a Si surface barrier \ensuremath{\Delta}E-E telescope followed by a HPGe detector (covering \ensuremath{\theta}\ensuremath{_{\textnormal{lab}}}=\ensuremath{\pm}23\ensuremath{^\circ}), both placed at}\\
\parbox[b][0.3cm]{17.7cm}{\ensuremath{\theta}\ensuremath{_{\textnormal{lab}}}=0\ensuremath{^\circ}.}\\
\parbox[b][0.3cm]{17.7cm}{\addtolength{\parindent}{-0.2in}\href{https://www.nndc.bnl.gov/nsr/nsrlink.jsp?2014Da01,B}{2014Da01}: \ensuremath{^{\textnormal{3}}}He(\ensuremath{^{\textnormal{20}}}Ne,\ensuremath{\alpha}\ensuremath{\gamma}) E=34 MeV. This study describes the scattering chamber used in (\href{https://www.nndc.bnl.gov/nsr/nsrlink.jsp?2006Ka50,B}{2006Ka50}) and (\href{https://www.nndc.bnl.gov/nsr/nsrlink.jsp?2008My01,B}{2008My01})}\\
\parbox[b][0.3cm]{17.7cm}{experiments and very briefly mentions the experimental setup of those measurements.}\\
\vspace{12pt}
\underline{$^{19}$Ne Levels}\\
\begin{longtable}{cccccc@{\extracolsep{\fill}}c}
\multicolumn{2}{c}{E(level)$^{}$}&J$^{\pi}$$^{{\hyperlink{NE11LEVEL2}{c}}}$&\multicolumn{2}{c}{T$_{1/2}$$^{{\hyperlink{NE11LEVEL3}{d}}}$}&Comments&\\[-.2cm]
\multicolumn{2}{c}{\hrulefill}&\hrulefill&\multicolumn{2}{c}{\hrulefill}&\hrulefill&
\endfirsthead
\multicolumn{1}{r@{}}{0}&\multicolumn{1}{@{}l}{\ensuremath{^{{\hyperlink{NE11LEVEL0}{a}}}}}&\multicolumn{1}{l}{1/2\ensuremath{^{+}}}&&&\parbox[t][0.3cm]{11.821051cm}{\raggedright E(level): Measured in (\href{https://www.nndc.bnl.gov/nsr/nsrlink.jsp?1970Bh02,B}{1970Bh02}, \href{https://www.nndc.bnl.gov/nsr/nsrlink.jsp?2006Ka50,B}{2006Ka50}, \href{https://www.nndc.bnl.gov/nsr/nsrlink.jsp?2008My01,B}{2008My01}).\vspace{0.1cm}}&\\
\multicolumn{1}{r@{}}{238}&\multicolumn{1}{@{.}l}{34 {\it 15}}&\multicolumn{1}{l}{5/2\ensuremath{^{+}}}&&&\parbox[t][0.3cm]{11.821051cm}{\raggedright E(level): From (\href{https://www.nndc.bnl.gov/nsr/nsrlink.jsp?1970Bh02,B}{1970Bh02}). See also E\ensuremath{_{\textnormal{x}}}=238 keV (\href{https://www.nndc.bnl.gov/nsr/nsrlink.jsp?2006Ka50,B}{2006Ka50}).\vspace{0.1cm}}&\\
\multicolumn{1}{r@{}}{275}&\multicolumn{1}{@{.}l}{30 {\it 20}}&\multicolumn{1}{l}{1/2\ensuremath{^{-}}}&\multicolumn{1}{r@{}}{42}&\multicolumn{1}{@{.}l}{6 ps {\it 21}}&\parbox[t][0.3cm]{11.821051cm}{\raggedright E(level): From (\href{https://www.nndc.bnl.gov/nsr/nsrlink.jsp?1970Bh02,B}{1970Bh02}). See also E\ensuremath{_{\textnormal{x}}}=275 keV (\href{https://www.nndc.bnl.gov/nsr/nsrlink.jsp?2006Ka50,B}{2006Ka50}, \href{https://www.nndc.bnl.gov/nsr/nsrlink.jsp?2008My01,B}{2008My01}).\vspace{0.1cm}}&\\
&&&&&\parbox[t][0.3cm]{11.821051cm}{\raggedright T\ensuremath{_{1/2}}: From \ensuremath{\tau}=61.4 ps \textit{30} (\href{https://www.nndc.bnl.gov/nsr/nsrlink.jsp?1970Bh02,B}{1970Bh02}).\vspace{0.1cm}}&\\
\multicolumn{1}{r@{}}{1508}&\multicolumn{1}{@{}l}{}&\multicolumn{1}{l}{5/2\ensuremath{^{-}}}&&&\parbox[t][0.3cm]{11.821051cm}{\raggedright E(level): From (\href{https://www.nndc.bnl.gov/nsr/nsrlink.jsp?2006Ka50,B}{2006Ka50}).\vspace{0.1cm}}&\\
\multicolumn{1}{r@{}}{1536}&\multicolumn{1}{@{}l}{\ensuremath{^{{\hyperlink{NE11LEVEL1}{b}}}}}&\multicolumn{1}{l}{3/2\ensuremath{^{+}}}&\multicolumn{1}{r@{}}{13}&\multicolumn{1}{@{.}l}{24 fs {\it +90\textminus87}}&\parbox[t][0.3cm]{11.821051cm}{\raggedright T\ensuremath{_{1/2}}: From \ensuremath{\tau}=19.1 fs \textit{+7{\textminus}6} (stat.) \textit{11} (sys.) (\href{https://www.nndc.bnl.gov/nsr/nsrlink.jsp?2008My01,B}{2008My01}).\vspace{0.1cm}}&\\
\multicolumn{1}{r@{}}{1615}&\multicolumn{1}{@{.}l}{28\ensuremath{^{{\hyperlink{NE11LEVEL0}{a}}}}}&\multicolumn{1}{l}{3/2\ensuremath{^{-}}}&&&\parbox[t][0.3cm]{11.821051cm}{\raggedright E(level): State observed in (\href{https://www.nndc.bnl.gov/nsr/nsrlink.jsp?2008My01,B}{2008My01}).\vspace{0.1cm}}&\\
\multicolumn{1}{r@{}}{2794}&\multicolumn{1}{@{.}l}{4\ensuremath{^{{\hyperlink{NE11LEVEL0}{a}}}}}&\multicolumn{1}{l}{9/2\ensuremath{^{+}}}&&&\parbox[t][0.3cm]{11.821051cm}{\raggedright E(level): State observed in (\href{https://www.nndc.bnl.gov/nsr/nsrlink.jsp?2008My01,B}{2008My01}).\vspace{0.1cm}}&\\
\multicolumn{1}{r@{}}{4035}&\multicolumn{1}{@{}l}{\ensuremath{^{{\hyperlink{NE11LEVEL1}{b}}}}}&\multicolumn{1}{l}{3/2\ensuremath{^{+}}}&\multicolumn{1}{r@{}}{4}&\multicolumn{1}{@{.}l}{78\ensuremath{^{{\hyperlink{NE11LEVEL4}{e}}}} fs {\it 115}}&\parbox[t][0.3cm]{11.821051cm}{\raggedright E(level): See also E\ensuremath{_{\textnormal{x}}}=4034 keV (\href{https://www.nndc.bnl.gov/nsr/nsrlink.jsp?2006Ka50,B}{2006Ka50}: See Fig. 7).\vspace{0.1cm}}&\\
&&&&&\parbox[t][0.3cm]{11.821051cm}{\raggedright T\ensuremath{_{1/2}}: From \ensuremath{\tau}=6.9 fs \textit{+15{\textminus}15} (stat.) \textit{7} (sys.) (\href{https://www.nndc.bnl.gov/nsr/nsrlink.jsp?2008My01,B}{2008My01}). See also (1) T\ensuremath{_{\textnormal{1/2}}}=7.6 fs\vspace{0.1cm}}&\\
&&&&&\parbox[t][0.3cm]{11.821051cm}{\raggedright {\ }{\ }{\ }\textit{+28{\textminus}21} deduced from \ensuremath{\tau}=11 fs \textit{+4{\textminus}3} at 1\ensuremath{\sigma} level (\href{https://www.nndc.bnl.gov/nsr/nsrlink.jsp?2006Ka50,B}{2006Ka50}). This value is not\vspace{0.1cm}}&\\
&&&&&\parbox[t][0.3cm]{11.821051cm}{\raggedright {\ }{\ }{\ }considered because the same group measured this lifetime more precisely in\vspace{0.1cm}}&\\
&&&&&\parbox[t][0.3cm]{11.821051cm}{\raggedright {\ }{\ }{\ }(\href{https://www.nndc.bnl.gov/nsr/nsrlink.jsp?2008My01,B}{2008My01}). (\href{https://www.nndc.bnl.gov/nsr/nsrlink.jsp?2006Ka50,B}{2006Ka50}) also reported \ensuremath{\tau}=11 fs \textit{+8{\textminus}7} at 95.45\% C.L. (2\ensuremath{\sigma}) resulting\vspace{0.1cm}}&\\
&&&&&\parbox[t][0.3cm]{11.821051cm}{\raggedright {\ }{\ }{\ }in T\ensuremath{_{\textnormal{1/2}}}=7.6 fs \textit{+55{\textminus}48}; (2) \ensuremath{\tau}=6.6 fs \textit{+24{\textminus}21} (stat.) \textit{7} (sys.) (\href{https://www.nndc.bnl.gov/nsr/nsrlink.jsp?2008My01,B}{2008My01}) deduced\vspace{0.1cm}}&\\
&&&&&\parbox[t][0.3cm]{11.821051cm}{\raggedright {\ }{\ }{\ }from DSAM for the \ensuremath{\gamma} ray from the \ensuremath{^{\textnormal{19}}}Ne*(4035)\ensuremath{\rightarrow}\ensuremath{^{\textnormal{19}}}Ne*(1536) decay; and (3)\vspace{0.1cm}}&\\
&&&&&\parbox[t][0.3cm]{11.821051cm}{\raggedright {\ }{\ }{\ }\ensuremath{\tau}=7.1 fs \textit{+19{\textminus}19} (stat.) \textit{6} (sys.) from DSAM for the \ensuremath{\gamma} ray emitted by the\vspace{0.1cm}}&\\
&&&&&\parbox[t][0.3cm]{11.821051cm}{\raggedright {\ }{\ }{\ }\ensuremath{^{\textnormal{19}}}Ne*(4035)\ensuremath{\rightarrow}\ensuremath{^{\textnormal{19}}}Ne\ensuremath{_{\textnormal{g.s.}}} decay (\href{https://www.nndc.bnl.gov/nsr/nsrlink.jsp?2008My01,B}{2008My01}).\vspace{0.1cm}}&\\
&&&&&\parbox[t][0.3cm]{11.821051cm}{\raggedright \ensuremath{\Gamma}\ensuremath{_{\ensuremath{\alpha}}}\ensuremath{<}200 \ensuremath{\mu}eV was mentioned by (\href{https://www.nndc.bnl.gov/nsr/nsrlink.jsp?2005Ka50,B}{2005Ka50}) at 99.73\% C.L.: (\href{https://www.nndc.bnl.gov/nsr/nsrlink.jsp?2005Ka50,B}{2005Ka50}) calculated\vspace{0.1cm}}&\\
&&&&&\parbox[t][0.3cm]{11.821051cm}{\raggedright {\ }{\ }{\ }the joint likelihood for the lifetime of this state using their deduced lifetime and that\vspace{0.1cm}}&\\
&&&&&\parbox[t][0.3cm]{11.821051cm}{\raggedright {\ }{\ }{\ }of (\href{https://www.nndc.bnl.gov/nsr/nsrlink.jsp?2005Ta28,B}{2005Ta28}). This distribution constrained the lifetime to 3 fs \ensuremath{<}\ensuremath{\tau}\ensuremath{<} 22 fs at the\vspace{0.1cm}}&\\
&&&&&\parbox[t][0.3cm]{11.821051cm}{\raggedright {\ }{\ }{\ }99.73\% C.L., peaking at 12 fs. Using the 3\ensuremath{\sigma} upper limit on the \ensuremath{\Gamma}\ensuremath{_{\ensuremath{\alpha}}}/\ensuremath{\Gamma} from\vspace{0.1cm}}&\\
&&&&&\parbox[t][0.3cm]{11.821051cm}{\raggedright {\ }{\ }{\ }(\href{https://www.nndc.bnl.gov/nsr/nsrlink.jsp?2003Da13,B}{2003Da13}) and the 3\ensuremath{\sigma} lower limit on \ensuremath{\tau}, (\href{https://www.nndc.bnl.gov/nsr/nsrlink.jsp?2006Ka50,B}{2006Ka50}) determined \ensuremath{\Gamma}\ensuremath{_{\ensuremath{\alpha}}} mentioned\vspace{0.1cm}}&\\
&&&&&\parbox[t][0.3cm]{11.821051cm}{\raggedright {\ }{\ }{\ }above.\vspace{0.1cm}}&\\
\multicolumn{1}{r@{}}{4144}&\multicolumn{1}{@{}l}{\ensuremath{^{{\hyperlink{NE11LEVEL1}{b}}}}}&\multicolumn{1}{l}{(7/2\ensuremath{^{-}})}&\multicolumn{1}{r@{}}{9}&\multicolumn{1}{@{.}l}{7 fs {\it +30\textminus29}}&\parbox[t][0.3cm]{11.821051cm}{\raggedright T\ensuremath{_{1/2}}: From \ensuremath{\tau}=14.0 fs \textit{+42{\textminus}40} (stat.) \textit{12} (sys.) (\href{https://www.nndc.bnl.gov/nsr/nsrlink.jsp?2008My01,B}{2008My01}).\vspace{0.1cm}}&\\
\multicolumn{1}{r@{}}{4200}&\multicolumn{1}{@{}l}{\ensuremath{^{{\hyperlink{NE11LEVEL1}{b}}}}}&\multicolumn{1}{l}{(9/2\ensuremath{^{-}})}&\multicolumn{1}{r@{}}{26}&\multicolumn{1}{@{ }l}{fs {\it +14\textminus7}}&\parbox[t][0.3cm]{11.821051cm}{\raggedright T\ensuremath{_{1/2}}: From \ensuremath{\tau}=38 fs \textit{+20{\textminus}10} (stat.) \textit{2} (sys.) (\href{https://www.nndc.bnl.gov/nsr/nsrlink.jsp?2008My01,B}{2008My01}).\vspace{0.1cm}}&\\
\multicolumn{1}{r@{}}{4378}&\multicolumn{1}{@{}l}{\ensuremath{^{{\hyperlink{NE11LEVEL1}{b}}}}}&\multicolumn{1}{l}{7/2\ensuremath{^{+}}}&\multicolumn{1}{r@{}}{$\leq$3}&\multicolumn{1}{@{.}l}{74 fs}&\parbox[t][0.3cm]{11.821051cm}{\raggedright T\ensuremath{_{1/2}}: From \ensuremath{\tau}\ensuremath{\leq}5.4 fs at 95\% C.L. (\href{https://www.nndc.bnl.gov/nsr/nsrlink.jsp?2008My01,B}{2008My01}).\vspace{0.1cm}}&\\
&&&&&\parbox[t][0.3cm]{11.821051cm}{\raggedright T\ensuremath{_{1/2}}: The upper 1\ensuremath{\sigma} statistical and systematic errors were found to be 1.4 and 0.6 fs,\vspace{0.1cm}}&\\
&&&&&\parbox[t][0.3cm]{11.821051cm}{\raggedright {\ }{\ }{\ }respectively (\href{https://www.nndc.bnl.gov/nsr/nsrlink.jsp?2008My01,B}{2008My01}).\vspace{0.1cm}}&\\
\multicolumn{1}{r@{}}{4548}&\multicolumn{1}{@{}l}{\ensuremath{^{{\hyperlink{NE11LEVEL1}{b}}}}}&\multicolumn{1}{l}{3/2\ensuremath{^{-}}}&\multicolumn{1}{r@{}}{13}&\multicolumn{1}{@{.}l}{0\ensuremath{^{{\hyperlink{NE11LEVEL4}{e}}}} fs {\it +26\textminus24}}&\parbox[t][0.3cm]{11.821051cm}{\raggedright T\ensuremath{_{1/2}}: From \ensuremath{\tau}=18.7 fs \textit{+30{\textminus}26} (stat.) \textit{22} (sys.) (\href{https://www.nndc.bnl.gov/nsr/nsrlink.jsp?2008My01,B}{2008My01}), which results in\vspace{0.1cm}}&\\
&&&&&\parbox[t][0.3cm]{11.821051cm}{\raggedright {\ }{\ }{\ }T\ensuremath{_{\textnormal{1/2}}}=12.96 fs \textit{+258{\textminus}240}. Other value: \ensuremath{\tau}=16.6 fs \textit{+44{\textminus}36} (stat.) \textit{16} (sys.) from DSAM\vspace{0.1cm}}&\\
&&&&&\parbox[t][0.3cm]{11.821051cm}{\raggedright {\ }{\ }{\ }for the \ensuremath{\gamma} ray emitted by the \ensuremath{^{\textnormal{19}}}Ne*(4548)\ensuremath{\rightarrow}\ensuremath{^{\textnormal{19}}}Ne*(275) decay; and \ensuremath{\tau}=19.9 fs \textit{+30{\textminus}36}\vspace{0.1cm}}&\\
&&&&&\parbox[t][0.3cm]{11.821051cm}{\raggedright {\ }{\ }{\ }(stat.) \textit{23} (sys.) from DSAM for the \ensuremath{\gamma} ray from the \ensuremath{^{\textnormal{19}}}Ne*(4548)\ensuremath{\rightarrow}\ensuremath{^{\textnormal{19}}}Ne\ensuremath{_{\textnormal{g.s.}}} decay\vspace{0.1cm}}&\\
&&&&&\parbox[t][0.3cm]{11.821051cm}{\raggedright {\ }{\ }{\ }(\href{https://www.nndc.bnl.gov/nsr/nsrlink.jsp?2008My01,B}{2008My01}).\vspace{0.1cm}}&\\
\end{longtable}
\begin{textblock}{29}(0,27.3)
Continued on next page (footnotes at end of table)
\end{textblock}
\clearpage
\begin{longtable}{cccccc@{\extracolsep{\fill}}c}
\\[-.4cm]
\multicolumn{7}{c}{{\bf \small \underline{\ensuremath{^{\textnormal{3}}}He(\ensuremath{^{\textnormal{20}}}Ne,\ensuremath{\alpha}\ensuremath{\gamma})\hspace{0.2in}\href{https://www.nndc.bnl.gov/nsr/nsrlink.jsp?1970Bh02,B}{1970Bh02},\href{https://www.nndc.bnl.gov/nsr/nsrlink.jsp?2008My01,B}{2008My01} (continued)}}}\\
\multicolumn{7}{c}{~}\\
\multicolumn{7}{c}{\underline{\ensuremath{^{19}}Ne Levels (continued)}}\\
\multicolumn{7}{c}{~}\\
\multicolumn{2}{c}{E(level)$^{}$}&J$^{\pi}$$^{{\hyperlink{NE11LEVEL2}{c}}}$&\multicolumn{2}{c}{T$_{1/2}$$^{{\hyperlink{NE11LEVEL3}{d}}}$}&Comments&\\[-.2cm]
\multicolumn{2}{c}{\hrulefill}&\hrulefill&\multicolumn{2}{c}{\hrulefill}&\hrulefill&
\endhead
\multicolumn{1}{r@{}}{4602}&\multicolumn{1}{@{}l}{\ensuremath{^{{\hyperlink{NE11LEVEL1}{b}}}}}&\multicolumn{1}{l}{5/2\ensuremath{^{+}}}&\multicolumn{1}{r@{}}{5}&\multicolumn{1}{@{.}l}{3 fs {\it +16\textminus15}}&\parbox[t][0.3cm]{12.695081cm}{\raggedright T\ensuremath{_{1/2}}: From \ensuremath{\tau}=7.6 fs \textit{+21{\textminus}20} (stat.) \textit{9} (sys.) (\href{https://www.nndc.bnl.gov/nsr/nsrlink.jsp?2008My01,B}{2008My01}), which results in T\ensuremath{_{\textnormal{1/2}}}=5.27 fs\vspace{0.1cm}}&\\
&&&&&\parbox[t][0.3cm]{12.695081cm}{\raggedright {\ }{\ }{\ }\textit{+159{\textminus}152}.\vspace{0.1cm}}&\\
\multicolumn{1}{r@{}}{4634}&\multicolumn{1}{@{}l}{\ensuremath{^{{\hyperlink{NE11LEVEL1}{b}}}}}&\multicolumn{1}{l}{13/2\ensuremath{^{+}}}&&&\parbox[t][0.3cm]{12.695081cm}{\raggedright Random coincidences with the 1836-keV \ensuremath{\gamma} ray emitted by the \ensuremath{^{\textnormal{88}}}Y source, used to monitor\vspace{0.1cm}}&\\
&&&&&\parbox[t][0.3cm]{12.695081cm}{\raggedright {\ }{\ }{\ }the electronics, prevented a determination of the lifetime of this level by (\href{https://www.nndc.bnl.gov/nsr/nsrlink.jsp?2008My01,B}{2008My01}).\vspace{0.1cm}}&\\
\end{longtable}
\parbox[b][0.3cm]{17.7cm}{\makebox[1ex]{\ensuremath{^{\hypertarget{NE11LEVEL0}{a}}}} From the \ensuremath{^{\textnormal{19}}}Ne Adopted Levels.}\\
\parbox[b][0.3cm]{17.7cm}{\makebox[1ex]{\ensuremath{^{\hypertarget{NE11LEVEL1}{b}}}} From (\href{https://www.nndc.bnl.gov/nsr/nsrlink.jsp?2008My01,B}{2008My01}: See Table I).}\\
\parbox[b][0.3cm]{17.7cm}{\makebox[1ex]{\ensuremath{^{\hypertarget{NE11LEVEL2}{c}}}} From the \ensuremath{^{\textnormal{19}}}Ne Adopted Levels.}\\
\parbox[b][0.3cm]{17.7cm}{\makebox[1ex]{\ensuremath{^{\hypertarget{NE11LEVEL3}{d}}}} Whenever separate statistical and systematic uncertainties in \ensuremath{\tau} (lifetime) are provided, these uncertainties are combined in the}\\
\parbox[b][0.3cm]{17.7cm}{{\ }{\ }reported, deduced half-lives.}\\
\parbox[b][0.3cm]{17.7cm}{\makebox[1ex]{\ensuremath{^{\hypertarget{NE11LEVEL4}{e}}}} Two \ensuremath{\gamma}-decay branches were observed for this state in (\href{https://www.nndc.bnl.gov/nsr/nsrlink.jsp?2008My01,B}{2008My01}). The lifetime measurements from both decay branches were}\\
\parbox[b][0.3cm]{17.7cm}{{\ }{\ }combined by the authors using a joint likelihood for the lifetime to determine the most likely value for \ensuremath{\tau}, which is presented in}\\
\parbox[b][0.3cm]{17.7cm}{{\ }{\ }Table I of (\href{https://www.nndc.bnl.gov/nsr/nsrlink.jsp?2008My01,B}{2008My01}) as the combined result. This value was used to deduce the half-life mentioned here from that study.}\\
\vspace{0.5cm}
\underline{$\gamma$($^{19}$Ne)}\\
\begin{longtable}{ccccccc@{}ccccc@{\extracolsep{\fill}}c}
\multicolumn{2}{c}{E\ensuremath{_{\gamma}}}&\multicolumn{2}{c}{E\ensuremath{_{i}}(level)}&J\ensuremath{^{\pi}_{i}}&\multicolumn{2}{c}{E\ensuremath{_{f}}}&J\ensuremath{^{\pi}_{f}}&Mult.&\multicolumn{2}{c}{\ensuremath{\alpha}\ensuremath{^{\hyperlink{NE11GAMMA3}{d}}}}&Comments&\\[-.2cm]
\multicolumn{2}{c}{\hrulefill}&\multicolumn{2}{c}{\hrulefill}&\hrulefill&\multicolumn{2}{c}{\hrulefill}&\hrulefill&\hrulefill&\multicolumn{2}{c}{\hrulefill}&\hrulefill&
\endfirsthead
\multicolumn{1}{r@{}}{238}&\multicolumn{1}{@{.}l}{34\ensuremath{^{\hyperlink{NE11GAMMA0}{a}}}}&\multicolumn{1}{r@{}}{238}&\multicolumn{1}{@{.}l}{34}&\multicolumn{1}{l}{5/2\ensuremath{^{+}}}&\multicolumn{1}{r@{}}{0}&\multicolumn{1}{@{}l}{}&\multicolumn{1}{@{}l}{1/2\ensuremath{^{+}}}&&&&&\\
\multicolumn{1}{r@{}}{275}&\multicolumn{1}{@{.}l}{30\ensuremath{^{\hyperlink{NE11GAMMA0}{a}}}}&\multicolumn{1}{r@{}}{275}&\multicolumn{1}{@{.}l}{30}&\multicolumn{1}{l}{1/2\ensuremath{^{-}}}&\multicolumn{1}{r@{}}{0}&\multicolumn{1}{@{}l}{}&\multicolumn{1}{@{}l}{1/2\ensuremath{^{+}}}&\multicolumn{1}{l}{E1}&\multicolumn{1}{r@{}}{1}&\multicolumn{1}{@{.}l}{40\ensuremath{\times10^{-4}} {\it 2}}&\parbox[t][0.3cm]{8.004341cm}{\raggedright B(E1)(W.u.)=0.00107 \textit{6}\vspace{0.1cm}}&\\
&&&&&&&&&&&\parbox[t][0.3cm]{8.004341cm}{\raggedright \ensuremath{\alpha}(K)=0.0001325 \textit{19}; \ensuremath{\alpha}(L)=7.34\ensuremath{\times}10\ensuremath{^{\textnormal{$-$6}}} \textit{10}\vspace{0.1cm}}&\\
&&&&&&&&&&&\parbox[t][0.3cm]{8.004341cm}{\raggedright Mult.: From (\href{https://www.nndc.bnl.gov/nsr/nsrlink.jsp?1970Bh02,B}{1970Bh02}).\vspace{0.1cm}}&\\
&&&&&&&&&&&\parbox[t][0.3cm]{8.004341cm}{\raggedright B(E1)(W.u.): Other values: (1) 0.00106 W.u. \textit{5} and\vspace{0.1cm}}&\\
&&&&&&&&&&&\parbox[t][0.3cm]{8.004341cm}{\raggedright {\ }{\ }{\ }B(E1)\ensuremath{_{^{\textnormal{19}}\textnormal{Ne}}}/B(E1)\ensuremath{_{^{\textnormal{19}}\textnormal{F}}}=0.95 \textit{6} based on \ensuremath{\tau}=61.4 ps \textit{30}\vspace{0.1cm}}&\\
&&&&&&&&&&&\parbox[t][0.3cm]{8.004341cm}{\raggedright {\ }{\ }{\ }(\href{https://www.nndc.bnl.gov/nsr/nsrlink.jsp?1970Bh02,B}{1970Bh02}). (2) 0.00107 W.u. \textit{+52{\textminus}7} (\href{https://www.nndc.bnl.gov/nsr/nsrlink.jsp?1970Bh02,B}{1970Bh02})\vspace{0.1cm}}&\\
&&&&&&&&&&&\parbox[t][0.3cm]{8.004341cm}{\raggedright {\ }{\ }{\ }determined based on \ensuremath{\tau}=61 ps \textit{+4{\textminus}20} deduced for this\vspace{0.1cm}}&\\
&&&&&&&&&&&\parbox[t][0.3cm]{8.004341cm}{\raggedright {\ }{\ }{\ }state by (\href{https://www.nndc.bnl.gov/nsr/nsrlink.jsp?1969Ni09,B}{1969Ni09}).\vspace{0.1cm}}&\\
&&&&&&&&&&&\parbox[t][0.3cm]{8.004341cm}{\raggedright The mean Doppler shift of the 275-keV \ensuremath{\gamma} ray was\vspace{0.1cm}}&\\
&&&&&&&&&&&\parbox[t][0.3cm]{8.004341cm}{\raggedright {\ }{\ }{\ }measured to be 12.03 keV \textit{10} or 4.37\% \textit{4} with no\vspace{0.1cm}}&\\
&&&&&&&&&&&\parbox[t][0.3cm]{8.004341cm}{\raggedright {\ }{\ }{\ }observable dependence on the target thickness\vspace{0.1cm}}&\\
&&&&&&&&&&&\parbox[t][0.3cm]{8.004341cm}{\raggedright {\ }{\ }{\ }(\href{https://www.nndc.bnl.gov/nsr/nsrlink.jsp?1970Bh02,B}{1970Bh02}).\vspace{0.1cm}}&\\
\multicolumn{1}{r@{}}{1233}&\multicolumn{1}{@{}l}{\ensuremath{^{\hyperlink{NE11GAMMA1}{b}}}}&\multicolumn{1}{r@{}}{1508}&\multicolumn{1}{@{}l}{}&\multicolumn{1}{l}{5/2\ensuremath{^{-}}}&\multicolumn{1}{r@{}}{275}&\multicolumn{1}{@{.}l}{30 }&\multicolumn{1}{@{}l}{1/2\ensuremath{^{-}}}&&&&&\\
\multicolumn{1}{r@{}}{1297}&\multicolumn{1}{@{.}l}{8\ensuremath{^{\hyperlink{NE11GAMMA2}{c}}}}&\multicolumn{1}{r@{}}{1536}&\multicolumn{1}{@{}l}{}&\multicolumn{1}{l}{3/2\ensuremath{^{+}}}&\multicolumn{1}{r@{}}{238}&\multicolumn{1}{@{.}l}{34 }&\multicolumn{1}{@{}l}{5/2\ensuremath{^{+}}}&&&&&\\
\multicolumn{1}{r@{}}{1839}&\multicolumn{1}{@{}l}{}&\multicolumn{1}{r@{}}{4634}&\multicolumn{1}{@{}l}{}&\multicolumn{1}{l}{13/2\ensuremath{^{+}}}&\multicolumn{1}{r@{}}{2794}&\multicolumn{1}{@{.}l}{4}&\multicolumn{1}{@{}l}{9/2\ensuremath{^{+}}}&&\multicolumn{1}{r@{}}{}&\multicolumn{1}{@{}l}{}&\parbox[t][0.3cm]{8.004341cm}{\raggedright E\ensuremath{_{\gamma}}: From Fig. 13 of (\href{https://www.nndc.bnl.gov/nsr/nsrlink.jsp?2008My01,B}{2008My01}).\vspace{0.1cm}}&\\
\multicolumn{1}{r@{}}{2498}&\multicolumn{1}{@{.}l}{5\ensuremath{^{\hyperlink{NE11GAMMA2}{c}}}}&\multicolumn{1}{r@{}}{4035}&\multicolumn{1}{@{}l}{}&\multicolumn{1}{l}{3/2\ensuremath{^{+}}}&\multicolumn{1}{r@{}}{1536}&\multicolumn{1}{@{}l}{}&\multicolumn{1}{@{}l}{3/2\ensuremath{^{+}}}&&&&&\\
\multicolumn{1}{r@{}}{2635}&\multicolumn{1}{@{.}l}{9\ensuremath{^{\hyperlink{NE11GAMMA2}{c}}}}&\multicolumn{1}{r@{}}{4144}&\multicolumn{1}{@{}l}{}&\multicolumn{1}{l}{(7/2\ensuremath{^{-}})}&\multicolumn{1}{r@{}}{1508}&\multicolumn{1}{@{}l}{}&\multicolumn{1}{@{}l}{5/2\ensuremath{^{-}}}&&&&&\\
\multicolumn{1}{r@{}}{2692}&\multicolumn{1}{@{.}l}{7\ensuremath{^{\hyperlink{NE11GAMMA2}{c}}}}&\multicolumn{1}{r@{}}{4200}&\multicolumn{1}{@{}l}{}&\multicolumn{1}{l}{(9/2\ensuremath{^{-}})}&\multicolumn{1}{r@{}}{1508}&\multicolumn{1}{@{}l}{}&\multicolumn{1}{@{}l}{5/2\ensuremath{^{-}}}&&&&&\\
\multicolumn{1}{r@{}}{4034}&\multicolumn{1}{@{.}l}{5\ensuremath{^{\hyperlink{NE11GAMMA2}{c}}}}&\multicolumn{1}{r@{}}{4035}&\multicolumn{1}{@{}l}{}&\multicolumn{1}{l}{3/2\ensuremath{^{+}}}&\multicolumn{1}{r@{}}{0}&\multicolumn{1}{@{}l}{}&\multicolumn{1}{@{}l}{1/2\ensuremath{^{+}}}&&&&\parbox[t][0.3cm]{8.004341cm}{\raggedright E\ensuremath{_{\gamma}}: This \ensuremath{\gamma} ray was also observed in (\href{https://www.nndc.bnl.gov/nsr/nsrlink.jsp?2006Ka50,B}{2006Ka50}), see Fig.\vspace{0.1cm}}&\\
&&&&&&&&&&&\parbox[t][0.3cm]{8.004341cm}{\raggedright {\ }{\ }{\ }7.\vspace{0.1cm}}&\\
\multicolumn{1}{r@{}}{4139}&\multicolumn{1}{@{.}l}{5\ensuremath{^{\hyperlink{NE11GAMMA2}{c}}}}&\multicolumn{1}{r@{}}{4378}&\multicolumn{1}{@{}l}{}&\multicolumn{1}{l}{7/2\ensuremath{^{+}}}&\multicolumn{1}{r@{}}{238}&\multicolumn{1}{@{.}l}{34 }&\multicolumn{1}{@{}l}{5/2\ensuremath{^{+}}}&&&&&\\
\multicolumn{1}{r@{}}{4272}&\multicolumn{1}{@{.}l}{6\ensuremath{^{\hyperlink{NE11GAMMA2}{c}}}}&\multicolumn{1}{r@{}}{4548}&\multicolumn{1}{@{}l}{}&\multicolumn{1}{l}{3/2\ensuremath{^{-}}}&\multicolumn{1}{r@{}}{275}&\multicolumn{1}{@{.}l}{30 }&\multicolumn{1}{@{}l}{1/2\ensuremath{^{-}}}&&&&&\\
\multicolumn{1}{r@{}}{4363}&\multicolumn{1}{@{.}l}{5\ensuremath{^{\hyperlink{NE11GAMMA2}{c}}}}&\multicolumn{1}{r@{}}{4602}&\multicolumn{1}{@{}l}{}&\multicolumn{1}{l}{5/2\ensuremath{^{+}}}&\multicolumn{1}{r@{}}{238}&\multicolumn{1}{@{.}l}{34 }&\multicolumn{1}{@{}l}{5/2\ensuremath{^{+}}}&&&&&\\
\multicolumn{1}{r@{}}{4547}&\multicolumn{1}{@{.}l}{7\ensuremath{^{\hyperlink{NE11GAMMA2}{c}}}}&\multicolumn{1}{r@{}}{4548}&\multicolumn{1}{@{}l}{}&\multicolumn{1}{l}{3/2\ensuremath{^{-}}}&\multicolumn{1}{r@{}}{0}&\multicolumn{1}{@{}l}{}&\multicolumn{1}{@{}l}{1/2\ensuremath{^{+}}}&&&&&\\
\end{longtable}
\parbox[b][0.3cm]{17.7cm}{\makebox[1ex]{\ensuremath{^{\hypertarget{NE11GAMMA0}{a}}}} \ensuremath{\gamma} ray observed in (\href{https://www.nndc.bnl.gov/nsr/nsrlink.jsp?1970Bh02,B}{1970Bh02}, \href{https://www.nndc.bnl.gov/nsr/nsrlink.jsp?2006Ka50,B}{2006Ka50}). These authors did not report E\ensuremath{_{\ensuremath{\gamma}}}. So, the \ensuremath{\gamma}-ray energy is from level-energy difference}\\
\parbox[b][0.3cm]{17.7cm}{{\ }{\ }corrected for recoil energy by the evaluator and based on the level-energies deduced by (\href{https://www.nndc.bnl.gov/nsr/nsrlink.jsp?1970Bh02,B}{1970Bh02}).}\\
\parbox[b][0.3cm]{17.7cm}{\makebox[1ex]{\ensuremath{^{\hypertarget{NE11GAMMA1}{b}}}} \ensuremath{\gamma} ray observed in (\href{https://www.nndc.bnl.gov/nsr/nsrlink.jsp?2006Ka50,B}{2006Ka50}), who did not report E\ensuremath{_{\ensuremath{\gamma}}}. The \ensuremath{\gamma}-ray energy is from level-energy difference, for the 1508 keV\ensuremath{\rightarrow}275}\\
\parbox[b][0.3cm]{17.7cm}{{\ }{\ }keV transition, corrected for recoil energy by the evaluator.}\\
\parbox[b][0.3cm]{17.7cm}{\makebox[1ex]{\ensuremath{^{\hypertarget{NE11GAMMA2}{c}}}} From (\href{https://www.nndc.bnl.gov/nsr/nsrlink.jsp?2005Ta28,B}{2005Ta28}: \ensuremath{^{\textnormal{17}}}O(\ensuremath{^{\textnormal{3}}}He,n\ensuremath{\gamma})): See Table I in (\href{https://www.nndc.bnl.gov/nsr/nsrlink.jsp?2008My01,B}{2008My01}).}\\
\parbox[b][0.3cm]{17.7cm}{\makebox[1ex]{\ensuremath{^{\hypertarget{NE11GAMMA3}{d}}}} Total theoretical internal conversion coefficients, calculated using the BrIcc code (\href{https://www.nndc.bnl.gov/nsr/nsrlink.jsp?2008Ki07,B}{2008Ki07}) with ``Frozen Orbitals''}\\
\parbox[b][0.3cm]{17.7cm}{{\ }{\ }approximation based on \ensuremath{\gamma}-ray energies, assigned multipolarities, and mixing ratios, unless otherwise specified.}\\
\vspace{0.5cm}
\clearpage
\begin{figure}[h]
\begin{center}
\includegraphics{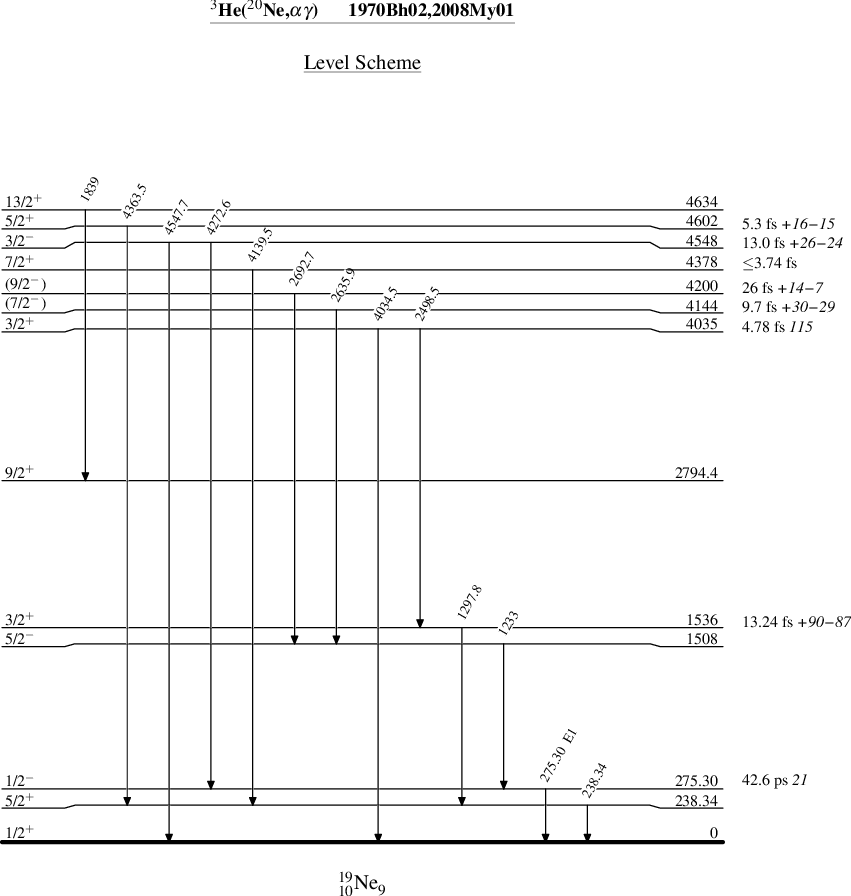}\\
\end{center}
\end{figure}
\clearpage
\subsection[\hspace{-0.2cm}\ensuremath{^{\textnormal{3}}}He(\ensuremath{^{\textnormal{20}}}Ne,\ensuremath{^{\textnormal{19}}}Ne),(\ensuremath{^{\textnormal{20}}}Ne,\ensuremath{\alpha})]{ }
\vspace{-27pt}
\vspace{0.3cm}
\hypertarget{NE12}{{\bf \small \underline{\ensuremath{^{\textnormal{3}}}He(\ensuremath{^{\textnormal{20}}}Ne,\ensuremath{^{\textnormal{19}}}Ne),(\ensuremath{^{\textnormal{20}}}Ne,\ensuremath{\alpha})\hspace{0.2in}\href{https://www.nndc.bnl.gov/nsr/nsrlink.jsp?2003Re16,B}{2003Re16},\href{https://www.nndc.bnl.gov/nsr/nsrlink.jsp?2003Re25,B}{2003Re25}}}}\\
\vspace{4pt}
\vspace{8pt}
\parbox[b][0.3cm]{17.7cm}{\addtolength{\parindent}{-0.2in}\ensuremath{^{\textnormal{20}}}Ne(\ensuremath{^{\textnormal{3}}}He,\ensuremath{\alpha}) reaction in inverse kinematics.}\\
\parbox[b][0.3cm]{17.7cm}{\addtolength{\parindent}{-0.2in}J\ensuremath{^{\ensuremath{\pi}}}(\ensuremath{^{\textnormal{3}}}He\ensuremath{_{\textnormal{g.s.}}})=1/2\ensuremath{^{\textnormal{+}}} and J\ensuremath{^{\ensuremath{\pi}}}(\ensuremath{^{\textnormal{20}}}Ne\ensuremath{_{\textnormal{g.s.}}})=0\ensuremath{^{\textnormal{+}}}.}\\
\parbox[b][0.3cm]{17.7cm}{\addtolength{\parindent}{-0.2in}\href{https://www.nndc.bnl.gov/nsr/nsrlink.jsp?2003Re16,B}{2003Re16}, \href{https://www.nndc.bnl.gov/nsr/nsrlink.jsp?2003Re25,B}{2003Re25}: \ensuremath{^{\textnormal{3}}}He(\ensuremath{^{\textnormal{20}}}Ne,\ensuremath{\alpha}), \ensuremath{^{\textnormal{3}}}He(\ensuremath{^{\textnormal{20}}}Ne,\ensuremath{^{\textnormal{19}}}Ne*\ensuremath{\rightarrow}\ensuremath{\alpha}+\ensuremath{^{\textnormal{15}}}O), \ensuremath{^{\textnormal{3}}}He(\ensuremath{^{\textnormal{20}}}Ne,\ensuremath{^{\textnormal{19}}}Ne*\ensuremath{\rightarrow}p+\ensuremath{^{\textnormal{18}}}F) E=98.3 MeV (at target center);}\\
\parbox[b][0.3cm]{17.7cm}{measured \ensuremath{\alpha}-particles from the reaction using a Si \ensuremath{\Delta}E-E telescope with a position sensitive \ensuremath{\Delta}E section at \ensuremath{\theta}\ensuremath{_{\textnormal{lab}}}=50\ensuremath{^\circ} covering}\\
\parbox[b][0.3cm]{17.7cm}{\ensuremath{\theta}\ensuremath{_{\textnormal{c.m.}}}=6.8\ensuremath{^\circ}{\textminus}17.2\ensuremath{^\circ}; momentum analyzed \ensuremath{^{\textnormal{19}}}Ne* recoils and their in-flight heavy decay residues, \ensuremath{^{\textnormal{15}}}O and \ensuremath{^{\textnormal{18}}}F ions, using an Enge}\\
\parbox[b][0.3cm]{17.7cm}{spectrograph placed at \ensuremath{\theta}\ensuremath{_{\textnormal{lab}}}=3.7\ensuremath{^\circ}; measured (in coincidence with the \ensuremath{\alpha}-particles) position, energy, B\ensuremath{\rho}, and timing of these heavy}\\
\parbox[b][0.3cm]{17.7cm}{ions using a focal plane detection system. Deduced \ensuremath{^{\textnormal{19}}}Ne excitation energy spectra, \ensuremath{^{\textnormal{19}}}Ne levels, and \ensuremath{\alpha}-decay branching ratios.}\\
\vspace{12pt}
\underline{$^{19}$Ne Levels}\\
\vspace{0.34cm}
\parbox[b][0.3cm]{17.7cm}{\addtolength{\parindent}{-0.254cm}\textit{Notes}:}\\
\parbox[b][0.3cm]{17.7cm}{\addtolength{\parindent}{-0.254cm}(1) The reported \ensuremath{\Gamma}\ensuremath{_{\ensuremath{\gamma}}} are either from the decay of the mirror level in \ensuremath{^{\textnormal{19}}}F or from theoretical calculations suggested by (\href{https://www.nndc.bnl.gov/nsr/nsrlink.jsp?2003Da03,B}{2003Da03})}\\
\parbox[b][0.3cm]{17.7cm}{and references therein.}\\
\parbox[b][0.3cm]{17.7cm}{\addtolength{\parindent}{-0.254cm}(2) (\href{https://www.nndc.bnl.gov/nsr/nsrlink.jsp?2003Re16,B}{2003Re16}): Observed an increase in \ensuremath{\Gamma}\ensuremath{_{\ensuremath{\gamma}}}/\ensuremath{\Gamma} at E\ensuremath{_{\textnormal{x}}}(\ensuremath{^{\textnormal{19}}}Ne)=6-7 MeV, indicating that there are states in this excitation energy}\\
\parbox[b][0.3cm]{17.7cm}{region, which decay with 10\% probability via \ensuremath{\gamma} emission. Those authors attributed these \ensuremath{\gamma} decays to high-spin states with J\ensuremath{^{\ensuremath{\pi}}}=7/2\ensuremath{^{-}},}\\
\parbox[b][0.3cm]{17.7cm}{9/2\ensuremath{^{\textnormal{+}}}, and 11/2\ensuremath{^{-}} based on an analogy with similar levels observed in this energy range in the mirror nucleus, \ensuremath{^{\textnormal{19}}}F, with \ensuremath{\Gamma}\ensuremath{_{\ensuremath{\gamma}}}/\ensuremath{\Gamma}}\\
\parbox[b][0.3cm]{17.7cm}{branching ratios between 2 and 20\% (\href{https://www.nndc.bnl.gov/nsr/nsrlink.jsp?1986Ca19,B}{1986Ca19}).}\\
\parbox[b][0.3cm]{17.7cm}{\addtolength{\parindent}{-0.254cm}(3) \ensuremath{\omega}\ensuremath{\gamma}=(2J\ensuremath{_{\textnormal{R}}}+1)\ensuremath{\Gamma}\ensuremath{_{\ensuremath{\gamma}}}B\ensuremath{_{\ensuremath{\alpha}}}/[(2J\ensuremath{_{\textnormal{p}}}+1)(2J\ensuremath{_{\textnormal{t}}}+1)(1+B\ensuremath{_{\ensuremath{\alpha}}})], where J\ensuremath{_{\textnormal{R}}}, J\ensuremath{_{\textnormal{p}}}, and J\ensuremath{_{\textnormal{t}}} are the spins of the final \ensuremath{^{\textnormal{19}}}Ne resonance, projectile, and}\\
\parbox[b][0.3cm]{17.7cm}{target, respectively; and B\ensuremath{_{\ensuremath{\alpha}}}=\ensuremath{\Gamma}\ensuremath{_{\ensuremath{\alpha}}}/\ensuremath{\Gamma} (\href{https://www.nndc.bnl.gov/nsr/nsrlink.jsp?2003Re25,B}{2003Re25}).}\\
\vspace{0.34cm}
\begin{longtable}{cccccccc@{\extracolsep{\fill}}c}
\multicolumn{2}{c}{E(level)$^{{\hyperlink{NE12LEVEL0}{a}}}$}&J$^{\pi}$$^{{\hyperlink{NE12LEVEL1}{b}}}$&\multicolumn{2}{c}{E\ensuremath{_{\textnormal{c.m.}}}(\ensuremath{^{\textnormal{15}}}O+\ensuremath{\alpha}) (keV)$^{{\hyperlink{NE12LEVEL2}{c}}}$}&\multicolumn{2}{c}{\ensuremath{\Gamma}\ensuremath{\alpha}/\ensuremath{\Gamma}$^{{\hyperlink{NE12LEVEL2}{c}}}$}&Comments&\\[-.2cm]
\multicolumn{2}{c}{\hrulefill}&\hrulefill&\multicolumn{2}{c}{\hrulefill}&\multicolumn{2}{c}{\hrulefill}&\hrulefill&
\endfirsthead
\multicolumn{1}{r@{}}{4032}&\multicolumn{1}{@{}l}{}&\multicolumn{1}{l}{3/2\ensuremath{^{+}}}&\multicolumn{1}{r@{}}{504}&\multicolumn{1}{@{ }l}{keV}&\multicolumn{1}{r@{}}{$<$6}&\multicolumn{1}{@{}l}{\ensuremath{\times10^{-4}}}&\parbox[t][0.3cm]{8.336079cm}{\raggedright E(level): Reported by (\href{https://www.nndc.bnl.gov/nsr/nsrlink.jsp?2003Re16,B}{2003Re16}, \href{https://www.nndc.bnl.gov/nsr/nsrlink.jsp?2003Re25,B}{2003Re25}).\vspace{0.1cm}}&\\
&&&&&&&\parbox[t][0.3cm]{8.336079cm}{\raggedright \ensuremath{\Gamma}\ensuremath{_{\ensuremath{\alpha}}}/\ensuremath{\Gamma}\ensuremath{\leq}6\ensuremath{\times}10\ensuremath{^{\textnormal{$-$4}}} was mentioned in the abstract and\vspace{0.1cm}}&\\
&&&&&&&\parbox[t][0.3cm]{8.336079cm}{\raggedright {\ }{\ }{\ }conclusion sections of (\href{https://www.nndc.bnl.gov/nsr/nsrlink.jsp?2003Re16,B}{2003Re16}). Table II of this study\vspace{0.1cm}}&\\
&&&&&&&\parbox[t][0.3cm]{8.336079cm}{\raggedright {\ }{\ }{\ }reports this value as \ensuremath{\Gamma}\ensuremath{_{\ensuremath{\alpha}}}/\ensuremath{\Gamma}\ensuremath{<}6\ensuremath{\times}10\ensuremath{^{\textnormal{$-$4}}}. Note that (\href{https://www.nndc.bnl.gov/nsr/nsrlink.jsp?2003Re25,B}{2003Re25})\vspace{0.1cm}}&\\
&&&&&&&\parbox[t][0.3cm]{8.336079cm}{\raggedright {\ }{\ }{\ }reports \ensuremath{\Gamma}\ensuremath{_{\ensuremath{\alpha}}}/\ensuremath{\Gamma}\ensuremath{\leq}7\ensuremath{\times}10\ensuremath{^{\textnormal{$-$4}}}. This value is also mentioned in\vspace{0.1cm}}&\\
&&&&&&&\parbox[t][0.3cm]{8.336079cm}{\raggedright {\ }{\ }{\ }(\href{https://www.nndc.bnl.gov/nsr/nsrlink.jsp?2003Re16,B}{2003Re16}), where the authors report that they obtained an\vspace{0.1cm}}&\\
&&&&&&&\parbox[t][0.3cm]{8.336079cm}{\raggedright {\ }{\ }{\ }upper limit of 7\ensuremath{\times}10\ensuremath{^{\textnormal{$-$4}}} for \ensuremath{\Gamma}\ensuremath{_{\ensuremath{\alpha}}}/\ensuremath{\Gamma} at 90\% C.L. However,\vspace{0.1cm}}&\\
&&&&&&&\parbox[t][0.3cm]{8.336079cm}{\raggedright {\ }{\ }{\ }this value was updated to 6\ensuremath{\times}10\ensuremath{^{\textnormal{$-$4}}} when they replaced the\vspace{0.1cm}}&\\
&&&&&&&\parbox[t][0.3cm]{8.336079cm}{\raggedright {\ }{\ }{\ }\ensuremath{\Delta}E-E telescope with an annular position sensitive Si\vspace{0.1cm}}&\\
&&&&&&&\parbox[t][0.3cm]{8.336079cm}{\raggedright {\ }{\ }{\ }detector during a test run.\vspace{0.1cm}}&\\
&&&&&&&\parbox[t][0.3cm]{8.336079cm}{\raggedright No \ensuremath{^{\textnormal{15}}}O+\ensuremath{\alpha} events were observed for this state during a\vspace{0.1cm}}&\\
&&&&&&&\parbox[t][0.3cm]{8.336079cm}{\raggedright {\ }{\ }{\ }3.5-day-long experiment (\href{https://www.nndc.bnl.gov/nsr/nsrlink.jsp?2003Re16,B}{2003Re16}).\vspace{0.1cm}}&\\
&&&&&&&\parbox[t][0.3cm]{8.336079cm}{\raggedright (\ensuremath{\omega}\ensuremath{\gamma})\ensuremath{_{\textnormal{(}\ensuremath{\alpha}\textnormal{,}\ensuremath{\gamma}\textnormal{)}}}=25 \ensuremath{\mu}eV deduced by (\href{https://www.nndc.bnl.gov/nsr/nsrlink.jsp?2003Re16,B}{2003Re16}). This value is\vspace{0.1cm}}&\\
&&&&&&&\parbox[t][0.3cm]{8.336079cm}{\raggedright {\ }{\ }{\ }obtained using \ensuremath{\Gamma}\ensuremath{_{\ensuremath{\gamma}}}=21 meV from (\href{https://www.nndc.bnl.gov/nsr/nsrlink.jsp?2000Ha26,B}{2000Ha26}): \ensuremath{\Gamma}\ensuremath{_{\ensuremath{\gamma}}}=12\vspace{0.1cm}}&\\
&&&&&&&\parbox[t][0.3cm]{8.336079cm}{\raggedright {\ }{\ }{\ }meV \textit{+9{\textminus}5}, see the \ensuremath{^{\textnormal{197}}}Au(\ensuremath{^{\textnormal{19}}}Ne, \ensuremath{^{\textnormal{19}}}Ne\ensuremath{'}) dataset, note that\vspace{0.1cm}}&\\
&&&&&&&\parbox[t][0.3cm]{8.336079cm}{\raggedright {\ }{\ }{\ }21 meV is obtained from adding 12 meV and +9 meV\vspace{0.1cm}}&\\
&&&&&&&\parbox[t][0.3cm]{8.336079cm}{\raggedright {\ }{\ }{\ }from the upper uncertainty band.\vspace{0.1cm}}&\\
\multicolumn{1}{r@{}}{4140}&\multicolumn{1}{@{}l}{}&\multicolumn{1}{l}{(7/2\ensuremath{^{-}})}&\multicolumn{1}{r@{}}{611}&\multicolumn{1}{@{ }l}{keV}&&&\parbox[t][0.3cm]{8.336079cm}{\raggedright E(level): Reported by (\href{https://www.nndc.bnl.gov/nsr/nsrlink.jsp?2003Re16,B}{2003Re16}, \href{https://www.nndc.bnl.gov/nsr/nsrlink.jsp?2003Re25,B}{2003Re25}).\vspace{0.1cm}}&\\
&&&&&&&\parbox[t][0.3cm]{8.336079cm}{\raggedright (\ensuremath{\omega}\ensuremath{\gamma})\ensuremath{_{\textnormal{(}\ensuremath{\alpha}\textnormal{,}\ensuremath{\gamma}\textnormal{)}}}=0.57 \ensuremath{\mu}eV deduced by (\href{https://www.nndc.bnl.gov/nsr/nsrlink.jsp?2003Re16,B}{2003Re16}) assuming\vspace{0.1cm}}&\\
&&&&&&&\parbox[t][0.3cm]{8.336079cm}{\raggedright {\ }{\ }{\ }C\ensuremath{^{\textnormal{2}}}S\ensuremath{_{\ensuremath{\alpha}}}=0.1.\vspace{0.1cm}}&\\
\multicolumn{1}{r@{}}{4196}&\multicolumn{1}{@{}l}{}&\multicolumn{1}{l}{(9/2\ensuremath{^{-}})}&\multicolumn{1}{r@{}}{668}&\multicolumn{1}{@{ }l}{keV}&&&\parbox[t][0.3cm]{8.336079cm}{\raggedright E(level): Reported by (\href{https://www.nndc.bnl.gov/nsr/nsrlink.jsp?2003Re16,B}{2003Re16}, \href{https://www.nndc.bnl.gov/nsr/nsrlink.jsp?2003Re25,B}{2003Re25}).\vspace{0.1cm}}&\\
&&&&&&&\parbox[t][0.3cm]{8.336079cm}{\raggedright (\ensuremath{\omega}\ensuremath{\gamma})\ensuremath{_{\textnormal{(}\ensuremath{\alpha}\textnormal{,}\ensuremath{\gamma}\textnormal{)}}}=8.9 \ensuremath{\mu}eV deduced by (\href{https://www.nndc.bnl.gov/nsr/nsrlink.jsp?2003Re16,B}{2003Re16}) assuming\vspace{0.1cm}}&\\
&&&&&&&\parbox[t][0.3cm]{8.336079cm}{\raggedright {\ }{\ }{\ }C\ensuremath{^{\textnormal{2}}}S\ensuremath{_{\ensuremath{\alpha}}}=0.1.\vspace{0.1cm}}&\\
\multicolumn{1}{r@{}}{4378}&\multicolumn{1}{@{}l}{}&\multicolumn{1}{l}{7/2\ensuremath{^{+}}}&\multicolumn{1}{r@{}}{850}&\multicolumn{1}{@{ }l}{keV}&\multicolumn{1}{r@{}}{0}&\multicolumn{1}{@{.}l}{016 {\it 5}}&\parbox[t][0.3cm]{8.336079cm}{\raggedright E(level): Reported by (\href{https://www.nndc.bnl.gov/nsr/nsrlink.jsp?2003Re16,B}{2003Re16}, \href{https://www.nndc.bnl.gov/nsr/nsrlink.jsp?2003Re25,B}{2003Re25}).\vspace{0.1cm}}&\\
&&&&&&&\parbox[t][0.3cm]{8.336079cm}{\raggedright \ensuremath{\Gamma}\ensuremath{\alpha}/\ensuremath{\Gamma}: The validity of this value may be\vspace{0.1cm}}&\\
&&&&&&&\parbox[t][0.3cm]{8.336079cm}{\raggedright {\ }{\ }{\ }questionable because the energy resolution of the\vspace{0.1cm}}&\\
&&&&&&&\parbox[t][0.3cm]{8.336079cm}{\raggedright {\ }{\ }{\ }(\href{https://www.nndc.bnl.gov/nsr/nsrlink.jsp?2003Re16,B}{2003Re16}) measurement was \ensuremath{\sim}220 keV, which makes this\vspace{0.1cm}}&\\
&&&&&&&\parbox[t][0.3cm]{8.336079cm}{\raggedright {\ }{\ }{\ }state unresolved from the 4548-keV state. (\href{https://www.nndc.bnl.gov/nsr/nsrlink.jsp?2007TaZX,B}{2007TaZX})\vspace{0.1cm}}&\\
&&&&&&&\parbox[t][0.3cm]{8.336079cm}{\raggedright {\ }{\ }{\ }also mentions that (\href{https://www.nndc.bnl.gov/nsr/nsrlink.jsp?2003Re16,B}{2003Re16}) could not resolve the states\vspace{0.1cm}}&\\
&&&&&&&\parbox[t][0.3cm]{8.336079cm}{\raggedright {\ }{\ }{\ }in this energy region in \ensuremath{^{\textnormal{19}}}Ne, and thus any contribution\vspace{0.1cm}}&\\
&&&&&&&\parbox[t][0.3cm]{8.336079cm}{\raggedright {\ }{\ }{\ }from higher lying states with large \ensuremath{\Gamma}\ensuremath{_{\ensuremath{\alpha}}}/\ensuremath{\Gamma} would\vspace{0.1cm}}&\\
&&&&&&&\parbox[t][0.3cm]{8.336079cm}{\raggedright {\ }{\ }{\ }contaminate the possible decay from this state measured by\vspace{0.1cm}}&\\
&&&&&&&\parbox[t][0.3cm]{8.336079cm}{\raggedright {\ }{\ }{\ }(\href{https://www.nndc.bnl.gov/nsr/nsrlink.jsp?2003Re16,B}{2003Re16}).\vspace{0.1cm}}&\\
\end{longtable}
\begin{textblock}{29}(0,27.3)
Continued on next page (footnotes at end of table)
\end{textblock}
\clearpage
\begin{longtable}{cccccccc@{\extracolsep{\fill}}c}
\\[-.4cm]
\multicolumn{9}{c}{{\bf \small \underline{\ensuremath{^{\textnormal{3}}}He(\ensuremath{^{\textnormal{20}}}Ne,\ensuremath{^{\textnormal{19}}}Ne),(\ensuremath{^{\textnormal{20}}}Ne,\ensuremath{\alpha})\hspace{0.2in}\href{https://www.nndc.bnl.gov/nsr/nsrlink.jsp?2003Re16,B}{2003Re16},\href{https://www.nndc.bnl.gov/nsr/nsrlink.jsp?2003Re25,B}{2003Re25} (continued)}}}\\
\multicolumn{9}{c}{~}\\
\multicolumn{9}{c}{\underline{\ensuremath{^{19}}Ne Levels (continued)}}\\
\multicolumn{9}{c}{~}\\
\multicolumn{2}{c}{E(level)$^{{\hyperlink{NE12LEVEL0}{a}}}$}&J$^{\pi}$$^{{\hyperlink{NE12LEVEL1}{b}}}$&\multicolumn{2}{c}{E\ensuremath{_{\textnormal{c.m.}}}(\ensuremath{^{\textnormal{15}}}O+\ensuremath{\alpha}) (keV)$^{{\hyperlink{NE12LEVEL2}{c}}}$}&\multicolumn{2}{c}{\ensuremath{\Gamma}\ensuremath{\alpha}/\ensuremath{\Gamma}$^{{\hyperlink{NE12LEVEL2}{c}}}$}&Comments&\\[-.2cm]
\multicolumn{2}{c}{\hrulefill}&\hrulefill&\multicolumn{2}{c}{\hrulefill}&\multicolumn{2}{c}{\hrulefill}&\hrulefill&
\endhead
&&&&&&&\parbox[t][0.3cm]{8.507279cm}{\raggedright (\ensuremath{\omega}\ensuremath{\gamma})\ensuremath{_{\textnormal{(}\ensuremath{\alpha}\textnormal{,}\ensuremath{\gamma}\textnormal{)}}}=29 meV \textit{11} deduced by (\href{https://www.nndc.bnl.gov/nsr/nsrlink.jsp?2003Re16,B}{2003Re16}) using\vspace{0.1cm}}&\\
&&&&&&&\parbox[t][0.3cm]{8.507279cm}{\raggedright {\ }{\ }{\ }\ensuremath{\Gamma}\ensuremath{_{\ensuremath{\alpha}}}/\ensuremath{\Gamma}\ensuremath{_{\ensuremath{\gamma}}}\ensuremath{\leq}1.5\ensuremath{\times}10\ensuremath{^{\textnormal{$-$3}}} (\href{https://www.nndc.bnl.gov/nsr/nsrlink.jsp?2003Re25,B}{2003Re25}) and \ensuremath{\Gamma}\ensuremath{_{\ensuremath{\gamma}}}=458 meV \textit{92},\vspace{0.1cm}}&\\
&&&&&&&\parbox[t][0.3cm]{8.507279cm}{\raggedright {\ }{\ }{\ }which was reported by (\href{https://www.nndc.bnl.gov/nsr/nsrlink.jsp?2003Da03,B}{2003Da03}). The latter value is from\vspace{0.1cm}}&\\
&&&&&&&\parbox[t][0.3cm]{8.507279cm}{\raggedright {\ }{\ }{\ }shell model calculations by B. A. Brown and is obtained via\vspace{0.1cm}}&\\
&&&&&&&\parbox[t][0.3cm]{8.507279cm}{\raggedright {\ }{\ }{\ }a priv. comm. with (\href{https://www.nndc.bnl.gov/nsr/nsrlink.jsp?2003Da03,B}{2003Da03}). A 1\ensuremath{\sigma} uncertainty of 20\%\vspace{0.1cm}}&\\
&&&&&&&\parbox[t][0.3cm]{8.507279cm}{\raggedright {\ }{\ }{\ }was assigned to the calculated \ensuremath{\Gamma}\ensuremath{_{\ensuremath{\gamma}}} width by (\href{https://www.nndc.bnl.gov/nsr/nsrlink.jsp?2003Da03,B}{2003Da03},\vspace{0.1cm}}&\\
&&&&&&&\parbox[t][0.3cm]{8.507279cm}{\raggedright {\ }{\ }{\ }\href{https://www.nndc.bnl.gov/nsr/nsrlink.jsp?2003Da13,B}{2003Da13}).\vspace{0.1cm}}&\\
\multicolumn{1}{r@{}}{4548}&\multicolumn{1}{@{}l}{}&\multicolumn{1}{l}{3/2\ensuremath{^{-}}}&\multicolumn{1}{r@{}}{1020}&\multicolumn{1}{@{ }l}{keV}&&&\parbox[t][0.3cm]{8.507279cm}{\raggedright E(level): Reported by (\href{https://www.nndc.bnl.gov/nsr/nsrlink.jsp?2003Re16,B}{2003Re16}).\vspace{0.1cm}}&\\
&&&&&&&\parbox[t][0.3cm]{8.507279cm}{\raggedright (\ensuremath{\omega}\ensuremath{\gamma})\ensuremath{_{\textnormal{(}\ensuremath{\alpha}\textnormal{,}\ensuremath{\gamma}\textnormal{)}}}=4.0 meV \textit{+35{\textminus}15} deduced by (\href{https://www.nndc.bnl.gov/nsr/nsrlink.jsp?2003Re16,B}{2003Re16}) using\vspace{0.1cm}}&\\
&&&&&&&\parbox[t][0.3cm]{8.507279cm}{\raggedright {\ }{\ }{\ }\ensuremath{\Gamma}\ensuremath{_{\ensuremath{\gamma}}}=39 meV \textit{+34{\textminus}15} (\href{https://www.nndc.bnl.gov/nsr/nsrlink.jsp?1990Ma05,B}{1990Ma05}). This value comes from\vspace{0.1cm}}&\\
&&&&&&&\parbox[t][0.3cm]{8.507279cm}{\raggedright {\ }{\ }{\ }the \ensuremath{^{\textnormal{19}}}F*(4556.1) mirror state reported by (\href{https://www.nndc.bnl.gov/nsr/nsrlink.jsp?1990Ma05,B}{1990Ma05}: See\vspace{0.1cm}}&\\
&&&&&&&\parbox[t][0.3cm]{8.507279cm}{\raggedright {\ }{\ }{\ }\ensuremath{^{\textnormal{19}}}F(\ensuremath{^{\textnormal{3}}}He,t) dataset). We highlight that Table II in\vspace{0.1cm}}&\\
&&&&&&&\parbox[t][0.3cm]{8.507279cm}{\raggedright {\ }{\ }{\ }(\href{https://www.nndc.bnl.gov/nsr/nsrlink.jsp?2003Re16,B}{2003Re16}) mentions that \ensuremath{\Gamma}\ensuremath{_{\ensuremath{\gamma}}} value but it has a misprint:\vspace{0.1cm}}&\\
&&&&&&&\parbox[t][0.3cm]{8.507279cm}{\raggedright {\ }{\ }{\ }\ensuremath{\Gamma}\ensuremath{_{\ensuremath{\gamma}}}=39.0 meV \textit{+340{\textminus}15} is reported instead of \ensuremath{\Gamma}\ensuremath{_{\ensuremath{\gamma}}}=39 meV\vspace{0.1cm}}&\\
&&&&&&&\parbox[t][0.3cm]{8.507279cm}{\raggedright {\ }{\ }{\ }\textit{+34{\textminus}15} (see also Table I in (\href{https://www.nndc.bnl.gov/nsr/nsrlink.jsp?2003Da03,B}{2003Da03}) and Table I in\vspace{0.1cm}}&\\
&&&&&&&\parbox[t][0.3cm]{8.507279cm}{\raggedright {\ }{\ }{\ }(\href{https://www.nndc.bnl.gov/nsr/nsrlink.jsp?2003Da13,B}{2003Da13}) that report the correct value).\vspace{0.1cm}}&\\
\multicolumn{1}{r@{}}{4600}&\multicolumn{1}{@{}l}{}&\multicolumn{1}{l}{5/2\ensuremath{^{+}}}&\multicolumn{1}{r@{}}{1071}&\multicolumn{1}{@{ }l}{keV}&&&\parbox[t][0.3cm]{8.507279cm}{\raggedright E(level): Reported by (\href{https://www.nndc.bnl.gov/nsr/nsrlink.jsp?2003Re16,B}{2003Re16}).\vspace{0.1cm}}&\\
&&&&&&&\parbox[t][0.3cm]{8.507279cm}{\raggedright (\ensuremath{\omega}\ensuremath{\gamma})\ensuremath{_{\textnormal{(}\ensuremath{\alpha}\textnormal{,}\ensuremath{\gamma}\textnormal{)}}}=90 meV \textit{50} obtained by (\href{https://www.nndc.bnl.gov/nsr/nsrlink.jsp?2003Re16,B}{2003Re16}) using\vspace{0.1cm}}&\\
&&&&&&&\parbox[t][0.3cm]{8.507279cm}{\raggedright {\ }{\ }{\ }\ensuremath{\Gamma}\ensuremath{_{\ensuremath{\gamma}}}=101 meV \textit{55} (\href{https://www.nndc.bnl.gov/nsr/nsrlink.jsp?1982Ki10,B}{1982Ki10}), who deduced \ensuremath{\tau}=6.5 fs \textit{35} for\vspace{0.1cm}}&\\
&&&&&&&\parbox[t][0.3cm]{8.507279cm}{\raggedright {\ }{\ }{\ }the \ensuremath{^{\textnormal{19}}}F*(4550) analog state using \ensuremath{^{\textnormal{15}}}N(\ensuremath{\alpha},\ensuremath{\gamma}).\vspace{0.1cm}}&\\
\multicolumn{1}{r@{}}{4712}&\multicolumn{1}{@{}l}{}&\multicolumn{1}{l}{5/2\ensuremath{^{-}}}&\multicolumn{1}{r@{}}{1183}&\multicolumn{1}{@{ }l}{keV}&&&\parbox[t][0.3cm]{8.507279cm}{\raggedright E(level): Reported by (\href{https://www.nndc.bnl.gov/nsr/nsrlink.jsp?2003Re16,B}{2003Re16}).\vspace{0.1cm}}&\\
&&&&&&&\parbox[t][0.3cm]{8.507279cm}{\raggedright (\ensuremath{\omega}\ensuremath{\gamma})\ensuremath{_{\textnormal{(}\ensuremath{\alpha}\textnormal{,}\ensuremath{\gamma}\textnormal{)}}}=110 meV \textit{21} determined by (\href{https://www.nndc.bnl.gov/nsr/nsrlink.jsp?2003Re16,B}{2003Re16}) using\vspace{0.1cm}}&\\
&&&&&&&\parbox[t][0.3cm]{8.507279cm}{\raggedright {\ }{\ }{\ }\ensuremath{\Gamma}\ensuremath{_{\ensuremath{\gamma}}}=43 meV \textit{8} (\href{https://www.nndc.bnl.gov/nsr/nsrlink.jsp?1990Ma05,B}{1990Ma05}), which was deduced for the\vspace{0.1cm}}&\\
&&&&&&&\parbox[t][0.3cm]{8.507279cm}{\raggedright {\ }{\ }{\ }\ensuremath{^{\textnormal{19}}}F*(4682.5) mirror state.\vspace{0.1cm}}&\\
\multicolumn{1}{r@{}}{5092}&\multicolumn{1}{@{}l}{}&\multicolumn{1}{l}{5/2\ensuremath{^{+}}}&\multicolumn{1}{r@{}}{1563}&\multicolumn{1}{@{ }l}{keV}&\multicolumn{1}{r@{}}{0}&\multicolumn{1}{@{.}l}{8 {\it 1}}&\parbox[t][0.3cm]{8.507279cm}{\raggedright E(level): Reported by (\href{https://www.nndc.bnl.gov/nsr/nsrlink.jsp?2003Re16,B}{2003Re16}).\vspace{0.1cm}}&\\
&&&&&&&\parbox[t][0.3cm]{8.507279cm}{\raggedright (\ensuremath{\omega}\ensuremath{\gamma})\ensuremath{_{\textnormal{(}\ensuremath{\alpha}\textnormal{,}\ensuremath{\gamma}\textnormal{)}}}=530 meV \textit{110} determined by (\href{https://www.nndc.bnl.gov/nsr/nsrlink.jsp?2003Re16,B}{2003Re16}) using\vspace{0.1cm}}&\\
&&&&&&&\parbox[t][0.3cm]{8.507279cm}{\raggedright {\ }{\ }{\ }\ensuremath{\Gamma}\ensuremath{_{\ensuremath{\gamma}}}=196 meV \textit{39} (\href{https://www.nndc.bnl.gov/nsr/nsrlink.jsp?2003Da03,B}{2003Da03}). This \ensuremath{\Gamma}\ensuremath{_{\ensuremath{\gamma}}} is from shell\vspace{0.1cm}}&\\
&&&&&&&\parbox[t][0.3cm]{8.507279cm}{\raggedright {\ }{\ }{\ }model calculations by B. A. Brown obtained via a priv.\vspace{0.1cm}}&\\
&&&&&&&\parbox[t][0.3cm]{8.507279cm}{\raggedright {\ }{\ }{\ }comm. with (\href{https://www.nndc.bnl.gov/nsr/nsrlink.jsp?2003Da03,B}{2003Da03}). A 1\ensuremath{\sigma} uncertainty of 20\% was\vspace{0.1cm}}&\\
&&&&&&&\parbox[t][0.3cm]{8.507279cm}{\raggedright {\ }{\ }{\ }assigned to the calculated \ensuremath{\Gamma}\ensuremath{_{\ensuremath{\gamma}}} width by (\href{https://www.nndc.bnl.gov/nsr/nsrlink.jsp?2003Da03,B}{2003Da03},\vspace{0.1cm}}&\\
&&&&&&&\parbox[t][0.3cm]{8.507279cm}{\raggedright {\ }{\ }{\ }\href{https://www.nndc.bnl.gov/nsr/nsrlink.jsp?2003Da13,B}{2003Da13}).\vspace{0.1cm}}&\\
\end{longtable}
\parbox[b][0.3cm]{17.7cm}{\makebox[1ex]{\ensuremath{^{\hypertarget{NE12LEVEL0}{a}}}} Level-energies are deduced from the center-of-mass \ensuremath{^{\textnormal{15}}}O+\ensuremath{\alpha} resonance energies from (\href{https://www.nndc.bnl.gov/nsr/nsrlink.jsp?2003Re16,B}{2003Re16}): E\ensuremath{_{\textnormal{x}}}=E\ensuremath{_{\ensuremath{\alpha}}}(c.m.)+Q{\textminus}value, where}\\
\parbox[b][0.3cm]{17.7cm}{{\ }{\ }Q=3528.5 keV is obtained using the \ensuremath{^{\textnormal{15}}}O, \ensuremath{^{\textnormal{4}}}He, and \ensuremath{^{\textnormal{19}}}Ne mass excesses from (\href{https://www.nndc.bnl.gov/nsr/nsrlink.jsp?2021Wa16,B}{2021Wa16}: AME-2020). The excitation energies}\\
\parbox[b][0.3cm]{17.7cm}{{\ }{\ }are rounded to the nearest integer values.}\\
\parbox[b][0.3cm]{17.7cm}{\makebox[1ex]{\ensuremath{^{\hypertarget{NE12LEVEL1}{b}}}} From the Adopted Levels of \ensuremath{^{\textnormal{19}}}Ne.}\\
\parbox[b][0.3cm]{17.7cm}{\makebox[1ex]{\ensuremath{^{\hypertarget{NE12LEVEL2}{c}}}} From (\href{https://www.nndc.bnl.gov/nsr/nsrlink.jsp?2003Re16,B}{2003Re16}).}\\
\vspace{0.5cm}
\clearpage
\subsection[\hspace{-0.2cm}\ensuremath{^{\textnormal{4}}}He(\ensuremath{^{\textnormal{15}}}O,\ensuremath{\alpha}):res]{ }
\vspace{-27pt}
\vspace{0.3cm}
\hypertarget{NE13}{{\bf \small \underline{\ensuremath{^{\textnormal{4}}}He(\ensuremath{^{\textnormal{15}}}O,\ensuremath{\alpha}):res\hspace{0.2in}\href{https://www.nndc.bnl.gov/nsr/nsrlink.jsp?2017To14,B}{2017To14}}}}\\
\vspace{4pt}
\vspace{8pt}
\parbox[b][0.3cm]{17.7cm}{\addtolength{\parindent}{-0.2in}\ensuremath{^{\textnormal{15}}}O(\ensuremath{\alpha},\ensuremath{\alpha}) resonant inelastic scattering in inverse kinematics.}\\
\parbox[b][0.3cm]{17.7cm}{\addtolength{\parindent}{-0.2in}J\ensuremath{^{\ensuremath{\pi}}}(\ensuremath{^{\textnormal{4}}}He\ensuremath{_{\textnormal{g.s.}}})=0\ensuremath{^{\textnormal{+}}} and J\ensuremath{^{\ensuremath{\pi}}}(\ensuremath{^{\textnormal{15}}}O\ensuremath{_{\textnormal{g.s.}}})=1/2\ensuremath{^{-}}.}\\
\parbox[b][0.3cm]{17.7cm}{\addtolength{\parindent}{-0.2in}\href{https://www.nndc.bnl.gov/nsr/nsrlink.jsp?2006Va06,B}{2006Va06}: \ensuremath{^{\textnormal{4}}}He(\ensuremath{^{\textnormal{15}}}O,\ensuremath{\alpha}) E=12.5 MeV; measured energies and TOF for elastically scattered \ensuremath{\alpha} particles using a position sensitive}\\
\parbox[b][0.3cm]{17.7cm}{annular Si detector covering \ensuremath{\theta}\ensuremath{_{\textnormal{lab}}}=12\ensuremath{^\circ}{\textminus}29\ensuremath{^\circ} corresponding to \ensuremath{\theta}\ensuremath{_{\textnormal{c.m.}}}=122\ensuremath{^\circ}{\textminus}156\ensuremath{^\circ}. Deduced the \ensuremath{\alpha}-partial width for the \ensuremath{^{\textnormal{19}}}Ne*(5351)}\\
\parbox[b][0.3cm]{17.7cm}{level using two independent analysis methods.}\\
\parbox[b][0.3cm]{17.7cm}{\addtolength{\parindent}{-0.2in}\href{https://www.nndc.bnl.gov/nsr/nsrlink.jsp?2017To14,B}{2017To14}: \ensuremath{^{\textnormal{4}}}He(\ensuremath{^{\textnormal{15}}}O,\ensuremath{\alpha}) E=28.5 MeV; measured energy and TOF of \ensuremath{\alpha} particles using a position sensitive Si detector at \ensuremath{\theta}\ensuremath{_{\textnormal{lab}}}=0\ensuremath{^\circ}}\\
\parbox[b][0.3cm]{17.7cm}{(see \ensuremath{\theta}\ensuremath{_{\textnormal{lab}}}=3\ensuremath{^\circ} reported by (\href{https://www.nndc.bnl.gov/nsr/nsrlink.jsp?2022Go03,B}{2022Go03}) from the EXFOR database). Deduced the \ensuremath{^{\textnormal{4}}}He(\ensuremath{^{\textnormal{15}}}O,\ensuremath{\alpha}) excitation function using Thick Target}\\
\parbox[b][0.3cm]{17.7cm}{Inverse Kinematics techniques with an energy resolution of \ensuremath{\approx}50 keV (FWHM) (\href{https://www.nndc.bnl.gov/nsr/nsrlink.jsp?2022Go03,B}{2022Go03} reports the resolution to be \ensuremath{\approx}30 keV).}\\
\parbox[b][0.3cm]{17.7cm}{Performed an R-matrix analysis of the data, including \ensuremath{\alpha}\ensuremath{_{\textnormal{0}}} and p\ensuremath{_{\textnormal{0}}} particle-decay channels. Deduced \ensuremath{^{\textnormal{19}}}Ne levels. Comparisons with}\\
\parbox[b][0.3cm]{17.7cm}{literature and mirror levels in \ensuremath{^{\textnormal{19}}}F are given. Comparison of the calculated and measured \ensuremath{\alpha}+\ensuremath{^{\textnormal{15}}}O rotational levels is given.}\\
\vspace{0.385cm}
\parbox[b][0.3cm]{17.7cm}{\addtolength{\parindent}{-0.2in}\textit{Related Experiments on the Properties of \ensuremath{^{19}}F* Mirror Levels}:}\\
\parbox[b][0.3cm]{17.7cm}{\addtolength{\parindent}{-0.2in}\href{https://www.nndc.bnl.gov/nsr/nsrlink.jsp?2019La08,B}{2019La08}: \ensuremath{^{\textnormal{4}}}He(\ensuremath{^{\textnormal{15}}}N,\ensuremath{\alpha}) E=40.23 MeV; measured E\ensuremath{_{\ensuremath{\alpha}}}, I\ensuremath{_{\ensuremath{\alpha}}}, differential \ensuremath{\sigma}(\ensuremath{\theta}) using \ensuremath{\Delta}E-E silicon telescopes for \ensuremath{\alpha} detection and Si}\\
\parbox[b][0.3cm]{17.7cm}{surface barrier detectors for \ensuremath{^{\textnormal{15}}}N-like recoils. Deduced \ensuremath{^{\textnormal{19}}}F levels, resonances, J, \ensuremath{\pi}, \ensuremath{\Gamma}\ensuremath{_{\textnormal{p}}}, \ensuremath{\Gamma}\ensuremath{_{\ensuremath{\alpha}}} and potential \ensuremath{\alpha}-cluster states using}\\
\parbox[b][0.3cm]{17.7cm}{R-matrix analysis with the AZURE2 code. Assigned \ensuremath{^{\textnormal{19}}}Ne-\ensuremath{^{\textnormal{19}}}F mirror states for the \ensuremath{^{\textnormal{19}}}Ne states with the excitation energies in the}\\
\parbox[b][0.3cm]{17.7cm}{E\ensuremath{_{\textnormal{x}}}=6-8.8 MeV region.}\\
\parbox[b][0.3cm]{17.7cm}{\addtolength{\parindent}{-0.2in}\href{https://www.nndc.bnl.gov/nsr/nsrlink.jsp?2022Go03,B}{2022Go03}: Analyzed the \ensuremath{^{\textnormal{19}}}F data of (\href{https://www.nndc.bnl.gov/nsr/nsrlink.jsp?1961Sm02,B}{1961Sm02}: \ensuremath{^{\textnormal{4}}}He(\ensuremath{^{\textnormal{15}}}N,\ensuremath{\alpha}) E\ensuremath{_{\textnormal{c.m.}}}=1.75-5.5 MeV); (\href{https://www.nndc.bnl.gov/nsr/nsrlink.jsp?2019La08,B}{2019La08}: \ensuremath{^{\textnormal{4}}}He(\ensuremath{^{\textnormal{15}}}N,\ensuremath{\alpha}) E=40.23 MeV); and}\\
\parbox[b][0.3cm]{17.7cm}{(\href{https://www.nndc.bnl.gov/nsr/nsrlink.jsp?2022Vo01,B}{2022Vo01}: \ensuremath{^{\textnormal{4}}}He(\ensuremath{^{\textnormal{15}}}N,\ensuremath{\alpha}) E=21 MeV); as well as the \ensuremath{^{\textnormal{19}}}Ne data of (\href{https://www.nndc.bnl.gov/nsr/nsrlink.jsp?2017To14,B}{2017To14}: \ensuremath{^{\textnormal{4}}}He(\ensuremath{^{\textnormal{15}}}O,\ensuremath{\alpha}) E=28.5 MeV) using R-matrix analysis}\\
\parbox[b][0.3cm]{17.7cm}{with the computer code AZURE. Deduced \ensuremath{^{\textnormal{19}}}Ne levels, J, \ensuremath{\pi}, \ensuremath{\Gamma}\ensuremath{_{\ensuremath{\alpha}}}; assigned mirror levels in \ensuremath{^{\textnormal{19}}}Ne-\ensuremath{^{\textnormal{19}}}F system; calculated the}\\
\parbox[b][0.3cm]{17.7cm}{Coulomb shift as a function of excitation energy for the \ensuremath{^{\textnormal{19}}}Ne levels. Evaluator highlights that this study presents the excitation}\\
\parbox[b][0.3cm]{17.7cm}{energies and \ensuremath{\alpha} partial widths of the \ensuremath{^{\textnormal{19}}}Ne states without any uncertainty. Considering that there are many states in the region of}\\
\parbox[b][0.3cm]{17.7cm}{interest to this work, for which there are discrepant J\ensuremath{^{\ensuremath{\pi}}} values and widths in the literature, we excluded the results of this work and}\\
\parbox[b][0.3cm]{17.7cm}{only used it as a guide.}\\
\vspace{0.385cm}
\parbox[b][0.3cm]{17.7cm}{\addtolength{\parindent}{-0.2in}\textit{Theory}:}\\
\parbox[b][0.3cm]{17.7cm}{\addtolength{\parindent}{-0.2in}\href{https://www.nndc.bnl.gov/nsr/nsrlink.jsp?2021Sa42,B}{2021Sa42}: \ensuremath{^{\textnormal{4}}}He(\ensuremath{^{\textnormal{15}}}O,\ensuremath{\alpha}); deduced level-energies and J\ensuremath{^{\ensuremath{\pi}}} assignments for the \ensuremath{^{\textnormal{19}}}Ne excited states; deduced lifetimes for the 4.14- and}\\
\parbox[b][0.3cm]{17.7cm}{4.2-MeV states. Calculations were done using the shifted Deng-Fan potential model; investigated the \ensuremath{\alpha}+\ensuremath{^{\textnormal{15}}}O cluster structure in}\\
\parbox[b][0.3cm]{17.7cm}{\ensuremath{^{\textnormal{19}}}Ne using the Nikiforov-Uvarov method to reproduce the rotational bands of \ensuremath{^{\textnormal{19}}}Ne with the \ensuremath{\alpha}+\ensuremath{^{\textnormal{15}}}O configuration corresponding to}\\
\parbox[b][0.3cm]{17.7cm}{N=8 band with negative parities and N=9 band with positive parities. Good agreement is found between the calculated and the}\\
\parbox[b][0.3cm]{17.7cm}{experimental results.}\\
\vspace{12pt}
\underline{$^{19}$Ne Levels}\\
\vspace{0.34cm}
\parbox[b][0.3cm]{17.7cm}{\addtolength{\parindent}{-0.254cm}\textit{Notes}:}\\
\parbox[b][0.3cm]{17.7cm}{\addtolength{\parindent}{-0.254cm}(1) \ensuremath{\theta}\ensuremath{^{\textnormal{2}}_{\ensuremath{\alpha}}} is the ratio of the \ensuremath{\alpha} particle reduced width to the Wigner single-particle limit. Note that \ensuremath{\theta}\ensuremath{_{\ensuremath{\alpha}}^{\textnormal{2}}}\ensuremath{>}0.1 indicates significant \ensuremath{\alpha}}\\
\parbox[b][0.3cm]{17.7cm}{clustering (\href{https://www.nndc.bnl.gov/nsr/nsrlink.jsp?2017To14,B}{2017To14}).}\\
\parbox[b][0.3cm]{17.7cm}{\addtolength{\parindent}{-0.254cm}(2) Mirror assignments are from Table V in (\href{https://www.nndc.bnl.gov/nsr/nsrlink.jsp?2019La08,B}{2019La08}). This study investigated \ensuremath{^{\textnormal{19}}}F mirror nucleus, and the J\ensuremath{^{\ensuremath{\pi}}} assignments for}\\
\parbox[b][0.3cm]{17.7cm}{the \ensuremath{^{\textnormal{19}}}F* states are based on a multi-level, multi-channel R-matrix analysis of \ensuremath{^{\textnormal{4}}}He(\ensuremath{^{\textnormal{15}}}N,\ensuremath{\alpha}).}\\
\vspace{0.34cm}
\begin{longtable}{cccccc@{\extracolsep{\fill}}c}
\multicolumn{2}{c}{E(level)$^{{\hyperlink{NE13LEVEL0}{a}}}$}&J$^{\pi}$$^{{\hyperlink{NE13LEVEL4}{e}}}$&\multicolumn{2}{c}{T\ensuremath{_{\textnormal{1/2}}} or \ensuremath{\Gamma}$^{{\hyperlink{NE13LEVEL3}{d}}}$}&Comments&\\[-.2cm]
\multicolumn{2}{c}{\hrulefill}&\hrulefill&\multicolumn{2}{c}{\hrulefill}&\hrulefill&
\endfirsthead
\multicolumn{1}{r@{}}{4134}&\multicolumn{1}{@{.}l}{6\ensuremath{^{{\hyperlink{NE13LEVEL2}{c}}}}}&\multicolumn{1}{l}{7/2\ensuremath{^{-}}}&\multicolumn{1}{r@{}}{9}&\multicolumn{1}{@{.}l}{94\ensuremath{^{{\hyperlink{NE13LEVEL2}{c}}}} fs}&\parbox[t][0.3cm]{12.876981cm}{\raggedright T\ensuremath{_{\textnormal{1/2}}}: From \ensuremath{\tau}=14.34 fs, which is calculated by (\href{https://www.nndc.bnl.gov/nsr/nsrlink.jsp?2021Sa42,B}{2021Sa42}) using \ensuremath{\Gamma}\ensuremath{_{\ensuremath{\alpha}}}=1.108\ensuremath{\times}10\ensuremath{^{\textnormal{$-$5}}} eV.\vspace{0.1cm}}&\\
&&&&&\parbox[t][0.3cm]{12.876981cm}{\raggedright \ensuremath{\Gamma}\ensuremath{_{\ensuremath{\gamma}}}=0.0459 eV: Calculated by (\href{https://www.nndc.bnl.gov/nsr/nsrlink.jsp?2021Sa42,B}{2021Sa42}).\vspace{0.1cm}}&\\
&&&&&\parbox[t][0.3cm]{12.876981cm}{\raggedright \ensuremath{\Gamma}\ensuremath{_{\textnormal{sp}}}=1.108\ensuremath{\times}10\ensuremath{^{\textnormal{$-$5}}} eV calculated by (\href{https://www.nndc.bnl.gov/nsr/nsrlink.jsp?2021Sa42,B}{2021Sa42}): \ensuremath{\Gamma}\ensuremath{_{\ensuremath{\alpha}}}=S\ensuremath{_{\ensuremath{\alpha}}}\ensuremath{\Gamma}\ensuremath{_{\textnormal{sp}}}, where \ensuremath{\Gamma}\ensuremath{_{\textnormal{sp}}} is the calculated\vspace{0.1cm}}&\\
&&&&&\parbox[t][0.3cm]{12.876981cm}{\raggedright {\ }{\ }{\ }single-particle \ensuremath{\alpha} width, and C\ensuremath{^{\textnormal{2}}}S\ensuremath{_{\ensuremath{\alpha}}}=1-4 is the experimentally obtained \ensuremath{\alpha}-stripping\vspace{0.1cm}}&\\
&&&&&\parbox[t][0.3cm]{12.876981cm}{\raggedright {\ }{\ }{\ }spectroscopic factor from (\href{https://www.nndc.bnl.gov/nsr/nsrlink.jsp?2009Ta09,B}{2009Ta09}: \ensuremath{^{\textnormal{19}}}F(\ensuremath{^{\textnormal{3}}}He,t)). These values lead to\vspace{0.1cm}}&\\
&&&&&\parbox[t][0.3cm]{12.876981cm}{\raggedright {\ }{\ }{\ }\ensuremath{\Gamma}\ensuremath{_{\ensuremath{\alpha}}}=(1.108-4.433)\ensuremath{\times}10\ensuremath{^{\textnormal{$-$5}}} eV calculated by (\href{https://www.nndc.bnl.gov/nsr/nsrlink.jsp?2021Sa42,B}{2021Sa42}).\vspace{0.1cm}}&\\
&&&&&\parbox[t][0.3cm]{12.876981cm}{\raggedright J\ensuremath{^{\pi}}: From (\href{https://www.nndc.bnl.gov/nsr/nsrlink.jsp?2020Ha31,B}{2020Ha31}: \ensuremath{^{\textnormal{19}}}F(\ensuremath{^{\textnormal{3}}}He,t\ensuremath{\gamma})) recommended by (\href{https://www.nndc.bnl.gov/nsr/nsrlink.jsp?2021Sa42,B}{2021Sa42}).\vspace{0.1cm}}&\\
\multicolumn{1}{r@{}}{4195}&\multicolumn{1}{@{.}l}{1\ensuremath{^{{\hyperlink{NE13LEVEL2}{c}}}}}&\multicolumn{1}{l}{9/2\ensuremath{^{-}}}&\multicolumn{1}{r@{}}{24}&\multicolumn{1}{@{.}l}{83\ensuremath{^{{\hyperlink{NE13LEVEL2}{c}}}} fs}&\parbox[t][0.3cm]{12.876981cm}{\raggedright T\ensuremath{_{\textnormal{1/2}}}: From \ensuremath{\tau}=35.82 fs calculated by (\href{https://www.nndc.bnl.gov/nsr/nsrlink.jsp?2021Sa42,B}{2021Sa42}).\vspace{0.1cm}}&\\
&&&&&\parbox[t][0.3cm]{12.876981cm}{\raggedright \ensuremath{\Gamma}\ensuremath{_{\ensuremath{\gamma}}}=0.0183 eV: Calculated by (\href{https://www.nndc.bnl.gov/nsr/nsrlink.jsp?2021Sa42,B}{2021Sa42}).\vspace{0.1cm}}&\\
\end{longtable}
\begin{textblock}{29}(0,27.3)
Continued on next page (footnotes at end of table)
\end{textblock}
\clearpage
\begin{longtable}{cccccc@{\extracolsep{\fill}}c}
\\[-.4cm]
\multicolumn{7}{c}{{\bf \small \underline{\ensuremath{^{\textnormal{4}}}He(\ensuremath{^{\textnormal{15}}}O,\ensuremath{\alpha}):res\hspace{0.2in}\href{https://www.nndc.bnl.gov/nsr/nsrlink.jsp?2017To14,B}{2017To14} (continued)}}}\\
\multicolumn{7}{c}{~}\\
\multicolumn{7}{c}{\underline{\ensuremath{^{19}}Ne Levels (continued)}}\\
\multicolumn{7}{c}{~}\\
\multicolumn{2}{c}{E(level)$^{{\hyperlink{NE13LEVEL0}{a}}}$}&J$^{\pi}$$^{{\hyperlink{NE13LEVEL4}{e}}}$&\multicolumn{2}{c}{T\ensuremath{_{\textnormal{1/2}}} or \ensuremath{\Gamma}$^{{\hyperlink{NE13LEVEL3}{d}}}$}&Comments&\\[-.2cm]
\multicolumn{2}{c}{\hrulefill}&\hrulefill&\multicolumn{2}{c}{\hrulefill}&\hrulefill&
\endhead
&&&&&\parbox[t][0.3cm]{11.697841cm}{\raggedright \ensuremath{\Gamma}\ensuremath{_{\ensuremath{\alpha}}}=2.769\ensuremath{\times}10\ensuremath{^{\textnormal{$-$5}}} eV \textit{+2154{\textminus}1539} calculated by (\href{https://www.nndc.bnl.gov/nsr/nsrlink.jsp?2021Sa42,B}{2021Sa42}).\vspace{0.1cm}}&\\
&&&&&\parbox[t][0.3cm]{11.697841cm}{\raggedright \ensuremath{\Gamma}\ensuremath{_{\textnormal{sp}}}=6.154\ensuremath{\times}10\ensuremath{^{\textnormal{$-$5}}} eV calculated by (\href{https://www.nndc.bnl.gov/nsr/nsrlink.jsp?2021Sa42,B}{2021Sa42}): \ensuremath{\Gamma}\ensuremath{_{\ensuremath{\alpha}}}=S\ensuremath{_{\ensuremath{\alpha}}}\ensuremath{\Gamma}\ensuremath{_{\textnormal{sp}}}, where \ensuremath{\Gamma}\ensuremath{_{\textnormal{sp}}} is the\vspace{0.1cm}}&\\
&&&&&\parbox[t][0.3cm]{11.697841cm}{\raggedright {\ }{\ }{\ }calculated single-particle \ensuremath{\alpha} width, and C\ensuremath{^{\textnormal{2}}}S\ensuremath{_{\ensuremath{\alpha}}}=0.45 \textit{+35{\textminus}25} (\href{https://www.nndc.bnl.gov/nsr/nsrlink.jsp?2009Ta09,B}{2009Ta09}: \ensuremath{^{\textnormal{19}}}F(\ensuremath{^{\textnormal{3}}}He,t)),\vspace{0.1cm}}&\\
&&&&&\parbox[t][0.3cm]{11.697841cm}{\raggedright {\ }{\ }{\ }which is the experimentally obtained \ensuremath{\alpha}-stripping spectroscopic factor.\vspace{0.1cm}}&\\
&&&&&\parbox[t][0.3cm]{11.697841cm}{\raggedright J\ensuremath{^{\pi}}: From (\href{https://www.nndc.bnl.gov/nsr/nsrlink.jsp?2020Ha31,B}{2020Ha31}) recommended by (\href{https://www.nndc.bnl.gov/nsr/nsrlink.jsp?2021Sa42,B}{2021Sa42}).\vspace{0.1cm}}&\\
\multicolumn{1}{r@{}}{5359}&\multicolumn{1}{@{ }l}{{\it 6}}&\multicolumn{1}{l}{1/2\ensuremath{^{+}}}&\multicolumn{1}{r@{}}{6}&\multicolumn{1}{@{.}l}{6 keV {\it 34}}&\parbox[t][0.3cm]{11.697841cm}{\raggedright \ensuremath{\Gamma}\ensuremath{\alpha}=6.6 keV \textit{34}\vspace{0.1cm}}&\\
&&&&&\parbox[t][0.3cm]{11.697841cm}{\raggedright E(level): See also E\ensuremath{_{\textnormal{x}}}=5.35 MeV (\href{https://www.nndc.bnl.gov/nsr/nsrlink.jsp?2006Va06,B}{2006Va06}) deduced from E\ensuremath{_{\textnormal{c.m.}}}(\ensuremath{^{\textnormal{15}}}O+\ensuremath{\alpha})=1.82 MeV\vspace{0.1cm}}&\\
&&&&&\parbox[t][0.3cm]{11.697841cm}{\raggedright {\ }{\ }{\ }(\href{https://www.nndc.bnl.gov/nsr/nsrlink.jsp?2006Va06,B}{2006Va06}) and the \ensuremath{\alpha}-separation energy of S\ensuremath{_{\ensuremath{\alpha}}}(\ensuremath{^{\textnormal{19}}}Ne)=3528.5 keV (\href{https://www.nndc.bnl.gov/nsr/nsrlink.jsp?2021Wa16,B}{2021Wa16}).\vspace{0.1cm}}&\\
&&&&&\parbox[t][0.3cm]{11.697841cm}{\raggedright \ensuremath{\Gamma}\ensuremath{_{\ensuremath{\alpha}}}: Unweighted average of \ensuremath{\Gamma}\ensuremath{_{\ensuremath{\alpha}}}=3.2 keV \textit{16} (\href{https://www.nndc.bnl.gov/nsr/nsrlink.jsp?2006Va06,B}{2006Va06}) and \ensuremath{\Gamma}\ensuremath{_{\ensuremath{\alpha}}}=10 keV \textit{3}\vspace{0.1cm}}&\\
&&&&&\parbox[t][0.3cm]{11.697841cm}{\raggedright {\ }{\ }{\ }(\href{https://www.nndc.bnl.gov/nsr/nsrlink.jsp?2017To14,B}{2017To14}), which are both determined from R-matrix analysis. Other value:\vspace{0.1cm}}&\\
&&&&&\parbox[t][0.3cm]{11.697841cm}{\raggedright {\ }{\ }{\ }(\href{https://www.nndc.bnl.gov/nsr/nsrlink.jsp?2006Va06,B}{2006Va06}) also deduced \ensuremath{\Gamma}\ensuremath{_{\ensuremath{\alpha}}}=2.9 keV \textit{18} from a global \ensuremath{\chi}\ensuremath{^{\textnormal{2}}} fit to modified energy\vspace{0.1cm}}&\\
&&&&&\parbox[t][0.3cm]{11.697841cm}{\raggedright {\ }{\ }{\ }bins of the \ensuremath{\alpha} spectrum.\vspace{0.1cm}}&\\
&&&&&\parbox[t][0.3cm]{11.697841cm}{\raggedright \ensuremath{\Gamma}: From \ensuremath{\Gamma}=\ensuremath{\Gamma}\ensuremath{_{\ensuremath{\alpha}}} (\href{https://www.nndc.bnl.gov/nsr/nsrlink.jsp?2006Va06,B}{2006Va06}, \href{https://www.nndc.bnl.gov/nsr/nsrlink.jsp?2017To14,B}{2017To14}).\vspace{0.1cm}}&\\
&&&&&\parbox[t][0.3cm]{11.697841cm}{\raggedright \ensuremath{\theta}\ensuremath{^{\textnormal{2}}_{\ensuremath{\alpha}}}=0.61 \textit{10} (\href{https://www.nndc.bnl.gov/nsr/nsrlink.jsp?2017To14,B}{2017To14}).\vspace{0.1cm}}&\\
\multicolumn{1}{r@{}}{5487}&\multicolumn{1}{@{ }l}{{\it 4}}&\multicolumn{1}{l}{3/2\ensuremath{^{+}}}&\multicolumn{1}{r@{}}{9}&\multicolumn{1}{@{ }l}{keV {\it 2}}&\parbox[t][0.3cm]{11.697841cm}{\raggedright \ensuremath{\Gamma}\ensuremath{\alpha}=9 keV \textit{2} (\href{https://www.nndc.bnl.gov/nsr/nsrlink.jsp?2017To14,B}{2017To14})\vspace{0.1cm}}&\\
&&&&&\parbox[t][0.3cm]{11.697841cm}{\raggedright \ensuremath{\theta}\ensuremath{^{\textnormal{2}}_{\ensuremath{\alpha}}}=0.39 \textit{4} (\href{https://www.nndc.bnl.gov/nsr/nsrlink.jsp?2017To14,B}{2017To14}).\vspace{0.1cm}}&\\
\multicolumn{1}{r@{}}{5704?}&\multicolumn{1}{@{}l}{\ensuremath{^{{\hyperlink{NE13LEVEL1}{b}}}} {\it 8}}&\multicolumn{1}{l}{5/2\ensuremath{^{-}}}&\multicolumn{1}{r@{}}{29}&\multicolumn{1}{@{ }l}{keV {\it 6}}&\parbox[t][0.3cm]{11.697841cm}{\raggedright \ensuremath{\Gamma}\ensuremath{\alpha}=29 keV \textit{6} (\href{https://www.nndc.bnl.gov/nsr/nsrlink.jsp?2017To14,B}{2017To14})\vspace{0.1cm}}&\\
&&&&&\parbox[t][0.3cm]{11.697841cm}{\raggedright E(level): Evaluator highlights that the most recent investigation of the \ensuremath{^{\textnormal{19}}}Ne-\ensuremath{^{\textnormal{19}}}F\vspace{0.1cm}}&\\
&&&&&\parbox[t][0.3cm]{11.697841cm}{\raggedright {\ }{\ }{\ }mirror study by (\href{https://www.nndc.bnl.gov/nsr/nsrlink.jsp?2022Go03,B}{2022Go03}), which included the data from (\href{https://www.nndc.bnl.gov/nsr/nsrlink.jsp?2017To14,B}{2017To14}) and has an\vspace{0.1cm}}&\\
&&&&&\parbox[t][0.3cm]{11.697841cm}{\raggedright {\ }{\ }{\ }improved understanding of \ensuremath{^{\textnormal{19}}}F, found that if they consider this level in their\vspace{0.1cm}}&\\
&&&&&\parbox[t][0.3cm]{11.697841cm}{\raggedright {\ }{\ }{\ }R-matrix analysis, it leads to a very poor fit. Therefore, this level was omitted from\vspace{0.1cm}}&\\
&&&&&\parbox[t][0.3cm]{11.697841cm}{\raggedright {\ }{\ }{\ }their analysis. This casts doubt on the existence of this state. We therefore, made\vspace{0.1cm}}&\\
&&&&&\parbox[t][0.3cm]{11.697841cm}{\raggedright {\ }{\ }{\ }the level tentative and did not adopt this state in the \ensuremath{^{\textnormal{19}}}Ne Adopted Levels.\vspace{0.1cm}}&\\
&&&&&\parbox[t][0.3cm]{11.697841cm}{\raggedright \ensuremath{\theta}\ensuremath{^{\textnormal{2}}_{\ensuremath{\alpha}}}=0.98 \textit{10} (\href{https://www.nndc.bnl.gov/nsr/nsrlink.jsp?2017To14,B}{2017To14}).\vspace{0.1cm}}&\\
\multicolumn{1}{r@{}}{5983}&\multicolumn{1}{@{ }l}{{\it 9}}&\multicolumn{1}{l}{3/2\ensuremath{^{-}}}&\multicolumn{1}{r@{}}{21}&\multicolumn{1}{@{ }l}{keV {\it 8}}&\parbox[t][0.3cm]{11.697841cm}{\raggedright \ensuremath{\Gamma}\ensuremath{\alpha}=21 keV \textit{8} (\href{https://www.nndc.bnl.gov/nsr/nsrlink.jsp?2017To14,B}{2017To14})\vspace{0.1cm}}&\\
&&&&&\parbox[t][0.3cm]{11.697841cm}{\raggedright Proposed mirror state: \ensuremath{^{\textnormal{19}}}F*(6088 keV, 3/2\ensuremath{^{-}}) (\href{https://www.nndc.bnl.gov/nsr/nsrlink.jsp?2019La08,B}{2019La08}). However, that \ensuremath{^{\textnormal{19}}}F* state\vspace{0.1cm}}&\\
&&&&&\parbox[t][0.3cm]{11.697841cm}{\raggedright {\ }{\ }{\ }has a total width that is twice as large (see \href{https://www.nndc.bnl.gov/nsr/nsrlink.jsp?2019La08,B}{2019La08}) as that of this \ensuremath{^{\textnormal{19}}}Ne* state.\vspace{0.1cm}}&\\
&&&&&\parbox[t][0.3cm]{11.697841cm}{\raggedright \ensuremath{\theta}\ensuremath{^{\textnormal{2}}_{\ensuremath{\alpha}}}=0.42 \textit{8} (\href{https://www.nndc.bnl.gov/nsr/nsrlink.jsp?2017To14,B}{2017To14}).\vspace{0.1cm}}&\\
\multicolumn{1}{r@{}}{6197}&\multicolumn{1}{@{}l}{\ensuremath{^{{\hyperlink{NE13LEVEL1}{b}}}} {\it 8}}&\multicolumn{1}{l}{(1/2\ensuremath{^{-}},1/2\ensuremath{^{+}})}&\multicolumn{1}{r@{}}{16}&\multicolumn{1}{@{ }l}{keV {\it 7}}&\parbox[t][0.3cm]{11.697841cm}{\raggedright \ensuremath{\Gamma}\ensuremath{\alpha}=16 keV \textit{5} (\href{https://www.nndc.bnl.gov/nsr/nsrlink.jsp?2017To14,B}{2017To14})\vspace{0.1cm}}&\\
&&&&&\parbox[t][0.3cm]{11.697841cm}{\raggedright \ensuremath{\Gamma}: It is not clear why the uncertainties in \ensuremath{\Gamma}\ensuremath{_{\textnormal{tot}}} and \ensuremath{\Gamma}\ensuremath{_{\ensuremath{\alpha}}} from (\href{https://www.nndc.bnl.gov/nsr/nsrlink.jsp?2017To14,B}{2017To14}) differ.\vspace{0.1cm}}&\\
&&&&&\parbox[t][0.3cm]{11.697841cm}{\raggedright \ensuremath{\theta}\ensuremath{^{\textnormal{2}}_{\ensuremath{\alpha}}}=0.14 \textit{2} for J\ensuremath{^{\ensuremath{\pi}}}=1/2\ensuremath{^{-}} (\href{https://www.nndc.bnl.gov/nsr/nsrlink.jsp?2017To14,B}{2017To14}).\vspace{0.1cm}}&\\
&&&&&\parbox[t][0.3cm]{11.697841cm}{\raggedright Proposed mirror state: \ensuremath{^{\textnormal{19}}}F*(6255 keV, 1/2\ensuremath{^{\textnormal{+}}}) (\href{https://www.nndc.bnl.gov/nsr/nsrlink.jsp?2019La08,B}{2019La08}).\vspace{0.1cm}}&\\
\multicolumn{1}{r@{}}{6279}&\multicolumn{1}{@{ }l}{{\it 2}}&\multicolumn{1}{l}{(5/2\ensuremath{^{+}})}&\multicolumn{1}{r@{}}{6}&\multicolumn{1}{@{ }l}{keV {\it 2}}&\parbox[t][0.3cm]{11.697841cm}{\raggedright \ensuremath{\Gamma}\ensuremath{\alpha}=6 keV \textit{2} (\href{https://www.nndc.bnl.gov/nsr/nsrlink.jsp?2017To14,B}{2017To14})\vspace{0.1cm}}&\\
&&&&&\parbox[t][0.3cm]{11.697841cm}{\raggedright \ensuremath{\Gamma},\ensuremath{\Gamma}\ensuremath{\alpha}: Deduced by (\href{https://www.nndc.bnl.gov/nsr/nsrlink.jsp?2017To14,B}{2017To14}) from R-matrix analysis assuming J\ensuremath{^{\ensuremath{\pi}}}=5/2\ensuremath{^{\textnormal{+}}}.\vspace{0.1cm}}&\\
&&&&&\parbox[t][0.3cm]{11.697841cm}{\raggedright \ensuremath{\theta}\ensuremath{^{\textnormal{2}}_{\ensuremath{\alpha}}}=0.27 \textit{5} (\href{https://www.nndc.bnl.gov/nsr/nsrlink.jsp?2017To14,B}{2017To14}) assuming J\ensuremath{^{\ensuremath{\pi}}}=5/2\ensuremath{^{\textnormal{+}}}.\vspace{0.1cm}}&\\
&&&&&\parbox[t][0.3cm]{11.697841cm}{\raggedright Proposed mirror state: \ensuremath{^{\textnormal{19}}}F*(6282 keV, 5/2\ensuremath{^{\textnormal{+}}}) (\href{https://www.nndc.bnl.gov/nsr/nsrlink.jsp?2019La08,B}{2019La08}).\vspace{0.1cm}}&\\
\multicolumn{1}{r@{}}{6395}&\multicolumn{1}{@{ }l}{{\it 5}}&\multicolumn{1}{l}{1/2\ensuremath{^{-}}}&\multicolumn{1}{r@{}}{181}&\multicolumn{1}{@{ }l}{keV {\it 58}}&\parbox[t][0.3cm]{11.697841cm}{\raggedright \ensuremath{\Gamma}\ensuremath{\alpha}=181 keV \textit{58} (\href{https://www.nndc.bnl.gov/nsr/nsrlink.jsp?2017To14,B}{2017To14})\vspace{0.1cm}}&\\
&&&&&\parbox[t][0.3cm]{11.697841cm}{\raggedright \ensuremath{\theta}\ensuremath{^{\textnormal{2}}_{\ensuremath{\alpha}}}=0.44 \textit{7} (\href{https://www.nndc.bnl.gov/nsr/nsrlink.jsp?2017To14,B}{2017To14}).\vspace{0.1cm}}&\\
\multicolumn{1}{r@{}}{7030}&\multicolumn{1}{@{}l}{\ensuremath{^{{\hyperlink{NE13LEVEL1}{b}}}} {\it 4}}&\multicolumn{1}{l}{7/2\ensuremath{^{+}}}&\multicolumn{1}{r@{}}{12}&\multicolumn{1}{@{ }l}{keV {\it 3}}&\parbox[t][0.3cm]{11.697841cm}{\raggedright \ensuremath{\Gamma}\ensuremath{\alpha}=12 keV \textit{3} (\href{https://www.nndc.bnl.gov/nsr/nsrlink.jsp?2017To14,B}{2017To14})\vspace{0.1cm}}&\\
&&&&&\parbox[t][0.3cm]{11.697841cm}{\raggedright \ensuremath{\theta}\ensuremath{^{\textnormal{2}}_{\ensuremath{\alpha}}}=0.17 \textit{2} (\href{https://www.nndc.bnl.gov/nsr/nsrlink.jsp?2017To14,B}{2017To14}).\vspace{0.1cm}}&\\
&&&&&\parbox[t][0.3cm]{11.697841cm}{\raggedright Proposed mirror state: \ensuremath{^{\textnormal{19}}}F*(7114 keV, 7/2\ensuremath{^{\textnormal{+}}}) (\href{https://www.nndc.bnl.gov/nsr/nsrlink.jsp?2019La08,B}{2019La08}).\vspace{0.1cm}}&\\
\multicolumn{1}{r@{}}{7153}&\multicolumn{1}{@{ }l}{{\it 9}}&\multicolumn{1}{l}{3/2\ensuremath{^{+}}}&\multicolumn{1}{r@{}}{252}&\multicolumn{1}{@{ }l}{keV {\it 46}}&\parbox[t][0.3cm]{11.697841cm}{\raggedright \ensuremath{\Gamma}\ensuremath{\alpha}=233 keV \textit{44} (\href{https://www.nndc.bnl.gov/nsr/nsrlink.jsp?2017To14,B}{2017To14})\vspace{0.1cm}}&\\
&&&&&\parbox[t][0.3cm]{11.697841cm}{\raggedright \ensuremath{\Gamma}\ensuremath{_{\textnormal{p}}}=19 keV \textit{14} (\href{https://www.nndc.bnl.gov/nsr/nsrlink.jsp?2017To14,B}{2017To14})\vspace{0.1cm}}&\\
&&&&&\parbox[t][0.3cm]{11.697841cm}{\raggedright \ensuremath{\Gamma}: (\href{https://www.nndc.bnl.gov/nsr/nsrlink.jsp?2017To14,B}{2017To14}) reports \ensuremath{\Gamma}=252 keV \textit{39}. This uncertainty is less than that deduced by\vspace{0.1cm}}&\\
&&&&&\parbox[t][0.3cm]{11.697841cm}{\raggedright {\ }{\ }{\ }summing \ensuremath{\Gamma}\ensuremath{_{\ensuremath{\alpha}}} and \ensuremath{\Gamma}\ensuremath{_{\textnormal{p}}} together. So, the evaluator changed the uncertainty in \ensuremath{\Gamma}\ensuremath{_{\textnormal{tot}}}\vspace{0.1cm}}&\\
&&&&&\parbox[t][0.3cm]{11.697841cm}{\raggedright {\ }{\ }{\ }accordingly.\vspace{0.1cm}}&\\
&&&&&\parbox[t][0.3cm]{11.697841cm}{\raggedright \ensuremath{\theta}\ensuremath{^{\textnormal{2}}_{\ensuremath{\alpha}}}=0.39 \textit{4} (\href{https://www.nndc.bnl.gov/nsr/nsrlink.jsp?2017To14,B}{2017To14}).\vspace{0.1cm}}&\\
&&&&&\parbox[t][0.3cm]{11.697841cm}{\raggedright Proposed mirror state: \ensuremath{^{\textnormal{19}}}F*(7262 keV, 3/2\ensuremath{^{\textnormal{+}}}) (\href{https://www.nndc.bnl.gov/nsr/nsrlink.jsp?2019La08,B}{2019La08}).\vspace{0.1cm}}&\\
\multicolumn{1}{r@{}}{7378}&\multicolumn{1}{@{ }l}{{\it 7}}&\multicolumn{1}{l}{7/2\ensuremath{^{+}}}&\multicolumn{1}{r@{}}{121}&\multicolumn{1}{@{ }l}{keV {\it 9}}&\parbox[t][0.3cm]{11.697841cm}{\raggedright \ensuremath{\Gamma}\ensuremath{\alpha}=121 keV \textit{9} (\href{https://www.nndc.bnl.gov/nsr/nsrlink.jsp?2017To14,B}{2017To14})\vspace{0.1cm}}&\\
&&&&&\parbox[t][0.3cm]{11.697841cm}{\raggedright \ensuremath{\theta}\ensuremath{^{\textnormal{2}}_{\ensuremath{\alpha}}}=0.44 \textit{2} (\href{https://www.nndc.bnl.gov/nsr/nsrlink.jsp?2017To14,B}{2017To14}).\vspace{0.1cm}}&\\
&&&&&\parbox[t][0.3cm]{11.697841cm}{\raggedright Proposed mirror state \ensuremath{^{\textnormal{19}}}F*(7560 keV, 7/2\ensuremath{^{\textnormal{+}}}) (\href{https://www.nndc.bnl.gov/nsr/nsrlink.jsp?2019La08,B}{2019La08}).\vspace{0.1cm}}&\\
\multicolumn{1}{r@{}}{7469}&\multicolumn{1}{@{ }l}{{\it 7}}&\multicolumn{1}{l}{5/2\ensuremath{^{+}}}&\multicolumn{1}{r@{}}{83}&\multicolumn{1}{@{ }l}{keV {\it 17}}&\parbox[t][0.3cm]{11.697841cm}{\raggedright \ensuremath{\Gamma}\ensuremath{\alpha}=83 keV \textit{17} (\href{https://www.nndc.bnl.gov/nsr/nsrlink.jsp?2017To14,B}{2017To14})\vspace{0.1cm}}&\\
&&&&&\parbox[t][0.3cm]{11.697841cm}{\raggedright \ensuremath{\theta}\ensuremath{^{\textnormal{2}}_{\ensuremath{\alpha}}}=0.34 \textit{3} (\href{https://www.nndc.bnl.gov/nsr/nsrlink.jsp?2017To14,B}{2017To14}).\vspace{0.1cm}}&\\
\end{longtable}
\begin{textblock}{29}(0,27.3)
Continued on next page (footnotes at end of table)
\end{textblock}
\clearpage
\begin{longtable}{cccccc@{\extracolsep{\fill}}c}
\\[-.4cm]
\multicolumn{7}{c}{{\bf \small \underline{\ensuremath{^{\textnormal{4}}}He(\ensuremath{^{\textnormal{15}}}O,\ensuremath{\alpha}):res\hspace{0.2in}\href{https://www.nndc.bnl.gov/nsr/nsrlink.jsp?2017To14,B}{2017To14} (continued)}}}\\
\multicolumn{7}{c}{~}\\
\multicolumn{7}{c}{\underline{\ensuremath{^{19}}Ne Levels (continued)}}\\
\multicolumn{7}{c}{~}\\
\multicolumn{2}{c}{E(level)$^{{\hyperlink{NE13LEVEL0}{a}}}$}&J$^{\pi}$$^{{\hyperlink{NE13LEVEL4}{e}}}$&\multicolumn{2}{c}{T\ensuremath{_{\textnormal{1/2}}} or \ensuremath{\Gamma}$^{{\hyperlink{NE13LEVEL3}{d}}}$}&Comments&\\[-.2cm]
\multicolumn{2}{c}{\hrulefill}&\hrulefill&\multicolumn{2}{c}{\hrulefill}&\hrulefill&
\endhead
&&&&&\parbox[t][0.3cm]{11.22276cm}{\raggedright Proposed mirror state: \ensuremath{^{\textnormal{19}}}F*(7539.6 keV, 5/2\ensuremath{^{\textnormal{+}}}) (\href{https://www.nndc.bnl.gov/nsr/nsrlink.jsp?2019La08,B}{2019La08}). However, that \ensuremath{^{\textnormal{19}}}F*\vspace{0.1cm}}&\\
&&&&&\parbox[t][0.3cm]{11.22276cm}{\raggedright {\ }{\ }{\ }state has a total width that is \ensuremath{\sim}7 times smaller than (see \href{https://www.nndc.bnl.gov/nsr/nsrlink.jsp?2019La08,B}{2019La08}) that of this\vspace{0.1cm}}&\\
&&&&&\parbox[t][0.3cm]{11.22276cm}{\raggedright {\ }{\ }{\ }\ensuremath{^{\textnormal{19}}}Ne* state.\vspace{0.1cm}}&\\
\multicolumn{1}{r@{}}{7568}&\multicolumn{1}{@{ }l}{{\it 27}}&\multicolumn{1}{l}{(3/2\ensuremath{^{+}},1/2\ensuremath{^{+}})}&\multicolumn{1}{r@{}}{774}&\multicolumn{1}{@{ }l}{keV {\it 144}}&\parbox[t][0.3cm]{11.22276cm}{\raggedright \ensuremath{\Gamma}\ensuremath{\alpha}=774 keV \textit{144} (\href{https://www.nndc.bnl.gov/nsr/nsrlink.jsp?2017To14,B}{2017To14})\vspace{0.1cm}}&\\
&&&&&\parbox[t][0.3cm]{11.22276cm}{\raggedright The authors mention that this state is either a different \ensuremath{\alpha}-resonance from that\vspace{0.1cm}}&\\
&&&&&\parbox[t][0.3cm]{11.22276cm}{\raggedright {\ }{\ }{\ }reported in (\href{https://www.nndc.bnl.gov/nsr/nsrlink.jsp?2009Da07,B}{2009Da07}: \ensuremath{^{\textnormal{1}}}H(\ensuremath{^{\textnormal{19}}}Ne,p)) at E\ensuremath{_{\textnormal{x}}}=7616 keV \textit{16}, or that the\vspace{0.1cm}}&\\
&&&&&\parbox[t][0.3cm]{11.22276cm}{\raggedright {\ }{\ }{\ }\ensuremath{^{\textnormal{19}}}Ne*(7568) may be an additional, unidentified level at a similar excitation\vspace{0.1cm}}&\\
&&&&&\parbox[t][0.3cm]{11.22276cm}{\raggedright {\ }{\ }{\ }energy.\vspace{0.1cm}}&\\
&&&&&\parbox[t][0.3cm]{11.22276cm}{\raggedright Proposed mirror state: \ensuremath{^{\textnormal{19}}}F*(7660.6 keV, 3/2\ensuremath{^{\textnormal{+}}}) (\href{https://www.nndc.bnl.gov/nsr/nsrlink.jsp?2019La08,B}{2019La08}).\vspace{0.1cm}}&\\
&&&&&\parbox[t][0.3cm]{11.22276cm}{\raggedright We highlight that (\href{https://www.nndc.bnl.gov/nsr/nsrlink.jsp?2022Go03,B}{2022Go03}) disputed the J\ensuremath{^{\ensuremath{\pi}}} assignments of (\href{https://www.nndc.bnl.gov/nsr/nsrlink.jsp?2017To14,B}{2017To14}) for this\vspace{0.1cm}}&\\
&&&&&\parbox[t][0.3cm]{11.22276cm}{\raggedright {\ }{\ }{\ }\ensuremath{^{\textnormal{19}}}Ne* state and reported that the existence of a J\ensuremath{^{\ensuremath{\pi}}}=3/2\ensuremath{^{\textnormal{+}}} state near 7.6 MeV in\vspace{0.1cm}}&\\
&&&&&\parbox[t][0.3cm]{11.22276cm}{\raggedright {\ }{\ }{\ }the \ensuremath{^{\textnormal{19}}}F* mirror nucleus (where they expected the mirror state) results in a very\vspace{0.1cm}}&\\
&&&&&\parbox[t][0.3cm]{11.22276cm}{\raggedright {\ }{\ }{\ }poor fit for their \ensuremath{^{\textnormal{19}}}F data.\vspace{0.1cm}}&\\
&&&&&\parbox[t][0.3cm]{11.22276cm}{\raggedright \ensuremath{\theta}\ensuremath{^{\textnormal{2}}_{\ensuremath{\alpha}}}=0.57 \textit{5} for J\ensuremath{^{\ensuremath{\pi}}}=3/2\ensuremath{^{\textnormal{+}}} (\href{https://www.nndc.bnl.gov/nsr/nsrlink.jsp?2017To14,B}{2017To14}).\vspace{0.1cm}}&\\
\multicolumn{1}{r@{}}{8022}&\multicolumn{1}{@{}l}{\ensuremath{^{{\hyperlink{NE13LEVEL1}{b}}}} {\it 4}}&&&&\parbox[t][0.3cm]{11.22276cm}{\raggedright \ensuremath{\Gamma}=\ensuremath{\Gamma}\ensuremath{_{\ensuremath{\alpha}}}=64 keV \textit{10} (\href{https://www.nndc.bnl.gov/nsr/nsrlink.jsp?2017To14,B}{2017To14}): From R-matrix.\vspace{0.1cm}}&\\
&&&&&\parbox[t][0.3cm]{11.22276cm}{\raggedright J=9/2\ensuremath{^{\textnormal{+}}} (\href{https://www.nndc.bnl.gov/nsr/nsrlink.jsp?2017To14,B}{2017To14}): From R-matrix.\vspace{0.1cm}}&\\
&&&&&\parbox[t][0.3cm]{11.22276cm}{\raggedright \ensuremath{\theta}\ensuremath{^{\textnormal{2}}_{\ensuremath{\alpha}}}=0.84 \textit{7} (\href{https://www.nndc.bnl.gov/nsr/nsrlink.jsp?2017To14,B}{2017To14}) for J\ensuremath{^{\ensuremath{\pi}}}=9/2\ensuremath{^{\textnormal{+}}}.\vspace{0.1cm}}&\\
&&&&&\parbox[t][0.3cm]{11.22276cm}{\raggedright Proposed mirror state: \ensuremath{^{\textnormal{19}}}F*(7929 keV, 7/2\ensuremath{^{\textnormal{+}}}) (\href{https://www.nndc.bnl.gov/nsr/nsrlink.jsp?2019La08,B}{2019La08}). However, that \ensuremath{^{\textnormal{19}}}F*\vspace{0.1cm}}&\\
&&&&&\parbox[t][0.3cm]{11.22276cm}{\raggedright {\ }{\ }{\ }state has a total width that is 4 times larger than (see \href{https://www.nndc.bnl.gov/nsr/nsrlink.jsp?2019La08,B}{2019La08}) that of this\vspace{0.1cm}}&\\
&&&&&\parbox[t][0.3cm]{11.22276cm}{\raggedright {\ }{\ }{\ }\ensuremath{^{\textnormal{19}}}Ne* state. Moreover, the J\ensuremath{^{\ensuremath{\pi}}} values for this proposed mirror pair are\vspace{0.1cm}}&\\
&&&&&\parbox[t][0.3cm]{11.22276cm}{\raggedright {\ }{\ }{\ }inconsistent. We highlight that in a more recent \ensuremath{^{\textnormal{19}}}F-\ensuremath{^{\textnormal{19}}}Ne mirror study by\vspace{0.1cm}}&\\
&&&&&\parbox[t][0.3cm]{11.22276cm}{\raggedright {\ }{\ }{\ }(\href{https://www.nndc.bnl.gov/nsr/nsrlink.jsp?2022Go03,B}{2022Go03}), the \ensuremath{^{\textnormal{4}}}He(\ensuremath{^{\textnormal{15}}}O,\ensuremath{\alpha}) data of (\href{https://www.nndc.bnl.gov/nsr/nsrlink.jsp?2017To14,B}{2017To14}) in addition to other similar\vspace{0.1cm}}&\\
&&&&&\parbox[t][0.3cm]{11.22276cm}{\raggedright {\ }{\ }{\ }data were reanalyzed using R-matrix and two states were found in \ensuremath{^{\textnormal{19}}}Ne within\vspace{0.1cm}}&\\
&&&&&\parbox[t][0.3cm]{11.22276cm}{\raggedright {\ }{\ }{\ }this energy range at 8053 keV with J\ensuremath{^{\ensuremath{\pi}}}=5/2\ensuremath{^{-}} and \ensuremath{\Gamma}\ensuremath{_{\ensuremath{\alpha}}}=120 keV; and 8103 keV\vspace{0.1cm}}&\\
&&&&&\parbox[t][0.3cm]{11.22276cm}{\raggedright {\ }{\ }{\ }with J\ensuremath{^{\ensuremath{\pi}}}=3/2\ensuremath{^{-}} and \ensuremath{\Gamma}\ensuremath{_{\ensuremath{\alpha}}}=48 keV. Due to inconsistent results from the same data,\vspace{0.1cm}}&\\
&&&&&\parbox[t][0.3cm]{11.22276cm}{\raggedright {\ }{\ }{\ }we did not adopt the \ensuremath{\Gamma}, \ensuremath{\Gamma}\ensuremath{_{\ensuremath{\alpha}}} and J\ensuremath{^{\ensuremath{\pi}}} from (\href{https://www.nndc.bnl.gov/nsr/nsrlink.jsp?2017To14,B}{2017To14}).\vspace{0.1cm}}&\\
\multicolumn{1}{r@{}}{8223}&\multicolumn{1}{@{}l}{\ensuremath{^{{\hyperlink{NE13LEVEL1}{b}}}} {\it 7}}&\multicolumn{1}{l}{5/2\ensuremath{^{+}}}&\multicolumn{1}{r@{}}{377}&\multicolumn{1}{@{ }l}{keV {\it 34}}&\parbox[t][0.3cm]{11.22276cm}{\raggedright \ensuremath{\Gamma}\ensuremath{\alpha}=377 keV \textit{34} (\href{https://www.nndc.bnl.gov/nsr/nsrlink.jsp?2017To14,B}{2017To14})\vspace{0.1cm}}&\\
&&&&&\parbox[t][0.3cm]{11.22276cm}{\raggedright \ensuremath{\theta}\ensuremath{^{\textnormal{2}}_{\ensuremath{\alpha}}}=0.51 \textit{2} (\href{https://www.nndc.bnl.gov/nsr/nsrlink.jsp?2017To14,B}{2017To14}).\vspace{0.1cm}}&\\
&&&&&\parbox[t][0.3cm]{11.22276cm}{\raggedright Proposed mirror state: \ensuremath{^{\textnormal{19}}}F*(8199 keV, 5/2\ensuremath{^{\textnormal{+}}}) (\href{https://www.nndc.bnl.gov/nsr/nsrlink.jsp?2019La08,B}{2019La08}). But, that \ensuremath{^{\textnormal{19}}}F* state has\vspace{0.1cm}}&\\
&&&&&\parbox[t][0.3cm]{11.22276cm}{\raggedright {\ }{\ }{\ }a total width that is an order of magnitude smaller (see \href{https://www.nndc.bnl.gov/nsr/nsrlink.jsp?2019La08,B}{2019La08}) than that of\vspace{0.1cm}}&\\
&&&&&\parbox[t][0.3cm]{11.22276cm}{\raggedright {\ }{\ }{\ }the \ensuremath{^{\textnormal{19}}}Ne* state.\vspace{0.1cm}}&\\
\multicolumn{1}{r@{}}{8428}&\multicolumn{1}{@{}l}{\ensuremath{^{{\hyperlink{NE13LEVEL1}{b}}}} {\it 2}}&\multicolumn{1}{l}{(13/2\ensuremath{^{-}},11/2\ensuremath{^{+}})}&\multicolumn{1}{r@{}}{4}&\multicolumn{1}{@{ }l}{keV {\it 1}}&\parbox[t][0.3cm]{11.22276cm}{\raggedright \ensuremath{\Gamma}\ensuremath{\alpha}=4 keV \textit{1} (\href{https://www.nndc.bnl.gov/nsr/nsrlink.jsp?2017To14,B}{2017To14})\vspace{0.1cm}}&\\
&&&&&\parbox[t][0.3cm]{11.22276cm}{\raggedright (\href{https://www.nndc.bnl.gov/nsr/nsrlink.jsp?2017To14,B}{2017To14}) reports that this state may be a member of the \ensuremath{^{\textnormal{15}}}O+\ensuremath{\alpha} rotational band\vspace{0.1cm}}&\\
&&&&&\parbox[t][0.3cm]{11.22276cm}{\raggedright {\ }{\ }{\ }(see Table II).\vspace{0.1cm}}&\\
&&&&&\parbox[t][0.3cm]{11.22276cm}{\raggedright \ensuremath{\theta}\ensuremath{^{\textnormal{2}}_{\ensuremath{\alpha}}}=0.31 \textit{4} for J\ensuremath{^{\ensuremath{\pi}}}=13/2\ensuremath{^{-}} (\href{https://www.nndc.bnl.gov/nsr/nsrlink.jsp?2017To14,B}{2017To14}).\vspace{0.1cm}}&\\
&&&&&\parbox[t][0.3cm]{11.22276cm}{\raggedright Proposed mirror state: \ensuremath{^{\textnormal{19}}}F*(8288 keV, 13/2\ensuremath{^{-}}) (\href{https://www.nndc.bnl.gov/nsr/nsrlink.jsp?2019La08,B}{2019La08}).\vspace{0.1cm}}&\\
\multicolumn{1}{r@{}}{8680}&\multicolumn{1}{@{}l}{\ensuremath{^{{\hyperlink{NE13LEVEL1}{b}}}} {\it 1}}&\multicolumn{1}{l}{(9/2\ensuremath{^{-}},7/2)}&\multicolumn{1}{r@{}}{3}&\multicolumn{1}{@{ }l}{keV {\it 1}}&\parbox[t][0.3cm]{11.22276cm}{\raggedright \ensuremath{\Gamma}\ensuremath{\alpha}=3 keV \textit{1} (\href{https://www.nndc.bnl.gov/nsr/nsrlink.jsp?2017To14,B}{2017To14})\vspace{0.1cm}}&\\
&&&&&\parbox[t][0.3cm]{11.22276cm}{\raggedright \ensuremath{\theta}\ensuremath{^{\textnormal{2}}_{\ensuremath{\alpha}}}=0.054 \textit{6} for J\ensuremath{^{\ensuremath{\pi}}}=9/2\ensuremath{^{-}} (\href{https://www.nndc.bnl.gov/nsr/nsrlink.jsp?2017To14,B}{2017To14}).\vspace{0.1cm}}&\\
&&&&&\parbox[t][0.3cm]{11.22276cm}{\raggedright J\ensuremath{^{\pi}}: (\href{https://www.nndc.bnl.gov/nsr/nsrlink.jsp?2017To14,B}{2017To14}) determined J\ensuremath{^{\ensuremath{\pi}}}=(9/2\ensuremath{^{-}}, 7/2\ensuremath{^{-}}) from R-matrix.\vspace{0.1cm}}&\\
&&&&&\parbox[t][0.3cm]{11.22276cm}{\raggedright Proposed mirror state: \ensuremath{^{\textnormal{19}}}F*(8943.9 keV, 7/2\ensuremath{^{\textnormal{+}}}) (\href{https://www.nndc.bnl.gov/nsr/nsrlink.jsp?2019La08,B}{2019La08}). Evaluator highlights\vspace{0.1cm}}&\\
&&&&&\parbox[t][0.3cm]{11.22276cm}{\raggedright {\ }{\ }{\ }that the \ensuremath{^{\textnormal{19}}}F* state has a positive parity as opposed to the negative parity\vspace{0.1cm}}&\\
&&&&&\parbox[t][0.3cm]{11.22276cm}{\raggedright {\ }{\ }{\ }assigned to this \ensuremath{^{\textnormal{19}}}Ne* state by (\href{https://www.nndc.bnl.gov/nsr/nsrlink.jsp?2017To14,B}{2017To14}). Therefore, we did not adopt the\vspace{0.1cm}}&\\
&&&&&\parbox[t][0.3cm]{11.22276cm}{\raggedright {\ }{\ }{\ }parity for J=7/2 here.\vspace{0.1cm}}&\\
\multicolumn{1}{r@{}}{8790?}&\multicolumn{1}{@{}l}{}&&&&\parbox[t][0.3cm]{11.22276cm}{\raggedright \ensuremath{\Gamma}=\ensuremath{\Gamma}\ensuremath{_{\ensuremath{\alpha}}}=4 keV \textit{1} (\href{https://www.nndc.bnl.gov/nsr/nsrlink.jsp?2017To14,B}{2017To14}) from R-matrix.\vspace{0.1cm}}&\\
&&&&&\parbox[t][0.3cm]{11.22276cm}{\raggedright J\ensuremath{^{\pi}}: (\href{https://www.nndc.bnl.gov/nsr/nsrlink.jsp?2017To14,B}{2017To14}) determined J\ensuremath{^{\ensuremath{\pi}}}=(11/2) from their R-matrix analysis.\vspace{0.1cm}}&\\
&&&&&\parbox[t][0.3cm]{11.22276cm}{\raggedright (\href{https://www.nndc.bnl.gov/nsr/nsrlink.jsp?2017To14,B}{2017To14}) reports that this state may be part of the \ensuremath{^{\textnormal{15}}}O+\ensuremath{\alpha} rotational band (see\vspace{0.1cm}}&\\
&&&&&\parbox[t][0.3cm]{11.22276cm}{\raggedright {\ }{\ }{\ }Table II).\vspace{0.1cm}}&\\
&&&&&\parbox[t][0.3cm]{11.22276cm}{\raggedright \ensuremath{\theta}\ensuremath{^{\textnormal{2}}_{\ensuremath{\alpha}}}=0.10 \textit{3} (\href{https://www.nndc.bnl.gov/nsr/nsrlink.jsp?2017To14,B}{2017To14}) for J=(11/2).\vspace{0.1cm}}&\\
&&&&&\parbox[t][0.3cm]{11.22276cm}{\raggedright (\href{https://www.nndc.bnl.gov/nsr/nsrlink.jsp?2019La08,B}{2019La08}) paired the \ensuremath{^{\textnormal{19}}}F*(8953 keV, 9/2\ensuremath{^{-}}) level with this \ensuremath{^{\textnormal{19}}}Ne level. However,\vspace{0.1cm}}&\\
&&&&&\parbox[t][0.3cm]{11.22276cm}{\raggedright {\ }{\ }{\ }the \ensuremath{^{\textnormal{19}}}F* level has J\ensuremath{^{\ensuremath{\pi}}}=9/2\ensuremath{^{-}} and a total width that is 5 times as large (see\vspace{0.1cm}}&\\
&&&&&\parbox[t][0.3cm]{11.22276cm}{\raggedright {\ }{\ }{\ }\href{https://www.nndc.bnl.gov/nsr/nsrlink.jsp?2019La08,B}{2019La08}) as that of the \ensuremath{^{\textnormal{19}}}Ne* level. These findings cast doubt on the results\vspace{0.1cm}}&\\
\end{longtable}
\begin{textblock}{29}(0,27.3)
Continued on next page (footnotes at end of table)
\end{textblock}
\clearpage
\begin{longtable}{cccccc@{\extracolsep{\fill}}c}
\\[-.4cm]
\multicolumn{7}{c}{{\bf \small \underline{\ensuremath{^{\textnormal{4}}}He(\ensuremath{^{\textnormal{15}}}O,\ensuremath{\alpha}):res\hspace{0.2in}\href{https://www.nndc.bnl.gov/nsr/nsrlink.jsp?2017To14,B}{2017To14} (continued)}}}\\
\multicolumn{7}{c}{~}\\
\multicolumn{7}{c}{\underline{\ensuremath{^{19}}Ne Levels (continued)}}\\
\multicolumn{7}{c}{~}\\
\multicolumn{2}{c}{E(level)$^{{\hyperlink{NE13LEVEL0}{a}}}$}&J$^{\pi}$$^{{\hyperlink{NE13LEVEL4}{e}}}$&\multicolumn{2}{c}{T\ensuremath{_{\textnormal{1/2}}} or \ensuremath{\Gamma}$^{{\hyperlink{NE13LEVEL3}{d}}}$}&Comments&\\[-.2cm]
\multicolumn{2}{c}{\hrulefill}&\hrulefill&\multicolumn{2}{c}{\hrulefill}&\hrulefill&
\endhead
&&&&&\parbox[t][0.3cm]{12.889821cm}{\raggedright {\ }{\ }{\ }of (\href{https://www.nndc.bnl.gov/nsr/nsrlink.jsp?2017To14,B}{2017To14}), and therefore, we did not adopt \ensuremath{\Gamma}, \ensuremath{\Gamma}\ensuremath{_{\ensuremath{\alpha}}} and J from the latter study.\vspace{0.1cm}}&\\
\end{longtable}
\parbox[b][0.3cm]{17.7cm}{\makebox[1ex]{\ensuremath{^{\hypertarget{NE13LEVEL0}{a}}}} From (\href{https://www.nndc.bnl.gov/nsr/nsrlink.jsp?2017To14,B}{2017To14}) unless otherwise noted. The uncertainties in the excitation energies from (\href{https://www.nndc.bnl.gov/nsr/nsrlink.jsp?2017To14,B}{2017To14}) are statistical only. An}\\
\parbox[b][0.3cm]{17.7cm}{{\ }{\ }additional \ensuremath{\approx}50 keV systematic uncertainty should be added in quadrature to each E\ensuremath{_{\textnormal{x}}}(\ensuremath{^{\textnormal{19}}}Ne) measured by (\href{https://www.nndc.bnl.gov/nsr/nsrlink.jsp?2017To14,B}{2017To14}: See Table}\\
\parbox[b][0.3cm]{17.7cm}{{\ }{\ }I).}\\
\parbox[b][0.3cm]{17.7cm}{\makebox[1ex]{\ensuremath{^{\hypertarget{NE13LEVEL1}{b}}}} State observed for the first time in (\href{https://www.nndc.bnl.gov/nsr/nsrlink.jsp?2017To14,B}{2017To14}).}\\
\parbox[b][0.3cm]{17.7cm}{\makebox[1ex]{\ensuremath{^{\hypertarget{NE13LEVEL2}{c}}}} From the theoretical results of (\href{https://www.nndc.bnl.gov/nsr/nsrlink.jsp?2021Sa42,B}{2021Sa42}).}\\
\parbox[b][0.3cm]{17.7cm}{\makebox[1ex]{\ensuremath{^{\hypertarget{NE13LEVEL3}{d}}}} From (\href{https://www.nndc.bnl.gov/nsr/nsrlink.jsp?2017To14,B}{2017To14}) unless otherwise noted.}\\
\parbox[b][0.3cm]{17.7cm}{\makebox[1ex]{\ensuremath{^{\hypertarget{NE13LEVEL4}{e}}}} From R-matrix analysis in (\href{https://www.nndc.bnl.gov/nsr/nsrlink.jsp?2017To14,B}{2017To14}) using the AZURE2 code unless otherwise noted.}\\
\vspace{0.5cm}
\clearpage
\subsection[\hspace{-0.2cm}\ensuremath{^{\textnormal{7}}}Li(\ensuremath{^{\textnormal{15}}}O,t\ensuremath{\gamma})]{ }
\vspace{-27pt}
\vspace{0.3cm}
\hypertarget{NE14}{{\bf \small \underline{\ensuremath{^{\textnormal{7}}}Li(\ensuremath{^{\textnormal{15}}}O,t\ensuremath{\gamma})\hspace{0.2in}\href{https://www.nndc.bnl.gov/nsr/nsrlink.jsp?2021As10,B}{2021As10}}}}\\
\vspace{4pt}
\vspace{8pt}
\parbox[b][0.3cm]{17.7cm}{\addtolength{\parindent}{-0.2in}\ensuremath{^{\textnormal{15}}}O(\ensuremath{^{\textnormal{7}}}Li,t) \ensuremath{\alpha}-transfer reaction in inverse kinematics.}\\
\parbox[b][0.3cm]{17.7cm}{\addtolength{\parindent}{-0.2in}J\ensuremath{^{\ensuremath{\pi}}}(\ensuremath{^{\textnormal{7}}}Li\ensuremath{_{\textnormal{g.s.}}})=3/2\ensuremath{^{-}} and J\ensuremath{^{\ensuremath{\pi}}}(\ensuremath{^{\textnormal{15}}}O\ensuremath{_{\textnormal{g.s.}}})=1/2\ensuremath{^{-}}.}\\
\parbox[b][0.3cm]{17.7cm}{\addtolength{\parindent}{-0.2in}\href{https://www.nndc.bnl.gov/nsr/nsrlink.jsp?2021As10,B}{2021As10}: \ensuremath{^{\textnormal{7}}}Li(\ensuremath{^{\textnormal{15}}}O,t\ensuremath{\gamma}) E=4.7 MeV/nucleon; measured t-\ensuremath{\gamma}-\ensuremath{^{\textnormal{19}}}Ne triple coincidence events; measured the energy and emission angle}\\
\parbox[b][0.3cm]{17.7cm}{of tritons using the MUGAST array that consisted of 13 position sensitive Si detectors, including 4 detectors of MUST2 array that}\\
\parbox[b][0.3cm]{17.7cm}{were placed at forward angles covering \ensuremath{\theta}\ensuremath{_{\textnormal{lab}}}=8\ensuremath{^\circ}{\textminus}50\ensuremath{^\circ}. The remaining detectors covered 70\% of \ensuremath{\phi} of the backward hemisphere. The}\\
\parbox[b][0.3cm]{17.7cm}{\ensuremath{\gamma} rays were measured using the AGATA array that consisted of 41 HPGe detectors placed in the backward angles surrounding the}\\
\parbox[b][0.3cm]{17.7cm}{MUGAST array. The rigidity and TOF of the \ensuremath{^{\textnormal{19}}}Ne recoils were measured using the large acceptance VAMOS++ spectrometer}\\
\parbox[b][0.3cm]{17.7cm}{placed at \ensuremath{\theta}\ensuremath{_{\textnormal{lab}}}=0\ensuremath{^\circ} covering an angular acceptance of \ensuremath{\pm}4.6\ensuremath{^\circ}. A partial energy spectrum for the \ensuremath{^{\textnormal{7}}}Li(\ensuremath{^{\textnormal{15}}}O,t\ensuremath{\gamma}) reaction was presented.}\\
\parbox[b][0.3cm]{17.7cm}{Obtained \ensuremath{^{\textnormal{19}}}Ne levels using the missing mass analysis techniques.}\\
\vspace{12pt}
\underline{$^{19}$Ne Levels}\\
\begin{longtable}{cccc@{\extracolsep{\fill}}c}
\multicolumn{2}{c}{E(level)$^{{\hyperlink{NE14LEVEL0}{a}}}$}&J$^{\pi}$$^{{\hyperlink{NE14LEVEL2}{c}}}$&Comments&\\[-.2cm]
\multicolumn{2}{c}{\hrulefill}&\hrulefill&\hrulefill&
\endfirsthead
\multicolumn{1}{r@{}}{0}&\multicolumn{1}{@{}l}{\ensuremath{^{{\hyperlink{NE14LEVEL1}{b}}}}}&\multicolumn{1}{l}{1/2\ensuremath{^{+}}}&&\\
\multicolumn{1}{r@{}}{238}&\multicolumn{1}{@{}l}{\ensuremath{^{{\hyperlink{NE14LEVEL1}{b}}}}}&\multicolumn{1}{l}{5/2\ensuremath{^{+}}}&&\\
\multicolumn{1}{r@{}}{275}&\multicolumn{1}{@{}l}{\ensuremath{^{{\hyperlink{NE14LEVEL1}{b}}}}}&\multicolumn{1}{l}{1/2\ensuremath{^{-}}}&\parbox[t][0.3cm]{14.85416cm}{\raggedright E(level): This state was directly populated in (\href{https://www.nndc.bnl.gov/nsr/nsrlink.jsp?2021As10,B}{2021As10}: See text).\vspace{0.1cm}}&\\
\multicolumn{1}{r@{}}{1507}&\multicolumn{1}{@{}l}{\ensuremath{^{{\hyperlink{NE14LEVEL1}{b}}}}}&\multicolumn{1}{l}{5/2\ensuremath{^{-}}}&&\\
\multicolumn{1}{r@{}}{1536}&\multicolumn{1}{@{}l}{\ensuremath{^{{\hyperlink{NE14LEVEL1}{b}}}}}&\multicolumn{1}{l}{3/2\ensuremath{^{+}}}&&\\
\multicolumn{1}{r@{}}{4033}&\multicolumn{1}{@{}l}{}&\multicolumn{1}{l}{3/2\ensuremath{^{+}}}&&\\
\multicolumn{1}{r@{}}{4139}&\multicolumn{1}{@{}l}{}&\multicolumn{1}{l}{(7/2\ensuremath{^{-}})}&&\\
\multicolumn{1}{r@{}}{4197}&\multicolumn{1}{@{.}l}{1}&\multicolumn{1}{l}{(9/2\ensuremath{^{-}})}&&\\
\end{longtable}
\parbox[b][0.3cm]{17.7cm}{\makebox[1ex]{\ensuremath{^{\hypertarget{NE14LEVEL0}{a}}}} Unless otherwise notes, these values are deduced from E\ensuremath{_{\ensuremath{\gamma}}} values from Fig. 18 of (\href{https://www.nndc.bnl.gov/nsr/nsrlink.jsp?2021As10,B}{2021As10}) with nuclear recoil corrections}\\
\parbox[b][0.3cm]{17.7cm}{{\ }{\ }applied.}\\
\parbox[b][0.3cm]{17.7cm}{\makebox[1ex]{\ensuremath{^{\hypertarget{NE14LEVEL1}{b}}}} From the \ensuremath{^{\textnormal{19}}}Ne Adopted Levels rounded to the nearest integer.}\\
\parbox[b][0.3cm]{17.7cm}{\makebox[1ex]{\ensuremath{^{\hypertarget{NE14LEVEL2}{c}}}} From the \ensuremath{^{\textnormal{19}}}Ne Adopted Levels.}\\
\vspace{0.5cm}
\underline{$\gamma$($^{19}$Ne)}\\
\begin{longtable}{ccccccccc@{}cc@{\extracolsep{\fill}}c}
\multicolumn{2}{c}{E\ensuremath{_{i}}(level)}&J\ensuremath{^{\pi}_{i}}&\multicolumn{2}{c}{E\ensuremath{_{\gamma}}\ensuremath{^{\hyperlink{NE14GAMMA0}{a}}}}&\multicolumn{2}{c}{I\ensuremath{_{\gamma}}\ensuremath{^{\hyperlink{NE14GAMMA0}{a}}}}&\multicolumn{2}{c}{E\ensuremath{_{f}}}&J\ensuremath{^{\pi}_{f}}&Comments&\\[-.2cm]
\multicolumn{2}{c}{\hrulefill}&\hrulefill&\multicolumn{2}{c}{\hrulefill}&\multicolumn{2}{c}{\hrulefill}&\multicolumn{2}{c}{\hrulefill}&\hrulefill&\hrulefill&
\endfirsthead
\multicolumn{1}{r@{}}{275}&\multicolumn{1}{@{}l}{}&\multicolumn{1}{l}{1/2\ensuremath{^{-}}}&\multicolumn{1}{r@{}}{274}&\multicolumn{1}{@{.}l}{99}&\multicolumn{1}{r@{}}{}&\multicolumn{1}{@{}l}{}&\multicolumn{1}{r@{}}{0}&\multicolumn{1}{@{}l}{}&\multicolumn{1}{@{}l}{1/2\ensuremath{^{+}}}&\parbox[t][0.3cm]{9.985801cm}{\raggedright E\ensuremath{_{\gamma}}: (\href{https://www.nndc.bnl.gov/nsr/nsrlink.jsp?2021As10,B}{2021As10}): This \ensuremath{\gamma} ray could either be from the direct population of\vspace{0.1cm}}&\\
&&&&&&&&&&\parbox[t][0.3cm]{9.985801cm}{\raggedright {\ }{\ }{\ }the 275-keV state, or by feeding the 275-keV state from the decay of\vspace{0.1cm}}&\\
&&&&&&&&&&\parbox[t][0.3cm]{9.985801cm}{\raggedright {\ }{\ }{\ }higher-lying \ensuremath{^{\textnormal{19}}}Ne states up to 10 MeV (see text). However,\vspace{0.1cm}}&\\
&&&&&&&&&&\parbox[t][0.3cm]{9.985801cm}{\raggedright {\ }{\ }{\ }(\href{https://www.nndc.bnl.gov/nsr/nsrlink.jsp?2021As10,B}{2021As10}) only presents a partial spectrum of \ensuremath{^{\textnormal{19}}}Ne. So, the energy of\vspace{0.1cm}}&\\
&&&&&&&&&&\parbox[t][0.3cm]{9.985801cm}{\raggedright {\ }{\ }{\ }this \ensuremath{\gamma} ray is taken from the \ensuremath{^{\textnormal{19}}}Ne Adopted Gammas.\vspace{0.1cm}}&\\
\multicolumn{1}{r@{}}{4033}&\multicolumn{1}{@{}l}{}&\multicolumn{1}{l}{3/2\ensuremath{^{+}}}&\multicolumn{1}{r@{}}{2497}&\multicolumn{1}{@{}l}{}&\multicolumn{1}{r@{}}{15}&\multicolumn{1}{@{}l}{\ensuremath{^{\hyperlink{NE14GAMMA1}{b}}} {\it 5}}&\multicolumn{1}{r@{}}{1536}&\multicolumn{1}{@{}l}{}&\multicolumn{1}{@{}l}{3/2\ensuremath{^{+}}}&&\\
&&&\multicolumn{1}{r@{}}{3758}&\multicolumn{1}{@{}l}{}&\multicolumn{1}{r@{}}{5}&\multicolumn{1}{@{}l}{\ensuremath{^{\hyperlink{NE14GAMMA1}{b}}} {\it 5}}&\multicolumn{1}{r@{}}{275}&\multicolumn{1}{@{}l}{}&\multicolumn{1}{@{}l}{1/2\ensuremath{^{-}}}&&\\
&&&\multicolumn{1}{r@{}}{4033}&\multicolumn{1}{@{}l}{}&\multicolumn{1}{r@{}}{80}&\multicolumn{1}{@{}l}{\ensuremath{^{\hyperlink{NE14GAMMA1}{b}}} {\it 15}}&\multicolumn{1}{r@{}}{0}&\multicolumn{1}{@{}l}{}&\multicolumn{1}{@{}l}{1/2\ensuremath{^{+}}}&&\\
\multicolumn{1}{r@{}}{4139}&\multicolumn{1}{@{}l}{}&\multicolumn{1}{l}{(7/2\ensuremath{^{-}})}&\multicolumn{1}{r@{}}{2632}&\multicolumn{1}{@{}l}{}&\multicolumn{1}{r@{}}{100}&\multicolumn{1}{@{}l}{}&\multicolumn{1}{r@{}}{1507}&\multicolumn{1}{@{}l}{}&\multicolumn{1}{@{}l}{5/2\ensuremath{^{-}}}&&\\
\multicolumn{1}{r@{}}{4197}&\multicolumn{1}{@{.}l}{1}&\multicolumn{1}{l}{(9/2\ensuremath{^{-}})}&\multicolumn{1}{r@{}}{2689}&\multicolumn{1}{@{.}l}{5}&\multicolumn{1}{r@{}}{80}&\multicolumn{1}{@{ }l}{{\it 5}}&\multicolumn{1}{r@{}}{1507}&\multicolumn{1}{@{}l}{}&\multicolumn{1}{@{}l}{5/2\ensuremath{^{-}}}&&\\
&&&\multicolumn{1}{r@{}}{3958}&\multicolumn{1}{@{.}l}{8\ensuremath{^{\hyperlink{NE14GAMMA2}{c}}}}&\multicolumn{1}{r@{}}{20}&\multicolumn{1}{@{ }l}{{\it 5}}&\multicolumn{1}{r@{}}{238}&\multicolumn{1}{@{}l}{}&\multicolumn{1}{@{}l}{5/2\ensuremath{^{+}}}&\parbox[t][0.3cm]{9.985801cm}{\raggedright E\ensuremath{_{\gamma}}: We highlight that the E\ensuremath{_{\ensuremath{\gamma}}}=3958.8 keV was measured by (\href{https://www.nndc.bnl.gov/nsr/nsrlink.jsp?1971Da31,B}{1971Da31}:\vspace{0.1cm}}&\\
&&&&&&&&&&\parbox[t][0.3cm]{9.985801cm}{\raggedright {\ }{\ }{\ }\ensuremath{^{\textnormal{17}}}O(\ensuremath{^{\textnormal{3}}}He,n\ensuremath{\gamma})) as a weak transition from the decay of the \ensuremath{^{\textnormal{19}}}Ne*(4197)\vspace{0.1cm}}&\\
&&&&&&&&&&\parbox[t][0.3cm]{9.985801cm}{\raggedright {\ }{\ }{\ }level to the \ensuremath{^{\textnormal{19}}}Ne*(238) level; however, (\href{https://www.nndc.bnl.gov/nsr/nsrlink.jsp?2020Ha31,B}{2020Ha31}: \ensuremath{^{\textnormal{19}}}F(\ensuremath{^{\textnormal{3}}}He,t\ensuremath{\gamma}))\vspace{0.1cm}}&\\
&&&&&&&&&&\parbox[t][0.3cm]{9.985801cm}{\raggedright {\ }{\ }{\ }claims that this transition is erroneous. From Fig. 18 of (\href{https://www.nndc.bnl.gov/nsr/nsrlink.jsp?2021As10,B}{2021As10}), it\vspace{0.1cm}}&\\
&&&&&&&&&&\parbox[t][0.3cm]{9.985801cm}{\raggedright {\ }{\ }{\ }is unclear if that study observed this \ensuremath{\gamma} ray or if it is simply being\vspace{0.1cm}}&\\
&&&&&&&&&&\parbox[t][0.3cm]{9.985801cm}{\raggedright {\ }{\ }{\ }reported from the literature.\vspace{0.1cm}}&\\
\end{longtable}
\parbox[b][0.3cm]{17.7cm}{\makebox[1ex]{\ensuremath{^{\hypertarget{NE14GAMMA0}{a}}}} From (\href{https://www.nndc.bnl.gov/nsr/nsrlink.jsp?2021As10,B}{2021As10}: See Fig. 18) and most likely reported from (\href{https://www.nndc.bnl.gov/nsr/nsrlink.jsp?1971Da31,B}{1971Da31}: \ensuremath{^{\textnormal{17}}}O(\ensuremath{^{\textnormal{3}}}He,n\ensuremath{\gamma})).}\\
\parbox[b][0.3cm]{17.7cm}{\makebox[1ex]{\ensuremath{^{\hypertarget{NE14GAMMA1}{b}}}} This value is from (\href{https://www.nndc.bnl.gov/nsr/nsrlink.jsp?1973Da31,B}{1973Da31}: \ensuremath{^{\textnormal{17}}}O(\ensuremath{^{\textnormal{3}}}He,n\ensuremath{\gamma})), which is cited by (\href{https://www.nndc.bnl.gov/nsr/nsrlink.jsp?2021As10,B}{2021As10}).}\\
\parbox[b][0.3cm]{17.7cm}{\makebox[1ex]{\ensuremath{^{\hypertarget{NE14GAMMA2}{c}}}} Placement of transition in the level scheme is uncertain.}\\
\vspace{0.5cm}
\clearpage
\begin{figure}[h]
\begin{center}
\includegraphics{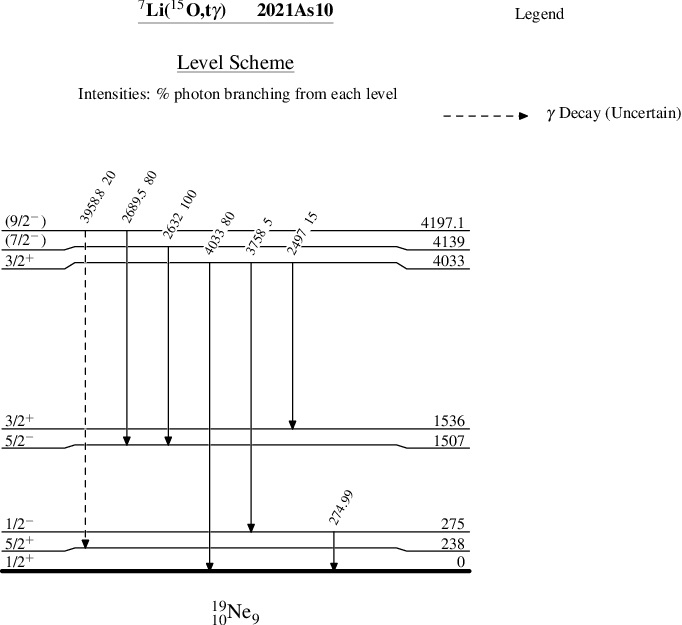}\\
\end{center}
\end{figure}
\clearpage
\subsection[\hspace{-0.2cm}\ensuremath{^{\textnormal{9}}}Be(\ensuremath{^{\textnormal{20}}}Ne,\ensuremath{^{\textnormal{19}}}Ne)]{ }
\vspace{-27pt}
\vspace{0.3cm}
\hypertarget{NE15}{{\bf \small \underline{\ensuremath{^{\textnormal{9}}}Be(\ensuremath{^{\textnormal{20}}}Ne,\ensuremath{^{\textnormal{19}}}Ne)\hspace{0.2in}\href{https://www.nndc.bnl.gov/nsr/nsrlink.jsp?1992Ge08,B}{1992Ge08},\href{https://www.nndc.bnl.gov/nsr/nsrlink.jsp?1997Kl06,B}{1997Kl06}}}}\\
\vspace{4pt}
\vspace{8pt}
\parbox[b][0.3cm]{17.7cm}{\addtolength{\parindent}{-0.2in}One-neutron knockout reaction in inverse kinematics.}\\
\parbox[b][0.3cm]{17.7cm}{\addtolength{\parindent}{-0.2in}J\ensuremath{^{\ensuremath{\pi}}}(\ensuremath{^{\textnormal{9}}}Be\ensuremath{_{\textnormal{g.s.}}})=3/2\ensuremath{^{-}} and J\ensuremath{^{\ensuremath{\pi}}}(\ensuremath{^{\textnormal{20}}}Ne\ensuremath{_{\textnormal{g.s.}}})=0\ensuremath{^{\textnormal{+}}}.}\\
\parbox[b][0.3cm]{17.7cm}{\addtolength{\parindent}{-0.2in}\href{https://www.nndc.bnl.gov/nsr/nsrlink.jsp?1992Ge08,B}{1992Ge08}, \href{https://www.nndc.bnl.gov/nsr/nsrlink.jsp?1997Kl06,B}{1997Kl06}: \ensuremath{^{\textnormal{9}}}Be(\ensuremath{^{\textnormal{20}}}Ne,\ensuremath{^{\textnormal{19}}}Ne) E=310 MeV/nucleon; produced a \ensuremath{^{\textnormal{19}}}Ne secondary beam from projectile fragmentation of a}\\
\parbox[b][0.3cm]{17.7cm}{\ensuremath{^{\textnormal{20}}}Ne primary beam on a \ensuremath{^{\textnormal{9}}}Be target placed at the entrance of the high resolution Forward Spectrometer (FRS) at GSI. Momentum}\\
\parbox[b][0.3cm]{17.7cm}{analyzed the \ensuremath{^{\textnormal{19}}}Ne beam using FRS; injected and accumulated (for 120 s) the \ensuremath{^{\textnormal{19}}}Ne beam in the GSI experimental storage ring}\\
\parbox[b][0.3cm]{17.7cm}{(ESR); measured the decay curve of \ensuremath{^{\textnormal{19}}}Ne using the ESR$'$s particle detector to measure the fully ionized \ensuremath{^{\textnormal{19}}}F decay products from}\\
\parbox[b][0.3cm]{17.7cm}{the in-flight \ensuremath{\beta}-decay of \ensuremath{^{\textnormal{19}}}Ne. Deduced the half-life of \ensuremath{^{\textnormal{19}}}Ne\ensuremath{_{\textnormal{g.s.}}} as T\ensuremath{_{\textnormal{1/2}}}=18.5 s \textit{6}. Evaluator notes that the \ensuremath{^{\textnormal{19}}}Ne beam in the ESR}\\
\parbox[b][0.3cm]{17.7cm}{was contaminated with \ensuremath{^{\textnormal{15}}}O by 11\%.}\\
\vspace{12pt}
\underline{$^{19}$Ne Levels}\\
\begin{longtable}{cccccc@{\extracolsep{\fill}}c}
\multicolumn{2}{c}{E(level)$^{{\hyperlink{NE15LEVEL0}{a}}}$}&J$^{\pi}$$^{{\hyperlink{NE15LEVEL0}{a}}}$&\multicolumn{2}{c}{T\ensuremath{_{\textnormal{1/2}}}$^{}$}&Comments&\\[-.2cm]
\multicolumn{2}{c}{\hrulefill}&\hrulefill&\multicolumn{2}{c}{\hrulefill}&\hrulefill&
\endfirsthead
\multicolumn{1}{r@{}}{0}&\multicolumn{1}{@{}l}{}&\multicolumn{1}{l}{1/2\ensuremath{^{+}}}&\multicolumn{1}{r@{}}{18}&\multicolumn{1}{@{.}l}{5 s {\it 6}}&\parbox[t][0.3cm]{13.416261cm}{\raggedright T\ensuremath{_{1/2}}: From (\href{https://www.nndc.bnl.gov/nsr/nsrlink.jsp?1992Ge08,B}{1992Ge08}, \href{https://www.nndc.bnl.gov/nsr/nsrlink.jsp?1997Kl06,B}{1997Kl06}).\vspace{0.1cm}}&\\
\end{longtable}
\parbox[b][0.3cm]{17.7cm}{\makebox[1ex]{\ensuremath{^{\hypertarget{NE15LEVEL0}{a}}}} From the \ensuremath{^{\textnormal{19}}}Ne Adopted Levels.}\\
\vspace{0.5cm}
\clearpage
\subsection[\hspace{-0.2cm}\ensuremath{^{\textnormal{10}}}B(\ensuremath{^{\textnormal{10}}}B,\ensuremath{^{\textnormal{19}}}Ne),(\ensuremath{^{\textnormal{11}}}B,\ensuremath{^{\textnormal{19}}}Ne)]{ }
\vspace{-27pt}
\vspace{0.3cm}
\hypertarget{NE16}{{\bf \small \underline{\ensuremath{^{\textnormal{10}}}B(\ensuremath{^{\textnormal{10}}}B,\ensuremath{^{\textnormal{19}}}Ne),(\ensuremath{^{\textnormal{11}}}B,\ensuremath{^{\textnormal{19}}}Ne)\hspace{0.2in}\href{https://www.nndc.bnl.gov/nsr/nsrlink.jsp?1976Hi05,B}{1976Hi05}}}}\\
\vspace{4pt}
\vspace{8pt}
\parbox[b][0.3cm]{17.7cm}{\addtolength{\parindent}{-0.2in}J\ensuremath{^{\ensuremath{\pi}}}(\ensuremath{^{\textnormal{10}}}B\ensuremath{_{\textnormal{g.s.}}})=3\ensuremath{^{\textnormal{+}}} and J\ensuremath{^{\ensuremath{\pi}}}(\ensuremath{^{\textnormal{11}}}B\ensuremath{_{\textnormal{g.s.}}})=3/2\ensuremath{^{-}}.}\\
\parbox[b][0.3cm]{17.7cm}{\addtolength{\parindent}{-0.2in}\href{https://www.nndc.bnl.gov/nsr/nsrlink.jsp?1976Hi05,B}{1976Hi05}: \ensuremath{^{\textnormal{10}}}B(\ensuremath{^{\textnormal{10}}}B,\ensuremath{^{\textnormal{19}}}Ne) E\ensuremath{_{\textnormal{c.m.}}}=1.84-3.66 MeV, and \ensuremath{^{\textnormal{11}}}B(\ensuremath{^{\textnormal{10}}}B,\ensuremath{^{\textnormal{19}}}Ne) E=1.61-3.94 MeV; measured \ensuremath{\gamma} rays emitted from the decay}\\
\parbox[b][0.3cm]{17.7cm}{of the populated residual nuclei, including \ensuremath{^{\textnormal{19}}}Ne, using two NaI detectors at \ensuremath{\theta}\ensuremath{_{\textnormal{lab}}}=0\ensuremath{^\circ} and 180\ensuremath{^\circ}. Deduced total \ensuremath{\sigma}(E) for the}\\
\parbox[b][0.3cm]{17.7cm}{\ensuremath{^{\textnormal{10}}}B+\ensuremath{^{\textnormal{10}}}B, \ensuremath{^{\textnormal{11}}}B+\ensuremath{^{\textnormal{10}}}B and \ensuremath{^{\textnormal{11}}}B+\ensuremath{^{\textnormal{11}}}B reactions at E\ensuremath{_{\textnormal{c.m.}}}=1.84-3.66 MeV, 1.61-3.94 MeV, and 1.56-3.65 MeV, respectively. Presented}\\
\parbox[b][0.3cm]{17.7cm}{optical model analysis.}\\
\vspace{12pt}
\underline{$^{19}$Ne Levels}\\
\begin{longtable}{cc@{\extracolsep{\fill}}c}
\multicolumn{2}{c}{E(level)$^{{\hyperlink{NE16LEVEL0}{a}}}$}&\\[-.2cm]
\multicolumn{2}{c}{\hrulefill}&
\endfirsthead
\multicolumn{1}{r@{}}{0}&\multicolumn{1}{@{}l}{}&\\
\multicolumn{1}{r@{}}{238}&\multicolumn{1}{@{}l}{}&\\
\multicolumn{1}{r@{}}{275}&\multicolumn{1}{@{}l}{}&\\
\end{longtable}
\parbox[b][0.3cm]{17.7cm}{\makebox[1ex]{\ensuremath{^{\hypertarget{NE16LEVEL0}{a}}}} From (\href{https://www.nndc.bnl.gov/nsr/nsrlink.jsp?1976Hi05,B}{1976Hi05}).}\\
\vspace{0.5cm}
\underline{$\gamma$($^{19}$Ne)}\\
\begin{longtable}{cccccc@{\extracolsep{\fill}}c}
\multicolumn{2}{c}{E\ensuremath{_{\gamma}}\ensuremath{^{\hyperlink{NE16GAMMA0}{a}}}}&\multicolumn{2}{c}{E\ensuremath{_{i}}(level)}&\multicolumn{2}{c}{E\ensuremath{_{f}}}&\\[-.2cm]
\multicolumn{2}{c}{\hrulefill}&\multicolumn{2}{c}{\hrulefill}&\multicolumn{2}{c}{\hrulefill}&
\endfirsthead
\multicolumn{1}{r@{}}{238}&\multicolumn{1}{@{}l}{}&\multicolumn{1}{r@{}}{238}&\multicolumn{1}{@{}l}{}&\multicolumn{1}{r@{}}{0}&\multicolumn{1}{@{}l}{}&\\
\multicolumn{1}{r@{}}{275}&\multicolumn{1}{@{}l}{}&\multicolumn{1}{r@{}}{275}&\multicolumn{1}{@{}l}{}&\multicolumn{1}{r@{}}{0}&\multicolumn{1}{@{}l}{}&\\
\end{longtable}
\parbox[b][0.3cm]{17.7cm}{\makebox[1ex]{\ensuremath{^{\hypertarget{NE16GAMMA0}{a}}}} From (\href{https://www.nndc.bnl.gov/nsr/nsrlink.jsp?1976Hi05,B}{1976Hi05}) measured via the \ensuremath{^{\textnormal{10}}}B+\ensuremath{^{\textnormal{11}}}B and \ensuremath{^{\textnormal{10}}}B+\ensuremath{^{\textnormal{10}}}B reactions.}\\
\vspace{0.5cm}
\begin{figure}[h]
\begin{center}
\includegraphics{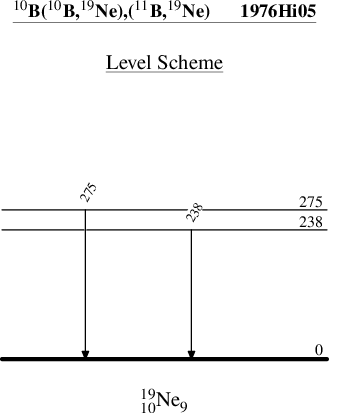}\\
\end{center}
\end{figure}
\clearpage
\subsection[\hspace{-0.2cm}\ensuremath{^{\textnormal{10}}}B(\ensuremath{^{\textnormal{14}}}N,\ensuremath{\alpha}n)]{ }
\vspace{-27pt}
\vspace{0.3cm}
\hypertarget{NE17}{{\bf \small \underline{\ensuremath{^{\textnormal{10}}}B(\ensuremath{^{\textnormal{14}}}N,\ensuremath{\alpha}n)\hspace{0.2in}\href{https://www.nndc.bnl.gov/nsr/nsrlink.jsp?1969Ni09,B}{1969Ni09},\href{https://www.nndc.bnl.gov/nsr/nsrlink.jsp?1978Wu05,B}{1978Wu05},\href{https://www.nndc.bnl.gov/nsr/nsrlink.jsp?1983De26,B}{1983De26}}}}\\
\vspace{4pt}
\vspace{8pt}
\parbox[b][0.3cm]{17.7cm}{\addtolength{\parindent}{-0.2in}J\ensuremath{^{\ensuremath{\pi}}}(\ensuremath{^{\textnormal{10}}}B\ensuremath{_{\textnormal{g.s.}}})=3\ensuremath{^{\textnormal{+}}} and J\ensuremath{^{\ensuremath{\pi}}}(\ensuremath{^{\textnormal{14}}}N\ensuremath{_{\textnormal{g.s.}}})=1\ensuremath{^{\textnormal{+}}}.}\\
\parbox[b][0.3cm]{17.7cm}{\addtolength{\parindent}{-0.2in}M. Fieher, P. Lehmann, A. Leveque and R. Pick, Compt. Rend. 241 (1955) 1946: Potentially \ensuremath{^{\textnormal{10}}}B(\ensuremath{^{\textnormal{14}}}N,\ensuremath{\alpha}n); measured the lifetime}\\
\parbox[b][0.3cm]{17.7cm}{of the \ensuremath{^{\textnormal{19}}}Ne*(275) level as \ensuremath{\tau}\ensuremath{<}5 ns (as cited by \href{https://www.nndc.bnl.gov/nsr/nsrlink.jsp?1969Ni09,B}{1969Ni09}: See Table I in that study).}\\
\parbox[b][0.3cm]{17.7cm}{\addtolength{\parindent}{-0.2in}\href{https://www.nndc.bnl.gov/nsr/nsrlink.jsp?1969Ni09,B}{1969Ni09}: \ensuremath{^{\textnormal{10}}}B(\ensuremath{^{\textnormal{14}}}N,\ensuremath{\alpha}n) E=15 MeV; measured Doppler-shift attenuation for the 275-keV \ensuremath{\gamma} ray from de-excitation of the}\\
\parbox[b][0.3cm]{17.7cm}{\ensuremath{^{\textnormal{19}}}Ne*(275) state using the plunger method. Measured \ensuremath{\gamma} rays using a Ge(Li) detector at \ensuremath{\theta}\ensuremath{_{\textnormal{lab}}}=0\ensuremath{^\circ}. Various beam energies and a}\\
\parbox[b][0.3cm]{17.7cm}{fixed flight path of 0.7 m was used. Deduced lifetime for the \ensuremath{^{\textnormal{19}}}Ne*(275) level.}\\
\parbox[b][0.3cm]{17.7cm}{\addtolength{\parindent}{-0.2in}\href{https://www.nndc.bnl.gov/nsr/nsrlink.jsp?1977Hi01,B}{1977Hi01}:\ensuremath{^{\textnormal{10}}}B(\ensuremath{^{\textnormal{14}}}N,\ensuremath{\alpha}n) E\ensuremath{_{\textnormal{c.m.}}}=2.64-5.97 MeV; measured total fusion cross section, measured prompt \ensuremath{\gamma} rays emitted by the various}\\
\parbox[b][0.3cm]{17.7cm}{residual nuclei, including \ensuremath{^{\textnormal{19}}}Ne, using two NaI detectors. Deduced absolute cross sections.}\\
\parbox[b][0.3cm]{17.7cm}{\addtolength{\parindent}{-0.2in}\href{https://www.nndc.bnl.gov/nsr/nsrlink.jsp?1978Wu05,B}{1978Wu05}: \ensuremath{^{\textnormal{10}}}B(\ensuremath{^{\textnormal{14}}}N,\ensuremath{\alpha}n) E\ensuremath{_{\textnormal{c.m.}}}=2.9-7.5 MeV; measured cross sections and \ensuremath{\gamma}-ray yields for the formation of various residual nuclei}\\
\parbox[b][0.3cm]{17.7cm}{formed by particle evaporation from the \ensuremath{^{\textnormal{24}}}Mg compound system, including \ensuremath{^{\textnormal{19}}}Ne, using a Ge(Li) detector at \ensuremath{\theta}\ensuremath{_{\textnormal{lab}}}=0\ensuremath{^\circ}. Deduced}\\
\parbox[b][0.3cm]{17.7cm}{partial and total fusion cross sections for the \ensuremath{^{\textnormal{10}}}B(\ensuremath{^{\textnormal{14}}}N,\ensuremath{\alpha}n) reaction as a function of incident energy using statistical model}\\
\parbox[b][0.3cm]{17.7cm}{calculations. Deduced summing-branching factor \ensuremath{\beta} and the bound-state fraction F (see text for definitions) for the populated}\\
\parbox[b][0.3cm]{17.7cm}{residual nuclei. Comparison with literature is given.}\\
\parbox[b][0.3cm]{17.7cm}{\addtolength{\parindent}{-0.2in}\href{https://www.nndc.bnl.gov/nsr/nsrlink.jsp?1983De26,B}{1983De26}: \ensuremath{^{\textnormal{10}}}B(\ensuremath{^{\textnormal{14}}}N,\ensuremath{\alpha}n) E\ensuremath{_{\textnormal{c.m.}}}=8-25.75 MeV; measured excitation functions for the production of A=10-23 reaction products using}\\
\parbox[b][0.3cm]{17.7cm}{two Ge(Li) detectors at \ensuremath{\theta}\ensuremath{_{\textnormal{lab}}}=55\ensuremath{^\circ} and \ensuremath{\theta}\ensuremath{_{\textnormal{lab}}}=125\ensuremath{^\circ}. Deduced total fusion \ensuremath{\sigma}(E)=380 mb \textit{5} and critical angular momentum for fusion.}\\
\vspace{12pt}
\underline{$^{19}$Ne Levels}\\
\vspace{0.34cm}
\parbox[b][0.3cm]{17.7cm}{\addtolength{\parindent}{-0.254cm}\textit{Notes}:}\\
\parbox[b][0.3cm]{17.7cm}{\addtolength{\parindent}{-0.254cm}(1) For partial and total \ensuremath{^{\textnormal{10}}}B(\ensuremath{^{\textnormal{14}}}N,\ensuremath{\alpha}n) fusion cross sections as a function of E\ensuremath{_{\textnormal{c.m.}}}, see Table 4 in (\href{https://www.nndc.bnl.gov/nsr/nsrlink.jsp?1978Wu05,B}{1978Wu05}). Note that those}\\
\parbox[b][0.3cm]{17.7cm}{cross sections have 3\% statistical uncertainty. The systematic uncertainty is 15\% at low energies and 20\% at high energies.}\\
\parbox[b][0.3cm]{17.7cm}{\addtolength{\parindent}{-0.254cm}(2) The summing-branching factor, \ensuremath{\beta}, given below from (\href{https://www.nndc.bnl.gov/nsr/nsrlink.jsp?1978Wu05,B}{1978Wu05}) represents the joint probability of formation of the residual}\\
\parbox[b][0.3cm]{17.7cm}{nucleus in the particular state emitting a \ensuremath{\gamma} ray. The de-exciting state is not fed by higher excited states. When the compound}\\
\parbox[b][0.3cm]{17.7cm}{nucleus is formed at sufficiently high excitation energy, the population of a low-lying state of spin J in an evaporation residue is}\\
\parbox[b][0.3cm]{17.7cm}{proportional to 2J+1. \ensuremath{\beta}\ensuremath{_{\textnormal{2J+1}}} shows the effects of this rule on the values of \ensuremath{\beta} (\href{https://www.nndc.bnl.gov/nsr/nsrlink.jsp?1978Wu05,B}{1978Wu05}).}\\
\parbox[b][0.3cm]{17.7cm}{\addtolength{\parindent}{-0.254cm}(3) Level-density parameters were deduced by (\href{https://www.nndc.bnl.gov/nsr/nsrlink.jsp?1978Wu05,B}{1978Wu05}) as T=3.85 MeV, E\ensuremath{_{\textnormal{0}}}={\textminus}6.05 MeV, U$'$=10.4 MeV, a=1.70 MeV\ensuremath{^{\textnormal{$-$1}}}, and}\\
\parbox[b][0.3cm]{17.7cm}{\ensuremath{\Delta}={\textminus}3.99. See (\href{https://www.nndc.bnl.gov/nsr/nsrlink.jsp?1977Ch10,B}{1977Ch10}) for definitions of these parameters.}\\
\vspace{0.34cm}
\begin{longtable}{cccccc@{\extracolsep{\fill}}c}
\multicolumn{2}{c}{E(level)$^{}$}&J$^{\pi}$$^{{\hyperlink{NE17LEVEL0}{a}}}$&\multicolumn{2}{c}{T$_{1/2}$$^{}$}&Comments&\\[-.2cm]
\multicolumn{2}{c}{\hrulefill}&\hrulefill&\multicolumn{2}{c}{\hrulefill}&\hrulefill&
\endfirsthead
\multicolumn{1}{r@{}}{0}&\multicolumn{1}{@{}l}{}&\multicolumn{1}{l}{1/2\ensuremath{^{+}}}&&&\parbox[t][0.3cm]{13.093121cm}{\raggedright E(level): From (\href{https://www.nndc.bnl.gov/nsr/nsrlink.jsp?1969Ni09,B}{1969Ni09}, \href{https://www.nndc.bnl.gov/nsr/nsrlink.jsp?1978Wu05,B}{1978Wu05}, \href{https://www.nndc.bnl.gov/nsr/nsrlink.jsp?1983De26,B}{1983De26}).\vspace{0.1cm}}&\\
&&&&&\parbox[t][0.3cm]{13.093121cm}{\raggedright Fusion cross section for the \ensuremath{^{\textnormal{10}}}B(\ensuremath{^{\textnormal{14}}}N,\ensuremath{^{\textnormal{19}}}Ne) reaction was extrapolated by (\href{https://www.nndc.bnl.gov/nsr/nsrlink.jsp?1983De26,B}{1983De26}) to\vspace{0.1cm}}&\\
&&&&&\parbox[t][0.3cm]{13.093121cm}{\raggedright {\ }{\ }{\ }E\ensuremath{_{\textnormal{c.m.}}}=7.5 MeV. The result was 19 mb \textit{1} (stat.). A systematic uncertainty of 12\% should be\vspace{0.1cm}}&\\
&&&&&\parbox[t][0.3cm]{13.093121cm}{\raggedright {\ }{\ }{\ }added in quadrature (see text).\vspace{0.1cm}}&\\
\multicolumn{1}{r@{}}{238}&\multicolumn{1}{@{}l}{}&\multicolumn{1}{l}{5/2\ensuremath{^{+}}}&&&\parbox[t][0.3cm]{13.093121cm}{\raggedright E(level): From (\href{https://www.nndc.bnl.gov/nsr/nsrlink.jsp?1978Wu05,B}{1978Wu05}, \href{https://www.nndc.bnl.gov/nsr/nsrlink.jsp?1983De26,B}{1983De26}).\vspace{0.1cm}}&\\
&&&&&\parbox[t][0.3cm]{13.093121cm}{\raggedright \ensuremath{\beta}=0.53 at E\ensuremath{_{\textnormal{c.m.}}}=2.5 MeV (\href{https://www.nndc.bnl.gov/nsr/nsrlink.jsp?1978Wu05,B}{1978Wu05}).\vspace{0.1cm}}&\\
&&&&&\parbox[t][0.3cm]{13.093121cm}{\raggedright \ensuremath{\beta}=0.529 at E\ensuremath{_{\textnormal{c.m.}}}=7.5 MeV (\href{https://www.nndc.bnl.gov/nsr/nsrlink.jsp?1978Wu05,B}{1978Wu05}).\vspace{0.1cm}}&\\
&&&&&\parbox[t][0.3cm]{13.093121cm}{\raggedright \ensuremath{\beta}\ensuremath{_{\textnormal{2J+1}}}=0.521 (\href{https://www.nndc.bnl.gov/nsr/nsrlink.jsp?1978Wu05,B}{1978Wu05}).\vspace{0.1cm}}&\\
\multicolumn{1}{r@{}}{275}&\multicolumn{1}{@{}l}{}&\multicolumn{1}{l}{1/2\ensuremath{^{-}}}&\multicolumn{1}{r@{}}{42}&\multicolumn{1}{@{ }l}{ps {\it +3\textminus14}}&\parbox[t][0.3cm]{13.093121cm}{\raggedright E(level): From (M. Fieher, P. Lehmann, A. Leveque and R. Pick, Compt. Rend. 241 (1955)\vspace{0.1cm}}&\\
&&&&&\parbox[t][0.3cm]{13.093121cm}{\raggedright {\ }{\ }{\ }1946; \href{https://www.nndc.bnl.gov/nsr/nsrlink.jsp?1969Ni09,B}{1969Ni09}; \href{https://www.nndc.bnl.gov/nsr/nsrlink.jsp?1978Wu05,B}{1978Wu05}; \href{https://www.nndc.bnl.gov/nsr/nsrlink.jsp?1983De26,B}{1983De26}).\vspace{0.1cm}}&\\
&&&&&\parbox[t][0.3cm]{13.093121cm}{\raggedright T\ensuremath{_{1/2}}: From \ensuremath{\tau}=61 ps \textit{+4{\textminus}20} (\href{https://www.nndc.bnl.gov/nsr/nsrlink.jsp?1969Ni09,B}{1969Ni09}), which leads to T\ensuremath{_{\textnormal{1/2}}}=42.3 ps \textit{+28{\textminus}139}. The quoted\vspace{0.1cm}}&\\
&&&&&\parbox[t][0.3cm]{13.093121cm}{\raggedright {\ }{\ }{\ }uncertainties are combined statistical and systematic uncertainties. See also \ensuremath{\tau}\ensuremath{<}5 ns (M. Fieher,\vspace{0.1cm}}&\\
&&&&&\parbox[t][0.3cm]{13.093121cm}{\raggedright {\ }{\ }{\ }P. Lehmann, A. Leveque and R. Pick, Compt. Rend. 241 (1955) 1946).\vspace{0.1cm}}&\\
&&&&&\parbox[t][0.3cm]{13.093121cm}{\raggedright \ensuremath{\beta}=0.228 at E\ensuremath{_{\textnormal{c.m.}}}=2.5 MeV (\href{https://www.nndc.bnl.gov/nsr/nsrlink.jsp?1978Wu05,B}{1978Wu05}).\vspace{0.1cm}}&\\
&&&&&\parbox[t][0.3cm]{13.093121cm}{\raggedright \ensuremath{\beta}=0.317 at E\ensuremath{_{\textnormal{c.m.}}}=7.5 MeV (\href{https://www.nndc.bnl.gov/nsr/nsrlink.jsp?1978Wu05,B}{1978Wu05}).\vspace{0.1cm}}&\\
&&&&&\parbox[t][0.3cm]{13.093121cm}{\raggedright \ensuremath{\beta}\ensuremath{_{\textnormal{2J+1}}}=0.251 (\href{https://www.nndc.bnl.gov/nsr/nsrlink.jsp?1978Wu05,B}{1978Wu05}).\vspace{0.1cm}}&\\
\multicolumn{1}{r@{}}{1511}&\multicolumn{1}{@{}l}{}&&&&\parbox[t][0.3cm]{13.093121cm}{\raggedright E(level): From (\href{https://www.nndc.bnl.gov/nsr/nsrlink.jsp?1978Wu05,B}{1978Wu05}), where E\ensuremath{_{\textnormal{x}}} is not reported. Evaluator deduced this energy from the\vspace{0.1cm}}&\\
&&&&&\parbox[t][0.3cm]{13.093121cm}{\raggedright {\ }{\ }{\ }sum of E\ensuremath{_{\ensuremath{\gamma}}} energies for the decay cascade and by considering negligible recoil energies.\vspace{0.1cm}}&\\
\end{longtable}
\parbox[b][0.3cm]{17.7cm}{\makebox[1ex]{\ensuremath{^{\hypertarget{NE17LEVEL0}{a}}}} From the \ensuremath{^{\textnormal{19}}}Ne Adopted Levels.}\\
\vspace{0.5cm}
\clearpage
\vspace{0.3cm}
\vspace*{-0.5cm}
{\bf \small \underline{\ensuremath{^{\textnormal{10}}}B(\ensuremath{^{\textnormal{14}}}N,\ensuremath{\alpha}n)\hspace{0.2in}\href{https://www.nndc.bnl.gov/nsr/nsrlink.jsp?1969Ni09,B}{1969Ni09},\href{https://www.nndc.bnl.gov/nsr/nsrlink.jsp?1978Wu05,B}{1978Wu05},\href{https://www.nndc.bnl.gov/nsr/nsrlink.jsp?1983De26,B}{1983De26} (continued)}}\\
\vspace{0.3cm}
\underline{$\gamma$($^{19}$Ne)}\\
\begin{longtable}{ccccccc@{}ccccc@{\extracolsep{\fill}}c}
\multicolumn{2}{c}{E\ensuremath{_{\gamma}}}&\multicolumn{2}{c}{E\ensuremath{_{i}}(level)}&J\ensuremath{^{\pi}_{i}}&\multicolumn{2}{c}{E\ensuremath{_{f}}}&J\ensuremath{^{\pi}_{f}}&Mult.&\multicolumn{2}{c}{\ensuremath{\alpha}\ensuremath{^{\hyperlink{NE17GAMMA0}{a}}}}&Comments&\\[-.2cm]
\multicolumn{2}{c}{\hrulefill}&\multicolumn{2}{c}{\hrulefill}&\hrulefill&\multicolumn{2}{c}{\hrulefill}&\hrulefill&\hrulefill&\multicolumn{2}{c}{\hrulefill}&\hrulefill&
\endfirsthead
\multicolumn{1}{r@{}}{238}&\multicolumn{1}{@{}l}{}&\multicolumn{1}{r@{}}{238}&\multicolumn{1}{@{}l}{}&\multicolumn{1}{l}{5/2\ensuremath{^{+}}}&\multicolumn{1}{r@{}}{0}&\multicolumn{1}{@{}l}{}&\multicolumn{1}{@{}l}{1/2\ensuremath{^{+}}}&&\multicolumn{1}{r@{}}{}&\multicolumn{1}{@{}l}{}&\parbox[t][0.3cm]{9.3611cm}{\raggedright E\ensuremath{_{\gamma}}: From (\href{https://www.nndc.bnl.gov/nsr/nsrlink.jsp?1978Wu05,B}{1978Wu05}, \href{https://www.nndc.bnl.gov/nsr/nsrlink.jsp?1983De26,B}{1983De26}).\vspace{0.1cm}}&\\
\multicolumn{1}{r@{}}{275}&\multicolumn{1}{@{}l}{}&\multicolumn{1}{r@{}}{275}&\multicolumn{1}{@{}l}{}&\multicolumn{1}{l}{1/2\ensuremath{^{-}}}&\multicolumn{1}{r@{}}{0}&\multicolumn{1}{@{}l}{}&\multicolumn{1}{@{}l}{1/2\ensuremath{^{+}}}&\multicolumn{1}{l}{E1}&\multicolumn{1}{r@{}}{1}&\multicolumn{1}{@{.}l}{40\ensuremath{\times10^{-4}} {\it 2}}&\parbox[t][0.3cm]{9.3611cm}{\raggedright B(E1)(W.u.)=0.00109 \textit{+20{\textminus}8}\vspace{0.1cm}}&\\
&&&&&&&&&&&\parbox[t][0.3cm]{9.3611cm}{\raggedright \ensuremath{\alpha}(K)=0.0001330 \textit{19}; \ensuremath{\alpha}(L)=7.36\ensuremath{\times}10\ensuremath{^{\textnormal{$-$6}}} \textit{10}\vspace{0.1cm}}&\\
&&&&&&&&&&&\parbox[t][0.3cm]{9.3611cm}{\raggedright E\ensuremath{_{\gamma}}: From (\href{https://www.nndc.bnl.gov/nsr/nsrlink.jsp?1969Ni09,B}{1969Ni09}): E\ensuremath{_{\ensuremath{\gamma}}} is not reported and is deduced from\vspace{0.1cm}}&\\
&&&&&&&&&&&\parbox[t][0.3cm]{9.3611cm}{\raggedright {\ }{\ }{\ }level-energy difference considering the negligible recoil energy.\vspace{0.1cm}}&\\
&&&&&&&&&&&\parbox[t][0.3cm]{9.3611cm}{\raggedright {\ }{\ }{\ }See also (\href{https://www.nndc.bnl.gov/nsr/nsrlink.jsp?1978Wu05,B}{1978Wu05}, \href{https://www.nndc.bnl.gov/nsr/nsrlink.jsp?1983De26,B}{1983De26}).\vspace{0.1cm}}&\\
&&&&&&&&&&&\parbox[t][0.3cm]{9.3611cm}{\raggedright Mult.: From (\href{https://www.nndc.bnl.gov/nsr/nsrlink.jsp?1969Ni09,B}{1969Ni09}).\vspace{0.1cm}}&\\
&&&&&&&&&&&\parbox[t][0.3cm]{9.3611cm}{\raggedright B(E1)(W.u.): See also 0.00107 W.u. \textit{+35{\textminus}12} (\href{https://www.nndc.bnl.gov/nsr/nsrlink.jsp?1969Ni09,B}{1969Ni09}), who reported\vspace{0.1cm}}&\\
&&&&&&&&&&&\parbox[t][0.3cm]{9.3611cm}{\raggedright {\ }{\ }{\ }that the equality of this E1 transition strength for the\vspace{0.1cm}}&\\
&&&&&&&&&&&\parbox[t][0.3cm]{9.3611cm}{\raggedright {\ }{\ }{\ }\ensuremath{^{\textnormal{19}}}Ne*(275)\ensuremath{\rightarrow}\ensuremath{^{\textnormal{19}}}Ne\ensuremath{_{\textnormal{g.s.}}}+\ensuremath{\gamma} transition to that of the\vspace{0.1cm}}&\\
&&&&&&&&&&&\parbox[t][0.3cm]{9.3611cm}{\raggedright {\ }{\ }{\ }\ensuremath{^{\textnormal{19}}}F*(109.9)\ensuremath{\rightarrow}\ensuremath{^{\textnormal{19}}}F\ensuremath{_{\textnormal{g.s.}}}+\ensuremath{\gamma} transition with B(E1)=0.001220 W.u. \textit{14}\vspace{0.1cm}}&\\
&&&&&&&&&&&\parbox[t][0.3cm]{9.3611cm}{\raggedright {\ }{\ }{\ }(\href{https://www.nndc.bnl.gov/nsr/nsrlink.jsp?1969Ni09,B}{1969Ni09}) indicates that these two states are mirror levels and\vspace{0.1cm}}&\\
&&&&&&&&&&&\parbox[t][0.3cm]{9.3611cm}{\raggedright {\ }{\ }{\ }confirms the level order inversion in this pair.\vspace{0.1cm}}&\\
\multicolumn{1}{r@{}}{1236}&\multicolumn{1}{@{}l}{}&\multicolumn{1}{r@{}}{1511}&\multicolumn{1}{@{}l}{}&&\multicolumn{1}{r@{}}{275}&\multicolumn{1}{@{}l}{}&\multicolumn{1}{@{}l}{1/2\ensuremath{^{-}}}&&\multicolumn{1}{r@{}}{}&\multicolumn{1}{@{}l}{}&\parbox[t][0.3cm]{9.3611cm}{\raggedright E\ensuremath{_{\gamma}}: From (\href{https://www.nndc.bnl.gov/nsr/nsrlink.jsp?1978Wu05,B}{1978Wu05}).\vspace{0.1cm}}&\\
\end{longtable}
\parbox[b][0.3cm]{17.7cm}{\makebox[1ex]{\ensuremath{^{\hypertarget{NE17GAMMA0}{a}}}} Total theoretical internal conversion coefficients, calculated using the BrIcc code (\href{https://www.nndc.bnl.gov/nsr/nsrlink.jsp?2008Ki07,B}{2008Ki07}) with ``Frozen Orbitals''}\\
\parbox[b][0.3cm]{17.7cm}{{\ }{\ }approximation based on \ensuremath{\gamma}-ray energies, assigned multipolarities, and mixing ratios, unless otherwise specified.}\\
\vspace{0.5cm}
\begin{figure}[h]
\begin{center}
\includegraphics{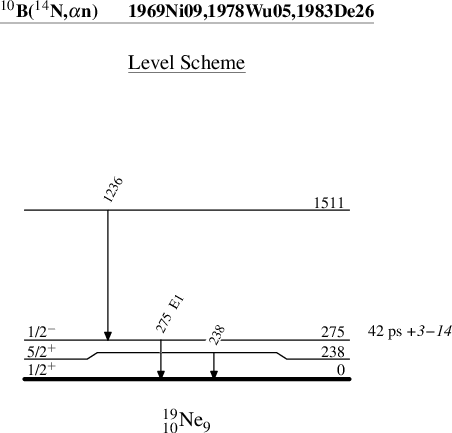}\\
\end{center}
\end{figure}
\clearpage
\subsection[\hspace{-0.2cm}\ensuremath{^{\textnormal{12}}}C(\ensuremath{^{\textnormal{12}}}C,n\ensuremath{\alpha})]{ }
\vspace{-27pt}
\vspace{0.3cm}
\hypertarget{NE18}{{\bf \small \underline{\ensuremath{^{\textnormal{12}}}C(\ensuremath{^{\textnormal{12}}}C,n\ensuremath{\alpha})\hspace{0.2in}\href{https://www.nndc.bnl.gov/nsr/nsrlink.jsp?1982Sa27,B}{1982Sa27}}}}\\
\vspace{4pt}
\vspace{8pt}
\parbox[b][0.3cm]{17.7cm}{\addtolength{\parindent}{-0.2in}J\ensuremath{^{\ensuremath{\pi}}}(\ensuremath{^{\textnormal{12}}}C\ensuremath{_{\textnormal{g.s.}}})=0\ensuremath{^{\textnormal{+}}}.}\\
\parbox[b][0.3cm]{17.7cm}{\addtolength{\parindent}{-0.2in}\href{https://www.nndc.bnl.gov/nsr/nsrlink.jsp?1982Sa27,B}{1982Sa27}: \ensuremath{^{\textnormal{12}}}C(\ensuremath{^{\textnormal{12}}}C,n\ensuremath{\alpha}) E\ensuremath{_{\textnormal{c.m.}}}=5.25-20 MeV and E\ensuremath{_{\textnormal{lab}}}=10.5-40 MeV; measured \ensuremath{\sigma}(E) for production of A=16-23 reaction products;}\\
\parbox[b][0.3cm]{17.7cm}{deduced total fusion \ensuremath{\sigma} vs. E using statistical analysis. The \ensuremath{\gamma} rays were measured using a Ge(Li) detector at \ensuremath{\theta}\ensuremath{_{\textnormal{lab}}}=55\ensuremath{^\circ}. Observed}\\
\parbox[b][0.3cm]{17.7cm}{evidence of anomalies in the \ensuremath{^{\textnormal{12}}}C+\ensuremath{^{\textnormal{12}}}C excitation function above the Coulomb barrier, whose source is of non-statistical origin.}\\
\vspace{12pt}
\underline{$^{19}$Ne Levels}\\
\vspace{0.34cm}
\parbox[b][0.3cm]{17.7cm}{\addtolength{\parindent}{-0.254cm}The cross section of \ensuremath{^{\textnormal{10}}}C(\ensuremath{^{\textnormal{10}}}C,n\ensuremath{\alpha})\ensuremath{^{\textnormal{19}}}Ne was estimated to be \ensuremath{\sigma}\ensuremath{<}10 mb.}\\
\vspace{0.34cm}
\begin{longtable}{cc@{\extracolsep{\fill}}c}
\multicolumn{2}{c}{E(level)$^{{\hyperlink{NE18LEVEL0}{a}}}$}&\\[-.2cm]
\multicolumn{2}{c}{\hrulefill}&
\endfirsthead
\multicolumn{1}{r@{}}{0}&\multicolumn{1}{@{}l}{}&\\
\multicolumn{1}{r@{}}{238}&\multicolumn{1}{@{.}l}{3}&\\
\multicolumn{1}{r@{}}{275}&\multicolumn{1}{@{.}l}{2}&\\
\end{longtable}
\parbox[b][0.3cm]{17.7cm}{\makebox[1ex]{\ensuremath{^{\hypertarget{NE18LEVEL0}{a}}}} From E\ensuremath{_{\ensuremath{\gamma}}}. Note that the recoil energy is negligible.}\\
\vspace{0.5cm}
\underline{$\gamma$($^{19}$Ne)}\\
\begin{longtable}{cccccc@{\extracolsep{\fill}}c}
\multicolumn{2}{c}{E\ensuremath{_{\gamma}}\ensuremath{^{\hyperlink{NE18GAMMA0}{a}}}}&\multicolumn{2}{c}{E\ensuremath{_{i}}(level)}&\multicolumn{2}{c}{E\ensuremath{_{f}}}&\\[-.2cm]
\multicolumn{2}{c}{\hrulefill}&\multicolumn{2}{c}{\hrulefill}&\multicolumn{2}{c}{\hrulefill}&
\endfirsthead
\multicolumn{1}{r@{}}{238}&\multicolumn{1}{@{.}l}{3}&\multicolumn{1}{r@{}}{238}&\multicolumn{1}{@{.}l}{3}&\multicolumn{1}{r@{}}{0}&\multicolumn{1}{@{}l}{}&\\
\multicolumn{1}{r@{}}{275}&\multicolumn{1}{@{.}l}{2}&\multicolumn{1}{r@{}}{275}&\multicolumn{1}{@{.}l}{2}&\multicolumn{1}{r@{}}{0}&\multicolumn{1}{@{}l}{}&\\
\end{longtable}
\parbox[b][0.3cm]{17.7cm}{\makebox[1ex]{\ensuremath{^{\hypertarget{NE18GAMMA0}{a}}}} From (\href{https://www.nndc.bnl.gov/nsr/nsrlink.jsp?1982Sa27,B}{1982Sa27}).}\\
\vspace{0.5cm}
\begin{figure}[h]
\begin{center}
\includegraphics{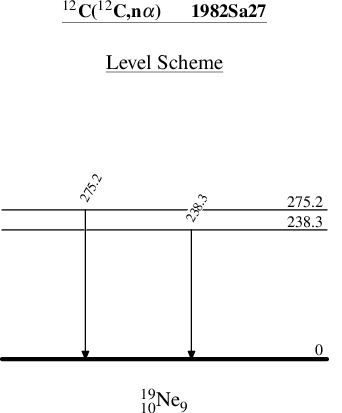}\\
\end{center}
\end{figure}
\clearpage
\subsection[\hspace{-0.2cm}\ensuremath{^{\textnormal{12}}}C(\ensuremath{^{\textnormal{20}}}Ne,\ensuremath{^{\textnormal{13}}}C)]{ }
\vspace{-27pt}
\vspace{0.3cm}
\hypertarget{NE19}{{\bf \small \underline{\ensuremath{^{\textnormal{12}}}C(\ensuremath{^{\textnormal{20}}}Ne,\ensuremath{^{\textnormal{13}}}C)\hspace{0.2in}\href{https://www.nndc.bnl.gov/nsr/nsrlink.jsp?1979Sh18,B}{1979Sh18}}}}\\
\vspace{4pt}
\vspace{8pt}
\parbox[b][0.3cm]{17.7cm}{\addtolength{\parindent}{-0.2in}J\ensuremath{^{\ensuremath{\pi}}}(\ensuremath{^{\textnormal{12}}}C\ensuremath{_{\textnormal{g.s.}}})=0\ensuremath{^{\textnormal{+}}} and J\ensuremath{^{\ensuremath{\pi}}}(\ensuremath{^{\textnormal{20}}}Ne\ensuremath{_{\textnormal{g.s.}}})=0\ensuremath{^{\textnormal{+}}}.}\\
\parbox[b][0.3cm]{17.7cm}{\addtolength{\parindent}{-0.2in}\href{https://www.nndc.bnl.gov/nsr/nsrlink.jsp?1979Sh18,B}{1979Sh18}: \ensuremath{^{\textnormal{12}}}C(\ensuremath{^{\textnormal{20}}}Ne,\ensuremath{^{\textnormal{13}}}C) E=50-80 MeV; measured reaction products with Z\ensuremath{>}3 using an ionization chamber backed by a position}\\
\parbox[b][0.3cm]{17.7cm}{sensitive Si \ensuremath{\Delta}E-E telescope; measured \ensuremath{\sigma}(E,\ensuremath{\theta}) at \ensuremath{\theta}\ensuremath{_{\textnormal{lab}}}=5\ensuremath{^\circ}{\textminus}40\ensuremath{^\circ} for E=60, 72.5, 74, 75.2 MeV; measured excitation function at}\\
\parbox[b][0.3cm]{17.7cm}{\ensuremath{\theta}\ensuremath{_{\textnormal{lab}}}=12\ensuremath{^\circ}{\textminus}21\ensuremath{^\circ}; deduced reaction mechanism.}\\
\vspace{12pt}
\underline{$^{19}$Ne Levels}\\
\begin{longtable}{ccc@{\extracolsep{\fill}}c}
\multicolumn{2}{c}{E(level)$^{}$}&Comments&\\[-.2cm]
\multicolumn{2}{c}{\hrulefill}&\hrulefill&
\endfirsthead
\multicolumn{1}{r@{}}{0}&\multicolumn{1}{@{}l}{}&\parbox[t][0.3cm]{16.24712cm}{\raggedright E(level): From (\href{https://www.nndc.bnl.gov/nsr/nsrlink.jsp?1979Sh18,B}{1979Sh18}).\vspace{0.1cm}}&\\
\end{longtable}
\clearpage
\subsection[\hspace{-0.2cm}\ensuremath{^{\textnormal{12}}}C(\ensuremath{^{\textnormal{20}}}Ne,\ensuremath{^{\textnormal{19}}}Ne)]{ }
\vspace{-27pt}
\vspace{0.3cm}
\hypertarget{NE20}{{\bf \small \underline{\ensuremath{^{\textnormal{12}}}C(\ensuremath{^{\textnormal{20}}}Ne,\ensuremath{^{\textnormal{19}}}Ne)\hspace{0.2in}\href{https://www.nndc.bnl.gov/nsr/nsrlink.jsp?2013Uj01,B}{2013Uj01},\href{https://www.nndc.bnl.gov/nsr/nsrlink.jsp?2017Fo24,B}{2017Fo24}}}}\\
\vspace{4pt}
\vspace{8pt}
\parbox[b][0.3cm]{17.7cm}{\addtolength{\parindent}{-0.2in}One neutron knockout reaction in inverse kinematics.}\\
\parbox[b][0.3cm]{17.7cm}{\addtolength{\parindent}{-0.2in}J\ensuremath{^{\ensuremath{\pi}}}(\ensuremath{^{\textnormal{12}}}C\ensuremath{_{\textnormal{g.s.}}})=0\ensuremath{^{\textnormal{+}}} and J\ensuremath{^{\ensuremath{\pi}}}(\ensuremath{^{\textnormal{20}}}Ne\ensuremath{_{\textnormal{g.s.}}})=0\ensuremath{^{\textnormal{+}}}.}\\
\parbox[b][0.3cm]{17.7cm}{\addtolength{\parindent}{-0.2in}\href{https://www.nndc.bnl.gov/nsr/nsrlink.jsp?2013Uj01,B}{2013Uj01}: \ensuremath{^{\textnormal{12}}}C(\ensuremath{^{\textnormal{20}}}Ne,\ensuremath{^{\textnormal{19}}}Ne) E=6 MeV/nucleon; implanted the \ensuremath{^{\textnormal{19}}}Ne beam into a 100 \ensuremath{\mu}m-thick niobium foil; measured \ensuremath{\beta}\ensuremath{^{\textnormal{+}}}}\\
\parbox[b][0.3cm]{17.7cm}{particles from the decay of \ensuremath{^{\textnormal{19}}}Ne using a plastic scintillator coupled to 2 PMTs, which operated in coincidence mode. Beam was on}\\
\parbox[b][0.3cm]{17.7cm}{for two half-lives followed by counting for 20 half-lives. Measurements were carried out in 1 h cycles. During each cycle, the Nb}\\
\parbox[b][0.3cm]{17.7cm}{target was cooled to 4 K (superconducting phase). Several implantations followed by counting were performed. The Nb target was}\\
\parbox[b][0.3cm]{17.7cm}{then warmed up to 16 K (metallic phase) and the implantation+counting was repeated. The authors were interested to measure}\\
\parbox[b][0.3cm]{17.7cm}{changes in the \ensuremath{^{\textnormal{19}}}Ne half-life due to possible superscreening effect at the superconducting phase. Measured the \ensuremath{^{\textnormal{19}}}Ne decay curve}\\
\parbox[b][0.3cm]{17.7cm}{and deduced the \ensuremath{^{\textnormal{19}}}Ne half-lives at T=4 K and T=16 K. The half-lives in these temperatures are only presented graphically (see}\\
\parbox[b][0.3cm]{17.7cm}{Fig. 3 of that study). These half-lives are consistent with one another at 1\ensuremath{\sigma} level with an average change of 0.95\% \textit{78}.}\\
\parbox[b][0.3cm]{17.7cm}{\addtolength{\parindent}{-0.2in}\href{https://www.nndc.bnl.gov/nsr/nsrlink.jsp?2017Fo24,B}{2017Fo24}: \ensuremath{^{\textnormal{12}}}C(\ensuremath{^{\textnormal{20}}}Ne,\ensuremath{^{\textnormal{19}}}Ne) E=95 MeV/nucleon, post-accelerated \ensuremath{^{\textnormal{19}}}Ne ions to E=5.0982 MeV/nucleon and implanted them into a}\\
\parbox[b][0.3cm]{17.7cm}{Pb target using 33 cycles of 6.4 s beam on, 440 s beam off. Measured the annihilation \ensuremath{\gamma} rays from \ensuremath{^{\textnormal{19}}}Ne(\ensuremath{\beta}\ensuremath{^{\textnormal{+}}}) decay in coincidence}\\
\parbox[b][0.3cm]{17.7cm}{using a plastic scintillator coupled to 2 PMTs. Deduced the half-life of \ensuremath{^{\textnormal{19}}}Ne\ensuremath{_{\textnormal{g.s.}}}. A good agreement with the results of (\href{https://www.nndc.bnl.gov/nsr/nsrlink.jsp?2012Tr06,B}{2012Tr06},}\\
\parbox[b][0.3cm]{17.7cm}{\href{https://www.nndc.bnl.gov/nsr/nsrlink.jsp?2013Uj01,B}{2013Uj01}) is achieved. However, the present result is inconsistent with that of (\href{https://www.nndc.bnl.gov/nsr/nsrlink.jsp?2014Br06,B}{2014Br06}) by 3.2\ensuremath{\sigma}. Discussed the impact of the}\\
\parbox[b][0.3cm]{17.7cm}{present half-life on the value of the V\ensuremath{_{\textnormal{ud}}} element of the CKM mixing matrix.}\\
\vspace{12pt}
\underline{$^{19}$Ne Levels}\\
\begin{longtable}{cccccc@{\extracolsep{\fill}}c}
\multicolumn{2}{c}{E(level)$^{{\hyperlink{NE20LEVEL0}{a}}}$}&J$^{\pi}$$^{{\hyperlink{NE20LEVEL0}{a}}}$&\multicolumn{2}{c}{T\ensuremath{_{\textnormal{1/2}}}$^{}$}&Comments&\\[-.2cm]
\multicolumn{2}{c}{\hrulefill}&\hrulefill&\multicolumn{2}{c}{\hrulefill}&\hrulefill&
\endfirsthead
\multicolumn{1}{r@{}}{0}&\multicolumn{1}{@{}l}{}&\multicolumn{1}{l}{1/2\ensuremath{^{+}}}&\multicolumn{1}{r@{}}{17}&\multicolumn{1}{@{.}l}{2565 s {\it 21}}&\parbox[t][0.3cm]{12.782821cm}{\raggedright T\ensuremath{_{1/2}}: Weighted average of T\ensuremath{_{\textnormal{1/2}}}=17.2569 s \textit{19} (stat.) \textit{9} (sys.) (\href{https://www.nndc.bnl.gov/nsr/nsrlink.jsp?2017Fo24,B}{2017Fo24}) and T\ensuremath{_{\textnormal{1/2}}}=17.254 s\vspace{0.1cm}}&\\
&&&&&\parbox[t][0.3cm]{12.782821cm}{\raggedright {\ }{\ }{\ }\textit{5} (\href{https://www.nndc.bnl.gov/nsr/nsrlink.jsp?2013Uj01,B}{2013Uj01}), where the systematic uncertainty is added in quadrature to the uncertainty in\vspace{0.1cm}}&\\
&&&&&\parbox[t][0.3cm]{12.782821cm}{\raggedright {\ }{\ }{\ }the weighted average.\vspace{0.1cm}}&\\
&&&&&\parbox[t][0.3cm]{12.782821cm}{\raggedright (\href{https://www.nndc.bnl.gov/nsr/nsrlink.jsp?2013Uj01,B}{2013Uj01}) reported that the difference in the electron screening potential energy between the\vspace{0.1cm}}&\\
&&&&&\parbox[t][0.3cm]{12.782821cm}{\raggedright {\ }{\ }{\ }two Nb phases (superconducting at 4 K vs. metallic at 16 K) that would induce an average\vspace{0.1cm}}&\\
&&&&&\parbox[t][0.3cm]{12.782821cm}{\raggedright {\ }{\ }{\ }half-life change (due to temperature difference) of 0.95\% \textit{78} is U\ensuremath{_{\textnormal{e}}}=110 eV \textit{90} eV for \ensuremath{^{\textnormal{19}}}Ne.\vspace{0.1cm}}&\\
&&&&&\parbox[t][0.3cm]{12.782821cm}{\raggedright {\ }{\ }{\ }No difference was observed in the half-life between the two phases within the limits of\vspace{0.1cm}}&\\
&&&&&\parbox[t][0.3cm]{12.782821cm}{\raggedright {\ }{\ }{\ }experimental accuracy (0.04\%) for the half-life (\href{https://www.nndc.bnl.gov/nsr/nsrlink.jsp?2013Uj01,B}{2013Uj01}). The obtained electron screening\vspace{0.1cm}}&\\
&&&&&\parbox[t][0.3cm]{12.782821cm}{\raggedright {\ }{\ }{\ }potential in a superconductor is well below the predicted value of 20 keV by the\vspace{0.1cm}}&\\
&&&&&\parbox[t][0.3cm]{12.782821cm}{\raggedright {\ }{\ }{\ }superscreening theoretical model developed by (\href{https://www.nndc.bnl.gov/nsr/nsrlink.jsp?1991St14,B}{1991St14}), which casts doubt on that model.\vspace{0.1cm}}&\\
\end{longtable}
\parbox[b][0.3cm]{17.7cm}{\makebox[1ex]{\ensuremath{^{\hypertarget{NE20LEVEL0}{a}}}} From the \ensuremath{^{\textnormal{19}}}Ne Adopted Levels.}\\
\vspace{0.5cm}
\clearpage
\subsection[\hspace{-0.2cm}\ensuremath{^{\textnormal{15}}}N(\ensuremath{^{\textnormal{12}}}C,\ensuremath{^{\textnormal{8}}}Li)]{ }
\vspace{-27pt}
\vspace{0.3cm}
\hypertarget{NE21}{{\bf \small \underline{\ensuremath{^{\textnormal{15}}}N(\ensuremath{^{\textnormal{12}}}C,\ensuremath{^{\textnormal{8}}}Li)\hspace{0.2in}\href{https://www.nndc.bnl.gov/nsr/nsrlink.jsp?1979Ra10,B}{1979Ra10}}}}\\
\vspace{4pt}
\vspace{8pt}
\parbox[b][0.3cm]{17.7cm}{\addtolength{\parindent}{-0.2in}J\ensuremath{^{\ensuremath{\pi}}}(\ensuremath{^{\textnormal{15}}}N\ensuremath{_{\textnormal{g.s.}}})=1/2\ensuremath{^{-}} and J\ensuremath{^{\ensuremath{\pi}}}(\ensuremath{^{\textnormal{12}}}C\ensuremath{_{\textnormal{g.s.}}})=0\ensuremath{^{\textnormal{+}}}.}\\
\parbox[b][0.3cm]{17.7cm}{\addtolength{\parindent}{-0.2in}\href{https://www.nndc.bnl.gov/nsr/nsrlink.jsp?1979Ra10,B}{1979Ra10}: \ensuremath{^{\textnormal{15}}}N(\ensuremath{^{\textnormal{12}}}C,\ensuremath{^{\textnormal{8}}}Li) E=115 MeV; a \ensuremath{^{\textnormal{12}}}C beam accelerated by the variable energy cyclotron of the Harwell Atomic Energy}\\
\parbox[b][0.3cm]{17.7cm}{Research Establishment bombarded a gas cell filled with 100 Torr of 99.5\% enriched \ensuremath{^{\textnormal{15}}}N gas. Measured reaction products using a}\\
\parbox[b][0.3cm]{17.7cm}{Si \ensuremath{\Delta}E-\ensuremath{\Delta}E-E telescope followed by a Si veto detector to reject the long-range particles. The angular coverage of the detection}\\
\parbox[b][0.3cm]{17.7cm}{system was for \ensuremath{\theta}\ensuremath{_{\textnormal{lab}}}=7\ensuremath{^\circ}{\textminus}12\ensuremath{^\circ}.}\\
\vspace{12pt}
\underline{$^{19}$Ne Levels}\\
\vspace{0.34cm}
\parbox[b][0.3cm]{17.7cm}{\addtolength{\parindent}{-0.254cm}(\href{https://www.nndc.bnl.gov/nsr/nsrlink.jsp?1979Ra10,B}{1979Ra10}) concluded that a direct \ensuremath{^{\textnormal{4}}}Li transfer mechanism, probably of a sequential nature, dominates this reaction.}\\
\vspace{0.34cm}
\begin{longtable}{cccc@{\extracolsep{\fill}}c}
\multicolumn{2}{c}{E(level)$^{{\hyperlink{NE21LEVEL0}{a}}}$}&J$^{\pi}$$^{{\hyperlink{NE21LEVEL1}{b}}}$&Comments&\\[-.2cm]
\multicolumn{2}{c}{\hrulefill}&\hrulefill&\hrulefill&
\endfirsthead
\multicolumn{1}{r@{}}{2.79\ensuremath{\times10^{3}}}&\multicolumn{1}{@{}l}{}&\multicolumn{1}{l}{9/2\ensuremath{^{+}}}&&\\
\multicolumn{1}{r@{}}{4.63\ensuremath{\times10^{3}}}&\multicolumn{1}{@{}l}{}&\multicolumn{1}{l}{13/2\ensuremath{^{+}}}&\parbox[t][0.3cm]{14.867001cm}{\raggedright E(level): Table 10 of (\href{https://www.nndc.bnl.gov/nsr/nsrlink.jsp?1979Ra10,B}{1979Ra10}) reports this state as 4.62 MeV.\vspace{0.1cm}}&\\
\end{longtable}
\parbox[b][0.3cm]{17.7cm}{\makebox[1ex]{\ensuremath{^{\hypertarget{NE21LEVEL0}{a}}}} From (\href{https://www.nndc.bnl.gov/nsr/nsrlink.jsp?1979Ra10,B}{1979Ra10}), where it was reported that to populate this state, a proton is transferred into the hole in the \textit{p}-shell to form an}\\
\parbox[b][0.3cm]{17.7cm}{{\ }{\ }intermediate \ensuremath{^{\textnormal{16}}}O\ensuremath{_{\textnormal{g.s.}}} with J\ensuremath{^{\ensuremath{\pi}}}=0\ensuremath{^{\textnormal{+}}}, and the cluster is then transferred into the \textit{sd}-shell to form a \ensuremath{^{\textnormal{3}}}He cluster state in \ensuremath{^{\textnormal{19}}}Ne.}\\
\parbox[b][0.3cm]{17.7cm}{\makebox[1ex]{\ensuremath{^{\hypertarget{NE21LEVEL1}{b}}}} From the \ensuremath{^{\textnormal{19}}}Ne Adopted Levels.}\\
\vspace{0.5cm}
\clearpage
\subsection[\hspace{-0.2cm}\ensuremath{^{\textnormal{15}}}O(\ensuremath{\alpha},\ensuremath{\gamma})]{ }
\vspace{-27pt}
\vspace{0.3cm}
\hypertarget{NE22}{{\bf \small \underline{\ensuremath{^{\textnormal{15}}}O(\ensuremath{\alpha},\ensuremath{\gamma})\hspace{0.2in}\href{https://www.nndc.bnl.gov/nsr/nsrlink.jsp?2011Da24,B}{2011Da24}}}}\\
\vspace{4pt}
\vspace{8pt}
\parbox[b][0.3cm]{17.7cm}{\addtolength{\parindent}{-0.2in}\textbf{Foreword:}}\\
\parbox[b][0.3cm]{17.7cm}{\addtolength{\parindent}{-0.2in}The \ensuremath{^{\textnormal{15}}}O(\ensuremath{\alpha},\ensuremath{\gamma}) reaction is not directly measured yet. This dataset presents a history of various studies which aimed to determine the}\\
\parbox[b][0.3cm]{17.7cm}{\ensuremath{^{\textnormal{15}}}O(\ensuremath{\alpha},\ensuremath{\gamma}) astrophysical reaction rate. The \ensuremath{^{\textnormal{19}}}Ne nuclear structure properties that are significant for determination of the \ensuremath{^{\textnormal{15}}}O(\ensuremath{\alpha},\ensuremath{\gamma})}\\
\parbox[b][0.3cm]{17.7cm}{astrophysical reaction rate are obtained from indirect measurements (listed below) and from information deduced from the mirror}\\
\parbox[b][0.3cm]{17.7cm}{states. Considering the significance of the \ensuremath{^{\textnormal{15}}}O(\ensuremath{\alpha},\ensuremath{\gamma}) reaction in astrophysics, this dataset is designed to present a summary of these}\\
\parbox[b][0.3cm]{17.7cm}{indirect studies, all in one reaction dataset. We note, however, that the results of these studies are already presented in other}\\
\parbox[b][0.3cm]{17.7cm}{reaction datasets. Therefore, we only list the excitation energies from the \ensuremath{^{\textnormal{19}}}Ne Adopted Levels for the states that were discussed in}\\
\parbox[b][0.3cm]{17.7cm}{the literature regarding this reaction rate.}\\
\vspace{0.385cm}
\parbox[b][0.3cm]{17.7cm}{\addtolength{\parindent}{-0.2in}\textit{Indirect Experimental Work:}}\\
\parbox[b][0.3cm]{17.7cm}{\addtolength{\parindent}{-0.2in}\href{https://www.nndc.bnl.gov/nsr/nsrlink.jsp?1987MaZQ,B}{1987MaZQ}, \href{https://www.nndc.bnl.gov/nsr/nsrlink.jsp?1987Ma31,B}{1987Ma31}: \ensuremath{^{\textnormal{15}}}N(\ensuremath{\alpha},\ensuremath{\gamma}) E=690, 700 keV; measured E\ensuremath{_{\ensuremath{\gamma}}}, I\ensuremath{_{\ensuremath{\gamma}}}, and thick target yield curve for the E\ensuremath{_{\ensuremath{\alpha}}^{\textnormal{c.m.}}}=536 keV}\\
\parbox[b][0.3cm]{17.7cm}{resonance in \ensuremath{^{\textnormal{19}}}F using a Ge(Li) detector at \ensuremath{\theta}\ensuremath{_{\textnormal{lab}}}=55\ensuremath{^\circ}. At both beam energies, the E\ensuremath{_{\ensuremath{\alpha}}^{\textnormal{c.m.}}}=542 keV resonance in \ensuremath{^{\textnormal{19}}}F remained}\\
\parbox[b][0.3cm]{17.7cm}{unobserved. Deduced \ensuremath{\omega}\ensuremath{\gamma}=97 \ensuremath{\mu}eV \textit{20} and \ensuremath{\Gamma}\ensuremath{_{\ensuremath{\alpha}}}=32 \ensuremath{\mu}eV \textit{7} for the E\ensuremath{_{\textnormal{c.m.}}}=536 keV resonance, and \ensuremath{\omega}\ensuremath{\gamma}\ensuremath{<}1\ensuremath{\times}10\ensuremath{^{\textnormal{$-$5}}} eV and \ensuremath{\Gamma}\ensuremath{_{\ensuremath{\alpha}}}\ensuremath{<}5}\\
\parbox[b][0.3cm]{17.7cm}{\ensuremath{\mu}eV for the E\ensuremath{_{\textnormal{c.m.}}}=542 keV resonance, corresponding to the \ensuremath{^{\textnormal{19}}}F*(4550, 4556) states, respectively. These levels were proposed to}\\
\parbox[b][0.3cm]{17.7cm}{be the mirror states to the \ensuremath{^{\textnormal{19}}}Ne*(4600, 4549) levels, respectively. Assumed \ensuremath{\Gamma}\ensuremath{_{\ensuremath{\gamma}}}\ensuremath{\approx}0.1 eV (\href{https://www.nndc.bnl.gov/nsr/nsrlink.jsp?1983Aj01,B}{1983Aj01}) for the aforementioned \ensuremath{^{\textnormal{19}}}Ne}\\
\parbox[b][0.3cm]{17.7cm}{states and deduced \ensuremath{\omega}\ensuremath{\gamma} for them. Obtained the \ensuremath{^{\textnormal{15}}}O(\ensuremath{\alpha},\ensuremath{\gamma}) reaction rate for 0.1-10 GK. Comparison with the theoretical rate of}\\
\parbox[b][0.3cm]{17.7cm}{(\href{https://www.nndc.bnl.gov/nsr/nsrlink.jsp?1986La07,B}{1986La07}) is discussed.}\\
\parbox[b][0.3cm]{17.7cm}{\addtolength{\parindent}{-0.2in}\href{https://www.nndc.bnl.gov/nsr/nsrlink.jsp?1995Ma28,B}{1995Ma28}, \href{https://www.nndc.bnl.gov/nsr/nsrlink.jsp?1996Ma07,B}{1996Ma07}: \ensuremath{^{\textnormal{15}}}N(\ensuremath{^{\textnormal{6}}}Li,d) E=22 MeV; measured deuterons$'$ angular distributions at \ensuremath{\theta}\ensuremath{_{\textnormal{lab}}}=7.5\ensuremath{^\circ}{\textminus}80\ensuremath{^\circ} using a spectrograph;}\\
\parbox[b][0.3cm]{17.7cm}{determined \ensuremath{\alpha}-spectroscopic factor for the \ensuremath{^{\textnormal{19}}}Ne*(4033) state assuming equality of spectroscopic factors for \ensuremath{^{\textnormal{19}}}Ne-\ensuremath{^{\textnormal{19}}}F mirror levels.}\\
\parbox[b][0.3cm]{17.7cm}{Deduced \ensuremath{\Gamma}\ensuremath{_{\ensuremath{\alpha}}} for the aforementioned \ensuremath{^{\textnormal{19}}}Ne state. Deduced the contribution of the \ensuremath{^{\textnormal{19}}}Ne*(4033) state to the \ensuremath{^{\textnormal{15}}}O(\ensuremath{\alpha},\ensuremath{\gamma}) astrophysical}\\
\parbox[b][0.3cm]{17.7cm}{rate at 0.2\ensuremath{\leq}T\ensuremath{\leq}1 GK. Comparison with the reaction rates of (\href{https://www.nndc.bnl.gov/nsr/nsrlink.jsp?1986La07,B}{1986La07},\href{https://www.nndc.bnl.gov/nsr/nsrlink.jsp?1990Ma05,B}{1990Ma05}) and the sensitivity of the reaction rate on r\ensuremath{_{\textnormal{0}\ensuremath{\alpha}}}}\\
\parbox[b][0.3cm]{17.7cm}{are discussed.}\\
\parbox[b][0.3cm]{17.7cm}{\addtolength{\parindent}{-0.2in}\href{https://www.nndc.bnl.gov/nsr/nsrlink.jsp?1995Wi26,B}{1995Wi26}: \ensuremath{^{\textnormal{15}}}N(\ensuremath{\alpha},\ensuremath{\gamma}) and \ensuremath{^{\textnormal{15}}}N(\ensuremath{\alpha},\ensuremath{\alpha}) E=0.65-2.65 MeV; measured E\ensuremath{_{\ensuremath{\gamma}}} and I\ensuremath{_{\ensuremath{\gamma}}} using a HPGe detector with a Compton suppression}\\
\parbox[b][0.3cm]{17.7cm}{BGO shielding at \ensuremath{\theta}\ensuremath{_{\textnormal{lab}}}=90\ensuremath{^\circ} covering \ensuremath{\theta}\ensuremath{_{\textnormal{lab}}}=60\ensuremath{^\circ}{\textminus}120\ensuremath{^\circ}; measured thick target yield curves for the \ensuremath{^{\textnormal{19}}}F*(4550, 4556) levels}\\
\parbox[b][0.3cm]{17.7cm}{corresponding to E\ensuremath{_{\ensuremath{\alpha}\textnormal{,c.m.}}}=536 and 542 keV resonances, respectively; measured elastically scattered \ensuremath{\alpha} particles using two Si surface}\\
\parbox[b][0.3cm]{17.7cm}{barrier detectors. Deduced \ensuremath{\Gamma}\ensuremath{_{\ensuremath{\alpha}}} and \ensuremath{\theta}\ensuremath{_{\ensuremath{\alpha}}^{\textnormal{2}}} (reduced width) for the previously mentioned resonances and \ensuremath{\omega}\ensuremath{\gamma}\ensuremath{_{\textnormal{(}\ensuremath{\alpha}\textnormal{,}\ensuremath{\gamma}\textnormal{)}}} for the 536-keV}\\
\parbox[b][0.3cm]{17.7cm}{resonance relative to the value measured by (\href{https://www.nndc.bnl.gov/nsr/nsrlink.jsp?1987Ma31,B}{1987Ma31}) for the same resonance. Assumed equality in \ensuremath{\theta}\ensuremath{_{\ensuremath{\alpha}}^{\textnormal{2}}} for \ensuremath{^{\textnormal{19}}}F and \ensuremath{^{\textnormal{19}}}Ne}\\
\parbox[b][0.3cm]{17.7cm}{analog states and deduced resonance parameters for the \ensuremath{^{\textnormal{19}}}Ne*(4379, 4549, 4600) levels using \ensuremath{\Gamma}\ensuremath{_{\ensuremath{\alpha}}}/\ensuremath{\Gamma} ratios from (\href{https://www.nndc.bnl.gov/nsr/nsrlink.jsp?1990Ma05,B}{1990Ma05}).}\\
\vspace{0.385cm}
\parbox[b][0.3cm]{17.7cm}{\addtolength{\parindent}{-0.2in}\textit{Experiments Already Presented in Other Datasets}:}\\
\parbox[b][0.3cm]{17.7cm}{\addtolength{\parindent}{-0.2in}\href{https://www.nndc.bnl.gov/nsr/nsrlink.jsp?1989MaZX,B}{1989MaZX}, \href{https://www.nndc.bnl.gov/nsr/nsrlink.jsp?1990Ma05,B}{1990Ma05}: \ensuremath{^{\textnormal{19}}}F(\ensuremath{^{\textnormal{3}}}He,t)\ensuremath{^{\textnormal{19}}}Ne*(\ensuremath{\alpha}) E=29.8 MeV; measured t-\ensuremath{\alpha} coincidence events; deduced the \ensuremath{^{\textnormal{19}}}Ne*(4033, 4140, 4197,}\\
\parbox[b][0.3cm]{17.7cm}{4379, 4549, 4600, 4635, 4712, 5092, 5351) levels. Deduced \ensuremath{\Gamma}\ensuremath{_{\ensuremath{\alpha}}}/\ensuremath{\Gamma} for most of these states. Obtained E\ensuremath{_{\textnormal{c.m.}}}, \ensuremath{\Gamma}\ensuremath{_{\ensuremath{\gamma}}}, \ensuremath{\Gamma}\ensuremath{_{\ensuremath{\alpha}}}, \ensuremath{\omega}\ensuremath{\gamma}\ensuremath{_{\textnormal{(}\ensuremath{\alpha}\textnormal{,}\ensuremath{\gamma}\textnormal{)}}},}\\
\parbox[b][0.3cm]{17.7cm}{and J\ensuremath{^{\ensuremath{\pi}}} for these states based on mirror levels in \ensuremath{^{\textnormal{19}}}F, determined \ensuremath{\Gamma}\ensuremath{_{\ensuremath{\alpha}}}/\ensuremath{\Gamma} from (\href{https://www.nndc.bnl.gov/nsr/nsrlink.jsp?1990Ma05,B}{1990Ma05}) and other literature. Deduced the}\\
\parbox[b][0.3cm]{17.7cm}{resonance contributions and the total \ensuremath{^{\textnormal{15}}}O(\ensuremath{\alpha},\ensuremath{\gamma}) reaction rate at T=0.1-10 GK. Discussed astrophysical implications.}\\
\parbox[b][0.3cm]{17.7cm}{\addtolength{\parindent}{-0.2in}\href{https://www.nndc.bnl.gov/nsr/nsrlink.jsp?2000Ha26,B}{2000Ha26}, \href{https://www.nndc.bnl.gov/nsr/nsrlink.jsp?2001Ha12,B}{2001Ha12}: \ensuremath{^{\textnormal{197}}}Au(\ensuremath{^{\textnormal{19}}}Ne,\ensuremath{^{\textnormal{19}}}Ne\ensuremath{'}) E=55 MeV/nucleon; estimated \ensuremath{\Gamma}\ensuremath{_{\ensuremath{\gamma}}}=12 meV \textit{+9{\textminus}5} for the \ensuremath{^{\textnormal{19}}}Ne*(4033 keV, 3/2\ensuremath{^{\textnormal{+}}}) level,}\\
\parbox[b][0.3cm]{17.7cm}{which plays an important role for the \ensuremath{^{\textnormal{15}}}O(\ensuremath{\alpha},\ensuremath{\gamma}) reaction rate at nova temperatures.}\\
\parbox[b][0.3cm]{17.7cm}{\addtolength{\parindent}{-0.2in}\href{https://www.nndc.bnl.gov/nsr/nsrlink.jsp?2003Da13,B}{2003Da13}, \href{https://www.nndc.bnl.gov/nsr/nsrlink.jsp?2003Da13,B}{2003Da13}: \ensuremath{^{\textnormal{1}}}H(\ensuremath{^{\textnormal{21}}}Ne,t) E=43 MeV/nucleon; deduced \ensuremath{\Gamma}\ensuremath{_{\ensuremath{\alpha}}}/\ensuremath{\Gamma} and \ensuremath{\Gamma}\ensuremath{_{\ensuremath{\alpha}}} for 6 important resonances. Deduced the \ensuremath{^{\textnormal{15}}}O(\ensuremath{\alpha},\ensuremath{\gamma})}\\
\parbox[b][0.3cm]{17.7cm}{direct capture rate and resonant rate contributions (for T\ensuremath{<}1.9 GK) corresponding to the 6 levels at 4033, 4379, 4549, 4600, 4712,}\\
\parbox[b][0.3cm]{17.7cm}{and 5092 keV. Performed hydrodynamic calculations of nova outbursts. Discussed comparison of \ensuremath{^{\textnormal{15}}}O(\ensuremath{\alpha},\ensuremath{\gamma}) and \ensuremath{^{\textnormal{15}}}O(\ensuremath{\beta}\ensuremath{^{\textnormal{+}}}) decay}\\
\parbox[b][0.3cm]{17.7cm}{rate. Concluded that no significant breakout from the hot CNO cycle into the rp-process takes place in novae via the \ensuremath{^{\textnormal{15}}}O(\ensuremath{\alpha},\ensuremath{\gamma})}\\
\parbox[b][0.3cm]{17.7cm}{reaction.}\\
\parbox[b][0.3cm]{17.7cm}{\addtolength{\parindent}{-0.2in}\href{https://www.nndc.bnl.gov/nsr/nsrlink.jsp?2003Re16,B}{2003Re16}, \href{https://www.nndc.bnl.gov/nsr/nsrlink.jsp?2003Re25,B}{2003Re25}: \ensuremath{^{\textnormal{3}}}He(\ensuremath{^{\textnormal{20}}}Ne,\ensuremath{\alpha}) E=98 MeV; deduced \ensuremath{\Gamma}\ensuremath{_{\ensuremath{\alpha}}}/\ensuremath{\Gamma} and \ensuremath{\omega}\ensuremath{\gamma}\ensuremath{_{\textnormal{(}\ensuremath{\alpha}\textnormal{,}\ensuremath{\gamma}\textnormal{)}}} for 6 important resonances. Determined the}\\
\parbox[b][0.3cm]{17.7cm}{\ensuremath{^{\textnormal{15}}}O(\ensuremath{\alpha},\ensuremath{\gamma}) resonant contributions and the total reaction rate for 0.1\ensuremath{<}T\ensuremath{<}1 GK corresponding to the \ensuremath{^{\textnormal{19}}}Ne*(4033, 4140, 4197, 4379,}\\
\parbox[b][0.3cm]{17.7cm}{4549, 4600, 4712, and 5092) levels. Discussed comparison of \ensuremath{^{\textnormal{15}}}O(\ensuremath{\alpha},\ensuremath{\gamma}) and \ensuremath{^{\textnormal{15}}}O(\ensuremath{\beta}\ensuremath{^{\textnormal{+}}}) decay rate for novae, type I X-ray bursts and}\\
\parbox[b][0.3cm]{17.7cm}{super massive stars. Provided comparison of different \ensuremath{^{\textnormal{15}}}O(\ensuremath{\alpha},\ensuremath{\gamma}) rates. Concluded that no significant breakout from the hot CNO}\\
\parbox[b][0.3cm]{17.7cm}{cycle into the rp-process in novae takes place via the \ensuremath{^{\textnormal{15}}}O(\ensuremath{\alpha},\ensuremath{\gamma}) reaction. However, this reaction was reported to likely have a}\\
\parbox[b][0.3cm]{17.7cm}{strong effect on the fate of super massive stars. They also mentioned that in type I X-ray bursts, the \ensuremath{^{\textnormal{15}}}O(\ensuremath{\alpha},\ensuremath{\gamma}) reaction is much}\\
\parbox[b][0.3cm]{17.7cm}{faster than the \ensuremath{^{\textnormal{15}}}O(\ensuremath{\beta}\ensuremath{^{\textnormal{+}}}) decay rate.}\\
\parbox[b][0.3cm]{17.7cm}{\addtolength{\parindent}{-0.2in}\href{https://www.nndc.bnl.gov/nsr/nsrlink.jsp?2005Ta28,B}{2005Ta28}: \ensuremath{^{\textnormal{17}}}O(\ensuremath{^{\textnormal{3}}}He,n\ensuremath{\gamma}) E=3 MeV; deduced the level-energies and lifetimes of the \ensuremath{^{\textnormal{19}}}Ne*(1.5-4.6 MeV) states, including those of}\\
\parbox[b][0.3cm]{17.7cm}{the \ensuremath{^{\textnormal{19}}}Ne*(4.03 MeV) state, which dominates the \ensuremath{^{\textnormal{15}}}O(\ensuremath{\alpha},\ensuremath{\gamma}) reaction rate at nova temperatures. The results reduce the uncertainty of}\\
\parbox[b][0.3cm]{17.7cm}{the \ensuremath{\gamma}-ray partial width of the 4.03-MeV state by a factor of 3, and are thus expected to improve the uncertainty in the \ensuremath{^{\textnormal{15}}}O(\ensuremath{\alpha},\ensuremath{\gamma})}\\
\parbox[b][0.3cm]{17.7cm}{reaction rate at these temperatures.}\\
\parbox[b][0.3cm]{17.7cm}{\addtolength{\parindent}{-0.2in}\href{https://www.nndc.bnl.gov/nsr/nsrlink.jsp?2006Ka50,B}{2006Ka50}: \ensuremath{^{\textnormal{3}}}He(\ensuremath{^{\textnormal{20}}}Ne,\ensuremath{\alpha}) E=34 MeV; measured lifetime of the \ensuremath{^{\textnormal{19}}}Ne*(4033) level. Constructed the joint likelihood for the lifetime}\\
\clearpage
\vspace{0.3cm}
{\bf \small \underline{\ensuremath{^{\textnormal{15}}}O(\ensuremath{\alpha},\ensuremath{\gamma})\hspace{0.2in}\href{https://www.nndc.bnl.gov/nsr/nsrlink.jsp?2011Da24,B}{2011Da24} (continued)}}\\
\vspace{0.3cm}
\parbox[b][0.3cm]{17.7cm}{using their result together with that of (\href{https://www.nndc.bnl.gov/nsr/nsrlink.jsp?2005Ta28,B}{2005Ta28}). From this, they deduced the 3\ensuremath{\sigma} lower limit on lifetime at the 99.73\% C.L.,}\\
\parbox[b][0.3cm]{17.7cm}{which was used together with the 3\ensuremath{\sigma} upper limit on the \ensuremath{\Gamma}\ensuremath{_{\ensuremath{\alpha}}}/\ensuremath{\Gamma} from (\href{https://www.nndc.bnl.gov/nsr/nsrlink.jsp?2003Da13,B}{2003Da13}) to obtain \ensuremath{\Gamma}\ensuremath{_{\ensuremath{\alpha}}}\ensuremath{<}200 \ensuremath{\mu}eV for the \ensuremath{^{\textnormal{19}}}Ne*(4033)}\\
\parbox[b][0.3cm]{17.7cm}{level. Discussed the implication of this width on the \ensuremath{^{\textnormal{15}}}O(\ensuremath{\alpha},\ensuremath{\gamma}) reaction rate.}\\
\parbox[b][0.3cm]{17.7cm}{\addtolength{\parindent}{-0.2in}\href{https://www.nndc.bnl.gov/nsr/nsrlink.jsp?2007TaZX,B}{2007TaZX}, \href{https://www.nndc.bnl.gov/nsr/nsrlink.jsp?2007Ta13,B}{2007Ta13}, \href{https://www.nndc.bnl.gov/nsr/nsrlink.jsp?2009Ta09,B}{2009Ta09}: \ensuremath{^{\textnormal{19}}}F(\ensuremath{^{\textnormal{3}}}He,t)\ensuremath{^{\textnormal{19}}}Ne*(\ensuremath{\alpha}) E=24 MeV; measured t-\ensuremath{\alpha} coincidence events; and deduced \ensuremath{\Gamma}\ensuremath{_{\ensuremath{\alpha}}}/\ensuremath{\Gamma}, \ensuremath{\Gamma}\ensuremath{_{\ensuremath{\alpha}}},}\\
\parbox[b][0.3cm]{17.7cm}{and \ensuremath{\omega}\ensuremath{\gamma}\ensuremath{_{\textnormal{(}\ensuremath{\alpha}\textnormal{,}\ensuremath{\gamma}\textnormal{)}}} for the \ensuremath{^{\textnormal{19}}}Ne*(4.03, 4.14+4.2 (unresolved), 4.38, 4.55, 4.60, 4.71, 5.09 MeV) levels. Deduced the \ensuremath{^{\textnormal{15}}}O(\ensuremath{\alpha},\ensuremath{\gamma}) reaction}\\
\parbox[b][0.3cm]{17.7cm}{rate at T=0.2-1.5 GK. Discussed the contribution of the 4.14-keV level to the reaction rate.}\\
\parbox[b][0.3cm]{17.7cm}{\addtolength{\parindent}{-0.2in}\href{https://www.nndc.bnl.gov/nsr/nsrlink.jsp?2008My01,B}{2008My01}: \ensuremath{^{\textnormal{3}}}He(\ensuremath{^{\textnormal{20}}}Ne,\ensuremath{\alpha}) E=34 MeV; measured lifetimes of 6 levels above the \ensuremath{^{\textnormal{15}}}O+\ensuremath{\alpha} threshold in \ensuremath{^{\textnormal{19}}}Ne and briefly discussed the}\\
\parbox[b][0.3cm]{17.7cm}{astrophysical implications of this measurement on the \ensuremath{^{\textnormal{15}}}O(\ensuremath{\alpha},\ensuremath{\gamma}) reaction rate.}\\
\parbox[b][0.3cm]{17.7cm}{\addtolength{\parindent}{-0.2in}\href{https://www.nndc.bnl.gov/nsr/nsrlink.jsp?2017To14,B}{2017To14}: \ensuremath{^{\textnormal{4}}}He(\ensuremath{^{\textnormal{15}}}O,\ensuremath{\alpha}) E=28.5 MeV; measured \ensuremath{\Gamma}\ensuremath{_{\ensuremath{\alpha}}}, \ensuremath{\Gamma}\ensuremath{_{\textnormal{p}}} (only for 1 state) and \ensuremath{\Gamma} for many of the \ensuremath{\alpha} resonances in \ensuremath{^{\textnormal{19}}}Ne.}\\
\parbox[b][0.3cm]{17.7cm}{Predicted \ensuremath{\alpha}+\ensuremath{^{\textnormal{15}}}O rotational structure. Mentioned that an enhanced \ensuremath{\alpha} structure in some of these states could explain the increased}\\
\parbox[b][0.3cm]{17.7cm}{contributions by some levels in the astrophysical \ensuremath{^{\textnormal{15}}}O(\ensuremath{\alpha},\ensuremath{\gamma}) reaction rate at temperatures relevant to type I X-ray bursts.}\\
\parbox[b][0.3cm]{17.7cm}{\addtolength{\parindent}{-0.2in}\href{https://www.nndc.bnl.gov/nsr/nsrlink.jsp?2019Ha14,B}{2019Ha14}: Deduced the fractional contributions of the 4.14- and 4.20-MeV states to the \ensuremath{^{\textnormal{15}}}O(\ensuremath{\alpha},\ensuremath{\gamma}) reaction rate at 0.2-2 GK}\\
\parbox[b][0.3cm]{17.7cm}{assuming J\ensuremath{^{\ensuremath{\pi}}}=7/2\ensuremath{^{-}} and 9/2\ensuremath{^{-}} for the \ensuremath{^{\textnormal{19}}}Ne*(4.14, 4.2-MeV) states, respectively. They recommended the \ensuremath{\alpha}-decay branching ratios of}\\
\parbox[b][0.3cm]{17.7cm}{\ensuremath{\Gamma}\ensuremath{_{\ensuremath{\alpha}}}/\ensuremath{\Gamma}=1.2\ensuremath{\times}10\ensuremath{^{\textnormal{$-$3}}} from (\href{https://www.nndc.bnl.gov/nsr/nsrlink.jsp?2009Ta09,B}{2009Ta09}).}\\
\vspace{0.385cm}
\parbox[b][0.3cm]{17.7cm}{\addtolength{\parindent}{-0.2in}\textit{Theoretical Work}:}\\
\parbox[b][0.3cm]{17.7cm}{\addtolength{\parindent}{-0.2in}R. V. Wagoner, Astrophys. J. Suppl. Ser., 18 (1969) 247: Calculated the parameterized \ensuremath{^{\textnormal{15}}}O(\ensuremath{\alpha},\ensuremath{\gamma}) reaction rate as a function of}\\
\parbox[b][0.3cm]{17.7cm}{temperature in GK.}\\
\parbox[b][0.3cm]{17.7cm}{\addtolength{\parindent}{-0.2in}R. K. Wallace and S. E. Woosley Astrophys. J. Suppl, 45 (1981) 389: Calculated the \ensuremath{^{\textnormal{15}}}O(\ensuremath{\alpha},\ensuremath{\gamma}) reaction rate using only the}\\
\parbox[b][0.3cm]{17.7cm}{resonances$'$ contributions. For those excitation energies that were experimentally unknown, the energies of their mirror level in \ensuremath{^{\textnormal{19}}}F}\\
\parbox[b][0.3cm]{17.7cm}{were used. Estimated the reduced \ensuremath{\alpha}-widths, J\ensuremath{^{\ensuremath{\pi}}}, L, \ensuremath{\Gamma}\ensuremath{_{\ensuremath{\alpha}}}, \ensuremath{\Gamma}\ensuremath{_{\ensuremath{\gamma}}}, and \ensuremath{\omega}\ensuremath{\gamma}\ensuremath{_{\textnormal{(}\ensuremath{\alpha}\textnormal{,}\ensuremath{\gamma}\textnormal{)}}} for the states involved. The \ensuremath{\gamma}-widths were mostly}\\
\parbox[b][0.3cm]{17.7cm}{from analog states in \ensuremath{^{\textnormal{19}}}F. Calculated the \ensuremath{\alpha}-widths from a barrier penetration model with a reduced \ensuremath{\alpha}-width of \ensuremath{\theta}\ensuremath{_{\ensuremath{\alpha}}^{\textnormal{2}}}=0.02 for all}\\
\parbox[b][0.3cm]{17.7cm}{\ensuremath{^{\textnormal{19}}}Ne states involved in the reaction rate calculation.}\\
\parbox[b][0.3cm]{17.7cm}{\addtolength{\parindent}{-0.2in}\href{https://www.nndc.bnl.gov/nsr/nsrlink.jsp?1986La07,B}{1986La07}: Deduced the reduced \ensuremath{\alpha}-widths for the \ensuremath{^{\textnormal{19}}}Ne states involved in the \ensuremath{^{\textnormal{15}}}O(\ensuremath{\alpha},\ensuremath{\gamma}) reaction rate with 504\ensuremath{\leq}E\ensuremath{_{\textnormal{c.m.}}}\ensuremath{\leq}3 MeV;}\\
\parbox[b][0.3cm]{17.7cm}{determined J\ensuremath{^{\ensuremath{\pi}}}, L, \ensuremath{\Gamma}\ensuremath{_{\ensuremath{\alpha}}}, \ensuremath{\Gamma}\ensuremath{_{\ensuremath{\gamma}}} and \ensuremath{\omega}\ensuremath{\gamma}\ensuremath{_{\textnormal{(}\ensuremath{\alpha}\textnormal{,}\ensuremath{\gamma}\textnormal{)}}} for these states; deduced the \ensuremath{^{\textnormal{15}}}O(\ensuremath{\alpha},\ensuremath{\gamma}) direct capture rate for 0.1-10 GK assuming E1}\\
\parbox[b][0.3cm]{17.7cm}{transitions; determined the total reaction rate for 0.1-10 GK; compared the \ensuremath{^{\textnormal{15}}}O(\ensuremath{\alpha},\ensuremath{\gamma}) rate with the previous rate and the \ensuremath{^{\textnormal{15}}}O(\ensuremath{\beta}\ensuremath{^{\textnormal{+}}})}\\
\parbox[b][0.3cm]{17.7cm}{decay rate; discussed the consequences of the new reaction rate on the CNO cycle and the rp-process nucleosynthesis.}\\
\parbox[b][0.3cm]{17.7cm}{\addtolength{\parindent}{-0.2in}\href{https://www.nndc.bnl.gov/nsr/nsrlink.jsp?1987De05,B}{1987De05}: Constructed fully antisymmetric wave functions of the \ensuremath{\alpha}+\ensuremath{^{\textnormal{15}}}N\ensuremath{\rightarrow}\ensuremath{^{\textnormal{19}}}F and \ensuremath{\alpha}+\ensuremath{^{\textnormal{15}}}O\ensuremath{\rightarrow}\ensuremath{^{\textnormal{19}}}Ne systems using the framework of}\\
\parbox[b][0.3cm]{17.7cm}{the Generator Coordinate Method (GCM). Calculated the radiative-capture cross section for \ensuremath{^{\textnormal{15}}}O(\ensuremath{\alpha},\ensuremath{\gamma}) at low energies. Using the}\\
\parbox[b][0.3cm]{17.7cm}{GCM formalism, they obtained three states (not presented) above the \ensuremath{\alpha}-threshold with J\ensuremath{^{\ensuremath{\pi}}}=9/2\ensuremath{^{-}}, 7/2\ensuremath{^{-}} and 13/2\ensuremath{^{\textnormal{+}}}. The other states}\\
\parbox[b][0.3cm]{17.7cm}{involved in the \ensuremath{^{\textnormal{15}}}O(\ensuremath{\alpha},\ensuremath{\gamma}) reaction rate were challenging to construct as they belong to K\ensuremath{^{\ensuremath{\pi}}}=3/2\ensuremath{^{\textnormal{+}}} and 3/2\ensuremath{^{-}} rotational bands arising}\\
\parbox[b][0.3cm]{17.7cm}{from a less asymmetric \ensuremath{^{\textnormal{7}}}Li+\ensuremath{^{\textnormal{12}}}C cluster configuration with small \ensuremath{\alpha}-widths.}\\
\parbox[b][0.3cm]{17.7cm}{\addtolength{\parindent}{-0.2in}\href{https://www.nndc.bnl.gov/nsr/nsrlink.jsp?1988Bu01,B}{1988Bu01}: Deduced direct capture and resonant contributions to the \ensuremath{^{\textnormal{15}}}O(\ensuremath{\alpha},\ensuremath{\gamma}) reaction rate. Parameterized the reaction rate using}\\
\parbox[b][0.3cm]{17.7cm}{the format presented by (\href{https://www.nndc.bnl.gov/nsr/nsrlink.jsp?1983Ha55,B}{1983Ha55}).}\\
\parbox[b][0.3cm]{17.7cm}{\addtolength{\parindent}{-0.2in}\href{https://www.nndc.bnl.gov/nsr/nsrlink.jsp?1988Ca26,B}{1988Ca26}: Deduced the direct capture (based on the work of \href{https://www.nndc.bnl.gov/nsr/nsrlink.jsp?1986La07,B}{1986La07}) and resonant contributions (for 23 resonances with}\\
\parbox[b][0.3cm]{17.7cm}{E\ensuremath{_{\textnormal{c.m.}}}=0.611-3 MeV from \href{https://www.nndc.bnl.gov/nsr/nsrlink.jsp?1986La07,B}{1986La07} and \href{https://www.nndc.bnl.gov/nsr/nsrlink.jsp?1987Ma31,B}{1987Ma31}) to the \ensuremath{^{\textnormal{15}}}O(\ensuremath{\alpha},\ensuremath{\gamma}) reaction rate for 0.001-10 GK. Provided analytical expression}\\
\parbox[b][0.3cm]{17.7cm}{of the reaction rate.}\\
\parbox[b][0.3cm]{17.7cm}{\addtolength{\parindent}{-0.2in}\href{https://www.nndc.bnl.gov/nsr/nsrlink.jsp?1996Ha26,B}{1996Ha26}: Adopted the reaction rate of (\href{https://www.nndc.bnl.gov/nsr/nsrlink.jsp?1995Ma28,B}{1995Ma28}) and presented the rate in tabular format for T=0.1-1 GK. (\href{https://www.nndc.bnl.gov/nsr/nsrlink.jsp?2010Cy01,B}{2010Cy01}) cited}\\
\parbox[b][0.3cm]{17.7cm}{that (\href{https://www.nndc.bnl.gov/nsr/nsrlink.jsp?1996Ha26,B}{1996Ha26}) found \ensuremath{\Gamma}\ensuremath{_{\ensuremath{\alpha}}}/\ensuremath{\Gamma}\ensuremath{\approx}1.2\ensuremath{\times}10\ensuremath{^{\textnormal{$-$4}}} for the \ensuremath{^{\textnormal{19}}}Ne*(4033) state.}\\
\parbox[b][0.3cm]{17.7cm}{\addtolength{\parindent}{-0.2in}\href{https://www.nndc.bnl.gov/nsr/nsrlink.jsp?1997De14,B}{1997De14}: Deduced resonance properties of the \ensuremath{^{\textnormal{19}}}Ne*(4379, 4549, 4600, 4712, 5092) states based on the data of (\href{https://www.nndc.bnl.gov/nsr/nsrlink.jsp?1996De07,B}{1996De07}:}\\
\parbox[b][0.3cm]{17.7cm}{\ensuremath{^{\textnormal{15}}}N(\ensuremath{^{\textnormal{7}}}Li,t) E=28 MeV; where the \ensuremath{\alpha} spectroscopic factors of \ensuremath{^{\textnormal{19}}}F levels were deduced from a DWBA analysis of the angular}\\
\parbox[b][0.3cm]{17.7cm}{distributions measured using a spectrograph) using the assumptions that \ensuremath{\Gamma}\ensuremath{_{\ensuremath{\gamma}}}(\ensuremath{^{\textnormal{19}}}Ne*)=\ensuremath{\Gamma}\ensuremath{_{\ensuremath{\gamma}}}(\ensuremath{^{\textnormal{19}}}F*)=\ensuremath{\Gamma}(\ensuremath{^{\textnormal{19}}}F*) [because}\\
\parbox[b][0.3cm]{17.7cm}{\ensuremath{\Gamma}\ensuremath{_{\ensuremath{\gamma}}}/\ensuremath{\Gamma}(\ensuremath{^{\textnormal{19}}}F*)\ensuremath{\approx}1 (\href{https://www.nndc.bnl.gov/nsr/nsrlink.jsp?1989Pr01,B}{1989Pr01})] and \ensuremath{\theta}\ensuremath{^{\textnormal{2}}_{\ensuremath{\alpha}}}(\ensuremath{^{\textnormal{19}}}F*)=\ensuremath{\theta}\ensuremath{^{\textnormal{2}}_{\ensuremath{\alpha}}}(\ensuremath{^{\textnormal{19}}}Ne*). Compared the results with the same properties from \ensuremath{^{\textnormal{19}}}F* mirror}\\
\parbox[b][0.3cm]{17.7cm}{levels. Analyzed uncertainties found in the literature concerning the deduced \ensuremath{^{\textnormal{19}}}Ne* resonance strengths when the above}\\
\parbox[b][0.3cm]{17.7cm}{assumptions were used. Concluded that the resonance strengths obtained under such assumptions remain uncertain because of the}\\
\parbox[b][0.3cm]{17.7cm}{questionable validity of these approximations particularly when the levels of interest have a weak \ensuremath{\alpha}-cluster structure. Concluded}\\
\parbox[b][0.3cm]{17.7cm}{that the \ensuremath{^{\textnormal{15}}}O(\ensuremath{\alpha},\ensuremath{\gamma}) reaction rates that rely on the data for \ensuremath{\alpha}-transfer on \ensuremath{^{\textnormal{15}}}N are uncertain to at least one order of magnitude.}\\
\parbox[b][0.3cm]{17.7cm}{\addtolength{\parindent}{-0.2in}\href{https://www.nndc.bnl.gov/nsr/nsrlink.jsp?2000Du09,B}{2000Du09}: Investigated the \ensuremath{^{\textnormal{15}}}O(\ensuremath{\alpha},\ensuremath{\gamma}) reaction rate by using a microscopic multi-cluster model for \ensuremath{^{\textnormal{15}}}O+\ensuremath{\alpha}. Calculated the \ensuremath{\alpha}+\ensuremath{^{\textnormal{15}}}O}\\
\parbox[b][0.3cm]{17.7cm}{wave functions using the Generator Coordinate Method. Deduced the rotational band structure, the E1 and E2 transition}\\
\parbox[b][0.3cm]{17.7cm}{probabilities, \ensuremath{^{\textnormal{19}}}Ne excitation energies, and the non-resonant \ensuremath{^{\textnormal{15}}}O(\ensuremath{\alpha},\ensuremath{\gamma}) capture cross section. Obtained the S-factor for E\ensuremath{_{\textnormal{c.m.}}}\ensuremath{<}4}\\
\parbox[b][0.3cm]{17.7cm}{MeV corresponding to all bound states of the 1/2\ensuremath{^{\textnormal{+}}_{\textnormal{1}}} and 1/2\ensuremath{^{-}_{\textnormal{1}}} bands with E1 transitions. Presented an analytical expression for the}\\
\parbox[b][0.3cm]{17.7cm}{\ensuremath{^{\textnormal{15}}}O(\ensuremath{\alpha},\ensuremath{\gamma}) reaction rate. A comparison with previous reaction rates is presented.}\\
\parbox[b][0.3cm]{17.7cm}{\addtolength{\parindent}{-0.2in}\href{https://www.nndc.bnl.gov/nsr/nsrlink.jsp?2001La16,B}{2001La16}, \href{https://www.nndc.bnl.gov/nsr/nsrlink.jsp?2002La29,B}{2002La29}: Discussed the \ensuremath{^{\textnormal{15}}}O(\ensuremath{\alpha},\ensuremath{\gamma}) reaction and its effect on the novae and type I X-ray bursts nucleosynthesis.}\\
\clearpage
\vspace{0.3cm}
{\bf \small \underline{\ensuremath{^{\textnormal{15}}}O(\ensuremath{\alpha},\ensuremath{\gamma})\hspace{0.2in}\href{https://www.nndc.bnl.gov/nsr/nsrlink.jsp?2011Da24,B}{2011Da24} (continued)}}\\
\vspace{0.3cm}
\parbox[b][0.3cm]{17.7cm}{Obtained upper limits on the \ensuremath{\alpha}-branching ratios of the \ensuremath{^{\textnormal{19}}}Ne*(4033, 4140, 4197) states and deduced the \ensuremath{\alpha}-branching ratios for the}\\
\parbox[b][0.3cm]{17.7cm}{\ensuremath{^{\textnormal{19}}}Ne*(4600, 5092, 5351/5424/5463, 6013/6092 keV) states.}\\
\parbox[b][0.3cm]{17.7cm}{\addtolength{\parindent}{-0.2in}\href{https://www.nndc.bnl.gov/nsr/nsrlink.jsp?2002Os05,B}{2002Os05}: Discussed the \ensuremath{^{\textnormal{15}}}O(\ensuremath{\alpha},\ensuremath{\gamma}) reaction and its effect on the novae and type I X-ray bursts nucleosynthesis. Obtained \ensuremath{^{\textnormal{19}}}Ne}\\
\parbox[b][0.3cm]{17.7cm}{level-energies and \ensuremath{\alpha}-branching ratio for the \ensuremath{^{\textnormal{19}}}Ne*(4.6 MeV) state.}\\
\parbox[b][0.3cm]{17.7cm}{\addtolength{\parindent}{-0.2in}\href{https://www.nndc.bnl.gov/nsr/nsrlink.jsp?2003Fo15,B}{2003Fo15}: Calculated the theoretical \ensuremath{\alpha} spectroscopic factor (S\ensuremath{_{\ensuremath{\alpha}}}) for the \ensuremath{^{\textnormal{19}}}Ne*(4.03 MeV) state using the experimental data of}\\
\parbox[b][0.3cm]{17.7cm}{(\href{https://www.nndc.bnl.gov/nsr/nsrlink.jsp?1995Ma28,B}{1995Ma28}, \href{https://www.nndc.bnl.gov/nsr/nsrlink.jsp?1996Ma07,B}{1996Ma07}). These data investigated the \ensuremath{^{\textnormal{19}}}F* analog level for the \ensuremath{^{\textnormal{19}}}Ne*(4.03 MeV) state. (\href{https://www.nndc.bnl.gov/nsr/nsrlink.jsp?2003Fo15,B}{2003Fo15}) deduced}\\
\parbox[b][0.3cm]{17.7cm}{S\ensuremath{_{\ensuremath{\alpha}}}=0.052 for the \ensuremath{^{\textnormal{19}}}Ne*(4.03 MeV) state, which resulted in a theoretical \ensuremath{\Gamma}\ensuremath{_{\ensuremath{\alpha}}}=7.5 \ensuremath{\mu}eV. The authors explained that their results}\\
\parbox[b][0.3cm]{17.7cm}{would reduce the \ensuremath{^{\textnormal{15}}}O(\ensuremath{\alpha},\ensuremath{\gamma}) reaction rate determined by (\href{https://www.nndc.bnl.gov/nsr/nsrlink.jsp?1995Ma28,B}{1995Ma28}) by a factor of 1.3.}\\
\parbox[b][0.3cm]{17.7cm}{\addtolength{\parindent}{-0.2in}\href{https://www.nndc.bnl.gov/nsr/nsrlink.jsp?2006Fi07,B}{2006Fi07}: Deduced the lower and upper limits, as well as the recommended \ensuremath{^{\textnormal{15}}}O(\ensuremath{\alpha},\ensuremath{\gamma}) reaction rate for T=0.1-1 GK using the}\\
\parbox[b][0.3cm]{17.7cm}{direct capture rates of (\href{https://www.nndc.bnl.gov/nsr/nsrlink.jsp?1986La07,B}{1986La07}, \href{https://www.nndc.bnl.gov/nsr/nsrlink.jsp?2000Du09,B}{2000Du09}) and the experimental resonance parameters from (\href{https://www.nndc.bnl.gov/nsr/nsrlink.jsp?1973Da31,B}{1973Da31}, \href{https://www.nndc.bnl.gov/nsr/nsrlink.jsp?1996Ma07,B}{1996Ma07}, \href{https://www.nndc.bnl.gov/nsr/nsrlink.jsp?2000Ha26,B}{2000Ha26},}\\
\parbox[b][0.3cm]{17.7cm}{\href{https://www.nndc.bnl.gov/nsr/nsrlink.jsp?2002Wi18,B}{2002Wi18}, \href{https://www.nndc.bnl.gov/nsr/nsrlink.jsp?2003Da13,B}{2003Da13}, \href{https://www.nndc.bnl.gov/nsr/nsrlink.jsp?2003Re16,B}{2003Re16}) to obtain the resonant rate and its uncertainty band. Performed a network nucleosynthesis}\\
\parbox[b][0.3cm]{17.7cm}{calculation for type I X-ray bursts using a time-dependent X-ray burst model. Discussed implications of the \ensuremath{^{\textnormal{15}}}O(\ensuremath{\alpha},\ensuremath{\gamma}) rate on the}\\
\parbox[b][0.3cm]{17.7cm}{X-ray bursts.}\\
\parbox[b][0.3cm]{17.7cm}{\addtolength{\parindent}{-0.2in}\href{https://www.nndc.bnl.gov/nsr/nsrlink.jsp?2011Da24,B}{2011Da24}: Deduced resonance parameters for the \ensuremath{^{\textnormal{19}}}Ne*(4.03, 4.14, 4.20, 4.38, 4.55, 4.60, 4.71, and 5.09 MeV) states based on the}\\
\parbox[b][0.3cm]{17.7cm}{evaluated (by the authors) available experimental data in the literature and from likelihood distributions constructed by the authors.}\\
\parbox[b][0.3cm]{17.7cm}{Evaluated the \ensuremath{^{\textnormal{15}}}O(\ensuremath{\alpha},\ensuremath{\gamma}) reaction rate (median rate and high and low rates with 99.73\% C.L.) at 0.1-2 GK using Monte Carlo}\\
\parbox[b][0.3cm]{17.7cm}{techniques. Investigated the impact of this rate on the nucleosynthesis of type I X-ray bursts using 3 hydrodynamic models.}\\
\parbox[b][0.3cm]{17.7cm}{\addtolength{\parindent}{-0.2in}\href{https://www.nndc.bnl.gov/nsr/nsrlink.jsp?2014Ot03,B}{2014Ot03}: Deduced \ensuremath{^{\textnormal{19}}}Ne levels, resonances, J\ensuremath{^{\ensuremath{\pi}}}, \ensuremath{\alpha}+\ensuremath{^{\textnormal{15}}}O(0, 1/2\ensuremath{^{-}}) rotational bands, and decay widths using a simple \ensuremath{\alpha}+\ensuremath{^{\textnormal{15}}}O}\\
\parbox[b][0.3cm]{17.7cm}{interaction potential deduced by (\href{https://www.nndc.bnl.gov/nsr/nsrlink.jsp?1990Kr16,B}{1990Kr16}) from \ensuremath{\alpha}+\ensuremath{^{\textnormal{16}}}O tuned to reproduce the \ensuremath{^{\textnormal{15}}}N(\ensuremath{\alpha},\ensuremath{\alpha}) elastic scattering data at E=6.85, 23.7,}\\
\parbox[b][0.3cm]{17.7cm}{28, 48.7, 54.1 MeV, where \ensuremath{\sigma}(\ensuremath{\theta}) was analyzed. Comparison with experimental results are given.}\\
\parbox[b][0.3cm]{17.7cm}{\addtolength{\parindent}{-0.2in}\href{https://www.nndc.bnl.gov/nsr/nsrlink.jsp?2015Pa46,B}{2015Pa46}: Favored the reaction rate deduced by (\href{https://www.nndc.bnl.gov/nsr/nsrlink.jsp?2011Da24,B}{2011Da24}) over those deduced by (\href{https://www.nndc.bnl.gov/nsr/nsrlink.jsp?2007Ta13,B}{2007Ta13}, \href{https://www.nndc.bnl.gov/nsr/nsrlink.jsp?2009Ta09,B}{2009Ta09}).}\\
\parbox[b][0.3cm]{17.7cm}{\addtolength{\parindent}{-0.2in}\href{https://www.nndc.bnl.gov/nsr/nsrlink.jsp?2016OtZZ,B}{2016OtZZ}: Calculated \ensuremath{^{\textnormal{19}}}Ne energy-levels and J\ensuremath{^{\ensuremath{\pi}}} using microscopic coupled-channel approach to \ensuremath{^{\textnormal{15}}}O+\ensuremath{\alpha} and \ensuremath{^{\textnormal{16}}}O+\ensuremath{^{\textnormal{3}}}He, and using}\\
\parbox[b][0.3cm]{17.7cm}{macroscopic approach based on \ensuremath{^{\textnormal{15}}}O+\ensuremath{\alpha} potential model; deduced several resonances above the \ensuremath{\alpha} threshold of weak-coupling}\\
\parbox[b][0.3cm]{17.7cm}{structure of \ensuremath{\alpha}-particle plus 1 hole inside the \ensuremath{^{\textnormal{16}}}O nucleus. \ensuremath{^{\textnormal{19}}}Ne levels were compared to data. The \ensuremath{^{\textnormal{19}}}Ne*(4033 keV, 3/2\ensuremath{^{\textnormal{+}}}) state,}\\
\parbox[b][0.3cm]{17.7cm}{which plays a significant role in the \ensuremath{^{\textnormal{15}}}O(\ensuremath{\alpha},\ensuremath{\gamma}) astrophysical reaction rate was not reproduced.}\\
\parbox[b][0.3cm]{17.7cm}{\addtolength{\parindent}{-0.2in}\href{https://www.nndc.bnl.gov/nsr/nsrlink.jsp?2021Sa42,B}{2021Sa42}: Deduced theoretical level-energies, J\ensuremath{^{\ensuremath{\pi}}} values, lifetimes, \ensuremath{\Gamma}\ensuremath{_{\ensuremath{\gamma}}}, \ensuremath{\Gamma}\ensuremath{_{\ensuremath{\alpha}}}, and \ensuremath{\Gamma}\ensuremath{_{\textnormal{sp}}} (the reduced transition probability) for the}\\
\parbox[b][0.3cm]{17.7cm}{\ensuremath{^{\textnormal{19}}}Ne*(4.14, 4.2 MeV) states in the frame of the shifted Deng-Fan potential model using the Nikiforov-Uvarov method.}\\
\vspace{0.385cm}
\parbox[b][0.3cm]{17.7cm}{\addtolength{\parindent}{-0.2in}\textit{See also}:}\\
\parbox[b][0.3cm]{17.7cm}{\addtolength{\parindent}{-0.2in}R. E. Taam and R. E. Picklum, Astrophys. J. 233 (1979) 327, \href{https://www.nndc.bnl.gov/nsr/nsrlink.jsp?1996Ku31,B}{1996Ku31}, \href{https://www.nndc.bnl.gov/nsr/nsrlink.jsp?1999Wi06,B}{1999Wi06}, \href{https://www.nndc.bnl.gov/nsr/nsrlink.jsp?2001Ch44,B}{2001Ch44}, \href{https://www.nndc.bnl.gov/nsr/nsrlink.jsp?2005Fi13,B}{2005Fi13}, \href{https://www.nndc.bnl.gov/nsr/nsrlink.jsp?2005Fi06,B}{2005Fi06}, \href{https://www.nndc.bnl.gov/nsr/nsrlink.jsp?2006Co26,B}{2006Co26},}\\
\parbox[b][0.3cm]{17.7cm}{\href{https://www.nndc.bnl.gov/nsr/nsrlink.jsp?2006Co27,B}{2006Co27}, \href{https://www.nndc.bnl.gov/nsr/nsrlink.jsp?2007Wi06,B}{2007Wi06}, \href{https://www.nndc.bnl.gov/nsr/nsrlink.jsp?2008Fi11,B}{2008Fi11}, \href{https://www.nndc.bnl.gov/nsr/nsrlink.jsp?2007Lu10,B}{2007Lu10}, \href{https://www.nndc.bnl.gov/nsr/nsrlink.jsp?2010Cy01,B}{2010Cy01}, \href{https://www.nndc.bnl.gov/nsr/nsrlink.jsp?2010Wi15,B}{2010Wi15}, and \href{https://www.nndc.bnl.gov/nsr/nsrlink.jsp?2016Cy01,B}{2016Cy01}.}\\
\vspace{12pt}
\underline{$^{19}$Ne Levels}\\
\begin{longtable}{cc@{\extracolsep{\fill}}c}
\multicolumn{2}{c}{E(level)$^{{\hyperlink{NE22LEVEL0}{a}}}$}&\\[-.2cm]
\multicolumn{2}{c}{\hrulefill}&
\endfirsthead
\multicolumn{1}{r@{}}{4034}&\multicolumn{1}{@{.}l}{5}&\\
\multicolumn{1}{r@{}}{4142}&\multicolumn{1}{@{.}l}{8}&\\
\multicolumn{1}{r@{}}{4199}&\multicolumn{1}{@{.}l}{5}&\\
\multicolumn{1}{r@{}}{4377}&\multicolumn{1}{@{.}l}{5}&\\
\multicolumn{1}{r@{}}{4548}&\multicolumn{1}{@{.}l}{6}&\\
\multicolumn{1}{r@{}}{4602}&\multicolumn{1}{@{.}l}{4}&\\
\multicolumn{1}{r@{}}{4708}&\multicolumn{1}{@{.}l}{5}&\\
\multicolumn{1}{r@{}}{5091}&\multicolumn{1}{@{.}l}{1}&\\
\end{longtable}
\parbox[b][0.3cm]{17.7cm}{\makebox[1ex]{\ensuremath{^{\hypertarget{NE22LEVEL0}{a}}}} From the \ensuremath{^{\textnormal{19}}}Ne Adopted Levels.}\\
\vspace{0.5cm}
\clearpage
\subsection[\hspace{-0.2cm}\ensuremath{^{\textnormal{16}}}O(\ensuremath{^{\textnormal{3}}}He,X)]{ }
\vspace{-27pt}
\vspace{0.3cm}
\hypertarget{NE23}{{\bf \small \underline{\ensuremath{^{\textnormal{16}}}O(\ensuremath{^{\textnormal{3}}}He,X)\hspace{0.2in}\href{https://www.nndc.bnl.gov/nsr/nsrlink.jsp?1959Br79,B}{1959Br79},\href{https://www.nndc.bnl.gov/nsr/nsrlink.jsp?1983Wa05,B}{1983Wa05}}}}\\
\vspace{4pt}
\vspace{8pt}
\parbox[b][0.3cm]{17.7cm}{\addtolength{\parindent}{-0.2in}J\ensuremath{^{\ensuremath{\pi}}}(\ensuremath{^{\textnormal{16}}}O\ensuremath{_{\textnormal{g.s.}}})=0\ensuremath{^{\textnormal{+}}} and J\ensuremath{^{\ensuremath{\pi}}}(\ensuremath{^{\textnormal{3}}}He)=1/2\ensuremath{^{\textnormal{+}}}.}\\
\parbox[b][0.3cm]{17.7cm}{\addtolength{\parindent}{-0.2in}\href{https://www.nndc.bnl.gov/nsr/nsrlink.jsp?1958Br86,B}{1958Br86}: \ensuremath{^{\textnormal{16}}}O(\ensuremath{^{\textnormal{3}}}He,\ensuremath{\alpha}) E=2.0-5.4 MeV; measured yield curves for the \ensuremath{^{\textnormal{19}}}Ne*\ensuremath{\rightarrow}\ensuremath{\alpha}\ensuremath{_{\textnormal{0}}}+\ensuremath{^{\textnormal{15}}}O\ensuremath{_{\textnormal{g.s.}}}, \ensuremath{^{\textnormal{19}}}Ne*\ensuremath{\rightarrow}p\ensuremath{_{\textnormal{0}}}+\ensuremath{^{\textnormal{18}}}F\ensuremath{_{\textnormal{g.s.}}}, and}\\
\parbox[b][0.3cm]{17.7cm}{\ensuremath{^{\textnormal{19}}}Ne*\ensuremath{\rightarrow}p\ensuremath{_{\textnormal{1,2,3,4,5}}}+\ensuremath{^{\textnormal{18}}}F* decays. All yield curves show pronounced resonance structure; in particular, at E\ensuremath{_{\textnormal{lab}}}\ensuremath{\approx}2.45 MeV,}\\
\parbox[b][0.3cm]{17.7cm}{corresponding to \ensuremath{^{\textnormal{19}}}Ne*(\ensuremath{\approx}10.503 MeV). Measured angular distributions at several energies.}\\
\parbox[b][0.3cm]{17.7cm}{\addtolength{\parindent}{-0.2in}\href{https://www.nndc.bnl.gov/nsr/nsrlink.jsp?1959Br79,B}{1959Br79}: \ensuremath{^{\textnormal{16}}}O(\ensuremath{^{\textnormal{3}}}He,\ensuremath{\alpha}) and \ensuremath{^{\textnormal{16}}}O(\ensuremath{^{\textnormal{3}}}He,p) E=2.1-3.1 MeV, and \ensuremath{^{\textnormal{16}}}O(\ensuremath{^{\textnormal{3}}}He,\ensuremath{\gamma}) E=2.42 MeV; measured \ensuremath{\sigma}(E\ensuremath{_{^{\textnormal{3}}\textnormal{He}}},\ensuremath{\theta}) of the \ensuremath{^{\textnormal{16}}}O+\ensuremath{^{\textnormal{3}}}He}\\
\parbox[b][0.3cm]{17.7cm}{capture reaction using a gridded ionization chamber placed at \ensuremath{\theta}\ensuremath{_{\textnormal{lab}}}=83\ensuremath{^\circ}, 90\ensuremath{^\circ}, 119\ensuremath{^\circ}, and 145\ensuremath{^\circ} with a resolution of \ensuremath{\Delta}E/E\ensuremath{\approx}0.9\% at}\\
\parbox[b][0.3cm]{17.7cm}{FWHM for 10-MeV \ensuremath{\alpha} particles. Measured angular distributions corresponding to \ensuremath{\alpha}\ensuremath{_{\textnormal{0}}} and p\ensuremath{_{\textnormal{0$-$7}}} and observed a resonance structure}\\
\parbox[b][0.3cm]{17.7cm}{near E\ensuremath{_{\textnormal{lab}}}=2.4 MeV in all these channels. Measured the \ensuremath{\gamma} rays emitted following the \ensuremath{^{\textnormal{3}}}He capture by \ensuremath{^{\textnormal{16}}}O using NaI(Tl) detectors}\\
\parbox[b][0.3cm]{17.7cm}{at \ensuremath{\theta}\ensuremath{_{\textnormal{lab}}}=15\ensuremath{^\circ} and 90\ensuremath{^\circ}. Deduced coefficients of Legendre polynomials from measured angular distributions; deduced cross sections for}\\
\parbox[b][0.3cm]{17.7cm}{\ensuremath{^{\textnormal{16}}}O(\ensuremath{^{\textnormal{3}}}He,\ensuremath{\alpha}) and \ensuremath{^{\textnormal{16}}}O(\ensuremath{^{\textnormal{3}}}He,p) at E\ensuremath{_{\textnormal{lab}}}=2.4 MeV; deduced \ensuremath{\Gamma}, and \ensuremath{\alpha}-, p- and \ensuremath{^{\textnormal{3}}}He-partial widths. Part of these data are also}\\
\parbox[b][0.3cm]{17.7cm}{presented in (\href{https://www.nndc.bnl.gov/nsr/nsrlink.jsp?1960Br40,B}{1960Br40}), see section 2.2.1.}\\
\parbox[b][0.3cm]{17.7cm}{\addtolength{\parindent}{-0.2in}\href{https://www.nndc.bnl.gov/nsr/nsrlink.jsp?1959Hi73,B}{1959Hi73}: \ensuremath{^{\textnormal{16}}}O(\ensuremath{^{\textnormal{3}}}He,\ensuremath{\alpha}) E=5.70, 5.89, 9.16 MeV; measured the \ensuremath{\sigma}(E\ensuremath{_{\ensuremath{\alpha}}},\ensuremath{\theta}) for the \ensuremath{\alpha}\ensuremath{_{\textnormal{0}}} group at all given energies at \ensuremath{\theta}\ensuremath{_{\textnormal{c.m.}}}=10\ensuremath{^\circ}{\textminus}130\ensuremath{^\circ}}\\
\parbox[b][0.3cm]{17.7cm}{and \ensuremath{\sigma}(E\ensuremath{_{\ensuremath{\alpha}}},\ensuremath{\theta}) for the \ensuremath{\alpha}\ensuremath{_{\textnormal{1}}} and \ensuremath{\alpha}\ensuremath{_{\textnormal{2}}} groups at 5.70 and 9.16 MeV incident energies and \ensuremath{\theta}\ensuremath{_{\textnormal{c.m.}}}=10\ensuremath{^\circ}{\textminus}100\ensuremath{^\circ}. Used DWBA calculations to}\\
\parbox[b][0.3cm]{17.7cm}{deduce L. The authors concluded that at E\ensuremath{\geq}6 MeV, the \ensuremath{^{\textnormal{16}}}O(\ensuremath{^{\textnormal{3}}}He,\ensuremath{\alpha}) reaction proceeds through direct reaction mechanism, while at}\\
\parbox[b][0.3cm]{17.7cm}{E\ensuremath{<}6 MeV, the compound nuclear reaction competes strongly with direct reaction mechanism. They also observed a resonance}\\
\parbox[b][0.3cm]{17.7cm}{corresponding to E\ensuremath{_{^{\textnormal{3}}\textnormal{He}}}=5.7 MeV (resonance energy not given).}\\
\parbox[b][0.3cm]{17.7cm}{\addtolength{\parindent}{-0.2in}\href{https://www.nndc.bnl.gov/nsr/nsrlink.jsp?1961Si09,B}{1961Si09}: \ensuremath{^{\textnormal{16}}}O(\ensuremath{^{\textnormal{3}}}He,\ensuremath{^{\textnormal{3}}}He) and \ensuremath{^{\textnormal{16}}}O(\ensuremath{^{\textnormal{3}}}He,\ensuremath{\alpha}) E=1-3 MeV; measured charged particles using a CsI(Tl) crystal with a resolution that}\\
\parbox[b][0.3cm]{17.7cm}{varied from \ensuremath{\Delta}E/E=8\% for 1.7-MeV protons to 22\% for 460-keV \ensuremath{\alpha}s. Measured differential cross sections at \ensuremath{\theta}\ensuremath{_{\textnormal{lab}}}=14\ensuremath{^\circ}{\textminus}166\ensuremath{^\circ}.}\\
\parbox[b][0.3cm]{17.7cm}{Observed the 10.46-MeV state that was measured by (\href{https://www.nndc.bnl.gov/nsr/nsrlink.jsp?1959Br79,B}{1959Br79}) and saw the evidence of the tail of a higher energy resonance in}\\
\parbox[b][0.3cm]{17.7cm}{the form of a rise in the cross section above E\ensuremath{_{\textnormal{lab}}}=2.9 MeV. No evidence was seen for the resonance observed by (\href{https://www.nndc.bnl.gov/nsr/nsrlink.jsp?1959Br79,B}{1959Br79}) at}\\
\parbox[b][0.3cm]{17.7cm}{E\ensuremath{_{\textnormal{lab}}}=2.425 MeV. There was evidence in the elastic scattering data that indicated destructive interference between the level at}\\
\parbox[b][0.3cm]{17.7cm}{E\ensuremath{_{\textnormal{x}}}=10.46 MeV and the tail of the higher energy resonance suggesting they both have even parities.}\\
\parbox[b][0.3cm]{17.7cm}{\addtolength{\parindent}{-0.2in}\href{https://www.nndc.bnl.gov/nsr/nsrlink.jsp?1961To03,B}{1961To03}: \ensuremath{^{\textnormal{16}}}O(\ensuremath{^{\textnormal{3}}}He,n) E\ensuremath{\leq}10 MeV. This work aimed at studying \ensuremath{^{\textnormal{12}}}C(\ensuremath{^{\textnormal{3}}}He,n) and \ensuremath{^{\textnormal{16}}}O(\ensuremath{^{\textnormal{3}}}He,n) by measuring neutrons and their}\\
\parbox[b][0.3cm]{17.7cm}{TOF using two long BF\ensuremath{_{\textnormal{3}}} counters. The authors claim that about 15 of the \ensuremath{^{\textnormal{19}}}Ne* resonances were populated in the region of}\\
\parbox[b][0.3cm]{17.7cm}{E\ensuremath{_{\textnormal{lab}}}=3.9-5.7 MeV, corresponding to the \ensuremath{^{\textnormal{19}}}Ne*(11.7-13.2 MeV) states with an average spacing of 0.1 MeV. These states are not}\\
\parbox[b][0.3cm]{17.7cm}{reported. A very small yield was observed at the neutron threshold energy (Q=8442.16 keV \textit{16} from \href{https://www.nndc.bnl.gov/nsr/nsrlink.jsp?2021Wa16,B}{2021Wa16}), which was}\\
\parbox[b][0.3cm]{17.7cm}{attributed to the absence of the \ensuremath{^{\textnormal{19}}}Ne* levels with J=1/2 assignment at this excitation, resulting in the almost complete inhibition of}\\
\parbox[b][0.3cm]{17.7cm}{the \textit{s}-wave neutron emission.}\\
\parbox[b][0.3cm]{17.7cm}{\addtolength{\parindent}{-0.2in}\href{https://www.nndc.bnl.gov/nsr/nsrlink.jsp?1965Al05,B}{1965Al05}: \ensuremath{^{\textnormal{16}}}O(\ensuremath{^{\textnormal{3}}}He,\ensuremath{\alpha}\ensuremath{_{\textnormal{0}}}) E=8, 8.5, 9, 9.42, and 10 MeV and \ensuremath{^{\textnormal{16}}}O(\ensuremath{^{\textnormal{3}}}He, \ensuremath{^{\textnormal{3}}}He) E=8.5, 9.42 MeV; measured \ensuremath{\sigma}(E\ensuremath{_{\ensuremath{\alpha}}},\ensuremath{\theta}) at \ensuremath{\theta}\ensuremath{_{\textnormal{lab}}}=0\ensuremath{^\circ}{\textminus}30\ensuremath{^\circ}}\\
\parbox[b][0.3cm]{17.7cm}{using a Buechner-Bainbridge magnetic spectrometer and at \ensuremath{\theta}\ensuremath{_{\textnormal{lab}}}=20\ensuremath{^\circ}{\textminus}174\ensuremath{^\circ} using an array of surface barrier detectors. Measured}\\
\parbox[b][0.3cm]{17.7cm}{\ensuremath{\sigma}(\ensuremath{\theta}) for elastic scattering to deduce entrance optical model parameters. Analyzed the \ensuremath{\alpha} angular distribution corresponding to}\\
\parbox[b][0.3cm]{17.7cm}{\ensuremath{^{\textnormal{15}}}O\ensuremath{_{\textnormal{g.s.}}} using a zero-range DWBA analysis with the JULIE code. Concluded that compound nuclear reaction plays a role in this}\\
\parbox[b][0.3cm]{17.7cm}{reaction.}\\
\parbox[b][0.3cm]{17.7cm}{\addtolength{\parindent}{-0.2in}\href{https://www.nndc.bnl.gov/nsr/nsrlink.jsp?1966Ha21,B}{1966Ha21}: \ensuremath{^{\textnormal{16}}}O(\ensuremath{^{\textnormal{3}}}He,\ensuremath{\alpha}) E=3-10 MeV; measured the absolute cross section of this reaction as a function of \ensuremath{^{\textnormal{3}}}He bombarding energy}\\
\parbox[b][0.3cm]{17.7cm}{by measuring the annihilation photons from the \ensuremath{\beta}\ensuremath{^{\textnormal{+}}} decay of \ensuremath{^{\textnormal{15}}}O using a NaI(Tl) detector. Cross sections (in mb) are provided in}\\
\parbox[b][0.3cm]{17.7cm}{Table II. Concluded that the \ensuremath{^{\textnormal{16}}}O(\ensuremath{^{\textnormal{3}}}He,\ensuremath{\alpha}) reaction at these energies proceeds via the direct reaction mechanism.}\\
\parbox[b][0.3cm]{17.7cm}{\addtolength{\parindent}{-0.2in}\href{https://www.nndc.bnl.gov/nsr/nsrlink.jsp?1967Ro10,B}{1967Ro10}: \ensuremath{^{\textnormal{16}}}O(\ensuremath{^{\textnormal{3}}}He,\ensuremath{^{\textnormal{3}}}He), \ensuremath{^{\textnormal{16}}}O(\ensuremath{^{\textnormal{3}}}He,p), \ensuremath{^{\textnormal{16}}}O(\ensuremath{^{\textnormal{3}}}He,\ensuremath{\alpha}\ensuremath{_{\textnormal{0}}}) E=3.5-6.5 MeV; measured excitation functions and angular distributions;}\\
\parbox[b][0.3cm]{17.7cm}{measured charged particles using two Si surface barrier detectors. Elastic scattering was measured at \ensuremath{\theta}\ensuremath{_{\textnormal{lab}}}=75\ensuremath{^\circ}{\textminus}165\ensuremath{^\circ}. Observed a}\\
\parbox[b][0.3cm]{17.7cm}{resonance at E\ensuremath{_{\textnormal{lab}}}(\ensuremath{^{\textnormal{3}}}He)=5.05 MeV in the elastic scattering data and in (\ensuremath{^{\textnormal{3}}}He,p) channel via p\ensuremath{_{\textnormal{0,1,5}}}. No evidence of this resonance}\\
\parbox[b][0.3cm]{17.7cm}{was observed on the (\ensuremath{^{\textnormal{3}}}He,\ensuremath{\alpha}\ensuremath{_{\textnormal{0}}}) channel. Deduced \ensuremath{^{\textnormal{19}}}Ne level-energy, \ensuremath{\Gamma}, J\ensuremath{^{\ensuremath{\pi}}} and L for the observed resonance, as well as \ensuremath{\sigma}(E\ensuremath{_{^{\textnormal{3}}\textnormal{He}}})}\\
\parbox[b][0.3cm]{17.7cm}{for the \ensuremath{^{\textnormal{16}}}O(\ensuremath{^{\textnormal{3}}}He,\ensuremath{\alpha}\ensuremath{_{\textnormal{0}}}) reaction (see Table 1).}\\
\parbox[b][0.3cm]{17.7cm}{\addtolength{\parindent}{-0.2in}\href{https://www.nndc.bnl.gov/nsr/nsrlink.jsp?1969Da08,B}{1969Da08}: \ensuremath{^{\textnormal{16}}}O(\ensuremath{^{\textnormal{3}}}He,\ensuremath{\alpha}\ensuremath{_{\textnormal{0}}}) E=4-9 MeV; measured reaction products using silicon surface barrier detectors placed at \ensuremath{\theta}\ensuremath{_{\textnormal{lab}}}=55\ensuremath{^\circ}, 125\ensuremath{^\circ}}\\
\parbox[b][0.3cm]{17.7cm}{and 165\ensuremath{^\circ}. Measured \ensuremath{\sigma}(E\ensuremath{_{^{\textnormal{3}}\textnormal{He}}},\ensuremath{\theta}). Analyzed the excitation function using the statistical fluctuation model of (\href{https://www.nndc.bnl.gov/nsr/nsrlink.jsp?1960Er03,B}{1960Er03}) and deduced}\\
\parbox[b][0.3cm]{17.7cm}{\ensuremath{\Gamma}=130 keV \textit{20} for the coherence width of the \ensuremath{^{\textnormal{19}}}Ne compound system in the excitation energy range of 12-15 MeV. Deduced}\\
\parbox[b][0.3cm]{17.7cm}{coefficients of the cross correlation and auto-correlation functions used to determine the cross section averaged over a number of}\\
\parbox[b][0.3cm]{17.7cm}{resonances. The findings of this study are in disagreement with the lack of observation of any resonances in the \ensuremath{\alpha}\ensuremath{_{\textnormal{0}}} channel by}\\
\parbox[b][0.3cm]{17.7cm}{(\href{https://www.nndc.bnl.gov/nsr/nsrlink.jsp?1967Ro10,B}{1967Ro10}).}\\
\parbox[b][0.3cm]{17.7cm}{\addtolength{\parindent}{-0.2in}\href{https://www.nndc.bnl.gov/nsr/nsrlink.jsp?1971OtZX,B}{1971OtZX}, \href{https://www.nndc.bnl.gov/nsr/nsrlink.jsp?1972Ot01,B}{1972Ot01}: \ensuremath{^{\textnormal{16}}}O(\ensuremath{^{\textnormal{3}}}He,\ensuremath{^{\textnormal{3}}}He) and \ensuremath{^{\textnormal{16}}}O(\ensuremath{^{\textnormal{3}}}He,\ensuremath{\alpha}) E=3.1-7.0; procured and analyzed \ensuremath{^{\textnormal{16}}}O(\ensuremath{^{\textnormal{3}}}He,\ensuremath{^{\textnormal{3}}}He) data obtained at E=2.7-4.0}\\
\parbox[b][0.3cm]{17.7cm}{MeV from (Rong-sheng Jin, Ph.D. Thesis, Ohio State University (1965), unpublished); measured the excitation function of the}\\
\parbox[b][0.3cm]{17.7cm}{\ensuremath{^{\textnormal{16}}}O(\ensuremath{^{\textnormal{3}}}He,\ensuremath{\alpha}) reaction at \ensuremath{\theta}\ensuremath{_{\textnormal{lab}}}=70\ensuremath{^\circ}, 80\ensuremath{^\circ}, 90\ensuremath{^\circ}, 100\ensuremath{^\circ}, 120\ensuremath{^\circ}, 130\ensuremath{^\circ}, 140\ensuremath{^\circ}, and 150\ensuremath{^\circ} and \ensuremath{\alpha} angular distributions at 6 energies between}\\
\parbox[b][0.3cm]{17.7cm}{4.45-5.5 MeV using a Si \ensuremath{\Delta}E-E telescope. Remeasured \ensuremath{\sigma}(E\ensuremath{_{\ensuremath{\alpha}}},\ensuremath{\theta}) at \ensuremath{\theta}\ensuremath{_{\textnormal{lab}}}=130\ensuremath{^\circ}, 140\ensuremath{^\circ}, and 150\ensuremath{^\circ} in 20 keV steps at 4-4.56 MeV to}\\
\parbox[b][0.3cm]{17.7cm}{study the region near E\ensuremath{_{\textnormal{lab}}}=4.3 MeV. Analyzed the elastic scattering data using optical model potential plus resonance analysis.}\\
\parbox[b][0.3cm]{17.7cm}{Analyzed the transfer data using Legendre polynomials analysis for angular distributions and a multi-channel multi-level R-matrix}\\
\clearpage
\vspace{0.3cm}
{\bf \small \underline{\ensuremath{^{\textnormal{16}}}O(\ensuremath{^{\textnormal{3}}}He,X)\hspace{0.2in}\href{https://www.nndc.bnl.gov/nsr/nsrlink.jsp?1959Br79,B}{1959Br79},\href{https://www.nndc.bnl.gov/nsr/nsrlink.jsp?1983Wa05,B}{1983Wa05} (continued)}}\\
\vspace{0.3cm}
\parbox[b][0.3cm]{17.7cm}{analysis using the MULTI code. Deduced \ensuremath{^{\textnormal{19}}}Ne resonances, \ensuremath{\Gamma}, and J\ensuremath{^{\ensuremath{\pi}}}. Compared the results with the predictions of an \ensuremath{\alpha}-particle}\\
\parbox[b][0.3cm]{17.7cm}{core-excited, threshold-state model.}\\
\parbox[b][0.3cm]{17.7cm}{\addtolength{\parindent}{-0.2in}\href{https://www.nndc.bnl.gov/nsr/nsrlink.jsp?1972WeZG,B}{1972WeZG}: \ensuremath{^{\textnormal{16}}}O(\ensuremath{^{\textnormal{3}}}He,\ensuremath{\alpha}); measured \ensuremath{\sigma}(\ensuremath{\theta}); deduced \ensuremath{^{\textnormal{19}}}Ne levels.}\\
\parbox[b][0.3cm]{17.7cm}{\addtolength{\parindent}{-0.2in}\href{https://www.nndc.bnl.gov/nsr/nsrlink.jsp?1972We01,B}{1972We01}: \ensuremath{^{\textnormal{16}}}O(\ensuremath{^{\textnormal{3}}}He,\ensuremath{\alpha}) and \ensuremath{^{\textnormal{16}}}O(\ensuremath{^{\textnormal{3}}}He,\ensuremath{^{\textnormal{3}}}He); measured \ensuremath{^{\textnormal{19}}}Ne states with \ensuremath{\alpha}+\ensuremath{^{\textnormal{15}}}O \ensuremath{\alpha}-cluster configuration within the E\ensuremath{_{\textnormal{x}}}(\ensuremath{^{\textnormal{19}}}Ne)\ensuremath{\sim}10-13}\\
\parbox[b][0.3cm]{17.7cm}{MeV region; analyzed data by means of R-matrix and optical potential plus resonance analyses. The excitation energies are}\\
\parbox[b][0.3cm]{17.7cm}{presented in low resolution graphical format. Concluded that the \ensuremath{\alpha}-particle threshold state must have L=0.}\\
\parbox[b][0.3cm]{17.7cm}{\addtolength{\parindent}{-0.2in}\href{https://www.nndc.bnl.gov/nsr/nsrlink.jsp?1973ShZB,B}{1973ShZB}: \ensuremath{^{\textnormal{16}}}O(\ensuremath{^{\textnormal{3}}}He,\ensuremath{\gamma}); measured \ensuremath{\sigma}(E\ensuremath{_{\ensuremath{\gamma}}},\ensuremath{\theta}).}\\
\parbox[b][0.3cm]{17.7cm}{\addtolength{\parindent}{-0.2in}\href{https://www.nndc.bnl.gov/nsr/nsrlink.jsp?1980ChZF,B}{1980ChZF}, \href{https://www.nndc.bnl.gov/nsr/nsrlink.jsp?1983Wa05,B}{1983Wa05}: \ensuremath{^{\textnormal{16}}}O(\ensuremath{^{\textnormal{3}}}He,\ensuremath{\gamma}) E=3-19 MeV at \ensuremath{\theta}\ensuremath{_{\textnormal{lab}}}=90\ensuremath{^\circ} and E=5-11 MeV at \ensuremath{\theta}\ensuremath{_{\textnormal{lab}}}=40\ensuremath{^\circ}; measured the excitation function and}\\
\parbox[b][0.3cm]{17.7cm}{angular distributions for the \ensuremath{^{\textnormal{3}}}He capture using a NaI(Tl) detector surrounded by a NE-102A plastic scintillator to reject the cosmic}\\
\parbox[b][0.3cm]{17.7cm}{rays. The resolution was 5\%, i.e., \ensuremath{\Delta}E\ensuremath{\sim}500 keV (FWHM). Angular distributions were fitted using Legendre polynomials up to P\ensuremath{_{\textnormal{3}}}}\\
\parbox[b][0.3cm]{17.7cm}{term. Deduced \ensuremath{^{\textnormal{19}}}Ne level-energies, J\ensuremath{^{\ensuremath{\pi}}}, \ensuremath{\Gamma}, \ensuremath{\Gamma}\ensuremath{_{\ensuremath{\gamma}}} and \ensuremath{\Gamma}\ensuremath{_{^{\textnormal{3}}\textnormal{He}}} using the Breit-Wigner analysis. Cluster model calculations were}\\
\parbox[b][0.3cm]{17.7cm}{performed by (\href{https://www.nndc.bnl.gov/nsr/nsrlink.jsp?1983Wa05,B}{1983Wa05}) using cosh potential, which could not reproduce the data. A shell model calculation was performed,}\\
\parbox[b][0.3cm]{17.7cm}{which reproduced the experimental excitation function.}\\
\vspace{0.385cm}
\parbox[b][0.3cm]{17.7cm}{\addtolength{\parindent}{-0.2in}\textit{Theory}:}\\
\parbox[b][0.3cm]{17.7cm}{\addtolength{\parindent}{-0.2in}\href{https://www.nndc.bnl.gov/nsr/nsrlink.jsp?1985Ch27,B}{1985Ch27}: \ensuremath{^{\textnormal{16}}}O(\ensuremath{^{\textnormal{3}}}He,\ensuremath{\gamma}) E not given; calculated \ensuremath{\sigma} vs. target excitation; deduced entrance and exit channel dependences.}\\
\parbox[b][0.3cm]{17.7cm}{\addtolength{\parindent}{-0.2in}\href{https://www.nndc.bnl.gov/nsr/nsrlink.jsp?1997Kh07,B}{1997Kh07}: \ensuremath{^{\textnormal{16}}}O(\ensuremath{^{\textnormal{3}}}He,\ensuremath{^{\textnormal{3}}}He) E=25-60 MeV; analyzed \ensuremath{\sigma}(\ensuremath{\theta}); deduced parameters and reaction \ensuremath{\sigma}. Used a complex folding potential.}\\
\parbox[b][0.3cm]{17.7cm}{\addtolength{\parindent}{-0.2in}\href{https://www.nndc.bnl.gov/nsr/nsrlink.jsp?2008Oh03,B}{2008Oh03}: \ensuremath{^{\textnormal{16}}}O(\ensuremath{^{\textnormal{3}}}He,\ensuremath{^{\textnormal{3}}}He) E=15, 25, 32, 40.9, 60 MeV; calculated \ensuremath{\sigma} and angular distributions using a central double folding}\\
\parbox[b][0.3cm]{17.7cm}{potential. Deduced phase shifts. Discussed \ensuremath{^{\textnormal{3}}}He clustering in \ensuremath{^{\textnormal{19}}}Ne as a vibrational mode, in which the relative motion within the}\\
\parbox[b][0.3cm]{17.7cm}{cluster is excited. Discussed the results of (\href{https://www.nndc.bnl.gov/nsr/nsrlink.jsp?1967Ro10,B}{1967Ro10}, \href{https://www.nndc.bnl.gov/nsr/nsrlink.jsp?1983Wa05,B}{1983Wa05}) within this framework.}\\
\parbox[b][0.3cm]{17.7cm}{\addtolength{\parindent}{-0.2in}\href{https://www.nndc.bnl.gov/nsr/nsrlink.jsp?2016OtZY,B}{2016OtZY}: \ensuremath{^{\textnormal{16}}}O(\ensuremath{^{\textnormal{3}}}He,\ensuremath{\alpha}) E not given; calculated negative parity and positive parity states, J, \ensuremath{\pi} using microscopic and macroscopic}\\
\parbox[b][0.3cm]{17.7cm}{potential models, adiabatic energy curves calculated by coupled channels for \ensuremath{^{\textnormal{3}}}He+\ensuremath{^{\textnormal{16}}}O, \ensuremath{\alpha}+\ensuremath{^{\textnormal{15}}}O, \ensuremath{^{\textnormal{5}}}He+\ensuremath{^{\textnormal{14}}}O; compared with data.}\\
\parbox[b][0.3cm]{17.7cm}{Confirmed the 5p-2h configuration of the 4.03-MeV state in \ensuremath{^{\textnormal{19}}}Ne originally proposed by (\href{https://www.nndc.bnl.gov/nsr/nsrlink.jsp?1978Fo26,B}{1978Fo26}).}\\
\vspace{0.385cm}
\vspace{12pt}

\raggedright\texttt{\ \ \  \underline{Legendre coefficients for \hspace{-0.04cm}\ensuremath{^{3}}He capture on \hspace{-0.04cm}\ensuremath{^{\textnormal{16}}}O (\href{https://www.nndc.bnl.gov/nsr/nsrlink.jsp?1983Wa05,B}{1983Wa05}):}}\\
\raggedright\texttt{\ \ \ \ \ \ \ \ \ \ \ \ \ \ \ \ \ \ \ \ {Decay mode: \hspace{-0.04cm}\ensuremath{\gamma}\ensuremath{_{\textnormal{0$-$2}}}}}\\
\raggedright\texttt{\ E(\ensuremath{^{\textnormal{3}}}He)(MeV)\ \ \ \ \ \ \ \ \ \ \ a1\ \ \ \ \ \ \ \ \ \ \ \ a2\ \ \ \ \ \ \ \ \ \ \ \ a3}\\
\raggedright\texttt{\ 6.42\ \ \ \ \ \ \ \ \ \ \ \ \ \ \ 0.05\ \textit{27}\ \ \ \ \ \ \ {\textminus}0.48\ \textit{45}\ \ \ \ \ \ {\textminus}0.42\ \textit{38}}\\
\raggedright\texttt{\ 7.66\ \ \ \ \ \ \ \ \ \ \ \ \ \ \ 0.35\ \textit{17}\ \ \ \ \ \ \ {\textminus}0.70\ \textit{28}\ \ \ \ \ \ {\textminus}0.17\ \textit{25}}\\
\raggedright\texttt{\ 8.41\ \ \ \ \ \ \ \ \ \ \ \ \ \ \ 0.52\ \textit{40}\ \ \ \ \ \ \ {\textminus}0.54\ \textit{59}\ \ \ \ \ \ {\textminus}0.79\ \textit{43}}\\
\raggedright\texttt{\ 9.33\ \ \ \ \ \ \ \ \ \ \ \ \ \ \ 0.08\ \textit{19}\ \ \ \ \ \ \ {\textminus}0.74\ \textit{31}}\\
\centering \texttt{}\\
\underline{$^{19}$Ne Levels}\\
\vspace{0.34cm}
\parbox[b][0.3cm]{17.7cm}{\addtolength{\parindent}{-0.254cm}\textit{Notes}:}\\
\parbox[b][0.3cm]{17.7cm}{\addtolength{\parindent}{-0.254cm}(1) (2J+1)\ensuremath{\Gamma}\ensuremath{_{^{\textnormal{3}}\textnormal{He}}}\ensuremath{\Gamma}\ensuremath{_{\ensuremath{\gamma}}} (keV\ensuremath{^{\textnormal{2}}}): (\href{https://www.nndc.bnl.gov/nsr/nsrlink.jsp?1983Wa05,B}{1983Wa05}) acknowledged that the values derived using the Breit-Wigner analysis may not be}\\
\parbox[b][0.3cm]{17.7cm}{reliable since such an analysis assumes that the resonances are isolated and are incoherent with the background. This assumption is}\\
\parbox[b][0.3cm]{17.7cm}{not likely to be valid for that study.}\\
\parbox[b][0.3cm]{17.7cm}{\addtolength{\parindent}{-0.254cm}(2) The \ensuremath{\gamma} rays corresponding to the \ensuremath{\gamma}\ensuremath{_{\textnormal{0,1,2}}} and \ensuremath{\gamma}\ensuremath{_{\textnormal{3,4,5}}} decays de-exciting \ensuremath{^{\textnormal{19}}}Ne* states were unresolved in (\href{https://www.nndc.bnl.gov/nsr/nsrlink.jsp?1983Wa05,B}{1983Wa05}). Moreover,}\\
\parbox[b][0.3cm]{17.7cm}{\ensuremath{\gamma}\ensuremath{_{\textnormal{6}}} label appears on the excitation function shown in Fig. 4 of (\href{https://www.nndc.bnl.gov/nsr/nsrlink.jsp?1983Wa05,B}{1983Wa05}) but there is no mention of the state that de-excited via}\\
\parbox[b][0.3cm]{17.7cm}{this transition.}\\
\parbox[b][0.3cm]{17.7cm}{\addtolength{\parindent}{-0.254cm}(3) (\href{https://www.nndc.bnl.gov/nsr/nsrlink.jsp?1983Wa05,B}{1983Wa05}) claims (see section 3.1) that there are indications that the \ensuremath{\gamma}\ensuremath{_{\textnormal{1}}} transition is responsible for the resonances observed}\\
\parbox[b][0.3cm]{17.7cm}{at E(\ensuremath{^{\textnormal{3}}}He)=7.65 and 9.26 MeV.}\\
\parbox[b][0.3cm]{17.7cm}{\addtolength{\parindent}{-0.254cm}(4) For the \ensuremath{\gamma}\ensuremath{_{\textnormal{0$-$2}}} decay channels, the angular distributions were fitted by (\href{https://www.nndc.bnl.gov/nsr/nsrlink.jsp?1983Wa05,B}{1983Wa05}) using the Legendre polynomials, whose}\\
\parbox[b][0.3cm]{17.7cm}{coefficients are given in the table presented above.}\\
\parbox[b][0.3cm]{17.7cm}{\addtolength{\parindent}{-0.254cm}(5) (\href{https://www.nndc.bnl.gov/nsr/nsrlink.jsp?1983Wa05,B}{1983Wa05}) did not report discrete \ensuremath{\gamma}-ray transitions or their energies.}\\
\parbox[b][0.3cm]{17.7cm}{\addtolength{\parindent}{-0.254cm}(6) (\href{https://www.nndc.bnl.gov/nsr/nsrlink.jsp?1969Da08,B}{1969Da08}) deduced \ensuremath{\Gamma}=130 keV \textit{20} for the coherence width of the \ensuremath{^{\textnormal{19}}}Ne compound system in the excitation energy range of}\\
\parbox[b][0.3cm]{17.7cm}{12-15 MeV.}\\
\parbox[b][0.3cm]{17.7cm}{\addtolength{\parindent}{-0.254cm}(7) (\href{https://www.nndc.bnl.gov/nsr/nsrlink.jsp?1959Br79,B}{1959Br79}) and (\href{https://www.nndc.bnl.gov/nsr/nsrlink.jsp?1972Ot01,B}{1972Ot01}) observed asymmetries in the \ensuremath{\alpha} angular distributions in forward and backward center-of-mass}\\
\parbox[b][0.3cm]{17.7cm}{angles, respectively. These were attributed to strong contributions from the direct reaction mechanism and strong compound nucleus}\\
\parbox[b][0.3cm]{17.7cm}{contributions to the \ensuremath{^{\textnormal{16}}}O(\ensuremath{^{\textnormal{3}}}He,\ensuremath{\alpha}) cross section, respectively.}\\
\parbox[b][0.3cm]{17.7cm}{\addtolength{\parindent}{-0.254cm}(8) The \ensuremath{\theta}\ensuremath{^{\textnormal{2}}} values reported here are the dimensionless reduced width calculated by (\href{https://www.nndc.bnl.gov/nsr/nsrlink.jsp?2008Oh03,B}{2008Oh03}) using a channel radius of 5 fm.}\\
\vspace{0.34cm}
\begin{textblock}{29}(0,27.3)
Continued on next page (footnotes at end of table)
\end{textblock}
\clearpage
\vspace{0.3cm}
{\bf \small \underline{\ensuremath{^{\textnormal{16}}}O(\ensuremath{^{\textnormal{3}}}He,X)\hspace{0.2in}\href{https://www.nndc.bnl.gov/nsr/nsrlink.jsp?1959Br79,B}{1959Br79},\href{https://www.nndc.bnl.gov/nsr/nsrlink.jsp?1983Wa05,B}{1983Wa05} (continued)}}\\
\vspace{0.3cm}
\underline{$^{19}$Ne Levels (continued)}\\
\begin{longtable}{ccccccccc@{\extracolsep{\fill}}c}
\multicolumn{2}{c}{E(level)$^{{\hyperlink{NE23LEVEL0}{a}}}$}&J$^{\pi}$$^{}$&\multicolumn{2}{c}{\ensuremath{\Gamma}$^{{\hyperlink{NE23LEVEL3}{d}}}$}&L$^{}$&\multicolumn{2}{c}{E(\ensuremath{^{\textnormal{3}}}He,lab) (MeV)$^{}$}&Comments&\\[-.2cm]
\multicolumn{2}{c}{\hrulefill}&\hrulefill&\multicolumn{2}{c}{\hrulefill}&\hrulefill&\multicolumn{2}{c}{\hrulefill}&\hrulefill&
\endfirsthead
\multicolumn{1}{r@{}}{10461}&\multicolumn{1}{@{ }l}{{\it 8}}&\multicolumn{1}{l}{1/2\ensuremath{^{{\hyperlink{NE23LEVEL4}{e}}}}}&\multicolumn{1}{r@{}}{355}&\multicolumn{1}{@{}l}{\ensuremath{^{{\hyperlink{NE23LEVEL2}{c}}}} keV}&\multicolumn{1}{l}{0}&\multicolumn{1}{r@{}}{2400}&\multicolumn{1}{@{ }l}{{\it 10}}&\parbox[t][0.3cm]{9.29712cm}{\raggedright \ensuremath{\Gamma}\ensuremath{\alpha}=135 keV (\href{https://www.nndc.bnl.gov/nsr/nsrlink.jsp?1959Br79,B}{1959Br79})\vspace{0.1cm}}&\\
&&&&&&&&\parbox[t][0.3cm]{9.29712cm}{\raggedright \ensuremath{\Gamma}\ensuremath{_{\textnormal{p}}}=170 keV (\href{https://www.nndc.bnl.gov/nsr/nsrlink.jsp?1959Br79,B}{1959Br79})\vspace{0.1cm}}&\\
&&&&&&&&\parbox[t][0.3cm]{9.29712cm}{\raggedright \ensuremath{\Gamma}\ensuremath{_{^{\textnormal{3}}\textnormal{He}}}=50 keV (\href{https://www.nndc.bnl.gov/nsr/nsrlink.jsp?1959Br79,B}{1959Br79}).\vspace{0.1cm}}&\\
&&&&&&&&\parbox[t][0.3cm]{9.29712cm}{\raggedright \ensuremath{\Gamma}\ensuremath{_{^{\textnormal{3}}\textnormal{He}}}: See Table 4 in (\href{https://www.nndc.bnl.gov/nsr/nsrlink.jsp?1959Br79,B}{1959Br79}). The ratio of the reduced partial\vspace{0.1cm}}&\\
&&&&&&&&\parbox[t][0.3cm]{9.29712cm}{\raggedright {\ }{\ }{\ }\ensuremath{^{\textnormal{3}}}He width to the Wigner limit was determined to be 0.67\vspace{0.1cm}}&\\
&&&&&&&&\parbox[t][0.3cm]{9.29712cm}{\raggedright {\ }{\ }{\ }(\href{https://www.nndc.bnl.gov/nsr/nsrlink.jsp?1959Br79,B}{1959Br79}: See Table 4) resulting in \ensuremath{\gamma}\ensuremath{^{\textnormal{2}}}(\ensuremath{^{\textnormal{3}}}He)=281 keV\vspace{0.1cm}}&\\
&&&&&&&&\parbox[t][0.3cm]{9.29712cm}{\raggedright {\ }{\ }{\ }(\href{https://www.nndc.bnl.gov/nsr/nsrlink.jsp?1960Br40,B}{1960Br40}, see Table 3) as cited by (\href{https://www.nndc.bnl.gov/nsr/nsrlink.jsp?1967Ro10,B}{1967Ro10}). The ratio to the\vspace{0.1cm}}&\\
&&&&&&&&\parbox[t][0.3cm]{9.29712cm}{\raggedright {\ }{\ }{\ }Wigner limit for the \ensuremath{\alpha}-partial width of 0.05 was obtained\vspace{0.1cm}}&\\
&&&&&&&&\parbox[t][0.3cm]{9.29712cm}{\raggedright {\ }{\ }{\ }(\href{https://www.nndc.bnl.gov/nsr/nsrlink.jsp?1959Br79,B}{1959Br79}: See Table 4) resulting in \ensuremath{\gamma}\ensuremath{^{\textnormal{2}}}(\ensuremath{^{\textnormal{4}}}He)=15 keV\vspace{0.1cm}}&\\
&&&&&&&&\parbox[t][0.3cm]{9.29712cm}{\raggedright {\ }{\ }{\ }(\href{https://www.nndc.bnl.gov/nsr/nsrlink.jsp?1960Br40,B}{1960Br40}, see Table 3) as cited by (\href{https://www.nndc.bnl.gov/nsr/nsrlink.jsp?1967Ro10,B}{1967Ro10}). See also\vspace{0.1cm}}&\\
&&&&&&&&\parbox[t][0.3cm]{9.29712cm}{\raggedright {\ }{\ }{\ }\ensuremath{\Gamma}\ensuremath{_{^{\textnormal{3}}\textnormal{He}}}=33 keV (\href{https://www.nndc.bnl.gov/nsr/nsrlink.jsp?1961Si09,B}{1961Si09}) and \ensuremath{\Gamma}\ensuremath{_{^{\textnormal{3}}\textnormal{He}}}/\ensuremath{\Gamma}=0.11 (\href{https://www.nndc.bnl.gov/nsr/nsrlink.jsp?1961Si09,B}{1961Si09}) from\vspace{0.1cm}}&\\
&&&&&&&&\parbox[t][0.3cm]{9.29712cm}{\raggedright {\ }{\ }{\ }the magnitude of the dip in the \ensuremath{\alpha} angular distribution data at\vspace{0.1cm}}&\\
&&&&&&&&\parbox[t][0.3cm]{9.29712cm}{\raggedright {\ }{\ }{\ }E(\ensuremath{^{\textnormal{3}}}He, lab)=2.373 MeV.\vspace{0.1cm}}&\\
&&&&&&&&\parbox[t][0.3cm]{9.29712cm}{\raggedright E(level): See also E\ensuremath{_{\textnormal{x}}}=10.436 MeV (\href{https://www.nndc.bnl.gov/nsr/nsrlink.jsp?1959Br79,B}{1959Br79}: See Table 4) and\vspace{0.1cm}}&\\
&&&&&&&&\parbox[t][0.3cm]{9.29712cm}{\raggedright {\ }{\ }{\ }E\ensuremath{_{\textnormal{x}}}=10428 keV associated with E(\ensuremath{^{\textnormal{3}}}He, lab)=2.36 MeV\vspace{0.1cm}}&\\
&&&&&&&&\parbox[t][0.3cm]{9.29712cm}{\raggedright {\ }{\ }{\ }(\href{https://www.nndc.bnl.gov/nsr/nsrlink.jsp?1961Si09,B}{1961Si09}). Those authors reported E\ensuremath{_{\textnormal{x}}}=10405 keV (see Table\vspace{0.1cm}}&\\
&&&&&&&&\parbox[t][0.3cm]{9.29712cm}{\raggedright {\ }{\ }{\ }III). Our value differs due to a revised \ensuremath{^{\textnormal{3}}}He separation energy for\vspace{0.1cm}}&\\
&&&&&&&&\parbox[t][0.3cm]{9.29712cm}{\raggedright {\ }{\ }{\ }\ensuremath{^{\textnormal{19}}}Ne\ensuremath{_{\textnormal{g.s.}}} from AME-2020.\vspace{0.1cm}}&\\
&&&&&&&&\parbox[t][0.3cm]{9.29712cm}{\raggedright E(\ensuremath{^{\textnormal{3}}}He,lab) (MeV): From (\href{https://www.nndc.bnl.gov/nsr/nsrlink.jsp?1959Br79,B}{1959Br79}), where the uncertainty is\vspace{0.1cm}}&\\
&&&&&&&&\parbox[t][0.3cm]{9.29712cm}{\raggedright {\ }{\ }{\ }reported in the text.\vspace{0.1cm}}&\\
&&&&&&&&\parbox[t][0.3cm]{9.29712cm}{\raggedright \ensuremath{\Gamma}: \ensuremath{\Gamma}=\ensuremath{\Gamma}\ensuremath{_{\textnormal{p}}}+\ensuremath{\Gamma}\ensuremath{_{\ensuremath{\alpha}}}+\ensuremath{\Gamma}\ensuremath{_{^{\textnormal{3}}\textnormal{He}}}. See also \ensuremath{\Gamma}=300 keV (\href{https://www.nndc.bnl.gov/nsr/nsrlink.jsp?1961Si09,B}{1961Si09}).\vspace{0.1cm}}&\\
&&&&&&&&\parbox[t][0.3cm]{9.29712cm}{\raggedright \ensuremath{\Gamma}\ensuremath{_{\textnormal{p}}}=\ensuremath{\sum}\ensuremath{\Gamma}\ensuremath{_{\textnormal{p}}}, where p\ensuremath{_{\textnormal{0,1,2,3,4,5,6,7}}} branches were observed by\vspace{0.1cm}}&\\
&&&&&&&&\parbox[t][0.3cm]{9.29712cm}{\raggedright {\ }{\ }{\ }(\href{https://www.nndc.bnl.gov/nsr/nsrlink.jsp?1959Br79,B}{1959Br79}).\vspace{0.1cm}}&\\
&&&&&&&&\parbox[t][0.3cm]{9.29712cm}{\raggedright (\href{https://www.nndc.bnl.gov/nsr/nsrlink.jsp?1961Si09,B}{1961Si09}) did not deduce \ensuremath{\Gamma}\ensuremath{_{\ensuremath{\alpha}}} and \ensuremath{\Gamma}\ensuremath{_{\textnormal{p}}}, but they acknowledged\vspace{0.1cm}}&\\
&&&&&&&&\parbox[t][0.3cm]{9.29712cm}{\raggedright {\ }{\ }{\ }that the values reported by (\href{https://www.nndc.bnl.gov/nsr/nsrlink.jsp?1959Br79,B}{1959Br79}) were not inconsistent with\vspace{0.1cm}}&\\
&&&&&&&&\parbox[t][0.3cm]{9.29712cm}{\raggedright {\ }{\ }{\ }their data.\vspace{0.1cm}}&\\
&&&&&&&&\parbox[t][0.3cm]{9.29712cm}{\raggedright J\ensuremath{^{\pi}}: (\href{https://www.nndc.bnl.gov/nsr/nsrlink.jsp?1961Si09,B}{1961Si09}) also supported this assignment.\vspace{0.1cm}}&\\
&&&&&&&&\parbox[t][0.3cm]{9.29712cm}{\raggedright L: From (\href{https://www.nndc.bnl.gov/nsr/nsrlink.jsp?1961Si09,B}{1961Si09}): The theoretical cross section was calculated\vspace{0.1cm}}&\\
&&&&&&&&\parbox[t][0.3cm]{9.29712cm}{\raggedright {\ }{\ }{\ }for a single level using the Wigner-Eisenbud one-level formalism.\vspace{0.1cm}}&\\
&&&&&&&&\parbox[t][0.3cm]{9.29712cm}{\raggedright {\ }{\ }{\ }Calculations assumed scattering of particles of spin 1/2 by the\vspace{0.1cm}}&\\
&&&&&&&&\parbox[t][0.3cm]{9.29712cm}{\raggedright {\ }{\ }{\ }target nuclei of spin 0. The calculated values were fitted to the\vspace{0.1cm}}&\\
&&&&&&&&\parbox[t][0.3cm]{9.29712cm}{\raggedright {\ }{\ }{\ }\ensuremath{^{\textnormal{16}}}O+\ensuremath{^{\textnormal{3}}}He elastic scattering data from (\href{https://www.nndc.bnl.gov/nsr/nsrlink.jsp?1961Si09,B}{1961Si09}).\vspace{0.1cm}}&\\
&&&&&&&&\parbox[t][0.3cm]{9.29712cm}{\raggedright (\href{https://www.nndc.bnl.gov/nsr/nsrlink.jsp?1959Br79,B}{1959Br79}): \ensuremath{\sigma}\ensuremath{<}0.8 \ensuremath{\mu}b corresponds to an E1 or M1 transition to\vspace{0.1cm}}&\\
&&&&&&&&\parbox[t][0.3cm]{9.29712cm}{\raggedright {\ }{\ }{\ }the \ensuremath{^{\textnormal{19}}}Ne\ensuremath{_{\textnormal{g.s.}}} with \ensuremath{\Gamma}\ensuremath{_{\ensuremath{\gamma}}}\ensuremath{\approx}0.7 eV. See also Table 3 in (\href{https://www.nndc.bnl.gov/nsr/nsrlink.jsp?1960Br40,B}{1960Br40}).\vspace{0.1cm}}&\\
&&&&&&&&\parbox[t][0.3cm]{9.29712cm}{\raggedright {\ }{\ }{\ }The observation of the \ensuremath{\gamma}-ray transition to the ground state was\vspace{0.1cm}}&\\
&&&&&&&&\parbox[t][0.3cm]{9.29712cm}{\raggedright {\ }{\ }{\ }not confirmed. However, the authors mentioned that the observed\vspace{0.1cm}}&\\
&&&&&&&&\parbox[t][0.3cm]{9.29712cm}{\raggedright {\ }{\ }{\ }yield in the region of interest, where this transition was expected,\vspace{0.1cm}}&\\
&&&&&&&&\parbox[t][0.3cm]{9.29712cm}{\raggedright {\ }{\ }{\ }was significantly higher at forward angles. This implied that the\vspace{0.1cm}}&\\
&&&&&&&&\parbox[t][0.3cm]{9.29712cm}{\raggedright {\ }{\ }{\ }observed yield could not be entirely attributed to the transitions\vspace{0.1cm}}&\\
&&&&&&&&\parbox[t][0.3cm]{9.29712cm}{\raggedright {\ }{\ }{\ }from the resonance, which indicated that the actual capture cross\vspace{0.1cm}}&\\
&&&&&&&&\parbox[t][0.3cm]{9.29712cm}{\raggedright {\ }{\ }{\ }section may be as much as an order of magnitude less than the\vspace{0.1cm}}&\\
&&&&&&&&\parbox[t][0.3cm]{9.29712cm}{\raggedright {\ }{\ }{\ }upper limit quoted.\vspace{0.1cm}}&\\
&&&&&&&&\parbox[t][0.3cm]{9.29712cm}{\raggedright Decay modes: \ensuremath{\alpha}\ensuremath{_{\textnormal{0}}}, \ensuremath{^{\textnormal{3}}}He, and p\ensuremath{_{\textnormal{0,1,2,3,4,5,6,7}}} (\href{https://www.nndc.bnl.gov/nsr/nsrlink.jsp?1959Br79,B}{1959Br79}).\vspace{0.1cm}}&\\
&&&&&&&&\parbox[t][0.3cm]{9.29712cm}{\raggedright The levels observed by (\href{https://www.nndc.bnl.gov/nsr/nsrlink.jsp?1972Ot01,B}{1972Ot01}) are not pure \ensuremath{\alpha}-particle states\vspace{0.1cm}}&\\
&&&&&&&&\parbox[t][0.3cm]{9.29712cm}{\raggedright {\ }{\ }{\ }since they have non-zero widths in other channels. The strengths\vspace{0.1cm}}&\\
&&&&&&&&\parbox[t][0.3cm]{9.29712cm}{\raggedright {\ }{\ }{\ }reported by (\href{https://www.nndc.bnl.gov/nsr/nsrlink.jsp?1972Ot01,B}{1972Ot01}) assume that the \ensuremath{\alpha}-particle width\vspace{0.1cm}}&\\
&&&&&&&&\parbox[t][0.3cm]{9.29712cm}{\raggedright {\ }{\ }{\ }dominates.\vspace{0.1cm}}&\\
&&&&&&&&\parbox[t][0.3cm]{9.29712cm}{\raggedright The R-matrix calculation of (\href{https://www.nndc.bnl.gov/nsr/nsrlink.jsp?1972Ot01,B}{1972Ot01}) used hard-sphere plus\vspace{0.1cm}}&\\
&&&&&&&&\parbox[t][0.3cm]{9.29712cm}{\raggedright {\ }{\ }{\ }Rutherford scattering phase shifts to represent the potential\vspace{0.1cm}}&\\
&&&&&&&&\parbox[t][0.3cm]{9.29712cm}{\raggedright {\ }{\ }{\ }scattering.\vspace{0.1cm}}&\\
\multicolumn{1}{r@{}}{10482}&\multicolumn{1}{@{ }l}{{\it 8}}&\multicolumn{1}{l}{5/2\ensuremath{^{{\hyperlink{NE23LEVEL4}{e}}}}}&\multicolumn{1}{r@{}}{45}&\multicolumn{1}{@{}l}{\ensuremath{^{{\hyperlink{NE23LEVEL2}{c}}}} keV}&&\multicolumn{1}{r@{}}{2425}&\multicolumn{1}{@{ }l}{{\it 10}}&\parbox[t][0.3cm]{9.29712cm}{\raggedright \ensuremath{\Gamma}\ensuremath{\alpha}=22.3 keV (\href{https://www.nndc.bnl.gov/nsr/nsrlink.jsp?1959Br79,B}{1959Br79})\vspace{0.1cm}}&\\
&&&&&&&&\parbox[t][0.3cm]{9.29712cm}{\raggedright \ensuremath{\Gamma}\ensuremath{_{\textnormal{p}}}=22.3 keV (\href{https://www.nndc.bnl.gov/nsr/nsrlink.jsp?1959Br79,B}{1959Br79})\vspace{0.1cm}}&\\
&&&&&&&&\parbox[t][0.3cm]{9.29712cm}{\raggedright \ensuremath{\Gamma}\ensuremath{_{^{\textnormal{3}}\textnormal{He}}}=0.45 keV (\href{https://www.nndc.bnl.gov/nsr/nsrlink.jsp?1959Br79,B}{1959Br79}).\vspace{0.1cm}}&\\
&&&&&&&&\parbox[t][0.3cm]{9.29712cm}{\raggedright \ensuremath{\Gamma}\ensuremath{_{^{\textnormal{3}}\textnormal{He}}}: See Table 4 in (\href{https://www.nndc.bnl.gov/nsr/nsrlink.jsp?1959Br79,B}{1959Br79}). The ratio of the reduced partial\vspace{0.1cm}}&\\
\end{longtable}
\begin{textblock}{29}(0,27.3)
Continued on next page (footnotes at end of table)
\end{textblock}
\clearpage
\begin{longtable}{ccccccccc@{\extracolsep{\fill}}c}
\\[-.4cm]
\multicolumn{10}{c}{{\bf \small \underline{\ensuremath{^{\textnormal{16}}}O(\ensuremath{^{\textnormal{3}}}He,X)\hspace{0.2in}\href{https://www.nndc.bnl.gov/nsr/nsrlink.jsp?1959Br79,B}{1959Br79},\href{https://www.nndc.bnl.gov/nsr/nsrlink.jsp?1983Wa05,B}{1983Wa05} (continued)}}}\\
\multicolumn{10}{c}{~}\\
\multicolumn{10}{c}{\underline{\ensuremath{^{19}}Ne Levels (continued)}}\\
\multicolumn{10}{c}{~}\\
\multicolumn{2}{c}{E(level)$^{{\hyperlink{NE23LEVEL0}{a}}}$}&J$^{\pi}$$^{}$&\multicolumn{2}{c}{\ensuremath{\Gamma}$^{{\hyperlink{NE23LEVEL3}{d}}}$}&L$^{}$&\multicolumn{2}{c}{E(\ensuremath{^{\textnormal{3}}}He,lab) (MeV)$^{}$}&Comments&\\[-.2cm]
\multicolumn{2}{c}{\hrulefill}&\hrulefill&\multicolumn{2}{c}{\hrulefill}&\hrulefill&\multicolumn{2}{c}{\hrulefill}&\hrulefill&
\endhead
&&&&&&&&\parbox[t][0.3cm]{7.4717007cm}{\raggedright {\ }{\ }{\ }\ensuremath{^{\textnormal{3}}}He width to the Wigner limit was determined to be\vspace{0.1cm}}&\\
&&&&&&&&\parbox[t][0.3cm]{7.4717007cm}{\raggedright {\ }{\ }{\ }0.03 (\href{https://www.nndc.bnl.gov/nsr/nsrlink.jsp?1959Br79,B}{1959Br79}: See Table 4) resulting in\vspace{0.1cm}}&\\
&&&&&&&&\parbox[t][0.3cm]{7.4717007cm}{\raggedright {\ }{\ }{\ }\ensuremath{\gamma}\ensuremath{^{\textnormal{2}}}(\ensuremath{^{\textnormal{3}}}He)=11 keV (\href{https://www.nndc.bnl.gov/nsr/nsrlink.jsp?1960Br40,B}{1960Br40}: See Table 3) as cited\vspace{0.1cm}}&\\
&&&&&&&&\parbox[t][0.3cm]{7.4717007cm}{\raggedright {\ }{\ }{\ }by (\href{https://www.nndc.bnl.gov/nsr/nsrlink.jsp?1967Ro10,B}{1967Ro10}). The ratio to the Wigner limit for\vspace{0.1cm}}&\\
&&&&&&&&\parbox[t][0.3cm]{7.4717007cm}{\raggedright {\ }{\ }{\ }the \ensuremath{\alpha}-partial width of 0.01 was obtained (\href{https://www.nndc.bnl.gov/nsr/nsrlink.jsp?1959Br79,B}{1959Br79}:\vspace{0.1cm}}&\\
&&&&&&&&\parbox[t][0.3cm]{7.4717007cm}{\raggedright {\ }{\ }{\ }See Table 4) resulting in \ensuremath{\gamma}\ensuremath{^{\textnormal{2}}}(\ensuremath{^{\textnormal{4}}}He)=3.5 keV\vspace{0.1cm}}&\\
&&&&&&&&\parbox[t][0.3cm]{7.4717007cm}{\raggedright {\ }{\ }{\ }(\href{https://www.nndc.bnl.gov/nsr/nsrlink.jsp?1960Br40,B}{1960Br40}: See Table 3) as cited by (\href{https://www.nndc.bnl.gov/nsr/nsrlink.jsp?1967Ro10,B}{1967Ro10}).\vspace{0.1cm}}&\\
&&&&&&&&\parbox[t][0.3cm]{7.4717007cm}{\raggedright E(level): See also E\ensuremath{_{\textnormal{x}}}=10.457 MeV (\href{https://www.nndc.bnl.gov/nsr/nsrlink.jsp?1959Br79,B}{1959Br79}: See\vspace{0.1cm}}&\\
&&&&&&&&\parbox[t][0.3cm]{7.4717007cm}{\raggedright {\ }{\ }{\ }Table 4) and E\ensuremath{_{\textnormal{x}}}=10503 keV associated with E(\ensuremath{^{\textnormal{3}}}He,\vspace{0.1cm}}&\\
&&&&&&&&\parbox[t][0.3cm]{7.4717007cm}{\raggedright {\ }{\ }{\ }lab)\ensuremath{\approx}2450 keV (\href{https://www.nndc.bnl.gov/nsr/nsrlink.jsp?1958Br86,B}{1958Br86}). These values differ\vspace{0.1cm}}&\\
&&&&&&&&\parbox[t][0.3cm]{7.4717007cm}{\raggedright {\ }{\ }{\ }because we used the revised \ensuremath{^{\textnormal{3}}}He separation energy\vspace{0.1cm}}&\\
&&&&&&&&\parbox[t][0.3cm]{7.4717007cm}{\raggedright {\ }{\ }{\ }for \ensuremath{^{\textnormal{19}}}Ne\ensuremath{_{\textnormal{g.s.}}} from AME-2020.\vspace{0.1cm}}&\\
&&&&&&&&\parbox[t][0.3cm]{7.4717007cm}{\raggedright E(\ensuremath{^{\textnormal{3}}}He,lab) (MeV): From (\href{https://www.nndc.bnl.gov/nsr/nsrlink.jsp?1959Br79,B}{1959Br79}), where the\vspace{0.1cm}}&\\
&&&&&&&&\parbox[t][0.3cm]{7.4717007cm}{\raggedright {\ }{\ }{\ }uncertainty is reported in the text.\vspace{0.1cm}}&\\
&&&&&&&&\parbox[t][0.3cm]{7.4717007cm}{\raggedright \ensuremath{\Gamma}\ensuremath{_{\textnormal{p}}}=\ensuremath{\sum}\ensuremath{\Gamma}\ensuremath{_{\textnormal{p}}}, where p\ensuremath{_{\textnormal{0,1,2,3,4,5,6,7}}} branches were\vspace{0.1cm}}&\\
&&&&&&&&\parbox[t][0.3cm]{7.4717007cm}{\raggedright {\ }{\ }{\ }observed by (\href{https://www.nndc.bnl.gov/nsr/nsrlink.jsp?1959Br79,B}{1959Br79}).\vspace{0.1cm}}&\\
&&&&&&&&\parbox[t][0.3cm]{7.4717007cm}{\raggedright \ensuremath{\Gamma}: \ensuremath{\Gamma}=\ensuremath{\Gamma}\ensuremath{_{\textnormal{p}}}+\ensuremath{\Gamma}\ensuremath{_{\ensuremath{\alpha}}}+\ensuremath{\Gamma}\ensuremath{_{^{\textnormal{3}}\textnormal{He}}}.\vspace{0.1cm}}&\\
&&&&&&&&\parbox[t][0.3cm]{7.4717007cm}{\raggedright (\href{https://www.nndc.bnl.gov/nsr/nsrlink.jsp?1959Br79,B}{1959Br79}) suggested that this state probably\vspace{0.1cm}}&\\
&&&&&&&&\parbox[t][0.3cm]{7.4717007cm}{\raggedright {\ }{\ }{\ }de-excites via a cascade of transitions to the lower\vspace{0.1cm}}&\\
&&&&&&&&\parbox[t][0.3cm]{7.4717007cm}{\raggedright {\ }{\ }{\ }levels rather than by a direct E2 transition if the\vspace{0.1cm}}&\\
&&&&&&&&\parbox[t][0.3cm]{7.4717007cm}{\raggedright {\ }{\ }{\ }parity is positive. If the parity is negative, the width\vspace{0.1cm}}&\\
&&&&&&&&\parbox[t][0.3cm]{7.4717007cm}{\raggedright {\ }{\ }{\ }for a direct M2 de-excitation would be expected to\vspace{0.1cm}}&\\
&&&&&&&&\parbox[t][0.3cm]{7.4717007cm}{\raggedright {\ }{\ }{\ }be negligible.\vspace{0.1cm}}&\\
&&&&&&&&\parbox[t][0.3cm]{7.4717007cm}{\raggedright Decay modes: \ensuremath{\alpha}\ensuremath{_{\textnormal{0}}}, \ensuremath{^{\textnormal{3}}}He, and p\ensuremath{_{\textnormal{0,1,2,3,4,5,6,7}}}, and (\ensuremath{\gamma})\vspace{0.1cm}}&\\
&&&&&&&&\parbox[t][0.3cm]{7.4717007cm}{\raggedright {\ }{\ }{\ }(\href{https://www.nndc.bnl.gov/nsr/nsrlink.jsp?1959Br79,B}{1959Br79}).\vspace{0.1cm}}&\\
\multicolumn{1}{r@{}}{11.51\ensuremath{\times10^{3}}}&\multicolumn{1}{@{ }l}{{\it 4}}&\multicolumn{1}{l}{(3/2\ensuremath{^{-}},1/2\ensuremath{^{-}})\ensuremath{^{{\hyperlink{NE23LEVEL5}{f}}}}}&\multicolumn{1}{r@{}}{24}&\multicolumn{1}{@{}l}{\ensuremath{^{{\hyperlink{NE23LEVEL5}{f}}}} keV {\it 24}}&\multicolumn{1}{l}{1}&\multicolumn{1}{r@{}}{3.65\ensuremath{\times10^{3}}}&\multicolumn{1}{@{}l}{\ensuremath{^{{\hyperlink{NE23LEVEL5}{f}}}} {\it 5}}&\parbox[t][0.3cm]{7.4717007cm}{\raggedright E(level): This level was observed in the elastic\vspace{0.1cm}}&\\
&&&&&&&&\parbox[t][0.3cm]{7.4717007cm}{\raggedright {\ }{\ }{\ }scattering excitation function of \ensuremath{^{\textnormal{16}}}O(\ensuremath{^{\textnormal{3}}}He,\ensuremath{^{\textnormal{3}}}He), as\vspace{0.1cm}}&\\
&&&&&&&&\parbox[t][0.3cm]{7.4717007cm}{\raggedright {\ }{\ }{\ }well as that of the \ensuremath{^{\textnormal{16}}}O(\ensuremath{^{\textnormal{3}}}He,\ensuremath{\alpha}) reaction (\href{https://www.nndc.bnl.gov/nsr/nsrlink.jsp?1972Ot01,B}{1972Ot01}).\vspace{0.1cm}}&\\
&&&&&&&&\parbox[t][0.3cm]{7.4717007cm}{\raggedright \ensuremath{\Gamma}: The uncertainty in \ensuremath{\Gamma} was reported as 25 keV\vspace{0.1cm}}&\\
&&&&&&&&\parbox[t][0.3cm]{7.4717007cm}{\raggedright {\ }{\ }{\ }(\href{https://www.nndc.bnl.gov/nsr/nsrlink.jsp?1972Ot01,B}{1972Ot01}), which we changed to 24 keV to avoid\vspace{0.1cm}}&\\
&&&&&&&&\parbox[t][0.3cm]{7.4717007cm}{\raggedright {\ }{\ }{\ }having a negative width. Other value: \ensuremath{\Gamma}=25 keV\vspace{0.1cm}}&\\
&&&&&&&&\parbox[t][0.3cm]{7.4717007cm}{\raggedright {\ }{\ }{\ }from the analysis of the \ensuremath{^{\textnormal{16}}}O(\ensuremath{^{\textnormal{3}}}He,\ensuremath{^{\textnormal{3}}}He) data in\vspace{0.1cm}}&\\
&&&&&&&&\parbox[t][0.3cm]{7.4717007cm}{\raggedright {\ }{\ }{\ }(\href{https://www.nndc.bnl.gov/nsr/nsrlink.jsp?1972Ot01,B}{1972Ot01}).\vspace{0.1cm}}&\\
&&&&&&&&\parbox[t][0.3cm]{7.4717007cm}{\raggedright J\ensuremath{^{\pi}},L: (\href{https://www.nndc.bnl.gov/nsr/nsrlink.jsp?1972Ot01,B}{1972Ot01}) modeled this state as an \ensuremath{\alpha}+\ensuremath{^{\textnormal{15}}}O\vspace{0.1cm}}&\\
&&&&&&&&\parbox[t][0.3cm]{7.4717007cm}{\raggedright {\ }{\ }{\ }configuration using an \ensuremath{\alpha}-particle, core-excited,\vspace{0.1cm}}&\\
&&&&&&&&\parbox[t][0.3cm]{7.4717007cm}{\raggedright {\ }{\ }{\ }threshold-state model. The prediction of this model\vspace{0.1cm}}&\\
&&&&&&&&\parbox[t][0.3cm]{7.4717007cm}{\raggedright {\ }{\ }{\ }was consistent with J\ensuremath{^{\ensuremath{\pi}}}=3/2\ensuremath{^{-}}. However, the optical\vspace{0.1cm}}&\\
&&&&&&&&\parbox[t][0.3cm]{7.4717007cm}{\raggedright {\ }{\ }{\ }model analysis of the measured \ensuremath{^{\textnormal{3}}}He angular\vspace{0.1cm}}&\\
&&&&&&&&\parbox[t][0.3cm]{7.4717007cm}{\raggedright {\ }{\ }{\ }distribution, from which L=1 was determined, could\vspace{0.1cm}}&\\
&&&&&&&&\parbox[t][0.3cm]{7.4717007cm}{\raggedright {\ }{\ }{\ }not rule out J\ensuremath{^{\ensuremath{\pi}}}=1/2\ensuremath{^{-}}.\vspace{0.1cm}}&\\
&&&&&&&&\parbox[t][0.3cm]{7.4717007cm}{\raggedright (\ensuremath{\Gamma}\ensuremath{_{^{\textnormal{3}}\textnormal{He}}}\ensuremath{\Gamma}\ensuremath{_{\ensuremath{\alpha}_{\textnormal{0}}}})\ensuremath{_{\textnormal{c.m.}}}=8.0\ensuremath{\times}10\ensuremath{^{\textnormal{1}}} (keV)\ensuremath{^{\textnormal{2}}} (\href{https://www.nndc.bnl.gov/nsr/nsrlink.jsp?1972Ot01,B}{1972Ot01}).\vspace{0.1cm}}&\\
&&&&&&&&\parbox[t][0.3cm]{7.4717007cm}{\raggedright Decay modes: \ensuremath{\alpha} and \ensuremath{^{\textnormal{3}}}He (\href{https://www.nndc.bnl.gov/nsr/nsrlink.jsp?1972Ot01,B}{1972Ot01}).\vspace{0.1cm}}&\\
\multicolumn{1}{r@{}}{12.08\ensuremath{\times10^{3}}?}&\multicolumn{1}{@{ }l}{{\it 4}}&\multicolumn{1}{l}{(3/2\ensuremath{^{+}},1/2\ensuremath{^{+}})\ensuremath{^{{\hyperlink{NE23LEVEL5}{f}}}}}&\multicolumn{1}{r@{}}{75}&\multicolumn{1}{@{}l}{\ensuremath{^{{\hyperlink{NE23LEVEL5}{f}}}} keV {\it 25}}&&\multicolumn{1}{r@{}}{4.32\ensuremath{\times10^{3}}}&\multicolumn{1}{@{}l}{\ensuremath{^{{\hyperlink{NE23LEVEL5}{f}}}} {\it 5}}&\parbox[t][0.3cm]{7.4717007cm}{\raggedright E(level): This state was observed as an unresolved\vspace{0.1cm}}&\\
&&&&&&&&\parbox[t][0.3cm]{7.4717007cm}{\raggedright {\ }{\ }{\ }shoulder to the 12.23-MeV state (\href{https://www.nndc.bnl.gov/nsr/nsrlink.jsp?1972Ot01,B}{1972Ot01}): See\vspace{0.1cm}}&\\
&&&&&&&&\parbox[t][0.3cm]{7.4717007cm}{\raggedright {\ }{\ }{\ }the discussion and Fig. 15, which seem to suggest\vspace{0.1cm}}&\\
&&&&&&&&\parbox[t][0.3cm]{7.4717007cm}{\raggedright {\ }{\ }{\ }that this state was considered to be tentative.\vspace{0.1cm}}&\\
&&&&&&&&\parbox[t][0.3cm]{7.4717007cm}{\raggedright J\ensuremath{^{\pi}}: (\href{https://www.nndc.bnl.gov/nsr/nsrlink.jsp?1972Ot01,B}{1972Ot01}): The prediction of an \ensuremath{\alpha}-particle,\vspace{0.1cm}}&\\
&&&&&&&&\parbox[t][0.3cm]{7.4717007cm}{\raggedright {\ }{\ }{\ }core-excited threshold-state model describing this\vspace{0.1cm}}&\\
&&&&&&&&\parbox[t][0.3cm]{7.4717007cm}{\raggedright {\ }{\ }{\ }state as an \ensuremath{^{\textnormal{15}}}O+\ensuremath{\alpha} configuration is consistent with\vspace{0.1cm}}&\\
&&&&&&&&\parbox[t][0.3cm]{7.4717007cm}{\raggedright {\ }{\ }{\ }J\ensuremath{^{\ensuremath{\pi}}}=3/2\ensuremath{^{\textnormal{+}}}. This assignment improved the fit to the\vspace{0.1cm}}&\\
&&&&&&&&\parbox[t][0.3cm]{7.4717007cm}{\raggedright {\ }{\ }{\ }measured (\ensuremath{^{\textnormal{3}}}He,\ensuremath{\alpha}) cross section in this energy\vspace{0.1cm}}&\\
&&&&&&&&\parbox[t][0.3cm]{7.4717007cm}{\raggedright {\ }{\ }{\ }region. But the authors pointed out that a J\ensuremath{^{\ensuremath{\pi}}}=1/2\ensuremath{^{\textnormal{+}}}\vspace{0.1cm}}&\\
\end{longtable}
\begin{textblock}{29}(0,27.3)
Continued on next page (footnotes at end of table)
\end{textblock}
\clearpage
\begin{longtable}{cccccccc@{\extracolsep{\fill}}c}
\\[-.4cm]
\multicolumn{9}{c}{{\bf \small \underline{\ensuremath{^{\textnormal{16}}}O(\ensuremath{^{\textnormal{3}}}He,X)\hspace{0.2in}\href{https://www.nndc.bnl.gov/nsr/nsrlink.jsp?1959Br79,B}{1959Br79},\href{https://www.nndc.bnl.gov/nsr/nsrlink.jsp?1983Wa05,B}{1983Wa05} (continued)}}}\\
\multicolumn{9}{c}{~}\\
\multicolumn{9}{c}{\underline{\ensuremath{^{19}}Ne Levels (continued)}}\\
\multicolumn{9}{c}{~}\\
\multicolumn{2}{c}{E(level)$^{{\hyperlink{NE23LEVEL0}{a}}}$}&J$^{\pi}$$^{}$&\multicolumn{2}{c}{\ensuremath{\Gamma}$^{{\hyperlink{NE23LEVEL3}{d}}}$}&\multicolumn{2}{c}{E(\ensuremath{^{\textnormal{3}}}He,lab) (MeV)$^{}$}&Comments&\\[-.2cm]
\multicolumn{2}{c}{\hrulefill}&\hrulefill&\multicolumn{2}{c}{\hrulefill}&\multicolumn{2}{c}{\hrulefill}&\hrulefill&
\endhead
&&&&&&&\parbox[t][0.3cm]{8.58218cm}{\raggedright {\ }{\ }{\ }assignment was not ruled out.\vspace{0.1cm}}&\\
&&&&&&&\parbox[t][0.3cm]{8.58218cm}{\raggedright (\ensuremath{\Gamma}\ensuremath{_{^{\textnormal{3}}\textnormal{He}}}\ensuremath{\Gamma}\ensuremath{_{\ensuremath{\alpha}_{\textnormal{0}}}})\ensuremath{_{\textnormal{c.m.}}}\ensuremath{<}3.73\ensuremath{\times}10\ensuremath{^{\textnormal{1}}} (keV)\ensuremath{^{\textnormal{2}}} (\href{https://www.nndc.bnl.gov/nsr/nsrlink.jsp?1972Ot01,B}{1972Ot01}).\vspace{0.1cm}}&\\
&&&&&&&\parbox[t][0.3cm]{8.58218cm}{\raggedright Decay modes: \ensuremath{\alpha} and \ensuremath{^{\textnormal{3}}}He (\href{https://www.nndc.bnl.gov/nsr/nsrlink.jsp?1972Ot01,B}{1972Ot01}).\vspace{0.1cm}}&\\
\multicolumn{1}{r@{}}{12.23\ensuremath{\times10^{3}}}&\multicolumn{1}{@{ }l}{{\it 4}}&\multicolumn{1}{l}{5/2\ensuremath{^{(+)}}\ensuremath{^{{\hyperlink{NE23LEVEL5}{f}}}}}&\multicolumn{1}{r@{}}{200}&\multicolumn{1}{@{}l}{\ensuremath{^{{\hyperlink{NE23LEVEL5}{f}}}} keV {\it 25}}&\multicolumn{1}{r@{}}{4.50\ensuremath{\times10^{3}}}&\multicolumn{1}{@{}l}{\ensuremath{^{{\hyperlink{NE23LEVEL5}{f}}}} {\it 5}}&\parbox[t][0.3cm]{8.58218cm}{\raggedright J\ensuremath{^{\pi}},\ensuremath{\Gamma},E(\ensuremath{^{\textnormal{3}}}He,lab) (MeV): Confirmed by the two-level analysis\vspace{0.1cm}}&\\
&&&&&&&\parbox[t][0.3cm]{8.58218cm}{\raggedright {\ }{\ }{\ }of the \ensuremath{\alpha} angular distribution data using Legendre\vspace{0.1cm}}&\\
&&&&&&&\parbox[t][0.3cm]{8.58218cm}{\raggedright {\ }{\ }{\ }polynomials (\href{https://www.nndc.bnl.gov/nsr/nsrlink.jsp?1972Ot01,B}{1972Ot01}). This analysis is unable to\vspace{0.1cm}}&\\
&&&&&&&\parbox[t][0.3cm]{8.58218cm}{\raggedright {\ }{\ }{\ }differentiate between positive or negative parities of this\vspace{0.1cm}}&\\
&&&&&&&\parbox[t][0.3cm]{8.58218cm}{\raggedright {\ }{\ }{\ }level and the \ensuremath{^{\textnormal{19}}}Ne*(12.50 MeV, 7/2\ensuremath{^{\textnormal{+}}}) state. It only yields\vspace{0.1cm}}&\\
&&&&&&&\parbox[t][0.3cm]{8.58218cm}{\raggedright {\ }{\ }{\ }identical parities for both levels. The authors chose positive\vspace{0.1cm}}&\\
&&&&&&&\parbox[t][0.3cm]{8.58218cm}{\raggedright {\ }{\ }{\ }parities for both states. (\href{https://www.nndc.bnl.gov/nsr/nsrlink.jsp?1972Ot01,B}{1972Ot01}) mentioned that the\vspace{0.1cm}}&\\
&&&&&&&\parbox[t][0.3cm]{8.58218cm}{\raggedright {\ }{\ }{\ }interference of these levels with the 1/2\ensuremath{^{\textnormal{+}}} state at 12.86 MeV\vspace{0.1cm}}&\\
&&&&&&&\parbox[t][0.3cm]{8.58218cm}{\raggedright {\ }{\ }{\ }would have caused considerable difficulty if the opposite\vspace{0.1cm}}&\\
&&&&&&&\parbox[t][0.3cm]{8.58218cm}{\raggedright {\ }{\ }{\ }parity had been assumed. They therefore considered the\vspace{0.1cm}}&\\
&&&&&&&\parbox[t][0.3cm]{8.58218cm}{\raggedright {\ }{\ }{\ }positive-parity assignments for both these states credible,\vspace{0.1cm}}&\\
&&&&&&&\parbox[t][0.3cm]{8.58218cm}{\raggedright {\ }{\ }{\ }though by no means definite. Thus, the evaluator made the\vspace{0.1cm}}&\\
&&&&&&&\parbox[t][0.3cm]{8.58218cm}{\raggedright {\ }{\ }{\ }parities tentative.\vspace{0.1cm}}&\\
&&&&&&&\parbox[t][0.3cm]{8.58218cm}{\raggedright J\ensuremath{^{\pi}}: See (\href{https://www.nndc.bnl.gov/nsr/nsrlink.jsp?1972Ot01,B}{1972Ot01}), where the prediction of an \ensuremath{\alpha}-particle\vspace{0.1cm}}&\\
&&&&&&&\parbox[t][0.3cm]{8.58218cm}{\raggedright {\ }{\ }{\ }core-excited, threshold-state model, which describes this\vspace{0.1cm}}&\\
&&&&&&&\parbox[t][0.3cm]{8.58218cm}{\raggedright {\ }{\ }{\ }state as an \ensuremath{^{\textnormal{15}}}O+\ensuremath{\alpha} configuration is consistent with J\ensuremath{^{\ensuremath{\pi}}}=5/2\ensuremath{^{\textnormal{+}}}.\vspace{0.1cm}}&\\
&&&&&&&\parbox[t][0.3cm]{8.58218cm}{\raggedright (\ensuremath{\Gamma}\ensuremath{_{^{\textnormal{3}}\textnormal{He}}}\ensuremath{\Gamma}\ensuremath{_{\ensuremath{\alpha}_{\textnormal{0}}}})\ensuremath{_{\textnormal{c.m.}}}=3.11\ensuremath{\times}10\ensuremath{^{\textnormal{3}}} (keV)\ensuremath{^{\textnormal{2}}} (\href{https://www.nndc.bnl.gov/nsr/nsrlink.jsp?1972Ot01,B}{1972Ot01}) from\vspace{0.1cm}}&\\
&&&&&&&\parbox[t][0.3cm]{8.58218cm}{\raggedright {\ }{\ }{\ }R-matrix analysis. See also (\ensuremath{\Gamma}\ensuremath{_{^{\textnormal{3}}\textnormal{He}}}\ensuremath{\Gamma}\ensuremath{_{\ensuremath{\alpha}_{\textnormal{0}}}})\ensuremath{_{\textnormal{c.m.}}}=3.045\ensuremath{\times}10\ensuremath{^{\textnormal{3}}}\vspace{0.1cm}}&\\
&&&&&&&\parbox[t][0.3cm]{8.58218cm}{\raggedright {\ }{\ }{\ }(keV)\ensuremath{^{\textnormal{2}}} (\href{https://www.nndc.bnl.gov/nsr/nsrlink.jsp?1972Ot01,B}{1972Ot01}) from a two-level analysis using Legendre\vspace{0.1cm}}&\\
&&&&&&&\parbox[t][0.3cm]{8.58218cm}{\raggedright {\ }{\ }{\ }polynomials (see Table 2).\vspace{0.1cm}}&\\
&&&&&&&\parbox[t][0.3cm]{8.58218cm}{\raggedright Decay modes: \ensuremath{\alpha} and \ensuremath{^{\textnormal{3}}}He (\href{https://www.nndc.bnl.gov/nsr/nsrlink.jsp?1972Ot01,B}{1972Ot01}).\vspace{0.1cm}}&\\
\multicolumn{1}{r@{}}{12.50\ensuremath{\times10^{3}}}&\multicolumn{1}{@{ }l}{{\it 4}}&\multicolumn{1}{l}{7/2\ensuremath{^{(+)}}\ensuremath{^{{\hyperlink{NE23LEVEL5}{f}}}}}&\multicolumn{1}{r@{}}{150}&\multicolumn{1}{@{}l}{\ensuremath{^{{\hyperlink{NE23LEVEL5}{f}}}} keV {\it 25}}&\multicolumn{1}{r@{}}{4.82\ensuremath{\times10^{3}}}&\multicolumn{1}{@{}l}{\ensuremath{^{{\hyperlink{NE23LEVEL5}{f}}}} {\it 5}}&\parbox[t][0.3cm]{8.58218cm}{\raggedright E(level): (\href{https://www.nndc.bnl.gov/nsr/nsrlink.jsp?1972Ot01,B}{1972Ot01}) first used a two-level analysis using\vspace{0.1cm}}&\\
&&&&&&&\parbox[t][0.3cm]{8.58218cm}{\raggedright {\ }{\ }{\ }Legendre polynomials to analyze the E\ensuremath{_{\textnormal{x}}}=12.23 MeV level\vspace{0.1cm}}&\\
&&&&&&&\parbox[t][0.3cm]{8.58218cm}{\raggedright {\ }{\ }{\ }(corresponding to the E(\ensuremath{^{\textnormal{3}}}He, lab)=4.50 MeV) and a\vspace{0.1cm}}&\\
&&&&&&&\parbox[t][0.3cm]{8.58218cm}{\raggedright {\ }{\ }{\ }shoulder peak at approximately E(\ensuremath{^{\textnormal{3}}}He, lab)=4.70 MeV, for\vspace{0.1cm}}&\\
&&&&&&&\parbox[t][0.3cm]{8.58218cm}{\raggedright {\ }{\ }{\ }which they determined E\ensuremath{_{\textnormal{x}}}=12.39 MeV. The data were\vspace{0.1cm}}&\\
&&&&&&&\parbox[t][0.3cm]{8.58218cm}{\raggedright {\ }{\ }{\ }insufficient to deduce reliable information on that shoulder\vspace{0.1cm}}&\\
&&&&&&&\parbox[t][0.3cm]{8.58218cm}{\raggedright {\ }{\ }{\ }structure in the excitation function. So, those authors\vspace{0.1cm}}&\\
&&&&&&&\parbox[t][0.3cm]{8.58218cm}{\raggedright {\ }{\ }{\ }remeasured the excitation function with better energy\vspace{0.1cm}}&\\
&&&&&&&\parbox[t][0.3cm]{8.58218cm}{\raggedright {\ }{\ }{\ }resolution and used R-matrix analysis instead, which\vspace{0.1cm}}&\\
&&&&&&&\parbox[t][0.3cm]{8.58218cm}{\raggedright {\ }{\ }{\ }substantially improved the fit. The R-matrix fit shifted the\vspace{0.1cm}}&\\
&&&&&&&\parbox[t][0.3cm]{8.58218cm}{\raggedright {\ }{\ }{\ }resonance energy from E\ensuremath{_{\textnormal{lab}}}(\ensuremath{^{\textnormal{3}}}He)=4.7 MeV to\vspace{0.1cm}}&\\
&&&&&&&\parbox[t][0.3cm]{8.58218cm}{\raggedright {\ }{\ }{\ }E\ensuremath{_{\textnormal{lab}}}(\ensuremath{^{\textnormal{3}}}He)=4.82 MeV \textit{5} corresponding to E\ensuremath{_{\textnormal{x}}}=12.50 MeV \textit{4}.\vspace{0.1cm}}&\\
&&&&&&&\parbox[t][0.3cm]{8.58218cm}{\raggedright {\ }{\ }{\ }(\href{https://www.nndc.bnl.gov/nsr/nsrlink.jsp?1972Ot01,B}{1972Ot01}) concluded that in addition to the E\ensuremath{_{\textnormal{x}}}=12.23{\textminus}MeV\vspace{0.1cm}}&\\
&&&&&&&\parbox[t][0.3cm]{8.58218cm}{\raggedright {\ }{\ }{\ }level, there is an \ensuremath{\alpha} resonance at E(\ensuremath{^{\textnormal{3}}}He, lab)=4.82 MeV \textit{5}\vspace{0.1cm}}&\\
&&&&&&&\parbox[t][0.3cm]{8.58218cm}{\raggedright {\ }{\ }{\ }(see Tables 2 and 3 in that study, respectively).\vspace{0.1cm}}&\\
&&&&&&&\parbox[t][0.3cm]{8.58218cm}{\raggedright {\ }{\ }{\ }Consequently, we adopted the improved result of E\ensuremath{_{\textnormal{x}}}=12.50\vspace{0.1cm}}&\\
&&&&&&&\parbox[t][0.3cm]{8.58218cm}{\raggedright {\ }{\ }{\ }MeV \textit{4} over the less reliable analysis that resulted in\vspace{0.1cm}}&\\
&&&&&&&\parbox[t][0.3cm]{8.58218cm}{\raggedright {\ }{\ }{\ }E\ensuremath{_{\textnormal{x}}}=12.39 MeV.\vspace{0.1cm}}&\\
&&&&&&&\parbox[t][0.3cm]{8.58218cm}{\raggedright \ensuremath{\Gamma}: See also \ensuremath{\Gamma}=180 keV deduced using the two-level analysis\vspace{0.1cm}}&\\
&&&&&&&\parbox[t][0.3cm]{8.58218cm}{\raggedright {\ }{\ }{\ }of the \ensuremath{\alpha} angular distribution data using Legendre\vspace{0.1cm}}&\\
&&&&&&&\parbox[t][0.3cm]{8.58218cm}{\raggedright {\ }{\ }{\ }polynomials (\href{https://www.nndc.bnl.gov/nsr/nsrlink.jsp?1972Ot01,B}{1972Ot01}).\vspace{0.1cm}}&\\
&&&&&&&\parbox[t][0.3cm]{8.58218cm}{\raggedright J\ensuremath{^{\pi}}: Confirmed by the two-level analysis of the \ensuremath{\alpha} angular\vspace{0.1cm}}&\\
&&&&&&&\parbox[t][0.3cm]{8.58218cm}{\raggedright {\ }{\ }{\ }distribution data using Legendre polynomials (\href{https://www.nndc.bnl.gov/nsr/nsrlink.jsp?1972Ot01,B}{1972Ot01}).\vspace{0.1cm}}&\\
&&&&&&&\parbox[t][0.3cm]{8.58218cm}{\raggedright {\ }{\ }{\ }This analysis is insensitive to parity. For the reasons\vspace{0.1cm}}&\\
&&&&&&&\parbox[t][0.3cm]{8.58218cm}{\raggedright {\ }{\ }{\ }mentioned for the 12.23-MeV state, which also hold for this\vspace{0.1cm}}&\\
&&&&&&&\parbox[t][0.3cm]{8.58218cm}{\raggedright {\ }{\ }{\ }state, the parity is considered tentative.\vspace{0.1cm}}&\\
&&&&&&&\parbox[t][0.3cm]{8.58218cm}{\raggedright J\ensuremath{^{\pi}}: (\href{https://www.nndc.bnl.gov/nsr/nsrlink.jsp?1972Ot01,B}{1972Ot01}): The prediction of an \ensuremath{\alpha}-particle, core-excited,\vspace{0.1cm}}&\\
&&&&&&&\parbox[t][0.3cm]{8.58218cm}{\raggedright {\ }{\ }{\ }threshold-state model, which describes this state as an\vspace{0.1cm}}&\\
&&&&&&&\parbox[t][0.3cm]{8.58218cm}{\raggedright {\ }{\ }{\ }\ensuremath{^{\textnormal{15}}}O+\ensuremath{\alpha} configuration is consistent with J\ensuremath{^{\ensuremath{\pi}}}=7/2\ensuremath{^{\textnormal{+}}}.\vspace{0.1cm}}&\\
&&&&&&&\parbox[t][0.3cm]{8.58218cm}{\raggedright (\ensuremath{\Gamma}\ensuremath{_{^{\textnormal{3}}\textnormal{He}}}\ensuremath{\Gamma}\ensuremath{_{\ensuremath{\alpha}_{\textnormal{0}}}})\ensuremath{_{\textnormal{c.m.}}}=1.14\ensuremath{\times}10\ensuremath{^{\textnormal{3}}} (keV)\ensuremath{^{\textnormal{2}}} (\href{https://www.nndc.bnl.gov/nsr/nsrlink.jsp?1972Ot01,B}{1972Ot01}) from\vspace{0.1cm}}&\\
&&&&&&&\parbox[t][0.3cm]{8.58218cm}{\raggedright {\ }{\ }{\ }R-matrix analysis. See also (\ensuremath{\Gamma}\ensuremath{_{^{\textnormal{3}}\textnormal{He}}}\ensuremath{\Gamma}\ensuremath{_{\ensuremath{\alpha}_{\textnormal{0}}}})\ensuremath{_{\textnormal{c.m.}}}=1.225\ensuremath{\times}10\ensuremath{^{\textnormal{3}}}\vspace{0.1cm}}&\\
\end{longtable}
\begin{textblock}{29}(0,27.3)
Continued on next page (footnotes at end of table)
\end{textblock}
\clearpage
\begin{longtable}{ccccccccc@{\extracolsep{\fill}}c}
\\[-.4cm]
\multicolumn{10}{c}{{\bf \small \underline{\ensuremath{^{\textnormal{16}}}O(\ensuremath{^{\textnormal{3}}}He,X)\hspace{0.2in}\href{https://www.nndc.bnl.gov/nsr/nsrlink.jsp?1959Br79,B}{1959Br79},\href{https://www.nndc.bnl.gov/nsr/nsrlink.jsp?1983Wa05,B}{1983Wa05} (continued)}}}\\
\multicolumn{10}{c}{~}\\
\multicolumn{10}{c}{\underline{\ensuremath{^{19}}Ne Levels (continued)}}\\
\multicolumn{10}{c}{~}\\
\multicolumn{2}{c}{E(level)$^{{\hyperlink{NE23LEVEL0}{a}}}$}&J$^{\pi}$$^{}$&\multicolumn{2}{c}{\ensuremath{\Gamma}$^{{\hyperlink{NE23LEVEL3}{d}}}$}&L$^{}$&\multicolumn{2}{c}{E(\ensuremath{^{\textnormal{3}}}He,lab) (MeV)$^{}$}&Comments&\\[-.2cm]
\multicolumn{2}{c}{\hrulefill}&\hrulefill&\multicolumn{2}{c}{\hrulefill}&\hrulefill&\multicolumn{2}{c}{\hrulefill}&\hrulefill&
\endhead
&&&&&&&&\parbox[t][0.3cm]{8.6829405cm}{\raggedright {\ }{\ }{\ }(keV)\ensuremath{^{\textnormal{2}}} (\href{https://www.nndc.bnl.gov/nsr/nsrlink.jsp?1972Ot01,B}{1972Ot01}) from the two-level analysis using\vspace{0.1cm}}&\\
&&&&&&&&\parbox[t][0.3cm]{8.6829405cm}{\raggedright {\ }{\ }{\ }Legendre polynomials (see Table 2).\vspace{0.1cm}}&\\
&&&&&&&&\parbox[t][0.3cm]{8.6829405cm}{\raggedright Decay modes: \ensuremath{\alpha} and \ensuremath{^{\textnormal{3}}}He (\href{https://www.nndc.bnl.gov/nsr/nsrlink.jsp?1972Ot01,B}{1972Ot01}).\vspace{0.1cm}}&\\
\multicolumn{1}{r@{}}{12.86\ensuremath{\times10^{3}}}&\multicolumn{1}{@{ }l}{{\it 4}}&\multicolumn{1}{l}{1/2\ensuremath{^{+}}}&\multicolumn{1}{r@{}}{160}&\multicolumn{1}{@{ }l}{keV {\it 25}}&\multicolumn{1}{l}{0}&\multicolumn{1}{r@{}}{5.25\ensuremath{\times10^{3}}}&\multicolumn{1}{@{ }l}{{\it 5}}&\parbox[t][0.3cm]{8.6829405cm}{\raggedright E(\ensuremath{^{\textnormal{3}}}He,lab) (MeV): From (\href{https://www.nndc.bnl.gov/nsr/nsrlink.jsp?1972Ot01,B}{1972Ot01}). See also E(\ensuremath{^{\textnormal{3}}}He,\vspace{0.1cm}}&\\
&&&&&&&&\parbox[t][0.3cm]{8.6829405cm}{\raggedright {\ }{\ }{\ }lab)=5.05 MeV \textit{5} (\href{https://www.nndc.bnl.gov/nsr/nsrlink.jsp?1967Ro10,B}{1967Ro10}).\vspace{0.1cm}}&\\
&&&&&&&&\parbox[t][0.3cm]{8.6829405cm}{\raggedright E(level): See also 12.69 MeV \textit{4} deduced from relativistic\vspace{0.1cm}}&\\
&&&&&&&&\parbox[t][0.3cm]{8.6829405cm}{\raggedright {\ }{\ }{\ }conversion of E(\ensuremath{^{\textnormal{3}}}He, lab)=5.05 MeV \textit{5} (\href{https://www.nndc.bnl.gov/nsr/nsrlink.jsp?1967Ro10,B}{1967Ro10}).\vspace{0.1cm}}&\\
&&&&&&&&\parbox[t][0.3cm]{8.6829405cm}{\raggedright E(level): (\href{https://www.nndc.bnl.gov/nsr/nsrlink.jsp?1972Ot01,B}{1972Ot01}) believed that this level corresponded to\vspace{0.1cm}}&\\
&&&&&&&&\parbox[t][0.3cm]{8.6829405cm}{\raggedright {\ }{\ }{\ }the 12.69-MeV level observed for the first time by\vspace{0.1cm}}&\\
&&&&&&&&\parbox[t][0.3cm]{8.6829405cm}{\raggedright {\ }{\ }{\ }(\href{https://www.nndc.bnl.gov/nsr/nsrlink.jsp?1967Ro10,B}{1967Ro10}). (\href{https://www.nndc.bnl.gov/nsr/nsrlink.jsp?1972Ot01,B}{1972Ot01}) attributed the discrepancy in the\vspace{0.1cm}}&\\
&&&&&&&&\parbox[t][0.3cm]{8.6829405cm}{\raggedright {\ }{\ }{\ }energy of this level with respect to that deduced from the\vspace{0.1cm}}&\\
&&&&&&&&\parbox[t][0.3cm]{8.6829405cm}{\raggedright {\ }{\ }{\ }resonance energy reported by (\href{https://www.nndc.bnl.gov/nsr/nsrlink.jsp?1967Ro10,B}{1967Ro10}) to the particular\vspace{0.1cm}}&\\
&&&&&&&&\parbox[t][0.3cm]{8.6829405cm}{\raggedright {\ }{\ }{\ }choice of background amplitude made in the analysis of\vspace{0.1cm}}&\\
&&&&&&&&\parbox[t][0.3cm]{8.6829405cm}{\raggedright {\ }{\ }{\ }(\href{https://www.nndc.bnl.gov/nsr/nsrlink.jsp?1967Ro10,B}{1967Ro10}). The evaluator recommends the resonance energy\vspace{0.1cm}}&\\
&&&&&&&&\parbox[t][0.3cm]{8.6829405cm}{\raggedright {\ }{\ }{\ }reported by (\href{https://www.nndc.bnl.gov/nsr/nsrlink.jsp?1972Ot01,B}{1972Ot01}) that is deduced from a multi-channel\vspace{0.1cm}}&\\
&&&&&&&&\parbox[t][0.3cm]{8.6829405cm}{\raggedright {\ }{\ }{\ }multi-level R-matrix analysis with a careful attention to the\vspace{0.1cm}}&\\
&&&&&&&&\parbox[t][0.3cm]{8.6829405cm}{\raggedright {\ }{\ }{\ }background.\vspace{0.1cm}}&\\
&&&&&&&&\parbox[t][0.3cm]{8.6829405cm}{\raggedright \ensuremath{\Gamma}: From a multi-channel, multi-level R-matrix analysis by\vspace{0.1cm}}&\\
&&&&&&&&\parbox[t][0.3cm]{8.6829405cm}{\raggedright {\ }{\ }{\ }(\href{https://www.nndc.bnl.gov/nsr/nsrlink.jsp?1972Ot01,B}{1972Ot01}). See also \ensuremath{\Gamma}=180 keV \textit{40} (\href{https://www.nndc.bnl.gov/nsr/nsrlink.jsp?1967Ro10,B}{1967Ro10}) deduced\vspace{0.1cm}}&\\
&&&&&&&&\parbox[t][0.3cm]{8.6829405cm}{\raggedright {\ }{\ }{\ }from a one-level Breit-Wigner analysis, where poor\vspace{0.1cm}}&\\
&&&&&&&&\parbox[t][0.3cm]{8.6829405cm}{\raggedright {\ }{\ }{\ }background analysis may have affected the result.\vspace{0.1cm}}&\\
&&&&&&&&\parbox[t][0.3cm]{8.6829405cm}{\raggedright \ensuremath{\Gamma}\ensuremath{_{^{\textnormal{3}}\textnormal{He}}}/\ensuremath{\Gamma}=0.43 \textit{3} (\href{https://www.nndc.bnl.gov/nsr/nsrlink.jsp?1967Ro10,B}{1967Ro10}): This result may be inaccurate\vspace{0.1cm}}&\\
&&&&&&&&\parbox[t][0.3cm]{8.6829405cm}{\raggedright {\ }{\ }{\ }due to the poor background analysis for the corresponding\vspace{0.1cm}}&\\
&&&&&&&&\parbox[t][0.3cm]{8.6829405cm}{\raggedright {\ }{\ }{\ }resonance.\vspace{0.1cm}}&\\
&&&&&&&&\parbox[t][0.3cm]{8.6829405cm}{\raggedright (\ensuremath{\Gamma}\ensuremath{_{^{\textnormal{3}}\textnormal{He}}}\ensuremath{\Gamma}\ensuremath{_{\ensuremath{\alpha}_{\textnormal{0}}}})\ensuremath{_{\textnormal{c.m.}}}=2.8\ensuremath{\times}10\ensuremath{^{\textnormal{3}}} (keV)\ensuremath{^{\textnormal{2}}} (\href{https://www.nndc.bnl.gov/nsr/nsrlink.jsp?1972Ot01,B}{1972Ot01}).\vspace{0.1cm}}&\\
&&&&&&&&\parbox[t][0.3cm]{8.6829405cm}{\raggedright \ensuremath{\gamma}\ensuremath{^{\textnormal{2}}}(\ensuremath{^{\textnormal{3}}}He)=80 \textit{20}: The \ensuremath{^{\textnormal{3}}}He reduced width from (\href{https://www.nndc.bnl.gov/nsr/nsrlink.jsp?1967Ro10,B}{1967Ro10}).\vspace{0.1cm}}&\\
&&&&&&&&\parbox[t][0.3cm]{8.6829405cm}{\raggedright L,J\ensuremath{^{\pi}}: From (\href{https://www.nndc.bnl.gov/nsr/nsrlink.jsp?1967Ro10,B}{1967Ro10}): Analysis of the \ensuremath{^{\textnormal{3}}}He angular\vspace{0.1cm}}&\\
&&&&&&&&\parbox[t][0.3cm]{8.6829405cm}{\raggedright {\ }{\ }{\ }distributions using Legendre polynomials together with the\vspace{0.1cm}}&\\
&&&&&&&&\parbox[t][0.3cm]{8.6829405cm}{\raggedright {\ }{\ }{\ }analysis of the difference between pure potential scattering\vspace{0.1cm}}&\\
&&&&&&&&\parbox[t][0.3cm]{8.6829405cm}{\raggedright {\ }{\ }{\ }and potential plus compound scattering extracted from the\vspace{0.1cm}}&\\
&&&&&&&&\parbox[t][0.3cm]{8.6829405cm}{\raggedright {\ }{\ }{\ }data and normalized to pure potential scattering ruled out all\vspace{0.1cm}}&\\
&&&&&&&&\parbox[t][0.3cm]{8.6829405cm}{\raggedright {\ }{\ }{\ }odd L-values and L=2 and L=4. So, the unique L=0\vspace{0.1cm}}&\\
&&&&&&&&\parbox[t][0.3cm]{8.6829405cm}{\raggedright {\ }{\ }{\ }assignment was deduced (considering that there was only one\vspace{0.1cm}}&\\
&&&&&&&&\parbox[t][0.3cm]{8.6829405cm}{\raggedright {\ }{\ }{\ }resonance observed), which resulted in J\ensuremath{^{\ensuremath{\pi}}}=1/2\ensuremath{^{\textnormal{+}}}. This\vspace{0.1cm}}&\\
&&&&&&&&\parbox[t][0.3cm]{8.6829405cm}{\raggedright {\ }{\ }{\ }assignment is supported by the multi-channel multi-level\vspace{0.1cm}}&\\
&&&&&&&&\parbox[t][0.3cm]{8.6829405cm}{\raggedright {\ }{\ }{\ }R-matrix analysis of (\href{https://www.nndc.bnl.gov/nsr/nsrlink.jsp?1972Ot01,B}{1972Ot01}).\vspace{0.1cm}}&\\
&&&&&&&&\parbox[t][0.3cm]{8.6829405cm}{\raggedright J\ensuremath{^{\pi}}: (\href{https://www.nndc.bnl.gov/nsr/nsrlink.jsp?1972Ot01,B}{1972Ot01}): The prediction of an \ensuremath{\alpha}-particle, core-excited,\vspace{0.1cm}}&\\
&&&&&&&&\parbox[t][0.3cm]{8.6829405cm}{\raggedright {\ }{\ }{\ }threshold-state model, which describes this state as an \ensuremath{^{\textnormal{15}}}O+\ensuremath{\alpha}\vspace{0.1cm}}&\\
&&&&&&&&\parbox[t][0.3cm]{8.6829405cm}{\raggedright {\ }{\ }{\ }configuration is consistent with J\ensuremath{^{\ensuremath{\pi}}}=1/2\ensuremath{^{\textnormal{+}}}.\vspace{0.1cm}}&\\
&&&&&&&&\parbox[t][0.3cm]{8.6829405cm}{\raggedright Decay modes: \ensuremath{^{\textnormal{3}}}He, p\ensuremath{_{\textnormal{0}}}, p\ensuremath{_{\textnormal{1}}}, p\ensuremath{_{\textnormal{5}}}, n\ensuremath{_{\textnormal{0}}} with lower limit partial\vspace{0.1cm}}&\\
&&&&&&&&\parbox[t][0.3cm]{8.6829405cm}{\raggedright {\ }{\ }{\ }widths of 0.40, 0.08, 0.06, 0.08, and 0.06, respectively; and\vspace{0.1cm}}&\\
&&&&&&&&\parbox[t][0.3cm]{8.6829405cm}{\raggedright {\ }{\ }{\ }with upper limit partial widths of 0.46, 0.09, 0.16, 0.10, and\vspace{0.1cm}}&\\
&&&&&&&&\parbox[t][0.3cm]{8.6829405cm}{\raggedright {\ }{\ }{\ }0.12, respectively. Therefore, 0.68\ensuremath{\Gamma}\ensuremath{\leq}\ensuremath{\sum}\ensuremath{\Gamma}\ensuremath{_{\textnormal{i}}}\ensuremath{\leq}0.97\ensuremath{\Gamma}\vspace{0.1cm}}&\\
&&&&&&&&\parbox[t][0.3cm]{8.6829405cm}{\raggedright {\ }{\ }{\ }(\href{https://www.nndc.bnl.gov/nsr/nsrlink.jsp?1967Ro10,B}{1967Ro10}). The proton and \ensuremath{^{\textnormal{3}}}He partial widths were deduced\vspace{0.1cm}}&\\
&&&&&&&&\parbox[t][0.3cm]{8.6829405cm}{\raggedright {\ }{\ }{\ }using the formalism outlined in (E. Vogt, in Nuclear\vspace{0.1cm}}&\\
&&&&&&&&\parbox[t][0.3cm]{8.6829405cm}{\raggedright {\ }{\ }{\ }reactions, ed. by P. M. Endt and M. Demeur (North-Holland\vspace{0.1cm}}&\\
&&&&&&&&\parbox[t][0.3cm]{8.6829405cm}{\raggedright {\ }{\ }{\ }Publ. Co., Amsterdam, 1960) chapter V, p. 215), while the\vspace{0.1cm}}&\\
&&&&&&&&\parbox[t][0.3cm]{8.6829405cm}{\raggedright {\ }{\ }{\ }neutron partial widths are estimated from the data of\vspace{0.1cm}}&\\
&&&&&&&&\parbox[t][0.3cm]{8.6829405cm}{\raggedright {\ }{\ }{\ }(\href{https://www.nndc.bnl.gov/nsr/nsrlink.jsp?1961To03,B}{1961To03}) at E\ensuremath{_{\textnormal{lab}}}=5 MeV.\vspace{0.1cm}}&\\
&&&&&&&&\parbox[t][0.3cm]{8.6829405cm}{\raggedright According to the theoretical work of (\href{https://www.nndc.bnl.gov/nsr/nsrlink.jsp?2008Oh03,B}{2008Oh03}), this state is\vspace{0.1cm}}&\\
&&&&&&&&\parbox[t][0.3cm]{8.6829405cm}{\raggedright {\ }{\ }{\ }the experimental evidence for the existence of an N=8 higher\vspace{0.1cm}}&\\
&&&&&&&&\parbox[t][0.3cm]{8.6829405cm}{\raggedright {\ }{\ }{\ }nodal (vibrational, \ensuremath{^{\textnormal{3}}}He cluster) state. This level is an L=0,\vspace{0.1cm}}&\\
&&&&&&&&\parbox[t][0.3cm]{8.6829405cm}{\raggedright {\ }{\ }{\ }N=8 nodal state, whose large \ensuremath{\Gamma}\ensuremath{_{^{\textnormal{3}}\textnormal{He}}}/\ensuremath{\Gamma} is in accordance with\vspace{0.1cm}}&\\
&&&&&&&&\parbox[t][0.3cm]{8.6829405cm}{\raggedright {\ }{\ }{\ }the characteristic of a higher nodal member state.\vspace{0.1cm}}&\\
\end{longtable}
\begin{textblock}{29}(0,27.3)
Continued on next page (footnotes at end of table)
\end{textblock}
\clearpage
\begin{longtable}{ccccc@{\extracolsep{\fill}}c}
\\[-.4cm]
\multicolumn{6}{c}{{\bf \small \underline{\ensuremath{^{\textnormal{16}}}O(\ensuremath{^{\textnormal{3}}}He,X)\hspace{0.2in}\href{https://www.nndc.bnl.gov/nsr/nsrlink.jsp?1959Br79,B}{1959Br79},\href{https://www.nndc.bnl.gov/nsr/nsrlink.jsp?1983Wa05,B}{1983Wa05} (continued)}}}\\
\multicolumn{6}{c}{~}\\
\multicolumn{6}{c}{\underline{\ensuremath{^{19}}Ne Levels (continued)}}\\
\multicolumn{6}{c}{~}\\
\multicolumn{2}{c}{E(level)$^{{\hyperlink{NE23LEVEL0}{a}}}$}&\multicolumn{2}{c}{\ensuremath{\Gamma}$^{{\hyperlink{NE23LEVEL3}{d}}}$}&Comments&\\[-.2cm]
\multicolumn{2}{c}{\hrulefill}&\multicolumn{2}{c}{\hrulefill}&\hrulefill&
\endhead
\multicolumn{1}{r@{}}{13.80\ensuremath{\times10^{3}}}&\multicolumn{1}{@{}l}{\ensuremath{^{{\hyperlink{NE23LEVEL1}{b}}}} {\it 25}}&\multicolumn{1}{r@{}}{0}&\multicolumn{1}{@{.}l}{67 MeV {\it 25}}&\parbox[t][0.3cm]{12.59004cm}{\raggedright E(level),\ensuremath{\Gamma}: From (\href{https://www.nndc.bnl.gov/nsr/nsrlink.jsp?1983Wa05,B}{1983Wa05}).\vspace{0.1cm}}&\\
&&&&\parbox[t][0.3cm]{12.59004cm}{\raggedright (2J+1)\ensuremath{\Gamma}\ensuremath{_{^{\textnormal{3}}\textnormal{He}}}\ensuremath{\Gamma}\ensuremath{_{\ensuremath{\gamma}}}=30 (keV\ensuremath{^{\textnormal{2}}}) \textit{17} (\href{https://www.nndc.bnl.gov/nsr/nsrlink.jsp?1983Wa05,B}{1983Wa05}).\vspace{0.1cm}}&\\
&&&&\parbox[t][0.3cm]{12.59004cm}{\raggedright Decay modes: \ensuremath{^{\textnormal{3}}}He and \ensuremath{\gamma}\ensuremath{_{\textnormal{0$-$2}}} (\href{https://www.nndc.bnl.gov/nsr/nsrlink.jsp?1983Wa05,B}{1983Wa05}: See section 3.1).\vspace{0.1cm}}&\\
&&&&\parbox[t][0.3cm]{12.59004cm}{\raggedright \ensuremath{\theta}\ensuremath{^{\textnormal{2}}}=0.20 (\href{https://www.nndc.bnl.gov/nsr/nsrlink.jsp?2008Oh03,B}{2008Oh03}) deduced assuming that the state is a member of L=2, N=8 higher\vspace{0.1cm}}&\\
&&&&\parbox[t][0.3cm]{12.59004cm}{\raggedright {\ }{\ }{\ }nodal band.\vspace{0.1cm}}&\\
\multicolumn{1}{r@{}}{14.88\ensuremath{\times10^{3}}}&\multicolumn{1}{@{}l}{\ensuremath{^{{\hyperlink{NE23LEVEL1}{b}}}} {\it 13}}&\multicolumn{1}{r@{}}{0}&\multicolumn{1}{@{.}l}{62 MeV {\it 13}}&\parbox[t][0.3cm]{12.59004cm}{\raggedright E(level),\ensuremath{\Gamma}: From (\href{https://www.nndc.bnl.gov/nsr/nsrlink.jsp?1983Wa05,B}{1983Wa05}).\vspace{0.1cm}}&\\
&&&&\parbox[t][0.3cm]{12.59004cm}{\raggedright (2J+1)\ensuremath{\Gamma}\ensuremath{_{^{\textnormal{3}}\textnormal{He}}}\ensuremath{\Gamma}\ensuremath{_{\ensuremath{\gamma}}}=89 (keV\ensuremath{^{\textnormal{2}}}) \textit{44} (\href{https://www.nndc.bnl.gov/nsr/nsrlink.jsp?1983Wa05,B}{1983Wa05}). This value was also reported as 89 (keV)\ensuremath{^{\textnormal{2}}} \textit{40}\vspace{0.1cm}}&\\
&&&&\parbox[t][0.3cm]{12.59004cm}{\raggedright {\ }{\ }{\ }in the text.\vspace{0.1cm}}&\\
&&&&\parbox[t][0.3cm]{12.59004cm}{\raggedright Decay modes: \ensuremath{^{\textnormal{3}}}He and \ensuremath{\gamma}\ensuremath{_{\textnormal{0$-$2}}} (\href{https://www.nndc.bnl.gov/nsr/nsrlink.jsp?1983Wa05,B}{1983Wa05}: See section 3.1).\vspace{0.1cm}}&\\
&&&&\parbox[t][0.3cm]{12.59004cm}{\raggedright \ensuremath{\theta}\ensuremath{^{\textnormal{2}}}=0.13 (\href{https://www.nndc.bnl.gov/nsr/nsrlink.jsp?2008Oh03,B}{2008Oh03}) deduced assuming that the state is a member of L=2, N=8 higher\vspace{0.1cm}}&\\
&&&&\parbox[t][0.3cm]{12.59004cm}{\raggedright {\ }{\ }{\ }nodal band.\vspace{0.1cm}}&\\
\multicolumn{1}{r@{}}{16.24\ensuremath{\times10^{3}}}&\multicolumn{1}{@{}l}{\ensuremath{^{{\hyperlink{NE23LEVEL1}{b}}}} {\it 13}}&\multicolumn{1}{r@{}}{0}&\multicolumn{1}{@{.}l}{40 MeV {\it 13}}&\parbox[t][0.3cm]{12.59004cm}{\raggedright E(level),\ensuremath{\Gamma}: From (\href{https://www.nndc.bnl.gov/nsr/nsrlink.jsp?1983Wa05,B}{1983Wa05}).\vspace{0.1cm}}&\\
&&&&\parbox[t][0.3cm]{12.59004cm}{\raggedright (2J+1)\ensuremath{\Gamma}\ensuremath{_{^{\textnormal{3}}\textnormal{He}}}\ensuremath{\Gamma}\ensuremath{_{\ensuremath{\gamma}}}=18 (keV\ensuremath{^{\textnormal{2}}}) \textit{4} (\href{https://www.nndc.bnl.gov/nsr/nsrlink.jsp?1983Wa05,B}{1983Wa05}).\vspace{0.1cm}}&\\
&&&&\parbox[t][0.3cm]{12.59004cm}{\raggedright Decay modes: \ensuremath{^{\textnormal{3}}}He and \ensuremath{\gamma}\ensuremath{_{\textnormal{0$-$2}}} (\href{https://www.nndc.bnl.gov/nsr/nsrlink.jsp?1983Wa05,B}{1983Wa05}: See section 3.1).\vspace{0.1cm}}&\\
&&&&\parbox[t][0.3cm]{12.59004cm}{\raggedright \ensuremath{\theta}\ensuremath{^{\textnormal{2}}}=0.07 (\href{https://www.nndc.bnl.gov/nsr/nsrlink.jsp?2008Oh03,B}{2008Oh03}) deduced assuming that the state is a member of L=2, N=8 higher\vspace{0.1cm}}&\\
&&&&\parbox[t][0.3cm]{12.59004cm}{\raggedright {\ }{\ }{\ }nodal band.\vspace{0.1cm}}&\\
\multicolumn{1}{r@{}}{18.4\ensuremath{\times10^{3}}}&\multicolumn{1}{@{ }l}{{\it 5}}&\multicolumn{1}{r@{}}{4}&\multicolumn{1}{@{.}l}{4 MeV {\it 5}}&\parbox[t][0.3cm]{12.59004cm}{\raggedright E(level),\ensuremath{\Gamma}: From (\href{https://www.nndc.bnl.gov/nsr/nsrlink.jsp?1983Wa05,B}{1983Wa05}).\vspace{0.1cm}}&\\
&&&&\parbox[t][0.3cm]{12.59004cm}{\raggedright E(level): This state was observed by (\href{https://www.nndc.bnl.gov/nsr/nsrlink.jsp?1983Wa05,B}{1983Wa05}) as an underlying structure on the \ensuremath{\gamma}\ensuremath{_{\textnormal{0$-$2}}}\vspace{0.1cm}}&\\
&&&&\parbox[t][0.3cm]{12.59004cm}{\raggedright {\ }{\ }{\ }peak. This structure appears to be more weakly populated under the \ensuremath{\gamma}\ensuremath{_{\textnormal{3$-$5}}} peak. No such\vspace{0.1cm}}&\\
&&&&\parbox[t][0.3cm]{12.59004cm}{\raggedright {\ }{\ }{\ }structure was observed under the \ensuremath{\gamma}\ensuremath{_{\textnormal{0}}} peak.\vspace{0.1cm}}&\\
&&&&\parbox[t][0.3cm]{12.59004cm}{\raggedright (2J+1)\ensuremath{\Gamma}\ensuremath{_{^{\textnormal{3}}\textnormal{He}}}\ensuremath{\Gamma}\ensuremath{_{\ensuremath{\gamma}}}=17.0\ensuremath{\times}10\ensuremath{^{\textnormal{3}}} (keV\ensuremath{^{\textnormal{2}}}) \textit{53} (\href{https://www.nndc.bnl.gov/nsr/nsrlink.jsp?1983Wa05,B}{1983Wa05}).\vspace{0.1cm}}&\\
&&&&\parbox[t][0.3cm]{12.59004cm}{\raggedright Decay modes: \ensuremath{^{\textnormal{3}}}He and \ensuremath{\gamma}\ensuremath{_{\textnormal{0$-$2}}} (\href{https://www.nndc.bnl.gov/nsr/nsrlink.jsp?1983Wa05,B}{1983Wa05}).\vspace{0.1cm}}&\\
&&&&\parbox[t][0.3cm]{12.59004cm}{\raggedright The results of (\href{https://www.nndc.bnl.gov/nsr/nsrlink.jsp?1983Wa05,B}{1983Wa05}) and their shell model calculations are consistent with a\vspace{0.1cm}}&\\
&&&&\parbox[t][0.3cm]{12.59004cm}{\raggedright {\ }{\ }{\ }predominantly giant dipole resonance character of this state, but it is populated much more\vspace{0.1cm}}&\\
&&&&\parbox[t][0.3cm]{12.59004cm}{\raggedright {\ }{\ }{\ }weakly in a \ensuremath{^{\textnormal{3}}}He capture than in a proton capture reaction.\vspace{0.1cm}}&\\
&&&&\parbox[t][0.3cm]{12.59004cm}{\raggedright \ensuremath{\sum}(2J+1)\ensuremath{\Gamma}\ensuremath{_{\ensuremath{\gamma}}}=286 keV (\href{https://www.nndc.bnl.gov/nsr/nsrlink.jsp?1983Wa05,B}{1983Wa05}): From the shell model calculations for an E\ensuremath{_{\ensuremath{\gamma}}}=18 MeV\vspace{0.1cm}}&\\
&&&&\parbox[t][0.3cm]{12.59004cm}{\raggedright {\ }{\ }{\ }and assuming that the transition is predominantly to the J\ensuremath{^{\ensuremath{\pi}}}=5/2\ensuremath{^{\textnormal{+}}} first excited state at 238\vspace{0.1cm}}&\\
&&&&\parbox[t][0.3cm]{12.59004cm}{\raggedright {\ }{\ }{\ }keV.\vspace{0.1cm}}&\\
&&&&\parbox[t][0.3cm]{12.59004cm}{\raggedright (\href{https://www.nndc.bnl.gov/nsr/nsrlink.jsp?2008Oh03,B}{2008Oh03}) reported that this state may be considered a candidate for the member state of\vspace{0.1cm}}&\\
&&&&\parbox[t][0.3cm]{12.59004cm}{\raggedright {\ }{\ }{\ }the N=8 higher nodal, vibrational \ensuremath{^{\textnormal{3}}}He cluster band with L=4.\vspace{0.1cm}}&\\
&&&&\parbox[t][0.3cm]{12.59004cm}{\raggedright \ensuremath{\theta}\ensuremath{^{\textnormal{2}}}=1.0 (\href{https://www.nndc.bnl.gov/nsr/nsrlink.jsp?2008Oh03,B}{2008Oh03}) deduced assuming that the state is a member of L=4, N=8 higher nodal\vspace{0.1cm}}&\\
&&&&\parbox[t][0.3cm]{12.59004cm}{\raggedright {\ }{\ }{\ }band.\vspace{0.1cm}}&\\
\end{longtable}
\parbox[b][0.3cm]{17.7cm}{\makebox[1ex]{\ensuremath{^{\hypertarget{NE23LEVEL0}{a}}}} Level-energies are deduced from relativistic conversion of the laboratory \ensuremath{^{\textnormal{16}}}O+\ensuremath{^{\textnormal{3}}}He resonance energy at E(\ensuremath{^{\textnormal{3}}}He,lab), listed in the}\\
\parbox[b][0.3cm]{17.7cm}{{\ }{\ }table, to the excitation energy and by using the \ensuremath{^{\textnormal{16}}}O, \ensuremath{^{\textnormal{3}}}He and \ensuremath{^{\textnormal{19}}}Ne masses from (\href{https://www.nndc.bnl.gov/nsr/nsrlink.jsp?2021Wa16,B}{2021Wa16}: AME-2020). E\ensuremath{_{\textnormal{x}}}=S\ensuremath{_{^{\textnormal{3}}\textnormal{He}}}+E\ensuremath{_{\textnormal{c.m.}}}}\\
\parbox[b][0.3cm]{17.7cm}{{\ }{\ }(relativistic).}\\
\parbox[b][0.3cm]{17.7cm}{\makebox[1ex]{\ensuremath{^{\hypertarget{NE23LEVEL1}{b}}}} (\href{https://www.nndc.bnl.gov/nsr/nsrlink.jsp?2008Oh03,B}{2008Oh03}) explains that this state should be regarded as a member of an N=8 higher nodal (vibrational state with \ensuremath{^{\textnormal{3}}}He cluster)}\\
\parbox[b][0.3cm]{17.7cm}{{\ }{\ }band. This state may be considered to be fragmented from the higher nodal L=2 state (\href{https://www.nndc.bnl.gov/nsr/nsrlink.jsp?2008Oh03,B}{2008Oh03}).}\\
\parbox[b][0.3cm]{17.7cm}{\makebox[1ex]{\ensuremath{^{\hypertarget{NE23LEVEL2}{c}}}} Obtained from the analysis of the corresponding \ensuremath{\alpha} angular distribution in (\href{https://www.nndc.bnl.gov/nsr/nsrlink.jsp?1959Br79,B}{1959Br79}), which also provided the given partial}\\
\parbox[b][0.3cm]{17.7cm}{{\ }{\ }widths.}\\
\parbox[b][0.3cm]{17.7cm}{\makebox[1ex]{\ensuremath{^{\hypertarget{NE23LEVEL3}{d}}}} Those total widths deduced from R-matrix analysis by (\href{https://www.nndc.bnl.gov/nsr/nsrlink.jsp?1972Ot01,B}{1972Ot01}) were estimated to be accurate to 25 keV.}\\
\parbox[b][0.3cm]{17.7cm}{\makebox[1ex]{\ensuremath{^{\hypertarget{NE23LEVEL4}{e}}}} Fitting the \ensuremath{\alpha} angular distributions measured by (\href{https://www.nndc.bnl.gov/nsr/nsrlink.jsp?1959Br79,B}{1959Br79}) with the Legendre polynomials revealed the necessity of a strong}\\
\parbox[b][0.3cm]{17.7cm}{{\ }{\ }negative P\ensuremath{_{\textnormal{2}}} and a non-zero P\ensuremath{_{\textnormal{4}}} contributions in the \ensuremath{\sigma}(\ensuremath{\theta}) for the \ensuremath{\alpha}\ensuremath{_{\textnormal{0}}} group. Considering that the entrance and exit channel spins}\\
\parbox[b][0.3cm]{17.7cm}{{\ }{\ }are both 1/2, if the reaction proceeds through a single, isolated resonance, it is impossible to obtain negative coefficients in the}\\
\parbox[b][0.3cm]{17.7cm}{{\ }{\ }Legendre polynomial expansion of the angular distribution of the reaction products (\href{https://www.nndc.bnl.gov/nsr/nsrlink.jsp?1959Br79,B}{1959Br79}). This means that instead of one}\\
\parbox[b][0.3cm]{17.7cm}{{\ }{\ }resonance, interference between at least two separate resonances is responsible for the data observed. The best fit to the measured}\\
\parbox[b][0.3cm]{17.7cm}{{\ }{\ }\ensuremath{\alpha} angular distributions was obtained when J=1/2 and 5/2 assignments were assumed for the two resonances involved. Parity}\\
\parbox[b][0.3cm]{17.7cm}{{\ }{\ }cannot be established from those angular distributions. However, they are symmetric about \ensuremath{\theta}\ensuremath{_{\textnormal{c.m.}}}=90\ensuremath{^\circ}. Furthermore, the obtained}\\
\parbox[b][0.3cm]{17.7cm}{{\ }{\ }Legendre coefficients of the best fits varied slowly with energy over the resonance, which indicates that the two resonances must}\\
\parbox[b][0.3cm]{17.7cm}{{\ }{\ }have the same parity. The authors assumed positive parities but they acknowledge that no conclusive evidence exists as to why}\\
\begin{textblock}{29}(0,27.3)
Continued on next page (footnotes at end of table)
\end{textblock}
\clearpage
\vspace*{-0.5cm}
{\bf \small \underline{\ensuremath{^{\textnormal{16}}}O(\ensuremath{^{\textnormal{3}}}He,X)\hspace{0.2in}\href{https://www.nndc.bnl.gov/nsr/nsrlink.jsp?1959Br79,B}{1959Br79},\href{https://www.nndc.bnl.gov/nsr/nsrlink.jsp?1983Wa05,B}{1983Wa05} (continued)}}\\
\vspace{0.3cm}
\underline{$^{19}$Ne Levels (continued)}\\
\vspace{0.3cm}
\parbox[b][0.3cm]{17.7cm}{{\ }{\ }parities should be positive. We therefore, did not adopt the parity for this state.}\\
\parbox[b][0.3cm]{17.7cm}{\makebox[1ex]{\ensuremath{^{\hypertarget{NE23LEVEL5}{f}}}} From the multi-channel multi-level R-matrix analysis of the \ensuremath{^{\textnormal{16}}}O(\ensuremath{^{\textnormal{3}}}He,\ensuremath{\alpha}) data of (\href{https://www.nndc.bnl.gov/nsr/nsrlink.jsp?1972Ot01,B}{1972Ot01}). A background pole was used in this}\\
\parbox[b][0.3cm]{17.7cm}{{\ }{\ }analysis at E\ensuremath{_{\textnormal{x}}}=11.39 MeV (deduced by the evaluator) associated with a resonance at E(\ensuremath{^{\textnormal{3}}}He, lab)=3.50 MeV with J\ensuremath{^{\ensuremath{\pi}}}=5/2\ensuremath{^{-}} and a}\\
\parbox[b][0.3cm]{17.7cm}{{\ }{\ }total width of \ensuremath{\Gamma}=1500 keV. The procedure of introducing this background level strongly affected the deduced resonance energies.}\\
\parbox[b][0.3cm]{17.7cm}{{\ }{\ }Thus, a confidence limit of \ensuremath{\pm}50 keV was placed on the reported resonance energies in the laboratory frame from that study. This}\\
\parbox[b][0.3cm]{17.7cm}{{\ }{\ }translate into a 40 keV uncertainty on the excitation energies that we obtained from (\href{https://www.nndc.bnl.gov/nsr/nsrlink.jsp?1972Ot01,B}{1972Ot01}).}\\
\vspace{0.5cm}
\clearpage
\subsection[\hspace{-0.2cm}\ensuremath{^{\textnormal{16}}}O(\ensuremath{\alpha},n),(\ensuremath{\alpha},n\ensuremath{\gamma})]{ }
\vspace{-27pt}
\vspace{0.3cm}
\hypertarget{NE24}{{\bf \small \underline{\ensuremath{^{\textnormal{16}}}O(\ensuremath{\alpha},n),(\ensuremath{\alpha},n\ensuremath{\gamma})\hspace{0.2in}\href{https://www.nndc.bnl.gov/nsr/nsrlink.jsp?1970Gi09,B}{1970Gi09},\href{https://www.nndc.bnl.gov/nsr/nsrlink.jsp?1981Ov01,B}{1981Ov01}}}}\\
\vspace{4pt}
\vspace{8pt}
\parbox[b][0.3cm]{17.7cm}{\addtolength{\parindent}{-0.2in}J\ensuremath{^{\ensuremath{\pi}}}(\ensuremath{^{\textnormal{16}}}O\ensuremath{_{\textnormal{g.s.}}})=0\ensuremath{^{\textnormal{+}}} and J\ensuremath{^{\ensuremath{\pi}}}(\ensuremath{\alpha})=0\ensuremath{^{\textnormal{+}}}.}\\
\parbox[b][0.3cm]{17.7cm}{\addtolength{\parindent}{-0.2in}\href{https://www.nndc.bnl.gov/nsr/nsrlink.jsp?1960Ja12,B}{1960Ja12}: \ensuremath{^{\textnormal{16}}}O(\ensuremath{\alpha},n) E=20.5 MeV; measured the half-life of \ensuremath{^{\textnormal{19}}}Ne using activation technique. A sample was irradiated for 1}\\
\parbox[b][0.3cm]{17.7cm}{half-life, transported to the counting station in 0.3 s, and counted using a plastic scintillator. Deduced T\ensuremath{_{\textnormal{1/2}}}(\ensuremath{^{\textnormal{19}}}Ne\ensuremath{_{\textnormal{g.s.}}})=16.72 s \textit{5}.}\\
\parbox[b][0.3cm]{17.7cm}{\addtolength{\parindent}{-0.2in}\href{https://www.nndc.bnl.gov/nsr/nsrlink.jsp?1969Ya05,B}{1969Ya05}: \ensuremath{^{\textnormal{16}}}O(\ensuremath{\alpha},n\ensuremath{\gamma}) E=20-50 MeV; measured E\ensuremath{_{\ensuremath{\gamma}}} for prompt and \ensuremath{\alpha} delayed \ensuremath{\gamma} rays using a Ge(Li) detector at \ensuremath{\theta}\ensuremath{_{\textnormal{lab}}}=126\ensuremath{^\circ}. The}\\
\parbox[b][0.3cm]{17.7cm}{\ensuremath{^{\textnormal{16}}}O(\ensuremath{\alpha},n) reaction was a contaminant in their data. Measured the half-life of the \ensuremath{^{\textnormal{19}}}Ne*(238 keV) level.}\\
\parbox[b][0.3cm]{17.7cm}{\addtolength{\parindent}{-0.2in}\href{https://www.nndc.bnl.gov/nsr/nsrlink.jsp?1970Gi09,B}{1970Gi09}: \ensuremath{^{\textnormal{16}}}O(\ensuremath{\alpha},n\ensuremath{\gamma}) E=15-19 MeV; measured neutrons and \ensuremath{\gamma}-rays in coincidence using an NE-213 liquid scintillator at \ensuremath{\theta}\ensuremath{_{\textnormal{lab}}}=0\ensuremath{^\circ}}\\
\parbox[b][0.3cm]{17.7cm}{and 150\ensuremath{^\circ} and a Ge(Li) detector at \ensuremath{\theta}\ensuremath{_{\textnormal{lab}}}=90\ensuremath{^\circ} and 30\ensuremath{^\circ}, respectively. Deduced the half-lives of the \ensuremath{^{\textnormal{19}}}Ne*(1508, 1536, 1615) states}\\
\parbox[b][0.3cm]{17.7cm}{using Doppler shift attenuation method.}\\
\parbox[b][0.3cm]{17.7cm}{\addtolength{\parindent}{-0.2in}\href{https://www.nndc.bnl.gov/nsr/nsrlink.jsp?1971It02,B}{1971It02}: \ensuremath{^{\textnormal{16}}}O(\ensuremath{\alpha},n\ensuremath{\gamma}) E=21-26 MeV; measured E\ensuremath{_{\ensuremath{\gamma}}} using a Ge(Li) detector that was placed at different angles between \ensuremath{\theta}\ensuremath{_{\textnormal{lab}}}=0\ensuremath{^\circ}}\\
\parbox[b][0.3cm]{17.7cm}{and 90\ensuremath{^\circ}. Deduced the \ensuremath{^{\textnormal{19}}}Ne level-energies and T\ensuremath{_{\textnormal{1/2}}} for the \ensuremath{^{\textnormal{19}}}Ne*(1508 and 1536) states using Doppler shift attenuation method.}\\
\parbox[b][0.3cm]{17.7cm}{Comparison with (\href{https://www.nndc.bnl.gov/nsr/nsrlink.jsp?1970Gi09,B}{1970Gi09}) is discussed.}\\
\parbox[b][0.3cm]{17.7cm}{\addtolength{\parindent}{-0.2in}\href{https://www.nndc.bnl.gov/nsr/nsrlink.jsp?1973De34,B}{1973De34}: \ensuremath{^{\textnormal{16}}}O(\ensuremath{\alpha},n) E=15.9-22.6 MeV; measured \ensuremath{\sigma}(E) using the activation technique and by detecting the annihilation \ensuremath{\gamma} rays}\\
\parbox[b][0.3cm]{17.7cm}{from the \ensuremath{\beta}-decay of \ensuremath{^{\textnormal{19}}}Ne. Observed an enhancement of the \ensuremath{\alpha} differential cross sections at backward angles. Normalized \ensuremath{\sigma}(E) for}\\
\parbox[b][0.3cm]{17.7cm}{the \ensuremath{^{\textnormal{16}}}O(\ensuremath{\alpha},n) reaction to that of the \ensuremath{^{\textnormal{12}}}C(\ensuremath{\alpha},n) reaction and obtained a normalization factor of 1.67\ensuremath{\times}10\ensuremath{^{\textnormal{$-$27}}} \textit{12}.}\\
\parbox[b][0.3cm]{17.7cm}{\addtolength{\parindent}{-0.2in}\href{https://www.nndc.bnl.gov/nsr/nsrlink.jsp?1973Gr29,B}{1973Gr29}: \ensuremath{^{\textnormal{16}}}O(\ensuremath{\alpha},n\ensuremath{\gamma}) E=15-26 and E=31 MeV; measured the excitation function of the \ensuremath{^{\textnormal{16}}}O(\ensuremath{\alpha},n) reaction using the activation}\\
\parbox[b][0.3cm]{17.7cm}{technique and by measuring the coincident annihilation \ensuremath{\gamma} rays from the \ensuremath{\beta}-decay of \ensuremath{^{\textnormal{19}}}Ne using 2 NaI(Tl) detectors. Energy}\\
\parbox[b][0.3cm]{17.7cm}{resolution was 40-60 keV (FWHM) for E\ensuremath{_{\ensuremath{\alpha}}}=15.5-26.8 MeV. Determined the absolute cross section of the \ensuremath{^{\textnormal{16}}}O(\ensuremath{\alpha},n) reaction as 44}\\
\parbox[b][0.3cm]{17.7cm}{mb \textit{4} at E\ensuremath{_{\textnormal{x}}}=23.7 MeV with a statistical uncertainty of \ensuremath{<}3\%.}\\
\parbox[b][0.3cm]{17.7cm}{\addtolength{\parindent}{-0.2in}\href{https://www.nndc.bnl.gov/nsr/nsrlink.jsp?1975SkZY,B}{1975SkZY}: \ensuremath{^{\textnormal{16}}}O(\ensuremath{\alpha},n) E=21.9 MeV; measured \ensuremath{\sigma}(E\ensuremath{_{\textnormal{n}}},\ensuremath{\theta}).}\\
\parbox[b][0.3cm]{17.7cm}{\addtolength{\parindent}{-0.2in}\href{https://www.nndc.bnl.gov/nsr/nsrlink.jsp?1978OvZZ,B}{1978OvZZ}, \href{https://www.nndc.bnl.gov/nsr/nsrlink.jsp?1981Ov01,B}{1981Ov01}: \ensuremath{^{\textnormal{16}}}O(\ensuremath{\alpha},n) E=40 and 41 MeV; measured neutrons using a TOF-spectrometer; measured the excitation}\\
\parbox[b][0.3cm]{17.7cm}{function of \ensuremath{^{\textnormal{16}}}O(\ensuremath{\alpha},n) and deduced \ensuremath{\sigma}(\ensuremath{\theta},E\ensuremath{_{\textnormal{n}}}). The resolution varied between \ensuremath{\Delta}E(FWHM)=360-500 keV depending on the target}\\
\parbox[b][0.3cm]{17.7cm}{used. Measured neutron angular distributions corresponding to the resolved \ensuremath{^{\textnormal{19}}}Ne states at \ensuremath{\theta}\ensuremath{_{\textnormal{lab}}}=0\ensuremath{^\circ}{\textminus}50\ensuremath{^\circ}. Deduced \ensuremath{^{\textnormal{19}}}Ne}\\
\parbox[b][0.3cm]{17.7cm}{level-energies, J, \ensuremath{\pi}, and spectroscopic factors using a zero-range DWBA analysis with DWUCK4.}\\
\parbox[b][0.3cm]{17.7cm}{\addtolength{\parindent}{-0.2in}\href{https://www.nndc.bnl.gov/nsr/nsrlink.jsp?1983Pi07,B}{1983Pi07}: \ensuremath{^{\textnormal{16}}}O(\ensuremath{\alpha},n\ensuremath{\gamma}) E=32 MeV; measured E\ensuremath{_{\ensuremath{\gamma}}} of low-energy prompt \ensuremath{\gamma} rays produced from activation using a shielded Ge}\\
\parbox[b][0.3cm]{17.7cm}{detector.}\\
\vspace{0.385cm}
\parbox[b][0.3cm]{17.7cm}{\addtolength{\parindent}{-0.2in}\textit{Theory}:}\\
\parbox[b][0.3cm]{17.7cm}{\addtolength{\parindent}{-0.2in}\href{https://www.nndc.bnl.gov/nsr/nsrlink.jsp?1977Gr18,B}{1977Gr18}: \ensuremath{^{\textnormal{16}}}O(\ensuremath{\alpha},n); performed Hauser-Feshbach calculations to determine \ensuremath{\sigma}(E) for this reaction. The agreement with the}\\
\parbox[b][0.3cm]{17.7cm}{measured excitation function is good within a factor of 2. The authors concluded that the \ensuremath{^{\textnormal{16}}}O(\ensuremath{\alpha},n) reaction cannot be described by}\\
\parbox[b][0.3cm]{17.7cm}{a pure statistical theory.}\\
\vspace{12pt}
\underline{$^{19}$Ne Levels}\\
\vspace{0.34cm}
\parbox[b][0.3cm]{17.7cm}{\addtolength{\parindent}{-0.254cm}(\href{https://www.nndc.bnl.gov/nsr/nsrlink.jsp?1981Ov01,B}{1981Ov01}) presents d\ensuremath{\sigma}/d\ensuremath{\Omega}\ensuremath{_{\textnormal{lab}}}(\ensuremath{\theta}\ensuremath{_{\textnormal{lab}}}=15\ensuremath{^\circ}) at E\ensuremath{_{\ensuremath{\alpha}}}=41 MeV and the integrated (over \ensuremath{\theta}\ensuremath{_{\textnormal{lab}}}=0\ensuremath{^\circ}{\textminus}60\ensuremath{^\circ}) cross section for each observed}\\
\parbox[b][0.3cm]{17.7cm}{\ensuremath{^{\textnormal{19}}}Ne state.}\\
\vspace{0.34cm}
\begin{longtable}{ccccccccc@{\extracolsep{\fill}}c}
\multicolumn{2}{c}{E(level)$^{{\hyperlink{NE24LEVEL0}{a}}}$}&J$^{\pi}$$^{}$&\multicolumn{2}{c}{T\ensuremath{_{\textnormal{1/2}}}$^{}$}&L$^{}$&\multicolumn{2}{c}{S\ensuremath{_{\textnormal{rel}}}$^{{\hyperlink{NE24LEVEL3}{d}}}$}&Comments&\\[-.2cm]
\multicolumn{2}{c}{\hrulefill}&\hrulefill&\multicolumn{2}{c}{\hrulefill}&\hrulefill&\multicolumn{2}{c}{\hrulefill}&\hrulefill&
\endfirsthead
\multicolumn{1}{r@{}}{0}&\multicolumn{1}{@{}l}{}&\multicolumn{1}{l}{1/2\ensuremath{^{+}}}&\multicolumn{1}{r@{}}{16}&\multicolumn{1}{@{.}l}{72 s {\it 5}}&&\multicolumn{1}{r@{}}{0}&\multicolumn{1}{@{.}l}{83\ensuremath{^{{\hyperlink{NE24LEVEL4}{e}}}}}&\parbox[t][0.3cm]{10.810101cm}{\raggedright E(level): From (\href{https://www.nndc.bnl.gov/nsr/nsrlink.jsp?1960Ja12,B}{1960Ja12}, \href{https://www.nndc.bnl.gov/nsr/nsrlink.jsp?1969Ya05,B}{1969Ya05}, \href{https://www.nndc.bnl.gov/nsr/nsrlink.jsp?1971It02,B}{1971It02}, \href{https://www.nndc.bnl.gov/nsr/nsrlink.jsp?1981Ov01,B}{1981Ov01}, \href{https://www.nndc.bnl.gov/nsr/nsrlink.jsp?1983Pi07,B}{1983Pi07}).\vspace{0.1cm}}&\\
&&&&&&&&\parbox[t][0.3cm]{10.810101cm}{\raggedright T\ensuremath{_{1/2}}: From (\href{https://www.nndc.bnl.gov/nsr/nsrlink.jsp?1960Ja12,B}{1960Ja12}).\vspace{0.1cm}}&\\
&&&&&&&&\parbox[t][0.3cm]{10.810101cm}{\raggedright J\ensuremath{^{\pi}}: From the \ensuremath{^{\textnormal{19}}}Ne Adopted Levels.\vspace{0.1cm}}&\\
\multicolumn{1}{r@{}}{238}&\multicolumn{1}{@{.}l}{5}&\multicolumn{1}{l}{(5/2\ensuremath{^{+}})\ensuremath{^{{\hyperlink{NE24LEVEL2}{c}}}}}&\multicolumn{1}{r@{}}{18}&\multicolumn{1}{@{ }l}{ns {\it 2}}&\multicolumn{1}{l}{(2)}&\multicolumn{1}{r@{}}{0}&\multicolumn{1}{@{.}l}{83\ensuremath{^{{\hyperlink{NE24LEVEL4}{e}}}}}&\parbox[t][0.3cm]{10.810101cm}{\raggedright E(level): From a least-squares fit to the measured E\ensuremath{_{\ensuremath{\gamma}}} (\href{https://www.nndc.bnl.gov/nsr/nsrlink.jsp?1969Ya05,B}{1969Ya05}). See also\vspace{0.1cm}}&\\
&&&&&&&&\parbox[t][0.3cm]{10.810101cm}{\raggedright {\ }{\ }{\ }E\ensuremath{_{\textnormal{x}}}=190 keV (\href{https://www.nndc.bnl.gov/nsr/nsrlink.jsp?1981Ov01,B}{1981Ov01}), where this is an unresolved doublet that consisted\vspace{0.1cm}}&\\
&&&&&&&&\parbox[t][0.3cm]{10.810101cm}{\raggedright {\ }{\ }{\ }of the \ensuremath{^{\textnormal{19}}}Ne\ensuremath{_{\textnormal{g.s.}}} and \ensuremath{^{\textnormal{19}}}Ne*(238) level; and 238 keV (\href{https://www.nndc.bnl.gov/nsr/nsrlink.jsp?1983Pi07,B}{1983Pi07}).\vspace{0.1cm}}&\\
&&&&&&&&\parbox[t][0.3cm]{10.810101cm}{\raggedright T\ensuremath{_{1/2}}: From Fig. 3 of (\href{https://www.nndc.bnl.gov/nsr/nsrlink.jsp?1969Ya05,B}{1969Ya05}). It is not clear if the uncertainty is measured\vspace{0.1cm}}&\\
&&&&&&&&\parbox[t][0.3cm]{10.810101cm}{\raggedright {\ }{\ }{\ }by those authors, or if it comes from the literature.\vspace{0.1cm}}&\\
&&&&&&&&\parbox[t][0.3cm]{10.810101cm}{\raggedright J\ensuremath{^{\pi}}: The zero-range DWBA analysis of (\href{https://www.nndc.bnl.gov/nsr/nsrlink.jsp?1981Ov01,B}{1981Ov01}) for the unresolved doublet\vspace{0.1cm}}&\\
&&&&&&&&\parbox[t][0.3cm]{10.810101cm}{\raggedright {\ }{\ }{\ }at 190 keV was best fitted using J\ensuremath{^{\ensuremath{\pi}}}=(1/2\ensuremath{^{\textnormal{+}}}+5/2\ensuremath{^{\textnormal{+}}}). Since the ground state has\vspace{0.1cm}}&\\
&&&&&&&&\parbox[t][0.3cm]{10.810101cm}{\raggedright {\ }{\ }{\ }an established J\ensuremath{^{\ensuremath{\pi}}}=1/2\ensuremath{^{\textnormal{+}}} assignment, the evaluator took J\ensuremath{^{\ensuremath{\pi}}}=(5/2\ensuremath{^{\textnormal{+}}}) for this\vspace{0.1cm}}&\\
&&&&&&&&\parbox[t][0.3cm]{10.810101cm}{\raggedright {\ }{\ }{\ }state.\vspace{0.1cm}}&\\
&&&&&&&&\parbox[t][0.3cm]{10.810101cm}{\raggedright L: From the zero-range DWBA analysis of (\href{https://www.nndc.bnl.gov/nsr/nsrlink.jsp?1981Ov01,B}{1981Ov01}: See text).\vspace{0.1cm}}&\\
\multicolumn{1}{r@{}}{273}&\multicolumn{1}{@{.}l}{9 {\it 7}}&&&&&&&\parbox[t][0.3cm]{10.810101cm}{\raggedright E(level): From a least-squares fit to the measured E\ensuremath{_{\ensuremath{\gamma}}} (\href{https://www.nndc.bnl.gov/nsr/nsrlink.jsp?1971It02,B}{1971It02}).\vspace{0.1cm}}&\\
\end{longtable}
\begin{textblock}{29}(0,27.3)
Continued on next page (footnotes at end of table)
\end{textblock}
\clearpage
\begin{longtable}{ccccccccc@{\extracolsep{\fill}}c}
\\[-.4cm]
\multicolumn{10}{c}{{\bf \small \underline{\ensuremath{^{\textnormal{16}}}O(\ensuremath{\alpha},n),(\ensuremath{\alpha},n\ensuremath{\gamma})\hspace{0.2in}\href{https://www.nndc.bnl.gov/nsr/nsrlink.jsp?1970Gi09,B}{1970Gi09},\href{https://www.nndc.bnl.gov/nsr/nsrlink.jsp?1981Ov01,B}{1981Ov01} (continued)}}}\\
\multicolumn{10}{c}{~}\\
\multicolumn{10}{c}{\underline{\ensuremath{^{19}}Ne Levels (continued)}}\\
\multicolumn{10}{c}{~}\\
\multicolumn{2}{c}{E(level)$^{{\hyperlink{NE24LEVEL0}{a}}}$}&J$^{\pi}$$^{}$&\multicolumn{2}{c}{T\ensuremath{_{\textnormal{1/2}}}$^{}$}&L$^{}$&\multicolumn{2}{c}{S\ensuremath{_{\textnormal{rel}}}$^{{\hyperlink{NE24LEVEL3}{d}}}$}&Comments&\\[-.2cm]
\multicolumn{2}{c}{\hrulefill}&\hrulefill&\multicolumn{2}{c}{\hrulefill}&\hrulefill&\multicolumn{2}{c}{\hrulefill}&\hrulefill&
\endhead
\multicolumn{1}{r@{}}{1506}&\multicolumn{1}{@{.}l}{1 {\it 8}}&&\multicolumn{1}{r@{}}{0}&\multicolumn{1}{@{.}l}{97\ensuremath{^{{\hyperlink{NE24LEVEL1}{b}}}} ps {\it +35\textminus42}}&&&&\parbox[t][0.3cm]{8.6915cm}{\raggedright E(level): From a least-squares fit to the measured E\ensuremath{_{\ensuremath{\gamma}}}\vspace{0.1cm}}&\\
&&&&&&&&\parbox[t][0.3cm]{8.6915cm}{\raggedright {\ }{\ }{\ }(\href{https://www.nndc.bnl.gov/nsr/nsrlink.jsp?1971It02,B}{1971It02}).\vspace{0.1cm}}&\\
&&&&&&&&\parbox[t][0.3cm]{8.6915cm}{\raggedright T\ensuremath{_{1/2}}: From \ensuremath{\tau}=1.4 ps \textit{+5{\textminus}6} (\href{https://www.nndc.bnl.gov/nsr/nsrlink.jsp?1971It02,B}{1971It02}). See also T\ensuremath{_{\textnormal{1/2}}}=2.84 ps\vspace{0.1cm}}&\\
&&&&&&&&\parbox[t][0.3cm]{8.6915cm}{\raggedright {\ }{\ }{\ }\textit{+243{\textminus}97} from \ensuremath{\tau}=4.1 ps \textit{+35{\textminus}14} (\href{https://www.nndc.bnl.gov/nsr/nsrlink.jsp?1970Gi09,B}{1970Gi09}). Those authors\vspace{0.1cm}}&\\
&&&&&&&&\parbox[t][0.3cm]{8.6915cm}{\raggedright {\ }{\ }{\ }reported \ensuremath{\Gamma}=0.17 meV \textit{8}.\vspace{0.1cm}}&\\
\multicolumn{1}{r@{}}{1536}&\multicolumn{1}{@{.}l}{5}&\multicolumn{1}{l}{(3/2\ensuremath{^{+}})\ensuremath{^{{\hyperlink{NE24LEVEL2}{c}}}}}&\multicolumn{1}{r@{}}{19}&\multicolumn{1}{@{}l}{\ensuremath{^{{\hyperlink{NE24LEVEL1}{b}}}} fs {\it 10}}&\multicolumn{1}{l}{(2)}&\multicolumn{1}{r@{}}{1}&\multicolumn{1}{@{.}l}{62}&\parbox[t][0.3cm]{8.6915cm}{\raggedright E(level): From a least-squares fit to the measured E\ensuremath{_{\ensuremath{\gamma}}}\vspace{0.1cm}}&\\
&&&&&&&&\parbox[t][0.3cm]{8.6915cm}{\raggedright {\ }{\ }{\ }(\href{https://www.nndc.bnl.gov/nsr/nsrlink.jsp?1970Gi09,B}{1970Gi09}). See also 1.55 MeV (\href{https://www.nndc.bnl.gov/nsr/nsrlink.jsp?1981Ov01,B}{1981Ov01}), where it is\vspace{0.1cm}}&\\
&&&&&&&&\parbox[t][0.3cm]{8.6915cm}{\raggedright {\ }{\ }{\ }likely that this state contains unresolved contributions from\vspace{0.1cm}}&\\
&&&&&&&&\parbox[t][0.3cm]{8.6915cm}{\raggedright {\ }{\ }{\ }other states.\vspace{0.1cm}}&\\
&&&&&&&&\parbox[t][0.3cm]{8.6915cm}{\raggedright T\ensuremath{_{1/2}}: From \ensuremath{\tau}=0.028 ps \textit{15} (\href{https://www.nndc.bnl.gov/nsr/nsrlink.jsp?1970Gi09,B}{1970Gi09}) which leads to\vspace{0.1cm}}&\\
&&&&&&&&\parbox[t][0.3cm]{8.6915cm}{\raggedright {\ }{\ }{\ }T\ensuremath{_{\textnormal{1/2}}}=19.4 fs \textit{104}. (\href{https://www.nndc.bnl.gov/nsr/nsrlink.jsp?1970Gi09,B}{1970Gi09}) reported \ensuremath{\Gamma}=24 meV \textit{+27{\textminus}8}.\vspace{0.1cm}}&\\
&&&&&&&&\parbox[t][0.3cm]{8.6915cm}{\raggedright J\ensuremath{^{\pi}}: The DWBA fit does not describe the data very well (see\vspace{0.1cm}}&\\
&&&&&&&&\parbox[t][0.3cm]{8.6915cm}{\raggedright {\ }{\ }{\ }Fig. 4 in \href{https://www.nndc.bnl.gov/nsr/nsrlink.jsp?1981Ov01,B}{1981Ov01}).\vspace{0.1cm}}&\\
&&&&&&&&\parbox[t][0.3cm]{8.6915cm}{\raggedright L: From the zero-range DWBA analysis of (\href{https://www.nndc.bnl.gov/nsr/nsrlink.jsp?1981Ov01,B}{1981Ov01}: See\vspace{0.1cm}}&\\
&&&&&&&&\parbox[t][0.3cm]{8.6915cm}{\raggedright {\ }{\ }{\ }text).\vspace{0.1cm}}&\\
&&&&&&&&\parbox[t][0.3cm]{8.6915cm}{\raggedright C=141 (\href{https://www.nndc.bnl.gov/nsr/nsrlink.jsp?1981Ov01,B}{1981Ov01}): The normalization constant.\vspace{0.1cm}}&\\
\multicolumn{1}{r@{}}{1614}&\multicolumn{1}{@{.}l}{1}&&\multicolumn{1}{r@{}}{125}&\multicolumn{1}{@{}l}{\ensuremath{^{{\hyperlink{NE24LEVEL1}{b}}}} fs {\it 42}}&&&&\parbox[t][0.3cm]{8.6915cm}{\raggedright E(level): From a least-squares fit to the measured E\ensuremath{_{\ensuremath{\gamma}}}\vspace{0.1cm}}&\\
&&&&&&&&\parbox[t][0.3cm]{8.6915cm}{\raggedright {\ }{\ }{\ }(\href{https://www.nndc.bnl.gov/nsr/nsrlink.jsp?1970Gi09,B}{1970Gi09}).\vspace{0.1cm}}&\\
&&&&&&&&\parbox[t][0.3cm]{8.6915cm}{\raggedright T\ensuremath{_{1/2}}: From \ensuremath{\tau}=0.18 ps \textit{6} (\href{https://www.nndc.bnl.gov/nsr/nsrlink.jsp?1970Gi09,B}{1970Gi09}). They reported \ensuremath{\Gamma}=3.7\vspace{0.1cm}}&\\
&&&&&&&&\parbox[t][0.3cm]{8.6915cm}{\raggedright {\ }{\ }{\ }meV \textit{+18{\textminus}9}.\vspace{0.1cm}}&\\
\multicolumn{1}{r@{}}{2.78\ensuremath{\times10^{3}}}&\multicolumn{1}{@{}l}{}&\multicolumn{1}{l}{(9/2\ensuremath{^{+}})\ensuremath{^{{\hyperlink{NE24LEVEL2}{c}}}}}&&&\multicolumn{1}{l}{(4)}&\multicolumn{1}{r@{}}{1}&\multicolumn{1}{@{.}l}{0}&\parbox[t][0.3cm]{8.6915cm}{\raggedright E(level),J\ensuremath{^{\pi}},L,S\ensuremath{_{\textnormal{rel}}}: From (\href{https://www.nndc.bnl.gov/nsr/nsrlink.jsp?1981Ov01,B}{1981Ov01}).\vspace{0.1cm}}&\\
&&&&&&&&\parbox[t][0.3cm]{8.6915cm}{\raggedright L: From the zero-range DWBA analysis of (\href{https://www.nndc.bnl.gov/nsr/nsrlink.jsp?1981Ov01,B}{1981Ov01}): See\vspace{0.1cm}}&\\
&&&&&&&&\parbox[t][0.3cm]{8.6915cm}{\raggedright {\ }{\ }{\ }text.\vspace{0.1cm}}&\\
&&&&&&&&\parbox[t][0.3cm]{8.6915cm}{\raggedright C=87 (\href{https://www.nndc.bnl.gov/nsr/nsrlink.jsp?1981Ov01,B}{1981Ov01}): The normalization constant.\vspace{0.1cm}}&\\
\multicolumn{1}{r@{}}{4.20\ensuremath{\times10^{3}}}&\multicolumn{1}{@{}l}{}&&&&&&&\parbox[t][0.3cm]{8.6915cm}{\raggedright E(level): From (\href{https://www.nndc.bnl.gov/nsr/nsrlink.jsp?1981Ov01,B}{1981Ov01}).\vspace{0.1cm}}&\\
\multicolumn{1}{r@{}}{4.63\ensuremath{\times10^{3}}}&\multicolumn{1}{@{}l}{}&\multicolumn{1}{l}{(13/2\ensuremath{^{+}})\ensuremath{^{{\hyperlink{NE24LEVEL2}{c}}}}}&&&&\multicolumn{1}{r@{}}{2}&\multicolumn{1}{@{.}l}{92}&\parbox[t][0.3cm]{8.6915cm}{\raggedright E(level),J\ensuremath{^{\pi}},S\ensuremath{_{\textnormal{rel}}}: From (\href{https://www.nndc.bnl.gov/nsr/nsrlink.jsp?1981Ov01,B}{1981Ov01}).\vspace{0.1cm}}&\\
&&&&&&&&\parbox[t][0.3cm]{8.6915cm}{\raggedright L: (\href{https://www.nndc.bnl.gov/nsr/nsrlink.jsp?1981Ov01,B}{1981Ov01}) did not provide L; however, the authors discuss\vspace{0.1cm}}&\\
&&&&&&&&\parbox[t][0.3cm]{8.6915cm}{\raggedright {\ }{\ }{\ }that the \ensuremath{^{\textnormal{16}}}O(\ensuremath{\alpha},n) reaction can be described by a \ensuremath{^{\textnormal{3}}}He-like\vspace{0.1cm}}&\\
&&&&&&&&\parbox[t][0.3cm]{8.6915cm}{\raggedright {\ }{\ }{\ }transfer and that \ensuremath{^{\textnormal{3}}}He would be in a (N,s,J)=(0,0,1/2) state.\vspace{0.1cm}}&\\
&&&&&&&&\parbox[t][0.3cm]{8.6915cm}{\raggedright {\ }{\ }{\ }Considering that the \ensuremath{^{\textnormal{16}}}O\ensuremath{_{\textnormal{g.s.}}} has a J\ensuremath{^{\ensuremath{\pi}}}=0\ensuremath{^{\textnormal{+}}} assignment, this\vspace{0.1cm}}&\\
&&&&&&&&\parbox[t][0.3cm]{8.6915cm}{\raggedright {\ }{\ }{\ }uniquely determines L=6.\vspace{0.1cm}}&\\
&&&&&&&&\parbox[t][0.3cm]{8.6915cm}{\raggedright A 10\% admixture of (\textit{sd})(\textit{fp})\ensuremath{^{\textnormal{2}}} was necessary to obtain\vspace{0.1cm}}&\\
&&&&&&&&\parbox[t][0.3cm]{8.6915cm}{\raggedright {\ }{\ }{\ }quantitative agreement between the calculated and\vspace{0.1cm}}&\\
&&&&&&&&\parbox[t][0.3cm]{8.6915cm}{\raggedright {\ }{\ }{\ }experimental spectroscopic factors (\href{https://www.nndc.bnl.gov/nsr/nsrlink.jsp?1981Ov01,B}{1981Ov01}).\vspace{0.1cm}}&\\
&&&&&&&&\parbox[t][0.3cm]{8.6915cm}{\raggedright C=254 (\href{https://www.nndc.bnl.gov/nsr/nsrlink.jsp?1981Ov01,B}{1981Ov01}): The normalization constant.\vspace{0.1cm}}&\\
\multicolumn{1}{r@{}}{5.43\ensuremath{\times10^{3}}}&\multicolumn{1}{@{}l}{}&\multicolumn{1}{l}{(7/2\ensuremath{^{+}})\ensuremath{^{{\hyperlink{NE24LEVEL2}{c}}}}}&&&\multicolumn{1}{l}{(4)}&\multicolumn{1}{r@{}}{1}&\multicolumn{1}{@{.}l}{63}&\parbox[t][0.3cm]{8.6915cm}{\raggedright E(level): From (\href{https://www.nndc.bnl.gov/nsr/nsrlink.jsp?1981Ov01,B}{1981Ov01}), where it is likely that this state\vspace{0.1cm}}&\\
&&&&&&&&\parbox[t][0.3cm]{8.6915cm}{\raggedright {\ }{\ }{\ }contains unresolved contributions from other states.\vspace{0.1cm}}&\\
&&&&&&&&\parbox[t][0.3cm]{8.6915cm}{\raggedright J\ensuremath{^{\pi}},L,S\ensuremath{_{\textnormal{rel}}}: From (\href{https://www.nndc.bnl.gov/nsr/nsrlink.jsp?1981Ov01,B}{1981Ov01}).\vspace{0.1cm}}&\\
&&&&&&&&\parbox[t][0.3cm]{8.6915cm}{\raggedright L: From the zero-range DWBA analysis of (\href{https://www.nndc.bnl.gov/nsr/nsrlink.jsp?1981Ov01,B}{1981Ov01}): See\vspace{0.1cm}}&\\
&&&&&&&&\parbox[t][0.3cm]{8.6915cm}{\raggedright {\ }{\ }{\ }text.\vspace{0.1cm}}&\\
&&&&&&&&\parbox[t][0.3cm]{8.6915cm}{\raggedright C=142 (\href{https://www.nndc.bnl.gov/nsr/nsrlink.jsp?1981Ov01,B}{1981Ov01}): The normalization constant.\vspace{0.1cm}}&\\
\multicolumn{1}{r@{}}{6.2\ensuremath{\times10^{3}}}&\multicolumn{1}{@{}l}{\ensuremath{^{{\hyperlink{NE24LEVEL5}{f}}}}}&&&&&&&&\\
\multicolumn{1}{r@{}}{6.80\ensuremath{\times10^{3}}}&\multicolumn{1}{@{}l}{\ensuremath{^{{\hyperlink{NE24LEVEL5}{f}}}}}&&&&&&&&\\
\multicolumn{1}{r@{}}{7.61\ensuremath{\times10^{3}}}&\multicolumn{1}{@{}l}{\ensuremath{^{{\hyperlink{NE24LEVEL5}{f}}}}}&&&&&&&&\\
\multicolumn{1}{r@{}}{8.42\ensuremath{\times10^{3}}}&\multicolumn{1}{@{}l}{\ensuremath{^{{\hyperlink{NE24LEVEL5}{f}}}}}&&&&&&&&\\
\multicolumn{1}{r@{}}{8.95\ensuremath{\times10^{3}}}&\multicolumn{1}{@{}l}{\ensuremath{^{{\hyperlink{NE24LEVEL5}{f}}}}}&&&&&&&&\\
\multicolumn{1}{r@{}}{9.23\ensuremath{\times10^{3}}}&\multicolumn{1}{@{}l}{\ensuremath{^{{\hyperlink{NE24LEVEL5}{f}}}}}&&&&&&&&\\
\multicolumn{1}{r@{}}{9.88\ensuremath{\times10^{3}}}&\multicolumn{1}{@{}l}{\ensuremath{^{{\hyperlink{NE24LEVEL5}{f}}}}}&&&&&&&&\\
\multicolumn{1}{r@{}}{10.40\ensuremath{\times10^{3}}}&\multicolumn{1}{@{}l}{\ensuremath{^{{\hyperlink{NE24LEVEL5}{f}}}}}&&&&&&&&\\
\multicolumn{1}{r@{}}{11.09\ensuremath{\times10^{3}}}&\multicolumn{1}{@{}l}{\ensuremath{^{{\hyperlink{NE24LEVEL5}{f}}}}}&&&&&&&&\\
\multicolumn{1}{r@{}}{12.49\ensuremath{\times10^{3}}}&\multicolumn{1}{@{}l}{\ensuremath{^{{\hyperlink{NE24LEVEL5}{f}}}}}&&&&&&&&\\
\end{longtable}
\begin{textblock}{29}(0,27.3)
Continued on next page (footnotes at end of table)
\end{textblock}
\clearpage
\vspace*{-0.5cm}
{\bf \small \underline{\ensuremath{^{\textnormal{16}}}O(\ensuremath{\alpha},n),(\ensuremath{\alpha},n\ensuremath{\gamma})\hspace{0.2in}\href{https://www.nndc.bnl.gov/nsr/nsrlink.jsp?1970Gi09,B}{1970Gi09},\href{https://www.nndc.bnl.gov/nsr/nsrlink.jsp?1981Ov01,B}{1981Ov01} (continued)}}\\
\vspace{0.3cm}
\underline{$^{19}$Ne Levels (continued)}\\
\vspace{0.3cm}
\parbox[b][0.3cm]{17.7cm}{\makebox[1ex]{\ensuremath{^{\hypertarget{NE24LEVEL0}{a}}}} From (\href{https://www.nndc.bnl.gov/nsr/nsrlink.jsp?1981Ov01,B}{1981Ov01}), where many of the levels for E\ensuremath{_{\textnormal{x}}}=1.55-12.49 MeV are unresolved states.}\\
\parbox[b][0.3cm]{17.7cm}{\makebox[1ex]{\ensuremath{^{\hypertarget{NE24LEVEL1}{b}}}} The uncertainties on the lifetimes deduced by (\href{https://www.nndc.bnl.gov/nsr/nsrlink.jsp?1970Gi09,B}{1970Gi09}) are combinations of systematic and statistical uncertainties.}\\
\parbox[b][0.3cm]{17.7cm}{\makebox[1ex]{\ensuremath{^{\hypertarget{NE24LEVEL2}{c}}}} From zero-range DWBA analysis using DWUCK4 by (\href{https://www.nndc.bnl.gov/nsr/nsrlink.jsp?1981Ov01,B}{1981Ov01}). They reported that the neutron angular distributions for the}\\
\parbox[b][0.3cm]{17.7cm}{{\ }{\ }J\ensuremath{^{\ensuremath{\pi}}}\ensuremath{\leq}9/2\ensuremath{^{\textnormal{+}}} members of the ground-state rotational band in \ensuremath{^{\textnormal{19}}}Ne exhibited a J-dependence supported by the DWBA predictions,}\\
\parbox[b][0.3cm]{17.7cm}{{\ }{\ }where the magnitudes of the theoretical distributions were more sensitive to the form factor parameters for the transitions with}\\
\parbox[b][0.3cm]{17.7cm}{{\ }{\ }J=L{\textminus}S. Thus, the shapes of the angular distributions for the states with J\ensuremath{^{\ensuremath{\pi}}}\ensuremath{\leq}9/2\ensuremath{^{\textnormal{+}}} at this E\ensuremath{_{\ensuremath{\alpha}}} energy are characterized by the}\\
\parbox[b][0.3cm]{17.7cm}{{\ }{\ }transferred J rather than the transferred L. The \ensuremath{^{\textnormal{16}}}O(\ensuremath{\alpha},n) reaction can be well described as the simple direct transfer of a \ensuremath{^{\textnormal{3}}}He-like}\\
\parbox[b][0.3cm]{17.7cm}{{\ }{\ }cluster with internal quantum numbers L=0 and s=1/2.}\\
\parbox[b][0.3cm]{17.7cm}{\makebox[1ex]{\ensuremath{^{\hypertarget{NE24LEVEL3}{d}}}} From (\href{https://www.nndc.bnl.gov/nsr/nsrlink.jsp?1981Ov01,B}{1981Ov01}): The relative spectroscopic factors are obtained from the zero-range DWBA analysis and are defined as the}\\
\parbox[b][0.3cm]{17.7cm}{{\ }{\ }normalization constant C relative to the C=87 deduced by (\href{https://www.nndc.bnl.gov/nsr/nsrlink.jsp?1981Ov01,B}{1981Ov01}) for the \ensuremath{^{\textnormal{19}}}Ne*(2.79 MeV, 9/2\ensuremath{^{\textnormal{+}}}) state.}\\
\parbox[b][0.3cm]{17.7cm}{\makebox[1ex]{\ensuremath{^{\hypertarget{NE24LEVEL4}{e}}}} For the \ensuremath{^{\textnormal{19}}}Ne\ensuremath{_{\textnormal{g.s.}}} and the \ensuremath{^{\textnormal{19}}}Ne*(238) level based on a normalization constant C=72 obtained for those states (\href{https://www.nndc.bnl.gov/nsr/nsrlink.jsp?1981Ov01,B}{1981Ov01}).}\\
\parbox[b][0.3cm]{17.7cm}{\makebox[1ex]{\ensuremath{^{\hypertarget{NE24LEVEL5}{f}}}} From (\href{https://www.nndc.bnl.gov/nsr/nsrlink.jsp?1981Ov01,B}{1981Ov01}).}\\
\vspace{0.5cm}
\underline{$\gamma$($^{19}$Ne)}\\
\begin{longtable}{ccccccc@{}ccccc@{\extracolsep{\fill}}c}
\multicolumn{2}{c}{E\ensuremath{_{\gamma}}}&\multicolumn{2}{c}{E\ensuremath{_{i}}(level)}&J\ensuremath{^{\pi}_{i}}&\multicolumn{2}{c}{E\ensuremath{_{f}}}&J\ensuremath{^{\pi}_{f}}&Mult.&\multicolumn{2}{c}{\ensuremath{\alpha}\ensuremath{^{\hyperlink{NE24GAMMA0}{a}}}}&Comments&\\[-.2cm]
\multicolumn{2}{c}{\hrulefill}&\multicolumn{2}{c}{\hrulefill}&\hrulefill&\multicolumn{2}{c}{\hrulefill}&\hrulefill&\hrulefill&\multicolumn{2}{c}{\hrulefill}&\hrulefill&
\endfirsthead
\multicolumn{1}{r@{}}{238}&\multicolumn{1}{@{.}l}{5}&\multicolumn{1}{r@{}}{238}&\multicolumn{1}{@{.}l}{5}&\multicolumn{1}{l}{(5/2\ensuremath{^{+}})}&\multicolumn{1}{r@{}}{0}&\multicolumn{1}{@{}l}{}&\multicolumn{1}{@{}l}{1/2\ensuremath{^{+}}}&\multicolumn{1}{l}{E2}&\multicolumn{1}{r@{}}{1}&\multicolumn{1}{@{.}l}{42\ensuremath{\times10^{-3}} {\it 2}}&\parbox[t][0.3cm]{8.094221cm}{\raggedright B(E2)(W.u.)=13.5 \textit{+17{\textminus}14}\vspace{0.1cm}}&\\
&&&&&&&&&&&\parbox[t][0.3cm]{8.094221cm}{\raggedright \ensuremath{\alpha}(K)=0.001343 \textit{19}; \ensuremath{\alpha}(L)=7.44\ensuremath{\times}10\ensuremath{^{\textnormal{$-$5}}} \textit{10}\vspace{0.1cm}}&\\
&&&&&&&&&&&\parbox[t][0.3cm]{8.094221cm}{\raggedright E\ensuremath{_{\gamma}}: From Figs. 4, 6, 12, 15, and 17 of (\href{https://www.nndc.bnl.gov/nsr/nsrlink.jsp?1969Ya05,B}{1969Ya05}). Note\vspace{0.1cm}}&\\
&&&&&&&&&&&\parbox[t][0.3cm]{8.094221cm}{\raggedright {\ }{\ }{\ }that Fig. 8 shows this \ensuremath{\gamma} ray with a quoted energy of\vspace{0.1cm}}&\\
&&&&&&&&&&&\parbox[t][0.3cm]{8.094221cm}{\raggedright {\ }{\ }{\ }238.6 keV. See also E\ensuremath{_{\ensuremath{\gamma}}}=238 keV (\href{https://www.nndc.bnl.gov/nsr/nsrlink.jsp?1983Pi07,B}{1983Pi07}).\vspace{0.1cm}}&\\
&&&&&&&&&&&\parbox[t][0.3cm]{8.094221cm}{\raggedright Mult.: From Fig. 3 of (\href{https://www.nndc.bnl.gov/nsr/nsrlink.jsp?1969Ya05,B}{1969Ya05}), where it is not clear if\vspace{0.1cm}}&\\
&&&&&&&&&&&\parbox[t][0.3cm]{8.094221cm}{\raggedright {\ }{\ }{\ }this value comes from the literature, or if those authors\vspace{0.1cm}}&\\
&&&&&&&&&&&\parbox[t][0.3cm]{8.094221cm}{\raggedright {\ }{\ }{\ }deduced it.\vspace{0.1cm}}&\\
\multicolumn{1}{r@{}}{273}&\multicolumn{1}{@{.}l}{9 {\it 7}}&\multicolumn{1}{r@{}}{273}&\multicolumn{1}{@{.}l}{9}&&\multicolumn{1}{r@{}}{0}&\multicolumn{1}{@{}l}{}&\multicolumn{1}{@{}l}{1/2\ensuremath{^{+}}}&&\multicolumn{1}{r@{}}{}&\multicolumn{1}{@{}l}{}&\parbox[t][0.3cm]{8.094221cm}{\raggedright E\ensuremath{_{\gamma}}: From (\href{https://www.nndc.bnl.gov/nsr/nsrlink.jsp?1971It02,B}{1971It02}).\vspace{0.1cm}}&\\
\multicolumn{1}{r@{}}{1232}&\multicolumn{1}{@{.}l}{2 {\it 2}}&\multicolumn{1}{r@{}}{1506}&\multicolumn{1}{@{.}l}{1}&&\multicolumn{1}{r@{}}{273}&\multicolumn{1}{@{.}l}{9 }&&&\multicolumn{1}{r@{}}{}&\multicolumn{1}{@{}l}{}&\parbox[t][0.3cm]{8.094221cm}{\raggedright E\ensuremath{_{\gamma}}: From (\href{https://www.nndc.bnl.gov/nsr/nsrlink.jsp?1971It02,B}{1971It02}).\vspace{0.1cm}}&\\
&&&&&&&&&&&\parbox[t][0.3cm]{8.094221cm}{\raggedright The observed Doppler shift for this \ensuremath{\gamma} ray was \ensuremath{\Delta}E\ensuremath{_{\ensuremath{\gamma}}}=7.70\vspace{0.1cm}}&\\
&&&&&&&&&&&\parbox[t][0.3cm]{8.094221cm}{\raggedright {\ }{\ }{\ }keV \textit{79} (\href{https://www.nndc.bnl.gov/nsr/nsrlink.jsp?1971It02,B}{1971It02}).\vspace{0.1cm}}&\\
&&&&&&&&&&&\parbox[t][0.3cm]{8.094221cm}{\raggedright F(\ensuremath{\tau})=0.37 \textit{4} measured at E\ensuremath{_{\ensuremath{\alpha}}}=20.6 MeV (\href{https://www.nndc.bnl.gov/nsr/nsrlink.jsp?1971It02,B}{1971It02}). See\vspace{0.1cm}}&\\
&&&&&&&&&&&\parbox[t][0.3cm]{8.094221cm}{\raggedright {\ }{\ }{\ }also F(\ensuremath{\tau})=0.17 \textit{9}: The unweighted average of 0.08 \textit{5} and\vspace{0.1cm}}&\\
&&&&&&&&&&&\parbox[t][0.3cm]{8.094221cm}{\raggedright {\ }{\ }{\ }0.26 \textit{2}, both of which are measured by (\href{https://www.nndc.bnl.gov/nsr/nsrlink.jsp?1970Gi09,B}{1970Gi09}) at\vspace{0.1cm}}&\\
&&&&&&&&&&&\parbox[t][0.3cm]{8.094221cm}{\raggedright {\ }{\ }{\ }E\ensuremath{_{\ensuremath{\alpha}}}=19 MeV.\vspace{0.1cm}}&\\
\multicolumn{1}{r@{}}{1298}&\multicolumn{1}{@{.}l}{0}&\multicolumn{1}{r@{}}{1536}&\multicolumn{1}{@{.}l}{5}&\multicolumn{1}{l}{(3/2\ensuremath{^{+}})}&\multicolumn{1}{r@{}}{238}&\multicolumn{1}{@{.}l}{5}&\multicolumn{1}{@{}l}{(5/2\ensuremath{^{+}})}&&\multicolumn{1}{r@{}}{}&\multicolumn{1}{@{}l}{}&\parbox[t][0.3cm]{8.094221cm}{\raggedright E\ensuremath{_{\gamma}}: From the \ensuremath{^{\textnormal{19}}}F(p,n\ensuremath{\gamma}) measurement by (\href{https://www.nndc.bnl.gov/nsr/nsrlink.jsp?1970Gi09,B}{1970Gi09}).\vspace{0.1cm}}&\\
&&&&&&&&&&&\parbox[t][0.3cm]{8.094221cm}{\raggedright F(\ensuremath{\tau})=0.91 \textit{5}: The weighted average of 0.89 \textit{10} and 0.92 \textit{5},\vspace{0.1cm}}&\\
&&&&&&&&&&&\parbox[t][0.3cm]{8.094221cm}{\raggedright {\ }{\ }{\ }both of which are measured by (\href{https://www.nndc.bnl.gov/nsr/nsrlink.jsp?1970Gi09,B}{1970Gi09}) at E\ensuremath{_{\ensuremath{\alpha}}}=19\vspace{0.1cm}}&\\
&&&&&&&&&&&\parbox[t][0.3cm]{8.094221cm}{\raggedright {\ }{\ }{\ }MeV.\vspace{0.1cm}}&\\
\multicolumn{1}{r@{}}{1340}&\multicolumn{1}{@{.}l}{1}&\multicolumn{1}{r@{}}{1614}&\multicolumn{1}{@{.}l}{1}&&\multicolumn{1}{r@{}}{273}&\multicolumn{1}{@{.}l}{9 }&&&\multicolumn{1}{r@{}}{}&\multicolumn{1}{@{}l}{}&\parbox[t][0.3cm]{8.094221cm}{\raggedright E\ensuremath{_{\gamma}}: From the \ensuremath{^{\textnormal{19}}}F(p,n\ensuremath{\gamma}) measurement by (\href{https://www.nndc.bnl.gov/nsr/nsrlink.jsp?1970Gi09,B}{1970Gi09}).\vspace{0.1cm}}&\\
&&&&&&&&&&&\parbox[t][0.3cm]{8.094221cm}{\raggedright F(\ensuremath{\tau})=0.69 \textit{11}: The unweighted average of 0.58 \textit{6} and 0.80\vspace{0.1cm}}&\\
&&&&&&&&&&&\parbox[t][0.3cm]{8.094221cm}{\raggedright {\ }{\ }{\ }\textit{5}, both of which are measured by (\href{https://www.nndc.bnl.gov/nsr/nsrlink.jsp?1970Gi09,B}{1970Gi09}) at E\ensuremath{_{\ensuremath{\alpha}}}=19\vspace{0.1cm}}&\\
&&&&&&&&&&&\parbox[t][0.3cm]{8.094221cm}{\raggedright {\ }{\ }{\ }MeV.\vspace{0.1cm}}&\\
\end{longtable}
\parbox[b][0.3cm]{17.7cm}{\makebox[1ex]{\ensuremath{^{\hypertarget{NE24GAMMA0}{a}}}} Total theoretical internal conversion coefficients, calculated using the BrIcc code (\href{https://www.nndc.bnl.gov/nsr/nsrlink.jsp?2008Ki07,B}{2008Ki07}) with ``Frozen Orbitals''}\\
\parbox[b][0.3cm]{17.7cm}{{\ }{\ }approximation based on \ensuremath{\gamma}-ray energies, assigned multipolarities, and mixing ratios, unless otherwise specified.}\\
\vspace{0.5cm}
\clearpage
\begin{figure}[h]
\begin{center}
\includegraphics{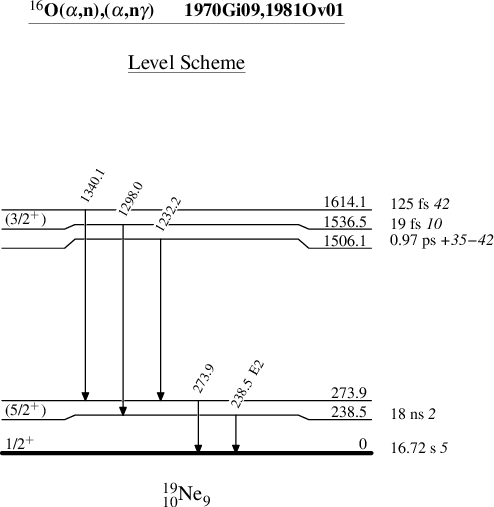}\\
\end{center}
\end{figure}
\clearpage
\subsection[\hspace{-0.2cm}\ensuremath{^{\textnormal{16}}}O(\ensuremath{^{\textnormal{6}}}Li,t)]{ }
\vspace{-27pt}
\vspace{0.3cm}
\hypertarget{NE25}{{\bf \small \underline{\ensuremath{^{\textnormal{16}}}O(\ensuremath{^{\textnormal{6}}}Li,t)\hspace{0.2in}\href{https://www.nndc.bnl.gov/nsr/nsrlink.jsp?1972Ga08,B}{1972Ga08},\href{https://www.nndc.bnl.gov/nsr/nsrlink.jsp?1979Ma26,B}{1979Ma26},\href{https://www.nndc.bnl.gov/nsr/nsrlink.jsp?2023Ma57,B}{2023Ma57}}}}\\
\vspace{4pt}
\vspace{8pt}
\parbox[b][0.3cm]{17.7cm}{\addtolength{\parindent}{-0.2in}\ensuremath{^{\textnormal{3}}}He pick-up reaction.}\\
\parbox[b][0.3cm]{17.7cm}{\addtolength{\parindent}{-0.2in}J\ensuremath{^{\ensuremath{\pi}}}(\ensuremath{^{\textnormal{16}}}O\ensuremath{_{\textnormal{g.s.}}})=0\ensuremath{^{\textnormal{+}}} and J\ensuremath{^{\ensuremath{\pi}}}(\ensuremath{^{\textnormal{6}}}Li\ensuremath{_{\textnormal{g.s.}}})=1\ensuremath{^{\textnormal{+}}}.}\\
\parbox[b][0.3cm]{17.7cm}{\addtolength{\parindent}{-0.2in}\href{https://www.nndc.bnl.gov/nsr/nsrlink.jsp?1971Bi06,B}{1971Bi06}, \href{https://www.nndc.bnl.gov/nsr/nsrlink.jsp?1972Ga08,B}{1972Ga08}: \ensuremath{^{\textnormal{16}}}O(\ensuremath{^{\textnormal{6}}}Li,t) and \ensuremath{^{\textnormal{16}}}O(\ensuremath{^{\textnormal{6}}}Li,\ensuremath{^{\textnormal{3}}}He) E=24 MeV; momentum analyzed the reaction products using a magnetic}\\
\parbox[b][0.3cm]{17.7cm}{spectrograph with a nuclear emulsion plate at its focal plane. Measured triton angular distributions \ensuremath{\sigma}(E\ensuremath{_{\textnormal{t}}}) at \ensuremath{\theta}\ensuremath{_{\textnormal{lab}}}=7.5\ensuremath{^\circ}{\textminus}82.5\ensuremath{^\circ}}\\
\parbox[b][0.3cm]{17.7cm}{(\href{https://www.nndc.bnl.gov/nsr/nsrlink.jsp?1972Ga08,B}{1972Ga08}). Energy resolution was 35 keV (FWHM) in (\href{https://www.nndc.bnl.gov/nsr/nsrlink.jsp?1972Ga08,B}{1972Ga08}). (\href{https://www.nndc.bnl.gov/nsr/nsrlink.jsp?1971Bi06,B}{1971Bi06}) presents differential cross sections (with 15\%}\\
\parbox[b][0.3cm]{17.7cm}{uncertainty) for each observed \ensuremath{^{\textnormal{19}}}Ne level. (\href{https://www.nndc.bnl.gov/nsr/nsrlink.jsp?1972Ga08,B}{1972Ga08}) performed a zero-range DWBA analysis assuming a three-nucleon-cluster}\\
\parbox[b][0.3cm]{17.7cm}{transfer. Deduced \ensuremath{^{\textnormal{19}}}Ne level-energies, J\ensuremath{^{\ensuremath{\pi}}} assignments, isobaric analog states, which were studied at the same energy and via the}\\
\parbox[b][0.3cm]{17.7cm}{\ensuremath{^{\textnormal{16}}}O(\ensuremath{^{\textnormal{6}}}Li,\ensuremath{^{\textnormal{3}}}He) reaction. (\href{https://www.nndc.bnl.gov/nsr/nsrlink.jsp?1972Ga08,B}{1972Ga08}) deduced the spectroscopic strengths for the strong transitions observed.}\\
\parbox[b][0.3cm]{17.7cm}{\addtolength{\parindent}{-0.2in}\href{https://www.nndc.bnl.gov/nsr/nsrlink.jsp?1971GaZO,B}{1971GaZO}: \ensuremath{^{\textnormal{16}}}O(\ensuremath{^{\textnormal{6}}}Li,t) and \ensuremath{^{\textnormal{16}}}O(\ensuremath{^{\textnormal{6}}}Li,\ensuremath{^{\textnormal{3}}}He) E=24 MeV; measured \ensuremath{\sigma}(\ensuremath{\theta}); deduced \ensuremath{^{\textnormal{19}}}Ne relative spectroscopic factors.}\\
\parbox[b][0.3cm]{17.7cm}{\addtolength{\parindent}{-0.2in}\href{https://www.nndc.bnl.gov/nsr/nsrlink.jsp?1971GaZY,B}{1971GaZY}: \ensuremath{^{\textnormal{16}}}O(\ensuremath{^{\textnormal{6}}}Li,t) and \ensuremath{^{\textnormal{16}}}O(\ensuremath{^{\textnormal{6}}}Li,\ensuremath{^{\textnormal{3}}}He) E not given; measured \ensuremath{\sigma}(E\ensuremath{_{\textnormal{t}}}); deduced \ensuremath{^{\textnormal{19}}}Ne analog states.}\\
\parbox[b][0.3cm]{17.7cm}{\addtolength{\parindent}{-0.2in}\href{https://www.nndc.bnl.gov/nsr/nsrlink.jsp?1971GoZX,B}{1971GoZX}, H. E. Gove and A. D. Panagiotou, Bull. Am. Phys. Soc. 16 (1971) 490: \ensuremath{^{\textnormal{16}}}O(\ensuremath{^{\textnormal{6}}}Li,t) E=36 MeV; measured \ensuremath{\sigma}(E\ensuremath{_{\textnormal{t}}},\ensuremath{\theta});}\\
\parbox[b][0.3cm]{17.7cm}{deduced \ensuremath{^{\textnormal{19}}}Ne level-energies and J\ensuremath{^{\ensuremath{\pi}}} assignments.}\\
\parbox[b][0.3cm]{17.7cm}{\addtolength{\parindent}{-0.2in}\href{https://www.nndc.bnl.gov/nsr/nsrlink.jsp?1972Pa29,B}{1972Pa29}: \ensuremath{^{\textnormal{16}}}O(\ensuremath{^{\textnormal{6}}}Li,t) E=35, 36 MeV; momentum analyzed the reaction products using a split-pole spectrograph; measured the}\\
\parbox[b][0.3cm]{17.7cm}{excitation function \ensuremath{\sigma}(E\ensuremath{_{\textnormal{t}}},\ensuremath{\theta}) at \ensuremath{\theta}\ensuremath{_{\textnormal{lab}}}=6\ensuremath{^\circ}{\textminus}39\ensuremath{^\circ}; deduced \ensuremath{^{\textnormal{19}}}Ne level-energies, J\ensuremath{^{\ensuremath{\pi}}}, and L. Comparison with analog states in \ensuremath{^{\textnormal{19}}}F are}\\
\parbox[b][0.3cm]{17.7cm}{presented. Discussed K\ensuremath{^{\ensuremath{\pi}}}=1/2\ensuremath{^{\textnormal{+}}} and 1/2\ensuremath{^{-}} bands in \ensuremath{^{\textnormal{19}}}F and \ensuremath{^{\textnormal{19}}}Ne.}\\
\parbox[b][0.3cm]{17.7cm}{\addtolength{\parindent}{-0.2in}\href{https://www.nndc.bnl.gov/nsr/nsrlink.jsp?1973Bi02,B}{1973Bi02}: \ensuremath{^{\textnormal{16}}}O(\ensuremath{^{\textnormal{6}}}Li,t) E not given; analyzed additional data (not clear if the data were from (\href{https://www.nndc.bnl.gov/nsr/nsrlink.jsp?1971Bi06,B}{1971Bi06}, \href{https://www.nndc.bnl.gov/nsr/nsrlink.jsp?1972Ga08,B}{1972Ga08}) or new data)}\\
\parbox[b][0.3cm]{17.7cm}{around E\ensuremath{_{\textnormal{x}}}(\ensuremath{^{\textnormal{19}}}Ne)=4.6 MeV and confirmed the existence of the 4.59-MeV tentative state found in (\href{https://www.nndc.bnl.gov/nsr/nsrlink.jsp?1971Bi06,B}{1971Bi06}), whose centroid}\\
\parbox[b][0.3cm]{17.7cm}{relative to that of the \ensuremath{^{\textnormal{19}}}Ne*(4625) state remained constant at \ensuremath{\theta}\ensuremath{_{\textnormal{lab}}}=7.5\ensuremath{^\circ}{\textminus}30\ensuremath{^\circ}. Deduced E\ensuremath{_{\textnormal{x}}}=4593 keV \textit{6} for this state. Comparison}\\
\parbox[b][0.3cm]{17.7cm}{with the \ensuremath{^{\textnormal{19}}}Ne*(4605) level observed by (D. Dehnhard and H. Ohnuma, John H. Williams Laboratory, Annual Report, 1971}\\
\parbox[b][0.3cm]{17.7cm}{(unpublished), p. 46) further supported the existence of this state. Possible mirror correspondences are discussed.}\\
\parbox[b][0.3cm]{17.7cm}{\addtolength{\parindent}{-0.2in}\href{https://www.nndc.bnl.gov/nsr/nsrlink.jsp?1973Bi07,B}{1973Bi07}: \ensuremath{^{\textnormal{16}}}O(\ensuremath{^{\textnormal{6}}}Li,t) and \ensuremath{^{\textnormal{16}}}O(\ensuremath{^{\textnormal{6}}}Li,\ensuremath{^{\textnormal{6}}}Li) E=22-24.6 MeV in 100 keV steps; measured the elastic scattering using a surface barrier}\\
\parbox[b][0.3cm]{17.7cm}{detector at \ensuremath{\theta}\ensuremath{_{\textnormal{lab}}}=15\ensuremath{^\circ} and the transfer reaction at \ensuremath{\theta}\ensuremath{_{\textnormal{lab}}}=7.5\ensuremath{^\circ} using a \ensuremath{\Delta}E-E telescope that consisted of Si surface barrier detectors.}\\
\parbox[b][0.3cm]{17.7cm}{Measured \ensuremath{\sigma}(E\ensuremath{_{^{\textnormal{6}}\textnormal{Li}}}) and \ensuremath{\sigma}(E\ensuremath{_{\textnormal{t}}}) for the \ensuremath{^{\textnormal{19}}}Ne*(0, 0.24+0.27, 1.51+1.54+1.62, and 2.79 MeV) states (+ sign indicates unresolved}\\
\parbox[b][0.3cm]{17.7cm}{states). Performed DWBA analysis using the same optical potentials used in (\href{https://www.nndc.bnl.gov/nsr/nsrlink.jsp?1972Ga08,B}{1972Ga08}) and confirmed that the nature of the}\\
\parbox[b][0.3cm]{17.7cm}{\ensuremath{^{\textnormal{16}}}O(\ensuremath{^{\textnormal{6}}}Li,t) reaction at this energy is predominantly direct mechanism.}\\
\parbox[b][0.3cm]{17.7cm}{\addtolength{\parindent}{-0.2in}\href{https://www.nndc.bnl.gov/nsr/nsrlink.jsp?1976WoZX,B}{1976WoZX}: \ensuremath{^{\textnormal{16}}}O(\ensuremath{^{\textnormal{6}}}Li,t) E=34 MeV; measured \ensuremath{\sigma}(E\ensuremath{_{\textnormal{t}}}); deduced reaction Q-value.}\\
\parbox[b][0.3cm]{17.7cm}{\addtolength{\parindent}{-0.2in}\href{https://www.nndc.bnl.gov/nsr/nsrlink.jsp?1977MaZB,B}{1977MaZB}, \href{https://www.nndc.bnl.gov/nsr/nsrlink.jsp?1977MaZR,B}{1977MaZR}: \ensuremath{^{\textnormal{16}}}O(\ensuremath{^{\textnormal{6}}}Li,t) E=44 MeV; measured \ensuremath{\sigma}(\ensuremath{\theta}).}\\
\parbox[b][0.3cm]{17.7cm}{\addtolength{\parindent}{-0.2in}\href{https://www.nndc.bnl.gov/nsr/nsrlink.jsp?1979Ma26,B}{1979Ma26}: \ensuremath{^{\textnormal{16}}}O(\ensuremath{^{\textnormal{6}}}Li,t) E=46 MeV; measured tritons using a Si surface barrier \ensuremath{\Delta}E-E telescope at \ensuremath{\theta}\ensuremath{_{\textnormal{lab}}}=15\ensuremath{^\circ} with energy resolution}\\
\parbox[b][0.3cm]{17.7cm}{of \ensuremath{\Delta}E(FWHM)=150 keV. Measured \ensuremath{\sigma}(\ensuremath{\theta}). Deduced \ensuremath{^{\textnormal{19}}}Ne level-energies and suggested triton cluster states outside a closed-shell}\\
\parbox[b][0.3cm]{17.7cm}{\ensuremath{^{\textnormal{16}}}O\ensuremath{_{\textnormal{g.s.}}} core using a folded-potential model.}\\
\parbox[b][0.3cm]{17.7cm}{\addtolength{\parindent}{-0.2in}\href{https://www.nndc.bnl.gov/nsr/nsrlink.jsp?1979MaZO,B}{1979MaZO}: \ensuremath{^{\textnormal{16}}}O(\ensuremath{^{\textnormal{6}}}Li,t) E=44 MeV; measured \ensuremath{\sigma}(\ensuremath{\theta}); deduced \ensuremath{^{\textnormal{19}}}Ne level-energies and their cluster structure. Performed DWBA and}\\
\parbox[b][0.3cm]{17.7cm}{Hauser-Feshbach analyses, and SU(3) shell model calculations.}\\
\parbox[b][0.3cm]{17.7cm}{\addtolength{\parindent}{-0.2in}\href{https://www.nndc.bnl.gov/nsr/nsrlink.jsp?1992RoZZ,B}{1992RoZZ}: \ensuremath{^{\textnormal{16}}}O(\ensuremath{^{\textnormal{6}}}Li,t) E not given; deduced \ensuremath{^{\textnormal{19}}}Ne levels.}\\
\parbox[b][0.3cm]{17.7cm}{\addtolength{\parindent}{-0.2in}\href{https://www.nndc.bnl.gov/nsr/nsrlink.jsp?1998Ut02,B}{1998Ut02}: \ensuremath{^{\textnormal{16}}}O(\ensuremath{^{\textnormal{6}}}Li,t) E=28 and 32 MeV; measured reaction products using a Browne-Buechner spectrograph at \ensuremath{\theta}\ensuremath{_{\textnormal{lab}}}=7.4\ensuremath{^\circ} and 14\ensuremath{^\circ}}\\
\parbox[b][0.3cm]{17.7cm}{and a split-pole spectrograph at \ensuremath{\theta}\ensuremath{_{\textnormal{lab}}}=8\ensuremath{^\circ} and 11\ensuremath{^\circ} (in a separate experiment). Obtained \ensuremath{^{\textnormal{19}}}Ne level-energies.}\\
\parbox[b][0.3cm]{17.7cm}{\addtolength{\parindent}{-0.2in}\href{https://www.nndc.bnl.gov/nsr/nsrlink.jsp?2023Ma57,B}{2023Ma57}: \ensuremath{^{\textnormal{6}}}Li(\ensuremath{^{\textnormal{16}}}O,\ensuremath{^{\textnormal{19}}}Ne*) E=96 MeV (inverse kinematics); measured \ensuremath{^{\textnormal{3}}}H-p and \ensuremath{^{\textnormal{3}}}H-\ensuremath{\alpha} coincidence events, where protons and}\\
\parbox[b][0.3cm]{17.7cm}{\ensuremath{\alpha}-particles were produced from \ensuremath{^{\textnormal{19}}}Ne* decay, using 4 pairs of \ensuremath{\Delta}E-\ensuremath{\Delta}E-E telescopes at \ensuremath{\theta}\ensuremath{_{\textnormal{lab}}}=\ensuremath{\pm}24\ensuremath{^\circ} and \ensuremath{\theta}\ensuremath{_{\textnormal{lab}}}=\ensuremath{\pm}47\ensuremath{^\circ}; and 4 pairs of}\\
\parbox[b][0.3cm]{17.7cm}{\ensuremath{\Delta}E-E telescopes at \ensuremath{\theta}\ensuremath{_{\textnormal{lab}}}=\ensuremath{\pm}70\ensuremath{^\circ} and \ensuremath{\theta}\ensuremath{_{\textnormal{lab}}}=\ensuremath{\pm}110\ensuremath{^\circ}. Deduced \ensuremath{^{\textnormal{19}}}Ne excitation energies up to 10.23 MeV using invariant mass analysis}\\
\parbox[b][0.3cm]{17.7cm}{and from the detected protons and \ensuremath{\alpha} particles and the determined momenta and energies of \ensuremath{^{\textnormal{15}}}O and \ensuremath{^{\textnormal{18}}}F decay products. Deduced}\\
\parbox[b][0.3cm]{17.7cm}{\ensuremath{\Gamma}\ensuremath{_{\ensuremath{\alpha}}}/\ensuremath{\Gamma} and \ensuremath{\Gamma}\ensuremath{_{\textnormal{p}}}/\ensuremath{\Gamma}\ensuremath{_{\ensuremath{\alpha}}} for some of these states. The experimental resolution was 300-350 keV (FWHM).}\\
\vspace{0.385cm}
\parbox[b][0.3cm]{17.7cm}{\addtolength{\parindent}{-0.2in}\textit{Theory}:}\\
\parbox[b][0.3cm]{17.7cm}{\addtolength{\parindent}{-0.2in}\href{https://www.nndc.bnl.gov/nsr/nsrlink.jsp?1978Pi06,B}{1978Pi06}: Calculated three-nucleon cluster strengths for the \ensuremath{^{\textnormal{19}}}Ne*(275, 1508, 1616, 4140, 4197) states belonging to the K\ensuremath{^{\ensuremath{\pi}}}=1/2\ensuremath{^{-}}}\\
\parbox[b][0.3cm]{17.7cm}{rotational band, which have 4p-1h characteristics.}\\
\parbox[b][0.3cm]{17.7cm}{\addtolength{\parindent}{-0.2in}\href{https://www.nndc.bnl.gov/nsr/nsrlink.jsp?1983Cu02,B}{1983Cu02}: Using weak coupling hypothesis, the authors compared the 3p-2h states in \ensuremath{^{\textnormal{17}}}O, measured via \ensuremath{^{\textnormal{14}}}C(\ensuremath{^{\textnormal{6}}}Li,t) at 34 MeV by}\\
\parbox[b][0.3cm]{17.7cm}{(\href{https://www.nndc.bnl.gov/nsr/nsrlink.jsp?1981Cu11,B}{1981Cu11}), with the 3p-0h states, belonging to the K\ensuremath{^{\ensuremath{\pi}}}=1/2\ensuremath{^{\textnormal{+}}} rotational band, in \ensuremath{^{\textnormal{19}}}Ne which were measured via \ensuremath{^{\textnormal{16}}}O(\ensuremath{^{\textnormal{6}}}Li,t) at 36}\\
\parbox[b][0.3cm]{17.7cm}{MeV by (\href{https://www.nndc.bnl.gov/nsr/nsrlink.jsp?1972Pa29,B}{1972Pa29}).}\\
\vspace{12pt}
\clearpage
\vspace{0.3cm}
\vspace*{-0.5cm}
{\bf \small \underline{\ensuremath{^{\textnormal{16}}}O(\ensuremath{^{\textnormal{6}}}Li,t)\hspace{0.2in}\href{https://www.nndc.bnl.gov/nsr/nsrlink.jsp?1972Ga08,B}{1972Ga08},\href{https://www.nndc.bnl.gov/nsr/nsrlink.jsp?1979Ma26,B}{1979Ma26},\href{https://www.nndc.bnl.gov/nsr/nsrlink.jsp?2023Ma57,B}{2023Ma57} (continued)}}\\
\vspace{0.3cm}
\underline{$^{19}$Ne Levels}\\
\vspace{0.34cm}
\parbox[b][0.3cm]{17.7cm}{\addtolength{\parindent}{-0.254cm}\textit{Notes}:}\\
\parbox[b][0.3cm]{17.7cm}{\addtolength{\parindent}{-0.254cm}(1) The given \ensuremath{\sigma}\ensuremath{_{\textnormal{exp}}}/\ensuremath{\sigma}\ensuremath{_{\textnormal{DWBA}}} from (\href{https://www.nndc.bnl.gov/nsr/nsrlink.jsp?1972Ga08,B}{1972Ga08}) are from set 3 in that study (see Table III), which produced the best DWBA fits.}\\
\parbox[b][0.3cm]{17.7cm}{\addtolength{\parindent}{-0.254cm}(2) The uncertainty in each of the reported cross sections by (\href{https://www.nndc.bnl.gov/nsr/nsrlink.jsp?1979Ma26,B}{1979Ma26}) is \ensuremath{\pm}1\%{\textminus}4\% statistical and \ensuremath{\sim}\ensuremath{\pm}10\% systematic.}\\
\parbox[b][0.3cm]{17.7cm}{\addtolength{\parindent}{-0.254cm}(3) (\href{https://www.nndc.bnl.gov/nsr/nsrlink.jsp?2023Ma57,B}{2023Ma57}) deduced \ensuremath{\Gamma}\ensuremath{_{\ensuremath{\alpha}}}/\ensuremath{\Gamma} values from the measured \ensuremath{\Gamma}\ensuremath{_{\textnormal{p}}}/\ensuremath{\Gamma}\ensuremath{_{\ensuremath{\alpha}}} and by taking \ensuremath{\Gamma}\ensuremath{_{\textnormal{tot}}} from literature (a reference is given for}\\
\parbox[b][0.3cm]{17.7cm}{each state) while assuming that those states have \ensuremath{\Gamma}\ensuremath{_{\ensuremath{\gamma}}}=0 as suggested by the evaluation of (\href{https://www.nndc.bnl.gov/nsr/nsrlink.jsp?2007Ne09,B}{2007Ne09}) and by (\href{https://www.nndc.bnl.gov/nsr/nsrlink.jsp?2019Ha08,B}{2019Ha08}:}\\
\parbox[b][0.3cm]{17.7cm}{\ensuremath{^{\textnormal{19}}}F(\ensuremath{^{\textnormal{3}}}He,t\ensuremath{\gamma})).}\\
\parbox[b][0.3cm]{17.7cm}{\addtolength{\parindent}{-0.254cm}(4) In addition to the statistical uncertainties listed here for \ensuremath{\Gamma}\ensuremath{_{\ensuremath{\alpha}}}/\ensuremath{\Gamma} and \ensuremath{\Gamma}\ensuremath{_{\textnormal{p}}}/\ensuremath{\Gamma}\ensuremath{_{\ensuremath{\alpha}}} from (\href{https://www.nndc.bnl.gov/nsr/nsrlink.jsp?2023Ma57,B}{2023Ma57}), there is an additional 15\%}\\
\parbox[b][0.3cm]{17.7cm}{systematic uncertainty (not listed here) due to possible variation of the background function used for the fits in (\href{https://www.nndc.bnl.gov/nsr/nsrlink.jsp?2023Ma57,B}{2023Ma57}).}\\
\vspace{0.34cm}
\begin{longtable}{ccccc@{\extracolsep{\fill}}c}
\multicolumn{2}{c}{E(level)$^{{\hyperlink{NE25LEVEL2}{c}}{\hyperlink{NE25LEVEL3}{d}}{\hyperlink{NE25LEVEL4}{e}}}$}&J$^{\pi}$$^{}$&L$^{}$&Comments&\\[-.2cm]
\multicolumn{2}{c}{\hrulefill}&\hrulefill&\hrulefill&\hrulefill&
\endfirsthead
\multicolumn{1}{r@{}}{0}&\multicolumn{1}{@{}l}{\ensuremath{^{{\hyperlink{NE25LEVEL0}{a}}{\hyperlink{NE25LEVEL5}{f}}}}}&\multicolumn{1}{l}{1/2\ensuremath{^{+}}\ensuremath{^{{\hyperlink{NE25LEVEL9}{j}}}}}&\multicolumn{1}{l}{0$^{{\hyperlink{NE25LEVEL9}{j}}}$}&\parbox[t][0.3cm]{12.1857605cm}{\raggedright E(level): From (\href{https://www.nndc.bnl.gov/nsr/nsrlink.jsp?1971Bi06,B}{1971Bi06}, \href{https://www.nndc.bnl.gov/nsr/nsrlink.jsp?1972Ga08,B}{1972Ga08}, \href{https://www.nndc.bnl.gov/nsr/nsrlink.jsp?1972Pa29,B}{1972Pa29}, \href{https://www.nndc.bnl.gov/nsr/nsrlink.jsp?1973Bi07,B}{1973Bi07}).\vspace{0.1cm}}&\\
&&&&\parbox[t][0.3cm]{12.1857605cm}{\raggedright J\ensuremath{^{\pi}}: From (\href{https://www.nndc.bnl.gov/nsr/nsrlink.jsp?1972Ga08,B}{1972Ga08}) based on zero-range DWBA analysis using DWUCK and assuming\vspace{0.1cm}}&\\
&&&&\parbox[t][0.3cm]{12.1857605cm}{\raggedright {\ }{\ }{\ }a three nucleon cluster transfer. L is not given by (\href{https://www.nndc.bnl.gov/nsr/nsrlink.jsp?1972Ga08,B}{1972Ga08}). See also J\ensuremath{^{\ensuremath{\pi}}}=1/2\ensuremath{^{\textnormal{+}}}\vspace{0.1cm}}&\\
&&&&\parbox[t][0.3cm]{12.1857605cm}{\raggedright {\ }{\ }{\ }(\href{https://www.nndc.bnl.gov/nsr/nsrlink.jsp?1972Pa29,B}{1972Pa29}).\vspace{0.1cm}}&\\
&&&&\parbox[t][0.3cm]{12.1857605cm}{\raggedright \ensuremath{\sigma}\ensuremath{_{\textnormal{exp}}}/\ensuremath{\sigma}\ensuremath{_{\textnormal{DWBA}}}=0.37 (\href{https://www.nndc.bnl.gov/nsr/nsrlink.jsp?1972Ga08,B}{1972Ga08}).\vspace{0.1cm}}&\\
\multicolumn{1}{r@{}}{241}&\multicolumn{1}{@{}l}{\ensuremath{^{{\hyperlink{NE25LEVEL0}{a}}{\hyperlink{NE25LEVEL5}{f}}}} {\it 40}}&\multicolumn{1}{l}{5/2\ensuremath{^{+}}\ensuremath{^{{\hyperlink{NE25LEVEL9}{j}}}}}&\multicolumn{1}{l}{2$^{{\hyperlink{NE25LEVEL9}{j}}}$}&\parbox[t][0.3cm]{12.1857605cm}{\raggedright E(level): From (\href{https://www.nndc.bnl.gov/nsr/nsrlink.jsp?1972Pa29,B}{1972Pa29}). See also 0.24 MeV (\href{https://www.nndc.bnl.gov/nsr/nsrlink.jsp?1971Bi06,B}{1971Bi06}, \href{https://www.nndc.bnl.gov/nsr/nsrlink.jsp?1972Ga08,B}{1972Ga08}, \href{https://www.nndc.bnl.gov/nsr/nsrlink.jsp?1973Bi07,B}{1973Bi07}:\vspace{0.1cm}}&\\
&&&&\parbox[t][0.3cm]{12.1857605cm}{\raggedright {\ }{\ }{\ }Unresolved from the 0.27 MeV state); and 0.23 MeV (\href{https://www.nndc.bnl.gov/nsr/nsrlink.jsp?1979Ma26,B}{1979Ma26}), which used this\vspace{0.1cm}}&\\
&&&&\parbox[t][0.3cm]{12.1857605cm}{\raggedright {\ }{\ }{\ }state as a calibration point.\vspace{0.1cm}}&\\
&&&&\parbox[t][0.3cm]{12.1857605cm}{\raggedright J\ensuremath{^{\pi}},L: From (\href{https://www.nndc.bnl.gov/nsr/nsrlink.jsp?1972Ga08,B}{1972Ga08}) based on zero-range DWBA analysis using DWUCK and\vspace{0.1cm}}&\\
&&&&\parbox[t][0.3cm]{12.1857605cm}{\raggedright {\ }{\ }{\ }assuming a three nucleon cluster transfer, see text for L. See also J\ensuremath{^{\ensuremath{\pi}}}=5/2\ensuremath{^{\textnormal{+}}} (\href{https://www.nndc.bnl.gov/nsr/nsrlink.jsp?1972Pa29,B}{1972Pa29}).\vspace{0.1cm}}&\\
&&&&\parbox[t][0.3cm]{12.1857605cm}{\raggedright d\ensuremath{\sigma}/d\ensuremath{\Omega}\ensuremath{_{\textnormal{c.m.}}}=63 \ensuremath{\mu}b/sr (\href{https://www.nndc.bnl.gov/nsr/nsrlink.jsp?1979Ma26,B}{1979Ma26}).\vspace{0.1cm}}&\\
&&&&\parbox[t][0.3cm]{12.1857605cm}{\raggedright \ensuremath{\sigma}\ensuremath{_{\textnormal{exp}}}/\ensuremath{\sigma}\ensuremath{_{\textnormal{DWBA}}}=0.36 (\href{https://www.nndc.bnl.gov/nsr/nsrlink.jsp?1972Ga08,B}{1972Ga08}).\vspace{0.1cm}}&\\
\multicolumn{1}{r@{}}{0.27\ensuremath{\times10^{3}}}&\multicolumn{1}{@{}l}{\ensuremath{^{{\hyperlink{NE25LEVEL1}{b}}{\hyperlink{NE25LEVEL6}{g}}}}}&&&\parbox[t][0.3cm]{12.1857605cm}{\raggedright E(level): From (\href{https://www.nndc.bnl.gov/nsr/nsrlink.jsp?1971Bi06,B}{1971Bi06}, \href{https://www.nndc.bnl.gov/nsr/nsrlink.jsp?1972Ga08,B}{1972Ga08}, \href{https://www.nndc.bnl.gov/nsr/nsrlink.jsp?1973Bi07,B}{1973Bi07}: Unresolved from the 0.24-MeV state).\vspace{0.1cm}}&\\
\multicolumn{1}{r@{}}{1.51\ensuremath{\times10^{3}}}&\multicolumn{1}{@{}l}{\ensuremath{^{{\hyperlink{NE25LEVEL1}{b}}{\hyperlink{NE25LEVEL6}{g}}}}}&\multicolumn{1}{l}{(5/2\ensuremath{^{-}})\ensuremath{^{{\hyperlink{NE25LEVEL12}{m}}}}}&&\parbox[t][0.3cm]{12.1857605cm}{\raggedright E(level): From (\href{https://www.nndc.bnl.gov/nsr/nsrlink.jsp?1971Bi06,B}{1971Bi06}, \href{https://www.nndc.bnl.gov/nsr/nsrlink.jsp?1972Ga08,B}{1972Ga08}, \href{https://www.nndc.bnl.gov/nsr/nsrlink.jsp?1973Bi07,B}{1973Bi07}: Unresolved from the 1.54- and\vspace{0.1cm}}&\\
&&&&\parbox[t][0.3cm]{12.1857605cm}{\raggedright {\ }{\ }{\ }1.62-MeV states).\vspace{0.1cm}}&\\
\multicolumn{1}{r@{}}{1540}&\multicolumn{1}{@{}l}{\ensuremath{^{{\hyperlink{NE25LEVEL0}{a}}{\hyperlink{NE25LEVEL5}{f}}{\hyperlink{NE25LEVEL13}{n}}}} {\it 20}}&\multicolumn{1}{l}{3/2\ensuremath{^{+}}\ensuremath{^{{\hyperlink{NE25LEVEL9}{j}}}}}&\multicolumn{1}{l}{2$^{{\hyperlink{NE25LEVEL9}{j}}}$}&\parbox[t][0.3cm]{12.1857605cm}{\raggedright E(level): See also 1.54 MeV (\href{https://www.nndc.bnl.gov/nsr/nsrlink.jsp?1971Bi06,B}{1971Bi06}, \href{https://www.nndc.bnl.gov/nsr/nsrlink.jsp?1972Ga08,B}{1972Ga08}, \href{https://www.nndc.bnl.gov/nsr/nsrlink.jsp?1973Bi07,B}{1973Bi07}: Unresolved from the 1.51-\vspace{0.1cm}}&\\
&&&&\parbox[t][0.3cm]{12.1857605cm}{\raggedright {\ }{\ }{\ }and 1.62-MeV states); and 1538 keV \textit{40} (\href{https://www.nndc.bnl.gov/nsr/nsrlink.jsp?1972Pa29,B}{1972Pa29}).\vspace{0.1cm}}&\\
&&&&\parbox[t][0.3cm]{12.1857605cm}{\raggedright J\ensuremath{^{\pi}},L: From (\href{https://www.nndc.bnl.gov/nsr/nsrlink.jsp?1972Ga08,B}{1972Ga08}) based on zero-range DWBA analysis using DWUCK and\vspace{0.1cm}}&\\
&&&&\parbox[t][0.3cm]{12.1857605cm}{\raggedright {\ }{\ }{\ }assuming a three nucleon cluster transfer, see text for L. See also J\ensuremath{^{\ensuremath{\pi}}}=3/2\ensuremath{^{\textnormal{+}}} (\href{https://www.nndc.bnl.gov/nsr/nsrlink.jsp?1971Bi06,B}{1971Bi06})\vspace{0.1cm}}&\\
&&&&\parbox[t][0.3cm]{12.1857605cm}{\raggedright {\ }{\ }{\ }based on comparison of the relative \ensuremath{^{\textnormal{16}}}O(\ensuremath{^{\textnormal{6}}}Li,\ensuremath{^{\textnormal{3}}}He) and \ensuremath{^{\textnormal{16}}}O(\ensuremath{^{\textnormal{6}}}Li,t) transition strengths\vspace{0.1cm}}&\\
&&&&\parbox[t][0.3cm]{12.1857605cm}{\raggedright {\ }{\ }{\ }populating the \ensuremath{^{\textnormal{19}}}F* and \ensuremath{^{\textnormal{19}}}Ne* analog states, respectively, and from mirror states\vspace{0.1cm}}&\\
&&&&\parbox[t][0.3cm]{12.1857605cm}{\raggedright {\ }{\ }{\ }analogy; and J\ensuremath{^{\ensuremath{\pi}}}=3/2\ensuremath{^{\textnormal{+}}} (\href{https://www.nndc.bnl.gov/nsr/nsrlink.jsp?1972Pa29,B}{1972Pa29}).\vspace{0.1cm}}&\\
&&&&\parbox[t][0.3cm]{12.1857605cm}{\raggedright L: From (\href{https://www.nndc.bnl.gov/nsr/nsrlink.jsp?1972Pa29,B}{1972Pa29}): See text.\vspace{0.1cm}}&\\
&&&&\parbox[t][0.3cm]{12.1857605cm}{\raggedright \ensuremath{\sigma}\ensuremath{_{\textnormal{exp}}}/\ensuremath{\sigma}\ensuremath{_{\textnormal{DWBA}}}=0.32 (\href{https://www.nndc.bnl.gov/nsr/nsrlink.jsp?1972Ga08,B}{1972Ga08}).\vspace{0.1cm}}&\\
\multicolumn{1}{r@{}}{1.61\ensuremath{\times10^{3}}}&\multicolumn{1}{@{}l}{\ensuremath{^{{\hyperlink{NE25LEVEL1}{b}}{\hyperlink{NE25LEVEL6}{g}}}} {\it 4}}&\multicolumn{1}{l}{(3/2\ensuremath{^{-}})\ensuremath{^{{\hyperlink{NE25LEVEL12}{m}}}}}&&\parbox[t][0.3cm]{12.1857605cm}{\raggedright E(level): From (\href{https://www.nndc.bnl.gov/nsr/nsrlink.jsp?1972Pa29,B}{1972Pa29}). See also 1.62 MeV (\href{https://www.nndc.bnl.gov/nsr/nsrlink.jsp?1971Bi06,B}{1971Bi06}, \href{https://www.nndc.bnl.gov/nsr/nsrlink.jsp?1972Ga08,B}{1972Ga08}, \href{https://www.nndc.bnl.gov/nsr/nsrlink.jsp?1973Bi07,B}{1973Bi07}:\vspace{0.1cm}}&\\
&&&&\parbox[t][0.3cm]{12.1857605cm}{\raggedright {\ }{\ }{\ }Unresolved from the 1.51- and 1.54-MeV states).\vspace{0.1cm}}&\\
\multicolumn{1}{r@{}}{2.77\ensuremath{\times10^{3}}}&\multicolumn{1}{@{}l}{\ensuremath{^{{\hyperlink{NE25LEVEL0}{a}}{\hyperlink{NE25LEVEL5}{f}}}} {\it 4}}&\multicolumn{1}{l}{9/2\ensuremath{^{+}}\ensuremath{^{{\hyperlink{NE25LEVEL9}{j}}{\hyperlink{NE25LEVEL12}{m}}}}}&\multicolumn{1}{l}{4$^{{\hyperlink{NE25LEVEL9}{j}}}$}&\parbox[t][0.3cm]{12.1857605cm}{\raggedright E(level): From (\href{https://www.nndc.bnl.gov/nsr/nsrlink.jsp?1972Pa29,B}{1972Pa29}). See also 2.79 MeV (\href{https://www.nndc.bnl.gov/nsr/nsrlink.jsp?1971Bi06,B}{1971Bi06}, \href{https://www.nndc.bnl.gov/nsr/nsrlink.jsp?1972Ga08,B}{1972Ga08}, \href{https://www.nndc.bnl.gov/nsr/nsrlink.jsp?1973Bi07,B}{1973Bi07}), and\vspace{0.1cm}}&\\
&&&&\parbox[t][0.3cm]{12.1857605cm}{\raggedright {\ }{\ }{\ }2.80 MeV (\href{https://www.nndc.bnl.gov/nsr/nsrlink.jsp?1979Ma26,B}{1979Ma26}), which was used as a calibration point.\vspace{0.1cm}}&\\
&&&&\parbox[t][0.3cm]{12.1857605cm}{\raggedright J\ensuremath{^{\pi}},L: From (\href{https://www.nndc.bnl.gov/nsr/nsrlink.jsp?1972Ga08,B}{1972Ga08}) based on zero-range DWBA analysis using DWUCK and\vspace{0.1cm}}&\\
&&&&\parbox[t][0.3cm]{12.1857605cm}{\raggedright {\ }{\ }{\ }assuming a three nucleon cluster transfer, see text for L. See also J\ensuremath{^{\ensuremath{\pi}}}=9/2\ensuremath{^{\textnormal{+}}} (\href{https://www.nndc.bnl.gov/nsr/nsrlink.jsp?1972Pa29,B}{1972Pa29}).\vspace{0.1cm}}&\\
&&&&\parbox[t][0.3cm]{12.1857605cm}{\raggedright d\ensuremath{\sigma}/d\ensuremath{\Omega}\ensuremath{_{\textnormal{c.m.}}}=148 \ensuremath{\mu}b/sr (\href{https://www.nndc.bnl.gov/nsr/nsrlink.jsp?1979Ma26,B}{1979Ma26}).\vspace{0.1cm}}&\\
&&&&\parbox[t][0.3cm]{12.1857605cm}{\raggedright \ensuremath{\sigma}\ensuremath{_{\textnormal{exp}}}/\ensuremath{\sigma}\ensuremath{_{\textnormal{DWBA}}}=0.33 (\href{https://www.nndc.bnl.gov/nsr/nsrlink.jsp?1972Ga08,B}{1972Ga08}).\vspace{0.1cm}}&\\
\multicolumn{1}{r@{}}{4.04\ensuremath{\times10^{3}}}&\multicolumn{1}{@{}l}{}&\multicolumn{1}{l}{(3/2\ensuremath{^{+}})\ensuremath{^{{\hyperlink{NE25LEVEL12}{m}}}}}&&\parbox[t][0.3cm]{12.1857605cm}{\raggedright E(level): From (\href{https://www.nndc.bnl.gov/nsr/nsrlink.jsp?1971Bi06,B}{1971Bi06}).\vspace{0.1cm}}&\\
\multicolumn{1}{r@{}}{4.15\ensuremath{\times10^{3}}}&\multicolumn{1}{@{}l}{\ensuremath{^{{\hyperlink{NE25LEVEL1}{b}}{\hyperlink{NE25LEVEL6}{g}}}} {\it 4}}&\multicolumn{1}{l}{(7/2\ensuremath{^{-}},9/2\ensuremath{^{-}})\ensuremath{^{{\hyperlink{NE25LEVEL12}{m}}}}}&&\parbox[t][0.3cm]{12.1857605cm}{\raggedright E(level): From (\href{https://www.nndc.bnl.gov/nsr/nsrlink.jsp?1972Pa29,B}{1972Pa29}). See also 4.14 MeV (\href{https://www.nndc.bnl.gov/nsr/nsrlink.jsp?1971Bi06,B}{1971Bi06}, \href{https://www.nndc.bnl.gov/nsr/nsrlink.jsp?1972Ga08,B}{1972Ga08}).\vspace{0.1cm}}&\\
\multicolumn{1}{r@{}}{4.21\ensuremath{\times10^{3}}}&\multicolumn{1}{@{}l}{\ensuremath{^{{\hyperlink{NE25LEVEL1}{b}}{\hyperlink{NE25LEVEL6}{g}}{\hyperlink{NE25LEVEL13}{n}}}} {\it 2}}&\multicolumn{1}{l}{(9/2\ensuremath{^{-}},7/2\ensuremath{^{-}})\ensuremath{^{{\hyperlink{NE25LEVEL12}{m}}}}}&&\parbox[t][0.3cm]{12.1857605cm}{\raggedright E(level): See also 4.20 MeV \textit{4} (\href{https://www.nndc.bnl.gov/nsr/nsrlink.jsp?1972Pa29,B}{1972Pa29}) and 4.20 MeV (\href{https://www.nndc.bnl.gov/nsr/nsrlink.jsp?1971Bi06,B}{1971Bi06}, \href{https://www.nndc.bnl.gov/nsr/nsrlink.jsp?1972Ga08,B}{1972Ga08}).\vspace{0.1cm}}&\\
\multicolumn{1}{r@{}}{4.38\ensuremath{\times10^{3}}}&\multicolumn{1}{@{}l}{\ensuremath{^{{\hyperlink{NE25LEVEL13}{n}}}} {\it 2}}&\multicolumn{1}{l}{(7/2\ensuremath{^{+}})\ensuremath{^{{\hyperlink{NE25LEVEL12}{m}}}}}&&\parbox[t][0.3cm]{12.1857605cm}{\raggedright E(level): See also 4.38 MeV (\href{https://www.nndc.bnl.gov/nsr/nsrlink.jsp?1971Bi06,B}{1971Bi06}) and 4.38 MeV \textit{4} (\href{https://www.nndc.bnl.gov/nsr/nsrlink.jsp?1972Pa29,B}{1972Pa29}).\vspace{0.1cm}}&\\
\multicolumn{1}{r@{}}{4.55\ensuremath{\times10^{3}}}&\multicolumn{1}{@{}l}{}&&&\parbox[t][0.3cm]{12.1857605cm}{\raggedright E(level): From (\href{https://www.nndc.bnl.gov/nsr/nsrlink.jsp?1971Bi06,B}{1971Bi06}).\vspace{0.1cm}}&\\
\multicolumn{1}{r@{}}{4593}&\multicolumn{1}{@{ }l}{{\it 6}}&&&\parbox[t][0.3cm]{12.1857605cm}{\raggedright E(level): From (\href{https://www.nndc.bnl.gov/nsr/nsrlink.jsp?1973Bi02,B}{1973Bi02}). See also a tentative state at 4.59 MeV (\href{https://www.nndc.bnl.gov/nsr/nsrlink.jsp?1971Bi06,B}{1971Bi06}).\vspace{0.1cm}}&\\
&&&&\parbox[t][0.3cm]{12.1857605cm}{\raggedright \ensuremath{\sigma}\ensuremath{_{\textnormal{max}}}(\ensuremath{\theta})=0.026 mb/sr (\href{https://www.nndc.bnl.gov/nsr/nsrlink.jsp?1973Bi02,B}{1973Bi02}).\vspace{0.1cm}}&\\
\multicolumn{1}{r@{}}{4.63\ensuremath{\times10^{3}}}&\multicolumn{1}{@{}l}{\ensuremath{^{{\hyperlink{NE25LEVEL0}{a}}{\hyperlink{NE25LEVEL5}{f}}}} {\it 2}}&\multicolumn{1}{l}{(13/2\ensuremath{^{+}})\ensuremath{^{{\hyperlink{NE25LEVEL9}{j}}}}}&\multicolumn{1}{l}{6$^{{\hyperlink{NE25LEVEL9}{j}}}$}&\parbox[t][0.3cm]{12.1857605cm}{\raggedright E(level): Weighted average of 4.61 MeV \textit{4} (\href{https://www.nndc.bnl.gov/nsr/nsrlink.jsp?1972Pa29,B}{1972Pa29}) and 4.64 MeV \textit{2} (\href{https://www.nndc.bnl.gov/nsr/nsrlink.jsp?1979Ma26,B}{1979Ma26}).\vspace{0.1cm}}&\\
&&&&\parbox[t][0.3cm]{12.1857605cm}{\raggedright {\ }{\ }{\ }See also 4.62 MeV (\href{https://www.nndc.bnl.gov/nsr/nsrlink.jsp?1971Bi06,B}{1971Bi06}: See also 4625 keV as cited by \href{https://www.nndc.bnl.gov/nsr/nsrlink.jsp?1973Bi02,B}{1973Bi02}) and 4.62\vspace{0.1cm}}&\\
\end{longtable}
\begin{textblock}{29}(0,27.3)
Continued on next page (footnotes at end of table)
\end{textblock}
\clearpage
\begin{longtable}{ccccc@{\extracolsep{\fill}}c}
\\[-.4cm]
\multicolumn{6}{c}{{\bf \small \underline{\ensuremath{^{\textnormal{16}}}O(\ensuremath{^{\textnormal{6}}}Li,t)\hspace{0.2in}\href{https://www.nndc.bnl.gov/nsr/nsrlink.jsp?1972Ga08,B}{1972Ga08},\href{https://www.nndc.bnl.gov/nsr/nsrlink.jsp?1979Ma26,B}{1979Ma26},\href{https://www.nndc.bnl.gov/nsr/nsrlink.jsp?2023Ma57,B}{2023Ma57} (continued)}}}\\
\multicolumn{6}{c}{~}\\
\multicolumn{6}{c}{\underline{\ensuremath{^{19}}Ne Levels (continued)}}\\
\multicolumn{6}{c}{~}\\
\multicolumn{2}{c}{E(level)$^{{\hyperlink{NE25LEVEL2}{c}}{\hyperlink{NE25LEVEL3}{d}}{\hyperlink{NE25LEVEL4}{e}}}$}&J$^{\pi}$$^{}$&L$^{}$&Comments&\\[-.2cm]
\multicolumn{2}{c}{\hrulefill}&\hrulefill&\hrulefill&\hrulefill&
\endhead
&&&&\parbox[t][0.3cm]{12.023121cm}{\raggedright {\ }{\ }{\ }MeV (\href{https://www.nndc.bnl.gov/nsr/nsrlink.jsp?1972Ga08,B}{1972Ga08}).\vspace{0.1cm}}&\\
&&&&\parbox[t][0.3cm]{12.023121cm}{\raggedright J\ensuremath{^{\pi}},L: From (\href{https://www.nndc.bnl.gov/nsr/nsrlink.jsp?1972Pa29,B}{1972Pa29}): The analysis of triton angular distribution populating this state\vspace{0.1cm}}&\\
&&&&\parbox[t][0.3cm]{12.023121cm}{\raggedright {\ }{\ }{\ }resulted in J\ensuremath{^{\ensuremath{\pi}}}=11/2\ensuremath{^{\textnormal{+}}} and 13/2\ensuremath{^{\textnormal{+}}}. The latter was selected on the basis of mirror levels\vspace{0.1cm}}&\\
&&&&\parbox[t][0.3cm]{12.023121cm}{\raggedright {\ }{\ }{\ }analysis and guided by the result of (\href{https://www.nndc.bnl.gov/nsr/nsrlink.jsp?1971Bi06,B}{1971Bi06}), which led to J\ensuremath{^{\ensuremath{\pi}}}=13/2\ensuremath{^{\textnormal{+}}} based on\vspace{0.1cm}}&\\
&&&&\parbox[t][0.3cm]{12.023121cm}{\raggedright {\ }{\ }{\ }comparison of the relative \ensuremath{^{\textnormal{16}}}O(\ensuremath{^{\textnormal{6}}}Li,\ensuremath{^{\textnormal{3}}}He) and \ensuremath{^{\textnormal{16}}}O(\ensuremath{^{\textnormal{6}}}Li,t) transition strengths populating\vspace{0.1cm}}&\\
&&&&\parbox[t][0.3cm]{12.023121cm}{\raggedright {\ }{\ }{\ }the \ensuremath{^{\textnormal{19}}}F* and \ensuremath{^{\textnormal{19}}}Ne* analog states, respectively, and from mirror states analogy.\vspace{0.1cm}}&\\
&&&&\parbox[t][0.3cm]{12.023121cm}{\raggedright d\ensuremath{\sigma}/d\ensuremath{\Omega}\ensuremath{_{\textnormal{c.m.}}}=182 \ensuremath{\mu}b/sr (\href{https://www.nndc.bnl.gov/nsr/nsrlink.jsp?1979Ma26,B}{1979Ma26}).\vspace{0.1cm}}&\\
\multicolumn{1}{r@{}}{4.74\ensuremath{\times10^{3}}}&\multicolumn{1}{@{ }l}{{\it 10}}&&&\parbox[t][0.3cm]{12.023121cm}{\raggedright E(level): From (\href{https://www.nndc.bnl.gov/nsr/nsrlink.jsp?2023Ma57,B}{2023Ma57}), where the uncertainty is systematic. These authors assumed\vspace{0.1cm}}&\\
&&&&\parbox[t][0.3cm]{12.023121cm}{\raggedright {\ }{\ }{\ }this state is the same as the E\ensuremath{_{\textnormal{x}}}=4712 keV \textit{10} state (\href{https://www.nndc.bnl.gov/nsr/nsrlink.jsp?2009Ta09,B}{2009Ta09}: \ensuremath{^{\textnormal{19}}}F(\ensuremath{^{\textnormal{3}}}He,t)). See also\vspace{0.1cm}}&\\
&&&&\parbox[t][0.3cm]{12.023121cm}{\raggedright {\ }{\ }{\ }4.71 MeV (\href{https://www.nndc.bnl.gov/nsr/nsrlink.jsp?1971Bi06,B}{1971Bi06}).\vspace{0.1cm}}&\\
&&&&\parbox[t][0.3cm]{12.023121cm}{\raggedright Decay mode: Predominantly \ensuremath{\alpha} (\href{https://www.nndc.bnl.gov/nsr/nsrlink.jsp?2023Ma57,B}{2023Ma57}).\vspace{0.1cm}}&\\
\multicolumn{1}{r@{}}{5.09\ensuremath{\times10^{3}}}&\multicolumn{1}{@{}l}{}&\multicolumn{1}{l}{(5/2\ensuremath{^{-}},7/2\ensuremath{^{-}})\ensuremath{^{{\hyperlink{NE25LEVEL12}{m}}}}}&&\parbox[t][0.3cm]{12.023121cm}{\raggedright E(level): From (\href{https://www.nndc.bnl.gov/nsr/nsrlink.jsp?1971Bi06,B}{1971Bi06}).\vspace{0.1cm}}&\\
\multicolumn{1}{r@{}}{5.35\ensuremath{\times10^{3}}}&\multicolumn{1}{@{}l}{}&&&\parbox[t][0.3cm]{12.023121cm}{\raggedright E(level): From (\href{https://www.nndc.bnl.gov/nsr/nsrlink.jsp?1971Bi06,B}{1971Bi06}).\vspace{0.1cm}}&\\
\multicolumn{1}{r@{}}{5.41\ensuremath{\times10^{3}}}&\multicolumn{1}{@{}l}{\ensuremath{^{{\hyperlink{NE25LEVEL0}{a}}{\hyperlink{NE25LEVEL5}{f}}}} {\it 4}}&\multicolumn{1}{l}{(7/2\ensuremath{^{+}})\ensuremath{^{{\hyperlink{NE25LEVEL9}{j}}{\hyperlink{NE25LEVEL12}{m}}}}}&\multicolumn{1}{l}{4$^{{\hyperlink{NE25LEVEL9}{j}}}$}&\parbox[t][0.3cm]{12.023121cm}{\raggedright E(level): From (\href{https://www.nndc.bnl.gov/nsr/nsrlink.jsp?1972Pa29,B}{1972Pa29}), who considered this state to be the same as 5.43 MeV state\vspace{0.1cm}}&\\
&&&&\parbox[t][0.3cm]{12.023121cm}{\raggedright {\ }{\ }{\ }measured by (\href{https://www.nndc.bnl.gov/nsr/nsrlink.jsp?1971Bi06,B}{1971Bi06}) based on the J\ensuremath{^{\ensuremath{\pi}}} assignments of these states. See also 5.44\vspace{0.1cm}}&\\
&&&&\parbox[t][0.3cm]{12.023121cm}{\raggedright {\ }{\ }{\ }MeV \textit{10} (sys.) (\href{https://www.nndc.bnl.gov/nsr/nsrlink.jsp?2023Ma57,B}{2023Ma57}); 5.43 MeV (\href{https://www.nndc.bnl.gov/nsr/nsrlink.jsp?1971Bi06,B}{1971Bi06}, \href{https://www.nndc.bnl.gov/nsr/nsrlink.jsp?1972Ga08,B}{1972Ga08}); 5.42 MeV\vspace{0.1cm}}&\\
&&&&\parbox[t][0.3cm]{12.023121cm}{\raggedright {\ }{\ }{\ }(\href{https://www.nndc.bnl.gov/nsr/nsrlink.jsp?1979Ma26,B}{1979Ma26}), which used it as a calibration point; and 5417 keV (\href{https://www.nndc.bnl.gov/nsr/nsrlink.jsp?1998Ut02,B}{1998Ut02}).\vspace{0.1cm}}&\\
&&&&\parbox[t][0.3cm]{12.023121cm}{\raggedright E(level): (\href{https://www.nndc.bnl.gov/nsr/nsrlink.jsp?1971Bi06,B}{1971Bi06}) mentioned that this state may be identified as the analog of the\vspace{0.1cm}}&\\
&&&&\parbox[t][0.3cm]{12.023121cm}{\raggedright {\ }{\ }{\ }\ensuremath{^{\textnormal{19}}}F*(5.47 MeV, 7/2\ensuremath{^{\textnormal{+}}}) level.\vspace{0.1cm}}&\\
&&&&\parbox[t][0.3cm]{12.023121cm}{\raggedright J\ensuremath{^{\pi}},L: See also J\ensuremath{^{\ensuremath{\pi}}}=7/2\ensuremath{^{\textnormal{+}}} with L=4 from (\href{https://www.nndc.bnl.gov/nsr/nsrlink.jsp?1972Ga08,B}{1972Ga08}), where incomplete angular\vspace{0.1cm}}&\\
&&&&\parbox[t][0.3cm]{12.023121cm}{\raggedright {\ }{\ }{\ }distributions deduced for this state prevented the authors to show the DWBA analysis\vspace{0.1cm}}&\\
&&&&\parbox[t][0.3cm]{12.023121cm}{\raggedright {\ }{\ }{\ }performed; and J\ensuremath{^{\ensuremath{\pi}}}=7/2\ensuremath{^{\textnormal{+}}} from (\href{https://www.nndc.bnl.gov/nsr/nsrlink.jsp?1972Pa29,B}{1972Pa29}), where the analysis of triton angular\vspace{0.1cm}}&\\
&&&&\parbox[t][0.3cm]{12.023121cm}{\raggedright {\ }{\ }{\ }distribution corresponding to this state resulted in J\ensuremath{^{\ensuremath{\pi}}}=7/2\ensuremath{^{\textnormal{+}}} and 9/2\ensuremath{^{\textnormal{+}}}. The latter was\vspace{0.1cm}}&\\
&&&&\parbox[t][0.3cm]{12.023121cm}{\raggedright {\ }{\ }{\ }selected on the basis of mirror level analysis and guided by the K\ensuremath{^{\ensuremath{\pi}}}=1/2\ensuremath{^{\textnormal{+}}} rotational\vspace{0.1cm}}&\\
&&&&\parbox[t][0.3cm]{12.023121cm}{\raggedright {\ }{\ }{\ }band in \ensuremath{^{\textnormal{19}}}Ne.\vspace{0.1cm}}&\\
&&&&\parbox[t][0.3cm]{12.023121cm}{\raggedright d\ensuremath{\sigma}/d\ensuremath{\Omega}\ensuremath{_{\textnormal{c.m.}}}=143 \ensuremath{\mu}b/sr (\href{https://www.nndc.bnl.gov/nsr/nsrlink.jsp?1979Ma26,B}{1979Ma26}).\vspace{0.1cm}}&\\
&&&&\parbox[t][0.3cm]{12.023121cm}{\raggedright \ensuremath{\sigma}\ensuremath{_{\textnormal{exp}}}/\ensuremath{\sigma}\ensuremath{_{\textnormal{DWBA}}}=0.25 (\href{https://www.nndc.bnl.gov/nsr/nsrlink.jsp?1972Ga08,B}{1972Ga08}).\vspace{0.1cm}}&\\
&&&&\parbox[t][0.3cm]{12.023121cm}{\raggedright Decay mode: Predominantly \ensuremath{\alpha} (\href{https://www.nndc.bnl.gov/nsr/nsrlink.jsp?2023Ma57,B}{2023Ma57}).\vspace{0.1cm}}&\\
\multicolumn{1}{r@{}}{6017}&\multicolumn{1}{@{}l}{}&&&\parbox[t][0.3cm]{12.023121cm}{\raggedright E(level): From (\href{https://www.nndc.bnl.gov/nsr/nsrlink.jsp?1998Ut02,B}{1998Ut02}).\vspace{0.1cm}}&\\
\multicolumn{1}{r@{}}{6.09\ensuremath{\times10^{3}}}&\multicolumn{1}{@{ }l}{{\it 2}}&&&\parbox[t][0.3cm]{12.023121cm}{\raggedright E(level): Weighted average of 6.08 MeV \textit{2} (\href{https://www.nndc.bnl.gov/nsr/nsrlink.jsp?1979Ma26,B}{1979Ma26}) and 6.12 MeV \textit{4} (\href{https://www.nndc.bnl.gov/nsr/nsrlink.jsp?1972Pa29,B}{1972Pa29}).\vspace{0.1cm}}&\\
&&&&\parbox[t][0.3cm]{12.023121cm}{\raggedright {\ }{\ }{\ }See also 6092 keV (\href{https://www.nndc.bnl.gov/nsr/nsrlink.jsp?1998Ut02,B}{1998Ut02}); and 6.02 MeV \textit{10} (sys.) (\href{https://www.nndc.bnl.gov/nsr/nsrlink.jsp?2023Ma57,B}{2023Ma57}), who considered\vspace{0.1cm}}&\\
&&&&\parbox[t][0.3cm]{12.023121cm}{\raggedright {\ }{\ }{\ }this state to be the same as the 6100-keV level from (\href{https://www.nndc.bnl.gov/nsr/nsrlink.jsp?2019Ha08,B}{2019Ha08}: \ensuremath{^{\textnormal{19}}}F(\ensuremath{^{\textnormal{3}}}He,t\ensuremath{\gamma})).\vspace{0.1cm}}&\\
&&&&\parbox[t][0.3cm]{12.023121cm}{\raggedright Decay mode: Predominantly \ensuremath{\alpha} (\href{https://www.nndc.bnl.gov/nsr/nsrlink.jsp?2023Ma57,B}{2023Ma57}).\vspace{0.1cm}}&\\
\multicolumn{1}{r@{}}{6140}&\multicolumn{1}{@{}l}{}&&&\parbox[t][0.3cm]{12.023121cm}{\raggedright E(level): From (\href{https://www.nndc.bnl.gov/nsr/nsrlink.jsp?1998Ut02,B}{1998Ut02}).\vspace{0.1cm}}&\\
\multicolumn{1}{r@{}}{6.28\ensuremath{\times10^{3}}}&\multicolumn{1}{@{ }l}{{\it 2}}&\multicolumn{1}{l}{(5/2\ensuremath{^{-}},7/2\ensuremath{^{-}})\ensuremath{^{{\hyperlink{NE25LEVEL9}{j}}}}}&\multicolumn{1}{l}{(3)$^{{\hyperlink{NE25LEVEL9}{j}}}$}&\parbox[t][0.3cm]{12.023121cm}{\raggedright E(level): Weighted average of 6.27 MeV \textit{4} (\href{https://www.nndc.bnl.gov/nsr/nsrlink.jsp?1972Pa29,B}{1972Pa29}) and 6.28 MeV \textit{2} (\href{https://www.nndc.bnl.gov/nsr/nsrlink.jsp?1979Ma26,B}{1979Ma26}).\vspace{0.1cm}}&\\
&&&&\parbox[t][0.3cm]{12.023121cm}{\raggedright {\ }{\ }{\ }See also 6287 keV (\href{https://www.nndc.bnl.gov/nsr/nsrlink.jsp?1998Ut02,B}{1998Ut02}); and 6.26 MeV \textit{10} (sys.) (\href{https://www.nndc.bnl.gov/nsr/nsrlink.jsp?2023Ma57,B}{2023Ma57}).\vspace{0.1cm}}&\\
&&&&\parbox[t][0.3cm]{12.023121cm}{\raggedright Decay mode: Predominantly \ensuremath{\alpha} (\href{https://www.nndc.bnl.gov/nsr/nsrlink.jsp?2023Ma57,B}{2023Ma57}).\vspace{0.1cm}}&\\
\multicolumn{1}{r@{}}{6.50\ensuremath{\times10^{3}}}&\multicolumn{1}{@{ }l}{{\it 10}}&&&\parbox[t][0.3cm]{12.023121cm}{\raggedright E(level): From (\href{https://www.nndc.bnl.gov/nsr/nsrlink.jsp?2023Ma57,B}{2023Ma57}), where the uncertainty is systematic. Evaluator notes that\vspace{0.1cm}}&\\
&&&&\parbox[t][0.3cm]{12.023121cm}{\raggedright {\ }{\ }{\ }(\href{https://www.nndc.bnl.gov/nsr/nsrlink.jsp?2023Ma57,B}{2023Ma57}) considered this state to be the same as the state with E\ensuremath{_{\textnormal{x}}}=6423 keV \textit{3}\vspace{0.1cm}}&\\
&&&&\parbox[t][0.3cm]{12.023121cm}{\raggedright {\ }{\ }{\ }from (\href{https://www.nndc.bnl.gov/nsr/nsrlink.jsp?2019Ha08,B}{2019Ha08}: \ensuremath{^{\textnormal{19}}}F(\ensuremath{^{\textnormal{3}}}He,t\ensuremath{\gamma})). See also 6434 keV (\href{https://www.nndc.bnl.gov/nsr/nsrlink.jsp?1998Ut02,B}{1998Ut02}: \ensuremath{^{\textnormal{16}}}O(\ensuremath{^{\textnormal{6}}}Li,t)). We\vspace{0.1cm}}&\\
&&&&\parbox[t][0.3cm]{12.023121cm}{\raggedright {\ }{\ }{\ }highlight that there are multiple states in the E\ensuremath{_{\textnormal{x}}}=6.4 MeV region in \ensuremath{^{\textnormal{19}}}Ne. Therefore,\vspace{0.1cm}}&\\
&&&&\parbox[t][0.3cm]{12.023121cm}{\raggedright {\ }{\ }{\ }the 6423-keV state from (\href{https://www.nndc.bnl.gov/nsr/nsrlink.jsp?2019Ha08,B}{2019Ha08}) and the 6434-keV state from (\href{https://www.nndc.bnl.gov/nsr/nsrlink.jsp?1998Ut02,B}{1998Ut02}:\vspace{0.1cm}}&\\
&&&&\parbox[t][0.3cm]{12.023121cm}{\raggedright {\ }{\ }{\ }\ensuremath{^{\textnormal{16}}}O(\ensuremath{^{\textnormal{6}}}Li,t)) may refer to two different states. But since the latter energy was reported\vspace{0.1cm}}&\\
&&&&\parbox[t][0.3cm]{12.023121cm}{\raggedright {\ }{\ }{\ }without an uncertainty and due to the lack of J\ensuremath{^{\ensuremath{\pi}}} assignment, we cannot make a\vspace{0.1cm}}&\\
&&&&\parbox[t][0.3cm]{12.023121cm}{\raggedright {\ }{\ }{\ }distinction.\vspace{0.1cm}}&\\
&&&&\parbox[t][0.3cm]{12.023121cm}{\raggedright Decay mode: Predominantly \ensuremath{\alpha}. However, decay via p emission is also energetically\vspace{0.1cm}}&\\
&&&&\parbox[t][0.3cm]{12.023121cm}{\raggedright {\ }{\ }{\ }allowed (\href{https://www.nndc.bnl.gov/nsr/nsrlink.jsp?2023Ma57,B}{2023Ma57}).\vspace{0.1cm}}&\\
\multicolumn{1}{r@{}}{6742}&\multicolumn{1}{@{}l}{}&&&\parbox[t][0.3cm]{12.023121cm}{\raggedright E(level): From (\href{https://www.nndc.bnl.gov/nsr/nsrlink.jsp?1998Ut02,B}{1998Ut02}).\vspace{0.1cm}}&\\
\multicolumn{1}{r@{}}{6.85\ensuremath{\times10^{3}}}&\multicolumn{1}{@{ }l}{{\it 2}}&\multicolumn{1}{l}{(9/2\ensuremath{^{-}},11/2\ensuremath{^{-}})\ensuremath{^{{\hyperlink{NE25LEVEL9}{j}}}}}&\multicolumn{1}{l}{(5)$^{{\hyperlink{NE25LEVEL9}{j}}}$}&\parbox[t][0.3cm]{12.023121cm}{\raggedright E(level): Weighted average of 6.83 MeV \textit{4} (\href{https://www.nndc.bnl.gov/nsr/nsrlink.jsp?1972Pa29,B}{1972Pa29}) and 6.85 MeV \textit{2} (\href{https://www.nndc.bnl.gov/nsr/nsrlink.jsp?1979Ma26,B}{1979Ma26}).\vspace{0.1cm}}&\\
&&&&\parbox[t][0.3cm]{12.023121cm}{\raggedright E(level): See also 6861 keV (\href{https://www.nndc.bnl.gov/nsr/nsrlink.jsp?1998Ut02,B}{1998Ut02}); and 6.89 MeV \textit{10} (sys.) (\href{https://www.nndc.bnl.gov/nsr/nsrlink.jsp?2023Ma57,B}{2023Ma57}). Those\vspace{0.1cm}}&\\
&&&&\parbox[t][0.3cm]{12.023121cm}{\raggedright {\ }{\ }{\ }authors assumed this state is the same as the E\ensuremath{_{\textnormal{x}}}=6853 keV \textit{3} level measured by\vspace{0.1cm}}&\\
&&&&\parbox[t][0.3cm]{12.023121cm}{\raggedright {\ }{\ }{\ }(\href{https://www.nndc.bnl.gov/nsr/nsrlink.jsp?2019Ha08,B}{2019Ha08}: \ensuremath{^{\textnormal{19}}}F(\ensuremath{^{\textnormal{3}}}He,t\ensuremath{\gamma})).\vspace{0.1cm}}&\\
&&&&\parbox[t][0.3cm]{12.023121cm}{\raggedright J\ensuremath{^{\pi}},L: (\href{https://www.nndc.bnl.gov/nsr/nsrlink.jsp?1972Pa29,B}{1972Pa29}): The analysis of triton angular distribution corresponding to this state\vspace{0.1cm}}&\\
&&&&\parbox[t][0.3cm]{12.023121cm}{\raggedright {\ }{\ }{\ }resulted in J\ensuremath{^{\ensuremath{\pi}}}=(9/2\ensuremath{^{-}}, 11/2\ensuremath{^{-}}, 9/2\ensuremath{^{\textnormal{+}}}, 11/2\ensuremath{^{\textnormal{+}}}, and 13/2\ensuremath{^{\textnormal{+}}}) with L=(5,6). The results with\vspace{0.1cm}}&\\
\end{longtable}
\begin{textblock}{29}(0,27.3)
Continued on next page (footnotes at end of table)
\end{textblock}
\clearpage
\begin{longtable}{cccccc@{\extracolsep{\fill}}c}
\\[-.4cm]
\multicolumn{7}{c}{{\bf \small \underline{\ensuremath{^{\textnormal{16}}}O(\ensuremath{^{\textnormal{6}}}Li,t)\hspace{0.2in}\href{https://www.nndc.bnl.gov/nsr/nsrlink.jsp?1972Ga08,B}{1972Ga08},\href{https://www.nndc.bnl.gov/nsr/nsrlink.jsp?1979Ma26,B}{1979Ma26},\href{https://www.nndc.bnl.gov/nsr/nsrlink.jsp?2023Ma57,B}{2023Ma57} (continued)}}}\\
\multicolumn{7}{c}{~}\\
\multicolumn{7}{c}{\underline{\ensuremath{^{19}}Ne Levels (continued)}}\\
\multicolumn{7}{c}{~}\\
\multicolumn{2}{c}{E(level)$^{{\hyperlink{NE25LEVEL2}{c}}{\hyperlink{NE25LEVEL3}{d}}{\hyperlink{NE25LEVEL4}{e}}}$}&J$^{\pi}$$^{}$&\multicolumn{2}{c}{\ensuremath{\theta}\ensuremath{^{\textnormal{2}}_{\ensuremath{\alpha}}}$^{{\hyperlink{NE25LEVEL10}{k}}}$}&Comments&\\[-.2cm]
\multicolumn{2}{c}{\hrulefill}&\hrulefill&\multicolumn{2}{c}{\hrulefill}&\hrulefill&
\endhead
&&&&&\parbox[t][0.3cm]{11.691421cm}{\raggedright {\ }{\ }{\ }J\ensuremath{^{\ensuremath{\pi}}}=(11/2\ensuremath{^{\textnormal{+}}}, 13/2\ensuremath{^{\textnormal{+}}}) deduced from L=6 were discarded by those authors based on the\vspace{0.1cm}}&\\
&&&&&\parbox[t][0.3cm]{11.691421cm}{\raggedright {\ }{\ }{\ }mirror level analysis, and the theoretical prediction by D. Strottman (priv. comm.,\vspace{0.1cm}}&\\
&&&&&\parbox[t][0.3cm]{11.691421cm}{\raggedright {\ }{\ }{\ }SU(3) model), who suggested that the 13/2\ensuremath{_{\textnormal{2}}^{\textnormal{+}}} level in \ensuremath{^{\textnormal{19}}}Ne lies at about 9.3 MeV.\vspace{0.1cm}}&\\
&&&&&\parbox[t][0.3cm]{11.691421cm}{\raggedright d\ensuremath{\sigma}/d\ensuremath{\Omega}\ensuremath{_{\textnormal{c.m.}}}=95 \ensuremath{\mu}b/sr (\href{https://www.nndc.bnl.gov/nsr/nsrlink.jsp?1979Ma26,B}{1979Ma26}).\vspace{0.1cm}}&\\
&&&&&\parbox[t][0.3cm]{11.691421cm}{\raggedright Decay mode: Predominantly \ensuremath{\alpha}. However, decay via p emission is also energetically\vspace{0.1cm}}&\\
&&&&&\parbox[t][0.3cm]{11.691421cm}{\raggedright {\ }{\ }{\ }allowed (\href{https://www.nndc.bnl.gov/nsr/nsrlink.jsp?2023Ma57,B}{2023Ma57}).\vspace{0.1cm}}&\\
\multicolumn{1}{r@{}}{7171}&\multicolumn{1}{@{}l}{}&&&&\parbox[t][0.3cm]{11.691421cm}{\raggedright E(level): From (\href{https://www.nndc.bnl.gov/nsr/nsrlink.jsp?1998Ut02,B}{1998Ut02}).\vspace{0.1cm}}&\\
\multicolumn{1}{r@{}}{7241}&\multicolumn{1}{@{}l}{}&&&&\parbox[t][0.3cm]{11.691421cm}{\raggedright E(level): From (\href{https://www.nndc.bnl.gov/nsr/nsrlink.jsp?1998Ut02,B}{1998Ut02}).\vspace{0.1cm}}&\\
&&&&&\parbox[t][0.3cm]{11.691421cm}{\raggedright E(level): See also 7.20 MeV \textit{4} (\href{https://www.nndc.bnl.gov/nsr/nsrlink.jsp?1972Pa29,B}{1972Pa29}); 7.21 MeV \textit{2} (\href{https://www.nndc.bnl.gov/nsr/nsrlink.jsp?1979Ma26,B}{1979Ma26}); and 7.22 MeV\vspace{0.1cm}}&\\
&&&&&\parbox[t][0.3cm]{11.691421cm}{\raggedright {\ }{\ }{\ }\textit{10} (sys.) (\href{https://www.nndc.bnl.gov/nsr/nsrlink.jsp?2023Ma57,B}{2023Ma57}).\vspace{0.1cm}}&\\
&&&&&\parbox[t][0.3cm]{11.691421cm}{\raggedright This state was observed to decay via \ensuremath{\alpha}+\ensuremath{^{\textnormal{15}}}O in (\href{https://www.nndc.bnl.gov/nsr/nsrlink.jsp?2023Ma57,B}{2023Ma57}). However, low statistics\vspace{0.1cm}}&\\
&&&&&\parbox[t][0.3cm]{11.691421cm}{\raggedright {\ }{\ }{\ }in the p+\ensuremath{^{\textnormal{18}}}F decay channel prevented those authors to extract the \ensuremath{\Gamma}\ensuremath{_{\textnormal{p}}}/\ensuremath{\Gamma}\ensuremath{_{\ensuremath{\alpha}}} for this\vspace{0.1cm}}&\\
&&&&&\parbox[t][0.3cm]{11.691421cm}{\raggedright {\ }{\ }{\ }state.\vspace{0.1cm}}&\\
\multicolumn{1}{r@{}}{7.61\ensuremath{\times10^{3}}}&\multicolumn{1}{@{}l}{\ensuremath{^{{\hyperlink{NE25LEVEL8}{i}}}} {\it 10}}&&\multicolumn{1}{r@{}}{7}&\multicolumn{1}{@{.}l}{3\ensuremath{\times10^{-3}} {\it 35}}&\parbox[t][0.3cm]{11.691421cm}{\raggedright \ensuremath{\Gamma}\ensuremath{_{\textnormal{p}}}/\ensuremath{\Gamma}\ensuremath{\alpha}=0.57 \textit{10} (\href{https://www.nndc.bnl.gov/nsr/nsrlink.jsp?2023Ma57,B}{2023Ma57})\vspace{0.1cm}}&\\
&&&&&\parbox[t][0.3cm]{11.691421cm}{\raggedright E(level): From (\href{https://www.nndc.bnl.gov/nsr/nsrlink.jsp?2023Ma57,B}{2023Ma57}), where the uncertainty is systematic. These authors\vspace{0.1cm}}&\\
&&&&&\parbox[t][0.3cm]{11.691421cm}{\raggedright {\ }{\ }{\ }considered this state to be the E\ensuremath{_{\textnormal{x}}}=7616 keV \textit{5} state measured by (\href{https://www.nndc.bnl.gov/nsr/nsrlink.jsp?2009Da07,B}{2009Da07}:\vspace{0.1cm}}&\\
&&&&&\parbox[t][0.3cm]{11.691421cm}{\raggedright {\ }{\ }{\ }\ensuremath{^{\textnormal{1}}}H(\ensuremath{^{\textnormal{19}}}Ne,p\ensuremath{'})).\vspace{0.1cm}}&\\
&&&&&\parbox[t][0.3cm]{11.691421cm}{\raggedright E(level): See also 7.55 MeV \textit{4} (\href{https://www.nndc.bnl.gov/nsr/nsrlink.jsp?1972Pa29,B}{1972Pa29}) and 7596 keV (\href{https://www.nndc.bnl.gov/nsr/nsrlink.jsp?1998Ut02,B}{1998Ut02}).\vspace{0.1cm}}&\\
&&&&&\parbox[t][0.3cm]{11.691421cm}{\raggedright \ensuremath{\Gamma}\ensuremath{_{\ensuremath{\alpha}}}/\ensuremath{\Gamma}=63.7\% \textit{41} deduced by (\href{https://www.nndc.bnl.gov/nsr/nsrlink.jsp?2023Ma57,B}{2023Ma57}) using \ensuremath{\Gamma}=21 keV \textit{10} from (\href{https://www.nndc.bnl.gov/nsr/nsrlink.jsp?2009Da07,B}{2009Da07}:\vspace{0.1cm}}&\\
&&&&&\parbox[t][0.3cm]{11.691421cm}{\raggedright {\ }{\ }{\ }\ensuremath{^{\textnormal{1}}}H(\ensuremath{^{\textnormal{19}}}Ne,p\ensuremath{'})).\vspace{0.1cm}}&\\
\multicolumn{1}{r@{}}{7.91\ensuremath{\times10^{3}}}&\multicolumn{1}{@{}l}{\ensuremath{^{{\hyperlink{NE25LEVEL8}{i}}}} {\it 4}}&\multicolumn{1}{l}{(1/2\ensuremath{^{+}})}&\multicolumn{1}{r@{}}{0}&\multicolumn{1}{@{.}l}{073 {\it 27}}&\parbox[t][0.3cm]{11.691421cm}{\raggedright \ensuremath{\Gamma}\ensuremath{_{\textnormal{p}}}/\ensuremath{\Gamma}\ensuremath{\alpha}=0.85 \textit{10} (\href{https://www.nndc.bnl.gov/nsr/nsrlink.jsp?2023Ma57,B}{2023Ma57})\vspace{0.1cm}}&\\
&&&&&\parbox[t][0.3cm]{11.691421cm}{\raggedright E(level): From (\href{https://www.nndc.bnl.gov/nsr/nsrlink.jsp?1972Pa29,B}{1972Pa29}).\vspace{0.1cm}}&\\
&&&&&\parbox[t][0.3cm]{11.691421cm}{\raggedright E(level): See also 7.85 MeV \textit{10} (sys.) (\href{https://www.nndc.bnl.gov/nsr/nsrlink.jsp?2023Ma57,B}{2023Ma57}): See the footnote on the excitation\vspace{0.1cm}}&\\
&&&&&\parbox[t][0.3cm]{11.691421cm}{\raggedright {\ }{\ }{\ }energy.\vspace{0.1cm}}&\\
&&&&&\parbox[t][0.3cm]{11.691421cm}{\raggedright J\ensuremath{^{\pi}}: From (\href{https://www.nndc.bnl.gov/nsr/nsrlink.jsp?2009Da07,B}{2009Da07}: \ensuremath{^{\textnormal{1}}}H(\ensuremath{^{\textnormal{19}}}Ne,p\ensuremath{'})) as cited by (\href{https://www.nndc.bnl.gov/nsr/nsrlink.jsp?2023Ma57,B}{2023Ma57}), who paired this state\vspace{0.1cm}}&\\
&&&&&\parbox[t][0.3cm]{11.691421cm}{\raggedright {\ }{\ }{\ }with the 7863 keV \textit{39} state measured by (\href{https://www.nndc.bnl.gov/nsr/nsrlink.jsp?2019Ha08,B}{2019Ha08}: \ensuremath{^{\textnormal{19}}}F(\ensuremath{^{\textnormal{3}}}He,t\ensuremath{\gamma})), whose J\ensuremath{^{\ensuremath{\pi}}} value\vspace{0.1cm}}&\\
&&&&&\parbox[t][0.3cm]{11.691421cm}{\raggedright {\ }{\ }{\ }was tentatively determined to be (1/2\ensuremath{^{\textnormal{+}}}) by (\href{https://www.nndc.bnl.gov/nsr/nsrlink.jsp?2009Da07,B}{2009Da07}).\vspace{0.1cm}}&\\
&&&&&\parbox[t][0.3cm]{11.691421cm}{\raggedright J\ensuremath{^{\pi}}: See also J\ensuremath{^{\ensuremath{\pi}}}=(11/2\ensuremath{^{\textnormal{+}}}) (\href{https://www.nndc.bnl.gov/nsr/nsrlink.jsp?1972Pa29,B}{1972Pa29}), who suggested that this state may be a possible\vspace{0.1cm}}&\\
&&&&&\parbox[t][0.3cm]{11.691421cm}{\raggedright {\ }{\ }{\ }candidate for being the J\ensuremath{^{\ensuremath{\pi}}}=11/2\ensuremath{^{\textnormal{+}}} member of the K\ensuremath{^{\ensuremath{\pi}}}=1/2\ensuremath{^{\textnormal{+}}} band in \ensuremath{^{\textnormal{19}}}Ne based on\vspace{0.1cm}}&\\
&&&&&\parbox[t][0.3cm]{11.691421cm}{\raggedright {\ }{\ }{\ }the systematics of the K\ensuremath{^{\ensuremath{\pi}}}=1/2\ensuremath{^{\textnormal{+}}} bands in \ensuremath{^{\textnormal{19}}}F and \ensuremath{^{\textnormal{19}}}Ne. However, evidence for the\vspace{0.1cm}}&\\
&&&&&\parbox[t][0.3cm]{11.691421cm}{\raggedright {\ }{\ }{\ }J\ensuremath{^{\ensuremath{\pi}}} assignment of this state from the later studies do not support J\ensuremath{^{\ensuremath{\pi}}}=(11/2\ensuremath{^{\textnormal{+}}}).\vspace{0.1cm}}&\\
&&&&&\parbox[t][0.3cm]{11.691421cm}{\raggedright \ensuremath{\Gamma}\ensuremath{_{\ensuremath{\alpha}}}/\ensuremath{\Gamma}=0.541 \textit{29} deduced by (\href{https://www.nndc.bnl.gov/nsr/nsrlink.jsp?2023Ma57,B}{2023Ma57}) using \ensuremath{\Gamma}=292 keV \textit{107} from (\href{https://www.nndc.bnl.gov/nsr/nsrlink.jsp?2009Da07,B}{2009Da07}:\vspace{0.1cm}}&\\
&&&&&\parbox[t][0.3cm]{11.691421cm}{\raggedright {\ }{\ }{\ }\ensuremath{^{\textnormal{1}}}H(\ensuremath{^{\textnormal{19}}}Ne,p\ensuremath{'})).\vspace{0.1cm}}&\\
\multicolumn{1}{r@{}}{8.08\ensuremath{\times10^{3}}}&\multicolumn{1}{@{}l}{\ensuremath{^{{\hyperlink{NE25LEVEL13}{n}}}} {\it 2}}&&&&&\\
\multicolumn{1}{r@{}}{8.14\ensuremath{\times10^{3}}}&\multicolumn{1}{@{}l}{\ensuremath{^{{\hyperlink{NE25LEVEL8}{i}}}} {\it 10}}&&\multicolumn{1}{r@{}}{7}&\multicolumn{1}{@{.}l}{8\ensuremath{\times10^{-3}} {\it 57}}&\parbox[t][0.3cm]{11.691421cm}{\raggedright \ensuremath{\Gamma}\ensuremath{_{\textnormal{p}}}/\ensuremath{\Gamma}\ensuremath{\alpha}=1.20 \textit{12} (\href{https://www.nndc.bnl.gov/nsr/nsrlink.jsp?2023Ma57,B}{2023Ma57})\vspace{0.1cm}}&\\
&&&&&\parbox[t][0.3cm]{11.691421cm}{\raggedright E(level): From (\href{https://www.nndc.bnl.gov/nsr/nsrlink.jsp?2023Ma57,B}{2023Ma57}), where the uncertainty is systematic. They paired this\vspace{0.1cm}}&\\
&&&&&\parbox[t][0.3cm]{11.691421cm}{\raggedright {\ }{\ }{\ }state with the E\ensuremath{_{\textnormal{x}}}=7974 keV \textit{10} state measured by (\href{https://www.nndc.bnl.gov/nsr/nsrlink.jsp?2009Da07,B}{2009Da07}: \ensuremath{^{\textnormal{1}}}H(\ensuremath{^{\textnormal{19}}}Ne,p\ensuremath{'})).\vspace{0.1cm}}&\\
&&&&&\parbox[t][0.3cm]{11.691421cm}{\raggedright This state may have an \ensuremath{\alpha}-cluster structure (\href{https://www.nndc.bnl.gov/nsr/nsrlink.jsp?2023Ma57,B}{2023Ma57}).\vspace{0.1cm}}&\\
&&&&&\parbox[t][0.3cm]{11.691421cm}{\raggedright \ensuremath{\Gamma}\ensuremath{_{\ensuremath{\alpha}}}/\ensuremath{\Gamma}=0.455 \textit{25} deduced by (\href{https://www.nndc.bnl.gov/nsr/nsrlink.jsp?2023Ma57,B}{2023Ma57}) using \ensuremath{\Gamma}=11 keV \textit{8} from (\href{https://www.nndc.bnl.gov/nsr/nsrlink.jsp?2009Da07,B}{2009Da07}:\vspace{0.1cm}}&\\
&&&&&\parbox[t][0.3cm]{11.691421cm}{\raggedright {\ }{\ }{\ }\ensuremath{^{\textnormal{1}}}H(\ensuremath{^{\textnormal{19}}}Ne,p\ensuremath{'})).\vspace{0.1cm}}&\\
\multicolumn{1}{r@{}}{8.45\ensuremath{\times10^{3}}}&\multicolumn{1}{@{}l}{\ensuremath{^{{\hyperlink{NE25LEVEL8}{i}}{\hyperlink{NE25LEVEL13}{n}}}} {\it 2}}&&\multicolumn{1}{r@{}}{0}&\multicolumn{1}{@{.}l}{46\ensuremath{^{{\hyperlink{NE25LEVEL11}{l}}}} {\it 12}}&\parbox[t][0.3cm]{11.691421cm}{\raggedright \ensuremath{\Gamma}\ensuremath{_{\textnormal{p}}}/\ensuremath{\Gamma}\ensuremath{\alpha}=0.59 \textit{6} (\href{https://www.nndc.bnl.gov/nsr/nsrlink.jsp?2023Ma57,B}{2023Ma57})\vspace{0.1cm}}&\\
&&&&&\parbox[t][0.3cm]{11.691421cm}{\raggedright E(level): See also 8.51 MeV \textit{10} (sys.) (\href{https://www.nndc.bnl.gov/nsr/nsrlink.jsp?2023Ma57,B}{2023Ma57}). They reported that this state may\vspace{0.1cm}}&\\
&&&&&\parbox[t][0.3cm]{11.691421cm}{\raggedright {\ }{\ }{\ }correspond to the \ensuremath{^{\textnormal{19}}}Ne*(8428 keV, (13/2\ensuremath{^{-}})) state measured by (\href{https://www.nndc.bnl.gov/nsr/nsrlink.jsp?2017To14,B}{2017To14}:\vspace{0.1cm}}&\\
&&&&&\parbox[t][0.3cm]{11.691421cm}{\raggedright {\ }{\ }{\ }\ensuremath{^{\textnormal{4}}}He(\ensuremath{^{\textnormal{15}}}O,\ensuremath{\alpha})).\vspace{0.1cm}}&\\
&&&&&\parbox[t][0.3cm]{11.691421cm}{\raggedright \ensuremath{\Gamma}\ensuremath{_{\ensuremath{\alpha}}}/\ensuremath{\Gamma}=0.629 \textit{24} deduced by (\href{https://www.nndc.bnl.gov/nsr/nsrlink.jsp?2023Ma57,B}{2023Ma57}) using \ensuremath{\Gamma}=4 keV \textit{1} from (\href{https://www.nndc.bnl.gov/nsr/nsrlink.jsp?2017To14,B}{2017To14}:\vspace{0.1cm}}&\\
&&&&&\parbox[t][0.3cm]{11.691421cm}{\raggedright {\ }{\ }{\ }\ensuremath{^{\textnormal{4}}}He(\ensuremath{^{\textnormal{15}}}O,\ensuremath{\alpha})).\vspace{0.1cm}}&\\
&&&&&\parbox[t][0.3cm]{11.691421cm}{\raggedright \ensuremath{\theta}\ensuremath{^{\textnormal{2}}_{\ensuremath{\alpha}}}: This result implies the existence of a strong \ensuremath{\alpha}+\ensuremath{^{\textnormal{15}}}O cluster configuration for\vspace{0.1cm}}&\\
&&&&&\parbox[t][0.3cm]{11.691421cm}{\raggedright {\ }{\ }{\ }this state (\href{https://www.nndc.bnl.gov/nsr/nsrlink.jsp?2023Ma57,B}{2023Ma57}).\vspace{0.1cm}}&\\
\multicolumn{1}{r@{}}{8.70\ensuremath{\times10^{3}}}&\multicolumn{1}{@{ }l}{{\it 4}}&&&&\parbox[t][0.3cm]{11.691421cm}{\raggedright E(level): From (\href{https://www.nndc.bnl.gov/nsr/nsrlink.jsp?1972Pa29,B}{1972Pa29}).\vspace{0.1cm}}&\\
\multicolumn{1}{r@{}}{8.89\ensuremath{\times10^{3}}}&\multicolumn{1}{@{}l}{\ensuremath{^{{\hyperlink{NE25LEVEL8}{i}}}} {\it 10}}&&\multicolumn{1}{r@{}}{0}&\multicolumn{1}{@{.}l}{12\ensuremath{^{{\hyperlink{NE25LEVEL11}{l}}}} {\it 3}}&\parbox[t][0.3cm]{11.691421cm}{\raggedright \ensuremath{\Gamma}\ensuremath{_{\textnormal{p}}}/\ensuremath{\Gamma}\ensuremath{\alpha}=2.37 \textit{23} (\href{https://www.nndc.bnl.gov/nsr/nsrlink.jsp?2023Ma57,B}{2023Ma57})\vspace{0.1cm}}&\\
&&&&&\parbox[t][0.3cm]{11.691421cm}{\raggedright E(level): From 8.89 MeV \textit{10} (sys.) (\href{https://www.nndc.bnl.gov/nsr/nsrlink.jsp?2023Ma57,B}{2023Ma57}). These authors paired this state with\vspace{0.1cm}}&\\
&&&&&\parbox[t][0.3cm]{11.691421cm}{\raggedright {\ }{\ }{\ }the tentative \ensuremath{^{\textnormal{19}}}Ne*(8790 keV) state with J=(11/2) measured by (\href{https://www.nndc.bnl.gov/nsr/nsrlink.jsp?2017To14,B}{2017To14}:\vspace{0.1cm}}&\\
&&&&&\parbox[t][0.3cm]{11.691421cm}{\raggedright {\ }{\ }{\ }\ensuremath{^{\textnormal{4}}}He(\ensuremath{^{\textnormal{15}}}O,\ensuremath{\alpha})).\vspace{0.1cm}}&\\
\end{longtable}
\begin{textblock}{29}(0,27.3)
Continued on next page (footnotes at end of table)
\end{textblock}
\clearpage
\begin{longtable}{ccc@{\extracolsep{\fill}}c}
\\[-.4cm]
\multicolumn{4}{c}{{\bf \small \underline{\ensuremath{^{\textnormal{16}}}O(\ensuremath{^{\textnormal{6}}}Li,t)\hspace{0.2in}\href{https://www.nndc.bnl.gov/nsr/nsrlink.jsp?1972Ga08,B}{1972Ga08},\href{https://www.nndc.bnl.gov/nsr/nsrlink.jsp?1979Ma26,B}{1979Ma26},\href{https://www.nndc.bnl.gov/nsr/nsrlink.jsp?2023Ma57,B}{2023Ma57} (continued)}}}\\
\multicolumn{4}{c}{~}\\
\multicolumn{4}{c}{\underline{\ensuremath{^{19}}Ne Levels (continued)}}\\
\multicolumn{4}{c}{~}\\
\multicolumn{2}{c}{E(level)$^{{\hyperlink{NE25LEVEL2}{c}}{\hyperlink{NE25LEVEL3}{d}}{\hyperlink{NE25LEVEL4}{e}}}$}&Comments&\\[-.2cm]
\multicolumn{2}{c}{\hrulefill}&\hrulefill&
\endhead
&&\parbox[t][0.3cm]{15.207081cm}{\raggedright \ensuremath{\Gamma}\ensuremath{_{\ensuremath{\alpha}}}/\ensuremath{\Gamma}=0.297 \textit{20} deduced by (\href{https://www.nndc.bnl.gov/nsr/nsrlink.jsp?2023Ma57,B}{2023Ma57}) using \ensuremath{\Gamma}=4 keV \textit{1} from (\href{https://www.nndc.bnl.gov/nsr/nsrlink.jsp?2017To14,B}{2017To14}: \ensuremath{^{\textnormal{4}}}He(\ensuremath{^{\textnormal{15}}}O,\ensuremath{\alpha})).\vspace{0.1cm}}&\\
&&\parbox[t][0.3cm]{15.207081cm}{\raggedright \ensuremath{\theta}\ensuremath{^{\textnormal{2}}_{\ensuremath{\alpha}}}: This result implies the existence of a strong \ensuremath{\alpha}+\ensuremath{^{\textnormal{15}}}O cluster configuration for this state (\href{https://www.nndc.bnl.gov/nsr/nsrlink.jsp?2023Ma57,B}{2023Ma57}).\vspace{0.1cm}}&\\
&&\parbox[t][0.3cm]{15.207081cm}{\raggedright d\ensuremath{\sigma}/d\ensuremath{\Omega}\ensuremath{_{\textnormal{c.m.}}}=321 \ensuremath{\mu}b/sr (\href{https://www.nndc.bnl.gov/nsr/nsrlink.jsp?1979Ma26,B}{1979Ma26}).\vspace{0.1cm}}&\\
\multicolumn{1}{r@{}}{8.94\ensuremath{\times10^{3}}}&\multicolumn{1}{@{}l}{\ensuremath{^{{\hyperlink{NE25LEVEL13}{n}}}} {\it 2}}&&\\
\multicolumn{1}{r@{}}{9.25\ensuremath{\times10^{3}}}&\multicolumn{1}{@{}l}{\ensuremath{^{{\hyperlink{NE25LEVEL8}{i}}}} {\it 10}}&\parbox[t][0.3cm]{15.207081cm}{\raggedright \ensuremath{\Gamma}\ensuremath{_{\textnormal{p}}}/\ensuremath{\Gamma}\ensuremath{\alpha}=1.31 \textit{23} (\href{https://www.nndc.bnl.gov/nsr/nsrlink.jsp?2023Ma57,B}{2023Ma57})\vspace{0.1cm}}&\\
&&\parbox[t][0.3cm]{15.207081cm}{\raggedright E(level): From (\href{https://www.nndc.bnl.gov/nsr/nsrlink.jsp?2023Ma57,B}{2023Ma57}), where the uncertainty is systematic. Note that (\href{https://www.nndc.bnl.gov/nsr/nsrlink.jsp?2023Ma57,B}{2023Ma57}) claimed that the 9.25\vspace{0.1cm}}&\\
&&\parbox[t][0.3cm]{15.207081cm}{\raggedright {\ }{\ }{\ }MeV state was first observed in their experiment. However, the evaluator associates this state with the known\vspace{0.1cm}}&\\
&&\parbox[t][0.3cm]{15.207081cm}{\raggedright {\ }{\ }{\ }state at 9240 keV \textit{20} (\href{https://www.nndc.bnl.gov/nsr/nsrlink.jsp?1972Ha03,B}{1972Ha03}).\vspace{0.1cm}}&\\
&&\parbox[t][0.3cm]{15.207081cm}{\raggedright Decay modes: p and \ensuremath{\alpha} (\href{https://www.nndc.bnl.gov/nsr/nsrlink.jsp?2023Ma57,B}{2023Ma57}).\vspace{0.1cm}}&\\
\multicolumn{1}{r@{}}{9.38\ensuremath{\times10^{3}}}&\multicolumn{1}{@{ }l}{{\it 4}}&\parbox[t][0.3cm]{15.207081cm}{\raggedright E(level): From (\href{https://www.nndc.bnl.gov/nsr/nsrlink.jsp?1972Pa29,B}{1972Pa29}).\vspace{0.1cm}}&\\
\multicolumn{1}{r@{}}{9.81\ensuremath{\times10^{3}}}&\multicolumn{1}{@{}l}{\ensuremath{^{{\hyperlink{NE25LEVEL13}{n}}}} {\it 2}}&\parbox[t][0.3cm]{15.207081cm}{\raggedright E(level): See also 9.77 MeV \textit{10} (\href{https://www.nndc.bnl.gov/nsr/nsrlink.jsp?2023Ma57,B}{2023Ma57}), who claimed that the 9.77 MeV state was first observed in their\vspace{0.1cm}}&\\
&&\parbox[t][0.3cm]{15.207081cm}{\raggedright {\ }{\ }{\ }experiment and paired this state with the E\ensuremath{_{\textnormal{x}}}=9788 keV \textit{13} level from (\href{https://www.nndc.bnl.gov/nsr/nsrlink.jsp?2011Ad24,B}{2011Ad24}: \ensuremath{^{\textnormal{2}}}H(\ensuremath{^{\textnormal{18}}}F,\ensuremath{^{\textnormal{19}}}Ne)).\vspace{0.1cm}}&\\
&&\parbox[t][0.3cm]{15.207081cm}{\raggedright (d\ensuremath{\sigma}/d\ensuremath{\Omega})\ensuremath{_{\textnormal{c.m.}}}=364 \ensuremath{\mu}b/sr (\href{https://www.nndc.bnl.gov/nsr/nsrlink.jsp?1979Ma26,B}{1979Ma26}).\vspace{0.1cm}}&\\
&&\parbox[t][0.3cm]{15.207081cm}{\raggedright Decay mode: \ensuremath{\alpha} (\href{https://www.nndc.bnl.gov/nsr/nsrlink.jsp?2023Ma57,B}{2023Ma57}).\vspace{0.1cm}}&\\
\multicolumn{1}{r@{}}{10.01\ensuremath{\times10^{3}}}&\multicolumn{1}{@{}l}{\ensuremath{^{{\hyperlink{NE25LEVEL13}{n}}}} {\it 2}}&\parbox[t][0.3cm]{15.207081cm}{\raggedright (d\ensuremath{\sigma}/d\ensuremath{\Omega})\ensuremath{_{\textnormal{c.m.}}}=246 \ensuremath{\mu}b/sr (\href{https://www.nndc.bnl.gov/nsr/nsrlink.jsp?1979Ma26,B}{1979Ma26}).\vspace{0.1cm}}&\\
\multicolumn{1}{r@{}}{10.23\ensuremath{\times10^{3}}}&\multicolumn{1}{@{}l}{\ensuremath{^{{\hyperlink{NE25LEVEL7}{h}}}} {\it 10}}&\parbox[t][0.3cm]{15.207081cm}{\raggedright E(level): From (\href{https://www.nndc.bnl.gov/nsr/nsrlink.jsp?2023Ma57,B}{2023Ma57}), where the uncertainty is systematic.\vspace{0.1cm}}&\\
&&\parbox[t][0.3cm]{15.207081cm}{\raggedright Decay mode: \ensuremath{\alpha} (\href{https://www.nndc.bnl.gov/nsr/nsrlink.jsp?2023Ma57,B}{2023Ma57}).\vspace{0.1cm}}&\\
\multicolumn{1}{r@{}}{11.08\ensuremath{\times10^{3}}}&\multicolumn{1}{@{}l}{\ensuremath{^{{\hyperlink{NE25LEVEL13}{n}}}} {\it 2}}&\parbox[t][0.3cm]{15.207081cm}{\raggedright (d\ensuremath{\sigma}/d\ensuremath{\Omega})\ensuremath{_{\textnormal{c.m.}}}=200 \ensuremath{\mu}b/sr (\href{https://www.nndc.bnl.gov/nsr/nsrlink.jsp?1979Ma26,B}{1979Ma26}).\vspace{0.1cm}}&\\
\multicolumn{1}{r@{}}{11.24\ensuremath{\times10^{3}}}&\multicolumn{1}{@{}l}{\ensuremath{^{{\hyperlink{NE25LEVEL13}{n}}}} {\it 2}}&\parbox[t][0.3cm]{15.207081cm}{\raggedright (d\ensuremath{\sigma}/d\ensuremath{\Omega})\ensuremath{_{\textnormal{c.m.}}}=200 \ensuremath{\mu}b/sr (\href{https://www.nndc.bnl.gov/nsr/nsrlink.jsp?1979Ma26,B}{1979Ma26}).\vspace{0.1cm}}&\\
\multicolumn{1}{r@{}}{11.40\ensuremath{\times10^{3}}}&\multicolumn{1}{@{}l}{\ensuremath{^{{\hyperlink{NE25LEVEL13}{n}}}} {\it 2}}&\parbox[t][0.3cm]{15.207081cm}{\raggedright (d\ensuremath{\sigma}/d\ensuremath{\Omega})\ensuremath{_{\textnormal{c.m.}}}=200 \ensuremath{\mu}b/sr (\href{https://www.nndc.bnl.gov/nsr/nsrlink.jsp?1979Ma26,B}{1979Ma26}).\vspace{0.1cm}}&\\
\multicolumn{1}{r@{}}{12.56\ensuremath{\times10^{3}}}&\multicolumn{1}{@{}l}{\ensuremath{^{{\hyperlink{NE25LEVEL13}{n}}}} {\it 2}}&\parbox[t][0.3cm]{15.207081cm}{\raggedright (d\ensuremath{\sigma}/d\ensuremath{\Omega})\ensuremath{_{\textnormal{c.m.}}}=273 \ensuremath{\mu}b/sr (\href{https://www.nndc.bnl.gov/nsr/nsrlink.jsp?1979Ma26,B}{1979Ma26}).\vspace{0.1cm}}&\\
\multicolumn{1}{r@{}}{13.1\ensuremath{\times10^{3}}}&\multicolumn{1}{@{}l}{\ensuremath{^{{\hyperlink{NE25LEVEL13}{n}}}} {\it 3}}&&\\
\multicolumn{1}{r@{}}{13.22\ensuremath{\times10^{3}}}&\multicolumn{1}{@{}l}{\ensuremath{^{{\hyperlink{NE25LEVEL13}{n}}}} {\it 3}}&&\\
\multicolumn{1}{r@{}}{14.18\ensuremath{\times10^{3}}}&\multicolumn{1}{@{}l}{\ensuremath{^{{\hyperlink{NE25LEVEL13}{n}}}} {\it 3}}&\parbox[t][0.3cm]{15.207081cm}{\raggedright (d\ensuremath{\sigma}/d\ensuremath{\Omega})\ensuremath{_{\textnormal{c.m.}}}=72 \ensuremath{\mu}b/sr (\href{https://www.nndc.bnl.gov/nsr/nsrlink.jsp?1979Ma26,B}{1979Ma26}).\vspace{0.1cm}}&\\
\multicolumn{1}{r@{}}{14.44\ensuremath{\times10^{3}}}&\multicolumn{1}{@{}l}{\ensuremath{^{{\hyperlink{NE25LEVEL13}{n}}}} {\it 3}}&&\\
\multicolumn{1}{r@{}}{14.78\ensuremath{\times10^{3}}}&\multicolumn{1}{@{}l}{\ensuremath{^{{\hyperlink{NE25LEVEL13}{n}}}} {\it 3}}&\parbox[t][0.3cm]{15.207081cm}{\raggedright (d\ensuremath{\sigma}/d\ensuremath{\Omega})\ensuremath{_{\textnormal{c.m.}}}=181 \ensuremath{\mu}b/sr (\href{https://www.nndc.bnl.gov/nsr/nsrlink.jsp?1979Ma26,B}{1979Ma26}).\vspace{0.1cm}}&\\
\end{longtable}
\parbox[b][0.3cm]{17.7cm}{\makebox[1ex]{\ensuremath{^{\hypertarget{NE25LEVEL0}{a}}}} Seq.(A): K\ensuremath{^{\ensuremath{\pi}}}=1/2\ensuremath{^{+}} g.s. band (\href{https://www.nndc.bnl.gov/nsr/nsrlink.jsp?1971Bi06,B}{1971Bi06}).}\\
\parbox[b][0.3cm]{17.7cm}{\makebox[1ex]{\ensuremath{^{\hypertarget{NE25LEVEL1}{b}}}} Seq.(B): K\ensuremath{^{\ensuremath{\pi}}}=1/2\ensuremath{^{-}} band (\href{https://www.nndc.bnl.gov/nsr/nsrlink.jsp?1971Bi06,B}{1971Bi06}).}\\
\parbox[b][0.3cm]{17.7cm}{\makebox[1ex]{\ensuremath{^{\hypertarget{NE25LEVEL2}{c}}}} The uncertainties in the excitation energies deduced by (\href{https://www.nndc.bnl.gov/nsr/nsrlink.jsp?1972Pa29,B}{1972Pa29}) were reported to be \ensuremath{\pm}40 keV. See text.}\\
\parbox[b][0.3cm]{17.7cm}{\makebox[1ex]{\ensuremath{^{\hypertarget{NE25LEVEL3}{d}}}} A 100-keV systematic uncertainty is estimated by (\href{https://www.nndc.bnl.gov/nsr/nsrlink.jsp?2023Ma57,B}{2023Ma57}) for all their measured excitation energies due to the energy}\\
\parbox[b][0.3cm]{17.7cm}{{\ }{\ }calibration of their Si detectors.}\\
\parbox[b][0.3cm]{17.7cm}{\makebox[1ex]{\ensuremath{^{\hypertarget{NE25LEVEL4}{e}}}} The energies reported by (\href{https://www.nndc.bnl.gov/nsr/nsrlink.jsp?1971Bi06,B}{1971Bi06}) are from (\href{https://www.nndc.bnl.gov/nsr/nsrlink.jsp?1970Ga18,B}{1970Ga18}: \ensuremath{^{\textnormal{20}}}Ne(\ensuremath{^{\textnormal{3}}}He,\ensuremath{\alpha})).}\\
\parbox[b][0.3cm]{17.7cm}{\makebox[1ex]{\ensuremath{^{\hypertarget{NE25LEVEL5}{f}}}} This state is a member of K\ensuremath{^{\ensuremath{\pi}}}=1/2\ensuremath{^{\textnormal{+}}} rotational band (\href{https://www.nndc.bnl.gov/nsr/nsrlink.jsp?1971Bi06,B}{1971Bi06}, \href{https://www.nndc.bnl.gov/nsr/nsrlink.jsp?1972Ga08,B}{1972Ga08}, \href{https://www.nndc.bnl.gov/nsr/nsrlink.jsp?1972Pa29,B}{1972Pa29}). (\href{https://www.nndc.bnl.gov/nsr/nsrlink.jsp?1971Bi06,B}{1971Bi06}) assumed this band to be based}\\
\parbox[b][0.3cm]{17.7cm}{{\ }{\ }on three particles and 0 holes in the 2\textit{s}{\textminus}1\textit{d} shell outside the closed 1\textit{s}{\textminus}1\textit{p} shell in the \ensuremath{^{\textnormal{16}}}O\ensuremath{_{\textnormal{g.s.}}} core.}\\
\parbox[b][0.3cm]{17.7cm}{\makebox[1ex]{\ensuremath{^{\hypertarget{NE25LEVEL6}{g}}}} This state is a member of K\ensuremath{^{\ensuremath{\pi}}}=1/2\ensuremath{^{-}} rotational band (\href{https://www.nndc.bnl.gov/nsr/nsrlink.jsp?1971Bi06,B}{1971Bi06}). (\href{https://www.nndc.bnl.gov/nsr/nsrlink.jsp?1978Pi06,B}{1978Pi06}) argued that core excitation through admixture of}\\
\parbox[b][0.3cm]{17.7cm}{{\ }{\ }2p-2h and 4p-4h configurations in the \ensuremath{^{\textnormal{16}}}O\ensuremath{_{\textnormal{g.s.}}} core suggested by (\href{https://www.nndc.bnl.gov/nsr/nsrlink.jsp?1971Bi06,B}{1971Bi06}) is insufficient to account for the relatively strong}\\
\parbox[b][0.3cm]{17.7cm}{{\ }{\ }population of these states observed by (\href{https://www.nndc.bnl.gov/nsr/nsrlink.jsp?1971Bi06,B}{1971Bi06}, \href{https://www.nndc.bnl.gov/nsr/nsrlink.jsp?1972Ga08,B}{1972Ga08}). Instead, (\href{https://www.nndc.bnl.gov/nsr/nsrlink.jsp?1978Pi06,B}{1978Pi06}) suggested that a 5\% admixture of \textit{sd}\ensuremath{^{\textnormal{2}}}\textit{pf} with}\\
\parbox[b][0.3cm]{17.7cm}{{\ }{\ }\textit{sd}\ensuremath{^{\textnormal{4}}}\textit{p}\ensuremath{~^{\textnormal{$-$1}}} could account very well for the experimentally observed \ensuremath{^{\textnormal{3}}}He cluster transfer strengths in (\href{https://www.nndc.bnl.gov/nsr/nsrlink.jsp?1971Bi06,B}{1971Bi06}, \href{https://www.nndc.bnl.gov/nsr/nsrlink.jsp?1972Ga08,B}{1972Ga08}).}\\
\parbox[b][0.3cm]{17.7cm}{\makebox[1ex]{\ensuremath{^{\hypertarget{NE25LEVEL7}{h}}}} Measured for the first time by (\href{https://www.nndc.bnl.gov/nsr/nsrlink.jsp?2023Ma57,B}{2023Ma57}).}\\
\parbox[b][0.3cm]{17.7cm}{\makebox[1ex]{\ensuremath{^{\hypertarget{NE25LEVEL8}{i}}}} (\href{https://www.nndc.bnl.gov/nsr/nsrlink.jsp?2023Ma57,B}{2023Ma57}) reconstructed this state from both the \ensuremath{\alpha}+\ensuremath{^{\textnormal{15}}}O and p+\ensuremath{^{\textnormal{18}}}F decay channels.}\\
\parbox[b][0.3cm]{17.7cm}{\makebox[1ex]{\ensuremath{^{\hypertarget{NE25LEVEL9}{j}}}} From (\href{https://www.nndc.bnl.gov/nsr/nsrlink.jsp?1972Pa29,B}{1972Pa29}): Based on simple comparisons of the shapes of the experimental triton angular distributions with the theoretical}\\
\parbox[b][0.3cm]{17.7cm}{{\ }{\ }angular distributions (using SU(3) shell model) for various L-transfers obtained through private communication between those}\\
\parbox[b][0.3cm]{17.7cm}{{\ }{\ }authors and D. Strottman. These calculations are not presented.}\\
\parbox[b][0.3cm]{17.7cm}{\makebox[1ex]{\ensuremath{^{\hypertarget{NE25LEVEL10}{k}}}} Dimensionless reduced \ensuremath{\alpha} width from (\href{https://www.nndc.bnl.gov/nsr/nsrlink.jsp?2023Ma57,B}{2023Ma57}). Note that an additional 15\% systematic uncertainty should be considered for}\\
\parbox[b][0.3cm]{17.7cm}{{\ }{\ }the \ensuremath{\theta}\ensuremath{^{\textnormal{2}}_{\ensuremath{\alpha}}} values, see text in (\href{https://www.nndc.bnl.gov/nsr/nsrlink.jsp?2023Ma57,B}{2023Ma57}).}\\
\parbox[b][0.3cm]{17.7cm}{\makebox[1ex]{\ensuremath{^{\hypertarget{NE25LEVEL11}{l}}}} This value is measured for the first time by (\href{https://www.nndc.bnl.gov/nsr/nsrlink.jsp?2023Ma57,B}{2023Ma57}).}\\
\parbox[b][0.3cm]{17.7cm}{\makebox[1ex]{\ensuremath{^{\hypertarget{NE25LEVEL12}{m}}}} From (\href{https://www.nndc.bnl.gov/nsr/nsrlink.jsp?1971Bi06,B}{1971Bi06}) based on comparison of the relative \ensuremath{^{\textnormal{16}}}O(\ensuremath{^{\textnormal{6}}}Li,\ensuremath{^{\textnormal{3}}}He) and \ensuremath{^{\textnormal{16}}}O(\ensuremath{^{\textnormal{6}}}Li,t) transition strengths populating the \ensuremath{^{\textnormal{19}}}F* and}\\
\parbox[b][0.3cm]{17.7cm}{{\ }{\ }\ensuremath{^{\textnormal{19}}}Ne* analog states, respectively, and from mirror states analogy. Since these are weak arguments, we made those assignments}\\
\begin{textblock}{29}(0,27.3)
Continued on next page (footnotes at end of table)
\end{textblock}
\clearpage
\vspace*{-0.5cm}
{\bf \small \underline{\ensuremath{^{\textnormal{16}}}O(\ensuremath{^{\textnormal{6}}}Li,t)\hspace{0.2in}\href{https://www.nndc.bnl.gov/nsr/nsrlink.jsp?1972Ga08,B}{1972Ga08},\href{https://www.nndc.bnl.gov/nsr/nsrlink.jsp?1979Ma26,B}{1979Ma26},\href{https://www.nndc.bnl.gov/nsr/nsrlink.jsp?2023Ma57,B}{2023Ma57} (continued)}}\\
\vspace{0.3cm}
\underline{$^{19}$Ne Levels (continued)}\\
\vspace{0.3cm}
\parbox[b][0.3cm]{17.7cm}{{\ }{\ }tentative.}\\
\parbox[b][0.3cm]{17.7cm}{\makebox[1ex]{\ensuremath{^{\hypertarget{NE25LEVEL13}{n}}}} From (\href{https://www.nndc.bnl.gov/nsr/nsrlink.jsp?1979Ma26,B}{1979Ma26}).}\\
\vspace{0.5cm}
\clearpage
\clearpage
\begin{figure}[h]
\begin{center}
\includegraphics{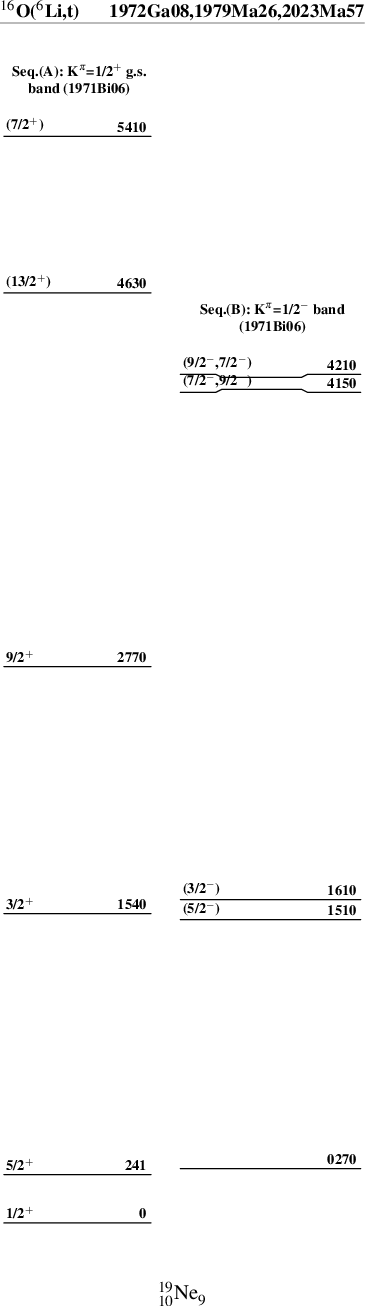}\\
\end{center}
\end{figure}
\clearpage
\subsection[\hspace{-0.2cm}\ensuremath{^{\textnormal{16}}}O(\ensuremath{^{\textnormal{10}}}B,\ensuremath{^{\textnormal{7}}}Li)]{ }
\vspace{-27pt}
\vspace{0.3cm}
\hypertarget{NE26}{{\bf \small \underline{\ensuremath{^{\textnormal{16}}}O(\ensuremath{^{\textnormal{10}}}B,\ensuremath{^{\textnormal{7}}}Li)\hspace{0.2in}\href{https://www.nndc.bnl.gov/nsr/nsrlink.jsp?1976Ha06,B}{1976Ha06}}}}\\
\vspace{4pt}
\vspace{8pt}
\parbox[b][0.3cm]{17.7cm}{\addtolength{\parindent}{-0.2in}\ensuremath{^{\textnormal{3}}}He transfer reaction.}\\
\parbox[b][0.3cm]{17.7cm}{\addtolength{\parindent}{-0.2in}J\ensuremath{^{\ensuremath{\pi}}}(\ensuremath{^{\textnormal{16}}}O\ensuremath{_{\textnormal{g.s.}}})=0\ensuremath{^{\textnormal{+}}} and J\ensuremath{^{\ensuremath{\pi}}}(\ensuremath{^{\textnormal{10}}}B\ensuremath{_{\textnormal{g.s.}}})=3\ensuremath{^{\textnormal{+}}}.}\\
\parbox[b][0.3cm]{17.7cm}{\addtolength{\parindent}{-0.2in}\href{https://www.nndc.bnl.gov/nsr/nsrlink.jsp?1975HaYZ,B}{1975HaYZ}: \ensuremath{^{\textnormal{16}}}O(\ensuremath{^{\textnormal{10}}}B,\ensuremath{^{\textnormal{7}}}Li) E=100 MeV; measured \ensuremath{\sigma}.}\\
\parbox[b][0.3cm]{17.7cm}{\addtolength{\parindent}{-0.2in}\href{https://www.nndc.bnl.gov/nsr/nsrlink.jsp?1975NaZF,B}{1975NaZF}: \ensuremath{^{\textnormal{16}}}O(\ensuremath{^{\textnormal{10}}}B,\ensuremath{^{\textnormal{7}}}Li); measured \ensuremath{\sigma}.}\\
\parbox[b][0.3cm]{17.7cm}{\addtolength{\parindent}{-0.2in}\href{https://www.nndc.bnl.gov/nsr/nsrlink.jsp?1976Ha06,B}{1976Ha06}: \ensuremath{^{\textnormal{16}}}O(\ensuremath{^{\textnormal{10}}}B,\ensuremath{^{\textnormal{7}}}Li) E=100 MeV; measured reaction products using a Si surface barrier \ensuremath{\Delta}E-E telescope with the resolution of}\\
\parbox[b][0.3cm]{17.7cm}{\ensuremath{\Delta}E(FWHM)=250-300 keV. Measured excitation function at \ensuremath{\theta}\ensuremath{_{\textnormal{lab}}}=10.8\ensuremath{^\circ}. Deduced \ensuremath{^{\textnormal{19}}}Ne levels. Performed deformed model}\\
\parbox[b][0.3cm]{17.7cm}{calculations to predict the energies of high-lying states. The results showed that these levels could be explained by (\textit{sd})\ensuremath{^{\textnormal{2}}}(\textit{fp})\ensuremath{^{\textnormal{1}}} and}\\
\parbox[b][0.3cm]{17.7cm}{(\textit{fp})\ensuremath{^{\textnormal{3}}} configurations outside an \ensuremath{^{\textnormal{16}}}O\ensuremath{_{\textnormal{g.s.}}} core. Mirror levels comparison and discussions on the \ensuremath{^{\textnormal{19}}}F rotational bands are provided.}\\
\parbox[b][0.3cm]{17.7cm}{\addtolength{\parindent}{-0.2in}\href{https://www.nndc.bnl.gov/nsr/nsrlink.jsp?1977HaZN,B}{1977HaZN}: \ensuremath{^{\textnormal{16}}}O(\ensuremath{^{\textnormal{10}}}B,\ensuremath{^{\textnormal{7}}}Li); measured \ensuremath{\sigma}; deduced \ensuremath{^{\textnormal{19}}}Ne levels, rotational band, and J\ensuremath{^{\ensuremath{\pi}}}.}\\
\vspace{12pt}
\underline{$^{19}$Ne Levels}\\
\begin{longtable}{cccccc@{\extracolsep{\fill}}c}
\multicolumn{2}{c}{E(level)$^{{\hyperlink{NE26LEVEL1}{b}}}$}&J$^{\pi}$$^{{\hyperlink{NE26LEVEL2}{c}}}$&\multicolumn{2}{c}{E\ensuremath{_{\textnormal{x}}}(\ensuremath{^{\textnormal{19}}}F*) Mirror (MeV)$^{{\hyperlink{NE26LEVEL3}{d}}}$}&Comments&\\[-.2cm]
\multicolumn{2}{c}{\hrulefill}&\hrulefill&\multicolumn{2}{c}{\hrulefill}&\hrulefill&
\endfirsthead
\multicolumn{1}{r@{}}{0}&\multicolumn{1}{@{}l}{\ensuremath{^{{\hyperlink{NE26LEVEL0}{a}}}}}&\multicolumn{1}{l}{[1/2\ensuremath{^{+}}]}&&&&\\
\multicolumn{1}{r@{}}{238}&\multicolumn{1}{@{}l}{\ensuremath{^{{\hyperlink{NE26LEVEL0}{a}}}}}&\multicolumn{1}{l}{[5/2\ensuremath{^{+}}]}&\multicolumn{1}{r@{}}{0}&\multicolumn{1}{@{.}l}{20}&&\\
\multicolumn{1}{r@{}}{1.54\ensuremath{\times10^{3}}}&\multicolumn{1}{@{}l}{\ensuremath{^{{\hyperlink{NE26LEVEL0}{a}}}}}&\multicolumn{1}{l}{[3/2\ensuremath{^{+}}]}&\multicolumn{1}{r@{}}{1}&\multicolumn{1}{@{.}l}{55}&&\\
\multicolumn{1}{r@{}}{2.79\ensuremath{\times10^{3}}}&\multicolumn{1}{@{}l}{\ensuremath{^{{\hyperlink{NE26LEVEL0}{a}}}}}&\multicolumn{1}{l}{[9/2\ensuremath{^{+}}]}&\multicolumn{1}{r@{}}{2}&\multicolumn{1}{@{.}l}{78}&&\\
\multicolumn{1}{r@{}}{4.20\ensuremath{\times10^{3}}}&\multicolumn{1}{@{}l}{}&\multicolumn{1}{l}{(7/2\ensuremath{^{-}},9/2\ensuremath{^{-}})}&\multicolumn{1}{r@{}}{4}&\multicolumn{1}{@{.}l}{03}&&\\
\multicolumn{1}{r@{}}{4.62\ensuremath{\times10^{3}}}&\multicolumn{1}{@{}l}{\ensuremath{^{{\hyperlink{NE26LEVEL0}{a}}}}}&\multicolumn{1}{l}{[13/2\ensuremath{^{+}}]}&\multicolumn{1}{r@{}}{4}&\multicolumn{1}{@{.}l}{65}&&\\
\multicolumn{1}{r@{}}{5.44\ensuremath{\times10^{3}}}&\multicolumn{1}{@{}l}{\ensuremath{^{{\hyperlink{NE26LEVEL0}{a}}}}}&\multicolumn{1}{l}{[7/2\ensuremath{^{+}}]}&\multicolumn{1}{r@{}}{5}&\multicolumn{1}{@{.}l}{47}&&\\
\multicolumn{1}{r@{}}{6.03\ensuremath{\times10^{3}}}&\multicolumn{1}{@{}l}{}&&&&&\\
\multicolumn{1}{r@{}}{6.28\ensuremath{\times10^{3}}}&\multicolumn{1}{@{}l}{}&&&&&\\
\multicolumn{1}{r@{}}{6.77\ensuremath{\times10^{3}}}&\multicolumn{1}{@{}l}{}&&&&&\\
\multicolumn{1}{r@{}}{7.64\ensuremath{\times10^{3}}}&\multicolumn{1}{@{}l}{}&&&&&\\
\multicolumn{1}{r@{}}{8.20\ensuremath{\times10^{3}}}&\multicolumn{1}{@{}l}{}&&&&&\\
\multicolumn{1}{r@{}}{8.41\ensuremath{\times10^{3}}}&\multicolumn{1}{@{}l}{}&&&&&\\
\multicolumn{1}{r@{}}{8.94\ensuremath{\times10^{3}}}&\multicolumn{1}{@{}l}{}&&\multicolumn{1}{r@{}}{8}&\multicolumn{1}{@{.}l}{98}&&\\
\multicolumn{1}{r@{}}{9.88\ensuremath{\times10^{3}}}&\multicolumn{1}{@{}l}{}&&&&\parbox[t][0.3cm]{10.092839cm}{\raggedright J\ensuremath{^{\pi}}: (\href{https://www.nndc.bnl.gov/nsr/nsrlink.jsp?1976Ha06,B}{1976Ha06}): In an attempt to locate the \ensuremath{^{\textnormal{19}}}Ne* state mirror to the\vspace{0.1cm}}&\\
&&&&&\parbox[t][0.3cm]{10.092839cm}{\raggedright {\ }{\ }{\ }\ensuremath{^{\textnormal{19}}}F*(10.42 MeV, J\ensuremath{^{\ensuremath{\pi}}}=13/2\ensuremath{^{\textnormal{+}}_{\textnormal{2}}}) level, a comparison was made for the\vspace{0.1cm}}&\\
&&&&&\parbox[t][0.3cm]{10.092839cm}{\raggedright {\ }{\ }{\ }theoretical decay widths of \ensuremath{^{\textnormal{19}}}F and \ensuremath{^{\textnormal{19}}}Ne mirror levels in the E\ensuremath{_{\textnormal{x}}}=8-12\vspace{0.1cm}}&\\
&&&&&\parbox[t][0.3cm]{10.092839cm}{\raggedright {\ }{\ }{\ }MeV region assuming that Coulomb penetrabilities and reduced widths\vspace{0.1cm}}&\\
&&&&&\parbox[t][0.3cm]{10.092839cm}{\raggedright {\ }{\ }{\ }were less than 10\% of the Wigner limits. The results indicated that a\vspace{0.1cm}}&\\
&&&&&\parbox[t][0.3cm]{10.092839cm}{\raggedright {\ }{\ }{\ }\ensuremath{^{\textnormal{19}}}Ne* level with E\ensuremath{_{\textnormal{x}}}=10-11 MeV and J\ensuremath{^{\ensuremath{\pi}}}\ensuremath{\leq}13/2\ensuremath{^{-}} or J\ensuremath{\leq}11/2 would be\vspace{0.1cm}}&\\
&&&&&\parbox[t][0.3cm]{10.092839cm}{\raggedright {\ }{\ }{\ }considerably (500 keV or more) broader than such a level in \ensuremath{^{\textnormal{19}}}F* due\vspace{0.1cm}}&\\
&&&&&\parbox[t][0.3cm]{10.092839cm}{\raggedright {\ }{\ }{\ }to the differences in proton decay of these nuclei in this region. A\vspace{0.1cm}}&\\
&&&&&\parbox[t][0.3cm]{10.092839cm}{\raggedright {\ }{\ }{\ }\ensuremath{^{\textnormal{19}}}Ne* level at E\ensuremath{_{\textnormal{x}}}=10 MeV was expected to have a small decay width\vspace{0.1cm}}&\\
&&&&&\parbox[t][0.3cm]{10.092839cm}{\raggedright {\ }{\ }{\ }of \ensuremath{\Gamma}\ensuremath{<}100 keV. Thus, the authors mentioned that the only reasonable\vspace{0.1cm}}&\\
&&&&&\parbox[t][0.3cm]{10.092839cm}{\raggedright {\ }{\ }{\ }explanation was that the \ensuremath{^{\textnormal{19}}}Ne*(9.88 MeV) state was a group of\vspace{0.1cm}}&\\
&&&&&\parbox[t][0.3cm]{10.092839cm}{\raggedright {\ }{\ }{\ }unresolved states, which consisted of the mirror levels to the \ensuremath{^{\textnormal{19}}}F*(9.87\vspace{0.1cm}}&\\
&&&&&\parbox[t][0.3cm]{10.092839cm}{\raggedright {\ }{\ }{\ }MeV) and \ensuremath{^{\textnormal{19}}}F*(10.42 MeV, J\ensuremath{^{\ensuremath{\pi}}}=13/2\ensuremath{^{\textnormal{+}}}) states.\vspace{0.1cm}}&\\
\multicolumn{1}{r@{}}{10.20\ensuremath{\times10^{3}}}&\multicolumn{1}{@{}l}{}&&&&&\\
\multicolumn{1}{r@{}}{11.09\ensuremath{\times10^{3}}}&\multicolumn{1}{@{}l}{}&\multicolumn{1}{l}{(13/2\ensuremath{^{-}},11/2)}&\multicolumn{1}{r@{}}{11}&\multicolumn{1}{@{.}l}{33}&\parbox[t][0.3cm]{10.092839cm}{\raggedright J\ensuremath{^{\pi}}: From (\href{https://www.nndc.bnl.gov/nsr/nsrlink.jsp?1976Ha06,B}{1976Ha06}) based on the theoretical decay widths and\vspace{0.1cm}}&\\
&&&&&\parbox[t][0.3cm]{10.092839cm}{\raggedright {\ }{\ }{\ }semi-classical transition strengths deduced at this excitation energy.\vspace{0.1cm}}&\\
\multicolumn{1}{r@{}}{12.48\ensuremath{\times10^{3}}}&\multicolumn{1}{@{}l}{}&&\multicolumn{1}{r@{}}{12}&\multicolumn{1}{@{.}l}{79}&&\\
\multicolumn{1}{r@{}}{14.17\ensuremath{\times10^{3}}}&\multicolumn{1}{@{}l}{}&&\multicolumn{1}{r@{}}{14}&\multicolumn{1}{@{.}l}{15}&&\\
\multicolumn{1}{r@{}}{14.61\ensuremath{\times10^{3}}}&\multicolumn{1}{@{}l}{}&&\multicolumn{1}{r@{}}{14}&\multicolumn{1}{@{.}l}{99}&&\\
\multicolumn{1}{r@{}}{15.40\ensuremath{\times10^{3}}}&\multicolumn{1}{@{}l}{}&&\multicolumn{1}{r@{}}{15}&\multicolumn{1}{@{.}l}{54}&&\\
\end{longtable}
\parbox[b][0.3cm]{17.7cm}{\makebox[1ex]{\ensuremath{^{\hypertarget{NE26LEVEL0}{a}}}} Seq.(A): K\ensuremath{^{\ensuremath{\pi}}}=1/2\ensuremath{^{+}} g.s. band (\href{https://www.nndc.bnl.gov/nsr/nsrlink.jsp?1976Ha06,B}{1976Ha06}).}\\
\parbox[b][0.3cm]{17.7cm}{\makebox[1ex]{\ensuremath{^{\hypertarget{NE26LEVEL1}{b}}}} From (\href{https://www.nndc.bnl.gov/nsr/nsrlink.jsp?1976Ha06,B}{1976Ha06}): The excitation energy uncertainty was \ensuremath{\pm}100 keV for high excitation energies.}\\
\parbox[b][0.3cm]{17.7cm}{\makebox[1ex]{\ensuremath{^{\hypertarget{NE26LEVEL2}{c}}}} From (\href{https://www.nndc.bnl.gov/nsr/nsrlink.jsp?1976Ha06,B}{1976Ha06}): Based on mirror level analysis; comparison of the \ensuremath{^{\textnormal{19}}}F-\ensuremath{^{\textnormal{19}}}Ne K\ensuremath{^{\ensuremath{\pi}}}=1/2\ensuremath{^{\textnormal{+}}} ground state rotational bands; and the}\\
\parbox[b][0.3cm]{17.7cm}{{\ }{\ }results of (\href{https://www.nndc.bnl.gov/nsr/nsrlink.jsp?1974Ts03,B}{1974Ts03}: \ensuremath{^{\textnormal{16}}}O(\ensuremath{^{\textnormal{7}}}Li,\ensuremath{\alpha})) and (\href{https://www.nndc.bnl.gov/nsr/nsrlink.jsp?1972Bi14,B}{1972Bi14}: \ensuremath{^{\textnormal{16}}}O(\ensuremath{^{\textnormal{6}}}Li,\ensuremath{^{\textnormal{3}}}He)) for the J\ensuremath{^{\ensuremath{\pi}}} assignments of \ensuremath{^{\textnormal{19}}}F* mirror levels and (\href{https://www.nndc.bnl.gov/nsr/nsrlink.jsp?1972Pa29,B}{1972Pa29}:}\\
\begin{textblock}{29}(0,27.3)
Continued on next page (footnotes at end of table)
\end{textblock}
\clearpage
\vspace*{-0.5cm}
{\bf \small \underline{\ensuremath{^{\textnormal{16}}}O(\ensuremath{^{\textnormal{10}}}B,\ensuremath{^{\textnormal{7}}}Li)\hspace{0.2in}\href{https://www.nndc.bnl.gov/nsr/nsrlink.jsp?1976Ha06,B}{1976Ha06} (continued)}}\\
\vspace{0.3cm}
\underline{$^{19}$Ne Levels (continued)}\\
\vspace{0.3cm}
\parbox[b][0.3cm]{17.7cm}{{\ }{\ }\ensuremath{^{\textnormal{16}}}O(\ensuremath{^{\textnormal{6}}}Li,t)) and (\href{https://www.nndc.bnl.gov/nsr/nsrlink.jsp?1972Ga08,B}{1972Ga08}: \ensuremath{^{\textnormal{16}}}O(\ensuremath{^{\textnormal{6}}}Li,t)) for the \ensuremath{^{\textnormal{19}}}Ne* levels.}\\
\parbox[b][0.3cm]{17.7cm}{\makebox[1ex]{\ensuremath{^{\hypertarget{NE26LEVEL3}{d}}}} Mirror levels are assigned by (\href{https://www.nndc.bnl.gov/nsr/nsrlink.jsp?1976Ha06,B}{1976Ha06}).}\\
\vspace{0.5cm}
\clearpage
\clearpage
\begin{figure}[h]
\begin{center}
\includegraphics{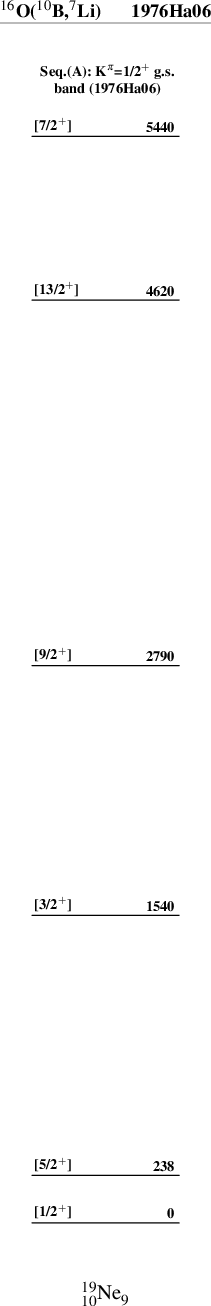}\\
\end{center}
\end{figure}
\clearpage
\subsection[\hspace{-0.2cm}\ensuremath{^{\textnormal{16}}}O(\ensuremath{^{\textnormal{11}}}B,\ensuremath{^{\textnormal{8}}}Li)]{ }
\vspace{-27pt}
\vspace{0.3cm}
\hypertarget{NE27}{{\bf \small \underline{\ensuremath{^{\textnormal{16}}}O(\ensuremath{^{\textnormal{11}}}B,\ensuremath{^{\textnormal{8}}}Li)\hspace{0.2in}\href{https://www.nndc.bnl.gov/nsr/nsrlink.jsp?1979Ra10,B}{1979Ra10},\href{https://www.nndc.bnl.gov/nsr/nsrlink.jsp?1981Go11,B}{1981Go11}}}}\\
\vspace{4pt}
\vspace{8pt}
\parbox[b][0.3cm]{17.7cm}{\addtolength{\parindent}{-0.2in}\ensuremath{^{\textnormal{3}}}He transfer reaction.}\\
\parbox[b][0.3cm]{17.7cm}{\addtolength{\parindent}{-0.2in}J\ensuremath{^{\ensuremath{\pi}}}(\ensuremath{^{\textnormal{16}}}O\ensuremath{_{\textnormal{g.s.}}})=0\ensuremath{^{\textnormal{+}}} and J\ensuremath{^{\ensuremath{\pi}}}(\ensuremath{^{\textnormal{11}}}B\ensuremath{_{\textnormal{g.s.}}})=3/2\ensuremath{^{-}}.}\\
\parbox[b][0.3cm]{17.7cm}{\addtolength{\parindent}{-0.2in}\href{https://www.nndc.bnl.gov/nsr/nsrlink.jsp?1979Ra10,B}{1979Ra10}: \ensuremath{^{\textnormal{16}}}O(\ensuremath{^{\textnormal{11}}}B,\ensuremath{^{\textnormal{8}}}Li) E=115 MeV; measured reaction products using a Si \ensuremath{\Delta}E-\ensuremath{\Delta}E-E telescope followed by a Si veto detector}\\
\parbox[b][0.3cm]{17.7cm}{to reject the long-range particles. The angular coverage of the detection system was for \ensuremath{\theta}\ensuremath{_{\textnormal{lab}}}=7\ensuremath{^\circ}{\textminus}12\ensuremath{^\circ}. Measured \ensuremath{\sigma}(\ensuremath{\theta}).}\\
\parbox[b][0.3cm]{17.7cm}{\addtolength{\parindent}{-0.2in}\href{https://www.nndc.bnl.gov/nsr/nsrlink.jsp?1981Go11,B}{1981Go11}: \ensuremath{^{\textnormal{16}}}O(\ensuremath{^{\textnormal{11}}}B,\ensuremath{^{\textnormal{8}}}Li) E=115 MeV; reanalyzed the data of (\href{https://www.nndc.bnl.gov/nsr/nsrlink.jsp?1979Ra10,B}{1979Ra10}) using an exact finite-range DWBA analysis (using the}\\
\parbox[b][0.3cm]{17.7cm}{LOLA code) and a parameterized cluster model potential for the bound-state form factors. Deduced J\ensuremath{^{\ensuremath{\pi}}} assignments for \ensuremath{^{\textnormal{19}}}Ne levels;}\\
\parbox[b][0.3cm]{17.7cm}{deduced shell model and experimental spectroscopic factors for cluster transfer to the \ensuremath{^{\textnormal{19}}}Ne* states.}\\
\vspace{12pt}
\underline{$^{19}$Ne Levels}\\
\vspace{0.34cm}
\parbox[b][0.3cm]{17.7cm}{\addtolength{\parindent}{-0.254cm}\textit{Notes}:}\\
\parbox[b][0.3cm]{17.7cm}{\addtolength{\parindent}{-0.254cm}(1) (\href{https://www.nndc.bnl.gov/nsr/nsrlink.jsp?1981Go11,B}{1981Go11}): N=\ensuremath{\sigma}\ensuremath{_{\textnormal{exp}}}/\ensuremath{\sigma}\ensuremath{_{\textnormal{DWBA}}} for populating cluster states with 2N+L=6.}\\
\parbox[b][0.3cm]{17.7cm}{\addtolength{\parindent}{-0.254cm}(2) (\href{https://www.nndc.bnl.gov/nsr/nsrlink.jsp?1981Go11,B}{1981Go11}): S\ensuremath{_{\textnormal{DWBA}}}=N/C\ensuremath{_{\textnormal{1}}^{\textnormal{2}}}S\ensuremath{_{\textnormal{1}}}C\ensuremath{_{\textnormal{2}}^{\textnormal{2}}}, where C\ensuremath{_{\textnormal{1}}} and C\ensuremath{_{\textnormal{2}}} are Clebsch-Gordan coefficients for the isospin coupling and S\ensuremath{_{\textnormal{1}}} is the}\\
\parbox[b][0.3cm]{17.7cm}{three nucleon spectroscopic factor.}\\
\parbox[b][0.3cm]{17.7cm}{\addtolength{\parindent}{-0.254cm}(3) (\href{https://www.nndc.bnl.gov/nsr/nsrlink.jsp?1981Go11,B}{1981Go11}): S\ensuremath{_{\textnormal{SM}}} is the shell model spectroscopic factor predicted with the plane wave interaction.}\\
\parbox[b][0.3cm]{17.7cm}{\addtolength{\parindent}{-0.254cm}(4) (\href{https://www.nndc.bnl.gov/nsr/nsrlink.jsp?1981Go11,B}{1981Go11}): S\ensuremath{_{\textnormal{sc}}} is the relative experimental spectroscopic factor, using semi-classical theory for the reaction dynamics,}\\
\parbox[b][0.3cm]{17.7cm}{normalized such that S\ensuremath{_{\textnormal{sc}}}= S\ensuremath{_{\textnormal{DWBA}}} for the \ensuremath{^{\textnormal{19}}}Ne*(2.79 MeV, 9/2\ensuremath{^{\textnormal{+}}}) state.}\\
\parbox[b][0.3cm]{17.7cm}{\addtolength{\parindent}{-0.254cm}(5) (\href{https://www.nndc.bnl.gov/nsr/nsrlink.jsp?1979Ra10,B}{1979Ra10}) concluded that a direct \ensuremath{^{\textnormal{3}}}He transfer mechanism, probably of a sequential nature, dominates this reaction.}\\
\vspace{0.34cm}
\begin{longtable}{cccccc@{\extracolsep{\fill}}c}
\multicolumn{2}{c}{E(level)$^{{\hyperlink{NE27LEVEL0}{a}}}$}&J$^{\pi}$$^{{\hyperlink{NE27LEVEL1}{b}}}$&\multicolumn{2}{c}{S\ensuremath{_{\textnormal{rel}}}$^{{\hyperlink{NE27LEVEL1}{b}}}$}&Comments&\\[-.2cm]
\multicolumn{2}{c}{\hrulefill}&\hrulefill&\multicolumn{2}{c}{\hrulefill}&\hrulefill&
\endfirsthead
\multicolumn{1}{r@{}}{2.79\ensuremath{\times10^{3}}}&\multicolumn{1}{@{}l}{}&\multicolumn{1}{l}{9/2\ensuremath{^{+}}}&\multicolumn{1}{r@{}}{1}&\multicolumn{1}{@{.}l}{00}&\parbox[t][0.3cm]{13.72014cm}{\raggedright N=0.019 \textit{6}, S\ensuremath{_{\textnormal{DWBA}}}=0.06, S\ensuremath{_{\textnormal{SM}}}=0.25, and S\ensuremath{_{\textnormal{sc}}}=0.04 (\href{https://www.nndc.bnl.gov/nsr/nsrlink.jsp?1981Go11,B}{1981Go11}). See \textit{Notes} for the definitions.\vspace{0.1cm}}&\\
\multicolumn{1}{r@{}}{4.64\ensuremath{\times10^{3}}}&\multicolumn{1}{@{}l}{}&\multicolumn{1}{l}{13/2\ensuremath{^{+}}}&\multicolumn{1}{r@{}}{2}&\multicolumn{1}{@{.}l}{89}&\parbox[t][0.3cm]{13.72014cm}{\raggedright E(level): See also 4.63 MeV (\href{https://www.nndc.bnl.gov/nsr/nsrlink.jsp?1979Ra10,B}{1979Ra10}), where Table 10 reports this energy as 4.62 MeV. Note\vspace{0.1cm}}&\\
&&&&&\parbox[t][0.3cm]{13.72014cm}{\raggedright {\ }{\ }{\ }also that (\href{https://www.nndc.bnl.gov/nsr/nsrlink.jsp?1981Go11,B}{1981Go11}) reports the energy as 4.63 MeV on Fig. 4.\vspace{0.1cm}}&\\
&&&&&\parbox[t][0.3cm]{13.72014cm}{\raggedright d\ensuremath{\sigma}/d\ensuremath{\Omega}=0.19 mb/sr (\href{https://www.nndc.bnl.gov/nsr/nsrlink.jsp?1979Ra10,B}{1979Ra10}), where the \ensuremath{^{\textnormal{8}}}Li angular distribution was featureless. Therefore, results\vspace{0.1cm}}&\\
&&&&&\parbox[t][0.3cm]{13.72014cm}{\raggedright {\ }{\ }{\ }obtained at a single angle (not given) can be taken as the representative of the given cross section.\vspace{0.1cm}}&\\
&&&&&\parbox[t][0.3cm]{13.72014cm}{\raggedright N=0.055 \textit{6}, S\ensuremath{_{\textnormal{DWBA}}}=0.17, S\ensuremath{_{\textnormal{SM}}}=0.22, and S\ensuremath{_{\textnormal{sc}}}=0.07 (\href{https://www.nndc.bnl.gov/nsr/nsrlink.jsp?1981Go11,B}{1981Go11}).\vspace{0.1cm}}&\\
\multicolumn{1}{r@{}}{5.42\ensuremath{\times10^{3}}}&\multicolumn{1}{@{}l}{}&\multicolumn{1}{l}{7/2\ensuremath{^{+}}}&\multicolumn{1}{r@{}}{1}&\multicolumn{1}{@{.}l}{37}&\parbox[t][0.3cm]{13.72014cm}{\raggedright E(level): See also 5.34 MeV (\href{https://www.nndc.bnl.gov/nsr/nsrlink.jsp?1979Ra10,B}{1979Ra10}). The energy of this level is also reported as 5.34 MeV on\vspace{0.1cm}}&\\
&&&&&\parbox[t][0.3cm]{13.72014cm}{\raggedright {\ }{\ }{\ }Fig. 4 of (\href{https://www.nndc.bnl.gov/nsr/nsrlink.jsp?1981Go11,B}{1981Go11}).\vspace{0.1cm}}&\\
&&&&&\parbox[t][0.3cm]{13.72014cm}{\raggedright N=0.026 \textit{13}, S\ensuremath{_{\textnormal{DWBA}}}=0.08, S\ensuremath{_{\textnormal{SM}}}=0.24, and S\ensuremath{_{\textnormal{sc}}}=0.09 (\href{https://www.nndc.bnl.gov/nsr/nsrlink.jsp?1981Go11,B}{1981Go11}).\vspace{0.1cm}}&\\
\multicolumn{1}{r@{}}{9.80\ensuremath{\times10^{3}}}&\multicolumn{1}{@{}l}{}&\multicolumn{1}{l}{13/2\ensuremath{^{+}}}&\multicolumn{1}{r@{}}{1}&\multicolumn{1}{@{.}l}{37}&\parbox[t][0.3cm]{13.72014cm}{\raggedright E(level): See also 9.75 MeV (\href{https://www.nndc.bnl.gov/nsr/nsrlink.jsp?1979Ra10,B}{1979Ra10}).\vspace{0.1cm}}&\\
&&&&&\parbox[t][0.3cm]{13.72014cm}{\raggedright J\ensuremath{^{\pi}}: (\href{https://www.nndc.bnl.gov/nsr/nsrlink.jsp?1981Go11,B}{1981Go11}) mentioned that this state is probably populated entirely due to the 13/2\ensuremath{^{\textnormal{+}}_{\textnormal{2}}} strength\vspace{0.1cm}}&\\
&&&&&\parbox[t][0.3cm]{13.72014cm}{\raggedright {\ }{\ }{\ }in \ensuremath{^{\textnormal{19}}}Ne. They suggested that the shell model prediction (by N. S. Godwin, D. Phil. Thesis, Oxford\vspace{0.1cm}}&\\
&&&&&\parbox[t][0.3cm]{13.72014cm}{\raggedright {\ }{\ }{\ }(1979), unpublished) for an 11/2\ensuremath{^{\textnormal{+}}} state at this energy is most likely erroneous. The DWBA\vspace{0.1cm}}&\\
&&&&&\parbox[t][0.3cm]{13.72014cm}{\raggedright {\ }{\ }{\ }analysis of (\href{https://www.nndc.bnl.gov/nsr/nsrlink.jsp?1981Go11,B}{1981Go11}) predicts cross sections of the same order of magnitude for the 11/2\ensuremath{^{\textnormal{+}}} and\vspace{0.1cm}}&\\
&&&&&\parbox[t][0.3cm]{13.72014cm}{\raggedright {\ }{\ }{\ }13/2\ensuremath{^{\textnormal{+}}} states, but no state is observed that can be identified with an 11/2\ensuremath{^{\textnormal{+}}} state. Therefore, that\vspace{0.1cm}}&\\
&&&&&\parbox[t][0.3cm]{13.72014cm}{\raggedright {\ }{\ }{\ }study suggests that the cluster strength for the 11/2\ensuremath{^{\textnormal{+}}_{\textnormal{1}}} in \ensuremath{^{\textnormal{19}}}Ne is highly fragmented.\vspace{0.1cm}}&\\
&&&&&\parbox[t][0.3cm]{13.72014cm}{\raggedright N=0.026 \textit{9}, S\ensuremath{_{\textnormal{DWBA}}}=0.08, S\ensuremath{_{\textnormal{SM}}}=0.13, and S\ensuremath{_{\textnormal{sc}}}=0.02 (\href{https://www.nndc.bnl.gov/nsr/nsrlink.jsp?1981Go11,B}{1981Go11}).\vspace{0.1cm}}&\\
\multicolumn{1}{r@{}}{12.5\ensuremath{\times10^{3}}}&\multicolumn{1}{@{}l}{}&&&&\parbox[t][0.3cm]{13.72014cm}{\raggedright E(level): See also 12.27 MeV (\href{https://www.nndc.bnl.gov/nsr/nsrlink.jsp?1979Ra10,B}{1979Ra10}).\vspace{0.1cm}}&\\
\end{longtable}
\parbox[b][0.3cm]{17.7cm}{\makebox[1ex]{\ensuremath{^{\hypertarget{NE27LEVEL0}{a}}}} From (\href{https://www.nndc.bnl.gov/nsr/nsrlink.jsp?1981Go11,B}{1981Go11}).}\\
\parbox[b][0.3cm]{17.7cm}{\makebox[1ex]{\ensuremath{^{\hypertarget{NE27LEVEL1}{b}}}} From the exact finite-range DWBA analysis in (\href{https://www.nndc.bnl.gov/nsr/nsrlink.jsp?1981Go11,B}{1981Go11}). The spectroscopic factors for the three nucleon transfer are relative to}\\
\parbox[b][0.3cm]{17.7cm}{{\ }{\ }that of the \ensuremath{^{\textnormal{19}}}Ne*(2.79 MeV, 9/2\ensuremath{^{\textnormal{+}}}) state.}\\
\vspace{0.5cm}
\clearpage
\subsection[\hspace{-0.2cm}\ensuremath{^{\textnormal{16}}}O(\ensuremath{^{\textnormal{12}}}C,\ensuremath{^{\textnormal{9}}}Be)]{ }
\vspace{-27pt}
\vspace{0.3cm}
\hypertarget{NE28}{{\bf \small \underline{\ensuremath{^{\textnormal{16}}}O(\ensuremath{^{\textnormal{12}}}C,\ensuremath{^{\textnormal{9}}}Be)\hspace{0.2in}\href{https://www.nndc.bnl.gov/nsr/nsrlink.jsp?1981Go11,B}{1981Go11},\href{https://www.nndc.bnl.gov/nsr/nsrlink.jsp?1988Kr11,B}{1988Kr11}}}}\\
\vspace{4pt}
\vspace{8pt}
\parbox[b][0.3cm]{17.7cm}{\addtolength{\parindent}{-0.2in}\ensuremath{^{\textnormal{3}}}He transfer reaction.}\\
\parbox[b][0.3cm]{17.7cm}{\addtolength{\parindent}{-0.2in}J\ensuremath{^{\ensuremath{\pi}}}(\ensuremath{^{\textnormal{16}}}O\ensuremath{_{\textnormal{g.s.}}})=0\ensuremath{^{\textnormal{+}}} and J\ensuremath{^{\ensuremath{\pi}}}(\ensuremath{^{\textnormal{12}}}C\ensuremath{_{\textnormal{g.s.}}})=0\ensuremath{^{\textnormal{+}}}.}\\
\parbox[b][0.3cm]{17.7cm}{\addtolength{\parindent}{-0.2in}\href{https://www.nndc.bnl.gov/nsr/nsrlink.jsp?1972Sc21,B}{1972Sc21}: \ensuremath{^{\textnormal{12}}}C(\ensuremath{^{\textnormal{12}}}C,\ensuremath{^{\textnormal{9}}}Be) E=114 MeV; measured reaction products using a Si \ensuremath{\Delta}E-\ensuremath{\Delta}E-E telescope followed by a Si veto detector}\\
\parbox[b][0.3cm]{17.7cm}{to reject the long-range particles. The angular coverage of the detection system was for \ensuremath{\theta}\ensuremath{_{\textnormal{lab}}}=7\ensuremath{^\circ}{\textminus}35\ensuremath{^\circ}.}\\
\parbox[b][0.3cm]{17.7cm}{\addtolength{\parindent}{-0.2in}\href{https://www.nndc.bnl.gov/nsr/nsrlink.jsp?1974An36,B}{1974An36}: \ensuremath{^{\textnormal{16}}}O(\ensuremath{^{\textnormal{12}}}C,\ensuremath{^{\textnormal{9}}}Be) E=114 MeV; measured reaction products using a \ensuremath{\Delta}E-\ensuremath{\Delta}E-E-AC telescope that consisted of Si surface}\\
\parbox[b][0.3cm]{17.7cm}{barrier detectors. An anti-coincidence detector (AC) was used for vetoing high energy particles. Energy resolution was}\\
\parbox[b][0.3cm]{17.7cm}{\ensuremath{\Delta}E(FWHM)=400 keV. Measured \ensuremath{\sigma}(E,\ensuremath{\theta}).}\\
\parbox[b][0.3cm]{17.7cm}{\addtolength{\parindent}{-0.2in}\href{https://www.nndc.bnl.gov/nsr/nsrlink.jsp?1977HaZN,B}{1977HaZN}: \ensuremath{^{\textnormal{16}}}O(\ensuremath{^{\textnormal{12}}}C,\ensuremath{^{\textnormal{9}}}Be); measured \ensuremath{\sigma}; deduced \ensuremath{^{\textnormal{19}}}Ne levels, K, J, and \ensuremath{\pi}.}\\
\parbox[b][0.3cm]{17.7cm}{\addtolength{\parindent}{-0.2in}\href{https://www.nndc.bnl.gov/nsr/nsrlink.jsp?1979Ra10,B}{1979Ra10}: \ensuremath{^{\textnormal{16}}}O(\ensuremath{^{\textnormal{12}}}C,\ensuremath{^{\textnormal{9}}}Be) E=115 MeV; measured reaction products using the same detector mentioned above covering \ensuremath{\theta}\ensuremath{_{\textnormal{lab}}}=7\ensuremath{^\circ}{\textminus}12\ensuremath{^\circ}.}\\
\parbox[b][0.3cm]{17.7cm}{\addtolength{\parindent}{-0.2in}\href{https://www.nndc.bnl.gov/nsr/nsrlink.jsp?1981Go11,B}{1981Go11}: \ensuremath{^{\textnormal{16}}}O(\ensuremath{^{\textnormal{12}}}C,\ensuremath{^{\textnormal{9}}}Be) E=115 MeV; reanalyzed the data of (\href{https://www.nndc.bnl.gov/nsr/nsrlink.jsp?1979Ra10,B}{1979Ra10}) using an exact finite-range DWBA analysis (using the}\\
\parbox[b][0.3cm]{17.7cm}{LOLA code) and a parameterized cluster model potential for the bound-state form factors. Deduced J\ensuremath{^{\ensuremath{\pi}}} assignments for the \ensuremath{^{\textnormal{19}}}Ne}\\
\parbox[b][0.3cm]{17.7cm}{levels; obtained shell model and experimental spectroscopic factors for cluster transfer to the \ensuremath{^{\textnormal{19}}}Ne states.}\\
\parbox[b][0.3cm]{17.7cm}{\addtolength{\parindent}{-0.2in}\href{https://www.nndc.bnl.gov/nsr/nsrlink.jsp?1988Kr11,B}{1988Kr11}: \ensuremath{^{\textnormal{16}}}O(\ensuremath{^{\textnormal{12}}}C,\ensuremath{^{\textnormal{9}}}Be) E=480 MeV; momentum analyzed the reaction products using the SPEG spectrometer and associated}\\
\parbox[b][0.3cm]{17.7cm}{detectors. An energy resolution of \ensuremath{\Delta}E(FWHM)=200 keV was achieved. Measured angular distributions of the reaction products at}\\
\parbox[b][0.3cm]{17.7cm}{\ensuremath{\theta}\ensuremath{_{\textnormal{lab}}}=0\ensuremath{^\circ}{\textminus}4\ensuremath{^\circ}. Deduced \ensuremath{^{\textnormal{19}}}Ne levels and J\ensuremath{^{\ensuremath{\pi}}} assignments using an exact finite-range DWBA analysis with the PTOLEMY computer}\\
\parbox[b][0.3cm]{17.7cm}{code and assuming the 0\textit{s} cluster approximation.}\\
\vspace{12pt}
\underline{$^{19}$Ne Levels}\\
\vspace{0.34cm}
\parbox[b][0.3cm]{17.7cm}{\addtolength{\parindent}{-0.254cm}\textit{Notes}:}\\
\parbox[b][0.3cm]{17.7cm}{\addtolength{\parindent}{-0.254cm}(1) (\href{https://www.nndc.bnl.gov/nsr/nsrlink.jsp?1981Go11,B}{1981Go11}): N=\ensuremath{\sigma}\ensuremath{_{\textnormal{exp}}}/\ensuremath{\sigma}\ensuremath{_{\textnormal{DWBA}}} for populating cluster states with 2N+L=6 in \ensuremath{^{\textnormal{19}}}Ne.}\\
\parbox[b][0.3cm]{17.7cm}{\addtolength{\parindent}{-0.254cm}(2) S\ensuremath{_{\textnormal{DWBA}}}=N/C\ensuremath{_{\textnormal{1}}^{\textnormal{2}}}S\ensuremath{_{\textnormal{1}}}C\ensuremath{_{\textnormal{2}}^{\textnormal{2}}}, where C\ensuremath{_{\textnormal{1}}} and C\ensuremath{_{\textnormal{2}}} are Clebsch-Gordan coefficients for the isospin coupling and S\ensuremath{_{\textnormal{1}}} is the three nucleon}\\
\parbox[b][0.3cm]{17.7cm}{spectroscopic factor from (\href{https://www.nndc.bnl.gov/nsr/nsrlink.jsp?1981Go11,B}{1981Go11}).}\\
\parbox[b][0.3cm]{17.7cm}{\addtolength{\parindent}{-0.254cm}(3) S\ensuremath{_{\textnormal{SM}}} is the shell model spectroscopic factor predicted with the plane wave interaction from (\href{https://www.nndc.bnl.gov/nsr/nsrlink.jsp?1981Go11,B}{1981Go11}).}\\
\parbox[b][0.3cm]{17.7cm}{\addtolength{\parindent}{-0.254cm}(4) S\ensuremath{_{\textnormal{sc}}} is the relative experimental spectroscopic factor, using semi-classical theory for the reaction dynamics, normalized such that}\\
\parbox[b][0.3cm]{17.7cm}{S\ensuremath{_{\textnormal{sc}}}= S\ensuremath{_{\textnormal{DWBA}}} for the \ensuremath{^{\textnormal{19}}}Ne*(2.79 MeV, 9/2\ensuremath{^{\textnormal{+}}}) state from (\href{https://www.nndc.bnl.gov/nsr/nsrlink.jsp?1981Go11,B}{1981Go11}).}\\
\parbox[b][0.3cm]{17.7cm}{\addtolength{\parindent}{-0.254cm}(5) The \ensuremath{^{\textnormal{16}}}O(\ensuremath{^{\textnormal{12}}}C,\ensuremath{^{\textnormal{9}}}Be) reaction preferentially populates high spin states with stretched configurations (\href{https://www.nndc.bnl.gov/nsr/nsrlink.jsp?1988Kr11,B}{1988Kr11}).}\\
\parbox[b][0.3cm]{17.7cm}{\addtolength{\parindent}{-0.254cm}(6) d\ensuremath{\sigma}/d\ensuremath{\Omega}\ensuremath{\sim}1 mb/sr at forward angles for \ensuremath{^{\textnormal{12}}}C(\ensuremath{^{\textnormal{12}}}C,n\ensuremath{\alpha})\ensuremath{^{\textnormal{19}}}Ne (\href{https://www.nndc.bnl.gov/nsr/nsrlink.jsp?1972Sc21,B}{1972Sc21}) at E\ensuremath{_{\textnormal{lab}}}=114 MeV.}\\
\vspace{0.34cm}
\begin{longtable}{cccccc@{\extracolsep{\fill}}c}
\multicolumn{2}{c}{E(level)$^{}$}&J$^{\pi}$$^{}$&\multicolumn{2}{c}{S\ensuremath{_{\textnormal{rel}}}$^{{\hyperlink{NE28LEVEL3}{d}}}$}&Comments&\\[-.2cm]
\multicolumn{2}{c}{\hrulefill}&\hrulefill&\multicolumn{2}{c}{\hrulefill}&\hrulefill&
\endfirsthead
\multicolumn{1}{r@{}}{0}&\multicolumn{1}{@{}l}{\ensuremath{^{{\hyperlink{NE28LEVEL0}{a}}}}}&&&&\parbox[t][0.3cm]{13.313541cm}{\raggedright E(level): From (\href{https://www.nndc.bnl.gov/nsr/nsrlink.jsp?1988Kr11,B}{1988Kr11}): Unresolved with the E\ensuremath{_{\textnormal{x}}}=0.23 MeV.\vspace{0.1cm}}&\\
\multicolumn{1}{r@{}}{0.24\ensuremath{\times10^{3}}}&\multicolumn{1}{@{}l}{}&\multicolumn{1}{l}{5/2\ensuremath{^{+}}\ensuremath{^{{\hyperlink{NE28LEVEL1}{b}}}}}&\multicolumn{1}{r@{}}{1}&\multicolumn{1}{@{.}l}{40}&\parbox[t][0.3cm]{13.313541cm}{\raggedright E(level): From (\href{https://www.nndc.bnl.gov/nsr/nsrlink.jsp?1979Ra10,B}{1979Ra10}, \href{https://www.nndc.bnl.gov/nsr/nsrlink.jsp?1981Go11,B}{1981Go11}). See also 0.23 MeV (\href{https://www.nndc.bnl.gov/nsr/nsrlink.jsp?1988Kr11,B}{1988Kr11}): Unresolved with the\vspace{0.1cm}}&\\
&&&&&\parbox[t][0.3cm]{13.313541cm}{\raggedright {\ }{\ }{\ }ground state.\vspace{0.1cm}}&\\
&&&&&\parbox[t][0.3cm]{13.313541cm}{\raggedright N=0.10 \textit{3}, S\ensuremath{_{\textnormal{DWBA}}}=0.07, S\ensuremath{_{\textnormal{SM}}}=0.25, and S\ensuremath{_{\textnormal{sc}}}=0.19 (\href{https://www.nndc.bnl.gov/nsr/nsrlink.jsp?1981Go11,B}{1981Go11}) at E\ensuremath{_{\textnormal{lab}}}=115 MeV.\vspace{0.1cm}}&\\
\multicolumn{1}{r@{}}{1.54\ensuremath{\times10^{3}}}&\multicolumn{1}{@{}l}{}&&&&\parbox[t][0.3cm]{13.313541cm}{\raggedright E(level): From (\href{https://www.nndc.bnl.gov/nsr/nsrlink.jsp?1979Ra10,B}{1979Ra10}, \href{https://www.nndc.bnl.gov/nsr/nsrlink.jsp?1981Go11,B}{1981Go11}). See also 1.5 MeV (\href{https://www.nndc.bnl.gov/nsr/nsrlink.jsp?1988Kr11,B}{1988Kr11}).\vspace{0.1cm}}&\\
\multicolumn{1}{r@{}}{2.79\ensuremath{\times10^{3}}}&\multicolumn{1}{@{}l}{\ensuremath{^{{\hyperlink{NE28LEVEL0}{a}}}}}&\multicolumn{1}{l}{9/2\ensuremath{^{+}}\ensuremath{^{{\hyperlink{NE28LEVEL1}{b}}{\hyperlink{NE28LEVEL2}{c}}}}}&\multicolumn{1}{r@{}}{1}&\multicolumn{1}{@{.}l}{00}&\parbox[t][0.3cm]{13.313541cm}{\raggedright E(level): From (\href{https://www.nndc.bnl.gov/nsr/nsrlink.jsp?1979Ra10,B}{1979Ra10}, \href{https://www.nndc.bnl.gov/nsr/nsrlink.jsp?1981Go11,B}{1981Go11}). See also 2.8 MeV (\href{https://www.nndc.bnl.gov/nsr/nsrlink.jsp?1988Kr11,B}{1988Kr11}).\vspace{0.1cm}}&\\
&&&&&\parbox[t][0.3cm]{13.313541cm}{\raggedright This state belongs to the (\textit{sd})\ensuremath{^{\textnormal{3}}}, 2N+L, ground state rotational band in \ensuremath{^{\textnormal{19}}}Ne (\href{https://www.nndc.bnl.gov/nsr/nsrlink.jsp?1988Kr11,B}{1988Kr11}).\vspace{0.1cm}}&\\
&&&&&\parbox[t][0.3cm]{13.313541cm}{\raggedright This state could have the (1\textit{d}\ensuremath{_{\textnormal{5/2}}})\ensuremath{^{\textnormal{2}}}(2\textit{s}\ensuremath{_{\textnormal{1/2}}}) stretched configuration (\href{https://www.nndc.bnl.gov/nsr/nsrlink.jsp?1988Kr11,B}{1988Kr11}).\vspace{0.1cm}}&\\
&&&&&\parbox[t][0.3cm]{13.313541cm}{\raggedright N=0.07 \textit{1}, S\ensuremath{_{\textnormal{DWBA}}}=0.05, S\ensuremath{_{\textnormal{SM}}}=0.25, and S\ensuremath{_{\textnormal{sc}}}=0.05 (\href{https://www.nndc.bnl.gov/nsr/nsrlink.jsp?1981Go11,B}{1981Go11}) at E\ensuremath{_{\textnormal{lab}}}=115 MeV.\vspace{0.1cm}}&\\
&&&&&\parbox[t][0.3cm]{13.313541cm}{\raggedright \ensuremath{\sigma}\ensuremath{_{\textnormal{exp}}}/\ensuremath{\sigma}\ensuremath{_{\textnormal{DWBA}}}=0.039 (\href{https://www.nndc.bnl.gov/nsr/nsrlink.jsp?1988Kr11,B}{1988Kr11}): See Fig. 15, E\ensuremath{_{\textnormal{lab}}}=115 MeV.\vspace{0.1cm}}&\\
\multicolumn{1}{r@{}}{4.64\ensuremath{\times10^{3}}}&\multicolumn{1}{@{}l}{\ensuremath{^{{\hyperlink{NE28LEVEL0}{a}}}}}&\multicolumn{1}{l}{13/2\ensuremath{^{+}}\ensuremath{^{{\hyperlink{NE28LEVEL1}{b}}{\hyperlink{NE28LEVEL2}{c}}}}}&\multicolumn{1}{r@{}}{3}&\multicolumn{1}{@{.}l}{22}&\parbox[t][0.3cm]{13.313541cm}{\raggedright E(level): From (\href{https://www.nndc.bnl.gov/nsr/nsrlink.jsp?1988Kr11,B}{1988Kr11}) and (\href{https://www.nndc.bnl.gov/nsr/nsrlink.jsp?1981Go11,B}{1981Go11}), where the energy of this level is reported as 4.63\vspace{0.1cm}}&\\
&&&&&\parbox[t][0.3cm]{13.313541cm}{\raggedright {\ }{\ }{\ }MeV on Fig. 4 and 4.64 MeV elsewhere. See also 4.63 MeV (\href{https://www.nndc.bnl.gov/nsr/nsrlink.jsp?1979Ra10,B}{1979Ra10}), where Table 10\vspace{0.1cm}}&\\
&&&&&\parbox[t][0.3cm]{13.313541cm}{\raggedright {\ }{\ }{\ }reports this state as 4.62 MeV; 4.6 MeV (\href{https://www.nndc.bnl.gov/nsr/nsrlink.jsp?1972Sc21,B}{1972Sc21}) and 4.60 MeV (\href{https://www.nndc.bnl.gov/nsr/nsrlink.jsp?1974An36,B}{1974An36}).\vspace{0.1cm}}&\\
&&&&&\parbox[t][0.3cm]{13.313541cm}{\raggedright J\ensuremath{^{\pi}}: The evaluator notes that the DWBA fit to the data from (\href{https://www.nndc.bnl.gov/nsr/nsrlink.jsp?1988Kr11,B}{1988Kr11}) is not a good fit.\vspace{0.1cm}}&\\
&&&&&\parbox[t][0.3cm]{13.313541cm}{\raggedright (\href{https://www.nndc.bnl.gov/nsr/nsrlink.jsp?1979Ra10,B}{1979Ra10}) reported the configuration of this state to be (\textit{sd})\ensuremath{^{\textnormal{3}}}.\vspace{0.1cm}}&\\
&&&&&\parbox[t][0.3cm]{13.313541cm}{\raggedright (\href{https://www.nndc.bnl.gov/nsr/nsrlink.jsp?1972Sc21,B}{1972Sc21}) identified this state to have a stretched configuration of (\textit{d}\ensuremath{_{\textnormal{5/2}}})\ensuremath{^{\textnormal{2}}_{\textnormal{13/2}~^{\textnormal{+}}}}. They considered\vspace{0.1cm}}&\\
&&&&&\parbox[t][0.3cm]{13.313541cm}{\raggedright {\ }{\ }{\ }this level to be the mirror of the \ensuremath{^{\textnormal{19}}}F*(4648) state from (\href{https://www.nndc.bnl.gov/nsr/nsrlink.jsp?1969Ja09,B}{1969Ja09}). However, (\href{https://www.nndc.bnl.gov/nsr/nsrlink.jsp?1988Kr11,B}{1988Kr11})\vspace{0.1cm}}&\\
&&&&&\parbox[t][0.3cm]{13.313541cm}{\raggedright {\ }{\ }{\ }suggested that this state could have the (1\textit{d}\ensuremath{_{\textnormal{5/2}}})\ensuremath{^{\textnormal{3}}} stretched configuration.\vspace{0.1cm}}&\\
&&&&&\parbox[t][0.3cm]{13.313541cm}{\raggedright This state belongs to the (\textit{sd})\ensuremath{^{\textnormal{3}}}, 2N+L, ground state rotational band in \ensuremath{^{\textnormal{19}}}Ne (\href{https://www.nndc.bnl.gov/nsr/nsrlink.jsp?1988Kr11,B}{1988Kr11}).\vspace{0.1cm}}&\\
&&&&&\parbox[t][0.3cm]{13.313541cm}{\raggedright S\ensuremath{_{\textnormal{rel}}}: From the unweighted average of 3.57 and 2.86 (\href{https://www.nndc.bnl.gov/nsr/nsrlink.jsp?1981Go11,B}{1981Go11}: See Table 6, E\ensuremath{_{\textnormal{lab}}}=115 MeV).\vspace{0.1cm}}&\\
&&&&&\parbox[t][0.3cm]{13.313541cm}{\raggedright {\ }{\ }{\ }(\href{https://www.nndc.bnl.gov/nsr/nsrlink.jsp?1981Go11,B}{1981Go11}) suggested that the inconsistencies in the spectroscopic factors could be due, in part,\vspace{0.1cm}}&\\
&&&&&\parbox[t][0.3cm]{13.313541cm}{\raggedright {\ }{\ }{\ }to a poor description of the radial wave function of the projectile cluster state, which was not\vspace{0.1cm}}&\\
\end{longtable}
\begin{textblock}{29}(0,27.3)
Continued on next page (footnotes at end of table)
\end{textblock}
\clearpage
\begin{longtable}{cccccc@{\extracolsep{\fill}}c}
\\[-.4cm]
\multicolumn{7}{c}{{\bf \small \underline{\ensuremath{^{\textnormal{16}}}O(\ensuremath{^{\textnormal{12}}}C,\ensuremath{^{\textnormal{9}}}Be)\hspace{0.2in}\href{https://www.nndc.bnl.gov/nsr/nsrlink.jsp?1981Go11,B}{1981Go11},\href{https://www.nndc.bnl.gov/nsr/nsrlink.jsp?1988Kr11,B}{1988Kr11} (continued)}}}\\
\multicolumn{7}{c}{~}\\
\multicolumn{7}{c}{\underline{\ensuremath{^{19}}Ne Levels (continued)}}\\
\multicolumn{7}{c}{~}\\
\multicolumn{2}{c}{E(level)$^{}$}&J$^{\pi}$$^{}$&\multicolumn{2}{c}{S\ensuremath{_{\textnormal{rel}}}$^{{\hyperlink{NE28LEVEL3}{d}}}$}&Comments&\\[-.2cm]
\multicolumn{2}{c}{\hrulefill}&\hrulefill&\multicolumn{2}{c}{\hrulefill}&\hrulefill&
\endhead
&&&&&\parbox[t][0.3cm]{12.710061cm}{\raggedright {\ }{\ }{\ }very well known.\vspace{0.1cm}}&\\
&&&&&\parbox[t][0.3cm]{12.710061cm}{\raggedright N=0.22 \textit{3} (\href{https://www.nndc.bnl.gov/nsr/nsrlink.jsp?1981Go11,B}{1981Go11}): From the weighted average of 0.25 \textit{4} and 0.20 \textit{4}, see Table 5,\vspace{0.1cm}}&\\
&&&&&\parbox[t][0.3cm]{12.710061cm}{\raggedright {\ }{\ }{\ }E\ensuremath{_{\textnormal{lab}}}=115 MeV.\vspace{0.1cm}}&\\
&&&&&\parbox[t][0.3cm]{12.710061cm}{\raggedright S\ensuremath{_{\textnormal{DWBA}}}=0.15 (\href{https://www.nndc.bnl.gov/nsr/nsrlink.jsp?1981Go11,B}{1981Go11}): The unweighted average of 0.17 and 0.13 (\href{https://www.nndc.bnl.gov/nsr/nsrlink.jsp?1981Go11,B}{1981Go11}: See Table\vspace{0.1cm}}&\\
&&&&&\parbox[t][0.3cm]{12.710061cm}{\raggedright {\ }{\ }{\ }5, E\ensuremath{_{\textnormal{lab}}}=115 MeV).\vspace{0.1cm}}&\\
&&&&&\parbox[t][0.3cm]{12.710061cm}{\raggedright S\ensuremath{_{\textnormal{SM}}}=0.22 and S\ensuremath{_{\textnormal{sc}}}=0.11 (\href{https://www.nndc.bnl.gov/nsr/nsrlink.jsp?1981Go11,B}{1981Go11}) at E\ensuremath{_{\textnormal{lab}}}=115 MeV.\vspace{0.1cm}}&\\
&&&&&\parbox[t][0.3cm]{12.710061cm}{\raggedright \ensuremath{\sigma}\ensuremath{_{\textnormal{exp}}}/\ensuremath{\sigma}\ensuremath{_{\textnormal{DWBA}}}=0.043 (\href{https://www.nndc.bnl.gov/nsr/nsrlink.jsp?1988Kr11,B}{1988Kr11}: See Fig. 15) at E\ensuremath{_{\textnormal{lab}}}=480 MeV.\vspace{0.1cm}}&\\
&&&&&\parbox[t][0.3cm]{12.710061cm}{\raggedright d\ensuremath{\sigma}/d\ensuremath{\Omega}\ensuremath{\sim}1 mb/sr at forward angles for \ensuremath{^{\textnormal{16}}}O(\ensuremath{^{\textnormal{12}}}C,\ensuremath{^{\textnormal{9}}}Be) (\href{https://www.nndc.bnl.gov/nsr/nsrlink.jsp?1972Sc21,B}{1972Sc21}) at E\ensuremath{_{\textnormal{lab}}}=114 MeV.\vspace{0.1cm}}&\\
\multicolumn{1}{r@{}}{8.9\ensuremath{\times10^{3}}}&\multicolumn{1}{@{}l}{}&\multicolumn{1}{l}{[11/2\ensuremath{^{-}}]}&&&\parbox[t][0.3cm]{12.710061cm}{\raggedright E(level): From (\href{https://www.nndc.bnl.gov/nsr/nsrlink.jsp?1988Kr11,B}{1988Kr11}).\vspace{0.1cm}}&\\
&&&&&\parbox[t][0.3cm]{12.710061cm}{\raggedright J\ensuremath{^{\pi}}: From mirror level analysis by (\href{https://www.nndc.bnl.gov/nsr/nsrlink.jsp?1988Kr11,B}{1988Kr11}).\vspace{0.1cm}}&\\
\multicolumn{1}{r@{}}{9.8\ensuremath{\times10^{3}}}&\multicolumn{1}{@{}l}{}&\multicolumn{1}{l}{(13/2\ensuremath{^{+}},11/2\ensuremath{^{+}})}&\multicolumn{1}{r@{}}{1}&\multicolumn{1}{@{.}l}{57}&\parbox[t][0.3cm]{12.710061cm}{\raggedright E(level): From (\href{https://www.nndc.bnl.gov/nsr/nsrlink.jsp?1981Go11,B}{1981Go11}, \href{https://www.nndc.bnl.gov/nsr/nsrlink.jsp?1988Kr11,B}{1988Kr11}). See also 9.75 MeV (\href{https://www.nndc.bnl.gov/nsr/nsrlink.jsp?1979Ra10,B}{1979Ra10}).\vspace{0.1cm}}&\\
&&&&&\parbox[t][0.3cm]{12.710061cm}{\raggedright J\ensuremath{^{\pi}}: From J\ensuremath{^{\ensuremath{\pi}}}=13/2\ensuremath{^{\textnormal{+}}} (\href{https://www.nndc.bnl.gov/nsr/nsrlink.jsp?1981Go11,B}{1981Go11}) deduced from the exact finite-range DWBA analysis (see\vspace{0.1cm}}&\\
&&&&&\parbox[t][0.3cm]{12.710061cm}{\raggedright {\ }{\ }{\ }Fig. 5b); and J\ensuremath{^{\ensuremath{\pi}}}=(11/2\ensuremath{^{\textnormal{+}}}) (\href{https://www.nndc.bnl.gov/nsr/nsrlink.jsp?1988Kr11,B}{1988Kr11}) deduced from the exact finite-range DWBA analysis.\vspace{0.1cm}}&\\
&&&&&\parbox[t][0.3cm]{12.710061cm}{\raggedright {\ }{\ }{\ }The evaluator notes that the DWBA fit to those data misses the data points at \ensuremath{\theta}\ensuremath{_{\textnormal{c.m.}}}\ensuremath{\leq}5\ensuremath{^\circ}.\vspace{0.1cm}}&\\
&&&&&\parbox[t][0.3cm]{12.710061cm}{\raggedright S\ensuremath{_{\textnormal{rel}}}: From (\href{https://www.nndc.bnl.gov/nsr/nsrlink.jsp?1981Go11,B}{1981Go11}) at E\ensuremath{_{\textnormal{lab}}}=115 MeV and for J\ensuremath{^{\ensuremath{\pi}}}=13/2\ensuremath{^{\textnormal{+}}}.\vspace{0.1cm}}&\\
&&&&&\parbox[t][0.3cm]{12.710061cm}{\raggedright N=0.11 \textit{3}, S\ensuremath{_{\textnormal{DWBA}}}=0.07, S\ensuremath{_{\textnormal{SM}}}=0.13, and S\ensuremath{_{\textnormal{sc}}}=0.04 (\href{https://www.nndc.bnl.gov/nsr/nsrlink.jsp?1981Go11,B}{1981Go11}) at E\ensuremath{_{\textnormal{lab}}}=115 MeV and for\vspace{0.1cm}}&\\
&&&&&\parbox[t][0.3cm]{12.710061cm}{\raggedright {\ }{\ }{\ }J\ensuremath{^{\ensuremath{\pi}}}=13/2\ensuremath{^{\textnormal{+}}}.\vspace{0.1cm}}&\\
&&&&&\parbox[t][0.3cm]{12.710061cm}{\raggedright \ensuremath{\sigma}\ensuremath{_{\textnormal{exp}}}/\ensuremath{\sigma}\ensuremath{_{\textnormal{DWBA}}}=0.105 (\href{https://www.nndc.bnl.gov/nsr/nsrlink.jsp?1988Kr11,B}{1988Kr11}: See Fig. 15) at E\ensuremath{_{\textnormal{lab}}}=480 MeV and deduced for J\ensuremath{^{\ensuremath{\pi}}}=(11/2\ensuremath{^{\textnormal{+}}}).\vspace{0.1cm}}&\\
\multicolumn{1}{r@{}}{12.3\ensuremath{\times10^{3}}}&\multicolumn{1}{@{}l}{}&\multicolumn{1}{l}{(17/2\ensuremath{^{-}})\ensuremath{^{{\hyperlink{NE28LEVEL2}{c}}}}}&&&\parbox[t][0.3cm]{12.710061cm}{\raggedright E(level): From (\href{https://www.nndc.bnl.gov/nsr/nsrlink.jsp?1988Kr11,B}{1988Kr11}). See also 12.27 MeV (\href{https://www.nndc.bnl.gov/nsr/nsrlink.jsp?1979Ra10,B}{1979Ra10}), and 12.5 MeV (\href{https://www.nndc.bnl.gov/nsr/nsrlink.jsp?1981Go11,B}{1981Go11}).\vspace{0.1cm}}&\\
&&&&&\parbox[t][0.3cm]{12.710061cm}{\raggedright J\ensuremath{^{\pi}}: The evaluator notes that the DWBA fit to the data from (\href{https://www.nndc.bnl.gov/nsr/nsrlink.jsp?1988Kr11,B}{1988Kr11}) misses the data\vspace{0.1cm}}&\\
&&&&&\parbox[t][0.3cm]{12.710061cm}{\raggedright {\ }{\ }{\ }points at \ensuremath{\theta}\ensuremath{_{\textnormal{c.m.}}}\ensuremath{\geq}8\ensuremath{^\circ}.\vspace{0.1cm}}&\\
&&&&&\parbox[t][0.3cm]{12.710061cm}{\raggedright \ensuremath{\sigma}\ensuremath{_{\textnormal{exp}}}/\ensuremath{\sigma}\ensuremath{_{\textnormal{DWBA}}}=0.001 (\href{https://www.nndc.bnl.gov/nsr/nsrlink.jsp?1988Kr11,B}{1988Kr11}: See Fig. 15) at E\ensuremath{_{\textnormal{lab}}}=480 MeV and deduced for J\ensuremath{^{\ensuremath{\pi}}}=(17/2\ensuremath{^{-}}).\vspace{0.1cm}}&\\
\end{longtable}
\parbox[b][0.3cm]{17.7cm}{\makebox[1ex]{\ensuremath{^{\hypertarget{NE28LEVEL0}{a}}}} Seq.(A): K\ensuremath{^{\ensuremath{\pi}}}=1/2\ensuremath{^{+}} g.s. band (\href{https://www.nndc.bnl.gov/nsr/nsrlink.jsp?1988Kr11,B}{1988Kr11}).}\\
\parbox[b][0.3cm]{17.7cm}{\makebox[1ex]{\ensuremath{^{\hypertarget{NE28LEVEL1}{b}}}} From the exact finite-range DWBA analysis of (\href{https://www.nndc.bnl.gov/nsr/nsrlink.jsp?1981Go11,B}{1981Go11}): See Fig. 5b.}\\
\parbox[b][0.3cm]{17.7cm}{\makebox[1ex]{\ensuremath{^{\hypertarget{NE28LEVEL2}{c}}}} From the exact finite-range DWBA analysis of (\href{https://www.nndc.bnl.gov/nsr/nsrlink.jsp?1988Kr11,B}{1988Kr11}), where the authors reported that the \ensuremath{^{\textnormal{9}}}Be angular distributions}\\
\parbox[b][0.3cm]{17.7cm}{{\ }{\ }corresponding to the \ensuremath{^{\textnormal{19}}}Ne states were spin-independent.}\\
\parbox[b][0.3cm]{17.7cm}{\makebox[1ex]{\ensuremath{^{\hypertarget{NE28LEVEL3}{d}}}} Spectroscopic factors for three nucleon transfer relative to that of the \ensuremath{^{\textnormal{19}}}Ne*(2.79 MeV, 9/2\ensuremath{^{\textnormal{+}}}) state. The values are from}\\
\parbox[b][0.3cm]{17.7cm}{{\ }{\ }(\href{https://www.nndc.bnl.gov/nsr/nsrlink.jsp?1981Go11,B}{1981Go11}).}\\
\vspace{0.5cm}
\clearpage
\clearpage
\begin{figure}[h]
\begin{center}
\includegraphics{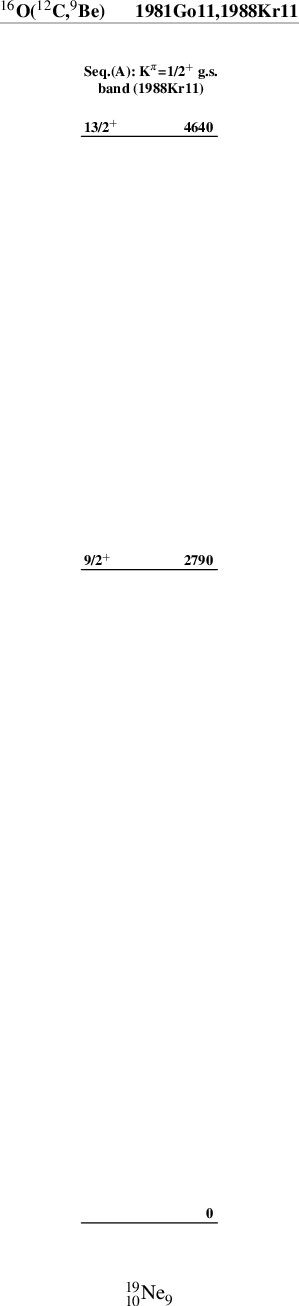}\\
\end{center}
\end{figure}
\clearpage
\subsection[\hspace{-0.2cm}\ensuremath{^{\textnormal{17}}}O(\ensuremath{^{\textnormal{3}}}He,n),(\ensuremath{^{\textnormal{3}}}He,n\ensuremath{\gamma})]{ }
\vspace{-27pt}
\vspace{0.3cm}
\hypertarget{NE29}{{\bf \small \underline{\ensuremath{^{\textnormal{17}}}O(\ensuremath{^{\textnormal{3}}}He,n),(\ensuremath{^{\textnormal{3}}}He,n\ensuremath{\gamma})\hspace{0.2in}\href{https://www.nndc.bnl.gov/nsr/nsrlink.jsp?1973Da31,B}{1973Da31},\href{https://www.nndc.bnl.gov/nsr/nsrlink.jsp?2005Ta28,B}{2005Ta28}}}}\\
\vspace{4pt}
\vspace{8pt}
\parbox[b][0.3cm]{17.7cm}{\addtolength{\parindent}{-0.2in}Two proton transfer reaction.}\\
\parbox[b][0.3cm]{17.7cm}{\addtolength{\parindent}{-0.2in}J\ensuremath{^{\ensuremath{\pi}}}(\ensuremath{^{\textnormal{17}}}O\ensuremath{_{\textnormal{g.s.}}})=5/2\ensuremath{^{\textnormal{+}}} and J\ensuremath{^{\ensuremath{\pi}}}(\ensuremath{^{\textnormal{3}}}He\ensuremath{_{\textnormal{g.s.}}})=1/2\ensuremath{^{\textnormal{+}}}.}\\
\parbox[b][0.3cm]{17.7cm}{\addtolength{\parindent}{-0.2in}\href{https://www.nndc.bnl.gov/nsr/nsrlink.jsp?1968Gu07,B}{1968Gu07}: \ensuremath{^{\textnormal{17}}}O(\ensuremath{^{\textnormal{3}}}He,n) E=3.076 MeV; measured neutron time-of-flight using NE-213 liquid scintillators; measured neutron angular}\\
\parbox[b][0.3cm]{17.7cm}{distributions at \ensuremath{\theta}\ensuremath{_{\textnormal{lab}}}=0\ensuremath{^\circ}{\textminus}110\ensuremath{^\circ}. Deduced \ensuremath{^{\textnormal{19}}}Ne level-energies and J\ensuremath{^{\ensuremath{\pi}}} assignments using a DWBA analysis.}\\
\parbox[b][0.3cm]{17.7cm}{\addtolength{\parindent}{-0.2in}\href{https://www.nndc.bnl.gov/nsr/nsrlink.jsp?1973Da31,B}{1973Da31}, \href{https://www.nndc.bnl.gov/nsr/nsrlink.jsp?1974DaYR,B}{1974DaYR}: \ensuremath{^{\textnormal{17}}}O(\ensuremath{^{\textnormal{3}}}He,n\ensuremath{\gamma}) E=3.0, 4.0, 5.0 MeV; measured n-\ensuremath{\gamma} coincidences using a liquid NE-213 scintillator at}\\
\parbox[b][0.3cm]{17.7cm}{\ensuremath{\theta}\ensuremath{_{\textnormal{lab}}}=0\ensuremath{^\circ}, 20\ensuremath{^\circ}, {\textminus}90\ensuremath{^\circ} and a Ge(Li) detector at \ensuremath{\theta}\ensuremath{_{\textnormal{lab}}}=90\ensuremath{^\circ}, 28\ensuremath{^\circ}, {\textminus}45\ensuremath{^\circ}. Measured E\ensuremath{_{\ensuremath{\gamma}}} (Ge detector covered \ensuremath{\theta}\ensuremath{_{\textnormal{lab}}}=88\ensuremath{^\circ}{\textminus}92\ensuremath{^\circ}), I\ensuremath{_{\ensuremath{\gamma}}},}\\
\parbox[b][0.3cm]{17.7cm}{\ensuremath{\gamma}-branching ratios (Ge detector covered \ensuremath{\theta}\ensuremath{_{\textnormal{lab}}}={\textminus}10\ensuremath{^\circ}{\textminus}80\ensuremath{^\circ}) and \ensuremath{^{\textnormal{19}}}Ne* lifetimes using the Doppler-shift attenuation method with a}\\
\parbox[b][0.3cm]{17.7cm}{timing resolution of 5.5 ns (FWHM). Deduced \ensuremath{^{\textnormal{19}}}Ne level-energies and lifetime limits. Assigned \ensuremath{^{\textnormal{19}}}F and \ensuremath{^{\textnormal{19}}}Ne mirror states on the}\\
\parbox[b][0.3cm]{17.7cm}{basis of \ensuremath{\gamma}-branching and discussed a state at \ensuremath{^{\textnormal{19}}}Ne*(4.78 MeV).}\\
\parbox[b][0.3cm]{17.7cm}{\addtolength{\parindent}{-0.2in}\href{https://www.nndc.bnl.gov/nsr/nsrlink.jsp?1973DaZP,B}{1973DaZP}: \ensuremath{^{\textnormal{17}}}O(\ensuremath{^{\textnormal{3}}}He,n); deduced \ensuremath{^{\textnormal{19}}}Ne levels.}\\
\parbox[b][0.3cm]{17.7cm}{\addtolength{\parindent}{-0.2in}\href{https://www.nndc.bnl.gov/nsr/nsrlink.jsp?2005Ta28,B}{2005Ta28}, \href{https://www.nndc.bnl.gov/nsr/nsrlink.jsp?2007TaZX,B}{2007TaZX}: \ensuremath{^{\textnormal{17}}}O(\ensuremath{^{\textnormal{3}}}He,n\ensuremath{\gamma}) E=3.0 MeV; measured n-\ensuremath{\gamma} coincidences using a HPGe detector placed at \ensuremath{\theta}\ensuremath{_{\textnormal{lab}}}=45\ensuremath{^\circ}, 135\ensuremath{^\circ}, or}\\
\parbox[b][0.3cm]{17.7cm}{28.5\ensuremath{^\circ} and a liquid scintillator at \ensuremath{\theta}\ensuremath{_{\textnormal{lab}}}=0\ensuremath{^\circ} or 90\ensuremath{^\circ} (with the Ge detector at 28.5\ensuremath{^\circ}). Deduced the \ensuremath{^{\textnormal{19}}}Ne*(1.5-4.6 MeV) level-energies}\\
\parbox[b][0.3cm]{17.7cm}{and lifetimes using the Doppler-shift attenuation method via a full line-shape analysis and GEANT4 simulations. Discussed}\\
\parbox[b][0.3cm]{17.7cm}{astrophysical implications for the \ensuremath{^{\textnormal{15}}}O(\ensuremath{\alpha},\ensuremath{\gamma}) reaction rate at nova temperatures.}\\
\parbox[b][0.3cm]{17.7cm}{\addtolength{\parindent}{-0.2in}\href{https://www.nndc.bnl.gov/nsr/nsrlink.jsp?2007HoZY,B}{2007HoZY}: \ensuremath{^{\textnormal{17}}}O(\ensuremath{^{\textnormal{3}}}He,n) E=4.2 MeV; measured neutron time-of-flight using three NE-213 and BC-501A liquid scintillators at}\\
\parbox[b][0.3cm]{17.7cm}{\ensuremath{\theta}\ensuremath{_{\textnormal{lab}}}=20\ensuremath{^\circ}. Discussed the measured NTOF spectrum.}\\
\vspace{12pt}
\underline{$^{19}$Ne Levels}\\
\vspace{0.34cm}
\parbox[b][0.3cm]{17.7cm}{\addtolength{\parindent}{-0.254cm}When an excitation energy is determined from a least-squares fit to \ensuremath{\gamma}-ray energies, evaluator assumed \ensuremath{\Delta}E\ensuremath{_{\ensuremath{\gamma}}}=1 where no}\\
\parbox[b][0.3cm]{17.7cm}{uncertainty in E\ensuremath{_{\ensuremath{\gamma}}} is given.}\\
\vspace{0.34cm}
\begin{longtable}{ccccccc@{\extracolsep{\fill}}c}
\multicolumn{2}{c}{E(level)$^{}$}&J$^{\pi}$$^{{\hyperlink{NE29LEVEL3}{d}}}$&\multicolumn{2}{c}{T\ensuremath{_{\textnormal{1/2}}}$^{{\hyperlink{NE29LEVEL1}{b}}}$}&L$^{{\hyperlink{NE29LEVEL3}{d}}}$&Comments&\\[-.2cm]
\multicolumn{2}{c}{\hrulefill}&\hrulefill&\multicolumn{2}{c}{\hrulefill}&\hrulefill&\hrulefill&
\endfirsthead
\multicolumn{1}{r@{}}{0}&\multicolumn{1}{@{}l}{\ensuremath{^{{\hyperlink{NE29LEVEL0}{a}}}}}&\multicolumn{1}{l}{1/2\ensuremath{^{+}}\ensuremath{^{{\hyperlink{NE29LEVEL0}{a}}}}}&&&&\parbox[t][0.3cm]{11.40484cm}{\raggedright E(level): State is populated in (\href{https://www.nndc.bnl.gov/nsr/nsrlink.jsp?1968Gu07,B}{1968Gu07}, \href{https://www.nndc.bnl.gov/nsr/nsrlink.jsp?1973Da31,B}{1973Da31}, \href{https://www.nndc.bnl.gov/nsr/nsrlink.jsp?2005Ta28,B}{2005Ta28}).\vspace{0.1cm}}&\\
\multicolumn{1}{r@{}}{238}&\multicolumn{1}{@{.}l}{18\ensuremath{^{{\hyperlink{NE29LEVEL0}{a}}}}}&\multicolumn{1}{l}{5/2\ensuremath{^{+}}\ensuremath{^{{\hyperlink{NE29LEVEL0}{a}}}}}&&&\multicolumn{1}{l}{0}&\parbox[t][0.3cm]{11.40484cm}{\raggedright E(level): Other values: 0.24 MeV \textit{7} (\href{https://www.nndc.bnl.gov/nsr/nsrlink.jsp?1968Gu07,B}{1968Gu07}): Unresolved from the ground and\vspace{0.1cm}}&\\
&&&&&&\parbox[t][0.3cm]{11.40484cm}{\raggedright {\ }{\ }{\ }0.28-MeV states; 0.24 MeV (\href{https://www.nndc.bnl.gov/nsr/nsrlink.jsp?1973Da31,B}{1973Da31}), which was also populated in\vspace{0.1cm}}&\\
&&&&&&\parbox[t][0.3cm]{11.40484cm}{\raggedright {\ }{\ }{\ }(\href{https://www.nndc.bnl.gov/nsr/nsrlink.jsp?2005Ta28,B}{2005Ta28}) but the energy is not reported.\vspace{0.1cm}}&\\
&&&&&&\parbox[t][0.3cm]{11.40484cm}{\raggedright J\ensuremath{^{\pi}},L: From DWBA analysis of (\href{https://www.nndc.bnl.gov/nsr/nsrlink.jsp?1968Gu07,B}{1968Gu07}) using a code called ``due to Yates''.\vspace{0.1cm}}&\\
&&&&&&\parbox[t][0.3cm]{11.40484cm}{\raggedright {\ }{\ }{\ }The evaluator notes that the DWBA fit does not describe the neutron angular\vspace{0.1cm}}&\\
&&&&&&\parbox[t][0.3cm]{11.40484cm}{\raggedright {\ }{\ }{\ }distribution at \ensuremath{\theta}\ensuremath{_{\textnormal{c.m.}}}\ensuremath{>}60\ensuremath{^\circ}.\vspace{0.1cm}}&\\
&&&&&&\parbox[t][0.3cm]{11.40484cm}{\raggedright J\ensuremath{^{\pi}}: See also J\ensuremath{^{\ensuremath{\pi}}}=5/2\ensuremath{^{\textnormal{+}}} (\href{https://www.nndc.bnl.gov/nsr/nsrlink.jsp?1973Da31,B}{1973Da31}) based on mirror levels analysis.\vspace{0.1cm}}&\\
\multicolumn{1}{r@{}}{275}&\multicolumn{1}{@{.}l}{06\ensuremath{^{{\hyperlink{NE29LEVEL0}{a}}}}}&\multicolumn{1}{l}{1/2\ensuremath{^{-}}\ensuremath{^{{\hyperlink{NE29LEVEL0}{a}}}}}&&&&\parbox[t][0.3cm]{11.40484cm}{\raggedright E(level): This state was populated in (\href{https://www.nndc.bnl.gov/nsr/nsrlink.jsp?1968Gu07,B}{1968Gu07}: Unresolved); and in (\href{https://www.nndc.bnl.gov/nsr/nsrlink.jsp?2005Ta28,B}{2005Ta28})\vspace{0.1cm}}&\\
&&&&&&\parbox[t][0.3cm]{11.40484cm}{\raggedright {\ }{\ }{\ }but they did not report its energy. Other value: 0.28 MeV (\href{https://www.nndc.bnl.gov/nsr/nsrlink.jsp?1973Da31,B}{1973Da31}).\vspace{0.1cm}}&\\
\multicolumn{1}{r@{}}{1507}&\multicolumn{1}{@{.}l}{51 {\it 35}}&\multicolumn{1}{l}{(5/2\ensuremath{^{-}})\ensuremath{^{{\hyperlink{NE29LEVEL2}{c}}}}}&\multicolumn{1}{r@{}}{1}&\multicolumn{1}{@{.}l}{18 ps {\it 21}}&&\parbox[t][0.3cm]{11.40484cm}{\raggedright E(level): From (\href{https://www.nndc.bnl.gov/nsr/nsrlink.jsp?2005Ta28,B}{2005Ta28}).\vspace{0.1cm}}&\\
&&&&&&\parbox[t][0.3cm]{11.40484cm}{\raggedright E(level): See also 1.55 MeV \textit{6} (\href{https://www.nndc.bnl.gov/nsr/nsrlink.jsp?1968Gu07,B}{1968Gu07}): An unresolved peak consisting of the\vspace{0.1cm}}&\\
&&&&&&\parbox[t][0.3cm]{11.40484cm}{\raggedright {\ }{\ }{\ }\ensuremath{^{\textnormal{19}}}Ne*(1.51, 1.54, 1.61 MeV) states; and 1.51 MeV (\href{https://www.nndc.bnl.gov/nsr/nsrlink.jsp?1973Da31,B}{1973Da31}).\vspace{0.1cm}}&\\
&&&&&&\parbox[t][0.3cm]{11.40484cm}{\raggedright T\ensuremath{_{1/2}}: From \ensuremath{\tau}=1.7\ensuremath{\times}10\ensuremath{^{\textnormal{3}}} fs \textit{3} (\href{https://www.nndc.bnl.gov/nsr/nsrlink.jsp?2005Ta28,B}{2005Ta28}).\vspace{0.1cm}}&\\
\multicolumn{1}{r@{}}{1536}&\multicolumn{1}{@{.}l}{1 {\it 8}}&\multicolumn{1}{l}{(3/2\ensuremath{^{+}})\ensuremath{^{{\hyperlink{NE29LEVEL2}{c}}}}}&\multicolumn{1}{r@{}}{11}&\multicolumn{1}{@{.}l}{1 fs {\it 28}}&&\parbox[t][0.3cm]{11.40484cm}{\raggedright E(level): From a least-squares fit to the E\ensuremath{_{\ensuremath{\gamma}}} from (\href{https://www.nndc.bnl.gov/nsr/nsrlink.jsp?2005Ta28,B}{2005Ta28}).\vspace{0.1cm}}&\\
&&&&&&\parbox[t][0.3cm]{11.40484cm}{\raggedright E(level): See also 1.55 MeV \textit{6} (\href{https://www.nndc.bnl.gov/nsr/nsrlink.jsp?1968Gu07,B}{1968Gu07}): An unresolved peak consisting of the\vspace{0.1cm}}&\\
&&&&&&\parbox[t][0.3cm]{11.40484cm}{\raggedright {\ }{\ }{\ }\ensuremath{^{\textnormal{19}}}Ne*(1.51, 1.54, 1.61 MeV) states; 1.54 MeV (\href{https://www.nndc.bnl.gov/nsr/nsrlink.jsp?1973Da31,B}{1973Da31}); and 1536.05 keV \textit{36}\vspace{0.1cm}}&\\
&&&&&&\parbox[t][0.3cm]{11.40484cm}{\raggedright {\ }{\ }{\ }(\href{https://www.nndc.bnl.gov/nsr/nsrlink.jsp?2005Ta28,B}{2005Ta28}).\vspace{0.1cm}}&\\
&&&&&&\parbox[t][0.3cm]{11.40484cm}{\raggedright T\ensuremath{_{1/2}}: From \ensuremath{\tau}=16 fs \textit{4} (\href{https://www.nndc.bnl.gov/nsr/nsrlink.jsp?2005Ta28,B}{2005Ta28}).\vspace{0.1cm}}&\\
\multicolumn{1}{r@{}}{1615}&\multicolumn{1}{@{.}l}{4 {\it 4}}&\multicolumn{1}{l}{(3/2\ensuremath{^{-}})\ensuremath{^{{\hyperlink{NE29LEVEL2}{c}}}}}&\multicolumn{1}{r@{}}{55}&\multicolumn{1}{@{ }l}{fs {\it 10}}&&\parbox[t][0.3cm]{11.40484cm}{\raggedright E(level): From (\href{https://www.nndc.bnl.gov/nsr/nsrlink.jsp?2005Ta28,B}{2005Ta28}).\vspace{0.1cm}}&\\
&&&&&&\parbox[t][0.3cm]{11.40484cm}{\raggedright E(level): See also 1.55 MeV \textit{6} (\href{https://www.nndc.bnl.gov/nsr/nsrlink.jsp?1968Gu07,B}{1968Gu07}): An unresolved peak consisting of the\vspace{0.1cm}}&\\
&&&&&&\parbox[t][0.3cm]{11.40484cm}{\raggedright {\ }{\ }{\ }\ensuremath{^{\textnormal{19}}}Ne*(1.51, 1.54, 1.61 MeV) states; and 1.62 MeV (\href{https://www.nndc.bnl.gov/nsr/nsrlink.jsp?1973Da31,B}{1973Da31}).\vspace{0.1cm}}&\\
&&&&&&\parbox[t][0.3cm]{11.40484cm}{\raggedright T\ensuremath{_{1/2}}: From \ensuremath{\tau}=80 fs \textit{15} (\href{https://www.nndc.bnl.gov/nsr/nsrlink.jsp?2005Ta28,B}{2005Ta28}) leading to T\ensuremath{_{\textnormal{1/2}}}=55.4 fs \textit{104}.\vspace{0.1cm}}&\\
\multicolumn{1}{r@{}}{2794}&\multicolumn{1}{@{.}l}{2 {\it 4}}&\multicolumn{1}{l}{(9/2\ensuremath{^{+}})}&\multicolumn{1}{r@{}}{69}&\multicolumn{1}{@{ }l}{fs {\it 8}}&\multicolumn{1}{l}{2}&\parbox[t][0.3cm]{11.40484cm}{\raggedright E(level): From (\href{https://www.nndc.bnl.gov/nsr/nsrlink.jsp?2005Ta28,B}{2005Ta28}).\vspace{0.1cm}}&\\
&&&&&&\parbox[t][0.3cm]{11.40484cm}{\raggedright E(level): See also 2.78 MeV \textit{5} (\href{https://www.nndc.bnl.gov/nsr/nsrlink.jsp?1968Gu07,B}{1968Gu07}) and 2.79 MeV (\href{https://www.nndc.bnl.gov/nsr/nsrlink.jsp?1973Da31,B}{1973Da31}).\vspace{0.1cm}}&\\
&&&&&&\parbox[t][0.3cm]{11.40484cm}{\raggedright T\ensuremath{_{1/2}}: From \ensuremath{\tau}=100 fs \textit{12} (\href{https://www.nndc.bnl.gov/nsr/nsrlink.jsp?2005Ta28,B}{2005Ta28}).\vspace{0.1cm}}&\\
&&&&&&\parbox[t][0.3cm]{11.40484cm}{\raggedright L: From DWBA analysis of (\href{https://www.nndc.bnl.gov/nsr/nsrlink.jsp?1968Gu07,B}{1968Gu07}) using a code called ``due to Yates''. This\vspace{0.1cm}}&\\
&&&&&&\parbox[t][0.3cm]{11.40484cm}{\raggedright {\ }{\ }{\ }led to L=2 and L=3; however, (\href{https://www.nndc.bnl.gov/nsr/nsrlink.jsp?1968Gu07,B}{1968Gu07}) preferred L=2 over L=3 due to a\vspace{0.1cm}}&\\
&&&&&&\parbox[t][0.3cm]{11.40484cm}{\raggedright {\ }{\ }{\ }better agreement between the measured neutron angular distribution and the\vspace{0.1cm}}&\\
&&&&&&\parbox[t][0.3cm]{11.40484cm}{\raggedright {\ }{\ }{\ }resulting DWBA curve at forward angles for L=2.\vspace{0.1cm}}&\\
\end{longtable}
\begin{textblock}{29}(0,27.3)
Continued on next page (footnotes at end of table)
\end{textblock}
\clearpage
\begin{longtable}{cccccc@{\extracolsep{\fill}}c}
\\[-.4cm]
\multicolumn{7}{c}{{\bf \small \underline{\ensuremath{^{\textnormal{17}}}O(\ensuremath{^{\textnormal{3}}}He,n),(\ensuremath{^{\textnormal{3}}}He,n\ensuremath{\gamma})\hspace{0.2in}\href{https://www.nndc.bnl.gov/nsr/nsrlink.jsp?1973Da31,B}{1973Da31},\href{https://www.nndc.bnl.gov/nsr/nsrlink.jsp?2005Ta28,B}{2005Ta28} (continued)}}}\\
\multicolumn{7}{c}{~}\\
\multicolumn{7}{c}{\underline{\ensuremath{^{19}}Ne Levels (continued)}}\\
\multicolumn{7}{c}{~}\\
\multicolumn{2}{c}{E(level)$^{}$}&J$^{\pi}$$^{{\hyperlink{NE29LEVEL3}{d}}}$&\multicolumn{2}{c}{T\ensuremath{_{\textnormal{1/2}}}$^{{\hyperlink{NE29LEVEL1}{b}}}$}&Comments&\\[-.2cm]
\multicolumn{2}{c}{\hrulefill}&\hrulefill&\multicolumn{2}{c}{\hrulefill}&\hrulefill&
\endhead
&&&&&\parbox[t][0.3cm]{11.655041cm}{\raggedright J\ensuremath{^{\pi}}: From the DWBA analysis of (\href{https://www.nndc.bnl.gov/nsr/nsrlink.jsp?1968Gu07,B}{1968Gu07}) and on the basis of a mirror levels\vspace{0.1cm}}&\\
&&&&&\parbox[t][0.3cm]{11.655041cm}{\raggedright {\ }{\ }{\ }analysis in the same study, (\href{https://www.nndc.bnl.gov/nsr/nsrlink.jsp?1968Gu07,B}{1968Gu07}) suggested J\ensuremath{^{\ensuremath{\pi}}}=(7/2\ensuremath{^{\textnormal{+}}},9/2\ensuremath{^{\textnormal{+}}}) with L=2. Later\vspace{0.1cm}}&\\
&&&&&\parbox[t][0.3cm]{11.655041cm}{\raggedright {\ }{\ }{\ }on, (\href{https://www.nndc.bnl.gov/nsr/nsrlink.jsp?1973Da31,B}{1973Da31}) assigned J\ensuremath{^{\ensuremath{\pi}}}=9/2\ensuremath{^{\textnormal{+}}} based on mirror levels analysis and by assuming\vspace{0.1cm}}&\\
&&&&&\parbox[t][0.3cm]{11.655041cm}{\raggedright {\ }{\ }{\ }the \ensuremath{^{\textnormal{19}}}F*(2.78 MeV, 9/2\ensuremath{^{\textnormal{+}}}) level as the mirror.\vspace{0.1cm}}&\\
\multicolumn{1}{r@{}}{3.70\ensuremath{\times10^{3}}?}&\multicolumn{1}{@{ }l}{{\it 4}}&&&&\parbox[t][0.3cm]{11.655041cm}{\raggedright E(level): From (\href{https://www.nndc.bnl.gov/nsr/nsrlink.jsp?1968Gu07,B}{1968Gu07}). The later studies did not observe this state, and thus this\vspace{0.1cm}}&\\
&&&&&\parbox[t][0.3cm]{11.655041cm}{\raggedright {\ }{\ }{\ }state is not included in the \ensuremath{^{\textnormal{19}}}Ne Adopted Levels.\vspace{0.1cm}}&\\
\multicolumn{1}{r@{}}{4034}&\multicolumn{1}{@{.}l}{6 {\it 7}}&\multicolumn{1}{l}{(3/2\ensuremath{^{+}})}&\multicolumn{1}{r@{}}{9}&\multicolumn{1}{@{ }l}{fs {\it +6\textminus4}}&\parbox[t][0.3cm]{11.655041cm}{\raggedright E(level): Weighted average of 4.01 MeV \textit{2} (\href{https://www.nndc.bnl.gov/nsr/nsrlink.jsp?1968Gu07,B}{1968Gu07}); 4032.9 keV \textit{24} (\href{https://www.nndc.bnl.gov/nsr/nsrlink.jsp?1973Da31,B}{1973Da31});\vspace{0.1cm}}&\\
&&&&&\parbox[t][0.3cm]{11.655041cm}{\raggedright {\ }{\ }{\ }and 4034.8 keV \textit{7} from the least-squares fit to the E\ensuremath{_{\ensuremath{\gamma}}} values from (\href{https://www.nndc.bnl.gov/nsr/nsrlink.jsp?2005Ta28,B}{2005Ta28}).\vspace{0.1cm}}&\\
&&&&&\parbox[t][0.3cm]{11.655041cm}{\raggedright E(level): See also 4034.5 keV \textit{8} reported by (\href{https://www.nndc.bnl.gov/nsr/nsrlink.jsp?2005Ta28,B}{2005Ta28}).\vspace{0.1cm}}&\\
&&&&&\parbox[t][0.3cm]{11.655041cm}{\raggedright T\ensuremath{_{1/2}}: From \ensuremath{\tau}=13 fs \textit{+9{\textminus}6} deduced at 1\ensuremath{\sigma} C.L. (\href{https://www.nndc.bnl.gov/nsr/nsrlink.jsp?2005Ta28,B}{2005Ta28}). See also T\ensuremath{_{\textnormal{1/2}}}=9 fs \textit{+11{\textminus}6}\vspace{0.1cm}}&\\
&&&&&\parbox[t][0.3cm]{11.655041cm}{\raggedright {\ }{\ }{\ }from \ensuremath{\tau}=13 fs \textit{+16{\textminus}9} obtained at 2\ensuremath{\sigma} C.L. (\href{https://www.nndc.bnl.gov/nsr/nsrlink.jsp?2005Ta28,B}{2005Ta28}); and T\ensuremath{_{\textnormal{1/2}}}\ensuremath{<}35 fs from \ensuremath{\tau}\ensuremath{<}0.05\vspace{0.1cm}}&\\
&&&&&\parbox[t][0.3cm]{11.655041cm}{\raggedright {\ }{\ }{\ }ps (\href{https://www.nndc.bnl.gov/nsr/nsrlink.jsp?1973Da31,B}{1973Da31}).\vspace{0.1cm}}&\\
&&&&&\parbox[t][0.3cm]{11.655041cm}{\raggedright J\ensuremath{^{\pi}}: From J\ensuremath{^{\ensuremath{\pi}}}=3/2\ensuremath{^{\textnormal{+}}} (\href{https://www.nndc.bnl.gov/nsr/nsrlink.jsp?2007TaZX,B}{2007TaZX}) based on comparison of the \ensuremath{\gamma}-ray decay scheme of\vspace{0.1cm}}&\\
&&&&&\parbox[t][0.3cm]{11.655041cm}{\raggedright {\ }{\ }{\ }this \ensuremath{^{\textnormal{19}}}Ne* state to those of the \ensuremath{^{\textnormal{19}}}F* levels in this vicinity and based on mirror\vspace{0.1cm}}&\\
&&&&&\parbox[t][0.3cm]{11.655041cm}{\raggedright {\ }{\ }{\ }level analysis in (\href{https://www.nndc.bnl.gov/nsr/nsrlink.jsp?1973Da31,B}{1973Da31}).\vspace{0.1cm}}&\\
&&&&&\parbox[t][0.3cm]{11.655041cm}{\raggedright (\href{https://www.nndc.bnl.gov/nsr/nsrlink.jsp?1973Da31,B}{1973Da31}) assigned the \ensuremath{^{\textnormal{19}}}F*(3.91 MeV, 3/2\ensuremath{^{\textnormal{+}}}) state as the mirror level of the\vspace{0.1cm}}&\\
&&&&&\parbox[t][0.3cm]{11.655041cm}{\raggedright {\ }{\ }{\ }\ensuremath{^{\textnormal{19}}}Ne*(4035) state.\vspace{0.1cm}}&\\
\multicolumn{1}{r@{}}{4143}&\multicolumn{1}{@{.}l}{2 {\it 11}}&\multicolumn{1}{l}{(7/2\ensuremath{^{-}})}&\multicolumn{1}{r@{}}{12}&\multicolumn{1}{@{.}l}{5 fs {\it +14\textminus21}}&\parbox[t][0.3cm]{11.655041cm}{\raggedright E(level): Weighted average of 4.13 MeV \textit{3} (\href{https://www.nndc.bnl.gov/nsr/nsrlink.jsp?1968Gu07,B}{1968Gu07}); 4140 keV \textit{4} (\href{https://www.nndc.bnl.gov/nsr/nsrlink.jsp?1973Da31,B}{1973Da31});\vspace{0.1cm}}&\\
&&&&&\parbox[t][0.3cm]{11.655041cm}{\raggedright {\ }{\ }{\ }and 4143.5 keV \textit{11} from the least-squares fit to the E\ensuremath{_{\ensuremath{\gamma}}} values from (\href{https://www.nndc.bnl.gov/nsr/nsrlink.jsp?2005Ta28,B}{2005Ta28}).\vspace{0.1cm}}&\\
&&&&&\parbox[t][0.3cm]{11.655041cm}{\raggedright E(level): See also 4143.5 keV \textit{6} reported by (\href{https://www.nndc.bnl.gov/nsr/nsrlink.jsp?2005Ta28,B}{2005Ta28}).\vspace{0.1cm}}&\\
&&&&&\parbox[t][0.3cm]{11.655041cm}{\raggedright T\ensuremath{_{1/2}}: From \ensuremath{\tau}=18 fs \textit{+2{\textminus}3} (\href{https://www.nndc.bnl.gov/nsr/nsrlink.jsp?2005Ta28,B}{2005Ta28}). See also T\ensuremath{_{\textnormal{1/2}}}\ensuremath{<}208 fs from \ensuremath{\tau}\ensuremath{<}0.30 ps\vspace{0.1cm}}&\\
&&&&&\parbox[t][0.3cm]{11.655041cm}{\raggedright {\ }{\ }{\ }(\href{https://www.nndc.bnl.gov/nsr/nsrlink.jsp?1973Da31,B}{1973Da31}).\vspace{0.1cm}}&\\
&&&&&\parbox[t][0.3cm]{11.655041cm}{\raggedright J\ensuremath{^{\pi}}: From (\href{https://www.nndc.bnl.gov/nsr/nsrlink.jsp?2005Ta28,B}{2005Ta28}, \href{https://www.nndc.bnl.gov/nsr/nsrlink.jsp?2007TaZX,B}{2007TaZX}) based on comparison of the \ensuremath{\gamma}-ray decay scheme of\vspace{0.1cm}}&\\
&&&&&\parbox[t][0.3cm]{11.655041cm}{\raggedright {\ }{\ }{\ }this state to those of the \ensuremath{^{\textnormal{19}}}F* levels in this vicinity.\vspace{0.1cm}}&\\
&&&&&\parbox[t][0.3cm]{11.655041cm}{\raggedright J\ensuremath{^{\pi}}: See also J\ensuremath{^{\ensuremath{\pi}}}=(9/2\ensuremath{^{-}}) in (\href{https://www.nndc.bnl.gov/nsr/nsrlink.jsp?1973Da31,B}{1973Da31}), who claimed that this state is formed by the\vspace{0.1cm}}&\\
&&&&&\parbox[t][0.3cm]{11.655041cm}{\raggedright {\ }{\ }{\ }coupling of a \textit{p}\ensuremath{_{\textnormal{1/2}}} hole with the J\ensuremath{^{\ensuremath{\pi}}}=4\ensuremath{^{\textnormal{+}}} member of the ground state rotational band\vspace{0.1cm}}&\\
&&&&&\parbox[t][0.3cm]{11.655041cm}{\raggedright {\ }{\ }{\ }of \ensuremath{^{\textnormal{20}}}Ne. (\href{https://www.nndc.bnl.gov/nsr/nsrlink.jsp?1973Da31,B}{1973Da31}) determined the \ensuremath{^{\textnormal{19}}}F*(4.03 MeV, 9/2\ensuremath{^{-}}) level as the mirror\vspace{0.1cm}}&\\
&&&&&\parbox[t][0.3cm]{11.655041cm}{\raggedright {\ }{\ }{\ }state.\vspace{0.1cm}}&\\
\multicolumn{1}{r@{}}{4199}&\multicolumn{1}{@{.}l}{6 {\it 10}}&\multicolumn{1}{l}{(9/2\ensuremath{^{-}})}&\multicolumn{1}{r@{}}{30}&\multicolumn{1}{@{ }l}{fs {\it +8\textminus6}}&\parbox[t][0.3cm]{11.655041cm}{\raggedright E(level): Weighted average (with external errors) of 4197.1 keV \textit{24} (\href{https://www.nndc.bnl.gov/nsr/nsrlink.jsp?1973Da31,B}{1973Da31}) and\vspace{0.1cm}}&\\
&&&&&\parbox[t][0.3cm]{11.655041cm}{\raggedright {\ }{\ }{\ }4200.0 keV \textit{10} from a least-squares fit to the E\ensuremath{_{\ensuremath{\gamma}}} values from (\href{https://www.nndc.bnl.gov/nsr/nsrlink.jsp?2005Ta28,B}{2005Ta28}).\vspace{0.1cm}}&\\
&&&&&\parbox[t][0.3cm]{11.655041cm}{\raggedright E(level): See also 4200.3 keV \textit{11} reported by (\href{https://www.nndc.bnl.gov/nsr/nsrlink.jsp?2005Ta28,B}{2005Ta28}).\vspace{0.1cm}}&\\
&&&&&\parbox[t][0.3cm]{11.655041cm}{\raggedright T\ensuremath{_{1/2}}: From \ensuremath{\tau}=43 fs \textit{+12{\textminus}9} (\href{https://www.nndc.bnl.gov/nsr/nsrlink.jsp?2005Ta28,B}{2005Ta28}). See also T\ensuremath{_{\textnormal{1/2}}}\ensuremath{<}243 fs from \ensuremath{\tau}\ensuremath{<}0.35 ps\vspace{0.1cm}}&\\
&&&&&\parbox[t][0.3cm]{11.655041cm}{\raggedright {\ }{\ }{\ }(\href{https://www.nndc.bnl.gov/nsr/nsrlink.jsp?1973Da31,B}{1973Da31}).\vspace{0.1cm}}&\\
&&&&&\parbox[t][0.3cm]{11.655041cm}{\raggedright J\ensuremath{^{\pi}}: From (\href{https://www.nndc.bnl.gov/nsr/nsrlink.jsp?2005Ta28,B}{2005Ta28}, \href{https://www.nndc.bnl.gov/nsr/nsrlink.jsp?2007TaZX,B}{2007TaZX}) based on comparison of the \ensuremath{\gamma}-ray decay scheme of\vspace{0.1cm}}&\\
&&&&&\parbox[t][0.3cm]{11.655041cm}{\raggedright {\ }{\ }{\ }this level to those of the \ensuremath{^{\textnormal{19}}}F* levels in this vicinity.\vspace{0.1cm}}&\\
&&&&&\parbox[t][0.3cm]{11.655041cm}{\raggedright J\ensuremath{^{\pi}}: See also J\ensuremath{^{\ensuremath{\pi}}}=(7/2\ensuremath{^{-}}) in (\href{https://www.nndc.bnl.gov/nsr/nsrlink.jsp?1973Da31,B}{1973Da31}), who claimed that this state is formed by the\vspace{0.1cm}}&\\
&&&&&\parbox[t][0.3cm]{11.655041cm}{\raggedright {\ }{\ }{\ }coupling of a \textit{p}\ensuremath{_{\textnormal{1/2}}} hole with the J\ensuremath{^{\ensuremath{\pi}}}=4\ensuremath{^{\textnormal{+}}} member of the ground state rotational band\vspace{0.1cm}}&\\
&&&&&\parbox[t][0.3cm]{11.655041cm}{\raggedright {\ }{\ }{\ }of \ensuremath{^{\textnormal{20}}}Ne. (\href{https://www.nndc.bnl.gov/nsr/nsrlink.jsp?1973Da31,B}{1973Da31}) determined the \ensuremath{^{\textnormal{19}}}F*(4.00 MeV, 7/2\ensuremath{^{-}}) level as the mirror\vspace{0.1cm}}&\\
&&&&&\parbox[t][0.3cm]{11.655041cm}{\raggedright {\ }{\ }{\ }state.\vspace{0.1cm}}&\\
\multicolumn{1}{r@{}}{4379}&\multicolumn{1}{@{.}l}{1 {\it 16}}&\multicolumn{1}{l}{(7/2\ensuremath{^{+}})\ensuremath{^{{\hyperlink{NE29LEVEL2}{c}}}}}&\multicolumn{1}{r@{}}{3}&\multicolumn{1}{@{.}l}{5 fs {\it +21\textminus14}}&\parbox[t][0.3cm]{11.655041cm}{\raggedright E(level): Weighted average of 4.36 MeV \textit{3} (\href{https://www.nndc.bnl.gov/nsr/nsrlink.jsp?1968Gu07,B}{1968Gu07}), 4379.1 keV \textit{22} (\href{https://www.nndc.bnl.gov/nsr/nsrlink.jsp?1973Da31,B}{1973Da31})\vspace{0.1cm}}&\\
&&&&&\parbox[t][0.3cm]{11.655041cm}{\raggedright {\ }{\ }{\ }and 4379.2 keV \textit{16} from a least-squares fit to the E\ensuremath{_{\ensuremath{\gamma}}} values from (\href{https://www.nndc.bnl.gov/nsr/nsrlink.jsp?2005Ta28,B}{2005Ta28}).\vspace{0.1cm}}&\\
&&&&&\parbox[t][0.3cm]{11.655041cm}{\raggedright E(level): See also 4377.8 keV \textit{6} reported by (\href{https://www.nndc.bnl.gov/nsr/nsrlink.jsp?2005Ta28,B}{2005Ta28}).\vspace{0.1cm}}&\\
&&&&&\parbox[t][0.3cm]{11.655041cm}{\raggedright T\ensuremath{_{1/2}}: From \ensuremath{\tau}=5 fs \textit{+3{\textminus}2} (\href{https://www.nndc.bnl.gov/nsr/nsrlink.jsp?2005Ta28,B}{2005Ta28}). See also T\ensuremath{_{\textnormal{1/2}}}\ensuremath{<}83 fs from \ensuremath{\tau}\ensuremath{<}0.12 ps\vspace{0.1cm}}&\\
&&&&&\parbox[t][0.3cm]{11.655041cm}{\raggedright {\ }{\ }{\ }(\href{https://www.nndc.bnl.gov/nsr/nsrlink.jsp?1973Da31,B}{1973Da31}).\vspace{0.1cm}}&\\
&&&&&\parbox[t][0.3cm]{11.655041cm}{\raggedright Mirror state was determined to be the \ensuremath{^{\textnormal{19}}}F*(4.38 MeV, 7/2\ensuremath{^{\textnormal{+}}}) level (\href{https://www.nndc.bnl.gov/nsr/nsrlink.jsp?1973Da31,B}{1973Da31}).\vspace{0.1cm}}&\\
&&&&&\parbox[t][0.3cm]{11.655041cm}{\raggedright J\ensuremath{^{\pi}}: From mirror analysis of (\href{https://www.nndc.bnl.gov/nsr/nsrlink.jsp?1973Da31,B}{1973Da31}) and based on J\ensuremath{^{\ensuremath{\pi}}}=7/2\ensuremath{^{\textnormal{+}}} deduced by\vspace{0.1cm}}&\\
&&&&&\parbox[t][0.3cm]{11.655041cm}{\raggedright {\ }{\ }{\ }(\href{https://www.nndc.bnl.gov/nsr/nsrlink.jsp?2007TaZX,B}{2007TaZX}) from comparison of the \ensuremath{\gamma}-ray decay scheme of this state and those of\vspace{0.1cm}}&\\
&&&&&\parbox[t][0.3cm]{11.655041cm}{\raggedright {\ }{\ }{\ }the \ensuremath{^{\textnormal{19}}}F* levels in this vicinity.\vspace{0.1cm}}&\\
\multicolumn{1}{r@{}}{4548}&\multicolumn{1}{@{.}l}{3 {\it 8}}&\multicolumn{1}{l}{(3/2\ensuremath{^{-}})}&\multicolumn{1}{r@{}}{10}&\multicolumn{1}{@{ }l}{fs {\it +8\textminus4}}&\parbox[t][0.3cm]{11.655041cm}{\raggedright E(level): From a least-squares fit to the E\ensuremath{_{\ensuremath{\gamma}}} values from (\href{https://www.nndc.bnl.gov/nsr/nsrlink.jsp?2005Ta28,B}{2005Ta28}).\vspace{0.1cm}}&\\
&&&&&\parbox[t][0.3cm]{11.655041cm}{\raggedright E(level): See also 4.54 MeV \textit{3} (\href{https://www.nndc.bnl.gov/nsr/nsrlink.jsp?1968Gu07,B}{1968Gu07}); 4549 keV \textit{4} (\href{https://www.nndc.bnl.gov/nsr/nsrlink.jsp?1973Da31,B}{1973Da31}); and 4547.7\vspace{0.1cm}}&\\
&&&&&\parbox[t][0.3cm]{11.655041cm}{\raggedright {\ }{\ }{\ }keV \textit{10} reported by (\href{https://www.nndc.bnl.gov/nsr/nsrlink.jsp?2005Ta28,B}{2005Ta28}).\vspace{0.1cm}}&\\
&&&&&\parbox[t][0.3cm]{11.655041cm}{\raggedright T\ensuremath{_{1/2}}: From \ensuremath{\tau}=15 fs \textit{+11{\textminus}5} (\href{https://www.nndc.bnl.gov/nsr/nsrlink.jsp?2005Ta28,B}{2005Ta28}), which leads to T\ensuremath{_{\textnormal{1/2}}}=10.4 fs \textit{+76{\textminus}35}. See also\vspace{0.1cm}}&\\
&&&&&\parbox[t][0.3cm]{11.655041cm}{\raggedright {\ }{\ }{\ }T\ensuremath{_{\textnormal{1/2}}}\ensuremath{<}55.4 fs from \ensuremath{\tau}\ensuremath{<}0.08 ps (\href{https://www.nndc.bnl.gov/nsr/nsrlink.jsp?1973Da31,B}{1973Da31}).\vspace{0.1cm}}&\\
&&&&&\parbox[t][0.3cm]{11.655041cm}{\raggedright J\ensuremath{^{\pi}}: From J\ensuremath{^{\ensuremath{\pi}}}=3/2\ensuremath{^{-}} (\href{https://www.nndc.bnl.gov/nsr/nsrlink.jsp?2007TaZX,B}{2007TaZX}) based on comparison of the \ensuremath{\gamma}-ray decay scheme of\vspace{0.1cm}}&\\
\end{longtable}
\begin{textblock}{29}(0,27.3)
Continued on next page (footnotes at end of table)
\end{textblock}
\clearpage
\begin{longtable}{cccccc@{\extracolsep{\fill}}c}
\\[-.4cm]
\multicolumn{7}{c}{{\bf \small \underline{\ensuremath{^{\textnormal{17}}}O(\ensuremath{^{\textnormal{3}}}He,n),(\ensuremath{^{\textnormal{3}}}He,n\ensuremath{\gamma})\hspace{0.2in}\href{https://www.nndc.bnl.gov/nsr/nsrlink.jsp?1973Da31,B}{1973Da31},\href{https://www.nndc.bnl.gov/nsr/nsrlink.jsp?2005Ta28,B}{2005Ta28} (continued)}}}\\
\multicolumn{7}{c}{~}\\
\multicolumn{7}{c}{\underline{\ensuremath{^{19}}Ne Levels (continued)}}\\
\multicolumn{7}{c}{~}\\
\multicolumn{2}{c}{E(level)$^{}$}&J$^{\pi}$$^{{\hyperlink{NE29LEVEL3}{d}}}$&\multicolumn{2}{c}{T\ensuremath{_{\textnormal{1/2}}}$^{{\hyperlink{NE29LEVEL1}{b}}}$}&Comments&\\[-.2cm]
\multicolumn{2}{c}{\hrulefill}&\hrulefill&\multicolumn{2}{c}{\hrulefill}&\hrulefill&
\endhead
&&&&&\parbox[t][0.3cm]{11.478651cm}{\raggedright {\ }{\ }{\ }this state to those of the \ensuremath{^{\textnormal{19}}}F* levels in this region.\vspace{0.1cm}}&\\
&&&&&\parbox[t][0.3cm]{11.478651cm}{\raggedright J\ensuremath{^{\pi}}: See also J\ensuremath{^{\ensuremath{\pi}}}=(1/2\ensuremath{^{-}}, 3/2\ensuremath{^{-}}) from (\href{https://www.nndc.bnl.gov/nsr/nsrlink.jsp?1973Da31,B}{1973Da31}) but it is not clear if these\vspace{0.1cm}}&\\
&&&&&\parbox[t][0.3cm]{11.478651cm}{\raggedright {\ }{\ }{\ }assignments were deduced based on the data of (\href{https://www.nndc.bnl.gov/nsr/nsrlink.jsp?1973Da31,B}{1973Da31}). They assigned the\vspace{0.1cm}}&\\
&&&&&\parbox[t][0.3cm]{11.478651cm}{\raggedright {\ }{\ }{\ }\ensuremath{^{\textnormal{19}}}F*(4556 keV, 3/2\ensuremath{^{-}}) state as the mirror level.\vspace{0.1cm}}&\\
\multicolumn{1}{r@{}}{4602}&\multicolumn{1}{@{.}l}{1 {\it 10}}&\multicolumn{1}{l}{(5/2\ensuremath{^{+}})\ensuremath{^{{\hyperlink{NE29LEVEL2}{c}}}}}&\multicolumn{1}{r@{}}{4}&\multicolumn{1}{@{.}l}{9 fs {\it +35\textminus28}}&\parbox[t][0.3cm]{11.478651cm}{\raggedright E(level): From a least-squares fit to the E\ensuremath{_{\ensuremath{\gamma}}} values from (\href{https://www.nndc.bnl.gov/nsr/nsrlink.jsp?2005Ta28,B}{2005Ta28}).\vspace{0.1cm}}&\\
&&&&&\parbox[t][0.3cm]{11.478651cm}{\raggedright E(level): See also 4.61 MeV \textit{4} (\href{https://www.nndc.bnl.gov/nsr/nsrlink.jsp?1968Gu07,B}{1968Gu07}); 4605 keV \textit{5} (\href{https://www.nndc.bnl.gov/nsr/nsrlink.jsp?1973Da31,B}{1973Da31}); and 4601.8\vspace{0.1cm}}&\\
&&&&&\parbox[t][0.3cm]{11.478651cm}{\raggedright {\ }{\ }{\ }keV \textit{8} reported in (\href{https://www.nndc.bnl.gov/nsr/nsrlink.jsp?2005Ta28,B}{2005Ta28}).\vspace{0.1cm}}&\\
&&&&&\parbox[t][0.3cm]{11.478651cm}{\raggedright T\ensuremath{_{1/2}}: From \ensuremath{\tau}=7 fs \textit{+5{\textminus}4} (\href{https://www.nndc.bnl.gov/nsr/nsrlink.jsp?2005Ta28,B}{2005Ta28}). See also T\ensuremath{_{\textnormal{1/2}}}\ensuremath{<}111 fs from \ensuremath{\tau}\ensuremath{<}0.16 ps\vspace{0.1cm}}&\\
&&&&&\parbox[t][0.3cm]{11.478651cm}{\raggedright {\ }{\ }{\ }(\href{https://www.nndc.bnl.gov/nsr/nsrlink.jsp?1973Da31,B}{1973Da31}).\vspace{0.1cm}}&\\
&&&&&\parbox[t][0.3cm]{11.478651cm}{\raggedright Mirror state was determined to be \ensuremath{^{\textnormal{19}}}F*(4550 MeV, 5/2\ensuremath{^{\textnormal{+}}}) (\href{https://www.nndc.bnl.gov/nsr/nsrlink.jsp?1973Da31,B}{1973Da31}).\vspace{0.1cm}}&\\
&&&&&\parbox[t][0.3cm]{11.478651cm}{\raggedright J\ensuremath{^{\pi}}: See also J\ensuremath{^{\ensuremath{\pi}}}=5/2\ensuremath{^{\textnormal{+}}} (\href{https://www.nndc.bnl.gov/nsr/nsrlink.jsp?2007TaZX,B}{2007TaZX}) based on comparison of the \ensuremath{\gamma}-ray decay scheme\vspace{0.1cm}}&\\
&&&&&\parbox[t][0.3cm]{11.478651cm}{\raggedright {\ }{\ }{\ }for this state to those of the \ensuremath{^{\textnormal{19}}}F* levels in this region.\vspace{0.1cm}}&\\
\multicolumn{1}{r@{}}{4634}&\multicolumn{1}{@{.}l}{1 {\it 9}}&\multicolumn{1}{l}{(13/2\ensuremath{^{+}})\ensuremath{^{{\hyperlink{NE29LEVEL2}{c}}}}}&\multicolumn{1}{r@{}}{$>$0}&\multicolumn{1}{@{.}l}{7 ps}&\parbox[t][0.3cm]{11.478651cm}{\raggedright E(level): Weighted average of 4634.0 keV \textit{9} (\href{https://www.nndc.bnl.gov/nsr/nsrlink.jsp?2005Ta28,B}{2005Ta28}) and 4635 keV \textit{4}\vspace{0.1cm}}&\\
&&&&&\parbox[t][0.3cm]{11.478651cm}{\raggedright {\ }{\ }{\ }(\href{https://www.nndc.bnl.gov/nsr/nsrlink.jsp?1973Da31,B}{1973Da31}).\vspace{0.1cm}}&\\
&&&&&\parbox[t][0.3cm]{11.478651cm}{\raggedright T\ensuremath{_{1/2}}: From \ensuremath{\tau}\ensuremath{>}1\ensuremath{\times}10\ensuremath{^{\textnormal{3}}} fs (\href{https://www.nndc.bnl.gov/nsr/nsrlink.jsp?2005Ta28,B}{2005Ta28}). See also T\ensuremath{_{\textnormal{1/2}}}\ensuremath{>}693 fs from \ensuremath{\tau}\ensuremath{>}1 ps (\href{https://www.nndc.bnl.gov/nsr/nsrlink.jsp?1973Da31,B}{1973Da31}).\vspace{0.1cm}}&\\
&&&&&\parbox[t][0.3cm]{11.478651cm}{\raggedright J\ensuremath{^{\pi}}: See also J\ensuremath{^{\ensuremath{\pi}}}\ensuremath{>}7/2 (\href{https://www.nndc.bnl.gov/nsr/nsrlink.jsp?1973Da31,B}{1973Da31}) based on the fact that the \ensuremath{\gamma} decay proceeded\vspace{0.1cm}}&\\
&&&&&\parbox[t][0.3cm]{11.478651cm}{\raggedright {\ }{\ }{\ }preferentially to a state with J=9/2 instead of nearby states with J\ensuremath{^{\ensuremath{\pi}}}=1/2\ensuremath{^{\ensuremath{\pm}}}, 3/2\ensuremath{^{\ensuremath{\pm}}},\vspace{0.1cm}}&\\
&&&&&\parbox[t][0.3cm]{11.478651cm}{\raggedright {\ }{\ }{\ }and 5/2\ensuremath{^{\ensuremath{\pm}}}. The authors reported that if J=(9/2, 11/2) assignments were selected, the\vspace{0.1cm}}&\\
&&&&&\parbox[t][0.3cm]{11.478651cm}{\raggedright {\ }{\ }{\ }\ensuremath{\gamma} ray observed from the decay of this state would be M1 or E1 transitions with\vspace{0.1cm}}&\\
&&&&&\parbox[t][0.3cm]{11.478651cm}{\raggedright {\ }{\ }{\ }strengths of B(M1)\ensuremath{<}0.005 W.u. and B(E1)\ensuremath{<}2\ensuremath{\times}10\ensuremath{^{\textnormal{$-$4}}} W.u., which would be too\vspace{0.1cm}}&\\
&&&&&\parbox[t][0.3cm]{11.478651cm}{\raggedright {\ }{\ }{\ }weak to be observed. Therefore, using this analogy and from the mirror levels\vspace{0.1cm}}&\\
&&&&&\parbox[t][0.3cm]{11.478651cm}{\raggedright {\ }{\ }{\ }analysis, (\href{https://www.nndc.bnl.gov/nsr/nsrlink.jsp?1973Da31,B}{1973Da31}) assigned J\ensuremath{^{\ensuremath{\pi}}}=13/2\ensuremath{^{\textnormal{+}}} to this state. See also J\ensuremath{^{\ensuremath{\pi}}}=13/2\ensuremath{^{\textnormal{+}}}\vspace{0.1cm}}&\\
&&&&&\parbox[t][0.3cm]{11.478651cm}{\raggedright {\ }{\ }{\ }(\href{https://www.nndc.bnl.gov/nsr/nsrlink.jsp?2007TaZX,B}{2007TaZX}) based on comparison of the \ensuremath{\gamma}-ray decay scheme of this state to those\vspace{0.1cm}}&\\
&&&&&\parbox[t][0.3cm]{11.478651cm}{\raggedright {\ }{\ }{\ }of the \ensuremath{^{\textnormal{19}}}F* levels in this vicinity.\vspace{0.1cm}}&\\
&&&&&\parbox[t][0.3cm]{11.478651cm}{\raggedright E(level): (\href{https://www.nndc.bnl.gov/nsr/nsrlink.jsp?1973Da31,B}{1973Da31}) assigned the \ensuremath{^{\textnormal{19}}}F*(4.65 MeV, 13/2\ensuremath{^{\textnormal{+}}}) level as the mirror state.\vspace{0.1cm}}&\\
\multicolumn{1}{r@{}}{4.69\ensuremath{\times10^{3}}}&\multicolumn{1}{@{ }l}{{\it 3}}&\multicolumn{1}{l}{(5/2\ensuremath{^{-}})\ensuremath{^{{\hyperlink{NE29LEVEL2}{c}}}}}&&&\parbox[t][0.3cm]{11.478651cm}{\raggedright E(level): From (\href{https://www.nndc.bnl.gov/nsr/nsrlink.jsp?1968Gu07,B}{1968Gu07}). Evaluator highlights that (\href{https://www.nndc.bnl.gov/nsr/nsrlink.jsp?1973Da31,B}{1973Da31}) suggested that a\vspace{0.1cm}}&\\
&&&&&\parbox[t][0.3cm]{11.478651cm}{\raggedright {\ }{\ }{\ }nearby state measured at 4783 keV \textit{20} by (\href{https://www.nndc.bnl.gov/nsr/nsrlink.jsp?1970Ga18,B}{1970Ga18}) is erroneous based on\vspace{0.1cm}}&\\
&&&&&\parbox[t][0.3cm]{11.478651cm}{\raggedright {\ }{\ }{\ }mirror levels analysis and the unpublished \ensuremath{^{\textnormal{19}}}F(\ensuremath{^{\textnormal{3}}}He,t) data taken at numerous\vspace{0.1cm}}&\\
&&&&&\parbox[t][0.3cm]{11.478651cm}{\raggedright {\ }{\ }{\ }angles by (D. Delmhard and H. Ohnuma, John H. Williams Laboratory, University\vspace{0.1cm}}&\\
&&&&&\parbox[t][0.3cm]{11.478651cm}{\raggedright {\ }{\ }{\ }of Minnesota, Annual report (1971) unpublished, and by priv. comm. with D.\vspace{0.1cm}}&\\
&&&&&\parbox[t][0.3cm]{11.478651cm}{\raggedright {\ }{\ }{\ }Dehnhard). (\href{https://www.nndc.bnl.gov/nsr/nsrlink.jsp?1973Da31,B}{1973Da31}) saw no evidence for transitions from the \ensuremath{^{\textnormal{19}}}Ne*(4.69\vspace{0.1cm}}&\\
&&&&&\parbox[t][0.3cm]{11.478651cm}{\raggedright {\ }{\ }{\ }MeV) and the 4.78-MeV states.\vspace{0.1cm}}&\\
&&&&&\parbox[t][0.3cm]{11.478651cm}{\raggedright J\ensuremath{^{\pi}}: (\href{https://www.nndc.bnl.gov/nsr/nsrlink.jsp?1973Da31,B}{1973Da31}) suggested the \ensuremath{^{\textnormal{19}}}F*(4.68 MeV, 5/2\ensuremath{^{-}}) level as the mirror to this\vspace{0.1cm}}&\\
&&&&&\parbox[t][0.3cm]{11.478651cm}{\raggedright {\ }{\ }{\ }state, which was not observed in their experiment.\vspace{0.1cm}}&\\
\multicolumn{1}{r@{}}{5096}&\multicolumn{1}{@{ }l}{{\it 10}}&\multicolumn{1}{l}{(5/2\ensuremath{^{+}})\ensuremath{^{{\hyperlink{NE29LEVEL2}{c}}}}}&&&\parbox[t][0.3cm]{11.478651cm}{\raggedright E(level): Weighted average of 5.09 MeV \textit{3} (\href{https://www.nndc.bnl.gov/nsr/nsrlink.jsp?1968Gu07,B}{1968Gu07}) and a tentative state at 5097\vspace{0.1cm}}&\\
&&&&&\parbox[t][0.3cm]{11.478651cm}{\raggedright {\ }{\ }{\ }keV \textit{10} (\href{https://www.nndc.bnl.gov/nsr/nsrlink.jsp?1973Da31,B}{1973Da31}).\vspace{0.1cm}}&\\
&&&&&\parbox[t][0.3cm]{11.478651cm}{\raggedright Mirror state was determined to be \ensuremath{^{\textnormal{19}}}F*(5.11 MeV, 5/2\ensuremath{^{\textnormal{+}}}) (\href{https://www.nndc.bnl.gov/nsr/nsrlink.jsp?1973Da31,B}{1973Da31}).\vspace{0.1cm}}&\\
\end{longtable}
\parbox[b][0.3cm]{17.7cm}{\makebox[1ex]{\ensuremath{^{\hypertarget{NE29LEVEL0}{a}}}} From the \ensuremath{^{\textnormal{19}}}Ne Adopted Levels.}\\
\parbox[b][0.3cm]{17.7cm}{\makebox[1ex]{\ensuremath{^{\hypertarget{NE29LEVEL1}{b}}}} From (\href{https://www.nndc.bnl.gov/nsr/nsrlink.jsp?2005Ta28,B}{2005Ta28}) at 1\ensuremath{\sigma} C.L. unless otherwise noted.}\\
\parbox[b][0.3cm]{17.7cm}{\makebox[1ex]{\ensuremath{^{\hypertarget{NE29LEVEL2}{c}}}} From (\href{https://www.nndc.bnl.gov/nsr/nsrlink.jsp?1973Da31,B}{1973Da31}) based on mirror level assignments (see Fig. 2 and the discussions). Since this evidence is weak, we have made}\\
\parbox[b][0.3cm]{17.7cm}{{\ }{\ }the assignment tentative.}\\
\parbox[b][0.3cm]{17.7cm}{\makebox[1ex]{\ensuremath{^{\hypertarget{NE29LEVEL3}{d}}}} For those states whose J\ensuremath{^{\ensuremath{\pi}}} assignment and L-transfer are recommended from (\href{https://www.nndc.bnl.gov/nsr/nsrlink.jsp?1968Gu07,B}{1968Gu07}), we note that those authors}\\
\parbox[b][0.3cm]{17.7cm}{{\ }{\ }acknowledged that the results of their DWBA analysis may not be reliable due to the contributions of the compound nuclear}\\
\parbox[b][0.3cm]{17.7cm}{{\ }{\ }reaction mechanism at E\ensuremath{_{\textnormal{lab}}}=3 MeV incident energy. However, the measured neutron angular distributions were forward peaked}\\
\parbox[b][0.3cm]{17.7cm}{{\ }{\ }implying that direct reaction mechanism may have played a role.}\\
\vspace{0.5cm}
\clearpage
\vspace{0.3cm}
\vspace*{-0.5cm}
{\bf \small \underline{\ensuremath{^{\textnormal{17}}}O(\ensuremath{^{\textnormal{3}}}He,n),(\ensuremath{^{\textnormal{3}}}He,n\ensuremath{\gamma})\hspace{0.2in}\href{https://www.nndc.bnl.gov/nsr/nsrlink.jsp?1973Da31,B}{1973Da31},\href{https://www.nndc.bnl.gov/nsr/nsrlink.jsp?2005Ta28,B}{2005Ta28} (continued)}}\\
\vspace{0.3cm}
\underline{$\gamma$($^{19}$Ne)}\\
\begin{longtable}{ccccccccc@{}ccccc@{\extracolsep{\fill}}c}
\multicolumn{2}{c}{E\ensuremath{_{i}}(level)}&J\ensuremath{^{\pi}_{i}}&\multicolumn{2}{c}{E\ensuremath{_{\gamma}}}&\multicolumn{2}{c}{I\ensuremath{_{\ensuremath{\gamma}}} (\%)\ensuremath{^{\hyperlink{NE29GAMMA4}{e}}}}&\multicolumn{2}{c}{E\ensuremath{_{f}}}&J\ensuremath{^{\pi}_{f}}&Mult.&\multicolumn{2}{c}{\ensuremath{\alpha}\ensuremath{^{\hyperlink{NE29GAMMA5}{f}}}}&Comments&\\[-.2cm]
\multicolumn{2}{c}{\hrulefill}&\hrulefill&\multicolumn{2}{c}{\hrulefill}&\multicolumn{2}{c}{\hrulefill}&\multicolumn{2}{c}{\hrulefill}&\hrulefill&\hrulefill&\multicolumn{2}{c}{\hrulefill}&\hrulefill&
\endfirsthead
\multicolumn{1}{r@{}}{238}&\multicolumn{1}{@{.}l}{18}&\multicolumn{1}{l}{5/2\ensuremath{^{+}}}&\multicolumn{1}{r@{}}{238}&\multicolumn{1}{@{.}l}{11\ensuremath{^{\hyperlink{NE29GAMMA3}{d}}}}&\multicolumn{1}{r@{}}{}&\multicolumn{1}{@{}l}{}&\multicolumn{1}{r@{}}{0}&\multicolumn{1}{@{}l}{}&\multicolumn{1}{@{}l}{1/2\ensuremath{^{+}}}&&&&&\\
\multicolumn{1}{r@{}}{275}&\multicolumn{1}{@{.}l}{06}&\multicolumn{1}{l}{1/2\ensuremath{^{-}}}&\multicolumn{1}{r@{}}{275}&\multicolumn{1}{@{.}l}{06\ensuremath{^{\hyperlink{NE29GAMMA3}{d}}}}&\multicolumn{1}{r@{}}{}&\multicolumn{1}{@{}l}{}&\multicolumn{1}{r@{}}{0}&\multicolumn{1}{@{}l}{}&\multicolumn{1}{@{}l}{1/2\ensuremath{^{+}}}&&&&&\\
\multicolumn{1}{r@{}}{1507}&\multicolumn{1}{@{.}l}{51}&\multicolumn{1}{l}{(5/2\ensuremath{^{-}})}&\multicolumn{1}{r@{}}{1232}&\multicolumn{1}{@{.}l}{41\ensuremath{^{\hyperlink{NE29GAMMA0}{a}}}}&\multicolumn{1}{r@{}}{}&\multicolumn{1}{@{}l}{}&\multicolumn{1}{r@{}}{275}&\multicolumn{1}{@{.}l}{06}&\multicolumn{1}{@{}l}{1/2\ensuremath{^{-}}}&&&&&\\
&&&\multicolumn{1}{r@{}}{1269}&\multicolumn{1}{@{.}l}{29\ensuremath{^{\hyperlink{NE29GAMMA0}{a}}}}&\multicolumn{1}{r@{}}{}&\multicolumn{1}{@{}l}{}&\multicolumn{1}{r@{}}{238}&\multicolumn{1}{@{.}l}{18}&\multicolumn{1}{@{}l}{5/2\ensuremath{^{+}}}&&&&&\\
\multicolumn{1}{r@{}}{1536}&\multicolumn{1}{@{.}l}{1}&\multicolumn{1}{l}{(3/2\ensuremath{^{+}})}&\multicolumn{1}{r@{}}{1297}&\multicolumn{1}{@{.}l}{8\ensuremath{^{\hyperlink{NE29GAMMA1}{b}}} {\it 4}}&\multicolumn{1}{r@{}}{}&\multicolumn{1}{@{}l}{}&\multicolumn{1}{r@{}}{238}&\multicolumn{1}{@{.}l}{18}&\multicolumn{1}{@{}l}{5/2\ensuremath{^{+}}}&&&&&\\
\multicolumn{1}{r@{}}{1615}&\multicolumn{1}{@{.}l}{4}&\multicolumn{1}{l}{(3/2\ensuremath{^{-}})}&\multicolumn{1}{r@{}}{1340}&\multicolumn{1}{@{.}l}{3\ensuremath{^{\hyperlink{NE29GAMMA0}{a}}}}&\multicolumn{1}{r@{}}{}&\multicolumn{1}{@{}l}{}&\multicolumn{1}{r@{}}{275}&\multicolumn{1}{@{.}l}{06}&\multicolumn{1}{@{}l}{1/2\ensuremath{^{-}}}&&&&&\\
\multicolumn{1}{r@{}}{2794}&\multicolumn{1}{@{.}l}{2}&\multicolumn{1}{l}{(9/2\ensuremath{^{+}})}&\multicolumn{1}{r@{}}{2555}&\multicolumn{1}{@{.}l}{8\ensuremath{^{\hyperlink{NE29GAMMA0}{a}}}}&\multicolumn{1}{r@{}}{}&\multicolumn{1}{@{}l}{}&\multicolumn{1}{r@{}}{238}&\multicolumn{1}{@{.}l}{18}&\multicolumn{1}{@{}l}{5/2\ensuremath{^{+}}}&&&&&\\
\multicolumn{1}{r@{}}{4034}&\multicolumn{1}{@{.}l}{6}&\multicolumn{1}{l}{(3/2\ensuremath{^{+}})}&\multicolumn{1}{r@{}}{2498}&\multicolumn{1}{@{.}l}{5\ensuremath{^{\hyperlink{NE29GAMMA1}{b}}} {\it 9}}&\multicolumn{1}{r@{}}{15}&\multicolumn{1}{@{ }l}{{\it 5}}&\multicolumn{1}{r@{}}{1536}&\multicolumn{1}{@{.}l}{1 }&\multicolumn{1}{@{}l}{(3/2\ensuremath{^{+}})}&&&&&\\
&&&\multicolumn{1}{r@{}}{3759}&\multicolumn{1}{@{.}l}{1\ensuremath{^{\hyperlink{NE29GAMMA0}{a}}}}&\multicolumn{1}{r@{}}{5}&\multicolumn{1}{@{ }l}{{\it 15}}&\multicolumn{1}{r@{}}{275}&\multicolumn{1}{@{.}l}{06}&\multicolumn{1}{@{}l}{1/2\ensuremath{^{-}}}&&&&&\\
&&&\multicolumn{1}{r@{}}{4034}&\multicolumn{1}{@{.}l}{5\ensuremath{^{\hyperlink{NE29GAMMA1}{b}}} {\it 8}}&\multicolumn{1}{r@{}}{80}&\multicolumn{1}{@{ }l}{{\it 15}}&\multicolumn{1}{r@{}}{0}&\multicolumn{1}{@{}l}{}&\multicolumn{1}{@{}l}{1/2\ensuremath{^{+}}}&&&&&\\
\multicolumn{1}{r@{}}{4143}&\multicolumn{1}{@{.}l}{2}&\multicolumn{1}{l}{(7/2\ensuremath{^{-}})}&\multicolumn{1}{r@{}}{2635}&\multicolumn{1}{@{.}l}{9\ensuremath{^{\hyperlink{NE29GAMMA1}{b}}} {\it 7}}&\multicolumn{1}{r@{}}{100}&\multicolumn{1}{@{}l}{}&\multicolumn{1}{r@{}}{1507}&\multicolumn{1}{@{.}l}{51 }&\multicolumn{1}{@{}l}{(5/2\ensuremath{^{-}})}&&&&&\\
\multicolumn{1}{r@{}}{4199}&\multicolumn{1}{@{.}l}{6}&\multicolumn{1}{l}{(9/2\ensuremath{^{-}})}&\multicolumn{1}{r@{}}{2692}&\multicolumn{1}{@{.}l}{7\ensuremath{^{\hyperlink{NE29GAMMA1}{b}}} {\it 11}}&\multicolumn{1}{r@{}}{80}&\multicolumn{1}{@{ }l}{{\it 5}}&\multicolumn{1}{r@{}}{1507}&\multicolumn{1}{@{.}l}{51 }&\multicolumn{1}{@{}l}{(5/2\ensuremath{^{-}})}&&&&&\\
&&&\multicolumn{1}{r@{}}{3961}&\multicolumn{1}{@{.}l}{0\ensuremath{^{\hyperlink{NE29GAMMA2}{c}}}}&\multicolumn{1}{r@{}}{20}&\multicolumn{1}{@{ }l}{{\it 5}}&\multicolumn{1}{r@{}}{238}&\multicolumn{1}{@{.}l}{18}&\multicolumn{1}{@{}l}{5/2\ensuremath{^{+}}}&&&&&\\
\multicolumn{1}{r@{}}{4379}&\multicolumn{1}{@{.}l}{1}&\multicolumn{1}{l}{(7/2\ensuremath{^{+}})}&\multicolumn{1}{r@{}}{1584}&\multicolumn{1}{@{.}l}{8\ensuremath{^{\hyperlink{NE29GAMMA0}{a}}}}&\multicolumn{1}{r@{}}{15}&\multicolumn{1}{@{ }l}{{\it 4}}&\multicolumn{1}{r@{}}{2794}&\multicolumn{1}{@{.}l}{2 }&\multicolumn{1}{@{}l}{(9/2\ensuremath{^{+}})}&&&&&\\
&&&\multicolumn{1}{r@{}}{4139}&\multicolumn{1}{@{.}l}{5\ensuremath{^{\hyperlink{NE29GAMMA1}{b}}} {\it 6}}&\multicolumn{1}{r@{}}{85}&\multicolumn{1}{@{ }l}{{\it 4}}&\multicolumn{1}{r@{}}{238}&\multicolumn{1}{@{.}l}{18}&\multicolumn{1}{@{}l}{5/2\ensuremath{^{+}}}&&&&&\\
\multicolumn{1}{r@{}}{4548}&\multicolumn{1}{@{.}l}{3}&\multicolumn{1}{l}{(3/2\ensuremath{^{-}})}&\multicolumn{1}{r@{}}{4272}&\multicolumn{1}{@{.}l}{6\ensuremath{^{\hyperlink{NE29GAMMA1}{b}}} {\it 10}}&\multicolumn{1}{r@{}}{65}&\multicolumn{1}{@{ }l}{{\it 25}}&\multicolumn{1}{r@{}}{275}&\multicolumn{1}{@{.}l}{06}&\multicolumn{1}{@{}l}{1/2\ensuremath{^{-}}}&&&&&\\
&&&\multicolumn{1}{r@{}}{4547}&\multicolumn{1}{@{.}l}{7\ensuremath{^{\hyperlink{NE29GAMMA1}{b}}} {\it 10}}&\multicolumn{1}{r@{}}{35}&\multicolumn{1}{@{ }l}{{\it 25}}&\multicolumn{1}{r@{}}{0}&\multicolumn{1}{@{}l}{}&\multicolumn{1}{@{}l}{1/2\ensuremath{^{+}}}&&&&&\\
\multicolumn{1}{r@{}}{4602}&\multicolumn{1}{@{.}l}{1}&\multicolumn{1}{l}{(5/2\ensuremath{^{+}})}&\multicolumn{1}{r@{}}{3065}&\multicolumn{1}{@{.}l}{7\ensuremath{^{\hyperlink{NE29GAMMA0}{a}}}}&\multicolumn{1}{r@{}}{10}&\multicolumn{1}{@{ }l}{{\it 5}}&\multicolumn{1}{r@{}}{1536}&\multicolumn{1}{@{.}l}{1 }&\multicolumn{1}{@{}l}{(3/2\ensuremath{^{+}})}&&&&&\\
&&&\multicolumn{1}{r@{}}{4363}&\multicolumn{1}{@{.}l}{5\ensuremath{^{\hyperlink{NE29GAMMA1}{b}}} {\it 8}}&\multicolumn{1}{r@{}}{90}&\multicolumn{1}{@{ }l}{{\it 5}}&\multicolumn{1}{r@{}}{238}&\multicolumn{1}{@{.}l}{18}&\multicolumn{1}{@{}l}{5/2\ensuremath{^{+}}}&&&&&\\
\multicolumn{1}{r@{}}{4634}&\multicolumn{1}{@{.}l}{1}&\multicolumn{1}{l}{(13/2\ensuremath{^{+}})}&\multicolumn{1}{r@{}}{1839}&\multicolumn{1}{@{.}l}{8\ensuremath{^{\hyperlink{NE29GAMMA0}{a}}}}&\multicolumn{1}{r@{}}{100}&\multicolumn{1}{@{}l}{}&\multicolumn{1}{r@{}}{2794}&\multicolumn{1}{@{.}l}{2 }&\multicolumn{1}{@{}l}{(9/2\ensuremath{^{+}})}&\multicolumn{1}{l}{E2}&\multicolumn{1}{r@{}}{2}&\multicolumn{1}{@{.}l}{40\ensuremath{\times10^{-4}} {\it 3}}&\parbox[t][0.3cm]{5.648381cm}{\raggedright B(E2)(W.u.)\ensuremath{<}13\vspace{0.1cm}}&\\
&&&&&&&&&&&&&\parbox[t][0.3cm]{5.648381cm}{\raggedright \ensuremath{\alpha}(K)=3.00\ensuremath{\times}10\ensuremath{^{\textnormal{$-$6}}} \textit{4}; \ensuremath{\alpha}(L)=1.658\ensuremath{\times}10\ensuremath{^{\textnormal{$-$7}}} \textit{23}\vspace{0.1cm}}&\\
&&&&&&&&&&&&&\parbox[t][0.3cm]{5.648381cm}{\raggedright \ensuremath{\alpha}(IPF)=0.0002366 \textit{33}\vspace{0.1cm}}&\\
&&&&&&&&&&&&&\parbox[t][0.3cm]{5.648381cm}{\raggedright Mult.: From (\href{https://www.nndc.bnl.gov/nsr/nsrlink.jsp?1973Da31,B}{1973Da31}) based on ruling\vspace{0.1cm}}&\\
&&&&&&&&&&&&&\parbox[t][0.3cm]{5.648381cm}{\raggedright {\ }{\ }{\ }out the possibility that this transition\vspace{0.1cm}}&\\
&&&&&&&&&&&&&\parbox[t][0.3cm]{5.648381cm}{\raggedright {\ }{\ }{\ }could be M1 or E1 (see the comment\vspace{0.1cm}}&\\
&&&&&&&&&&&&&\parbox[t][0.3cm]{5.648381cm}{\raggedright {\ }{\ }{\ }on the J\ensuremath{^{\ensuremath{\pi}}} assignment of the 4634-keV\vspace{0.1cm}}&\\
&&&&&&&&&&&&&\parbox[t][0.3cm]{5.648381cm}{\raggedright {\ }{\ }{\ }level).\vspace{0.1cm}}&\\
\end{longtable}
\parbox[b][0.3cm]{17.7cm}{\makebox[1ex]{\ensuremath{^{\hypertarget{NE29GAMMA0}{a}}}} This \ensuremath{\gamma} ray was measured by (\href{https://www.nndc.bnl.gov/nsr/nsrlink.jsp?1973Da31,B}{1973Da31}, \href{https://www.nndc.bnl.gov/nsr/nsrlink.jsp?2005Ta28,B}{2005Ta28}). But they did not report E\ensuremath{_{\ensuremath{\gamma}}}, so we obtained it from the level-energy}\\
\parbox[b][0.3cm]{17.7cm}{{\ }{\ }differences corrected for the nuclear recoil energy.}\\
\parbox[b][0.3cm]{17.7cm}{\makebox[1ex]{\ensuremath{^{\hypertarget{NE29GAMMA1}{b}}}} From (\href{https://www.nndc.bnl.gov/nsr/nsrlink.jsp?2005Ta28,B}{2005Ta28}). We highlight that (\href{https://www.nndc.bnl.gov/nsr/nsrlink.jsp?1973Da31,B}{1973Da31}, \href{https://www.nndc.bnl.gov/nsr/nsrlink.jsp?2005Ta28,B}{2005Ta28}) did not report the energy of the observed transitions; however, Table}\\
\parbox[b][0.3cm]{17.7cm}{{\ }{\ }I in (\href{https://www.nndc.bnl.gov/nsr/nsrlink.jsp?2008My01,B}{2008My01}: \ensuremath{^{\textnormal{3}}}He(\ensuremath{^{\textnormal{20}}}Ne,\ensuremath{\alpha}\ensuremath{\gamma})) cites some of the \ensuremath{\gamma}-ray energies measured by (\href{https://www.nndc.bnl.gov/nsr/nsrlink.jsp?2005Ta28,B}{2005Ta28}). Those values are reported here.}\\
\parbox[b][0.3cm]{17.7cm}{\makebox[1ex]{\ensuremath{^{\hypertarget{NE29GAMMA2}{c}}}} This \ensuremath{\gamma} ray was measured by (\href{https://www.nndc.bnl.gov/nsr/nsrlink.jsp?1973Da31,B}{1973Da31}). But they did not report E\ensuremath{_{\ensuremath{\gamma}}}, so we obtained it from the level-energy differences}\\
\parbox[b][0.3cm]{17.7cm}{{\ }{\ }corrected for the nuclear recoil energy. We note that (\href{https://www.nndc.bnl.gov/nsr/nsrlink.jsp?2020Ha31,B}{2020Ha31}: \ensuremath{^{\textnormal{19}}}F(\ensuremath{^{\textnormal{3}}}He,t\ensuremath{\gamma})) disputed the existence of this branch.}\\
\parbox[b][0.3cm]{17.7cm}{\makebox[1ex]{\ensuremath{^{\hypertarget{NE29GAMMA3}{d}}}} From the \ensuremath{^{\textnormal{19}}}Ne Adopted Gammas.}\\
\parbox[b][0.3cm]{17.7cm}{\makebox[1ex]{\ensuremath{^{\hypertarget{NE29GAMMA4}{e}}}} From (\href{https://www.nndc.bnl.gov/nsr/nsrlink.jsp?1973Da31,B}{1973Da31}: See Fig. 2).}\\
\parbox[b][0.3cm]{17.7cm}{\makebox[1ex]{\ensuremath{^{\hypertarget{NE29GAMMA5}{f}}}} Total theoretical internal conversion coefficients, calculated using the BrIcc code (\href{https://www.nndc.bnl.gov/nsr/nsrlink.jsp?2008Ki07,B}{2008Ki07}) with ``Frozen Orbitals''}\\
\parbox[b][0.3cm]{17.7cm}{{\ }{\ }approximation based on \ensuremath{\gamma}-ray energies, assigned multipolarities, and mixing ratios, unless otherwise specified.}\\
\vspace{0.5cm}
\clearpage
\begin{figure}[h]
\begin{center}
\includegraphics{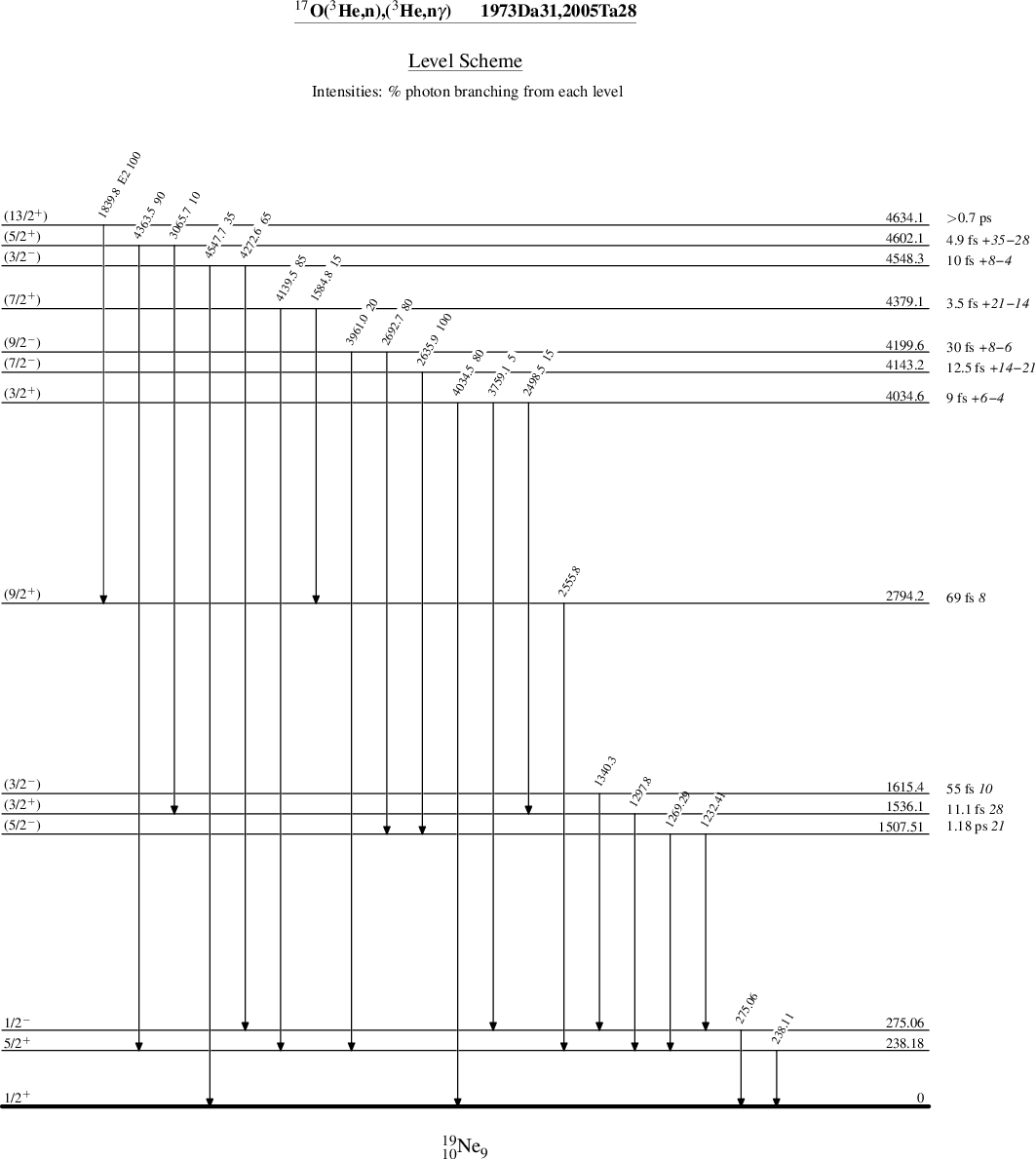}\\
\end{center}
\end{figure}
\clearpage
\subsection[\hspace{-0.2cm}\ensuremath{^{\textnormal{18}}}O(p,\ensuremath{\pi}\ensuremath{^{-}})]{ }
\vspace{-27pt}
\vspace{0.3cm}
\hypertarget{NE30}{{\bf \small \underline{\ensuremath{^{\textnormal{18}}}O(p,\ensuremath{\pi}\ensuremath{^{-}})\hspace{0.2in}\href{https://www.nndc.bnl.gov/nsr/nsrlink.jsp?1982Vi05,B}{1982Vi05},\href{https://www.nndc.bnl.gov/nsr/nsrlink.jsp?1986Ke04,B}{1986Ke04}}}}\\
\vspace{4pt}
\vspace{8pt}
\parbox[b][0.3cm]{17.7cm}{\addtolength{\parindent}{-0.2in}Double charge exchange reaction.}\\
\parbox[b][0.3cm]{17.7cm}{\addtolength{\parindent}{-0.2in}J\ensuremath{^{\ensuremath{\pi}}}(\ensuremath{^{\textnormal{18}}}O\ensuremath{_{\textnormal{g.s.}}})=0\ensuremath{^{\textnormal{+}}} and J\ensuremath{^{\ensuremath{\pi}}}(\ensuremath{\pi}\ensuremath{^{-}})=0\ensuremath{^{-}}.}\\
\parbox[b][0.3cm]{17.7cm}{\addtolength{\parindent}{-0.2in}\href{https://www.nndc.bnl.gov/nsr/nsrlink.jsp?1982Vi05,B}{1982Vi05}, \href{https://www.nndc.bnl.gov/nsr/nsrlink.jsp?1983Vi01,B}{1983Vi01}: \ensuremath{^{\textnormal{18}}}O(p,\ensuremath{\pi}\ensuremath{^{-}}) E=191, 205, 206 MeV; measured the energies and emission angles of the \ensuremath{\pi}\ensuremath{^{-}} particles using a}\\
\parbox[b][0.3cm]{17.7cm}{quadrupole-quadrupole split-pole magnetic spectrometer (aka a broad-range pion spectrometer) and its associated focal plane}\\
\parbox[b][0.3cm]{17.7cm}{detectors with an energy resolution of \ensuremath{\Delta}E(FWHM)\ensuremath{>}150 keV. Measured momentum spectra and double-differential cross sections.}\\
\parbox[b][0.3cm]{17.7cm}{Populated a few high-spin excited states in \ensuremath{^{\textnormal{19}}}Ne within the E\ensuremath{_{\textnormal{x}}}=0-15 MeV region, including the \ensuremath{^{\textnormal{19}}}Ne*(4.64 MeV) state and a state}\\
\parbox[b][0.3cm]{17.7cm}{at \ensuremath{\sim}10 MeV, which is clearly visible in the spectra.}\\
\parbox[b][0.3cm]{17.7cm}{\addtolength{\parindent}{-0.2in}\href{https://www.nndc.bnl.gov/nsr/nsrlink.jsp?1983KeZX,B}{1983KeZX}, \href{https://www.nndc.bnl.gov/nsr/nsrlink.jsp?1986Ke04,B}{1986Ke04}: \ensuremath{^{\textnormal{18}}}O(pol. p,\ensuremath{\pi}\ensuremath{^{-}}) E=200, 201 MeV; measured energies and emission angles of the outgoing negative pion}\\
\parbox[b][0.3cm]{17.7cm}{reaction products using the broad-range pion spectrometer, mentioned above, which was capable of measuring \ensuremath{\Delta}E-\ensuremath{\Delta}E-E and TOF.}\\
\parbox[b][0.3cm]{17.7cm}{Measured the angular distributions of the differential cross sections and analyzing powers vs. \ensuremath{\theta} for the \ensuremath{^{\textnormal{18}}}O(p,\ensuremath{\pi}\ensuremath{^{-}}) reaction}\\
\parbox[b][0.3cm]{17.7cm}{populating the \ensuremath{^{\textnormal{19}}}Ne*(2.8, 4.6, 10 MeV) levels. Discussed J\ensuremath{^{\ensuremath{\pi}}} assignments of these states.}\\
\vspace{0.385cm}
\parbox[b][0.3cm]{17.7cm}{\addtolength{\parindent}{-0.2in}\textit{Theory}:}\\
\parbox[b][0.3cm]{17.7cm}{\addtolength{\parindent}{-0.2in}\href{https://www.nndc.bnl.gov/nsr/nsrlink.jsp?1984BeZZ,B}{1984BeZZ}: \ensuremath{^{\textnormal{18}}}O(p,\ensuremath{\pi}\ensuremath{^{-}}) E not given; analyzed data; deduced \ensuremath{^{\textnormal{19}}}Ne certain level enhancements using two-nucleon model.}\\
\parbox[b][0.3cm]{17.7cm}{\addtolength{\parindent}{-0.2in}\href{https://www.nndc.bnl.gov/nsr/nsrlink.jsp?1987Vi13,B}{1987Vi13}: \ensuremath{^{\textnormal{18}}}O(p,\ensuremath{\pi}\ensuremath{^{-}}); discussed the analyzing powers.}\\
\vspace{12pt}
\underline{$^{19}$Ne Levels}\\
\begin{longtable}{cccc@{\extracolsep{\fill}}c}
\multicolumn{2}{c}{E(level)$^{{\hyperlink{NE30LEVEL0}{a}}}$}&J$^{\pi}$$^{}$&Comments&\\[-.2cm]
\multicolumn{2}{c}{\hrulefill}&\hrulefill&\hrulefill&
\endfirsthead
\multicolumn{1}{r@{}}{0}&\multicolumn{1}{@{}l}{}&&\parbox[t][0.3cm]{14.691521cm}{\raggedright E(level): From (\href{https://www.nndc.bnl.gov/nsr/nsrlink.jsp?1986Ke04,B}{1986Ke04}): Weakly populated.\vspace{0.1cm}}&\\
&&&\parbox[t][0.3cm]{14.691521cm}{\raggedright E(level): A state near the ground state was also weakly populated in (\href{https://www.nndc.bnl.gov/nsr/nsrlink.jsp?1982Vi05,B}{1982Vi05}, \href{https://www.nndc.bnl.gov/nsr/nsrlink.jsp?1983Vi01,B}{1983Vi01}), but the authors\vspace{0.1cm}}&\\
&&&\parbox[t][0.3cm]{14.691521cm}{\raggedright {\ }{\ }{\ }did not report its energy.\vspace{0.1cm}}&\\
\multicolumn{1}{r@{}}{2.8\ensuremath{\times10^{3}}}&\multicolumn{1}{@{}l}{}&\multicolumn{1}{l}{9/2\ensuremath{^{+}}}&\parbox[t][0.3cm]{14.691521cm}{\raggedright E(level): From (\href{https://www.nndc.bnl.gov/nsr/nsrlink.jsp?1986Ke04,B}{1986Ke04}): Weakly populated.\vspace{0.1cm}}&\\
&&&\parbox[t][0.3cm]{14.691521cm}{\raggedright E(level): A state near 2 MeV was also weakly populated in (\href{https://www.nndc.bnl.gov/nsr/nsrlink.jsp?1982Vi05,B}{1982Vi05}, \href{https://www.nndc.bnl.gov/nsr/nsrlink.jsp?1983Vi01,B}{1983Vi01}), but the authors did not\vspace{0.1cm}}&\\
&&&\parbox[t][0.3cm]{14.691521cm}{\raggedright {\ }{\ }{\ }report its energy. Since this state has a high spin, the evaluator speculates that the same state was\vspace{0.1cm}}&\\
&&&\parbox[t][0.3cm]{14.691521cm}{\raggedright {\ }{\ }{\ }populated in (\href{https://www.nndc.bnl.gov/nsr/nsrlink.jsp?1982Vi05,B}{1982Vi05}, \href{https://www.nndc.bnl.gov/nsr/nsrlink.jsp?1983Vi01,B}{1983Vi01}).\vspace{0.1cm}}&\\
&&&\parbox[t][0.3cm]{14.691521cm}{\raggedright J\ensuremath{^{\pi}}: From the \ensuremath{^{\textnormal{19}}}Ne Adopted Levels.\vspace{0.1cm}}&\\
&&&\parbox[t][0.3cm]{14.691521cm}{\raggedright (\href{https://www.nndc.bnl.gov/nsr/nsrlink.jsp?1986Ke04,B}{1986Ke04}) pointed out that the angular distribution of the differential cross section of this state peaked at\vspace{0.1cm}}&\\
&&&\parbox[t][0.3cm]{14.691521cm}{\raggedright {\ }{\ }{\ }backward angles in contrast to the forward peaking associated with the measured angular distributions of\vspace{0.1cm}}&\\
&&&\parbox[t][0.3cm]{14.691521cm}{\raggedright {\ }{\ }{\ }the \ensuremath{^{\textnormal{19}}}Ne*(4.6, 10 MeV) states in the same study. It was unclear to those authors if this was a nuclear\vspace{0.1cm}}&\\
&&&\parbox[t][0.3cm]{14.691521cm}{\raggedright {\ }{\ }{\ }structure effect or a reaction mechanism effect.\vspace{0.1cm}}&\\
\multicolumn{1}{r@{}}{4.6\ensuremath{\times10^{3}}}&\multicolumn{1}{@{}l}{\ensuremath{^{{\hyperlink{NE30LEVEL1}{b}}}}}&\multicolumn{1}{l}{13/2\ensuremath{^{+}}}&\parbox[t][0.3cm]{14.691521cm}{\raggedright E(level): This state was populated and mentioned in (\href{https://www.nndc.bnl.gov/nsr/nsrlink.jsp?1982Vi05,B}{1982Vi05}: E\ensuremath{_{\textnormal{x}}}=4.64 MeV) and (\href{https://www.nndc.bnl.gov/nsr/nsrlink.jsp?1986Ke04,B}{1986Ke04}), where the\vspace{0.1cm}}&\\
&&&\parbox[t][0.3cm]{14.691521cm}{\raggedright {\ }{\ }{\ }energy is reported to be 4.6 MeV and 4.64 MeV.\vspace{0.1cm}}&\\
&&&\parbox[t][0.3cm]{14.691521cm}{\raggedright J\ensuremath{^{\pi}}: From the \ensuremath{^{\textnormal{19}}}Ne Adopted Levels. (\href{https://www.nndc.bnl.gov/nsr/nsrlink.jsp?1986Ke04,B}{1986Ke04}) mentioned that the pattern of the measured analyzing\vspace{0.1cm}}&\\
&&&\parbox[t][0.3cm]{14.691521cm}{\raggedright {\ }{\ }{\ }power (A(\ensuremath{\theta})) for this state supports the J\ensuremath{^{\ensuremath{\pi}}}=13/2\ensuremath{^{\textnormal{+}}} assignment. Also, a shell model calculation by B. H.\vspace{0.1cm}}&\\
&&&\parbox[t][0.3cm]{14.691521cm}{\raggedright {\ }{\ }{\ }Wildenthal (priv. comm. with \href{https://www.nndc.bnl.gov/nsr/nsrlink.jsp?1986Ke04,B}{1986Ke04}) estimated that the first J\ensuremath{^{\ensuremath{\pi}}}=13/2\ensuremath{^{\textnormal{+}}} state in the A=19 system\vspace{0.1cm}}&\\
&&&\parbox[t][0.3cm]{14.691521cm}{\raggedright {\ }{\ }{\ }occurred at 4.8 MeV, which corresponds closely to the strong transition observed in this energy region in\vspace{0.1cm}}&\\
&&&\parbox[t][0.3cm]{14.691521cm}{\raggedright {\ }{\ }{\ }the \ensuremath{^{\textnormal{18}}}O(p,\ensuremath{\pi}\ensuremath{^{-}}) spectrum obtained by (\href{https://www.nndc.bnl.gov/nsr/nsrlink.jsp?1986Ke04,B}{1986Ke04}).\vspace{0.1cm}}&\\
&&&\parbox[t][0.3cm]{14.691521cm}{\raggedright d\ensuremath{\sigma}/d\ensuremath{\Omega}\ensuremath{_{\textnormal{lab}}}(\ensuremath{\theta}\ensuremath{_{\textnormal{lab}}}=30\ensuremath{^\circ})=20.0 nb/sr \textit{15} (\href{https://www.nndc.bnl.gov/nsr/nsrlink.jsp?1983Vi01,B}{1983Vi01}), where the uncertainty comes from the statistical and\vspace{0.1cm}}&\\
&&&\parbox[t][0.3cm]{14.691521cm}{\raggedright {\ }{\ }{\ }background subtraction uncertainties.\vspace{0.1cm}}&\\
\multicolumn{1}{r@{}}{6.4\ensuremath{\times10^{3}}}&\multicolumn{1}{@{}l}{}&&\parbox[t][0.3cm]{14.691521cm}{\raggedright E(level): A strong peak is clearly visible around this energy in the spectra obtained in (\href{https://www.nndc.bnl.gov/nsr/nsrlink.jsp?1982Vi05,B}{1982Vi05},\vspace{0.1cm}}&\\
&&&\parbox[t][0.3cm]{14.691521cm}{\raggedright {\ }{\ }{\ }\href{https://www.nndc.bnl.gov/nsr/nsrlink.jsp?1986Ke04,B}{1986Ke04}) but the authors did not report its energy.\vspace{0.1cm}}&\\
&&&\parbox[t][0.3cm]{14.691521cm}{\raggedright (\href{https://www.nndc.bnl.gov/nsr/nsrlink.jsp?1983Vi01,B}{1983Vi01}) suggested that this state probably corresponds to the mirror of the known \ensuremath{^{\textnormal{19}}}F*(6.50 MeV,\vspace{0.1cm}}&\\
&&&\parbox[t][0.3cm]{14.691521cm}{\raggedright {\ }{\ }{\ }11/2\ensuremath{^{\textnormal{+}}}) level.\vspace{0.1cm}}&\\
&&&\parbox[t][0.3cm]{14.691521cm}{\raggedright This state is presumed by (\href{https://www.nndc.bnl.gov/nsr/nsrlink.jsp?1982Vi05,B}{1982Vi05}, \href{https://www.nndc.bnl.gov/nsr/nsrlink.jsp?1983Vi01,B}{1983Vi01}, \href{https://www.nndc.bnl.gov/nsr/nsrlink.jsp?1986Ke04,B}{1986Ke04}) to have a predominantly 2p-1h configuration\vspace{0.1cm}}&\\
&&&\parbox[t][0.3cm]{14.691521cm}{\raggedright {\ }{\ }{\ }with respect to the \ensuremath{^{\textnormal{18}}}O\ensuremath{_{\textnormal{g.s.}}} configuration because of the two nucleon nature of the (p,\ensuremath{\pi}\ensuremath{^{-}}) reaction\vspace{0.1cm}}&\\
&&&\parbox[t][0.3cm]{14.691521cm}{\raggedright {\ }{\ }{\ }mechanism. Therefore, (\href{https://www.nndc.bnl.gov/nsr/nsrlink.jsp?1986Ke04,B}{1986Ke04}) expected this final state to have a relatively pure (\textit{sd})\ensuremath{^{\textnormal{n}}} configuration.\vspace{0.1cm}}&\\
\multicolumn{1}{r@{}}{10.0\ensuremath{\times10^{3}}}&\multicolumn{1}{@{}l}{\ensuremath{^{{\hyperlink{NE30LEVEL1}{b}}}}}&&\parbox[t][0.3cm]{14.691521cm}{\raggedright A strongly populated peak at or near this energy is populated in (\href{https://www.nndc.bnl.gov/nsr/nsrlink.jsp?1982Vi05,B}{1982Vi05}) but the authors did not report\vspace{0.1cm}}&\\
&&&\parbox[t][0.3cm]{14.691521cm}{\raggedright {\ }{\ }{\ }its energy. This state is also populated and mentioned in (\href{https://www.nndc.bnl.gov/nsr/nsrlink.jsp?1986Ke04,B}{1986Ke04}).\vspace{0.1cm}}&\\
&&&\parbox[t][0.3cm]{14.691521cm}{\raggedright J\ensuremath{^{\pi}}: (\href{https://www.nndc.bnl.gov/nsr/nsrlink.jsp?1976Ha06,B}{1976Ha06}) concluded that the analog of the \ensuremath{^{\textnormal{19}}}F*(10.42 MeV, 13/2\ensuremath{^{\textnormal{+}}}) level was expected to be a\vspace{0.1cm}}&\\
&&&\parbox[t][0.3cm]{14.691521cm}{\raggedright {\ }{\ }{\ }relatively narrow state with \ensuremath{\Gamma}\ensuremath{<}100 keV at E\ensuremath{_{\textnormal{x}}}\ensuremath{\sim}10 MeV in \ensuremath{^{\textnormal{19}}}Ne. (\href{https://www.nndc.bnl.gov/nsr/nsrlink.jsp?1983Vi01,B}{1983Vi01}) mentioned that several\vspace{0.1cm}}&\\
&&&\parbox[t][0.3cm]{14.691521cm}{\raggedright {\ }{\ }{\ }high-spin states, particularly the \ensuremath{^{\textnormal{19}}}F*(10.4 MeV, 13/2\ensuremath{^{\textnormal{+}}_{\textnormal{2}}}) level from (\href{https://www.nndc.bnl.gov/nsr/nsrlink.jsp?1983Aj01,B}{1983Aj01}), could be the mirror of\vspace{0.1cm}}&\\
&&&\parbox[t][0.3cm]{14.691521cm}{\raggedright {\ }{\ }{\ }this state. Therefore, (\href{https://www.nndc.bnl.gov/nsr/nsrlink.jsp?1986Ke04,B}{1986Ke04}) concluded that the observed \ensuremath{^{\textnormal{19}}}Ne state at 10 MeV, which was populated\vspace{0.1cm}}&\\
&&&\parbox[t][0.3cm]{14.691521cm}{\raggedright {\ }{\ }{\ }strongly by the \ensuremath{^{\textnormal{18}}}O(p,\ensuremath{\pi}\ensuremath{^{-}}) reaction may have a J\ensuremath{^{\ensuremath{\pi}}}=13/2\ensuremath{^{\textnormal{+}}} assignment. The measured (by \href{https://www.nndc.bnl.gov/nsr/nsrlink.jsp?1986Ke04,B}{1986Ke04})\vspace{0.1cm}}&\\
\end{longtable}
\begin{textblock}{29}(0,27.3)
Continued on next page (footnotes at end of table)
\end{textblock}
\clearpage
\begin{longtable}{cccc@{\extracolsep{\fill}}c}
\\[-.4cm]
\multicolumn{5}{c}{{\bf \small \underline{\ensuremath{^{\textnormal{18}}}O(p,\ensuremath{\pi}\ensuremath{^{-}})\hspace{0.2in}\href{https://www.nndc.bnl.gov/nsr/nsrlink.jsp?1982Vi05,B}{1982Vi05},\href{https://www.nndc.bnl.gov/nsr/nsrlink.jsp?1986Ke04,B}{1986Ke04} (continued)}}}\\
\multicolumn{5}{c}{~}\\
\multicolumn{5}{c}{\underline{\ensuremath{^{19}}Ne Levels (continued)}}\\
\multicolumn{5}{c}{~}\\
\multicolumn{2}{c}{E(level)$^{{\hyperlink{NE30LEVEL0}{a}}}$}&J$^{\pi}$$^{}$&Comments&\\[-.2cm]
\multicolumn{2}{c}{\hrulefill}&\hrulefill&\hrulefill&
\endhead
&&&\parbox[t][0.3cm]{15.333521cm}{\raggedright {\ }{\ }{\ }pattern of the analyzing power for this state may or may not support this J\ensuremath{^{\ensuremath{\pi}}} assignment. (\href{https://www.nndc.bnl.gov/nsr/nsrlink.jsp?1986Ke04,B}{1986Ke04})\vspace{0.1cm}}&\\
&&&\parbox[t][0.3cm]{15.333521cm}{\raggedright {\ }{\ }{\ }acknowledged that this argument was less convincing for this state because of the poorer statistics. A shell\vspace{0.1cm}}&\\
&&&\parbox[t][0.3cm]{15.333521cm}{\raggedright {\ }{\ }{\ }model calculation by B. H. Wildenthal (priv. comm. with \href{https://www.nndc.bnl.gov/nsr/nsrlink.jsp?1986Ke04,B}{1986Ke04}) estimated that the second J\ensuremath{^{\ensuremath{\pi}}}=13/2\ensuremath{^{\textnormal{+}}} state\vspace{0.1cm}}&\\
&&&\parbox[t][0.3cm]{15.333521cm}{\raggedright {\ }{\ }{\ }in the A=19 system occurred at 9.9 MeV, which corresponds closely to the transition observed in this energy\vspace{0.1cm}}&\\
&&&\parbox[t][0.3cm]{15.333521cm}{\raggedright {\ }{\ }{\ }region in the \ensuremath{^{\textnormal{18}}}O(p,\ensuremath{\pi}\ensuremath{^{-}}) spectrum obtained by (\href{https://www.nndc.bnl.gov/nsr/nsrlink.jsp?1986Ke04,B}{1986Ke04}). Even though all this evidence may point to a\vspace{0.1cm}}&\\
&&&\parbox[t][0.3cm]{15.333521cm}{\raggedright {\ }{\ }{\ }J\ensuremath{^{\ensuremath{\pi}}}=13/2\ensuremath{^{\textnormal{+}}} assignment, we think the evidence is weak, and therefore we did not adopt J\ensuremath{^{\ensuremath{\pi}}}=13/2\ensuremath{^{\textnormal{+}}} for this state.\vspace{0.1cm}}&\\
\end{longtable}
\parbox[b][0.3cm]{17.7cm}{\makebox[1ex]{\ensuremath{^{\hypertarget{NE30LEVEL0}{a}}}} From (\href{https://www.nndc.bnl.gov/nsr/nsrlink.jsp?1983Vi01,B}{1983Vi01}) unless otherwise noted. Note that (\href{https://www.nndc.bnl.gov/nsr/nsrlink.jsp?1982Vi05,B}{1982Vi05}, \href{https://www.nndc.bnl.gov/nsr/nsrlink.jsp?1983Vi01,B}{1983Vi01}) mentioned that the relative excitation energies for a}\\
\parbox[b][0.3cm]{17.7cm}{{\ }{\ }given target were accurate to \ensuremath{\sim}\ensuremath{\pm}100 keV over a 10-MeV excitation energy range.}\\
\parbox[b][0.3cm]{17.7cm}{\makebox[1ex]{\ensuremath{^{\hypertarget{NE30LEVEL1}{b}}}} This state is presumed by (\href{https://www.nndc.bnl.gov/nsr/nsrlink.jsp?1982Vi05,B}{1982Vi05}, \href{https://www.nndc.bnl.gov/nsr/nsrlink.jsp?1983Vi01,B}{1983Vi01}, \href{https://www.nndc.bnl.gov/nsr/nsrlink.jsp?1986Ke04,B}{1986Ke04}) to have a predominantly 2p-1h configuration with respect to the}\\
\parbox[b][0.3cm]{17.7cm}{{\ }{\ }\ensuremath{^{\textnormal{18}}}O\ensuremath{_{\textnormal{g.s.}}} configuration because of the two nucleon nature of the (p,\ensuremath{\pi}\ensuremath{^{-}}) reaction mechanism. Therefore, (\href{https://www.nndc.bnl.gov/nsr/nsrlink.jsp?1986Ke04,B}{1986Ke04}) expected this}\\
\parbox[b][0.3cm]{17.7cm}{{\ }{\ }final state to have a relatively pure (\textit{sd})\ensuremath{^{\textnormal{n}}} configuration.}\\
\vspace{0.5cm}
\clearpage
\subsection[\hspace{-0.2cm}\ensuremath{^{\textnormal{19}}}F(p,n),(p,n\ensuremath{\gamma}),(d,2n\ensuremath{\gamma})]{ }
\vspace{-27pt}
\vspace{0.3cm}
\hypertarget{NE31}{{\bf \small \underline{\ensuremath{^{\textnormal{19}}}F(p,n),(p,n\ensuremath{\gamma}),(d,2n\ensuremath{\gamma})\hspace{0.2in}\href{https://www.nndc.bnl.gov/nsr/nsrlink.jsp?1970Gi09,B}{1970Gi09},\href{https://www.nndc.bnl.gov/nsr/nsrlink.jsp?1977Le03,B}{1977Le03}}}}\\
\vspace{4pt}
\vspace{8pt}
\parbox[b][0.3cm]{17.7cm}{\addtolength{\parindent}{-0.2in}Charge exchange reaction.}\\
\parbox[b][0.3cm]{17.7cm}{\addtolength{\parindent}{-0.2in}J\ensuremath{^{\ensuremath{\pi}}}(\ensuremath{^{\textnormal{19}}}F\ensuremath{_{\textnormal{g.s.}}})=1/2\ensuremath{^{\textnormal{+}}} and J\ensuremath{^{\ensuremath{\pi}}}(p)=1/2\ensuremath{^{\textnormal{+}}}.}\\
\parbox[b][0.3cm]{17.7cm}{\addtolength{\parindent}{-0.2in}\href{https://www.nndc.bnl.gov/nsr/nsrlink.jsp?1939Cr04,B}{1939Cr04}: \ensuremath{^{\textnormal{19}}}F(p,n) E=5.3 MeV; measured the ground-state reaction threshold. Results are reported to be ``\textit{about 4.2 MeV}'' by the}\\
\parbox[b][0.3cm]{17.7cm}{authors and 4.18 MeV \textit{25} as cited by (\href{https://www.nndc.bnl.gov/nsr/nsrlink.jsp?1952Wi27,B}{1952Wi27}).}\\
\parbox[b][0.3cm]{17.7cm}{\addtolength{\parindent}{-0.2in}\href{https://www.nndc.bnl.gov/nsr/nsrlink.jsp?1955Ma84,B}{1955Ma84}: \ensuremath{^{\textnormal{19}}}F(p,n) E=4.2-5.9 MeV; measured the ground-state threshold as E\ensuremath{_{\textnormal{thresh}}}=4235 keV \textit{5} (corresponding to Q={\textminus}4022 keV}\\
\parbox[b][0.3cm]{17.7cm}{\textit{5}) using two paraffin-moderated BF\ensuremath{_{\textnormal{3}}} counters, one of which was sensitive to slow neutrons. Measured the ratio of slow neutrons}\\
\parbox[b][0.3cm]{17.7cm}{detected by these detectors and found two \ensuremath{^{\textnormal{19}}}Ne* levels at E\ensuremath{_{\textnormal{thresh}}}=4489 keV \textit{5} and 4530 keV \textit{5} corresponding to E\ensuremath{_{\textnormal{x}}}=241 keV \textit{4}}\\
\parbox[b][0.3cm]{17.7cm}{and 280 keV \textit{4} and to Q={\textminus}4263 keV \textit{5} and Q={\textminus}4302 keV \textit{5}, respectively. No other excited states appeared with E\ensuremath{_{\textnormal{x}}}\ensuremath{<}1.5 MeV.}\\
\parbox[b][0.3cm]{17.7cm}{Measured (d\ensuremath{\sigma}/d\ensuremath{\Omega})\ensuremath{_{\textnormal{lab}}}=13 mb/sr for \ensuremath{^{\textnormal{19}}}F(p,n) relative to \ensuremath{^{\textnormal{7}}}Li(p,n) at 4.74 MeV and \ensuremath{\theta}\ensuremath{_{\textnormal{lab}}}=0\ensuremath{^\circ}{\textminus}10\ensuremath{^\circ}.}\\
\parbox[b][0.3cm]{17.7cm}{\addtolength{\parindent}{-0.2in}\href{https://www.nndc.bnl.gov/nsr/nsrlink.jsp?1955Ki28,B}{1955Ki28}: \ensuremath{^{\textnormal{19}}}F(p,n); repeated the energy calibration of the 5.5-MV Van de Graaff accelerator at the Oak Ridge National Laboratory}\\
\parbox[b][0.3cm]{17.7cm}{using various resonance measurements after installation of an NMR probe. As a result, the ground-state threshold for the \ensuremath{^{\textnormal{19}}}F(p,n)}\\
\parbox[b][0.3cm]{17.7cm}{reaction was redetermined to be E\ensuremath{_{\textnormal{thresh}}}=4240 keV \textit{8} (see also E\ensuremath{_{\textnormal{thresh}}}=4240 keV \textit{5} (Chapman and Bichsel, Universitv of Maryland}\\
\parbox[b][0.3cm]{17.7cm}{Technical Report No. 100 (revised), \textit{as quoted by} J. B. Marion and T. W. Bonner, \textit{Fast neutron physics}, edited by J. B. Marion and}\\
\parbox[b][0.3cm]{17.7cm}{J. L. Fowler (Academic Press, Inc., New York), Chap. 5, 1959)).}\\
\parbox[b][0.3cm]{17.7cm}{\addtolength{\parindent}{-0.2in}\href{https://www.nndc.bnl.gov/nsr/nsrlink.jsp?1952Wi27,B}{1952Wi27}: \ensuremath{^{\textnormal{19}}}F(p,n) E=0.15-5.4 MeV; measured neutrons using a long counter sensitive to low energy neutrons with E\ensuremath{_{\textnormal{n}}}\ensuremath{\approx}10 keV}\\
\parbox[b][0.3cm]{17.7cm}{placed at \ensuremath{\theta}\ensuremath{_{\textnormal{lab}}}=0\ensuremath{^\circ}. Measured the ground-state threshold for the \ensuremath{^{\textnormal{19}}}F(p,n) reaction relative to a \ensuremath{^{\textnormal{20}}}Ne resonance at 935.3 keV}\\
\parbox[b][0.3cm]{17.7cm}{(measured using a H\ensuremath{_{\textnormal{2}}^{\textnormal{+}}} beam) and at 486 keV (measured using a H\ensuremath{_{\textnormal{3}}^{\textnormal{+}}}) beam. The resulting relative threshold is E\ensuremath{_{\textnormal{thresh}}}=4253 keV}\\
\parbox[b][0.3cm]{17.7cm}{\textit{5}. This value was used to calculate the \ensuremath{^{\textnormal{19}}}Ne-\ensuremath{^{\textnormal{19}}}F mass difference as 0.00349 u and to determine the positron end point energy for}\\
\parbox[b][0.3cm]{17.7cm}{the \ensuremath{\beta}\ensuremath{^{\textnormal{+}}} decay of \ensuremath{^{\textnormal{19}}}Ne\ensuremath{_{\textnormal{g.s.}}} as 2.24 MeV.}\\
\parbox[b][0.3cm]{17.7cm}{\addtolength{\parindent}{-0.2in}\href{https://www.nndc.bnl.gov/nsr/nsrlink.jsp?1957Ba09,B}{1957Ba09}: \ensuremath{^{\textnormal{19}}}F(p,n\ensuremath{\gamma}); measured E\ensuremath{_{\ensuremath{\gamma}}}=242 keV \textit{5} and E\ensuremath{_{\ensuremath{\gamma}}}=281 keV \textit{8} and lifetimes of \ensuremath{\tau}=18 ns \textit{2} and \ensuremath{\tau}\ensuremath{<}5 ns for the first two excited}\\
\parbox[b][0.3cm]{17.7cm}{states of \ensuremath{^{\textnormal{19}}}Ne, reported by (\href{https://www.nndc.bnl.gov/nsr/nsrlink.jsp?1955Ma84,B}{1955Ma84}), respectively.}\\
\parbox[b][0.3cm]{17.7cm}{\addtolength{\parindent}{-0.2in}\href{https://www.nndc.bnl.gov/nsr/nsrlink.jsp?1959Br06,B}{1959Br06}: \ensuremath{^{\textnormal{19}}}F(p,n) E=2-10 MeV; measured thresholds of the (p,n) reactions on a number of nuclei, including \ensuremath{^{\textnormal{19}}}F using the}\\
\parbox[b][0.3cm]{17.7cm}{counter ratio technique. Deduced the Q-values and \ensuremath{\beta}\ensuremath{^{\textnormal{+}}} end-point energies for the measured (p,n) reactions. Measured slow-neutron}\\
\parbox[b][0.3cm]{17.7cm}{excitation curve as a function of proton bombarding energy and obtained E\ensuremath{_{\textnormal{thresh}}}=4229 keV \textit{6} for the \ensuremath{^{\textnormal{19}}}F(p,n) reaction. From this}\\
\parbox[b][0.3cm]{17.7cm}{result, the authors deduced Q(\ensuremath{^{\textnormal{19}}}F(p,n))={\textminus}4016 keV \textit{6} (based on tabulated masses by \href{https://www.nndc.bnl.gov/nsr/nsrlink.jsp?1958Kr73,B}{1958Kr73}), E\ensuremath{_{\ensuremath{\beta}^{\textnormal{+}}}}=2211 keV \textit{6} (based on}\\
\parbox[b][0.3cm]{17.7cm}{neutron-hydrogen atom mass difference of 0.7830 MeV \textit{9} by (\href{https://www.nndc.bnl.gov/nsr/nsrlink.jsp?1958Kr73,B}{1958Kr73}), and \ensuremath{^{\textnormal{19}}}Ne mass of 19.0079168 u \textit{64} assuming 1}\\
\parbox[b][0.3cm]{17.7cm}{u=931.143 MeV \textit{10}).}\\
\parbox[b][0.3cm]{17.7cm}{\addtolength{\parindent}{-0.2in}\href{https://www.nndc.bnl.gov/nsr/nsrlink.jsp?1959Gi47,B}{1959Gi47}: \ensuremath{^{\textnormal{19}}}F(p,n) E=4.235-5 MeV; embedded their target in the center of a highly pure graphite sphere (to fully absorb \ensuremath{\beta}\ensuremath{^{\textnormal{+}}}}\\
\parbox[b][0.3cm]{17.7cm}{particles) and measured neutrons using an array of 8 BF\ensuremath{_{\textnormal{3}}} counters placed at the surface of the graphite sphere. Measured absolute}\\
\parbox[b][0.3cm]{17.7cm}{cross section as a function of proton energy for the \ensuremath{^{\textnormal{19}}}F(p,n) reaction. Deduced E\ensuremath{_{\textnormal{thresh}}}(\ensuremath{^{\textnormal{19}}}F(p,n))=4235 keV \textit{5}.}\\
\parbox[b][0.3cm]{17.7cm}{\addtolength{\parindent}{-0.2in}\href{https://www.nndc.bnl.gov/nsr/nsrlink.jsp?1961Be13,B}{1961Be13}: \ensuremath{^{\textnormal{19}}}F(p,n) E\ensuremath{\sim}4.22-4.24 MeV; measured neutrons using a paraffin-moderated \ensuremath{^{\textnormal{10}}}BF\ensuremath{_{\textnormal{3}}} proportional counter; measured the}\\
\parbox[b][0.3cm]{17.7cm}{ground-state threshold as E\ensuremath{_{\textnormal{thresh}}}=4233.3 keV \textit{20} using a 180\ensuremath{^\circ} spectrometer. Deduced a Q-value of {\textminus}4019.5 keV \textit{20} for the}\\
\parbox[b][0.3cm]{17.7cm}{\ensuremath{^{\textnormal{19}}}F(p,n) reaction.}\\
\parbox[b][0.3cm]{17.7cm}{\addtolength{\parindent}{-0.2in}\href{https://www.nndc.bnl.gov/nsr/nsrlink.jsp?1961Ry04,B}{1961Ry04}: \ensuremath{^{\textnormal{19}}}F(p,n) E=4.2 MeV; measured the threshold energy for the \ensuremath{^{\textnormal{19}}}F(p,n)\ensuremath{^{\textnormal{19}}}Ne\ensuremath{_{\textnormal{g.s.}}} using a 180\ensuremath{^\circ} spectrometer; obtained}\\
\parbox[b][0.3cm]{17.7cm}{E\ensuremath{_{\textnormal{thresh}}}(\ensuremath{^{\textnormal{19}}}Ne\ensuremath{_{\textnormal{g.s.}}})=4234.7 keV \textit{10}.}\\
\parbox[b][0.3cm]{17.7cm}{\addtolength{\parindent}{-0.2in}\href{https://www.nndc.bnl.gov/nsr/nsrlink.jsp?1962Fr09,B}{1962Fr09}: \ensuremath{^{\textnormal{19}}}F(p,n) E=5.2, 6.2-6.9 MeV; measured ground-state threshold energy and Q-value of the \ensuremath{^{\textnormal{19}}}F(p,n)\ensuremath{^{\textnormal{19}}}Ne\ensuremath{_{\textnormal{g.s.}}} reaction}\\
\parbox[b][0.3cm]{17.7cm}{using the counter ratio technique and a \ensuremath{^{\textnormal{3}}}He ionization chamber. Deduced E\ensuremath{_{\textnormal{thresh}}}=4233 keV \textit{5} and Q-value={\textminus}4019 keV \textit{5}.}\\
\parbox[b][0.3cm]{17.7cm}{Measured \ensuremath{\sigma}(E;E\ensuremath{_{\textnormal{n}}}) excitation function using both techniques and observed n\ensuremath{_{\textnormal{0}}}, n\ensuremath{_{\textnormal{1+2}}} and n\ensuremath{_{\textnormal{3+4}}} neutron groups. Deduced \ensuremath{^{\textnormal{19}}}Ne}\\
\parbox[b][0.3cm]{17.7cm}{excitation energies corresponding to these neutron groups using each method. Determined \ensuremath{^{\textnormal{19}}}Ne excited levels at E\ensuremath{_{\textnormal{x}}}=1506 keV \textit{5},}\\
\parbox[b][0.3cm]{17.7cm}{1538 keV \textit{4}, and 1612 keV \textit{5}.}\\
\parbox[b][0.3cm]{17.7cm}{\addtolength{\parindent}{-0.2in}\href{https://www.nndc.bnl.gov/nsr/nsrlink.jsp?1963Je04,B}{1963Je04}: \ensuremath{^{\textnormal{19}}}F(p,n\ensuremath{\gamma}) E=4.9-11 MeV; using activation technique, the authors measured coincident annihilation radiations from the}\\
\parbox[b][0.3cm]{17.7cm}{\ensuremath{^{\textnormal{19}}}Ne\ensuremath{_{\textnormal{g.s.}}} decay via the fast-slow technique using two NaI(Tl) detectors. Measured the excitation function. Deduced \ensuremath{^{\textnormal{20}}}Ne levels}\\
\parbox[b][0.3cm]{17.7cm}{from proton resonances observed in the excitation function. Deduced \ensuremath{\sigma}\ensuremath{_{\textnormal{absolute}}}(\ensuremath{^{\textnormal{19}}}F(p,n))=27 mb \textit{4} at E\ensuremath{_{\textnormal{p}}}=5.58 MeV.}\\
\parbox[b][0.3cm]{17.7cm}{\addtolength{\parindent}{-0.2in}\href{https://www.nndc.bnl.gov/nsr/nsrlink.jsp?1963Gi09,B}{1963Gi09}: \ensuremath{^{\textnormal{19}}}F(p,n\ensuremath{\gamma}) E=5.4 MeV; measured the relative intensities of internal conversion electrons and of \ensuremath{\gamma} rays emitted from}\\
\parbox[b][0.3cm]{17.7cm}{transitions to \ensuremath{^{\textnormal{19}}}Ne\ensuremath{_{\textnormal{g.s.}}} from the \ensuremath{^{\textnormal{19}}}Ne*(238, 275) levels. The internal conversion electron spectrum was measured for electron}\\
\parbox[b][0.3cm]{17.7cm}{energies between 100-350 keV using a magnetic-lens-type \ensuremath{\beta}-ray spectrometer. The background subtracted spectrum shows 2 peaks}\\
\parbox[b][0.3cm]{17.7cm}{associated with the 241 keV \textit{4} and 271 keV \textit{4} K-electron converted \ensuremath{\gamma} rays from the decays of the \ensuremath{^{\textnormal{19}}}Ne*(238, 275) levels. From}\\
\parbox[b][0.3cm]{17.7cm}{the measured relative intensities between the two \ensuremath{\gamma} rays and between the two conversion electrons, the authors concluded that the}\\
\parbox[b][0.3cm]{17.7cm}{238-keV level has a larger total angular momentum than the 275-keV state, and thus confirming that the spin sequence of these two}\\
\parbox[b][0.3cm]{17.7cm}{levels is inverted with respect to that in the mirror nucleus, \ensuremath{^{\textnormal{19}}}F.}\\
\parbox[b][0.3cm]{17.7cm}{\addtolength{\parindent}{-0.2in}\href{https://www.nndc.bnl.gov/nsr/nsrlink.jsp?1965Va23,B}{1965Va23}: \ensuremath{^{\textnormal{19}}}F(p,n) E=18.5 MeV to 5.6 GeV; measured reaction products and deduced \ensuremath{\sigma} and \ensuremath{\sigma}(E).}\\
\parbox[b][0.3cm]{17.7cm}{\addtolength{\parindent}{-0.2in}\href{https://www.nndc.bnl.gov/nsr/nsrlink.jsp?1965We05,B}{1965We05}: \ensuremath{^{\textnormal{19}}}F(p,n) E=7.5, 8.6, 9.75 MeV; measured neutrons TOF spectra over a shielded neutron flight path of 10.3 m. The}\\
\clearpage
\vspace{0.3cm}
{\bf \small \underline{\ensuremath{^{\textnormal{19}}}F(p,n),(p,n\ensuremath{\gamma}),(d,2n\ensuremath{\gamma})\hspace{0.2in}\href{https://www.nndc.bnl.gov/nsr/nsrlink.jsp?1970Gi09,B}{1970Gi09},\href{https://www.nndc.bnl.gov/nsr/nsrlink.jsp?1977Le03,B}{1977Le03} (continued)}}\\
\vspace{0.3cm}
\parbox[b][0.3cm]{17.7cm}{spectra were measured at \ensuremath{\theta}\ensuremath{_{\textnormal{lab}}}=30\ensuremath{^\circ} for E=7.5 MeV, at \ensuremath{\theta}\ensuremath{_{\textnormal{lab}}}=120\ensuremath{^\circ} for E=8.6 MeV, and at \ensuremath{\theta}\ensuremath{_{\textnormal{lab}}}=30\ensuremath{^\circ}, 60\ensuremath{^\circ}, and 90\ensuremath{^\circ} for E=9.75 MeV.}\\
\parbox[b][0.3cm]{17.7cm}{Measured \ensuremath{\sigma}(E;E\ensuremath{_{\textnormal{n}}},\ensuremath{\theta}). Observed n\ensuremath{_{\textnormal{0}}}, n\ensuremath{_{\textnormal{1+2}}} (unresolved), n\ensuremath{_{\textnormal{3+4}}} (unresolved), n\ensuremath{_{\textnormal{5}}}, and n\ensuremath{_{\textnormal{6}}} groups. The latter group was only observed}\\
\parbox[b][0.3cm]{17.7cm}{at E=9.75 MeV and corresponds to the \ensuremath{^{\textnormal{19}}}Ne*(2.78 MeV) state, which was observed here for first time. Deduced \ensuremath{^{\textnormal{19}}}Ne excitation}\\
\parbox[b][0.3cm]{17.7cm}{energies. The results are consistent with those of (\href{https://www.nndc.bnl.gov/nsr/nsrlink.jsp?1962Fr09,B}{1962Fr09}).}\\
\parbox[b][0.3cm]{17.7cm}{\addtolength{\parindent}{-0.2in}\href{https://www.nndc.bnl.gov/nsr/nsrlink.jsp?1967Dr08,B}{1967Dr08}: \ensuremath{^{\textnormal{19}}}F(p,n\ensuremath{_{\textnormal{0+1+2+3?}}}) E=6.8 MeV; deduced nuclear properties.}\\
\parbox[b][0.3cm]{17.7cm}{\addtolength{\parindent}{-0.2in}\href{https://www.nndc.bnl.gov/nsr/nsrlink.jsp?1968Ri08,B}{1968Ri08}: \ensuremath{^{\textnormal{19}}}F(p,n\ensuremath{\gamma}) E=4.324-6.010 MeV; measured \ensuremath{\sigma}(E) using the activation technique. Measured annihilation \ensuremath{\gamma} rays in}\\
\parbox[b][0.3cm]{17.7cm}{coincidence using two NaI(T1) detectors; deduced absolute \ensuremath{\sigma}(E). Comparison with previously measured cross sections is provided.}\\
\parbox[b][0.3cm]{17.7cm}{\addtolength{\parindent}{-0.2in}\href{https://www.nndc.bnl.gov/nsr/nsrlink.jsp?1968Go10,B}{1968Go10}: \ensuremath{^{\textnormal{19}}}F(p,n\ensuremath{\gamma}) E=5.5 MeV; measured T\ensuremath{_{\textnormal{1/2}}}(\ensuremath{^{\textnormal{19}}}Ne\ensuremath{_{\textnormal{g.s.}}})=17.36 s \textit{6} by detection, in coincidence, of annihilation radiations}\\
\parbox[b][0.3cm]{17.7cm}{following the \ensuremath{^{\textnormal{19}}}Ne(\ensuremath{\beta}\ensuremath{^{\textnormal{+}}})\ensuremath{^{\textnormal{19}}}F decay using two NaI(Tl) detectors facing each other.}\\
\parbox[b][0.3cm]{17.7cm}{\addtolength{\parindent}{-0.2in}\href{https://www.nndc.bnl.gov/nsr/nsrlink.jsp?1969Bl02,B}{1969Bl02}: \ensuremath{^{\textnormal{19}}}F(p,n\ensuremath{\gamma}) E=4.4-6.1 MeV; measured \ensuremath{\sigma}(E;E\ensuremath{_{\ensuremath{\gamma}}},\ensuremath{\theta},H,t) for \ensuremath{^{\textnormal{19}}}F(p,n\ensuremath{\gamma}) at E\ensuremath{_{\textnormal{p}}}=4.4-6.1 MeV using a Ge(Li) detector at}\\
\parbox[b][0.3cm]{17.7cm}{\ensuremath{\theta}\ensuremath{_{\textnormal{lab}}}=0\ensuremath{^\circ}. Deduced \ensuremath{^{\textnormal{19}}}Ne* level- and \ensuremath{\gamma}-ray energies for the first two excited states. The excitation function showed a resonance at}\\
\parbox[b][0.3cm]{17.7cm}{E\ensuremath{_{\textnormal{p}}}=5.25 MeV. Simultaneously measured the time-dependent intensities of the \ensuremath{\gamma} rays from the decays of the \ensuremath{^{\textnormal{19}}}Ne*(238) and}\\
\parbox[b][0.3cm]{17.7cm}{\ensuremath{^{\textnormal{19}}}F*(197) levels. These were populated via (p,n) and (p,p\ensuremath{'}) reactions, respectively, on a \ensuremath{^{\textnormal{19}}}F target and were measured at E\ensuremath{_{\textnormal{p}}}=5.25}\\
\parbox[b][0.3cm]{17.7cm}{MeV using pulsed beam differential delay constant angle method and by using 2 NaI(Tl) detectors at \ensuremath{\theta}\ensuremath{_{\textnormal{lab}}}=25\ensuremath{^\circ} and {\textminus}65\ensuremath{^\circ}. Measured}\\
\parbox[b][0.3cm]{17.7cm}{spin precession spectra at E\ensuremath{_{\textnormal{p}}}=5.25 MeV and B=34 kG. Deduced the g-factor of \textit{g}={\textminus}0.296 \textit{3} for the 238-keV level. Comparison with}\\
\parbox[b][0.3cm]{17.7cm}{various models$'$ predictions are discussed. Measured T\ensuremath{_{\textnormal{1/2}}}=17.7 ns \textit{7} and T\ensuremath{_{\textnormal{1/2}}}\ensuremath{<}0.3 ns for the \ensuremath{^{\textnormal{19}}}Ne*(238, 275) levels, respectively,}\\
\parbox[b][0.3cm]{17.7cm}{from n-\ensuremath{\gamma} coincidence events at E\ensuremath{_{\textnormal{p}}}=6.2 MeV.}\\
\parbox[b][0.3cm]{17.7cm}{\addtolength{\parindent}{-0.2in}\href{https://www.nndc.bnl.gov/nsr/nsrlink.jsp?1969Ov01,B}{1969Ov01}: \ensuremath{^{\textnormal{19}}}F(p,n) E=2-20 MeV; measured neutrons using a BF\ensuremath{_{\textnormal{3}}} long counter; measured \ensuremath{\sigma}(E). Deduced E\ensuremath{_{\textnormal{thresh}}}=4233.7 keV \textit{7}}\\
\parbox[b][0.3cm]{17.7cm}{for the ground state.}\\
\parbox[b][0.3cm]{17.7cm}{\addtolength{\parindent}{-0.2in}\href{https://www.nndc.bnl.gov/nsr/nsrlink.jsp?1970Gi09,B}{1970Gi09}: \ensuremath{^{\textnormal{19}}}F(p,n\ensuremath{\gamma}) E=8.4-10.4 MeV; measured n-\ensuremath{\gamma} coincidences using an NE-213 liquid scintillator at \ensuremath{\theta}\ensuremath{_{\textnormal{lab}}}=0\ensuremath{^\circ} and 90\ensuremath{^\circ}, and a}\\
\parbox[b][0.3cm]{17.7cm}{Ge(Li) detector placed at \ensuremath{\theta}\ensuremath{_{\textnormal{lab}}}=90\ensuremath{^\circ}, 30\ensuremath{^\circ}, and 125\ensuremath{^\circ} to detect the neutrons and \ensuremath{\gamma} rays, respectively. Measured E\ensuremath{_{\ensuremath{\gamma}}}, I\ensuremath{_{\ensuremath{\gamma}}}, \ensuremath{\gamma}-ray angular}\\
\parbox[b][0.3cm]{17.7cm}{correlations (\ensuremath{\sigma}(E;E\ensuremath{_{\textnormal{n}}},E\ensuremath{_{\ensuremath{\gamma}}},\ensuremath{\theta}\ensuremath{_{\textnormal{n}\ensuremath{\gamma}}})) of the emitted \ensuremath{\gamma} rays. Deduced \ensuremath{^{\textnormal{19}}}Ne levels, J, \ensuremath{\pi}, and branching ratios. Comparison of \ensuremath{\gamma}-ray and}\\
\parbox[b][0.3cm]{17.7cm}{level-energies with those deduced by (\href{https://www.nndc.bnl.gov/nsr/nsrlink.jsp?1967Ol05,B}{1967Ol05}) are discussed. Deduced the lifetime of the \ensuremath{^{\textnormal{19}}}Ne*(2795) state using DSAM.}\\
\parbox[b][0.3cm]{17.7cm}{\addtolength{\parindent}{-0.2in}\href{https://www.nndc.bnl.gov/nsr/nsrlink.jsp?1971It02,B}{1971It02}: \ensuremath{^{\textnormal{19}}}F(p,n\ensuremath{\gamma}) E=21-26 MeV; measured E\ensuremath{_{\ensuremath{\gamma}}} using a Ge(Li) detector that was placed at different angles between \ensuremath{\theta}\ensuremath{_{\textnormal{lab}}}=0\ensuremath{^\circ} and}\\
\parbox[b][0.3cm]{17.7cm}{90\ensuremath{^\circ}. Deduced \ensuremath{^{\textnormal{19}}}Ne level-energies and T\ensuremath{_{\textnormal{1/2}}} for the \ensuremath{^{\textnormal{19}}}Ne*(1508, 1536) levels using DSAM. Comparison with (\href{https://www.nndc.bnl.gov/nsr/nsrlink.jsp?1970Gi09,B}{1970Gi09}) is}\\
\parbox[b][0.3cm]{17.7cm}{discussed.}\\
\parbox[b][0.3cm]{17.7cm}{\addtolength{\parindent}{-0.2in}\href{https://www.nndc.bnl.gov/nsr/nsrlink.jsp?1972Ku24,B}{1972Ku24}: \ensuremath{^{\textnormal{19}}}F(p,n) and \ensuremath{^{\textnormal{19}}}F(p,n\ensuremath{\gamma}) E=5.82-5.98 MeV; measured the total cross section of \ensuremath{^{\textnormal{19}}}F(p,n) using the activation technique}\\
\parbox[b][0.3cm]{17.7cm}{by measuring (in coincidence) the annihilation photons emitted following the \ensuremath{\beta} decay of \ensuremath{^{\textnormal{19}}}Ne. Deduced \ensuremath{\sigma}(E)\ensuremath{_{\textnormal{tot}}}=40-50 mb with an}\\
\parbox[b][0.3cm]{17.7cm}{uncertainty of 20\% based on the results of (\href{https://www.nndc.bnl.gov/nsr/nsrlink.jsp?1968Ri08,B}{1968Ri08}). They also measured differential cross sections of \ensuremath{^{\textnormal{19}}}F(p,n\ensuremath{_{\textnormal{0}}}) and}\\
\parbox[b][0.3cm]{17.7cm}{\ensuremath{^{\textnormal{19}}}F(p,n\ensuremath{_{\textnormal{1}}}+n\ensuremath{_{\textnormal{2}}}) (unresolved) at \ensuremath{\theta}\ensuremath{_{\textnormal{lab}}}=0\ensuremath{^\circ}, 90\ensuremath{^\circ}, 150\ensuremath{^\circ}, and 163\ensuremath{^\circ} using a pulsed beam and neutron time-of-flight technique. Populated}\\
\parbox[b][0.3cm]{17.7cm}{the T=2 \ensuremath{^{\textnormal{20}}}Ne*(18430) excited state corresponding to the \ensuremath{^{\textnormal{19}}}F+p resonance at E\ensuremath{_{\textnormal{p}}}=5879 keV \textit{7}, which decays to the \ensuremath{^{\textnormal{19}}}Ne*(0, 238,}\\
\parbox[b][0.3cm]{17.7cm}{275) levels among many other decay paths. The n\ensuremath{_{\textnormal{1}}} and n\ensuremath{_{\textnormal{2}}} decays to the \ensuremath{^{\textnormal{19}}}Ne*(238, 275) levels were unresolved.}\\
\parbox[b][0.3cm]{17.7cm}{\addtolength{\parindent}{-0.2in}\href{https://www.nndc.bnl.gov/nsr/nsrlink.jsp?1972Sh08,B}{1972Sh08}, \href{https://www.nndc.bnl.gov/nsr/nsrlink.jsp?1974Sh06,B}{1974Sh06}: \ensuremath{^{\textnormal{19}}}F(p,n) E=2-60; measured neutrons using a long counter; measured the ground-state Q-value as part of the}\\
\parbox[b][0.3cm]{17.7cm}{calibration of an analyzing magnet. The result is not reported. The proton energy reported here is not specific to the \ensuremath{^{\textnormal{19}}}F(p,n)}\\
\parbox[b][0.3cm]{17.7cm}{reaction.}\\
\parbox[b][0.3cm]{17.7cm}{\addtolength{\parindent}{-0.2in}\href{https://www.nndc.bnl.gov/nsr/nsrlink.jsp?1974DeZX,B}{1974DeZX}: \ensuremath{^{\textnormal{19}}}F(p,n\ensuremath{\gamma}); measured \ensuremath{\sigma}(E\ensuremath{_{\ensuremath{\gamma}}}), and n-\ensuremath{\gamma} coincidence events; deduced \ensuremath{^{\textnormal{19}}}Ne levels and \ensuremath{\gamma} branching ratios.}\\
\parbox[b][0.3cm]{17.7cm}{\addtolength{\parindent}{-0.2in}\href{https://www.nndc.bnl.gov/nsr/nsrlink.jsp?1977Le03,B}{1977Le03}: \ensuremath{^{\textnormal{19}}}F(p,n\ensuremath{\gamma}) E=12 MeV; measured n-\ensuremath{\gamma} and \ensuremath{\gamma}-\ensuremath{\gamma} coincidence events using a Ge(Li) detector with a resolution of 4 keV at}\\
\parbox[b][0.3cm]{17.7cm}{E\ensuremath{_{\ensuremath{\gamma}}}=1.33 MeV placed at \ensuremath{\theta}\ensuremath{_{\textnormal{lab}}}=90\ensuremath{^\circ} and two NE-213 liquid scintillators placed symmetrically at \ensuremath{\theta}\ensuremath{_{\textnormal{lab}}}=\ensuremath{\pm}36\ensuremath{^\circ}; measured E\ensuremath{_{\ensuremath{\gamma}}} and n-\ensuremath{\gamma}}\\
\parbox[b][0.3cm]{17.7cm}{TOF with 4 ns timing resolution; deduced lifetimes of the \ensuremath{^{\textnormal{19}}}Ne*(1.51, 1.54, 1.62, 2.79 MeV) levels using DSAM. Deduced B(\ensuremath{\lambda}).}\\
\parbox[b][0.3cm]{17.7cm}{Comparison with previously measured and theoretical lifetimes and transition strengths are discussed.}\\
\parbox[b][0.3cm]{17.7cm}{\addtolength{\parindent}{-0.2in}\href{https://www.nndc.bnl.gov/nsr/nsrlink.jsp?1983RaZU,B}{1983RaZU}: \ensuremath{^{\textnormal{19}}}F(p,n) E=160 MeV; measured \ensuremath{\sigma}(\ensuremath{\theta}); deduced \ensuremath{^{\textnormal{19}}}Ne levels, Gamow-Teller transition strength distribution. Discussed}\\
\parbox[b][0.3cm]{17.7cm}{sum rule comparisons and TOF.}\\
\parbox[b][0.3cm]{17.7cm}{\addtolength{\parindent}{-0.2in}\href{https://www.nndc.bnl.gov/nsr/nsrlink.jsp?1983Wi04,B}{1983Wi04}: \ensuremath{^{\textnormal{19}}}F(p,n) E\ensuremath{\approx}threshold; measured yield vs. E\ensuremath{_{\textnormal{p}}} as part of the calibration of a dual 90\ensuremath{^\circ} analyzing magnet system.}\\
\parbox[b][0.3cm]{17.7cm}{\addtolength{\parindent}{-0.2in}\href{https://www.nndc.bnl.gov/nsr/nsrlink.jsp?1984Ra22,B}{1984Ra22}: \ensuremath{^{\textnormal{19}}}F(p,n) E=120, 160 MeV; neutron-TOF was measured using a 100-m flight path and three plastic scintillators covering}\\
\parbox[b][0.3cm]{17.7cm}{\ensuremath{\theta}\ensuremath{_{\textnormal{lab}}}=0\ensuremath{^\circ}{\textminus}25\ensuremath{^\circ} and a resolution of 800-900 ps. Measured neutron angular distributions at \ensuremath{\theta}\ensuremath{_{\textnormal{lab}}}=0\ensuremath{^\circ}, 2.5\ensuremath{^\circ}, and 6.2\ensuremath{^\circ} at E\ensuremath{_{\textnormal{p}}}=120 MeV and}\\
\parbox[b][0.3cm]{17.7cm}{at \ensuremath{\theta}\ensuremath{_{\textnormal{lab}}}=0\ensuremath{^\circ}, 2.5\ensuremath{^\circ}, 5\ensuremath{^\circ}, 7.5\ensuremath{^\circ}, 10\ensuremath{^\circ}, and 15\ensuremath{^\circ} at E\ensuremath{_{\textnormal{p}}}=160 MeV. Deduced \ensuremath{^{\textnormal{19}}}Ne levels, Gamow-Teller transition strength distribution, and}\\
\parbox[b][0.3cm]{17.7cm}{transfer orbital angular momenta using DWBA analysis. This study provides d\ensuremath{\sigma}/d\ensuremath{\Omega}\ensuremath{_{\textnormal{c.m.}}}(\ensuremath{\theta}\ensuremath{_{\textnormal{c.m.}}}=0\ensuremath{^\circ}) for populated \ensuremath{^{\textnormal{19}}}Ne states at}\\
\parbox[b][0.3cm]{17.7cm}{E\ensuremath{_{\textnormal{p}}}=120 and 160 MeV.}\\
\parbox[b][0.3cm]{17.7cm}{\addtolength{\parindent}{-0.2in}\href{https://www.nndc.bnl.gov/nsr/nsrlink.jsp?1984TaZS,B}{1984TaZS}: \ensuremath{^{\textnormal{19}}}F(p,n); spin transfer measurements.}\\
\parbox[b][0.3cm]{17.7cm}{\addtolength{\parindent}{-0.2in}\href{https://www.nndc.bnl.gov/nsr/nsrlink.jsp?1985Ba66,B}{1985Ba66}: \ensuremath{^{\textnormal{19}}}F(p,n) E\ensuremath{\leq}7 MeV; measured neutrons using a NE-213 liquid scintillator; measured \ensuremath{\sigma}(E\ensuremath{_{\textnormal{n}}}) at \ensuremath{\theta}\ensuremath{_{\textnormal{lab}}}=0\ensuremath{^\circ} using thick}\\
\parbox[b][0.3cm]{17.7cm}{target yield curve method; populated n\ensuremath{_{\textnormal{0}}} and n\ensuremath{_{\textnormal{1+2}}} (unresolved) neutron groups.}\\
\parbox[b][0.3cm]{17.7cm}{\addtolength{\parindent}{-0.2in}\href{https://www.nndc.bnl.gov/nsr/nsrlink.jsp?1985Wa24,B}{1985Wa24}: \ensuremath{^{\textnormal{19}}}F(p,n) E=160 MeV; measured \ensuremath{\sigma}(\ensuremath{\theta}) and \ensuremath{\sigma}(\ensuremath{\theta}\ensuremath{_{\textnormal{n}}},E\ensuremath{_{\textnormal{n}}}).}\\
\parbox[b][0.3cm]{17.7cm}{\addtolength{\parindent}{-0.2in}\href{https://www.nndc.bnl.gov/nsr/nsrlink.jsp?1987Ra23,B}{1987Ra23}: \ensuremath{^{\textnormal{19}}}F(p,n\ensuremath{\gamma}) E=7, 9 MeV; measured proton induced \ensuremath{\gamma} ray yields from a variety of thick targets using a Ge(Li) detector}\\
\clearpage
\vspace{0.3cm}
{\bf \small \underline{\ensuremath{^{\textnormal{19}}}F(p,n),(p,n\ensuremath{\gamma}),(d,2n\ensuremath{\gamma})\hspace{0.2in}\href{https://www.nndc.bnl.gov/nsr/nsrlink.jsp?1970Gi09,B}{1970Gi09},\href{https://www.nndc.bnl.gov/nsr/nsrlink.jsp?1977Le03,B}{1977Le03} (continued)}}\\
\vspace{0.3cm}
\parbox[b][0.3cm]{17.7cm}{with a resolution of 1.9 keV at E\ensuremath{_{\ensuremath{\gamma}}}=1.33 MeV placed at \ensuremath{\theta}\ensuremath{_{\textnormal{lab}}}=55\ensuremath{^\circ}. Measured neutrons in coincidence with \ensuremath{\gamma} rays using a BF\ensuremath{_{\textnormal{3}}}}\\
\parbox[b][0.3cm]{17.7cm}{counter. Provided absolute \ensuremath{\gamma} ray yields and relative normalized neutron yields from various targets at E\ensuremath{_{\textnormal{p}}}=7 and 9 MeV.}\\
\parbox[b][0.3cm]{17.7cm}{\addtolength{\parindent}{-0.2in}\href{https://www.nndc.bnl.gov/nsr/nsrlink.jsp?1990HuZY,B}{1990HuZY}: \ensuremath{^{\textnormal{19}}}F(pol. p,n), E=120, 160 MeV; measured transverse polarization transfer coefficient.}\\
\parbox[b][0.3cm]{17.7cm}{\addtolength{\parindent}{-0.2in}\href{https://www.nndc.bnl.gov/nsr/nsrlink.jsp?1990Wa10,B}{1990Wa10}: \ensuremath{^{\textnormal{19}}}F(p,n\ensuremath{\gamma}) E=4.24-28 MeV; a target located at \ensuremath{\theta}\ensuremath{_{\textnormal{lab}}}=45\ensuremath{^\circ} was bombarded for 20 s. After activation, the annihilation \ensuremath{\gamma}}\\
\parbox[b][0.3cm]{17.7cm}{rays emitted from the decay of \ensuremath{^{\textnormal{19}}}Ne were measured in coincidence for 1 hour using two plastic scintillators facing each other.}\\
\parbox[b][0.3cm]{17.7cm}{Measured energy integrated cross section from E\ensuremath{_{\textnormal{thresh}}} to 7 MeV. Calculated \ensuremath{^{\textnormal{19}}}Ne production yield useful for production of}\\
\parbox[b][0.3cm]{17.7cm}{radioactive beams.}\\
\parbox[b][0.3cm]{17.7cm}{\addtolength{\parindent}{-0.2in}\href{https://www.nndc.bnl.gov/nsr/nsrlink.jsp?1993HuZT,B}{1993HuZT}: \ensuremath{^{\textnormal{19}}}F(p,n) E not given; measured residuals atomic yields per incident proton; deduced radioactive beam production.}\\
\vspace{0.385cm}
\parbox[b][0.3cm]{17.7cm}{\addtolength{\parindent}{-0.2in}\textit{The \ensuremath{^{19}}F(d,2n) Studies}:}\\
\parbox[b][0.3cm]{17.7cm}{\addtolength{\parindent}{-0.2in}\href{https://www.nndc.bnl.gov/nsr/nsrlink.jsp?1954Na29,B}{1954Na29}: \ensuremath{^{\textnormal{19}}}F(d,2n) E=28 MeV; activation technique using cycles that consisted of 1 min. irradiation followed by 10 s counting}\\
\parbox[b][0.3cm]{17.7cm}{using a Geiger counter. Measured the \ensuremath{^{\textnormal{19}}}Ne decay curve and deduced T\ensuremath{_{\textnormal{1/2}}}=19 s \textit{1}.}\\
\parbox[b][0.3cm]{17.7cm}{\addtolength{\parindent}{-0.2in}\href{https://www.nndc.bnl.gov/nsr/nsrlink.jsp?1984Pi07,B}{1984Pi07}: \ensuremath{^{\textnormal{19}}}F(p,n\ensuremath{\gamma}) E=8 MeV and \ensuremath{^{\textnormal{19}}}F(d,2n\ensuremath{\gamma}) E=16 MeV; measured E\ensuremath{_{\ensuremath{\gamma}}} for prompt \ensuremath{\gamma} rays with E\ensuremath{<}600 keV using a Ge-detector.}\\
\vspace{0.385cm}
\parbox[b][0.3cm]{17.7cm}{\addtolength{\parindent}{-0.2in}\textit{The \ensuremath{^{19}}F(p,n) Studies with Relevant Information on the \ensuremath{^{\textnormal{19}}}Ne(\ensuremath{\beta}\ensuremath{^{\textnormal{+}}})\ensuremath{^{\textnormal{19}}}F Decay and \ensuremath{^{\textnormal{19}}}Ne Lifetime}:}\\
\parbox[b][0.3cm]{17.7cm}{\addtolength{\parindent}{-0.2in}\href{https://www.nndc.bnl.gov/nsr/nsrlink.jsp?1939Fo01,B}{1939Fo01}: \ensuremath{^{\textnormal{19}}}F(p,n)\ensuremath{^{\textnormal{19}}}Ne(\ensuremath{\beta}) E not given. This is the study in which \ensuremath{^{\textnormal{19}}}Ne was first identified. The details are not provided. The}\\
\parbox[b][0.3cm]{17.7cm}{authors measured the half-life of \ensuremath{^{\textnormal{19}}}Ne\ensuremath{_{\textnormal{g.s.}}} as T\ensuremath{_{\textnormal{1/2}}}=20 s and measured the positron decay curve of \ensuremath{^{\textnormal{19}}}Ne using a cloud chamber and}\\
\parbox[b][0.3cm]{17.7cm}{obtained the \ensuremath{\beta}\ensuremath{^{\textnormal{+}}} end-point energy of E\ensuremath{_{\ensuremath{\beta}^{\textnormal{+}}}}=2.5 MeV.}\\
\parbox[b][0.3cm]{17.7cm}{\addtolength{\parindent}{-0.2in}\href{https://www.nndc.bnl.gov/nsr/nsrlink.jsp?1939Wh02,B}{1939Wh02}: \ensuremath{^{\textnormal{19}}}F(p,n)\ensuremath{^{\textnormal{19}}}Ne(\ensuremath{\beta}) E\ensuremath{\leq}6 MeV; used activation technique; samples were moved to a Lauritsen electroscope using a}\\
\parbox[b][0.3cm]{17.7cm}{moving tape; measured \ensuremath{\beta}\ensuremath{^{\textnormal{+}}} particles from the \ensuremath{^{\textnormal{19}}}Ne\ensuremath{_{\textnormal{g.s.}}} decay using a hydrogen-filled Wilson cloud chamber placed in a uniform}\\
\parbox[b][0.3cm]{17.7cm}{magnetic field with a strength of 600 G. Measured absorption curve of positrons and annihilation radiation from \ensuremath{^{\textnormal{19}}}Ne; deduced}\\
\parbox[b][0.3cm]{17.7cm}{T\ensuremath{_{\textnormal{1/2}}}(\ensuremath{^{\textnormal{19}}}Ne\ensuremath{_{\textnormal{g.s.}}})=20.3 s \textit{5}, \ensuremath{\beta}\ensuremath{^{\textnormal{+}}} end-point energy of 2.2 MeV, and E\ensuremath{_{\textnormal{thresh}}}=4.18 MeV \textit{25} for the \ensuremath{^{\textnormal{19}}}F(p,n) reaction.}\\
\parbox[b][0.3cm]{17.7cm}{\addtolength{\parindent}{-0.2in}\href{https://www.nndc.bnl.gov/nsr/nsrlink.jsp?1949Sh25,B}{1949Sh25}: \ensuremath{^{\textnormal{19}}}F(p,n)\ensuremath{^{\textnormal{19}}}Ne(\ensuremath{\beta}) E not given; measured the decay curve of \ensuremath{^{\textnormal{19}}}Ne and deduced T\ensuremath{_{\textnormal{1/2}}}(\ensuremath{^{\textnormal{19}}}Ne\ensuremath{_{\textnormal{g.s.}}})=18.2 s \textit{6}. The authors}\\
\parbox[b][0.3cm]{17.7cm}{acknowledged that the half-life they determined was lower than that deduced by (\href{https://www.nndc.bnl.gov/nsr/nsrlink.jsp?1939Wh02,B}{1939Wh02}: T\ensuremath{_{\textnormal{1/2}}}=20.3 s \textit{5}) and mentioned that}\\
\parbox[b][0.3cm]{17.7cm}{the longer half-life obtained by (\href{https://www.nndc.bnl.gov/nsr/nsrlink.jsp?1939Wh02,B}{1939Wh02}) could be due to the presence of a small amount of impurity in the solid target used by}\\
\parbox[b][0.3cm]{17.7cm}{(\href{https://www.nndc.bnl.gov/nsr/nsrlink.jsp?1939Wh02,B}{1939Wh02}). (\href{https://www.nndc.bnl.gov/nsr/nsrlink.jsp?1949Sh25,B}{1949Sh25}) deduced an end-point energy of 2.3 MeV \textit{1} for the positrons from the decay of \ensuremath{^{\textnormal{19}}}Ne.}\\
\parbox[b][0.3cm]{17.7cm}{\addtolength{\parindent}{-0.2in}\href{https://www.nndc.bnl.gov/nsr/nsrlink.jsp?1951Bl75,B}{1951Bl75}: \ensuremath{^{\textnormal{19}}}F(p,n) E=4.2-6.8 MeV; measured the absolute (p,n) cross section using activation technique. Deduced \ensuremath{^{\textnormal{19}}}Ne half-life}\\
\parbox[b][0.3cm]{17.7cm}{as 18.6 s \textit{4} and the positrons end-point energy as 2.3 MeV.}\\
\parbox[b][0.3cm]{17.7cm}{\addtolength{\parindent}{-0.2in}\href{https://www.nndc.bnl.gov/nsr/nsrlink.jsp?1952Sc15,B}{1952Sc15}: \ensuremath{^{\textnormal{19}}}F(p,n)\ensuremath{^{\textnormal{19}}}Ne(\ensuremath{\beta}) E not given; measured \ensuremath{\beta}\ensuremath{^{\textnormal{+}}} decay curve of \ensuremath{^{\textnormal{19}}}Ne\ensuremath{_{\textnormal{g.s.}}} using a 180\ensuremath{^\circ} beta-ray spectrometer and a NaI}\\
\parbox[b][0.3cm]{17.7cm}{detector for detecting (in coincidence) possible \ensuremath{\beta}-delayed \ensuremath{\gamma} rays. The activated sample was transferred to the counting station in}\\
\parbox[b][0.3cm]{17.7cm}{\ensuremath{\sim}30 s. The sample was then counted for 2 half-lives and allowed to decay entirely without counting. Deduced T\ensuremath{_{\textnormal{1/2}}}=18.5 s \textit{5} and a}\\
\parbox[b][0.3cm]{17.7cm}{\ensuremath{\beta}\ensuremath{^{\textnormal{+}}} end-point energy of 2.18 MeV \textit{3} for \ensuremath{^{\textnormal{19}}}Ne\ensuremath{_{\textnormal{g.s.}}}. The authors only observed the annihilation \ensuremath{\gamma} rays and no \ensuremath{\gamma} rays with energies}\\
\parbox[b][0.3cm]{17.7cm}{above the 511-keV were observed.}\\
\parbox[b][0.3cm]{17.7cm}{\addtolength{\parindent}{-0.2in}\href{https://www.nndc.bnl.gov/nsr/nsrlink.jsp?1958We25,B}{1958We25}, \href{https://www.nndc.bnl.gov/nsr/nsrlink.jsp?1960Wa04,B}{1960Wa04}: \ensuremath{^{\textnormal{19}}}F(p,n)\ensuremath{^{\textnormal{19}}}Ne(\ensuremath{\beta}\ensuremath{^{\textnormal{+}}}) E=1-3 MeV; measured \ensuremath{\beta}\ensuremath{^{\textnormal{+}}} particles using a 180\ensuremath{^\circ} deflection single-focusing spectrometer}\\
\parbox[b][0.3cm]{17.7cm}{with two gas proportional counters in the focal plane, which measured the \ensuremath{\beta} rays in coincidence. Beam was on for 3 half-lives and}\\
\parbox[b][0.3cm]{17.7cm}{then turned off. Counting lasted for up to 12 half-lives. Deduced T\ensuremath{_{\textnormal{1/2}}}(\ensuremath{^{\textnormal{19}}}Ne)=19.5 s \textit{10}, an end-point energy of E\ensuremath{_{\ensuremath{\beta}}}=2.24 MeV \textit{1},}\\
\parbox[b][0.3cm]{17.7cm}{and a Coulomb energy difference for \ensuremath{^{\textnormal{19}}}Ne-\ensuremath{^{\textnormal{19}}}F mirrors of 4.04 MeV \textit{1}.}\\
\parbox[b][0.3cm]{17.7cm}{\addtolength{\parindent}{-0.2in}\href{https://www.nndc.bnl.gov/nsr/nsrlink.jsp?1962Ea02,B}{1962Ea02}: \ensuremath{^{\textnormal{19}}}F(p,n) E\ensuremath{\approx}4.8 MeV; measured the \ensuremath{^{\textnormal{19}}}Ne\ensuremath{_{\textnormal{g.s.}}} decay curve; measured the annihilation photons following the \ensuremath{\beta}\ensuremath{^{\textnormal{+}}} decay of}\\
\parbox[b][0.3cm]{17.7cm}{\ensuremath{^{\textnormal{19}}}Ne\ensuremath{_{\textnormal{g.s.}}} to \ensuremath{^{\textnormal{19}}}F\ensuremath{_{\textnormal{g.s.}}} using a NaI(Tl) detector. Deduced the half-life of \ensuremath{^{\textnormal{19}}}Ne as T\ensuremath{_{\textnormal{1/2}}}=17.43 s \textit{6}. This work also cites}\\
\parbox[b][0.3cm]{17.7cm}{T\ensuremath{_{\textnormal{1/2}}}(\ensuremath{^{\textnormal{19}}}Ne\ensuremath{_{\textnormal{g.s.}}})=20.3 s \textit{5} measured by (W. B. Herrmannsfehlt, R. J. Burman, P. Stahelin, J. S. Allen, and T. II. Braid, Bull. Amer.}\\
\parbox[b][0.3cm]{17.7cm}{Phys. Soc., 4 (1959) 77).}\\
\parbox[b][0.3cm]{17.7cm}{\addtolength{\parindent}{-0.2in}\href{https://www.nndc.bnl.gov/nsr/nsrlink.jsp?1974Ma31,B}{1974Ma31}, \href{https://www.nndc.bnl.gov/nsr/nsrlink.jsp?1975MaXA,B}{1975MaXA}: \ensuremath{^{\textnormal{19}}}F(p,n) E=11 MeV; an activated sample was kept inside a beryllium rabbit by a Ta foil and was}\\
\parbox[b][0.3cm]{17.7cm}{transported by an air shuttle to a Ge(Li) detector for counting, the duration of which was 34.8 s. Attempted to measure the very}\\
\parbox[b][0.3cm]{17.7cm}{weak 1357-keV \ensuremath{\beta}-delayed \ensuremath{\gamma} rays emitted from the \ensuremath{^{\textnormal{19}}}Ne\ensuremath{_{\textnormal{g.s.}}}\ensuremath{\rightarrow}\ensuremath{^{\textnormal{19}}}F*(1554)\ensuremath{\rightarrow}\ensuremath{^{\textnormal{19}}}F*(197)+\ensuremath{\gamma} decay. Deduced an upper limit (at 2\ensuremath{\sigma}}\\
\parbox[b][0.3cm]{17.7cm}{C.L.) of 3\ensuremath{\times}10\ensuremath{^{\textnormal{$-$5}}} for the branching ratio of the \ensuremath{^{\textnormal{19}}}Ne\ensuremath{_{\textnormal{g.s.}}}\ensuremath{\rightarrow}\ensuremath{^{\textnormal{19}}}F*(1554) decay. The authors reported that this implies a log \textit{ft}\ensuremath{>}5.6.}\\
\parbox[b][0.3cm]{17.7cm}{\addtolength{\parindent}{-0.2in}\href{https://www.nndc.bnl.gov/nsr/nsrlink.jsp?1974Wi14,B}{1974Wi14}: \ensuremath{^{\textnormal{19}}}F(p,n) E=7 MeV; deduced the half-life of \ensuremath{^{\textnormal{19}}}Ne\ensuremath{_{\textnormal{g.s.}}} by multiscaling \ensuremath{\beta}\ensuremath{^{\textnormal{+}}} particles measured using a plastic scintillator.}\\
\parbox[b][0.3cm]{17.7cm}{Beam was on for 5 seconds. The activated samples were transported to the counting area. The counting started with a 3-seconds}\\
\parbox[b][0.3cm]{17.7cm}{delay. Deduced T\ensuremath{_{\textnormal{1/2}}}=17.36 s \textit{6} and obtained an \textit{ft} value of 1728.4 s \textit{67}. Determined a Gamow-Teller matrix elements of}\\
\parbox[b][0.3cm]{17.7cm}{R\ensuremath{_{\textnormal{e}}}\ensuremath{<}\ensuremath{\sigma}\ensuremath{>}=1.6006 \textit{42}, where R\ensuremath{_{\textnormal{e}}} is the magnitude of the ratio of the axial to vector coupling constant for \ensuremath{^{\textnormal{19}}}Ne.}\\
\parbox[b][0.3cm]{17.7cm}{\addtolength{\parindent}{-0.2in}\href{https://www.nndc.bnl.gov/nsr/nsrlink.jsp?1975Az01,B}{1975Az01}: \ensuremath{^{\textnormal{19}}}F(p,n) E=13 MeV; deduced half-life of \ensuremath{^{\textnormal{19}}}Ne\ensuremath{_{\textnormal{g.s.}}} by preparing activated samples, which were encapsulated in high}\\
\parbox[b][0.3cm]{17.7cm}{purity Be cylinders with very low oxygen content. Transported the samples to the counting station, in 2 s, where a plastic (NE-102)}\\
\parbox[b][0.3cm]{17.7cm}{scintillator coupled to a PMT measured the \ensuremath{\beta}\ensuremath{^{\textnormal{+}}} particles from the decay of \ensuremath{^{\textnormal{19}}}Ne\ensuremath{_{\textnormal{g.s.}}}. Data accumulation occurred with a delay of}\\
\parbox[b][0.3cm]{17.7cm}{20-40 s. Obtained T\ensuremath{_{\textnormal{1/2}}}(\ensuremath{^{\textnormal{19}}}Ne\ensuremath{_{\textnormal{g.s.}}})=17.219 s \textit{17}. Deduced an end-point energy of E\ensuremath{_{\ensuremath{\beta}^{\textnormal{+}}}}=2216.2 keV \textit{4} and a \textit{ft} value of 1714.3 \textit{60}.}\\
\clearpage
\vspace{0.3cm}
{\bf \small \underline{\ensuremath{^{\textnormal{19}}}F(p,n),(p,n\ensuremath{\gamma}),(d,2n\ensuremath{\gamma})\hspace{0.2in}\href{https://www.nndc.bnl.gov/nsr/nsrlink.jsp?1970Gi09,B}{1970Gi09},\href{https://www.nndc.bnl.gov/nsr/nsrlink.jsp?1977Le03,B}{1977Le03} (continued)}}\\
\vspace{0.3cm}
\parbox[b][0.3cm]{17.7cm}{\addtolength{\parindent}{-0.2in}\href{https://www.nndc.bnl.gov/nsr/nsrlink.jsp?1975FrZY,B}{1975FrZY}, \href{https://www.nndc.bnl.gov/nsr/nsrlink.jsp?1975Fr15,B}{1975Fr15}: \ensuremath{^{\textnormal{19}}}F(p,n) E=12 MeV; produced an activated, gaseous sample from SF\ensuremath{_{\textnormal{6}}} target; purified the activated sample}\\
\parbox[b][0.3cm]{17.7cm}{by LN\ensuremath{_{\textnormal{2}}} cooling; the radioactive gas diffused into a mylar cell with thin walls kept in an Al-chamber. This chamber was under}\\
\parbox[b][0.3cm]{17.7cm}{vacuum and in a 3-kG magnetic field to reduce the in-flight \ensuremath{\beta}\ensuremath{^{\textnormal{+}}} particles from the \ensuremath{^{\textnormal{19}}}Ne decay. A shielded, collimated Ge(Li)}\\
\parbox[b][0.3cm]{17.7cm}{detector at \ensuremath{\theta}\ensuremath{_{\textnormal{lab}}}=0\ensuremath{^\circ} measured the 1357-keV \ensuremath{\beta}-delayed \ensuremath{\gamma} rays from the \ensuremath{^{\textnormal{19}}}Ne\ensuremath{_{\textnormal{g.s.}}}\ensuremath{\rightarrow}\ensuremath{^{\textnormal{19}}}F*(1554 keV, 3/2\ensuremath{^{\textnormal{+}}})\ensuremath{\rightarrow}\ensuremath{^{\textnormal{19}}}F*(197 keV, 5/2\ensuremath{^{\textnormal{+}}})}\\
\parbox[b][0.3cm]{17.7cm}{decay relative to the 511-keV annihilation \ensuremath{\gamma} rays from the \ensuremath{^{\textnormal{19}}}Ne\ensuremath{_{\textnormal{g.s.}}}\ensuremath{\rightarrow}\ensuremath{^{\textnormal{19}}}F\ensuremath{_{\textnormal{g.s.}}} decay. Measured I\ensuremath{_{\ensuremath{\gamma}\textnormal{=1357 keV}}}/I\ensuremath{_{\ensuremath{\gamma}\textnormal{=511 keV}}}. Deduced}\\
\parbox[b][0.3cm]{17.7cm}{a branching ratio of 8.2\ensuremath{\times}10\ensuremath{^{\textnormal{$-$6}}} \textit{20} for the \ensuremath{^{\textnormal{19}}}Ne\ensuremath{_{\textnormal{g.s.}}}\ensuremath{\rightarrow}\ensuremath{^{\textnormal{19}}}F*(1554 keV, 3/2\ensuremath{^{\textnormal{+}}}) decay branch. However, due to the \ensuremath{^{\textnormal{19}}}O contamination,}\\
\parbox[b][0.3cm]{17.7cm}{which can also produce the 1357-keV \ensuremath{\gamma} ray from \ensuremath{^{\textnormal{19}}}O\ensuremath{_{\textnormal{g.s.}}}\ensuremath{\rightarrow}\ensuremath{^{\textnormal{19}}}F*(1554)\ensuremath{\rightarrow}\ensuremath{^{\textnormal{19}}}F\ensuremath{_{\textnormal{g.s.}}} decay, they reported the resulting branching ratio}\\
\parbox[b][0.3cm]{17.7cm}{as an upper limit.}\\
\parbox[b][0.3cm]{17.7cm}{\addtolength{\parindent}{-0.2in}\href{https://www.nndc.bnl.gov/nsr/nsrlink.jsp?1976Al07,B}{1976Al07}: \ensuremath{^{\textnormal{19}}}F(p,n) E=7 MeV; activated a sample by proton bombardment (for 5-8 seconds) at a low incident energy to eliminate}\\
\parbox[b][0.3cm]{17.7cm}{the possibility of production of \ensuremath{^{\textnormal{19}}}O; surrounded the transport system with pure graphite to absorb all positrons from the in-flight}\\
\parbox[b][0.3cm]{17.7cm}{decay of \ensuremath{^{\textnormal{19}}}Ne to reduce normalization uncertainties; measured (for 15 seconds) the 1357-keV \ensuremath{\beta}-delayed \ensuremath{\gamma} rays from the}\\
\parbox[b][0.3cm]{17.7cm}{\ensuremath{^{\textnormal{19}}}Ne\ensuremath{_{\textnormal{g.s.}}}(\ensuremath{\beta}\ensuremath{^{\textnormal{+}}})\ensuremath{^{\textnormal{19}}}F*(1554)(\ensuremath{\gamma})\ensuremath{^{\textnormal{19}}}F*(197) decay using a Ge(Li) detector; measured E\ensuremath{_{\ensuremath{\gamma}}} and I\ensuremath{_{\ensuremath{\gamma}}}; also measured the annihilation photons}\\
\parbox[b][0.3cm]{17.7cm}{from the \ensuremath{^{\textnormal{19}}}Ne decay. Deduced \ensuremath{\beta}-branching ratio of 2.1\ensuremath{\times}10\ensuremath{^{\textnormal{$-$5}}} \textit{3} for the above mentioned branch. Determined log \textit{ft}=3.237 \textit{1} and}\\
\parbox[b][0.3cm]{17.7cm}{log \textit{ft}=5.72 \textit{6} for the \ensuremath{^{\textnormal{19}}}Ne\ensuremath{_{\textnormal{g.s.}}}(\ensuremath{\beta}\ensuremath{^{\textnormal{+}}})\ensuremath{^{\textnormal{19}}}F\ensuremath{_{\textnormal{g.s.}}} and \ensuremath{^{\textnormal{19}}}Ne\ensuremath{_{\textnormal{g.s.}}}\ensuremath{\rightarrow}\ensuremath{^{\textnormal{19}}}F*(1554) decay branches, respectively.}\\
\parbox[b][0.3cm]{17.7cm}{\addtolength{\parindent}{-0.2in}\href{https://www.nndc.bnl.gov/nsr/nsrlink.jsp?1981Ad05,B}{1981Ad05}, \href{https://www.nndc.bnl.gov/nsr/nsrlink.jsp?1983Ad03,B}{1983Ad03}: \ensuremath{^{\textnormal{19}}}F(p,n)\ensuremath{^{\textnormal{19}}}Ne(\ensuremath{\beta}\ensuremath{^{\textnormal{+}}}) E=6.4 MeV; produced \ensuremath{^{\textnormal{19}}}Ne by irradiation cycles of \ensuremath{\sim}20 s; transported \ensuremath{^{\textnormal{19}}}Ne through a}\\
\parbox[b][0.3cm]{17.7cm}{series of purifying traps to a counting cell, where the annihilation \ensuremath{\gamma} rays were measured (in cycles of \ensuremath{\sim}20 s) in coincidence with}\\
\parbox[b][0.3cm]{17.7cm}{the 110-keV \ensuremath{\gamma} rays from the \ensuremath{^{\textnormal{19}}}Ne\ensuremath{_{\textnormal{g.s., 1/2}^{\textnormal{+}}}}\ensuremath{\rightarrow}\ensuremath{^{\textnormal{19}}}F*(109.9 keV, 1/2\ensuremath{^{-}}) first forbidden decay. The annihilation \ensuremath{\gamma} rays were measured}\\
\parbox[b][0.3cm]{17.7cm}{using two NaI detectors at each end of the counting cell and the 110-keV \ensuremath{\gamma} rays were detected using a Ge(Li) detector on top of}\\
\parbox[b][0.3cm]{17.7cm}{the cell. Deduced a branching ratio of 1.20\ensuremath{\times}10\ensuremath{^{\textnormal{$-$4}}} \textit{20} for the first forbidden decay. (\href{https://www.nndc.bnl.gov/nsr/nsrlink.jsp?1983Ad03,B}{1983Ad03}) measured the singles 1357-keV \ensuremath{\gamma}}\\
\parbox[b][0.3cm]{17.7cm}{rays from \ensuremath{^{\textnormal{19}}}Ne\ensuremath{\rightarrow}\ensuremath{^{\textnormal{19}}}F*(1554 keV, 3/2\ensuremath{^{\textnormal{+}}}) decay using the Ge(Li) detector and deduced a branching ratio of 2.34\ensuremath{\times}10\ensuremath{^{\textnormal{$-$5}}} \textit{30} for this}\\
\parbox[b][0.3cm]{17.7cm}{latter decay. Determined \textit{ft} values. Discussed parity mixing effects and calculated B(\ensuremath{\lambda}) using shell model for Gamow-Teller}\\
\parbox[b][0.3cm]{17.7cm}{transitions.}\\
\parbox[b][0.3cm]{17.7cm}{\addtolength{\parindent}{-0.2in}\href{https://www.nndc.bnl.gov/nsr/nsrlink.jsp?2023By02,B}{2023By02}: \ensuremath{^{\textnormal{19}}}F(p,n)\ensuremath{^{\textnormal{19}}}Ne(\ensuremath{\beta}\ensuremath{^{\textnormal{+}}}) E=12 MeV; transported \ensuremath{^{\textnormal{19}}}Ne particles into a decay cell (a wave-guide) inside a 7 T superconducting}\\
\parbox[b][0.3cm]{17.7cm}{solenoid magnet. The \ensuremath{\beta}\ensuremath{^{\textnormal{+}}} particles from the \ensuremath{^{\textnormal{19}}}Ne decay were trapped and underwent cyclotron trajectories. Measured for the first}\\
\parbox[b][0.3cm]{17.7cm}{time the cyclotron radiation emission from MeV-scale \ensuremath{\beta}\ensuremath{^{\textnormal{+}}} particles from the decay of \ensuremath{^{\textnormal{19}}}Ne at different magnetic fields. Measured}\\
\parbox[b][0.3cm]{17.7cm}{the total radiated power of \ensuremath{\beta}$'$s vs. magnetic field by measuring the cyclotron frequencies. The results are consistent with}\\
\parbox[b][0.3cm]{17.7cm}{simulations and with the Larmor formula. Discussed applications of cyclotron radiation emission spectroscopy to physics beyond the}\\
\parbox[b][0.3cm]{17.7cm}{standard model.}\\
\vspace{0.385cm}
\parbox[b][0.3cm]{17.7cm}{\addtolength{\parindent}{-0.2in}\textit{\ensuremath{\beta}\ensuremath{^{+}}-\ensuremath{\nu} Angular Correlation Measurements and Fundamental Symmetry Results via the \ensuremath{^{\textnormal{19}}}F(p,n)\ensuremath{^{\textnormal{19}}}Ne(\ensuremath{\beta}\ensuremath{^{\textnormal{+}}})\ensuremath{^{\textnormal{19}}}F Reaction}:}\\
\parbox[b][0.3cm]{17.7cm}{\addtolength{\parindent}{-0.2in}\href{https://www.nndc.bnl.gov/nsr/nsrlink.jsp?1954Al29,B}{1954Al29}: \ensuremath{^{\textnormal{19}}}F(p,n)\ensuremath{^{\textnormal{19}}}Ne(\ensuremath{\beta}\ensuremath{^{\textnormal{+}}}) E not given; measured \ensuremath{\beta}\ensuremath{^{\textnormal{+}}}-\ensuremath{^{\textnormal{19}}}F coincidence events using a terphenyl crystal attached to a PMT to}\\
\parbox[b][0.3cm]{17.7cm}{detect \ensuremath{\beta}\ensuremath{^{\textnormal{+}}} particles at \ensuremath{\theta}\ensuremath{_{\textnormal{lab}}}=0\ensuremath{^\circ}. The \ensuremath{^{\textnormal{19}}}Ne recoils were defused into a silver-magnesium electron multiplier chamber filled with Ne}\\
\parbox[b][0.3cm]{17.7cm}{gas at \ensuremath{\theta}\ensuremath{_{\textnormal{lab}}}=180\ensuremath{^\circ}, where the \ensuremath{^{\textnormal{19}}}F negatively charged ions were detected. Measured \ensuremath{^{\textnormal{19}}}F and \ensuremath{\beta}\ensuremath{^{\textnormal{+}}} energy spectra and deduced the}\\
\parbox[b][0.3cm]{17.7cm}{electron-neutrino angular correlation coefficient of \ensuremath{\alpha}={\textminus}0.8 \textit{4}. This value was later disputed by (\href{https://www.nndc.bnl.gov/nsr/nsrlink.jsp?1957Al29,B}{1957Al29}).}\\
\parbox[b][0.3cm]{17.7cm}{\addtolength{\parindent}{-0.2in}\href{https://www.nndc.bnl.gov/nsr/nsrlink.jsp?1955Mb02,B}{1955Mb02}: \ensuremath{^{\textnormal{19}}}F(p,n)\ensuremath{^{\textnormal{19}}}Ne(\ensuremath{\beta}\ensuremath{^{\textnormal{+}}}) E=6 MeV; measured \ensuremath{\beta}\ensuremath{^{\textnormal{+}}}-\ensuremath{^{\textnormal{19}}}F coincidence events using an electrostatic spectrometer and a}\\
\parbox[b][0.3cm]{17.7cm}{scintillator; measured TOF distributions for \ensuremath{^{\textnormal{19}}}F from which the \ensuremath{^{\textnormal{19}}}Ne decay curve was deduced. Using a fit considering pure scalar,}\\
\parbox[b][0.3cm]{17.7cm}{vector, tensor and axial vector interactions, the authors deduced \ensuremath{\lambda}={\textminus}0.21 \textit{8} (stat.) (see Equation 3); and coupling constant of}\\
\parbox[b][0.3cm]{17.7cm}{C\ensuremath{_{\textnormal{S}}^{\textnormal{2}}}/C\ensuremath{_{\textnormal{T}}^{\textnormal{2}}}=1.1 \textit{3} (stat.). From these results, the authors reported that \ensuremath{^{\textnormal{19}}}Ne\ensuremath{_{\textnormal{g.s.}}} may have a probable nuclear configuration of (\textit{d}\ensuremath{_{\textnormal{5/2}}})\ensuremath{^{\textnormal{3}}}}\\
\parbox[b][0.3cm]{17.7cm}{with T=1/2 and not (\textit{s}\ensuremath{_{\textnormal{1/2}}})\ensuremath{^{\textnormal{3}}} with T=1/2 as suggested by Peaslee through private communication. An electron-neutrino angular}\\
\parbox[b][0.3cm]{17.7cm}{correlation coefficient of \ensuremath{\alpha}={\textminus}0.21 \textit{8} was deduced.}\\
\parbox[b][0.3cm]{17.7cm}{\addtolength{\parindent}{-0.2in}\href{https://www.nndc.bnl.gov/nsr/nsrlink.jsp?1957Al29,B}{1957Al29}: \ensuremath{^{\textnormal{19}}}F(p,n)\ensuremath{^{\textnormal{19}}}Ne(\ensuremath{\beta}\ensuremath{^{\textnormal{+}}}) E not given; measured \ensuremath{\beta}\ensuremath{^{\textnormal{+}}}-\ensuremath{^{\textnormal{19}}}F coincidence events using a terphenyl crystal attached to a PMT to}\\
\parbox[b][0.3cm]{17.7cm}{detect \ensuremath{\beta}\ensuremath{^{\textnormal{+}}} particles in a brass vacuum chamber lined with lucite at \ensuremath{\theta}\ensuremath{_{\textnormal{lab}}}=0\ensuremath{^\circ}. The \ensuremath{^{\textnormal{19}}}Ne recoils were negatively charged and defused}\\
\parbox[b][0.3cm]{17.7cm}{into a silver-magnesium electron multiplier chamber at \ensuremath{\theta}\ensuremath{_{\textnormal{lab}}}=180\ensuremath{^\circ}, with which the \ensuremath{^{\textnormal{19}}}F recoils were counted. Deduced T\ensuremath{_{\textnormal{1/2}}}=18.3 s}\\
\parbox[b][0.3cm]{17.7cm}{\textit{5} for \ensuremath{^{\textnormal{19}}}Ne\ensuremath{_{\textnormal{g.s.}}}, and an end-point energy of 2.23 MeV \textit{5} for the \ensuremath{\beta}\ensuremath{^{\textnormal{+}}} particles. Deduced the electron-neutrino angular correlation}\\
\parbox[b][0.3cm]{17.7cm}{coefficient of \ensuremath{\alpha}={\textminus}0.15 \textit{20} (stat.).}\\
\parbox[b][0.3cm]{17.7cm}{\addtolength{\parindent}{-0.2in}\href{https://www.nndc.bnl.gov/nsr/nsrlink.jsp?1957Go94,B}{1957Go94}, and M. L. Good and E. J. Lauer, University of California, Radiation Laboratory Report-3450 (unpublished, circa 1957):}\\
\parbox[b][0.3cm]{17.7cm}{\ensuremath{^{\textnormal{19}}}F(p,n)\ensuremath{^{\textnormal{19}}}Ne(\ensuremath{\beta}\ensuremath{^{\textnormal{+}}}) E=32 MeV; measured the \ensuremath{\beta}\ensuremath{^{\textnormal{+}}} particle energy using a scintillator; measured the \ensuremath{^{\textnormal{19}}}F recoil ions$'$ kinetic energy}\\
\parbox[b][0.3cm]{17.7cm}{by measuring the coincidence time-of-flight using an electron multiplier tube. Deduced an electron-neutrino angular correlation}\\
\parbox[b][0.3cm]{17.7cm}{coefficient of \ensuremath{\alpha}=+0.14 \textit{13}. The authors acknowledge that their result could be biased toward more positive values of \ensuremath{\alpha} due to not}\\
\parbox[b][0.3cm]{17.7cm}{accounting for the neutral \ensuremath{^{\textnormal{19}}}F ions.}\\
\parbox[b][0.3cm]{17.7cm}{\addtolength{\parindent}{-0.2in}\href{https://www.nndc.bnl.gov/nsr/nsrlink.jsp?1957Pe12,B}{1957Pe12}: \ensuremath{^{\textnormal{19}}}F(p,n)\ensuremath{^{\textnormal{19}}}Ne(\ensuremath{\beta}\ensuremath{\gamma}) E not given; measured \ensuremath{\beta}\ensuremath{^{\textnormal{+}}} particles using a stilbene crystal and \ensuremath{\gamma} rays using two NaI(Tl) crystals.}\\
\parbox[b][0.3cm]{17.7cm}{Measured the decay curve of \ensuremath{^{\textnormal{19}}}Ne from which they deduced T\ensuremath{_{\textnormal{1/2}}}=17.7 s \textit{1} for \ensuremath{^{\textnormal{19}}}Ne\ensuremath{_{\textnormal{g.s.}}}. Note that (\href{https://www.nndc.bnl.gov/nsr/nsrlink.jsp?1959Al10,B}{1959Al10}) mistakenly reported}\\
\parbox[b][0.3cm]{17.7cm}{this half-life as 17.7 s \textit{2} (see Table I in \href{https://www.nndc.bnl.gov/nsr/nsrlink.jsp?1959Al10,B}{1959Al10}).}\\
\clearpage
\vspace{0.3cm}
{\bf \small \underline{\ensuremath{^{\textnormal{19}}}F(p,n),(p,n\ensuremath{\gamma}),(d,2n\ensuremath{\gamma})\hspace{0.2in}\href{https://www.nndc.bnl.gov/nsr/nsrlink.jsp?1970Gi09,B}{1970Gi09},\href{https://www.nndc.bnl.gov/nsr/nsrlink.jsp?1977Le03,B}{1977Le03} (continued)}}\\
\vspace{0.3cm}
\parbox[b][0.3cm]{17.7cm}{\addtolength{\parindent}{-0.2in}\href{https://www.nndc.bnl.gov/nsr/nsrlink.jsp?1959Al10,B}{1959Al10}: \ensuremath{^{\textnormal{19}}}F(p,n)\ensuremath{^{\textnormal{19}}}Ne(\ensuremath{\beta}\ensuremath{^{\textnormal{+}}}); measured the energy spectrum of the \ensuremath{^{\textnormal{19}}}F decay products using an electrostatic spectrometer made of}\\
\parbox[b][0.3cm]{17.7cm}{electron multiplier AgMgNi alloy. Deduced T\ensuremath{_{\textnormal{1/2}}}(\ensuremath{^{\textnormal{19}}}Ne)=17.4 s \textit{2} and an electron-neutrino angular correlation coefficient of \ensuremath{\alpha}=0.00}\\
\parbox[b][0.3cm]{17.7cm}{\textit{8}. See also (\href{https://www.nndc.bnl.gov/nsr/nsrlink.jsp?2015Li47,B}{2015Li47}).}\\
\parbox[b][0.3cm]{17.7cm}{\addtolength{\parindent}{-0.2in}\href{https://www.nndc.bnl.gov/nsr/nsrlink.jsp?1974Ca17,B}{1974Ca17}: \ensuremath{^{\textnormal{19}}}F(p,n)\ensuremath{^{\textnormal{19}}}Ne(\ensuremath{\beta}\ensuremath{^{\textnormal{+}}}) E=15 MeV; polarized \ensuremath{^{\textnormal{19}}}Ne using a Stern-Gerlach magnet; trapped polarized \ensuremath{^{\textnormal{19}}}Ne for 4 seconds;}\\
\parbox[b][0.3cm]{17.7cm}{measured the \ensuremath{\beta}\ensuremath{^{\textnormal{+}}}-\ensuremath{^{\textnormal{19}}}F\ensuremath{^{-}} coincidence events, while \ensuremath{^{\textnormal{19}}}Ne was still polarized, using a Si(Li) detector to detect \ensuremath{\beta}\ensuremath{^{\textnormal{+}}} particles and an}\\
\parbox[b][0.3cm]{17.7cm}{electron multiplier for detecting \ensuremath{^{\textnormal{19}}}F negative ions; deduced time reversal coefficient D=0.002 \textit{4} and the relative phase angle}\\
\parbox[b][0.3cm]{17.7cm}{between axial-vector (A) and vector (V) couplings: \ensuremath{\phi}\ensuremath{_{\textnormal{A,V}}}(\ensuremath{^{\textnormal{19}}}Ne)=180.2\ensuremath{^\circ} \textit{4}.}\\
\parbox[b][0.3cm]{17.7cm}{\addtolength{\parindent}{-0.2in}\href{https://www.nndc.bnl.gov/nsr/nsrlink.jsp?1975Ca28,B}{1975Ca28}, \href{https://www.nndc.bnl.gov/nsr/nsrlink.jsp?1985Ca45,B}{1985Ca45}: \ensuremath{^{\textnormal{19}}}F(p,n)\ensuremath{^{\textnormal{19}}}Ne(\ensuremath{\beta}\ensuremath{^{\textnormal{+}}}) E=12 MeV; transported \ensuremath{^{\textnormal{19}}}Ne into the hot ion source cavity of the atomic-beam apparatus}\\
\parbox[b][0.3cm]{17.7cm}{at Princeton. The thermalized \ensuremath{^{\textnormal{19}}}Ne atoms effused out of the ion source and were polarized by three orbit defining slits and a}\\
\parbox[b][0.3cm]{17.7cm}{Stern-Gerlach magnet. The polarized \ensuremath{^{\textnormal{19}}}Ne particles were trapped in a cell and the \ensuremath{\beta}\ensuremath{^{\textnormal{+}}} particles from their decay were measured}\\
\parbox[b][0.3cm]{17.7cm}{using two plastic scintillators, which measured the up-down \ensuremath{\beta} asymmetry. Deduced the energy dependence of the angular}\\
\parbox[b][0.3cm]{17.7cm}{correlation between the initial nuclear spin of the polarized \ensuremath{^{\textnormal{19}}}Ne atoms and the direction of the emitted positrons from the \ensuremath{^{\textnormal{19}}}Ne}\\
\parbox[b][0.3cm]{17.7cm}{decay. Deduced a \ensuremath{\beta}-asymmetry parameter of A(0)={\textminus}3.91\% \textit{14} for the \ensuremath{^{\textnormal{19}}}Ne decay.}\\
\parbox[b][0.3cm]{17.7cm}{\addtolength{\parindent}{-0.2in}\href{https://www.nndc.bnl.gov/nsr/nsrlink.jsp?1977BaZZ,B}{1977BaZZ}, \href{https://www.nndc.bnl.gov/nsr/nsrlink.jsp?1977Ba08,B}{1977Ba08}: \ensuremath{^{\textnormal{19}}}F(p,n)\ensuremath{^{\textnormal{19}}}Ne(\ensuremath{\beta}\ensuremath{^{\textnormal{+}}}) E=12 MeV; produced, polarized and trapped a \ensuremath{^{\textnormal{19}}}Ne beam; measured \ensuremath{\beta}\ensuremath{^{\textnormal{+}}}-\ensuremath{^{\textnormal{19}}}F\ensuremath{^{-}}}\\
\parbox[b][0.3cm]{17.7cm}{coincidence events from the decay of \ensuremath{^{\textnormal{19}}}Ne using two plastic scintillators facing one another at the ends of the trapping cell to}\\
\parbox[b][0.3cm]{17.7cm}{detect the \ensuremath{\beta} rays and two MCP detectors placed perpendicular to the scintillators to detect the drifting \ensuremath{^{\textnormal{19}}}F\ensuremath{^{-}} ions. Measured the}\\
\parbox[b][0.3cm]{17.7cm}{time-reversal correlation and deduced D\ensuremath{_{\textnormal{exp}}}={\textminus}0.0005 \textit{10} (stat.).}\\
\parbox[b][0.3cm]{17.7cm}{\addtolength{\parindent}{-0.2in}\href{https://www.nndc.bnl.gov/nsr/nsrlink.jsp?1983Sc32,B}{1983Sc32}: \ensuremath{^{\textnormal{19}}}F(p,n)\ensuremath{^{\textnormal{19}}}Ne(\ensuremath{\beta}\ensuremath{^{\textnormal{+}}}) E=12 MeV; polarized \ensuremath{^{\textnormal{19}}}Ne using a Stern-Gerlach magnet; trapped the polarized \ensuremath{^{\textnormal{19}}}Ne for 6.5 s;}\\
\parbox[b][0.3cm]{17.7cm}{studied the \ensuremath{\beta}\ensuremath{^{\textnormal{+}}} asymmetry by measuring the counting-rate asymmetry (as the \ensuremath{^{\textnormal{19}}}Ne spin depolarized) on both sides of the trap using}\\
\parbox[b][0.3cm]{17.7cm}{gas proportional counters backed by scintillators and using by beta polarimetry. From the measure \ensuremath{\beta} polarization and asymmetry,}\\
\parbox[b][0.3cm]{17.7cm}{the authors deduced time reversal odd angular correlation magnitude R={\textminus}0.079 \textit{53}; deduced time reversal invariance, scalar, axial}\\
\parbox[b][0.3cm]{17.7cm}{vector coupling imaginary interference limit as Im(C\ensuremath{_{\textnormal{S}}}C\ensuremath{_{\textnormal{A}}}*)=0.19 \textit{13}.}\\
\parbox[b][0.3cm]{17.7cm}{\addtolength{\parindent}{-0.2in}\href{https://www.nndc.bnl.gov/nsr/nsrlink.jsp?1984Ha01,B}{1984Ha01}: \ensuremath{^{\textnormal{19}}}F(p,n)\ensuremath{^{\textnormal{19}}}Ne(\ensuremath{\beta}) E not given; polarized \ensuremath{^{\textnormal{19}}}Ne beam; trapped in a cell; measured \ensuremath{\beta}-\ensuremath{^{\textnormal{19}}}F\ensuremath{^{-}} coincidence events; deduced}\\
\parbox[b][0.3cm]{17.7cm}{asymmetry parameter D=+0.0004 \textit{8} (the article$'$s abstract reports D=+0.0040 \textit{8}); discussed time reversal symmetry$'$s validity.}\\
\parbox[b][0.3cm]{17.7cm}{\addtolength{\parindent}{-0.2in}\href{https://www.nndc.bnl.gov/nsr/nsrlink.jsp?1993Sa32,B}{1993Sa32}: \ensuremath{^{\textnormal{19}}}F(p,n)\ensuremath{^{\textnormal{19}}}Ne(\ensuremath{\beta}) E=12 MeV; produced and polarized a \ensuremath{^{\textnormal{19}}}Ne atomic beam, which was captured for 6.5 s in a cryogenic}\\
\parbox[b][0.3cm]{17.7cm}{cell with a plastic scintillator at each end and a HPGe detector placed above; measured \ensuremath{\beta}-\ensuremath{\gamma} coincidence events from the \ensuremath{^{\textnormal{19}}}Ne(g.s.,}\\
\parbox[b][0.3cm]{17.7cm}{1/2\ensuremath{^{\textnormal{+}}})\ensuremath{\rightarrow}\ensuremath{^{\textnormal{19}}}F*(109.9 keV, 1/2\ensuremath{^{-}})\ensuremath{\rightarrow}\ensuremath{^{\textnormal{19}}}F\ensuremath{_{\textnormal{g.s.}}} first forbidden decay. Deduced \ensuremath{\beta} asymmetry parameter (A\ensuremath{_{\ensuremath{\beta}}}=17\% \textit{12}) and the branching}\\
\parbox[b][0.3cm]{17.7cm}{ratio (BR=1.13\ensuremath{\times}10\ensuremath{^{\textnormal{$-$4}}} \textit{9}) for this decay, as well as the \ensuremath{\beta} asymmetry parameter for the \ensuremath{^{\textnormal{19}}}Ne\ensuremath{_{\textnormal{g.s.}}}\ensuremath{\rightarrow}\ensuremath{^{\textnormal{19}}}F\ensuremath{_{\textnormal{g.s.}}} decay (not reported).}\\
\parbox[b][0.3cm]{17.7cm}{Determined vector form factor coefficients for the first forbidden \ensuremath{\beta} decay.}\\
\parbox[b][0.3cm]{17.7cm}{\addtolength{\parindent}{-0.2in}D. Combs, G. Jones, W. Anderson, F. Calaprice, L. Hayen, and A. Young, \textit{A look into mirrors: A measurement of the}}\\
\parbox[b][0.3cm]{17.7cm}{\textit{\ensuremath{\beta}-asymmetry in \ensuremath{^{19}}Ne decay and searches for new physics}, arXiv:2009.13700v2 [nucl-ex] 20 Nov 2020 (unpublished):}\\
\parbox[b][0.3cm]{17.7cm}{\ensuremath{^{\textnormal{19}}}F(p,n)\ensuremath{^{\textnormal{19}}}Ne(\ensuremath{\beta}) E=12 MeV; reanalyzed the data of an earlier experiment by G. Jones (Ph.D. Thesis, Princeton University, (1996),}\\
\parbox[b][0.3cm]{17.7cm}{unpublished), where \ensuremath{^{\textnormal{19}}}Ne was produced, polarized and trapped in a decay cell for 3.5 seconds; measured the positrons from \ensuremath{^{\textnormal{19}}}Ne}\\
\parbox[b][0.3cm]{17.7cm}{decay using two back to back Si(Li) quadrant detectors at the ends of the cell. The 2020-authors carried out a Monte Carlo}\\
\parbox[b][0.3cm]{17.7cm}{simulation to correctly reconstruct the events that back scattered from one detector into the other. Deduced the zero-intercept of the}\\
\parbox[b][0.3cm]{17.7cm}{\ensuremath{\beta}-asymmetry as A\ensuremath{_{\textnormal{0}}}={\textminus}0.03871 \textit{26} (stat.) \textit{+65{\textminus}87} (sys.) and obtained the Fermi-to-Gamow-Teller mixing ratio of \ensuremath{\rho}=1.6014 \textit{8} (stat.)}\\
\parbox[b][0.3cm]{17.7cm}{\textit{+21{\textminus}28} (sys.). The authors determined \ensuremath{\vert}V\ensuremath{_{\textnormal{ud}}}\ensuremath{\vert}=0.9739 \textit{13} for \ensuremath{^{\textnormal{19}}}Ne. Note that these results are not peer-reviewed and are}\\
\parbox[b][0.3cm]{17.7cm}{unpublished.}\\
\vspace{0.385cm}
\parbox[b][0.3cm]{17.7cm}{\addtolength{\parindent}{-0.2in}\textit{Measurements of \ensuremath{^{19}}Ne Nuclear Magnetic Moment via the \ensuremath{^{\textnormal{19}}}F(p,n)\ensuremath{^{\textnormal{19}}}Ne(\ensuremath{\beta}\ensuremath{^{\textnormal{+}}})\ensuremath{^{\textnormal{19}}}F Reaction}:}\\
\parbox[b][0.3cm]{17.7cm}{\addtolength{\parindent}{-0.2in}\href{https://www.nndc.bnl.gov/nsr/nsrlink.jsp?1963Co22,B}{1963Co22}, \href{https://www.nndc.bnl.gov/nsr/nsrlink.jsp?1963Do15,B}{1963Do15}: \ensuremath{^{\textnormal{19}}}F(p,n)\ensuremath{^{\textnormal{19}}}Ne(\ensuremath{\beta}\ensuremath{^{\textnormal{+}}}) E=13 MeV; polarized atomic \ensuremath{^{\textnormal{19}}}Ne beam using a Stern-Gerlach magnet; Nuclear}\\
\parbox[b][0.3cm]{17.7cm}{Magnetic Resonance (NMR) transitions were induced on the atoms in flight; trapped these atoms for 12 seconds at a time in a cell}\\
\parbox[b][0.3cm]{17.7cm}{with a Geiger counter at each end; measured \ensuremath{\beta} asymmetry parameter (A\ensuremath{_{\ensuremath{\beta}}}) for \ensuremath{^{\textnormal{19}}}Ne by observing the reversal of beta-decay}\\
\parbox[b][0.3cm]{17.7cm}{asymmetry, which occurred when polarized \ensuremath{^{\textnormal{19}}}Ne nuclei underwent a NMR reorientation. Deduced A\ensuremath{_{\ensuremath{\beta}}}(\ensuremath{^{\textnormal{19}}}Ne)={\textminus}0.057 \textit{5}. Measured}\\
\parbox[b][0.3cm]{17.7cm}{the \ensuremath{^{\textnormal{19}}}Ne nuclear magnetic moment as \ensuremath{\mu}={\textminus}1.886 \ensuremath{\mu}\ensuremath{_{\textnormal{N}}} \textit{1}.}\\
\parbox[b][0.3cm]{17.7cm}{\addtolength{\parindent}{-0.2in}\href{https://www.nndc.bnl.gov/nsr/nsrlink.jsp?1982Ma39,B}{1982Ma39}: \ensuremath{^{\textnormal{19}}}F(p,n)\ensuremath{^{\textnormal{19}}}Ne(\ensuremath{\beta}\ensuremath{^{\textnormal{+}}}) E=12 MeV; produced, polarized, and trapped a \ensuremath{^{\textnormal{19}}}Ne atomic beam; induced NMR transitions while}\\
\parbox[b][0.3cm]{17.7cm}{the atoms were trapped in a cryogenic cell for 2.5 s inside a magnet; two plastic scintillators at the end of the magnet measured \ensuremath{\beta}\ensuremath{^{\textnormal{+}}}}\\
\parbox[b][0.3cm]{17.7cm}{particles from the \ensuremath{^{\textnormal{19}}}Ne\ensuremath{\rightarrow}\ensuremath{^{\textnormal{19}}}F ground-state decay. Deduced \ensuremath{\mu}={\textminus}1.88542 \ensuremath{\mu}\ensuremath{_{\textnormal{N}}} \textit{8} for the \ensuremath{^{\textnormal{19}}}Ne nuclear magnetic moment.}\\
\vspace{0.385cm}
\parbox[b][0.3cm]{17.7cm}{\addtolength{\parindent}{-0.2in}\textit{Theory}:}\\
\parbox[b][0.3cm]{17.7cm}{\addtolength{\parindent}{-0.2in}\href{https://www.nndc.bnl.gov/nsr/nsrlink.jsp?1966Ma60,B}{1966Ma60}: Revised the E\ensuremath{_{\textnormal{thresh}}}(\ensuremath{^{\textnormal{19}}}Ne\ensuremath{_{\textnormal{g.s.}}})=4233.2 keV \textit{20} deduced by (\href{https://www.nndc.bnl.gov/nsr/nsrlink.jsp?1961Be13,B}{1961Be13}) and E\ensuremath{_{\textnormal{thresh}}}(\ensuremath{^{\textnormal{19}}}Ne\ensuremath{_{\textnormal{g.s.}}})=4234.7 keV \textit{10} deduced}\\
\parbox[b][0.3cm]{17.7cm}{by (\href{https://www.nndc.bnl.gov/nsr/nsrlink.jsp?1961Ry04,B}{1961Ry04}) and recommended E\ensuremath{_{\textnormal{thresh}}}(\ensuremath{^{\textnormal{19}}}Ne\ensuremath{_{\textnormal{g.s.}}})=4234.3 keV \textit{8}. This result is also published in (\href{https://www.nndc.bnl.gov/nsr/nsrlink.jsp?1966Ma75,B}{1966Ma75}).}\\
\parbox[b][0.3cm]{17.7cm}{\addtolength{\parindent}{-0.2in}\href{https://www.nndc.bnl.gov/nsr/nsrlink.jsp?1974Do18,B}{1974Do18}: Suggested to study neutral weak currents through \ensuremath{^{\textnormal{19}}}F(\ensuremath{\nu}\ensuremath{_{\textnormal{e}}},\ensuremath{\nu}$'$\ensuremath{_{\textnormal{e}}})\ensuremath{^{\textnormal{19}}}F*(1554)(\ensuremath{\gamma})\ensuremath{^{\textnormal{19}}}F*(197)(\ensuremath{\gamma})\ensuremath{^{\textnormal{19}}}F\ensuremath{_{\textnormal{g.s.}}}. These authors}\\
\clearpage
\vspace{0.3cm}
{\bf \small \underline{\ensuremath{^{\textnormal{19}}}F(p,n),(p,n\ensuremath{\gamma}),(d,2n\ensuremath{\gamma})\hspace{0.2in}\href{https://www.nndc.bnl.gov/nsr/nsrlink.jsp?1970Gi09,B}{1970Gi09},\href{https://www.nndc.bnl.gov/nsr/nsrlink.jsp?1977Le03,B}{1977Le03} (continued)}}\\
\vspace{0.3cm}
\parbox[b][0.3cm]{17.7cm}{predicted a production cross section for reactor-based \ensuremath{\nu}\ensuremath{_{\textnormal{e}}} (electron neutrino) of 6.3\ensuremath{\times}10\ensuremath{^{\textnormal{$-$44}}}B cm\ensuremath{^{\textnormal{2}}}/nucleon, where B is the branching}\\
\parbox[b][0.3cm]{17.7cm}{for the \ensuremath{^{\textnormal{19}}}Ne\ensuremath{_{\textnormal{g.s.}}} decay to the \ensuremath{^{\textnormal{19}}}F*(1554) level. The authors used the single-particle Nilsson model with a K=1/2\ensuremath{^{\textnormal{+}}} band and four}\\
\parbox[b][0.3cm]{17.7cm}{adjustable parameters and calculated B to be 1.0\ensuremath{\times}10\ensuremath{^{\textnormal{$-$4}}} \textit{9}, implying a log \textit{ft}=5.0 \textit{+10{\textminus}3} for the \ensuremath{^{\textnormal{19}}}Ne\ensuremath{\rightarrow}\ensuremath{^{\textnormal{19}}}F*(1554) decay branch.}\\
\parbox[b][0.3cm]{17.7cm}{\addtolength{\parindent}{-0.2in}\href{https://www.nndc.bnl.gov/nsr/nsrlink.jsp?1976Fr13,B}{1976Fr13}: Discussed the results of (\href{https://www.nndc.bnl.gov/nsr/nsrlink.jsp?1969Ov01,B}{1969Ov01}) for calibrations of analyzing magnets in use with tandem accelerators.}\\
\parbox[b][0.3cm]{17.7cm}{\addtolength{\parindent}{-0.2in}\href{https://www.nndc.bnl.gov/nsr/nsrlink.jsp?1993Go15,B}{1993Go15}: \ensuremath{^{\textnormal{19}}}F(p,n) E\ensuremath{\approx}4.5-10.5 MeV; analyzed \ensuremath{\sigma}(E); deduced possibility of generating radioactive beams by cyclotrons and}\\
\parbox[b][0.3cm]{17.7cm}{magnetic separators.}\\
\parbox[b][0.3cm]{17.7cm}{\addtolength{\parindent}{-0.2in}\href{https://www.nndc.bnl.gov/nsr/nsrlink.jsp?1994Ga49,B}{1994Ga49}: \ensuremath{^{\textnormal{19}}}F(p,n) E=1 GeV; analyzed \ensuremath{\sigma}(\ensuremath{\theta}), mass dependences; deduced resonance phenomena related features.}\\
\parbox[b][0.3cm]{17.7cm}{\addtolength{\parindent}{-0.2in}\href{https://www.nndc.bnl.gov/nsr/nsrlink.jsp?1999An35,B}{1999An35}: Deduced the astrophysical S-factor for the \ensuremath{^{\textnormal{19}}}F(p,n) reaction at E\ensuremath{_{\textnormal{c.m.}}}\ensuremath{\approx}4-11 MeV based on the results of (\href{https://www.nndc.bnl.gov/nsr/nsrlink.jsp?1959Gi47,B}{1959Gi47},}\\
\parbox[b][0.3cm]{17.7cm}{\href{https://www.nndc.bnl.gov/nsr/nsrlink.jsp?1963Je04,B}{1963Je04}, \href{https://www.nndc.bnl.gov/nsr/nsrlink.jsp?1968Ri08,B}{1968Ri08}, \href{https://www.nndc.bnl.gov/nsr/nsrlink.jsp?1990Wa10,B}{1990Wa10}); deduced the \ensuremath{^{\textnormal{19}}}F(p,n) reaction rate for T=0.7-2.5 GK; provided an analytical approximation for}\\
\parbox[b][0.3cm]{17.7cm}{the reaction rate vs. temperature.}\\
\vspace{12pt}
\underline{$^{19}$Ne Levels}\\
\vspace{0.34cm}
\parbox[b][0.3cm]{17.7cm}{\addtolength{\parindent}{-0.254cm}\textit{Notes}:}\\
\parbox[b][0.3cm]{17.7cm}{\addtolength{\parindent}{-0.254cm}(1) The excitation energies reported from (\href{https://www.nndc.bnl.gov/nsr/nsrlink.jsp?1970Gi09,B}{1970Gi09}, \href{https://www.nndc.bnl.gov/nsr/nsrlink.jsp?1977Le03,B}{1977Le03}) were deduced by those authors from the recoil corrected \ensuremath{\gamma}-ray}\\
\parbox[b][0.3cm]{17.7cm}{energies measured by those authors.}\\
\parbox[b][0.3cm]{17.7cm}{\addtolength{\parindent}{-0.254cm}(2) The magnetic moments reported by (\href{https://www.nndc.bnl.gov/nsr/nsrlink.jsp?1969Bl02,B}{1969Bl02}) are from (\href{https://www.nndc.bnl.gov/nsr/nsrlink.jsp?1964Dr06,B}{1964Dr06}) and are cited by (\href{https://www.nndc.bnl.gov/nsr/nsrlink.jsp?1969Bl02,B}{1969Bl02}: See Table 1).}\\
\parbox[b][0.3cm]{17.7cm}{\addtolength{\parindent}{-0.254cm}(3) When an excitation energy is determined from a least-squares fit to \ensuremath{\gamma}-ray energies, evaluator assumed \ensuremath{\Delta}E\ensuremath{_{\ensuremath{\gamma}}}=1 where no}\\
\parbox[b][0.3cm]{17.7cm}{uncertainty in E\ensuremath{_{\ensuremath{\gamma}}} is given.}\\
\vspace{0.34cm}
\begin{longtable}{ccccccc@{\extracolsep{\fill}}c}
\multicolumn{2}{c}{E(level)$^{}$}&J$^{\pi}$$^{}$&\multicolumn{2}{c}{T$_{1/2}$$^{}$}&\ensuremath{\Delta}L$^{{\hyperlink{NE31LEVEL0}{a}}}$&Comments&\\[-.2cm]
\multicolumn{2}{c}{\hrulefill}&\hrulefill&\multicolumn{2}{c}{\hrulefill}&\hrulefill&\hrulefill&
\endfirsthead
\multicolumn{1}{r@{}}{0}&\multicolumn{1}{@{}l}{}&\multicolumn{1}{l}{1/2\ensuremath{^{+}}}&\multicolumn{1}{r@{}}{17}&\multicolumn{1}{@{.}l}{35 s {\it 4}}&\multicolumn{1}{l}{0}&\parbox[t][0.3cm]{12.365701cm}{\raggedright \ensuremath{\mu}={\textminus}1.88542 \textit{8} (\href{https://www.nndc.bnl.gov/nsr/nsrlink.jsp?1982Ma39,B}{1982Ma39})\vspace{0.1cm}}&\\
&&&&&&\parbox[t][0.3cm]{12.365701cm}{\raggedright T=1/2 (\href{https://www.nndc.bnl.gov/nsr/nsrlink.jsp?1984Ra22,B}{1984Ra22})\vspace{0.1cm}}&\\
&&&&&&\parbox[t][0.3cm]{12.365701cm}{\raggedright g=0.742 (\href{https://www.nndc.bnl.gov/nsr/nsrlink.jsp?1969Bl02,B}{1969Bl02})\vspace{0.1cm}}&\\
&&&&&&\parbox[t][0.3cm]{12.365701cm}{\raggedright E(level): From (\href{https://www.nndc.bnl.gov/nsr/nsrlink.jsp?1955Ma84,B}{1955Ma84}, \href{https://www.nndc.bnl.gov/nsr/nsrlink.jsp?1957Ba09,B}{1957Ba09}, \href{https://www.nndc.bnl.gov/nsr/nsrlink.jsp?1963Gi09,B}{1963Gi09}, \href{https://www.nndc.bnl.gov/nsr/nsrlink.jsp?1965We05,B}{1965We05}, \href{https://www.nndc.bnl.gov/nsr/nsrlink.jsp?1969Bl02,B}{1969Bl02}, \href{https://www.nndc.bnl.gov/nsr/nsrlink.jsp?1970Gi09,B}{1970Gi09},\vspace{0.1cm}}&\\
&&&&&&\parbox[t][0.3cm]{12.365701cm}{\raggedright {\ }{\ }{\ }\href{https://www.nndc.bnl.gov/nsr/nsrlink.jsp?1971It02,B}{1971It02}, \href{https://www.nndc.bnl.gov/nsr/nsrlink.jsp?1974Ma31,B}{1974Ma31}, \href{https://www.nndc.bnl.gov/nsr/nsrlink.jsp?1974Wi14,B}{1974Wi14}, \href{https://www.nndc.bnl.gov/nsr/nsrlink.jsp?1975Az01,B}{1975Az01}, \href{https://www.nndc.bnl.gov/nsr/nsrlink.jsp?1975Fr15,B}{1975Fr15}, \href{https://www.nndc.bnl.gov/nsr/nsrlink.jsp?1976Al07,B}{1976Al07}, \href{https://www.nndc.bnl.gov/nsr/nsrlink.jsp?1977Le03,B}{1977Le03}, \href{https://www.nndc.bnl.gov/nsr/nsrlink.jsp?1981Ad05,B}{1981Ad05},\vspace{0.1cm}}&\\
&&&&&&\parbox[t][0.3cm]{12.365701cm}{\raggedright {\ }{\ }{\ }\href{https://www.nndc.bnl.gov/nsr/nsrlink.jsp?1983Ad03,B}{1983Ad03}, \href{https://www.nndc.bnl.gov/nsr/nsrlink.jsp?1984Pi07,B}{1984Pi07}, \href{https://www.nndc.bnl.gov/nsr/nsrlink.jsp?1984Ra22,B}{1984Ra22}, \href{https://www.nndc.bnl.gov/nsr/nsrlink.jsp?1985Ba66,B}{1985Ba66}, and \href{https://www.nndc.bnl.gov/nsr/nsrlink.jsp?1987Ra23,B}{1987Ra23}).\vspace{0.1cm}}&\\
&&&&&&\parbox[t][0.3cm]{12.365701cm}{\raggedright T\ensuremath{_{1/2}}: Unweighted average of (1) 17.4 s \textit{2} (\href{https://www.nndc.bnl.gov/nsr/nsrlink.jsp?1959Al10,B}{1959Al10}); (2) 17.43 s \textit{6} (\href{https://www.nndc.bnl.gov/nsr/nsrlink.jsp?1962Ea02,B}{1962Ea02}), which\vspace{0.1cm}}&\\
&&&&&&\parbox[t][0.3cm]{12.365701cm}{\raggedright {\ }{\ }{\ }also mentioned T\ensuremath{_{\textnormal{1/2}}}=20.3 s \textit{5} from (W. B. Herrmannsfehlt, R. J. Burman, P. Stahelin, J.\vspace{0.1cm}}&\\
&&&&&&\parbox[t][0.3cm]{12.365701cm}{\raggedright {\ }{\ }{\ }S. Allen, and T. II. Braid, Bull. Amer. Phys. Soc., 4 (1959) 77); (3) 17.36 s \textit{6}\vspace{0.1cm}}&\\
&&&&&&\parbox[t][0.3cm]{12.365701cm}{\raggedright {\ }{\ }{\ }(\href{https://www.nndc.bnl.gov/nsr/nsrlink.jsp?1968Go10,B}{1968Go10}), which deduced a weighted average of T\ensuremath{_{\textnormal{1/2}}}=17.90 s \textit{16} from T\ensuremath{_{\textnormal{1/2}}}=17.4 s \textit{2}\vspace{0.1cm}}&\\
&&&&&&\parbox[t][0.3cm]{12.365701cm}{\raggedright {\ }{\ }{\ }(K. Way \textit{et al}. (ed.), Nuclear Data Sheets, NRC 60-1-41); T\ensuremath{_{\textnormal{1/2}}}=20.3 s \textit{5} (\href{https://www.nndc.bnl.gov/nsr/nsrlink.jsp?1962Ea02,B}{1962Ea02});\vspace{0.1cm}}&\\
&&&&&&\parbox[t][0.3cm]{12.365701cm}{\raggedright {\ }{\ }{\ }and their result; (4) 17.36 s \textit{6} (\href{https://www.nndc.bnl.gov/nsr/nsrlink.jsp?1974Wi14,B}{1974Wi14}); and (5) 17.219 s \textit{17} (\href{https://www.nndc.bnl.gov/nsr/nsrlink.jsp?1975Az01,B}{1975Az01}).\vspace{0.1cm}}&\\
&&&&&&\parbox[t][0.3cm]{12.365701cm}{\raggedright T\ensuremath{_{1/2}}: See also excluded half-life measurements of: 20 s (\href{https://www.nndc.bnl.gov/nsr/nsrlink.jsp?1939Fo01,B}{1939Fo01}); 20.3 s \textit{5}\vspace{0.1cm}}&\\
&&&&&&\parbox[t][0.3cm]{12.365701cm}{\raggedright {\ }{\ }{\ }(\href{https://www.nndc.bnl.gov/nsr/nsrlink.jsp?1939Wh02,B}{1939Wh02}); 18.2 s \textit{6} (\href{https://www.nndc.bnl.gov/nsr/nsrlink.jsp?1949Sh25,B}{1949Sh25}); 18.6 s \textit{4} (\href{https://www.nndc.bnl.gov/nsr/nsrlink.jsp?1951Bl75,B}{1951Bl75}); 18.5 s \textit{5} (\href{https://www.nndc.bnl.gov/nsr/nsrlink.jsp?1952Sc15,B}{1952Sc15}); 19 s \textit{1}\vspace{0.1cm}}&\\
&&&&&&\parbox[t][0.3cm]{12.365701cm}{\raggedright {\ }{\ }{\ }(\href{https://www.nndc.bnl.gov/nsr/nsrlink.jsp?1954Na29,B}{1954Na29}); 18.3 s \textit{5} (\href{https://www.nndc.bnl.gov/nsr/nsrlink.jsp?1957Al29,B}{1957Al29}); 17.7 s \textit{1} (\href{https://www.nndc.bnl.gov/nsr/nsrlink.jsp?1957Pe12,B}{1957Pe12}); and 19.5 s \textit{10} (\href{https://www.nndc.bnl.gov/nsr/nsrlink.jsp?1958We25,B}{1958We25},\vspace{0.1cm}}&\\
&&&&&&\parbox[t][0.3cm]{12.365701cm}{\raggedright {\ }{\ }{\ }\href{https://www.nndc.bnl.gov/nsr/nsrlink.jsp?1960Wa04,B}{1960Wa04}).\vspace{0.1cm}}&\\
&&&&&&\parbox[t][0.3cm]{12.365701cm}{\raggedright J\ensuremath{^{\pi}}: From the \ensuremath{^{\textnormal{19}}}Ne Adopted Levels.\vspace{0.1cm}}&\\
&&&&&&\parbox[t][0.3cm]{12.365701cm}{\raggedright J\ensuremath{^{\pi}}: See also (\href{https://www.nndc.bnl.gov/nsr/nsrlink.jsp?1957Ba09,B}{1957Ba09}): J\ensuremath{^{\ensuremath{\pi}}}=1/2\ensuremath{^{\textnormal{+}}}; (\href{https://www.nndc.bnl.gov/nsr/nsrlink.jsp?1962Fr09,B}{1962Fr09}), who reported that \ensuremath{^{\textnormal{19}}}Ne\ensuremath{_{\textnormal{g.s.}}} may be of\vspace{0.1cm}}&\\
&&&&&&\parbox[t][0.3cm]{12.365701cm}{\raggedright {\ }{\ }{\ }positive parity nature using mirror level analysis and by comparison of level spacing in\vspace{0.1cm}}&\\
&&&&&&\parbox[t][0.3cm]{12.365701cm}{\raggedright {\ }{\ }{\ }\ensuremath{^{\textnormal{19}}}F and \ensuremath{^{\textnormal{19}}}Ne; (\href{https://www.nndc.bnl.gov/nsr/nsrlink.jsp?1963Gi09,B}{1963Gi09}): J\ensuremath{^{\ensuremath{\pi}}}=1/2\ensuremath{^{\textnormal{+}}}; (\href{https://www.nndc.bnl.gov/nsr/nsrlink.jsp?1970Gi09,B}{1970Gi09}): J\ensuremath{^{\ensuremath{\pi}}}=1/2\ensuremath{^{\textnormal{+}}}; and (\href{https://www.nndc.bnl.gov/nsr/nsrlink.jsp?1977Le03,B}{1977Le03}): J\ensuremath{^{\ensuremath{\pi}}}=1/2\ensuremath{^{\textnormal{+}}}.\vspace{0.1cm}}&\\
&&&&&&\parbox[t][0.3cm]{12.365701cm}{\raggedright {\ }{\ }{\ }These J\ensuremath{^{\ensuremath{\pi}}} assignments were taken from literature.\vspace{0.1cm}}&\\
&&&&&&\parbox[t][0.3cm]{12.365701cm}{\raggedright \ensuremath{\mu}: From \ensuremath{\beta}-NMR using polarized \ensuremath{^{\textnormal{19}}}Ne (\href{https://www.nndc.bnl.gov/nsr/nsrlink.jsp?1982Ma39,B}{1982Ma39}).\vspace{0.1cm}}&\\
&&&&&&\parbox[t][0.3cm]{12.365701cm}{\raggedright \ensuremath{\mu}: See also {\textminus}1.886 \ensuremath{\mu}\ensuremath{_{\textnormal{N}}} \textit{1} (\href{https://www.nndc.bnl.gov/nsr/nsrlink.jsp?1963Co22,B}{1963Co22}) using \ensuremath{\beta}-NMR; and {\textminus}1.886 \ensuremath{\mu}\ensuremath{_{\textnormal{N}}} (\href{https://www.nndc.bnl.gov/nsr/nsrlink.jsp?1969Bl02,B}{1969Bl02}) deduced\vspace{0.1cm}}&\\
&&&&&&\parbox[t][0.3cm]{12.365701cm}{\raggedright {\ }{\ }{\ }using Time Dependent Perturbed Angular Distribution method, which was used to obtain\vspace{0.1cm}}&\\
&&&&&&\parbox[t][0.3cm]{12.365701cm}{\raggedright {\ }{\ }{\ }the isoscalar g-factor (see below).\vspace{0.1cm}}&\\
&&&&&&\parbox[t][0.3cm]{12.365701cm}{\raggedright Isoscalar g-factor=0.742 (\href{https://www.nndc.bnl.gov/nsr/nsrlink.jsp?1969Bl02,B}{1969Bl02}: See Table 3) deduced using pulsed beam differential\vspace{0.1cm}}&\\
&&&&&&\parbox[t][0.3cm]{12.365701cm}{\raggedright {\ }{\ }{\ }delay constant angle method.\vspace{0.1cm}}&\\
&&&&&&\parbox[t][0.3cm]{12.365701cm}{\raggedright Coulomb energy difference for \ensuremath{^{\textnormal{19}}}Ne-\ensuremath{^{\textnormal{19}}}F mirrors: 4.04 MeV \textit{1} (\href{https://www.nndc.bnl.gov/nsr/nsrlink.jsp?1960Wa04,B}{1960Wa04}).\vspace{0.1cm}}&\\
&&&&&&\parbox[t][0.3cm]{12.365701cm}{\raggedright B(GT)\ensuremath{_{\textnormal{g.s.}}}=1.64 \textit{2} (\href{https://www.nndc.bnl.gov/nsr/nsrlink.jsp?1984Ra22,B}{1984Ra22}). See also B(GT)\ensuremath{_{\textnormal{g.s.}}}=1.635 \textit{16} (\href{https://www.nndc.bnl.gov/nsr/nsrlink.jsp?1984Ra22,B}{1984Ra22}) deduced from\vspace{0.1cm}}&\\
&&&&&&\parbox[t][0.3cm]{12.365701cm}{\raggedright {\ }{\ }{\ }(g\ensuremath{_{\textnormal{A}}}/g\ensuremath{_{\textnormal{v}}})*B(GT)\ensuremath{_{\textnormal{g.s.}}}=2.544 \textit{16}. B(GT)\ensuremath{_{\textnormal{g.s.}}} is the Gamow-Teller strength for the ground\vspace{0.1cm}}&\\
&&&&&&\parbox[t][0.3cm]{12.365701cm}{\raggedright {\ }{\ }{\ }state decay transition. (\href{https://www.nndc.bnl.gov/nsr/nsrlink.jsp?1984Ra22,B}{1984Ra22}) reported that the excited states in \ensuremath{^{\textnormal{19}}}Ne contribute\vspace{0.1cm}}&\\
&&&&&&\parbox[t][0.3cm]{12.365701cm}{\raggedright {\ }{\ }{\ }33\% \textit{6} to the total sum strength of B(GT)= 1.97 \textit{6}. They deduced a quenching factor of\vspace{0.1cm}}&\\
&&&&&&\parbox[t][0.3cm]{12.365701cm}{\raggedright {\ }{\ }{\ }0.66 for the observed B(GT)\ensuremath{_{\textnormal{sum}}}.\vspace{0.1cm}}&\\
\end{longtable}
\begin{textblock}{29}(0,27.3)
Continued on next page (footnotes at end of table)
\end{textblock}
\clearpage
\begin{longtable}{ccccccc@{\extracolsep{\fill}}c}
\\[-.4cm]
\multicolumn{8}{c}{{\bf \small \underline{\ensuremath{^{\textnormal{19}}}F(p,n),(p,n\ensuremath{\gamma}),(d,2n\ensuremath{\gamma})\hspace{0.2in}\href{https://www.nndc.bnl.gov/nsr/nsrlink.jsp?1970Gi09,B}{1970Gi09},\href{https://www.nndc.bnl.gov/nsr/nsrlink.jsp?1977Le03,B}{1977Le03} (continued)}}}\\
\multicolumn{8}{c}{~}\\
\multicolumn{8}{c}{\underline{\ensuremath{^{19}}Ne Levels (continued)}}\\
\multicolumn{8}{c}{~}\\
\multicolumn{2}{c}{E(level)$^{}$}&J$^{\pi}$$^{}$&\multicolumn{2}{c}{T$_{1/2}$$^{}$}&\ensuremath{\Delta}L$^{{\hyperlink{NE31LEVEL0}{a}}}$&Comments&\\[-.2cm]
\multicolumn{2}{c}{\hrulefill}&\hrulefill&\multicolumn{2}{c}{\hrulefill}&\hrulefill&\hrulefill&
\endhead
&&&&&&\parbox[t][0.3cm]{12.29294cm}{\raggedright \ensuremath{\beta}\ensuremath{^{\textnormal{+}}}-\ensuremath{\nu} angular correlation coefficient: \ensuremath{\alpha}={\textminus}0.07 \textit{8}: Weighted average (with external errors)\vspace{0.1cm}}&\\
&&&&&&\parbox[t][0.3cm]{12.29294cm}{\raggedright {\ }{\ }{\ }of {\textminus}0.21 \textit{8} (\href{https://www.nndc.bnl.gov/nsr/nsrlink.jsp?1955Mb02,B}{1955Mb02}); {\textminus}0.15 \textit{20} (stat.) (\href{https://www.nndc.bnl.gov/nsr/nsrlink.jsp?1957Al29,B}{1957Al29}); +0.14 \textit{13} (\href{https://www.nndc.bnl.gov/nsr/nsrlink.jsp?1957Go94,B}{1957Go94}); and 0.00 \textit{8}\vspace{0.1cm}}&\\
&&&&&&\parbox[t][0.3cm]{12.29294cm}{\raggedright {\ }{\ }{\ }(\href{https://www.nndc.bnl.gov/nsr/nsrlink.jsp?1959Al10,B}{1959Al10}).\vspace{0.1cm}}&\\
&&&&&&\parbox[t][0.3cm]{12.29294cm}{\raggedright \ensuremath{\beta}\ensuremath{^{\textnormal{+}}}-\ensuremath{\nu} angular correlation coefficient: See also (1) \ensuremath{\alpha}={\textminus}0.8 \textit{4} (\href{https://www.nndc.bnl.gov/nsr/nsrlink.jsp?1954Al29,B}{1954Al29}), where their result\vspace{0.1cm}}&\\
&&&&&&\parbox[t][0.3cm]{12.29294cm}{\raggedright {\ }{\ }{\ }was deemed unreliable by (\href{https://www.nndc.bnl.gov/nsr/nsrlink.jsp?1957Al29,B}{1957Al29}); and (2) \ensuremath{\alpha}=0.14 \textit{15}, which is an unpublished\vspace{0.1cm}}&\\
&&&&&&\parbox[t][0.3cm]{12.29294cm}{\raggedright {\ }{\ }{\ }value by (M. L. Good and E. J. Lauer, University of California, Radiation Laboratory\vspace{0.1cm}}&\\
&&&&&&\parbox[t][0.3cm]{12.29294cm}{\raggedright {\ }{\ }{\ }Report-3450 (unpublished), circa 1957), and is cited by (\href{https://www.nndc.bnl.gov/nsr/nsrlink.jsp?1957Al29,B}{1957Al29}).\vspace{0.1cm}}&\\
&&&&&&\parbox[t][0.3cm]{12.29294cm}{\raggedright \ensuremath{\lambda}={\textminus}0.21 \textit{8} (stat.) (\href{https://www.nndc.bnl.gov/nsr/nsrlink.jsp?1955Mb02,B}{1955Mb02}), see Equation 3 in that study.\vspace{0.1cm}}&\\
&&&&&&\parbox[t][0.3cm]{12.29294cm}{\raggedright C\ensuremath{_{\textnormal{S}}^{\textnormal{2}}}/C\ensuremath{_{\textnormal{T}}^{\textnormal{2}}}=1.1 \textit{3} (stat.) (\href{https://www.nndc.bnl.gov/nsr/nsrlink.jsp?1955Mb02,B}{1955Mb02}).\vspace{0.1cm}}&\\
&&&&&&\parbox[t][0.3cm]{12.29294cm}{\raggedright (\href{https://www.nndc.bnl.gov/nsr/nsrlink.jsp?1955Mb02,B}{1955Mb02}) reported that \ensuremath{^{\textnormal{19}}}Ne\ensuremath{_{\textnormal{g.s.}}} may have a probable nuclear configuration of (\textit{d}\ensuremath{_{\textnormal{5/2}}})\ensuremath{^{\textnormal{3}}}\vspace{0.1cm}}&\\
&&&&&&\parbox[t][0.3cm]{12.29294cm}{\raggedright {\ }{\ }{\ }with T=1/2 and not (\textit{s}\ensuremath{_{\textnormal{1/2}}})\ensuremath{^{\textnormal{3}}} with T=1/2 as suggested by (D. C. Peaslee through private\vspace{0.1cm}}&\\
&&&&&&\parbox[t][0.3cm]{12.29294cm}{\raggedright {\ }{\ }{\ }communication by D. R. Maxson \textit{et al}). See also (D. C. Peaslee, Phys. Rev. 89 (1953)\vspace{0.1cm}}&\\
&&&&&&\parbox[t][0.3cm]{12.29294cm}{\raggedright {\ }{\ }{\ }1148).\vspace{0.1cm}}&\\
&&&&&&\parbox[t][0.3cm]{12.29294cm}{\raggedright A\ensuremath{_{\ensuremath{\beta}}}={\textminus}0.057 \textit{5} (\href{https://www.nndc.bnl.gov/nsr/nsrlink.jsp?1963Co22,B}{1963Co22}): \ensuremath{\beta} asymmetry parameter. See also the zero-intercept of the \ensuremath{\beta}\vspace{0.1cm}}&\\
&&&&&&\parbox[t][0.3cm]{12.29294cm}{\raggedright {\ }{\ }{\ }asymmetry parameter: A(0)={\textminus}0.0391 \textit{14} (\href{https://www.nndc.bnl.gov/nsr/nsrlink.jsp?1975Ca28,B}{1975Ca28}); and the unpublished value of\vspace{0.1cm}}&\\
&&&&&&\parbox[t][0.3cm]{12.29294cm}{\raggedright {\ }{\ }{\ }A\ensuremath{_{\textnormal{0}}}={\textminus}0.03871 \textit{26} (stat.) \textit{+65{\textminus}87} (sys.) from (D. Combs, G. Jones, W. Anderson, F.\vspace{0.1cm}}&\\
&&&&&&\parbox[t][0.3cm]{12.29294cm}{\raggedright {\ }{\ }{\ }Calaprice, L. Hayen, and A. Young, \textit{A look into mirrors: A measurement of the}\vspace{0.1cm}}&\\
&&&&&&\parbox[t][0.3cm]{12.29294cm}{\raggedright {\ }{\ }{\ }\ensuremath{\beta}-asymmetry in \ensuremath{^{\textnormal{19}}}Ne decay and searches for new physics, arXiv:2009.13700v2\vspace{0.1cm}}&\\
&&&&&&\parbox[t][0.3cm]{12.29294cm}{\raggedright {\ }{\ }{\ }[nucl-ex] 20 Nov 2020, unpublished). Other value: A\ensuremath{_{\ensuremath{\beta}}}=0.17 \textit{12} (\href{https://www.nndc.bnl.gov/nsr/nsrlink.jsp?1993Sa32,B}{1993Sa32}): Average \ensuremath{\beta}\vspace{0.1cm}}&\\
&&&&&&\parbox[t][0.3cm]{12.29294cm}{\raggedright {\ }{\ }{\ }asymmetry of the \ensuremath{^{\textnormal{19}}}Ne(g.s., 1/2\ensuremath{^{\textnormal{+}}})\ensuremath{\rightarrow}\ensuremath{^{\textnormal{19}}}F*(109.9 keV, 1/2\ensuremath{^{-}})\ensuremath{\rightarrow}\ensuremath{^{\textnormal{19}}}F\ensuremath{_{\textnormal{g.s.}}} first-forbidden\vspace{0.1cm}}&\\
&&&&&&\parbox[t][0.3cm]{12.29294cm}{\raggedright {\ }{\ }{\ }transition.\vspace{0.1cm}}&\\
&&&&&&\parbox[t][0.3cm]{12.29294cm}{\raggedright Fermi-to-Gamow-Teller mixing ratio: \ensuremath{\rho}=1.6014 \textit{8} (stat.) \textit{+21{\textminus}28} (sys.) (unpublished) and\vspace{0.1cm}}&\\
&&&&&&\parbox[t][0.3cm]{12.29294cm}{\raggedright {\ }{\ }{\ }\ensuremath{\vert}V\ensuremath{_{\textnormal{ud}}}\ensuremath{\vert}=0.9739 \textit{13} (unpublished) from (D. Combs, G. Jones, W. Anderson, F. Calaprice,\vspace{0.1cm}}&\\
&&&&&&\parbox[t][0.3cm]{12.29294cm}{\raggedright {\ }{\ }{\ }L. Hayen, and A. Young, \textit{A look into mirrors: A measurement of the \ensuremath{\beta}-asymmetry in}\vspace{0.1cm}}&\\
&&&&&&\parbox[t][0.3cm]{12.29294cm}{\raggedright {\ }{\ }{\ }\ensuremath{^{\textnormal{19}}}Ne decay and searches for new physics, arXiv:2009.13700v2 [nucl-ex] 20 Nov 2020,\vspace{0.1cm}}&\\
&&&&&&\parbox[t][0.3cm]{12.29294cm}{\raggedright {\ }{\ }{\ }unpublished).\vspace{0.1cm}}&\\
&&&&&&\parbox[t][0.3cm]{12.29294cm}{\raggedright (K. Wildermuth and Y. C. Tang, Phys. Rev. Lett. 6 (1961) 17) attributed the \ensuremath{^{\textnormal{19}}}Ne\ensuremath{_{\textnormal{g.s.}}} to\vspace{0.1cm}}&\\
&&&&&&\parbox[t][0.3cm]{12.29294cm}{\raggedright {\ }{\ }{\ }an unexcited \ensuremath{^{\textnormal{16}}}O core plus a \ensuremath{^{\textnormal{3}}}He cluster in relative motion of orbital angular\vspace{0.1cm}}&\\
&&&&&&\parbox[t][0.3cm]{12.29294cm}{\raggedright {\ }{\ }{\ }momentum L=0.\vspace{0.1cm}}&\\
&&&&&&\parbox[t][0.3cm]{12.29294cm}{\raggedright \textit{ft}=1721 s \textit{7}: Weighted average (with external errors) of 1728.4 s \textit{67} (\href{https://www.nndc.bnl.gov/nsr/nsrlink.jsp?1974Wi14,B}{1974Wi14}) and\vspace{0.1cm}}&\\
&&&&&&\parbox[t][0.3cm]{12.29294cm}{\raggedright {\ }{\ }{\ }1714.3 s \textit{60} (\href{https://www.nndc.bnl.gov/nsr/nsrlink.jsp?1975Az01,B}{1975Az01}).\vspace{0.1cm}}&\\
&&&&&&\parbox[t][0.3cm]{12.29294cm}{\raggedright Time reversal coefficient: See D=0.002 \textit{4} (\href{https://www.nndc.bnl.gov/nsr/nsrlink.jsp?1974Ca17,B}{1974Ca17}); D={\textminus}0.0005 \textit{10} (stat.) (\href{https://www.nndc.bnl.gov/nsr/nsrlink.jsp?1977Ba08,B}{1977Ba08});\vspace{0.1cm}}&\\
&&&&&&\parbox[t][0.3cm]{12.29294cm}{\raggedright {\ }{\ }{\ }and D=+0.0004 \textit{8} (\href{https://www.nndc.bnl.gov/nsr/nsrlink.jsp?1984Ha01,B}{1984Ha01}: See also D=+0.0040 \textit{8} in the abstract).\vspace{0.1cm}}&\\
&&&&&&\parbox[t][0.3cm]{12.29294cm}{\raggedright Time reversal odd angular correlation magnitude: R={\textminus}0.079 \textit{53} (\href{https://www.nndc.bnl.gov/nsr/nsrlink.jsp?1983Sc32,B}{1983Sc32}).\vspace{0.1cm}}&\\
&&&&&&\parbox[t][0.3cm]{12.29294cm}{\raggedright Im(C\ensuremath{_{\textnormal{S}}}C\ensuremath{_{\textnormal{A}}}*)=0.19 \textit{13} (\href{https://www.nndc.bnl.gov/nsr/nsrlink.jsp?1983Sc32,B}{1983Sc32}): Time reversal invariance, scalar, axial vector coupling\vspace{0.1cm}}&\\
&&&&&&\parbox[t][0.3cm]{12.29294cm}{\raggedright {\ }{\ }{\ }imaginary interference limit.\vspace{0.1cm}}&\\
&&&&&&\parbox[t][0.3cm]{12.29294cm}{\raggedright \ensuremath{\phi}\ensuremath{_{\textnormal{A,V}}}(\ensuremath{^{\textnormal{19}}}Ne)=180.2\ensuremath{^\circ} \textit{4} (\href{https://www.nndc.bnl.gov/nsr/nsrlink.jsp?1974Ca17,B}{1974Ca17}): Relative phase angle between axial-vector (A) and\vspace{0.1cm}}&\\
&&&&&&\parbox[t][0.3cm]{12.29294cm}{\raggedright {\ }{\ }{\ }vector (V) couplings.\vspace{0.1cm}}&\\
&&&&&&\parbox[t][0.3cm]{12.29294cm}{\raggedright R\ensuremath{_{\textnormal{e}}}\ensuremath{<}\ensuremath{\sigma}\ensuremath{>}=1.6006 \textit{42} (\href{https://www.nndc.bnl.gov/nsr/nsrlink.jsp?1974Wi14,B}{1974Wi14}): Gamow-Teller matrix element, where R\ensuremath{_{\textnormal{e}}} is the magnitude\vspace{0.1cm}}&\\
&&&&&&\parbox[t][0.3cm]{12.29294cm}{\raggedright {\ }{\ }{\ }of the ratio of the axial to vector coupling constant for \ensuremath{^{\textnormal{19}}}Ne.\vspace{0.1cm}}&\\
&&&&&&\parbox[t][0.3cm]{12.29294cm}{\raggedright Decay mode: \ensuremath{\beta}\ensuremath{^{\textnormal{+}}}.\vspace{0.1cm}}&\\
&&&&&&\parbox[t][0.3cm]{12.29294cm}{\raggedright See (\href{https://www.nndc.bnl.gov/nsr/nsrlink.jsp?1974Ma31,B}{1974Ma31}, \href{https://www.nndc.bnl.gov/nsr/nsrlink.jsp?1976Al07,B}{1976Al07}, \href{https://www.nndc.bnl.gov/nsr/nsrlink.jsp?1981Ad05,B}{1981Ad05}, \href{https://www.nndc.bnl.gov/nsr/nsrlink.jsp?1983Ad03,B}{1983Ad03}, \href{https://www.nndc.bnl.gov/nsr/nsrlink.jsp?1993Sa32,B}{1993Sa32}) for the \ensuremath{^{\textnormal{19}}}Ne(\ensuremath{\beta}\ensuremath{^{\textnormal{+}}}) decay\vspace{0.1cm}}&\\
&&&&&&\parbox[t][0.3cm]{12.29294cm}{\raggedright {\ }{\ }{\ }branches and their branching ratios. Note that (\href{https://www.nndc.bnl.gov/nsr/nsrlink.jsp?1976Al07,B}{1976Al07}) disputed the results of\vspace{0.1cm}}&\\
&&&&&&\parbox[t][0.3cm]{12.29294cm}{\raggedright {\ }{\ }{\ }(\href{https://www.nndc.bnl.gov/nsr/nsrlink.jsp?1975Fr15,B}{1975Fr15}).\vspace{0.1cm}}&\\
&&&&&&\parbox[t][0.3cm]{12.29294cm}{\raggedright d\ensuremath{\sigma}/d\ensuremath{\Omega}\ensuremath{_{\textnormal{lab}}}(\ensuremath{\theta}\ensuremath{_{\textnormal{lab}}}=163\ensuremath{^\circ})=0.8 mb/sr \textit{4} at E\ensuremath{_{\textnormal{p}}}=5.879 MeV for the \ensuremath{^{\textnormal{19}}}F(p,n\ensuremath{_{\textnormal{0}}}) channel\vspace{0.1cm}}&\\
&&&&&&\parbox[t][0.3cm]{12.29294cm}{\raggedright {\ }{\ }{\ }(\href{https://www.nndc.bnl.gov/nsr/nsrlink.jsp?1972Ku24,B}{1972Ku24}).\vspace{0.1cm}}&\\
\multicolumn{1}{r@{}}{238}&\multicolumn{1}{@{.}l}{1 {\it 2}}&\multicolumn{1}{l}{5/2\ensuremath{^{+}}}&\multicolumn{1}{r@{}}{17}&\multicolumn{1}{@{.}l}{7 ns {\it 7}}&\multicolumn{1}{l}{2}&\parbox[t][0.3cm]{12.29294cm}{\raggedright \ensuremath{\mu}={\textminus}0.740 (\href{https://www.nndc.bnl.gov/nsr/nsrlink.jsp?1969Bl02,B}{1969Bl02})\vspace{0.1cm}}&\\
&&&&&&\parbox[t][0.3cm]{12.29294cm}{\raggedright g={\textminus}0.296 \textit{3} (\href{https://www.nndc.bnl.gov/nsr/nsrlink.jsp?1969Bl02,B}{1969Bl02})\vspace{0.1cm}}&\\
&&&&&&\parbox[t][0.3cm]{12.29294cm}{\raggedright E(level): From a least-squares fit to E\ensuremath{_{\ensuremath{\gamma}}} values from (\href{https://www.nndc.bnl.gov/nsr/nsrlink.jsp?1957Ba09,B}{1957Ba09}, \href{https://www.nndc.bnl.gov/nsr/nsrlink.jsp?1963Gi09,B}{1963Gi09}, \href{https://www.nndc.bnl.gov/nsr/nsrlink.jsp?1969Bl02,B}{1969Bl02},\vspace{0.1cm}}&\\
&&&&&&\parbox[t][0.3cm]{12.29294cm}{\raggedright {\ }{\ }{\ }\href{https://www.nndc.bnl.gov/nsr/nsrlink.jsp?1970Gi09,B}{1970Gi09}, \href{https://www.nndc.bnl.gov/nsr/nsrlink.jsp?1971It02,B}{1971It02}).\vspace{0.1cm}}&\\
&&&&&&\parbox[t][0.3cm]{12.29294cm}{\raggedright E(level): See also 241 keV \textit{4} (\href{https://www.nndc.bnl.gov/nsr/nsrlink.jsp?1955Ma84,B}{1955Ma84}); 241 keV (\href{https://www.nndc.bnl.gov/nsr/nsrlink.jsp?1962Fr09,B}{1962Fr09}), which was from an\vspace{0.1cm}}&\\
&&&&&&\parbox[t][0.3cm]{12.29294cm}{\raggedright {\ }{\ }{\ }unresolved n\ensuremath{_{\textnormal{1+2}}} neutron groups; 0.25 MeV \textit{2} (\href{https://www.nndc.bnl.gov/nsr/nsrlink.jsp?1965We05,B}{1965We05}): Unresolved n\ensuremath{_{\textnormal{1+2}}} neutron\vspace{0.1cm}}&\\
&&&&&&\parbox[t][0.3cm]{12.29294cm}{\raggedright {\ }{\ }{\ }groups; 0.24 MeV (\href{https://www.nndc.bnl.gov/nsr/nsrlink.jsp?1972Ku24,B}{1972Ku24}): Unresolved from the 275-keV state; 238 keV\vspace{0.1cm}}&\\
&&&&&&\parbox[t][0.3cm]{12.29294cm}{\raggedright {\ }{\ }{\ }(\href{https://www.nndc.bnl.gov/nsr/nsrlink.jsp?1984Pi07,B}{1984Pi07}); 238 keV (\href{https://www.nndc.bnl.gov/nsr/nsrlink.jsp?1985Ba66,B}{1985Ba66}): Unresolved from the 275-keV level; and 0.24 MeV\vspace{0.1cm}}&\\
\end{longtable}
\begin{textblock}{29}(0,27.3)
Continued on next page (footnotes at end of table)
\end{textblock}
\clearpage
\begin{longtable}{cccccc@{\extracolsep{\fill}}c}
\\[-.4cm]
\multicolumn{7}{c}{{\bf \small \underline{\ensuremath{^{\textnormal{19}}}F(p,n),(p,n\ensuremath{\gamma}),(d,2n\ensuremath{\gamma})\hspace{0.2in}\href{https://www.nndc.bnl.gov/nsr/nsrlink.jsp?1970Gi09,B}{1970Gi09},\href{https://www.nndc.bnl.gov/nsr/nsrlink.jsp?1977Le03,B}{1977Le03} (continued)}}}\\
\multicolumn{7}{c}{~}\\
\multicolumn{7}{c}{\underline{\ensuremath{^{19}}Ne Levels (continued)}}\\
\multicolumn{7}{c}{~}\\
\multicolumn{2}{c}{E(level)$^{}$}&J$^{\pi}$$^{}$&\multicolumn{2}{c}{T$_{1/2}$$^{}$}&Comments&\\[-.2cm]
\multicolumn{2}{c}{\hrulefill}&\hrulefill&\multicolumn{2}{c}{\hrulefill}&\hrulefill&
\endhead
&&&&&\parbox[t][0.3cm]{12.604291cm}{\raggedright {\ }{\ }{\ }(\href{https://www.nndc.bnl.gov/nsr/nsrlink.jsp?1984Ra22,B}{1984Ra22}): Unresolved from the ground state at \ensuremath{\theta}\ensuremath{_{\textnormal{lab}}}\ensuremath{>}10\ensuremath{^\circ}.\vspace{0.1cm}}&\\
&&&&&\parbox[t][0.3cm]{12.604291cm}{\raggedright T\ensuremath{_{1/2}}: Weighted average of (1) 18.0 ns \textit{21} (\href{https://www.nndc.bnl.gov/nsr/nsrlink.jsp?1957Ba09,B}{1957Ba09}) from \ensuremath{\tau}=26 ns \textit{3} as cited by\vspace{0.1cm}}&\\
&&&&&\parbox[t][0.3cm]{12.604291cm}{\raggedright {\ }{\ }{\ }(\href{https://www.nndc.bnl.gov/nsr/nsrlink.jsp?1967Be14,B}{1967Be14}) (see also T\ensuremath{_{\textnormal{1/2}}}=18 ns \textit{2} as cited by \href{https://www.nndc.bnl.gov/nsr/nsrlink.jsp?1955Ma84,B}{1955Ma84}); and (2) 17.7 ns \textit{7} (\href{https://www.nndc.bnl.gov/nsr/nsrlink.jsp?1969Bl02,B}{1969Bl02}),\vspace{0.1cm}}&\\
&&&&&\parbox[t][0.3cm]{12.604291cm}{\raggedright {\ }{\ }{\ }which also reports T\ensuremath{_{\textnormal{1/2}}}=17.7 ns \textit{5} in section 4.2.1.\vspace{0.1cm}}&\\
&&&&&\parbox[t][0.3cm]{12.604291cm}{\raggedright J\ensuremath{^{\pi}}: From the \ensuremath{^{\textnormal{19}}}Ne Adopted Levels.\vspace{0.1cm}}&\\
&&&&&\parbox[t][0.3cm]{12.604291cm}{\raggedright J\ensuremath{^{\pi}}: See also (1) J\ensuremath{^{\ensuremath{\pi}}}=(5/2\ensuremath{^{\textnormal{+}}}) (\href{https://www.nndc.bnl.gov/nsr/nsrlink.jsp?1962Fr09,B}{1962Fr09}), who reported that this level may be of positive\vspace{0.1cm}}&\\
&&&&&\parbox[t][0.3cm]{12.604291cm}{\raggedright {\ }{\ }{\ }parity nature using mirror level analysis and by comparison of level spacing in \ensuremath{^{\textnormal{19}}}F and\vspace{0.1cm}}&\\
&&&&&\parbox[t][0.3cm]{12.604291cm}{\raggedright {\ }{\ }{\ }\ensuremath{^{\textnormal{19}}}Ne; (2) J\ensuremath{^{\ensuremath{\pi}}}=5/2\ensuremath{^{\textnormal{+}}} (\href{https://www.nndc.bnl.gov/nsr/nsrlink.jsp?1963Gi09,B}{1963Gi09}): See the comment below on the ratios of 271 keV to 241\vspace{0.1cm}}&\\
&&&&&\parbox[t][0.3cm]{12.604291cm}{\raggedright {\ }{\ }{\ }keV \ensuremath{\gamma} rays and internal conversion coefficients; (3) J\ensuremath{^{\ensuremath{\pi}}}=5/2\ensuremath{^{\textnormal{+}}} (\href{https://www.nndc.bnl.gov/nsr/nsrlink.jsp?1969Bl02,B}{1969Bl02}), who reported\vspace{0.1cm}}&\\
&&&&&\parbox[t][0.3cm]{12.604291cm}{\raggedright {\ }{\ }{\ }that a non-zero a\ensuremath{_{\textnormal{4}}} angular correlation coefficient means that J\ensuremath{\geq}5/2 from L\ensuremath{_{\ensuremath{\gamma}}}\ensuremath{\geq}2; (4)\vspace{0.1cm}}&\\
&&&&&\parbox[t][0.3cm]{12.604291cm}{\raggedright {\ }{\ }{\ }J\ensuremath{^{\ensuremath{\pi}}}=5/2\ensuremath{^{\textnormal{+}}} (\href{https://www.nndc.bnl.gov/nsr/nsrlink.jsp?1970Gi09,B}{1970Gi09}) from mirror level analysis guided by the level$'$s lifetime from\vspace{0.1cm}}&\\
&&&&&\parbox[t][0.3cm]{12.604291cm}{\raggedright {\ }{\ }{\ }(\href{https://www.nndc.bnl.gov/nsr/nsrlink.jsp?1959Aj76,B}{1959Aj76}), relative internal conversion coefficient from (\href{https://www.nndc.bnl.gov/nsr/nsrlink.jsp?1963Gi09,B}{1963Gi09}), and the significant\vspace{0.1cm}}&\\
&&&&&\parbox[t][0.3cm]{12.604291cm}{\raggedright {\ }{\ }{\ }non-zero a\ensuremath{_{\textnormal{4}}} angular correlation coefficient measured in (\href{https://www.nndc.bnl.gov/nsr/nsrlink.jsp?1970Gi09,B}{1970Gi09}) for the\vspace{0.1cm}}&\\
&&&&&\parbox[t][0.3cm]{12.604291cm}{\raggedright {\ }{\ }{\ }\ensuremath{^{\textnormal{19}}}Ne*(238)\ensuremath{\rightarrow}\ensuremath{^{\textnormal{19}}}Ne\ensuremath{_{\textnormal{g.s.}}} transition; and (5) J\ensuremath{^{\ensuremath{\pi}}}=5/2\ensuremath{^{\textnormal{+}}} (\href{https://www.nndc.bnl.gov/nsr/nsrlink.jsp?1957Ba09,B}{1957Ba09}, \href{https://www.nndc.bnl.gov/nsr/nsrlink.jsp?1977Le03,B}{1977Le03}) assumed from\vspace{0.1cm}}&\\
&&&&&\parbox[t][0.3cm]{12.604291cm}{\raggedright {\ }{\ }{\ }literature.\vspace{0.1cm}}&\\
&&&&&\parbox[t][0.3cm]{12.604291cm}{\raggedright \ensuremath{\mu}: From the Time Dependent Perturbed Angular Distribution method. (\href{https://www.nndc.bnl.gov/nsr/nsrlink.jsp?1969Bl02,B}{1969Bl02}).\vspace{0.1cm}}&\\
&&&&&\parbox[t][0.3cm]{12.604291cm}{\raggedright (\href{https://www.nndc.bnl.gov/nsr/nsrlink.jsp?1969Bl02,B}{1969Bl02}) also determined the isoscalar g-factor: {\textminus}0.296 \textit{3} from a Larmor frequency ratio of\vspace{0.1cm}}&\\
&&&&&\parbox[t][0.3cm]{12.604291cm}{\raggedright {\ }{\ }{\ }\ensuremath{\omega}\ensuremath{_{\textnormal{B}}}(Ne)/\ensuremath{\omega}\ensuremath{_{\textnormal{B}}}(F)={\textminus}0.2054 \textit{15} (\href{https://www.nndc.bnl.gov/nsr/nsrlink.jsp?1969Bl02,B}{1969Bl02}), which is multiplied by the \ensuremath{^{\textnormal{19}}}F \textit{g}{\textminus}factor=+1.442 \textit{3}\vspace{0.1cm}}&\\
&&&&&\parbox[t][0.3cm]{12.604291cm}{\raggedright {\ }{\ }{\ }(\href{https://www.nndc.bnl.gov/nsr/nsrlink.jsp?1969Bl18,B}{1969Bl18}). The sign for the frequency ratio was obtained from the phases of the spin\vspace{0.1cm}}&\\
&&&&&\parbox[t][0.3cm]{12.604291cm}{\raggedright {\ }{\ }{\ }precession spectra at t=0 (\href{https://www.nndc.bnl.gov/nsr/nsrlink.jsp?1969Bl02,B}{1969Bl02}). See also g=0.573 \textit{3} (\href{https://www.nndc.bnl.gov/nsr/nsrlink.jsp?1969Bl02,B}{1969Bl02}: See Table 3).\vspace{0.1cm}}&\\
&&&&&\parbox[t][0.3cm]{12.604291cm}{\raggedright d\ensuremath{\sigma}/d\ensuremath{\Omega}\ensuremath{_{\textnormal{lab}}}(\ensuremath{\theta}\ensuremath{_{\textnormal{lab}}}=163\ensuremath{^\circ})=0.50 mb/sr \textit{25} at E\ensuremath{_{\textnormal{p}}}=5.879 MeV for the \ensuremath{^{\textnormal{19}}}F(p,n\ensuremath{_{\textnormal{1}}}+n\ensuremath{_{\textnormal{2}}}) channel\vspace{0.1cm}}&\\
&&&&&\parbox[t][0.3cm]{12.604291cm}{\raggedright {\ }{\ }{\ }(\href{https://www.nndc.bnl.gov/nsr/nsrlink.jsp?1972Ku24,B}{1972Ku24}).\vspace{0.1cm}}&\\
&&&&&\parbox[t][0.3cm]{12.604291cm}{\raggedright (K. Wildermuth and Y. C. Tang, Phys. Rev. Lett. 6 (1961) 17) attributed this state to an\vspace{0.1cm}}&\\
&&&&&\parbox[t][0.3cm]{12.604291cm}{\raggedright {\ }{\ }{\ }unexcited \ensuremath{^{\textnormal{16}}}O core plus a \ensuremath{^{\textnormal{3}}}He cluster in relative motion of orbital angular momentum\vspace{0.1cm}}&\\
&&&&&\parbox[t][0.3cm]{12.604291cm}{\raggedright {\ }{\ }{\ }L=1.\vspace{0.1cm}}&\\
\multicolumn{1}{r@{}}{275}&\multicolumn{1}{@{.}l}{1 {\it 2}}&\multicolumn{1}{l}{1/2\ensuremath{^{-}}}&\multicolumn{1}{r@{}}{$<$0}&\multicolumn{1}{@{.}l}{3 ns}&\parbox[t][0.3cm]{12.604291cm}{\raggedright E(level): From a least-squares fit to the E\ensuremath{_{\ensuremath{\gamma}}} values from (\href{https://www.nndc.bnl.gov/nsr/nsrlink.jsp?1957Ba09,B}{1957Ba09}, \href{https://www.nndc.bnl.gov/nsr/nsrlink.jsp?1963Gi09,B}{1963Gi09}, \href{https://www.nndc.bnl.gov/nsr/nsrlink.jsp?1969Bl02,B}{1969Bl02},\vspace{0.1cm}}&\\
&&&&&\parbox[t][0.3cm]{12.604291cm}{\raggedright {\ }{\ }{\ }\href{https://www.nndc.bnl.gov/nsr/nsrlink.jsp?1970Gi09,B}{1970Gi09}).\vspace{0.1cm}}&\\
&&&&&\parbox[t][0.3cm]{12.604291cm}{\raggedright E(level): See also 280 keV \textit{4} (\href{https://www.nndc.bnl.gov/nsr/nsrlink.jsp?1955Ma84,B}{1955Ma84}); 280 keV (\href{https://www.nndc.bnl.gov/nsr/nsrlink.jsp?1962Fr09,B}{1962Fr09}): Unresolved n\ensuremath{_{\textnormal{1+2}}} neutron\vspace{0.1cm}}&\\
&&&&&\parbox[t][0.3cm]{12.604291cm}{\raggedright {\ }{\ }{\ }group; 0.28 MeV (\href{https://www.nndc.bnl.gov/nsr/nsrlink.jsp?1972Ku24,B}{1972Ku24}): Unresolved from the 238-keV level; 275 keV (\href{https://www.nndc.bnl.gov/nsr/nsrlink.jsp?1984Pi07,B}{1984Pi07});\vspace{0.1cm}}&\\
&&&&&\parbox[t][0.3cm]{12.604291cm}{\raggedright {\ }{\ }{\ }and 275 keV (\href{https://www.nndc.bnl.gov/nsr/nsrlink.jsp?1985Ba66,B}{1985Ba66}): Unresolved from the 238-keV level.\vspace{0.1cm}}&\\
&&&&&\parbox[t][0.3cm]{12.604291cm}{\raggedright T\ensuremath{_{1/2}}: From (\href{https://www.nndc.bnl.gov/nsr/nsrlink.jsp?1969Bl02,B}{1969Bl02}).\vspace{0.1cm}}&\\
&&&&&\parbox[t][0.3cm]{12.604291cm}{\raggedright T\ensuremath{_{1/2}}: See also T\ensuremath{_{\textnormal{1/2}}}\ensuremath{<}3.5 ns (\href{https://www.nndc.bnl.gov/nsr/nsrlink.jsp?1957Ba09,B}{1957Ba09}) from \ensuremath{\tau}\ensuremath{<}5 ns as cited by (\href{https://www.nndc.bnl.gov/nsr/nsrlink.jsp?1955Ma84,B}{1955Ma84}).\vspace{0.1cm}}&\\
&&&&&\parbox[t][0.3cm]{12.604291cm}{\raggedright J\ensuremath{^{\pi}}: From the relative internal conversion coefficient from (\href{https://www.nndc.bnl.gov/nsr/nsrlink.jsp?1963Gi09,B}{1963Gi09}). The isotropy of the\vspace{0.1cm}}&\\
&&&&&\parbox[t][0.3cm]{12.604291cm}{\raggedright {\ }{\ }{\ }\ensuremath{^{\textnormal{19}}}Ne*(275)\ensuremath{\rightarrow}\ensuremath{^{\textnormal{19}}}Ne\ensuremath{_{\textnormal{g.s.}}} transition reported by (\href{https://www.nndc.bnl.gov/nsr/nsrlink.jsp?1970Gi09,B}{1970Gi09}) supports the assigned J\ensuremath{^{\ensuremath{\pi}}} value.\vspace{0.1cm}}&\\
&&&&&\parbox[t][0.3cm]{12.604291cm}{\raggedright {\ }{\ }{\ }Moreover, (\href{https://www.nndc.bnl.gov/nsr/nsrlink.jsp?1970Gi09,B}{1970Gi09}) assigned J\ensuremath{^{\ensuremath{\pi}}}=1/2\ensuremath{^{-}} from mirror level analysis guided by the level$'$s\vspace{0.1cm}}&\\
&&&&&\parbox[t][0.3cm]{12.604291cm}{\raggedright {\ }{\ }{\ }lifetime from (\href{https://www.nndc.bnl.gov/nsr/nsrlink.jsp?1959Aj76,B}{1959Aj76}).\vspace{0.1cm}}&\\
&&&&&\parbox[t][0.3cm]{12.604291cm}{\raggedright J\ensuremath{^{\pi}}: See also (\href{https://www.nndc.bnl.gov/nsr/nsrlink.jsp?1962Fr09,B}{1962Fr09}): J\ensuremath{^{\ensuremath{\pi}}}=(1/2\ensuremath{^{-}}) from mirror analysis; (\href{https://www.nndc.bnl.gov/nsr/nsrlink.jsp?1963Gi09,B}{1963Gi09}): J\ensuremath{^{\ensuremath{\pi}}}=1/2\ensuremath{^{-}}; and\vspace{0.1cm}}&\\
&&&&&\parbox[t][0.3cm]{12.604291cm}{\raggedright {\ }{\ }{\ }(\href{https://www.nndc.bnl.gov/nsr/nsrlink.jsp?1957Ba09,B}{1957Ba09}, \href{https://www.nndc.bnl.gov/nsr/nsrlink.jsp?1969Bl02,B}{1969Bl02}, \href{https://www.nndc.bnl.gov/nsr/nsrlink.jsp?1977Le03,B}{1977Le03}): J\ensuremath{^{\ensuremath{\pi}}}=1/2\ensuremath{^{-}} assumed from literature.\vspace{0.1cm}}&\\
&&&&&\parbox[t][0.3cm]{12.604291cm}{\raggedright d\ensuremath{\sigma}/d\ensuremath{\Omega}\ensuremath{_{\textnormal{lab}}}(\ensuremath{\theta}\ensuremath{_{\textnormal{lab}}}=163\ensuremath{^\circ})=0.50 mb/sr \textit{25} at E\ensuremath{_{\textnormal{p}}}=5.879 MeV for the \ensuremath{^{\textnormal{19}}}F(p,n\ensuremath{_{\textnormal{1}}}+n\ensuremath{_{\textnormal{2}}}) channel\vspace{0.1cm}}&\\
&&&&&\parbox[t][0.3cm]{12.604291cm}{\raggedright {\ }{\ }{\ }(\href{https://www.nndc.bnl.gov/nsr/nsrlink.jsp?1972Ku24,B}{1972Ku24}).\vspace{0.1cm}}&\\
&&&&&\parbox[t][0.3cm]{12.604291cm}{\raggedright (K. Wildermuth and Y. C. Tang, Phys. Rev. Lett. 6 (1961) 17) attributed this state to an\vspace{0.1cm}}&\\
&&&&&\parbox[t][0.3cm]{12.604291cm}{\raggedright {\ }{\ }{\ }unexcited \ensuremath{^{\textnormal{15}}}O core plus an \ensuremath{\alpha} cluster in relative oscillation of orbital angular momentum\vspace{0.1cm}}&\\
&&&&&\parbox[t][0.3cm]{12.604291cm}{\raggedright {\ }{\ }{\ }L=0. The study by Wildermuth \textit{et al}. predicts that this level in \ensuremath{^{\textnormal{19}}}Ne would have a larger\vspace{0.1cm}}&\\
&&&&&\parbox[t][0.3cm]{12.604291cm}{\raggedright {\ }{\ }{\ }excitation energy than its mirror state in \ensuremath{^{\textnormal{19}}}F.\vspace{0.1cm}}&\\
\multicolumn{1}{r@{}}{1507}&\multicolumn{1}{@{.}l}{9 {\it 3}}&\multicolumn{1}{l}{5/2\ensuremath{^{-}}}&\multicolumn{1}{r@{}}{2}&\multicolumn{1}{@{.}l}{1 ps {\it +83\textminus14}}&\parbox[t][0.3cm]{12.604291cm}{\raggedright E(level): Weighted average of 1507.9 keV \textit{3} from a least-squares fit to the E\ensuremath{_{\ensuremath{\gamma}}} values from\vspace{0.1cm}}&\\
&&&&&\parbox[t][0.3cm]{12.604291cm}{\raggedright {\ }{\ }{\ }(\href{https://www.nndc.bnl.gov/nsr/nsrlink.jsp?1970Gi09,B}{1970Gi09}) and 1507.8 keV \textit{6} (\href{https://www.nndc.bnl.gov/nsr/nsrlink.jsp?1977Le03,B}{1977Le03}).\vspace{0.1cm}}&\\
&&&&&\parbox[t][0.3cm]{12.604291cm}{\raggedright E(level): See also (1) 1506 keV \textit{5} (\href{https://www.nndc.bnl.gov/nsr/nsrlink.jsp?1962Fr09,B}{1962Fr09}): Weighted average of E\ensuremath{_{\textnormal{x}}}=1509 keV \textit{+8{\textminus}6}\vspace{0.1cm}}&\\
&&&&&\parbox[t][0.3cm]{12.604291cm}{\raggedright {\ }{\ }{\ }using \ensuremath{^{\textnormal{3}}}He ionization chamber; and E\ensuremath{_{\textnormal{x}}}=1504 keV \textit{6} using the counter ratio method and\vspace{0.1cm}}&\\
&&&&&\parbox[t][0.3cm]{12.604291cm}{\raggedright {\ }{\ }{\ }corresponds to the unresolved n\ensuremath{_{\textnormal{3+4}}} neutron groups. The n\ensuremath{_{\textnormal{3}}} neutron group was significantly\vspace{0.1cm}}&\\
&&&&&\parbox[t][0.3cm]{12.604291cm}{\raggedright {\ }{\ }{\ }wider than a single group and arose from the excitation of two unresolved levels in \ensuremath{^{\textnormal{19}}}Ne.\vspace{0.1cm}}&\\
&&&&&\parbox[t][0.3cm]{12.604291cm}{\raggedright {\ }{\ }{\ }Uncertainty in E\ensuremath{_{\textnormal{x}}} is from the statistical and systematic uncertainties combined; (2) 1.51\vspace{0.1cm}}&\\
&&&&&\parbox[t][0.3cm]{12.604291cm}{\raggedright {\ }{\ }{\ }MeV \textit{2} (\href{https://www.nndc.bnl.gov/nsr/nsrlink.jsp?1965We05,B}{1965We05}): Unresolved n\ensuremath{_{\textnormal{3+4}}} neutron groups and determined from the difference\vspace{0.1cm}}&\\
&&&&&\parbox[t][0.3cm]{12.604291cm}{\raggedright {\ }{\ }{\ }in the center-of-mass energies of the n\ensuremath{_{\textnormal{0}}} and n\ensuremath{_{\textnormal{1,2}}} neutron groups; and (3) 1507.56 keV\vspace{0.1cm}}&\\
\end{longtable}
\begin{textblock}{29}(0,27.3)
Continued on next page (footnotes at end of table)
\end{textblock}
\clearpage
\begin{longtable}{ccccccc@{\extracolsep{\fill}}c}
\\[-.4cm]
\multicolumn{8}{c}{{\bf \small \underline{\ensuremath{^{\textnormal{19}}}F(p,n),(p,n\ensuremath{\gamma}),(d,2n\ensuremath{\gamma})\hspace{0.2in}\href{https://www.nndc.bnl.gov/nsr/nsrlink.jsp?1970Gi09,B}{1970Gi09},\href{https://www.nndc.bnl.gov/nsr/nsrlink.jsp?1977Le03,B}{1977Le03} (continued)}}}\\
\multicolumn{8}{c}{~}\\
\multicolumn{8}{c}{\underline{\ensuremath{^{19}}Ne Levels (continued)}}\\
\multicolumn{8}{c}{~}\\
\multicolumn{2}{c}{E(level)$^{}$}&J$^{\pi}$$^{}$&\multicolumn{2}{c}{T$_{1/2}$$^{}$}&\ensuremath{\Delta}L$^{{\hyperlink{NE31LEVEL0}{a}}}$&Comments&\\[-.2cm]
\multicolumn{2}{c}{\hrulefill}&\hrulefill&\multicolumn{2}{c}{\hrulefill}&\hrulefill&\hrulefill&
\endhead
&&&&&&\parbox[t][0.3cm]{11.49258cm}{\raggedright {\ }{\ }{\ }(\href{https://www.nndc.bnl.gov/nsr/nsrlink.jsp?1987Ra23,B}{1987Ra23}).\vspace{0.1cm}}&\\
&&&&&&\parbox[t][0.3cm]{11.49258cm}{\raggedright E(level): Other value: 1509 keV \textit{15} (\href{https://www.nndc.bnl.gov/nsr/nsrlink.jsp?1965We05,B}{1965We05}) from the measured\vspace{0.1cm}}&\\
&&&&&&\parbox[t][0.3cm]{11.49258cm}{\raggedright {\ }{\ }{\ }Q(\ensuremath{^{\textnormal{19}}}F(p,n)\ensuremath{^{\textnormal{19}}}Ne*(1.51 MeV))={\textminus}5.529 MeV (\href{https://www.nndc.bnl.gov/nsr/nsrlink.jsp?1965We05,B}{1965We05}) relative to\vspace{0.1cm}}&\\
&&&&&&\parbox[t][0.3cm]{11.49258cm}{\raggedright {\ }{\ }{\ }Q(\ensuremath{^{\textnormal{27}}}A1(p,n)\ensuremath{^{\textnormal{27}}}Si\ensuremath{_{\textnormal{g.s.}}})={\textminus}5.589 MeV (\href{https://www.nndc.bnl.gov/nsr/nsrlink.jsp?1965We05,B}{1965We05}) and by subtracting the\vspace{0.1cm}}&\\
&&&&&&\parbox[t][0.3cm]{11.49258cm}{\raggedright {\ }{\ }{\ }Q(\ensuremath{^{\textnormal{19}}}F(p,n)\ensuremath{^{\textnormal{19}}}Ne\ensuremath{_{\textnormal{g.s.}}})={\textminus}4.020 MeV (\href{https://www.nndc.bnl.gov/nsr/nsrlink.jsp?1961Be13,B}{1961Be13}). Evaluator updated E\ensuremath{_{\textnormal{x}}}=1507 keV\vspace{0.1cm}}&\\
&&&&&&\parbox[t][0.3cm]{11.49258cm}{\raggedright {\ }{\ }{\ }\textit{15} using Q={\textminus}5.529 MeV \textit{15} (\href{https://www.nndc.bnl.gov/nsr/nsrlink.jsp?1965We05,B}{1965We05}) and Q(\ensuremath{^{\textnormal{19}}}F(p,n)\ensuremath{^{\textnormal{19}}}Ne\ensuremath{_{\textnormal{g.s.}}})={\textminus}4021.85 keV\vspace{0.1cm}}&\\
&&&&&&\parbox[t][0.3cm]{11.49258cm}{\raggedright {\ }{\ }{\ }\textit{16} (\href{https://www.nndc.bnl.gov/nsr/nsrlink.jsp?2021Wa16,B}{2021Wa16}).\vspace{0.1cm}}&\\
&&&&&&\parbox[t][0.3cm]{11.49258cm}{\raggedright T\ensuremath{_{1/2}}: From (\href{https://www.nndc.bnl.gov/nsr/nsrlink.jsp?1977Le03,B}{1977Le03}) from \ensuremath{\tau}=3 ps \textit{+12{\textminus}2}, where the lower uncertainty was\vspace{0.1cm}}&\\
&&&&&&\parbox[t][0.3cm]{11.49258cm}{\raggedright {\ }{\ }{\ }reported as \textit{1.5} ps.\vspace{0.1cm}}&\\
&&&&&&\parbox[t][0.3cm]{11.49258cm}{\raggedright J\ensuremath{^{\pi}}: From J\ensuremath{^{\ensuremath{\pi}}}=5/2\ensuremath{^{-}} (\href{https://www.nndc.bnl.gov/nsr/nsrlink.jsp?1970Gi09,B}{1970Gi09}): The non-zero a\ensuremath{_{\textnormal{4}}} angular correlation coefficient for\vspace{0.1cm}}&\\
&&&&&&\parbox[t][0.3cm]{11.49258cm}{\raggedright {\ }{\ }{\ }the \ensuremath{^{\textnormal{19}}}Ne*(1508)\ensuremath{\rightarrow}\ensuremath{^{\textnormal{19}}}Ne*(275) transition requires J\ensuremath{\geq}5/2 for this level. From the\vspace{0.1cm}}&\\
&&&&&&\parbox[t][0.3cm]{11.49258cm}{\raggedright {\ }{\ }{\ }E2 multipolarity for the 1233-keV transition, (\href{https://www.nndc.bnl.gov/nsr/nsrlink.jsp?1970Gi09,B}{1970Gi09}) assigned J\ensuremath{^{\ensuremath{\pi}}}=5/2\ensuremath{^{-}}.\vspace{0.1cm}}&\\
&&&&&&\parbox[t][0.3cm]{11.49258cm}{\raggedright J\ensuremath{^{\pi}}: See also (\href{https://www.nndc.bnl.gov/nsr/nsrlink.jsp?1962Fr09,B}{1962Fr09}), who reported that this state may be of negative parity\vspace{0.1cm}}&\\
&&&&&&\parbox[t][0.3cm]{11.49258cm}{\raggedright {\ }{\ }{\ }nature and may correspond to excitation of a \textit{p}-shell neutron using mirror level\vspace{0.1cm}}&\\
&&&&&&\parbox[t][0.3cm]{11.49258cm}{\raggedright {\ }{\ }{\ }analysis and by comparison of level spacing in \ensuremath{^{\textnormal{19}}}F and \ensuremath{^{\textnormal{19}}}Ne; and J\ensuremath{^{\ensuremath{\pi}}}=5/2\ensuremath{^{-}}\vspace{0.1cm}}&\\
&&&&&&\parbox[t][0.3cm]{11.49258cm}{\raggedright {\ }{\ }{\ }(\href{https://www.nndc.bnl.gov/nsr/nsrlink.jsp?1977Le03,B}{1977Le03}) assumed from literature.\vspace{0.1cm}}&\\
\multicolumn{1}{r@{}}{1536}&\multicolumn{1}{@{.}l}{2 {\it 3}}&\multicolumn{1}{l}{3/2\ensuremath{^{+}}}&\multicolumn{1}{r@{}}{22}&\multicolumn{1}{@{ }l}{fs {\it +12\textminus11}}&\multicolumn{1}{l}{(0+2)}&\parbox[t][0.3cm]{11.49258cm}{\raggedright E(level): Weighted average of 1536.1 keV \textit{3} from a least-squares fit to the E\ensuremath{_{\ensuremath{\gamma}}}\vspace{0.1cm}}&\\
&&&&&&\parbox[t][0.3cm]{11.49258cm}{\raggedright {\ }{\ }{\ }values from (\href{https://www.nndc.bnl.gov/nsr/nsrlink.jsp?1970Gi09,B}{1970Gi09}, \href{https://www.nndc.bnl.gov/nsr/nsrlink.jsp?1971It02,B}{1971It02}) and 1536.6 keV \textit{6} (\href{https://www.nndc.bnl.gov/nsr/nsrlink.jsp?1977Le03,B}{1977Le03}).\vspace{0.1cm}}&\\
&&&&&&\parbox[t][0.3cm]{11.49258cm}{\raggedright E(level): See also (1) 1538 keV \textit{4} (\href{https://www.nndc.bnl.gov/nsr/nsrlink.jsp?1962Fr09,B}{1962Fr09}): Weighted average of E\ensuremath{_{\textnormal{x}}}=1540 keV \textit{6}\vspace{0.1cm}}&\\
&&&&&&\parbox[t][0.3cm]{11.49258cm}{\raggedright {\ }{\ }{\ }using \ensuremath{^{\textnormal{3}}}He ionization chamber; and E\ensuremath{_{\textnormal{x}}}=1537 keV \textit{6} using the counter ratio method\vspace{0.1cm}}&\\
&&&&&&\parbox[t][0.3cm]{11.49258cm}{\raggedright {\ }{\ }{\ }and corresponds to the unresolved n\ensuremath{_{\textnormal{3+4}}} neutron groups. The n\ensuremath{_{\textnormal{3}}} neutron group was\vspace{0.1cm}}&\\
&&&&&&\parbox[t][0.3cm]{11.49258cm}{\raggedright {\ }{\ }{\ }significantly wider than a single group and arose from the excitation of two\vspace{0.1cm}}&\\
&&&&&&\parbox[t][0.3cm]{11.49258cm}{\raggedright {\ }{\ }{\ }unresolved levels in \ensuremath{^{\textnormal{19}}}Ne. Uncertainty in E\ensuremath{_{\textnormal{x}}} is from the statistical and systematic\vspace{0.1cm}}&\\
&&&&&&\parbox[t][0.3cm]{11.49258cm}{\raggedright {\ }{\ }{\ }uncertainties combined; (2) 1.51 MeV \textit{2} (\href{https://www.nndc.bnl.gov/nsr/nsrlink.jsp?1965We05,B}{1965We05}); and (3) 1.54 MeV\vspace{0.1cm}}&\\
&&&&&&\parbox[t][0.3cm]{11.49258cm}{\raggedright {\ }{\ }{\ }(\href{https://www.nndc.bnl.gov/nsr/nsrlink.jsp?1984Ra22,B}{1984Ra22}).\vspace{0.1cm}}&\\
&&&&&&\parbox[t][0.3cm]{11.49258cm}{\raggedright T\ensuremath{_{1/2}}: From 22.2 fs \textit{+125{\textminus}111} deduced from \ensuremath{\tau}=32 fs \textit{+18{\textminus}16}, which is the weighted\vspace{0.1cm}}&\\
&&&&&&\parbox[t][0.3cm]{11.49258cm}{\raggedright {\ }{\ }{\ }average of (1) \ensuremath{\tau}=28 fs \textit{+18{\textminus}16} (\href{https://www.nndc.bnl.gov/nsr/nsrlink.jsp?1971It02,B}{1971It02}), where this value is, in turn, the\vspace{0.1cm}}&\\
&&&&&&\parbox[t][0.3cm]{11.49258cm}{\raggedright {\ }{\ }{\ }weighted average of \ensuremath{\tau}=27 fs \textit{+12{\textminus}13} and \ensuremath{\tau}=29 fs \textit{+17{\textminus}16} obtained using a\vspace{0.1cm}}&\\
&&&&&&\parbox[t][0.3cm]{11.49258cm}{\raggedright {\ }{\ }{\ }0.4-mg/cm\ensuremath{^{\textnormal{2}}}-thick CaF\ensuremath{_{\textnormal{2}}} target on Ni and on Au backings, respectively; and (2)\vspace{0.1cm}}&\\
&&&&&&\parbox[t][0.3cm]{11.49258cm}{\raggedright {\ }{\ }{\ }\ensuremath{\tau}=42 fs \textit{27} (\href{https://www.nndc.bnl.gov/nsr/nsrlink.jsp?1977Le03,B}{1977Le03}).\vspace{0.1cm}}&\\
&&&&&&\parbox[t][0.3cm]{11.49258cm}{\raggedright J\ensuremath{^{\pi}}: From (\href{https://www.nndc.bnl.gov/nsr/nsrlink.jsp?1970Gi09,B}{1970Gi09}), where the a\ensuremath{_{\textnormal{2}}} angular correlation coefficient for the\vspace{0.1cm}}&\\
&&&&&&\parbox[t][0.3cm]{11.49258cm}{\raggedright {\ }{\ }{\ }\ensuremath{^{\textnormal{19}}}Ne*(1536)\ensuremath{\rightarrow}\ensuremath{^{\textnormal{19}}}Ne*(238) transition requires J\ensuremath{\geq}3/2 for this level. From mirror\vspace{0.1cm}}&\\
&&&&&&\parbox[t][0.3cm]{11.49258cm}{\raggedright {\ }{\ }{\ }level analysis and based on the deduced transition strength, those authors assigned\vspace{0.1cm}}&\\
&&&&&&\parbox[t][0.3cm]{11.49258cm}{\raggedright {\ }{\ }{\ }J\ensuremath{^{\ensuremath{\pi}}}=3/2\ensuremath{^{\textnormal{+}}}.\vspace{0.1cm}}&\\
&&&&&&\parbox[t][0.3cm]{11.49258cm}{\raggedright J\ensuremath{^{\pi}}: See also (\href{https://www.nndc.bnl.gov/nsr/nsrlink.jsp?1962Fr09,B}{1962Fr09}), who reported that \ensuremath{^{\textnormal{19}}}Ne*(241) may be of positive parity\vspace{0.1cm}}&\\
&&&&&&\parbox[t][0.3cm]{11.49258cm}{\raggedright {\ }{\ }{\ }nature using mirror level analysis and by comparison of level spacing in \ensuremath{^{\textnormal{19}}}F and\vspace{0.1cm}}&\\
&&&&&&\parbox[t][0.3cm]{11.49258cm}{\raggedright {\ }{\ }{\ }\ensuremath{^{\textnormal{19}}}Ne. Other value: (\href{https://www.nndc.bnl.gov/nsr/nsrlink.jsp?1977Le03,B}{1977Le03}) assumed J\ensuremath{^{\ensuremath{\pi}}}=3/2\ensuremath{^{\textnormal{+}}} from literature.\vspace{0.1cm}}&\\
&&&&&&\parbox[t][0.3cm]{11.49258cm}{\raggedright B(GT)=0.045 \textit{15} (\href{https://www.nndc.bnl.gov/nsr/nsrlink.jsp?1984Ra22,B}{1984Ra22}).\vspace{0.1cm}}&\\
\multicolumn{1}{r@{}}{1615}&\multicolumn{1}{@{.}l}{4 {\it 3}}&\multicolumn{1}{l}{(3/2\ensuremath{^{-}})}&\multicolumn{1}{r@{}}{90}&\multicolumn{1}{@{ }l}{fs {\it 24}}&&\parbox[t][0.3cm]{11.49258cm}{\raggedright E(level): Weighted average of 1615.3 \textit{4} from a least-squares fit to the E\ensuremath{_{\ensuremath{\gamma}}}=1340.1\vspace{0.1cm}}&\\
&&&&&&\parbox[t][0.3cm]{11.49258cm}{\raggedright {\ }{\ }{\ }keV \textit{4} and E\ensuremath{_{\ensuremath{\gamma}}}=1615.4 keV \textit{7} transitions in (\href{https://www.nndc.bnl.gov/nsr/nsrlink.jsp?1970Gi09,B}{1970Gi09}) and 1615.7 keV \textit{6}\vspace{0.1cm}}&\\
&&&&&&\parbox[t][0.3cm]{11.49258cm}{\raggedright {\ }{\ }{\ }(\href{https://www.nndc.bnl.gov/nsr/nsrlink.jsp?1977Le03,B}{1977Le03}).\vspace{0.1cm}}&\\
&&&&&&\parbox[t][0.3cm]{11.49258cm}{\raggedright E(level): See also 1612 keV \textit{5} (\href{https://www.nndc.bnl.gov/nsr/nsrlink.jsp?1962Fr09,B}{1962Fr09}): Weighted average of E\ensuremath{_{\textnormal{x}}}=1613 keV \textit{8}\vspace{0.1cm}}&\\
&&&&&&\parbox[t][0.3cm]{11.49258cm}{\raggedright {\ }{\ }{\ }using \ensuremath{^{\textnormal{3}}}He ionization chamber; and E\ensuremath{_{\textnormal{x}}}=1612 keV \textit{7} using the counter ratio method\vspace{0.1cm}}&\\
&&&&&&\parbox[t][0.3cm]{11.49258cm}{\raggedright {\ }{\ }{\ }and corresponds to the unresolved n\ensuremath{_{\textnormal{3+4}}} neutron groups. The n\ensuremath{_{\textnormal{3}}} neutron group was\vspace{0.1cm}}&\\
&&&&&&\parbox[t][0.3cm]{11.49258cm}{\raggedright {\ }{\ }{\ }significantly wider than a single group and arose from the excitation of two\vspace{0.1cm}}&\\
&&&&&&\parbox[t][0.3cm]{11.49258cm}{\raggedright {\ }{\ }{\ }unresolved levels in \ensuremath{^{\textnormal{19}}}Ne. Uncertainty in E\ensuremath{_{\textnormal{x}}} is from the statistical and systematic\vspace{0.1cm}}&\\
&&&&&&\parbox[t][0.3cm]{11.49258cm}{\raggedright {\ }{\ }{\ }uncertainties combined. Other value: 1.62 MeV \textit{2} (\href{https://www.nndc.bnl.gov/nsr/nsrlink.jsp?1965We05,B}{1965We05}): Unresolved n\ensuremath{_{\textnormal{5}}}\vspace{0.1cm}}&\\
&&&&&&\parbox[t][0.3cm]{11.49258cm}{\raggedright {\ }{\ }{\ }neutron group.\vspace{0.1cm}}&\\
&&&&&&\parbox[t][0.3cm]{11.49258cm}{\raggedright T\ensuremath{_{1/2}}: From \ensuremath{\tau}=130 fs \textit{35} (\href{https://www.nndc.bnl.gov/nsr/nsrlink.jsp?1977Le03,B}{1977Le03}).\vspace{0.1cm}}&\\
&&&&&&\parbox[t][0.3cm]{11.49258cm}{\raggedright J\ensuremath{^{\pi}}: From (\href{https://www.nndc.bnl.gov/nsr/nsrlink.jsp?1970Gi09,B}{1970Gi09}): Analysis of the \ensuremath{\gamma}-ray angular correlations shows that J=1/2,\vspace{0.1cm}}&\\
&&&&&&\parbox[t][0.3cm]{11.49258cm}{\raggedright {\ }{\ }{\ }3/2, or 5/2. J\ensuremath{^{\ensuremath{\pi}}}=3/2\ensuremath{^{-}} was deduced based on mirror level analysis. Since this is a\vspace{0.1cm}}&\\
&&&&&&\parbox[t][0.3cm]{11.49258cm}{\raggedright {\ }{\ }{\ }weak argument, we made the assignment tentative.\vspace{0.1cm}}&\\
&&&&&&\parbox[t][0.3cm]{11.49258cm}{\raggedright J\ensuremath{^{\pi}}: See also (\href{https://www.nndc.bnl.gov/nsr/nsrlink.jsp?1962Fr09,B}{1962Fr09}), who reported that this state may be of negative parity\vspace{0.1cm}}&\\
\end{longtable}
\begin{textblock}{29}(0,27.3)
Continued on next page (footnotes at end of table)
\end{textblock}
\clearpage
\begin{longtable}{ccccccc@{\extracolsep{\fill}}c}
\\[-.4cm]
\multicolumn{8}{c}{{\bf \small \underline{\ensuremath{^{\textnormal{19}}}F(p,n),(p,n\ensuremath{\gamma}),(d,2n\ensuremath{\gamma})\hspace{0.2in}\href{https://www.nndc.bnl.gov/nsr/nsrlink.jsp?1970Gi09,B}{1970Gi09},\href{https://www.nndc.bnl.gov/nsr/nsrlink.jsp?1977Le03,B}{1977Le03} (continued)}}}\\
\multicolumn{8}{c}{~}\\
\multicolumn{8}{c}{\underline{\ensuremath{^{19}}Ne Levels (continued)}}\\
\multicolumn{8}{c}{~}\\
\multicolumn{2}{c}{E(level)$^{}$}&J$^{\pi}$$^{}$&\multicolumn{2}{c}{T$_{1/2}$$^{}$}&\ensuremath{\Delta}L$^{{\hyperlink{NE31LEVEL0}{a}}}$&Comments&\\[-.2cm]
\multicolumn{2}{c}{\hrulefill}&\hrulefill&\multicolumn{2}{c}{\hrulefill}&\hrulefill&\hrulefill&
\endhead
&&&&&&\parbox[t][0.3cm]{11.449781cm}{\raggedright {\ }{\ }{\ }nature and may correspond to excitation of a \textit{p}-shell neutron using mirror level\vspace{0.1cm}}&\\
&&&&&&\parbox[t][0.3cm]{11.449781cm}{\raggedright {\ }{\ }{\ }analysis and by comparison of level spacing in \ensuremath{^{\textnormal{19}}}F and \ensuremath{^{\textnormal{19}}}Ne. Other value:\vspace{0.1cm}}&\\
&&&&&&\parbox[t][0.3cm]{11.449781cm}{\raggedright {\ }{\ }{\ }(\href{https://www.nndc.bnl.gov/nsr/nsrlink.jsp?1977Le03,B}{1977Le03}) assumed J\ensuremath{^{\ensuremath{\pi}}}=3/2\ensuremath{^{-}} from literature.\vspace{0.1cm}}&\\
\multicolumn{1}{r@{}}{2794}&\multicolumn{1}{@{.}l}{7 {\it 5}}&\multicolumn{1}{l}{(9/2\ensuremath{^{+}})}&\multicolumn{1}{r@{}}{97}&\multicolumn{1}{@{ }l}{fs {\it 24}}&&\parbox[t][0.3cm]{11.449781cm}{\raggedright E(level): Weighted average of 2.78 MeV \textit{3} (\href{https://www.nndc.bnl.gov/nsr/nsrlink.jsp?1965We05,B}{1965We05}), which corresponds to the\vspace{0.1cm}}&\\
&&&&&&\parbox[t][0.3cm]{11.449781cm}{\raggedright {\ }{\ }{\ }n\ensuremath{_{\textnormal{6}}} neutron group that is only observed at E\ensuremath{_{\textnormal{lab}}}=9.75 MeV; 2794.7 keV \textit{5} from a\vspace{0.1cm}}&\\
&&&&&&\parbox[t][0.3cm]{11.449781cm}{\raggedright {\ }{\ }{\ }least-squares fit to the E\ensuremath{_{\ensuremath{\gamma}}} from (\href{https://www.nndc.bnl.gov/nsr/nsrlink.jsp?1970Gi09,B}{1970Gi09}); and 2794.7 keV \textit{6} (\href{https://www.nndc.bnl.gov/nsr/nsrlink.jsp?1977Le03,B}{1977Le03}).\vspace{0.1cm}}&\\
&&&&&&\parbox[t][0.3cm]{11.449781cm}{\raggedright E(level): This state was observed for the first time in (\href{https://www.nndc.bnl.gov/nsr/nsrlink.jsp?1965We05,B}{1965We05}) and was\vspace{0.1cm}}&\\
&&&&&&\parbox[t][0.3cm]{11.449781cm}{\raggedright {\ }{\ }{\ }considered as the analog of the \ensuremath{^{\textnormal{19}}}F*(2.78 MeV) level.\vspace{0.1cm}}&\\
&&&&&&\parbox[t][0.3cm]{11.449781cm}{\raggedright T\ensuremath{_{1/2}}: From \ensuremath{\tau}=140 fs \textit{35} (\href{https://www.nndc.bnl.gov/nsr/nsrlink.jsp?1977Le03,B}{1977Le03}).\vspace{0.1cm}}&\\
&&&&&&\parbox[t][0.3cm]{11.449781cm}{\raggedright T\ensuremath{_{1/2}}: See also 229 fs \textit{90} (\href{https://www.nndc.bnl.gov/nsr/nsrlink.jsp?1970Gi09,B}{1970Gi09}) from \ensuremath{\tau}=0.33 ps \textit{13}. (\href{https://www.nndc.bnl.gov/nsr/nsrlink.jsp?1977Le03,B}{1977Le03}) explains in\vspace{0.1cm}}&\\
&&&&&&\parbox[t][0.3cm]{11.449781cm}{\raggedright {\ }{\ }{\ }details the inconsistency between their measured lifetime and that of (\href{https://www.nndc.bnl.gov/nsr/nsrlink.jsp?1970Gi09,B}{1970Gi09})\vspace{0.1cm}}&\\
&&&&&&\parbox[t][0.3cm]{11.449781cm}{\raggedright {\ }{\ }{\ }and concludes that the experimental determination of F(\ensuremath{\tau}) is the reason behind\vspace{0.1cm}}&\\
&&&&&&\parbox[t][0.3cm]{11.449781cm}{\raggedright {\ }{\ }{\ }this discrepancy (see \href{https://www.nndc.bnl.gov/nsr/nsrlink.jsp?1977Le03,B}{1977Le03} for details), which has caused the lifetime\vspace{0.1cm}}&\\
&&&&&&\parbox[t][0.3cm]{11.449781cm}{\raggedright {\ }{\ }{\ }measured by (\href{https://www.nndc.bnl.gov/nsr/nsrlink.jsp?1970Gi09,B}{1970Gi09}) for this state to be erroneous. See also \ensuremath{\Gamma}\ensuremath{_{\ensuremath{\gamma}\textnormal{,tot}}}=2.0 meV\vspace{0.1cm}}&\\
&&&&&&\parbox[t][0.3cm]{11.449781cm}{\raggedright {\ }{\ }{\ }\textit{+13{\textminus}6} (\href{https://www.nndc.bnl.gov/nsr/nsrlink.jsp?1970Gi09,B}{1970Gi09}).\vspace{0.1cm}}&\\
&&&&&&\parbox[t][0.3cm]{11.449781cm}{\raggedright J\ensuremath{^{\pi}}: From J\ensuremath{^{\ensuremath{\pi}}}=9/2\ensuremath{^{\textnormal{+}}} (\href{https://www.nndc.bnl.gov/nsr/nsrlink.jsp?1970Gi09,B}{1970Gi09}) from \ensuremath{\gamma} ray angular correlations coefficients, which\vspace{0.1cm}}&\\
&&&&&&\parbox[t][0.3cm]{11.449781cm}{\raggedright {\ }{\ }{\ }are consistent with J\ensuremath{\geq}5/2. The measured lifetime further restricts the multipolarity\vspace{0.1cm}}&\\
&&&&&&\parbox[t][0.3cm]{11.449781cm}{\raggedright {\ }{\ }{\ }of the transition to E1, M1 or E2 implying that the possible spin parity\vspace{0.1cm}}&\\
&&&&&&\parbox[t][0.3cm]{11.449781cm}{\raggedright {\ }{\ }{\ }assignments are J\ensuremath{^{\ensuremath{\pi}}}=5/2\ensuremath{^{\textnormal{+}}}, 7/2\ensuremath{^{\textnormal{+}}}, or 9/2\ensuremath{^{\textnormal{+}}}. The latter assignment was chosen based\vspace{0.1cm}}&\\
&&&&&&\parbox[t][0.3cm]{11.449781cm}{\raggedright {\ }{\ }{\ }on the J\ensuremath{^{\ensuremath{\pi}}} value of the mirror level. Since this is not a strong argument, we made\vspace{0.1cm}}&\\
&&&&&&\parbox[t][0.3cm]{11.449781cm}{\raggedright {\ }{\ }{\ }the assignment tentative.\vspace{0.1cm}}&\\
&&&&&&\parbox[t][0.3cm]{11.449781cm}{\raggedright J\ensuremath{^{\pi}}: See also (\href{https://www.nndc.bnl.gov/nsr/nsrlink.jsp?1977Le03,B}{1977Le03}), which assumed J\ensuremath{^{\ensuremath{\pi}}}=9/2\ensuremath{^{\textnormal{+}}}.\vspace{0.1cm}}&\\
\multicolumn{1}{r@{}}{5.4\ensuremath{\times10^{3}}}&\multicolumn{1}{@{}l}{\ensuremath{^{{\hyperlink{NE31LEVEL1}{b}}}}}&&&&\multicolumn{1}{l}{0}&&\\
\multicolumn{1}{r@{}}{6.2\ensuremath{\times10^{3}}}&\multicolumn{1}{@{}l}{\ensuremath{^{{\hyperlink{NE31LEVEL1}{b}}}}}&&&&\multicolumn{1}{l}{(0+1)}&&\\
\multicolumn{1}{r@{}}{7.1\ensuremath{\times10^{3}}}&\multicolumn{1}{@{}l}{\ensuremath{^{{\hyperlink{NE31LEVEL1}{b}}}}}&&&&\multicolumn{1}{l}{(0+1)}&&\\
\multicolumn{1}{r@{}}{7.7\ensuremath{\times10^{3}}}&\multicolumn{1}{@{}l}{\ensuremath{^{{\hyperlink{NE31LEVEL1}{b}}}}}&&&&\multicolumn{1}{l}{(0+1)}&\parbox[t][0.3cm]{11.449781cm}{\raggedright B(GT)\ensuremath{\sim}0.035 (\href{https://www.nndc.bnl.gov/nsr/nsrlink.jsp?1984Ra22,B}{1984Ra22}).\vspace{0.1cm}}&\\
\multicolumn{1}{r@{}}{8.60\ensuremath{\times10^{3}}}&\multicolumn{1}{@{}l}{\ensuremath{^{{\hyperlink{NE31LEVEL1}{b}}}}}&&&&\multicolumn{1}{l}{(0)}&&\\
\multicolumn{1}{r@{}}{10.2\ensuremath{\times10^{3}}}&\multicolumn{1}{@{}l}{\ensuremath{^{{\hyperlink{NE31LEVEL1}{b}}}}}&&&&\multicolumn{1}{l}{(1)}&&\\
\multicolumn{1}{r@{}}{11.0\ensuremath{\times10^{3}}}&\multicolumn{1}{@{}l}{\ensuremath{^{{\hyperlink{NE31LEVEL1}{b}}}}}&&&&\multicolumn{1}{l}{0}&&\\
\multicolumn{1}{r@{}}{12.1\ensuremath{\times10^{3}}}&\multicolumn{1}{@{}l}{\ensuremath{^{{\hyperlink{NE31LEVEL1}{b}}}}}&&&&\multicolumn{1}{l}{(0+2)}&&\\
\end{longtable}
\parbox[b][0.3cm]{17.7cm}{\makebox[1ex]{\ensuremath{^{\hypertarget{NE31LEVEL0}{a}}}} From DWBA analysis of (\href{https://www.nndc.bnl.gov/nsr/nsrlink.jsp?1984Ra22,B}{1984Ra22}).}\\
\parbox[b][0.3cm]{17.7cm}{\makebox[1ex]{\ensuremath{^{\hypertarget{NE31LEVEL1}{b}}}} From (\href{https://www.nndc.bnl.gov/nsr/nsrlink.jsp?1984Ra22,B}{1984Ra22}).}\\
\vspace{0.5cm}
\clearpage
\vspace{0.3cm}
\begin{landscape}
\vspace*{-0.5cm}
{\bf \small \underline{\ensuremath{^{\textnormal{19}}}F(p,n),(p,n\ensuremath{\gamma}),(d,2n\ensuremath{\gamma})\hspace{0.2in}\href{https://www.nndc.bnl.gov/nsr/nsrlink.jsp?1970Gi09,B}{1970Gi09},\href{https://www.nndc.bnl.gov/nsr/nsrlink.jsp?1977Le03,B}{1977Le03} (continued)}}\\
\vspace{0.3cm}
\underline{$\gamma$($^{19}$Ne)}\\
\vspace{0.34cm}
\parbox[b][0.3cm]{22.5cm}{\addtolength{\parindent}{-0.254cm}\textit{Notes}:}\\
\parbox[b][0.3cm]{22.5cm}{\addtolength{\parindent}{-0.254cm}(1) When a\ensuremath{_{\textnormal{4}}} is not provided by (\href{https://www.nndc.bnl.gov/nsr/nsrlink.jsp?1970Gi09,B}{1970Gi09}), it is because the Legendre polynomial fit to that \ensuremath{\gamma}-ray$'$s angular distribution was not substantially improved by including}\\
\parbox[b][0.3cm]{22.5cm}{the Legendre polynomials of order higher than two.}\\
\parbox[b][0.3cm]{22.5cm}{\addtolength{\parindent}{-0.254cm}(2) The \ensuremath{\gamma} ray angular correlation coefficients reported by (\href{https://www.nndc.bnl.gov/nsr/nsrlink.jsp?1970Gi09,B}{1970Gi09}) are not corrected for the effect of the finite size of the Ge(Li) detector used in that study.}\\
\parbox[b][0.3cm]{22.5cm}{\addtolength{\parindent}{-0.254cm}(3) The uncertainties in the transition strengths reported by (\href{https://www.nndc.bnl.gov/nsr/nsrlink.jsp?1970Gi09,B}{1970Gi09}) reflect the uncertainties in the lifetime measurements (see \href{https://www.nndc.bnl.gov/nsr/nsrlink.jsp?1970Gi09,B}{1970Gi09}: \ensuremath{^{\textnormal{16}}}O(\ensuremath{\alpha},n\ensuremath{\gamma}) and \ensuremath{^{\textnormal{19}}}F(p,n)}\\
\parbox[b][0.3cm]{22.5cm}{datasets) and in the branching ratios determined by (\href{https://www.nndc.bnl.gov/nsr/nsrlink.jsp?1970Gi09,B}{1970Gi09}).}\\
\vspace{0.34cm}
\begin{longtable}{ccccccccc@{}ccccccc@{\extracolsep{\fill}}c}
\multicolumn{2}{c}{E\ensuremath{_{i}}(level)}&J\ensuremath{^{\pi}_{i}}&\multicolumn{2}{c}{E\ensuremath{_{\gamma}}}&\multicolumn{2}{c}{I\ensuremath{_{\ensuremath{\gamma}}} (\%)\ensuremath{^{\hyperlink{NE31GAMMA3}{d}}}}&\multicolumn{2}{c}{E\ensuremath{_{f}}}&J\ensuremath{^{\pi}_{f}}&Mult.&\multicolumn{2}{c}{\ensuremath{\delta}\ensuremath{^{\hyperlink{NE31GAMMA4}{e}}}}&\multicolumn{2}{c}{\ensuremath{\alpha}\ensuremath{^{\hyperlink{NE31GAMMA5}{f}}}}&Comments&\\[-.2cm]
\multicolumn{2}{c}{\hrulefill}&\hrulefill&\multicolumn{2}{c}{\hrulefill}&\multicolumn{2}{c}{\hrulefill}&\multicolumn{2}{c}{\hrulefill}&\hrulefill&\hrulefill&\multicolumn{2}{c}{\hrulefill}&\multicolumn{2}{c}{\hrulefill}&\hrulefill&
\endfirsthead
\multicolumn{1}{r@{}}{238}&\multicolumn{1}{@{.}l}{1}&\multicolumn{1}{l}{5/2\ensuremath{^{+}}}&\multicolumn{1}{r@{}}{238}&\multicolumn{1}{@{.}l}{1 {\it 2}}&\multicolumn{1}{r@{}}{}&\multicolumn{1}{@{}l}{}&\multicolumn{1}{r@{}}{0}&\multicolumn{1}{@{}l}{}&\multicolumn{1}{@{}l}{1/2\ensuremath{^{+}}}&\multicolumn{1}{l}{E2}&\multicolumn{1}{r@{}}{}&\multicolumn{1}{@{}l}{}&\multicolumn{1}{r@{}}{1}&\multicolumn{1}{@{.}l}{43\ensuremath{\times10^{-3}} {\it 2}}&\parbox[t][0.3cm]{10.275222cm}{\raggedright B(E2)(W.u.)=13.9 \textit{6}\vspace{0.1cm}}&\\
&&&&&&&&&&&&&&&\parbox[t][0.3cm]{10.275222cm}{\raggedright \ensuremath{\alpha}(K)=0.001352 \textit{19}; \ensuremath{\alpha}(L)=7.49\ensuremath{\times}10\ensuremath{^{\textnormal{$-$5}}} \textit{11}\vspace{0.1cm}}&\\
&&&&&&&&&&&&&&&\parbox[t][0.3cm]{10.275222cm}{\raggedright E\ensuremath{_{\gamma}}: Weighted average of 242 keV \textit{5} (\href{https://www.nndc.bnl.gov/nsr/nsrlink.jsp?1957Ba09,B}{1957Ba09}); 241 keV \textit{4} (\href{https://www.nndc.bnl.gov/nsr/nsrlink.jsp?1963Gi09,B}{1963Gi09});\vspace{0.1cm}}&\\
&&&&&&&&&&&&&&&\parbox[t][0.3cm]{10.275222cm}{\raggedright {\ }{\ }{\ }239 keV \textit{2} (\href{https://www.nndc.bnl.gov/nsr/nsrlink.jsp?1969Bl02,B}{1969Bl02}); 238.2 keV \textit{2} (\href{https://www.nndc.bnl.gov/nsr/nsrlink.jsp?1970Gi09,B}{1970Gi09}); and 236.8 keV \textit{7}\vspace{0.1cm}}&\\
&&&&&&&&&&&&&&&\parbox[t][0.3cm]{10.275222cm}{\raggedright {\ }{\ }{\ }(\href{https://www.nndc.bnl.gov/nsr/nsrlink.jsp?1971It02,B}{1971It02}). See also 238 keV (\href{https://www.nndc.bnl.gov/nsr/nsrlink.jsp?1984Pi07,B}{1984Pi07}).\vspace{0.1cm}}&\\
&&&&&&&&&&&&&&&\parbox[t][0.3cm]{10.275222cm}{\raggedright Mult.: From (\href{https://www.nndc.bnl.gov/nsr/nsrlink.jsp?1969Bl02,B}{1969Bl02}, \href{https://www.nndc.bnl.gov/nsr/nsrlink.jsp?1970Gi09,B}{1970Gi09}).\vspace{0.1cm}}&\\
&&&&&&&&&&&&&&&\parbox[t][0.3cm]{10.275222cm}{\raggedright (\href{https://www.nndc.bnl.gov/nsr/nsrlink.jsp?1963Gi09,B}{1963Gi09}) measured the relative intensities of the 238- and 275-keV \ensuremath{\gamma}\vspace{0.1cm}}&\\
&&&&&&&&&&&&&&&\parbox[t][0.3cm]{10.275222cm}{\raggedright {\ }{\ }{\ }rays and deduced a ratio of \ensuremath{\sim}15:1 assuming isotropic emissions of the \ensuremath{\gamma}\vspace{0.1cm}}&\\
&&&&&&&&&&&&&&&\parbox[t][0.3cm]{10.275222cm}{\raggedright {\ }{\ }{\ }rays and conversion electrons. (\href{https://www.nndc.bnl.gov/nsr/nsrlink.jsp?1963Gi09,B}{1963Gi09}) also measured the relative\vspace{0.1cm}}&\\
&&&&&&&&&&&&&&&\parbox[t][0.3cm]{10.275222cm}{\raggedright {\ }{\ }{\ }intensity of the 275-keV internal conversion electrons to that the 238-keV\vspace{0.1cm}}&\\
&&&&&&&&&&&&&&&\parbox[t][0.3cm]{10.275222cm}{\raggedright {\ }{\ }{\ }internal conversion electrons and obtained a ratio of 1:5. From these\vspace{0.1cm}}&\\
&&&&&&&&&&&&&&&\parbox[t][0.3cm]{10.275222cm}{\raggedright {\ }{\ }{\ }ratios, (\href{https://www.nndc.bnl.gov/nsr/nsrlink.jsp?1963Gi09,B}{1963Gi09}) concluded that the 238-keV transitions$'$ internal\vspace{0.1cm}}&\\
&&&&&&&&&&&&&&&\parbox[t][0.3cm]{10.275222cm}{\raggedright {\ }{\ }{\ }conversion coefficient is ten times or more larger than that for the\vspace{0.1cm}}&\\
&&&&&&&&&&&&&&&\parbox[t][0.3cm]{10.275222cm}{\raggedright {\ }{\ }{\ }275-keV transition. This result indicates that the 238-keV state has a\vspace{0.1cm}}&\\
&&&&&&&&&&&&&&&\parbox[t][0.3cm]{10.275222cm}{\raggedright {\ }{\ }{\ }larger total angular momentum than the 275-keV level.\vspace{0.1cm}}&\\
&&&&&&&&&&&&&&&\parbox[t][0.3cm]{10.275222cm}{\raggedright a\ensuremath{_{\textnormal{2}}}=+0.42 \textit{4} and a\ensuremath{_{\textnormal{4}}}={\textminus}0.11 \textit{5} (\href{https://www.nndc.bnl.gov/nsr/nsrlink.jsp?1970Gi09,B}{1970Gi09}): Angular correlation coefficients.\vspace{0.1cm}}&\\
&&&&&&&&&&&&&&&\parbox[t][0.3cm]{10.275222cm}{\raggedright {\ }{\ }{\ }See also a\ensuremath{_{\textnormal{2}}}=0.21 \textit{3} and a\ensuremath{_{\textnormal{4}}}=0.07 \textit{2} (\href{https://www.nndc.bnl.gov/nsr/nsrlink.jsp?1969Bl02,B}{1969Bl02}), who reported that these\vspace{0.1cm}}&\\
&&&&&&&&&&&&&&&\parbox[t][0.3cm]{10.275222cm}{\raggedright {\ }{\ }{\ }values have to possibly be increased by corrections (not provided) for\vspace{0.1cm}}&\\
&&&&&&&&&&&&&&&\parbox[t][0.3cm]{10.275222cm}{\raggedright {\ }{\ }{\ }fast relaxation processes of the spin precession.\vspace{0.1cm}}&\\
\multicolumn{1}{r@{}}{275}&\multicolumn{1}{@{.}l}{1}&\multicolumn{1}{l}{1/2\ensuremath{^{-}}}&\multicolumn{1}{r@{}}{275}&\multicolumn{1}{@{.}l}{1 {\it 2}}&\multicolumn{1}{r@{}}{}&\multicolumn{1}{@{}l}{}&\multicolumn{1}{r@{}}{0}&\multicolumn{1}{@{}l}{}&\multicolumn{1}{@{}l}{1/2\ensuremath{^{+}}}&\multicolumn{1}{l}{E1}&\multicolumn{1}{r@{}}{}&\multicolumn{1}{@{}l}{}&\multicolumn{1}{r@{}}{1}&\multicolumn{1}{@{.}l}{40\ensuremath{\times10^{-4}} {\it 2}}&\parbox[t][0.3cm]{10.275222cm}{\raggedright B(E1)(W.u.)\ensuremath{>}1.5\ensuremath{\times}10\ensuremath{^{\textnormal{$-$4}}}\vspace{0.1cm}}&\\
&&&&&&&&&&&&&&&\parbox[t][0.3cm]{10.275222cm}{\raggedright \ensuremath{\alpha}(K)=0.0001328 \textit{19}; \ensuremath{\alpha}(L)=7.36\ensuremath{\times}10\ensuremath{^{\textnormal{$-$6}}} \textit{10}\vspace{0.1cm}}&\\
&&&&&&&&&&&&&&&\parbox[t][0.3cm]{10.275222cm}{\raggedright E\ensuremath{_{\gamma}}: From (\href{https://www.nndc.bnl.gov/nsr/nsrlink.jsp?1970Gi09,B}{1970Gi09}).\vspace{0.1cm}}&\\
&&&&&&&&&&&&&&&\parbox[t][0.3cm]{10.275222cm}{\raggedright E\ensuremath{_{\gamma}}: See also 281 keV \textit{8} (\href{https://www.nndc.bnl.gov/nsr/nsrlink.jsp?1957Ba09,B}{1957Ba09}); 271 keV \textit{4} (\href{https://www.nndc.bnl.gov/nsr/nsrlink.jsp?1963Gi09,B}{1963Gi09}); and 276 keV\vspace{0.1cm}}&\\
&&&&&&&&&&&&&&&\parbox[t][0.3cm]{10.275222cm}{\raggedright {\ }{\ }{\ }\textit{2} (\href{https://www.nndc.bnl.gov/nsr/nsrlink.jsp?1969Bl02,B}{1969Bl02}). See also 275 keV (\href{https://www.nndc.bnl.gov/nsr/nsrlink.jsp?1984Pi07,B}{1984Pi07}: \ensuremath{^{\textnormal{19}}}F(p,n\ensuremath{\gamma}) and \ensuremath{^{\textnormal{19}}}F(d,2n\ensuremath{\gamma})).\vspace{0.1cm}}&\\
&&&&&&&&&&&&&&&\parbox[t][0.3cm]{10.275222cm}{\raggedright Mult.: From (\href{https://www.nndc.bnl.gov/nsr/nsrlink.jsp?1969Bl02,B}{1969Bl02}, \href{https://www.nndc.bnl.gov/nsr/nsrlink.jsp?1970Gi09,B}{1970Gi09}).\vspace{0.1cm}}&\\
&&&&&&&&&&&&&&&\parbox[t][0.3cm]{10.275222cm}{\raggedright a\ensuremath{_{\textnormal{2}}}=0.00 \textit{2} (\href{https://www.nndc.bnl.gov/nsr/nsrlink.jsp?1970Gi09,B}{1970Gi09}): Angular correlation coefficient.\vspace{0.1cm}}&\\
&&&&&&&&&&&&&&&\parbox[t][0.3cm]{10.275222cm}{\raggedright B(E1)(W.u.): See also B(E1)\ensuremath{>}1.2\ensuremath{\times}10\ensuremath{^{\textnormal{$-$4}}} W.u. (\href{https://www.nndc.bnl.gov/nsr/nsrlink.jsp?1969Bl02,B}{1969Bl02}).\vspace{0.1cm}}&\\
\multicolumn{1}{r@{}}{1507}&\multicolumn{1}{@{.}l}{9}&\multicolumn{1}{l}{5/2\ensuremath{^{-}}}&\multicolumn{1}{r@{}}{1232}&\multicolumn{1}{@{.}l}{8 {\it 3}}&\multicolumn{1}{r@{}}{88}&\multicolumn{1}{@{ }l}{{\it 3}}&\multicolumn{1}{r@{}}{275}&\multicolumn{1}{@{.}l}{1 }&\multicolumn{1}{@{}l}{1/2\ensuremath{^{-}}}&\multicolumn{1}{l}{E2+M3}&\multicolumn{1}{r@{}}{$-$0}&\multicolumn{1}{@{.}l}{1 {\it 2}}&\multicolumn{1}{r@{}}{2}&\multicolumn{1}{@{.}l}{22\ensuremath{\times10^{-5}} {\it 4}}&\parbox[t][0.3cm]{10.275222cm}{\raggedright B(E2)(W.u.)=3\ensuremath{\times}10\ensuremath{^{\textnormal{1}}} \textit{+6{\textminus}2}\vspace{0.1cm}}&\\
&&&&&&&&&&&&&&&\parbox[t][0.3cm]{10.275222cm}{\raggedright \ensuremath{\alpha}(K)=7.0\ensuremath{\times}10\ensuremath{^{\textnormal{$-$6}}} \textit{8}; \ensuremath{\alpha}(L)=3.9\ensuremath{\times}10\ensuremath{^{\textnormal{$-$7}}} \textit{5}\vspace{0.1cm}}&\\
&&&&&&&&&&&&&&&\parbox[t][0.3cm]{10.275222cm}{\raggedright \ensuremath{\alpha}(IPF)=1.48\ensuremath{\times}10\ensuremath{^{\textnormal{$-$5}}} \textit{11}\vspace{0.1cm}}&\\
&&&&&&&&&&&&&&&\parbox[t][0.3cm]{10.275222cm}{\raggedright E\ensuremath{_{\gamma}}: From (\href{https://www.nndc.bnl.gov/nsr/nsrlink.jsp?1970Gi09,B}{1970Gi09}). See also 1236 keV (\href{https://www.nndc.bnl.gov/nsr/nsrlink.jsp?1987Ra23,B}{1987Ra23}).\vspace{0.1cm}}&\\
\end{longtable}
\clearpage
\begin{longtable}{ccccccccc@{}ccccc@{\extracolsep{\fill}}c}
\\[-.4cm]
\multicolumn{15}{c}{{\bf \small \underline{\ensuremath{^{\textnormal{19}}}F(p,n),(p,n\ensuremath{\gamma}),(d,2n\ensuremath{\gamma})\hspace{0.2in}\href{https://www.nndc.bnl.gov/nsr/nsrlink.jsp?1970Gi09,B}{1970Gi09},\href{https://www.nndc.bnl.gov/nsr/nsrlink.jsp?1977Le03,B}{1977Le03} (continued)}}}\\
\multicolumn{15}{c}{~}\\
\multicolumn{15}{c}{\underline{$\gamma$($^{19}$Ne) (continued)}}\\
\multicolumn{15}{c}{~~~}\\
\multicolumn{2}{c}{E\ensuremath{_{i}}(level)}&J\ensuremath{^{\pi}_{i}}&\multicolumn{2}{c}{E\ensuremath{_{\gamma}}}&\multicolumn{2}{c}{I\ensuremath{_{\ensuremath{\gamma}}} (\%)\ensuremath{^{\hyperlink{NE31GAMMA3}{d}}}}&\multicolumn{2}{c}{E\ensuremath{_{f}}}&J\ensuremath{^{\pi}_{f}}&Mult.&\multicolumn{2}{c}{\ensuremath{\alpha}\ensuremath{^{\hyperlink{NE31GAMMA5}{f}}}}&Comments&\\[-.2cm]
\multicolumn{2}{c}{\hrulefill}&\hrulefill&\multicolumn{2}{c}{\hrulefill}&\multicolumn{2}{c}{\hrulefill}&\multicolumn{2}{c}{\hrulefill}&\hrulefill&\hrulefill&\multicolumn{2}{c}{\hrulefill}&\hrulefill&
\endhead
&&&&&&&&&&&&&\parbox[t][0.3cm]{11.467021cm}{\raggedright Mult.: From the transition strength and lifetime measurement of (\href{https://www.nndc.bnl.gov/nsr/nsrlink.jsp?1970Gi09,B}{1970Gi09}) using\vspace{0.1cm}}&\\
&&&&&&&&&&&&&\parbox[t][0.3cm]{11.467021cm}{\raggedright {\ }{\ }{\ }the \ensuremath{^{\textnormal{16}}}O(\ensuremath{\alpha},n\ensuremath{\gamma}) reaction, which led to exclusion of the M2 and higher\vspace{0.1cm}}&\\
&&&&&&&&&&&&&\parbox[t][0.3cm]{11.467021cm}{\raggedright {\ }{\ }{\ }multipolarities for this transition. See also E2 assumed by (\href{https://www.nndc.bnl.gov/nsr/nsrlink.jsp?1977Le03,B}{1977Le03}).\vspace{0.1cm}}&\\
&&&&&&&&&&&&&\parbox[t][0.3cm]{11.467021cm}{\raggedright a\ensuremath{_{\textnormal{2}}}=+0.46 \textit{8} and a\ensuremath{_{\textnormal{4}}}={\textminus}0.34 \textit{11} (\href{https://www.nndc.bnl.gov/nsr/nsrlink.jsp?1970Gi09,B}{1970Gi09}).\vspace{0.1cm}}&\\
&&&&&&&&&&&&&\parbox[t][0.3cm]{11.467021cm}{\raggedright B(E2)(W.u.): See also 21 W.u. \textit{10} (\href{https://www.nndc.bnl.gov/nsr/nsrlink.jsp?1970Gi09,B}{1970Gi09}) and 41 W.u. \textit{+22{\textminus}11} (\href{https://www.nndc.bnl.gov/nsr/nsrlink.jsp?1977Le03,B}{1977Le03}), which is\vspace{0.1cm}}&\\
&&&&&&&&&&&&&\parbox[t][0.3cm]{11.467021cm}{\raggedright {\ }{\ }{\ }deduced using the mixing ratio from (\href{https://www.nndc.bnl.gov/nsr/nsrlink.jsp?1970Gi09,B}{1970Gi09}).\vspace{0.1cm}}&\\
&&&&&&&&&&&&&\parbox[t][0.3cm]{11.467021cm}{\raggedright F(\ensuremath{\tau})=0.20 \textit{4}: Weighted average of 0.18 \textit{5} and 0.22 \textit{6}, both measured by (\href{https://www.nndc.bnl.gov/nsr/nsrlink.jsp?1970Gi09,B}{1970Gi09})\vspace{0.1cm}}&\\
&&&&&&&&&&&&&\parbox[t][0.3cm]{11.467021cm}{\raggedright {\ }{\ }{\ }at E\ensuremath{_{\textnormal{p}}}=8.75 and 10.4 MeV, respectively. See also the Doppler shift of 0.55 keV \textit{50}\vspace{0.1cm}}&\\
&&&&&&&&&&&&&\parbox[t][0.3cm]{11.467021cm}{\raggedright {\ }{\ }{\ }(\href{https://www.nndc.bnl.gov/nsr/nsrlink.jsp?1977Le03,B}{1977Le03}: See Fig. 2).\vspace{0.1cm}}&\\
\multicolumn{1}{r@{}}{1507}&\multicolumn{1}{@{.}l}{9}&\multicolumn{1}{l}{5/2\ensuremath{^{-}}}&\multicolumn{1}{r@{}}{1269}&\multicolumn{1}{@{.}l}{7}&\multicolumn{1}{r@{}}{12}&\multicolumn{1}{@{}l}{\ensuremath{^{\hyperlink{NE31GAMMA2}{c}}} {\it 3}}&\multicolumn{1}{r@{}}{238}&\multicolumn{1}{@{.}l}{1 }&\multicolumn{1}{@{}l}{5/2\ensuremath{^{+}}}&\multicolumn{1}{l}{E1}&\multicolumn{1}{r@{}}{1}&\multicolumn{1}{@{.}l}{09\ensuremath{\times10^{-4}} {\it 2}}&\parbox[t][0.3cm]{11.467021cm}{\raggedright B(E1)(W.u.)=3\ensuremath{\times}10\ensuremath{^{\textnormal{$-$5}}} \textit{+6{\textminus}2}\vspace{0.1cm}}&\\
&&&&&&&&&&&&&\parbox[t][0.3cm]{11.467021cm}{\raggedright \ensuremath{\alpha}(K)=3.22\ensuremath{\times}10\ensuremath{^{\textnormal{$-$6}}} \textit{5}; \ensuremath{\alpha}(L)=1.781\ensuremath{\times}10\ensuremath{^{\textnormal{$-$7}}} \textit{25}\vspace{0.1cm}}&\\
&&&&&&&&&&&&&\parbox[t][0.3cm]{11.467021cm}{\raggedright \ensuremath{\alpha}(IPF)=0.0001052 \textit{15}\vspace{0.1cm}}&\\
&&&&&&&&&&&&&\parbox[t][0.3cm]{11.467021cm}{\raggedright E\ensuremath{_{\gamma}}: Deduced by evaluator from level-energy difference corrected for recoil energy\vspace{0.1cm}}&\\
&&&&&&&&&&&&&\parbox[t][0.3cm]{11.467021cm}{\raggedright {\ }{\ }{\ }based on the (\href{https://www.nndc.bnl.gov/nsr/nsrlink.jsp?1970Gi09,B}{1970Gi09}) results, where this transition was unresolved from the\vspace{0.1cm}}&\\
&&&&&&&&&&&&&\parbox[t][0.3cm]{11.467021cm}{\raggedright {\ }{\ }{\ }\ensuremath{^{\textnormal{19}}}Ne*(1536)\ensuremath{\rightarrow}\ensuremath{^{\textnormal{19}}}Ne*(275) transition.\vspace{0.1cm}}&\\
&&&&&&&&&&&&&\parbox[t][0.3cm]{11.467021cm}{\raggedright Mult.: From (\href{https://www.nndc.bnl.gov/nsr/nsrlink.jsp?1970Gi09,B}{1970Gi09}). See also E1 assumed in (\href{https://www.nndc.bnl.gov/nsr/nsrlink.jsp?1977Le03,B}{1977Le03}), where this transition\vspace{0.1cm}}&\\
&&&&&&&&&&&&&\parbox[t][0.3cm]{11.467021cm}{\raggedright {\ }{\ }{\ }was not observed.\vspace{0.1cm}}&\\
&&&&&&&&&&&&&\parbox[t][0.3cm]{11.467021cm}{\raggedright B(E1)(W.u.): See also 2.0\ensuremath{\times}10\ensuremath{^{\textnormal{$-$5}}} W.u. \textit{10} (\href{https://www.nndc.bnl.gov/nsr/nsrlink.jsp?1970Gi09,B}{1970Gi09}) and 4.0\ensuremath{\times}10\ensuremath{^{\textnormal{$-$5}}} W.u. \textit{+22{\textminus}14}\vspace{0.1cm}}&\\
&&&&&&&&&&&&&\parbox[t][0.3cm]{11.467021cm}{\raggedright {\ }{\ }{\ }(\href{https://www.nndc.bnl.gov/nsr/nsrlink.jsp?1977Le03,B}{1977Le03}).\vspace{0.1cm}}&\\
&&&\multicolumn{1}{r@{}}{1507}&\multicolumn{1}{@{.}l}{8\ensuremath{^{\hyperlink{NE31GAMMA0}{a}}}}&\multicolumn{1}{r@{}}{$<$3}&\multicolumn{1}{@{}l}{\ensuremath{^{\hyperlink{NE31GAMMA1}{b}}}}&\multicolumn{1}{r@{}}{0}&\multicolumn{1}{@{}l}{}&\multicolumn{1}{@{}l}{1/2\ensuremath{^{+}}}&\multicolumn{1}{l}{M2}&\multicolumn{1}{r@{}}{3}&\multicolumn{1}{@{.}l}{08\ensuremath{\times10^{-5}} {\it 4}}&\parbox[t][0.3cm]{11.467021cm}{\raggedright \ensuremath{\alpha}(K)=6.29\ensuremath{\times}10\ensuremath{^{\textnormal{$-$6}}} \textit{9}; \ensuremath{\alpha}(L)=3.48\ensuremath{\times}10\ensuremath{^{\textnormal{$-$7}}} \textit{5}\vspace{0.1cm}}&\\
&&&&&&&&&&&&&\parbox[t][0.3cm]{11.467021cm}{\raggedright \ensuremath{\alpha}(IPF)=2.414\ensuremath{\times}10\ensuremath{^{\textnormal{$-$5}}} \textit{34}\vspace{0.1cm}}&\\
&&&&&&&&&&&&&\parbox[t][0.3cm]{11.467021cm}{\raggedright Mult.: From (\href{https://www.nndc.bnl.gov/nsr/nsrlink.jsp?1970Gi09,B}{1970Gi09}).\vspace{0.1cm}}&\\
&&&&&&&&&&&&&\parbox[t][0.3cm]{11.467021cm}{\raggedright B(M2)\ensuremath{<}9 W.u. (\href{https://www.nndc.bnl.gov/nsr/nsrlink.jsp?1970Gi09,B}{1970Gi09}).\vspace{0.1cm}}&\\
\multicolumn{1}{r@{}}{1536}&\multicolumn{1}{@{.}l}{2}&\multicolumn{1}{l}{3/2\ensuremath{^{+}}}&\multicolumn{1}{r@{}}{1260}&\multicolumn{1}{@{.}l}{7}&\multicolumn{1}{r@{}}{5}&\multicolumn{1}{@{}l}{\ensuremath{^{\hyperlink{NE31GAMMA2}{c}}} {\it 3}}&\multicolumn{1}{r@{}}{275}&\multicolumn{1}{@{.}l}{1 }&\multicolumn{1}{@{}l}{1/2\ensuremath{^{-}}}&\multicolumn{1}{l}{E1}&\multicolumn{1}{r@{}}{1}&\multicolumn{1}{@{.}l}{02\ensuremath{\times10^{-4}} {\it 1}}&\parbox[t][0.3cm]{11.467021cm}{\raggedright B(E1)(W.u.)=0.0011 \textit{+13{\textminus}6}\vspace{0.1cm}}&\\
&&&&&&&&&&&&&\parbox[t][0.3cm]{11.467021cm}{\raggedright \ensuremath{\alpha}(K)=3.26\ensuremath{\times}10\ensuremath{^{\textnormal{$-$6}}} \textit{5}; \ensuremath{\alpha}(L)=1.803\ensuremath{\times}10\ensuremath{^{\textnormal{$-$7}}} \textit{25}\vspace{0.1cm}}&\\
&&&&&&&&&&&&&\parbox[t][0.3cm]{11.467021cm}{\raggedright \ensuremath{\alpha}(IPF)=9.90\ensuremath{\times}10\ensuremath{^{\textnormal{$-$5}}} \textit{14}\vspace{0.1cm}}&\\
&&&&&&&&&&&&&\parbox[t][0.3cm]{11.467021cm}{\raggedright E\ensuremath{_{\gamma}}: Deduced by evaluator from level-energy difference corrected for recoil energy\vspace{0.1cm}}&\\
&&&&&&&&&&&&&\parbox[t][0.3cm]{11.467021cm}{\raggedright {\ }{\ }{\ }based on the (\href{https://www.nndc.bnl.gov/nsr/nsrlink.jsp?1970Gi09,B}{1970Gi09}) results, where this transition was unresolved from the\vspace{0.1cm}}&\\
&&&&&&&&&&&&&\parbox[t][0.3cm]{11.467021cm}{\raggedright {\ }{\ }{\ }\ensuremath{^{\textnormal{19}}}Ne*(1508)\ensuremath{\rightarrow}\ensuremath{^{\textnormal{19}}}Ne*(238) transition.\vspace{0.1cm}}&\\
&&&&&&&&&&&&&\parbox[t][0.3cm]{11.467021cm}{\raggedright Mult.: From (\href{https://www.nndc.bnl.gov/nsr/nsrlink.jsp?1970Gi09,B}{1970Gi09}).\vspace{0.1cm}}&\\
&&&&&&&&&&&&&\parbox[t][0.3cm]{11.467021cm}{\raggedright B(E1)(W.u.): See also 1.2\ensuremath{\times}10\ensuremath{^{\textnormal{$-$3}}} W.u. \textit{+20{\textminus}8} (\href{https://www.nndc.bnl.gov/nsr/nsrlink.jsp?1970Gi09,B}{1970Gi09}) and 1.1\ensuremath{\times}10\ensuremath{^{\textnormal{$-$3}}} W.u. \textit{+9{\textminus}7}\vspace{0.1cm}}&\\
&&&&&&&&&&&&&\parbox[t][0.3cm]{11.467021cm}{\raggedright {\ }{\ }{\ }(\href{https://www.nndc.bnl.gov/nsr/nsrlink.jsp?1977Le03,B}{1977Le03}).\vspace{0.1cm}}&\\
&&&\multicolumn{1}{r@{}}{1298}&\multicolumn{1}{@{.}l}{0 {\it 3}}&\multicolumn{1}{r@{}}{95}&\multicolumn{1}{@{ }l}{{\it 3}}&\multicolumn{1}{r@{}}{238}&\multicolumn{1}{@{.}l}{1 }&\multicolumn{1}{@{}l}{5/2\ensuremath{^{+}}}&\multicolumn{1}{l}{M1}&\multicolumn{1}{r@{}}{2}&\multicolumn{1}{@{.}l}{429\ensuremath{\times10^{-5}} {\it 34}}&\parbox[t][0.3cm]{11.467021cm}{\raggedright B(M1)(W.u.)=0.42 \textit{+40{\textminus}15}\vspace{0.1cm}}&\\
&&&&&&&&&&&&&\parbox[t][0.3cm]{11.467021cm}{\raggedright \ensuremath{\alpha}(K)=4.64\ensuremath{\times}10\ensuremath{^{\textnormal{$-$6}}} \textit{6}; \ensuremath{\alpha}(L)=2.57\ensuremath{\times}10\ensuremath{^{\textnormal{$-$7}}} \textit{4}\vspace{0.1cm}}&\\
&&&&&&&&&&&&&\parbox[t][0.3cm]{11.467021cm}{\raggedright \ensuremath{\alpha}(IPF)=1.939\ensuremath{\times}10\ensuremath{^{\textnormal{$-$5}}} \textit{28}\vspace{0.1cm}}&\\
&&&&&&&&&&&&&\parbox[t][0.3cm]{11.467021cm}{\raggedright E\ensuremath{_{\gamma}}: Weighted average of 1298.0 keV \textit{4} (\href{https://www.nndc.bnl.gov/nsr/nsrlink.jsp?1970Gi09,B}{1970Gi09}) and 1297.9 keV \textit{4} (\href{https://www.nndc.bnl.gov/nsr/nsrlink.jsp?1971It02,B}{1971It02}).\vspace{0.1cm}}&\\
&&&&&&&&&&&&&\parbox[t][0.3cm]{11.467021cm}{\raggedright Mult.: From (\href{https://www.nndc.bnl.gov/nsr/nsrlink.jsp?1970Gi09,B}{1970Gi09}). See also M1 assumed in (\href{https://www.nndc.bnl.gov/nsr/nsrlink.jsp?1977Le03,B}{1977Le03}).\vspace{0.1cm}}&\\
&&&&&&&&&&&&&\parbox[t][0.3cm]{11.467021cm}{\raggedright a\ensuremath{_{\textnormal{2}}}={\textminus}0.15 \textit{8} (\href{https://www.nndc.bnl.gov/nsr/nsrlink.jsp?1970Gi09,B}{1970Gi09}).\vspace{0.1cm}}&\\
&&&&&&&&&&&&&\parbox[t][0.3cm]{11.467021cm}{\raggedright B(M1)(W.u.): See also 0.50 W.u. \textit{+56{\textminus}17} (\href{https://www.nndc.bnl.gov/nsr/nsrlink.jsp?1970Gi09,B}{1970Gi09}) and 4.5\ensuremath{\times}10\ensuremath{^{\textnormal{$-$1}}} W.u. \textit{+26{\textminus}12}\vspace{0.1cm}}&\\
&&&&&&&&&&&&&\parbox[t][0.3cm]{11.467021cm}{\raggedright {\ }{\ }{\ }(\href{https://www.nndc.bnl.gov/nsr/nsrlink.jsp?1977Le03,B}{1977Le03}). The reported B(M1) deduced by (\href{https://www.nndc.bnl.gov/nsr/nsrlink.jsp?1977Le03,B}{1977Le03}) was determined using a\vspace{0.1cm}}&\\
\end{longtable}
\clearpage
\begin{longtable}{ccccccccc@{}ccccc@{\extracolsep{\fill}}c}
\\[-.4cm]
\multicolumn{15}{c}{{\bf \small \underline{\ensuremath{^{\textnormal{19}}}F(p,n),(p,n\ensuremath{\gamma}),(d,2n\ensuremath{\gamma})\hspace{0.2in}\href{https://www.nndc.bnl.gov/nsr/nsrlink.jsp?1970Gi09,B}{1970Gi09},\href{https://www.nndc.bnl.gov/nsr/nsrlink.jsp?1977Le03,B}{1977Le03} (continued)}}}\\
\multicolumn{15}{c}{~}\\
\multicolumn{15}{c}{\underline{$\gamma$($^{19}$Ne) (continued)}}\\
\multicolumn{15}{c}{~~~}\\
\multicolumn{2}{c}{E\ensuremath{_{i}}(level)}&J\ensuremath{^{\pi}_{i}}&\multicolumn{2}{c}{E\ensuremath{_{\gamma}}}&\multicolumn{2}{c}{I\ensuremath{_{\ensuremath{\gamma}}} (\%)\ensuremath{^{\hyperlink{NE31GAMMA3}{d}}}}&\multicolumn{2}{c}{E\ensuremath{_{f}}}&J\ensuremath{^{\pi}_{f}}&Mult.&\multicolumn{2}{c}{\ensuremath{\alpha}\ensuremath{^{\hyperlink{NE31GAMMA5}{f}}}}&Comments&\\[-.2cm]
\multicolumn{2}{c}{\hrulefill}&\hrulefill&\multicolumn{2}{c}{\hrulefill}&\multicolumn{2}{c}{\hrulefill}&\multicolumn{2}{c}{\hrulefill}&\hrulefill&\hrulefill&\multicolumn{2}{c}{\hrulefill}&\hrulefill&
\endhead
&&&&&&&&&&&&&\parbox[t][0.3cm]{11.45418cm}{\raggedright {\ }{\ }{\ }mixing ratio of \ensuremath{\delta}=0 assuming an uncertainty on \ensuremath{\delta} corresponding to B(E2)\ensuremath{<}100\vspace{0.1cm}}&\\
&&&&&&&&&&&&&\parbox[t][0.3cm]{11.45418cm}{\raggedright {\ }{\ }{\ }W.u., which was suggested by (\href{https://www.nndc.bnl.gov/nsr/nsrlink.jsp?1974En05,B}{1974En05}).\vspace{0.1cm}}&\\
&&&&&&&&&&&&&\parbox[t][0.3cm]{11.45418cm}{\raggedright \ensuremath{\Gamma}\ensuremath{_{\ensuremath{\gamma}}}=0.02 eV estimated by (\href{https://www.nndc.bnl.gov/nsr/nsrlink.jsp?1970Gi09,B}{1970Gi09}).\vspace{0.1cm}}&\\
&&&&&&&&&&&&&\parbox[t][0.3cm]{11.45418cm}{\raggedright F(\ensuremath{\tau})=0.96 \textit{5}: Weighted average of 0.92 \textit{9} and 0.97 \textit{5} both measured by (\href{https://www.nndc.bnl.gov/nsr/nsrlink.jsp?1970Gi09,B}{1970Gi09})\vspace{0.1cm}}&\\
&&&&&&&&&&&&&\parbox[t][0.3cm]{11.45418cm}{\raggedright {\ }{\ }{\ }at E\ensuremath{_{\textnormal{p}}}=8.75 and 10.4 MeV, respectively.\vspace{0.1cm}}&\\
&&&&&&&&&&&&&\parbox[t][0.3cm]{11.45418cm}{\raggedright See also F(\ensuremath{\tau})=0.89 \textit{3} (\href{https://www.nndc.bnl.gov/nsr/nsrlink.jsp?1971It02,B}{1971It02}): Weighted average of 0.91 \textit{10}; 0.88 \textit{3}; and 0.89 \textit{3}\vspace{0.1cm}}&\\
&&&&&&&&&&&&&\parbox[t][0.3cm]{11.45418cm}{\raggedright {\ }{\ }{\ }for E\ensuremath{_{\textnormal{p}}}=6.35 MeV and a recoil velocity of 0.6\% of the speed of light measured\vspace{0.1cm}}&\\
&&&&&&&&&&&&&\parbox[t][0.3cm]{11.45418cm}{\raggedright {\ }{\ }{\ }using a 11.3-mg/cm\ensuremath{^{\textnormal{2}}}-thick CaF\ensuremath{_{\textnormal{2}}} target on Au backing, and a 0.4-mg/cm\ensuremath{^{\textnormal{2}}}-thick\vspace{0.1cm}}&\\
&&&&&&&&&&&&&\parbox[t][0.3cm]{11.45418cm}{\raggedright {\ }{\ }{\ }CaF\ensuremath{_{\textnormal{2}}} target on Ni and on Au backings, respectively. The F(\ensuremath{\tau})=0.91 \textit{10} resulted in\vspace{0.1cm}}&\\
&&&&&&&&&&&&&\parbox[t][0.3cm]{11.45418cm}{\raggedright {\ }{\ }{\ }the lifetime of the \ensuremath{^{\textnormal{19}}}Ne*(1536) level to be \ensuremath{\tau}\ensuremath{<}76 fs (\href{https://www.nndc.bnl.gov/nsr/nsrlink.jsp?1971It02,B}{1971It02}). Using the thick\vspace{0.1cm}}&\\
&&&&&&&&&&&&&\parbox[t][0.3cm]{11.45418cm}{\raggedright {\ }{\ }{\ }target, the \ensuremath{\gamma}-ray energy was shifted by \ensuremath{\Delta}E\ensuremath{_{\ensuremath{\gamma}}}=6.56 keV. Using the thin target\vspace{0.1cm}}&\\
&&&&&&&&&&&&&\parbox[t][0.3cm]{11.45418cm}{\raggedright {\ }{\ }{\ }with Ni and Au backings resulted in \ensuremath{\Delta}E\ensuremath{_{\ensuremath{\gamma}}}=6.67 keV \textit{20} and 6.74 keV \textit{20},\vspace{0.1cm}}&\\
&&&&&&&&&&&&&\parbox[t][0.3cm]{11.45418cm}{\raggedright {\ }{\ }{\ }respectively. See also F(\ensuremath{\tau})=87\% \textit{7} (\href{https://www.nndc.bnl.gov/nsr/nsrlink.jsp?1977Le03,B}{1977Le03}), where a Doppler shift of 8.3 keV\vspace{0.1cm}}&\\
&&&&&&&&&&&&&\parbox[t][0.3cm]{11.45418cm}{\raggedright {\ }{\ }{\ }\textit{5} was measured for this \ensuremath{\gamma} ray transition, see Fig. 2.\vspace{0.1cm}}&\\
\multicolumn{1}{r@{}}{1536}&\multicolumn{1}{@{.}l}{2}&\multicolumn{1}{l}{3/2\ensuremath{^{+}}}&\multicolumn{1}{r@{}}{1535}&\multicolumn{1}{@{.}l}{7}&\multicolumn{1}{r@{}}{$<$6}&\multicolumn{1}{@{}l}{\ensuremath{^{\hyperlink{NE31GAMMA1}{b}}}}&\multicolumn{1}{r@{}}{0}&\multicolumn{1}{@{}l}{}&\multicolumn{1}{@{}l}{1/2\ensuremath{^{+}}}&\multicolumn{1}{l}{M1}&\multicolumn{1}{r@{}}{7}&\multicolumn{1}{@{.}l}{69\ensuremath{\times10^{-5}} {\it 11}}&\parbox[t][0.3cm]{11.45418cm}{\raggedright B(M1)(W.u.)\ensuremath{<}0.033\vspace{0.1cm}}&\\
&&&&&&&&&&&&&\parbox[t][0.3cm]{11.45418cm}{\raggedright \ensuremath{\alpha}(K)=3.48\ensuremath{\times}10\ensuremath{^{\textnormal{$-$6}}} \textit{5}; \ensuremath{\alpha}(L)=1.927\ensuremath{\times}10\ensuremath{^{\textnormal{$-$7}}} \textit{27}\vspace{0.1cm}}&\\
&&&&&&&&&&&&&\parbox[t][0.3cm]{11.45418cm}{\raggedright \ensuremath{\alpha}(IPF)=7.33\ensuremath{\times}10\ensuremath{^{\textnormal{$-$5}}} \textit{10}\vspace{0.1cm}}&\\
&&&&&&&&&&&&&\parbox[t][0.3cm]{11.45418cm}{\raggedright E\ensuremath{_{\gamma}}: From (\href{https://www.nndc.bnl.gov/nsr/nsrlink.jsp?1970Gi09,B}{1970Gi09}: See Fig. 1) measured at E\ensuremath{_{\textnormal{p}}}=8.48 MeV; however, this energy\vspace{0.1cm}}&\\
&&&&&&&&&&&&&\parbox[t][0.3cm]{11.45418cm}{\raggedright {\ }{\ }{\ }is not reported in their Table 1 of (\href{https://www.nndc.bnl.gov/nsr/nsrlink.jsp?1970Gi09,B}{1970Gi09}) potentially due to low statistics.\vspace{0.1cm}}&\\
&&&&&&&&&&&&&\parbox[t][0.3cm]{11.45418cm}{\raggedright Mult.: From (\href{https://www.nndc.bnl.gov/nsr/nsrlink.jsp?1970Gi09,B}{1970Gi09}).\vspace{0.1cm}}&\\
&&&&&&&&&&&&&\parbox[t][0.3cm]{11.45418cm}{\raggedright B(M1)(W.u.): See also B(M1)\ensuremath{<}4.1\ensuremath{\times}10\ensuremath{^{\textnormal{$-$2}}} W.u. (\href{https://www.nndc.bnl.gov/nsr/nsrlink.jsp?1970Gi09,B}{1970Gi09}).\vspace{0.1cm}}&\\
\multicolumn{1}{r@{}}{1615}&\multicolumn{1}{@{.}l}{4}&\multicolumn{1}{l}{(3/2\ensuremath{^{-}})}&\multicolumn{1}{r@{}}{1340}&\multicolumn{1}{@{.}l}{1 {\it 4}}&\multicolumn{1}{r@{}}{70}&\multicolumn{1}{@{ }l}{{\it 4}}&\multicolumn{1}{r@{}}{275}&\multicolumn{1}{@{.}l}{1 }&\multicolumn{1}{@{}l}{1/2\ensuremath{^{-}}}&\multicolumn{1}{l}{M1}&\multicolumn{1}{r@{}}{3}&\multicolumn{1}{@{.}l}{12\ensuremath{\times10^{-5}} {\it 4}}&\parbox[t][0.3cm]{11.45418cm}{\raggedright B(M1)(W.u.)=0.059 \textit{+22{\textminus}13}\vspace{0.1cm}}&\\
&&&&&&&&&&&&&\parbox[t][0.3cm]{11.45418cm}{\raggedright \ensuremath{\alpha}(K)=4.39\ensuremath{\times}10\ensuremath{^{\textnormal{$-$6}}} \textit{6}; \ensuremath{\alpha}(L)=2.430\ensuremath{\times}10\ensuremath{^{\textnormal{$-$7}}} \textit{34}\vspace{0.1cm}}&\\
&&&&&&&&&&&&&\parbox[t][0.3cm]{11.45418cm}{\raggedright \ensuremath{\alpha}(IPF)=2.66\ensuremath{\times}10\ensuremath{^{\textnormal{$-$5}}} \textit{4}\vspace{0.1cm}}&\\
&&&&&&&&&&&&&\parbox[t][0.3cm]{11.45418cm}{\raggedright E\ensuremath{_{\gamma}}: From (\href{https://www.nndc.bnl.gov/nsr/nsrlink.jsp?1970Gi09,B}{1970Gi09}).\vspace{0.1cm}}&\\
&&&&&&&&&&&&&\parbox[t][0.3cm]{11.45418cm}{\raggedright Mult.: From (\href{https://www.nndc.bnl.gov/nsr/nsrlink.jsp?1970Gi09,B}{1970Gi09}). See also M1 assumed by (\href{https://www.nndc.bnl.gov/nsr/nsrlink.jsp?1977Le03,B}{1977Le03}).\vspace{0.1cm}}&\\
&&&&&&&&&&&&&\parbox[t][0.3cm]{11.45418cm}{\raggedright a\ensuremath{_{\textnormal{2}}}=0.00 \textit{8} (\href{https://www.nndc.bnl.gov/nsr/nsrlink.jsp?1970Gi09,B}{1970Gi09}).\vspace{0.1cm}}&\\
&&&&&&&&&&&&&\parbox[t][0.3cm]{11.45418cm}{\raggedright B(M1)(W.u.): See also 5.1\ensuremath{\times}10\ensuremath{^{\textnormal{$-$2}}} W.u. \textit{+26{\textminus}13} (\href{https://www.nndc.bnl.gov/nsr/nsrlink.jsp?1970Gi09,B}{1970Gi09}) and 6.5\ensuremath{\times}10\ensuremath{^{\textnormal{$-$2}}} W.u. \textit{19}\vspace{0.1cm}}&\\
&&&&&&&&&&&&&\parbox[t][0.3cm]{11.45418cm}{\raggedright {\ }{\ }{\ }(\href{https://www.nndc.bnl.gov/nsr/nsrlink.jsp?1977Le03,B}{1977Le03}). This value was determined using a mixing ratio of \ensuremath{\delta}=0 assuming an\vspace{0.1cm}}&\\
&&&&&&&&&&&&&\parbox[t][0.3cm]{11.45418cm}{\raggedright {\ }{\ }{\ }uncertainty on \ensuremath{\delta} corresponding to B(E2)\ensuremath{<}100 W.u., which was suggested by\vspace{0.1cm}}&\\
&&&&&&&&&&&&&\parbox[t][0.3cm]{11.45418cm}{\raggedright {\ }{\ }{\ }(\href{https://www.nndc.bnl.gov/nsr/nsrlink.jsp?1974En05,B}{1974En05}).\vspace{0.1cm}}&\\
&&&&&&&&&&&&&\parbox[t][0.3cm]{11.45418cm}{\raggedright F(\ensuremath{\tau})=0.69 \textit{4}: Weighted average of 0.65 \textit{7} and 0.71 \textit{5} both measured by (\href{https://www.nndc.bnl.gov/nsr/nsrlink.jsp?1970Gi09,B}{1970Gi09})\vspace{0.1cm}}&\\
&&&&&&&&&&&&&\parbox[t][0.3cm]{11.45418cm}{\raggedright {\ }{\ }{\ }at E\ensuremath{_{\textnormal{p}}}=8.75 and 10.4 MeV, respectively. See also F(\ensuremath{\tau})=66\% \textit{7} (\href{https://www.nndc.bnl.gov/nsr/nsrlink.jsp?1977Le03,B}{1977Le03}), where\vspace{0.1cm}}&\\
&&&&&&&&&&&&&\parbox[t][0.3cm]{11.45418cm}{\raggedright {\ }{\ }{\ }a Doppler shift of 6.6 keV \textit{5} was measured, see Fig. 2.\vspace{0.1cm}}&\\
&&&\multicolumn{1}{r@{}}{1377}&\multicolumn{1}{@{.}l}{1}&\multicolumn{1}{r@{}}{10}&\multicolumn{1}{@{}l}{\ensuremath{^{\hyperlink{NE31GAMMA1}{b}}} {\it 3}}&\multicolumn{1}{r@{}}{238}&\multicolumn{1}{@{.}l}{1 }&\multicolumn{1}{@{}l}{5/2\ensuremath{^{+}}}&\multicolumn{1}{l}{E1}&\multicolumn{1}{r@{}}{1}&\multicolumn{1}{@{.}l}{81\ensuremath{\times10^{-4}} {\it 3}}&\parbox[t][0.3cm]{11.45418cm}{\raggedright B(E1)(W.u.)=3.4\ensuremath{\times}10\ensuremath{^{\textnormal{$-$4}}} \textit{+16{\textminus}11}\vspace{0.1cm}}&\\
&&&&&&&&&&&&&\parbox[t][0.3cm]{11.45418cm}{\raggedright \ensuremath{\alpha}(K)=2.80\ensuremath{\times}10\ensuremath{^{\textnormal{$-$6}}} \textit{4}; \ensuremath{\alpha}(L)=1.552\ensuremath{\times}10\ensuremath{^{\textnormal{$-$7}}} \textit{22}\vspace{0.1cm}}&\\
&&&&&&&&&&&&&\parbox[t][0.3cm]{11.45418cm}{\raggedright \ensuremath{\alpha}(IPF)=0.0001783 \textit{25}\vspace{0.1cm}}&\\
&&&&&&&&&&&&&\parbox[t][0.3cm]{11.45418cm}{\raggedright E\ensuremath{_{\gamma}}: From (\href{https://www.nndc.bnl.gov/nsr/nsrlink.jsp?1970Gi09,B}{1970Gi09}: See Fig. 1) measured at E\ensuremath{_{\textnormal{p}}}=8.48 MeV; however, this energy\vspace{0.1cm}}&\\
&&&&&&&&&&&&&\parbox[t][0.3cm]{11.45418cm}{\raggedright {\ }{\ }{\ }is not reported in their Table 1 potentially due to low statistics.\vspace{0.1cm}}&\\
\end{longtable}
\clearpage
\begin{longtable}{ccccccccc@{}ccccccc@{\extracolsep{\fill}}c}
\\[-.4cm]
\multicolumn{17}{c}{{\bf \small \underline{\ensuremath{^{\textnormal{19}}}F(p,n),(p,n\ensuremath{\gamma}),(d,2n\ensuremath{\gamma})\hspace{0.2in}\href{https://www.nndc.bnl.gov/nsr/nsrlink.jsp?1970Gi09,B}{1970Gi09},\href{https://www.nndc.bnl.gov/nsr/nsrlink.jsp?1977Le03,B}{1977Le03} (continued)}}}\\
\multicolumn{17}{c}{~}\\
\multicolumn{17}{c}{\underline{$\gamma$($^{19}$Ne) (continued)}}\\
\multicolumn{17}{c}{~~~}\\
\multicolumn{2}{c}{E\ensuremath{_{i}}(level)}&J\ensuremath{^{\pi}_{i}}&\multicolumn{2}{c}{E\ensuremath{_{\gamma}}}&\multicolumn{2}{c}{I\ensuremath{_{\ensuremath{\gamma}}} (\%)\ensuremath{^{\hyperlink{NE31GAMMA3}{d}}}}&\multicolumn{2}{c}{E\ensuremath{_{f}}}&J\ensuremath{^{\pi}_{f}}&Mult.&\multicolumn{2}{c}{\ensuremath{\delta}\ensuremath{^{\hyperlink{NE31GAMMA4}{e}}}}&\multicolumn{2}{c}{\ensuremath{\alpha}\ensuremath{^{\hyperlink{NE31GAMMA5}{f}}}}&Comments&\\[-.2cm]
\multicolumn{2}{c}{\hrulefill}&\hrulefill&\multicolumn{2}{c}{\hrulefill}&\multicolumn{2}{c}{\hrulefill}&\multicolumn{2}{c}{\hrulefill}&\hrulefill&\hrulefill&\multicolumn{2}{c}{\hrulefill}&\multicolumn{2}{c}{\hrulefill}&\hrulefill&
\endhead
&&&&&&&&&&&&&&&\parbox[t][0.3cm]{8.6787815cm}{\raggedright Mult.: From (\href{https://www.nndc.bnl.gov/nsr/nsrlink.jsp?1970Gi09,B}{1970Gi09}). See also E1 assumed by (\href{https://www.nndc.bnl.gov/nsr/nsrlink.jsp?1977Le03,B}{1977Le03}),\vspace{0.1cm}}&\\
&&&&&&&&&&&&&&&\parbox[t][0.3cm]{8.6787815cm}{\raggedright {\ }{\ }{\ }where this transition was not observed.\vspace{0.1cm}}&\\
&&&&&&&&&&&&&&&\parbox[t][0.3cm]{8.6787815cm}{\raggedright B(E1)(W.u.): See also 2.9\ensuremath{\times}10\ensuremath{^{\textnormal{$-$4}}} W.u. \textit{+17{\textminus}11} (\href{https://www.nndc.bnl.gov/nsr/nsrlink.jsp?1970Gi09,B}{1970Gi09}) and\vspace{0.1cm}}&\\
&&&&&&&&&&&&&&&\parbox[t][0.3cm]{8.6787815cm}{\raggedright {\ }{\ }{\ }3.7\ensuremath{\times}10\ensuremath{^{\textnormal{$-$4}}} W.u. \textit{14} (\href{https://www.nndc.bnl.gov/nsr/nsrlink.jsp?1977Le03,B}{1977Le03}).\vspace{0.1cm}}&\\
\multicolumn{1}{r@{}}{1615}&\multicolumn{1}{@{.}l}{4}&\multicolumn{1}{l}{(3/2\ensuremath{^{-}})}&\multicolumn{1}{r@{}}{1615}&\multicolumn{1}{@{.}l}{4 {\it 7}}&\multicolumn{1}{r@{}}{20}&\multicolumn{1}{@{ }l}{{\it 3}}&\multicolumn{1}{r@{}}{0}&\multicolumn{1}{@{}l}{}&\multicolumn{1}{@{}l}{1/2\ensuremath{^{+}}}&\multicolumn{1}{l}{E1}&\multicolumn{1}{r@{}}{}&\multicolumn{1}{@{}l}{}&\multicolumn{1}{r@{}}{3}&\multicolumn{1}{@{.}l}{61\ensuremath{\times10^{-4}} {\it 5}}&\parbox[t][0.3cm]{8.6787815cm}{\raggedright B(E1)(W.u.)=4.2\ensuremath{\times}10\ensuremath{^{\textnormal{$-$4}}} \textit{+17{\textminus}10}\vspace{0.1cm}}&\\
&&&&&&&&&&&&&&&\parbox[t][0.3cm]{8.6787815cm}{\raggedright \ensuremath{\alpha}(K)=2.169\ensuremath{\times}10\ensuremath{^{\textnormal{$-$6}}} \textit{30}; \ensuremath{\alpha}(L)=1.201\ensuremath{\times}10\ensuremath{^{\textnormal{$-$7}}} \textit{17}\vspace{0.1cm}}&\\
&&&&&&&&&&&&&&&\parbox[t][0.3cm]{8.6787815cm}{\raggedright \ensuremath{\alpha}(IPF)=0.000359 \textit{5}\vspace{0.1cm}}&\\
&&&&&&&&&&&&&&&\parbox[t][0.3cm]{8.6787815cm}{\raggedright E\ensuremath{_{\gamma}}: From (\href{https://www.nndc.bnl.gov/nsr/nsrlink.jsp?1970Gi09,B}{1970Gi09}).\vspace{0.1cm}}&\\
&&&&&&&&&&&&&&&\parbox[t][0.3cm]{8.6787815cm}{\raggedright Mult.: From (\href{https://www.nndc.bnl.gov/nsr/nsrlink.jsp?1970Gi09,B}{1970Gi09}). See also E1 assumed by (\href{https://www.nndc.bnl.gov/nsr/nsrlink.jsp?1977Le03,B}{1977Le03}),\vspace{0.1cm}}&\\
&&&&&&&&&&&&&&&\parbox[t][0.3cm]{8.6787815cm}{\raggedright {\ }{\ }{\ }where this transition was not observed.\vspace{0.1cm}}&\\
&&&&&&&&&&&&&&&\parbox[t][0.3cm]{8.6787815cm}{\raggedright a\ensuremath{_{\textnormal{2}}}={\textminus}0.21 \textit{20} (\href{https://www.nndc.bnl.gov/nsr/nsrlink.jsp?1970Gi09,B}{1970Gi09}).\vspace{0.1cm}}&\\
&&&&&&&&&&&&&&&\parbox[t][0.3cm]{8.6787815cm}{\raggedright B(E1)(W.u.): See also 3.6\ensuremath{\times}10\ensuremath{^{\textnormal{$-$4}}} W.u. \textit{+19{\textminus}10} (\href{https://www.nndc.bnl.gov/nsr/nsrlink.jsp?1970Gi09,B}{1970Gi09}) and\vspace{0.1cm}}&\\
&&&&&&&&&&&&&&&\parbox[t][0.3cm]{8.6787815cm}{\raggedright {\ }{\ }{\ }4.6\ensuremath{\times}10\ensuremath{^{\textnormal{$-$4}}} W.u. \textit{+15{\textminus}11} (\href{https://www.nndc.bnl.gov/nsr/nsrlink.jsp?1977Le03,B}{1977Le03}).\vspace{0.1cm}}&\\
\multicolumn{1}{r@{}}{2794}&\multicolumn{1}{@{.}l}{7}&\multicolumn{1}{l}{(9/2\ensuremath{^{+}})}&\multicolumn{1}{r@{}}{1179}&\multicolumn{1}{@{.}l}{3\ensuremath{^{\hyperlink{NE31GAMMA0}{a}}}}&\multicolumn{1}{r@{}}{$<$10}&\multicolumn{1}{@{}l}{\ensuremath{^{\hyperlink{NE31GAMMA1}{b}}}}&\multicolumn{1}{r@{}}{1615}&\multicolumn{1}{@{.}l}{4 }&\multicolumn{1}{@{}l}{(3/2\ensuremath{^{-}})}&&&&&&&\\
&&&\multicolumn{1}{r@{}}{1258}&\multicolumn{1}{@{.}l}{9\ensuremath{^{\hyperlink{NE31GAMMA0}{a}}}}&\multicolumn{1}{r@{}}{$<$10}&\multicolumn{1}{@{}l}{\ensuremath{^{\hyperlink{NE31GAMMA1}{b}}}}&\multicolumn{1}{r@{}}{1536}&\multicolumn{1}{@{.}l}{2 }&\multicolumn{1}{@{}l}{3/2\ensuremath{^{+}}}&&&&&&&\\
&&&\multicolumn{1}{r@{}}{1286}&\multicolumn{1}{@{.}l}{8\ensuremath{^{\hyperlink{NE31GAMMA0}{a}}}}&\multicolumn{1}{r@{}}{$<$12}&\multicolumn{1}{@{}l}{\ensuremath{^{\hyperlink{NE31GAMMA1}{b}}}}&\multicolumn{1}{r@{}}{1507}&\multicolumn{1}{@{.}l}{9 }&\multicolumn{1}{@{}l}{5/2\ensuremath{^{-}}}&&&&&&&\\
&&&\multicolumn{1}{r@{}}{2519}&\multicolumn{1}{@{.}l}{4\ensuremath{^{\hyperlink{NE31GAMMA0}{a}}}}&\multicolumn{1}{r@{}}{$<$10}&\multicolumn{1}{@{}l}{\ensuremath{^{\hyperlink{NE31GAMMA1}{b}}}}&\multicolumn{1}{r@{}}{275}&\multicolumn{1}{@{.}l}{1 }&\multicolumn{1}{@{}l}{1/2\ensuremath{^{-}}}&&&&&&&\\
&&&\multicolumn{1}{r@{}}{2556}&\multicolumn{1}{@{.}l}{2 {\it 15}}&\multicolumn{1}{r@{}}{100}&\multicolumn{1}{@{}l}{}&\multicolumn{1}{r@{}}{238}&\multicolumn{1}{@{.}l}{1 }&\multicolumn{1}{@{}l}{5/2\ensuremath{^{+}}}&\multicolumn{1}{l}{E2+M3}&\multicolumn{1}{r@{}}{$-$0}&\multicolumn{1}{@{.}l}{1 {\it 5}}&\multicolumn{1}{r@{}}{0}&\multicolumn{1}{@{.}l}{00058 {\it 10}}&\parbox[t][0.3cm]{8.6787815cm}{\raggedright B(E2)(W.u.)=14 \textit{+10{\textminus}4}\vspace{0.1cm}}&\\
&&&&&&&&&&&&&&&\parbox[t][0.3cm]{8.6787815cm}{\raggedright \ensuremath{\alpha}(K)=1.7\ensuremath{\times}10\ensuremath{^{\textnormal{$-$6}}} \textit{4}; \ensuremath{\alpha}(L)=9.4\ensuremath{\times}10\ensuremath{^{\textnormal{$-$8}}} \textit{22}\vspace{0.1cm}}&\\
&&&&&&&&&&&&&&&\parbox[t][0.3cm]{8.6787815cm}{\raggedright \ensuremath{\alpha}(IPF)=0.00058 \textit{10}\vspace{0.1cm}}&\\
&&&&&&&&&&&&&&&\parbox[t][0.3cm]{8.6787815cm}{\raggedright E\ensuremath{_{\gamma}}: From (\href{https://www.nndc.bnl.gov/nsr/nsrlink.jsp?1970Gi09,B}{1970Gi09}).\vspace{0.1cm}}&\\
&&&&&&&&&&&&&&&\parbox[t][0.3cm]{8.6787815cm}{\raggedright Mult.: From (\href{https://www.nndc.bnl.gov/nsr/nsrlink.jsp?1970Gi09,B}{1970Gi09}). See also E2 assumed by (\href{https://www.nndc.bnl.gov/nsr/nsrlink.jsp?1977Le03,B}{1977Le03}).\vspace{0.1cm}}&\\
&&&&&&&&&&&&&&&\parbox[t][0.3cm]{8.6787815cm}{\raggedright a\ensuremath{_{\textnormal{2}}}=+0.16 \textit{13} and a\ensuremath{_{\textnormal{4}}}={\textminus}0.55 \textit{18} (\href{https://www.nndc.bnl.gov/nsr/nsrlink.jsp?1970Gi09,B}{1970Gi09}).\vspace{0.1cm}}&\\
&&&&&&&&&&&&&&&\parbox[t][0.3cm]{8.6787815cm}{\raggedright B(E2)(W.u.): See also 7.4 W.u. \textit{+48{\textminus}22} (\href{https://www.nndc.bnl.gov/nsr/nsrlink.jsp?1970Gi09,B}{1970Gi09}) and 18 W.u.\vspace{0.1cm}}&\\
&&&&&&&&&&&&&&&\parbox[t][0.3cm]{8.6787815cm}{\raggedright {\ }{\ }{\ }\textit{+6{\textminus}4} (\href{https://www.nndc.bnl.gov/nsr/nsrlink.jsp?1977Le03,B}{1977Le03}) using I\ensuremath{_{\ensuremath{\gamma}}} from (\href{https://www.nndc.bnl.gov/nsr/nsrlink.jsp?1970Gi09,B}{1970Gi09}). (\href{https://www.nndc.bnl.gov/nsr/nsrlink.jsp?1977Le03,B}{1977Le03})\vspace{0.1cm}}&\\
&&&&&&&&&&&&&&&\parbox[t][0.3cm]{8.6787815cm}{\raggedright {\ }{\ }{\ }used the mixing ratio of (\href{https://www.nndc.bnl.gov/nsr/nsrlink.jsp?1970Gi09,B}{1970Gi09}) to deduce the transition\vspace{0.1cm}}&\\
&&&&&&&&&&&&&&&\parbox[t][0.3cm]{8.6787815cm}{\raggedright {\ }{\ }{\ }strength given here. Unlike the transition strength deduced for\vspace{0.1cm}}&\\
&&&&&&&&&&&&&&&\parbox[t][0.3cm]{8.6787815cm}{\raggedright {\ }{\ }{\ }this transition by (\href{https://www.nndc.bnl.gov/nsr/nsrlink.jsp?1970Gi09,B}{1970Gi09}), the transition strength obtained\vspace{0.1cm}}&\\
&&&&&&&&&&&&&&&\parbox[t][0.3cm]{8.6787815cm}{\raggedright {\ }{\ }{\ }by (\href{https://www.nndc.bnl.gov/nsr/nsrlink.jsp?1977Le03,B}{1977Le03}) for the 2556-keV \ensuremath{\gamma} ray is consistent with the\vspace{0.1cm}}&\\
&&&&&&&&&&&&&&&\parbox[t][0.3cm]{8.6787815cm}{\raggedright {\ }{\ }{\ }theoretical estimation of B(E2)=22 W.u. (\href{https://www.nndc.bnl.gov/nsr/nsrlink.jsp?1969Be93,B}{1969Be93}) as cited\vspace{0.1cm}}&\\
&&&&&&&&&&&&&&&\parbox[t][0.3cm]{8.6787815cm}{\raggedright {\ }{\ }{\ }by (\href{https://www.nndc.bnl.gov/nsr/nsrlink.jsp?1977Le03,B}{1977Le03}).\vspace{0.1cm}}&\\
&&&&&&&&&&&&&&&\parbox[t][0.3cm]{8.6787815cm}{\raggedright F(\ensuremath{\tau})=0.57 \textit{6} measured at E\ensuremath{_{\textnormal{p}}}=10.4 MeV (\href{https://www.nndc.bnl.gov/nsr/nsrlink.jsp?1970Gi09,B}{1970Gi09}). See also\vspace{0.1cm}}&\\
&&&&&&&&&&&&&&&\parbox[t][0.3cm]{8.6787815cm}{\raggedright {\ }{\ }{\ }F(\ensuremath{\tau})=65\% \textit{5} (\href{https://www.nndc.bnl.gov/nsr/nsrlink.jsp?1977Le03,B}{1977Le03}), where a Doppler shift of 11.4 keV\vspace{0.1cm}}&\\
&&&&&&&&&&&&&&&\parbox[t][0.3cm]{8.6787815cm}{\raggedright {\ }{\ }{\ }\textit{4} was measured, see Fig. 2.\vspace{0.1cm}}&\\
&&&\multicolumn{1}{r@{}}{2794}&\multicolumn{1}{@{.}l}{5\ensuremath{^{\hyperlink{NE31GAMMA0}{a}}}}&\multicolumn{1}{r@{}}{$<$10}&\multicolumn{1}{@{}l}{\ensuremath{^{\hyperlink{NE31GAMMA1}{b}}}}&\multicolumn{1}{r@{}}{0}&\multicolumn{1}{@{}l}{}&\multicolumn{1}{@{}l}{1/2\ensuremath{^{+}}}&&&&&&&\\
\end{longtable}
\parbox[b][0.3cm]{22.5cm}{\makebox[1ex]{\ensuremath{^{\hypertarget{NE31GAMMA0}{a}}}} This transition was not observed by (\href{https://www.nndc.bnl.gov/nsr/nsrlink.jsp?1970Gi09,B}{1970Gi09}). The evaluator deduced its energy from the level-energy difference corrected for the nuclear recoil energy.}\\
\parbox[b][0.3cm]{22.5cm}{\makebox[1ex]{\ensuremath{^{\hypertarget{NE31GAMMA1}{b}}}} From (\href{https://www.nndc.bnl.gov/nsr/nsrlink.jsp?1970Gi09,B}{1970Gi09}): Deduced the anticipated position in the coincident \ensuremath{\gamma}-ray spectrum for all the possible decay modes of each level from the measured level-energies.}\\
\clearpage
\vspace*{-0.5cm}
{\bf \small \underline{\ensuremath{^{\textnormal{19}}}F(p,n),(p,n\ensuremath{\gamma}),(d,2n\ensuremath{\gamma})\hspace{0.2in}\href{https://www.nndc.bnl.gov/nsr/nsrlink.jsp?1970Gi09,B}{1970Gi09},\href{https://www.nndc.bnl.gov/nsr/nsrlink.jsp?1977Le03,B}{1977Le03} (continued)}}\\
\vspace{0.3cm}
\underline{$\gamma$($^{19}$Ne) (continued)}\\
\vspace{0.3cm}
\parbox[b][0.3cm]{22.5cm}{{\ }{\ }To obtain the relative intensities of these weaker transitions, the spectra used to obtain the angular distributions of the prominent observed transitions were summed}\\
\parbox[b][0.3cm]{22.5cm}{{\ }{\ }and corrected for the effects of angular correlations and the relative efficiency of the Ge(Li) detector as a function of energy.}\\
\parbox[b][0.3cm]{22.5cm}{\makebox[1ex]{\ensuremath{^{\hypertarget{NE31GAMMA2}{c}}}} From (\href{https://www.nndc.bnl.gov/nsr/nsrlink.jsp?1970Gi09,B}{1970Gi09}): Determined the relative intensity of this unresolved \ensuremath{\gamma} ray by comparing the deduced centroid with the one found in the measured spectrum.}\\
\parbox[b][0.3cm]{22.5cm}{\makebox[1ex]{\ensuremath{^{\hypertarget{NE31GAMMA3}{d}}}} From (\href{https://www.nndc.bnl.gov/nsr/nsrlink.jsp?1970Gi09,B}{1970Gi09}).}\\
\parbox[b][0.3cm]{22.5cm}{\makebox[1ex]{\ensuremath{^{\hypertarget{NE31GAMMA4}{e}}}} From (\href{https://www.nndc.bnl.gov/nsr/nsrlink.jsp?1970Gi09,B}{1970Gi09}), where the sign of the mixing ratio was determined based on the convention of (\href{https://www.nndc.bnl.gov/nsr/nsrlink.jsp?1967Ro21,B}{1967Ro21}).}\\
\parbox[b][0.3cm]{22.5cm}{\makebox[1ex]{\ensuremath{^{\hypertarget{NE31GAMMA5}{f}}}} Total theoretical internal conversion coefficients, calculated using the BrIcc code (\href{https://www.nndc.bnl.gov/nsr/nsrlink.jsp?2008Ki07,B}{2008Ki07}) with ``Frozen Orbitals'' approximation based on \ensuremath{\gamma}-ray energies, assigned}\\
\parbox[b][0.3cm]{22.5cm}{{\ }{\ }multipolarities, and mixing ratios, unless otherwise specified.}\\
\vspace{0.5cm}
\end{landscape}\clearpage
\clearpage
\begin{figure}[h]
\begin{center}
\includegraphics{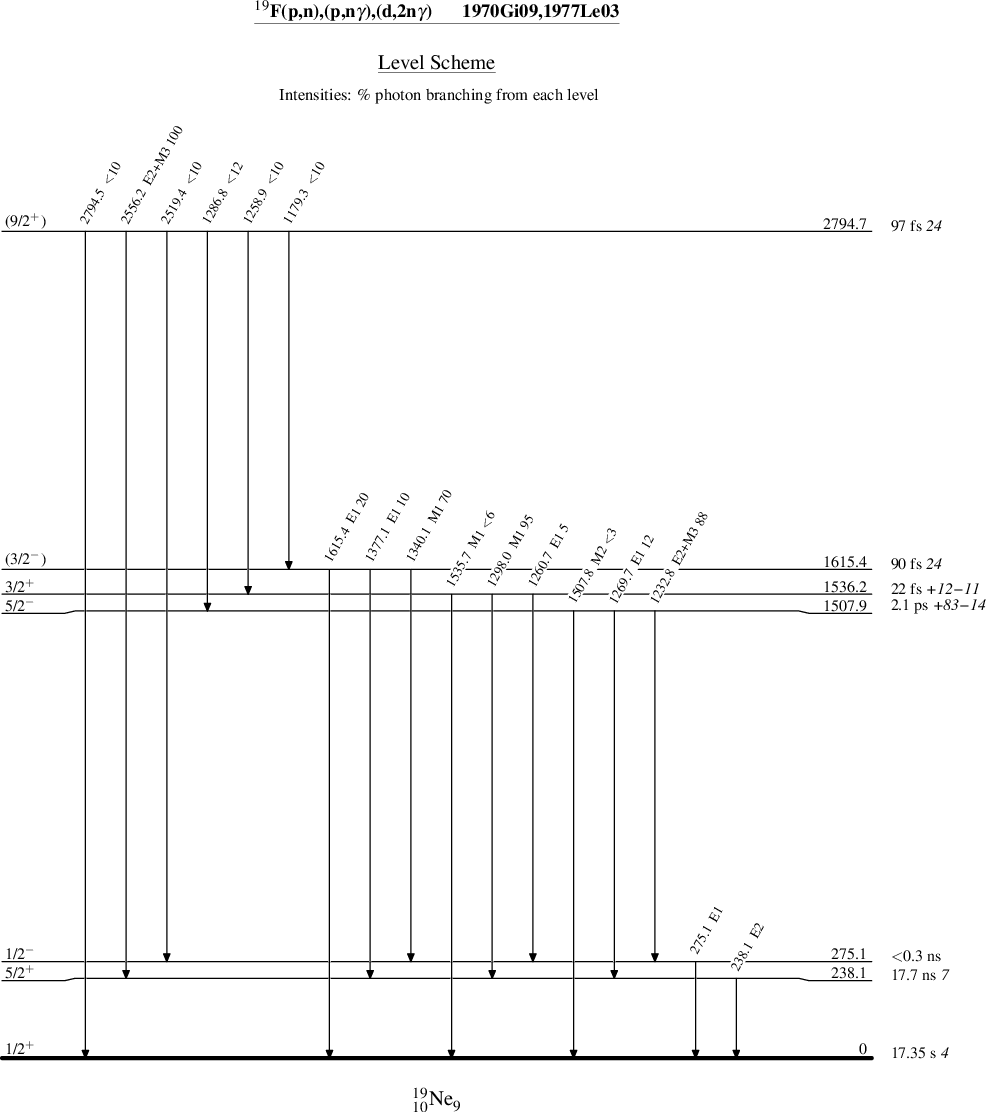}\\
\end{center}
\end{figure}
\clearpage
\subsection[\hspace{-0.2cm}\ensuremath{^{\textnormal{19}}}F(\ensuremath{^{\textnormal{3}}}He,t)]{ }
\vspace{-27pt}
\vspace{0.3cm}
\hypertarget{NE32}{{\bf \small \underline{\ensuremath{^{\textnormal{19}}}F(\ensuremath{^{\textnormal{3}}}He,t)\hspace{0.2in}\href{https://www.nndc.bnl.gov/nsr/nsrlink.jsp?1998Ut02,B}{1998Ut02},\href{https://www.nndc.bnl.gov/nsr/nsrlink.jsp?2013La01,B}{2013La01},\href{https://www.nndc.bnl.gov/nsr/nsrlink.jsp?2019Ka15,B}{2019Ka15}}}}\\
\vspace{4pt}
\vspace{8pt}
\parbox[b][0.3cm]{17.7cm}{\addtolength{\parindent}{-0.2in}Charge exchange reaction.}\\
\parbox[b][0.3cm]{17.7cm}{\addtolength{\parindent}{-0.2in}J\ensuremath{^{\ensuremath{\pi}}}(\ensuremath{^{\textnormal{19}}}F\ensuremath{_{\textnormal{g.s.}}})=1/2\ensuremath{^{\textnormal{+}}} and J\ensuremath{^{\ensuremath{\pi}}}(\ensuremath{^{\textnormal{3}}}He\ensuremath{_{\textnormal{g.s.}}})=1/2\ensuremath{^{\textnormal{+}}}.}\\
\parbox[b][0.3cm]{17.7cm}{\addtolength{\parindent}{-0.2in}\href{https://www.nndc.bnl.gov/nsr/nsrlink.jsp?1970Sc05,B}{1970Sc05}: \ensuremath{^{\textnormal{19}}}F(\ensuremath{^{\textnormal{3}}}He,t) E=26 MeV; measured tritons angular distributions using a \ensuremath{\Delta}E-E telescope covering \ensuremath{\theta}\ensuremath{_{\textnormal{lab}}}=10\ensuremath{^\circ}{\textminus}60\ensuremath{^\circ}.}\\
\parbox[b][0.3cm]{17.7cm}{Resolution was 50 keV (FWHM). Deduced \ensuremath{^{\textnormal{19}}}Ne levels, and L.}\\
\parbox[b][0.3cm]{17.7cm}{\addtolength{\parindent}{-0.2in}\href{https://www.nndc.bnl.gov/nsr/nsrlink.jsp?1989MaZX,B}{1989MaZX}, \href{https://www.nndc.bnl.gov/nsr/nsrlink.jsp?1990Ma05,B}{1990Ma05}: \ensuremath{^{\textnormal{19}}}F(\ensuremath{^{\textnormal{3}}}He,t)\ensuremath{^{\textnormal{19}}}Ne*(\ensuremath{\alpha}) E=29.8 MeV; momentum analyzed tritons using a Q3D spectrograph and its focal}\\
\parbox[b][0.3cm]{17.7cm}{plane detector system at \ensuremath{\theta}\ensuremath{_{\textnormal{lab}}}=0\ensuremath{^\circ}. Measured t-\ensuremath{\alpha} coincidence events using three Si surface barrier detectors at \ensuremath{\theta}\ensuremath{_{\textnormal{lab}}}={\textminus}80\ensuremath{^\circ}, 120\ensuremath{^\circ} and}\\
\parbox[b][0.3cm]{17.7cm}{155\ensuremath{^\circ} to measure energies and TOF of \ensuremath{\alpha}-particles. Deduced the \ensuremath{^{\textnormal{19}}}Ne*(4033, 4140, 4197, 4379, 4549, 4600, 4635, 4712, 5092,}\\
\parbox[b][0.3cm]{17.7cm}{5351) levels. Deduced \ensuremath{\Gamma}\ensuremath{_{\ensuremath{\alpha}}}/\ensuremath{\Gamma} for most of these states. Deduced the \ensuremath{^{\textnormal{15}}}O(\ensuremath{\alpha},\ensuremath{\gamma}) reaction rate at T=0.1-10 GK.}\\
\parbox[b][0.3cm]{17.7cm}{\addtolength{\parindent}{-0.2in}\href{https://www.nndc.bnl.gov/nsr/nsrlink.jsp?1991Ja04,B}{1991Ja04}: \ensuremath{^{\textnormal{19}}}F(\ensuremath{^{\textnormal{3}}}He,t) E=200 MeV; momentum analyzed tritons using a K-600 spectrograph at \ensuremath{\theta}\ensuremath{_{\textnormal{lab}}}=0\ensuremath{^\circ}. Resolution was 50 keV}\\
\parbox[b][0.3cm]{17.7cm}{(FWHM). Measured \ensuremath{\sigma}(E\ensuremath{_{\textnormal{t}}}).}\\
\parbox[b][0.3cm]{17.7cm}{\addtolength{\parindent}{-0.2in}\href{https://www.nndc.bnl.gov/nsr/nsrlink.jsp?1992RoZZ,B}{1992RoZZ}: \ensuremath{^{\textnormal{19}}}F(\ensuremath{^{\textnormal{3}}}He,t) E not given; deduced \ensuremath{^{\textnormal{19}}}Ne levels.}\\
\parbox[b][0.3cm]{17.7cm}{\addtolength{\parindent}{-0.2in}\href{https://www.nndc.bnl.gov/nsr/nsrlink.jsp?1993UtZZ,B}{1993UtZZ}, \href{https://www.nndc.bnl.gov/nsr/nsrlink.jsp?1998Ut02,B}{1998Ut02}: \ensuremath{^{\textnormal{19}}}F(\ensuremath{^{\textnormal{3}}}He,t), \ensuremath{^{\textnormal{19}}}F(\ensuremath{^{\textnormal{3}}}He,t)\ensuremath{^{\textnormal{19}}}Ne*(p) and \ensuremath{^{\textnormal{19}}}F(\ensuremath{^{\textnormal{3}}}He,t)\ensuremath{^{\textnormal{19}}}Ne*(\ensuremath{\alpha}) E=29.8 MeV; momentum analyzed the reaction}\\
\parbox[b][0.3cm]{17.7cm}{products using a Browne-Buechner spectrograph; measured triton angular distributions at \ensuremath{\theta}\ensuremath{_{\textnormal{lab}}}=0\ensuremath{^\circ}, 5\ensuremath{^\circ}, 10\ensuremath{^\circ}, and 15\ensuremath{^\circ}. Energy}\\
\parbox[b][0.3cm]{17.7cm}{resolution was \ensuremath{\Delta}E(FWHM)=24 keV. In a separate experiment at the same energy, the authors measured \ensuremath{^{\textnormal{3}}}H-p and \ensuremath{^{\textnormal{3}}}H-\ensuremath{\alpha}}\\
\parbox[b][0.3cm]{17.7cm}{coincidences using a Q3D spectrograph at \ensuremath{\theta}\ensuremath{_{\textnormal{lab}}}=0\ensuremath{^\circ} to momentum analyze the tritons and 3 Si surface barrier detectors at \ensuremath{\theta}\ensuremath{_{\textnormal{lab}}}=90\ensuremath{^\circ},}\\
\parbox[b][0.3cm]{17.7cm}{110\ensuremath{^\circ}, and 145\ensuremath{^\circ} to measure the protons and \ensuremath{\alpha}-particles resulting from the decay of the \ensuremath{^{\textnormal{19}}}Ne* levels. Deduced \ensuremath{^{\textnormal{19}}}Ne level-energies}\\
\parbox[b][0.3cm]{17.7cm}{and branching ratios. Deduced the \ensuremath{^{\textnormal{18}}}F(p,\ensuremath{\alpha}) and \ensuremath{^{\textnormal{18}}}F(p,\ensuremath{\gamma}) reaction rates and discussed astrophysical implications.}\\
\parbox[b][0.3cm]{17.7cm}{\addtolength{\parindent}{-0.2in}K. Kumagai, M.Sc. Thesis, Tohoku University (1999), unpublished, \href{https://www.nndc.bnl.gov/nsr/nsrlink.jsp?2002Ku12,B}{2002Ku12}: \ensuremath{^{\textnormal{19}}}F(\ensuremath{^{\textnormal{3}}}He,t)\ensuremath{^{\textnormal{19}}}Ne*(\ensuremath{\alpha}) E=30 MeV; momentum}\\
\parbox[b][0.3cm]{17.7cm}{analyzed the tritons using a QDD spectrograph. Measured t-\ensuremath{\alpha} coincidence events, where the \ensuremath{\alpha} particles were measured at backward}\\
\parbox[b][0.3cm]{17.7cm}{angles using 4 position sensitive Si detectors covering 11\% of 4\ensuremath{\pi}. Measured \ensuremath{\Gamma}\ensuremath{_{\ensuremath{\alpha}}}/\ensuremath{\Gamma} branching ratios of a few \ensuremath{^{\textnormal{19}}}Ne* states with}\\
\parbox[b][0.3cm]{17.7cm}{E\ensuremath{_{\textnormal{x}}}\ensuremath{>}5 MeV. The branching ratios for the \ensuremath{^{\textnormal{19}}}Ne*(4033, 4379) could not be measured due to very small \ensuremath{\alpha} yield from the decay of}\\
\parbox[b][0.3cm]{17.7cm}{those states. The results are not presented by (\href{https://www.nndc.bnl.gov/nsr/nsrlink.jsp?2002Ku12,B}{2002Ku12}), which only presents a brief description of the experiment.}\\
\parbox[b][0.3cm]{17.7cm}{\addtolength{\parindent}{-0.2in}\href{https://www.nndc.bnl.gov/nsr/nsrlink.jsp?2004Vi05,B}{2004Vi05}: \ensuremath{^{\textnormal{19}}}F(\ensuremath{^{\textnormal{3}}}He,t)\ensuremath{^{\textnormal{19}}}Ne*(\ensuremath{\alpha}) and \ensuremath{^{\textnormal{19}}}F(\ensuremath{^{\textnormal{3}}}He,t)\ensuremath{^{\textnormal{19}}}Ne*(p) E=25 MeV; momentum analyzed and detected tritons using an Enge}\\
\parbox[b][0.3cm]{17.7cm}{spectrograph and its focal plane detectors placed at \ensuremath{\theta}\ensuremath{_{\textnormal{lab}}}=0\ensuremath{^\circ}; measured protons and \ensuremath{\alpha} particles from the decay of \ensuremath{^{\textnormal{19}}}Ne* states in}\\
\parbox[b][0.3cm]{17.7cm}{coincidence with tritons using the Yale Lamp Shade Array. This was an array of 5 position sensitive Si detectors covering}\\
\parbox[b][0.3cm]{17.7cm}{\ensuremath{\theta}\ensuremath{_{\textnormal{lab}}}=130\ensuremath{^\circ}{\textminus}165\ensuremath{^\circ}. Deduced \ensuremath{^{\textnormal{19}}}Ne levels. Measured \ensuremath{\alpha} angular distributions corresponding to the \ensuremath{^{\textnormal{19}}}Ne*(6742, 6861) levels. Measured}\\
\parbox[b][0.3cm]{17.7cm}{\ensuremath{\Gamma}\ensuremath{_{\ensuremath{\alpha}}}/\ensuremath{\Gamma} for the \ensuremath{^{\textnormal{19}}}Ne*(4379, 4549, 4600, 4712, 5092, 6742, 6861, 7076) states and \ensuremath{\Gamma}\ensuremath{_{\textnormal{p}}}/\ensuremath{\Gamma} for the \ensuremath{^{\textnormal{19}}}Ne*(6742, 6861, 7076) states.}\\
\parbox[b][0.3cm]{17.7cm}{\addtolength{\parindent}{-0.2in}\href{https://www.nndc.bnl.gov/nsr/nsrlink.jsp?2007TaZX,B}{2007TaZX}, \href{https://www.nndc.bnl.gov/nsr/nsrlink.jsp?2007Ta13,B}{2007Ta13}, \href{https://www.nndc.bnl.gov/nsr/nsrlink.jsp?2009Ta09,B}{2009Ta09}: \ensuremath{^{\textnormal{19}}}F(\ensuremath{^{\textnormal{3}}}He,t)\ensuremath{^{\textnormal{19}}}Ne*(\ensuremath{\alpha}) E=24 MeV; target at \ensuremath{\theta}\ensuremath{_{\textnormal{lab}}}=30\ensuremath{^\circ}; measured t-\ensuremath{\alpha} coincidence events using the}\\
\parbox[b][0.3cm]{17.7cm}{TwinSol (to separate and momentum analyze the tritons detected by a position sensitive \ensuremath{\Delta}E-E telescope covering \ensuremath{\theta}\ensuremath{_{\textnormal{lab}}}=2\ensuremath{^\circ}{\textminus}7.5\ensuremath{^\circ})}\\
\parbox[b][0.3cm]{17.7cm}{and the LESA position sensitive Si array (covering \ensuremath{\theta}\ensuremath{_{\textnormal{lab}}}=60\ensuremath{^\circ}{\textminus}150\ensuremath{^\circ} to measure the energy and TOF of the \ensuremath{\alpha}-particles from the}\\
\parbox[b][0.3cm]{17.7cm}{decays of unbound \ensuremath{^{\textnormal{19}}}Ne* states). Deduced the \ensuremath{^{\textnormal{19}}}Ne*(4.03, 4.14+4.2, 4.38, 4.55, 4.60, 4.71, 5.09 MeV) levels. Measured \ensuremath{\alpha}-decay}\\
\parbox[b][0.3cm]{17.7cm}{angular distributions for the 4.55, 4.60, 4.71, and 5.09 MeV states. Deduced \ensuremath{\Gamma}\ensuremath{_{\ensuremath{\alpha}}}/\ensuremath{\Gamma} branching ratios for most levels observed.}\\
\parbox[b][0.3cm]{17.7cm}{Deduced \ensuremath{\Gamma}\ensuremath{_{\ensuremath{\alpha}}} based on the branching ratios and lifetimes measured in (\href{https://www.nndc.bnl.gov/nsr/nsrlink.jsp?2005Ta28,B}{2005Ta28}: \ensuremath{^{\textnormal{17}}}O(\ensuremath{^{\textnormal{3}}}He,n\ensuremath{\gamma})). Deduced the \ensuremath{^{\textnormal{15}}}O(\ensuremath{\alpha},\ensuremath{\gamma}) reaction}\\
\parbox[b][0.3cm]{17.7cm}{rate at T=0.2-1.5 GK. Astrophysical implications are discussed.}\\
\parbox[b][0.3cm]{17.7cm}{\addtolength{\parindent}{-0.2in}\href{https://www.nndc.bnl.gov/nsr/nsrlink.jsp?2013La01,B}{2013La01}: \ensuremath{^{\textnormal{19}}}F(\ensuremath{^{\textnormal{3}}}He,t) E=25 MeV; momentum analyzed the tritons using a Q3D spectrograph and its focal plane detection system}\\
\parbox[b][0.3cm]{17.7cm}{placed at \ensuremath{\theta}\ensuremath{_{\textnormal{lab}}}=10\ensuremath{^\circ}{\textminus}50\ensuremath{^\circ}. Energy resolution\ensuremath{\sim}14 keV (FWHM). Deduced \ensuremath{^{\textnormal{19}}}Ne* states. Observed an unresolved group near the proton}\\
\parbox[b][0.3cm]{17.7cm}{threshold, which was best fitted using 3 peaks with fixed widths. As a result, a new state was found at E\ensuremath{_{\textnormal{x}}}=6440 keV \textit{3} (stat.) \textit{2}}\\
\parbox[b][0.3cm]{17.7cm}{(sys.). Measured triton angular distributions and deduced L and J\ensuremath{^{\ensuremath{\pi}}} values using two-step finite-range DWBA analysis via}\\
\parbox[b][0.3cm]{17.7cm}{FRESCO. Performed multi-channel R-matrix calculations using the DREAM code. Deduced the \ensuremath{^{\textnormal{18}}}F(p,\ensuremath{\alpha}) reaction rate at nova}\\
\parbox[b][0.3cm]{17.7cm}{temperatures and used it to carry out nova nucleosynthesis calculations.}\\
\parbox[b][0.3cm]{17.7cm}{\addtolength{\parindent}{-0.2in}\href{https://www.nndc.bnl.gov/nsr/nsrlink.jsp?2015Pa46,B}{2015Pa46}: \ensuremath{^{\textnormal{19}}}F(\ensuremath{^{\textnormal{3}}}He,t) E=25 MeV; momentum analyzed tritons using a Q3D spectrograph and its focal plane detector system placed}\\
\parbox[b][0.3cm]{17.7cm}{at \ensuremath{\theta}\ensuremath{_{\textnormal{lab}}}=10\ensuremath{^\circ}{\textminus}50\ensuremath{^\circ}. Energy resolution was typically 10 to 15 keV (FWHM); measured triton angular distributions and deduced J\ensuremath{^{\ensuremath{\pi}}}}\\
\parbox[b][0.3cm]{17.7cm}{values using a finite-range coupled-channels analysis via the FRESCO code. Those authors did not report their excitation energies}\\
\parbox[b][0.3cm]{17.7cm}{and instead quoted the values from (\href{https://www.nndc.bnl.gov/nsr/nsrlink.jsp?2009Ta09,B}{2009Ta09}) and (\href{https://www.nndc.bnl.gov/nsr/nsrlink.jsp?2011Da24,B}{2011Da24}: evaluation of \ensuremath{^{\textnormal{15}}}O(\ensuremath{\alpha},\ensuremath{\gamma})). They discussed that the 6.29-MeV state}\\
\parbox[b][0.3cm]{17.7cm}{in \ensuremath{^{\textnormal{19}}}Ne is either a doublet or a single broad state. In addition, those authors mentioned that one or more new levels may exist near}\\
\parbox[b][0.3cm]{17.7cm}{the 6.86-MeV state.}\\
\parbox[b][0.3cm]{17.7cm}{\addtolength{\parindent}{-0.2in}\href{https://www.nndc.bnl.gov/nsr/nsrlink.jsp?2019Ka15,B}{2019Ka15}: \ensuremath{^{\textnormal{19}}}F(\ensuremath{^{\textnormal{3}}}He,t) E=140 MeV/nucleon; momentum analyzed tritons using the Grand Raiden spectrometer in dispersion}\\
\parbox[b][0.3cm]{17.7cm}{matched mode placed at \ensuremath{\theta}\ensuremath{_{\textnormal{lab}}}=0\ensuremath{^\circ} with an acceptance of \ensuremath{<}2.5\ensuremath{^\circ}. Resolutions were 41 keV and \ensuremath{\leq}0.3\ensuremath{^\circ} (FWHM for both). Measured}\\
\parbox[b][0.3cm]{17.7cm}{triton angular distributions at 5 angular ranges in the \ensuremath{\theta}\ensuremath{_{\textnormal{lab}}}=0\ensuremath{^\circ}{\textminus}2\ensuremath{^\circ} range. (\ensuremath{^{\textnormal{3}}}He,t) reactions measured at extreme forward angles and}\\
\parbox[b][0.3cm]{17.7cm}{in intermediate energies are described by one pion charge-exchange mechanism for Gamow-Teller (GT) \ensuremath{\Delta}L=0 transitions. The goal}\\
\parbox[b][0.3cm]{17.7cm}{therefore was to identify \ensuremath{^{\textnormal{19}}}Ne* states associated with \ensuremath{\Delta}L=0 transitions corresponding to J\ensuremath{^{\ensuremath{\pi}}}=1/2\ensuremath{^{\textnormal{+}}} or 3/2\ensuremath{^{\textnormal{+}}}. Deduced \ensuremath{^{\textnormal{19}}}Ne levels}\\
\parbox[b][0.3cm]{17.7cm}{and determined the ones associated with \ensuremath{\Delta}L=0 GT transitions. Deduced the astrophysical S-factor for the \ensuremath{^{\textnormal{18}}}F(p,\ensuremath{\alpha}) reaction at}\\
\parbox[b][0.3cm]{17.7cm}{E\ensuremath{_{\textnormal{c.m.}}}\ensuremath{\leq}1 MeV and discussed the implications.}\\
\clearpage
\vspace{0.3cm}
{\bf \small \underline{\ensuremath{^{\textnormal{19}}}F(\ensuremath{^{\textnormal{3}}}He,t)\hspace{0.2in}\href{https://www.nndc.bnl.gov/nsr/nsrlink.jsp?1998Ut02,B}{1998Ut02},\href{https://www.nndc.bnl.gov/nsr/nsrlink.jsp?2013La01,B}{2013La01},\href{https://www.nndc.bnl.gov/nsr/nsrlink.jsp?2019Ka15,B}{2019Ka15} (continued)}}\\
\vspace{0.3cm}
\parbox[b][0.3cm]{17.7cm}{\addtolength{\parindent}{-0.2in}\href{https://www.nndc.bnl.gov/nsr/nsrlink.jsp?2021Ri04,B}{2021Ri04}: \ensuremath{^{\textnormal{19}}}F(\ensuremath{^{\textnormal{3}}}He,t)\ensuremath{^{\textnormal{19}}}Ne*(p) and \ensuremath{^{\textnormal{19}}}F(\ensuremath{^{\textnormal{3}}}He,t)\ensuremath{^{\textnormal{19}}}Ne*(\ensuremath{\alpha}) E=25 MeV; momentum analyzed and measured the light reaction products}\\
\parbox[b][0.3cm]{17.7cm}{using an Enge spectrograph and its focal plane detector at \ensuremath{\theta}\ensuremath{_{\textnormal{lab,effective}}}=12\ensuremath{^\circ}; measured \ensuremath{^{\textnormal{3}}}H-p and \ensuremath{^{\textnormal{3}}}H-\ensuremath{\alpha} coincidence events, where}\\
\parbox[b][0.3cm]{17.7cm}{protons and \ensuremath{\alpha}s were emitted from the decay of the \ensuremath{^{\textnormal{19}}}Ne* states, using an array of 6 position sensitive Si detectors, which}\\
\parbox[b][0.3cm]{17.7cm}{constructed 3 walls of 2 detectors, placed at \ensuremath{\theta}\ensuremath{_{\textnormal{lab}}}={\textminus}135\ensuremath{^\circ}, 113\ensuremath{^\circ}, and 155\ensuremath{^\circ} covering \ensuremath{\theta}\ensuremath{_{\textnormal{c.m.}}}=90\ensuremath{^\circ}{\textminus}172\ensuremath{^\circ}. Peak width resolution was \ensuremath{\approx}85}\\
\parbox[b][0.3cm]{17.7cm}{keV (FWHM). Measured angular correlation distributions between the tritons and protons or \ensuremath{\alpha} particles; measured \ensuremath{\alpha} decay}\\
\parbox[b][0.3cm]{17.7cm}{branching ratios for the \ensuremath{^{\textnormal{19}}}Ne* unbound states; deduced \ensuremath{^{\textnormal{19}}}Ne levels and J\ensuremath{^{\ensuremath{\pi}}} values; discussed tentative evidence for the}\\
\parbox[b][0.3cm]{17.7cm}{\ensuremath{^{\textnormal{19}}}Ne*(6008) state with \ensuremath{\Gamma}=124 keV; deduced the \ensuremath{^{\textnormal{18}}}F(p,\ensuremath{\alpha}) rate and its S-factor and discussed astrophysical implications.}\\
\parbox[b][0.3cm]{17.7cm}{\addtolength{\parindent}{-0.2in}\href{https://www.nndc.bnl.gov/nsr/nsrlink.jsp?2025PhZZ,B}{2025PhZZ}: \ensuremath{^{\textnormal{19}}}F(\ensuremath{^{\textnormal{3}}}He,t)\ensuremath{^{\textnormal{19}}}Ne*(a) and \ensuremath{^{\textnormal{19}}}F(\ensuremath{^{\textnormal{3}}}He,t)\ensuremath{^{\textnormal{19}}}Ne*(p) E=24 MeV; momentum analyzed and measured light reaction products}\\
\parbox[b][0.3cm]{17.7cm}{using an Enge spectrograph and its associated focal plane detector at \ensuremath{\theta}\ensuremath{_{\textnormal{lab}}}=3\ensuremath{^\circ}; measured t-\ensuremath{\alpha} and t-p coincidence events using an}\\
\parbox[b][0.3cm]{17.7cm}{array of 5 position sensitive Si detectors called Silicon Array for Branching Ratio Experiments (SABRE), which was mounted in}\\
\parbox[b][0.3cm]{17.7cm}{the target chamber. Peak width resolution was \ensuremath{\approx}31 keV (FWHM). Measured angular correlation distributions between the tritons}\\
\parbox[b][0.3cm]{17.7cm}{and protons or \ensuremath{\alpha} particles; measured \ensuremath{\alpha} decay and p decay branching ratios for the \ensuremath{^{\textnormal{19}}}Ne* unbound states; deduced \ensuremath{^{\textnormal{19}}}Ne levels, J\ensuremath{^{\ensuremath{\pi}}}}\\
\parbox[b][0.3cm]{17.7cm}{values, and L-transfer for the \ensuremath{^{\textnormal{15}}}O+\ensuremath{\alpha} and \ensuremath{^{\textnormal{18}}}F+p decay products; deduced the \ensuremath{^{\textnormal{18}}}F(p,\ensuremath{\alpha}) S-factor and discussed astrophysical}\\
\parbox[b][0.3cm]{17.7cm}{implications.}\\
\vspace{12pt}
\underline{$^{19}$Ne Levels}\\
\vspace{0.34cm}
\parbox[b][0.3cm]{17.7cm}{\addtolength{\parindent}{-0.254cm}\textit{Notes}:}\\
\parbox[b][0.3cm]{17.7cm}{\addtolength{\parindent}{-0.254cm}(1) \ensuremath{\Gamma}\ensuremath{_{\ensuremath{\alpha}}} values quoted here from (\href{https://www.nndc.bnl.gov/nsr/nsrlink.jsp?2009Ta09,B}{2009Ta09}) are deduced from \ensuremath{\Gamma}\ensuremath{_{\ensuremath{\alpha}}}/\ensuremath{\Gamma} measured by (\href{https://www.nndc.bnl.gov/nsr/nsrlink.jsp?2009Ta09,B}{2009Ta09}) and \ensuremath{\tau} (lifetime) measured by}\\
\parbox[b][0.3cm]{17.7cm}{(\href{https://www.nndc.bnl.gov/nsr/nsrlink.jsp?2005Ta28,B}{2005Ta28}: \ensuremath{^{\textnormal{17}}}O(\ensuremath{^{\textnormal{3}}}He,n\ensuremath{\gamma})) except when noted otherwise.}\\
\parbox[b][0.3cm]{17.7cm}{\addtolength{\parindent}{-0.254cm}(2) \ensuremath{\Gamma}\ensuremath{_{\textnormal{p}}} and \ensuremath{\Gamma}\ensuremath{_{\ensuremath{\alpha}}} values reported by (\href{https://www.nndc.bnl.gov/nsr/nsrlink.jsp?2013La01,B}{2013La01}) are deduced using R-matrix analysis.}\\
\parbox[b][0.3cm]{17.7cm}{\addtolength{\parindent}{-0.254cm}(3) Evaluator notes that in the (\href{https://www.nndc.bnl.gov/nsr/nsrlink.jsp?2009Ta09,B}{2009Ta09}) study, there appears to be a large background in the coincident events corresponding to}\\
\parbox[b][0.3cm]{17.7cm}{populations of \ensuremath{^{\textnormal{19}}}Ne* states below E\ensuremath{_{\textnormal{x}}}=4.55 MeV (see Fig. 6 in that study). This background was not vigorously characterized,}\\
\parbox[b][0.3cm]{17.7cm}{which makes the deduced \ensuremath{\alpha} decay branching ratios less reliable. (\href{https://www.nndc.bnl.gov/nsr/nsrlink.jsp?2011Da24,B}{2011Da24}: Evaluation of \ensuremath{^{\textnormal{15}}}O(\ensuremath{\alpha},\ensuremath{\gamma})) has brought these facts into}\\
\parbox[b][0.3cm]{17.7cm}{attention. The coincidence timing spectra for the \ensuremath{^{\textnormal{19}}}Ne* \ensuremath{\alpha} decaying states below E\ensuremath{_{\textnormal{x}}}=4.55 MeV in (\href{https://www.nndc.bnl.gov/nsr/nsrlink.jsp?2009Ta09,B}{2009Ta09}) does not reveal a}\\
\parbox[b][0.3cm]{17.7cm}{convincing evidence for an \ensuremath{\alpha} decay peak from the 4.03, 4.14, 4.20, or 4.38 MeV states. In contrast, the \ensuremath{\Gamma}\ensuremath{_{\ensuremath{\alpha}}}/\ensuremath{\Gamma} values for states}\\
\parbox[b][0.3cm]{17.7cm}{lying at and above 4.55 MeV appear reliable as far as the background determination and signal-to-noise ratio are concerned.}\\
\parbox[b][0.3cm]{17.7cm}{\addtolength{\parindent}{-0.254cm}(4) We note that the L (angular momentum) values listed from (\href{https://www.nndc.bnl.gov/nsr/nsrlink.jsp?2025PhZZ,B}{2025PhZZ}) refer to the orbital angular momentum transfer for the}\\
\parbox[b][0.3cm]{17.7cm}{\ensuremath{\alpha} particles or for protons from the decay of \ensuremath{^{\textnormal{19}}}Ne* states to the \ensuremath{^{\textnormal{15}}}O(g.s., 1/2\ensuremath{^{-}}) or \ensuremath{^{\textnormal{18}}}F(g.s., 1\ensuremath{^{\textnormal{+}}}) levels, respectively. We also}\\
\parbox[b][0.3cm]{17.7cm}{highlight that the branching ratios listed in the comments from (\href{https://www.nndc.bnl.gov/nsr/nsrlink.jsp?2025PhZZ,B}{2025PhZZ}) are the lowest reported L-value. Those authors pointed}\\
\parbox[b][0.3cm]{17.7cm}{out to us that the branching ratios for the E\ensuremath{_{\textnormal{x}}}(\ensuremath{^{\textnormal{19}}}Ne*)\ensuremath{>}7.1 MeV levels are less reliable due to the low statistics and the complicated}\\
\parbox[b][0.3cm]{17.7cm}{level schemes.}\\
\vspace{0.34cm}
\begin{longtable}{ccccc@{\extracolsep{\fill}}c}
\multicolumn{2}{c}{E(level)$^{}$}&J$^{\pi}$$^{}$&L$^{}$&Comments&\\[-.2cm]
\multicolumn{2}{c}{\hrulefill}&\hrulefill&\hrulefill&\hrulefill&
\endfirsthead
\multicolumn{1}{r@{}}{0}&\multicolumn{1}{@{}l}{}&\multicolumn{1}{l}{1/2\ensuremath{^{+}}}&\multicolumn{1}{l}{0}&\parbox[t][0.3cm]{14.659601cm}{\raggedright E(level): Observed in (\href{https://www.nndc.bnl.gov/nsr/nsrlink.jsp?1970Sc05,B}{1970Sc05}, \href{https://www.nndc.bnl.gov/nsr/nsrlink.jsp?1991Ja04,B}{1991Ja04}).\vspace{0.1cm}}&\\
&&&&\parbox[t][0.3cm]{14.659601cm}{\raggedright E(level): This is the isobaric analog state (IAS) for \ensuremath{^{\textnormal{19}}}F\ensuremath{_{\textnormal{g.s.}}} (\href{https://www.nndc.bnl.gov/nsr/nsrlink.jsp?1991Ja04,B}{1991Ja04}).\vspace{0.1cm}}&\\
&&&&\parbox[t][0.3cm]{14.659601cm}{\raggedright L: From (\href{https://www.nndc.bnl.gov/nsr/nsrlink.jsp?1970Sc05,B}{1970Sc05}) from comparison of the triton angular distribution of this state with those of 0\ensuremath{^{\textnormal{+}}}\ensuremath{\rightarrow}0\ensuremath{^{\textnormal{+}}}\vspace{0.1cm}}&\\
&&&&\parbox[t][0.3cm]{14.659601cm}{\raggedright {\ }{\ }{\ }transitions from (\ensuremath{^{\textnormal{3}}}He,t) reactions on even-A target nuclei with J\ensuremath{^{\ensuremath{\pi}}}=0\ensuremath{^{\textnormal{+}}}. The L-transfer for such transitions\vspace{0.1cm}}&\\
&&&&\parbox[t][0.3cm]{14.659601cm}{\raggedright {\ }{\ }{\ }is unique. (\href{https://www.nndc.bnl.gov/nsr/nsrlink.jsp?1970Sc05,B}{1970Sc05}) reported that the transition populating the \ensuremath{^{\textnormal{19}}}Ne\ensuremath{_{\textnormal{g.s.}}} proceeds primarily via L=0\vspace{0.1cm}}&\\
&&&&\parbox[t][0.3cm]{14.659601cm}{\raggedright {\ }{\ }{\ }although L=0 and L=2 are both allowed.\vspace{0.1cm}}&\\
&&&&\parbox[t][0.3cm]{14.659601cm}{\raggedright J\ensuremath{^{\pi}}: From (\href{https://www.nndc.bnl.gov/nsr/nsrlink.jsp?1970Sc05,B}{1970Sc05}) deduced from L=0.\vspace{0.1cm}}&\\
&&&&\parbox[t][0.3cm]{14.659601cm}{\raggedright \ensuremath{^{\textnormal{19}}}F\ensuremath{_{\textnormal{g.s.}}}(\ensuremath{^{\textnormal{3}}}He,t)\ensuremath{^{\textnormal{19}}}Ne\ensuremath{_{\textnormal{g.s.}}} reaction is the non-spin-flip charge-exchange reaction that leads to the population of\vspace{0.1cm}}&\\
&&&&\parbox[t][0.3cm]{14.659601cm}{\raggedright {\ }{\ }{\ }the \ensuremath{^{\textnormal{19}}}Ne\ensuremath{_{\textnormal{g.s.}}} (IAS).\vspace{0.1cm}}&\\
&&&&\parbox[t][0.3cm]{14.659601cm}{\raggedright d\ensuremath{\sigma}/d\ensuremath{\Omega}\ensuremath{_{\textnormal{lab}}}(\ensuremath{\theta}\ensuremath{_{\textnormal{lab}}}=0\ensuremath{^\circ},E=200 MeV)=3.68 mb/sr \textit{2} (\href{https://www.nndc.bnl.gov/nsr/nsrlink.jsp?1991Ja04,B}{1991Ja04}) assuming an isotropic angular distribution.\vspace{0.1cm}}&\\
\multicolumn{1}{r@{}}{241}&\multicolumn{1}{@{}l}{}&\multicolumn{1}{l}{5/2\ensuremath{^{+}}}&\multicolumn{1}{l}{2}&\parbox[t][0.3cm]{14.659601cm}{\raggedright E(level): From (\href{https://www.nndc.bnl.gov/nsr/nsrlink.jsp?1970Sc05,B}{1970Sc05}).\vspace{0.1cm}}&\\
&&&&\parbox[t][0.3cm]{14.659601cm}{\raggedright L: From comparison of the triton angular distribution populating this state with those of 0\ensuremath{^{\textnormal{+}}}\ensuremath{\rightarrow}2\ensuremath{^{\textnormal{+}}} transitions\vspace{0.1cm}}&\\
&&&&\parbox[t][0.3cm]{14.659601cm}{\raggedright {\ }{\ }{\ }on even-A target nuclei with J\ensuremath{^{\ensuremath{\pi}}}=0\ensuremath{^{\textnormal{+}}} populated by (\ensuremath{^{\textnormal{3}}}He,t). The L-transfer for the transition populating the\vspace{0.1cm}}&\\
&&&&\parbox[t][0.3cm]{14.659601cm}{\raggedright {\ }{\ }{\ }even-J states in this case is unique. The transition populating \ensuremath{^{\textnormal{19}}}Ne*(241) level proceeds primarily via L=2\vspace{0.1cm}}&\\
&&&&\parbox[t][0.3cm]{14.659601cm}{\raggedright {\ }{\ }{\ }although L=2 and L=4 are allowed (\href{https://www.nndc.bnl.gov/nsr/nsrlink.jsp?1970Sc05,B}{1970Sc05}).\vspace{0.1cm}}&\\
&&&&\parbox[t][0.3cm]{14.659601cm}{\raggedright J\ensuremath{^{\pi}}: From (\href{https://www.nndc.bnl.gov/nsr/nsrlink.jsp?1970Sc05,B}{1970Sc05}) deduced from L=2.\vspace{0.1cm}}&\\
\multicolumn{1}{r@{}}{275}&\multicolumn{1}{@{}l}{}&\multicolumn{1}{l}{1/2\ensuremath{^{-}}}&&\parbox[t][0.3cm]{14.659601cm}{\raggedright E(level): From (\href{https://www.nndc.bnl.gov/nsr/nsrlink.jsp?1991Ja04,B}{1991Ja04}).\vspace{0.1cm}}&\\
&&&&\parbox[t][0.3cm]{14.659601cm}{\raggedright J\ensuremath{^{\pi}}: From the \ensuremath{^{\textnormal{19}}}Ne Adopted Levels and also assumed by (\href{https://www.nndc.bnl.gov/nsr/nsrlink.jsp?1991Ja04,B}{1991Ja04}).\vspace{0.1cm}}&\\
&&&&\parbox[t][0.3cm]{14.659601cm}{\raggedright d\ensuremath{\sigma}/d\ensuremath{\Omega}\ensuremath{_{\textnormal{lab}}}(\ensuremath{\theta}\ensuremath{_{\textnormal{lab}}}=0\ensuremath{^\circ},E=200 MeV)=0.302 mb/sr \textit{4} (\href{https://www.nndc.bnl.gov/nsr/nsrlink.jsp?1991Ja04,B}{1991Ja04}) assuming an isotropic angular distribution.\vspace{0.1cm}}&\\
\end{longtable}
\begin{textblock}{29}(0,27.3)
Continued on next page (footnotes at end of table)
\end{textblock}
\clearpage
\begin{longtable}{ccccc@{\extracolsep{\fill}}c}
\\[-.4cm]
\multicolumn{6}{c}{{\bf \small \underline{\ensuremath{^{\textnormal{19}}}F(\ensuremath{^{\textnormal{3}}}He,t)\hspace{0.2in}\href{https://www.nndc.bnl.gov/nsr/nsrlink.jsp?1998Ut02,B}{1998Ut02},\href{https://www.nndc.bnl.gov/nsr/nsrlink.jsp?2013La01,B}{2013La01},\href{https://www.nndc.bnl.gov/nsr/nsrlink.jsp?2019Ka15,B}{2019Ka15} (continued)}}}\\
\multicolumn{6}{c}{~}\\
\multicolumn{6}{c}{\underline{\ensuremath{^{19}}Ne Levels (continued)}}\\
\multicolumn{6}{c}{~}\\
\multicolumn{2}{c}{E(level)$^{}$}&J$^{\pi}$$^{}$&L$^{}$&Comments&\\[-.2cm]
\multicolumn{2}{c}{\hrulefill}&\hrulefill&\hrulefill&\hrulefill&
\endhead
\multicolumn{1}{r@{}}{1540}&\multicolumn{1}{@{}l}{}&\multicolumn{1}{l}{3/2\ensuremath{^{+}}}&\multicolumn{1}{l}{2}&\parbox[t][0.3cm]{13.409841cm}{\raggedright E(level): From (\href{https://www.nndc.bnl.gov/nsr/nsrlink.jsp?1970Sc05,B}{1970Sc05}).\vspace{0.1cm}}&\\
&&&&\parbox[t][0.3cm]{13.409841cm}{\raggedright L: From (\href{https://www.nndc.bnl.gov/nsr/nsrlink.jsp?1970Sc05,B}{1970Sc05}): Reported that this transition proceeds primarily via the lowest allowed L\vspace{0.1cm}}&\\
&&&&\parbox[t][0.3cm]{13.409841cm}{\raggedright {\ }{\ }{\ }value which does not require an accompanying spin flip, i.e., the transition proceeds primarily\vspace{0.1cm}}&\\
&&&&\parbox[t][0.3cm]{13.409841cm}{\raggedright {\ }{\ }{\ }via L=2 rather than L=0 (\href{https://www.nndc.bnl.gov/nsr/nsrlink.jsp?1970Sc05,B}{1970Sc05}).\vspace{0.1cm}}&\\
&&&&\parbox[t][0.3cm]{13.409841cm}{\raggedright J\ensuremath{^{\pi}}: From (\href{https://www.nndc.bnl.gov/nsr/nsrlink.jsp?1970Sc05,B}{1970Sc05}) deduced from L=2.\vspace{0.1cm}}&\\
\multicolumn{1}{r@{}}{2.78\ensuremath{\times10^{3}}}&\multicolumn{1}{@{}l}{}&\multicolumn{1}{l}{(9/2\ensuremath{^{+}})}&&\parbox[t][0.3cm]{13.409841cm}{\raggedright E(level): From (\href{https://www.nndc.bnl.gov/nsr/nsrlink.jsp?1970Sc05,B}{1970Sc05}).\vspace{0.1cm}}&\\
&&&&\parbox[t][0.3cm]{13.409841cm}{\raggedright J\ensuremath{^{\pi}}: From triton angular distribution populating this state (\href{https://www.nndc.bnl.gov/nsr/nsrlink.jsp?1970Sc05,B}{1970Sc05}: See Fig. 1).\vspace{0.1cm}}&\\
\multicolumn{1}{r@{}}{4033}&\multicolumn{1}{@{}l}{}&\multicolumn{1}{l}{3/2\ensuremath{^{+}}}&&\parbox[t][0.3cm]{13.409841cm}{\raggedright E(level): From (\href{https://www.nndc.bnl.gov/nsr/nsrlink.jsp?1990Ma05,B}{1990Ma05}, \href{https://www.nndc.bnl.gov/nsr/nsrlink.jsp?2004Vi05,B}{2004Vi05}: See Fig. 1).\vspace{0.1cm}}&\\
&&&&\parbox[t][0.3cm]{13.409841cm}{\raggedright E(level): See also 4.03 MeV (\href{https://www.nndc.bnl.gov/nsr/nsrlink.jsp?2007TaZX,B}{2007TaZX}: See Fig. 2) and (\href{https://www.nndc.bnl.gov/nsr/nsrlink.jsp?2007Ta13,B}{2007Ta13}); 4034.5 keV (\href{https://www.nndc.bnl.gov/nsr/nsrlink.jsp?2009Ta09,B}{2009Ta09})\vspace{0.1cm}}&\\
&&&&\parbox[t][0.3cm]{13.409841cm}{\raggedright {\ }{\ }{\ }taken from (\href{https://www.nndc.bnl.gov/nsr/nsrlink.jsp?2005Ta28,B}{2005Ta28}: \ensuremath{^{\textnormal{15}}}O(\ensuremath{^{\textnormal{3}}}He,n\ensuremath{\gamma})); and 4034 keV (\href{https://www.nndc.bnl.gov/nsr/nsrlink.jsp?2015Pa46,B}{2015Pa46}) quoted from (\href{https://www.nndc.bnl.gov/nsr/nsrlink.jsp?2009Ta09,B}{2009Ta09}) and\vspace{0.1cm}}&\\
&&&&\parbox[t][0.3cm]{13.409841cm}{\raggedright {\ }{\ }{\ }(\href{https://www.nndc.bnl.gov/nsr/nsrlink.jsp?2011Da24,B}{2011Da24}: Evaluation of \ensuremath{^{\textnormal{15}}}O(\ensuremath{\alpha},\ensuremath{\gamma})). This state was used as a calibration point in (\href{https://www.nndc.bnl.gov/nsr/nsrlink.jsp?2025PhZZ,B}{2025PhZZ}).\vspace{0.1cm}}&\\
&&&&\parbox[t][0.3cm]{13.409841cm}{\raggedright \ensuremath{\Gamma}\ensuremath{_{\ensuremath{\alpha}}}/\ensuremath{\Gamma}=2.9\ensuremath{\times}10\ensuremath{^{\textnormal{$-$4}}} \textit{21} measured for the first time by (\href{https://www.nndc.bnl.gov/nsr/nsrlink.jsp?2007TaZX,B}{2007TaZX}, \href{https://www.nndc.bnl.gov/nsr/nsrlink.jsp?2007Ta13,B}{2007Ta13}, \href{https://www.nndc.bnl.gov/nsr/nsrlink.jsp?2009Ta09,B}{2009Ta09}). Note that\vspace{0.1cm}}&\\
&&&&\parbox[t][0.3cm]{13.409841cm}{\raggedright {\ }{\ }{\ }the statistics under the kinematically corrected TOF peak corresponding to the t-\ensuremath{\alpha} coincidence\vspace{0.1cm}}&\\
&&&&\parbox[t][0.3cm]{13.409841cm}{\raggedright {\ }{\ }{\ }events and the number of detected t-\ensuremath{\alpha} coincidence events are very limited in (\href{https://www.nndc.bnl.gov/nsr/nsrlink.jsp?2009Ta09,B}{2009Ta09}). There\vspace{0.1cm}}&\\
&&&&\parbox[t][0.3cm]{13.409841cm}{\raggedright {\ }{\ }{\ }is also a significant background in the energy region of interest for this state. (\href{https://www.nndc.bnl.gov/nsr/nsrlink.jsp?2009Ta09,B}{2009Ta09})\vspace{0.1cm}}&\\
&&&&\parbox[t][0.3cm]{13.409841cm}{\raggedright {\ }{\ }{\ }claimed that a net \ensuremath{\alpha}-decay events, at the level of 1.4\ensuremath{\sigma} statistical significance was observed.\vspace{0.1cm}}&\\
&&&&\parbox[t][0.3cm]{13.409841cm}{\raggedright {\ }{\ }{\ }Using a Bayesian approach, those authors also placed an upper limit of \ensuremath{\Gamma}\ensuremath{_{\ensuremath{\alpha}}}/\ensuremath{\Gamma}\ensuremath{<}3.9\ensuremath{\times}10\ensuremath{^{\textnormal{$-$4}}} at 90\%\vspace{0.1cm}}&\\
&&&&\parbox[t][0.3cm]{13.409841cm}{\raggedright {\ }{\ }{\ }C.L. See also \ensuremath{\Gamma}\ensuremath{_{\ensuremath{\alpha}}}/\ensuremath{\Gamma}\ensuremath{<}0.03 from private communication of (\href{https://www.nndc.bnl.gov/nsr/nsrlink.jsp?2009Ta09,B}{2009Ta09}: See Table I) with T.\vspace{0.1cm}}&\\
&&&&\parbox[t][0.3cm]{13.409841cm}{\raggedright {\ }{\ }{\ }Motobayashi.\vspace{0.1cm}}&\\
&&&&\parbox[t][0.3cm]{13.409841cm}{\raggedright \ensuremath{\Gamma}\ensuremath{_{\ensuremath{\alpha}}}=17 \ensuremath{\mu}eV \textit{13} deduced by (\href{https://www.nndc.bnl.gov/nsr/nsrlink.jsp?2009Ta09,B}{2009Ta09}) using \ensuremath{\Gamma}\ensuremath{_{\ensuremath{\alpha}}}/\ensuremath{\Gamma} from (\href{https://www.nndc.bnl.gov/nsr/nsrlink.jsp?2009Ta09,B}{2009Ta09}) and \ensuremath{\tau}=13 fs \textit{+9{\textminus}6} from\vspace{0.1cm}}&\\
&&&&\parbox[t][0.3cm]{13.409841cm}{\raggedright {\ }{\ }{\ }(\href{https://www.nndc.bnl.gov/nsr/nsrlink.jsp?2005Ta28,B}{2005Ta28}: \ensuremath{^{\textnormal{15}}}O(\ensuremath{^{\textnormal{3}}}He,n\ensuremath{\gamma})).\vspace{0.1cm}}&\\
&&&&\parbox[t][0.3cm]{13.409841cm}{\raggedright \ensuremath{\Gamma}\ensuremath{_{\ensuremath{\alpha}}}: See also \ensuremath{\Gamma}\ensuremath{_{\ensuremath{\alpha}}}=0.0072 meV (\href{https://www.nndc.bnl.gov/nsr/nsrlink.jsp?1990Ma05,B}{1990Ma05}) deduced, see Table 1, based on a reduced \ensuremath{\alpha}-particle\vspace{0.1cm}}&\\
&&&&\parbox[t][0.3cm]{13.409841cm}{\raggedright {\ }{\ }{\ }width of 0.06 calculated by H. T. Fortune in 1986 (from a private communication of H. T.\vspace{0.1cm}}&\\
&&&&\parbox[t][0.3cm]{13.409841cm}{\raggedright {\ }{\ }{\ }Fortune and the authors of \href{https://www.nndc.bnl.gov/nsr/nsrlink.jsp?1990Ma05,B}{1990Ma05}).\vspace{0.1cm}}&\\
&&&&\parbox[t][0.3cm]{13.409841cm}{\raggedright \ensuremath{\Gamma}\ensuremath{_{\ensuremath{\gamma}}}=73 meV \textit{41} (\href{https://www.nndc.bnl.gov/nsr/nsrlink.jsp?1990Ma05,B}{1990Ma05}: See Table 1) deduced from the mirror state: \ensuremath{^{\textnormal{19}}}F*(3908).\vspace{0.1cm}}&\\
&&&&\parbox[t][0.3cm]{13.409841cm}{\raggedright J\ensuremath{^{\pi}}: From (1) J\ensuremath{^{\ensuremath{\pi}}}=3/2\ensuremath{^{\textnormal{+}}} (\href{https://www.nndc.bnl.gov/nsr/nsrlink.jsp?2015Pa46,B}{2015Pa46}) from finite-range coupled-channels analysis via FRESCO (see\vspace{0.1cm}}&\\
&&&&\parbox[t][0.3cm]{13.409841cm}{\raggedright {\ }{\ }{\ }Fig. 3), but J\ensuremath{^{\ensuremath{\pi}}}=3/2\ensuremath{^{-}} was also an acceptable fit albeit worse than that obtained for J\ensuremath{^{\ensuremath{\pi}}}=3/2\ensuremath{^{\textnormal{+}}}; and\vspace{0.1cm}}&\\
&&&&\parbox[t][0.3cm]{13.409841cm}{\raggedright {\ }{\ }{\ }(2) J\ensuremath{^{\ensuremath{\pi}}}=3/2\ensuremath{^{\textnormal{+}}} (\href{https://www.nndc.bnl.gov/nsr/nsrlink.jsp?2009Ta09,B}{2009Ta09}) from comparison of the \ensuremath{\gamma}-ray decay scheme of this \ensuremath{^{\textnormal{19}}}Ne state with its\vspace{0.1cm}}&\\
&&&&\parbox[t][0.3cm]{13.409841cm}{\raggedright {\ }{\ }{\ }mirror state in \ensuremath{^{\textnormal{19}}}F.\vspace{0.1cm}}&\\
&&&&\parbox[t][0.3cm]{13.409841cm}{\raggedright J\ensuremath{^{\pi}}: See also J\ensuremath{^{\ensuremath{\pi}}}=3/2\ensuremath{^{\textnormal{+}}} assumed by (\href{https://www.nndc.bnl.gov/nsr/nsrlink.jsp?1990Ma05,B}{1990Ma05}).\vspace{0.1cm}}&\\
\multicolumn{1}{r@{}}{4140}&\multicolumn{1}{@{}l}{}&\multicolumn{1}{l}{(9/2\ensuremath{^{-}},7/2)\ensuremath{^{{\hyperlink{NE32LEVEL2}{c}}}}}&&\parbox[t][0.3cm]{13.409841cm}{\raggedright E(level): From (\href{https://www.nndc.bnl.gov/nsr/nsrlink.jsp?1990Ma05,B}{1990Ma05}: See Fig. 2).\vspace{0.1cm}}&\\
&&&&\parbox[t][0.3cm]{13.409841cm}{\raggedright E(level): See also 4.14 MeV (\href{https://www.nndc.bnl.gov/nsr/nsrlink.jsp?2007TaZX,B}{2007TaZX}, \href{https://www.nndc.bnl.gov/nsr/nsrlink.jsp?2007Ta13,B}{2007Ta13}: See Fig. 2, unresolved from the \ensuremath{^{\textnormal{19}}}Ne*(4.2\vspace{0.1cm}}&\\
&&&&\parbox[t][0.3cm]{13.409841cm}{\raggedright {\ }{\ }{\ }MeV) level); 4143.5 keV (\href{https://www.nndc.bnl.gov/nsr/nsrlink.jsp?2009Ta09,B}{2009Ta09}) quoted from (\href{https://www.nndc.bnl.gov/nsr/nsrlink.jsp?2005Ta28,B}{2005Ta28}: \ensuremath{^{\textnormal{17}}}O(\ensuremath{^{\textnormal{3}}}He,n\ensuremath{\gamma})) and unresolved\vspace{0.1cm}}&\\
&&&&\parbox[t][0.3cm]{13.409841cm}{\raggedright {\ }{\ }{\ }from the \ensuremath{^{\textnormal{19}}}Ne*(4.2 MeV) level; and 4143 keV (\href{https://www.nndc.bnl.gov/nsr/nsrlink.jsp?2015Pa46,B}{2015Pa46}) quoted from (\href{https://www.nndc.bnl.gov/nsr/nsrlink.jsp?2009Ta09,B}{2009Ta09}) and\vspace{0.1cm}}&\\
&&&&\parbox[t][0.3cm]{13.409841cm}{\raggedright {\ }{\ }{\ }(\href{https://www.nndc.bnl.gov/nsr/nsrlink.jsp?2011Da24,B}{2011Da24}: Evaluation of \ensuremath{^{\textnormal{15}}}O(\ensuremath{\alpha},\ensuremath{\gamma})). This state was used as a calibration point in (\href{https://www.nndc.bnl.gov/nsr/nsrlink.jsp?2025PhZZ,B}{2025PhZZ}).\vspace{0.1cm}}&\\
&&&&\parbox[t][0.3cm]{13.409841cm}{\raggedright E(level): (\href{https://www.nndc.bnl.gov/nsr/nsrlink.jsp?2007TaZX,B}{2007TaZX}, \href{https://www.nndc.bnl.gov/nsr/nsrlink.jsp?2007Ta13,B}{2007Ta13}) mentioned that the 4.14-MeV state may have a possible \ensuremath{\alpha}-cluster\vspace{0.1cm}}&\\
&&&&\parbox[t][0.3cm]{13.409841cm}{\raggedright {\ }{\ }{\ }configuration.\vspace{0.1cm}}&\\
&&&&\parbox[t][0.3cm]{13.409841cm}{\raggedright J\ensuremath{^{\pi}}: From (1) J\ensuremath{^{\ensuremath{\pi}}}=9/2\ensuremath{^{-}}, 7/2\ensuremath{^{\textnormal{+}}} (\href{https://www.nndc.bnl.gov/nsr/nsrlink.jsp?2015Pa46,B}{2015Pa46}) from finite-range coupled-channels analysis via FRESCO.\vspace{0.1cm}}&\\
&&&&\parbox[t][0.3cm]{13.409841cm}{\raggedright {\ }{\ }{\ }J\ensuremath{^{\ensuremath{\pi}}}=9/2\ensuremath{^{-}} was recommended by (\href{https://www.nndc.bnl.gov/nsr/nsrlink.jsp?2015Pa46,B}{2015Pa46}) based on prior studies. J\ensuremath{^{\ensuremath{\pi}}}=7/2\ensuremath{^{-}} was a poorer fit to\vspace{0.1cm}}&\\
&&&&\parbox[t][0.3cm]{13.409841cm}{\raggedright {\ }{\ }{\ }the data of (\href{https://www.nndc.bnl.gov/nsr/nsrlink.jsp?2015Pa46,B}{2015Pa46}) particularly at forward angles. (2) J\ensuremath{^{\ensuremath{\pi}}}=7/2\ensuremath{^{-}} (\href{https://www.nndc.bnl.gov/nsr/nsrlink.jsp?2009Ta09,B}{2009Ta09}) from comparison\vspace{0.1cm}}&\\
&&&&\parbox[t][0.3cm]{13.409841cm}{\raggedright {\ }{\ }{\ }of the \ensuremath{\gamma}-ray decay scheme of this \ensuremath{^{\textnormal{19}}}Ne state with its proposed mirror state in \ensuremath{^{\textnormal{19}}}F.\vspace{0.1cm}}&\\
&&&&\parbox[t][0.3cm]{13.409841cm}{\raggedright J\ensuremath{^{\pi}}: See also J\ensuremath{^{\ensuremath{\pi}}}=9/2\ensuremath{^{-}} assumed by (\href{https://www.nndc.bnl.gov/nsr/nsrlink.jsp?1990Ma05,B}{1990Ma05}).\vspace{0.1cm}}&\\
&&&&\parbox[t][0.3cm]{13.409841cm}{\raggedright \ensuremath{\Gamma}\ensuremath{_{\ensuremath{\alpha}}}/\ensuremath{\Gamma}=1.2\ensuremath{\times}10\ensuremath{^{\textnormal{$-$3}}} \textit{5} measured for the first time by (\href{https://www.nndc.bnl.gov/nsr/nsrlink.jsp?2007TaZX,B}{2007TaZX}, \href{https://www.nndc.bnl.gov/nsr/nsrlink.jsp?2007Ta13,B}{2007Ta13}, \href{https://www.nndc.bnl.gov/nsr/nsrlink.jsp?2009Ta09,B}{2009Ta09}) and is the\vspace{0.1cm}}&\\
&&&&\parbox[t][0.3cm]{13.409841cm}{\raggedright {\ }{\ }{\ }combined branching ratio for the unresolved doublet observed at 4.14+4.2 MeV (\href{https://www.nndc.bnl.gov/nsr/nsrlink.jsp?2007Ta13,B}{2007Ta13},\vspace{0.1cm}}&\\
&&&&\parbox[t][0.3cm]{13.409841cm}{\raggedright {\ }{\ }{\ }\href{https://www.nndc.bnl.gov/nsr/nsrlink.jsp?2009Ta09,B}{2009Ta09}). From a comparison of the simulated decays of these states, (\href{https://www.nndc.bnl.gov/nsr/nsrlink.jsp?2007Ta13,B}{2007Ta13}, \href{https://www.nndc.bnl.gov/nsr/nsrlink.jsp?2009Ta09,B}{2009Ta09})\vspace{0.1cm}}&\\
&&&&\parbox[t][0.3cm]{13.409841cm}{\raggedright {\ }{\ }{\ }thought that the \ensuremath{\alpha} decay events are more likely from the 4.14-MeV state. However, after\vspace{0.1cm}}&\\
&&&&\parbox[t][0.3cm]{13.409841cm}{\raggedright {\ }{\ }{\ }calculating spectroscopic factors for these two states, the result favored the decay from the\vspace{0.1cm}}&\\
&&&&\parbox[t][0.3cm]{13.409841cm}{\raggedright {\ }{\ }{\ }4.2-MeV state. The uncertainties in the spectroscopic factors were too large to draw any\vspace{0.1cm}}&\\
&&&&\parbox[t][0.3cm]{13.409841cm}{\raggedright {\ }{\ }{\ }conclusions, and more data are needed to clarify this.\vspace{0.1cm}}&\\
&&&&\parbox[t][0.3cm]{13.409841cm}{\raggedright \ensuremath{\Gamma}\ensuremath{_{\ensuremath{\alpha}}}=44 \ensuremath{\mu}eV \textit{20} deduced by (\href{https://www.nndc.bnl.gov/nsr/nsrlink.jsp?2009Ta09,B}{2009Ta09}) using \ensuremath{\Gamma}\ensuremath{_{\ensuremath{\alpha}}}/\ensuremath{\Gamma} from (\href{https://www.nndc.bnl.gov/nsr/nsrlink.jsp?2009Ta09,B}{2009Ta09}) and \ensuremath{\tau}=18 fs \textit{+2{\textminus}3} from\vspace{0.1cm}}&\\
&&&&\parbox[t][0.3cm]{13.409841cm}{\raggedright {\ }{\ }{\ }(\href{https://www.nndc.bnl.gov/nsr/nsrlink.jsp?2005Ta28,B}{2005Ta28}: \ensuremath{^{\textnormal{15}}}O(\ensuremath{^{\textnormal{3}}}He,n\ensuremath{\gamma})) and assuming (1) J\ensuremath{^{\ensuremath{\pi}}}=7/2\ensuremath{^{-}} (\href{https://www.nndc.bnl.gov/nsr/nsrlink.jsp?2009Ta09,B}{2009Ta09}) and (2) that the \ensuremath{\alpha} decay\vspace{0.1cm}}&\\
&&&&\parbox[t][0.3cm]{13.409841cm}{\raggedright {\ }{\ }{\ }observed is solely from either the 4.20-MeV state or the 4.14-MeV state, and therefore this \ensuremath{\Gamma}\ensuremath{_{\ensuremath{\alpha}}}\vspace{0.1cm}}&\\
&&&&\parbox[t][0.3cm]{13.409841cm}{\raggedright {\ }{\ }{\ }value should not be used for both of these states together.\vspace{0.1cm}}&\\
\multicolumn{1}{r@{}}{4.20\ensuremath{\times10^{3}}}&\multicolumn{1}{@{}l}{}&\multicolumn{1}{l}{(7/2\ensuremath{^{-}},9/2\ensuremath{^{+}})\ensuremath{^{{\hyperlink{NE32LEVEL2}{c}}}}}&&\parbox[t][0.3cm]{13.409841cm}{\raggedright E(level): From (\href{https://www.nndc.bnl.gov/nsr/nsrlink.jsp?2007Ta13,B}{2007Ta13}: See Fig. 2, unresolved from the \ensuremath{^{\textnormal{19}}}Ne*(4.14 MeV) level).\vspace{0.1cm}}&\\
\end{longtable}
\begin{textblock}{29}(0,27.3)
Continued on next page (footnotes at end of table)
\end{textblock}
\clearpage
\begin{longtable}{cccc@{\extracolsep{\fill}}c}
\\[-.4cm]
\multicolumn{5}{c}{{\bf \small \underline{\ensuremath{^{\textnormal{19}}}F(\ensuremath{^{\textnormal{3}}}He,t)\hspace{0.2in}\href{https://www.nndc.bnl.gov/nsr/nsrlink.jsp?1998Ut02,B}{1998Ut02},\href{https://www.nndc.bnl.gov/nsr/nsrlink.jsp?2013La01,B}{2013La01},\href{https://www.nndc.bnl.gov/nsr/nsrlink.jsp?2019Ka15,B}{2019Ka15} (continued)}}}\\
\multicolumn{5}{c}{~}\\
\multicolumn{5}{c}{\underline{\ensuremath{^{19}}Ne Levels (continued)}}\\
\multicolumn{5}{c}{~}\\
\multicolumn{2}{c}{E(level)$^{}$}&J$^{\pi}$$^{}$&Comments&\\[-.2cm]
\multicolumn{2}{c}{\hrulefill}&\hrulefill&\hrulefill&
\endhead
&&&\parbox[t][0.3cm]{15.021081cm}{\raggedright E(level): See also 4197 keV (\href{https://www.nndc.bnl.gov/nsr/nsrlink.jsp?1990Ma05,B}{1990Ma05}: See Fig. 2); 4200.3 (\href{https://www.nndc.bnl.gov/nsr/nsrlink.jsp?2009Ta09,B}{2009Ta09}) quoted from (\href{https://www.nndc.bnl.gov/nsr/nsrlink.jsp?2005Ta28,B}{2005Ta28}:\vspace{0.1cm}}&\\
&&&\parbox[t][0.3cm]{15.021081cm}{\raggedright {\ }{\ }{\ }\ensuremath{^{\textnormal{15}}}O(\ensuremath{^{\textnormal{3}}}He,n\ensuremath{\gamma})) and unresolved from the \ensuremath{^{\textnormal{19}}}Ne*(4.14 MeV) level; and 4200 keV (\href{https://www.nndc.bnl.gov/nsr/nsrlink.jsp?2015Pa46,B}{2015Pa46}) quoted from\vspace{0.1cm}}&\\
&&&\parbox[t][0.3cm]{15.021081cm}{\raggedright {\ }{\ }{\ }(\href{https://www.nndc.bnl.gov/nsr/nsrlink.jsp?2009Ta09,B}{2009Ta09}) and (\href{https://www.nndc.bnl.gov/nsr/nsrlink.jsp?2011Da24,B}{2011Da24}: Evaluation of \ensuremath{^{\textnormal{15}}}O(\ensuremath{\alpha},\ensuremath{\gamma})). This state was used as a calibration point in\vspace{0.1cm}}&\\
&&&\parbox[t][0.3cm]{15.021081cm}{\raggedright {\ }{\ }{\ }(\href{https://www.nndc.bnl.gov/nsr/nsrlink.jsp?2025PhZZ,B}{2025PhZZ}).\vspace{0.1cm}}&\\
&&&\parbox[t][0.3cm]{15.021081cm}{\raggedright J\ensuremath{^{\pi}}: From (1) J\ensuremath{^{\ensuremath{\pi}}}=7/2\ensuremath{^{-}}, 9/2\ensuremath{^{\textnormal{+}}} (\href{https://www.nndc.bnl.gov/nsr/nsrlink.jsp?2015Pa46,B}{2015Pa46}) from a finite-range coupled-channels analysis via FRESCO, where\vspace{0.1cm}}&\\
&&&\parbox[t][0.3cm]{15.021081cm}{\raggedright {\ }{\ }{\ }J\ensuremath{^{\ensuremath{\pi}}}=7/2\ensuremath{^{-}} was recommended based on prior studies. A J\ensuremath{^{\ensuremath{\pi}}}=9/2\ensuremath{^{-}} gives a significantly worse fit to the data at\vspace{0.1cm}}&\\
&&&\parbox[t][0.3cm]{15.021081cm}{\raggedright {\ }{\ }{\ }the lowest measured angles. (2) J\ensuremath{^{\ensuremath{\pi}}}=9/2\ensuremath{^{-}} (\href{https://www.nndc.bnl.gov/nsr/nsrlink.jsp?2009Ta09,B}{2009Ta09}) from comparison of the \ensuremath{\gamma}-ray decay scheme of this\vspace{0.1cm}}&\\
&&&\parbox[t][0.3cm]{15.021081cm}{\raggedright {\ }{\ }{\ }\ensuremath{^{\textnormal{19}}}Ne state with its proposed mirror state in \ensuremath{^{\textnormal{19}}}F. See also J\ensuremath{^{\ensuremath{\pi}}}=7/2\ensuremath{^{-}} assumed by (\href{https://www.nndc.bnl.gov/nsr/nsrlink.jsp?1990Ma05,B}{1990Ma05}).\vspace{0.1cm}}&\\
&&&\parbox[t][0.3cm]{15.021081cm}{\raggedright \ensuremath{\Gamma}\ensuremath{_{\ensuremath{\alpha}}}/\ensuremath{\Gamma}=1.2\ensuremath{\times}10\ensuremath{^{\textnormal{$-$3}}} \textit{5} measured for the first time by (\href{https://www.nndc.bnl.gov/nsr/nsrlink.jsp?2007TaZX,B}{2007TaZX}, \href{https://www.nndc.bnl.gov/nsr/nsrlink.jsp?2007Ta13,B}{2007Ta13}, \href{https://www.nndc.bnl.gov/nsr/nsrlink.jsp?2009Ta09,B}{2009Ta09}) and is the combined\vspace{0.1cm}}&\\
&&&\parbox[t][0.3cm]{15.021081cm}{\raggedright {\ }{\ }{\ }branching ratio for the unresolved doublet observed at 4.14+4.2 MeV (\href{https://www.nndc.bnl.gov/nsr/nsrlink.jsp?2007Ta13,B}{2007Ta13}, \href{https://www.nndc.bnl.gov/nsr/nsrlink.jsp?2009Ta09,B}{2009Ta09}). See the above\vspace{0.1cm}}&\\
&&&\parbox[t][0.3cm]{15.021081cm}{\raggedright {\ }{\ }{\ }mentioned comment on the \ensuremath{\Gamma}\ensuremath{_{\ensuremath{\alpha}}}/\ensuremath{\Gamma} for the 4140-keV level.\vspace{0.1cm}}&\\
&&&\parbox[t][0.3cm]{15.021081cm}{\raggedright \ensuremath{\Gamma}\ensuremath{_{\ensuremath{\alpha}}}=18 \ensuremath{\mu}eV \textit{9} deduced by (\href{https://www.nndc.bnl.gov/nsr/nsrlink.jsp?2009Ta09,B}{2009Ta09}) using \ensuremath{\Gamma}\ensuremath{_{\ensuremath{\alpha}}}/\ensuremath{\Gamma} from (\href{https://www.nndc.bnl.gov/nsr/nsrlink.jsp?2009Ta09,B}{2009Ta09}) and \ensuremath{\tau}=43 fs \textit{+12{\textminus}9} from (\href{https://www.nndc.bnl.gov/nsr/nsrlink.jsp?2005Ta28,B}{2005Ta28}:\vspace{0.1cm}}&\\
&&&\parbox[t][0.3cm]{15.021081cm}{\raggedright {\ }{\ }{\ }\ensuremath{^{\textnormal{15}}}O(\ensuremath{^{\textnormal{3}}}He,n\ensuremath{\gamma})) and assuming (1) J\ensuremath{^{\ensuremath{\pi}}}=9/2\ensuremath{^{-}} (\href{https://www.nndc.bnl.gov/nsr/nsrlink.jsp?2009Ta09,B}{2009Ta09}) and (2) that the \ensuremath{\alpha} decay observed is solely from either\vspace{0.1cm}}&\\
&&&\parbox[t][0.3cm]{15.021081cm}{\raggedright {\ }{\ }{\ }the 4.20-MeV state or the 4.14-MeV state, and therefore this \ensuremath{\Gamma}\ensuremath{_{\ensuremath{\alpha}}} value should not be used for both of these\vspace{0.1cm}}&\\
&&&\parbox[t][0.3cm]{15.021081cm}{\raggedright {\ }{\ }{\ }states together.\vspace{0.1cm}}&\\
\multicolumn{1}{r@{}}{4379}&\multicolumn{1}{@{}l}{}&\multicolumn{1}{l}{(7/2\ensuremath{^{+}})}&\parbox[t][0.3cm]{15.021081cm}{\raggedright E(level): From (\href{https://www.nndc.bnl.gov/nsr/nsrlink.jsp?1990Ma05,B}{1990Ma05}, \href{https://www.nndc.bnl.gov/nsr/nsrlink.jsp?2004Vi05,B}{2004Vi05}: See Fig. 1).\vspace{0.1cm}}&\\
&&&\parbox[t][0.3cm]{15.021081cm}{\raggedright E(level): See also 4.38 MeV (\href{https://www.nndc.bnl.gov/nsr/nsrlink.jsp?2007TaZX,B}{2007TaZX}, \href{https://www.nndc.bnl.gov/nsr/nsrlink.jsp?2007Ta13,B}{2007Ta13}: See Fig. 2); 4377.8 keV (\href{https://www.nndc.bnl.gov/nsr/nsrlink.jsp?2009Ta09,B}{2009Ta09}) quoted from\vspace{0.1cm}}&\\
&&&\parbox[t][0.3cm]{15.021081cm}{\raggedright {\ }{\ }{\ }(\href{https://www.nndc.bnl.gov/nsr/nsrlink.jsp?2005Ta28,B}{2005Ta28}: \ensuremath{^{\textnormal{15}}}O(\ensuremath{^{\textnormal{3}}}He,n\ensuremath{\gamma})); 4378 keV (\href{https://www.nndc.bnl.gov/nsr/nsrlink.jsp?2015Pa46,B}{2015Pa46}) quoted from (\href{https://www.nndc.bnl.gov/nsr/nsrlink.jsp?2009Ta09,B}{2009Ta09}) and (\href{https://www.nndc.bnl.gov/nsr/nsrlink.jsp?2011Da24,B}{2011Da24}: \ensuremath{^{\textnormal{15}}}O(\ensuremath{\alpha},\ensuremath{\gamma})\vspace{0.1cm}}&\\
&&&\parbox[t][0.3cm]{15.021081cm}{\raggedright {\ }{\ }{\ }evaluation); and 4378 keV (\href{https://www.nndc.bnl.gov/nsr/nsrlink.jsp?2021Ri04,B}{2021Ri04}) used as a calibration point.\vspace{0.1cm}}&\\
&&&\parbox[t][0.3cm]{15.021081cm}{\raggedright \ensuremath{\Gamma}\ensuremath{_{\ensuremath{\alpha}}}/\ensuremath{\Gamma}: See (1) \ensuremath{\Gamma}\ensuremath{_{\ensuremath{\alpha}}}/\ensuremath{\Gamma}=0.044 \textit{32} (\href{https://www.nndc.bnl.gov/nsr/nsrlink.jsp?1990Ma05,B}{1990Ma05}): Evaluator notes that the TAC spectrum representing the t-\ensuremath{\alpha}\vspace{0.1cm}}&\\
&&&\parbox[t][0.3cm]{15.021081cm}{\raggedright {\ }{\ }{\ }coincidence events is flat and reveals no peak. Also the Si detector spectrum showing the \ensuremath{\alpha} decay events has\vspace{0.1cm}}&\\
&&&\parbox[t][0.3cm]{15.021081cm}{\raggedright {\ }{\ }{\ }a large background. Therefore, the deduced branching ratio is less reliable, see also remarks by (\href{https://www.nndc.bnl.gov/nsr/nsrlink.jsp?2011Da24,B}{2011Da24}:\vspace{0.1cm}}&\\
&&&\parbox[t][0.3cm]{15.021081cm}{\raggedright {\ }{\ }{\ }Evaluation of \ensuremath{^{\textnormal{15}}}O(\ensuremath{\alpha},\ensuremath{\gamma})); (2) \ensuremath{\Gamma}\ensuremath{_{\ensuremath{\alpha}}}/\ensuremath{\Gamma}\ensuremath{>}0.0027 (\href{https://www.nndc.bnl.gov/nsr/nsrlink.jsp?2004Vi05,B}{2004Vi05}) at 1\ensuremath{\sigma} C.L., this result is tentative with a probability\vspace{0.1cm}}&\\
&&&\parbox[t][0.3cm]{15.021081cm}{\raggedright {\ }{\ }{\ }of \ensuremath{>}90\%. The lower limit is placed due to a potential loss of some \ensuremath{\alpha} particles whose signals were below the\vspace{0.1cm}}&\\
&&&\parbox[t][0.3cm]{15.021081cm}{\raggedright {\ }{\ }{\ }detection threshold; (3) \ensuremath{\Gamma}\ensuremath{_{\ensuremath{\alpha}}}/\ensuremath{\Gamma}=1.2\ensuremath{\times}10\ensuremath{^{\textnormal{$-$3}}} \textit{3} (\href{https://www.nndc.bnl.gov/nsr/nsrlink.jsp?2007Ta13,B}{2007Ta13}, \href{https://www.nndc.bnl.gov/nsr/nsrlink.jsp?2007TaZX,B}{2007TaZX}, \href{https://www.nndc.bnl.gov/nsr/nsrlink.jsp?2009Ta09,B}{2009Ta09}). Those authors claimed to\vspace{0.1cm}}&\\
&&&\parbox[t][0.3cm]{15.021081cm}{\raggedright {\ }{\ }{\ }have measured this branching ratio for the first time and that their value differs from 0.044 \textit{32} (\href{https://www.nndc.bnl.gov/nsr/nsrlink.jsp?1990Ma05,B}{1990Ma05})\vspace{0.1cm}}&\\
&&&\parbox[t][0.3cm]{15.021081cm}{\raggedright {\ }{\ }{\ }and 16\ensuremath{\times}10\ensuremath{^{\textnormal{$-$3}}} \textit{5} (\href{https://www.nndc.bnl.gov/nsr/nsrlink.jsp?2003Re16,B}{2003Re16}: \ensuremath{^{\textnormal{3}}}He(\ensuremath{^{\textnormal{20}}}Ne,\ensuremath{\alpha})) because the latter two studies were handicapped by poor statistics\vspace{0.1cm}}&\\
&&&\parbox[t][0.3cm]{15.021081cm}{\raggedright {\ }{\ }{\ }and lack of experimental resolutions, respectively. (\href{https://www.nndc.bnl.gov/nsr/nsrlink.jsp?2009Ta09,B}{2009Ta09}) deduced \ensuremath{\Gamma}\ensuremath{_{\ensuremath{\alpha}}}=16\ensuremath{\times}10\ensuremath{^{\textnormal{$-$5}}} eV \textit{+11{\textminus}7} using their\vspace{0.1cm}}&\\
&&&\parbox[t][0.3cm]{15.021081cm}{\raggedright {\ }{\ }{\ }\ensuremath{\Gamma}\ensuremath{_{\ensuremath{\alpha}}}/\ensuremath{\Gamma} and \ensuremath{\tau}=5 fs \textit{+3{\textminus}2} from (\href{https://www.nndc.bnl.gov/nsr/nsrlink.jsp?2005Ta28,B}{2005Ta28}: \ensuremath{^{\textnormal{15}}}O(\ensuremath{^{\textnormal{3}}}He,n\ensuremath{\gamma})); (4) \ensuremath{\Gamma}\ensuremath{_{\ensuremath{\alpha}}}/\ensuremath{\Gamma}\ensuremath{<}0.04 from a private communication\vspace{0.1cm}}&\\
&&&\parbox[t][0.3cm]{15.021081cm}{\raggedright {\ }{\ }{\ }with T. Motobayashi and (\href{https://www.nndc.bnl.gov/nsr/nsrlink.jsp?2009Ta09,B}{2009Ta09}: See Table I).\vspace{0.1cm}}&\\
&&&\parbox[t][0.3cm]{15.021081cm}{\raggedright \ensuremath{\Gamma}\ensuremath{_{\ensuremath{\gamma}}}\ensuremath{>}60 meV (\href{https://www.nndc.bnl.gov/nsr/nsrlink.jsp?1990Ma05,B}{1990Ma05}: See Table 1) deduced from the analog state in \ensuremath{^{\textnormal{19}}}F.\vspace{0.1cm}}&\\
&&&\parbox[t][0.3cm]{15.021081cm}{\raggedright J\ensuremath{^{\pi}}: From comparison of the \ensuremath{\gamma}-ray decay scheme of this \ensuremath{^{\textnormal{19}}}Ne with its mirror state in \ensuremath{^{\textnormal{19}}}F by (\href{https://www.nndc.bnl.gov/nsr/nsrlink.jsp?2009Ta09,B}{2009Ta09}).\vspace{0.1cm}}&\\
&&&\parbox[t][0.3cm]{15.021081cm}{\raggedright {\ }{\ }{\ }Since this is a weak argument, we made the assignment tantative.\vspace{0.1cm}}&\\
&&&\parbox[t][0.3cm]{15.021081cm}{\raggedright J\ensuremath{^{\pi}}: See also J\ensuremath{^{\ensuremath{\pi}}}=7/2\ensuremath{^{\textnormal{+}}} assumed by (\href{https://www.nndc.bnl.gov/nsr/nsrlink.jsp?2004Vi05,B}{2004Vi05}) from (\href{https://www.nndc.bnl.gov/nsr/nsrlink.jsp?1995Ti07,B}{1995Ti07}); and J\ensuremath{^{\ensuremath{\pi}}}=7/2\ensuremath{^{\textnormal{+}}} assumed by (\href{https://www.nndc.bnl.gov/nsr/nsrlink.jsp?1990Ma05,B}{1990Ma05}).\vspace{0.1cm}}&\\
\multicolumn{1}{r@{}}{4549}&\multicolumn{1}{@{}l}{}&\multicolumn{1}{l}{3/2\ensuremath{^{-}}}&\parbox[t][0.3cm]{15.021081cm}{\raggedright \ensuremath{\Gamma}\ensuremath{\alpha}/\ensuremath{\Gamma}=0.07 \textit{2}\vspace{0.1cm}}&\\
&&&\parbox[t][0.3cm]{15.021081cm}{\raggedright E(level): From (\href{https://www.nndc.bnl.gov/nsr/nsrlink.jsp?1990Ma05,B}{1990Ma05}, \href{https://www.nndc.bnl.gov/nsr/nsrlink.jsp?2004Vi05,B}{2004Vi05}: See Fig. 1).\vspace{0.1cm}}&\\
&&&\parbox[t][0.3cm]{15.021081cm}{\raggedright E(level): See also 4.55 MeV (\href{https://www.nndc.bnl.gov/nsr/nsrlink.jsp?2007TaZX,B}{2007TaZX}, \href{https://www.nndc.bnl.gov/nsr/nsrlink.jsp?2007Ta13,B}{2007Ta13}: See Fig. 2), where this state is a resolved member of a\vspace{0.1cm}}&\\
&&&\parbox[t][0.3cm]{15.021081cm}{\raggedright {\ }{\ }{\ }close lying triplet; 4547.7 keV (\href{https://www.nndc.bnl.gov/nsr/nsrlink.jsp?2009Ta09,B}{2009Ta09}) quoted from (\href{https://www.nndc.bnl.gov/nsr/nsrlink.jsp?2005Ta28,B}{2005Ta28}: \ensuremath{^{\textnormal{15}}}O(\ensuremath{^{\textnormal{3}}}He,n\ensuremath{\gamma})), where this level is a\vspace{0.1cm}}&\\
&&&\parbox[t][0.3cm]{15.021081cm}{\raggedright {\ }{\ }{\ }resolved member of a triplet; 4548 keV (\href{https://www.nndc.bnl.gov/nsr/nsrlink.jsp?2015Pa46,B}{2015Pa46}) quoted from (\href{https://www.nndc.bnl.gov/nsr/nsrlink.jsp?2009Ta09,B}{2009Ta09}) and (\href{https://www.nndc.bnl.gov/nsr/nsrlink.jsp?2011Da24,B}{2011Da24}: Evaluation of\vspace{0.1cm}}&\\
&&&\parbox[t][0.3cm]{15.021081cm}{\raggedright {\ }{\ }{\ }\ensuremath{^{\textnormal{15}}}O(\ensuremath{\alpha},\ensuremath{\gamma})); and 4548 keV (\href{https://www.nndc.bnl.gov/nsr/nsrlink.jsp?2021Ri04,B}{2021Ri04}: See Fig. 3), where the state is populated from focal plane singles.\vspace{0.1cm}}&\\
&&&\parbox[t][0.3cm]{15.021081cm}{\raggedright \ensuremath{\Gamma}\ensuremath{_{\ensuremath{\alpha}}}/\ensuremath{\Gamma}: Weighted average of 0.07 \textit{3} (\href{https://www.nndc.bnl.gov/nsr/nsrlink.jsp?1990Ma05,B}{1990Ma05}); 0.06 \textit{4} (\href{https://www.nndc.bnl.gov/nsr/nsrlink.jsp?2004Vi05,B}{2004Vi05}) at 1\ensuremath{\sigma} level; 0.07 2 (\href{https://www.nndc.bnl.gov/nsr/nsrlink.jsp?2007TaZX,B}{2007TaZX},\vspace{0.1cm}}&\\
&&&\parbox[t][0.3cm]{15.021081cm}{\raggedright {\ }{\ }{\ }\href{https://www.nndc.bnl.gov/nsr/nsrlink.jsp?2007Ta13,B}{2007Ta13}, \href{https://www.nndc.bnl.gov/nsr/nsrlink.jsp?2009Ta09,B}{2009Ta09}); and 0.09 \textit{+4{\textminus}2} (private communication of (\href{https://www.nndc.bnl.gov/nsr/nsrlink.jsp?2009Ta09,B}{2009Ta09}) with T. Motobayashi): See\vspace{0.1cm}}&\\
&&&\parbox[t][0.3cm]{15.021081cm}{\raggedright {\ }{\ }{\ }Table I in (\href{https://www.nndc.bnl.gov/nsr/nsrlink.jsp?2009Ta09,B}{2009Ta09}).\vspace{0.1cm}}&\\
&&&\parbox[t][0.3cm]{15.021081cm}{\raggedright \ensuremath{\Gamma}\ensuremath{_{\ensuremath{\alpha}}}=3.5 meV \textit{+20{\textminus}17} deduced by (\href{https://www.nndc.bnl.gov/nsr/nsrlink.jsp?2009Ta09,B}{2009Ta09}) using their \ensuremath{\Gamma}\ensuremath{_{\ensuremath{\alpha}}}/\ensuremath{\Gamma} (see above) and \ensuremath{\tau}=15 fs \textit{+11{\textminus}5} from\vspace{0.1cm}}&\\
&&&\parbox[t][0.3cm]{15.021081cm}{\raggedright {\ }{\ }{\ }(\href{https://www.nndc.bnl.gov/nsr/nsrlink.jsp?2005Ta28,B}{2005Ta28}: \ensuremath{^{\textnormal{15}}}O(\ensuremath{^{\textnormal{3}}}He,n\ensuremath{\gamma})).\vspace{0.1cm}}&\\
&&&\parbox[t][0.3cm]{15.021081cm}{\raggedright \ensuremath{\Gamma}\ensuremath{_{\ensuremath{\gamma}}}=39 meV \textit{+34{\textminus}15} (\href{https://www.nndc.bnl.gov/nsr/nsrlink.jsp?1990Ma05,B}{1990Ma05}) deduced from analog state in \ensuremath{^{\textnormal{19}}}F (see Table 1).\vspace{0.1cm}}&\\
&&&\parbox[t][0.3cm]{15.021081cm}{\raggedright J\ensuremath{^{\pi}}: From (1) J\ensuremath{^{\ensuremath{\pi}}}=3/2\ensuremath{^{-}} (\href{https://www.nndc.bnl.gov/nsr/nsrlink.jsp?2009Ta09,B}{2009Ta09}) from comparison of the \ensuremath{\gamma}-ray decay scheme of this state in \ensuremath{^{\textnormal{19}}}Ne with that\vspace{0.1cm}}&\\
&&&\parbox[t][0.3cm]{15.021081cm}{\raggedright {\ }{\ }{\ }of the mirror state in \ensuremath{^{\textnormal{19}}}F; and (2) J\ensuremath{^{\ensuremath{\pi}}}=3/2\ensuremath{^{-}} (\href{https://www.nndc.bnl.gov/nsr/nsrlink.jsp?2015Pa46,B}{2015Pa46}) from a finite-range coupled-channels analysis via\vspace{0.1cm}}&\\
&&&\parbox[t][0.3cm]{15.021081cm}{\raggedright {\ }{\ }{\ }FRESCO, where J\ensuremath{^{\ensuremath{\pi}}}=3/2\ensuremath{^{\textnormal{+}}} and J\ensuremath{^{\ensuremath{\pi}}}=3/2\ensuremath{^{-}} both provide reasonable fits to the triton angular distribution data, but\vspace{0.1cm}}&\\
&&&\parbox[t][0.3cm]{15.021081cm}{\raggedright {\ }{\ }{\ }J\ensuremath{^{\ensuremath{\pi}}}=3/2\ensuremath{^{-}} was recommended based on prior studies. J\ensuremath{^{\ensuremath{\pi}}}=1/2\ensuremath{^{-}} is a poor fit to the data in that study.\vspace{0.1cm}}&\\
&&&\parbox[t][0.3cm]{15.021081cm}{\raggedright J\ensuremath{^{\pi}}: See also (1) J\ensuremath{^{\ensuremath{\pi}}}=3/2\ensuremath{^{-}} assumed by (\href{https://www.nndc.bnl.gov/nsr/nsrlink.jsp?1990Ma05,B}{1990Ma05}); and (2) J\ensuremath{^{\ensuremath{\pi}}}=(1/2\ensuremath{^{-}}) reported by (\href{https://www.nndc.bnl.gov/nsr/nsrlink.jsp?2004Vi05,B}{2004Vi05}) and taken from\vspace{0.1cm}}&\\
&&&\parbox[t][0.3cm]{15.021081cm}{\raggedright {\ }{\ }{\ }(\href{https://www.nndc.bnl.gov/nsr/nsrlink.jsp?1973Da31,B}{1973Da31}: \ensuremath{^{\textnormal{15}}}O(\ensuremath{^{\textnormal{3}}}He,n\ensuremath{\gamma})).\vspace{0.1cm}}&\\
\multicolumn{1}{r@{}}{4600}&\multicolumn{1}{@{}l}{}&\multicolumn{1}{l}{5/2\ensuremath{^{+}}}&\parbox[t][0.3cm]{15.021081cm}{\raggedright \ensuremath{\Gamma}\ensuremath{\alpha}/\ensuremath{\Gamma}=0.240 \textit{17}\vspace{0.1cm}}&\\
\end{longtable}
\begin{textblock}{29}(0,27.3)
Continued on next page (footnotes at end of table)
\end{textblock}
\clearpage
\begin{longtable}{cccc@{\extracolsep{\fill}}c}
\\[-.4cm]
\multicolumn{5}{c}{{\bf \small \underline{\ensuremath{^{\textnormal{19}}}F(\ensuremath{^{\textnormal{3}}}He,t)\hspace{0.2in}\href{https://www.nndc.bnl.gov/nsr/nsrlink.jsp?1998Ut02,B}{1998Ut02},\href{https://www.nndc.bnl.gov/nsr/nsrlink.jsp?2013La01,B}{2013La01},\href{https://www.nndc.bnl.gov/nsr/nsrlink.jsp?2019Ka15,B}{2019Ka15} (continued)}}}\\
\multicolumn{5}{c}{~}\\
\multicolumn{5}{c}{\underline{\ensuremath{^{19}}Ne Levels (continued)}}\\
\multicolumn{5}{c}{~}\\
\multicolumn{2}{c}{E(level)$^{}$}&J$^{\pi}$$^{}$&Comments&\\[-.2cm]
\multicolumn{2}{c}{\hrulefill}&\hrulefill&\hrulefill&
\endhead
&&&\parbox[t][0.3cm]{14.807081cm}{\raggedright E(level): From (\href{https://www.nndc.bnl.gov/nsr/nsrlink.jsp?1990Ma05,B}{1990Ma05}).\vspace{0.1cm}}&\\
&&&\parbox[t][0.3cm]{14.807081cm}{\raggedright E(level): See also 4600 keV (\href{https://www.nndc.bnl.gov/nsr/nsrlink.jsp?2004Vi05,B}{2004Vi05}: See Fig. 1) unresolved from the E\ensuremath{_{\textnormal{x}}}=4635 keV state. However, the\vspace{0.1cm}}&\\
&&&\parbox[t][0.3cm]{14.807081cm}{\raggedright {\ }{\ }{\ }latter state is not expected to emit significant numbers of \ensuremath{\alpha} particles due to its L=7 angular momentum\vspace{0.1cm}}&\\
&&&\parbox[t][0.3cm]{14.807081cm}{\raggedright {\ }{\ }{\ }barrier (\href{https://www.nndc.bnl.gov/nsr/nsrlink.jsp?2004Vi05,B}{2004Vi05}). Therefore, those authors attributed all of the measured \ensuremath{\alpha} particles to the decay of the\vspace{0.1cm}}&\\
&&&\parbox[t][0.3cm]{14.807081cm}{\raggedright {\ }{\ }{\ }\ensuremath{^{\textnormal{19}}}Ne*(4600) state; 4.60 MeV (\href{https://www.nndc.bnl.gov/nsr/nsrlink.jsp?2007Ta13,B}{2007Ta13}), which is a resolved member of a close lying triplet (see Fig. 2);\vspace{0.1cm}}&\\
&&&\parbox[t][0.3cm]{14.807081cm}{\raggedright {\ }{\ }{\ }4.60 MeV (\href{https://www.nndc.bnl.gov/nsr/nsrlink.jsp?2007TaZX,B}{2007TaZX}); 4601.8 keV (\href{https://www.nndc.bnl.gov/nsr/nsrlink.jsp?2009Ta09,B}{2009Ta09}) quoted from (\href{https://www.nndc.bnl.gov/nsr/nsrlink.jsp?2005Ta28,B}{2005Ta28}: \ensuremath{^{\textnormal{15}}}O(\ensuremath{^{\textnormal{3}}}He,n\ensuremath{\gamma})), where this level\vspace{0.1cm}}&\\
&&&\parbox[t][0.3cm]{14.807081cm}{\raggedright {\ }{\ }{\ }was a resolved member of a close lying triplet; 4602 keV (\href{https://www.nndc.bnl.gov/nsr/nsrlink.jsp?2015Pa46,B}{2015Pa46}) quoted from (\href{https://www.nndc.bnl.gov/nsr/nsrlink.jsp?2009Ta09,B}{2009Ta09}) and\vspace{0.1cm}}&\\
&&&\parbox[t][0.3cm]{14.807081cm}{\raggedright {\ }{\ }{\ }(\href{https://www.nndc.bnl.gov/nsr/nsrlink.jsp?2011Da24,B}{2011Da24}: Evaluation of \ensuremath{^{\textnormal{15}}}O(\ensuremath{\alpha},\ensuremath{\gamma})); and 4602 (\href{https://www.nndc.bnl.gov/nsr/nsrlink.jsp?2021Ri04,B}{2021Ri04}: See Fig. 3), where this state was populated\vspace{0.1cm}}&\\
&&&\parbox[t][0.3cm]{14.807081cm}{\raggedright {\ }{\ }{\ }from focal plane singles. This level was used as a calibration point in (\href{https://www.nndc.bnl.gov/nsr/nsrlink.jsp?2025PhZZ,B}{2025PhZZ}).\vspace{0.1cm}}&\\
&&&\parbox[t][0.3cm]{14.807081cm}{\raggedright \ensuremath{\Gamma}\ensuremath{_{\ensuremath{\alpha}}}/\ensuremath{\Gamma}: Weighted average (with external, combined weights) of 0.25 \textit{4} (\href{https://www.nndc.bnl.gov/nsr/nsrlink.jsp?1990Ma05,B}{1990Ma05}); 0.208 \textit{26} (\href{https://www.nndc.bnl.gov/nsr/nsrlink.jsp?2004Vi05,B}{2004Vi05}) at\vspace{0.1cm}}&\\
&&&\parbox[t][0.3cm]{14.807081cm}{\raggedright {\ }{\ }{\ }1\ensuremath{\sigma} level; 0.26 \textit{3} (\href{https://www.nndc.bnl.gov/nsr/nsrlink.jsp?2007TaZX,B}{2007TaZX}, \href{https://www.nndc.bnl.gov/nsr/nsrlink.jsp?2009Ta09,B}{2009Ta09}) and 0.29 \textit{+6{\textminus}4} from a private communication with T. Motobayashi\vspace{0.1cm}}&\\
&&&\parbox[t][0.3cm]{14.807081cm}{\raggedright {\ }{\ }{\ }and (\href{https://www.nndc.bnl.gov/nsr/nsrlink.jsp?2009Ta09,B}{2009Ta09}: See Table I).\vspace{0.1cm}}&\\
&&&\parbox[t][0.3cm]{14.807081cm}{\raggedright \ensuremath{\Gamma}\ensuremath{_{\ensuremath{\alpha}}}=24 meV \textit{+33{\textminus}10} deduced by (\href{https://www.nndc.bnl.gov/nsr/nsrlink.jsp?2009Ta09,B}{2009Ta09}) using their \ensuremath{\Gamma}\ensuremath{_{\ensuremath{\alpha}}}/\ensuremath{\Gamma} (see above) and \ensuremath{\tau}=7 fs \textit{+5{\textminus}4} from\vspace{0.1cm}}&\\
&&&\parbox[t][0.3cm]{14.807081cm}{\raggedright {\ }{\ }{\ }(\href{https://www.nndc.bnl.gov/nsr/nsrlink.jsp?2005Ta28,B}{2005Ta28}: \ensuremath{^{\textnormal{15}}}O(\ensuremath{^{\textnormal{3}}}He,n\ensuremath{\gamma})). Other value: \ensuremath{\Gamma}\ensuremath{_{\ensuremath{\alpha}}}=88 meV \textit{18} suggested by (\href{https://www.nndc.bnl.gov/nsr/nsrlink.jsp?1990Ma05,B}{1990Ma05}: See Table 1) from\vspace{0.1cm}}&\\
&&&\parbox[t][0.3cm]{14.807081cm}{\raggedright {\ }{\ }{\ }(\href{https://www.nndc.bnl.gov/nsr/nsrlink.jsp?1987Ma31,B}{1987Ma31}: \ensuremath{^{\textnormal{15}}}N(\ensuremath{\alpha},\ensuremath{\gamma})) for the mirror level in \ensuremath{^{\textnormal{19}}}F.\vspace{0.1cm}}&\\
&&&\parbox[t][0.3cm]{14.807081cm}{\raggedright \ensuremath{\Gamma}\ensuremath{_{\ensuremath{\gamma}}}\ensuremath{>}13 meV (\href{https://www.nndc.bnl.gov/nsr/nsrlink.jsp?1990Ma05,B}{1990Ma05}) deduced from analog state in \ensuremath{^{\textnormal{19}}}F (see Table 1 in that study).\vspace{0.1cm}}&\\
&&&\parbox[t][0.3cm]{14.807081cm}{\raggedright J\ensuremath{^{\pi}}: From (1) J\ensuremath{^{\ensuremath{\pi}}}=3/2\ensuremath{^{-}}, 5/2\ensuremath{^{\textnormal{+}}} (\href{https://www.nndc.bnl.gov/nsr/nsrlink.jsp?2015Pa46,B}{2015Pa46}) from a finite-range coupled-channels analysis via FRESCO, where\vspace{0.1cm}}&\\
&&&\parbox[t][0.3cm]{14.807081cm}{\raggedright {\ }{\ }{\ }J\ensuremath{^{\ensuremath{\pi}}}=5/2\ensuremath{^{\textnormal{+}}} was recommended based on prior studies; and (2) J\ensuremath{^{\ensuremath{\pi}}}=5/2\ensuremath{^{\textnormal{+}}} (\href{https://www.nndc.bnl.gov/nsr/nsrlink.jsp?2009Ta09,B}{2009Ta09}) from comparison of the\vspace{0.1cm}}&\\
&&&\parbox[t][0.3cm]{14.807081cm}{\raggedright {\ }{\ }{\ }\ensuremath{\gamma}-ray decay scheme of this \ensuremath{^{\textnormal{19}}}Ne state with that of the proposed mirror state in \ensuremath{^{\textnormal{19}}}F.\vspace{0.1cm}}&\\
&&&\parbox[t][0.3cm]{14.807081cm}{\raggedright J\ensuremath{^{\pi}}: See also J\ensuremath{^{\ensuremath{\pi}}}=5/2\ensuremath{^{\textnormal{+}}} assumed by (\href{https://www.nndc.bnl.gov/nsr/nsrlink.jsp?1990Ma05,B}{1990Ma05}); and J\ensuremath{^{\ensuremath{\pi}}}=3/2\ensuremath{^{-}} assumed by (\href{https://www.nndc.bnl.gov/nsr/nsrlink.jsp?2004Vi05,B}{2004Vi05}) and taken from\vspace{0.1cm}}&\\
&&&\parbox[t][0.3cm]{14.807081cm}{\raggedright {\ }{\ }{\ }(\href{https://www.nndc.bnl.gov/nsr/nsrlink.jsp?1995Ti07,B}{1995Ti07}).\vspace{0.1cm}}&\\
\multicolumn{1}{r@{}}{4635}&\multicolumn{1}{@{}l}{}&\multicolumn{1}{l}{(13/2\ensuremath{^{+}})}&\parbox[t][0.3cm]{14.807081cm}{\raggedright E(level): From (\href{https://www.nndc.bnl.gov/nsr/nsrlink.jsp?1990Ma05,B}{1990Ma05}, see Fig. 2).\vspace{0.1cm}}&\\
&&&\parbox[t][0.3cm]{14.807081cm}{\raggedright E(level): See also (1) 4634 keV (\href{https://www.nndc.bnl.gov/nsr/nsrlink.jsp?2009Ta09,B}{2009Ta09}) quoted from (\href{https://www.nndc.bnl.gov/nsr/nsrlink.jsp?2005Ta28,B}{2005Ta28}: \ensuremath{^{\textnormal{15}}}O(\ensuremath{^{\textnormal{3}}}He,n\ensuremath{\gamma})). This state was not\vspace{0.1cm}}&\\
&&&\parbox[t][0.3cm]{14.807081cm}{\raggedright {\ }{\ }{\ }populated in (\href{https://www.nndc.bnl.gov/nsr/nsrlink.jsp?2009Ta09,B}{2009Ta09}); (2) 4634 keV (\href{https://www.nndc.bnl.gov/nsr/nsrlink.jsp?2015Pa46,B}{2015Pa46}) quoted from (\href{https://www.nndc.bnl.gov/nsr/nsrlink.jsp?2011Da24,B}{2011Da24}: Evaluation of \ensuremath{^{\textnormal{15}}}O(\ensuremath{\alpha},\ensuremath{\gamma})); and\vspace{0.1cm}}&\\
&&&\parbox[t][0.3cm]{14.807081cm}{\raggedright {\ }{\ }{\ }(3) 4634 keV (\href{https://www.nndc.bnl.gov/nsr/nsrlink.jsp?2021Ri04,B}{2021Ri04}: See Fig. 3). This state was used as a calibration point in (\href{https://www.nndc.bnl.gov/nsr/nsrlink.jsp?2025PhZZ,B}{2025PhZZ}).\vspace{0.1cm}}&\\
&&&\parbox[t][0.3cm]{14.807081cm}{\raggedright J\ensuremath{^{\pi}}: From J\ensuremath{^{\ensuremath{\pi}}}=13/2\ensuremath{^{\textnormal{+}}} (\href{https://www.nndc.bnl.gov/nsr/nsrlink.jsp?2009Ta09,B}{2009Ta09}) from comparison of the \ensuremath{\gamma}-ray decay scheme of this state in \ensuremath{^{\textnormal{19}}}Ne with that\vspace{0.1cm}}&\\
&&&\parbox[t][0.3cm]{14.807081cm}{\raggedright {\ }{\ }{\ }of the proposed mirror state in \ensuremath{^{\textnormal{19}}}F. Since this is a weak argument, we made the assignment tentative.\vspace{0.1cm}}&\\
&&&\parbox[t][0.3cm]{14.807081cm}{\raggedright J\ensuremath{^{\pi}}: See also J\ensuremath{^{\ensuremath{\pi}}}=13/2\ensuremath{^{\textnormal{+}}} assumed by (\href{https://www.nndc.bnl.gov/nsr/nsrlink.jsp?1990Ma05,B}{1990Ma05}).\vspace{0.1cm}}&\\
\multicolumn{1}{r@{}}{4712}&\multicolumn{1}{@{ }l}{{\it 10}}&\multicolumn{1}{l}{(5/2\ensuremath{^{-}})}&\parbox[t][0.3cm]{14.807081cm}{\raggedright \ensuremath{\Gamma}\ensuremath{\alpha}/\ensuremath{\Gamma}=0.75 \textit{7}\vspace{0.1cm}}&\\
&&&\parbox[t][0.3cm]{14.807081cm}{\raggedright E(level): From (\href{https://www.nndc.bnl.gov/nsr/nsrlink.jsp?2009Ta09,B}{2009Ta09}: See Table II), where this state is a resolved member of a close lying triplet.\vspace{0.1cm}}&\\
&&&\parbox[t][0.3cm]{14.807081cm}{\raggedright E(level): See also 4712 keV (\href{https://www.nndc.bnl.gov/nsr/nsrlink.jsp?1990Ma05,B}{1990Ma05}, \href{https://www.nndc.bnl.gov/nsr/nsrlink.jsp?2004Vi05,B}{2004Vi05}: See Fig. 1); 4.71 MeV (\href{https://www.nndc.bnl.gov/nsr/nsrlink.jsp?2007TaZX,B}{2007TaZX}, \href{https://www.nndc.bnl.gov/nsr/nsrlink.jsp?2007Ta13,B}{2007Ta13}: See Fig.\vspace{0.1cm}}&\\
&&&\parbox[t][0.3cm]{14.807081cm}{\raggedright {\ }{\ }{\ }2), where this level is a resolved member of a close lying triplet; 4712 keV (\href{https://www.nndc.bnl.gov/nsr/nsrlink.jsp?2015Pa46,B}{2015Pa46}) quoted from\vspace{0.1cm}}&\\
&&&\parbox[t][0.3cm]{14.807081cm}{\raggedright {\ }{\ }{\ }(\href{https://www.nndc.bnl.gov/nsr/nsrlink.jsp?2009Ta09,B}{2009Ta09}) and (\href{https://www.nndc.bnl.gov/nsr/nsrlink.jsp?2011Da24,B}{2011Da24}: \ensuremath{^{\textnormal{15}}}O(\ensuremath{\alpha},\ensuremath{\gamma}) evaluation); 4712 keV (\href{https://www.nndc.bnl.gov/nsr/nsrlink.jsp?2021Ri04,B}{2021Ri04}: See Fig. 3); and 4709 keV \textit{3}\vspace{0.1cm}}&\\
&&&\parbox[t][0.3cm]{14.807081cm}{\raggedright {\ }{\ }{\ }(\href{https://www.nndc.bnl.gov/nsr/nsrlink.jsp?2025PhZZ,B}{2025PhZZ}: Unpublished).\vspace{0.1cm}}&\\
&&&\parbox[t][0.3cm]{14.807081cm}{\raggedright \ensuremath{\Gamma}\ensuremath{_{\ensuremath{\alpha}}}/\ensuremath{\Gamma}: Weighted average of 0.82 \textit{15} (\href{https://www.nndc.bnl.gov/nsr/nsrlink.jsp?1990Ma05,B}{1990Ma05}); 0.69 \textit{+11{\textminus}14} (\href{https://www.nndc.bnl.gov/nsr/nsrlink.jsp?2004Vi05,B}{2004Vi05}) at 1\ensuremath{\sigma} level; 0.80 \textit{15} (\href{https://www.nndc.bnl.gov/nsr/nsrlink.jsp?2009Ta09,B}{2009Ta09});\vspace{0.1cm}}&\\
&&&\parbox[t][0.3cm]{14.807081cm}{\raggedright {\ }{\ }{\ }and 0.67 \textit{+23{\textminus}14} from a private communication between T. Motobayashi with (\href{https://www.nndc.bnl.gov/nsr/nsrlink.jsp?2009Ta09,B}{2009Ta09}: See Table I).\vspace{0.1cm}}&\\
&&&\parbox[t][0.3cm]{14.807081cm}{\raggedright {\ }{\ }{\ }Other value: \ensuremath{\Gamma}\ensuremath{_{\ensuremath{\alpha}}}/\ensuremath{\Gamma}=0.64 \textit{24} (\href{https://www.nndc.bnl.gov/nsr/nsrlink.jsp?2025PhZZ,B}{2025PhZZ}: Unpublished).\vspace{0.1cm}}&\\
&&&\parbox[t][0.3cm]{14.807081cm}{\raggedright \ensuremath{\Gamma}\ensuremath{_{\ensuremath{\alpha}}}=200 meV \textit{70} deduced by (\href{https://www.nndc.bnl.gov/nsr/nsrlink.jsp?2009Ta09,B}{2009Ta09}) using their \ensuremath{\Gamma}\ensuremath{_{\ensuremath{\alpha}}}/\ensuremath{\Gamma} (see above) and by adopting (1) the \ensuremath{\tau}=15.4 fs\vspace{0.1cm}}&\\
&&&\parbox[t][0.3cm]{14.807081cm}{\raggedright {\ }{\ }{\ }\textit{30} from (\href{https://www.nndc.bnl.gov/nsr/nsrlink.jsp?1995Ti07,B}{1995Ti07}, see Table 19.9) from the \ensuremath{^{\textnormal{19}}}F*(4682.5, 5/2\ensuremath{^{-}}) mirror level, and (2) \ensuremath{\Gamma}\ensuremath{_{\ensuremath{\gamma}}}=40.7 meV \textit{81}\vspace{0.1cm}}&\\
&&&\parbox[t][0.3cm]{14.807081cm}{\raggedright {\ }{\ }{\ }from (\href{https://www.nndc.bnl.gov/nsr/nsrlink.jsp?1972Ro01,B}{1972Ro01}: \ensuremath{^{\textnormal{15}}}N(\ensuremath{\alpha},\ensuremath{\gamma})), which was also measured for the proposed mirror state in \ensuremath{^{\textnormal{19}}}F. See also\vspace{0.1cm}}&\\
&&&\parbox[t][0.3cm]{14.807081cm}{\raggedright {\ }{\ }{\ }\ensuremath{\Gamma}\ensuremath{_{\ensuremath{\alpha}}}=420 meV \textit{70} suggested bu (\href{https://www.nndc.bnl.gov/nsr/nsrlink.jsp?1990Ma05,B}{1990Ma05}) from (\href{https://www.nndc.bnl.gov/nsr/nsrlink.jsp?1972Ro01,B}{1972Ro01}: \ensuremath{^{\textnormal{15}}}N(\ensuremath{\alpha},\ensuremath{\gamma})) from the mirror state in \ensuremath{^{\textnormal{19}}}F.\vspace{0.1cm}}&\\
&&&\parbox[t][0.3cm]{14.807081cm}{\raggedright \ensuremath{\Gamma}\ensuremath{_{\ensuremath{\gamma}}}=43 meV \textit{8} (\href{https://www.nndc.bnl.gov/nsr/nsrlink.jsp?1990Ma05,B}{1990Ma05}) deduced from the analog state in \ensuremath{^{\textnormal{19}}}F (see Table 1 of that study).\vspace{0.1cm}}&\\
&&&\parbox[t][0.3cm]{14.807081cm}{\raggedright \ensuremath{\Gamma}=35 keV (\href{https://www.nndc.bnl.gov/nsr/nsrlink.jsp?2025PhZZ,B}{2025PhZZ}: Unpublished), where the experimental resolution was 31 keV.\vspace{0.1cm}}&\\
&&&\parbox[t][0.3cm]{14.807081cm}{\raggedright J\ensuremath{^{\pi}}: From (1) J\ensuremath{^{\ensuremath{\pi}}}=5/2\ensuremath{^{-}} (\href{https://www.nndc.bnl.gov/nsr/nsrlink.jsp?2009Ta09,B}{2009Ta09}) from comparison of the \ensuremath{\gamma}-ray decay scheme of this state in \ensuremath{^{\textnormal{19}}}Ne with\vspace{0.1cm}}&\\
&&&\parbox[t][0.3cm]{14.807081cm}{\raggedright {\ }{\ }{\ }that of the proposed mirror state in \ensuremath{^{\textnormal{19}}}F; (2) J\ensuremath{^{\ensuremath{\pi}}}=(5/2\ensuremath{^{-}}) (\href{https://www.nndc.bnl.gov/nsr/nsrlink.jsp?2015Pa46,B}{2015Pa46}) from finite-range coupled-channels\vspace{0.1cm}}&\\
&&&\parbox[t][0.3cm]{14.807081cm}{\raggedright {\ }{\ }{\ }analysis via the FRESCO, where J\ensuremath{^{\ensuremath{\pi}}}=5/2\ensuremath{^{-}} and J\ensuremath{^{\ensuremath{\pi}}}=7/2\ensuremath{^{-}} both provide reasonable fits to the triton angular\vspace{0.1cm}}&\\
&&&\parbox[t][0.3cm]{14.807081cm}{\raggedright {\ }{\ }{\ }distribution data, but J\ensuremath{^{\ensuremath{\pi}}}=5/2\ensuremath{^{-}} was recommended based on prior studies and because it was a better fit to\vspace{0.1cm}}&\\
&&&\parbox[t][0.3cm]{14.807081cm}{\raggedright {\ }{\ }{\ }their data; and (3) J\ensuremath{^{\ensuremath{\pi}}}=3/2\ensuremath{^{-}}, 5/2\ensuremath{^{-}} with L=2 (\href{https://www.nndc.bnl.gov/nsr/nsrlink.jsp?2025PhZZ,B}{2025PhZZ}) from \ensuremath{\alpha} angular correlations.\vspace{0.1cm}}&\\
&&&\parbox[t][0.3cm]{14.807081cm}{\raggedright J\ensuremath{^{\pi}}: See also J\ensuremath{^{\ensuremath{\pi}}}=5/2\ensuremath{^{-}} assumed by (\href{https://www.nndc.bnl.gov/nsr/nsrlink.jsp?1990Ma05,B}{1990Ma05}) and (\href{https://www.nndc.bnl.gov/nsr/nsrlink.jsp?2004Vi05,B}{2004Vi05}).\vspace{0.1cm}}&\\
\multicolumn{1}{r@{}}{5092}&\multicolumn{1}{@{ }l}{{\it 6}}&\multicolumn{1}{l}{(5/2\ensuremath{^{+}})}&\parbox[t][0.3cm]{14.807081cm}{\raggedright \ensuremath{\Gamma}\ensuremath{\alpha}/\ensuremath{\Gamma}=0.86 \textit{4}\vspace{0.1cm}}&\\
&&&\parbox[t][0.3cm]{14.807081cm}{\raggedright E(level): From (\href{https://www.nndc.bnl.gov/nsr/nsrlink.jsp?2009Ta09,B}{2009Ta09}: See Table II).\vspace{0.1cm}}&\\
\end{longtable}
\begin{textblock}{29}(0,27.3)
Continued on next page (footnotes at end of table)
\end{textblock}
\clearpage
\begin{longtable}{ccccc@{\extracolsep{\fill}}c}
\\[-.4cm]
\multicolumn{6}{c}{{\bf \small \underline{\ensuremath{^{\textnormal{19}}}F(\ensuremath{^{\textnormal{3}}}He,t)\hspace{0.2in}\href{https://www.nndc.bnl.gov/nsr/nsrlink.jsp?1998Ut02,B}{1998Ut02},\href{https://www.nndc.bnl.gov/nsr/nsrlink.jsp?2013La01,B}{2013La01},\href{https://www.nndc.bnl.gov/nsr/nsrlink.jsp?2019Ka15,B}{2019Ka15} (continued)}}}\\
\multicolumn{6}{c}{~}\\
\multicolumn{6}{c}{\underline{\ensuremath{^{19}}Ne Levels (continued)}}\\
\multicolumn{6}{c}{~}\\
\multicolumn{2}{c}{E(level)$^{}$}&J$^{\pi}$$^{}$&L$^{}$&Comments&\\[-.2cm]
\multicolumn{2}{c}{\hrulefill}&\hrulefill&\hrulefill&\hrulefill&
\endhead
&&&&\parbox[t][0.3cm]{13.773641cm}{\raggedright E(level): See also 5092 keV (\href{https://www.nndc.bnl.gov/nsr/nsrlink.jsp?1990Ma05,B}{1990Ma05}, \href{https://www.nndc.bnl.gov/nsr/nsrlink.jsp?2004Vi05,B}{2004Vi05}: See Fig. 1); 5.09 MeV (\href{https://www.nndc.bnl.gov/nsr/nsrlink.jsp?2007TaZX,B}{2007TaZX}, \href{https://www.nndc.bnl.gov/nsr/nsrlink.jsp?2007Ta13,B}{2007Ta13}:\vspace{0.1cm}}&\\
&&&&\parbox[t][0.3cm]{13.773641cm}{\raggedright {\ }{\ }{\ }See Fig. 2); 5092 keV (\href{https://www.nndc.bnl.gov/nsr/nsrlink.jsp?2015Pa46,B}{2015Pa46}) quoted from (\href{https://www.nndc.bnl.gov/nsr/nsrlink.jsp?2009Ta09,B}{2009Ta09}) and (\href{https://www.nndc.bnl.gov/nsr/nsrlink.jsp?2011Da24,B}{2011Da24}: Evaluation of\vspace{0.1cm}}&\\
&&&&\parbox[t][0.3cm]{13.773641cm}{\raggedright {\ }{\ }{\ }\ensuremath{^{\textnormal{15}}}O(\ensuremath{\alpha},\ensuremath{\gamma})); 5092 keV (\href{https://www.nndc.bnl.gov/nsr/nsrlink.jsp?2021Ri04,B}{2021Ri04}) used as a calibration point; and 5092 keV \textit{3} (\href{https://www.nndc.bnl.gov/nsr/nsrlink.jsp?2025PhZZ,B}{2025PhZZ}:\vspace{0.1cm}}&\\
&&&&\parbox[t][0.3cm]{13.773641cm}{\raggedright {\ }{\ }{\ }Unpublished).\vspace{0.1cm}}&\\
&&&&\parbox[t][0.3cm]{13.773641cm}{\raggedright \ensuremath{\Gamma}\ensuremath{>}22 meV (\href{https://www.nndc.bnl.gov/nsr/nsrlink.jsp?1990Ma05,B}{1990Ma05}: See Table 1) deduced for the \ensuremath{^{\textnormal{19}}}F* mirror level. Other value: \ensuremath{\Gamma}=31 keV\vspace{0.1cm}}&\\
&&&&\parbox[t][0.3cm]{13.773641cm}{\raggedright {\ }{\ }{\ }(\href{https://www.nndc.bnl.gov/nsr/nsrlink.jsp?2025PhZZ,B}{2025PhZZ}: Unpublished) dominated by the experimental resolution of 31 keV.\vspace{0.1cm}}&\\
&&&&\parbox[t][0.3cm]{13.773641cm}{\raggedright \ensuremath{\Gamma}\ensuremath{_{\ensuremath{\alpha}}}/\ensuremath{\Gamma}: Weighted average (with external errors) of 0.90 \textit{9} (\href{https://www.nndc.bnl.gov/nsr/nsrlink.jsp?1990Ma05,B}{1990Ma05}); 0.75 \textit{+6{\textminus}7} (\href{https://www.nndc.bnl.gov/nsr/nsrlink.jsp?2004Vi05,B}{2004Vi05}) at 1\ensuremath{\sigma}\vspace{0.1cm}}&\\
&&&&\parbox[t][0.3cm]{13.773641cm}{\raggedright {\ }{\ }{\ }level; 0.87 \textit{3} (\href{https://www.nndc.bnl.gov/nsr/nsrlink.jsp?2007TaZX,B}{2007TaZX}, \href{https://www.nndc.bnl.gov/nsr/nsrlink.jsp?2009Ta09,B}{2009Ta09}); and 1.11 \textit{+17{\textminus}13} from private communication of T.\vspace{0.1cm}}&\\
&&&&\parbox[t][0.3cm]{13.773641cm}{\raggedright {\ }{\ }{\ }Motobayashi with (\href{https://www.nndc.bnl.gov/nsr/nsrlink.jsp?2009Ta09,B}{2009Ta09}: See Table I). Other value: 0.67 \textit{6} (\href{https://www.nndc.bnl.gov/nsr/nsrlink.jsp?2025PhZZ,B}{2025PhZZ}: Unpublished).\vspace{0.1cm}}&\\
&&&&\parbox[t][0.3cm]{13.773641cm}{\raggedright J\ensuremath{^{\pi}}: From J\ensuremath{^{\ensuremath{\pi}}}=5/2\ensuremath{^{\textnormal{+}}} (\href{https://www.nndc.bnl.gov/nsr/nsrlink.jsp?2009Ta09,B}{2009Ta09}) from comparison of the \ensuremath{\gamma}-ray decay scheme of this state in \ensuremath{^{\textnormal{19}}}Ne\vspace{0.1cm}}&\\
&&&&\parbox[t][0.3cm]{13.773641cm}{\raggedright {\ }{\ }{\ }with that of the proposed mirror state in \ensuremath{^{\textnormal{19}}}F. This evidence is weak, and thus we made the\vspace{0.1cm}}&\\
&&&&\parbox[t][0.3cm]{13.773641cm}{\raggedright {\ }{\ }{\ }assignment tentative.\vspace{0.1cm}}&\\
&&&&\parbox[t][0.3cm]{13.773641cm}{\raggedright J\ensuremath{^{\pi}}: See also J\ensuremath{^{\ensuremath{\pi}}}=5/2\ensuremath{^{\textnormal{+}}} assumed by (\href{https://www.nndc.bnl.gov/nsr/nsrlink.jsp?1990Ma05,B}{1990Ma05}) and (\href{https://www.nndc.bnl.gov/nsr/nsrlink.jsp?2004Vi05,B}{2004Vi05}); and J\ensuremath{^{\ensuremath{\pi}}}=3/2\ensuremath{^{-}}, 5/2, 7/2\ensuremath{^{\textnormal{+}}} (\href{https://www.nndc.bnl.gov/nsr/nsrlink.jsp?2025PhZZ,B}{2025PhZZ})\vspace{0.1cm}}&\\
&&&&\parbox[t][0.3cm]{13.773641cm}{\raggedright {\ }{\ }{\ }from \ensuremath{\alpha} angular correlations with L=2,3. Those authors preferred L=3 leading to J\ensuremath{^{\ensuremath{\pi}}}=5/2\ensuremath{^{\textnormal{+}}}, 7/2\ensuremath{^{\textnormal{+}}};\vspace{0.1cm}}&\\
&&&&\parbox[t][0.3cm]{13.773641cm}{\raggedright {\ }{\ }{\ }however, the data cannot distinguish between those L-values.\vspace{0.1cm}}&\\
\multicolumn{1}{r@{}}{5345}&\multicolumn{1}{@{ }l}{{\it 5}}&&\multicolumn{1}{l}{0}&\parbox[t][0.3cm]{13.773641cm}{\raggedright \%\ensuremath{\alpha}=100 (\href{https://www.nndc.bnl.gov/nsr/nsrlink.jsp?1990Ma05,B}{1990Ma05},\href{https://www.nndc.bnl.gov/nsr/nsrlink.jsp?2009Ta09,B}{2009Ta09},\href{https://www.nndc.bnl.gov/nsr/nsrlink.jsp?2025PhZZ,B}{2025PhZZ})\vspace{0.1cm}}&\\
&&&&\parbox[t][0.3cm]{13.773641cm}{\raggedright E(level): From (\href{https://www.nndc.bnl.gov/nsr/nsrlink.jsp?2019Ka15,B}{2019Ka15}).\vspace{0.1cm}}&\\
&&&&\parbox[t][0.3cm]{13.773641cm}{\raggedright E(level): See also 5351 keV (\href{https://www.nndc.bnl.gov/nsr/nsrlink.jsp?1990Ma05,B}{1990Ma05}: See Fig. 2); 5351 keV (\href{https://www.nndc.bnl.gov/nsr/nsrlink.jsp?2004Vi05,B}{2004Vi05}: See Fig. 1); 5.35 MeV\vspace{0.1cm}}&\\
&&&&\parbox[t][0.3cm]{13.773641cm}{\raggedright {\ }{\ }{\ }(\href{https://www.nndc.bnl.gov/nsr/nsrlink.jsp?2007Ta13,B}{2007Ta13}: See Fig. 2), where this level is a resolved member of a close lying doublet; 5.35 MeV\vspace{0.1cm}}&\\
&&&&\parbox[t][0.3cm]{13.773641cm}{\raggedright {\ }{\ }{\ }(\href{https://www.nndc.bnl.gov/nsr/nsrlink.jsp?2009Ta09,B}{2009Ta09}), where this state is a resolved member of a close lying doublet; 5351 keV (\href{https://www.nndc.bnl.gov/nsr/nsrlink.jsp?2021Ri04,B}{2021Ri04}:\vspace{0.1cm}}&\\
&&&&\parbox[t][0.3cm]{13.773641cm}{\raggedright {\ }{\ }{\ }See Fig. 3); and 5352 keV \textit{3} (\href{https://www.nndc.bnl.gov/nsr/nsrlink.jsp?2025PhZZ,B}{2025PhZZ}: Unpublished).\vspace{0.1cm}}&\\
&&&&\parbox[t][0.3cm]{13.773641cm}{\raggedright \ensuremath{\Gamma}\ensuremath{_{\ensuremath{\gamma}}}/\ensuremath{\Gamma}\ensuremath{<}0.001 (\href{https://www.nndc.bnl.gov/nsr/nsrlink.jsp?1990Ma05,B}{1990Ma05}): Estimated from the mirror level.\vspace{0.1cm}}&\\
&&&&\parbox[t][0.3cm]{13.773641cm}{\raggedright L: From (\href{https://www.nndc.bnl.gov/nsr/nsrlink.jsp?2019Ka15,B}{2019Ka15}), who deduced \ensuremath{\Delta}L=0.\vspace{0.1cm}}&\\
&&&&\parbox[t][0.3cm]{13.773641cm}{\raggedright (\href{https://www.nndc.bnl.gov/nsr/nsrlink.jsp?1990Ma05,B}{1990Ma05}, \href{https://www.nndc.bnl.gov/nsr/nsrlink.jsp?2004Vi05,B}{2004Vi05}, \href{https://www.nndc.bnl.gov/nsr/nsrlink.jsp?2009Ta09,B}{2009Ta09}) reported that this state has an isotropic angular distribution in the\vspace{0.1cm}}&\\
&&&&\parbox[t][0.3cm]{13.773641cm}{\raggedright {\ }{\ }{\ }center-of-mass frame for its decay products, and that it is known to decay almost entirely by \ensuremath{\alpha}\vspace{0.1cm}}&\\
&&&&\parbox[t][0.3cm]{13.773641cm}{\raggedright {\ }{\ }{\ }emission. These findings are confirmed by (\href{https://www.nndc.bnl.gov/nsr/nsrlink.jsp?2025PhZZ,B}{2025PhZZ}).\vspace{0.1cm}}&\\
&&&&\parbox[t][0.3cm]{13.773641cm}{\raggedright \ensuremath{\Gamma}=39 keV (\href{https://www.nndc.bnl.gov/nsr/nsrlink.jsp?2025PhZZ,B}{2025PhZZ}: Unpublished), where the experimental resolution was 31 keV.\vspace{0.1cm}}&\\
\multicolumn{1}{r@{}}{5.42\ensuremath{\times10^{3}}}&\multicolumn{1}{@{}l}{}&&&\parbox[t][0.3cm]{13.773641cm}{\raggedright E(level): From (\href{https://www.nndc.bnl.gov/nsr/nsrlink.jsp?2007Ta13,B}{2007Ta13}, \href{https://www.nndc.bnl.gov/nsr/nsrlink.jsp?2009Ta09,B}{2009Ta09}), where this state was a resolved member of a close lying\vspace{0.1cm}}&\\
&&&&\parbox[t][0.3cm]{13.773641cm}{\raggedright {\ }{\ }{\ }doublet.\vspace{0.1cm}}&\\
&&&&\parbox[t][0.3cm]{13.773641cm}{\raggedright E(level): See also 5424 keV (\href{https://www.nndc.bnl.gov/nsr/nsrlink.jsp?2021Ri04,B}{2021Ri04}: See Fig. 3); and 5428 keV \textit{3} (\href{https://www.nndc.bnl.gov/nsr/nsrlink.jsp?2025PhZZ,B}{2025PhZZ}: Unpublished).\vspace{0.1cm}}&\\
&&&&\parbox[t][0.3cm]{13.773641cm}{\raggedright \ensuremath{\Gamma}=39 keV (\href{https://www.nndc.bnl.gov/nsr/nsrlink.jsp?2025PhZZ,B}{2025PhZZ}: Unpublished), where the experimental resolution was 31 keV.\vspace{0.1cm}}&\\
&&&&\parbox[t][0.3cm]{13.773641cm}{\raggedright \ensuremath{\Gamma}\ensuremath{_{\ensuremath{\alpha}}}/\ensuremath{\Gamma}=0.71 \textit{11} (\href{https://www.nndc.bnl.gov/nsr/nsrlink.jsp?2025PhZZ,B}{2025PhZZ}: Unpublished).\vspace{0.1cm}}&\\
&&&&\parbox[t][0.3cm]{13.773641cm}{\raggedright J\ensuremath{^{\pi}}: See J\ensuremath{^{\ensuremath{\pi}}}=5/2\ensuremath{^{\textnormal{+}}}, 7/2, 9/2\ensuremath{^{-}} with L=3,4 (\href{https://www.nndc.bnl.gov/nsr/nsrlink.jsp?2025PhZZ,B}{2025PhZZ}: Unpublished), where the authors favored L=3\vspace{0.1cm}}&\\
&&&&\parbox[t][0.3cm]{13.773641cm}{\raggedright {\ }{\ }{\ }leading to J\ensuremath{^{\ensuremath{\pi}}}=5/2\ensuremath{^{\textnormal{+}}}, 7/2\ensuremath{^{\textnormal{+}}}. We note that the L-values cannot be distinguished.\vspace{0.1cm}}&\\
\multicolumn{1}{r@{}}{5486}&\multicolumn{1}{@{ }l}{{\it 15}}&&\multicolumn{1}{l}{0}&\parbox[t][0.3cm]{13.773641cm}{\raggedright E(level),L: From (\href{https://www.nndc.bnl.gov/nsr/nsrlink.jsp?2019Ka15,B}{2019Ka15}).\vspace{0.1cm}}&\\
&&&&\parbox[t][0.3cm]{13.773641cm}{\raggedright E(level): See also 5463 keV from (\href{https://www.nndc.bnl.gov/nsr/nsrlink.jsp?2021Ri04,B}{2021Ri04}: See Fig. 3); and 5491 keV \textit{3} (\href{https://www.nndc.bnl.gov/nsr/nsrlink.jsp?2025PhZZ,B}{2025PhZZ}: Unpublished).\vspace{0.1cm}}&\\
&&&&\parbox[t][0.3cm]{13.773641cm}{\raggedright \ensuremath{\Gamma}=51 keV (\href{https://www.nndc.bnl.gov/nsr/nsrlink.jsp?2025PhZZ,B}{2025PhZZ}: Unpublished), where the experimental resolution was 31 keV.\vspace{0.1cm}}&\\
&&&&\parbox[t][0.3cm]{13.773641cm}{\raggedright \ensuremath{\Gamma}\ensuremath{_{\ensuremath{\alpha}}}/\ensuremath{\Gamma}=0.71 \textit{4} (\href{https://www.nndc.bnl.gov/nsr/nsrlink.jsp?2025PhZZ,B}{2025PhZZ}: Unpublished).\vspace{0.1cm}}&\\
&&&&\parbox[t][0.3cm]{13.773641cm}{\raggedright J\ensuremath{^{\pi}}: J\ensuremath{^{\ensuremath{\pi}}}=3/2\ensuremath{^{-}}, 5/2, 7/2\ensuremath{^{\textnormal{+}}} with L=2,3 (\href{https://www.nndc.bnl.gov/nsr/nsrlink.jsp?2025PhZZ,B}{2025PhZZ}). Those authors favored L=2 but the L-values are\vspace{0.1cm}}&\\
&&&&\parbox[t][0.3cm]{13.773641cm}{\raggedright {\ }{\ }{\ }indistinguishable.\vspace{0.1cm}}&\\
\multicolumn{1}{r@{}}{5539}&\multicolumn{1}{@{}l}{}&&&\parbox[t][0.3cm]{13.773641cm}{\raggedright E(level): From (\href{https://www.nndc.bnl.gov/nsr/nsrlink.jsp?2021Ri04,B}{2021Ri04}: See Fig. 3).\vspace{0.1cm}}&\\
&&&&\parbox[t][0.3cm]{13.773641cm}{\raggedright E(level): See also 5548 keV \textit{3} (\href{https://www.nndc.bnl.gov/nsr/nsrlink.jsp?2025PhZZ,B}{2025PhZZ}: Unpublished).\vspace{0.1cm}}&\\
&&&&\parbox[t][0.3cm]{13.773641cm}{\raggedright \ensuremath{\Gamma}=32 keV (\href{https://www.nndc.bnl.gov/nsr/nsrlink.jsp?2025PhZZ,B}{2025PhZZ}: Unpublished), where the experimental resolution was 31 keV.\vspace{0.1cm}}&\\
&&&&\parbox[t][0.3cm]{13.773641cm}{\raggedright \ensuremath{\Gamma}\ensuremath{_{\ensuremath{\alpha}}}/\ensuremath{\Gamma}=0.72 \textit{11} (\href{https://www.nndc.bnl.gov/nsr/nsrlink.jsp?2025PhZZ,B}{2025PhZZ}).\vspace{0.1cm}}&\\
&&&&\parbox[t][0.3cm]{13.773641cm}{\raggedright J\ensuremath{^{\pi}}: J\ensuremath{^{\ensuremath{\pi}}}=1/2\ensuremath{^{\textnormal{+}}}, 3/2, 5/2\ensuremath{^{-}} with L=1,2 (\href{https://www.nndc.bnl.gov/nsr/nsrlink.jsp?2025PhZZ,B}{2025PhZZ}). They favored L=1 leading to J\ensuremath{^{\ensuremath{\pi}}}=1/2\ensuremath{^{\textnormal{+}}} and 3/2\ensuremath{^{\textnormal{+}}}. We\vspace{0.1cm}}&\\
&&&&\parbox[t][0.3cm]{13.773641cm}{\raggedright {\ }{\ }{\ }note that L=1 and L=2 are indistinguishable.\vspace{0.1cm}}&\\
\multicolumn{1}{r@{}}{5824}&\multicolumn{1}{@{ }l}{{\it 10}}&\multicolumn{1}{l}{(1/2\ensuremath{^{+}})}&\multicolumn{1}{l}{0}&\parbox[t][0.3cm]{13.773641cm}{\raggedright E(level): From (\href{https://www.nndc.bnl.gov/nsr/nsrlink.jsp?2019Ka15,B}{2019Ka15}).\vspace{0.1cm}}&\\
&&&&\parbox[t][0.3cm]{13.773641cm}{\raggedright E(level): See also E\ensuremath{_{\textnormal{x}}}=5832 keV (\href{https://www.nndc.bnl.gov/nsr/nsrlink.jsp?2021Ri04,B}{2021Ri04}: See Fig. 3); and 5826 keV \textit{3} (\href{https://www.nndc.bnl.gov/nsr/nsrlink.jsp?2025PhZZ,B}{2025PhZZ}: Unpublished).\vspace{0.1cm}}&\\
&&&&\parbox[t][0.3cm]{13.773641cm}{\raggedright (\href{https://www.nndc.bnl.gov/nsr/nsrlink.jsp?2019Ka15,B}{2019Ka15}) reported the \ensuremath{^{\textnormal{19}}}F*(5.94 MeV) level as the likely mirror state for this \ensuremath{^{\textnormal{19}}}Ne* state.\vspace{0.1cm}}&\\
&&&&\parbox[t][0.3cm]{13.773641cm}{\raggedright \ensuremath{\Gamma}=33 keV (\href{https://www.nndc.bnl.gov/nsr/nsrlink.jsp?2025PhZZ,B}{2025PhZZ}: Unpublished), where the experimental resolution was 31 keV.\vspace{0.1cm}}&\\
&&&&\parbox[t][0.3cm]{13.773641cm}{\raggedright \ensuremath{\Gamma}\ensuremath{_{\ensuremath{\alpha}}}/\ensuremath{\Gamma}=1.10 \textit{8} (\href{https://www.nndc.bnl.gov/nsr/nsrlink.jsp?2025PhZZ,B}{2025PhZZ}).\vspace{0.1cm}}&\\
&&&&\parbox[t][0.3cm]{13.773641cm}{\raggedright L: From (\href{https://www.nndc.bnl.gov/nsr/nsrlink.jsp?2019Ka15,B}{2019Ka15}).\vspace{0.1cm}}&\\
&&&&\parbox[t][0.3cm]{13.773641cm}{\raggedright J\ensuremath{^{\pi}}: From (\href{https://www.nndc.bnl.gov/nsr/nsrlink.jsp?2019Ka15,B}{2019Ka15}) from \ensuremath{\Delta}L=0 leading to J\ensuremath{^{\ensuremath{\pi}}}=1/2\ensuremath{^{\textnormal{+}}}, 3/2\ensuremath{^{\textnormal{+}}} and by comparison with the \ensuremath{^{\textnormal{19}}}F*(5938,\vspace{0.1cm}}&\\
&&&&\parbox[t][0.3cm]{13.773641cm}{\raggedright {\ }{\ }{\ }1/2\ensuremath{^{\textnormal{+}}}) mirror state. Since this J\ensuremath{^{\ensuremath{\pi}}} assignment is guided by mirror analysis, we made the assignment\vspace{0.1cm}}&\\
\end{longtable}
\begin{textblock}{29}(0,27.3)
Continued on next page (footnotes at end of table)
\end{textblock}
\clearpage
\begin{longtable}{ccccccc@{\extracolsep{\fill}}c}
\\[-.4cm]
\multicolumn{8}{c}{{\bf \small \underline{\ensuremath{^{\textnormal{19}}}F(\ensuremath{^{\textnormal{3}}}He,t)\hspace{0.2in}\href{https://www.nndc.bnl.gov/nsr/nsrlink.jsp?1998Ut02,B}{1998Ut02},\href{https://www.nndc.bnl.gov/nsr/nsrlink.jsp?2013La01,B}{2013La01},\href{https://www.nndc.bnl.gov/nsr/nsrlink.jsp?2019Ka15,B}{2019Ka15} (continued)}}}\\
\multicolumn{8}{c}{~}\\
\multicolumn{8}{c}{\underline{\ensuremath{^{19}}Ne Levels (continued)}}\\
\multicolumn{8}{c}{~}\\
\multicolumn{2}{c}{E(level)$^{}$}&J$^{\pi}$$^{}$&\multicolumn{2}{c}{\ensuremath{\Gamma}$^{}$}&L$^{}$&Comments&\\[-.2cm]
\multicolumn{2}{c}{\hrulefill}&\hrulefill&\multicolumn{2}{c}{\hrulefill}&\hrulefill&\hrulefill&
\endhead
&&&&&&\parbox[t][0.3cm]{11.015361cm}{\raggedright {\ }{\ }{\ }tentative.\vspace{0.1cm}}&\\
&&&&&&\parbox[t][0.3cm]{11.015361cm}{\raggedright J\ensuremath{^{\pi}}: See also J\ensuremath{^{\ensuremath{\pi}}}=1/2\ensuremath{^{\textnormal{+}}}, 3/2, 5/2, 7/2\ensuremath{^{\textnormal{+}}} with L=1,2,3 (\href{https://www.nndc.bnl.gov/nsr/nsrlink.jsp?2025PhZZ,B}{2025PhZZ}), where the\vspace{0.1cm}}&\\
&&&&&&\parbox[t][0.3cm]{11.015361cm}{\raggedright {\ }{\ }{\ }authors favored L=1 even though the L-values are indistinguishable.\vspace{0.1cm}}&\\
\multicolumn{1}{r@{}}{6008?}&\multicolumn{1}{@{ }l}{{\it 20}}&\multicolumn{1}{l}{(1/2,3/2\ensuremath{^{+}})}&\multicolumn{1}{r@{}}{124}&\multicolumn{1}{@{ }l}{keV {\it 25}}&&\parbox[t][0.3cm]{11.015361cm}{\raggedright E(level): From (\href{https://www.nndc.bnl.gov/nsr/nsrlink.jsp?2021Ri04,B}{2021Ri04}): This tentative, new state was not considered by\vspace{0.1cm}}&\\
&&&&&&\parbox[t][0.3cm]{11.015361cm}{\raggedright {\ }{\ }{\ }those authors to be the same state as the 6014-keV level. (\href{https://www.nndc.bnl.gov/nsr/nsrlink.jsp?2021Ri04,B}{2021Ri04}) observed\vspace{0.1cm}}&\\
&&&&&&\parbox[t][0.3cm]{11.015361cm}{\raggedright {\ }{\ }{\ }\ensuremath{\approx}30 excess counts in the triton singles focal plane spectrum and the t-\ensuremath{\alpha}\vspace{0.1cm}}&\\
&&&&&&\parbox[t][0.3cm]{11.015361cm}{\raggedright {\ }{\ }{\ }coincidence spectrum in this energy region. They claimed to have achieved a\vspace{0.1cm}}&\\
&&&&&&\parbox[t][0.3cm]{11.015361cm}{\raggedright {\ }{\ }{\ }better fit if an additional, new, broad state was added to the fit, which already\vspace{0.1cm}}&\\
&&&&&&\parbox[t][0.3cm]{11.015361cm}{\raggedright {\ }{\ }{\ }included the 6014-keV state. Evaluator cautions that this new, tentative state at\vspace{0.1cm}}&\\
&&&&&&\parbox[t][0.3cm]{11.015361cm}{\raggedright {\ }{\ }{\ }6008-keV is degenerate with another state at 6014 keV, i.e., the centroid of\vspace{0.1cm}}&\\
&&&&&&\parbox[t][0.3cm]{11.015361cm}{\raggedright {\ }{\ }{\ }either states at 6008 and 6014 keV falls within the uncertainty in the centroid\vspace{0.1cm}}&\\
&&&&&&\parbox[t][0.3cm]{11.015361cm}{\raggedright {\ }{\ }{\ }of the other peak, and both states appear to have comparable widths thus\vspace{0.1cm}}&\\
&&&&&&\parbox[t][0.3cm]{11.015361cm}{\raggedright {\ }{\ }{\ }having a significant overlap. Moreover, the improved fit still has a reduced \ensuremath{\chi}\ensuremath{^{\textnormal{2}}}\vspace{0.1cm}}&\\
&&&&&&\parbox[t][0.3cm]{11.015361cm}{\raggedright {\ }{\ }{\ }of nearly 2, which does not imply an overall good fit. (\href{https://www.nndc.bnl.gov/nsr/nsrlink.jsp?2021Ri04,B}{2021Ri04})\vspace{0.1cm}}&\\
&&&&&&\parbox[t][0.3cm]{11.015361cm}{\raggedright {\ }{\ }{\ }acknowledged that it is not clear whether the 6008-keV state can be associated\vspace{0.1cm}}&\\
&&&&&&\parbox[t][0.3cm]{11.015361cm}{\raggedright {\ }{\ }{\ }with the broad \ensuremath{^{\textnormal{19}}}Ne state predicted theoretically by (\href{https://www.nndc.bnl.gov/nsr/nsrlink.jsp?2007Du09,B}{2007Du09}) using the\vspace{0.1cm}}&\\
&&&&&&\parbox[t][0.3cm]{11.015361cm}{\raggedright {\ }{\ }{\ }Generator Coordinate Method.\vspace{0.1cm}}&\\
&&&&&&\parbox[t][0.3cm]{11.015361cm}{\raggedright E(level): We note that the latest experiment by (\href{https://www.nndc.bnl.gov/nsr/nsrlink.jsp?2025PhZZ,B}{2025PhZZ}) does not see this\vspace{0.1cm}}&\\
&&&&&&\parbox[t][0.3cm]{11.015361cm}{\raggedright {\ }{\ }{\ }state.\vspace{0.1cm}}&\\
&&&&&&\parbox[t][0.3cm]{11.015361cm}{\raggedright \ensuremath{\Gamma}: From (\href{https://www.nndc.bnl.gov/nsr/nsrlink.jsp?2021Ri04,B}{2021Ri04}).\vspace{0.1cm}}&\\
&&&&&&\parbox[t][0.3cm]{11.015361cm}{\raggedright J\ensuremath{^{\pi}}: From (\href{https://www.nndc.bnl.gov/nsr/nsrlink.jsp?2021Ri04,B}{2021Ri04}) obtained by comparison of \ensuremath{\Gamma} with the Wigner limit.\vspace{0.1cm}}&\\
\multicolumn{1}{r@{}}{6014}&\multicolumn{1}{@{ }l}{{\it 3}}&\multicolumn{1}{l}{3/2\ensuremath{^{-}}}&&&\multicolumn{1}{l}{\ensuremath{\geq}1}&\parbox[t][0.3cm]{11.015361cm}{\raggedright E(level): From 6014 keV \textit{2} (stat.) \textit{2} (sys.) (\href{https://www.nndc.bnl.gov/nsr/nsrlink.jsp?2013La01,B}{2013La01}).\vspace{0.1cm}}&\\
&&&&&&\parbox[t][0.3cm]{11.015361cm}{\raggedright E(level): See also 6016 keV (\href{https://www.nndc.bnl.gov/nsr/nsrlink.jsp?1998Ut02,B}{1998Ut02}: See Fig. 1); 6.01 MeV (\href{https://www.nndc.bnl.gov/nsr/nsrlink.jsp?2007Ta13,B}{2007Ta13}: See\vspace{0.1cm}}&\\
&&&&&&\parbox[t][0.3cm]{11.015361cm}{\raggedright {\ }{\ }{\ }Fig. 2); 6.01 MeV (\href{https://www.nndc.bnl.gov/nsr/nsrlink.jsp?2009Ta09,B}{2009Ta09}: See Fig. 4); 6007 keV \textit{10} (\href{https://www.nndc.bnl.gov/nsr/nsrlink.jsp?2019Ka15,B}{2019Ka15}); 6014\vspace{0.1cm}}&\\
&&&&&&\parbox[t][0.3cm]{11.015361cm}{\raggedright {\ }{\ }{\ }keV (\href{https://www.nndc.bnl.gov/nsr/nsrlink.jsp?2021Ri04,B}{2021Ri04}) used as a calibration point; and 6012 keV \textit{3} (\href{https://www.nndc.bnl.gov/nsr/nsrlink.jsp?2025PhZZ,B}{2025PhZZ}:\vspace{0.1cm}}&\\
&&&&&&\parbox[t][0.3cm]{11.015361cm}{\raggedright {\ }{\ }{\ }Unpublished).\vspace{0.1cm}}&\\
&&&&&&\parbox[t][0.3cm]{11.015361cm}{\raggedright \ensuremath{\Gamma}=37 keV (\href{https://www.nndc.bnl.gov/nsr/nsrlink.jsp?2025PhZZ,B}{2025PhZZ}: Unpublished), where the experimental resolution was 31\vspace{0.1cm}}&\\
&&&&&&\parbox[t][0.3cm]{11.015361cm}{\raggedright {\ }{\ }{\ }keV.\vspace{0.1cm}}&\\
&&&&&&\parbox[t][0.3cm]{11.015361cm}{\raggedright \ensuremath{\Gamma}\ensuremath{_{\ensuremath{\alpha}}}/\ensuremath{\Gamma}=0.86 \textit{6} (\href{https://www.nndc.bnl.gov/nsr/nsrlink.jsp?2025PhZZ,B}{2025PhZZ}).\vspace{0.1cm}}&\\
&&&&&&\parbox[t][0.3cm]{11.015361cm}{\raggedright L: From (\href{https://www.nndc.bnl.gov/nsr/nsrlink.jsp?2019Ka15,B}{2019Ka15}).\vspace{0.1cm}}&\\
&&&&&&\parbox[t][0.3cm]{11.015361cm}{\raggedright J\ensuremath{^{\pi}}: From J\ensuremath{^{\ensuremath{\pi}}}=3/2\ensuremath{^{-}} (\href{https://www.nndc.bnl.gov/nsr/nsrlink.jsp?2013La01,B}{2013La01}) from a two-step finite-range DWBA analysis via\vspace{0.1cm}}&\\
&&&&&&\parbox[t][0.3cm]{11.015361cm}{\raggedright {\ }{\ }{\ }FRESCO; and J\ensuremath{^{\ensuremath{\pi}}}=3/2\ensuremath{^{-}} (\href{https://www.nndc.bnl.gov/nsr/nsrlink.jsp?2019Ka15,B}{2019Ka15}) from \ensuremath{\Delta}L\ensuremath{\geq}1 guided by mirror analysis.\vspace{0.1cm}}&\\
&&&&&&\parbox[t][0.3cm]{11.015361cm}{\raggedright J\ensuremath{^{\pi}}: See also J\ensuremath{^{\ensuremath{\pi}}}=3/2\ensuremath{^{-}}, 5/2, 7/2\ensuremath{^{\textnormal{+}}} with L=2,3 (\href{https://www.nndc.bnl.gov/nsr/nsrlink.jsp?2025PhZZ,B}{2025PhZZ}), where L=2 was\vspace{0.1cm}}&\\
&&&&&&\parbox[t][0.3cm]{11.015361cm}{\raggedright {\ }{\ }{\ }favored even though both L values describe the data.\vspace{0.1cm}}&\\
\multicolumn{1}{r@{}}{6072}&\multicolumn{1}{@{ }l}{{\it 3}}&\multicolumn{1}{l}{(3/2\ensuremath{^{+}},5/2\ensuremath{^{-}})}&&&&\parbox[t][0.3cm]{11.015361cm}{\raggedright E(level): From E\ensuremath{_{\textnormal{x}}}=6072 keV \textit{2} (stat.) \textit{2} keV (sys.) (\href{https://www.nndc.bnl.gov/nsr/nsrlink.jsp?2013La01,B}{2013La01}). Other values:\vspace{0.1cm}}&\\
&&&&&&\parbox[t][0.3cm]{11.015361cm}{\raggedright {\ }{\ }{\ }6078 keV (\href{https://www.nndc.bnl.gov/nsr/nsrlink.jsp?1998Ut02,B}{1998Ut02}: See Fig. 1); and 6067 keV \textit{3} (\href{https://www.nndc.bnl.gov/nsr/nsrlink.jsp?2025PhZZ,B}{2025PhZZ}: Unpublished).\vspace{0.1cm}}&\\
&&&&&&\parbox[t][0.3cm]{11.015361cm}{\raggedright Evaluator notes that Table I in (\href{https://www.nndc.bnl.gov/nsr/nsrlink.jsp?2013La01,B}{2013La01}) erroneously (confirmed by our\vspace{0.1cm}}&\\
&&&&&&\parbox[t][0.3cm]{11.015361cm}{\raggedright {\ }{\ }{\ }private communication with the authors) lists \ensuremath{\Gamma}\ensuremath{_{\textnormal{p}}}=\ensuremath{\Gamma}=0.143 keV for this\vspace{0.1cm}}&\\
&&&&&&\parbox[t][0.3cm]{11.015361cm}{\raggedright {\ }{\ }{\ }proton bound state.\vspace{0.1cm}}&\\
&&&&&&\parbox[t][0.3cm]{11.015361cm}{\raggedright \ensuremath{\Gamma}=37 keV (\href{https://www.nndc.bnl.gov/nsr/nsrlink.jsp?2025PhZZ,B}{2025PhZZ}: Unpublished), where the experimental resolution was 31\vspace{0.1cm}}&\\
&&&&&&\parbox[t][0.3cm]{11.015361cm}{\raggedright {\ }{\ }{\ }keV.\vspace{0.1cm}}&\\
&&&&&&\parbox[t][0.3cm]{11.015361cm}{\raggedright \ensuremath{\Gamma}\ensuremath{_{\ensuremath{\alpha}}}=6\ensuremath{\times}10\ensuremath{^{\textnormal{$-$4}}} keV (\href{https://www.nndc.bnl.gov/nsr/nsrlink.jsp?2013La01,B}{2013La01}).\vspace{0.1cm}}&\\
&&&&&&\parbox[t][0.3cm]{11.015361cm}{\raggedright \ensuremath{\Gamma}\ensuremath{_{\ensuremath{\alpha}}}/\ensuremath{\Gamma}=0.79 \textit{7} (\href{https://www.nndc.bnl.gov/nsr/nsrlink.jsp?2025PhZZ,B}{2025PhZZ}).\vspace{0.1cm}}&\\
&&&&&&\parbox[t][0.3cm]{11.015361cm}{\raggedright J\ensuremath{^{\pi}}: From a two-step finite-range DWBA analysis via FRESCO performed by\vspace{0.1cm}}&\\
&&&&&&\parbox[t][0.3cm]{11.015361cm}{\raggedright {\ }{\ }{\ }(\href{https://www.nndc.bnl.gov/nsr/nsrlink.jsp?2013La01,B}{2013La01}).\vspace{0.1cm}}&\\
&&&&&&\parbox[t][0.3cm]{11.015361cm}{\raggedright J\ensuremath{^{\pi}}: See also J\ensuremath{^{\ensuremath{\pi}}}=1/2\ensuremath{^{\textnormal{+}}}, 3/2, 5/2\ensuremath{^{-}} with L=1,2 (\href{https://www.nndc.bnl.gov/nsr/nsrlink.jsp?2025PhZZ,B}{2025PhZZ}), where L=1 was\vspace{0.1cm}}&\\
&&&&&&\parbox[t][0.3cm]{11.015361cm}{\raggedright {\ }{\ }{\ }favored even though both L-values are indistinguishable.\vspace{0.1cm}}&\\
\multicolumn{1}{r@{}}{6097}&\multicolumn{1}{@{ }l}{{\it 4}}&\multicolumn{1}{l}{(7/2,9/2)\ensuremath{^{+}}}&&&&\parbox[t][0.3cm]{11.015361cm}{\raggedright E(level): From E\ensuremath{_{\textnormal{x}}}=6097 keV \textit{3} (stat.) \textit{2} keV (sys.) (\href{https://www.nndc.bnl.gov/nsr/nsrlink.jsp?2013La01,B}{2013La01}). Other values:\vspace{0.1cm}}&\\
&&&&&&\parbox[t][0.3cm]{11.015361cm}{\raggedright {\ }{\ }{\ }6107 keV (\href{https://www.nndc.bnl.gov/nsr/nsrlink.jsp?1998Ut02,B}{1998Ut02}: See Fig. 1); and 6092 keV \textit{3} (\href{https://www.nndc.bnl.gov/nsr/nsrlink.jsp?2025PhZZ,B}{2025PhZZ}: Unpublished).\vspace{0.1cm}}&\\
&&&&&&\parbox[t][0.3cm]{11.015361cm}{\raggedright \ensuremath{\Gamma}=37 keV (\href{https://www.nndc.bnl.gov/nsr/nsrlink.jsp?2025PhZZ,B}{2025PhZZ}: Unpublished), where the experimental resolution was 31\vspace{0.1cm}}&\\
&&&&&&\parbox[t][0.3cm]{11.015361cm}{\raggedright {\ }{\ }{\ }keV.\vspace{0.1cm}}&\\
&&&&&&\parbox[t][0.3cm]{11.015361cm}{\raggedright \ensuremath{\Gamma}\ensuremath{_{\ensuremath{\alpha}}}/\ensuremath{\Gamma}=0.66 \textit{23} (\href{https://www.nndc.bnl.gov/nsr/nsrlink.jsp?2025PhZZ,B}{2025PhZZ}).\vspace{0.1cm}}&\\
&&&&&&\parbox[t][0.3cm]{11.015361cm}{\raggedright J\ensuremath{^{\pi}}: From a two-step finite-range DWBA analysis via FRESCO performed by\vspace{0.1cm}}&\\
&&&&&&\parbox[t][0.3cm]{11.015361cm}{\raggedright {\ }{\ }{\ }(\href{https://www.nndc.bnl.gov/nsr/nsrlink.jsp?2013La01,B}{2013La01}).\vspace{0.1cm}}&\\
&&&&&&\parbox[t][0.3cm]{11.015361cm}{\raggedright J\ensuremath{^{\pi}}: See also J\ensuremath{^{\ensuremath{\pi}}}=5/2\ensuremath{^{\textnormal{+}}}, 7/2, 9/2, 11/2\ensuremath{^{\textnormal{+}}} with L=3,4,5 (\href{https://www.nndc.bnl.gov/nsr/nsrlink.jsp?2025PhZZ,B}{2025PhZZ}), where L=4\vspace{0.1cm}}&\\
&&&&&&\parbox[t][0.3cm]{11.015361cm}{\raggedright {\ }{\ }{\ }was favored even though all those L-values describe the data.\vspace{0.1cm}}&\\
\end{longtable}
\begin{textblock}{29}(0,27.3)
Continued on next page (footnotes at end of table)
\end{textblock}
\clearpage
\begin{longtable}{ccccc@{\extracolsep{\fill}}c}
\\[-.4cm]
\multicolumn{6}{c}{{\bf \small \underline{\ensuremath{^{\textnormal{19}}}F(\ensuremath{^{\textnormal{3}}}He,t)\hspace{0.2in}\href{https://www.nndc.bnl.gov/nsr/nsrlink.jsp?1998Ut02,B}{1998Ut02},\href{https://www.nndc.bnl.gov/nsr/nsrlink.jsp?2013La01,B}{2013La01},\href{https://www.nndc.bnl.gov/nsr/nsrlink.jsp?2019Ka15,B}{2019Ka15} (continued)}}}\\
\multicolumn{6}{c}{~}\\
\multicolumn{6}{c}{\underline{\ensuremath{^{19}}Ne Levels (continued)}}\\
\multicolumn{6}{c}{~}\\
\multicolumn{2}{c}{E(level)$^{}$}&J$^{\pi}$$^{}$&L$^{}$&Comments&\\[-.2cm]
\multicolumn{2}{c}{\hrulefill}&\hrulefill&\hrulefill&\hrulefill&
\endhead
\multicolumn{1}{r@{}}{6132}&\multicolumn{1}{@{}l}{\ensuremath{^{{\hyperlink{NE32LEVEL0}{a}}}} {\it 4}}&\multicolumn{1}{l}{(1/2\ensuremath{^{+}},3/2\ensuremath{^{+}})}&\multicolumn{1}{l}{0}&\parbox[t][0.3cm]{13.497581cm}{\raggedright E(level): Weighted average of 6132 keV \textit{3} (stat.) \textit{2} (sys.) (\href{https://www.nndc.bnl.gov/nsr/nsrlink.jsp?2013La01,B}{2013La01}) and 6130 keV \textit{5}\vspace{0.1cm}}&\\
&&&&\parbox[t][0.3cm]{13.497581cm}{\raggedright {\ }{\ }{\ }(\href{https://www.nndc.bnl.gov/nsr/nsrlink.jsp?2019Ka15,B}{2019Ka15}). Other values: 6138 keV (\href{https://www.nndc.bnl.gov/nsr/nsrlink.jsp?1998Ut02,B}{1998Ut02}: See Fig. 1); 6132 keV (\href{https://www.nndc.bnl.gov/nsr/nsrlink.jsp?2015Pa46,B}{2015Pa46}); 6131 keV\vspace{0.1cm}}&\\
&&&&\parbox[t][0.3cm]{13.497581cm}{\raggedright {\ }{\ }{\ }(\href{https://www.nndc.bnl.gov/nsr/nsrlink.jsp?2021Ri04,B}{2021Ri04}: See Fig. 3); and 6133 keV \textit{3} (\href{https://www.nndc.bnl.gov/nsr/nsrlink.jsp?2025PhZZ,B}{2025PhZZ}: Unpublished).\vspace{0.1cm}}&\\
&&&&\parbox[t][0.3cm]{13.497581cm}{\raggedright \ensuremath{\Gamma}=37 keV (\href{https://www.nndc.bnl.gov/nsr/nsrlink.jsp?2025PhZZ,B}{2025PhZZ}: Unpublished), where the experimental resolution was 31 keV.\vspace{0.1cm}}&\\
&&&&\parbox[t][0.3cm]{13.497581cm}{\raggedright \ensuremath{\Gamma}\ensuremath{_{\ensuremath{\alpha}}}=7\ensuremath{\times}10\ensuremath{^{\textnormal{$-$4}}} keV (\href{https://www.nndc.bnl.gov/nsr/nsrlink.jsp?2013La01,B}{2013La01}).\vspace{0.1cm}}&\\
&&&&\parbox[t][0.3cm]{13.497581cm}{\raggedright \ensuremath{\Gamma}\ensuremath{_{\ensuremath{\alpha}}}/\ensuremath{\Gamma}=0.86 \textit{5} (\href{https://www.nndc.bnl.gov/nsr/nsrlink.jsp?2025PhZZ,B}{2025PhZZ}).\vspace{0.1cm}}&\\
&&&&\parbox[t][0.3cm]{13.497581cm}{\raggedright L: From (\href{https://www.nndc.bnl.gov/nsr/nsrlink.jsp?2019Ka15,B}{2019Ka15}).\vspace{0.1cm}}&\\
&&&&\parbox[t][0.3cm]{13.497581cm}{\raggedright J\ensuremath{^{\pi}}: From (1) J\ensuremath{^{\ensuremath{\pi}}}=(3/2\ensuremath{^{\textnormal{+}}}, 5/2\ensuremath{^{-}}) (\href{https://www.nndc.bnl.gov/nsr/nsrlink.jsp?2013La01,B}{2013La01}, \href{https://www.nndc.bnl.gov/nsr/nsrlink.jsp?2015Pa46,B}{2015Pa46}) from a two-step finite-range DWBA analysis\vspace{0.1cm}}&\\
&&&&\parbox[t][0.3cm]{13.497581cm}{\raggedright {\ }{\ }{\ }via FRESCO; and (2) J\ensuremath{^{\ensuremath{\pi}}}=(1/2\ensuremath{^{\textnormal{+}}}, 3/2\ensuremath{^{\textnormal{+}}}) (\href{https://www.nndc.bnl.gov/nsr/nsrlink.jsp?2019Ka15,B}{2019Ka15}) deduced from \ensuremath{\Delta}L=0. This study favored\vspace{0.1cm}}&\\
&&&&\parbox[t][0.3cm]{13.497581cm}{\raggedright {\ }{\ }{\ }J\ensuremath{^{\ensuremath{\pi}}}=1/2\ensuremath{^{\textnormal{+}}} based on the Coulomb energy shift between this \ensuremath{^{\textnormal{19}}}Ne* state and its proposed mirror\vspace{0.1cm}}&\\
&&&&\parbox[t][0.3cm]{13.497581cm}{\raggedright {\ }{\ }{\ }level at \ensuremath{^{\textnormal{19}}}F*(6255, 1/2\ensuremath{^{\textnormal{+}}}). If this \ensuremath{^{\textnormal{19}}}Ne* state is assumed to be a J\ensuremath{^{\ensuremath{\pi}}}=3/2\ensuremath{^{\textnormal{+}}} state, then the\vspace{0.1cm}}&\\
&&&&\parbox[t][0.3cm]{13.497581cm}{\raggedright {\ }{\ }{\ }Coulomb energy difference would be 365 keV, which (\href{https://www.nndc.bnl.gov/nsr/nsrlink.jsp?2019Ka15,B}{2019Ka15}) deemed to be too large, but\vspace{0.1cm}}&\\
&&&&\parbox[t][0.3cm]{13.497581cm}{\raggedright {\ }{\ }{\ }they did not rule out the J\ensuremath{^{\ensuremath{\pi}}}=3/2\ensuremath{^{\textnormal{+}}} assignment.\vspace{0.1cm}}&\\
&&&&\parbox[t][0.3cm]{13.497581cm}{\raggedright J\ensuremath{^{\pi}}: See also (1) (\href{https://www.nndc.bnl.gov/nsr/nsrlink.jsp?2021Ka51,B}{2021Ka51}), which supports the J\ensuremath{^{\ensuremath{\pi}}}=1/2\ensuremath{^{\textnormal{+}}} assignment due to the presence of the\vspace{0.1cm}}&\\
&&&&\parbox[t][0.3cm]{13.497581cm}{\raggedright {\ }{\ }{\ }strong J\ensuremath{^{\ensuremath{\pi}}}=1/2\ensuremath{^{\textnormal{+}}} resonance in \ensuremath{^{\textnormal{19}}}Ne observed by (\href{https://www.nndc.bnl.gov/nsr/nsrlink.jsp?2002La29,B}{2002La29}: \ensuremath{^{\textnormal{2}}}H(\ensuremath{^{\textnormal{18}}}Ne,\ensuremath{^{\textnormal{19}}}Ne)) near \ensuremath{\sim}6.1 MeV; (2)\vspace{0.1cm}}&\\
&&&&\parbox[t][0.3cm]{13.497581cm}{\raggedright {\ }{\ }{\ }(\href{https://www.nndc.bnl.gov/nsr/nsrlink.jsp?2021Ri04,B}{2021Ri04}), which recommended J\ensuremath{^{\ensuremath{\pi}}}=3/2\ensuremath{^{\textnormal{+}}} based on the literature; and (3) J\ensuremath{^{\ensuremath{\pi}}}=1/2\ensuremath{^{\textnormal{+}}}, 3/2, 5/2\ensuremath{^{-}}\vspace{0.1cm}}&\\
&&&&\parbox[t][0.3cm]{13.497581cm}{\raggedright {\ }{\ }{\ }with L=1,2 (\href{https://www.nndc.bnl.gov/nsr/nsrlink.jsp?2025PhZZ,B}{2025PhZZ}), where they favored L=1 even though both values describe the data.\vspace{0.1cm}}&\\
&&&&\parbox[t][0.3cm]{13.497581cm}{\raggedright Evaluator notes that Table I in (\href{https://www.nndc.bnl.gov/nsr/nsrlink.jsp?2013La01,B}{2013La01}) erroneously (confirmed by our private communication\vspace{0.1cm}}&\\
&&&&\parbox[t][0.3cm]{13.497581cm}{\raggedright {\ }{\ }{\ }with the authors) lists \ensuremath{\Gamma}\ensuremath{_{\textnormal{p}}}=\ensuremath{\Gamma}=0.143 keV for this proton bound state.\vspace{0.1cm}}&\\
\multicolumn{1}{r@{}}{6282}&\multicolumn{1}{@{ }l}{{\it 2}}&\multicolumn{1}{l}{(1/2,3/2,5/2)}&&\parbox[t][0.3cm]{13.497581cm}{\raggedright \ensuremath{\Gamma}\ensuremath{\alpha}/\ensuremath{\Gamma}=0.92 \textit{11} (\href{https://www.nndc.bnl.gov/nsr/nsrlink.jsp?2021Ri04,B}{2021Ri04})\vspace{0.1cm}}&\\
&&&&\parbox[t][0.3cm]{13.497581cm}{\raggedright E(level): From (\href{https://www.nndc.bnl.gov/nsr/nsrlink.jsp?2015Pa46,B}{2015Pa46}).\vspace{0.1cm}}&\\
&&&&\parbox[t][0.3cm]{13.497581cm}{\raggedright E(level): (\href{https://www.nndc.bnl.gov/nsr/nsrlink.jsp?2015Pa46,B}{2015Pa46}) reported that this state could be a single state, or possibly an unresolved\vspace{0.1cm}}&\\
&&&&\parbox[t][0.3cm]{13.497581cm}{\raggedright {\ }{\ }{\ }doublet, which is composed of the \ensuremath{^{\textnormal{19}}}Ne*(6282) and \ensuremath{^{\textnormal{19}}}Ne*(6295) states. Those authors\vspace{0.1cm}}&\\
&&&&\parbox[t][0.3cm]{13.497581cm}{\raggedright {\ }{\ }{\ }mentioned that a fit that consisted of a triplet at 6.4 MeV, an assumed single state at 6.86 MeV,\vspace{0.1cm}}&\\
&&&&\parbox[t][0.3cm]{13.497581cm}{\raggedright {\ }{\ }{\ }and an assumed single broad state (with \ensuremath{\Gamma}\ensuremath{\approx}16 keV) at 6.29 MeV was as good as a fit which\vspace{0.1cm}}&\\
&&&&\parbox[t][0.3cm]{13.497581cm}{\raggedright {\ }{\ }{\ }consisted of the \ensuremath{^{\textnormal{19}}}Ne*(6282, 6295, 6416, 6440, 6459, 6851, 6864) levels, where they considered\vspace{0.1cm}}&\\
&&&&\parbox[t][0.3cm]{13.497581cm}{\raggedright {\ }{\ }{\ }a doublet at \ensuremath{\sim}6.29 MeV, a triplet at 6.4 MeV and another doublet at 6851 keV and 6864 keV.\vspace{0.1cm}}&\\
&&&&\parbox[t][0.3cm]{13.497581cm}{\raggedright {\ }{\ }{\ }Those authors mentioned that the angular distribution of the tritons populating the peak around\vspace{0.1cm}}&\\
&&&&\parbox[t][0.3cm]{13.497581cm}{\raggedright {\ }{\ }{\ }6.29 MeV did not show any evidence indicating the contribution of two components, which\vspace{0.1cm}}&\\
&&&&\parbox[t][0.3cm]{13.497581cm}{\raggedright {\ }{\ }{\ }supports the identification of this peak as a single state, although the contributing levels may\vspace{0.1cm}}&\\
&&&&\parbox[t][0.3cm]{13.497581cm}{\raggedright {\ }{\ }{\ }simply have similar spin and parity. (\href{https://www.nndc.bnl.gov/nsr/nsrlink.jsp?2013La01,B}{2013La01}, \href{https://www.nndc.bnl.gov/nsr/nsrlink.jsp?2019Ka15,B}{2019Ka15}, \href{https://www.nndc.bnl.gov/nsr/nsrlink.jsp?2021Ri04,B}{2021Ri04}) also reported that the peak\vspace{0.1cm}}&\\
&&&&\parbox[t][0.3cm]{13.497581cm}{\raggedright {\ }{\ }{\ }around 6.29 MeV may be an unresolved peak. We considered this peak to be a member of a\vspace{0.1cm}}&\\
&&&&\parbox[t][0.3cm]{13.497581cm}{\raggedright {\ }{\ }{\ }close lying doublet (see the Adopted Levels for more convincing arguments).\vspace{0.1cm}}&\\
&&&&\parbox[t][0.3cm]{13.497581cm}{\raggedright E(level): See also 6290 keV (\href{https://www.nndc.bnl.gov/nsr/nsrlink.jsp?1998Ut02,B}{1998Ut02}: See Fig. 1); 6289 keV \textit{3} (stat.) \textit{2} (sys.) (\href{https://www.nndc.bnl.gov/nsr/nsrlink.jsp?2013La01,B}{2013La01}); 6288\vspace{0.1cm}}&\\
&&&&\parbox[t][0.3cm]{13.497581cm}{\raggedright {\ }{\ }{\ }keV \textit{5} (\href{https://www.nndc.bnl.gov/nsr/nsrlink.jsp?2019Ka15,B}{2019Ka15}); 6289 keV (\href{https://www.nndc.bnl.gov/nsr/nsrlink.jsp?2021Ri04,B}{2021Ri04}: See Fig. 3); and 6285 keV \textit{3} (\href{https://www.nndc.bnl.gov/nsr/nsrlink.jsp?2025PhZZ,B}{2025PhZZ}:\vspace{0.1cm}}&\\
&&&&\parbox[t][0.3cm]{13.497581cm}{\raggedright {\ }{\ }{\ }Unpublished).\vspace{0.1cm}}&\\
&&&&\parbox[t][0.3cm]{13.497581cm}{\raggedright J\ensuremath{^{\pi}}: Results are inconsistent: (1) J\ensuremath{^{\ensuremath{\pi}}}=(5/2\ensuremath{^{-}}, 11/2\ensuremath{^{-}}) (\href{https://www.nndc.bnl.gov/nsr/nsrlink.jsp?2013La01,B}{2013La01}), where the triton angular distribution\vspace{0.1cm}}&\\
&&&&\parbox[t][0.3cm]{13.497581cm}{\raggedright {\ }{\ }{\ }could not be reproduced by any assignment assuming this state is a single peak; however, those\vspace{0.1cm}}&\\
&&&&\parbox[t][0.3cm]{13.497581cm}{\raggedright {\ }{\ }{\ }with J\ensuremath{\geq}3/2 are preferred, this study argued that this level does not have a low-spin. (\href{https://www.nndc.bnl.gov/nsr/nsrlink.jsp?2015Pa46,B}{2015Pa46})\vspace{0.1cm}}&\\
&&&&\parbox[t][0.3cm]{13.497581cm}{\raggedright {\ }{\ }{\ }reanalyzed the data of (\href{https://www.nndc.bnl.gov/nsr/nsrlink.jsp?2013La01,B}{2013La01}) using a finite-range coupled-channels calculation and reported\vspace{0.1cm}}&\\
&&&&\parbox[t][0.3cm]{13.497581cm}{\raggedright {\ }{\ }{\ }consistent J\ensuremath{^{\ensuremath{\pi}}} values with those of (\href{https://www.nndc.bnl.gov/nsr/nsrlink.jsp?2013La01,B}{2013La01}); (2) J\ensuremath{^{\ensuremath{\pi}}}=(1/2\ensuremath{^{\textnormal{+}}}, 3/2\ensuremath{^{\textnormal{+}}}, 5/2\ensuremath{^{\textnormal{+}}}) (\href{https://www.nndc.bnl.gov/nsr/nsrlink.jsp?2019Ka15,B}{2019Ka15}), which\vspace{0.1cm}}&\\
&&&&\parbox[t][0.3cm]{13.497581cm}{\raggedright {\ }{\ }{\ }reported that their data are not compatible with a pure \ensuremath{\Delta}L=0 transition because the triton angular\vspace{0.1cm}}&\\
&&&&\parbox[t][0.3cm]{13.497581cm}{\raggedright {\ }{\ }{\ }distribution increases noticeably in strength at more backward angles, relative to that of known\vspace{0.1cm}}&\\
&&&&\parbox[t][0.3cm]{13.497581cm}{\raggedright {\ }{\ }{\ }Gamow-Teller \ensuremath{\Delta}L=0 transitions. If this state is a doublet with at least one member being a\vspace{0.1cm}}&\\
&&&&\parbox[t][0.3cm]{13.497581cm}{\raggedright {\ }{\ }{\ }non-Gamow-Teller state, a \ensuremath{\Delta}L=0 component would be allowed; (3) J=11/2 was preferred by\vspace{0.1cm}}&\\
&&&&\parbox[t][0.3cm]{13.497581cm}{\raggedright {\ }{\ }{\ }(\href{https://www.nndc.bnl.gov/nsr/nsrlink.jsp?2021Ri04,B}{2021Ri04}) from triton-\ensuremath{\alpha} angular correlation data assuming this state is a member of a doublet\vspace{0.1cm}}&\\
&&&&\parbox[t][0.3cm]{13.497581cm}{\raggedright {\ }{\ }{\ }and using a Legendre polynomial with k\ensuremath{_{\textnormal{max}}}=10. Assuming that this state is a single state at\vspace{0.1cm}}&\\
&&&&\parbox[t][0.3cm]{13.497581cm}{\raggedright {\ }{\ }{\ }6289 keV, J\ensuremath{\geq}7/2 is deduced from the best fit using a Legendre polynomial with k\ensuremath{_{\textnormal{max}}}=6.\vspace{0.1cm}}&\\
&&&&\parbox[t][0.3cm]{13.497581cm}{\raggedright {\ }{\ }{\ }(\href{https://www.nndc.bnl.gov/nsr/nsrlink.jsp?2021Ri04,B}{2021Ri04}) ruled out the possibility that the 6289-keV state is a single state with J=1/2 based on\vspace{0.1cm}}&\\
&&&&\parbox[t][0.3cm]{13.497581cm}{\raggedright {\ }{\ }{\ }a fit with k\ensuremath{_{\textnormal{max}}}=0 (isotropic), which does not describe the observed triton-\ensuremath{\alpha} angular distribution.\vspace{0.1cm}}&\\
&&&&\parbox[t][0.3cm]{13.497581cm}{\raggedright {\ }{\ }{\ }(\href{https://www.nndc.bnl.gov/nsr/nsrlink.jsp?2021Ri04,B}{2021Ri04}) reported that either only one member of the doublet with higher spin is populated in\vspace{0.1cm}}&\\
&&&&\parbox[t][0.3cm]{13.497581cm}{\raggedright {\ }{\ }{\ }their experiment, or both are populated to a non-negligible proportion and the \ensuremath{\alpha}-particle decay\vspace{0.1cm}}&\\
&&&&\parbox[t][0.3cm]{13.497581cm}{\raggedright {\ }{\ }{\ }measurements are mixing from both resulting in angular correlations that do not represent that of\vspace{0.1cm}}&\\
&&&&\parbox[t][0.3cm]{13.497581cm}{\raggedright {\ }{\ }{\ }a single state. (\href{https://www.nndc.bnl.gov/nsr/nsrlink.jsp?2025PhZZ,B}{2025PhZZ}) deduced J\ensuremath{^{\ensuremath{\pi}}}=1/2\ensuremath{^{\textnormal{+}}}, 3/2, 5/2\ensuremath{^{-}} with L=1,2 and preferred L=1 even\vspace{0.1cm}}&\\
\end{longtable}
\begin{textblock}{29}(0,27.3)
Continued on next page (footnotes at end of table)
\end{textblock}
\clearpage
\begin{longtable}{cccccc@{\extracolsep{\fill}}c}
\\[-.4cm]
\multicolumn{7}{c}{{\bf \small \underline{\ensuremath{^{\textnormal{19}}}F(\ensuremath{^{\textnormal{3}}}He,t)\hspace{0.2in}\href{https://www.nndc.bnl.gov/nsr/nsrlink.jsp?1998Ut02,B}{1998Ut02},\href{https://www.nndc.bnl.gov/nsr/nsrlink.jsp?2013La01,B}{2013La01},\href{https://www.nndc.bnl.gov/nsr/nsrlink.jsp?2019Ka15,B}{2019Ka15} (continued)}}}\\
\multicolumn{7}{c}{~}\\
\multicolumn{7}{c}{\underline{\ensuremath{^{19}}Ne Levels (continued)}}\\
\multicolumn{7}{c}{~}\\
\multicolumn{2}{c}{E(level)$^{}$}&J$^{\pi}$$^{}$&\multicolumn{2}{c}{\ensuremath{\Gamma}$^{}$}&Comments&\\[-.2cm]
\multicolumn{2}{c}{\hrulefill}&\hrulefill&\multicolumn{2}{c}{\hrulefill}&\hrulefill&
\endhead
&&&&&\parbox[t][0.3cm]{10.931721cm}{\raggedright {\ }{\ }{\ }though both L-values describe the data well.\vspace{0.1cm}}&\\
&&&&&\parbox[t][0.3cm]{10.931721cm}{\raggedright \ensuremath{\Gamma}=33 keV (\href{https://www.nndc.bnl.gov/nsr/nsrlink.jsp?2025PhZZ,B}{2025PhZZ}: Unpublished), where the experimental resolution was 31\vspace{0.1cm}}&\\
&&&&&\parbox[t][0.3cm]{10.931721cm}{\raggedright {\ }{\ }{\ }keV.\vspace{0.1cm}}&\\
&&&&&\parbox[t][0.3cm]{10.931721cm}{\raggedright \ensuremath{\Gamma}\ensuremath{_{\ensuremath{\alpha}}}/\ensuremath{\Gamma}: Obtained from (\href{https://www.nndc.bnl.gov/nsr/nsrlink.jsp?2021Ri04,B}{2021Ri04}) assuming J\ensuremath{\geq}(7/2) and that this state is a\vspace{0.1cm}}&\\
&&&&&\parbox[t][0.3cm]{10.931721cm}{\raggedright {\ }{\ }{\ }member of a doublet. The uncertainty is a quadratic sum of statistical and\vspace{0.1cm}}&\\
&&&&&\parbox[t][0.3cm]{10.931721cm}{\raggedright {\ }{\ }{\ }systematic uncertainties. Other value: 0.88 \textit{8} (\href{https://www.nndc.bnl.gov/nsr/nsrlink.jsp?2025PhZZ,B}{2025PhZZ}).\vspace{0.1cm}}&\\
\multicolumn{1}{r@{}}{6295}&\multicolumn{1}{@{ }l}{{\it 2}}&&&&\parbox[t][0.3cm]{10.931721cm}{\raggedright E(level): From (\href{https://www.nndc.bnl.gov/nsr/nsrlink.jsp?2015Pa46,B}{2015Pa46}), where this level is an unresolved member of the\vspace{0.1cm}}&\\
&&&&&\parbox[t][0.3cm]{10.931721cm}{\raggedright {\ }{\ }{\ }\ensuremath{^{\textnormal{19}}}Ne*(6282, 6295) doublet.\vspace{0.1cm}}&\\
\multicolumn{1}{r@{}}{6417}&\multicolumn{1}{@{}l}{\ensuremath{^{{\hyperlink{NE32LEVEL0}{a}}}} {\it 4}}&\multicolumn{1}{l}{(3/2\ensuremath{^{-}},5/2\ensuremath{^{+}})\ensuremath{^{{\hyperlink{NE32LEVEL1}{b}}}}}&&&\parbox[t][0.3cm]{10.931721cm}{\raggedright E(level): Weighted average of 6419 keV \textit{6} (\href{https://www.nndc.bnl.gov/nsr/nsrlink.jsp?1998Ut02,B}{1998Ut02}) and 6416 keV \textit{3} (stat.) \textit{2}\vspace{0.1cm}}&\\
&&&&&\parbox[t][0.3cm]{10.931721cm}{\raggedright {\ }{\ }{\ }(sys.) (\href{https://www.nndc.bnl.gov/nsr/nsrlink.jsp?2013La01,B}{2013La01}), who assumed this state is the previously identified\vspace{0.1cm}}&\\
&&&&&\parbox[t][0.3cm]{10.931721cm}{\raggedright {\ }{\ }{\ }\ensuremath{^{\textnormal{19}}}Ne*(6419 keV, 3/2\ensuremath{^{-}}) state by (\href{https://www.nndc.bnl.gov/nsr/nsrlink.jsp?2011Ad05,B}{2011Ad05}: \ensuremath{^{\textnormal{2}}}H(\ensuremath{^{\textnormal{18}}}F,\ensuremath{^{\textnormal{19}}}Ne)).\vspace{0.1cm}}&\\
&&&&&\parbox[t][0.3cm]{10.931721cm}{\raggedright E(level): See also 6146 keV (\href{https://www.nndc.bnl.gov/nsr/nsrlink.jsp?2015Pa46,B}{2015Pa46}), which considered this level to be a\vspace{0.1cm}}&\\
&&&&&\parbox[t][0.3cm]{10.931721cm}{\raggedright {\ }{\ }{\ }member of a possible unresolved triplet at \ensuremath{^{\textnormal{19}}}Ne*(6416, 6440, 6459); 6421\vspace{0.1cm}}&\\
&&&&&\parbox[t][0.3cm]{10.931721cm}{\raggedright {\ }{\ }{\ }keV \textit{10} (\href{https://www.nndc.bnl.gov/nsr/nsrlink.jsp?2019Ka15,B}{2019Ka15}), who reported that more than one unresolved state may\vspace{0.1cm}}&\\
&&&&&\parbox[t][0.3cm]{10.931721cm}{\raggedright {\ }{\ }{\ }contribute to this peak; 6419 keV (\href{https://www.nndc.bnl.gov/nsr/nsrlink.jsp?2021Ri04,B}{2021Ri04}: See Fig. 3); and 6416 keV \textit{3}\vspace{0.1cm}}&\\
&&&&&\parbox[t][0.3cm]{10.931721cm}{\raggedright {\ }{\ }{\ }(\href{https://www.nndc.bnl.gov/nsr/nsrlink.jsp?2025PhZZ,B}{2025PhZZ}: Unpublished).\vspace{0.1cm}}&\\
&&&&&\parbox[t][0.3cm]{10.931721cm}{\raggedright J\ensuremath{^{\pi}}: From J\ensuremath{^{\ensuremath{\pi}}}=(3/2\ensuremath{^{-}}, 5/2\ensuremath{^{\textnormal{+}}}) (\href{https://www.nndc.bnl.gov/nsr/nsrlink.jsp?2013La01,B}{2013La01}) from a two-step finite-range DWBA\vspace{0.1cm}}&\\
&&&&&\parbox[t][0.3cm]{10.931721cm}{\raggedright {\ }{\ }{\ }analysis via FRESCO.\vspace{0.1cm}}&\\
&&&&&\parbox[t][0.3cm]{10.931721cm}{\raggedright J\ensuremath{^{\pi}}: See also (1) J\ensuremath{^{\ensuremath{\pi}}}=(3/2\ensuremath{^{-}}) (\href{https://www.nndc.bnl.gov/nsr/nsrlink.jsp?2019Ka15,B}{2019Ka15}) taken from (\href{https://www.nndc.bnl.gov/nsr/nsrlink.jsp?2013La01,B}{2013La01}) and (\href{https://www.nndc.bnl.gov/nsr/nsrlink.jsp?2011Ad05,B}{2011Ad05}:\vspace{0.1cm}}&\\
&&&&&\parbox[t][0.3cm]{10.931721cm}{\raggedright {\ }{\ }{\ }\ensuremath{^{\textnormal{2}}}H(\ensuremath{^{\textnormal{18}}}F,\ensuremath{^{\textnormal{19}}}Ne)); (2) J\ensuremath{^{\ensuremath{\pi}}}=3/2\ensuremath{^{\textnormal{+}}} (\href{https://www.nndc.bnl.gov/nsr/nsrlink.jsp?1998Ut02,B}{1998Ut02}) from mirror analysis; (3)\vspace{0.1cm}}&\\
&&&&&\parbox[t][0.3cm]{10.931721cm}{\raggedright {\ }{\ }{\ }(\href{https://www.nndc.bnl.gov/nsr/nsrlink.jsp?2015Pa46,B}{2015Pa46}) reported J\ensuremath{^{\ensuremath{\pi}}} consistent with (\href{https://www.nndc.bnl.gov/nsr/nsrlink.jsp?2013La01,B}{2013La01}), but did not explicitly\vspace{0.1cm}}&\\
&&&&&\parbox[t][0.3cm]{10.931721cm}{\raggedright {\ }{\ }{\ }give the values.\vspace{0.1cm}}&\\
&&&&&\parbox[t][0.3cm]{10.931721cm}{\raggedright \ensuremath{\Gamma}\ensuremath{_{\textnormal{p}}}=4.7\ensuremath{\times}10\ensuremath{^{\textnormal{$-$50}}} keV for J\ensuremath{^{\ensuremath{\pi}}}=3/2\ensuremath{^{-}}, and \ensuremath{\Gamma}\ensuremath{_{\textnormal{p}}}=1.2\ensuremath{\times}10\ensuremath{^{\textnormal{$-$51}}} keV for J\ensuremath{^{\ensuremath{\pi}}}=5/2\ensuremath{^{\textnormal{+}}}\vspace{0.1cm}}&\\
&&&&&\parbox[t][0.3cm]{10.931721cm}{\raggedright {\ }{\ }{\ }(\href{https://www.nndc.bnl.gov/nsr/nsrlink.jsp?2013La01,B}{2013La01}).\vspace{0.1cm}}&\\
&&&&&\parbox[t][0.3cm]{10.931721cm}{\raggedright \ensuremath{\Gamma}\ensuremath{_{\ensuremath{\alpha}}}=0.5 keV for J\ensuremath{^{\ensuremath{\pi}}}=3/2\ensuremath{^{-}}, and \ensuremath{\Gamma}\ensuremath{_{\ensuremath{\alpha}}}=0.126 keV for J\ensuremath{^{\ensuremath{\pi}}}=5/2\ensuremath{^{\textnormal{+}}} (\href{https://www.nndc.bnl.gov/nsr/nsrlink.jsp?2013La01,B}{2013La01}).\vspace{0.1cm}}&\\
&&&&&\parbox[t][0.3cm]{10.931721cm}{\raggedright We highlight that (\href{https://www.nndc.bnl.gov/nsr/nsrlink.jsp?2025PhZZ,B}{2025PhZZ}) used two different procedures to fit the\vspace{0.1cm}}&\\
&&&&&\parbox[t][0.3cm]{10.931721cm}{\raggedright {\ }{\ }{\ }unresolved region around 6.4 MeV. Since those data are not peer-reviewed,\vspace{0.1cm}}&\\
&&&&&\parbox[t][0.3cm]{10.931721cm}{\raggedright {\ }{\ }{\ }we refrained using their results from this region until the data are published.\vspace{0.1cm}}&\\
\multicolumn{1}{r@{}}{6438}&\multicolumn{1}{@{}l}{\ensuremath{^{{\hyperlink{NE32LEVEL0}{a}}}} {\it 4}}&\multicolumn{1}{l}{(1/2\ensuremath{^{-}},3/2\ensuremath{^{-}},11/2\ensuremath{^{+}})\ensuremath{^{{\hyperlink{NE32LEVEL1}{b}}}}}&\multicolumn{1}{r@{}}{216}&\multicolumn{1}{@{ }l}{keV {\it 19}}&\parbox[t][0.3cm]{10.931721cm}{\raggedright E(level): Weighted average (with external errors) of 6437 keV \textit{9} (\href{https://www.nndc.bnl.gov/nsr/nsrlink.jsp?1998Ut02,B}{1998Ut02});\vspace{0.1cm}}&\\
&&&&&\parbox[t][0.3cm]{10.931721cm}{\raggedright {\ }{\ }{\ }6440 keV \textit{3} (stat.) \textit{2} (sys.) (\href{https://www.nndc.bnl.gov/nsr/nsrlink.jsp?2013La01,B}{2013La01}), who claimed that they observed this\vspace{0.1cm}}&\\
&&&&&\parbox[t][0.3cm]{10.931721cm}{\raggedright {\ }{\ }{\ }state for the first time; and 6421 keV \textit{10} (\href{https://www.nndc.bnl.gov/nsr/nsrlink.jsp?2019Ka15,B}{2019Ka15}), which reported that\vspace{0.1cm}}&\\
&&&&&\parbox[t][0.3cm]{10.931721cm}{\raggedright {\ }{\ }{\ }more than one unresolved state may contribute to the observed peak.\vspace{0.1cm}}&\\
&&&&&\parbox[t][0.3cm]{10.931721cm}{\raggedright E(level): See also 6440 keV (\href{https://www.nndc.bnl.gov/nsr/nsrlink.jsp?2015Pa46,B}{2015Pa46}): Possible unresolved \ensuremath{^{\textnormal{19}}}Ne*(6416,\vspace{0.1cm}}&\\
&&&&&\parbox[t][0.3cm]{10.931721cm}{\raggedright {\ }{\ }{\ }6440, 6459) triplet; 6437 keV (\href{https://www.nndc.bnl.gov/nsr/nsrlink.jsp?2021Ri04,B}{2021Ri04}: See Fig. 3); and 6440 keV \textit{3}\vspace{0.1cm}}&\\
&&&&&\parbox[t][0.3cm]{10.931721cm}{\raggedright {\ }{\ }{\ }(\href{https://www.nndc.bnl.gov/nsr/nsrlink.jsp?2025PhZZ,B}{2025PhZZ}: Unpublished).\vspace{0.1cm}}&\\
&&&&&\parbox[t][0.3cm]{10.931721cm}{\raggedright \ensuremath{\Gamma}: From (\href{https://www.nndc.bnl.gov/nsr/nsrlink.jsp?1998Ut02,B}{1998Ut02}).\vspace{0.1cm}}&\\
&&&&&\parbox[t][0.3cm]{10.931721cm}{\raggedright J\ensuremath{^{\pi}}: From (1) J\ensuremath{^{\ensuremath{\pi}}}=(11/2\ensuremath{^{\textnormal{+}}}) (\href{https://www.nndc.bnl.gov/nsr/nsrlink.jsp?2013La01,B}{2013La01}) from a two-step finite-range DWBA\vspace{0.1cm}}&\\
&&&&&\parbox[t][0.3cm]{10.931721cm}{\raggedright {\ }{\ }{\ }analysis via FRESCO, where angular distributions were not consistent with\vspace{0.1cm}}&\\
&&&&&\parbox[t][0.3cm]{10.931721cm}{\raggedright {\ }{\ }{\ }J\ensuremath{^{\ensuremath{\pi}}}=3/2\ensuremath{^{\textnormal{+}}} or 5/2\ensuremath{^{\textnormal{+}}}. Data are best fitted with J\ensuremath{^{\ensuremath{\pi}}}=11/2\ensuremath{^{\textnormal{+}}} supported by the\vspace{0.1cm}}&\\
&&&&&\parbox[t][0.3cm]{10.931721cm}{\raggedright {\ }{\ }{\ }evaluation in (\href{https://www.nndc.bnl.gov/nsr/nsrlink.jsp?2007Ne09,B}{2007Ne09}). (2) J\ensuremath{^{\ensuremath{\pi}}}=(3/2\ensuremath{^{-}}) (\href{https://www.nndc.bnl.gov/nsr/nsrlink.jsp?2019Ka15,B}{2019Ka15}) recommended by\vspace{0.1cm}}&\\
&&&&&\parbox[t][0.3cm]{10.931721cm}{\raggedright {\ }{\ }{\ }(\href{https://www.nndc.bnl.gov/nsr/nsrlink.jsp?2011Ad05,B}{2011Ad05}: \ensuremath{^{\textnormal{2}}}H(\ensuremath{^{\textnormal{18}}}F,\ensuremath{^{\textnormal{19}}}Ne)) for E\ensuremath{_{\textnormal{x}}}\ensuremath{\sim}6.42 MeV; and (3) J\ensuremath{^{\ensuremath{\pi}}}=1/2\ensuremath{^{-}} (\href{https://www.nndc.bnl.gov/nsr/nsrlink.jsp?1998Ut02,B}{1998Ut02})\vspace{0.1cm}}&\\
&&&&&\parbox[t][0.3cm]{10.931721cm}{\raggedright {\ }{\ }{\ }from mirror analysis.\vspace{0.1cm}}&\\
&&&&&\parbox[t][0.3cm]{10.931721cm}{\raggedright J\ensuremath{^{\pi}}: See also (\href{https://www.nndc.bnl.gov/nsr/nsrlink.jsp?2015Pa46,B}{2015Pa46}), which reported that J\ensuremath{^{\ensuremath{\pi}}} values are consistent with\vspace{0.1cm}}&\\
&&&&&\parbox[t][0.3cm]{10.931721cm}{\raggedright {\ }{\ }{\ }(\href{https://www.nndc.bnl.gov/nsr/nsrlink.jsp?2013La01,B}{2013La01}), but did not explicitly give the values.\vspace{0.1cm}}&\\
\multicolumn{1}{r@{}}{6457}&\multicolumn{1}{@{}l}{\ensuremath{^{{\hyperlink{NE32LEVEL0}{a}}}} {\it 4}}&\multicolumn{1}{l}{(5/2\ensuremath{^{-}})}&&&\parbox[t][0.3cm]{10.931721cm}{\raggedright E(level): Weighted average of (1) 6450 keV \textit{6} (\href{https://www.nndc.bnl.gov/nsr/nsrlink.jsp?1998Ut02,B}{1998Ut02}), where in Fig. 1, the\vspace{0.1cm}}&\\
&&&&&\parbox[t][0.3cm]{10.931721cm}{\raggedright {\ }{\ }{\ }energy is reported to be 6449 keV, and (2) 6459 keV \textit{3} (stat.) \textit{2} (sys.)\vspace{0.1cm}}&\\
&&&&&\parbox[t][0.3cm]{10.931721cm}{\raggedright {\ }{\ }{\ }(\href{https://www.nndc.bnl.gov/nsr/nsrlink.jsp?2013La01,B}{2013La01}).\vspace{0.1cm}}&\\
&&&&&\parbox[t][0.3cm]{10.931721cm}{\raggedright E(level): See also 6459 keV (\href{https://www.nndc.bnl.gov/nsr/nsrlink.jsp?2015Pa46,B}{2015Pa46}): Possible unresolved \ensuremath{^{\textnormal{19}}}Ne*(6416,\vspace{0.1cm}}&\\
&&&&&\parbox[t][0.3cm]{10.931721cm}{\raggedright {\ }{\ }{\ }6440, 6459) triplet; 6448 keV (\href{https://www.nndc.bnl.gov/nsr/nsrlink.jsp?2021Ri04,B}{2021Ri04}: See Fig. 3); and 6459 keV \textit{3}\vspace{0.1cm}}&\\
&&&&&\parbox[t][0.3cm]{10.931721cm}{\raggedright {\ }{\ }{\ }(\href{https://www.nndc.bnl.gov/nsr/nsrlink.jsp?2025PhZZ,B}{2025PhZZ}: Unpublished).\vspace{0.1cm}}&\\
&&&&&\parbox[t][0.3cm]{10.931721cm}{\raggedright (\href{https://www.nndc.bnl.gov/nsr/nsrlink.jsp?2013La01,B}{2013La01}) deduced \ensuremath{\Gamma}\ensuremath{_{\textnormal{p}}}=8.4\ensuremath{\times}10\ensuremath{^{\textnormal{$-$14}}} keV and \ensuremath{\Gamma}\ensuremath{_{\ensuremath{\alpha}}}=\ensuremath{\Gamma}=5.5 keV from the\vspace{0.1cm}}&\\
&&&&&\parbox[t][0.3cm]{10.931721cm}{\raggedright {\ }{\ }{\ }R-matrix analysis of the \ensuremath{^{\textnormal{18}}}F(p,\ensuremath{\alpha}) astrophysical S-factor.\vspace{0.1cm}}&\\
&&&&&\parbox[t][0.3cm]{10.931721cm}{\raggedright J\ensuremath{^{\pi}}: From (\href{https://www.nndc.bnl.gov/nsr/nsrlink.jsp?2013La01,B}{2013La01}): J\ensuremath{^{\ensuremath{\pi}}}=(5/2\ensuremath{^{-}}) from the two-step finite-range DWBA analysis\vspace{0.1cm}}&\\
&&&&&\parbox[t][0.3cm]{10.931721cm}{\raggedright {\ }{\ }{\ }via FRESCO. A J\ensuremath{^{\ensuremath{\pi}}}=3/2\ensuremath{^{\textnormal{+}}} assignment cannot reproduce the low and high angle\vspace{0.1cm}}&\\
&&&&&\parbox[t][0.3cm]{10.931721cm}{\raggedright {\ }{\ }{\ }data simultaneously and is therefore excluded.\vspace{0.1cm}}&\\
\end{longtable}
\begin{textblock}{29}(0,27.3)
Continued on next page (footnotes at end of table)
\end{textblock}
\clearpage
\begin{longtable}{ccccc@{\extracolsep{\fill}}c}
\\[-.4cm]
\multicolumn{6}{c}{{\bf \small \underline{\ensuremath{^{\textnormal{19}}}F(\ensuremath{^{\textnormal{3}}}He,t)\hspace{0.2in}\href{https://www.nndc.bnl.gov/nsr/nsrlink.jsp?1998Ut02,B}{1998Ut02},\href{https://www.nndc.bnl.gov/nsr/nsrlink.jsp?2013La01,B}{2013La01},\href{https://www.nndc.bnl.gov/nsr/nsrlink.jsp?2019Ka15,B}{2019Ka15} (continued)}}}\\
\multicolumn{6}{c}{~}\\
\multicolumn{6}{c}{\underline{\ensuremath{^{19}}Ne Levels (continued)}}\\
\multicolumn{6}{c}{~}\\
\multicolumn{2}{c}{E(level)$^{}$}&J$^{\pi}$$^{}$&L$^{}$&Comments&\\[-.2cm]
\multicolumn{2}{c}{\hrulefill}&\hrulefill&\hrulefill&\hrulefill&
\endhead
&&&&\parbox[t][0.3cm]{14.205921cm}{\raggedright J\ensuremath{^{\pi}}: See also J\ensuremath{^{\ensuremath{\pi}}}=3/2\ensuremath{^{\textnormal{+}}} (\href{https://www.nndc.bnl.gov/nsr/nsrlink.jsp?1998Ut02,B}{1998Ut02}) from mirror analysis; and (\href{https://www.nndc.bnl.gov/nsr/nsrlink.jsp?2015Pa46,B}{2015Pa46}), which reported that J\ensuremath{^{\ensuremath{\pi}}} values\vspace{0.1cm}}&\\
&&&&\parbox[t][0.3cm]{14.205921cm}{\raggedright {\ }{\ }{\ }are consistent with (\href{https://www.nndc.bnl.gov/nsr/nsrlink.jsp?2013La01,B}{2013La01}), but did not explicitly give the values.\vspace{0.1cm}}&\\
&&&&\parbox[t][0.3cm]{14.205921cm}{\raggedright \ensuremath{\theta}\ensuremath{^{\textnormal{2}}_{\textnormal{p}}}=0.014: Reduced proton width from (\href{https://www.nndc.bnl.gov/nsr/nsrlink.jsp?2013La01,B}{2013La01}) and deduced for J\ensuremath{^{\ensuremath{\pi}}}=(5/2\ensuremath{^{-}}).\vspace{0.1cm}}&\\
\multicolumn{1}{r@{}}{6700}&\multicolumn{1}{@{ }l}{{\it 4}}&&&\parbox[t][0.3cm]{14.205921cm}{\raggedright E(level): From 6700 keV \textit{3} (stat.) \textit{2} (sys.) (\href{https://www.nndc.bnl.gov/nsr/nsrlink.jsp?2013La01,B}{2013La01}).\vspace{0.1cm}}&\\
&&&&\parbox[t][0.3cm]{14.205921cm}{\raggedright E(level): See also 6696 keV \textit{6} (\href{https://www.nndc.bnl.gov/nsr/nsrlink.jsp?1998Ut02,B}{1998Ut02}); 6700 keV (\href{https://www.nndc.bnl.gov/nsr/nsrlink.jsp?2015Pa46,B}{2015Pa46}); and 6700 keV (\href{https://www.nndc.bnl.gov/nsr/nsrlink.jsp?2021Ri04,B}{2021Ri04}: See Fig. 3).\vspace{0.1cm}}&\\
&&&&\parbox[t][0.3cm]{14.205921cm}{\raggedright J\ensuremath{^{\pi}}: See J\ensuremath{^{\ensuremath{\pi}}}=5/2\ensuremath{^{\textnormal{+}}} (\href{https://www.nndc.bnl.gov/nsr/nsrlink.jsp?1998Ut02,B}{1998Ut02}) from mirror analysis. Since this is a weak argument, which may not be\vspace{0.1cm}}&\\
&&&&\parbox[t][0.3cm]{14.205921cm}{\raggedright {\ }{\ }{\ }reliable anymore, we did not adopt the J\ensuremath{^{\ensuremath{\pi}}} value.\vspace{0.1cm}}&\\
\multicolumn{1}{r@{}}{6742}&\multicolumn{1}{@{}l}{\ensuremath{^{{\hyperlink{NE32LEVEL0}{a}}}} {\it 3}}&\multicolumn{1}{l}{3/2\ensuremath{^{-}}}&\multicolumn{1}{l}{\ensuremath{\geq}1}&\parbox[t][0.3cm]{14.205921cm}{\raggedright \ensuremath{\Gamma}\ensuremath{\alpha}/\ensuremath{\Gamma}=0.94 \textit{+5{\textminus}3}\vspace{0.1cm}}&\\
&&&&\parbox[t][0.3cm]{14.205921cm}{\raggedright E(level): Weighted average of 6741 keV \textit{6} (\href{https://www.nndc.bnl.gov/nsr/nsrlink.jsp?1998Ut02,B}{1998Ut02}), which was observed for the first time, and 6742\vspace{0.1cm}}&\\
&&&&\parbox[t][0.3cm]{14.205921cm}{\raggedright {\ }{\ }{\ }keV \textit{2} (stat.) \textit{2} (sys.) (\href{https://www.nndc.bnl.gov/nsr/nsrlink.jsp?2013La01,B}{2013La01}).\vspace{0.1cm}}&\\
&&&&\parbox[t][0.3cm]{14.205921cm}{\raggedright E(level): See also 6742 keV (\href{https://www.nndc.bnl.gov/nsr/nsrlink.jsp?2004Vi05,B}{2004Vi05}: See Fig. 1); 6742 keV (\href{https://www.nndc.bnl.gov/nsr/nsrlink.jsp?2015Pa46,B}{2015Pa46}); 6753 keV \textit{10} (\href{https://www.nndc.bnl.gov/nsr/nsrlink.jsp?2019Ka15,B}{2019Ka15}),\vspace{0.1cm}}&\\
&&&&\parbox[t][0.3cm]{14.205921cm}{\raggedright {\ }{\ }{\ }where more than one unresolved state may contribute to this peak; 6742 keV (\href{https://www.nndc.bnl.gov/nsr/nsrlink.jsp?2021Ri04,B}{2021Ri04}) used as a\vspace{0.1cm}}&\\
&&&&\parbox[t][0.3cm]{14.205921cm}{\raggedright {\ }{\ }{\ }calibration point; and 6740 keV \textit{3} (\href{https://www.nndc.bnl.gov/nsr/nsrlink.jsp?2025PhZZ,B}{2025PhZZ}: Unpublished).\vspace{0.1cm}}&\\
&&&&\parbox[t][0.3cm]{14.205921cm}{\raggedright \ensuremath{\Gamma}=\ensuremath{\Gamma}\ensuremath{_{\ensuremath{\alpha}}}=5.2 keV and \ensuremath{\Gamma}\ensuremath{_{\textnormal{p}}}=2.22 eV were suggested by (\href{https://www.nndc.bnl.gov/nsr/nsrlink.jsp?2013La01,B}{2013La01}). The former is taken from the\vspace{0.1cm}}&\\
&&&&\parbox[t][0.3cm]{14.205921cm}{\raggedright {\ }{\ }{\ }evaluation of (\href{https://www.nndc.bnl.gov/nsr/nsrlink.jsp?2007Ne09,B}{2007Ne09}) and is scaled from the \ensuremath{^{\textnormal{19}}}F*(6787). The latter is from (\href{https://www.nndc.bnl.gov/nsr/nsrlink.jsp?2002Bb02,B}{2002Bb02}). See also\vspace{0.1cm}}&\\
&&&&\parbox[t][0.3cm]{14.205921cm}{\raggedright {\ }{\ }{\ }\ensuremath{\Gamma}=33 keV (\href{https://www.nndc.bnl.gov/nsr/nsrlink.jsp?2025PhZZ,B}{2025PhZZ}), where the experimental resolution was 31 keV at FWHM.\vspace{0.1cm}}&\\
&&&&\parbox[t][0.3cm]{14.205921cm}{\raggedright \ensuremath{\Gamma}\ensuremath{_{\ensuremath{\alpha}}}/\ensuremath{\Gamma}: Weighted average of 1.04 \textit{8} (\href{https://www.nndc.bnl.gov/nsr/nsrlink.jsp?1998Ut02,B}{1998Ut02}); 0.901 \textit{+74{\textminus}31} (\href{https://www.nndc.bnl.gov/nsr/nsrlink.jsp?2004Vi05,B}{2004Vi05}) at 1\ensuremath{\sigma} level; and 0.92 \textit{9}\vspace{0.1cm}}&\\
&&&&\parbox[t][0.3cm]{14.205921cm}{\raggedright {\ }{\ }{\ }(\href{https://www.nndc.bnl.gov/nsr/nsrlink.jsp?2021Ri04,B}{2021Ri04}), where the uncertainty is a quadratic sum of statistical and systematic uncertainties. Other\vspace{0.1cm}}&\\
&&&&\parbox[t][0.3cm]{14.205921cm}{\raggedright {\ }{\ }{\ }value: 0.90 \textit{12} (\href{https://www.nndc.bnl.gov/nsr/nsrlink.jsp?2025PhZZ,B}{2025PhZZ}).\vspace{0.1cm}}&\\
&&&&\parbox[t][0.3cm]{14.205921cm}{\raggedright \ensuremath{\Gamma}\ensuremath{_{\textnormal{p}}}/\ensuremath{\Gamma}\ensuremath{>}0.003 (\href{https://www.nndc.bnl.gov/nsr/nsrlink.jsp?2004Vi05,B}{2004Vi05}). This value is tentative with a probability of \ensuremath{>}90\% (\href{https://www.nndc.bnl.gov/nsr/nsrlink.jsp?2004Vi05,B}{2004Vi05}). The lower\vspace{0.1cm}}&\\
&&&&\parbox[t][0.3cm]{14.205921cm}{\raggedright {\ }{\ }{\ }limit is placed due to a probable loss of some \ensuremath{\alpha} particles whose signals were below the detection\vspace{0.1cm}}&\\
&&&&\parbox[t][0.3cm]{14.205921cm}{\raggedright {\ }{\ }{\ }threshold in (\href{https://www.nndc.bnl.gov/nsr/nsrlink.jsp?2004Vi05,B}{2004Vi05}). Other value: 0.013 \textit{7} (\href{https://www.nndc.bnl.gov/nsr/nsrlink.jsp?2025PhZZ,B}{2025PhZZ}) for L\ensuremath{_{\textnormal{rel}}}(p+\ensuremath{^{\textnormal{18}}}F)=1 from the proton angular\vspace{0.1cm}}&\\
&&&&\parbox[t][0.3cm]{14.205921cm}{\raggedright {\ }{\ }{\ }correlations from the \ensuremath{^{\textnormal{19}}}Ne*(p) decay.\vspace{0.1cm}}&\\
&&&&\parbox[t][0.3cm]{14.205921cm}{\raggedright J\ensuremath{^{\pi}}: From (1) J\ensuremath{^{\ensuremath{\pi}}}=3/2\ensuremath{^{-}} (\href{https://www.nndc.bnl.gov/nsr/nsrlink.jsp?1998Ut02,B}{1998Ut02}) from relative populations and angular distributions measured for\vspace{0.1cm}}&\\
&&&&\parbox[t][0.3cm]{14.205921cm}{\raggedright {\ }{\ }{\ }\ensuremath{^{\textnormal{20}}}Ne(d,\ensuremath{^{\textnormal{3}}}He), \ensuremath{^{\textnormal{20}}}Ne(d,t), \ensuremath{^{\textnormal{16}}}O(\ensuremath{^{\textnormal{6}}}Li,\ensuremath{^{\textnormal{3}}}He), and \ensuremath{^{\textnormal{16}}}O(\ensuremath{^{\textnormal{6}}}Li,t) mirror reactions; (2) J\ensuremath{^{\ensuremath{\pi}}}=3/2\ensuremath{^{-}} (\href{https://www.nndc.bnl.gov/nsr/nsrlink.jsp?2004Vi05,B}{2004Vi05}) from\vspace{0.1cm}}&\\
&&&&\parbox[t][0.3cm]{14.205921cm}{\raggedright {\ }{\ }{\ }\ensuremath{\alpha} angular distributions, see Fig. 2, reported that J\ensuremath{^{\ensuremath{\pi}}}=1/2\ensuremath{^{-}} would imply an isotropic correlation function\vspace{0.1cm}}&\\
&&&&\parbox[t][0.3cm]{14.205921cm}{\raggedright {\ }{\ }{\ }for the decay \ensuremath{\alpha} particles, which did not fit the observed \ensuremath{\alpha} angular distribution. J\ensuremath{^{\ensuremath{\pi}}}=3/2\ensuremath{^{-}} resulted in a\vspace{0.1cm}}&\\
&&&&\parbox[t][0.3cm]{14.205921cm}{\raggedright {\ }{\ }{\ }much better fit to the \ensuremath{\alpha} angular distribution data; (3) J\ensuremath{^{\ensuremath{\pi}}}=(3/2\ensuremath{^{-}}) (\href{https://www.nndc.bnl.gov/nsr/nsrlink.jsp?2013La01,B}{2013La01}) from a two-step\vspace{0.1cm}}&\\
&&&&\parbox[t][0.3cm]{14.205921cm}{\raggedright {\ }{\ }{\ }finite-range DWBA analysis via FRESCO; (4) J\ensuremath{^{\ensuremath{\pi}}}=3/2\ensuremath{^{-}} (\href{https://www.nndc.bnl.gov/nsr/nsrlink.jsp?2019Ka15,B}{2019Ka15}) from mirror analysis; and (4)\vspace{0.1cm}}&\\
&&&&\parbox[t][0.3cm]{14.205921cm}{\raggedright {\ }{\ }{\ }J\ensuremath{^{\ensuremath{\pi}}}=(3/2\ensuremath{^{-}}) (\href{https://www.nndc.bnl.gov/nsr/nsrlink.jsp?2021Ri04,B}{2021Ri04}) was consistent with their triton-\ensuremath{\alpha} angular correlation, but their best fit was\vspace{0.1cm}}&\\
&&&&\parbox[t][0.3cm]{14.205921cm}{\raggedright {\ }{\ }{\ }deduced using J\ensuremath{\geq}5/2.\vspace{0.1cm}}&\\
&&&&\parbox[t][0.3cm]{14.205921cm}{\raggedright J\ensuremath{^{\pi}}: See also (\href{https://www.nndc.bnl.gov/nsr/nsrlink.jsp?2015Pa46,B}{2015Pa46}), which reported that J\ensuremath{^{\ensuremath{\pi}}} value is consistent with that of (\href{https://www.nndc.bnl.gov/nsr/nsrlink.jsp?2013La01,B}{2013La01}), but did not\vspace{0.1cm}}&\\
&&&&\parbox[t][0.3cm]{14.205921cm}{\raggedright {\ }{\ }{\ }explicitly give the value. Other value: J\ensuremath{^{\ensuremath{\pi}}}=3/2\ensuremath{^{-}}, 5/2, 7/2\ensuremath{^{\textnormal{+}}} with L=2,3, where L=2 was preferred even\vspace{0.1cm}}&\\
&&&&\parbox[t][0.3cm]{14.205921cm}{\raggedright {\ }{\ }{\ }though both L-values describe the data.\vspace{0.1cm}}&\\
&&&&\parbox[t][0.3cm]{14.205921cm}{\raggedright L: From (\href{https://www.nndc.bnl.gov/nsr/nsrlink.jsp?2019Ka15,B}{2019Ka15}).\vspace{0.1cm}}&\\
\multicolumn{1}{r@{}}{6862}&\multicolumn{1}{@{}l}{\ensuremath{^{{\hyperlink{NE32LEVEL0}{a}}}} {\it 3}}&\multicolumn{1}{l}{7/2\ensuremath{^{-}}}&&\parbox[t][0.3cm]{14.205921cm}{\raggedright \ensuremath{\Gamma}\ensuremath{\alpha}/\ensuremath{\Gamma}=0.925 \textit{+26{\textminus}28}\vspace{0.1cm}}&\\
&&&&\parbox[t][0.3cm]{14.205921cm}{\raggedright E(level): Weighted average of 6861 keV \textit{6} (\href{https://www.nndc.bnl.gov/nsr/nsrlink.jsp?1998Ut02,B}{1998Ut02}), which observed this state for the first time, and\vspace{0.1cm}}&\\
&&&&\parbox[t][0.3cm]{14.205921cm}{\raggedright {\ }{\ }{\ }6862 keV \textit{2} (stat.) \textit{2} (sys.) (\href{https://www.nndc.bnl.gov/nsr/nsrlink.jsp?2013La01,B}{2013La01}).\vspace{0.1cm}}&\\
&&&&\parbox[t][0.3cm]{14.205921cm}{\raggedright E(level): See also (1) 6861 keV (\href{https://www.nndc.bnl.gov/nsr/nsrlink.jsp?2004Vi05,B}{2004Vi05}: See Fig. 1); (2) 6851 keV \textit{4} and 6864 keV \textit{1} unresolved\vspace{0.1cm}}&\\
&&&&\parbox[t][0.3cm]{14.205921cm}{\raggedright {\ }{\ }{\ }doublet (\href{https://www.nndc.bnl.gov/nsr/nsrlink.jsp?2015Pa46,B}{2015Pa46}), where the evidence for the existence of a doublet is inconclusive. We therefore,\vspace{0.1cm}}&\\
&&&&\parbox[t][0.3cm]{14.205921cm}{\raggedright {\ }{\ }{\ }did not consider this doublet. (3) 6864 keV (\href{https://www.nndc.bnl.gov/nsr/nsrlink.jsp?2021Ri04,B}{2021Ri04}) used as a calibration point; and (4) 6859 keV\vspace{0.1cm}}&\\
&&&&\parbox[t][0.3cm]{14.205921cm}{\raggedright {\ }{\ }{\ }\textit{3} (\href{https://www.nndc.bnl.gov/nsr/nsrlink.jsp?2025PhZZ,B}{2025PhZZ}: Unpublished).\vspace{0.1cm}}&\\
&&&&\parbox[t][0.3cm]{14.205921cm}{\raggedright \ensuremath{\Gamma}=\ensuremath{\Gamma}\ensuremath{_{\ensuremath{\alpha}}}=1.2 keV and \ensuremath{\Gamma}\ensuremath{_{\textnormal{p}}}=1.1\ensuremath{\times}10\ensuremath{^{\textnormal{$-$5}}} keV were suggested by (\href{https://www.nndc.bnl.gov/nsr/nsrlink.jsp?2013La01,B}{2013La01}). These values are taken from\vspace{0.1cm}}&\\
&&&&\parbox[t][0.3cm]{14.205921cm}{\raggedright {\ }{\ }{\ }the evaluation of (\href{https://www.nndc.bnl.gov/nsr/nsrlink.jsp?2007Ne09,B}{2007Ne09}) and are scaled from the mirror state in \ensuremath{^{\textnormal{19}}}F. (\href{https://www.nndc.bnl.gov/nsr/nsrlink.jsp?2025PhZZ,B}{2025PhZZ}) reported \ensuremath{\Gamma}=31\vspace{0.1cm}}&\\
&&&&\parbox[t][0.3cm]{14.205921cm}{\raggedright {\ }{\ }{\ }keV, which is entirely dominated by their energy resolution of 31 keV at FWHM.\vspace{0.1cm}}&\\
&&&&\parbox[t][0.3cm]{14.205921cm}{\raggedright \ensuremath{\Gamma}\ensuremath{_{\ensuremath{\alpha}}}/\ensuremath{\Gamma}: Weighted average of 0.96 \textit{8} (\href{https://www.nndc.bnl.gov/nsr/nsrlink.jsp?1998Ut02,B}{1998Ut02}); 0.932 \textit{+28{\textminus}31} (\href{https://www.nndc.bnl.gov/nsr/nsrlink.jsp?2004Vi05,B}{2004Vi05}) at 1 \ensuremath{\sigma} level; and 0.81 \textit{9}\vspace{0.1cm}}&\\
&&&&\parbox[t][0.3cm]{14.205921cm}{\raggedright {\ }{\ }{\ }(\href{https://www.nndc.bnl.gov/nsr/nsrlink.jsp?2021Ri04,B}{2021Ri04}), where the uncertainty is the quadratic sum of statistical and systematic uncertainties. Other\vspace{0.1cm}}&\\
&&&&\parbox[t][0.3cm]{14.205921cm}{\raggedright {\ }{\ }{\ }value: 0.90 \textit{11} (\href{https://www.nndc.bnl.gov/nsr/nsrlink.jsp?2025PhZZ,B}{2025PhZZ}).\vspace{0.1cm}}&\\
&&&&\parbox[t][0.3cm]{14.205921cm}{\raggedright \ensuremath{\Gamma}\ensuremath{_{\textnormal{p}}}/\ensuremath{\Gamma}: See \ensuremath{\Gamma}\ensuremath{_{\textnormal{p}}}/\ensuremath{\Gamma}\ensuremath{<}0.025 (\href{https://www.nndc.bnl.gov/nsr/nsrlink.jsp?1998Ut02,B}{1998Ut02}); and \ensuremath{\Gamma}\ensuremath{_{\textnormal{p}}}/\ensuremath{\Gamma}\ensuremath{>}0.007 (\href{https://www.nndc.bnl.gov/nsr/nsrlink.jsp?2004Vi05,B}{2004Vi05}) at 1\ensuremath{\sigma} level. This latter value is\vspace{0.1cm}}&\\
&&&&\parbox[t][0.3cm]{14.205921cm}{\raggedright {\ }{\ }{\ }tentative with a probability of \ensuremath{>}90\%. The lower limit is placed due to a probable loss of some \ensuremath{\alpha}\vspace{0.1cm}}&\\
&&&&\parbox[t][0.3cm]{14.205921cm}{\raggedright {\ }{\ }{\ }particles whose signals were below the detection threshold.\vspace{0.1cm}}&\\
&&&&\parbox[t][0.3cm]{14.205921cm}{\raggedright J\ensuremath{^{\pi}}: From (1) J\ensuremath{^{\ensuremath{\pi}}}=7/2\ensuremath{^{-}} (\href{https://www.nndc.bnl.gov/nsr/nsrlink.jsp?1998Ut02,B}{1998Ut02}) from relative populations and angular distributions measured for\vspace{0.1cm}}&\\
&&&&\parbox[t][0.3cm]{14.205921cm}{\raggedright {\ }{\ }{\ }\ensuremath{^{\textnormal{20}}}Ne(d,\ensuremath{^{\textnormal{3}}}He), \ensuremath{^{\textnormal{20}}}Ne(d,t), \ensuremath{^{\textnormal{16}}}O(\ensuremath{^{\textnormal{6}}}Li,\ensuremath{^{\textnormal{3}}}He), and \ensuremath{^{\textnormal{16}}}O(\ensuremath{^{\textnormal{6}}}Li,t) mirror reactions; (2) J\ensuremath{^{\ensuremath{\pi}}}=7/2\ensuremath{^{-}} (\href{https://www.nndc.bnl.gov/nsr/nsrlink.jsp?2004Vi05,B}{2004Vi05}) from\vspace{0.1cm}}&\\
\end{longtable}
\begin{textblock}{29}(0,27.3)
Continued on next page (footnotes at end of table)
\end{textblock}
\clearpage
\begin{longtable}{ccccccc@{\extracolsep{\fill}}c}
\\[-.4cm]
\multicolumn{8}{c}{{\bf \small \underline{\ensuremath{^{\textnormal{19}}}F(\ensuremath{^{\textnormal{3}}}He,t)\hspace{0.2in}\href{https://www.nndc.bnl.gov/nsr/nsrlink.jsp?1998Ut02,B}{1998Ut02},\href{https://www.nndc.bnl.gov/nsr/nsrlink.jsp?2013La01,B}{2013La01},\href{https://www.nndc.bnl.gov/nsr/nsrlink.jsp?2019Ka15,B}{2019Ka15} (continued)}}}\\
\multicolumn{8}{c}{~}\\
\multicolumn{8}{c}{\underline{\ensuremath{^{19}}Ne Levels (continued)}}\\
\multicolumn{8}{c}{~}\\
\multicolumn{2}{c}{E(level)$^{}$}&J$^{\pi}$$^{}$&\multicolumn{2}{c}{\ensuremath{\Gamma}$^{}$}&L$^{}$&Comments&\\[-.2cm]
\multicolumn{2}{c}{\hrulefill}&\hrulefill&\multicolumn{2}{c}{\hrulefill}&\hrulefill&\hrulefill&
\endhead
&&&&&&\parbox[t][0.3cm]{12.256561cm}{\raggedright {\ }{\ }{\ }measured \ensuremath{\alpha} angular distributions (see Fig. 2); (3) J\ensuremath{^{\ensuremath{\pi}}}=(7/2\ensuremath{^{-}}) (\href{https://www.nndc.bnl.gov/nsr/nsrlink.jsp?2013La01,B}{2013La01}) from a\vspace{0.1cm}}&\\
&&&&&&\parbox[t][0.3cm]{12.256561cm}{\raggedright {\ }{\ }{\ }two-step finite-range DWBA analysis via FRESCO; and (4) J\ensuremath{^{\ensuremath{\pi}}}=7/2\ensuremath{^{-}} (\href{https://www.nndc.bnl.gov/nsr/nsrlink.jsp?2021Ri04,B}{2021Ri04}) from\vspace{0.1cm}}&\\
&&&&&&\parbox[t][0.3cm]{12.256561cm}{\raggedright {\ }{\ }{\ }the extracted triton-\ensuremath{\alpha} angular correlation data fitted using even Legendre polynomials up\vspace{0.1cm}}&\\
&&&&&&\parbox[t][0.3cm]{12.256561cm}{\raggedright {\ }{\ }{\ }to maximum order of 6.\vspace{0.1cm}}&\\
&&&&&&\parbox[t][0.3cm]{12.256561cm}{\raggedright J\ensuremath{^{\pi}}: See also (\href{https://www.nndc.bnl.gov/nsr/nsrlink.jsp?2015Pa46,B}{2015Pa46}), who reported J\ensuremath{^{\ensuremath{\pi}}} value is consistent with that of (\href{https://www.nndc.bnl.gov/nsr/nsrlink.jsp?2013La01,B}{2013La01}),\vspace{0.1cm}}&\\
&&&&&&\parbox[t][0.3cm]{12.256561cm}{\raggedright {\ }{\ }{\ }but did not give the value. Other values: J\ensuremath{^{\ensuremath{\pi}}}=7/2\ensuremath{^{-}}, 9/2\ensuremath{^{-}} with L=4 (\href{https://www.nndc.bnl.gov/nsr/nsrlink.jsp?2025PhZZ,B}{2025PhZZ}) from the\vspace{0.1cm}}&\\
&&&&&&\parbox[t][0.3cm]{12.256561cm}{\raggedright {\ }{\ }{\ }\ensuremath{\alpha} angular correlations.\vspace{0.1cm}}&\\
\multicolumn{1}{r@{}}{6968}&\multicolumn{1}{@{}l}{}&&&&&\parbox[t][0.3cm]{12.256561cm}{\raggedright E(level): From (\href{https://www.nndc.bnl.gov/nsr/nsrlink.jsp?2021Ri04,B}{2021Ri04}: See Fig. 3).\vspace{0.1cm}}&\\
\multicolumn{1}{r@{}}{7076}&\multicolumn{1}{@{ }l}{{\it 9}}&\multicolumn{1}{l}{3/2\ensuremath{^{+}}}&\multicolumn{1}{r@{}}{43}&\multicolumn{1}{@{ }l}{keV {\it 6}}&\multicolumn{1}{l}{0}&\parbox[t][0.3cm]{12.256561cm}{\raggedright \ensuremath{\Gamma}\ensuremath{_{\textnormal{p}}}/\ensuremath{\Gamma}=0.37 \textit{4} (\href{https://www.nndc.bnl.gov/nsr/nsrlink.jsp?1998Ut02,B}{1998Ut02})\vspace{0.1cm}}&\\
&&&&&&\parbox[t][0.3cm]{12.256561cm}{\raggedright \ensuremath{\Gamma}\ensuremath{\alpha}/\ensuremath{\Gamma}=0.63 \textit{6}\vspace{0.1cm}}&\\
&&&&&&\parbox[t][0.3cm]{12.256561cm}{\raggedright E(level): Weighted average (with external errors) of 7088 keV \textit{10} (\href{https://www.nndc.bnl.gov/nsr/nsrlink.jsp?2019Ka15,B}{2019Ka15}) and 7070\vspace{0.1cm}}&\\
&&&&&&\parbox[t][0.3cm]{12.256561cm}{\raggedright {\ }{\ }{\ }keV \textit{7} (\href{https://www.nndc.bnl.gov/nsr/nsrlink.jsp?1998Ut02,B}{1998Ut02}).\vspace{0.1cm}}&\\
&&&&&&\parbox[t][0.3cm]{12.256561cm}{\raggedright E(level): Various evidence in (\href{https://www.nndc.bnl.gov/nsr/nsrlink.jsp?1998Ut02,B}{1998Ut02}) suggest that this state is a single state: Centroid\vspace{0.1cm}}&\\
&&&&&&\parbox[t][0.3cm]{12.256561cm}{\raggedright {\ }{\ }{\ }position moved as kinematically expected in the spectra obtained at \ensuremath{\theta}\ensuremath{_{\textnormal{lab}}}=0\ensuremath{^\circ}{\textminus}15\ensuremath{^\circ}. Peak\vspace{0.1cm}}&\\
&&&&&&\parbox[t][0.3cm]{12.256561cm}{\raggedright {\ }{\ }{\ }shape remained the same. No improvement in the fit was observed if this peak was\vspace{0.1cm}}&\\
&&&&&&\parbox[t][0.3cm]{12.256561cm}{\raggedright {\ }{\ }{\ }fitted as an unresolved doublet. No significant difference in the proton/\ensuremath{\alpha}-particle\vspace{0.1cm}}&\\
&&&&&&\parbox[t][0.3cm]{12.256561cm}{\raggedright {\ }{\ }{\ }branching ratio was observed if the peak were treated as an unresolved doublet.\vspace{0.1cm}}&\\
&&&&&&\parbox[t][0.3cm]{12.256561cm}{\raggedright E(level): See also 7070 keV (\href{https://www.nndc.bnl.gov/nsr/nsrlink.jsp?2004Vi05,B}{2004Vi05}: See Fig. 1); 7076 keV (\href{https://www.nndc.bnl.gov/nsr/nsrlink.jsp?2015Pa46,B}{2015Pa46}); 7074 keV\vspace{0.1cm}}&\\
&&&&&&\parbox[t][0.3cm]{12.256561cm}{\raggedright {\ }{\ }{\ }(\href{https://www.nndc.bnl.gov/nsr/nsrlink.jsp?2021Ri04,B}{2021Ri04}) used as a calibration point and where the energy is presented as 7074 keV\vspace{0.1cm}}&\\
&&&&&&\parbox[t][0.3cm]{12.256561cm}{\raggedright {\ }{\ }{\ }on Fig. 3 but reported as 7076 keV in the text; 7060 keV and 7088 keV (\href{https://www.nndc.bnl.gov/nsr/nsrlink.jsp?1993UtZZ,B}{1993UtZZ});\vspace{0.1cm}}&\\
&&&&&&\parbox[t][0.3cm]{12.256561cm}{\raggedright {\ }{\ }{\ }and 7074 keV \textit{3} (\href{https://www.nndc.bnl.gov/nsr/nsrlink.jsp?2025PhZZ,B}{2025PhZZ}: Unpublished).\vspace{0.1cm}}&\\
&&&&&&\parbox[t][0.3cm]{12.256561cm}{\raggedright \ensuremath{\Gamma}: Weighted average of 44 keV \textit{6} (\href{https://www.nndc.bnl.gov/nsr/nsrlink.jsp?2019Ka15,B}{2019Ka15}) and 39 keV \textit{10} (\href{https://www.nndc.bnl.gov/nsr/nsrlink.jsp?1998Ut02,B}{1998Ut02}). The latter\vspace{0.1cm}}&\\
&&&&&&\parbox[t][0.3cm]{12.256561cm}{\raggedright {\ }{\ }{\ }study reported that the total observed width for this state was 46 keV, including an\vspace{0.1cm}}&\\
&&&&&&\parbox[t][0.3cm]{12.256561cm}{\raggedright {\ }{\ }{\ }experimental resolution of 24 keV. Note that when the instrumental resolution is\vspace{0.1cm}}&\\
&&&&&&\parbox[t][0.3cm]{12.256561cm}{\raggedright {\ }{\ }{\ }convoluted with a natural Lorentzian line shape, the widths do not add in quadrature.\vspace{0.1cm}}&\\
&&&&&&\parbox[t][0.3cm]{12.256561cm}{\raggedright {\ }{\ }{\ }Other value: \ensuremath{\Gamma}=58 keV (\href{https://www.nndc.bnl.gov/nsr/nsrlink.jsp?2025PhZZ,B}{2025PhZZ}), where the experimental resolution was 31 keV\vspace{0.1cm}}&\\
&&&&&&\parbox[t][0.3cm]{12.256561cm}{\raggedright {\ }{\ }{\ }(FWHM).\vspace{0.1cm}}&\\
&&&&&&\parbox[t][0.3cm]{12.256561cm}{\raggedright \ensuremath{\Gamma}\ensuremath{_{\textnormal{p}}}/\ensuremath{\Gamma}: Other values: 0.387 \textit{16} (\href{https://www.nndc.bnl.gov/nsr/nsrlink.jsp?2004Vi05,B}{2004Vi05}) at 1\ensuremath{\sigma} level; 0.58 \textit{8} (\href{https://www.nndc.bnl.gov/nsr/nsrlink.jsp?1993UtZZ,B}{1993UtZZ}); and 0.36 \textit{2}\vspace{0.1cm}}&\\
&&&&&&\parbox[t][0.3cm]{12.256561cm}{\raggedright {\ }{\ }{\ }(\href{https://www.nndc.bnl.gov/nsr/nsrlink.jsp?2025PhZZ,B}{2025PhZZ}) for L\ensuremath{_{\textnormal{rel}}}(p+\ensuremath{^{\textnormal{18}}}F)=0. \textit{Note}: The sum of the adopted \ensuremath{\Gamma}\ensuremath{_{\ensuremath{\alpha}}}/\ensuremath{\Gamma} and a weighted\vspace{0.1cm}}&\\
&&&&&&\parbox[t][0.3cm]{12.256561cm}{\raggedright {\ }{\ }{\ }average of the listed \ensuremath{\Gamma}\ensuremath{_{\textnormal{p}}}/\ensuremath{\Gamma} values would be slightly more than one. We therefore,\vspace{0.1cm}}&\\
&&&&&&\parbox[t][0.3cm]{12.256561cm}{\raggedright {\ }{\ }{\ }adopted \ensuremath{\Gamma}\ensuremath{_{\textnormal{p}}}/\ensuremath{\Gamma} value from (\href{https://www.nndc.bnl.gov/nsr/nsrlink.jsp?1998Ut02,B}{1998Ut02}).\vspace{0.1cm}}&\\
&&&&&&\parbox[t][0.3cm]{12.256561cm}{\raggedright \ensuremath{\Gamma}\ensuremath{_{\ensuremath{\alpha}}}/\ensuremath{\Gamma}: Weighted average of 0.62 \textit{7} (\href{https://www.nndc.bnl.gov/nsr/nsrlink.jsp?2021Ri04,B}{2021Ri04}), where the uncertainty is the quadratic\vspace{0.1cm}}&\\
&&&&&&\parbox[t][0.3cm]{12.256561cm}{\raggedright {\ }{\ }{\ }sum of statistical and systematic uncertainties, and 0.64 \textit{6} (\href{https://www.nndc.bnl.gov/nsr/nsrlink.jsp?1998Ut02,B}{1998Ut02}).\vspace{0.1cm}}&\\
&&&&&&\parbox[t][0.3cm]{12.256561cm}{\raggedright \ensuremath{\Gamma}\ensuremath{_{\ensuremath{\alpha}}}/\ensuremath{\Gamma}: See also (1) 0.613 (\href{https://www.nndc.bnl.gov/nsr/nsrlink.jsp?2004Vi05,B}{2004Vi05}) at 1\ensuremath{\sigma} level and obtained from a simultaneous fit\vspace{0.1cm}}&\\
&&&&&&\parbox[t][0.3cm]{12.256561cm}{\raggedright {\ }{\ }{\ }assuming \ensuremath{\Gamma}=\ensuremath{\Gamma}\ensuremath{_{\textnormal{p}}}+\ensuremath{\Gamma}\ensuremath{_{\ensuremath{\alpha}}}; and (2) \ensuremath{\Gamma}\ensuremath{_{\ensuremath{\alpha}}}/\ensuremath{\Gamma}=0.69 \textit{4} (\href{https://www.nndc.bnl.gov/nsr/nsrlink.jsp?2025PhZZ,B}{2025PhZZ}).\vspace{0.1cm}}&\\
&&&&&&\parbox[t][0.3cm]{12.256561cm}{\raggedright J\ensuremath{^{\pi}}: From (1) J\ensuremath{^{\ensuremath{\pi}}}=3/2\ensuremath{^{\textnormal{+}}} (\href{https://www.nndc.bnl.gov/nsr/nsrlink.jsp?2015Pa46,B}{2015Pa46}) from a finite-range coupled-channels analysis via\vspace{0.1cm}}&\\
&&&&&&\parbox[t][0.3cm]{12.256561cm}{\raggedright {\ }{\ }{\ }FRESCO; (2) J\ensuremath{^{\ensuremath{\pi}}}=3/2\ensuremath{^{\textnormal{+}}} (\href{https://www.nndc.bnl.gov/nsr/nsrlink.jsp?2019Ka15,B}{2019Ka15}) from \ensuremath{\Delta}L=0 and based on established value from\vspace{0.1cm}}&\\
&&&&&&\parbox[t][0.3cm]{12.256561cm}{\raggedright {\ }{\ }{\ }the literature; and (3) J\ensuremath{^{\ensuremath{\pi}}}=3/2\ensuremath{^{\textnormal{+}}} (\href{https://www.nndc.bnl.gov/nsr/nsrlink.jsp?2021Ri04,B}{2021Ri04}) from triton-\ensuremath{\alpha} angular correlation consistent\vspace{0.1cm}}&\\
&&&&&&\parbox[t][0.3cm]{12.256561cm}{\raggedright {\ }{\ }{\ }with an already established J\ensuremath{^{\ensuremath{\pi}}} from literature, which constrains the orbital angular\vspace{0.1cm}}&\\
&&&&&&\parbox[t][0.3cm]{12.256561cm}{\raggedright {\ }{\ }{\ }momentum of the \ensuremath{\alpha}-particle emitted from the decay of this state to L=1 implying\vspace{0.1cm}}&\\
&&&&&&\parbox[t][0.3cm]{12.256561cm}{\raggedright {\ }{\ }{\ }k\ensuremath{_{\textnormal{max}}}=2.\vspace{0.1cm}}&\\
&&&&&&\parbox[t][0.3cm]{12.256561cm}{\raggedright J\ensuremath{^{\pi}}: See also J\ensuremath{^{\ensuremath{\pi}}}=(1/2\ensuremath{^{\textnormal{+}}}, 3/2\ensuremath{^{\textnormal{+}}}) (\href{https://www.nndc.bnl.gov/nsr/nsrlink.jsp?1998Ut02,B}{1998Ut02}) from L=(0) based on a comparison of \ensuremath{\Gamma}\ensuremath{_{\textnormal{p}}}=14\vspace{0.1cm}}&\\
&&&&&&\parbox[t][0.3cm]{12.256561cm}{\raggedright {\ }{\ }{\ }keV \textit{4} with the single-particle Wigner limits for this state deduced for various choices of\vspace{0.1cm}}&\\
&&&&&&\parbox[t][0.3cm]{12.256561cm}{\raggedright {\ }{\ }{\ }J\ensuremath{^{\ensuremath{\pi}}} assignments, which favored a J\ensuremath{^{\ensuremath{\pi}}}=1/2\ensuremath{^{\textnormal{+}}} and 3/2\ensuremath{^{\textnormal{+}}} with L=0 assignment; J\ensuremath{^{\ensuremath{\pi}}}=3/2\ensuremath{^{\textnormal{+}}}\vspace{0.1cm}}&\\
&&&&&&\parbox[t][0.3cm]{12.256561cm}{\raggedright {\ }{\ }{\ }(\href{https://www.nndc.bnl.gov/nsr/nsrlink.jsp?2004Vi05,B}{2004Vi05}) taken from (a) (\href{https://www.nndc.bnl.gov/nsr/nsrlink.jsp?1996Re05,B}{1996Re05}: \ensuremath{^{\textnormal{18}}}F(p,\ensuremath{\alpha})), (b) (\href{https://www.nndc.bnl.gov/nsr/nsrlink.jsp?2001Gr01,B}{2001Gr01}: \ensuremath{^{\textnormal{1}}}H(\ensuremath{^{\textnormal{18}}}F,p), (c)\vspace{0.1cm}}&\\
&&&&&&\parbox[t][0.3cm]{12.256561cm}{\raggedright {\ }{\ }{\ }\ensuremath{^{\textnormal{1}}}H(\ensuremath{^{\textnormal{18}}}F,\ensuremath{\alpha})), and (d) (\href{https://www.nndc.bnl.gov/nsr/nsrlink.jsp?2001Ba49,B}{2001Ba49}: \ensuremath{^{\textnormal{1}}}H(\ensuremath{^{\textnormal{18}}}F,p) and \ensuremath{^{\textnormal{1}}}H(\ensuremath{^{\textnormal{18}}}F,\ensuremath{\alpha})). Other values: J\ensuremath{^{\ensuremath{\pi}}}=1/2\ensuremath{^{\textnormal{+}}},\vspace{0.1cm}}&\\
&&&&&&\parbox[t][0.3cm]{12.256561cm}{\raggedright {\ }{\ }{\ }3/2, 5/2\ensuremath{^{-}} with L=1,2, where L=1 was preferred even though L=2 was a better fit.\vspace{0.1cm}}&\\
&&&&&&\parbox[t][0.3cm]{12.256561cm}{\raggedright L: From (\href{https://www.nndc.bnl.gov/nsr/nsrlink.jsp?1998Ut02,B}{1998Ut02}: L=(0)) and (\href{https://www.nndc.bnl.gov/nsr/nsrlink.jsp?2019Ka15,B}{2019Ka15}: \ensuremath{\Delta}L=0).\vspace{0.1cm}}&\\
\multicolumn{1}{r@{}}{7173}&\multicolumn{1}{@{ }l}{{\it 5}}&\multicolumn{1}{l}{11/2}&&&&\parbox[t][0.3cm]{12.256561cm}{\raggedright E(level): From (\href{https://www.nndc.bnl.gov/nsr/nsrlink.jsp?1998Ut02,B}{1998Ut02}). See also 7173 keV (\href{https://www.nndc.bnl.gov/nsr/nsrlink.jsp?2015Pa46,B}{2015Pa46}, \href{https://www.nndc.bnl.gov/nsr/nsrlink.jsp?2021Ri04,B}{2021Ri04}: See Fig. 3).\vspace{0.1cm}}&\\
&&&&&&\parbox[t][0.3cm]{12.256561cm}{\raggedright J\ensuremath{^{\pi}}: From J\ensuremath{^{\ensuremath{\pi}}}=11/2\ensuremath{^{\textnormal{+}}}, 11/2\ensuremath{^{-}} (\href{https://www.nndc.bnl.gov/nsr/nsrlink.jsp?2015Pa46,B}{2015Pa46}) from a finite-range coupled-channels calculation\vspace{0.1cm}}&\\
&&&&&&\parbox[t][0.3cm]{12.256561cm}{\raggedright {\ }{\ }{\ }using FRESCO. These assignments both provide reasonable fits to the triton angular\vspace{0.1cm}}&\\
&&&&&&\parbox[t][0.3cm]{12.256561cm}{\raggedright {\ }{\ }{\ }distribution data, so J=11/2 was recommended; and J\ensuremath{^{\ensuremath{\pi}}}=(11/2\ensuremath{^{-}}) (\href{https://www.nndc.bnl.gov/nsr/nsrlink.jsp?1998Ut02,B}{1998Ut02}) from mirror\vspace{0.1cm}}&\\
&&&&&&\parbox[t][0.3cm]{12.256561cm}{\raggedright {\ }{\ }{\ }analysis, and thus the evaluator made the assignment tentative.\vspace{0.1cm}}&\\
&&&&&&\parbox[t][0.3cm]{12.256561cm}{\raggedright J\ensuremath{^{\pi}}: See also J\ensuremath{^{\ensuremath{\pi}}}=(11/2\ensuremath{^{-}}) based on the mirror levels analysis in the evaluation of\vspace{0.1cm}}&\\
&&&&&&\parbox[t][0.3cm]{12.256561cm}{\raggedright {\ }{\ }{\ }(\href{https://www.nndc.bnl.gov/nsr/nsrlink.jsp?2007Ne09,B}{2007Ne09}).\vspace{0.1cm}}&\\
\multicolumn{1}{r@{}}{7238}&\multicolumn{1}{@{ }l}{{\it 6}}&\multicolumn{1}{l}{(3/2\ensuremath{^{+}})}&&&&\parbox[t][0.3cm]{12.256561cm}{\raggedright E(level): From (\href{https://www.nndc.bnl.gov/nsr/nsrlink.jsp?1998Ut02,B}{1998Ut02}), who observed this state for the first time. See also 7238 keV\vspace{0.1cm}}&\\
\end{longtable}
\begin{textblock}{29}(0,27.3)
Continued on next page (footnotes at end of table)
\end{textblock}
\clearpage
\begin{longtable}{cccccc@{\extracolsep{\fill}}c}
\\[-.4cm]
\multicolumn{7}{c}{{\bf \small \underline{\ensuremath{^{\textnormal{19}}}F(\ensuremath{^{\textnormal{3}}}He,t)\hspace{0.2in}\href{https://www.nndc.bnl.gov/nsr/nsrlink.jsp?1998Ut02,B}{1998Ut02},\href{https://www.nndc.bnl.gov/nsr/nsrlink.jsp?2013La01,B}{2013La01},\href{https://www.nndc.bnl.gov/nsr/nsrlink.jsp?2019Ka15,B}{2019Ka15} (continued)}}}\\
\multicolumn{7}{c}{~}\\
\multicolumn{7}{c}{\underline{\ensuremath{^{19}}Ne Levels (continued)}}\\
\multicolumn{7}{c}{~}\\
\multicolumn{2}{c}{E(level)$^{}$}&J$^{\pi}$$^{}$&\multicolumn{2}{c}{\ensuremath{\Gamma}$^{}$}&Comments&\\[-.2cm]
\multicolumn{2}{c}{\hrulefill}&\hrulefill&\multicolumn{2}{c}{\hrulefill}&\hrulefill&
\endhead
&&&&&\parbox[t][0.3cm]{12.874841cm}{\raggedright {\ }{\ }{\ }(\href{https://www.nndc.bnl.gov/nsr/nsrlink.jsp?2015Pa46,B}{2015Pa46}).\vspace{0.1cm}}&\\
&&&&&\parbox[t][0.3cm]{12.874841cm}{\raggedright J\ensuremath{^{\pi}}: From (\href{https://www.nndc.bnl.gov/nsr/nsrlink.jsp?1998Ut02,B}{1998Ut02}): Based on relative populations and angular distributions measured for\vspace{0.1cm}}&\\
&&&&&\parbox[t][0.3cm]{12.874841cm}{\raggedright {\ }{\ }{\ }\ensuremath{^{\textnormal{20}}}Ne(d,\ensuremath{^{\textnormal{3}}}He), \ensuremath{^{\textnormal{20}}}Ne(d,t), \ensuremath{^{\textnormal{16}}}O(\ensuremath{^{\textnormal{6}}}Li,\ensuremath{^{\textnormal{3}}}He), and \ensuremath{^{\textnormal{16}}}O(\ensuremath{^{\textnormal{6}}}Li,t) mirror reactions. Since this is a weak\vspace{0.1cm}}&\\
&&&&&\parbox[t][0.3cm]{12.874841cm}{\raggedright {\ }{\ }{\ }argument, we considered the assignment tentative.\vspace{0.1cm}}&\\
\multicolumn{1}{r@{}}{7420}&\multicolumn{1}{@{}l}{}&\multicolumn{1}{l}{7/2\ensuremath{^{+}}}&&&\parbox[t][0.3cm]{12.874841cm}{\raggedright \ensuremath{\Gamma}\ensuremath{\alpha}/\ensuremath{\Gamma}=0.76 \textit{12} (\href{https://www.nndc.bnl.gov/nsr/nsrlink.jsp?2021Ri04,B}{2021Ri04})\vspace{0.1cm}}&\\
&&&&&\parbox[t][0.3cm]{12.874841cm}{\raggedright E(level): From (\href{https://www.nndc.bnl.gov/nsr/nsrlink.jsp?2021Ri04,B}{2021Ri04}). This state was observed in their focal plane singles and the\vspace{0.1cm}}&\\
&&&&&\parbox[t][0.3cm]{12.874841cm}{\raggedright {\ }{\ }{\ }triton-\ensuremath{\alpha} coincidence events.\vspace{0.1cm}}&\\
&&&&&\parbox[t][0.3cm]{12.874841cm}{\raggedright J\ensuremath{^{\pi}}: From triton-\ensuremath{\alpha} angular correlation in (\href{https://www.nndc.bnl.gov/nsr/nsrlink.jsp?2021Ri04,B}{2021Ri04}) fitted using even Legendre polynomials\vspace{0.1cm}}&\\
&&&&&\parbox[t][0.3cm]{12.874841cm}{\raggedright {\ }{\ }{\ }up to k\ensuremath{_{\textnormal{max}}}=6.\vspace{0.1cm}}&\\
&&&&&\parbox[t][0.3cm]{12.874841cm}{\raggedright \ensuremath{\Gamma}\ensuremath{_{\ensuremath{\alpha}}}/\ensuremath{\Gamma}: The uncertainty is the quadratic sum of statistical and systematic uncertainties\vspace{0.1cm}}&\\
&&&&&\parbox[t][0.3cm]{12.874841cm}{\raggedright {\ }{\ }{\ }(\href{https://www.nndc.bnl.gov/nsr/nsrlink.jsp?2021Ri04,B}{2021Ri04}).\vspace{0.1cm}}&\\
\multicolumn{1}{r@{}}{7500}&\multicolumn{1}{@{ }l}{{\it 9}}&\multicolumn{1}{l}{5/2\ensuremath{^{+}}}&\multicolumn{1}{r@{}}{16}&\multicolumn{1}{@{ }l}{keV {\it 16}}&\parbox[t][0.3cm]{12.874841cm}{\raggedright \ensuremath{\Gamma}\ensuremath{\alpha}/\ensuremath{\Gamma}=0.47 \textit{6} (\href{https://www.nndc.bnl.gov/nsr/nsrlink.jsp?2021Ri04,B}{2021Ri04})\vspace{0.1cm}}&\\
&&&&&\parbox[t][0.3cm]{12.874841cm}{\raggedright \ensuremath{\Gamma}\ensuremath{_{\textnormal{p}}}/\ensuremath{\Gamma}=0.66 \textit{7} (\href{https://www.nndc.bnl.gov/nsr/nsrlink.jsp?2021Ri04,B}{2021Ri04})\vspace{0.1cm}}&\\
&&&&&\parbox[t][0.3cm]{12.874841cm}{\raggedright E(level): From (\href{https://www.nndc.bnl.gov/nsr/nsrlink.jsp?1998Ut02,B}{1998Ut02}).\vspace{0.1cm}}&\\
&&&&&\parbox[t][0.3cm]{12.874841cm}{\raggedright E(level): See also 7499 keV (\href{https://www.nndc.bnl.gov/nsr/nsrlink.jsp?2021Ri04,B}{2021Ri04}), where the state$'$s energy was presented as 7499 keV\vspace{0.1cm}}&\\
&&&&&\parbox[t][0.3cm]{12.874841cm}{\raggedright {\ }{\ }{\ }on Fig. 3 but reported as 7500 keV in the text; 7500 keV (\href{https://www.nndc.bnl.gov/nsr/nsrlink.jsp?1993UtZZ,B}{1993UtZZ}); and 7484 keV \textit{3}\vspace{0.1cm}}&\\
&&&&&\parbox[t][0.3cm]{12.874841cm}{\raggedright {\ }{\ }{\ }(\href{https://www.nndc.bnl.gov/nsr/nsrlink.jsp?2025PhZZ,B}{2025PhZZ}: Unpublished).\vspace{0.1cm}}&\\
&&&&&\parbox[t][0.3cm]{12.874841cm}{\raggedright \ensuremath{\Gamma}: From (\href{https://www.nndc.bnl.gov/nsr/nsrlink.jsp?1998Ut02,B}{1998Ut02}).\vspace{0.1cm}}&\\
&&&&&\parbox[t][0.3cm]{12.874841cm}{\raggedright \ensuremath{\Gamma}: See also \ensuremath{\Gamma}=27 keV, which is dominated by the experimental resolution of 31 keV at\vspace{0.1cm}}&\\
&&&&&\parbox[t][0.3cm]{12.874841cm}{\raggedright {\ }{\ }{\ }FWHM.\vspace{0.1cm}}&\\
&&&&&\parbox[t][0.3cm]{12.874841cm}{\raggedright \ensuremath{\Gamma}\ensuremath{_{\ensuremath{\alpha}}}/\ensuremath{\Gamma} and \ensuremath{\Gamma}\ensuremath{_{\textnormal{p}}}/\ensuremath{\Gamma}: See also \ensuremath{\Gamma}\ensuremath{_{\ensuremath{\alpha}}}/\ensuremath{\Gamma}=0.16 \textit{2} (\href{https://www.nndc.bnl.gov/nsr/nsrlink.jsp?1998Ut02,B}{1998Ut02}); \ensuremath{\Gamma}\ensuremath{_{\textnormal{p}}}/\ensuremath{\Gamma}=0.84 \textit{4} (\href{https://www.nndc.bnl.gov/nsr/nsrlink.jsp?1998Ut02,B}{1998Ut02});\vspace{0.1cm}}&\\
&&&&&\parbox[t][0.3cm]{12.874841cm}{\raggedright {\ }{\ }{\ }\ensuremath{\Gamma}\ensuremath{_{\textnormal{p}}}/\ensuremath{\Gamma}\ensuremath{_{\ensuremath{\alpha}}}=2.7 \textit{9} (\href{https://www.nndc.bnl.gov/nsr/nsrlink.jsp?1993UtZZ,B}{1993UtZZ}); \ensuremath{\Gamma}\ensuremath{_{\textnormal{p}}}/\ensuremath{\Gamma}=0.69 \textit{9} (\href{https://www.nndc.bnl.gov/nsr/nsrlink.jsp?2025PhZZ,B}{2025PhZZ}) for L\ensuremath{_{\textnormal{rel}}}(p+\ensuremath{^{\textnormal{18}}}F)=2; and \ensuremath{\Gamma}\ensuremath{_{\ensuremath{\alpha}}}/\ensuremath{\Gamma}=0.05\vspace{0.1cm}}&\\
&&&&&\parbox[t][0.3cm]{12.874841cm}{\raggedright {\ }{\ }{\ }\textit{3} (\href{https://www.nndc.bnl.gov/nsr/nsrlink.jsp?2025PhZZ,B}{2025PhZZ}). Branching ratios deduced by (\href{https://www.nndc.bnl.gov/nsr/nsrlink.jsp?2021Ri04,B}{2021Ri04}) are inconsistent with those deduced\vspace{0.1cm}}&\\
&&&&&\parbox[t][0.3cm]{12.874841cm}{\raggedright {\ }{\ }{\ }by (\href{https://www.nndc.bnl.gov/nsr/nsrlink.jsp?1998Ut02,B}{1998Ut02}). (\href{https://www.nndc.bnl.gov/nsr/nsrlink.jsp?2021Ri04,B}{2021Ri04}) reported that in the procedure followed by (\href{https://www.nndc.bnl.gov/nsr/nsrlink.jsp?1998Ut02,B}{1998Ut02}), the\vspace{0.1cm}}&\\
&&&&&\parbox[t][0.3cm]{12.874841cm}{\raggedright {\ }{\ }{\ }\ensuremath{\sum}P(m)=1 relation between the magnetic sub-states population is ignored. This can lead to\vspace{0.1cm}}&\\
&&&&&\parbox[t][0.3cm]{12.874841cm}{\raggedright {\ }{\ }{\ }erroneous shape of the angular correlation function resulting in biased determination of the\vspace{0.1cm}}&\\
&&&&&\parbox[t][0.3cm]{12.874841cm}{\raggedright {\ }{\ }{\ }branching ratios by (\href{https://www.nndc.bnl.gov/nsr/nsrlink.jsp?1998Ut02,B}{1998Ut02}). Furthermore, the angular correlation data of (\href{https://www.nndc.bnl.gov/nsr/nsrlink.jsp?2021Ri04,B}{2021Ri04}) for\vspace{0.1cm}}&\\
&&&&&\parbox[t][0.3cm]{12.874841cm}{\raggedright {\ }{\ }{\ }this state could be contaminated by the decay of the \ensuremath{^{\textnormal{19}}}Ne*(7531) state. (\href{https://www.nndc.bnl.gov/nsr/nsrlink.jsp?2021Ri04,B}{2021Ri04}) deduced\vspace{0.1cm}}&\\
&&&&&\parbox[t][0.3cm]{12.874841cm}{\raggedright {\ }{\ }{\ }branching ratios for the combined \ensuremath{^{\textnormal{19}}}Ne*(7500) and \ensuremath{^{\textnormal{19}}}Ne*(7531) from the data of\vspace{0.1cm}}&\\
&&&&&\parbox[t][0.3cm]{12.874841cm}{\raggedright {\ }{\ }{\ }(\href{https://www.nndc.bnl.gov/nsr/nsrlink.jsp?1998Ut02,B}{1998Ut02}) and obtained \ensuremath{\Gamma}\ensuremath{_{\ensuremath{\alpha}}}/\ensuremath{\Gamma}\ensuremath{_{\textnormal{7500+7531}}}=0.27 \textit{3} and \ensuremath{\Gamma}\ensuremath{_{\textnormal{p}}}/\ensuremath{\Gamma}\ensuremath{_{\textnormal{7500+7531}}}=0.73 \textit{4}, which are\vspace{0.1cm}}&\\
&&&&&\parbox[t][0.3cm]{12.874841cm}{\raggedright {\ }{\ }{\ }in better agreement with the results of (\href{https://www.nndc.bnl.gov/nsr/nsrlink.jsp?2021Ri04,B}{2021Ri04}).\vspace{0.1cm}}&\\
&&&&&\parbox[t][0.3cm]{12.874841cm}{\raggedright J\ensuremath{^{\pi}}: From (\href{https://www.nndc.bnl.gov/nsr/nsrlink.jsp?2021Ri04,B}{2021Ri04}) based on triton-\ensuremath{\alpha} and triton-p angular correlations fitted using even\vspace{0.1cm}}&\\
&&&&&\parbox[t][0.3cm]{12.874841cm}{\raggedright {\ }{\ }{\ }Legendre polynomials up to k\ensuremath{_{\textnormal{max}}}=4.\vspace{0.1cm}}&\\
&&&&&\parbox[t][0.3cm]{12.874841cm}{\raggedright J\ensuremath{^{\pi}}: See also J\ensuremath{^{\ensuremath{\pi}}}=3/2\ensuremath{^{-}}, 5/2, 7/2, 9/2\ensuremath{^{-}} from the \ensuremath{\alpha} angular correlations with L\ensuremath{_{\textnormal{rel}}}(\ensuremath{\alpha}+\ensuremath{^{\textnormal{15}}}O)=2,3,4\vspace{0.1cm}}&\\
&&&&&\parbox[t][0.3cm]{12.874841cm}{\raggedright {\ }{\ }{\ }and J\ensuremath{^{\ensuremath{\pi}}}=1/2, 3/2, 5/2, 7/2, 9/2\ensuremath{^{-}} with L\ensuremath{_{\textnormal{rel}}}(p+\ensuremath{^{\textnormal{18}}}F)=1,2,3. Those authors preferred L=2 for\vspace{0.1cm}}&\\
&&&&&\parbox[t][0.3cm]{12.874841cm}{\raggedright {\ }{\ }{\ }protons and L=3 for \ensuremath{\alpha}s; however, the data can be fitted with all the listed L-values.\vspace{0.1cm}}&\\
\multicolumn{1}{r@{}}{7531}&\multicolumn{1}{@{ }l}{{\it 11}}&&\multicolumn{1}{r@{}}{31}&\multicolumn{1}{@{ }l}{keV {\it 16}}&\parbox[t][0.3cm]{12.874841cm}{\raggedright \ensuremath{\Gamma}\ensuremath{\alpha}/\ensuremath{\Gamma}=0.67 \textit{8} (\href{https://www.nndc.bnl.gov/nsr/nsrlink.jsp?1998Ut02,B}{1998Ut02})\vspace{0.1cm}}&\\
&&&&&\parbox[t][0.3cm]{12.874841cm}{\raggedright \ensuremath{\Gamma}\ensuremath{_{\textnormal{p}}}/\ensuremath{\Gamma}=0.33 \textit{6} (\href{https://www.nndc.bnl.gov/nsr/nsrlink.jsp?1998Ut02,B}{1998Ut02})\vspace{0.1cm}}&\\
&&&&&\parbox[t][0.3cm]{12.874841cm}{\raggedright E(level): From (\href{https://www.nndc.bnl.gov/nsr/nsrlink.jsp?1998Ut02,B}{1998Ut02}).\vspace{0.1cm}}&\\
&&&&&\parbox[t][0.3cm]{12.874841cm}{\raggedright E(level): See also 7531 keV (\href{https://www.nndc.bnl.gov/nsr/nsrlink.jsp?1993UtZZ,B}{1993UtZZ}, \href{https://www.nndc.bnl.gov/nsr/nsrlink.jsp?2021Ri04,B}{2021Ri04}: See Fig. 3); and 7509 keV \textit{3} (\href{https://www.nndc.bnl.gov/nsr/nsrlink.jsp?2025PhZZ,B}{2025PhZZ}:\vspace{0.1cm}}&\\
&&&&&\parbox[t][0.3cm]{12.874841cm}{\raggedright {\ }{\ }{\ }Unpublished).\vspace{0.1cm}}&\\
&&&&&\parbox[t][0.3cm]{12.874841cm}{\raggedright \ensuremath{\Gamma}: From (\href{https://www.nndc.bnl.gov/nsr/nsrlink.jsp?1998Ut02,B}{1998Ut02}). Other value: \ensuremath{\Gamma}=55 keV (\href{https://www.nndc.bnl.gov/nsr/nsrlink.jsp?2025PhZZ,B}{2025PhZZ}), where the experimental resolution\vspace{0.1cm}}&\\
&&&&&\parbox[t][0.3cm]{12.874841cm}{\raggedright {\ }{\ }{\ }was 31 keV at FWHM.\vspace{0.1cm}}&\\
&&&&&\parbox[t][0.3cm]{12.874841cm}{\raggedright See also \ensuremath{\Gamma}\ensuremath{_{\textnormal{p}}}/\ensuremath{\Gamma}\ensuremath{_{\ensuremath{\alpha}}}=0.29 \textit{9} (\href{https://www.nndc.bnl.gov/nsr/nsrlink.jsp?1993UtZZ,B}{1993UtZZ}); \ensuremath{\Gamma}\ensuremath{_{\textnormal{p}}}/\ensuremath{\Gamma}=0.26 \textit{5} (\href{https://www.nndc.bnl.gov/nsr/nsrlink.jsp?2025PhZZ,B}{2025PhZZ}) and \ensuremath{\Gamma}\ensuremath{_{\ensuremath{\alpha}}}/\ensuremath{\Gamma}=0.55 \textit{26}\vspace{0.1cm}}&\\
&&&&&\parbox[t][0.3cm]{12.874841cm}{\raggedright {\ }{\ }{\ }(\href{https://www.nndc.bnl.gov/nsr/nsrlink.jsp?2025PhZZ,B}{2025PhZZ}).\vspace{0.1cm}}&\\
&&&&&\parbox[t][0.3cm]{12.874841cm}{\raggedright J\ensuremath{^{\pi}}: See J\ensuremath{^{\ensuremath{\pi}}}=3/2\ensuremath{^{-}}, 5/2, 7/2, 9/2\ensuremath{^{-}} with L\ensuremath{_{\textnormal{rel}}}(\ensuremath{\alpha}+\ensuremath{^{\textnormal{15}}}O)=2,3,4 from the \ensuremath{\alpha} angular correlations and\vspace{0.1cm}}&\\
&&&&&\parbox[t][0.3cm]{12.874841cm}{\raggedright {\ }{\ }{\ }J\ensuremath{^{\ensuremath{\pi}}}=1/2, 3/2, 5/2, 7/2, 9/2\ensuremath{^{-}} with L\ensuremath{_{\textnormal{rel}}}(p+\ensuremath{^{\textnormal{18}}}F)=1,2,3. Those authors preferred L=2 for protons\vspace{0.1cm}}&\\
&&&&&\parbox[t][0.3cm]{12.874841cm}{\raggedright {\ }{\ }{\ }and L=3 for \ensuremath{\alpha}s; however, the data can be fitted with all the listed L-values.\vspace{0.1cm}}&\\
\multicolumn{1}{r@{}}{7558?}&\multicolumn{1}{@{}l}{}&&&&\parbox[t][0.3cm]{12.874841cm}{\raggedright E(level): From (\href{https://www.nndc.bnl.gov/nsr/nsrlink.jsp?2021Ri04,B}{2021Ri04}: See Fig. 3).\vspace{0.1cm}}&\\
\multicolumn{1}{r@{}}{7608}&\multicolumn{1}{@{ }l}{{\it 11}}&&\multicolumn{1}{r@{}}{45}&\multicolumn{1}{@{ }l}{keV {\it 16}}&\parbox[t][0.3cm]{12.874841cm}{\raggedright \ensuremath{\Gamma}\ensuremath{\alpha}/\ensuremath{\Gamma}=0.97 \textit{4} (\href{https://www.nndc.bnl.gov/nsr/nsrlink.jsp?1998Ut02,B}{1998Ut02})\vspace{0.1cm}}&\\
&&&&&\parbox[t][0.3cm]{12.874841cm}{\raggedright \ensuremath{\Gamma}\ensuremath{_{\textnormal{p}}}/\ensuremath{\Gamma}=0.04 \textit{2} (\href{https://www.nndc.bnl.gov/nsr/nsrlink.jsp?1998Ut02,B}{1998Ut02})\vspace{0.1cm}}&\\
&&&&&\parbox[t][0.3cm]{12.874841cm}{\raggedright T=3/2\vspace{0.1cm}}&\\
&&&&&\parbox[t][0.3cm]{12.874841cm}{\raggedright E(level): From (\href{https://www.nndc.bnl.gov/nsr/nsrlink.jsp?1998Ut02,B}{1998Ut02}). See also 7615 keV (\href{https://www.nndc.bnl.gov/nsr/nsrlink.jsp?2021Ri04,B}{2021Ri04}: See Fig. 3); and 7579 keV \textit{3}\vspace{0.1cm}}&\\
&&&&&\parbox[t][0.3cm]{12.874841cm}{\raggedright {\ }{\ }{\ }(\href{https://www.nndc.bnl.gov/nsr/nsrlink.jsp?2025PhZZ,B}{2025PhZZ}: Unpublished).\vspace{0.1cm}}&\\
\end{longtable}
\begin{textblock}{29}(0,27.3)
Continued on next page (footnotes at end of table)
\end{textblock}
\clearpage
\begin{longtable}{ccccccc@{\extracolsep{\fill}}c}
\\[-.4cm]
\multicolumn{8}{c}{{\bf \small \underline{\ensuremath{^{\textnormal{19}}}F(\ensuremath{^{\textnormal{3}}}He,t)\hspace{0.2in}\href{https://www.nndc.bnl.gov/nsr/nsrlink.jsp?1998Ut02,B}{1998Ut02},\href{https://www.nndc.bnl.gov/nsr/nsrlink.jsp?2013La01,B}{2013La01},\href{https://www.nndc.bnl.gov/nsr/nsrlink.jsp?2019Ka15,B}{2019Ka15} (continued)}}}\\
\multicolumn{8}{c}{~}\\
\multicolumn{8}{c}{\underline{\ensuremath{^{19}}Ne Levels (continued)}}\\
\multicolumn{8}{c}{~}\\
\multicolumn{2}{c}{E(level)$^{}$}&J$^{\pi}$$^{}$&\multicolumn{2}{c}{\ensuremath{\Gamma}$^{}$}&L$^{}$&Comments&\\[-.2cm]
\multicolumn{2}{c}{\hrulefill}&\hrulefill&\multicolumn{2}{c}{\hrulefill}&\hrulefill&\hrulefill&
\endhead
&&&&&&\parbox[t][0.3cm]{11.370601cm}{\raggedright \ensuremath{\Gamma}: From (\href{https://www.nndc.bnl.gov/nsr/nsrlink.jsp?1998Ut02,B}{1998Ut02}). See also \ensuremath{\Gamma}=45 keV (\href{https://www.nndc.bnl.gov/nsr/nsrlink.jsp?2025PhZZ,B}{2025PhZZ}).\vspace{0.1cm}}&\\
&&&&&&\parbox[t][0.3cm]{11.370601cm}{\raggedright (\href{https://www.nndc.bnl.gov/nsr/nsrlink.jsp?2009Mu07,B}{2009Mu07}: \ensuremath{^{\textnormal{1}}}H(\ensuremath{^{\textnormal{18}}}F,\ensuremath{\alpha}) and \ensuremath{^{\textnormal{1}}}H(\ensuremath{^{\textnormal{18}}}F,p)) reported that the absence of a distinct\vspace{0.1cm}}&\\
&&&&&&\parbox[t][0.3cm]{11.370601cm}{\raggedright {\ }{\ }{\ }feature corresponding to this state in their data, especially in the \ensuremath{^{\textnormal{18}}}F(p,\ensuremath{\alpha})\vspace{0.1cm}}&\\
&&&&&&\parbox[t][0.3cm]{11.370601cm}{\raggedright {\ }{\ }{\ }channel, is inconsistent with the \ensuremath{\Gamma}=46 keV \textit{16} reported by (\href{https://www.nndc.bnl.gov/nsr/nsrlink.jsp?1998Ut02,B}{1998Ut02}).\vspace{0.1cm}}&\\
&&&&&&\parbox[t][0.3cm]{11.370601cm}{\raggedright \ensuremath{\Gamma}\ensuremath{_{\ensuremath{\alpha}}}/\ensuremath{\Gamma}=0.84 \textit{12} (\href{https://www.nndc.bnl.gov/nsr/nsrlink.jsp?2025PhZZ,B}{2025PhZZ}), where they mention that this state decayed\vspace{0.1cm}}&\\
&&&&&&\parbox[t][0.3cm]{11.370601cm}{\raggedright {\ }{\ }{\ }exclusively via \ensuremath{\alpha} emission (see p. 120).\vspace{0.1cm}}&\\
&&&&&&\parbox[t][0.3cm]{11.370601cm}{\raggedright J\ensuremath{^{\pi}}: See J\ensuremath{^{\ensuremath{\pi}}}=3/2\ensuremath{^{-}}, 5/2, 7/2\ensuremath{^{\textnormal{+}}} with L=2,3 from \ensuremath{\alpha} angular correlations. Those authors\vspace{0.1cm}}&\\
&&&&&&\parbox[t][0.3cm]{11.370601cm}{\raggedright {\ }{\ }{\ }preferred L=2, which is indistinguishable from L=3.\vspace{0.1cm}}&\\
\multicolumn{1}{r@{}}{7644}&\multicolumn{1}{@{ }l}{{\it 12}}&&\multicolumn{1}{r@{}}{43}&\multicolumn{1}{@{ }l}{keV {\it 16}}&&\parbox[t][0.3cm]{11.370601cm}{\raggedright \ensuremath{\Gamma}\ensuremath{\alpha}/\ensuremath{\Gamma}=0.37 \textit{6} (\href{https://www.nndc.bnl.gov/nsr/nsrlink.jsp?1998Ut02,B}{1998Ut02})\vspace{0.1cm}}&\\
&&&&&&\parbox[t][0.3cm]{11.370601cm}{\raggedright \ensuremath{\Gamma}\ensuremath{_{\textnormal{p}}}/\ensuremath{\Gamma}=0.64 \textit{4} (\href{https://www.nndc.bnl.gov/nsr/nsrlink.jsp?1998Ut02,B}{1998Ut02})\vspace{0.1cm}}&\\
&&&&&&\parbox[t][0.3cm]{11.370601cm}{\raggedright E(level): From (\href{https://www.nndc.bnl.gov/nsr/nsrlink.jsp?1998Ut02,B}{1998Ut02}).\vspace{0.1cm}}&\\
&&&&&&\parbox[t][0.3cm]{11.370601cm}{\raggedright E(level): See also 7621 keV \textit{5} (\href{https://www.nndc.bnl.gov/nsr/nsrlink.jsp?2019Ka15,B}{2019Ka15}), where more than one unresolved state\vspace{0.1cm}}&\\
&&&&&&\parbox[t][0.3cm]{11.370601cm}{\raggedright {\ }{\ }{\ }may contribute to the observed peak; 7629 keV (\href{https://www.nndc.bnl.gov/nsr/nsrlink.jsp?1993UtZZ,B}{1993UtZZ}); 7644 keV\vspace{0.1cm}}&\\
&&&&&&\parbox[t][0.3cm]{11.370601cm}{\raggedright {\ }{\ }{\ }(\href{https://www.nndc.bnl.gov/nsr/nsrlink.jsp?2021Ri04,B}{2021Ri04}: See Fig. 3); and 7615 keV \textit{3} (\href{https://www.nndc.bnl.gov/nsr/nsrlink.jsp?2025PhZZ,B}{2025PhZZ}: Unpublished).\vspace{0.1cm}}&\\
&&&&&&\parbox[t][0.3cm]{11.370601cm}{\raggedright \ensuremath{\Gamma}: From (\href{https://www.nndc.bnl.gov/nsr/nsrlink.jsp?1998Ut02,B}{1998Ut02}). Other value: \ensuremath{\Gamma}=48 keV (\href{https://www.nndc.bnl.gov/nsr/nsrlink.jsp?2025PhZZ,B}{2025PhZZ}).\vspace{0.1cm}}&\\
&&&&&&\parbox[t][0.3cm]{11.370601cm}{\raggedright \ensuremath{\Gamma}\ensuremath{_{\ensuremath{\alpha}}}/\ensuremath{\Gamma}: See also 0.41 \textit{4} (\href{https://www.nndc.bnl.gov/nsr/nsrlink.jsp?2025PhZZ,B}{2025PhZZ}).\vspace{0.1cm}}&\\
&&&&&&\parbox[t][0.3cm]{11.370601cm}{\raggedright \ensuremath{\Gamma}\ensuremath{_{\textnormal{p}}}/\ensuremath{\Gamma}: See also 0.45 \textit{2} (\href{https://www.nndc.bnl.gov/nsr/nsrlink.jsp?2025PhZZ,B}{2025PhZZ}).\vspace{0.1cm}}&\\
&&&&&&\parbox[t][0.3cm]{11.370601cm}{\raggedright J\ensuremath{^{\pi}}: The data of (\href{https://www.nndc.bnl.gov/nsr/nsrlink.jsp?2019Ka15,B}{2019Ka15}) for this state is not consistent with a pure \ensuremath{\Delta}L=0\vspace{0.1cm}}&\\
&&&&&&\parbox[t][0.3cm]{11.370601cm}{\raggedright {\ }{\ }{\ }transition. (\href{https://www.nndc.bnl.gov/nsr/nsrlink.jsp?2019Ka15,B}{2019Ka15}) assumed this state to have J=3/2. See also\vspace{0.1cm}}&\\
&&&&&&\parbox[t][0.3cm]{11.370601cm}{\raggedright {\ }{\ }{\ }J\ensuremath{^{\ensuremath{\pi}}}=1/2\ensuremath{^{\textnormal{+}}},3/2,5/2\ensuremath{^{-}} with L\ensuremath{_{\textnormal{rel}}}(\ensuremath{\alpha}+\ensuremath{^{\textnormal{15}}}O)=1,2 and J\ensuremath{^{\ensuremath{\pi}}}=1/2, 3/2, 5/2, 7/2\ensuremath{^{-}} with\vspace{0.1cm}}&\\
&&&&&&\parbox[t][0.3cm]{11.370601cm}{\raggedright {\ }{\ }{\ }L\ensuremath{_{\textnormal{rel}}}(p+\ensuremath{^{\textnormal{18}}}F)=0,1,2. However, (\href{https://www.nndc.bnl.gov/nsr/nsrlink.jsp?2025PhZZ,B}{2025PhZZ}) favored L\ensuremath{_{\ensuremath{\alpha}}}=1 and L\ensuremath{_{\textnormal{p}}}=0 leading to\vspace{0.1cm}}&\\
&&&&&&\parbox[t][0.3cm]{11.370601cm}{\raggedright {\ }{\ }{\ }J\ensuremath{^{\ensuremath{\pi}}}=1/2\ensuremath{^{\textnormal{+}}} and 3/2\ensuremath{^{\textnormal{+}}}.\vspace{0.1cm}}&\\
\multicolumn{1}{r@{}}{7700}&\multicolumn{1}{@{}l}{}&&&&&\parbox[t][0.3cm]{11.370601cm}{\raggedright E(level): From (\href{https://www.nndc.bnl.gov/nsr/nsrlink.jsp?2021Ri04,B}{2021Ri04}: See Fig. 3).\vspace{0.1cm}}&\\
\multicolumn{1}{r@{}}{7757}&\multicolumn{1}{@{}l}{}&&&&&\parbox[t][0.3cm]{11.370601cm}{\raggedright E(level): From (\href{https://www.nndc.bnl.gov/nsr/nsrlink.jsp?2021Ri04,B}{2021Ri04}: See Fig. 3).\vspace{0.1cm}}&\\
\multicolumn{1}{r@{}}{7816}&\multicolumn{1}{@{ }l}{{\it 11}}&\multicolumn{1}{l}{(1/2\ensuremath{^{+}},3/2\ensuremath{^{+}})}&\multicolumn{1}{r@{}}{22}&\multicolumn{1}{@{ }l}{keV {\it 16}}&\multicolumn{1}{l}{0}&\parbox[t][0.3cm]{11.370601cm}{\raggedright \ensuremath{\Gamma}\ensuremath{\alpha}/\ensuremath{\Gamma}=0.19 \textit{9} (\href{https://www.nndc.bnl.gov/nsr/nsrlink.jsp?1998Ut02,B}{1998Ut02})\vspace{0.1cm}}&\\
&&&&&&\parbox[t][0.3cm]{11.370601cm}{\raggedright \ensuremath{\Gamma}\ensuremath{_{\textnormal{p}}}/\ensuremath{\Gamma}=0.81 \textit{11} (\href{https://www.nndc.bnl.gov/nsr/nsrlink.jsp?1998Ut02,B}{1998Ut02})\vspace{0.1cm}}&\\
&&&&&&\parbox[t][0.3cm]{11.370601cm}{\raggedright E(level): Weighted average of 7.79 MeV \textit{3} (\href{https://www.nndc.bnl.gov/nsr/nsrlink.jsp?2019Ka15,B}{2019Ka15}) and 7819 keV \textit{11}\vspace{0.1cm}}&\\
&&&&&&\parbox[t][0.3cm]{11.370601cm}{\raggedright {\ }{\ }{\ }(\href{https://www.nndc.bnl.gov/nsr/nsrlink.jsp?1998Ut02,B}{1998Ut02}).\vspace{0.1cm}}&\\
&&&&&&\parbox[t][0.3cm]{11.370601cm}{\raggedright E(level): See also (1) a tentative state at 7819 keV from (\href{https://www.nndc.bnl.gov/nsr/nsrlink.jsp?2021Ri04,B}{2021Ri04}), which was\vspace{0.1cm}}&\\
&&&&&&\parbox[t][0.3cm]{11.370601cm}{\raggedright {\ }{\ }{\ }considered tentative since part of this state was outside the detection region, and\vspace{0.1cm}}&\\
&&&&&&\parbox[t][0.3cm]{11.370601cm}{\raggedright {\ }{\ }{\ }therefore data were incomplete; and (2) 7795 keV \textit{3} (\href{https://www.nndc.bnl.gov/nsr/nsrlink.jsp?2025PhZZ,B}{2025PhZZ}: Unpublished).\vspace{0.1cm}}&\\
&&&&&&\parbox[t][0.3cm]{11.370601cm}{\raggedright J\ensuremath{^{\pi}}: From (\href{https://www.nndc.bnl.gov/nsr/nsrlink.jsp?2019Ka15,B}{2019Ka15}: \ensuremath{^{\textnormal{19}}}F(\ensuremath{^{\textnormal{3}}}He,t)) with \ensuremath{\Delta}L=0: Consistent with a pure \ensuremath{\Delta}L=0\vspace{0.1cm}}&\\
&&&&&&\parbox[t][0.3cm]{11.370601cm}{\raggedright {\ }{\ }{\ }Gamow-Teller transition.\vspace{0.1cm}}&\\
&&&&&&\parbox[t][0.3cm]{11.370601cm}{\raggedright J\ensuremath{^{\pi}}: See also J\ensuremath{^{\ensuremath{\pi}}}=1/2, 3/2, 5/2, 7/2, 9/2\ensuremath{^{-}} with L\ensuremath{_{\textnormal{rel}}}(p+\ensuremath{^{\textnormal{18}}}F)=2,3 and J\ensuremath{^{\ensuremath{\pi}}}=1/2\ensuremath{^{\textnormal{+}}}, 3/2\ensuremath{^{\textnormal{+}}}\vspace{0.1cm}}&\\
&&&&&&\parbox[t][0.3cm]{11.370601cm}{\raggedright {\ }{\ }{\ }with L\ensuremath{_{\textnormal{rel}}}(\ensuremath{\alpha}+\ensuremath{^{\textnormal{15}}}O)=1 from the proton and \ensuremath{\alpha} angular correlations in (\href{https://www.nndc.bnl.gov/nsr/nsrlink.jsp?2025PhZZ,B}{2025PhZZ}:\vspace{0.1cm}}&\\
&&&&&&\parbox[t][0.3cm]{11.370601cm}{\raggedright {\ }{\ }{\ }Unpublished).\vspace{0.1cm}}&\\
&&&&&&\parbox[t][0.3cm]{11.370601cm}{\raggedright See also \ensuremath{\Gamma}\ensuremath{_{\ensuremath{\alpha}}}/\ensuremath{\Gamma}=0.40 \textit{2} and \ensuremath{\Gamma}\ensuremath{_{\textnormal{p}}}/\ensuremath{\Gamma}=0.66 \textit{6} (\href{https://www.nndc.bnl.gov/nsr/nsrlink.jsp?2025PhZZ,B}{2025PhZZ}: Unpublished).\vspace{0.1cm}}&\\
&&&&&&\parbox[t][0.3cm]{11.370601cm}{\raggedright \ensuremath{\Gamma}: From (\href{https://www.nndc.bnl.gov/nsr/nsrlink.jsp?1998Ut02,B}{1998Ut02}).\vspace{0.1cm}}&\\
&&&&&&\parbox[t][0.3cm]{11.370601cm}{\raggedright \ensuremath{\Gamma}: See also \ensuremath{\Gamma}=130 keV \textit{10} (\href{https://www.nndc.bnl.gov/nsr/nsrlink.jsp?2019Ka15,B}{2019Ka15}) and \ensuremath{\Gamma}=36 keV (FWHM) (\href{https://www.nndc.bnl.gov/nsr/nsrlink.jsp?2025PhZZ,B}{2025PhZZ}:\vspace{0.1cm}}&\\
&&&&&&\parbox[t][0.3cm]{11.370601cm}{\raggedright {\ }{\ }{\ }Unpublished). (\href{https://www.nndc.bnl.gov/nsr/nsrlink.jsp?1998Ut02,B}{1998Ut02}) performed a high statistics t-p and t-\ensuremath{\alpha} coincidence run\vspace{0.1cm}}&\\
&&&&&&\parbox[t][0.3cm]{11.370601cm}{\raggedright {\ }{\ }{\ }using the Princeton QDDD spectrograph placed at \ensuremath{\theta}\ensuremath{_{\textnormal{lab}}}=0\ensuremath{^\circ} to measure tritons and\vspace{0.1cm}}&\\
&&&&&&\parbox[t][0.3cm]{11.370601cm}{\raggedright {\ }{\ }{\ }a Si surface barrier detector at \ensuremath{\theta}\ensuremath{_{\textnormal{lab}}}=145\ensuremath{^\circ} to measure protons and \ensuremath{\alpha} particles\vspace{0.1cm}}&\\
&&&&&&\parbox[t][0.3cm]{11.370601cm}{\raggedright {\ }{\ }{\ }from the decay of this state (see Fig. 2 of that study). Because of this dedicated\vspace{0.1cm}}&\\
&&&&&&\parbox[t][0.3cm]{11.370601cm}{\raggedright {\ }{\ }{\ }run (see Sinan Utku, Ph.D. Thesis, Yale University, 1994) and in light of the\vspace{0.1cm}}&\\
&&&&&&\parbox[t][0.3cm]{11.370601cm}{\raggedright {\ }{\ }{\ }unpublished results from (\href{https://www.nndc.bnl.gov/nsr/nsrlink.jsp?2025PhZZ,B}{2025PhZZ}), we accepted the branching ratios and the\vspace{0.1cm}}&\\
&&&&&&\parbox[t][0.3cm]{11.370601cm}{\raggedright {\ }{\ }{\ }width reported by (\href{https://www.nndc.bnl.gov/nsr/nsrlink.jsp?1998Ut02,B}{1998Ut02}). However, due to the inconsistencies between the\vspace{0.1cm}}&\\
&&&&&&\parbox[t][0.3cm]{11.370601cm}{\raggedright {\ }{\ }{\ }reported results, we cannot rule out the possibility of the existence of a\vspace{0.1cm}}&\\
&&&&&&\parbox[t][0.3cm]{11.370601cm}{\raggedright {\ }{\ }{\ }close-lying doublet in this region. Albeit, considering that the state was mainly\vspace{0.1cm}}&\\
&&&&&&\parbox[t][0.3cm]{11.370601cm}{\raggedright {\ }{\ }{\ }populated from the \ensuremath{^{\textnormal{19}}}F+\ensuremath{^{\textnormal{3}}}He interaction, this is unlikely. We also note that\vspace{0.1cm}}&\\
&&&&&&\parbox[t][0.3cm]{11.370601cm}{\raggedright {\ }{\ }{\ }(\href{https://www.nndc.bnl.gov/nsr/nsrlink.jsp?2019Ka15,B}{2019Ka15}) used the branching ratios measured by (\href{https://www.nndc.bnl.gov/nsr/nsrlink.jsp?1998Ut02,B}{1998Ut02}) to compute the\vspace{0.1cm}}&\\
&&&&&&\parbox[t][0.3cm]{11.370601cm}{\raggedright {\ }{\ }{\ }contribution of their 7.79-MeV state to the astrophysical \ensuremath{^{\textnormal{18}}}F(p,\ensuremath{\alpha}) reaction rate.\vspace{0.1cm}}&\\
&&&&&&\parbox[t][0.3cm]{11.370601cm}{\raggedright {\ }{\ }{\ }This favors the hypothesis that those authors considered that their 7.79-MeV,\vspace{0.1cm}}&\\
&&&&&&\parbox[t][0.3cm]{11.370601cm}{\raggedright {\ }{\ }{\ }\ensuremath{\Gamma}=130 keV state is the same as the 7819-keV, \ensuremath{\Gamma}=22 keV state measured by\vspace{0.1cm}}&\\
&&&&&&\parbox[t][0.3cm]{11.370601cm}{\raggedright {\ }{\ }{\ }(\href{https://www.nndc.bnl.gov/nsr/nsrlink.jsp?1998Ut02,B}{1998Ut02}).\vspace{0.1cm}}&\\
\end{longtable}
\begin{textblock}{29}(0,27.3)
Continued on next page (footnotes at end of table)
\end{textblock}
\clearpage
\vspace*{-0.5cm}
{\bf \small \underline{\ensuremath{^{\textnormal{19}}}F(\ensuremath{^{\textnormal{3}}}He,t)\hspace{0.2in}\href{https://www.nndc.bnl.gov/nsr/nsrlink.jsp?1998Ut02,B}{1998Ut02},\href{https://www.nndc.bnl.gov/nsr/nsrlink.jsp?2013La01,B}{2013La01},\href{https://www.nndc.bnl.gov/nsr/nsrlink.jsp?2019Ka15,B}{2019Ka15} (continued)}}\\
\vspace{0.3cm}
\underline{$^{19}$Ne Levels (continued)}\\
\vspace{0.3cm}
\parbox[b][0.3cm]{17.7cm}{\makebox[1ex]{\ensuremath{^{\hypertarget{NE32LEVEL0}{a}}}} The 2-keV systematic uncertainty recommended by (\href{https://www.nndc.bnl.gov/nsr/nsrlink.jsp?2013La01,B}{2013La01}) is added in quadrature to the statistical uncertainty in the}\\
\parbox[b][0.3cm]{17.7cm}{{\ }{\ }weighted average.}\\
\parbox[b][0.3cm]{17.7cm}{\makebox[1ex]{\ensuremath{^{\hypertarget{NE32LEVEL1}{b}}}} The triton angular distributions measured by (\href{https://www.nndc.bnl.gov/nsr/nsrlink.jsp?2019Ka15,B}{2019Ka15}) are inconsistent with a pure \ensuremath{\Delta}L=0 Gamow-Teller transition since the}\\
\parbox[b][0.3cm]{17.7cm}{{\ }{\ }cross section increases at backward angle; however, the strength of this state at backward angles does not rise as much as that of}\\
\parbox[b][0.3cm]{17.7cm}{{\ }{\ }a known J\ensuremath{^{\ensuremath{\pi}}}=3/2\ensuremath{^{-}} state. The data of (\href{https://www.nndc.bnl.gov/nsr/nsrlink.jsp?2019Ka15,B}{2019Ka15}) are consistent with unobserved states whose strengths are concentrated on the}\\
\parbox[b][0.3cm]{17.7cm}{{\ }{\ }more backward angles at excitation energies above the 6421 keV peak. This would be consistent with \ensuremath{\Delta}L\ensuremath{\geq}1 for the two}\\
\parbox[b][0.3cm]{17.7cm}{{\ }{\ }higher-lying states at 6438 keV and 6457 keV states. (\href{https://www.nndc.bnl.gov/nsr/nsrlink.jsp?2019Ka15,B}{2019Ka15}) also mentioned that this peak could be a candidate for a}\\
\parbox[b][0.3cm]{17.7cm}{{\ }{\ }J\ensuremath{^{\ensuremath{\pi}}}=3/2\ensuremath{^{\textnormal{+}}} state based on mirror levels analysis, but it would require an additional member since it is not fitted with pure \ensuremath{\Delta}L=0.}\\
\parbox[b][0.3cm]{17.7cm}{\makebox[1ex]{\ensuremath{^{\hypertarget{NE32LEVEL2}{c}}}} (\href{https://www.nndc.bnl.gov/nsr/nsrlink.jsp?2009Ta09,B}{2009Ta09}) suggested that J\ensuremath{^{\ensuremath{\pi}}} values for the 4143-keV and 4200-keV levels should be reversed from those accepted in}\\
\parbox[b][0.3cm]{17.7cm}{{\ }{\ }(\href{https://www.nndc.bnl.gov/nsr/nsrlink.jsp?1995Ti07,B}{1995Ti07}), which listed the order as \ensuremath{^{\textnormal{19}}}Ne*(4143 keV, (9/2\ensuremath{^{-}})) and \ensuremath{^{\textnormal{19}}}Ne*(4200 keV, (7/2\ensuremath{^{-}})) based on the \ensuremath{^{\textnormal{16}}}O(\ensuremath{^{\textnormal{6}}}Li,t) and}\\
\parbox[b][0.3cm]{17.7cm}{{\ }{\ }\ensuremath{^{\textnormal{17}}}O(\ensuremath{^{\textnormal{3}}}He,n\ensuremath{\gamma}) reactions. (\href{https://www.nndc.bnl.gov/nsr/nsrlink.jsp?2009Ta09,B}{2009Ta09}) calculated single-particle \ensuremath{\alpha}-widths for these two states. Based on the deduced spectroscopic}\\
\parbox[b][0.3cm]{17.7cm}{{\ }{\ }factors, as well as a comparison of the decay schemes of these two states with those of \ensuremath{^{\textnormal{19}}}F* states in this vicinity, they favored a}\\
\parbox[b][0.3cm]{17.7cm}{{\ }{\ }reversal of the J values for these two states.}\\
\vspace{0.5cm}
\clearpage
\subsection[\hspace{-0.2cm}\ensuremath{^{\textnormal{19}}}F(\ensuremath{^{\textnormal{3}}}He,t\ensuremath{\gamma})]{ }
\vspace{-27pt}
\vspace{0.3cm}
\hypertarget{NE33}{{\bf \small \underline{\ensuremath{^{\textnormal{19}}}F(\ensuremath{^{\textnormal{3}}}He,t\ensuremath{\gamma})\hspace{0.2in}\href{https://www.nndc.bnl.gov/nsr/nsrlink.jsp?2019Ha08,B}{2019Ha08},\href{https://www.nndc.bnl.gov/nsr/nsrlink.jsp?2020Ha31,B}{2020Ha31}}}}\\
\vspace{4pt}
\vspace{8pt}
\parbox[b][0.3cm]{17.7cm}{\addtolength{\parindent}{-0.2in}Charge exchange reaction.}\\
\parbox[b][0.3cm]{17.7cm}{\addtolength{\parindent}{-0.2in}J\ensuremath{^{\ensuremath{\pi}}}(\ensuremath{^{\textnormal{19}}}F\ensuremath{_{\textnormal{g.s.}}})=1/2\ensuremath{^{\textnormal{+}}} and J\ensuremath{^{\ensuremath{\pi}}}(\ensuremath{^{\textnormal{3}}}He\ensuremath{_{\textnormal{g.s.}}})=1/2\ensuremath{^{\textnormal{+}}}.}\\
\parbox[b][0.3cm]{17.7cm}{\addtolength{\parindent}{-0.2in}\href{https://www.nndc.bnl.gov/nsr/nsrlink.jsp?2019Ha08,B}{2019Ha08}, \href{https://www.nndc.bnl.gov/nsr/nsrlink.jsp?2019Ha14,B}{2019Ha14}, \href{https://www.nndc.bnl.gov/nsr/nsrlink.jsp?2020Ha31,B}{2020Ha31}: \ensuremath{^{\textnormal{19}}}F(\ensuremath{^{\textnormal{3}}}He,t\ensuremath{\gamma}) E=30 MeV; measured tritons-\ensuremath{\gamma}-\ensuremath{\gamma} coincidences using the GODDESS array that}\\
\parbox[b][0.3cm]{17.7cm}{consisted of six \ensuremath{\Delta}E-E position sensitive telescopes of ORRUBA covering \ensuremath{\theta}\ensuremath{_{\textnormal{lab}}}\ensuremath{\approx}18\ensuremath{^\circ}{\textminus}162\ensuremath{^\circ} to measure the tritons, and the}\\
\parbox[b][0.3cm]{17.7cm}{Gammasphere to measure the \ensuremath{\gamma} rays and \ensuremath{\gamma}-\ensuremath{\gamma} coincidence events from the \ensuremath{^{\textnormal{19}}}Ne* de-exciting states in the E\ensuremath{_{\textnormal{x}}}=0-6.9 MeV region.}\\
\parbox[b][0.3cm]{17.7cm}{Measured E\ensuremath{_{\ensuremath{\gamma}}} and I\ensuremath{_{\ensuremath{\gamma}}} for the observed transitions; reconstructed the \ensuremath{^{\textnormal{19}}}Ne levels from triple coincidence events. Energy resolution}\\
\parbox[b][0.3cm]{17.7cm}{was 15 keV (FWHM).}\\
\parbox[b][0.3cm]{17.7cm}{\addtolength{\parindent}{-0.2in}(\href{https://www.nndc.bnl.gov/nsr/nsrlink.jsp?2019Ha08,B}{2019Ha08}) discussed a sub-threshold, J\ensuremath{^{\ensuremath{\pi}}}=(11/2)\ensuremath{^{\textnormal{+}}} state at 6291.6 keV \textit{9}, as well as a close lying doublet at 6423 and 6441 keV}\\
\parbox[b][0.3cm]{17.7cm}{proposed to be J\ensuremath{^{\ensuremath{\pi}}}=3/2\ensuremath{^{\textnormal{+}}}. These states are near the proton threshold and could play a significant role in the \ensuremath{^{\textnormal{18}}}F(p,\ensuremath{\alpha}) reaction rate at}\\
\parbox[b][0.3cm]{17.7cm}{nova temperatures. The authors deduced the astrophysical S-factor at E\ensuremath{_{\textnormal{c.m.}}}\ensuremath{<}1 MeV using R-matrix via AZURE2 code and used it}\\
\parbox[b][0.3cm]{17.7cm}{to calculate the \ensuremath{^{\textnormal{18}}}F(p,\ensuremath{\alpha}) reaction rate at T=0.05-0.4 GK; performed nucleosynthesis calculations.}\\
\parbox[b][0.3cm]{17.7cm}{\addtolength{\parindent}{-0.2in}(\href{https://www.nndc.bnl.gov/nsr/nsrlink.jsp?2019Ha14,B}{2019Ha14}) discussed the E\ensuremath{_{\textnormal{x}}}=3.8-4.4 MeV region and the \ensuremath{\gamma} decay of the \ensuremath{^{\textnormal{19}}}Ne*(4.14, 4.20, 4.378, 4.602 MeV) states. Based on}\\
\parbox[b][0.3cm]{17.7cm}{the low spin-parity of the \ensuremath{^{\textnormal{19}}}Ne*(1616 keV, 3/2\ensuremath{^{-}}) state to which the 4.14-MeV state de-excites, the multipolarity of the latter}\\
\parbox[b][0.3cm]{17.7cm}{transition, and because in this vicinity in \ensuremath{^{\textnormal{19}}}F there are only one state with J\ensuremath{^{\ensuremath{\pi}}}=7/2\ensuremath{^{-}} and another with J\ensuremath{^{\ensuremath{\pi}}}=9/2\ensuremath{^{-}}, the authors suggest}\\
\parbox[b][0.3cm]{17.7cm}{that the previously accepted J\ensuremath{^{\ensuremath{\pi}}} values of the \ensuremath{^{\textnormal{19}}}Ne*(4.14, 4.20 MeV) states, recommended by (\href{https://www.nndc.bnl.gov/nsr/nsrlink.jsp?1995Ti07,B}{1995Ti07}), should be reversed to}\\
\parbox[b][0.3cm]{17.7cm}{7/2\ensuremath{^{-}} and 9/2\ensuremath{^{-}}, respectively.}\\
\parbox[b][0.3cm]{17.7cm}{\addtolength{\parindent}{-0.2in}(\href{https://www.nndc.bnl.gov/nsr/nsrlink.jsp?2020Ha31,B}{2020Ha31}) further expands on the analysis details and results and presents the entire observed \ensuremath{^{\textnormal{19}}}Ne decay scheme.}\\
\vspace{12pt}
\underline{$^{19}$Ne Levels}\\
\begin{longtable}{cccccc@{\extracolsep{\fill}}c}
\multicolumn{2}{c}{E(level)$^{{\hyperlink{NE33LEVEL0}{a}}}$}&J$^{\pi}$$^{{\hyperlink{NE33LEVEL1}{b}}}$&\multicolumn{2}{c}{E\ensuremath{_{\textnormal{x}}}(\ensuremath{^{\textnormal{19}}}F) Mirror (keV)$^{{\hyperlink{NE33LEVEL3}{d}}}$}&Comments&\\[-.2cm]
\multicolumn{2}{c}{\hrulefill}&\hrulefill&\multicolumn{2}{c}{\hrulefill}&\hrulefill&
\endfirsthead
\multicolumn{1}{r@{}}{0}&\multicolumn{1}{@{}l}{}&\multicolumn{1}{l}{1/2\ensuremath{^{+}}}&&&&\\
\multicolumn{1}{r@{}}{238}&\multicolumn{1}{@{.}l}{64 {\it 25}}&\multicolumn{1}{l}{5/2\ensuremath{^{+}}}&\multicolumn{1}{r@{}}{197}&\multicolumn{1}{@{}l}{}&&\\
\multicolumn{1}{r@{}}{275}&\multicolumn{1}{@{.}l}{45 {\it 25}}&\multicolumn{1}{l}{1/2\ensuremath{^{-}}}&\multicolumn{1}{r@{}}{109}&\multicolumn{1}{@{}l}{}&&\\
\multicolumn{1}{r@{}}{1507}&\multicolumn{1}{@{.}l}{7 {\it 3}}&\multicolumn{1}{l}{5/2\ensuremath{^{-}}}&\multicolumn{1}{r@{}}{1345}&\multicolumn{1}{@{}l}{}&&\\
\multicolumn{1}{r@{}}{1536}&\multicolumn{1}{@{.}l}{0 {\it 4}}&\multicolumn{1}{l}{3/2\ensuremath{^{+}}}&\multicolumn{1}{r@{}}{1554}&\multicolumn{1}{@{}l}{}&&\\
\multicolumn{1}{r@{}}{1615}&\multicolumn{1}{@{.}l}{6 {\it 5}}&\multicolumn{1}{l}{3/2\ensuremath{^{-}}}&\multicolumn{1}{r@{}}{1458}&\multicolumn{1}{@{}l}{}&&\\
\multicolumn{1}{r@{}}{2794}&\multicolumn{1}{@{.}l}{8 {\it 7}}&\multicolumn{1}{l}{9/2\ensuremath{^{+}}}&\multicolumn{1}{r@{}}{2779}&\multicolumn{1}{@{}l}{}&&\\
\multicolumn{1}{r@{}}{4034}&\multicolumn{1}{@{.}l}{7 {\it 9}}&\multicolumn{1}{l}{3/2\ensuremath{^{+}}}&\multicolumn{1}{r@{}}{3908}&\multicolumn{1}{@{}l}{}&&\\
\multicolumn{1}{r@{}}{4142}&\multicolumn{1}{@{.}l}{9 {\it 7}}&\multicolumn{1}{l}{(7/2\ensuremath{^{-}})\ensuremath{^{{\hyperlink{NE33LEVEL2}{c}}}}}&\multicolumn{1}{r@{}}{3998}&\multicolumn{1}{@{}l}{}&\parbox[t][0.3cm]{10.8311405cm}{\raggedright J\ensuremath{^{\pi}}: See also J\ensuremath{^{\ensuremath{\pi}}}=(9/2\ensuremath{^{-}}) (\href{https://www.nndc.bnl.gov/nsr/nsrlink.jsp?1995Ti07,B}{1995Ti07}).\vspace{0.1cm}}&\\
&&&&&\parbox[t][0.3cm]{10.8311405cm}{\raggedright The decay scheme presented by (\href{https://www.nndc.bnl.gov/nsr/nsrlink.jsp?2019Ha14,B}{2019Ha14}, \href{https://www.nndc.bnl.gov/nsr/nsrlink.jsp?2020Ha31,B}{2020Ha31}) indicates J\ensuremath{^{\ensuremath{\pi}}}=7/2\ensuremath{^{-}}\vspace{0.1cm}}&\\
&&&&&\parbox[t][0.3cm]{10.8311405cm}{\raggedright {\ }{\ }{\ }making this state the mirror of the \ensuremath{^{\textnormal{19}}}F*(3998) level.\vspace{0.1cm}}&\\
\multicolumn{1}{r@{}}{4200}&\multicolumn{1}{@{.}l}{1 {\it 11}}&\multicolumn{1}{l}{(9/2\ensuremath{^{-}})\ensuremath{^{{\hyperlink{NE33LEVEL2}{c}}}}}&\multicolumn{1}{r@{}}{4032}&\multicolumn{1}{@{}l}{}&\parbox[t][0.3cm]{10.8311405cm}{\raggedright J\ensuremath{^{\pi}}: See also J\ensuremath{^{\ensuremath{\pi}}}=(7/2\ensuremath{^{-}}) (\href{https://www.nndc.bnl.gov/nsr/nsrlink.jsp?1995Ti07,B}{1995Ti07}).\vspace{0.1cm}}&\\
&&&&&\parbox[t][0.3cm]{10.8311405cm}{\raggedright The decay scheme presented by (\href{https://www.nndc.bnl.gov/nsr/nsrlink.jsp?2019Ha14,B}{2019Ha14}, \href{https://www.nndc.bnl.gov/nsr/nsrlink.jsp?2020Ha31,B}{2020Ha31}) indicates J\ensuremath{^{\ensuremath{\pi}}}=9/2\ensuremath{^{-}}\vspace{0.1cm}}&\\
&&&&&\parbox[t][0.3cm]{10.8311405cm}{\raggedright {\ }{\ }{\ }making this state the mirror of the \ensuremath{^{\textnormal{19}}}F*(4032) level.\vspace{0.1cm}}&\\
&&&&&\parbox[t][0.3cm]{10.8311405cm}{\raggedright The weak \ensuremath{\gamma} decay transition to the \ensuremath{^{\textnormal{19}}}Ne*(238) level observed by (\href{https://www.nndc.bnl.gov/nsr/nsrlink.jsp?1973Da31,B}{1973Da31}:\vspace{0.1cm}}&\\
&&&&&\parbox[t][0.3cm]{10.8311405cm}{\raggedright {\ }{\ }{\ }\ensuremath{^{\textnormal{17}}}O(\ensuremath{^{\textnormal{3}}}He,n\ensuremath{\gamma})) from this state is not observed by (\href{https://www.nndc.bnl.gov/nsr/nsrlink.jsp?2019Ha14,B}{2019Ha14}, \href{https://www.nndc.bnl.gov/nsr/nsrlink.jsp?2020Ha31,B}{2020Ha31}).\vspace{0.1cm}}&\\
&&&&&\parbox[t][0.3cm]{10.8311405cm}{\raggedright {\ }{\ }{\ }(\href{https://www.nndc.bnl.gov/nsr/nsrlink.jsp?2019Ha14,B}{2019Ha14}) reports that there is no evidence for the 3961-keV \ensuremath{\gamma}-transition\vspace{0.1cm}}&\\
&&&&&\parbox[t][0.3cm]{10.8311405cm}{\raggedright {\ }{\ }{\ }de-exciting this state, and that it should have appeared in their triton-gated\vspace{0.1cm}}&\\
&&&&&\parbox[t][0.3cm]{10.8311405cm}{\raggedright {\ }{\ }{\ }\ensuremath{\gamma}-ray spectrum if it were to exist. As stated before, this transition was\vspace{0.1cm}}&\\
&&&&&\parbox[t][0.3cm]{10.8311405cm}{\raggedright {\ }{\ }{\ }reported by (\href{https://www.nndc.bnl.gov/nsr/nsrlink.jsp?1973Da31,B}{1973Da31}) as a weak transition in their spectra when only\vspace{0.1cm}}&\\
&&&&&\parbox[t][0.3cm]{10.8311405cm}{\raggedright {\ }{\ }{\ }gating on neutrons with no excitation energy gate and no \ensuremath{\gamma}-\ensuremath{\gamma} coincidences.\vspace{0.1cm}}&\\
&&&&&\parbox[t][0.3cm]{10.8311405cm}{\raggedright {\ }{\ }{\ }Therefore, (\href{https://www.nndc.bnl.gov/nsr/nsrlink.jsp?2019Ha14,B}{2019Ha14}) reported that it is likely that the previously observed\vspace{0.1cm}}&\\
&&&&&\parbox[t][0.3cm]{10.8311405cm}{\raggedright {\ }{\ }{\ }weak transition at 3961 keV was incorrectly placed by (\href{https://www.nndc.bnl.gov/nsr/nsrlink.jsp?1973Da31,B}{1973Da31}) as\vspace{0.1cm}}&\\
&&&&&\parbox[t][0.3cm]{10.8311405cm}{\raggedright {\ }{\ }{\ }depopulating the 4199.8-keV state.\vspace{0.1cm}}&\\
\multicolumn{1}{r@{}}{4377}&\multicolumn{1}{@{.}l}{6 {\it 10}}&\multicolumn{1}{l}{7/2\ensuremath{^{+}}}&\multicolumn{1}{r@{}}{4377}&\multicolumn{1}{@{}l}{}&&\\
\multicolumn{1}{r@{}}{4547}&\multicolumn{1}{@{.}l}{6 {\it 10}}&\multicolumn{1}{l}{3/2\ensuremath{^{-}}}&\multicolumn{1}{r@{}}{4556}&\multicolumn{1}{@{}l}{}&\parbox[t][0.3cm]{10.8311405cm}{\raggedright Two weak \ensuremath{\gamma} decay transitions from this state to the 1507- and 1536-keV states\vspace{0.1cm}}&\\
&&&&&\parbox[t][0.3cm]{10.8311405cm}{\raggedright {\ }{\ }{\ }were reported by (\href{https://www.nndc.bnl.gov/nsr/nsrlink.jsp?2020Ha31,B}{2020Ha31}), which should have been observed in\vspace{0.1cm}}&\\
&&&&&\parbox[t][0.3cm]{10.8311405cm}{\raggedright {\ }{\ }{\ }(\href{https://www.nndc.bnl.gov/nsr/nsrlink.jsp?1973Da31,B}{1973Da31}: \ensuremath{^{\textnormal{17}}}O(\ensuremath{^{\textnormal{3}}}He,n\ensuremath{\gamma})) but were not.\vspace{0.1cm}}&\\
\multicolumn{1}{r@{}}{4603}&\multicolumn{1}{@{.}l}{2 {\it 9}}&\multicolumn{1}{l}{5/2\ensuremath{^{+}}}&\multicolumn{1}{r@{}}{4549}&\multicolumn{1}{@{}l}{}&&\\
\multicolumn{1}{r@{}}{4634}&\multicolumn{1}{@{.}l}{6 {\it 8}}&\multicolumn{1}{l}{13/2\ensuremath{^{+}}}&\multicolumn{1}{r@{}}{4648}&\multicolumn{1}{@{}l}{}&&\\
\multicolumn{1}{r@{}}{4708}&\multicolumn{1}{@{.}l}{8 {\it 17}}&\multicolumn{1}{l}{5/2\ensuremath{^{-}}}&\multicolumn{1}{r@{}}{4682}&\multicolumn{1}{@{}l}{}&&\\
\multicolumn{1}{r@{}}{5091}&\multicolumn{1}{@{ }l}{{\it 3}}&\multicolumn{1}{l}{5/2\ensuremath{^{+}}}&\multicolumn{1}{r@{}}{5107}&\multicolumn{1}{@{}l}{}&&\\
\multicolumn{1}{r@{}}{6099}&\multicolumn{1}{@{.}l}{6 {\it 17}}&\multicolumn{1}{l}{(7/2\ensuremath{^{+}})}&\multicolumn{1}{r@{}}{6070}&\multicolumn{1}{@{}l}{}&&\\
\multicolumn{1}{r@{}}{6292}&\multicolumn{1}{@{.}l}{5 {\it 9}}&\multicolumn{1}{l}{(11/2\ensuremath{^{+}})}&\multicolumn{1}{r@{}}{6500}&\multicolumn{1}{@{}l}{}&\parbox[t][0.3cm]{10.8311405cm}{\raggedright J\ensuremath{^{\pi}}: See also (\href{https://www.nndc.bnl.gov/nsr/nsrlink.jsp?2019Ha08,B}{2019Ha08}).\vspace{0.1cm}}&\\
\end{longtable}
\begin{textblock}{29}(0,27.3)
Continued on next page (footnotes at end of table)
\end{textblock}
\clearpage
\begin{longtable}{cccccc@{\extracolsep{\fill}}c}
\\[-.4cm]
\multicolumn{7}{c}{{\bf \small \underline{\ensuremath{^{\textnormal{19}}}F(\ensuremath{^{\textnormal{3}}}He,t\ensuremath{\gamma})\hspace{0.2in}\href{https://www.nndc.bnl.gov/nsr/nsrlink.jsp?2019Ha08,B}{2019Ha08},\href{https://www.nndc.bnl.gov/nsr/nsrlink.jsp?2020Ha31,B}{2020Ha31} (continued)}}}\\
\multicolumn{7}{c}{~}\\
\multicolumn{7}{c}{\underline{\ensuremath{^{19}}Ne Levels (continued)}}\\
\multicolumn{7}{c}{~}\\
\multicolumn{2}{c}{E(level)$^{{\hyperlink{NE33LEVEL0}{a}}}$}&J$^{\pi}$$^{{\hyperlink{NE33LEVEL1}{b}}}$&\multicolumn{2}{c}{E\ensuremath{_{\textnormal{x}}}(\ensuremath{^{\textnormal{19}}}F) Mirror (keV)$^{{\hyperlink{NE33LEVEL3}{d}}}$}&Comments&\\[-.2cm]
\multicolumn{2}{c}{\hrulefill}&\hrulefill&\multicolumn{2}{c}{\hrulefill}&\hrulefill&
\endhead
&&&&&\parbox[t][0.3cm]{11.22062cm}{\raggedright We highlight a discrepancy in E\ensuremath{_{\textnormal{x}}} from (\href{https://www.nndc.bnl.gov/nsr/nsrlink.jsp?2019Ha08,B}{2019Ha08}) and (\href{https://www.nndc.bnl.gov/nsr/nsrlink.jsp?2020Ha31,B}{2020Ha31}): Fig. 1 in\vspace{0.1cm}}&\\
&&&&&\parbox[t][0.3cm]{11.22062cm}{\raggedright {\ }{\ }{\ }(\href{https://www.nndc.bnl.gov/nsr/nsrlink.jsp?2019Ha08,B}{2019Ha08}) shows the energy of this state as 6291.6 keV \textit{9}, while E\ensuremath{_{\textnormal{x}}}=6291.7\vspace{0.1cm}}&\\
&&&&&\parbox[t][0.3cm]{11.22062cm}{\raggedright {\ }{\ }{\ }keV \textit{9} is mentioned in the text. (\href{https://www.nndc.bnl.gov/nsr/nsrlink.jsp?2020Ha31,B}{2020Ha31}) reports E\ensuremath{_{\textnormal{x}}}=6291.6 keV \textit{9}.\vspace{0.1cm}}&\\
\multicolumn{1}{r@{}}{6424}&\multicolumn{1}{@{ }l}{{\it 3}}&\multicolumn{1}{l}{(3/2\ensuremath{^{+}})}&\multicolumn{1}{r@{}}{6527}&\multicolumn{1}{@{}l}{}&\parbox[t][0.3cm]{11.22062cm}{\raggedright J\ensuremath{^{\pi}}: See also (\href{https://www.nndc.bnl.gov/nsr/nsrlink.jsp?2019Ha08,B}{2019Ha08}).\vspace{0.1cm}}&\\
&&&&&\parbox[t][0.3cm]{11.22062cm}{\raggedright J\ensuremath{^{\pi}}: The J\ensuremath{^{\ensuremath{\pi}}}=7/2\ensuremath{^{\textnormal{+}}} assignment, based on multipolarity of the observed transitions, is\vspace{0.1cm}}&\\
&&&&&\parbox[t][0.3cm]{11.22062cm}{\raggedright {\ }{\ }{\ }ruled out in (\href{https://www.nndc.bnl.gov/nsr/nsrlink.jsp?2019Ha08,B}{2019Ha08}) because in that case the decay scheme of this \ensuremath{^{\textnormal{19}}}Ne\vspace{0.1cm}}&\\
&&&&&\parbox[t][0.3cm]{11.22062cm}{\raggedright {\ }{\ }{\ }state would be different from that of its proposed mirror level in \ensuremath{^{\textnormal{19}}}F.\vspace{0.1cm}}&\\
\multicolumn{1}{r@{}}{6442}&\multicolumn{1}{@{ }l}{{\it 3}}&\multicolumn{1}{l}{(3/2\ensuremath{^{+}})}&\multicolumn{1}{r@{}}{6497}&\multicolumn{1}{@{}l}{}&\parbox[t][0.3cm]{11.22062cm}{\raggedright J\ensuremath{^{\pi}}: See also (\href{https://www.nndc.bnl.gov/nsr/nsrlink.jsp?2019Ha08,B}{2019Ha08}).\vspace{0.1cm}}&\\
&&&&&\parbox[t][0.3cm]{11.22062cm}{\raggedright J\ensuremath{^{\pi}}: Because this state decays to the 238- and 1616-keV states, which have\vspace{0.1cm}}&\\
&&&&&\parbox[t][0.3cm]{11.22062cm}{\raggedright {\ }{\ }{\ }J\ensuremath{^{\ensuremath{\pi}}}=5/2\ensuremath{^{\textnormal{+}}} and 3/2\ensuremath{^{-}}, respectively, this state most likely has a J\ensuremath{\leq}7/2 (\href{https://www.nndc.bnl.gov/nsr/nsrlink.jsp?2020Ha31,B}{2020Ha31}).\vspace{0.1cm}}&\\
&&&&&\parbox[t][0.3cm]{11.22062cm}{\raggedright {\ }{\ }{\ }These authors already ruled out the J\ensuremath{^{\ensuremath{\pi}}}=11/2\ensuremath{^{\textnormal{+}}} assignment suggested by\vspace{0.1cm}}&\\
&&&&&\parbox[t][0.3cm]{11.22062cm}{\raggedright {\ }{\ }{\ }(\href{https://www.nndc.bnl.gov/nsr/nsrlink.jsp?2013La01,B}{2013La01}: \ensuremath{^{\textnormal{19}}}F(\ensuremath{^{\textnormal{3}}}He,t)). Therefore, by comparing the observed \ensuremath{\gamma}-ray decay\vspace{0.1cm}}&\\
&&&&&\parbox[t][0.3cm]{11.22062cm}{\raggedright {\ }{\ }{\ }scheme of this state with that in this energy region in \ensuremath{^{\textnormal{19}}}F, a J\ensuremath{^{\ensuremath{\pi}}}=3/2\ensuremath{^{\textnormal{+}}} was\vspace{0.1cm}}&\\
&&&&&\parbox[t][0.3cm]{11.22062cm}{\raggedright {\ }{\ }{\ }assigned to this state by (\href{https://www.nndc.bnl.gov/nsr/nsrlink.jsp?2020Ha31,B}{2020Ha31}).\vspace{0.1cm}}&\\
&&&&&\parbox[t][0.3cm]{11.22062cm}{\raggedright Note that the \ensuremath{\gamma}-ray transition from the \ensuremath{^{\textnormal{19}}}F*(6527 keV, 3/2\ensuremath{^{\textnormal{+}}}) level to the\vspace{0.1cm}}&\\
&&&&&\parbox[t][0.3cm]{11.22062cm}{\raggedright {\ }{\ }{\ }\ensuremath{^{\textnormal{19}}}F*(1458) is not observed. This is why (\href{https://www.nndc.bnl.gov/nsr/nsrlink.jsp?2020Ha31,B}{2020Ha31}) assigned the \ensuremath{^{\textnormal{19}}}F*(6497)\vspace{0.1cm}}&\\
&&&&&\parbox[t][0.3cm]{11.22062cm}{\raggedright {\ }{\ }{\ }as the mirror to the \ensuremath{^{\textnormal{19}}}Ne*(6441) state.\vspace{0.1cm}}&\\
\multicolumn{1}{r@{}}{6739}&\multicolumn{1}{@{ }l}{{\it 7}}&\multicolumn{1}{l}{3/2\ensuremath{^{-}}}&\multicolumn{1}{r@{}}{6787}&\multicolumn{1}{@{}l}{}&\parbox[t][0.3cm]{11.22062cm}{\raggedright J\ensuremath{^{\pi}}: This assignment was tentative in (\href{https://www.nndc.bnl.gov/nsr/nsrlink.jsp?2019Ha08,B}{2019Ha08}), which precedes that of\vspace{0.1cm}}&\\
&&&&&\parbox[t][0.3cm]{11.22062cm}{\raggedright {\ }{\ }{\ }(\href{https://www.nndc.bnl.gov/nsr/nsrlink.jsp?2020Ha31,B}{2020Ha31}).\vspace{0.1cm}}&\\
\multicolumn{1}{r@{}}{6853}&\multicolumn{1}{@{ }l}{{\it 4}}&\multicolumn{1}{l}{(7/2\ensuremath{^{-}})}&\multicolumn{1}{r@{}}{6927}&\multicolumn{1}{@{}l}{}&&\\
\end{longtable}
\parbox[b][0.3cm]{17.7cm}{\makebox[1ex]{\ensuremath{^{\hypertarget{NE33LEVEL0}{a}}}} From a least-squares fit to the \ensuremath{\gamma}-ray energies given in (\href{https://www.nndc.bnl.gov/nsr/nsrlink.jsp?2020Ha31,B}{2020Ha31}). See also the earlier work of (\href{https://www.nndc.bnl.gov/nsr/nsrlink.jsp?2019Ha08,B}{2019Ha08}, \href{https://www.nndc.bnl.gov/nsr/nsrlink.jsp?2019Ha14,B}{2019Ha14}), where}\\
\parbox[b][0.3cm]{17.7cm}{{\ }{\ }many of the excitation energies are deduced from an analysis, which is presented in (\href{https://www.nndc.bnl.gov/nsr/nsrlink.jsp?2020Ha31,B}{2020Ha31}). We did not take those energies}\\
\parbox[b][0.3cm]{17.7cm}{{\ }{\ }because nuclear recoil corrections were not applied in (\href{https://www.nndc.bnl.gov/nsr/nsrlink.jsp?2019Ha08,B}{2019Ha08}, \href{https://www.nndc.bnl.gov/nsr/nsrlink.jsp?2019Ha14,B}{2019Ha14}, \href{https://www.nndc.bnl.gov/nsr/nsrlink.jsp?2020Ha31,B}{2020Ha31}). So, a systematic error may be present}\\
\parbox[b][0.3cm]{17.7cm}{{\ }{\ }in their reported excited energies. Our least-squares fit to E\ensuremath{_{\ensuremath{\gamma}}} values correct for the nuclear recoils.}\\
\parbox[b][0.3cm]{17.7cm}{\makebox[1ex]{\ensuremath{^{\hypertarget{NE33LEVEL1}{b}}}} From (\href{https://www.nndc.bnl.gov/nsr/nsrlink.jsp?2020Ha31,B}{2020Ha31}) deduced from comparison of the observed \ensuremath{\gamma}-ray decay scheme of each of the \ensuremath{^{\textnormal{19}}}Ne* states with that of the}\\
\parbox[b][0.3cm]{17.7cm}{{\ }{\ }proposed mirror state in \ensuremath{^{\textnormal{19}}}F*. See also the earlier work of (\href{https://www.nndc.bnl.gov/nsr/nsrlink.jsp?2019Ha08,B}{2019Ha08}, \href{https://www.nndc.bnl.gov/nsr/nsrlink.jsp?2019Ha14,B}{2019Ha14}).}\\
\parbox[b][0.3cm]{17.7cm}{\makebox[1ex]{\ensuremath{^{\hypertarget{NE33LEVEL2}{c}}}} (\href{https://www.nndc.bnl.gov/nsr/nsrlink.jsp?2019Ha14,B}{2019Ha14}) suggested J\ensuremath{^{\ensuremath{\pi}}}=(7/2\ensuremath{^{-}}) and (9/2\ensuremath{^{-}}) for the \ensuremath{^{\textnormal{19}}}Ne*(4141.8) and \ensuremath{^{\textnormal{19}}}Ne*(4199.8) states, respectively. These are reversed}\\
\parbox[b][0.3cm]{17.7cm}{{\ }{\ }when compared with the prior \ensuremath{^{\textnormal{19}}}Ne evaluations based on \ensuremath{^{\textnormal{16}}}O(\ensuremath{^{\textnormal{6}}}Li,t) and \ensuremath{^{\textnormal{17}}}O(\ensuremath{^{\textnormal{3}}}He,\ensuremath{\gamma}) results in (\href{https://www.nndc.bnl.gov/nsr/nsrlink.jsp?1973Da31,B}{1973Da31}), or when compared}\\
\parbox[b][0.3cm]{17.7cm}{{\ }{\ }with the \ensuremath{^{\textnormal{19}}}F(\ensuremath{^{\textnormal{3}}}He,t) reaction study by (\href{https://www.nndc.bnl.gov/nsr/nsrlink.jsp?2015Pa46,B}{2015Pa46}). The results of (\href{https://www.nndc.bnl.gov/nsr/nsrlink.jsp?2019Ha14,B}{2019Ha14}, \href{https://www.nndc.bnl.gov/nsr/nsrlink.jsp?2020Ha31,B}{2020Ha31}) provide evidence that favors the reversal}\\
\parbox[b][0.3cm]{17.7cm}{{\ }{\ }of the J values for these states, in agreement with the findings of (\href{https://www.nndc.bnl.gov/nsr/nsrlink.jsp?2009Ta09,B}{2009Ta09}: \ensuremath{^{\textnormal{19}}}F(\ensuremath{^{\textnormal{3}}}He,t\ensuremath{\gamma})). The evaluation of the astrophysical}\\
\parbox[b][0.3cm]{17.7cm}{{\ }{\ }\ensuremath{^{\textnormal{15}}}O(\ensuremath{\alpha},\ensuremath{\gamma}) reaction rate by (\href{https://www.nndc.bnl.gov/nsr/nsrlink.jsp?2011Da24,B}{2011Da24}) agreed with the swapping of the spin assignments based on the reduced transition}\\
\parbox[b][0.3cm]{17.7cm}{{\ }{\ }probabilities for these two states; however, (\href{https://www.nndc.bnl.gov/nsr/nsrlink.jsp?2011Da24,B}{2011Da24}) recommended the tentative J\ensuremath{^{\ensuremath{\pi}}} assignments adopted by the last A=19}\\
\parbox[b][0.3cm]{17.7cm}{{\ }{\ }ENSDF evaluation of (\href{https://www.nndc.bnl.gov/nsr/nsrlink.jsp?1995Ti07,B}{1995Ti07}) based on the decay branching ratios of these two levels.}\\
\parbox[b][0.3cm]{17.7cm}{\makebox[1ex]{\ensuremath{^{\hypertarget{NE33LEVEL3}{d}}}} From (\href{https://www.nndc.bnl.gov/nsr/nsrlink.jsp?2020Ha31,B}{2020Ha31}).}\\
\vspace{0.5cm}
\underline{$\gamma$($^{19}$Ne)}\\
\begin{longtable}{ccccccccc@{}c@{\extracolsep{\fill}}c}
\multicolumn{2}{c}{E\ensuremath{_{i}}(level)}&J\ensuremath{^{\pi}_{i}}&\multicolumn{2}{c}{E\ensuremath{_{\gamma}}\ensuremath{^{\hyperlink{NE33GAMMA0}{a}}}}&\multicolumn{2}{c}{I\ensuremath{_{\ensuremath{\gamma}}} (\%)\ensuremath{^{\hyperlink{NE33GAMMA0}{a}}}}&\multicolumn{2}{c}{E\ensuremath{_{f}}}&J\ensuremath{^{\pi}_{f}}&\\[-.2cm]
\multicolumn{2}{c}{\hrulefill}&\hrulefill&\multicolumn{2}{c}{\hrulefill}&\multicolumn{2}{c}{\hrulefill}&\multicolumn{2}{c}{\hrulefill}&\hrulefill&
\endfirsthead
\multicolumn{1}{r@{}}{238}&\multicolumn{1}{@{.}l}{64}&\multicolumn{1}{l}{5/2\ensuremath{^{+}}}&\multicolumn{1}{r@{}}{238}&\multicolumn{1}{@{.}l}{4 {\it 3}}&\multicolumn{1}{r@{}}{100}&\multicolumn{1}{@{}l}{}&\multicolumn{1}{r@{}}{0}&\multicolumn{1}{@{}l}{}&\multicolumn{1}{@{}l}{1/2\ensuremath{^{+}}}&\\
\multicolumn{1}{r@{}}{275}&\multicolumn{1}{@{.}l}{45}&\multicolumn{1}{l}{1/2\ensuremath{^{-}}}&\multicolumn{1}{r@{}}{275}&\multicolumn{1}{@{.}l}{4 {\it 3}}&\multicolumn{1}{r@{}}{100}&\multicolumn{1}{@{}l}{}&\multicolumn{1}{r@{}}{0}&\multicolumn{1}{@{}l}{}&\multicolumn{1}{@{}l}{1/2\ensuremath{^{+}}}&\\
\multicolumn{1}{r@{}}{1507}&\multicolumn{1}{@{.}l}{7}&\multicolumn{1}{l}{5/2\ensuremath{^{-}}}&\multicolumn{1}{r@{}}{1232}&\multicolumn{1}{@{.}l}{4 {\it 3}}&\multicolumn{1}{r@{}}{82}&\multicolumn{1}{@{ }l}{{\it 3}}&\multicolumn{1}{r@{}}{275}&\multicolumn{1}{@{.}l}{45 }&\multicolumn{1}{@{}l}{1/2\ensuremath{^{-}}}&\\
&&&\multicolumn{1}{r@{}}{1269}&\multicolumn{1}{@{.}l}{0 {\it 3}}&\multicolumn{1}{r@{}}{18}&\multicolumn{1}{@{ }l}{{\it 3}}&\multicolumn{1}{r@{}}{238}&\multicolumn{1}{@{.}l}{64 }&\multicolumn{1}{@{}l}{5/2\ensuremath{^{+}}}&\\
\multicolumn{1}{r@{}}{1536}&\multicolumn{1}{@{.}l}{0}&\multicolumn{1}{l}{3/2\ensuremath{^{+}}}&\multicolumn{1}{r@{}}{1260}&\multicolumn{1}{@{.}l}{4 {\it 16}}&\multicolumn{1}{r@{}}{5}&\multicolumn{1}{@{.}l}{6 {\it 20}}&\multicolumn{1}{r@{}}{275}&\multicolumn{1}{@{.}l}{45 }&\multicolumn{1}{@{}l}{1/2\ensuremath{^{-}}}&\\
&&&\multicolumn{1}{r@{}}{1297}&\multicolumn{1}{@{.}l}{1 {\it 4}}&\multicolumn{1}{r@{}}{92}&\multicolumn{1}{@{ }l}{{\it 2}}&\multicolumn{1}{r@{}}{238}&\multicolumn{1}{@{.}l}{64 }&\multicolumn{1}{@{}l}{5/2\ensuremath{^{+}}}&\\
&&&\multicolumn{1}{r@{}}{1536}&\multicolumn{1}{@{.}l}{8 {\it 10}}&\multicolumn{1}{r@{}}{2}&\multicolumn{1}{@{.}l}{2 {\it 20}}&\multicolumn{1}{r@{}}{0}&\multicolumn{1}{@{}l}{}&\multicolumn{1}{@{}l}{1/2\ensuremath{^{+}}}&\\
\multicolumn{1}{r@{}}{1615}&\multicolumn{1}{@{.}l}{6}&\multicolumn{1}{l}{3/2\ensuremath{^{-}}}&\multicolumn{1}{r@{}}{1339}&\multicolumn{1}{@{.}l}{5 {\it 6}}&\multicolumn{1}{r@{}}{76}&\multicolumn{1}{@{ }l}{{\it 1}}&\multicolumn{1}{r@{}}{275}&\multicolumn{1}{@{.}l}{45 }&\multicolumn{1}{@{}l}{1/2\ensuremath{^{-}}}&\\
&&&\multicolumn{1}{r@{}}{1377}&\multicolumn{1}{@{.}l}{2 {\it 14}}&\multicolumn{1}{r@{}}{5}&\multicolumn{1}{@{ }l}{{\it 1}}&\multicolumn{1}{r@{}}{238}&\multicolumn{1}{@{.}l}{64 }&\multicolumn{1}{@{}l}{5/2\ensuremath{^{+}}}&\\
&&&\multicolumn{1}{r@{}}{1616}&\multicolumn{1}{@{.}l}{4 {\it 7}}&\multicolumn{1}{r@{}}{19}&\multicolumn{1}{@{ }l}{{\it 1}}&\multicolumn{1}{r@{}}{0}&\multicolumn{1}{@{}l}{}&\multicolumn{1}{@{}l}{1/2\ensuremath{^{+}}}&\\
\multicolumn{1}{r@{}}{2794}&\multicolumn{1}{@{.}l}{8}&\multicolumn{1}{l}{9/2\ensuremath{^{+}}}&\multicolumn{1}{r@{}}{2555}&\multicolumn{1}{@{.}l}{8 {\it 6}}&\multicolumn{1}{r@{}}{100}&\multicolumn{1}{@{}l}{}&\multicolumn{1}{r@{}}{238}&\multicolumn{1}{@{.}l}{64 }&\multicolumn{1}{@{}l}{5/2\ensuremath{^{+}}}&\\
\multicolumn{1}{r@{}}{4034}&\multicolumn{1}{@{.}l}{7}&\multicolumn{1}{l}{3/2\ensuremath{^{+}}}&\multicolumn{1}{r@{}}{2498}&\multicolumn{1}{@{.}l}{1 {\it 12}}&\multicolumn{1}{r@{}}{13}&\multicolumn{1}{@{ }l}{{\it 6}}&\multicolumn{1}{r@{}}{1536}&\multicolumn{1}{@{.}l}{0 }&\multicolumn{1}{@{}l}{3/2\ensuremath{^{+}}}&\\
\end{longtable}
\begin{textblock}{29}(0,27.3)
Continued on next page (footnotes at end of table)
\end{textblock}
\clearpage
\begin{longtable}{ccccccccc@{}ccccc@{\extracolsep{\fill}}c}
\\[-.4cm]
\multicolumn{15}{c}{{\bf \small \underline{\ensuremath{^{\textnormal{19}}}F(\ensuremath{^{\textnormal{3}}}He,t\ensuremath{\gamma})\hspace{0.2in}\href{https://www.nndc.bnl.gov/nsr/nsrlink.jsp?2019Ha08,B}{2019Ha08},\href{https://www.nndc.bnl.gov/nsr/nsrlink.jsp?2020Ha31,B}{2020Ha31} (continued)}}}\\
\multicolumn{15}{c}{~}\\
\multicolumn{15}{c}{\underline{$\gamma$($^{19}$Ne) (continued)}}\\
\multicolumn{15}{c}{~~~}\\
\multicolumn{2}{c}{E\ensuremath{_{i}}(level)}&J\ensuremath{^{\pi}_{i}}&\multicolumn{2}{c}{E\ensuremath{_{\gamma}}\ensuremath{^{\hyperlink{NE33GAMMA0}{a}}}}&\multicolumn{2}{c}{I\ensuremath{_{\ensuremath{\gamma}}} (\%)\ensuremath{^{\hyperlink{NE33GAMMA0}{a}}}}&\multicolumn{2}{c}{E\ensuremath{_{f}}}&J\ensuremath{^{\pi}_{f}}&Mult.&\multicolumn{2}{c}{\ensuremath{\alpha}\ensuremath{^{\hyperlink{NE33GAMMA3}{d}}}}&Comments&\\[-.2cm]
\multicolumn{2}{c}{\hrulefill}&\hrulefill&\multicolumn{2}{c}{\hrulefill}&\multicolumn{2}{c}{\hrulefill}&\multicolumn{2}{c}{\hrulefill}&\hrulefill&\hrulefill&\multicolumn{2}{c}{\hrulefill}&\hrulefill&
\endhead
\multicolumn{1}{r@{}}{4034}&\multicolumn{1}{@{.}l}{7}&\multicolumn{1}{l}{3/2\ensuremath{^{+}}}&\multicolumn{1}{r@{}}{3759}&\multicolumn{1}{@{.}l}{4 {\it 32}}&\multicolumn{1}{r@{}}{28}&\multicolumn{1}{@{ }l}{{\it 6}}&\multicolumn{1}{r@{}}{275}&\multicolumn{1}{@{.}l}{45 }&\multicolumn{1}{@{}l}{1/2\ensuremath{^{-}}}&&\multicolumn{1}{r@{}}{}&\multicolumn{1}{@{}l}{}&&\\
&&&\multicolumn{1}{r@{}}{4034}&\multicolumn{1}{@{.}l}{7 {\it 13}}&\multicolumn{1}{r@{}}{59}&\multicolumn{1}{@{ }l}{{\it 6}}&\multicolumn{1}{r@{}}{0}&\multicolumn{1}{@{}l}{}&\multicolumn{1}{@{}l}{1/2\ensuremath{^{+}}}&&\multicolumn{1}{r@{}}{}&\multicolumn{1}{@{}l}{}&&\\
\multicolumn{1}{r@{}}{4142}&\multicolumn{1}{@{.}l}{9}&\multicolumn{1}{l}{(7/2\ensuremath{^{-}})}&\multicolumn{1}{r@{}}{2527}&\multicolumn{1}{@{.}l}{2 {\it 10}}&\multicolumn{1}{r@{}}{14}&\multicolumn{1}{@{ }l}{{\it 4}}&\multicolumn{1}{r@{}}{1615}&\multicolumn{1}{@{.}l}{6 }&\multicolumn{1}{@{}l}{3/2\ensuremath{^{-}}}&&\multicolumn{1}{r@{}}{}&\multicolumn{1}{@{}l}{}&\parbox[t][0.3cm]{5.4857407cm}{\raggedright E\ensuremath{_{\gamma}}: This transition was first reported in\vspace{0.1cm}}&\\
&&&&&&&&&&&&&\parbox[t][0.3cm]{5.4857407cm}{\raggedright {\ }{\ }{\ }(\href{https://www.nndc.bnl.gov/nsr/nsrlink.jsp?2019Ha14,B}{2019Ha14}). See also (\href{https://www.nndc.bnl.gov/nsr/nsrlink.jsp?2020Ha31,B}{2020Ha31}).\vspace{0.1cm}}&\\
&&&&&&&&&&&&&\parbox[t][0.3cm]{5.4857407cm}{\raggedright I\ensuremath{_{\ensuremath{\gamma}}} (\%): See also (\href{https://www.nndc.bnl.gov/nsr/nsrlink.jsp?2019Ha14,B}{2019Ha14}).\vspace{0.1cm}}&\\
&&&\multicolumn{1}{r@{}}{2635}&\multicolumn{1}{@{.}l}{8 {\it 8}}&\multicolumn{1}{r@{}}{68}&\multicolumn{1}{@{ }l}{{\it 4}}&\multicolumn{1}{r@{}}{1507}&\multicolumn{1}{@{.}l}{7 }&\multicolumn{1}{@{}l}{5/2\ensuremath{^{-}}}&\multicolumn{1}{l}{M1}&\multicolumn{1}{r@{}}{5}&\multicolumn{1}{@{.}l}{09\ensuremath{\times10^{-4}} {\it 7}}&\parbox[t][0.3cm]{5.4857407cm}{\raggedright \ensuremath{\alpha}(K)=1.481\ensuremath{\times}10\ensuremath{^{\textnormal{$-$6}}} \textit{21}; \ensuremath{\alpha}(L)=8.20\ensuremath{\times}10\ensuremath{^{\textnormal{$-$8}}}\vspace{0.1cm}}&\\
&&&&&&&&&&&&&\parbox[t][0.3cm]{5.4857407cm}{\raggedright {\ }{\ }{\ }\textit{11}\vspace{0.1cm}}&\\
&&&&&&&&&&&&&\parbox[t][0.3cm]{5.4857407cm}{\raggedright \ensuremath{\alpha}(IPF)=0.000508 \textit{7}\vspace{0.1cm}}&\\
&&&&&&&&&&&&&\parbox[t][0.3cm]{5.4857407cm}{\raggedright E\ensuremath{_{\gamma}},I\ensuremath{_{\ensuremath{\gamma}}} (\%): See also (\href{https://www.nndc.bnl.gov/nsr/nsrlink.jsp?2019Ha14,B}{2019Ha14}).\vspace{0.1cm}}&\\
&&&&&&&&&&&&&\parbox[t][0.3cm]{5.4857407cm}{\raggedright Mult.: From (\href{https://www.nndc.bnl.gov/nsr/nsrlink.jsp?2019Ha14,B}{2019Ha14}).\vspace{0.1cm}}&\\
&&&\multicolumn{1}{r@{}}{3897}&\multicolumn{1}{@{.}l}{5 {\it 21}}&\multicolumn{1}{r@{}}{18}&\multicolumn{1}{@{ }l}{{\it 4}}&\multicolumn{1}{r@{}}{238}&\multicolumn{1}{@{.}l}{64 }&\multicolumn{1}{@{}l}{5/2\ensuremath{^{+}}}&&\multicolumn{1}{r@{}}{}&\multicolumn{1}{@{}l}{}&\parbox[t][0.3cm]{5.4857407cm}{\raggedright E\ensuremath{_{\gamma}}: This transition was first reported in\vspace{0.1cm}}&\\
&&&&&&&&&&&&&\parbox[t][0.3cm]{5.4857407cm}{\raggedright {\ }{\ }{\ }(\href{https://www.nndc.bnl.gov/nsr/nsrlink.jsp?2019Ha14,B}{2019Ha14}). See also (\href{https://www.nndc.bnl.gov/nsr/nsrlink.jsp?2020Ha31,B}{2020Ha31}).\vspace{0.1cm}}&\\
&&&&&&&&&&&&&\parbox[t][0.3cm]{5.4857407cm}{\raggedright I\ensuremath{_{\ensuremath{\gamma}}} (\%): See also (\href{https://www.nndc.bnl.gov/nsr/nsrlink.jsp?2019Ha14,B}{2019Ha14}).\vspace{0.1cm}}&\\
\multicolumn{1}{r@{}}{4200}&\multicolumn{1}{@{.}l}{1}&\multicolumn{1}{l}{(9/2\ensuremath{^{-}})}&\multicolumn{1}{r@{}}{2692}&\multicolumn{1}{@{.}l}{2 {\it 10}}&\multicolumn{1}{r@{}}{100}&\multicolumn{1}{@{}l}{}&\multicolumn{1}{r@{}}{1507}&\multicolumn{1}{@{.}l}{7 }&\multicolumn{1}{@{}l}{5/2\ensuremath{^{-}}}&\multicolumn{1}{l}{E2}&\multicolumn{1}{r@{}}{6}&\multicolumn{1}{@{.}l}{48\ensuremath{\times10^{-4}} {\it 9}}&\parbox[t][0.3cm]{5.4857407cm}{\raggedright \ensuremath{\alpha}(K)=1.543\ensuremath{\times}10\ensuremath{^{\textnormal{$-$6}}} \textit{22}; \ensuremath{\alpha}(L)=8.54\ensuremath{\times}10\ensuremath{^{\textnormal{$-$8}}}\vspace{0.1cm}}&\\
&&&&&&&&&&&&&\parbox[t][0.3cm]{5.4857407cm}{\raggedright {\ }{\ }{\ }\textit{12}\vspace{0.1cm}}&\\
&&&&&&&&&&&&&\parbox[t][0.3cm]{5.4857407cm}{\raggedright \ensuremath{\alpha}(IPF)=0.000647 \textit{9}\vspace{0.1cm}}&\\
&&&&&&&&&&&&&\parbox[t][0.3cm]{5.4857407cm}{\raggedright E\ensuremath{_{\gamma}},I\ensuremath{_{\ensuremath{\gamma}}} (\%): See also (\href{https://www.nndc.bnl.gov/nsr/nsrlink.jsp?2019Ha14,B}{2019Ha14}).\vspace{0.1cm}}&\\
&&&&&&&&&&&&&\parbox[t][0.3cm]{5.4857407cm}{\raggedright Mult.: From (\href{https://www.nndc.bnl.gov/nsr/nsrlink.jsp?2019Ha14,B}{2019Ha14}).\vspace{0.1cm}}&\\
\multicolumn{1}{r@{}}{4377}&\multicolumn{1}{@{.}l}{6}&\multicolumn{1}{l}{7/2\ensuremath{^{+}}}&\multicolumn{1}{r@{}}{1582}&\multicolumn{1}{@{.}l}{1 {\it 10}}&\multicolumn{1}{r@{}}{15}&\multicolumn{1}{@{ }l}{{\it 2}}&\multicolumn{1}{r@{}}{2794}&\multicolumn{1}{@{.}l}{8 }&\multicolumn{1}{@{}l}{9/2\ensuremath{^{+}}}&&\multicolumn{1}{r@{}}{}&\multicolumn{1}{@{}l}{}&&\\
&&&\multicolumn{1}{r@{}}{4139}&\multicolumn{1}{@{.}l}{7 {\it 14}}&\multicolumn{1}{r@{}}{85}&\multicolumn{1}{@{ }l}{{\it 2}}&\multicolumn{1}{r@{}}{238}&\multicolumn{1}{@{.}l}{64 }&\multicolumn{1}{@{}l}{5/2\ensuremath{^{+}}}&&\multicolumn{1}{r@{}}{}&\multicolumn{1}{@{}l}{}&&\\
\multicolumn{1}{r@{}}{4547}&\multicolumn{1}{@{.}l}{6}&\multicolumn{1}{l}{3/2\ensuremath{^{-}}}&\multicolumn{1}{r@{}}{3010}&\multicolumn{1}{@{.}l}{7\ensuremath{^{\hyperlink{NE33GAMMA1}{b}}} {\it 22}}&\multicolumn{1}{r@{}}{34}&\multicolumn{1}{@{ }l}{{\it 4}}&\multicolumn{1}{r@{}}{1536}&\multicolumn{1}{@{.}l}{0 }&\multicolumn{1}{@{}l}{3/2\ensuremath{^{+}}}&&&&&\\
&&&\multicolumn{1}{r@{}}{3046}&\multicolumn{1}{@{.}l}{5\ensuremath{^{\hyperlink{NE33GAMMA1}{b}}} {\it 35}}&\multicolumn{1}{r@{}}{7}&\multicolumn{1}{@{.}l}{1 {\it 36}}&\multicolumn{1}{r@{}}{1507}&\multicolumn{1}{@{.}l}{7 }&\multicolumn{1}{@{}l}{5/2\ensuremath{^{-}}}&&&&&\\
&&&\multicolumn{1}{r@{}}{4269}&\multicolumn{1}{@{.}l}{0 {\it 20}}&\multicolumn{1}{r@{}}{29}&\multicolumn{1}{@{ }l}{{\it 4}}&\multicolumn{1}{r@{}}{275}&\multicolumn{1}{@{.}l}{45 }&\multicolumn{1}{@{}l}{1/2\ensuremath{^{-}}}&&\multicolumn{1}{r@{}}{}&\multicolumn{1}{@{}l}{}&&\\
&&&\multicolumn{1}{r@{}}{4547}&\multicolumn{1}{@{.}l}{4 {\it 14}}&\multicolumn{1}{r@{}}{31}&\multicolumn{1}{@{ }l}{{\it 4}}&\multicolumn{1}{r@{}}{0}&\multicolumn{1}{@{}l}{}&\multicolumn{1}{@{}l}{1/2\ensuremath{^{+}}}&&\multicolumn{1}{r@{}}{}&\multicolumn{1}{@{}l}{}&&\\
\multicolumn{1}{r@{}}{4603}&\multicolumn{1}{@{.}l}{2}&\multicolumn{1}{l}{5/2\ensuremath{^{+}}}&\multicolumn{1}{r@{}}{2987}&\multicolumn{1}{@{.}l}{4\ensuremath{^{\hyperlink{NE33GAMMA1}{b}}} {\it 21}}&\multicolumn{1}{r@{}}{6}&\multicolumn{1}{@{.}l}{8 {\it 14}}&\multicolumn{1}{r@{}}{1615}&\multicolumn{1}{@{.}l}{6 }&\multicolumn{1}{@{}l}{3/2\ensuremath{^{-}}}&&&&&\\
&&&\multicolumn{1}{r@{}}{4364}&\multicolumn{1}{@{.}l}{1 {\it 12}}&\multicolumn{1}{r@{}}{85}&\multicolumn{1}{@{ }l}{{\it 1}}&\multicolumn{1}{r@{}}{238}&\multicolumn{1}{@{.}l}{64 }&\multicolumn{1}{@{}l}{5/2\ensuremath{^{+}}}&&\multicolumn{1}{r@{}}{}&\multicolumn{1}{@{}l}{}&&\\
&&&\multicolumn{1}{r@{}}{4602}&\multicolumn{1}{@{.}l}{3\ensuremath{^{\hyperlink{NE33GAMMA1}{b}}} {\it 17}}&\multicolumn{1}{r@{}}{8}&\multicolumn{1}{@{.}l}{6 {\it 14}}&\multicolumn{1}{r@{}}{0}&\multicolumn{1}{@{}l}{}&\multicolumn{1}{@{}l}{1/2\ensuremath{^{+}}}&&&&&\\
\multicolumn{1}{r@{}}{4634}&\multicolumn{1}{@{.}l}{6}&\multicolumn{1}{l}{13/2\ensuremath{^{+}}}&\multicolumn{1}{r@{}}{1839}&\multicolumn{1}{@{.}l}{5 {\it 4}}&\multicolumn{1}{r@{}}{100}&\multicolumn{1}{@{}l}{}&\multicolumn{1}{r@{}}{2794}&\multicolumn{1}{@{.}l}{8 }&\multicolumn{1}{@{}l}{9/2\ensuremath{^{+}}}&&\multicolumn{1}{r@{}}{}&\multicolumn{1}{@{}l}{}&&\\
\multicolumn{1}{r@{}}{4708}&\multicolumn{1}{@{.}l}{8}&\multicolumn{1}{l}{5/2\ensuremath{^{-}}}&\multicolumn{1}{r@{}}{3094}&\multicolumn{1}{@{.}l}{0\ensuremath{^{\hyperlink{NE33GAMMA1}{b}}} {\it 35}}&\multicolumn{1}{r@{}}{28}&\multicolumn{1}{@{.}l}{5\ensuremath{^{\hyperlink{NE33GAMMA2}{c}}} {\it 90}}&\multicolumn{1}{r@{}}{1615}&\multicolumn{1}{@{.}l}{6 }&\multicolumn{1}{@{}l}{3/2\ensuremath{^{-}}}&&&&&\\
&&&\multicolumn{1}{r@{}}{3200}&\multicolumn{1}{@{.}l}{5\ensuremath{^{\hyperlink{NE33GAMMA1}{b}}} {\it 18}}&\multicolumn{1}{r@{}}{71}&\multicolumn{1}{@{.}l}{5\ensuremath{^{\hyperlink{NE33GAMMA2}{c}}} {\it 90}}&\multicolumn{1}{r@{}}{1507}&\multicolumn{1}{@{.}l}{7 }&\multicolumn{1}{@{}l}{5/2\ensuremath{^{-}}}&&&&&\\
\multicolumn{1}{r@{}}{5091}&\multicolumn{1}{@{}l}{}&\multicolumn{1}{l}{5/2\ensuremath{^{+}}}&\multicolumn{1}{r@{}}{4852}&\multicolumn{1}{@{}l}{\ensuremath{^{\hyperlink{NE33GAMMA1}{b}}} {\it 3}}&\multicolumn{1}{r@{}}{100}&\multicolumn{1}{@{}l}{}&\multicolumn{1}{r@{}}{238}&\multicolumn{1}{@{.}l}{64 }&\multicolumn{1}{@{}l}{5/2\ensuremath{^{+}}}&&&&\parbox[t][0.3cm]{5.4857407cm}{\raggedright E\ensuremath{_{\gamma}}: (\href{https://www.nndc.bnl.gov/nsr/nsrlink.jsp?2020Ha31,B}{2020Ha31}) reported that this\vspace{0.1cm}}&\\
&&&&&&&&&&&&&\parbox[t][0.3cm]{5.4857407cm}{\raggedright {\ }{\ }{\ }transition is not observed for the\vspace{0.1cm}}&\\
&&&&&&&&&&&&&\parbox[t][0.3cm]{5.4857407cm}{\raggedright {\ }{\ }{\ }mirror level in \ensuremath{^{\textnormal{19}}}F.\vspace{0.1cm}}&\\
\multicolumn{1}{r@{}}{6099}&\multicolumn{1}{@{.}l}{6}&\multicolumn{1}{l}{(7/2\ensuremath{^{+}})}&\multicolumn{1}{r@{}}{4590}&\multicolumn{1}{@{}l}{\ensuremath{^{\hyperlink{NE33GAMMA1}{b}}} {\it 2}}&\multicolumn{1}{r@{}}{53}&\multicolumn{1}{@{ }l}{{\it 7}}&\multicolumn{1}{r@{}}{1507}&\multicolumn{1}{@{.}l}{7 }&\multicolumn{1}{@{}l}{5/2\ensuremath{^{-}}}&&&&&\\
&&&\multicolumn{1}{r@{}}{5863}&\multicolumn{1}{@{}l}{\ensuremath{^{\hyperlink{NE33GAMMA1}{b}}} {\it 3}}&\multicolumn{1}{r@{}}{47}&\multicolumn{1}{@{ }l}{{\it 7}}&\multicolumn{1}{r@{}}{238}&\multicolumn{1}{@{.}l}{64 }&\multicolumn{1}{@{}l}{5/2\ensuremath{^{+}}}&&&&&\\
\multicolumn{1}{r@{}}{6292}&\multicolumn{1}{@{.}l}{5}&\multicolumn{1}{l}{(11/2\ensuremath{^{+}})}&\multicolumn{1}{r@{}}{1657}&\multicolumn{1}{@{.}l}{6\ensuremath{^{\hyperlink{NE33GAMMA1}{b}}} {\it 6}}&\multicolumn{1}{r@{}}{32}&\multicolumn{1}{@{ }l}{{\it 3}}&\multicolumn{1}{r@{}}{4634}&\multicolumn{1}{@{.}l}{6 }&\multicolumn{1}{@{}l}{13/2\ensuremath{^{+}}}&&&&\parbox[t][0.3cm]{5.4857407cm}{\raggedright E\ensuremath{_{\gamma}},I\ensuremath{_{\ensuremath{\gamma}}} (\%): See also (\href{https://www.nndc.bnl.gov/nsr/nsrlink.jsp?2019Ha08,B}{2019Ha08}).\vspace{0.1cm}}&\\
&&&\multicolumn{1}{r@{}}{3498}&\multicolumn{1}{@{}l}{\ensuremath{^{\hyperlink{NE33GAMMA1}{b}}} {\it 1}}&\multicolumn{1}{r@{}}{68}&\multicolumn{1}{@{ }l}{{\it 3}}&\multicolumn{1}{r@{}}{2794}&\multicolumn{1}{@{.}l}{8 }&\multicolumn{1}{@{}l}{9/2\ensuremath{^{+}}}&&&&\parbox[t][0.3cm]{5.4857407cm}{\raggedright E\ensuremath{_{\gamma}},I\ensuremath{_{\ensuremath{\gamma}}} (\%): See also (\href{https://www.nndc.bnl.gov/nsr/nsrlink.jsp?2019Ha08,B}{2019Ha08}).\vspace{0.1cm}}&\\
\multicolumn{1}{r@{}}{6424}&\multicolumn{1}{@{}l}{}&\multicolumn{1}{l}{(3/2\ensuremath{^{+}})}&\multicolumn{1}{r@{}}{4913}&\multicolumn{1}{@{}l}{\ensuremath{^{\hyperlink{NE33GAMMA1}{b}}} {\it 5}}&\multicolumn{1}{r@{}}{42}&\multicolumn{1}{@{ }l}{{\it 11}}&\multicolumn{1}{r@{}}{1507}&\multicolumn{1}{@{.}l}{7 }&\multicolumn{1}{@{}l}{5/2\ensuremath{^{-}}}&&&&\parbox[t][0.3cm]{5.4857407cm}{\raggedright E\ensuremath{_{\gamma}},I\ensuremath{_{\ensuremath{\gamma}}} (\%): See also (\href{https://www.nndc.bnl.gov/nsr/nsrlink.jsp?2019Ha08,B}{2019Ha08}).\vspace{0.1cm}}&\\
&&&\multicolumn{1}{r@{}}{6147}&\multicolumn{1}{@{}l}{\ensuremath{^{\hyperlink{NE33GAMMA1}{b}}} {\it 6}}&\multicolumn{1}{r@{}}{27}&\multicolumn{1}{@{ }l}{{\it 11}}&\multicolumn{1}{r@{}}{275}&\multicolumn{1}{@{.}l}{45 }&\multicolumn{1}{@{}l}{1/2\ensuremath{^{-}}}&&&&\parbox[t][0.3cm]{5.4857407cm}{\raggedright E\ensuremath{_{\gamma}},I\ensuremath{_{\ensuremath{\gamma}}} (\%): See also (\href{https://www.nndc.bnl.gov/nsr/nsrlink.jsp?2019Ha08,B}{2019Ha08}).\vspace{0.1cm}}&\\
&&&&&&&&&&&&&\parbox[t][0.3cm]{5.4857407cm}{\raggedright Because of this transition to the\vspace{0.1cm}}&\\
&&&&&&&&&&&&&\parbox[t][0.3cm]{5.4857407cm}{\raggedright {\ }{\ }{\ }\ensuremath{^{\textnormal{19}}}Ne*(275 keV, 1/2\ensuremath{^{-}}) level,\vspace{0.1cm}}&\\
&&&&&&&&&&&&&\parbox[t][0.3cm]{5.4857407cm}{\raggedright {\ }{\ }{\ }(\href{https://www.nndc.bnl.gov/nsr/nsrlink.jsp?2020Ha31,B}{2020Ha31}) proposed a low\vspace{0.1cm}}&\\
&&&&&&&&&&&&&\parbox[t][0.3cm]{5.4857407cm}{\raggedright {\ }{\ }{\ }spin-parity (J\ensuremath{\leq}5/2) for the 6423-keV\vspace{0.1cm}}&\\
&&&&&&&&&&&&&\parbox[t][0.3cm]{5.4857407cm}{\raggedright {\ }{\ }{\ }state in \ensuremath{^{\textnormal{19}}}Ne. The decay scheme of\vspace{0.1cm}}&\\
&&&&&&&&&&&&&\parbox[t][0.3cm]{5.4857407cm}{\raggedright {\ }{\ }{\ }the two J\ensuremath{^{\ensuremath{\pi}}}=3/2\ensuremath{^{\textnormal{+}}} \ensuremath{^{\textnormal{19}}}F*(6497, 6527)\vspace{0.1cm}}&\\
&&&&&&&&&&&&&\parbox[t][0.3cm]{5.4857407cm}{\raggedright {\ }{\ }{\ }levels are similar to the transitions\vspace{0.1cm}}&\\
&&&&&&&&&&&&&\parbox[t][0.3cm]{5.4857407cm}{\raggedright {\ }{\ }{\ }observed for the \ensuremath{^{\textnormal{19}}}Ne*(6423) state.\vspace{0.1cm}}&\\
&&&&&&&&&&&&&\parbox[t][0.3cm]{5.4857407cm}{\raggedright {\ }{\ }{\ }Therefore, in the absence of an\vspace{0.1cm}}&\\
&&&&&&&&&&&&&\parbox[t][0.3cm]{5.4857407cm}{\raggedright {\ }{\ }{\ }obvious 3/2\ensuremath{^{-}} or 5/2\ensuremath{^{\textnormal{+}}} mirror level in\vspace{0.1cm}}&\\
\end{longtable}
\begin{textblock}{29}(0,27.3)
Continued on next page (footnotes at end of table)
\end{textblock}
\clearpage
\begin{longtable}{ccccccccc@{}cc@{\extracolsep{\fill}}c}
\\[-.4cm]
\multicolumn{12}{c}{{\bf \small \underline{\ensuremath{^{\textnormal{19}}}F(\ensuremath{^{\textnormal{3}}}He,t\ensuremath{\gamma})\hspace{0.2in}\href{https://www.nndc.bnl.gov/nsr/nsrlink.jsp?2019Ha08,B}{2019Ha08},\href{https://www.nndc.bnl.gov/nsr/nsrlink.jsp?2020Ha31,B}{2020Ha31} (continued)}}}\\
\multicolumn{12}{c}{~}\\
\multicolumn{12}{c}{\underline{$\gamma$($^{19}$Ne) (continued)}}\\
\multicolumn{12}{c}{~~~}\\
\multicolumn{2}{c}{E\ensuremath{_{i}}(level)}&J\ensuremath{^{\pi}_{i}}&\multicolumn{2}{c}{E\ensuremath{_{\gamma}}\ensuremath{^{\hyperlink{NE33GAMMA0}{a}}}}&\multicolumn{2}{c}{I\ensuremath{_{\ensuremath{\gamma}}} (\%)\ensuremath{^{\hyperlink{NE33GAMMA0}{a}}}}&\multicolumn{2}{c}{E\ensuremath{_{f}}}&J\ensuremath{^{\pi}_{f}}&Comments&\\[-.2cm]
\multicolumn{2}{c}{\hrulefill}&\hrulefill&\multicolumn{2}{c}{\hrulefill}&\multicolumn{2}{c}{\hrulefill}&\multicolumn{2}{c}{\hrulefill}&\hrulefill&\hrulefill&
\endhead
&&&&&&&&&&\parbox[t][0.3cm]{9.275321cm}{\raggedright {\ }{\ }{\ }\ensuremath{^{\textnormal{19}}}F as suggested by (\href{https://www.nndc.bnl.gov/nsr/nsrlink.jsp?2013La01,B}{2013La01}: \ensuremath{^{\textnormal{19}}}F(\ensuremath{^{\textnormal{3}}}He,t)) and (\href{https://www.nndc.bnl.gov/nsr/nsrlink.jsp?2011Ad05,B}{2011Ad05}:\vspace{0.1cm}}&\\
&&&&&&&&&&\parbox[t][0.3cm]{9.275321cm}{\raggedright {\ }{\ }{\ }\ensuremath{^{\textnormal{2}}}H(\ensuremath{^{\textnormal{18}}}F,\ensuremath{^{\textnormal{19}}}Ne)), (\href{https://www.nndc.bnl.gov/nsr/nsrlink.jsp?2020Ha31,B}{2020Ha31}) assigned a J\ensuremath{^{\ensuremath{\pi}}}=3/2\ensuremath{^{\textnormal{+}}} to the\vspace{0.1cm}}&\\
&&&&&&&&&&\parbox[t][0.3cm]{9.275321cm}{\raggedright {\ }{\ }{\ }\ensuremath{^{\textnormal{19}}}Ne*(6423) level.\vspace{0.1cm}}&\\
\multicolumn{1}{r@{}}{6424}&\multicolumn{1}{@{}l}{}&\multicolumn{1}{l}{(3/2\ensuremath{^{+}})}&\multicolumn{1}{r@{}}{6425}&\multicolumn{1}{@{}l}{\ensuremath{^{\hyperlink{NE33GAMMA1}{b}}} {\it 5}}&\multicolumn{1}{r@{}}{31}&\multicolumn{1}{@{ }l}{{\it 11}}&\multicolumn{1}{r@{}}{0}&\multicolumn{1}{@{}l}{}&\multicolumn{1}{@{}l}{1/2\ensuremath{^{+}}}&\parbox[t][0.3cm]{9.275321cm}{\raggedright E\ensuremath{_{\gamma}},I\ensuremath{_{\ensuremath{\gamma}}} (\%): See also (\href{https://www.nndc.bnl.gov/nsr/nsrlink.jsp?2019Ha08,B}{2019Ha08}).\vspace{0.1cm}}&\\
\multicolumn{1}{r@{}}{6442}&\multicolumn{1}{@{}l}{}&\multicolumn{1}{l}{(3/2\ensuremath{^{+}})}&\multicolumn{1}{r@{}}{4828}&\multicolumn{1}{@{}l}{\ensuremath{^{\hyperlink{NE33GAMMA1}{b}}} {\it 4}}&\multicolumn{1}{r@{}}{38}&\multicolumn{1}{@{ }l}{{\it 12}}&\multicolumn{1}{r@{}}{1615}&\multicolumn{1}{@{.}l}{6 }&\multicolumn{1}{@{}l}{3/2\ensuremath{^{-}}}&\parbox[t][0.3cm]{9.275321cm}{\raggedright E\ensuremath{_{\gamma}},I\ensuremath{_{\ensuremath{\gamma}}} (\%): See also (\href{https://www.nndc.bnl.gov/nsr/nsrlink.jsp?2019Ha08,B}{2019Ha08}).\vspace{0.1cm}}&\\
&&&&&&&&&&\parbox[t][0.3cm]{9.275321cm}{\raggedright I\ensuremath{_{\ensuremath{\gamma}}} (\%): The branching ratio of 32\% \textit{12} is also mentioned in the\vspace{0.1cm}}&\\
&&&&&&&&&&\parbox[t][0.3cm]{9.275321cm}{\raggedright {\ }{\ }{\ }text in (\href{https://www.nndc.bnl.gov/nsr/nsrlink.jsp?2020Ha31,B}{2020Ha31}), which is most likely erroneous since the sum\vspace{0.1cm}}&\\
&&&&&&&&&&\parbox[t][0.3cm]{9.275321cm}{\raggedright {\ }{\ }{\ }of the \ensuremath{\gamma} ray branching ratios from this state would not add to\vspace{0.1cm}}&\\
&&&&&&&&&&\parbox[t][0.3cm]{9.275321cm}{\raggedright {\ }{\ }{\ }100\% in that case.\vspace{0.1cm}}&\\
&&&\multicolumn{1}{r@{}}{6200}&\multicolumn{1}{@{}l}{\ensuremath{^{\hyperlink{NE33GAMMA1}{b}}} {\it 4}}&\multicolumn{1}{r@{}}{62}&\multicolumn{1}{@{ }l}{{\it 12}}&\multicolumn{1}{r@{}}{238}&\multicolumn{1}{@{.}l}{64 }&\multicolumn{1}{@{}l}{5/2\ensuremath{^{+}}}&\parbox[t][0.3cm]{9.275321cm}{\raggedright E\ensuremath{_{\gamma}},I\ensuremath{_{\ensuremath{\gamma}}} (\%): See also (\href{https://www.nndc.bnl.gov/nsr/nsrlink.jsp?2019Ha08,B}{2019Ha08}).\vspace{0.1cm}}&\\
\multicolumn{1}{r@{}}{6739}&\multicolumn{1}{@{}l}{}&\multicolumn{1}{l}{3/2\ensuremath{^{-}}}&\multicolumn{1}{r@{}}{5123}&\multicolumn{1}{@{}l}{\ensuremath{^{\hyperlink{NE33GAMMA1}{b}}} {\it 7}}&\multicolumn{1}{r@{}}{100}&\multicolumn{1}{@{}l}{}&\multicolumn{1}{r@{}}{1615}&\multicolumn{1}{@{.}l}{6 }&\multicolumn{1}{@{}l}{3/2\ensuremath{^{-}}}&\parbox[t][0.3cm]{9.275321cm}{\raggedright E\ensuremath{_{\gamma}}: This \ensuremath{\gamma} ray transition has limited statistics (see Fig. 8(i) in\vspace{0.1cm}}&\\
&&&&&&&&&&\parbox[t][0.3cm]{9.275321cm}{\raggedright {\ }{\ }{\ }(\href{https://www.nndc.bnl.gov/nsr/nsrlink.jsp?2020Ha31,B}{2020Ha31})).\vspace{0.1cm}}&\\
\multicolumn{1}{r@{}}{6853}&\multicolumn{1}{@{}l}{}&\multicolumn{1}{l}{(7/2\ensuremath{^{-}})}&\multicolumn{1}{r@{}}{2653}&\multicolumn{1}{@{}l}{\ensuremath{^{\hyperlink{NE33GAMMA1}{b}}} {\it 3}}&\multicolumn{1}{r@{}}{100}&\multicolumn{1}{@{}l}{}&\multicolumn{1}{r@{}}{4200}&\multicolumn{1}{@{.}l}{1 }&\multicolumn{1}{@{}l}{(9/2\ensuremath{^{-}})}&&\\
\end{longtable}
\parbox[b][0.3cm]{17.7cm}{\makebox[1ex]{\ensuremath{^{\hypertarget{NE33GAMMA0}{a}}}} From (\href{https://www.nndc.bnl.gov/nsr/nsrlink.jsp?2020Ha31,B}{2020Ha31}).}\\
\parbox[b][0.3cm]{17.7cm}{\makebox[1ex]{\ensuremath{^{\hypertarget{NE33GAMMA1}{b}}}} This transition was newly observed in (\href{https://www.nndc.bnl.gov/nsr/nsrlink.jsp?2020Ha31,B}{2020Ha31}).}\\
\parbox[b][0.3cm]{17.7cm}{\makebox[1ex]{\ensuremath{^{\hypertarget{NE33GAMMA2}{c}}}} From private communication between the authors of (\href{https://www.nndc.bnl.gov/nsr/nsrlink.jsp?2020Ha31,B}{2020Ha31}) and J. Kelley (TUNL, December-2020). This communication led}\\
\parbox[b][0.3cm]{17.7cm}{{\ }{\ }to the correction of the branching ratios, whose original sum was greater than 100\%.}\\
\parbox[b][0.3cm]{17.7cm}{\makebox[1ex]{\ensuremath{^{\hypertarget{NE33GAMMA3}{d}}}} Total theoretical internal conversion coefficients, calculated using the BrIcc code (\href{https://www.nndc.bnl.gov/nsr/nsrlink.jsp?2008Ki07,B}{2008Ki07}) with ``Frozen Orbitals''}\\
\parbox[b][0.3cm]{17.7cm}{{\ }{\ }approximation based on \ensuremath{\gamma}-ray energies, assigned multipolarities, and mixing ratios, unless otherwise specified.}\\
\vspace{0.5cm}
\clearpage
\begin{figure}[h]
\begin{center}
\includegraphics{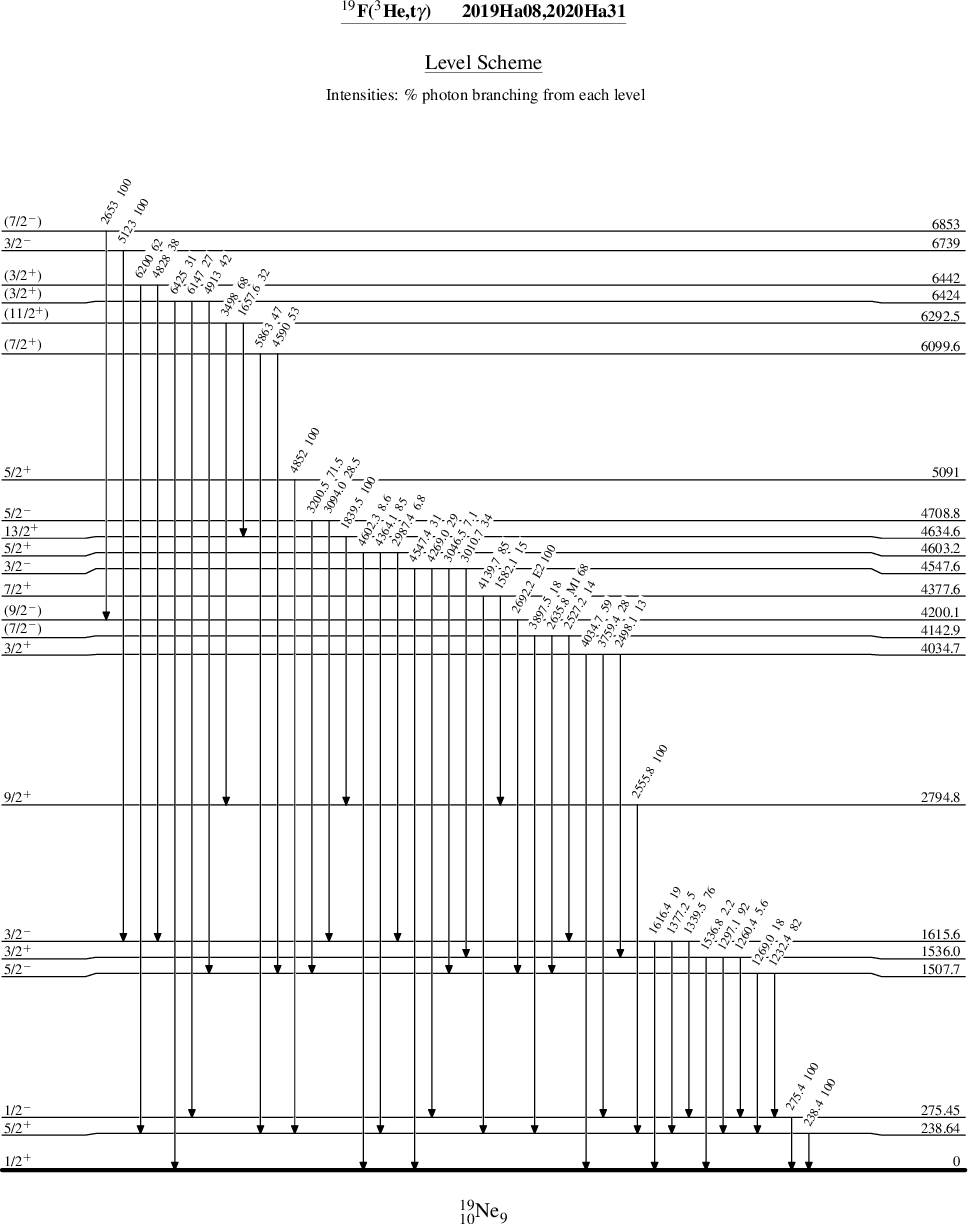}\\
\end{center}
\end{figure}
\clearpage
\subsection[\hspace{-0.2cm}\ensuremath{^{\textnormal{19}}}F(\ensuremath{^{\textnormal{6}}}Li,\ensuremath{^{\textnormal{6}}}He)]{ }
\vspace{-27pt}
\vspace{0.3cm}
\hypertarget{NE34}{{\bf \small \underline{\ensuremath{^{\textnormal{19}}}F(\ensuremath{^{\textnormal{6}}}Li,\ensuremath{^{\textnormal{6}}}He)\hspace{0.2in}\href{https://www.nndc.bnl.gov/nsr/nsrlink.jsp?1974Ga11,B}{1974Ga11}}}}\\
\vspace{4pt}
\vspace{8pt}
\parbox[b][0.3cm]{17.7cm}{\addtolength{\parindent}{-0.2in}Charge exchange reaction.}\\
\parbox[b][0.3cm]{17.7cm}{\addtolength{\parindent}{-0.2in}J\ensuremath{^{\ensuremath{\pi}}}(\ensuremath{^{\textnormal{6}}}Li\ensuremath{_{\textnormal{g.s.}}})=1\ensuremath{^{\textnormal{+}}} and J\ensuremath{^{\ensuremath{\pi}}}(\ensuremath{^{\textnormal{6}}}He\ensuremath{_{\textnormal{g.s.}}})=0\ensuremath{^{\textnormal{+}}}.}\\
\parbox[b][0.3cm]{17.7cm}{\addtolength{\parindent}{-0.2in}\href{https://www.nndc.bnl.gov/nsr/nsrlink.jsp?1974Ga11,B}{1974Ga11}: \ensuremath{^{\textnormal{19}}}F(\ensuremath{^{\textnormal{6}}}Li,\ensuremath{^{\textnormal{6}}}He) E=34 MeV; momentum analyzed and identified the reaction products using an Enge split-pole}\\
\parbox[b][0.3cm]{17.7cm}{spectrograph and its focal plane spark counter. Measured the excitation energies of \ensuremath{^{\textnormal{19}}}Ne up to E\ensuremath{_{\textnormal{x}}}=4.4 MeV and observed the}\\
\parbox[b][0.3cm]{17.7cm}{low-lying members of the K\ensuremath{^{\ensuremath{\pi}}}=1/2\ensuremath{^{\textnormal{+}}} ground state band. Measured the \ensuremath{^{\textnormal{6}}}He angular distributions at \ensuremath{\theta}\ensuremath{_{\textnormal{c.m.}}}=5\ensuremath{^\circ}{\textminus}30\ensuremath{^\circ} in steps of 5\ensuremath{^\circ}.}\\
\parbox[b][0.3cm]{17.7cm}{Deduced the J\ensuremath{^{\ensuremath{\pi}}} values using a single-step DWBA analysis.}\\
\vspace{12pt}
\underline{$^{19}$Ne Levels}\\
\begin{longtable}{ccccc@{\extracolsep{\fill}}c}
\multicolumn{2}{c}{E(level)$^{{\hyperlink{NE34LEVEL1}{b}}}$}&J$^{\pi}$$^{{\hyperlink{NE34LEVEL4}{e}}}$&L$^{{\hyperlink{NE34LEVEL4}{e}}}$&Comments&\\[-.2cm]
\multicolumn{2}{c}{\hrulefill}&\hrulefill&\hrulefill&\hrulefill&
\endfirsthead
\multicolumn{1}{r@{}}{0}&\multicolumn{1}{@{}l}{\ensuremath{^{{\hyperlink{NE34LEVEL0}{a}}{\hyperlink{NE34LEVEL2}{c}}}}}&\multicolumn{1}{l}{1/2\ensuremath{^{+}}}&\multicolumn{1}{l}{0}&&\\
\multicolumn{1}{r@{}}{238}&\multicolumn{1}{@{}l}{\ensuremath{^{{\hyperlink{NE34LEVEL0}{a}}{\hyperlink{NE34LEVEL2}{c}}}}}&\multicolumn{1}{l}{5/2\ensuremath{^{+}}}&\multicolumn{1}{l}{2}&&\\
\multicolumn{1}{r@{}}{1508?}&\multicolumn{1}{@{}l}{\ensuremath{^{{\hyperlink{NE34LEVEL3}{d}}}}}&&&&\\
\multicolumn{1}{r@{}}{1536}&\multicolumn{1}{@{}l}{\ensuremath{^{{\hyperlink{NE34LEVEL0}{a}}{\hyperlink{NE34LEVEL2}{c}}}}}&\multicolumn{1}{l}{3/2\ensuremath{^{+}}}&\multicolumn{1}{l}{2}&&\\
\multicolumn{1}{r@{}}{1615}&\multicolumn{1}{@{}l}{\ensuremath{^{{\hyperlink{NE34LEVEL3}{d}}}}}&&&&\\
\multicolumn{1}{r@{}}{2794}&\multicolumn{1}{@{}l}{\ensuremath{^{{\hyperlink{NE34LEVEL0}{a}}{\hyperlink{NE34LEVEL2}{c}}}}}&\multicolumn{1}{l}{9/2\ensuremath{^{+}}\ensuremath{^{{\hyperlink{NE34LEVEL5}{f}}}}}&\multicolumn{1}{l}{4$^{{\hyperlink{NE34LEVEL5}{f}}}$}&&\\
\multicolumn{1}{r@{}}{4368}&\multicolumn{1}{@{ }l}{{\it 10}}&\multicolumn{1}{l}{7/2\ensuremath{^{+}}\ensuremath{^{{\hyperlink{NE34LEVEL5}{f}}}}}&\multicolumn{1}{l}{4$^{{\hyperlink{NE34LEVEL5}{f}}}$}&\parbox[t][0.3cm]{13.966241cm}{\raggedright E(level): From (\href{https://www.nndc.bnl.gov/nsr/nsrlink.jsp?1974Ga11,B}{1974Ga11}): See Fig. 6.\vspace{0.1cm}}&\\
\end{longtable}
\parbox[b][0.3cm]{17.7cm}{\makebox[1ex]{\ensuremath{^{\hypertarget{NE34LEVEL0}{a}}}} Seq.(A): K\ensuremath{^{\ensuremath{\pi}}}=1/2\ensuremath{^{+}} g.s. band (\href{https://www.nndc.bnl.gov/nsr/nsrlink.jsp?1974Ga11,B}{1974Ga11}).}\\
\parbox[b][0.3cm]{17.7cm}{\makebox[1ex]{\ensuremath{^{\hypertarget{NE34LEVEL1}{b}}}} From (\href{https://www.nndc.bnl.gov/nsr/nsrlink.jsp?1974Ga11,B}{1974Ga11}), where the excitation energies were taken from (\href{https://www.nndc.bnl.gov/nsr/nsrlink.jsp?1972Aj02,B}{1972Aj02}: An older evaluation of A=19 nuclei), unless}\\
\parbox[b][0.3cm]{17.7cm}{{\ }{\ }otherwise noted.}\\
\parbox[b][0.3cm]{17.7cm}{\makebox[1ex]{\ensuremath{^{\hypertarget{NE34LEVEL2}{c}}}} This state is a member of the K\ensuremath{^{\ensuremath{\pi}}}=1/2\ensuremath{^{\textnormal{+}}} ground state rotational band (\href{https://www.nndc.bnl.gov/nsr/nsrlink.jsp?1974Ga11,B}{1974Ga11}).}\\
\parbox[b][0.3cm]{17.7cm}{\makebox[1ex]{\ensuremath{^{\hypertarget{NE34LEVEL3}{d}}}} This state was populated very weakly in (\href{https://www.nndc.bnl.gov/nsr/nsrlink.jsp?1974Ga11,B}{1974Ga11}): See Fig. 1 and text.}\\
\parbox[b][0.3cm]{17.7cm}{\makebox[1ex]{\ensuremath{^{\hypertarget{NE34LEVEL4}{e}}}} From the single-step DWBA analysis of (\href{https://www.nndc.bnl.gov/nsr/nsrlink.jsp?1974Ga11,B}{1974Ga11}).}\\
\parbox[b][0.3cm]{17.7cm}{\makebox[1ex]{\ensuremath{^{\hypertarget{NE34LEVEL5}{f}}}} (\href{https://www.nndc.bnl.gov/nsr/nsrlink.jsp?1974Ga11,B}{1974Ga11}) acknowledged that the \ensuremath{^{\textnormal{6}}}He angular distribution corresponding to the population of this state was not well represented}\\
\parbox[b][0.3cm]{17.7cm}{{\ }{\ }by an L=4 DWBA distribution. The authors attributed this effect to the fact that they neglected two-step processes in their DWBA}\\
\parbox[b][0.3cm]{17.7cm}{{\ }{\ }analysis, which may have played a significant role for populating this state.}\\
\vspace{0.5cm}
\clearpage
\clearpage
\begin{figure}[h]
\begin{center}
\includegraphics{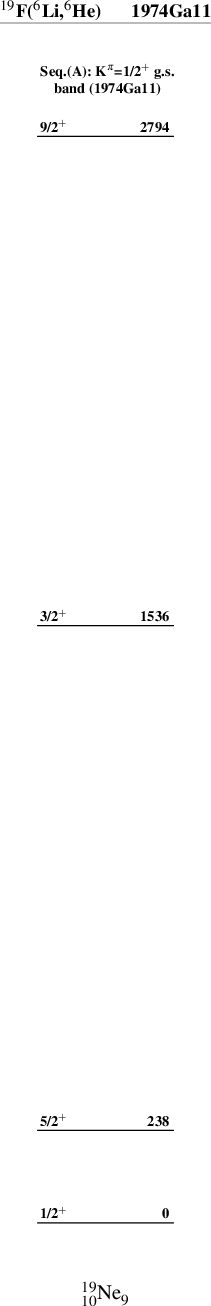}\\
\end{center}
\end{figure}
\clearpage
\subsection[\hspace{-0.2cm}\ensuremath{^{\textnormal{20}}}Ne(\ensuremath{\gamma},n)]{ }
\vspace{-27pt}
\vspace{0.3cm}
\hypertarget{NE35}{{\bf \small \underline{\ensuremath{^{\textnormal{20}}}Ne(\ensuremath{\gamma},n)\hspace{0.2in}\href{https://www.nndc.bnl.gov/nsr/nsrlink.jsp?1972Le33,B}{1972Le33},\href{https://www.nndc.bnl.gov/nsr/nsrlink.jsp?1981Al05,B}{1981Al05}}}}\\
\vspace{4pt}
\vspace{8pt}
\parbox[b][0.3cm]{17.7cm}{\addtolength{\parindent}{-0.2in}Photon induced reaction.}\\
\parbox[b][0.3cm]{17.7cm}{\addtolength{\parindent}{-0.2in}J\ensuremath{^{\ensuremath{\pi}}}(\ensuremath{^{\textnormal{20}}}Ne\ensuremath{_{\textnormal{g.s.}}})=0\ensuremath{^{\textnormal{+}}}.}\\
\parbox[b][0.3cm]{17.7cm}{\addtolength{\parindent}{-0.2in}\href{https://www.nndc.bnl.gov/nsr/nsrlink.jsp?1954Fe16,B}{1954Fe16}: \ensuremath{^{\textnormal{nat}}}Ne(\ensuremath{\gamma},n) E not given; measured \ensuremath{\sigma}(E\ensuremath{_{\ensuremath{\gamma}}}). Gaseous targets. Deduced parameters of a giant dipole resonance in \ensuremath{^{\textnormal{20}}}Ne,}\\
\parbox[b][0.3cm]{17.7cm}{which decayed via neutron emission to \ensuremath{^{\textnormal{19}}}Ne.}\\
\parbox[b][0.3cm]{17.7cm}{\addtolength{\parindent}{-0.2in}\href{https://www.nndc.bnl.gov/nsr/nsrlink.jsp?1972Le33,B}{1972Le33}: \ensuremath{^{\textnormal{20}}}Ne(\ensuremath{\gamma},n)\ensuremath{^{\textnormal{19}}}Ne(\ensuremath{\beta}) E not given; trapped \ensuremath{^{\textnormal{19}}}Ne (using consecutive cycles of 20 s beam on, followed by 50 s counting) in}\\
\parbox[b][0.3cm]{17.7cm}{a central proportional counter, which was set up in anti-coincidence with a surrounding plastic scintillator; measured K/\ensuremath{\beta}\ensuremath{^{\textnormal{+}}} ratio}\\
\parbox[b][0.3cm]{17.7cm}{(where K refers to K-electron capture) and deduced K/\ensuremath{\beta}\ensuremath{^{\textnormal{+}}}=9.6\ensuremath{\times}10\ensuremath{^{\textnormal{$-$4}}} \textit{3}. Theoretical K/\ensuremath{\beta}\ensuremath{^{\textnormal{+}}} ratio for \ensuremath{^{\textnormal{19}}}Ne was computed following}\\
\parbox[b][0.3cm]{17.7cm}{the work of (\href{https://www.nndc.bnl.gov/nsr/nsrlink.jsp?1963Ba21,B}{1963Ba21}, \href{https://www.nndc.bnl.gov/nsr/nsrlink.jsp?1963Ba72,B}{1963Ba72}, \href{https://www.nndc.bnl.gov/nsr/nsrlink.jsp?1970Va29,B}{1970Va29}). A better agreement between the experimental results and the calculations of}\\
\parbox[b][0.3cm]{17.7cm}{(\href{https://www.nndc.bnl.gov/nsr/nsrlink.jsp?1970Va29,B}{1970Va29}) was found.}\\
\parbox[b][0.3cm]{17.7cm}{\addtolength{\parindent}{-0.2in}\href{https://www.nndc.bnl.gov/nsr/nsrlink.jsp?1974Ve06,B}{1974Ve06}: \ensuremath{^{\textnormal{nat}}}Ne(\ensuremath{\gamma},n) E=16-26 MeV; a quasi-monochromatic photon beam was created, using the annihilation-in-flight technique}\\
\parbox[b][0.3cm]{17.7cm}{from a continuously variable monochromatic positron beam. Measured \ensuremath{\sigma}(\ensuremath{\gamma},n)+\ensuremath{\sigma}(\ensuremath{\gamma},pn) and \ensuremath{\sigma}(\ensuremath{\gamma},2n) as a function of photon energy}\\
\parbox[b][0.3cm]{17.7cm}{using a 250 liter gadolinium loaded liquid scintillator to detect neutrons. Deduced total photoneutron \ensuremath{\sigma} and integrated \ensuremath{\sigma}.}\\
\parbox[b][0.3cm]{17.7cm}{\addtolength{\parindent}{-0.2in}\href{https://www.nndc.bnl.gov/nsr/nsrlink.jsp?1974WoZS,B}{1974WoZS}: \ensuremath{^{\textnormal{20}}}Ne(\ensuremath{\gamma},n); measured \ensuremath{\sigma}(E\ensuremath{_{\ensuremath{\gamma}}};E\ensuremath{_{\textnormal{n}}}). Deduced \ensuremath{^{\textnormal{19}}}Ne levels.}\\
\parbox[b][0.3cm]{17.7cm}{\addtolength{\parindent}{-0.2in}\href{https://www.nndc.bnl.gov/nsr/nsrlink.jsp?1975Wo06,B}{1975Wo06}: \ensuremath{^{\textnormal{nat}}}Ne(\ensuremath{\gamma},n\ensuremath{_{\textnormal{0}}}+n\ensuremath{_{\textnormal{1}}}+n\ensuremath{_{\textnormal{2}}}) E=19-32 MeV in steps of 1 MeV; measured photoneutron energy distributions from bombarding a}\\
\parbox[b][0.3cm]{17.7cm}{liquid \ensuremath{^{\textnormal{nat}}}Ne target with photons. Neutrons were measured using a neutron time-of-flight spectrometer at \ensuremath{\theta}\ensuremath{_{\textnormal{lab}}}=90\ensuremath{^\circ}. Neutron}\\
\parbox[b][0.3cm]{17.7cm}{resolution was 11 keV at E\ensuremath{_{\textnormal{n}}}=1 MeV and 300 keV at E\ensuremath{_{\textnormal{n}}}=9 MeV. Measured d\ensuremath{\sigma}/d\ensuremath{\Omega}\ensuremath{_{\textnormal{lab}}}(\ensuremath{\gamma},n\ensuremath{_{\textnormal{0}}}+n\ensuremath{_{\textnormal{1}}}+n\ensuremath{_{\textnormal{2}}},\ensuremath{\theta}\ensuremath{_{\textnormal{lab}}}=90\ensuremath{^\circ}). Observed}\\
\parbox[b][0.3cm]{17.7cm}{photoneutron groups corresponding to photon absorption populating the \ensuremath{^{\textnormal{20}}}Ne*(18.8, 19.1, 20.1, 22.0, 23.0, and 24.8 MeV) states.}\\
\parbox[b][0.3cm]{17.7cm}{These are the strength distribution of a giant dipole resonance in \ensuremath{^{\textnormal{20}}}Ne at \ensuremath{\sim}20 MeV.}\\
\parbox[b][0.3cm]{17.7cm}{\addtolength{\parindent}{-0.2in}\href{https://www.nndc.bnl.gov/nsr/nsrlink.jsp?1981Al05,B}{1981Al05}: \ensuremath{^{\textnormal{20}}}Ne(\ensuremath{\gamma},n) E=16-29 MeV in steps of 100 keV; measured photoneutron energy distributions and cross section from}\\
\parbox[b][0.3cm]{17.7cm}{bombarding an enriched \ensuremath{^{\textnormal{20}}}Ne gas target with a photon beam. Neutrons were measured using a BF\ensuremath{_{\textnormal{3}}} counter. Observed photoneutron}\\
\parbox[b][0.3cm]{17.7cm}{groups corresponding to the neutron decay of the \ensuremath{^{\textnormal{20}}}Ne*(17.78, 19.0, 20.15, 22.6, 24.9, 27.5 MeV) states, which are part of the}\\
\parbox[b][0.3cm]{17.7cm}{strength distribution of a giant dipole resonance in \ensuremath{^{\textnormal{20}}}Ne, whose properties were deduced.}\\
\vspace{0.385cm}
\parbox[b][0.3cm]{17.7cm}{\addtolength{\parindent}{-0.2in}\textit{Theory}:}\\
\parbox[b][0.3cm]{17.7cm}{\addtolength{\parindent}{-0.2in}\href{https://www.nndc.bnl.gov/nsr/nsrlink.jsp?2000Va24,B}{2000Va24}: \ensuremath{^{\textnormal{nat}}}Ne(\ensuremath{\gamma},n) E\ensuremath{<}30 MeV; analyzed \ensuremath{\sigma}.}\\
\parbox[b][0.3cm]{17.7cm}{\addtolength{\parindent}{-0.2in}\href{https://www.nndc.bnl.gov/nsr/nsrlink.jsp?2002Va05,B}{2002Va05}: \ensuremath{^{\textnormal{nat}}}Ne(\ensuremath{\gamma},n) E=16-28 MeV; analyzed data from literature and deduced \ensuremath{\sigma}.}\\
\vspace{12pt}
\underline{$^{19}$Ne Levels}\\
\vspace{0.34cm}
\parbox[b][0.3cm]{17.7cm}{\addtolength{\parindent}{-0.254cm}\textit{Notes}:}\\
\parbox[b][0.3cm]{17.7cm}{\addtolength{\parindent}{-0.254cm}(1) (\href{https://www.nndc.bnl.gov/nsr/nsrlink.jsp?1975Wo06,B}{1975Wo06}) deduced the lower and upper limits on the total cross section integrated over photon energy of 18-31 MeV for}\\
\parbox[b][0.3cm]{17.7cm}{production of neutrons produced at \ensuremath{\theta}\ensuremath{_{\textnormal{lab}}}=90\ensuremath{^\circ} from the decay to \ensuremath{^{\textnormal{19}}}Ne. These limits are \ensuremath{\sim}56.9 MeV.mb and 76.8 MeV.mb,}\\
\parbox[b][0.3cm]{17.7cm}{respectively. To obtain these values, (\href{https://www.nndc.bnl.gov/nsr/nsrlink.jsp?1975Wo06,B}{1975Wo06}) assumed that all neutron-decays that did not proceed through n\ensuremath{_{\textnormal{0}}}, n\ensuremath{_{\textnormal{1}}}, or n\ensuremath{_{\textnormal{2}}} left}\\
\parbox[b][0.3cm]{17.7cm}{the \ensuremath{^{\textnormal{19}}}Ne residual nucleus at the 1.508 MeV excited state.}\\
\parbox[b][0.3cm]{17.7cm}{\addtolength{\parindent}{-0.254cm}(2) (\href{https://www.nndc.bnl.gov/nsr/nsrlink.jsp?1975Wo06,B}{1975Wo06}) found that above 18 MeV photon energy, the (\ensuremath{\gamma},n) reaction has an integrated strength of 37.2 MeV.mb, which is}\\
\parbox[b][0.3cm]{17.7cm}{approximately twice that (16 MeV.mb) of the (\ensuremath{\gamma},p) reaction.}\\
\parbox[b][0.3cm]{17.7cm}{\addtolength{\parindent}{-0.254cm}(3) The \ensuremath{^{\textnormal{20}}}Ne* states with E\ensuremath{_{\textnormal{x}}}\ensuremath{>}21 MeV decay to a number of \ensuremath{^{\textnormal{19}}}Ne* states. (\href{https://www.nndc.bnl.gov/nsr/nsrlink.jsp?1981Al05,B}{1981Al05}) did not determine which excited states}\\
\parbox[b][0.3cm]{17.7cm}{were populated from the decay of those states.}\\
\vspace{0.34cm}
\begin{longtable}{cccc@{\extracolsep{\fill}}c}
\multicolumn{2}{c}{E(level)$^{}$}&J$^{\pi}$$^{{\hyperlink{NE35LEVEL0}{a}}}$&Comments&\\[-.2cm]
\multicolumn{2}{c}{\hrulefill}&\hrulefill&\hrulefill&
\endfirsthead
\multicolumn{1}{r@{}}{0}&\multicolumn{1}{@{}l}{\ensuremath{^{{\hyperlink{NE35LEVEL1}{b}}}}}&\multicolumn{1}{l}{1/2\ensuremath{^{+}}}&\parbox[t][0.3cm]{15.256481cm}{\raggedright \ensuremath{\varepsilon}\ensuremath{_{\textnormal{K}}}/\ensuremath{\beta}\ensuremath{^{\textnormal{+}}}=9.6\ensuremath{\times}10\ensuremath{^{\textnormal{$-$4}}} \textit{3} (\href{https://www.nndc.bnl.gov/nsr/nsrlink.jsp?1972Le33,B}{1972Le33}): \ensuremath{\varepsilon}\ensuremath{_{\textnormal{K}}} refers to the coefficient of the K-electron capture.\vspace{0.1cm}}&\\
\multicolumn{1}{r@{}}{238}&\multicolumn{1}{@{}l}{\ensuremath{^{{\hyperlink{NE35LEVEL1}{b}}}}}&\multicolumn{1}{l}{5/2\ensuremath{^{+}}}&&\\
\multicolumn{1}{r@{}}{275}&\multicolumn{1}{@{}l}{\ensuremath{^{{\hyperlink{NE35LEVEL1}{b}}}}}&\multicolumn{1}{l}{1/2\ensuremath{^{-}}}&&\\
\multicolumn{1}{r@{}}{1536}&\multicolumn{1}{@{}l}{}&\multicolumn{1}{l}{3/2\ensuremath{^{+}}}&\parbox[t][0.3cm]{15.256481cm}{\raggedright E(level): From (\href{https://www.nndc.bnl.gov/nsr/nsrlink.jsp?1981Al05,B}{1981Al05}).\vspace{0.1cm}}&\\
\end{longtable}
\parbox[b][0.3cm]{17.7cm}{\makebox[1ex]{\ensuremath{^{\hypertarget{NE35LEVEL0}{a}}}} From the \ensuremath{^{\textnormal{19}}}Ne Adopted Levels.}\\
\parbox[b][0.3cm]{17.7cm}{\makebox[1ex]{\ensuremath{^{\hypertarget{NE35LEVEL1}{b}}}} From (\href{https://www.nndc.bnl.gov/nsr/nsrlink.jsp?1975Wo06,B}{1975Wo06}, \href{https://www.nndc.bnl.gov/nsr/nsrlink.jsp?1981Al05,B}{1981Al05}). The \ensuremath{^{\textnormal{19}}}Ne*(0, 238, 275) states were unresolved in (\href{https://www.nndc.bnl.gov/nsr/nsrlink.jsp?1975Wo06,B}{1975Wo06}).}\\
\vspace{0.5cm}
\clearpage
\subsection[\hspace{-0.2cm}\ensuremath{^{\textnormal{20}}}Ne(p,d)]{ }
\vspace{-27pt}
\vspace{0.3cm}
\hypertarget{NE36}{{\bf \small \underline{\ensuremath{^{\textnormal{20}}}Ne(p,d)\hspace{0.2in}\href{https://www.nndc.bnl.gov/nsr/nsrlink.jsp?2015Do10,B}{2015Do10},\href{https://www.nndc.bnl.gov/nsr/nsrlink.jsp?2017Ba42,B}{2017Ba42}}}}\\
\vspace{4pt}
\vspace{8pt}
\parbox[b][0.3cm]{17.7cm}{\addtolength{\parindent}{-0.2in}One neutron transfer reaction.}\\
\parbox[b][0.3cm]{17.7cm}{\addtolength{\parindent}{-0.2in}J\ensuremath{^{\ensuremath{\pi}}}(\ensuremath{^{\textnormal{20}}}Ne\ensuremath{_{\textnormal{g.s.}}})=0\ensuremath{^{\textnormal{+}}} and J\ensuremath{^{\ensuremath{\pi}}}(p)=1/2\ensuremath{^{\textnormal{+}}}.}\\
\parbox[b][0.3cm]{17.7cm}{\addtolength{\parindent}{-0.2in}\href{https://www.nndc.bnl.gov/nsr/nsrlink.jsp?1970MaZP,B}{1970MaZP}: \ensuremath{^{\textnormal{20}}}Ne(p,d) E=40 MeV; measured \ensuremath{\sigma}(\ensuremath{\theta}); deduced optical model parameters, \ensuremath{^{\textnormal{19}}}Ne levels, J, \ensuremath{\pi}, S, and \ensuremath{\beta}(L).}\\
\parbox[b][0.3cm]{17.7cm}{\addtolength{\parindent}{-0.2in}\href{https://www.nndc.bnl.gov/nsr/nsrlink.jsp?2014Pa58,B}{2014Pa58}, \href{https://www.nndc.bnl.gov/nsr/nsrlink.jsp?2015BaZQ,B}{2015BaZQ}, \href{https://www.nndc.bnl.gov/nsr/nsrlink.jsp?2015Ba51,B}{2015Ba51}, \href{https://www.nndc.bnl.gov/nsr/nsrlink.jsp?2017Ba42,B}{2017Ba42}: \ensuremath{^{\textnormal{20}}}Ne(p,d) E=30 MeV; beam impinged on the JENSA windowless jet gas target;}\\
\parbox[b][0.3cm]{17.7cm}{\ensuremath{^{\textnormal{nat}}}Ne gas used as target; measured the reaction products using the SIDAR Si array, which consisted of \ensuremath{\Delta}E-E telescopes covering}\\
\parbox[b][0.3cm]{17.7cm}{\ensuremath{\theta}\ensuremath{_{\textnormal{lab}}}=18\ensuremath{^\circ}{\textminus}53\ensuremath{^\circ}; measured deuteron angular distributions for strongly populated \ensuremath{^{\textnormal{19}}}Ne states; analyzed these data using finite-range}\\
\parbox[b][0.3cm]{17.7cm}{DWBA analysis performed using TWOFNR18. Assigned J\ensuremath{^{\ensuremath{\pi}}}=1/2\ensuremath{^{\textnormal{+}}} to the \ensuremath{^{\textnormal{19}}}Ne*(6282) state corresponding to a sub-threshold}\\
\parbox[b][0.3cm]{17.7cm}{resonance at E\ensuremath{_{\textnormal{c.m.}}}(\ensuremath{^{\textnormal{18}}}F+p)={\textminus}128 keV. Obtained the \ensuremath{^{\textnormal{18}}}F(p,\ensuremath{\alpha}) reaction rate; discussed the astrophysical implications.}\\
\parbox[b][0.3cm]{17.7cm}{\addtolength{\parindent}{-0.2in}\href{https://www.nndc.bnl.gov/nsr/nsrlink.jsp?2015Do10,B}{2015Do10}: \ensuremath{^{\textnormal{1}}}H(\ensuremath{^{\textnormal{20}}}Ne,d)\ensuremath{^{\textnormal{19}}}Ne*\ensuremath{\rightarrow}\ensuremath{\alpha}+\ensuremath{^{\textnormal{15}}}O E=50 MeV/nucleon; beam was injected into a storage ring while utilizing electron cooling;}\\
\parbox[b][0.3cm]{17.7cm}{an ultra pure hydrogen gas-jet target was bombarded by the beam; measured deuterons using a position sensitive \ensuremath{\Delta}E-E telescope;}\\
\parbox[b][0.3cm]{17.7cm}{measured d-\ensuremath{^{\textnormal{19}}}Ne and d-\ensuremath{^{\textnormal{15}}}O coincidence events using an array of 6 silicon pin diode detectors to detect the heavy reaction residues}\\
\parbox[b][0.3cm]{17.7cm}{and decay products. Deduced the \ensuremath{^{\textnormal{19}}}Ne excitation function. Energy resolution was \ensuremath{\approx}260 keV (FWHM).}\\
\parbox[b][0.3cm]{17.7cm}{\addtolength{\parindent}{-0.2in}\href{https://www.nndc.bnl.gov/nsr/nsrlink.jsp?2017Ch32,B}{2017Ch32}: \ensuremath{^{\textnormal{20}}}Ne(p,d) E not given; measured excitation energy spectrum using the JENSA gas-jet target with \ensuremath{^{\textnormal{nat}}}Ne gas. Results}\\
\parbox[b][0.3cm]{17.7cm}{and experimental setup are not discussed.}\\
\vspace{0.385cm}
\parbox[b][0.3cm]{17.7cm}{\addtolength{\parindent}{-0.2in}\textit{Theory}:}\\
\parbox[b][0.3cm]{17.7cm}{\addtolength{\parindent}{-0.2in}\href{https://www.nndc.bnl.gov/nsr/nsrlink.jsp?2018Ge07,B}{2018Ge07}: \ensuremath{^{\textnormal{20}}}Ne(p,d), \ensuremath{^{\textnormal{24}}}Mg(p,\ensuremath{^{\textnormal{19}}}Ne), \ensuremath{^{\textnormal{28}}}Si(p,\ensuremath{^{\textnormal{19}}}Ne) E=0.1-10 GeV/nucleon; analyzed production \ensuremath{\sigma}(E) with benchmark}\\
\parbox[b][0.3cm]{17.7cm}{parametrizations and compared with experimental data.}\\
\vspace{12pt}
\underline{$^{19}$Ne Levels}\\
\vspace{0.34cm}
\parbox[b][0.3cm]{17.7cm}{\addtolength{\parindent}{-0.254cm}The results from (\href{https://www.nndc.bnl.gov/nsr/nsrlink.jsp?2015Ba51,B}{2015Ba51}) are identical to those of (\href{https://www.nndc.bnl.gov/nsr/nsrlink.jsp?2017Ba42,B}{2017Ba42}).}\\
\vspace{0.34cm}
\begin{longtable}{ccccc@{\extracolsep{\fill}}c}
\multicolumn{2}{c}{E(level)$^{{\hyperlink{NE36LEVEL1}{b}}}$}&J$^{\pi}$$^{{\hyperlink{NE36LEVEL3}{d}}}$&L$^{{\hyperlink{NE36LEVEL2}{c}}}$&Comments&\\[-.2cm]
\multicolumn{2}{c}{\hrulefill}&\hrulefill&\hrulefill&\hrulefill&
\endfirsthead
\multicolumn{1}{r@{}}{0}&\multicolumn{1}{@{}l}{\ensuremath{^{{\hyperlink{NE36LEVEL0}{a}}}}}&\multicolumn{1}{l}{1/2\ensuremath{^{+}}}&\multicolumn{1}{l}{0}&\parbox[t][0.3cm]{12.79352cm}{\raggedright E(level): The energy obtained after calibration was E\ensuremath{_{\textnormal{x}}}=2 keV \textit{2} (\href{https://www.nndc.bnl.gov/nsr/nsrlink.jsp?2017Ba42,B}{2017Ba42}). The ground state\vspace{0.1cm}}&\\
&&&&\parbox[t][0.3cm]{12.79352cm}{\raggedright {\ }{\ }{\ }was also populated in (\href{https://www.nndc.bnl.gov/nsr/nsrlink.jsp?2015Do10,B}{2015Do10}).\vspace{0.1cm}}&\\
\multicolumn{1}{r@{}}{255}&\multicolumn{1}{@{ }l}{{\it 2}}&\multicolumn{1}{l}{(1/2\ensuremath{^{-}},3/2,5/2\ensuremath{^{+}})}&\multicolumn{1}{l}{1+2}&\parbox[t][0.3cm]{12.79352cm}{\raggedright E(level): Unresolved doublet that consists of the \ensuremath{^{\textnormal{19}}}Ne*(238, 275) states (\href{https://www.nndc.bnl.gov/nsr/nsrlink.jsp?2017Ba42,B}{2017Ba42}). These\vspace{0.1cm}}&\\
&&&&\parbox[t][0.3cm]{12.79352cm}{\raggedright {\ }{\ }{\ }states were also populated and unresolved in (\href{https://www.nndc.bnl.gov/nsr/nsrlink.jsp?2015Do10,B}{2015Do10}).\vspace{0.1cm}}&\\
&&&&\parbox[t][0.3cm]{12.79352cm}{\raggedright L: The deuteron angular distribution in (\href{https://www.nndc.bnl.gov/nsr/nsrlink.jsp?2017Ba42,B}{2017Ba42}: See Fig. 4) deviated from the DWBA\vspace{0.1cm}}&\\
&&&&\parbox[t][0.3cm]{12.79352cm}{\raggedright {\ }{\ }{\ }analysis with L=1+2 at \ensuremath{\theta}\ensuremath{_{\textnormal{c.m.}}}\ensuremath{\geq}45\ensuremath{^\circ}.\vspace{0.1cm}}&\\
\multicolumn{1}{r@{}}{1524}&\multicolumn{1}{@{ }l}{{\it 2}}&&&\parbox[t][0.3cm]{12.79352cm}{\raggedright E(level): Unresolved doublet that consists of the \ensuremath{^{\textnormal{19}}}Ne*(1507.6, 1536) states (\href{https://www.nndc.bnl.gov/nsr/nsrlink.jsp?2017Ba42,B}{2017Ba42}).\vspace{0.1cm}}&\\
&&&&\parbox[t][0.3cm]{12.79352cm}{\raggedright {\ }{\ }{\ }These states were also populated and unresolved in (\href{https://www.nndc.bnl.gov/nsr/nsrlink.jsp?2015Do10,B}{2015Do10}).\vspace{0.1cm}}&\\
\multicolumn{1}{r@{}}{1604}&\multicolumn{1}{@{ }l}{{\it 3}}&\multicolumn{1}{l}{(1/2\ensuremath{^{-}},3/2\ensuremath{^{-}})}&\multicolumn{1}{l}{1}&&\\
\multicolumn{1}{r@{}}{2792}&\multicolumn{1}{@{}l}{\ensuremath{^{{\hyperlink{NE36LEVEL0}{a}}}}}&\multicolumn{1}{l}{(7/2\ensuremath{^{+}},9/2\ensuremath{^{+}})}&\multicolumn{1}{l}{4}&\parbox[t][0.3cm]{12.79352cm}{\raggedright E(level): See also 2795 keV (\href{https://www.nndc.bnl.gov/nsr/nsrlink.jsp?2015Do10,B}{2015Do10}).\vspace{0.1cm}}&\\
&&&&\parbox[t][0.3cm]{12.79352cm}{\raggedright \ensuremath{\Gamma}(FWHM)\ensuremath{\approx}260 keV (\href{https://www.nndc.bnl.gov/nsr/nsrlink.jsp?2015Do10,B}{2015Do10}) dominated by the experimental energy resolution.\vspace{0.1cm}}&\\
\multicolumn{1}{r@{}}{4035}&\multicolumn{1}{@{ }l}{{\it 4}}&&&\parbox[t][0.3cm]{12.79352cm}{\raggedright E(level): See also \ensuremath{^{\textnormal{19}}}Ne*(4033, 4140, 4197) unresolved triplet states in (\href{https://www.nndc.bnl.gov/nsr/nsrlink.jsp?2015Do10,B}{2015Do10}).\vspace{0.1cm}}&\\
\multicolumn{1}{r@{}}{4153}&\multicolumn{1}{@{ }l}{{\it 4}}&&&&\\
\multicolumn{1}{r@{}}{4371}&\multicolumn{1}{@{ }l}{{\it 3}}&&&&\\
\multicolumn{1}{r@{}}{4556}&\multicolumn{1}{@{ }l}{{\it 3}}&&&&\\
\multicolumn{1}{r@{}}{5090}&\multicolumn{1}{@{ }l}{{\it 6}}&\multicolumn{1}{l}{(3/2\ensuremath{^{+}},5/2\ensuremath{^{+}})}&\multicolumn{1}{l}{2}&&\\
\multicolumn{1}{r@{}}{5424}&\multicolumn{1}{@{ }l}{{\it 7}}&&&&\\
\multicolumn{1}{r@{}}{5529}&\multicolumn{1}{@{ }l}{{\it 10}}&&&&\\
\multicolumn{1}{r@{}}{6017}&\multicolumn{1}{@{ }l}{{\it 3}}&&&&\\
\multicolumn{1}{r@{}}{6101}&\multicolumn{1}{@{ }l}{{\it 4}}&&&&\\
\multicolumn{1}{r@{}}{6282}&\multicolumn{1}{@{ }l}{{\it 3}}&&&\parbox[t][0.3cm]{12.79352cm}{\raggedright J\ensuremath{^{\pi}},L: (\href{https://www.nndc.bnl.gov/nsr/nsrlink.jsp?2017Ba42,B}{2017Ba42}) assigned J\ensuremath{^{\ensuremath{\pi}}}=1/2\ensuremath{^{\textnormal{+}}} with L=0 to this state and unambiguously ruled out\vspace{0.1cm}}&\\
&&&&\parbox[t][0.3cm]{12.79352cm}{\raggedright {\ }{\ }{\ }J\ensuremath{^{\ensuremath{\pi}}}=3/2\ensuremath{^{\textnormal{+}}} for this level claiming that the deuteron angular distribution populating this state is\vspace{0.1cm}}&\\
&&&&\parbox[t][0.3cm]{12.79352cm}{\raggedright {\ }{\ }{\ }inconsistent with L=2 (which could lead to J\ensuremath{^{\ensuremath{\pi}}}=3/2\ensuremath{^{\textnormal{+}}} or 5/2\ensuremath{^{\textnormal{+}}}). However, evaluator cautions\vspace{0.1cm}}&\\
&&&&\parbox[t][0.3cm]{12.79352cm}{\raggedright {\ }{\ }{\ }that the shape of L=0 angular distribution for this level from (\href{https://www.nndc.bnl.gov/nsr/nsrlink.jsp?2017Ba42,B}{2017Ba42}) is inconsistent\vspace{0.1cm}}&\\
&&&&\parbox[t][0.3cm]{12.79352cm}{\raggedright {\ }{\ }{\ }with the other L=0 distributions from that study and with what is generally expected for an\vspace{0.1cm}}&\\
&&&&\parbox[t][0.3cm]{12.79352cm}{\raggedright {\ }{\ }{\ }L=0 angular distribution (a deep first minimum). The presented DWBA calculation\vspace{0.1cm}}&\\
&&&&\parbox[t][0.3cm]{12.79352cm}{\raggedright {\ }{\ }{\ }resembles an L=1 or 2 distribution. For this reason, we cast doubt on the DWBA result for\vspace{0.1cm}}&\\
&&&&\parbox[t][0.3cm]{12.79352cm}{\raggedright {\ }{\ }{\ }this state from that study.\vspace{0.1cm}}&\\
\multicolumn{1}{r@{}}{6438}&\multicolumn{1}{@{ }l}{{\it 2}}&&&&\\
\multicolumn{1}{r@{}}{6742}&\multicolumn{1}{@{}l}{\ensuremath{^{{\hyperlink{NE36LEVEL0}{a}}}}}&\multicolumn{1}{l}{(1/2\ensuremath{^{-}},3/2\ensuremath{^{-}})}&\multicolumn{1}{l}{1}&&\\
\end{longtable}
\begin{textblock}{29}(0,27.3)
Continued on next page (footnotes at end of table)
\end{textblock}
\clearpage
\begin{longtable}{cc@{\extracolsep{\fill}}c}
\\[-.4cm]
\multicolumn{3}{c}{{\bf \small \underline{\ensuremath{^{\textnormal{20}}}Ne(p,d)\hspace{0.2in}\href{https://www.nndc.bnl.gov/nsr/nsrlink.jsp?2015Do10,B}{2015Do10},\href{https://www.nndc.bnl.gov/nsr/nsrlink.jsp?2017Ba42,B}{2017Ba42} (continued)}}}\\
\multicolumn{3}{c}{~}\\
\multicolumn{3}{c}{\underline{\ensuremath{^{19}}Ne Levels (continued)}}\\
\multicolumn{3}{c}{~}\\
\multicolumn{2}{c}{E(level)$^{{\hyperlink{NE36LEVEL1}{b}}}$}&\\[-.2cm]
\multicolumn{2}{c}{\hrulefill}&
\endhead
\multicolumn{1}{r@{}}{6865}&\multicolumn{1}{@{ }l}{{\it 3}}&\\
\multicolumn{1}{r@{}}{7067}&\multicolumn{1}{@{ }l}{{\it 2}}&\\
\end{longtable}
\parbox[b][0.3cm]{17.7cm}{\makebox[1ex]{\ensuremath{^{\hypertarget{NE36LEVEL0}{a}}}} This state was used as an internal calibration point in (\href{https://www.nndc.bnl.gov/nsr/nsrlink.jsp?2017Ba42,B}{2017Ba42}).}\\
\parbox[b][0.3cm]{17.7cm}{\makebox[1ex]{\ensuremath{^{\hypertarget{NE36LEVEL1}{b}}}} From (\href{https://www.nndc.bnl.gov/nsr/nsrlink.jsp?2017Ba42,B}{2017Ba42}). Each excitation energy quoted here has an additional 3 keV systematic uncertainty, which should be added in}\\
\parbox[b][0.3cm]{17.7cm}{{\ }{\ }quadrature.}\\
\parbox[b][0.3cm]{17.7cm}{\makebox[1ex]{\ensuremath{^{\hypertarget{NE36LEVEL2}{c}}}} From the finite-range DWBA analysis of (\href{https://www.nndc.bnl.gov/nsr/nsrlink.jsp?2017Ba42,B}{2017Ba42}) performed using the TWOFNR18 code.}\\
\parbox[b][0.3cm]{17.7cm}{\makebox[1ex]{\ensuremath{^{\hypertarget{NE36LEVEL3}{d}}}} Not provided by (\href{https://www.nndc.bnl.gov/nsr/nsrlink.jsp?2017Ba42,B}{2017Ba42}). The given J\ensuremath{^{\ensuremath{\pi}}} values are deduced by the evaluator from the given L-values.}\\
\vspace{0.5cm}
\clearpage
\subsection[\hspace{-0.2cm}\ensuremath{^{\textnormal{20}}}Ne(d,t)]{ }
\vspace{-27pt}
\vspace{0.3cm}
\hypertarget{NE37}{{\bf \small \underline{\ensuremath{^{\textnormal{20}}}Ne(d,t)\hspace{0.2in}\href{https://www.nndc.bnl.gov/nsr/nsrlink.jsp?1998Ut02,B}{1998Ut02},\href{https://www.nndc.bnl.gov/nsr/nsrlink.jsp?2002Ku12,B}{2002Ku12}}}}\\
\vspace{4pt}
\vspace{8pt}
\parbox[b][0.3cm]{17.7cm}{\addtolength{\parindent}{-0.2in}One neutron transfer reaction.}\\
\parbox[b][0.3cm]{17.7cm}{\addtolength{\parindent}{-0.2in}J\ensuremath{^{\ensuremath{\pi}}}(\ensuremath{^{\textnormal{20}}}Ne\ensuremath{_{\textnormal{g.s.}}})=0\ensuremath{^{\textnormal{+}}} and J\ensuremath{^{\ensuremath{\pi}}}(\ensuremath{^{\textnormal{2}}}H\ensuremath{_{\textnormal{g.s.}}})=1\ensuremath{^{\textnormal{+}}}.}\\
\parbox[b][0.3cm]{17.7cm}{\addtolength{\parindent}{-0.2in}\href{https://www.nndc.bnl.gov/nsr/nsrlink.jsp?1998Ut02,B}{1998Ut02}: \ensuremath{^{\textnormal{20}}}Ne(d,t) E=30 MeV; measured the reaction products using 7 Si surface barrier detectors covering \ensuremath{\theta}\ensuremath{_{\textnormal{lab}}}=12.5\ensuremath{^\circ}{\textminus}45\ensuremath{^\circ}. In a}\\
\parbox[b][0.3cm]{17.7cm}{separate experiment, those authors used a split-pole spectrograph at \ensuremath{\theta}\ensuremath{_{\textnormal{lab}}}=20\ensuremath{^\circ} to obtain a high resolution spectrum. Deduced \ensuremath{^{\textnormal{19}}}Ne}\\
\parbox[b][0.3cm]{17.7cm}{level-energies.}\\
\parbox[b][0.3cm]{17.7cm}{\addtolength{\parindent}{-0.2in}K. Kumagai, M.Sc. Thesis, Tohoku University (1999), unpublished, \href{https://www.nndc.bnl.gov/nsr/nsrlink.jsp?2002Ku12,B}{2002Ku12}: \ensuremath{^{\textnormal{20}}}Ne(d,t) and \ensuremath{^{\textnormal{20}}}Ne(d,\ensuremath{^{\textnormal{3}}}He); studied the}\\
\parbox[b][0.3cm]{17.7cm}{single-particle nature of those \ensuremath{^{\textnormal{19}}}Ne* states with E\ensuremath{_{\textnormal{x}}}\ensuremath{<}5 MeV, as well as their analog levels in \ensuremath{^{\textnormal{19}}}F using the DWBA analysis of the}\\
\parbox[b][0.3cm]{17.7cm}{triton angular distributions. The experimental setup is not described in (\href{https://www.nndc.bnl.gov/nsr/nsrlink.jsp?2002Ku12,B}{2002Ku12}), where a brief summary of the results for the}\\
\parbox[b][0.3cm]{17.7cm}{4033-keV state is given.}\\
\vspace{12pt}
\underline{$^{19}$Ne Levels}\\
\begin{longtable}{ccccccc@{\extracolsep{\fill}}c}
\multicolumn{2}{c}{E(level)$^{{\hyperlink{NE37LEVEL0}{a}}}$}&J$^{\pi}$$^{}$&L$^{}$&\multicolumn{2}{c}{S\ensuremath{_{\textnormal{n}}}$^{}$}&Comments&\\[-.2cm]
\multicolumn{2}{c}{\hrulefill}&\hrulefill&\hrulefill&\multicolumn{2}{c}{\hrulefill}&\hrulefill&
\endfirsthead
\multicolumn{1}{r@{}}{4033}&\multicolumn{1}{@{}l}{}&\multicolumn{1}{l}{(3/2\ensuremath{^{+}})}&\multicolumn{1}{l}{(2)}&\multicolumn{1}{r@{}}{0}&\multicolumn{1}{@{.}l}{04}&\parbox[t][0.3cm]{13.039801cm}{\raggedright E(level): From (\href{https://www.nndc.bnl.gov/nsr/nsrlink.jsp?2002Ku12,B}{2002Ku12}): State was very weakly populated suggesting an insignificant \textit{d}\ensuremath{_{\textnormal{3/2}}}\vspace{0.1cm}}&\\
&&&&&&\parbox[t][0.3cm]{13.039801cm}{\raggedright {\ }{\ }{\ }single-particle component (see text).\vspace{0.1cm}}&\\
&&&&&&\parbox[t][0.3cm]{13.039801cm}{\raggedright J\ensuremath{^{\pi}},L: From the DWBA analysis of (\href{https://www.nndc.bnl.gov/nsr/nsrlink.jsp?2002Ku12,B}{2002Ku12}). Considering that this state was populated very\vspace{0.1cm}}&\\
&&&&&&\parbox[t][0.3cm]{13.039801cm}{\raggedright {\ }{\ }{\ }weakly and because the DWBA results are not presented in (\href{https://www.nndc.bnl.gov/nsr/nsrlink.jsp?2002Ku12,B}{2002Ku12}), the evaluator\vspace{0.1cm}}&\\
&&&&&&\parbox[t][0.3cm]{13.039801cm}{\raggedright {\ }{\ }{\ }considered them tentative.\vspace{0.1cm}}&\\
&&&&&&\parbox[t][0.3cm]{13.039801cm}{\raggedright S\ensuremath{_{\textnormal{n}}}: This is the neutron spectroscopic factor from the \ensuremath{^{\textnormal{20}}}Ne(d,t) reaction obtained from the\vspace{0.1cm}}&\\
&&&&&&\parbox[t][0.3cm]{13.039801cm}{\raggedright {\ }{\ }{\ }DWBA analysis of (\href{https://www.nndc.bnl.gov/nsr/nsrlink.jsp?2002Ku12,B}{2002Ku12}, see text).\vspace{0.1cm}}&\\
&&&&&&\parbox[t][0.3cm]{13.039801cm}{\raggedright (\href{https://www.nndc.bnl.gov/nsr/nsrlink.jsp?2002Ku12,B}{2002Ku12}) concluded that this state has a 5p-2h configuration due to its very weak excitation\vspace{0.1cm}}&\\
&&&&&&\parbox[t][0.3cm]{13.039801cm}{\raggedright {\ }{\ }{\ }in the \ensuremath{^{\textnormal{20}}}Ne(d,t), but strong excitation in the \ensuremath{^{\textnormal{21}}}Ne(p,t) reactions.\vspace{0.1cm}}&\\
\multicolumn{1}{r@{}}{4549}&\multicolumn{1}{@{}l}{}&&&&&&\\
\multicolumn{1}{r@{}}{4600}&\multicolumn{1}{@{}l}{}&&&&&&\\
\multicolumn{1}{r@{}}{6013}&\multicolumn{1}{@{}l}{}&&&&&&\\
\multicolumn{1}{r@{}}{6741}&\multicolumn{1}{@{}l}{}&&&&&&\\
\end{longtable}
\parbox[b][0.3cm]{17.7cm}{\makebox[1ex]{\ensuremath{^{\hypertarget{NE37LEVEL0}{a}}}} From Fig. 5 of (\href{https://www.nndc.bnl.gov/nsr/nsrlink.jsp?1998Ut02,B}{1998Ut02}) unless otherwise noted. Excitation energies from (\href{https://www.nndc.bnl.gov/nsr/nsrlink.jsp?1998Ut02,B}{1998Ut02}) are measured at \ensuremath{\theta}\ensuremath{_{\textnormal{lab}}}=20\ensuremath{^\circ}.}\\
\vspace{0.5cm}
\clearpage
\subsection[\hspace{-0.2cm}\ensuremath{^{\textnormal{20}}}Ne(\ensuremath{^{\textnormal{3}}}He,\ensuremath{\alpha}),(\ensuremath{^{\textnormal{3}}}He,\ensuremath{\alpha}\ensuremath{\gamma})]{ }
\vspace{-27pt}
\vspace{0.3cm}
\hypertarget{NE38}{{\bf \small \underline{\ensuremath{^{\textnormal{20}}}Ne(\ensuremath{^{\textnormal{3}}}He,\ensuremath{\alpha}),(\ensuremath{^{\textnormal{3}}}He,\ensuremath{\alpha}\ensuremath{\gamma})\hspace{0.2in}\href{https://www.nndc.bnl.gov/nsr/nsrlink.jsp?1970Ga18,B}{1970Ga18},\href{https://www.nndc.bnl.gov/nsr/nsrlink.jsp?1972Ha03,B}{1972Ha03},\href{https://www.nndc.bnl.gov/nsr/nsrlink.jsp?2023Po03,B}{2023Po03}}}}\\
\vspace{4pt}
\vspace{8pt}
\parbox[b][0.3cm]{17.7cm}{\addtolength{\parindent}{-0.2in}One neutron transfer reaction.}\\
\parbox[b][0.3cm]{17.7cm}{\addtolength{\parindent}{-0.2in}J\ensuremath{^{\ensuremath{\pi}}}(\ensuremath{^{\textnormal{20}}}Ne\ensuremath{_{\textnormal{g.s.}}})=0\ensuremath{^{\textnormal{+}}} and J\ensuremath{^{\ensuremath{\pi}}}(\ensuremath{^{\textnormal{3}}}He\ensuremath{_{\textnormal{g.s.}}})=1/2\ensuremath{^{\textnormal{+}}}.}\\
\parbox[b][0.3cm]{17.7cm}{\addtolength{\parindent}{-0.2in}\href{https://www.nndc.bnl.gov/nsr/nsrlink.jsp?1967Be14,B}{1967Be14}: \ensuremath{^{\textnormal{20}}}Ne(\ensuremath{^{\textnormal{3}}}He,\ensuremath{\alpha}\ensuremath{\gamma}) E=3.1-3.3 MeV; measured \ensuremath{\gamma}-\ensuremath{^{\textnormal{3}}}He coincidence events using a pulsed beam (every 350 ns) and 2 NaI(Tl)}\\
\parbox[b][0.3cm]{17.7cm}{scintillators placed at \ensuremath{\theta}\ensuremath{_{\textnormal{lab}}}=\ensuremath{\pm}90\ensuremath{^\circ} to detect the \ensuremath{\gamma} rays from the \ensuremath{^{\textnormal{19}}}Ne*(241, 280) levels. The detector at \ensuremath{\theta}\ensuremath{_{\textnormal{lab}}}={\textminus}90\ensuremath{^\circ} was used as a}\\
\parbox[b][0.3cm]{17.7cm}{veto for the annihilation \ensuremath{\gamma} rays from \ensuremath{\beta}-decay of \ensuremath{^{\textnormal{19}}}Ne. Measured lifetimes of the \ensuremath{^{\textnormal{19}}}Ne*(241, 280) states as \ensuremath{\tau}=26.6 ns \textit{12} and \ensuremath{\tau}\ensuremath{\leq}5}\\
\parbox[b][0.3cm]{17.7cm}{ns, respectively.}\\
\parbox[b][0.3cm]{17.7cm}{\addtolength{\parindent}{-0.2in}\href{https://www.nndc.bnl.gov/nsr/nsrlink.jsp?1967Gr04,B}{1967Gr04}: \ensuremath{^{\textnormal{20}}}Ne(\ensuremath{^{\textnormal{3}}}He,\ensuremath{\alpha}) E=5.90, 6 MeV; measured \ensuremath{\alpha} particles from the reaction using a Si surface barrier detector placed at}\\
\parbox[b][0.3cm]{17.7cm}{\ensuremath{\theta}\ensuremath{_{\textnormal{lab}}}=130\ensuremath{^\circ}, 145\ensuremath{^\circ}, 165\ensuremath{^\circ}, and 167.5\ensuremath{^\circ}. Deduced \ensuremath{^{\textnormal{19}}}Ne level-energies for the \ensuremath{^{\textnormal{19}}}Ne*(4013, 4152, 4344, 4547, 4689, 5077) states. The}\\
\parbox[b][0.3cm]{17.7cm}{uncertainties in these level-energies were assigned to be \ensuremath{\pm}15 keV. Preliminary mirror level analysis was performed.}\\
\parbox[b][0.3cm]{17.7cm}{\addtolength{\parindent}{-0.2in}\href{https://www.nndc.bnl.gov/nsr/nsrlink.jsp?1967Ol05,B}{1967Ol05}: \ensuremath{^{\textnormal{20}}}Ne(\ensuremath{^{\textnormal{3}}}He,\ensuremath{\alpha}\ensuremath{\gamma}) E\ensuremath{_{\textnormal{eff}}}=6.1, 6.3 MeV; measured \ensuremath{\gamma} rays from decay of the \ensuremath{^{\textnormal{19}}}Ne excited states using a Ge(Li) detector at}\\
\parbox[b][0.3cm]{17.7cm}{\ensuremath{\theta}\ensuremath{_{\textnormal{lab}}}=90\ensuremath{^\circ}. Deduced \ensuremath{^{\textnormal{19}}}Ne levels. Deduced decay modes for the observed states by measuring \ensuremath{\gamma}-\ensuremath{\gamma} coincidence events using two}\\
\parbox[b][0.3cm]{17.7cm}{NaI(Tl) and a NaI(Tl) and a Ge(Li) detector placed at \ensuremath{\theta}\ensuremath{_{\textnormal{lab}}}=\ensuremath{\pm}90\ensuremath{^\circ}. Mirror level analysis was performed for the measured bound}\\
\parbox[b][0.3cm]{17.7cm}{states of \ensuremath{^{\textnormal{19}}}Ne.}\\
\parbox[b][0.3cm]{17.7cm}{\addtolength{\parindent}{-0.2in}\href{https://www.nndc.bnl.gov/nsr/nsrlink.jsp?1969Ba62,B}{1969Ba62}: \ensuremath{^{\textnormal{20}}}Ne(\ensuremath{^{\textnormal{3}}}He,\ensuremath{^{\textnormal{3}}}He) and \ensuremath{^{\textnormal{20}}}Ne(\ensuremath{^{\textnormal{3}}}He,\ensuremath{\alpha}) E=10, 15 MeV; measured the \ensuremath{^{\textnormal{3}}}He and \ensuremath{\alpha} angular distributions populating the}\\
\parbox[b][0.3cm]{17.7cm}{\ensuremath{^{\textnormal{20}}}Ne\ensuremath{_{\textnormal{g.s.}}} and \ensuremath{^{\textnormal{19}}}Ne(0, 278) levels, respectively, at \ensuremath{\theta}\ensuremath{_{\textnormal{c.m.}}}=20\ensuremath{^\circ}{\textminus}100\ensuremath{^\circ}; deduced the entrance channel$'$s optical model parameters from}\\
\parbox[b][0.3cm]{17.7cm}{the elastic scattering data; using these parameters, they performed a zero-range DWBA analysis for the (\ensuremath{^{\textnormal{3}}}He,\ensuremath{\alpha}) data to determine}\\
\parbox[b][0.3cm]{17.7cm}{the L and J\ensuremath{^{\ensuremath{\pi}}} values for the \ensuremath{^{\textnormal{19}}}Ne*(278) state. Discussed the DWBA fits to the \ensuremath{\alpha} angular distribution corresponding to the}\\
\parbox[b][0.3cm]{17.7cm}{\ensuremath{^{\textnormal{19}}}Ne*(278) state with L=2, 3, 4 and 5.}\\
\parbox[b][0.3cm]{17.7cm}{\addtolength{\parindent}{-0.2in}\href{https://www.nndc.bnl.gov/nsr/nsrlink.jsp?1970Ar25,B}{1970Ar25}: \ensuremath{^{\textnormal{20}}}Ne(\ensuremath{^{\textnormal{3}}}He,\ensuremath{\alpha}) E\ensuremath{\approx}35 MeV; measured \ensuremath{\sigma}(E\ensuremath{_{\ensuremath{\alpha}}},\ensuremath{\theta}). Deduced \ensuremath{^{\textnormal{19}}}Ne levels, J, \ensuremath{\pi}, L, and S.}\\
\parbox[b][0.3cm]{17.7cm}{\addtolength{\parindent}{-0.2in}\href{https://www.nndc.bnl.gov/nsr/nsrlink.jsp?1970Ga18,B}{1970Ga18}: \ensuremath{^{\textnormal{20}}}Ne(\ensuremath{^{\textnormal{3}}}He,\ensuremath{\alpha}) E=15 MeV; momentum analyzed and measured light reaction products using a magnetic spectrograph and}\\
\parbox[b][0.3cm]{17.7cm}{its focal plane detector placed at \ensuremath{\theta}\ensuremath{_{\textnormal{lab}}}=7.5\ensuremath{^\circ}{\textminus}87\ensuremath{^\circ} with a \ensuremath{\Delta}\ensuremath{\theta}=3.75\ensuremath{^\circ} interval. Deduced numerous \ensuremath{^{\textnormal{19}}}Ne levels from ground state to}\\
\parbox[b][0.3cm]{17.7cm}{E\ensuremath{_{\textnormal{x}}}=7064 MeV, including 16 new states with E\ensuremath{_{\textnormal{x}}}=4-7 MeV. Measured the \ensuremath{\alpha} angular distributions corresponding to the states}\\
\parbox[b][0.3cm]{17.7cm}{observed. Performed a zero-range DWBA analysis using the JULIE code and deduced L, J\ensuremath{^{\ensuremath{\pi}}}, and S for strong transitions. Discussed}\\
\parbox[b][0.3cm]{17.7cm}{and suggested mirror levels and members of K\ensuremath{^{\ensuremath{\pi}}}=1/2\ensuremath{^{-}} and K\ensuremath{^{\ensuremath{\pi}}}=1/2\ensuremath{^{\textnormal{+}}} rotational bands.}\\
\parbox[b][0.3cm]{17.7cm}{\addtolength{\parindent}{-0.2in}\href{https://www.nndc.bnl.gov/nsr/nsrlink.jsp?1970Ke24,B}{1970Ke24}; \href{https://www.nndc.bnl.gov/nsr/nsrlink.jsp?1971HaYA,B}{1971HaYA}; \href{https://www.nndc.bnl.gov/nsr/nsrlink.jsp?1972Ha03,B}{1972Ha03}: \ensuremath{^{\textnormal{20}}}Ne(\ensuremath{^{\textnormal{3}}}He,\ensuremath{\alpha}) E=18 MeV; measured \ensuremath{\alpha}-particles from the reaction using a surface barrier}\\
\parbox[b][0.3cm]{17.7cm}{detector placed at \ensuremath{\theta}\ensuremath{_{\textnormal{lab}}}=27.8\ensuremath{^\circ}, 32.8\ensuremath{^\circ}, 37.5\ensuremath{^\circ}, 47.5\ensuremath{^\circ}, 52.5\ensuremath{^\circ}, and 57.5\ensuremath{^\circ}; measured \ensuremath{\sigma}(E\ensuremath{_{\ensuremath{\alpha}}},\ensuremath{\theta}). Deduced 38 \ensuremath{^{\textnormal{19}}}Ne levels, 21 of which are}\\
\parbox[b][0.3cm]{17.7cm}{in E\ensuremath{_{\textnormal{x}}}=7.1\ensuremath{\sim}10.6 MeV range; performed zero-range DWBA analysis with zero lower cutoff radius using JULIE and extracted J, \ensuremath{\pi},}\\
\parbox[b][0.3cm]{17.7cm}{and relative spectroscopic factors for some of the observed states; compared the S\ensuremath{_{\textnormal{rel,exp}}} with the predictions of the weak coupling}\\
\parbox[b][0.3cm]{17.7cm}{model using the relative spectroscopic factor sum rule.}\\
\parbox[b][0.3cm]{17.7cm}{\addtolength{\parindent}{-0.2in}\href{https://www.nndc.bnl.gov/nsr/nsrlink.jsp?2023Po03,B}{2023Po03}: \ensuremath{^{\textnormal{20}}}Ne(\ensuremath{^{\textnormal{3}}}He,\ensuremath{\alpha}) E=21 MeV; momentum analyzed and detected the \ensuremath{\alpha} particles from the reaction using an Enge split-pole}\\
\parbox[b][0.3cm]{17.7cm}{spectrograph together with its focal plane detector placed at \ensuremath{\theta}\ensuremath{_{\textnormal{lab}}}=9\ensuremath{^\circ}, 12\ensuremath{^\circ}, 20\ensuremath{^\circ}, 22\ensuremath{^\circ}, 25\ensuremath{^\circ}, and 27\ensuremath{^\circ}. Measured \ensuremath{\alpha} angular distributions.}\\
\parbox[b][0.3cm]{17.7cm}{Deduced \ensuremath{^{\textnormal{19}}}Ne levels and J, \ensuremath{\pi}, and L using a finite-range DWBA analysis via the FRESCO code. Mirror levels were assigned to}\\
\parbox[b][0.3cm]{17.7cm}{the \ensuremath{^{\textnormal{19}}}Ne* states below 6 MeV excitation energy. Deduced the astrophysical S-factor for the \ensuremath{^{\textnormal{18}}}F(p,\ensuremath{\alpha}) reaction at E\ensuremath{_{\textnormal{c.m.}}}\ensuremath{<}1.7 MeV}\\
\parbox[b][0.3cm]{17.7cm}{using R-matrix analysis via AZURE2 with radius of 4.53 fm; deduced the \ensuremath{^{\textnormal{18}}}F(p,\ensuremath{\alpha}) reaction rate at T\ensuremath{\leq}0.5 GK; discussed the}\\
\parbox[b][0.3cm]{17.7cm}{interferences between resonances and the astrophysical implications.}\\
\vspace{0.385cm}
\parbox[b][0.3cm]{17.7cm}{\addtolength{\parindent}{-0.2in}\textit{Theory}:}\\
\parbox[b][0.3cm]{17.7cm}{\addtolength{\parindent}{-0.2in}\href{https://www.nndc.bnl.gov/nsr/nsrlink.jsp?1972En03,B}{1972En03}: Calculated \ensuremath{^{\textnormal{19}}}Ne \ensuremath{\gamma}-transition rates, B(\ensuremath{\lambda}), \ensuremath{\delta} (mixing ratios), level-widths, and spectroscopic factors for the levels}\\
\parbox[b][0.3cm]{17.7cm}{populated in the \ensuremath{^{\textnormal{20}}}Ne(\ensuremath{^{\textnormal{3}}}He,\ensuremath{\alpha}) reaction (based on the data of \href{https://www.nndc.bnl.gov/nsr/nsrlink.jsp?1970Ga18,B}{1970Ga18}) using the weak coupling model.}\\
\parbox[b][0.3cm]{17.7cm}{\addtolength{\parindent}{-0.2in}\href{https://www.nndc.bnl.gov/nsr/nsrlink.jsp?1972Ga14,B}{1972Ga14}: Calculated \ensuremath{^{\textnormal{19}}}F and \ensuremath{^{\textnormal{19}}}Ne mirror levels with isospin T=1/2 and up to E\ensuremath{_{\textnormal{x}}}=6 MeV; calculated B(M1), B(E2), and S based}\\
\parbox[b][0.3cm]{17.7cm}{on the Coriolis-mixing amplitudes.}\\
\parbox[b][0.3cm]{17.7cm}{\addtolength{\parindent}{-0.2in}\href{https://www.nndc.bnl.gov/nsr/nsrlink.jsp?1974Ga28,B}{1974Ga28}: Performed finite range, non-local DWBA calculations using the optical model parameters set 1 of (\href{https://www.nndc.bnl.gov/nsr/nsrlink.jsp?1970Ga18,B}{1970Ga18}) and}\\
\parbox[b][0.3cm]{17.7cm}{corrected the spectroscopic factors that were obtained by (\href{https://www.nndc.bnl.gov/nsr/nsrlink.jsp?1970Ga18,B}{1970Ga18}).}\\
\vspace{12pt}
\underline{$^{19}$Ne Levels}\\
\begin{longtable}{ccccccc@{\extracolsep{\fill}}c}
\multicolumn{2}{c}{E(level)$^{{\hyperlink{NE38LEVEL0}{a}}}$}&J$^{\pi}$$^{}$&L$^{}$&\multicolumn{2}{c}{C\ensuremath{^{\textnormal{2}}}S\ensuremath{_{\textnormal{exp}}}$^{{\hyperlink{NE38LEVEL5}{f}}}$}&Comments&\\[-.2cm]
\multicolumn{2}{c}{\hrulefill}&\hrulefill&\hrulefill&\multicolumn{2}{c}{\hrulefill}&\hrulefill&
\endfirsthead
\multicolumn{1}{r@{}}{0}&\multicolumn{1}{@{}l}{}&\multicolumn{1}{l}{1/2\ensuremath{^{+}}}&\multicolumn{1}{l}{0}&\multicolumn{1}{r@{}}{0}&\multicolumn{1}{@{.}l}{12}&\parbox[t][0.3cm]{12.789421cm}{\raggedright E(level): From (\href{https://www.nndc.bnl.gov/nsr/nsrlink.jsp?1967Gr04,B}{1967Gr04}, \href{https://www.nndc.bnl.gov/nsr/nsrlink.jsp?1967Ol05,B}{1967Ol05}, \href{https://www.nndc.bnl.gov/nsr/nsrlink.jsp?1969Ba62,B}{1969Ba62}, \href{https://www.nndc.bnl.gov/nsr/nsrlink.jsp?1970Ga18,B}{1970Ga18}, \href{https://www.nndc.bnl.gov/nsr/nsrlink.jsp?1972Ha03,B}{1972Ha03}: See Fig. 4).\vspace{0.1cm}}&\\
&&&&&&\parbox[t][0.3cm]{12.789421cm}{\raggedright J\ensuremath{^{\pi}}: From the zero-range DWBA analyses of (\href{https://www.nndc.bnl.gov/nsr/nsrlink.jsp?1969Ba62,B}{1969Ba62}: L=0) and (\href{https://www.nndc.bnl.gov/nsr/nsrlink.jsp?1970Ga18,B}{1970Ga18}, \href{https://www.nndc.bnl.gov/nsr/nsrlink.jsp?1972Ha03,B}{1972Ha03});\vspace{0.1cm}}&\\
&&&&&&\parbox[t][0.3cm]{12.789421cm}{\raggedright {\ }{\ }{\ }and the mirror analysis of (\href{https://www.nndc.bnl.gov/nsr/nsrlink.jsp?1967Ol05,B}{1967Ol05}). We note that the last two studies used JULIE code\vspace{0.1cm}}&\\
&&&&&&\parbox[t][0.3cm]{12.789421cm}{\raggedright {\ }{\ }{\ }and performed DWBA with zero lower cutoff radius. In both cases, the agreement between\vspace{0.1cm}}&\\
&&&&&&\parbox[t][0.3cm]{12.789421cm}{\raggedright {\ }{\ }{\ }the DWBA fits and the data is not good.\vspace{0.1cm}}&\\
\end{longtable}
\begin{textblock}{29}(0,27.3)
Continued on next page (footnotes at end of table)
\end{textblock}
\clearpage
\begin{longtable}{ccccccccc@{\extracolsep{\fill}}c}
\\[-.4cm]
\multicolumn{10}{c}{{\bf \small \underline{\ensuremath{^{\textnormal{20}}}Ne(\ensuremath{^{\textnormal{3}}}He,\ensuremath{\alpha}),(\ensuremath{^{\textnormal{3}}}He,\ensuremath{\alpha}\ensuremath{\gamma})\hspace{0.2in}\href{https://www.nndc.bnl.gov/nsr/nsrlink.jsp?1970Ga18,B}{1970Ga18},\href{https://www.nndc.bnl.gov/nsr/nsrlink.jsp?1972Ha03,B}{1972Ha03},\href{https://www.nndc.bnl.gov/nsr/nsrlink.jsp?2023Po03,B}{2023Po03} (continued)}}}\\
\multicolumn{10}{c}{~}\\
\multicolumn{10}{c}{\underline{\ensuremath{^{19}}Ne Levels (continued)}}\\
\multicolumn{10}{c}{~}\\
\multicolumn{2}{c}{E(level)$^{{\hyperlink{NE38LEVEL0}{a}}}$}&J$^{\pi}$$^{}$&\multicolumn{2}{c}{T$_{1/2}$$^{}$}&L$^{}$&\multicolumn{2}{c}{C\ensuremath{^{\textnormal{2}}}S\ensuremath{_{\textnormal{exp}}}$^{{\hyperlink{NE38LEVEL5}{f}}}$}&Comments&\\[-.2cm]
\multicolumn{2}{c}{\hrulefill}&\hrulefill&\multicolumn{2}{c}{\hrulefill}&\hrulefill&\multicolumn{2}{c}{\hrulefill}&\hrulefill&
\endhead
&&&&&&&&\parbox[t][0.3cm]{9.42461cm}{\raggedright L: From the zero-range DWBA analyses of (\href{https://www.nndc.bnl.gov/nsr/nsrlink.jsp?1969Ba62,B}{1969Ba62}), where the\vspace{0.1cm}}&\\
&&&&&&&&\parbox[t][0.3cm]{9.42461cm}{\raggedright {\ }{\ }{\ }analysis not is presented, but L=0 is mentioned; (\href{https://www.nndc.bnl.gov/nsr/nsrlink.jsp?1970Ga18,B}{1970Ga18}: L=0);\vspace{0.1cm}}&\\
&&&&&&&&\parbox[t][0.3cm]{9.42461cm}{\raggedright {\ }{\ }{\ }and (\href{https://www.nndc.bnl.gov/nsr/nsrlink.jsp?1972Ha03,B}{1972Ha03}: See Table II).\vspace{0.1cm}}&\\
&&&&&&&&\parbox[t][0.3cm]{9.42461cm}{\raggedright C\ensuremath{^{\textnormal{2}}}S: See also 0.20 calculated by (\href{https://www.nndc.bnl.gov/nsr/nsrlink.jsp?1974Ga28,B}{1974Ga28}: Tables 4 and 5).\vspace{0.1cm}}&\\
\multicolumn{1}{r@{}}{238}&\multicolumn{1}{@{.}l}{4 {\it 3}}&\multicolumn{1}{l}{5/2\ensuremath{^{+}}}&\multicolumn{1}{r@{}}{18}&\multicolumn{1}{@{.}l}{44 ns {\it 83}}&\multicolumn{1}{l}{2}&\multicolumn{1}{r@{}}{1}&\multicolumn{1}{@{.}l}{04}&\parbox[t][0.3cm]{9.42461cm}{\raggedright E(level): Weighted average of 238.4 keV \textit{3} from a least-squares fit\vspace{0.1cm}}&\\
&&&&&&&&\parbox[t][0.3cm]{9.42461cm}{\raggedright {\ }{\ }{\ }to the E\ensuremath{_{\ensuremath{\gamma}}} value from (\href{https://www.nndc.bnl.gov/nsr/nsrlink.jsp?1967Ol05,B}{1967Ol05}); 238 keV \textit{10} (\href{https://www.nndc.bnl.gov/nsr/nsrlink.jsp?1970Ga18,B}{1970Ga18}); and\vspace{0.1cm}}&\\
&&&&&&&&\parbox[t][0.3cm]{9.42461cm}{\raggedright {\ }{\ }{\ }239.7 keV \textit{20} (\href{https://www.nndc.bnl.gov/nsr/nsrlink.jsp?1972Ha03,B}{1972Ha03}).\vspace{0.1cm}}&\\
&&&&&&&&\parbox[t][0.3cm]{9.42461cm}{\raggedright E(level): See also 241 keV (\href{https://www.nndc.bnl.gov/nsr/nsrlink.jsp?1967Be14,B}{1967Be14}) and 241 keV \textit{15}\vspace{0.1cm}}&\\
&&&&&&&&\parbox[t][0.3cm]{9.42461cm}{\raggedright {\ }{\ }{\ }(\href{https://www.nndc.bnl.gov/nsr/nsrlink.jsp?1967Gr04,B}{1967Gr04}), which is an unresolved doublet.\vspace{0.1cm}}&\\
&&&&&&&&\parbox[t][0.3cm]{9.42461cm}{\raggedright T\ensuremath{_{1/2}}: From \ensuremath{\tau}=26.6 ns \textit{12} (\href{https://www.nndc.bnl.gov/nsr/nsrlink.jsp?1967Be14,B}{1967Be14}).\vspace{0.1cm}}&\\
&&&&&&&&\parbox[t][0.3cm]{9.42461cm}{\raggedright J\ensuremath{^{\pi}},L: From the zero-range DWBA analyses of (1) (\href{https://www.nndc.bnl.gov/nsr/nsrlink.jsp?1970Ga18,B}{1970Ga18}):\vspace{0.1cm}}&\\
&&&&&&&&\parbox[t][0.3cm]{9.42461cm}{\raggedright {\ }{\ }{\ }L=2, J\ensuremath{^{\ensuremath{\pi}}}=5/2\ensuremath{^{\textnormal{+}}} using JULIE and performed with zero lower cutoff\vspace{0.1cm}}&\\
&&&&&&&&\parbox[t][0.3cm]{9.42461cm}{\raggedright {\ }{\ }{\ }radius; and (2) (\href{https://www.nndc.bnl.gov/nsr/nsrlink.jsp?1972Ha03,B}{1972Ha03}): L=(2) and J\ensuremath{^{\ensuremath{\pi}}}=(5/2\ensuremath{^{\textnormal{+}}}) using JULIE\vspace{0.1cm}}&\\
&&&&&&&&\parbox[t][0.3cm]{9.42461cm}{\raggedright {\ }{\ }{\ }with zero lower cutoff radius, see Table II. We made those J\ensuremath{^{\ensuremath{\pi}}} and\vspace{0.1cm}}&\\
&&&&&&&&\parbox[t][0.3cm]{9.42461cm}{\raggedright {\ }{\ }{\ }L assignments tentative since the DWBA fit is not presented.\vspace{0.1cm}}&\\
&&&&&&&&\parbox[t][0.3cm]{9.42461cm}{\raggedright J\ensuremath{^{\pi}}: See also J\ensuremath{^{\ensuremath{\pi}}}=5/2\ensuremath{^{\textnormal{+}}} from the mirror analysis of (\href{https://www.nndc.bnl.gov/nsr/nsrlink.jsp?1967Ol05,B}{1967Ol05}).\vspace{0.1cm}}&\\
&&&&&&&&\parbox[t][0.3cm]{9.42461cm}{\raggedright C\ensuremath{^{\textnormal{2}}}S: See also 0.95 calculated by (\href{https://www.nndc.bnl.gov/nsr/nsrlink.jsp?1974Ga28,B}{1974Ga28}: Table 5).\vspace{0.1cm}}&\\
\multicolumn{1}{r@{}}{274}&\multicolumn{1}{@{.}l}{8 {\it 3}}&\multicolumn{1}{l}{1/2\ensuremath{^{-}}}&\multicolumn{1}{r@{}}{$\leq$3}&\multicolumn{1}{@{.}l}{5 ns}&\multicolumn{1}{l}{1}&\multicolumn{1}{r@{}}{1}&\multicolumn{1}{@{.}l}{96}&\parbox[t][0.3cm]{9.42461cm}{\raggedright E(level): Weighted average of 274.8 keV \textit{3} from a least-squares fit\vspace{0.1cm}}&\\
&&&&&&&&\parbox[t][0.3cm]{9.42461cm}{\raggedright {\ }{\ }{\ }to the E\ensuremath{_{\ensuremath{\gamma}}} value from (\href{https://www.nndc.bnl.gov/nsr/nsrlink.jsp?1967Ol05,B}{1967Ol05}); 273 keV \textit{10} (\href{https://www.nndc.bnl.gov/nsr/nsrlink.jsp?1970Ga18,B}{1970Ga18}); and\vspace{0.1cm}}&\\
&&&&&&&&\parbox[t][0.3cm]{9.42461cm}{\raggedright {\ }{\ }{\ }276.6 keV \textit{20} (\href{https://www.nndc.bnl.gov/nsr/nsrlink.jsp?1972Ha03,B}{1972Ha03}).\vspace{0.1cm}}&\\
&&&&&&&&\parbox[t][0.3cm]{9.42461cm}{\raggedright E(level): See also 280 keV (\href{https://www.nndc.bnl.gov/nsr/nsrlink.jsp?1967Be14,B}{1967Be14}) and 280 keV \textit{15}\vspace{0.1cm}}&\\
&&&&&&&&\parbox[t][0.3cm]{9.42461cm}{\raggedright {\ }{\ }{\ }(\href{https://www.nndc.bnl.gov/nsr/nsrlink.jsp?1967Gr04,B}{1967Gr04}), which is an unresolved doublet.\vspace{0.1cm}}&\\
&&&&&&&&\parbox[t][0.3cm]{9.42461cm}{\raggedright T\ensuremath{_{1/2}}: From \ensuremath{\tau}\ensuremath{\leq}5 ns (\href{https://www.nndc.bnl.gov/nsr/nsrlink.jsp?1967Be14,B}{1967Be14}).\vspace{0.1cm}}&\\
&&&&&&&&\parbox[t][0.3cm]{9.42461cm}{\raggedright J\ensuremath{^{\pi}},L: From the zero-range DWBA analyses of (1) (\href{https://www.nndc.bnl.gov/nsr/nsrlink.jsp?1970Ga18,B}{1970Ga18}): L=1\vspace{0.1cm}}&\\
&&&&&&&&\parbox[t][0.3cm]{9.42461cm}{\raggedright {\ }{\ }{\ }from 1\textit{p}\ensuremath{_{\textnormal{3/2}}} orbital using JULIE and performed with zero lower\vspace{0.1cm}}&\\
&&&&&&&&\parbox[t][0.3cm]{9.42461cm}{\raggedright {\ }{\ }{\ }cutoff radius; and (2) (\href{https://www.nndc.bnl.gov/nsr/nsrlink.jsp?1972Ha03,B}{1972Ha03}): L=(1) and J\ensuremath{^{\ensuremath{\pi}}}=(1/2\ensuremath{^{-}}) using\vspace{0.1cm}}&\\
&&&&&&&&\parbox[t][0.3cm]{9.42461cm}{\raggedright {\ }{\ }{\ }JULIE with zero lower cutoff radius, see Table II. We made those\vspace{0.1cm}}&\\
&&&&&&&&\parbox[t][0.3cm]{9.42461cm}{\raggedright {\ }{\ }{\ }J\ensuremath{^{\ensuremath{\pi}}} and L assignments tentative since the DWBA fit is not\vspace{0.1cm}}&\\
&&&&&&&&\parbox[t][0.3cm]{9.42461cm}{\raggedright {\ }{\ }{\ }presented.\vspace{0.1cm}}&\\
&&&&&&&&\parbox[t][0.3cm]{9.42461cm}{\raggedright J\ensuremath{^{\pi}}: See also J\ensuremath{^{\ensuremath{\pi}}}=1/2\ensuremath{^{-}} from the mirror analysis of (\href{https://www.nndc.bnl.gov/nsr/nsrlink.jsp?1967Ol05,B}{1967Ol05}).\vspace{0.1cm}}&\\
&&&&&&&&\parbox[t][0.3cm]{9.42461cm}{\raggedright C\ensuremath{^{\textnormal{2}}}S: See also 1.8 calculated by (\href{https://www.nndc.bnl.gov/nsr/nsrlink.jsp?1974Ga28,B}{1974Ga28}: Table 5).\vspace{0.1cm}}&\\
\multicolumn{1}{r@{}}{1504}&\multicolumn{1}{@{.}l}{0\ensuremath{^{{\hyperlink{NE38LEVEL1}{b}}}} {\it 30}}&\multicolumn{1}{l}{(5/2\ensuremath{^{-}})\ensuremath{^{{\hyperlink{NE38LEVEL4}{e}}}}}&&&\multicolumn{1}{l}{(3)$^{{\hyperlink{NE38LEVEL4}{e}}}$}&&&\parbox[t][0.3cm]{9.42461cm}{\raggedright E(level): From (\href{https://www.nndc.bnl.gov/nsr/nsrlink.jsp?1972Ha03,B}{1972Ha03}).\vspace{0.1cm}}&\\
&&&&&&&&\parbox[t][0.3cm]{9.42461cm}{\raggedright E(level): See also 1500.8 keV \textit{12} from a least-squares fit to the E\ensuremath{_{\ensuremath{\gamma}}}\vspace{0.1cm}}&\\
&&&&&&&&\parbox[t][0.3cm]{9.42461cm}{\raggedright {\ }{\ }{\ }values from (\href{https://www.nndc.bnl.gov/nsr/nsrlink.jsp?1967Ol05,B}{1967Ol05}). The E\ensuremath{_{\ensuremath{\gamma}}} and E\ensuremath{_{\textnormal{x}}} values reported by this\vspace{0.1cm}}&\\
&&&&&&&&\parbox[t][0.3cm]{9.42461cm}{\raggedright {\ }{\ }{\ }study seem to be a few keV off compared to the literature values.\vspace{0.1cm}}&\\
&&&&&&&&\parbox[t][0.3cm]{9.42461cm}{\raggedright {\ }{\ }{\ }It seems that a systematic uncertainty may have been unaccounted\vspace{0.1cm}}&\\
&&&&&&&&\parbox[t][0.3cm]{9.42461cm}{\raggedright {\ }{\ }{\ }for in that study. Other value: 1524 keV \textit{20} (\href{https://www.nndc.bnl.gov/nsr/nsrlink.jsp?1970Ga18,B}{1970Ga18}): An\vspace{0.1cm}}&\\
&&&&&&&&\parbox[t][0.3cm]{9.42461cm}{\raggedright {\ }{\ }{\ }unresolved doublet.\vspace{0.1cm}}&\\
&&&&&&&&\parbox[t][0.3cm]{9.42461cm}{\raggedright J\ensuremath{^{\pi}},L: From the zero-range DWBA analyses of (1) (\href{https://www.nndc.bnl.gov/nsr/nsrlink.jsp?1970Ga18,B}{1970Ga18}):\vspace{0.1cm}}&\\
&&&&&&&&\parbox[t][0.3cm]{9.42461cm}{\raggedright {\ }{\ }{\ }J\ensuremath{^{\ensuremath{\pi}}}=(5/2\ensuremath{^{-}}, 3/2\ensuremath{^{\textnormal{+}}}) and L=(2) using JULIE and performed with zero\vspace{0.1cm}}&\\
&&&&&&&&\parbox[t][0.3cm]{9.42461cm}{\raggedright {\ }{\ }{\ }lower cutoff radius, see the footnote; and (2) (\href{https://www.nndc.bnl.gov/nsr/nsrlink.jsp?1972Ha03,B}{1972Ha03}): L=(3)\vspace{0.1cm}}&\\
&&&&&&&&\parbox[t][0.3cm]{9.42461cm}{\raggedright {\ }{\ }{\ }and J\ensuremath{^{\ensuremath{\pi}}}=(5/2\ensuremath{^{-}}) using JULIE with zero lower cutoff radius, see\vspace{0.1cm}}&\\
&&&&&&&&\parbox[t][0.3cm]{9.42461cm}{\raggedright {\ }{\ }{\ }Table II. We made the J\ensuremath{^{\ensuremath{\pi}}} and L assignments from that study\vspace{0.1cm}}&\\
&&&&&&&&\parbox[t][0.3cm]{9.42461cm}{\raggedright {\ }{\ }{\ }tentative since the DWBA fit is not presented.\vspace{0.1cm}}&\\
&&&&&&&&\parbox[t][0.3cm]{9.42461cm}{\raggedright J\ensuremath{^{\pi}}: See also J\ensuremath{^{\ensuremath{\pi}}}=5/2\ensuremath{^{-}} from the mirror analysis of (\href{https://www.nndc.bnl.gov/nsr/nsrlink.jsp?1967Ol05,B}{1967Ol05}).\vspace{0.1cm}}&\\
\multicolumn{1}{r@{}}{1532}&\multicolumn{1}{@{.}l}{2\ensuremath{^{{\hyperlink{NE38LEVEL1}{b}}}} {\it 21}}&\multicolumn{1}{l}{(3/2\ensuremath{^{+}})\ensuremath{^{{\hyperlink{NE38LEVEL4}{e}}}}}&&&\multicolumn{1}{l}{(2)$^{{\hyperlink{NE38LEVEL4}{e}}}$}&\multicolumn{1}{r@{}}{0}&\multicolumn{1}{@{.}l}{73}&\parbox[t][0.3cm]{9.42461cm}{\raggedright E(level): Weighted average of 1532.4 keV \textit{30} (\href{https://www.nndc.bnl.gov/nsr/nsrlink.jsp?1972Ha03,B}{1972Ha03}) and 1532\vspace{0.1cm}}&\\
&&&&&&&&\parbox[t][0.3cm]{9.42461cm}{\raggedright {\ }{\ }{\ }keV \textit{3} (\href{https://www.nndc.bnl.gov/nsr/nsrlink.jsp?2023Po03,B}{2023Po03}).\vspace{0.1cm}}&\\
&&&&&&&&\parbox[t][0.3cm]{9.42461cm}{\raggedright E(level): See also 1541.6 keV \textit{12} from a least-squares fit to the E\ensuremath{_{\ensuremath{\gamma}}}\vspace{0.1cm}}&\\
&&&&&&&&\parbox[t][0.3cm]{9.42461cm}{\raggedright {\ }{\ }{\ }values from (\href{https://www.nndc.bnl.gov/nsr/nsrlink.jsp?1967Ol05,B}{1967Ol05}). This study may have a systematic\vspace{0.1cm}}&\\
&&&&&&&&\parbox[t][0.3cm]{9.42461cm}{\raggedright {\ }{\ }{\ }uncertainty that was not considered. Other value: 1524 keV \textit{20}\vspace{0.1cm}}&\\
&&&&&&&&\parbox[t][0.3cm]{9.42461cm}{\raggedright {\ }{\ }{\ }(\href{https://www.nndc.bnl.gov/nsr/nsrlink.jsp?1970Ga18,B}{1970Ga18}): An unresolved doublet.\vspace{0.1cm}}&\\
&&&&&&&&\parbox[t][0.3cm]{9.42461cm}{\raggedright J\ensuremath{^{\pi}},L: From the zero-range DWBA analyses of (1) (\href{https://www.nndc.bnl.gov/nsr/nsrlink.jsp?1970Ga18,B}{1970Ga18}):\vspace{0.1cm}}&\\
&&&&&&&&\parbox[t][0.3cm]{9.42461cm}{\raggedright {\ }{\ }{\ }J\ensuremath{^{\ensuremath{\pi}}}=(5/2\ensuremath{^{-}}, 3/2\ensuremath{^{\textnormal{+}}}) and L=(2) using JULIE and performed with zero\vspace{0.1cm}}&\\
&&&&&&&&\parbox[t][0.3cm]{9.42461cm}{\raggedright {\ }{\ }{\ }lower cutoff radius, see footnote; and (2) (\href{https://www.nndc.bnl.gov/nsr/nsrlink.jsp?1972Ha03,B}{1972Ha03}): L=(2) and\vspace{0.1cm}}&\\
&&&&&&&&\parbox[t][0.3cm]{9.42461cm}{\raggedright {\ }{\ }{\ }J\ensuremath{^{\ensuremath{\pi}}}=(3/2\ensuremath{^{\textnormal{+}}}) using JULIE with zero lower cutoff radius, see Table II.\vspace{0.1cm}}&\\
\end{longtable}
\begin{textblock}{29}(0,27.3)
Continued on next page (footnotes at end of table)
\end{textblock}
\clearpage
\begin{longtable}{ccccccc@{\extracolsep{\fill}}c}
\\[-.4cm]
\multicolumn{8}{c}{{\bf \small \underline{\ensuremath{^{\textnormal{20}}}Ne(\ensuremath{^{\textnormal{3}}}He,\ensuremath{\alpha}),(\ensuremath{^{\textnormal{3}}}He,\ensuremath{\alpha}\ensuremath{\gamma})\hspace{0.2in}\href{https://www.nndc.bnl.gov/nsr/nsrlink.jsp?1970Ga18,B}{1970Ga18},\href{https://www.nndc.bnl.gov/nsr/nsrlink.jsp?1972Ha03,B}{1972Ha03},\href{https://www.nndc.bnl.gov/nsr/nsrlink.jsp?2023Po03,B}{2023Po03} (continued)}}}\\
\multicolumn{8}{c}{~}\\
\multicolumn{8}{c}{\underline{\ensuremath{^{19}}Ne Levels (continued)}}\\
\multicolumn{8}{c}{~}\\
\multicolumn{2}{c}{E(level)$^{{\hyperlink{NE38LEVEL0}{a}}}$}&J$^{\pi}$$^{}$&L$^{}$&\multicolumn{2}{c}{C\ensuremath{^{\textnormal{2}}}S\ensuremath{_{\textnormal{exp}}}$^{{\hyperlink{NE38LEVEL5}{f}}}$}&Comments&\\[-.2cm]
\multicolumn{2}{c}{\hrulefill}&\hrulefill&\hrulefill&\multicolumn{2}{c}{\hrulefill}&\hrulefill&
\endhead
&&&&&&\parbox[t][0.3cm]{11.154461cm}{\raggedright {\ }{\ }{\ }We made those J\ensuremath{^{\ensuremath{\pi}}} and L assignments tentative since the DWBA fit is not\vspace{0.1cm}}&\\
&&&&&&\parbox[t][0.3cm]{11.154461cm}{\raggedright {\ }{\ }{\ }presented.\vspace{0.1cm}}&\\
&&&&&&\parbox[t][0.3cm]{11.154461cm}{\raggedright J\ensuremath{^{\pi}}: See also J\ensuremath{^{\ensuremath{\pi}}}=3/2\ensuremath{^{\textnormal{+}}} from the mirror analysis of (\href{https://www.nndc.bnl.gov/nsr/nsrlink.jsp?1967Ol05,B}{1967Ol05}).\vspace{0.1cm}}&\\
&&&&&&\parbox[t][0.3cm]{11.154461cm}{\raggedright C\ensuremath{^{\textnormal{2}}}S: See also 0.70 calculated by (\href{https://www.nndc.bnl.gov/nsr/nsrlink.jsp?1974Ga28,B}{1974Ga28}: Table 5).\vspace{0.1cm}}&\\
\multicolumn{1}{r@{}}{1612}&\multicolumn{1}{@{.}l}{5 {\it 21}}&\multicolumn{1}{l}{(3/2\ensuremath{^{-}})}&\multicolumn{1}{l}{(1)}&\multicolumn{1}{r@{}}{0}&\multicolumn{1}{@{.}l}{21}&\parbox[t][0.3cm]{11.154461cm}{\raggedright E(level): Weighted average of 1620 keV \textit{15} (\href{https://www.nndc.bnl.gov/nsr/nsrlink.jsp?1967Gr04,B}{1967Gr04}); 1615 keV \textit{10}\vspace{0.1cm}}&\\
&&&&&&\parbox[t][0.3cm]{11.154461cm}{\raggedright {\ }{\ }{\ }(\href{https://www.nndc.bnl.gov/nsr/nsrlink.jsp?1970Ga18,B}{1970Ga18}); 1611.5 keV \textit{30} (\href{https://www.nndc.bnl.gov/nsr/nsrlink.jsp?1972Ha03,B}{1972Ha03}); and 1613 keV \textit{3} (\href{https://www.nndc.bnl.gov/nsr/nsrlink.jsp?2023Po03,B}{2023Po03}).\vspace{0.1cm}}&\\
&&&&&&\parbox[t][0.3cm]{11.154461cm}{\raggedright E(level): See also 1607.1 keV \textit{12} from a least-squares fit to the E\ensuremath{_{\ensuremath{\gamma}}} values from\vspace{0.1cm}}&\\
&&&&&&\parbox[t][0.3cm]{11.154461cm}{\raggedright {\ }{\ }{\ }(\href{https://www.nndc.bnl.gov/nsr/nsrlink.jsp?1967Ol05,B}{1967Ol05}). This study may carry a systematic uncertainty.\vspace{0.1cm}}&\\
&&&&&&\parbox[t][0.3cm]{11.154461cm}{\raggedright J\ensuremath{^{\pi}},L: From the zero-range DWBA analyses of (1) (\href{https://www.nndc.bnl.gov/nsr/nsrlink.jsp?1970Ga18,B}{1970Ga18}): J\ensuremath{^{\ensuremath{\pi}}}=(3/2\ensuremath{^{-}}) and\vspace{0.1cm}}&\\
&&&&&&\parbox[t][0.3cm]{11.154461cm}{\raggedright {\ }{\ }{\ }L=(1) using JULIE and performed with zero lower cutoff radius, L-transfer\vspace{0.1cm}}&\\
&&&&&&\parbox[t][0.3cm]{11.154461cm}{\raggedright {\ }{\ }{\ }from the 1\textit{p}\ensuremath{_{\textnormal{3/2}}} orbital, where the DWBA fit does not describe the data at\vspace{0.1cm}}&\\
&&&&&&\parbox[t][0.3cm]{11.154461cm}{\raggedright {\ }{\ }{\ }\ensuremath{\theta}\ensuremath{_{\textnormal{c.m.}}}\ensuremath{>}30\ensuremath{^\circ}, J\ensuremath{^{\ensuremath{\pi}}}=(3/2\ensuremath{^{-}}) assignment is guided by the mirror level analysis; and (2)\vspace{0.1cm}}&\\
&&&&&&\parbox[t][0.3cm]{11.154461cm}{\raggedright {\ }{\ }{\ }(\href{https://www.nndc.bnl.gov/nsr/nsrlink.jsp?1972Ha03,B}{1972Ha03}): L=(1) and J\ensuremath{^{\ensuremath{\pi}}}=(3/2\ensuremath{^{-}}) using JULIE with zero lower cutoff radius,\vspace{0.1cm}}&\\
&&&&&&\parbox[t][0.3cm]{11.154461cm}{\raggedright {\ }{\ }{\ }see Table II. We made those J\ensuremath{^{\ensuremath{\pi}}} and L assignments tentative since the DWBA\vspace{0.1cm}}&\\
&&&&&&\parbox[t][0.3cm]{11.154461cm}{\raggedright {\ }{\ }{\ }fit is not presented.\vspace{0.1cm}}&\\
&&&&&&\parbox[t][0.3cm]{11.154461cm}{\raggedright J\ensuremath{^{\pi}}: See also J\ensuremath{^{\ensuremath{\pi}}}=3/2\ensuremath{^{\textnormal{+}}} from the mirror analysis of (\href{https://www.nndc.bnl.gov/nsr/nsrlink.jsp?1967Ol05,B}{1967Ol05}).\vspace{0.1cm}}&\\
&&&&&&\parbox[t][0.3cm]{11.154461cm}{\raggedright C\ensuremath{^{\textnormal{2}}}S: See also 0.21 calculated by (\href{https://www.nndc.bnl.gov/nsr/nsrlink.jsp?1974Ga28,B}{1974Ga28}: Table 5).\vspace{0.1cm}}&\\
\multicolumn{1}{r@{}}{2791}&\multicolumn{1}{@{.}l}{3 {\it 30}}&\multicolumn{1}{l}{(9/2\ensuremath{^{+}})}&\multicolumn{1}{l}{4,5}&&&\parbox[t][0.3cm]{11.154461cm}{\raggedright Weighted average of 2778 keV \textit{15} (\href{https://www.nndc.bnl.gov/nsr/nsrlink.jsp?1967Gr04,B}{1967Gr04}); 2793 keV \textit{10} (\href{https://www.nndc.bnl.gov/nsr/nsrlink.jsp?1970Ga18,B}{1970Ga18}); and\vspace{0.1cm}}&\\
&&&&&&\parbox[t][0.3cm]{11.154461cm}{\raggedright {\ }{\ }{\ }2791.7 keV \textit{30} (\href{https://www.nndc.bnl.gov/nsr/nsrlink.jsp?1972Ha03,B}{1972Ha03}).\vspace{0.1cm}}&\\
&&&&&&\parbox[t][0.3cm]{11.154461cm}{\raggedright E(level): See also 2775.8 keV \textit{34} reported by (\href{https://www.nndc.bnl.gov/nsr/nsrlink.jsp?1967Ol05,B}{1967Ol05}), which likely suffers\vspace{0.1cm}}&\\
&&&&&&\parbox[t][0.3cm]{11.154461cm}{\raggedright {\ }{\ }{\ }from systematic uncertainties that are unaccounted for; 2.78 MeV (\href{https://www.nndc.bnl.gov/nsr/nsrlink.jsp?1969Ba62,B}{1969Ba62});\vspace{0.1cm}}&\\
&&&&&&\parbox[t][0.3cm]{11.154461cm}{\raggedright {\ }{\ }{\ }and 2794.4 (\href{https://www.nndc.bnl.gov/nsr/nsrlink.jsp?2023Po03,B}{2023Po03}), which was a calibration point in that study.\vspace{0.1cm}}&\\
&&&&&&\parbox[t][0.3cm]{11.154461cm}{\raggedright J\ensuremath{^{\pi}},L: From the zero-range DWBA analyses of (1) (\href{https://www.nndc.bnl.gov/nsr/nsrlink.jsp?1969Ba62,B}{1969Ba62}): J=9/2, L=4,5,\vspace{0.1cm}}&\\
&&&&&&\parbox[t][0.3cm]{11.154461cm}{\raggedright {\ }{\ }{\ }where the DWBA fits with L=4 and L=5 are satisfactory fits to the measured \ensuremath{\alpha}\vspace{0.1cm}}&\\
&&&&&&\parbox[t][0.3cm]{11.154461cm}{\raggedright {\ }{\ }{\ }angular distribution data corresponding to this state. However, the fits are not\vspace{0.1cm}}&\\
&&&&&&\parbox[t][0.3cm]{11.154461cm}{\raggedright {\ }{\ }{\ }ideal. The authors certainly ruled out L=2 and L=3 and assigned J=(9/2\ensuremath{^{\textnormal{+}}}) based\vspace{0.1cm}}&\\
&&&&&&\parbox[t][0.3cm]{11.154461cm}{\raggedright {\ }{\ }{\ }on mirror analysis out of all other allowed spins deduced from L=4 and L=5;\vspace{0.1cm}}&\\
&&&&&&\parbox[t][0.3cm]{11.154461cm}{\raggedright {\ }{\ }{\ }and (2) (\href{https://www.nndc.bnl.gov/nsr/nsrlink.jsp?1972Ha03,B}{1972Ha03}): J\ensuremath{^{\ensuremath{\pi}}}=9/2\ensuremath{^{\textnormal{+}}} and L=4 using JULIE with zero lower cutoff\vspace{0.1cm}}&\\
&&&&&&\parbox[t][0.3cm]{11.154461cm}{\raggedright {\ }{\ }{\ }radius, where the DWBA fits do not describe the data very well.\vspace{0.1cm}}&\\
\multicolumn{1}{r@{}}{3841?}&\multicolumn{1}{@{ }l}{{\it 15}}&&&&&\parbox[t][0.3cm]{11.154461cm}{\raggedright E(level): From (\href{https://www.nndc.bnl.gov/nsr/nsrlink.jsp?1967Gr04,B}{1967Gr04}): This state was only observed at two angles:\vspace{0.1cm}}&\\
&&&&&&\parbox[t][0.3cm]{11.154461cm}{\raggedright {\ }{\ }{\ }\ensuremath{\theta}\ensuremath{_{\textnormal{lab}}}=165\ensuremath{^\circ} and 167.5\ensuremath{^\circ} at E=5.9 MeV. The statistics and the resolution in the\vspace{0.1cm}}&\\
&&&&&&\parbox[t][0.3cm]{11.154461cm}{\raggedright {\ }{\ }{\ }spectra at other angles were not sufficient to show this level as a kinematically\vspace{0.1cm}}&\\
&&&&&&\parbox[t][0.3cm]{11.154461cm}{\raggedright {\ }{\ }{\ }consistent state.\vspace{0.1cm}}&\\
&&&&&&\parbox[t][0.3cm]{11.154461cm}{\raggedright E(level): Evaluator notes that (\href{https://www.nndc.bnl.gov/nsr/nsrlink.jsp?1970Ga18,B}{1970Ga18}) searched for this tentative state and did\vspace{0.1cm}}&\\
&&&&&&\parbox[t][0.3cm]{11.154461cm}{\raggedright {\ }{\ }{\ }not observe it. The later studies suggest that this tentative state does not exist.\vspace{0.1cm}}&\\
&&&&&&\parbox[t][0.3cm]{11.154461cm}{\raggedright {\ }{\ }{\ }We therefore did not include it in the \ensuremath{^{\textnormal{19}}}Ne Adopted Levels.\vspace{0.1cm}}&\\
\multicolumn{1}{r@{}}{4034}&\multicolumn{1}{@{.}l}{7\ensuremath{^{{\hyperlink{NE38LEVEL2}{c}}{\hyperlink{NE38LEVEL8}{i}}}} {\it 23}}&\multicolumn{1}{l}{(3/2\ensuremath{^{+}},5/2\ensuremath{^{+}})}&\multicolumn{1}{l}{2}&&&\parbox[t][0.3cm]{11.154461cm}{\raggedright E(level): See also 4013 keV \textit{15} (\href{https://www.nndc.bnl.gov/nsr/nsrlink.jsp?1967Gr04,B}{1967Gr04}); and 4036 keV \textit{10} (\href{https://www.nndc.bnl.gov/nsr/nsrlink.jsp?1970Ga18,B}{1970Ga18}).\vspace{0.1cm}}&\\
&&&&&&\parbox[t][0.3cm]{11.154461cm}{\raggedright J\ensuremath{^{\pi}},L: From the zero-range DWBA analysis of (\href{https://www.nndc.bnl.gov/nsr/nsrlink.jsp?1970Ga18,B}{1970Ga18}) using JULIE and\vspace{0.1cm}}&\\
&&&&&&\parbox[t][0.3cm]{11.154461cm}{\raggedright {\ }{\ }{\ }performed with zero lower cutoff radius. The agreement between the DWBA\vspace{0.1cm}}&\\
&&&&&&\parbox[t][0.3cm]{11.154461cm}{\raggedright {\ }{\ }{\ }and the data is not good. The authors favored J\ensuremath{^{\ensuremath{\pi}}}=3/2\ensuremath{^{\textnormal{+}}} over J\ensuremath{^{\ensuremath{\pi}}}=5/2\ensuremath{^{\textnormal{+}}} based on\vspace{0.1cm}}&\\
&&&&&&\parbox[t][0.3cm]{11.154461cm}{\raggedright {\ }{\ }{\ }their mirror levels analysis.\vspace{0.1cm}}&\\
&&&&&&\parbox[t][0.3cm]{11.154461cm}{\raggedright See C\ensuremath{^{\textnormal{2}}}S\ensuremath{\leq}0.1 calculated by (\href{https://www.nndc.bnl.gov/nsr/nsrlink.jsp?1974Ga28,B}{1974Ga28}: Table 5).\vspace{0.1cm}}&\\
\multicolumn{1}{r@{}}{4142}&\multicolumn{1}{@{.}l}{1\ensuremath{^{{\hyperlink{NE38LEVEL2}{c}}{\hyperlink{NE38LEVEL8}{i}}}} {\it 22}}&\multicolumn{1}{l}{(9/2\ensuremath{^{-}})\ensuremath{^{{\hyperlink{NE38LEVEL6}{g}}}}}&&&&\parbox[t][0.3cm]{11.154461cm}{\raggedright E(level): See also 4152 keV \textit{15} (\href{https://www.nndc.bnl.gov/nsr/nsrlink.jsp?1967Gr04,B}{1967Gr04}); 4160 keV \textit{20} (\href{https://www.nndc.bnl.gov/nsr/nsrlink.jsp?1967Ol05,B}{1967Ol05}), which\vspace{0.1cm}}&\\
&&&&&&\parbox[t][0.3cm]{11.154461cm}{\raggedright {\ }{\ }{\ }likely suffers from systematics uncertainties; and 4142 keV \textit{10} (\href{https://www.nndc.bnl.gov/nsr/nsrlink.jsp?1970Ga18,B}{1970Ga18}).\vspace{0.1cm}}&\\
&&&&&&\parbox[t][0.3cm]{11.154461cm}{\raggedright J\ensuremath{^{\pi}}: From the zero-range DWBA analysis of (\href{https://www.nndc.bnl.gov/nsr/nsrlink.jsp?1970Ga18,B}{1970Ga18}) using JULIE and\vspace{0.1cm}}&\\
&&&&&&\parbox[t][0.3cm]{11.154461cm}{\raggedright {\ }{\ }{\ }performed with zero lower cutoff radius. L is not mentioned.\vspace{0.1cm}}&\\
\multicolumn{1}{r@{}}{4200}&\multicolumn{1}{@{}l}{\ensuremath{^{{\hyperlink{NE38LEVEL3}{d}}}} {\it 10}}&\multicolumn{1}{l}{(7/2\ensuremath{^{-}})\ensuremath{^{{\hyperlink{NE38LEVEL6}{g}}}}}&&&&\parbox[t][0.3cm]{11.154461cm}{\raggedright E(level): From (\href{https://www.nndc.bnl.gov/nsr/nsrlink.jsp?1970Ga18,B}{1970Ga18}).\vspace{0.1cm}}&\\
&&&&&&\parbox[t][0.3cm]{11.154461cm}{\raggedright E(level): See also 4199.8 keV (\href{https://www.nndc.bnl.gov/nsr/nsrlink.jsp?2023Po03,B}{2023Po03}): Used as a calibration point.\vspace{0.1cm}}&\\
&&&&&&\parbox[t][0.3cm]{11.154461cm}{\raggedright J\ensuremath{^{\pi}}: From the zero-range DWBA analysis of (\href{https://www.nndc.bnl.gov/nsr/nsrlink.jsp?1970Ga18,B}{1970Ga18}) using JULIE and\vspace{0.1cm}}&\\
&&&&&&\parbox[t][0.3cm]{11.154461cm}{\raggedright {\ }{\ }{\ }performed with zero lower cutoff radius. L is not mentioned.\vspace{0.1cm}}&\\
\multicolumn{1}{r@{}}{4379}&\multicolumn{1}{@{}l}{\ensuremath{^{{\hyperlink{NE38LEVEL2}{c}}}} {\it 10}}&\multicolumn{1}{l}{(7/2\ensuremath{^{+}})}&&&&\parbox[t][0.3cm]{11.154461cm}{\raggedright E(level): From (\href{https://www.nndc.bnl.gov/nsr/nsrlink.jsp?1970Ga18,B}{1970Ga18}).\vspace{0.1cm}}&\\
&&&&&&\parbox[t][0.3cm]{11.154461cm}{\raggedright E(level): See also 4344 keV \textit{15} (\href{https://www.nndc.bnl.gov/nsr/nsrlink.jsp?1967Gr04,B}{1967Gr04}).\vspace{0.1cm}}&\\
&&&&&&\parbox[t][0.3cm]{11.154461cm}{\raggedright (\href{https://www.nndc.bnl.gov/nsr/nsrlink.jsp?1970Ga18,B}{1970Ga18}) assigned the \ensuremath{^{\textnormal{19}}}F*(4.38 MeV, 7/2\ensuremath{^{\textnormal{+}}}) state as the mirror to this \ensuremath{^{\textnormal{19}}}Ne\vspace{0.1cm}}&\\
&&&&&&\parbox[t][0.3cm]{11.154461cm}{\raggedright {\ }{\ }{\ }level.\vspace{0.1cm}}&\\
\end{longtable}
\begin{textblock}{29}(0,27.3)
Continued on next page (footnotes at end of table)
\end{textblock}
\clearpage
\begin{longtable}{ccccccc@{\extracolsep{\fill}}c}
\\[-.4cm]
\multicolumn{8}{c}{{\bf \small \underline{\ensuremath{^{\textnormal{20}}}Ne(\ensuremath{^{\textnormal{3}}}He,\ensuremath{\alpha}),(\ensuremath{^{\textnormal{3}}}He,\ensuremath{\alpha}\ensuremath{\gamma})\hspace{0.2in}\href{https://www.nndc.bnl.gov/nsr/nsrlink.jsp?1970Ga18,B}{1970Ga18},\href{https://www.nndc.bnl.gov/nsr/nsrlink.jsp?1972Ha03,B}{1972Ha03},\href{https://www.nndc.bnl.gov/nsr/nsrlink.jsp?2023Po03,B}{2023Po03} (continued)}}}\\
\multicolumn{8}{c}{~}\\
\multicolumn{8}{c}{\underline{\ensuremath{^{19}}Ne Levels (continued)}}\\
\multicolumn{8}{c}{~}\\
\multicolumn{2}{c}{E(level)$^{{\hyperlink{NE38LEVEL0}{a}}}$}&J$^{\pi}$$^{}$&L$^{}$&\multicolumn{2}{c}{C\ensuremath{^{\textnormal{2}}}S\ensuremath{_{\textnormal{exp}}}$^{{\hyperlink{NE38LEVEL5}{f}}}$}&Comments&\\[-.2cm]
\multicolumn{2}{c}{\hrulefill}&\hrulefill&\hrulefill&\multicolumn{2}{c}{\hrulefill}&\hrulefill&
\endhead
&&&&&&\parbox[t][0.3cm]{11.518261cm}{\raggedright J\ensuremath{^{\pi}}: From (\href{https://www.nndc.bnl.gov/nsr/nsrlink.jsp?1970Ga18,B}{1970Ga18}) based on mirror level analysis. Since this is a weak argument,\vspace{0.1cm}}&\\
&&&&&&\parbox[t][0.3cm]{11.518261cm}{\raggedright {\ }{\ }{\ }we made the assignment tentative.\vspace{0.1cm}}&\\
\multicolumn{1}{r@{}}{4550}&\multicolumn{1}{@{}l}{\ensuremath{^{{\hyperlink{NE38LEVEL2}{c}}}} {\it 10}}&\multicolumn{1}{l}{(1/2\ensuremath{^{-}},3/2\ensuremath{^{-}})}&\multicolumn{1}{l}{1}&&&\parbox[t][0.3cm]{11.518261cm}{\raggedright E(level): Weighted average of 4547 keV \textit{15} (\href{https://www.nndc.bnl.gov/nsr/nsrlink.jsp?1967Gr04,B}{1967Gr04}) and 4551 keV \textit{10}\vspace{0.1cm}}&\\
&&&&&&\parbox[t][0.3cm]{11.518261cm}{\raggedright {\ }{\ }{\ }(\href{https://www.nndc.bnl.gov/nsr/nsrlink.jsp?1970Ga18,B}{1970Ga18}). Other value: 4547.7 keV (\href{https://www.nndc.bnl.gov/nsr/nsrlink.jsp?2023Po03,B}{2023Po03}): Used as a calibration point.\vspace{0.1cm}}&\\
&&&&&&\parbox[t][0.3cm]{11.518261cm}{\raggedright E(level): (\href{https://www.nndc.bnl.gov/nsr/nsrlink.jsp?1970Ga18,B}{1970Ga18}) reported that this state may be a possible mirror to the\vspace{0.1cm}}&\\
&&&&&&\parbox[t][0.3cm]{11.518261cm}{\raggedright {\ }{\ }{\ }\ensuremath{^{\textnormal{19}}}F*(4.56 MeV, 3/2\ensuremath{^{-}}) state.\vspace{0.1cm}}&\\
&&&&&&\parbox[t][0.3cm]{11.518261cm}{\raggedright J\ensuremath{^{\pi}},L: From the zero-range DWBA analysis of (\href{https://www.nndc.bnl.gov/nsr/nsrlink.jsp?1970Ga18,B}{1970Ga18}) using JULIE and\vspace{0.1cm}}&\\
&&&&&&\parbox[t][0.3cm]{11.518261cm}{\raggedright {\ }{\ }{\ }performed with zero lower cutoff radius. L-transfer is from the 1\textit{p}\ensuremath{_{\textnormal{3/2}}} or 1\textit{p}\ensuremath{_{\textnormal{1/2}}}\vspace{0.1cm}}&\\
&&&&&&\parbox[t][0.3cm]{11.518261cm}{\raggedright {\ }{\ }{\ }orbital (\href{https://www.nndc.bnl.gov/nsr/nsrlink.jsp?1970Ga18,B}{1970Ga18}). The DWBA fits do not describe the data at \ensuremath{\theta}\ensuremath{_{\textnormal{c.m.}}}\ensuremath{>}50\ensuremath{^\circ}.\vspace{0.1cm}}&\\
&&&&&&\parbox[t][0.3cm]{11.518261cm}{\raggedright C\ensuremath{^{\textnormal{2}}}S: See the calculated value of 0.57 (\href{https://www.nndc.bnl.gov/nsr/nsrlink.jsp?1974Ga28,B}{1974Ga28}: Table 5).\vspace{0.1cm}}&\\
\multicolumn{1}{r@{}}{4601}&\multicolumn{1}{@{.}l}{7\ensuremath{^{{\hyperlink{NE38LEVEL8}{i}}}} {\it 25}}&&&&&&\\
\multicolumn{1}{r@{}}{4633}&\multicolumn{1}{@{}l}{\ensuremath{^{{\hyperlink{NE38LEVEL3}{d}}{\hyperlink{NE38LEVEL8}{i}}}} {\it 3}}&&&&&\parbox[t][0.3cm]{11.518261cm}{\raggedright E(level): See also 4625 keV \textit{10} (\href{https://www.nndc.bnl.gov/nsr/nsrlink.jsp?1970Ga18,B}{1970Ga18}).\vspace{0.1cm}}&\\
\multicolumn{1}{r@{}}{4705}&\multicolumn{1}{@{}l}{\ensuremath{^{{\hyperlink{NE38LEVEL2}{c}}}} {\it 11}}&&&&&\parbox[t][0.3cm]{11.518261cm}{\raggedright Weighted average (with external errors) of 4689 keV \textit{15} (\href{https://www.nndc.bnl.gov/nsr/nsrlink.jsp?1967Gr04,B}{1967Gr04}) and 4712 keV\vspace{0.1cm}}&\\
&&&&&&\parbox[t][0.3cm]{11.518261cm}{\raggedright {\ }{\ }{\ }\textit{10} (\href{https://www.nndc.bnl.gov/nsr/nsrlink.jsp?1970Ga18,B}{1970Ga18}). See also 4708.8 keV (\href{https://www.nndc.bnl.gov/nsr/nsrlink.jsp?2023Po03,B}{2023Po03}: Used as a calibration point).\vspace{0.1cm}}&\\
&&&&&&\parbox[t][0.3cm]{11.518261cm}{\raggedright L: (\href{https://www.nndc.bnl.gov/nsr/nsrlink.jsp?1970Ga18,B}{1970Ga18}) suggested that the \ensuremath{\alpha} angular distribution corresponding to this state\vspace{0.1cm}}&\\
&&&&&&\parbox[t][0.3cm]{11.518261cm}{\raggedright {\ }{\ }{\ }has a characteristic shape of L=2. However, no DWBA fit was performed for this\vspace{0.1cm}}&\\
&&&&&&\parbox[t][0.3cm]{11.518261cm}{\raggedright {\ }{\ }{\ }weakly populated state.\vspace{0.1cm}}&\\
&&&&&&\parbox[t][0.3cm]{11.518261cm}{\raggedright (\href{https://www.nndc.bnl.gov/nsr/nsrlink.jsp?1970Ga18,B}{1970Ga18}) suggested that this state may be the possible mirror state to the\vspace{0.1cm}}&\\
&&&&&&\parbox[t][0.3cm]{11.518261cm}{\raggedright {\ }{\ }{\ }\ensuremath{^{\textnormal{19}}}F*(4.55 MeV, 5/2\ensuremath{^{\textnormal{+}}}) level.\vspace{0.1cm}}&\\
\multicolumn{1}{r@{}}{4783}&\multicolumn{1}{@{}l}{\ensuremath{^{{\hyperlink{NE38LEVEL3}{d}}}} {\it 20}}&&&&&\parbox[t][0.3cm]{11.518261cm}{\raggedright E(level): From (\href{https://www.nndc.bnl.gov/nsr/nsrlink.jsp?1970Ga18,B}{1970Ga18}).\vspace{0.1cm}}&\\
\multicolumn{1}{r@{}}{5090}&\multicolumn{1}{@{.}l}{3 {\it 25}}&&&&&\parbox[t][0.3cm]{11.518261cm}{\raggedright E(level): Weighted average of 5077 keV \textit{15} (\href{https://www.nndc.bnl.gov/nsr/nsrlink.jsp?1967Gr04,B}{1967Gr04}), where this state was free of\vspace{0.1cm}}&\\
&&&&&&\parbox[t][0.3cm]{11.518261cm}{\raggedright {\ }{\ }{\ }contamination by the \ensuremath{^{\textnormal{20}}}Ne(\ensuremath{^{\textnormal{3}}}He,p) reaction only at a single scattering angle; 5093\vspace{0.1cm}}&\\
&&&&&&\parbox[t][0.3cm]{11.518261cm}{\raggedright {\ }{\ }{\ }keV \textit{10} (\href{https://www.nndc.bnl.gov/nsr/nsrlink.jsp?1970Ga18,B}{1970Ga18}); 5086 keV \textit{10} (\href{https://www.nndc.bnl.gov/nsr/nsrlink.jsp?1972Ha03,B}{1972Ha03}); and 5090.8 keV \textit{25} (\href{https://www.nndc.bnl.gov/nsr/nsrlink.jsp?2023Po03,B}{2023Po03}).\vspace{0.1cm}}&\\
&&&&&&\parbox[t][0.3cm]{11.518261cm}{\raggedright L: (\href{https://www.nndc.bnl.gov/nsr/nsrlink.jsp?1970Ga18,B}{1970Ga18}) suggested that the \ensuremath{\alpha} angular distribution corresponding to this state\vspace{0.1cm}}&\\
&&&&&&\parbox[t][0.3cm]{11.518261cm}{\raggedright {\ }{\ }{\ }has a characteristic shape of L=2. However, no DWBA fit was performed for this\vspace{0.1cm}}&\\
&&&&&&\parbox[t][0.3cm]{11.518261cm}{\raggedright {\ }{\ }{\ }weakly populated state.\vspace{0.1cm}}&\\
\multicolumn{1}{r@{}}{5351}&\multicolumn{1}{@{}l}{\ensuremath{^{{\hyperlink{NE38LEVEL3}{d}}}} {\it 3}}&\multicolumn{1}{l}{(1/2\ensuremath{^{+}})}&\multicolumn{1}{l}{(0)}&\multicolumn{1}{r@{}}{0}&\multicolumn{1}{@{.}l}{01}&\parbox[t][0.3cm]{11.518261cm}{\raggedright E(level): Weighted average of 5351 keV \textit{10} (\href{https://www.nndc.bnl.gov/nsr/nsrlink.jsp?1970Ga18,B}{1970Ga18}) and 5351 keV \textit{3}\vspace{0.1cm}}&\\
&&&&&&\parbox[t][0.3cm]{11.518261cm}{\raggedright {\ }{\ }{\ }(\href{https://www.nndc.bnl.gov/nsr/nsrlink.jsp?2023Po03,B}{2023Po03}).\vspace{0.1cm}}&\\
&&&&&&\parbox[t][0.3cm]{11.518261cm}{\raggedright E(level): (\href{https://www.nndc.bnl.gov/nsr/nsrlink.jsp?1970Ga18,B}{1970Ga18}) reported that this state may be a possible mirror to the\vspace{0.1cm}}&\\
&&&&&&\parbox[t][0.3cm]{11.518261cm}{\raggedright {\ }{\ }{\ }\ensuremath{^{\textnormal{19}}}F*(5.34 MeV, 1/2\ensuremath{^{\textnormal{(+)}}}) level.\vspace{0.1cm}}&\\
&&&&&&\parbox[t][0.3cm]{11.518261cm}{\raggedright J\ensuremath{^{\pi}},L: From the zero-range DWBA analysis of (\href{https://www.nndc.bnl.gov/nsr/nsrlink.jsp?1970Ga18,B}{1970Ga18}) using JULIE and\vspace{0.1cm}}&\\
&&&&&&\parbox[t][0.3cm]{11.518261cm}{\raggedright {\ }{\ }{\ }performed with zero lower cutoff radius. The agreement between the DWBA and\vspace{0.1cm}}&\\
&&&&&&\parbox[t][0.3cm]{11.518261cm}{\raggedright {\ }{\ }{\ }the data is not good. So, the evaluator made L and J\ensuremath{^{\ensuremath{\pi}}} values tentative, whereas\vspace{0.1cm}}&\\
&&&&&&\parbox[t][0.3cm]{11.518261cm}{\raggedright {\ }{\ }{\ }(\href{https://www.nndc.bnl.gov/nsr/nsrlink.jsp?1970Ga18,B}{1970Ga18}) assigned firm values.\vspace{0.1cm}}&\\
\multicolumn{1}{r@{}}{5424}&\multicolumn{1}{@{}l}{\ensuremath{^{{\hyperlink{NE38LEVEL3}{d}}}} {\it 3}}&&&&&\parbox[t][0.3cm]{11.518261cm}{\raggedright E(level): Weighted average of 5423 keV \textit{10} (\href{https://www.nndc.bnl.gov/nsr/nsrlink.jsp?1972Ha03,B}{1972Ha03}) and 5424 keV \textit{3}\vspace{0.1cm}}&\\
&&&&&&\parbox[t][0.3cm]{11.518261cm}{\raggedright {\ }{\ }{\ }(\href{https://www.nndc.bnl.gov/nsr/nsrlink.jsp?2023Po03,B}{2023Po03}).\vspace{0.1cm}}&\\
&&&&&&\parbox[t][0.3cm]{11.518261cm}{\raggedright E(level): See also 5426 keV \textit{10} (\href{https://www.nndc.bnl.gov/nsr/nsrlink.jsp?1970Ga18,B}{1970Ga18}): A member of an unresolved doublet\vspace{0.1cm}}&\\
&&&&&&\parbox[t][0.3cm]{11.518261cm}{\raggedright {\ }{\ }{\ }with another level at E\ensuremath{_{\textnormal{x}}}=5463 keV \textit{20}.\vspace{0.1cm}}&\\
\multicolumn{1}{r@{}}{5463}&\multicolumn{1}{@{}l}{\ensuremath{^{{\hyperlink{NE38LEVEL3}{d}}}} {\it 20}}&&&&&\parbox[t][0.3cm]{11.518261cm}{\raggedright E(level): From (\href{https://www.nndc.bnl.gov/nsr/nsrlink.jsp?1970Ga18,B}{1970Ga18}): This state is reported to be a member of an unresolved\vspace{0.1cm}}&\\
&&&&&&\parbox[t][0.3cm]{11.518261cm}{\raggedright {\ }{\ }{\ }doublet with another level at E\ensuremath{_{\textnormal{x}}}=5426 keV \textit{10}.\vspace{0.1cm}}&\\
\multicolumn{1}{r@{}}{5543}&\multicolumn{1}{@{}l}{\ensuremath{^{{\hyperlink{NE38LEVEL3}{d}}}} {\it 3}}&&&&&\parbox[t][0.3cm]{11.518261cm}{\raggedright E(level): Weighted average of 5545 keV \textit{10} (\href{https://www.nndc.bnl.gov/nsr/nsrlink.jsp?1970Ga18,B}{1970Ga18}); 5517 keV \textit{20} (\href{https://www.nndc.bnl.gov/nsr/nsrlink.jsp?1972Ha03,B}{1972Ha03});\vspace{0.1cm}}&\\
&&&&&&\parbox[t][0.3cm]{11.518261cm}{\raggedright {\ }{\ }{\ }and 5543 keV \textit{3} (\href{https://www.nndc.bnl.gov/nsr/nsrlink.jsp?2023Po03,B}{2023Po03}).\vspace{0.1cm}}&\\
\multicolumn{1}{r@{}}{5832}&\multicolumn{1}{@{}l}{\ensuremath{^{{\hyperlink{NE38LEVEL3}{d}}}} {\it 10}}&&&&&\parbox[t][0.3cm]{11.518261cm}{\raggedright E(level): Weighted average of 5831 keV \textit{10} (\href{https://www.nndc.bnl.gov/nsr/nsrlink.jsp?1970Ga18,B}{1970Ga18}) and 5837 keV \textit{20}\vspace{0.1cm}}&\\
&&&&&&\parbox[t][0.3cm]{11.518261cm}{\raggedright {\ }{\ }{\ }(\href{https://www.nndc.bnl.gov/nsr/nsrlink.jsp?1972Ha03,B}{1972Ha03}).\vspace{0.1cm}}&\\
&&&&&&\parbox[t][0.3cm]{11.518261cm}{\raggedright E(level): See also 5828 keV (\href{https://www.nndc.bnl.gov/nsr/nsrlink.jsp?2023Po03,B}{2023Po03}): Used as a calibration point.\vspace{0.1cm}}&\\
\multicolumn{1}{r@{}}{6014}&\multicolumn{1}{@{}l}{\ensuremath{^{{\hyperlink{NE38LEVEL3}{d}}}} {\it 10}}&\multicolumn{1}{l}{(1/2\ensuremath{^{-}},3/2\ensuremath{^{-}})}&\multicolumn{1}{l}{1}&\multicolumn{1}{r@{}}{3}&\multicolumn{1}{@{.}l}{62}&\parbox[t][0.3cm]{11.518261cm}{\raggedright E(level): From (\href{https://www.nndc.bnl.gov/nsr/nsrlink.jsp?1972Ha03,B}{1972Ha03}).\vspace{0.1cm}}&\\
&&&&&&\parbox[t][0.3cm]{11.518261cm}{\raggedright E(level): See also 6012 keV \textit{10} (\href{https://www.nndc.bnl.gov/nsr/nsrlink.jsp?1970Ga18,B}{1970Ga18}), which is a member of an unresolved\vspace{0.1cm}}&\\
&&&&&&\parbox[t][0.3cm]{11.518261cm}{\raggedright {\ }{\ }{\ }doublet with another level at E\ensuremath{_{\textnormal{x}}}=6089 keV \textit{10}; and 6014.7 (\href{https://www.nndc.bnl.gov/nsr/nsrlink.jsp?2023Po03,B}{2023Po03}), which was\vspace{0.1cm}}&\\
&&&&&&\parbox[t][0.3cm]{11.518261cm}{\raggedright {\ }{\ }{\ }used as a calibration point.\vspace{0.1cm}}&\\
&&&&&&\parbox[t][0.3cm]{11.518261cm}{\raggedright E(level): (\href{https://www.nndc.bnl.gov/nsr/nsrlink.jsp?2023Po03,B}{2023Po03}) reported that this state is thought to be an unresolved doublet\vspace{0.1cm}}&\\
&&&&&&\parbox[t][0.3cm]{11.518261cm}{\raggedright {\ }{\ }{\ }by (\href{https://www.nndc.bnl.gov/nsr/nsrlink.jsp?2021Ri04,B}{2021Ri04}: \ensuremath{^{\textnormal{19}}}F(\ensuremath{^{\textnormal{3}}}He,t)) that is comprised of two states: One broad state at\vspace{0.1cm}}&\\
&&&&&&\parbox[t][0.3cm]{11.518261cm}{\raggedright {\ }{\ }{\ }E\ensuremath{_{\textnormal{x}}}\ensuremath{\approx}6008 keV with \ensuremath{\Gamma}=124 keV \textit{25}, and another state at E\ensuremath{_{\textnormal{x}}}\ensuremath{\approx}6014 keV. The\vspace{0.1cm}}&\\
&&&&&&\parbox[t][0.3cm]{11.518261cm}{\raggedright {\ }{\ }{\ }former state was not observed in (\href{https://www.nndc.bnl.gov/nsr/nsrlink.jsp?2023Po03,B}{2023Po03}). There is also no indication that the\vspace{0.1cm}}&\\
\end{longtable}
\begin{textblock}{29}(0,27.3)
Continued on next page (footnotes at end of table)
\end{textblock}
\clearpage
\begin{longtable}{ccccc@{\extracolsep{\fill}}c}
\\[-.4cm]
\multicolumn{6}{c}{{\bf \small \underline{\ensuremath{^{\textnormal{20}}}Ne(\ensuremath{^{\textnormal{3}}}He,\ensuremath{\alpha}),(\ensuremath{^{\textnormal{3}}}He,\ensuremath{\alpha}\ensuremath{\gamma})\hspace{0.2in}\href{https://www.nndc.bnl.gov/nsr/nsrlink.jsp?1970Ga18,B}{1970Ga18},\href{https://www.nndc.bnl.gov/nsr/nsrlink.jsp?1972Ha03,B}{1972Ha03},\href{https://www.nndc.bnl.gov/nsr/nsrlink.jsp?2023Po03,B}{2023Po03} (continued)}}}\\
\multicolumn{6}{c}{~}\\
\multicolumn{6}{c}{\underline{\ensuremath{^{19}}Ne Levels (continued)}}\\
\multicolumn{6}{c}{~}\\
\multicolumn{2}{c}{E(level)$^{{\hyperlink{NE38LEVEL0}{a}}}$}&J$^{\pi}$$^{}$&L$^{}$&Comments&\\[-.2cm]
\multicolumn{2}{c}{\hrulefill}&\hrulefill&\hrulefill&\hrulefill&
\endhead
&&&&\parbox[t][0.3cm]{12.665121cm}{\raggedright {\ }{\ }{\ }observed peak in the spectra of (\href{https://www.nndc.bnl.gov/nsr/nsrlink.jsp?2023Po03,B}{2023Po03}) is broader than a single peak. The \ensuremath{\alpha} angular\vspace{0.1cm}}&\\
&&&&\parbox[t][0.3cm]{12.665121cm}{\raggedright {\ }{\ }{\ }distribution obtained for this state by (\href{https://www.nndc.bnl.gov/nsr/nsrlink.jsp?2023Po03,B}{2023Po03}) indicates that only a single state was\vspace{0.1cm}}&\\
&&&&\parbox[t][0.3cm]{12.665121cm}{\raggedright {\ }{\ }{\ }strongly populated in (\href{https://www.nndc.bnl.gov/nsr/nsrlink.jsp?2023Po03,B}{2023Po03}).\vspace{0.1cm}}&\\
&&&&\parbox[t][0.3cm]{12.665121cm}{\raggedright (\href{https://www.nndc.bnl.gov/nsr/nsrlink.jsp?2023Po03,B}{2023Po03}) considered this state to be the analog of the \ensuremath{^{\textnormal{19}}}F*(6088 keV, 3/2\ensuremath{^{-}}) level.\vspace{0.1cm}}&\\
&&&&\parbox[t][0.3cm]{12.665121cm}{\raggedright J\ensuremath{^{\pi}},L: From (i) the zero-range DWBA analyses (using JULIE and with zero lower cutoff\vspace{0.1cm}}&\\
&&&&\parbox[t][0.3cm]{12.665121cm}{\raggedright {\ }{\ }{\ }radius) of (1) (\href{https://www.nndc.bnl.gov/nsr/nsrlink.jsp?1970Ga18,B}{1970Ga18}): J\ensuremath{^{\ensuremath{\pi}}}=(3/2\ensuremath{^{-}}, 1/2\ensuremath{^{-}}) and L=(1) from the 1\textit{p}\ensuremath{_{\textnormal{3/2}}} or 1\textit{p}\ensuremath{_{\textnormal{1/2}}} orbital. The\vspace{0.1cm}}&\\
&&&&\parbox[t][0.3cm]{12.665121cm}{\raggedright {\ }{\ }{\ }DWBA fits do not describe the data at \ensuremath{\theta}\ensuremath{_{\textnormal{c.m.}}}\ensuremath{\geq}60\ensuremath{^\circ}. (\href{https://www.nndc.bnl.gov/nsr/nsrlink.jsp?1970Ga18,B}{1970Ga18}) favored J\ensuremath{^{\ensuremath{\pi}}}=3/2\ensuremath{^{-}} based on\vspace{0.1cm}}&\\
&&&&\parbox[t][0.3cm]{12.665121cm}{\raggedright {\ }{\ }{\ }the spectroscopic factor sum rules; and (2) (\href{https://www.nndc.bnl.gov/nsr/nsrlink.jsp?1972Ha03,B}{1972Ha03}): J\ensuremath{^{\ensuremath{\pi}}}=(1/2\ensuremath{^{-}}, 3/2\ensuremath{^{-}}) and L=1, where\vspace{0.1cm}}&\\
&&&&\parbox[t][0.3cm]{12.665121cm}{\raggedright {\ }{\ }{\ }those authors reported no preference for J\ensuremath{^{\ensuremath{\pi}}}=3/2\ensuremath{^{-}} or J\ensuremath{^{\ensuremath{\pi}}}=1/2\ensuremath{^{-}} based on the spectroscopic\vspace{0.1cm}}&\\
&&&&\parbox[t][0.3cm]{12.665121cm}{\raggedright {\ }{\ }{\ }factor sum rule; and (ii) from the finite-range DWBA analysis of (\href{https://www.nndc.bnl.gov/nsr/nsrlink.jsp?2023Po03,B}{2023Po03}) using\vspace{0.1cm}}&\\
&&&&\parbox[t][0.3cm]{12.665121cm}{\raggedright {\ }{\ }{\ }FRESCO. They assigned J\ensuremath{^{\ensuremath{\pi}}}=(1/2\ensuremath{^{-}}, 3/2\ensuremath{^{-}}) with L=1.\vspace{0.1cm}}&\\
&&&&\parbox[t][0.3cm]{12.665121cm}{\raggedright C\ensuremath{^{\textnormal{2}}}S\ensuremath{_{\textnormal{exp}}}: Obtained for J\ensuremath{^{\ensuremath{\pi}}}=(3/2\ensuremath{^{-}}).\vspace{0.1cm}}&\\
&&&&\parbox[t][0.3cm]{12.665121cm}{\raggedright C\ensuremath{^{\textnormal{2}}}S: See also 1.4 calculated by (\href{https://www.nndc.bnl.gov/nsr/nsrlink.jsp?1974Ga28,B}{1974Ga28}: Table 5) for J\ensuremath{^{\ensuremath{\pi}}}=(3/2\ensuremath{^{-}}).\vspace{0.1cm}}&\\
\multicolumn{1}{r@{}}{6092}&\multicolumn{1}{@{}l}{\ensuremath{^{{\hyperlink{NE38LEVEL3}{d}}{\hyperlink{NE38LEVEL8}{i}}}} {\it 3}}&\multicolumn{1}{l}{(5/2\ensuremath{^{-}},7/2,9/2\ensuremath{^{+}})}&\multicolumn{1}{l}{3,4}&\parbox[t][0.3cm]{12.665121cm}{\raggedright E(level): See also 6104 keV \textit{15} (\href{https://www.nndc.bnl.gov/nsr/nsrlink.jsp?1972Ha03,B}{1972Ha03}); and 6089 keV \textit{10} (\href{https://www.nndc.bnl.gov/nsr/nsrlink.jsp?1970Ga18,B}{1970Ga18}), which is a\vspace{0.1cm}}&\\
&&&&\parbox[t][0.3cm]{12.665121cm}{\raggedright {\ }{\ }{\ }member of an unresolved doublet with another level at E\ensuremath{_{\textnormal{x}}}=6012 keV \textit{10}.\vspace{0.1cm}}&\\
&&&&\parbox[t][0.3cm]{12.665121cm}{\raggedright L,J\ensuremath{^{\pi}}: From finite-range DWBA analysis in (\href{https://www.nndc.bnl.gov/nsr/nsrlink.jsp?2023Po03,B}{2023Po03}) using FRESCO. The DWBA fits of\vspace{0.1cm}}&\\
&&&&\parbox[t][0.3cm]{12.665121cm}{\raggedright {\ }{\ }{\ }(\href{https://www.nndc.bnl.gov/nsr/nsrlink.jsp?2023Po03,B}{2023Po03}) do not describe the data well. Those authors reported that \ensuremath{\sigma}(E\ensuremath{_{\ensuremath{\alpha}}},\ensuremath{\theta}) is flat and\vspace{0.1cm}}&\\
&&&&\parbox[t][0.3cm]{12.665121cm}{\raggedright {\ }{\ }{\ }suggested that this would indicate a high spin state. The experimenters mentioned that L=3\vspace{0.1cm}}&\\
&&&&\parbox[t][0.3cm]{12.665121cm}{\raggedright {\ }{\ }{\ }is the best fit and L=4 is the next best fit. Evaluator notes that this is most likely a\vspace{0.1cm}}&\\
&&&&\parbox[t][0.3cm]{12.665121cm}{\raggedright {\ }{\ }{\ }misprint because the reduced \ensuremath{\chi}\ensuremath{^{\textnormal{2}}} should clearly be lower for L=4 than L=3. So most likely,\vspace{0.1cm}}&\\
&&&&\parbox[t][0.3cm]{12.665121cm}{\raggedright {\ }{\ }{\ }the best fit was achieved by L=4.\vspace{0.1cm}}&\\
\multicolumn{1}{r@{}}{6133}&\multicolumn{1}{@{}l}{\ensuremath{^{{\hyperlink{NE38LEVEL3}{d}}{\hyperlink{NE38LEVEL8}{i}}}} {\it 3}}&\multicolumn{1}{l}{3/2\ensuremath{^{+}}}&\multicolumn{1}{l}{2}&\parbox[t][0.3cm]{12.665121cm}{\raggedright E(level): Other value: 6149 keV \textit{20} (\href{https://www.nndc.bnl.gov/nsr/nsrlink.jsp?1970Ga18,B}{1970Ga18}).\vspace{0.1cm}}&\\
&&&&\parbox[t][0.3cm]{12.665121cm}{\raggedright L,J\ensuremath{^{\pi}}: From finite-range DWBA analysis in (\href{https://www.nndc.bnl.gov/nsr/nsrlink.jsp?2023Po03,B}{2023Po03}) using FRESCO and guided by the\vspace{0.1cm}}&\\
&&&&\parbox[t][0.3cm]{12.665121cm}{\raggedright {\ }{\ }{\ }results of (\href{https://www.nndc.bnl.gov/nsr/nsrlink.jsp?2019Ka15,B}{2019Ka15}: \ensuremath{^{\textnormal{19}}}F(\ensuremath{^{\textnormal{3}}}He,t)) for this state.\vspace{0.1cm}}&\\
&&&&\parbox[t][0.3cm]{12.665121cm}{\raggedright \ensuremath{\Gamma}\ensuremath{_{\ensuremath{\alpha}}}=0.74 keV (\href{https://www.nndc.bnl.gov/nsr/nsrlink.jsp?2019Ka15,B}{2019Ka15}: \ensuremath{^{\textnormal{19}}}F(\ensuremath{^{\textnormal{3}}}He,t)) and ANC=6 fm\ensuremath{^{\textnormal{$-$1/2}}} (\href{https://www.nndc.bnl.gov/nsr/nsrlink.jsp?2019Ka15,B}{2019Ka15}) were recommended\vspace{0.1cm}}&\\
&&&&\parbox[t][0.3cm]{12.665121cm}{\raggedright {\ }{\ }{\ }by and used in the R-matrix analysis of (\href{https://www.nndc.bnl.gov/nsr/nsrlink.jsp?2023Po03,B}{2023Po03}).\vspace{0.1cm}}&\\
\multicolumn{1}{r@{}}{6285}&\multicolumn{1}{@{}l}{\ensuremath{^{{\hyperlink{NE38LEVEL3}{d}}}} {\it 3}}&\multicolumn{1}{l}{1/2\ensuremath{^{+}}\&7/2\ensuremath{^{+}}}&\multicolumn{1}{l}{0+4}&\parbox[t][0.3cm]{12.665121cm}{\raggedright E(level): Weighted average of 6290 keV \textit{10} (\href{https://www.nndc.bnl.gov/nsr/nsrlink.jsp?1970Ga18,B}{1970Ga18}); 6289 keV \textit{10} (\href{https://www.nndc.bnl.gov/nsr/nsrlink.jsp?1972Ha03,B}{1972Ha03}); and\vspace{0.1cm}}&\\
&&&&\parbox[t][0.3cm]{12.665121cm}{\raggedright {\ }{\ }{\ }6284 keV \textit{3} (\href{https://www.nndc.bnl.gov/nsr/nsrlink.jsp?2023Po03,B}{2023Po03}).\vspace{0.1cm}}&\\
&&&&\parbox[t][0.3cm]{12.665121cm}{\raggedright E(level): (\href{https://www.nndc.bnl.gov/nsr/nsrlink.jsp?2023Po03,B}{2023Po03}) reported that this state is likely a closely spaced doublet consisting of a\vspace{0.1cm}}&\\
&&&&\parbox[t][0.3cm]{12.665121cm}{\raggedright {\ }{\ }{\ }low- and a high-spin state suggested by (\href{https://www.nndc.bnl.gov/nsr/nsrlink.jsp?2015Pa46,B}{2015Pa46}: \ensuremath{^{\textnormal{19}}}F(\ensuremath{^{\textnormal{3}}}He,t)) and (\href{https://www.nndc.bnl.gov/nsr/nsrlink.jsp?2020Ha31,B}{2020Ha31}:\vspace{0.1cm}}&\\
&&&&\parbox[t][0.3cm]{12.665121cm}{\raggedright {\ }{\ }{\ }\ensuremath{^{\textnormal{19}}}F(\ensuremath{^{\textnormal{3}}}He,t\ensuremath{\gamma})) with a significant impact on the interference patterns in the \ensuremath{^{\textnormal{18}}}F(p,\ensuremath{\alpha}) cross\vspace{0.1cm}}&\\
&&&&\parbox[t][0.3cm]{12.665121cm}{\raggedright {\ }{\ }{\ }section. In the measurement of (\href{https://www.nndc.bnl.gov/nsr/nsrlink.jsp?2023Po03,B}{2023Po03}), this state was obscured by contaminant states\vspace{0.1cm}}&\\
&&&&\parbox[t][0.3cm]{12.665121cm}{\raggedright {\ }{\ }{\ }and was only measured at 4 angles higher than \ensuremath{\theta}\ensuremath{_{\textnormal{c.m.}}}=20\ensuremath{^\circ}. No broadening of this state was\vspace{0.1cm}}&\\
&&&&\parbox[t][0.3cm]{12.665121cm}{\raggedright {\ }{\ }{\ }observed. (\href{https://www.nndc.bnl.gov/nsr/nsrlink.jsp?2023Po03,B}{2023Po03}) suggested the \ensuremath{^{\textnormal{19}}}F*(6.255 MeV, 1/2\ensuremath{^{\textnormal{+}}}) and \ensuremath{^{\textnormal{19}}}F*(6.330 MeV, 7/2\ensuremath{^{\textnormal{+}}})\vspace{0.1cm}}&\\
&&&&\parbox[t][0.3cm]{12.665121cm}{\raggedright {\ }{\ }{\ }levels as possible mirror states for the members of this doublet.\vspace{0.1cm}}&\\
&&&&\parbox[t][0.3cm]{12.665121cm}{\raggedright \ensuremath{\Gamma}\ensuremath{_{\ensuremath{\alpha}}}=2.51 keV was deduced by (\href{https://www.nndc.bnl.gov/nsr/nsrlink.jsp?2023Po03,B}{2023Po03}) for the higher energy member of this doublet\vspace{0.1cm}}&\\
&&&&\parbox[t][0.3cm]{12.665121cm}{\raggedright {\ }{\ }{\ }assuming J\ensuremath{^{\ensuremath{\pi}}}=7/2\ensuremath{^{\textnormal{+}}} for that state.\vspace{0.1cm}}&\\
&&&&\parbox[t][0.3cm]{12.665121cm}{\raggedright J\ensuremath{^{\pi}},L: From finite-range DWBA analysis in (\href{https://www.nndc.bnl.gov/nsr/nsrlink.jsp?2023Po03,B}{2023Po03}): J\ensuremath{^{\ensuremath{\pi}}}=1/2\ensuremath{^{\textnormal{+}}}+7/2\ensuremath{^{\textnormal{+}}} with L=0+4 using\vspace{0.1cm}}&\\
&&&&\parbox[t][0.3cm]{12.665121cm}{\raggedright {\ }{\ }{\ }FRESCO. The \ensuremath{\alpha} angular distribution measured by (\href{https://www.nndc.bnl.gov/nsr/nsrlink.jsp?2023Po03,B}{2023Po03}) corresponding to this state\vspace{0.1cm}}&\\
&&&&\parbox[t][0.3cm]{12.665121cm}{\raggedright {\ }{\ }{\ }could not be described by a single angular momentum transfer but rather by L=0+4 with\vspace{0.1cm}}&\\
&&&&\parbox[t][0.3cm]{12.665121cm}{\raggedright {\ }{\ }{\ }the 40\% \textit{10} + 60\% \textit{10} proportions, respectively. This observation supports an unresolved\vspace{0.1cm}}&\\
&&&&\parbox[t][0.3cm]{12.665121cm}{\raggedright {\ }{\ }{\ }doublet hypothesis. Angular momenta of L\ensuremath{\geq}4 for the higher energy member of the doublet\vspace{0.1cm}}&\\
&&&&\parbox[t][0.3cm]{12.665121cm}{\raggedright {\ }{\ }{\ }also fit the data of (\href{https://www.nndc.bnl.gov/nsr/nsrlink.jsp?2023Po03,B}{2023Po03}) well. The J\ensuremath{^{\ensuremath{\pi}}} value of the high spin member of the doublet\vspace{0.1cm}}&\\
&&&&\parbox[t][0.3cm]{12.665121cm}{\raggedright {\ }{\ }{\ }was assigned to be (11/2\ensuremath{^{\textnormal{+}}}) from the observed \ensuremath{\gamma}-rays following the decay of this state\vspace{0.1cm}}&\\
&&&&\parbox[t][0.3cm]{12.665121cm}{\raggedright {\ }{\ }{\ }(\href{https://www.nndc.bnl.gov/nsr/nsrlink.jsp?2020Ha31,B}{2020Ha31}: \ensuremath{^{\textnormal{19}}}F(\ensuremath{^{\textnormal{3}}}He,t\ensuremath{\gamma})). (\href{https://www.nndc.bnl.gov/nsr/nsrlink.jsp?2023Po03,B}{2023Po03}) recommended (based on mirror level analysis)\vspace{0.1cm}}&\\
&&&&\parbox[t][0.3cm]{12.665121cm}{\raggedright {\ }{\ }{\ }J\ensuremath{^{\ensuremath{\pi}}}=1/2\ensuremath{^{\textnormal{+}}} and J\ensuremath{^{\ensuremath{\pi}}}\ensuremath{\geq}7/2\ensuremath{^{\textnormal{+}}} for the members of this doublet to investigate the maximum effect\vspace{0.1cm}}&\\
&&&&\parbox[t][0.3cm]{12.665121cm}{\raggedright {\ }{\ }{\ }of this state on the \ensuremath{^{\textnormal{18}}}F(p,\ensuremath{\alpha}) reaction rate.\vspace{0.1cm}}&\\
&&&&\parbox[t][0.3cm]{12.665121cm}{\raggedright J\ensuremath{^{\pi}},L: See also J\ensuremath{^{\ensuremath{\pi}}}=(1/2, 3/2\ensuremath{^{\textnormal{+}}}) and L=(0,1,2) from the zero-range DWBA analysis of\vspace{0.1cm}}&\\
&&&&\parbox[t][0.3cm]{12.665121cm}{\raggedright {\ }{\ }{\ }(\href{https://www.nndc.bnl.gov/nsr/nsrlink.jsp?1972Ha03,B}{1972Ha03}) using JULIE with zero lower cutoff radius, see Table II. The DWBA fits do\vspace{0.1cm}}&\\
&&&&\parbox[t][0.3cm]{12.665121cm}{\raggedright {\ }{\ }{\ }not describe the data well. Therefore, the evaluator assumed these results are tentative.\vspace{0.1cm}}&\\
&&&&\parbox[t][0.3cm]{12.665121cm}{\raggedright ANC\ensuremath{<}16 fm\ensuremath{^{\textnormal{$-$1/2}}} (\href{https://www.nndc.bnl.gov/nsr/nsrlink.jsp?2023Po03,B}{2023Po03}): For the higher energy member of this doublet assuming\vspace{0.1cm}}&\\
&&&&\parbox[t][0.3cm]{12.665121cm}{\raggedright {\ }{\ }{\ }J\ensuremath{^{\ensuremath{\pi}}}=7/2\ensuremath{^{\textnormal{+}}}.\vspace{0.1cm}}&\\
\multicolumn{1}{r@{}}{6438}&\multicolumn{1}{@{}l}{\ensuremath{^{{\hyperlink{NE38LEVEL3}{d}}}} {\it 10}}&&&\parbox[t][0.3cm]{12.665121cm}{\raggedright E(level): From (\href{https://www.nndc.bnl.gov/nsr/nsrlink.jsp?1972Ha03,B}{1972Ha03}). See also 6433 keV \textit{20} (\href{https://www.nndc.bnl.gov/nsr/nsrlink.jsp?1970Ga18,B}{1970Ga18}): An unresolved doublet.\vspace{0.1cm}}&\\
\end{longtable}
\begin{textblock}{29}(0,27.3)
Continued on next page (footnotes at end of table)
\end{textblock}
\clearpage
\begin{longtable}{ccccccc@{\extracolsep{\fill}}c}
\\[-.4cm]
\multicolumn{8}{c}{{\bf \small \underline{\ensuremath{^{\textnormal{20}}}Ne(\ensuremath{^{\textnormal{3}}}He,\ensuremath{\alpha}),(\ensuremath{^{\textnormal{3}}}He,\ensuremath{\alpha}\ensuremath{\gamma})\hspace{0.2in}\href{https://www.nndc.bnl.gov/nsr/nsrlink.jsp?1970Ga18,B}{1970Ga18},\href{https://www.nndc.bnl.gov/nsr/nsrlink.jsp?1972Ha03,B}{1972Ha03},\href{https://www.nndc.bnl.gov/nsr/nsrlink.jsp?2023Po03,B}{2023Po03} (continued)}}}\\
\multicolumn{8}{c}{~}\\
\multicolumn{8}{c}{\underline{\ensuremath{^{19}}Ne Levels (continued)}}\\
\multicolumn{8}{c}{~}\\
\multicolumn{2}{c}{E(level)$^{{\hyperlink{NE38LEVEL0}{a}}}$}&J$^{\pi}$$^{}$&L$^{}$&\multicolumn{2}{c}{C\ensuremath{^{\textnormal{2}}}S\ensuremath{_{\textnormal{exp}}}$^{{\hyperlink{NE38LEVEL5}{f}}}$}&Comments&\\[-.2cm]
\multicolumn{2}{c}{\hrulefill}&\hrulefill&\hrulefill&\multicolumn{2}{c}{\hrulefill}&\hrulefill&
\endhead
\multicolumn{1}{r@{}}{6739}&\multicolumn{1}{@{}l}{\ensuremath{^{{\hyperlink{NE38LEVEL3}{d}}}} {\it 3}}&\multicolumn{1}{l}{(1/2\ensuremath{^{-}},3/2\ensuremath{^{-}})}&\multicolumn{1}{l}{1}&\multicolumn{1}{r@{}}{3}&\multicolumn{1}{@{.}l}{62}&\parbox[t][0.3cm]{11.278581cm}{\raggedright E(level): Weighted average of 6744 keV \textit{10} (\href{https://www.nndc.bnl.gov/nsr/nsrlink.jsp?1970Ga18,B}{1970Ga18}); 6741 keV \textit{10}\vspace{0.1cm}}&\\
&&&&&&\parbox[t][0.3cm]{11.278581cm}{\raggedright {\ }{\ }{\ }(\href{https://www.nndc.bnl.gov/nsr/nsrlink.jsp?1972Ha03,B}{1972Ha03}); and 6738 keV \textit{3} (\href{https://www.nndc.bnl.gov/nsr/nsrlink.jsp?2023Po03,B}{2023Po03}).\vspace{0.1cm}}&\\
&&&&&&\parbox[t][0.3cm]{11.278581cm}{\raggedright E(level): The mirror level was suggested to be \ensuremath{^{\textnormal{19}}}F*(6.787 MeV, 3/2\ensuremath{^{-}}) (\href{https://www.nndc.bnl.gov/nsr/nsrlink.jsp?2023Po03,B}{2023Po03}).\vspace{0.1cm}}&\\
&&&&&&\parbox[t][0.3cm]{11.278581cm}{\raggedright J\ensuremath{^{\pi}},L: From (i) the zero-range DWBA analyses (using JULIE and with zero cutoff\vspace{0.1cm}}&\\
&&&&&&\parbox[t][0.3cm]{11.278581cm}{\raggedright {\ }{\ }{\ }radius) of (1) (\href{https://www.nndc.bnl.gov/nsr/nsrlink.jsp?1970Ga18,B}{1970Ga18}): J\ensuremath{^{\ensuremath{\pi}}}=(1/2\ensuremath{^{-}}, 3/2\ensuremath{^{-}}) and L=1 from the 1\textit{p}\ensuremath{_{\textnormal{3/2}}} or 1\textit{p}\ensuremath{_{\textnormal{1/2}}}\vspace{0.1cm}}&\\
&&&&&&\parbox[t][0.3cm]{11.278581cm}{\raggedright {\ }{\ }{\ }orbital. The DWBA fits do not describe the data at \ensuremath{\theta}\ensuremath{_{\textnormal{c.m.}}}\ensuremath{\geq}30\ensuremath{^\circ}. Those authors\vspace{0.1cm}}&\\
&&&&&&\parbox[t][0.3cm]{11.278581cm}{\raggedright {\ }{\ }{\ }favored J\ensuremath{^{\ensuremath{\pi}}}=3/2\ensuremath{^{-}} based on the spectroscopic factor sum rules; and (2)\vspace{0.1cm}}&\\
&&&&&&\parbox[t][0.3cm]{11.278581cm}{\raggedright {\ }{\ }{\ }(\href{https://www.nndc.bnl.gov/nsr/nsrlink.jsp?1972Ha03,B}{1972Ha03}): J\ensuremath{^{\ensuremath{\pi}}}=(1/2\ensuremath{^{-}}, 3/2\ensuremath{^{-}}) and L=1. They reported no preference for J\ensuremath{^{\ensuremath{\pi}}}=3/2\ensuremath{^{-}}\vspace{0.1cm}}&\\
&&&&&&\parbox[t][0.3cm]{11.278581cm}{\raggedright {\ }{\ }{\ }or J\ensuremath{^{\ensuremath{\pi}}}=1/2\ensuremath{^{-}} based on the spectroscopic factor sum rule; and (ii) from the\vspace{0.1cm}}&\\
&&&&&&\parbox[t][0.3cm]{11.278581cm}{\raggedright {\ }{\ }{\ }finite-range DWBA analysis of (\href{https://www.nndc.bnl.gov/nsr/nsrlink.jsp?2023Po03,B}{2023Po03}): J\ensuremath{^{\ensuremath{\pi}}}=3/2\ensuremath{^{-}} and L=1 using FRESCO.\vspace{0.1cm}}&\\
&&&&&&\parbox[t][0.3cm]{11.278581cm}{\raggedright {\ }{\ }{\ }The DWBA fit was obtained using 3 data points at high angles. Results are\vspace{0.1cm}}&\\
&&&&&&\parbox[t][0.3cm]{11.278581cm}{\raggedright {\ }{\ }{\ }guided by the findings of (\href{https://www.nndc.bnl.gov/nsr/nsrlink.jsp?2004Vi05,B}{2004Vi05}: \ensuremath{^{\textnormal{19}}}F(\ensuremath{^{\textnormal{3}}}He,t)), (\href{https://www.nndc.bnl.gov/nsr/nsrlink.jsp?2019Ka15,B}{2019Ka15}: \ensuremath{^{\textnormal{19}}}F(\ensuremath{^{\textnormal{3}}}He,t)),\vspace{0.1cm}}&\\
&&&&&&\parbox[t][0.3cm]{11.278581cm}{\raggedright {\ }{\ }{\ }(\href{https://www.nndc.bnl.gov/nsr/nsrlink.jsp?2019Ha08,B}{2019Ha08}: \ensuremath{^{\textnormal{19}}}F(\ensuremath{^{\textnormal{3}}}He,t\ensuremath{\gamma})), and the evaluation of (\href{https://www.nndc.bnl.gov/nsr/nsrlink.jsp?2007Ne09,B}{2007Ne09}).\vspace{0.1cm}}&\\
&&&&&&\parbox[t][0.3cm]{11.278581cm}{\raggedright C\ensuremath{^{\textnormal{2}}}S\ensuremath{_{\textnormal{exp}}}: Obtained for J\ensuremath{^{\ensuremath{\pi}}}=(3/2\ensuremath{^{-}}).\vspace{0.1cm}}&\\
&&&&&&\parbox[t][0.3cm]{11.278581cm}{\raggedright See also C\ensuremath{^{\textnormal{2}}}S=1.5 calculated by (\href{https://www.nndc.bnl.gov/nsr/nsrlink.jsp?1974Ga28,B}{1974Ga28}: Table 5) for J\ensuremath{^{\ensuremath{\pi}}}=3/2\ensuremath{^{-}}. Note that this\vspace{0.1cm}}&\\
&&&&&&\parbox[t][0.3cm]{11.278581cm}{\raggedright {\ }{\ }{\ }theoretical value is obtained when the DWBA cross section normalization is\vspace{0.1cm}}&\\
&&&&&&\parbox[t][0.3cm]{11.278581cm}{\raggedright {\ }{\ }{\ }made at the second maximum. If the normalization is made at the first cross\vspace{0.1cm}}&\\
&&&&&&\parbox[t][0.3cm]{11.278581cm}{\raggedright {\ }{\ }{\ }section$'$s maximum, then the computed C\ensuremath{^{\textnormal{2}}}S\ensuremath{_{\textnormal{theo}}} will be 1.7 times larger\vspace{0.1cm}}&\\
&&&&&&\parbox[t][0.3cm]{11.278581cm}{\raggedright {\ }{\ }{\ }(\href{https://www.nndc.bnl.gov/nsr/nsrlink.jsp?1974Ga28,B}{1974Ga28}): See footnote d in Table 5.\vspace{0.1cm}}&\\
\multicolumn{1}{r@{}}{6861}&\multicolumn{1}{@{}l}{\ensuremath{^{{\hyperlink{NE38LEVEL3}{d}}}} {\it 4}}&\multicolumn{1}{l}{7/2\ensuremath{^{-}}}&\multicolumn{1}{l}{3}&&&\parbox[t][0.3cm]{11.278581cm}{\raggedright E(level): Weighted average of 6866 keV \textit{10} (\href{https://www.nndc.bnl.gov/nsr/nsrlink.jsp?1970Ga18,B}{1970Ga18}); 6858 keV \textit{10}\vspace{0.1cm}}&\\
&&&&&&\parbox[t][0.3cm]{11.278581cm}{\raggedright {\ }{\ }{\ }(\href{https://www.nndc.bnl.gov/nsr/nsrlink.jsp?1972Ha03,B}{1972Ha03}); and 6861 keV \textit{4} (\href{https://www.nndc.bnl.gov/nsr/nsrlink.jsp?2023Po03,B}{2023Po03}).\vspace{0.1cm}}&\\
&&&&&&\parbox[t][0.3cm]{11.278581cm}{\raggedright E(level): The mirror level was suggested to be \ensuremath{^{\textnormal{19}}}F*(6.927 MeV) (\href{https://www.nndc.bnl.gov/nsr/nsrlink.jsp?2023Po03,B}{2023Po03}).\vspace{0.1cm}}&\\
&&&&&&\parbox[t][0.3cm]{11.278581cm}{\raggedright J\ensuremath{^{\pi}},L: From finite-range DWBA analysis in (\href{https://www.nndc.bnl.gov/nsr/nsrlink.jsp?2023Po03,B}{2023Po03}) using FRESCO, which\vspace{0.1cm}}&\\
&&&&&&\parbox[t][0.3cm]{11.278581cm}{\raggedright {\ }{\ }{\ }confirms the result of (\href{https://www.nndc.bnl.gov/nsr/nsrlink.jsp?2013La01,B}{2013La01}: \ensuremath{^{\textnormal{19}}}F(\ensuremath{^{\textnormal{3}}}He,t)).\vspace{0.1cm}}&\\
&&&&&&\parbox[t][0.3cm]{11.278581cm}{\raggedright J\ensuremath{^{\pi}}: See also J\ensuremath{^{\ensuremath{\pi}}}=(1/2, 3/2\ensuremath{^{\textnormal{+}}}) with L=(0,1,2) from the zero-range DWBA analysis\vspace{0.1cm}}&\\
&&&&&&\parbox[t][0.3cm]{11.278581cm}{\raggedright {\ }{\ }{\ }with zero lower cutoff radius using JULIE in (\href{https://www.nndc.bnl.gov/nsr/nsrlink.jsp?1972Ha03,B}{1972Ha03}): See Table II and Fig.\vspace{0.1cm}}&\\
&&&&&&\parbox[t][0.3cm]{11.278581cm}{\raggedright {\ }{\ }{\ }5, where the DWBA fits do not describe the data well.\vspace{0.1cm}}&\\
\multicolumn{1}{r@{}}{7067}&\multicolumn{1}{@{}l}{\ensuremath{^{{\hyperlink{NE38LEVEL3}{d}}}} {\it 10}}&\multicolumn{1}{l}{3/2\ensuremath{^{+}}}&\multicolumn{1}{l}{2}&&&\parbox[t][0.3cm]{11.278581cm}{\raggedright E(level): Weighted average of 7064 keV \textit{20} (\href{https://www.nndc.bnl.gov/nsr/nsrlink.jsp?1970Ga18,B}{1970Ga18}) and 7068 keV \textit{10}\vspace{0.1cm}}&\\
&&&&&&\parbox[t][0.3cm]{11.278581cm}{\raggedright {\ }{\ }{\ }(\href{https://www.nndc.bnl.gov/nsr/nsrlink.jsp?1972Ha03,B}{1972Ha03}). See also 7074.7 keV (\href{https://www.nndc.bnl.gov/nsr/nsrlink.jsp?2023Po03,B}{2023Po03}): Used as a calibration point.\vspace{0.1cm}}&\\
&&&&&&\parbox[t][0.3cm]{11.278581cm}{\raggedright E(level): The mirror level was suggested to be \ensuremath{^{\textnormal{19}}}F*(7.262 MeV, 3/2\ensuremath{^{\textnormal{+}}})\vspace{0.1cm}}&\\
&&&&&&\parbox[t][0.3cm]{11.278581cm}{\raggedright {\ }{\ }{\ }(\href{https://www.nndc.bnl.gov/nsr/nsrlink.jsp?2023Po03,B}{2023Po03}).\vspace{0.1cm}}&\\
&&&&&&\parbox[t][0.3cm]{11.278581cm}{\raggedright J\ensuremath{^{\pi}},L: From finite-range DWBA analysis in (\href{https://www.nndc.bnl.gov/nsr/nsrlink.jsp?2023Po03,B}{2023Po03}) using FRESCO. The\vspace{0.1cm}}&\\
&&&&&&\parbox[t][0.3cm]{11.278581cm}{\raggedright {\ }{\ }{\ }DWBA fit was obtained using only 3 data points at \ensuremath{\theta}\ensuremath{_{\textnormal{c.m.}}}\ensuremath{\geq}25\ensuremath{^\circ}.\vspace{0.1cm}}&\\
\multicolumn{1}{r@{}}{7178?}&\multicolumn{1}{@{}l}{\ensuremath{^{{\hyperlink{NE38LEVEL7}{h}}}} {\it 15}}&&&&&&\\
\multicolumn{1}{r@{}}{7253}&\multicolumn{1}{@{}l}{\ensuremath{^{{\hyperlink{NE38LEVEL7}{h}}}} {\it 10}}&&&&&&\\
\multicolumn{1}{r@{}}{7326?}&\multicolumn{1}{@{}l}{\ensuremath{^{{\hyperlink{NE38LEVEL7}{h}}}} {\it 15}}&&&&&&\\
\multicolumn{1}{r@{}}{7549}&\multicolumn{1}{@{}l}{\ensuremath{^{{\hyperlink{NE38LEVEL8}{i}}}} {\it 5}}&&&&&\parbox[t][0.3cm]{11.278581cm}{\raggedright E(level): See also 7531 keV \textit{15} (\href{https://www.nndc.bnl.gov/nsr/nsrlink.jsp?1972Ha03,B}{1972Ha03}): A tentative state.\vspace{0.1cm}}&\\
\multicolumn{1}{r@{}}{7605}&\multicolumn{1}{@{}l}{\ensuremath{^{{\hyperlink{NE38LEVEL8}{i}}}} {\it 5}}&&&&&\parbox[t][0.3cm]{11.278581cm}{\raggedright E(level): See also 7614 keV \textit{20} (\href{https://www.nndc.bnl.gov/nsr/nsrlink.jsp?1972Ha03,B}{1972Ha03}).\vspace{0.1cm}}&\\
&&&&&&\parbox[t][0.3cm]{11.278581cm}{\raggedright (\href{https://www.nndc.bnl.gov/nsr/nsrlink.jsp?1972Ha03,B}{1972Ha03}) reported that this state may be the J\ensuremath{^{\ensuremath{\pi}}}=3/2\ensuremath{^{\textnormal{+}}}, T=3/2 state reported by\vspace{0.1cm}}&\\
&&&&&&\parbox[t][0.3cm]{11.278581cm}{\raggedright {\ }{\ }{\ }(\href{https://www.nndc.bnl.gov/nsr/nsrlink.jsp?1969Ha38,B}{1969Ha38}: \ensuremath{^{\textnormal{21}}}Ne(p,t)) at E\ensuremath{_{\textnormal{x}}}=7620 keV \textit{25}. Assuming this is the case, the\vspace{0.1cm}}&\\
&&&&&&\parbox[t][0.3cm]{11.278581cm}{\raggedright {\ }{\ }{\ }DWBA analysis of (\href{https://www.nndc.bnl.gov/nsr/nsrlink.jsp?1972Ha03,B}{1972Ha03}), which is not presented, places an upper limit on\vspace{0.1cm}}&\\
&&&&&&\parbox[t][0.3cm]{11.278581cm}{\raggedright {\ }{\ }{\ }the isospin impurity of this state at 2-4\%. Those authors also reported that the \ensuremath{\alpha}\vspace{0.1cm}}&\\
&&&&&&\parbox[t][0.3cm]{11.278581cm}{\raggedright {\ }{\ }{\ }angular distribution corresponding to this state (also not presented) is isotropic.\vspace{0.1cm}}&\\
\multicolumn{1}{r@{}}{7700}&\multicolumn{1}{@{}l}{\ensuremath{^{{\hyperlink{NE38LEVEL7}{h}}}} {\it 10}}&&&&&\parbox[t][0.3cm]{11.278581cm}{\raggedright E(level): See also 7706 keV (\href{https://www.nndc.bnl.gov/nsr/nsrlink.jsp?2023Po03,B}{2023Po03}): Used as a calibration point.\vspace{0.1cm}}&\\
\multicolumn{1}{r@{}}{7788}&\multicolumn{1}{@{}l}{\ensuremath{^{{\hyperlink{NE38LEVEL7}{h}}}} {\it 10}}&&&&&\parbox[t][0.3cm]{11.278581cm}{\raggedright E(level): This state was considered tentative by (\href{https://www.nndc.bnl.gov/nsr/nsrlink.jsp?1972Ha03,B}{1972Ha03}). However, (\href{https://www.nndc.bnl.gov/nsr/nsrlink.jsp?2023Po03,B}{2023Po03})\vspace{0.1cm}}&\\
&&&&&&\parbox[t][0.3cm]{11.278581cm}{\raggedright {\ }{\ }{\ }also observed this state and used the E\ensuremath{_{\textnormal{x}}}=7788 keV as a calibration point.\vspace{0.1cm}}&\\
&&&&&&\parbox[t][0.3cm]{11.278581cm}{\raggedright {\ }{\ }{\ }Therefore, we removed the tentative status.\vspace{0.1cm}}&\\
\multicolumn{1}{r@{}}{7970}&\multicolumn{1}{@{}l}{\ensuremath{^{{\hyperlink{NE38LEVEL8}{i}}}} {\it 4}}&&&&&\parbox[t][0.3cm]{11.278581cm}{\raggedright E(level): See also 7994 keV \textit{15} (\href{https://www.nndc.bnl.gov/nsr/nsrlink.jsp?1972Ha03,B}{1972Ha03}).\vspace{0.1cm}}&\\
\multicolumn{1}{r@{}}{8057}&\multicolumn{1}{@{}l}{\ensuremath{^{{\hyperlink{NE38LEVEL8}{i}}}} {\it 5}}&&&&&\parbox[t][0.3cm]{11.278581cm}{\raggedright E(level): See also 8063 keV \textit{15} (\href{https://www.nndc.bnl.gov/nsr/nsrlink.jsp?1972Ha03,B}{1972Ha03}).\vspace{0.1cm}}&\\
\multicolumn{1}{r@{}}{8236}&\multicolumn{1}{@{}l}{\ensuremath{^{{\hyperlink{NE38LEVEL7}{h}}}} {\it 10}}&\multicolumn{1}{l}{(1/2,3/2\ensuremath{^{+}})}&\multicolumn{1}{l}{(0,1,2)}&&&\parbox[t][0.3cm]{11.278581cm}{\raggedright E(level): See also 8236 keV (\href{https://www.nndc.bnl.gov/nsr/nsrlink.jsp?2023Po03,B}{2023Po03}): Used as a calibration point.\vspace{0.1cm}}&\\
&&&&&&\parbox[t][0.3cm]{11.278581cm}{\raggedright L,J\ensuremath{^{\pi}}: From a zero-range DWBA analysis with zero lower cutoff radius using\vspace{0.1cm}}&\\
&&&&&&\parbox[t][0.3cm]{11.278581cm}{\raggedright {\ }{\ }{\ }JULIE in (\href{https://www.nndc.bnl.gov/nsr/nsrlink.jsp?1972Ha03,B}{1972Ha03}: See Table II). The DWBA fits do not describe the data\vspace{0.1cm}}&\\
\end{longtable}
\begin{textblock}{29}(0,27.3)
Continued on next page (footnotes at end of table)
\end{textblock}
\clearpage
\begin{longtable}{ccc@{\extracolsep{\fill}}c}
\\[-.4cm]
\multicolumn{4}{c}{{\bf \small \underline{\ensuremath{^{\textnormal{20}}}Ne(\ensuremath{^{\textnormal{3}}}He,\ensuremath{\alpha}),(\ensuremath{^{\textnormal{3}}}He,\ensuremath{\alpha}\ensuremath{\gamma})\hspace{0.2in}\href{https://www.nndc.bnl.gov/nsr/nsrlink.jsp?1970Ga18,B}{1970Ga18},\href{https://www.nndc.bnl.gov/nsr/nsrlink.jsp?1972Ha03,B}{1972Ha03},\href{https://www.nndc.bnl.gov/nsr/nsrlink.jsp?2023Po03,B}{2023Po03} (continued)}}}\\
\multicolumn{4}{c}{~}\\
\multicolumn{4}{c}{\underline{\ensuremath{^{19}}Ne Levels (continued)}}\\
\multicolumn{4}{c}{~}\\
\multicolumn{2}{c}{E(level)$^{{\hyperlink{NE38LEVEL0}{a}}}$}&Comments&\\[-.2cm]
\multicolumn{2}{c}{\hrulefill}&\hrulefill&
\endhead
&&\parbox[t][0.3cm]{15.802001cm}{\raggedright {\ }{\ }{\ }well.\vspace{0.1cm}}&\\
\multicolumn{1}{r@{}}{8440}&\multicolumn{1}{@{}l}{\ensuremath{^{{\hyperlink{NE38LEVEL7}{h}}}} {\it 10}}&&\\
\multicolumn{1}{r@{}}{8523}&\multicolumn{1}{@{}l}{\ensuremath{^{{\hyperlink{NE38LEVEL7}{h}}}} {\it 10}}&&\\
\multicolumn{1}{r@{}}{8810?}&\multicolumn{1}{@{}l}{\ensuremath{^{{\hyperlink{NE38LEVEL7}{h}}}} {\it 25}}&&\\
\multicolumn{1}{r@{}}{8915}&\multicolumn{1}{@{}l}{\ensuremath{^{{\hyperlink{NE38LEVEL7}{h}}}} {\it 10}}&&\\
\multicolumn{1}{r@{}}{9013}&\multicolumn{1}{@{}l}{\ensuremath{^{{\hyperlink{NE38LEVEL7}{h}}}} {\it 10}}&&\\
\multicolumn{1}{r@{}}{9100?}&\multicolumn{1}{@{}l}{\ensuremath{^{{\hyperlink{NE38LEVEL7}{h}}}} {\it 20}}&&\\
\multicolumn{1}{r@{}}{9240}&\multicolumn{1}{@{}l}{\ensuremath{^{{\hyperlink{NE38LEVEL7}{h}}}} {\it 20}}&&\\
\multicolumn{1}{r@{}}{9489}&\multicolumn{1}{@{}l}{\ensuremath{^{{\hyperlink{NE38LEVEL7}{h}}}} {\it 25}}&&\\
\multicolumn{1}{r@{}}{9886}&\multicolumn{1}{@{}l}{\ensuremath{^{{\hyperlink{NE38LEVEL7}{h}}}} {\it 50}}&&\\
\multicolumn{1}{r@{}}{10407}&\multicolumn{1}{@{}l}{\ensuremath{^{{\hyperlink{NE38LEVEL7}{h}}}} {\it 30}}&&\\
\multicolumn{1}{r@{}}{10613}&\multicolumn{1}{@{}l}{\ensuremath{^{{\hyperlink{NE38LEVEL7}{h}}}} {\it 20}}&&\\
\end{longtable}
\parbox[b][0.3cm]{17.7cm}{\makebox[1ex]{\ensuremath{^{\hypertarget{NE38LEVEL0}{a}}}} The excitation energies quoted from (\href{https://www.nndc.bnl.gov/nsr/nsrlink.jsp?1972Ha03,B}{1972Ha03}) are the ``corrected energies'' given in Table I, which are deduced by those}\\
\parbox[b][0.3cm]{17.7cm}{{\ }{\ }authors after pulse height corrections were made.}\\
\parbox[b][0.3cm]{17.7cm}{\makebox[1ex]{\ensuremath{^{\hypertarget{NE38LEVEL1}{b}}}} See also 1521 keV \textit{20} (\href{https://www.nndc.bnl.gov/nsr/nsrlink.jsp?1967Gr04,B}{1967Gr04}): Unresolved doublet that consisted of E\ensuremath{_{\textnormal{x}}}=1507 keV \textit{20} (see Table I, measured at \ensuremath{\theta}\ensuremath{_{\textnormal{lab}}}=167.5\ensuremath{^\circ})}\\
\parbox[b][0.3cm]{17.7cm}{{\ }{\ }and E\ensuremath{_{\textnormal{x}}}=1535 keV \textit{20} (see Table I, measured at \ensuremath{\theta}\ensuremath{_{\textnormal{lab}}}=165\ensuremath{^\circ}). The energies of the constituent members of this doublet are}\\
\parbox[b][0.3cm]{17.7cm}{{\ }{\ }measured at angles where they seemed to have been separated from one another.}\\
\parbox[b][0.3cm]{17.7cm}{\makebox[1ex]{\ensuremath{^{\hypertarget{NE38LEVEL2}{c}}}} This state was first observed by (\href{https://www.nndc.bnl.gov/nsr/nsrlink.jsp?1967Gr04,B}{1967Gr04}).}\\
\parbox[b][0.3cm]{17.7cm}{\makebox[1ex]{\ensuremath{^{\hypertarget{NE38LEVEL3}{d}}}} This state was first observed by (\href{https://www.nndc.bnl.gov/nsr/nsrlink.jsp?1970Ga18,B}{1970Ga18}).}\\
\parbox[b][0.3cm]{17.7cm}{\makebox[1ex]{\ensuremath{^{\hypertarget{NE38LEVEL4}{e}}}} (\href{https://www.nndc.bnl.gov/nsr/nsrlink.jsp?1970Ga18,B}{1970Ga18}) assigned J\ensuremath{^{\ensuremath{\pi}}}=(5/2\ensuremath{^{-}}) to the \ensuremath{^{\textnormal{19}}}Ne*(1.51 MeV) member of the observed doublet and J\ensuremath{^{\ensuremath{\pi}}}=(3/2\ensuremath{^{\textnormal{+}}}) to the \ensuremath{^{\textnormal{19}}}Ne*(1.54}\\
\parbox[b][0.3cm]{17.7cm}{{\ }{\ }MeV) member based on mirror analysis for the former and DWBA analysis with L=2 for the latter. The DWBA fits shown in}\\
\parbox[b][0.3cm]{17.7cm}{{\ }{\ }(\href{https://www.nndc.bnl.gov/nsr/nsrlink.jsp?1970Ga18,B}{1970Ga18}) do not describe the data. The authors reported that at forward angles, the strength was nearly all contained by the}\\
\parbox[b][0.3cm]{17.7cm}{{\ }{\ }1.54-MeV member of the observed doublet. DWBA fit with L=3 was not performed.}\\
\parbox[b][0.3cm]{17.7cm}{\makebox[1ex]{\ensuremath{^{\hypertarget{NE38LEVEL5}{f}}}} From (\href{https://www.nndc.bnl.gov/nsr/nsrlink.jsp?1972En03,B}{1972En03}): See Table 15, data taken from (\href{https://www.nndc.bnl.gov/nsr/nsrlink.jsp?1970Ga18,B}{1970Ga18}). The quoted, calculated values from (\href{https://www.nndc.bnl.gov/nsr/nsrlink.jsp?1974Ga28,B}{1974Ga28}) are deduced with}\\
\parbox[b][0.3cm]{17.7cm}{{\ }{\ }the absolute DWBA normalization factor of 10.2, which is obtained from the finite-range, non-local DWBA analysis of}\\
\parbox[b][0.3cm]{17.7cm}{{\ }{\ }(\href{https://www.nndc.bnl.gov/nsr/nsrlink.jsp?1974Ga28,B}{1974Ga28}), see set 1 in Table 3 and Table 5 in that study.}\\
\parbox[b][0.3cm]{17.7cm}{\makebox[1ex]{\ensuremath{^{\hypertarget{NE38LEVEL6}{g}}}} Evaluator highlights that (\href{https://www.nndc.bnl.gov/nsr/nsrlink.jsp?1970Ga18,B}{1970Ga18}) initially assigned J\ensuremath{^{\ensuremath{\pi}}}=7/2\ensuremath{^{-}} to the \ensuremath{^{\textnormal{19}}}Ne*(4142) state and J\ensuremath{^{\ensuremath{\pi}}}=9/2\ensuremath{^{-}} to the \ensuremath{^{\textnormal{19}}}Ne*(4200) state}\\
\parbox[b][0.3cm]{17.7cm}{{\ }{\ }based on a comparison of the rotational bands in \ensuremath{^{\textnormal{19}}}F and \ensuremath{^{\textnormal{19}}}Ne mirror nuclei. Based on their DWBA analysis; however, they}\\
\parbox[b][0.3cm]{17.7cm}{{\ }{\ }suggested swapping the spin assignments of the \ensuremath{^{\textnormal{19}}}Ne*(4142, 4200) states. The previous evaluation by (\href{https://www.nndc.bnl.gov/nsr/nsrlink.jsp?1995Ti07,B}{1995Ti07}) tentatively}\\
\parbox[b][0.3cm]{17.7cm}{{\ }{\ }adopted this latter suggestion.}\\
\parbox[b][0.3cm]{17.7cm}{\makebox[1ex]{\ensuremath{^{\hypertarget{NE38LEVEL7}{h}}}} From (\href{https://www.nndc.bnl.gov/nsr/nsrlink.jsp?1972Ha03,B}{1972Ha03}).}\\
\parbox[b][0.3cm]{17.7cm}{\makebox[1ex]{\ensuremath{^{\hypertarget{NE38LEVEL8}{i}}}} From (\href{https://www.nndc.bnl.gov/nsr/nsrlink.jsp?2023Po03,B}{2023Po03}).}\\
\vspace{0.5cm}
\underline{$\gamma$($^{19}$Ne)}\\
\begin{longtable}{ccccccc@{}cc@{\extracolsep{\fill}}c}
\multicolumn{2}{c}{E\ensuremath{_{\gamma}}\ensuremath{^{\hyperlink{NE38GAMMA0}{a}}}}&\multicolumn{2}{c}{E\ensuremath{_{i}}(level)}&J\ensuremath{^{\pi}_{i}}&\multicolumn{2}{c}{E\ensuremath{_{f}}}&J\ensuremath{^{\pi}_{f}}&Comments&\\[-.2cm]
\multicolumn{2}{c}{\hrulefill}&\multicolumn{2}{c}{\hrulefill}&\hrulefill&\multicolumn{2}{c}{\hrulefill}&\hrulefill&\hrulefill&
\endfirsthead
\multicolumn{1}{r@{}}{238}&\multicolumn{1}{@{.}l}{4 {\it 3}}&\multicolumn{1}{r@{}}{238}&\multicolumn{1}{@{.}l}{4}&\multicolumn{1}{l}{5/2\ensuremath{^{+}}}&\multicolumn{1}{r@{}}{0}&\multicolumn{1}{@{}l}{}&\multicolumn{1}{@{}l}{1/2\ensuremath{^{+}}}&\parbox[t][0.3cm]{10.387941cm}{\raggedright E\ensuremath{_{\gamma}}: See also 241 keV (\href{https://www.nndc.bnl.gov/nsr/nsrlink.jsp?1967Be14,B}{1967Be14}).\vspace{0.1cm}}&\\
\multicolumn{1}{r@{}}{274}&\multicolumn{1}{@{.}l}{8 {\it 3}}&\multicolumn{1}{r@{}}{274}&\multicolumn{1}{@{.}l}{8}&\multicolumn{1}{l}{1/2\ensuremath{^{-}}}&\multicolumn{1}{r@{}}{0}&\multicolumn{1}{@{}l}{}&\multicolumn{1}{@{}l}{1/2\ensuremath{^{+}}}&\parbox[t][0.3cm]{10.387941cm}{\raggedright E\ensuremath{_{\gamma}}: See also 280 keV (\href{https://www.nndc.bnl.gov/nsr/nsrlink.jsp?1967Be14,B}{1967Be14}).\vspace{0.1cm}}&\\
\multicolumn{1}{r@{}}{1226}&\multicolumn{1}{@{.}l}{0 {\it 11}}&\multicolumn{1}{r@{}}{1504}&\multicolumn{1}{@{.}l}{0}&\multicolumn{1}{l}{(5/2\ensuremath{^{-}})}&\multicolumn{1}{r@{}}{274}&\multicolumn{1}{@{.}l}{8 }&\multicolumn{1}{@{}l}{1/2\ensuremath{^{-}}}&&\\
\multicolumn{1}{r@{}}{1303}&\multicolumn{1}{@{.}l}{2 {\it 11}}&\multicolumn{1}{r@{}}{1532}&\multicolumn{1}{@{.}l}{2}&\multicolumn{1}{l}{(3/2\ensuremath{^{+}})}&\multicolumn{1}{r@{}}{238}&\multicolumn{1}{@{.}l}{4 }&\multicolumn{1}{@{}l}{5/2\ensuremath{^{+}}}&&\\
\multicolumn{1}{r@{}}{1332}&\multicolumn{1}{@{.}l}{2 {\it 11}}&\multicolumn{1}{r@{}}{1612}&\multicolumn{1}{@{.}l}{5}&\multicolumn{1}{l}{(3/2\ensuremath{^{-}})}&\multicolumn{1}{r@{}}{274}&\multicolumn{1}{@{.}l}{8 }&\multicolumn{1}{@{}l}{1/2\ensuremath{^{-}}}&&\\
\multicolumn{1}{r@{}}{2537}&\multicolumn{1}{@{.}l}{2 {\it 34}}&\multicolumn{1}{r@{}}{2791}&\multicolumn{1}{@{.}l}{3}&\multicolumn{1}{l}{(9/2\ensuremath{^{+}})}&\multicolumn{1}{r@{}}{238}&\multicolumn{1}{@{.}l}{4 }&\multicolumn{1}{@{}l}{5/2\ensuremath{^{+}}}&&\\
\multicolumn{1}{r@{}}{2.66\ensuremath{\times10^{3}}}&\multicolumn{1}{@{ }l}{{\it 2}}&\multicolumn{1}{r@{}}{4142}&\multicolumn{1}{@{.}l}{1}&\multicolumn{1}{l}{(9/2\ensuremath{^{-}})}&\multicolumn{1}{r@{}}{1504}&\multicolumn{1}{@{.}l}{0 }&\multicolumn{1}{@{}l}{(5/2\ensuremath{^{-}})}&&\\
\end{longtable}
\parbox[b][0.3cm]{17.7cm}{\makebox[1ex]{\ensuremath{^{\hypertarget{NE38GAMMA0}{a}}}} From (\href{https://www.nndc.bnl.gov/nsr/nsrlink.jsp?1967Ol05,B}{1967Ol05}). We note that this study seems to have a systematic uncertainty that is not accounted for.}\\
\vspace{0.5cm}
\clearpage
\begin{figure}[h]
\begin{center}
\includegraphics{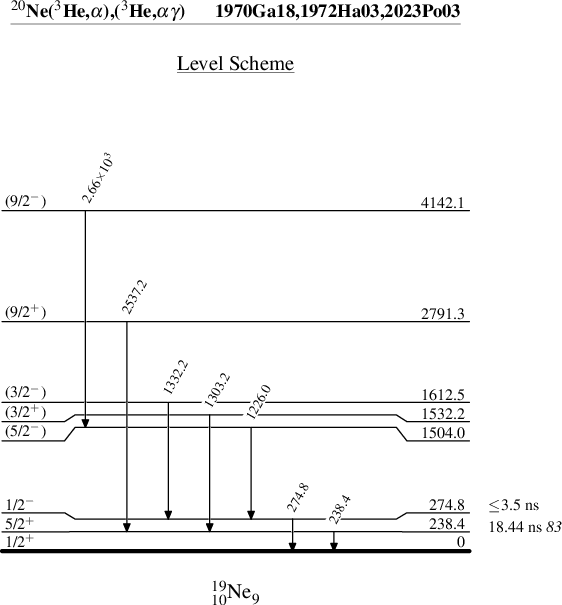}\\
\end{center}
\end{figure}
\clearpage
\subsection[\hspace{-0.2cm}\ensuremath{^{\textnormal{197}}}Au(\ensuremath{^{\textnormal{19}}}Ne,\ensuremath{^{\textnormal{19}}}Ne\ensuremath{'}\ensuremath{\gamma}):coulex]{ }
\vspace{-27pt}
\vspace{0.3cm}
\hypertarget{NE39}{{\bf \small \underline{\ensuremath{^{\textnormal{197}}}Au(\ensuremath{^{\textnormal{19}}}Ne,\ensuremath{^{\textnormal{19}}}Ne\ensuremath{'}\ensuremath{\gamma}):coulex\hspace{0.2in}\href{https://www.nndc.bnl.gov/nsr/nsrlink.jsp?2000Ha26,B}{2000Ha26}}}}\\
\vspace{4pt}
\vspace{8pt}
\parbox[b][0.3cm]{17.7cm}{\addtolength{\parindent}{-0.2in}Coulomb excitation experiment.}\\
\parbox[b][0.3cm]{17.7cm}{\addtolength{\parindent}{-0.2in}J\ensuremath{^{\ensuremath{\pi}}}(\ensuremath{^{\textnormal{197}}}Au\ensuremath{_{\textnormal{g.s.}}})=3/2\ensuremath{^{\textnormal{+}}} and J\ensuremath{^{\ensuremath{\pi}}}(\ensuremath{^{\textnormal{19}}}Ne\ensuremath{_{\textnormal{g.s.}}})=1/2\ensuremath{^{\textnormal{+}}}.}\\
\parbox[b][0.3cm]{17.7cm}{\addtolength{\parindent}{-0.2in}This dataset also includes data from \ensuremath{^{\textnormal{208}}}Pb(\ensuremath{^{\textnormal{19}}}Ne,\ensuremath{^{\textnormal{19}}}Ne\ensuremath{'}):coulex: J\ensuremath{^{\ensuremath{\pi}}}(\ensuremath{^{\textnormal{208}}}Pb\ensuremath{_{\textnormal{g.s.}}})=0\ensuremath{^{\textnormal{+}}}.}\\
\parbox[b][0.3cm]{17.7cm}{\addtolength{\parindent}{-0.2in}\href{https://www.nndc.bnl.gov/nsr/nsrlink.jsp?2000Ha26,B}{2000Ha26}, \href{https://www.nndc.bnl.gov/nsr/nsrlink.jsp?2001Ha12,B}{2001Ha12}: \ensuremath{^{\textnormal{197}}}Au(\ensuremath{^{\textnormal{19}}}Ne,\ensuremath{^{\textnormal{19}}}Ne\ensuremath{'}\ensuremath{\gamma}) E=55 MeV/nucleon; measured position and energy of inelastically scattered \ensuremath{^{\textnormal{19}}}Ne*}\\
\parbox[b][0.3cm]{17.7cm}{ions in coincidence with \ensuremath{\gamma} rays from the \ensuremath{^{\textnormal{19}}}Ne* decays following projectile Coulomb excitation. A fast-slow phoswich detector was}\\
\parbox[b][0.3cm]{17.7cm}{used to identify the \ensuremath{^{\textnormal{19}}}Ne beam ions at \ensuremath{\theta}\ensuremath{_{\textnormal{lab}}}\ensuremath{<}5\ensuremath{^\circ}, while a NaI(Tl) array was used to measure E\ensuremath{_{\ensuremath{\gamma}}} and I\ensuremath{_{\ensuremath{\gamma}}}. These authors deduced}\\
\parbox[b][0.3cm]{17.7cm}{cross sections for Coulomb excitation from the \ensuremath{^{\textnormal{19}}}Ne\ensuremath{_{\textnormal{g.s.}}}. Deduced B(\ensuremath{\sigma}\ensuremath{\lambda}) for the measured \ensuremath{^{\textnormal{19}}}Ne transitions. The \ensuremath{\gamma}-ray partial}\\
\parbox[b][0.3cm]{17.7cm}{width of the \ensuremath{^{\textnormal{19}}}Ne*(4033) state was determined. Discussed comparisons with mirror states and shell model calculations.}\\
\vspace{0.385cm}
\parbox[b][0.3cm]{17.7cm}{\addtolength{\parindent}{-0.2in}\textit{Theory}:}\\
\parbox[b][0.3cm]{17.7cm}{\addtolength{\parindent}{-0.2in}\href{https://www.nndc.bnl.gov/nsr/nsrlink.jsp?2007Be54,B}{2007Be54}: \ensuremath{^{\textnormal{208}}}Pb(\ensuremath{^{\textnormal{19}}}Ne,\ensuremath{^{\textnormal{19}}}Ne\ensuremath{'}) E=10, 20, 30, 50, 100, 200, 500 MeV/nucleon; calculated the cross sections for projectile Coulomb}\\
\parbox[b][0.3cm]{17.7cm}{excitation at the given incident energies; discussed retardation effects; Deduced B(E2) for \ensuremath{^{\textnormal{19}}}Ne*(238).}\\
\vspace{12pt}
\underline{$^{19}$Ne Levels}\\
\vspace{0.34cm}
\parbox[b][0.3cm]{17.7cm}{\addtolength{\parindent}{-0.254cm}(\href{https://www.nndc.bnl.gov/nsr/nsrlink.jsp?2000Ha26,B}{2000Ha26}): The dominant systematic uncertainties in the cross sections come from ambiguous angular distributions for mixed}\\
\parbox[b][0.3cm]{17.7cm}{M1/E2 transitions. These systematic uncertainties are negligible except for the \ensuremath{^{\textnormal{19}}}Ne*(1536) state.}\\
\vspace{0.34cm}
\begin{longtable}{cccc@{\extracolsep{\fill}}c}
\multicolumn{2}{c}{E(level)$^{{\hyperlink{NE39LEVEL0}{a}}}$}&J$^{\pi}$$^{{\hyperlink{NE39LEVEL2}{c}}}$&Comments&\\[-.2cm]
\multicolumn{2}{c}{\hrulefill}&\hrulefill&\hrulefill&
\endfirsthead
\multicolumn{1}{r@{}}{0}&\multicolumn{1}{@{}l}{\ensuremath{^{{\hyperlink{NE39LEVEL1}{b}}}}}&\multicolumn{1}{l}{1/2\ensuremath{^{+}}}&&\\
\multicolumn{1}{r@{}}{238}&\multicolumn{1}{@{}l}{\ensuremath{^{{\hyperlink{NE39LEVEL1}{b}}}}}&\multicolumn{1}{l}{5/2\ensuremath{^{+}}}&\parbox[t][0.3cm]{15.02536cm}{\raggedright B(E2,\ensuremath{\uparrow})=119 e\ensuremath{^{\textnormal{2}}}fm\ensuremath{^{\textnormal{4}}} (\href{https://www.nndc.bnl.gov/nsr/nsrlink.jsp?2007Be54,B}{2007Be54}): Calculated.\vspace{0.1cm}}&\\
\multicolumn{1}{r@{}}{275}&\multicolumn{1}{@{}l}{\ensuremath{^{{\hyperlink{NE39LEVEL1}{b}}}}}&\multicolumn{1}{l}{1/2\ensuremath{^{-}}}&&\\
\multicolumn{1}{r@{}}{1536}&\multicolumn{1}{@{}l}{}&\multicolumn{1}{l}{3/2\ensuremath{^{+}}}&\parbox[t][0.3cm]{15.02536cm}{\raggedright B(E2)\ensuremath{\uparrow}=0.0079 \textit{18} (\href{https://www.nndc.bnl.gov/nsr/nsrlink.jsp?2000Ha26,B}{2000Ha26})\vspace{0.1cm}}&\\
&&&\parbox[t][0.3cm]{15.02536cm}{\raggedright B(E2)\ensuremath{\uparrow}: From B(E2,\ensuremath{\uparrow})=79 e\ensuremath{^{\textnormal{2}}}fm\ensuremath{^{\textnormal{4}}} \textit{1} (stat.) \textit{18} (sys.) (\href{https://www.nndc.bnl.gov/nsr/nsrlink.jsp?2000Ha26,B}{2000Ha26}).\vspace{0.1cm}}&\\
&&&\parbox[t][0.3cm]{15.02536cm}{\raggedright Based on the data for the \ensuremath{^{\textnormal{19}}}F*(1554) mirror state from (\href{https://www.nndc.bnl.gov/nsr/nsrlink.jsp?1985Br15,B}{1985Br15}: \ensuremath{^{\textnormal{19}}}F(e,e\ensuremath{'})), (\href{https://www.nndc.bnl.gov/nsr/nsrlink.jsp?2000Ha26,B}{2000Ha26}) expected that the\vspace{0.1cm}}&\\
&&&\parbox[t][0.3cm]{15.02536cm}{\raggedright {\ }{\ }{\ }M1 contribution to the excitation cross section for the \ensuremath{^{\textnormal{19}}}Ne*(1536) state would be three orders of magnitude\vspace{0.1cm}}&\\
&&&\parbox[t][0.3cm]{15.02536cm}{\raggedright {\ }{\ }{\ }lower than that for the E2 excitation. The deduced E2 transition strength translates to T\ensuremath{_{\textnormal{1/2}}}=57 fs \textit{+16{\textminus}10}\vspace{0.1cm}}&\\
&&&\parbox[t][0.3cm]{15.02536cm}{\raggedright {\ }{\ }{\ }(using a Monte Carlo technique together with Java-RULER computer code) for the 1536-keV state, which is\vspace{0.1cm}}&\\
&&&\parbox[t][0.3cm]{15.02536cm}{\raggedright {\ }{\ }{\ }inconsistent with all the lifetimes measured for this state (see the Adopted Levels).\vspace{0.1cm}}&\\
&&&\parbox[t][0.3cm]{15.02536cm}{\raggedright \ensuremath{\sigma}\ensuremath{_{\textnormal{exc}}}=23.6 mb \textit{3} (stat.) \textit{31} (sys.) (\href{https://www.nndc.bnl.gov/nsr/nsrlink.jsp?2000Ha26,B}{2000Ha26}).\vspace{0.1cm}}&\\
\multicolumn{1}{r@{}}{1616}&\multicolumn{1}{@{}l}{}&\multicolumn{1}{l}{3/2\ensuremath{^{-}}}&\parbox[t][0.3cm]{15.02536cm}{\raggedright B(E1)\ensuremath{\uparrow}=18\ensuremath{\times}10\ensuremath{^{\textnormal{$-$6}}} \textit{4} (\href{https://www.nndc.bnl.gov/nsr/nsrlink.jsp?2000Ha26,B}{2000Ha26})\vspace{0.1cm}}&\\
&&&\parbox[t][0.3cm]{15.02536cm}{\raggedright B(E1)\ensuremath{\uparrow}: From B(E1,\ensuremath{\uparrow})=18\ensuremath{\times}10\ensuremath{^{\textnormal{$-$4}}} e\ensuremath{^{\textnormal{2}}}fm\ensuremath{^{\textnormal{2}}} \textit{3} (stat.) \textit{2} (sys.) (\href{https://www.nndc.bnl.gov/nsr/nsrlink.jsp?2000Ha26,B}{2000Ha26}), from which we deduced T\ensuremath{_{\textnormal{1/2}}}=21 fs\vspace{0.1cm}}&\\
&&&\parbox[t][0.3cm]{15.02536cm}{\raggedright {\ }{\ }{\ }\textit{+6{\textminus}4} for this state using a Monte-Carlo technique together with Java-RULER. This half-life is not consistent\vspace{0.1cm}}&\\
&&&\parbox[t][0.3cm]{15.02536cm}{\raggedright {\ }{\ }{\ }with those deduced from (\href{https://www.nndc.bnl.gov/nsr/nsrlink.jsp?1970Gi09,B}{1970Gi09}, \href{https://www.nndc.bnl.gov/nsr/nsrlink.jsp?1977Le03,B}{1977Le03}, \href{https://www.nndc.bnl.gov/nsr/nsrlink.jsp?2005Ta28,B}{2005Ta28}), see the Adopted Levels.\vspace{0.1cm}}&\\
&&&\parbox[t][0.3cm]{15.02536cm}{\raggedright \textit{Note}: (\href{https://www.nndc.bnl.gov/nsr/nsrlink.jsp?2000Ha26,B}{2000Ha26}) argued that the lifetime measurements by (\href{https://www.nndc.bnl.gov/nsr/nsrlink.jsp?1970Gi09,B}{1970Gi09}, \href{https://www.nndc.bnl.gov/nsr/nsrlink.jsp?1977Le03,B}{1977Le03}) using DSAM following\vspace{0.1cm}}&\\
&&&\parbox[t][0.3cm]{15.02536cm}{\raggedright {\ }{\ }{\ }light-ion fusion-evaporation reactions (\ensuremath{^{\textnormal{16}}}O(\ensuremath{\alpha},n\ensuremath{\gamma}) and \ensuremath{^{\textnormal{19}}}F(p,n\ensuremath{\gamma})) would not properly account for the\vspace{0.1cm}}&\\
&&&\parbox[t][0.3cm]{15.02536cm}{\raggedright {\ }{\ }{\ }side-feeding of this state and would result in a larger apparent lifetime for this state. Hence, (\href{https://www.nndc.bnl.gov/nsr/nsrlink.jsp?2000Ha26,B}{2000Ha26})\vspace{0.1cm}}&\\
&&&\parbox[t][0.3cm]{15.02536cm}{\raggedright {\ }{\ }{\ }recommended that the transition strengths obtained from the lifetime measurements by (\href{https://www.nndc.bnl.gov/nsr/nsrlink.jsp?1970Gi09,B}{1970Gi09}, \href{https://www.nndc.bnl.gov/nsr/nsrlink.jsp?1977Le03,B}{1977Le03})\vspace{0.1cm}}&\\
&&&\parbox[t][0.3cm]{15.02536cm}{\raggedright {\ }{\ }{\ }should be regarded as lower limits. But the (\href{https://www.nndc.bnl.gov/nsr/nsrlink.jsp?1970Gi09,B}{1970Gi09}: \ensuremath{^{\textnormal{16}}}O(\ensuremath{\alpha},n\ensuremath{\gamma})) and (\href{https://www.nndc.bnl.gov/nsr/nsrlink.jsp?1977Le03,B}{1977Le03}: \ensuremath{^{\textnormal{19}}}F(p,n\ensuremath{\gamma})) technique\vspace{0.1cm}}&\\
&&&\parbox[t][0.3cm]{15.02536cm}{\raggedright {\ }{\ }{\ }observed n-\ensuremath{\gamma} coincidence events, and their neutron time-of-flight measurements$'$ constraint should eliminate\vspace{0.1cm}}&\\
&&&\parbox[t][0.3cm]{15.02536cm}{\raggedright {\ }{\ }{\ }any side-feeding.\vspace{0.1cm}}&\\
&&&\parbox[t][0.3cm]{15.02536cm}{\raggedright \ensuremath{\sigma}\ensuremath{_{\textnormal{exc}}}=2.1 mb \textit{3} (stat.) (\href{https://www.nndc.bnl.gov/nsr/nsrlink.jsp?2000Ha26,B}{2000Ha26}).\vspace{0.1cm}}&\\
\multicolumn{1}{r@{}}{4033}&\multicolumn{1}{@{}l}{}&\multicolumn{1}{l}{3/2\ensuremath{^{+}}}&\parbox[t][0.3cm]{15.02536cm}{\raggedright \ensuremath{\Gamma}\ensuremath{_{\ensuremath{\gamma}}}=12\ensuremath{\times}10\ensuremath{^{\textnormal{$-$3}}} eV \textit{+9{\textminus}5} (\href{https://www.nndc.bnl.gov/nsr/nsrlink.jsp?2000Ha26,B}{2000Ha26})\vspace{0.1cm}}&\\
&&&\parbox[t][0.3cm]{15.02536cm}{\raggedright B(M1)\ensuremath{\uparrow}\ensuremath{<}9\ensuremath{\times}10\ensuremath{^{\textnormal{$-$3}}} (\href{https://www.nndc.bnl.gov/nsr/nsrlink.jsp?2000Ha26,B}{2000Ha26})\vspace{0.1cm}}&\\
&&&\parbox[t][0.3cm]{15.02536cm}{\raggedright B(E2)\ensuremath{\uparrow}\ensuremath{<}6.4\ensuremath{\times}10\ensuremath{^{\textnormal{$-$5}}} (\href{https://www.nndc.bnl.gov/nsr/nsrlink.jsp?2000Ha26,B}{2000Ha26})\vspace{0.1cm}}&\\
&&&\parbox[t][0.3cm]{15.02536cm}{\raggedright B(M1)\ensuremath{\uparrow}: From B(M1,\ensuremath{\uparrow})\ensuremath{<}0.90 \ensuremath{\mu}\ensuremath{_{\textnormal{N}}^{\textnormal{2}}} (at 2\ensuremath{\sigma}) (\href{https://www.nndc.bnl.gov/nsr/nsrlink.jsp?2000Ha26,B}{2000Ha26}), where the 2\ensuremath{\sigma} upper limit on B(M1) assumes an\vspace{0.1cm}}&\\
&&&\parbox[t][0.3cm]{15.02536cm}{\raggedright {\ }{\ }{\ }unmixed transition.\vspace{0.1cm}}&\\
&&&\parbox[t][0.3cm]{15.02536cm}{\raggedright B(E2)\ensuremath{\uparrow}: From B(E2,\ensuremath{\uparrow})\ensuremath{<}0.64 e\ensuremath{^{\textnormal{2}}}fm\ensuremath{^{\textnormal{4}}} (at 2\ensuremath{\sigma}) (\href{https://www.nndc.bnl.gov/nsr/nsrlink.jsp?2000Ha26,B}{2000Ha26}), where the 2\ensuremath{\sigma} upper limit on B(E2) assumes an\vspace{0.1cm}}&\\
&&&\parbox[t][0.3cm]{15.02536cm}{\raggedright {\ }{\ }{\ }unmixed transition.\vspace{0.1cm}}&\\
&&&\parbox[t][0.3cm]{15.02536cm}{\raggedright E(level): The excitation of this state was not positively identified in (\href{https://www.nndc.bnl.gov/nsr/nsrlink.jsp?2000Ha26,B}{2000Ha26}), see Fig. 1.\vspace{0.1cm}}&\\
&&&\parbox[t][0.3cm]{15.02536cm}{\raggedright T\ensuremath{_{1/2}}: Evaluator deduced T\ensuremath{_{\textnormal{1/2}}}=78 fs \textit{+22{\textminus}15} and T\ensuremath{_{\textnormal{1/2}}}\ensuremath{<}12.8 fs from B(M1)\ensuremath{<}9\ensuremath{\times}10\ensuremath{^{\textnormal{$-$3}}} e\ensuremath{^{\textnormal{2}}}b and B(E2)\ensuremath{<}6.4\ensuremath{\times}10\ensuremath{^{\textnormal{$-$5}}}\vspace{0.1cm}}&\\
&&&\parbox[t][0.3cm]{15.02536cm}{\raggedright {\ }{\ }{\ }e\ensuremath{^{\textnormal{2}}}b\ensuremath{^{\textnormal{2}}}, respectively, using a Monte Carlo technique together with Java-RULER computer code.\vspace{0.1cm}}&\\
&&&\parbox[t][0.3cm]{15.02536cm}{\raggedright \ensuremath{\Gamma}\ensuremath{_{\ensuremath{\gamma}}}: Deduced based on a mixing ratio of \ensuremath{\delta}=+0.14 deduced by (\href{https://www.nndc.bnl.gov/nsr/nsrlink.jsp?2000Ha26,B}{2000Ha26}) using shell model, the upper limit\vspace{0.1cm}}&\\
\end{longtable}
\begin{textblock}{29}(0,27.3)
Continued on next page (footnotes at end of table)
\end{textblock}
\clearpage
\begin{longtable}{cccc@{\extracolsep{\fill}}c}
\\[-.4cm]
\multicolumn{5}{c}{{\bf \small \underline{\ensuremath{^{\textnormal{197}}}Au(\ensuremath{^{\textnormal{19}}}Ne,\ensuremath{^{\textnormal{19}}}Ne\ensuremath{'}\ensuremath{\gamma}):coulex\hspace{0.2in}\href{https://www.nndc.bnl.gov/nsr/nsrlink.jsp?2000Ha26,B}{2000Ha26} (continued)}}}\\
\multicolumn{5}{c}{~}\\
\multicolumn{5}{c}{\underline{\ensuremath{^{19}}Ne Levels (continued)}}\\
\multicolumn{5}{c}{~}\\
\multicolumn{2}{c}{E(level)$^{{\hyperlink{NE39LEVEL0}{a}}}$}&J$^{\pi}$$^{{\hyperlink{NE39LEVEL2}{c}}}$&Comments&\\[-.2cm]
\multicolumn{2}{c}{\hrulefill}&\hrulefill&\hrulefill&
\endhead
&&&\parbox[t][0.3cm]{15.02536cm}{\raggedright {\ }{\ }{\ }lifetime (T\ensuremath{_{\textnormal{1/2}}}\ensuremath{<}35 fs) from (\href{https://www.nndc.bnl.gov/nsr/nsrlink.jsp?1973Da31,B}{1973Da31}: \ensuremath{^{\textnormal{17}}}O(\ensuremath{^{\textnormal{3}}}He,n\ensuremath{\gamma})) based on DSAM, and the upper limit cross sections\vspace{0.1cm}}&\\
&&&\parbox[t][0.3cm]{15.02536cm}{\raggedright {\ }{\ }{\ }deduced by (\href{https://www.nndc.bnl.gov/nsr/nsrlink.jsp?2000Ha26,B}{2000Ha26}). However, this \ensuremath{\Gamma}\ensuremath{_{\ensuremath{\gamma}}} value does not take into account the uncertainty in \ensuremath{\delta}, which is\vspace{0.1cm}}&\\
&&&\parbox[t][0.3cm]{15.02536cm}{\raggedright {\ }{\ }{\ }hard to estimate. See also \ensuremath{\Gamma}\ensuremath{_{\ensuremath{\gamma}}}=22 meV calculated from shell model using 5p-2h configuration; and \ensuremath{\Gamma}\ensuremath{_{\ensuremath{\gamma}}}=45\vspace{0.1cm}}&\\
&&&\parbox[t][0.3cm]{15.02536cm}{\raggedright {\ }{\ }{\ }meV \textit{+200{\textminus}33} deduced by (\href{https://www.nndc.bnl.gov/nsr/nsrlink.jsp?2000Ha26,B}{2000Ha26}) for a pure M1 transition by combining the Coulomb excitation data of\vspace{0.1cm}}&\\
&&&\parbox[t][0.3cm]{15.02536cm}{\raggedright {\ }{\ }{\ }(\href{https://www.nndc.bnl.gov/nsr/nsrlink.jsp?2000Ha26,B}{2000Ha26}) with those of the DSAM from (\href{https://www.nndc.bnl.gov/nsr/nsrlink.jsp?1973Da31,B}{1973Da31}: \ensuremath{^{\textnormal{17}}}O(\ensuremath{^{\textnormal{3}}}He,n\ensuremath{\gamma})).\vspace{0.1cm}}&\\
&&&\parbox[t][0.3cm]{15.02536cm}{\raggedright \ensuremath{\sigma}\ensuremath{_{\textnormal{exc}}}={\textminus}0.21 mb \textit{19} (stat.) for M1 transition (\href{https://www.nndc.bnl.gov/nsr/nsrlink.jsp?2000Ha26,B}{2000Ha26}). This results in \ensuremath{\Gamma}\ensuremath{_{\ensuremath{\gamma}}}\ensuremath{<}430 meV for a pure M1 transition\vspace{0.1cm}}&\\
&&&\parbox[t][0.3cm]{15.02536cm}{\raggedright {\ }{\ }{\ }(\href{https://www.nndc.bnl.gov/nsr/nsrlink.jsp?2000Ha26,B}{2000Ha26}).\vspace{0.1cm}}&\\
&&&\parbox[t][0.3cm]{15.02536cm}{\raggedright \ensuremath{\sigma}\ensuremath{_{\textnormal{exc}}}={\textminus}0.19 mb \textit{17} (stat.) for E2 transition (\href{https://www.nndc.bnl.gov/nsr/nsrlink.jsp?2000Ha26,B}{2000Ha26}). This yields \ensuremath{\Gamma}\ensuremath{_{\ensuremath{\gamma}}}\ensuremath{<}0.34 meV for a pure E2 transition\vspace{0.1cm}}&\\
&&&\parbox[t][0.3cm]{15.02536cm}{\raggedright {\ }{\ }{\ }(\href{https://www.nndc.bnl.gov/nsr/nsrlink.jsp?2000Ha26,B}{2000Ha26}). These authors reported that a dominant M1 transition (\ensuremath{\delta}=0) is consistent with the result of the\vspace{0.1cm}}&\\
&&&\parbox[t][0.3cm]{15.02536cm}{\raggedright {\ }{\ }{\ }DASM in (\href{https://www.nndc.bnl.gov/nsr/nsrlink.jsp?1973Da31,B}{1973Da31}: \ensuremath{^{\textnormal{17}}}O(\ensuremath{^{\textnormal{3}}}He,n\ensuremath{\gamma})), where a reported 2\ensuremath{\sigma} upper limit on lifetime corresponds to a \ensuremath{\Gamma}\ensuremath{_{\ensuremath{\gamma}}}\ensuremath{>}6.6\vspace{0.1cm}}&\\
&&&\parbox[t][0.3cm]{15.02536cm}{\raggedright {\ }{\ }{\ }meV. But a dominant E2 transition (\ensuremath{\delta}=\ensuremath{\infty}) is inconsistent with the results of (\href{https://www.nndc.bnl.gov/nsr/nsrlink.jsp?1973Da31,B}{1973Da31}).\vspace{0.1cm}}&\\
&&&\parbox[t][0.3cm]{15.02536cm}{\raggedright Evaluator notes that \ensuremath{\sigma}\ensuremath{_{\textnormal{exc}}}\ensuremath{<}0.16 mb is also reported in (\href{https://www.nndc.bnl.gov/nsr/nsrlink.jsp?2000Ha26,B}{2000Ha26}: See Fig. 1) but it is not clear if this is for\vspace{0.1cm}}&\\
&&&\parbox[t][0.3cm]{15.02536cm}{\raggedright {\ }{\ }{\ }M1 or E2 transition.\vspace{0.1cm}}&\\
&&&\parbox[t][0.3cm]{15.02536cm}{\raggedright (\href{https://www.nndc.bnl.gov/nsr/nsrlink.jsp?2000Ha26,B}{2000Ha26}) reported this state to be of 5p-2h configuration.\vspace{0.1cm}}&\\
\multicolumn{1}{r@{}}{4600}&\multicolumn{1}{@{}l}{}&\multicolumn{1}{l}{5/2\ensuremath{^{+}}}&\parbox[t][0.3cm]{15.02536cm}{\raggedright B(E2)\ensuremath{\uparrow}=0.0020 \textit{3} (\href{https://www.nndc.bnl.gov/nsr/nsrlink.jsp?2000Ha26,B}{2000Ha26})\vspace{0.1cm}}&\\
&&&\parbox[t][0.3cm]{15.02536cm}{\raggedright B(E2)\ensuremath{\uparrow}: From B(E2,\ensuremath{\uparrow})=20 e\ensuremath{^{\textnormal{2}}}fm\ensuremath{^{\textnormal{4}}} \textit{2} (stat.) \textit{2} (sys.) (\href{https://www.nndc.bnl.gov/nsr/nsrlink.jsp?2000Ha26,B}{2000Ha26}).\vspace{0.1cm}}&\\
&&&\parbox[t][0.3cm]{15.02536cm}{\raggedright We deduced T\ensuremath{_{\textnormal{1/2}}}=3.5 fs \textit{+8{\textminus}7} from B(E2,\ensuremath{\uparrow})=20 e\ensuremath{^{\textnormal{2}}}fm\ensuremath{^{\textnormal{4}}} \textit{2} (stat.) \textit{2} (sys.) (\href{https://www.nndc.bnl.gov/nsr/nsrlink.jsp?2000Ha26,B}{2000Ha26}) using a Monte-Carlo\vspace{0.1cm}}&\\
&&&\parbox[t][0.3cm]{15.02536cm}{\raggedright {\ }{\ }{\ }technique together with Java-RULER.\vspace{0.1cm}}&\\
&&&\parbox[t][0.3cm]{15.02536cm}{\raggedright \ensuremath{\sigma}\ensuremath{_{\textnormal{exc}}}=4.2 mb \textit{3} (stat.) (\href{https://www.nndc.bnl.gov/nsr/nsrlink.jsp?2000Ha26,B}{2000Ha26}).\vspace{0.1cm}}&\\
\end{longtable}
\parbox[b][0.3cm]{17.7cm}{\makebox[1ex]{\ensuremath{^{\hypertarget{NE39LEVEL0}{a}}}} From (\href{https://www.nndc.bnl.gov/nsr/nsrlink.jsp?2000Ha26,B}{2000Ha26}).}\\
\parbox[b][0.3cm]{17.7cm}{\makebox[1ex]{\ensuremath{^{\hypertarget{NE39LEVEL1}{b}}}} The cross section and reduced transition probability for populating this state were not measured by (\href{https://www.nndc.bnl.gov/nsr/nsrlink.jsp?2000Ha26,B}{2000Ha26}) due to the energy}\\
\parbox[b][0.3cm]{17.7cm}{{\ }{\ }threshold setting during the experiment.}\\
\parbox[b][0.3cm]{17.7cm}{\makebox[1ex]{\ensuremath{^{\hypertarget{NE39LEVEL2}{c}}}} From the \ensuremath{^{\textnormal{19}}}Ne Adopted Levels.}\\
\vspace{0.5cm}
\underline{$\gamma$($^{19}$Ne)}\\
\begin{longtable}{ccccccccc@{}ccccccc@{\extracolsep{\fill}}c}
\multicolumn{2}{c}{E\ensuremath{_{i}}(level)}&J\ensuremath{^{\pi}_{i}}&\multicolumn{2}{c}{E\ensuremath{_{\gamma}}\ensuremath{^{\hyperlink{NE39GAMMA0}{a}}}}&\multicolumn{2}{c}{I\ensuremath{_{\gamma}}\ensuremath{^{\hyperlink{NE39GAMMA3}{d}}}}&\multicolumn{2}{c}{E\ensuremath{_{f}}}&J\ensuremath{^{\pi}_{f}}&Mult.&\multicolumn{2}{c}{\ensuremath{\delta}}&\multicolumn{2}{c}{\ensuremath{\alpha}\ensuremath{^{\hyperlink{NE39GAMMA4}{e}}}}&Comments&\\[-.2cm]
\multicolumn{2}{c}{\hrulefill}&\hrulefill&\multicolumn{2}{c}{\hrulefill}&\multicolumn{2}{c}{\hrulefill}&\multicolumn{2}{c}{\hrulefill}&\hrulefill&\hrulefill&\multicolumn{2}{c}{\hrulefill}&\multicolumn{2}{c}{\hrulefill}&\hrulefill&
\endfirsthead
\multicolumn{1}{r@{}}{238}&\multicolumn{1}{@{}l}{}&\multicolumn{1}{l}{5/2\ensuremath{^{+}}}&\multicolumn{1}{r@{}}{238}&\multicolumn{1}{@{}l}{\ensuremath{^{\hyperlink{NE39GAMMA1}{b}\hyperlink{NE39GAMMA5}{f}}}}&\multicolumn{1}{r@{}}{}&\multicolumn{1}{@{}l}{}&\multicolumn{1}{r@{}}{0}&\multicolumn{1}{@{}l}{}&\multicolumn{1}{@{}l}{1/2\ensuremath{^{+}}}&&&&&&&\\
\multicolumn{1}{r@{}}{275}&\multicolumn{1}{@{}l}{}&\multicolumn{1}{l}{1/2\ensuremath{^{-}}}&\multicolumn{1}{r@{}}{275}&\multicolumn{1}{@{}l}{\ensuremath{^{\hyperlink{NE39GAMMA1}{b}\hyperlink{NE39GAMMA5}{f}}}}&\multicolumn{1}{r@{}}{}&\multicolumn{1}{@{}l}{}&\multicolumn{1}{r@{}}{0}&\multicolumn{1}{@{}l}{}&\multicolumn{1}{@{}l}{1/2\ensuremath{^{+}}}&&&&&&&\\
\multicolumn{1}{r@{}}{1536}&\multicolumn{1}{@{}l}{}&\multicolumn{1}{l}{3/2\ensuremath{^{+}}}&\multicolumn{1}{r@{}}{1261}&\multicolumn{1}{@{}l}{}&\multicolumn{1}{r@{}}{5}&\multicolumn{1}{@{}l}{}&\multicolumn{1}{r@{}}{275}&\multicolumn{1}{@{}l}{}&\multicolumn{1}{@{}l}{1/2\ensuremath{^{-}}}&&&&\multicolumn{1}{r@{}}{}&\multicolumn{1}{@{}l}{}&&\\
&&&\multicolumn{1}{r@{}}{1298}&\multicolumn{1}{@{}l}{}&\multicolumn{1}{r@{}}{100}&\multicolumn{1}{@{}l}{}&\multicolumn{1}{r@{}}{238}&\multicolumn{1}{@{}l}{}&\multicolumn{1}{@{}l}{5/2\ensuremath{^{+}}}&&&&\multicolumn{1}{r@{}}{}&\multicolumn{1}{@{}l}{}&&\\
\multicolumn{1}{r@{}}{1616}&\multicolumn{1}{@{}l}{}&\multicolumn{1}{l}{3/2\ensuremath{^{-}}}&\multicolumn{1}{r@{}}{1341}&\multicolumn{1}{@{}l}{}&\multicolumn{1}{r@{}}{100}&\multicolumn{1}{@{}l}{}&\multicolumn{1}{r@{}}{275}&\multicolumn{1}{@{}l}{}&\multicolumn{1}{@{}l}{1/2\ensuremath{^{-}}}&&&&\multicolumn{1}{r@{}}{}&\multicolumn{1}{@{}l}{}&&\\
&&&\multicolumn{1}{r@{}}{1378}&\multicolumn{1}{@{}l}{}&\multicolumn{1}{r@{}}{14}&\multicolumn{1}{@{}l}{}&\multicolumn{1}{r@{}}{238}&\multicolumn{1}{@{}l}{}&\multicolumn{1}{@{}l}{5/2\ensuremath{^{+}}}&&&&\multicolumn{1}{r@{}}{}&\multicolumn{1}{@{}l}{}&&\\
&&&\multicolumn{1}{r@{}}{1616}&\multicolumn{1}{@{}l}{}&\multicolumn{1}{r@{}}{29}&\multicolumn{1}{@{}l}{}&\multicolumn{1}{r@{}}{0}&\multicolumn{1}{@{}l}{}&\multicolumn{1}{@{}l}{1/2\ensuremath{^{+}}}&&&&\multicolumn{1}{r@{}}{}&\multicolumn{1}{@{}l}{}&&\\
\multicolumn{1}{r@{}}{4033}&\multicolumn{1}{@{}l}{}&\multicolumn{1}{l}{3/2\ensuremath{^{+}}}&\multicolumn{1}{r@{}}{2497}&\multicolumn{1}{@{}l}{\ensuremath{^{\hyperlink{NE39GAMMA2}{c}}}}&\multicolumn{1}{r@{}}{19}&\multicolumn{1}{@{}l}{}&\multicolumn{1}{r@{}}{1536}&\multicolumn{1}{@{}l}{}&\multicolumn{1}{@{}l}{3/2\ensuremath{^{+}}}&&&&&&&\\
&&&\multicolumn{1}{r@{}}{3758}&\multicolumn{1}{@{}l}{\ensuremath{^{\hyperlink{NE39GAMMA2}{c}}}}&\multicolumn{1}{r@{}}{6}&\multicolumn{1}{@{}l}{}&\multicolumn{1}{r@{}}{275}&\multicolumn{1}{@{}l}{}&\multicolumn{1}{@{}l}{1/2\ensuremath{^{-}}}&&&&&&&\\
&&&\multicolumn{1}{r@{}}{4033}&\multicolumn{1}{@{}l}{\ensuremath{^{\hyperlink{NE39GAMMA2}{c}}}}&\multicolumn{1}{r@{}}{100}&\multicolumn{1}{@{}l}{}&\multicolumn{1}{r@{}}{0}&\multicolumn{1}{@{}l}{}&\multicolumn{1}{@{}l}{1/2\ensuremath{^{+}}}&\multicolumn{1}{l}{M1+E2}&\multicolumn{1}{r@{}}{$<$0}&\multicolumn{1}{@{.}l}{23}&\multicolumn{1}{r@{}}{1}&\multicolumn{1}{@{.}l}{03\ensuremath{\times10^{-3}} {\it 2}}&\parbox[t][0.3cm]{6.435171cm}{\raggedright \ensuremath{\alpha}(K)=8.09\ensuremath{\times}10\ensuremath{^{\textnormal{$-$7}}} \textit{11}; \ensuremath{\alpha}(L)=4.48\ensuremath{\times}10\ensuremath{^{\textnormal{$-$8}}} \textit{6}\vspace{0.1cm}}&\\
&&&&&&&&&&&&&&&\parbox[t][0.3cm]{6.435171cm}{\raggedright \ensuremath{\alpha}(IPF)=0.001034 \textit{15}\vspace{0.1cm}}&\\
&&&&&&&&&&&&&&&\parbox[t][0.3cm]{6.435171cm}{\raggedright Mult.,\ensuremath{\delta}: From (\href{https://www.nndc.bnl.gov/nsr/nsrlink.jsp?2000Ha26,B}{2000Ha26}): Deduced \ensuremath{\vert}\ensuremath{\delta}\ensuremath{\vert}\ensuremath{<}0.23\vspace{0.1cm}}&\\
&&&&&&&&&&&&&&&\parbox[t][0.3cm]{6.435171cm}{\raggedright {\ }{\ }{\ }(theoretical) from the analysis of \ensuremath{\Gamma}\ensuremath{_{\ensuremath{\gamma}}} vs. \ensuremath{\delta}\vspace{0.1cm}}&\\
&&&&&&&&&&&&&&&\parbox[t][0.3cm]{6.435171cm}{\raggedright {\ }{\ }{\ }(see Fig. 3). They also reported \ensuremath{\delta}=+0.14\vspace{0.1cm}}&\\
&&&&&&&&&&&&&&&\parbox[t][0.3cm]{6.435171cm}{\raggedright {\ }{\ }{\ }calculated using shell model for a 5p-2h\vspace{0.1cm}}&\\
&&&&&&&&&&&&&&&\parbox[t][0.3cm]{6.435171cm}{\raggedright {\ }{\ }{\ }configuration. For \ensuremath{\delta}=+0.14, the measured\vspace{0.1cm}}&\\
&&&&&&&&&&&&&&&\parbox[t][0.3cm]{6.435171cm}{\raggedright {\ }{\ }{\ }cross section places simultaneous limits of\vspace{0.1cm}}&\\
&&&&&&&&&&&&&&&\parbox[t][0.3cm]{6.435171cm}{\raggedright {\ }{\ }{\ }B(M1,\ensuremath{\uparrow})\ensuremath{<}0.035 \ensuremath{\mu}\ensuremath{^{\textnormal{2}}_{\textnormal{N}}} and B(E2,\ensuremath{\uparrow})\ensuremath{<}0.61\vspace{0.1cm}}&\\
&&&&&&&&&&&&&&&\parbox[t][0.3cm]{6.435171cm}{\raggedright {\ }{\ }{\ }e\ensuremath{^{\textnormal{2}}}fm\ensuremath{^{\textnormal{4}}} (\href{https://www.nndc.bnl.gov/nsr/nsrlink.jsp?2000Ha26,B}{2000Ha26}).\vspace{0.1cm}}&\\
\multicolumn{1}{r@{}}{4600}&\multicolumn{1}{@{}l}{}&\multicolumn{1}{l}{5/2\ensuremath{^{+}}}&\multicolumn{1}{r@{}}{3064}&\multicolumn{1}{@{}l}{}&\multicolumn{1}{r@{}}{11}&\multicolumn{1}{@{}l}{}&\multicolumn{1}{r@{}}{1536}&\multicolumn{1}{@{}l}{}&\multicolumn{1}{@{}l}{3/2\ensuremath{^{+}}}&&&&\multicolumn{1}{r@{}}{}&\multicolumn{1}{@{}l}{}&&\\
&&&\multicolumn{1}{r@{}}{4362}&\multicolumn{1}{@{}l}{}&\multicolumn{1}{r@{}}{100}&\multicolumn{1}{@{}l}{}&\multicolumn{1}{r@{}}{238}&\multicolumn{1}{@{}l}{}&\multicolumn{1}{@{}l}{5/2\ensuremath{^{+}}}&&&&\multicolumn{1}{r@{}}{}&\multicolumn{1}{@{}l}{}&&\\
\end{longtable}
\parbox[b][0.3cm]{17.7cm}{\makebox[1ex]{\ensuremath{^{\hypertarget{NE39GAMMA0}{a}}}} The \ensuremath{\gamma} ray energies are not provided in (\href{https://www.nndc.bnl.gov/nsr/nsrlink.jsp?2000Ha26,B}{2000Ha26}). Therefore, they are deduced from level-energy differences. The observed \ensuremath{\gamma}}\\
\begin{textblock}{29}(0,27.3)
Continued on next page (footnotes at end of table)
\end{textblock}
\clearpage
\vspace*{-0.5cm}
{\bf \small \underline{\ensuremath{^{\textnormal{197}}}Au(\ensuremath{^{\textnormal{19}}}Ne,\ensuremath{^{\textnormal{19}}}Ne\ensuremath{'}\ensuremath{\gamma}):coulex\hspace{0.2in}\href{https://www.nndc.bnl.gov/nsr/nsrlink.jsp?2000Ha26,B}{2000Ha26} (continued)}}\\
\vspace{0.3cm}
\underline{$\gamma$($^{19}$Ne) (continued)}\\
\vspace{0.3cm}
\parbox[b][0.3cm]{17.7cm}{{\ }{\ }rays are indicated in Fig. 1 of (\href{https://www.nndc.bnl.gov/nsr/nsrlink.jsp?2000Ha26,B}{2000Ha26}).}\\
\parbox[b][0.3cm]{17.7cm}{\makebox[1ex]{\ensuremath{^{\hypertarget{NE39GAMMA1}{b}}}} This \ensuremath{\gamma} ray was most likely not observed due to energy thresholds in (\href{https://www.nndc.bnl.gov/nsr/nsrlink.jsp?2000Ha26,B}{2000Ha26}), see Fig. 1 and the caption of Table 1.}\\
\parbox[b][0.3cm]{17.7cm}{\makebox[1ex]{\ensuremath{^{\hypertarget{NE39GAMMA2}{c}}}} This transition was not positively identified in (\href{https://www.nndc.bnl.gov/nsr/nsrlink.jsp?2000Ha26,B}{2000Ha26}), see Fig. 1.}\\
\parbox[b][0.3cm]{17.7cm}{\makebox[1ex]{\ensuremath{^{\hypertarget{NE39GAMMA3}{d}}}} Relative branching ratios normalized to 100 for the strongest transition from (\href{https://www.nndc.bnl.gov/nsr/nsrlink.jsp?2000Ha26,B}{2000Ha26}: See Fig. 1). These values are not}\\
\parbox[b][0.3cm]{17.7cm}{{\ }{\ }measured by (\href{https://www.nndc.bnl.gov/nsr/nsrlink.jsp?2000Ha26,B}{2000Ha26}) and come from the evaluation of (\href{https://www.nndc.bnl.gov/nsr/nsrlink.jsp?1995Ti07,B}{1995Ti07}).}\\
\parbox[b][0.3cm]{17.7cm}{\makebox[1ex]{\ensuremath{^{\hypertarget{NE39GAMMA4}{e}}}} Total theoretical internal conversion coefficients, calculated using the BrIcc code (\href{https://www.nndc.bnl.gov/nsr/nsrlink.jsp?2008Ki07,B}{2008Ki07}) with ``Frozen Orbitals''}\\
\parbox[b][0.3cm]{17.7cm}{{\ }{\ }approximation based on \ensuremath{\gamma}-ray energies, assigned multipolarities, and mixing ratios, unless otherwise specified.}\\
\parbox[b][0.3cm]{17.7cm}{\makebox[1ex]{\ensuremath{^{\hypertarget{NE39GAMMA5}{f}}}} Placement of transition in the level scheme is uncertain.}\\
\vspace{0.5cm}
\clearpage
\begin{figure}[h]
\begin{center}
\includegraphics{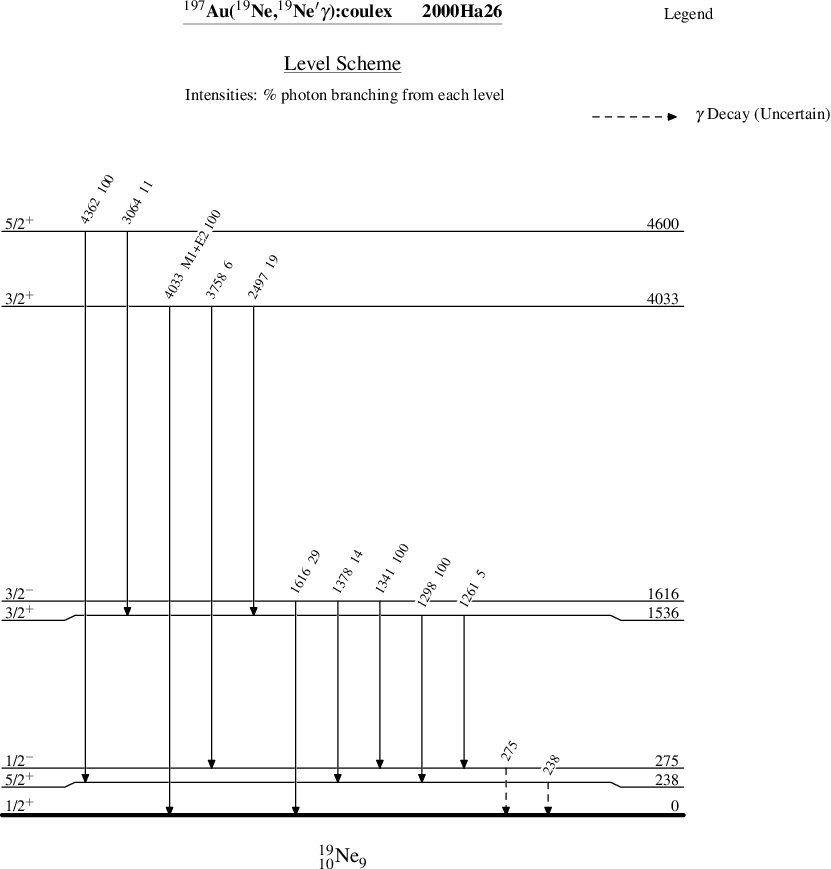}\\
\end{center}
\end{figure}
\clearpage
\subsection[\hspace{-0.2cm}Si(p,\ensuremath{^{\textnormal{19}}}Ne)]{ }
\vspace{-27pt}
\vspace{0.3cm}
\hypertarget{NE40}{{\bf \small \underline{Si(p,\ensuremath{^{\textnormal{19}}}Ne)\hspace{0.2in}\href{https://www.nndc.bnl.gov/nsr/nsrlink.jsp?2012Tr06,B}{2012Tr06},\href{https://www.nndc.bnl.gov/nsr/nsrlink.jsp?2012Tr09,B}{2012Tr09}}}}\\
\vspace{4pt}
\vspace{8pt}
\parbox[b][0.3cm]{17.7cm}{\addtolength{\parindent}{-0.2in}Spallation reaction.}\\
\parbox[b][0.3cm]{17.7cm}{\addtolength{\parindent}{-0.2in}J\ensuremath{^{\ensuremath{\pi}}}(\ensuremath{^{\textnormal{28}}}Si\ensuremath{_{\textnormal{g.s.}}})=0\ensuremath{^{\textnormal{+}}} and J\ensuremath{^{\ensuremath{\pi}}}(p)=1/2\ensuremath{^{\textnormal{+}}}.}\\
\parbox[b][0.3cm]{17.7cm}{\addtolength{\parindent}{-0.2in}\href{https://www.nndc.bnl.gov/nsr/nsrlink.jsp?2012Tr06,B}{2012Tr06}, \href{https://www.nndc.bnl.gov/nsr/nsrlink.jsp?2012Tr09,B}{2012Tr09}: SiC(p,\ensuremath{^{\textnormal{19}}}Ne) E\ensuremath{<}480 MeV. \ensuremath{^{\textnormal{19}}}Ne ions with E\ensuremath{\sim}37 keV were implanted into a thin aluminized mylar tape placed}\\
\parbox[b][0.3cm]{17.7cm}{at the center of the 8\ensuremath{\pi} \ensuremath{\gamma}-ray spectrometer. The lifetime of the \ensuremath{^{\textnormal{19}}}Ne\ensuremath{_{\textnormal{g.s.}}} was measured using consecutive cycles of background}\\
\parbox[b][0.3cm]{17.7cm}{measurement (2 s), \ensuremath{^{\textnormal{19}}}Ne implantation (\ensuremath{\sim}1 s), counting (300 s), and tape removal (1 s). The SCEPTAR scintillator array (20 plastic}\\
\parbox[b][0.3cm]{17.7cm}{scintillators) surrounded the target and provided 80\% of 4\ensuremath{\pi} coverage for detecting the \ensuremath{\beta}\ensuremath{^{\textnormal{+}}} particles emitted from the decay of}\\
\parbox[b][0.3cm]{17.7cm}{\ensuremath{^{\textnormal{19}}}Ne\ensuremath{_{\textnormal{g.s.}}}. \ensuremath{\beta}-\ensuremath{\gamma} coincidence events were measured using SCEPTAR and the 8\ensuremath{\pi} spectrometer. Various systematic uncertainties and}\\
\parbox[b][0.3cm]{17.7cm}{corrections are discussed. A statistical uncertainty of \ensuremath{\approx}0.002 s is obtained, while a systematic uncertainty of \ensuremath{\approx}0.007 s dominates}\\
\parbox[b][0.3cm]{17.7cm}{the total uncertainty. The net result of T\ensuremath{_{\textnormal{1/2}}}=17.262 s \textit{7} (sys.) is deduced, which disagrees with the result of (\href{https://www.nndc.bnl.gov/nsr/nsrlink.jsp?1975Az01,B}{1975Az01}: \ensuremath{^{\textnormal{19}}}F(p,n),}\\
\parbox[b][0.3cm]{17.7cm}{T\ensuremath{_{\textnormal{1/2}}}=17.219 s \textit{17}) by 2.5\ensuremath{\sigma}.}\\
\vspace{12pt}
\underline{$^{19}$Ne Levels}\\
\begin{longtable}{cccccc@{\extracolsep{\fill}}c}
\multicolumn{2}{c}{E(level)$^{}$}&J$^{\pi}$$^{{\hyperlink{NE40LEVEL0}{a}}}$&\multicolumn{2}{c}{T\ensuremath{_{\textnormal{1/2}}}$^{}$}&Comments&\\[-.2cm]
\multicolumn{2}{c}{\hrulefill}&\hrulefill&\multicolumn{2}{c}{\hrulefill}&\hrulefill&
\endfirsthead
\multicolumn{1}{r@{}}{0}&\multicolumn{1}{@{}l}{}&\multicolumn{1}{l}{1/2\ensuremath{^{+}}}&\multicolumn{1}{r@{}}{17}&\multicolumn{1}{@{.}l}{262 s {\it 7}}&\parbox[t][0.3cm]{13.33066cm}{\raggedright T=1/2 (\href{https://www.nndc.bnl.gov/nsr/nsrlink.jsp?2012Tr06,B}{2012Tr06})\vspace{0.1cm}}&\\
&&&&&\parbox[t][0.3cm]{13.33066cm}{\raggedright T\ensuremath{_{1/2}}: From (\href{https://www.nndc.bnl.gov/nsr/nsrlink.jsp?2012Tr06,B}{2012Tr06}), where the uncertainty is systematic. See also the preliminary report in\vspace{0.1cm}}&\\
&&&&&\parbox[t][0.3cm]{13.33066cm}{\raggedright {\ }{\ }{\ }(\href{https://www.nndc.bnl.gov/nsr/nsrlink.jsp?2012Tr09,B}{2012Tr09}).\vspace{0.1cm}}&\\
&&&&&\parbox[t][0.3cm]{13.33066cm}{\raggedright \textit{Ft}=1721.3 s \textit{12} (\href{https://www.nndc.bnl.gov/nsr/nsrlink.jsp?2012Tr06,B}{2012Tr06}).\vspace{0.1cm}}&\\
&&&&&\parbox[t][0.3cm]{13.33066cm}{\raggedright A\ensuremath{_{\ensuremath{\beta}}^{\textnormal{SM}}}={\textminus}0.0416 \textit{7}: The standard model beta asymmetry parameter deduced by (\href{https://www.nndc.bnl.gov/nsr/nsrlink.jsp?2012Tr06,B}{2012Tr06}). This\vspace{0.1cm}}&\\
&&&&&\parbox[t][0.3cm]{13.33066cm}{\raggedright {\ }{\ }{\ }can be compared with the measured value of A\ensuremath{_{\ensuremath{\beta}}^{\textnormal{exp}}}={\textminus}0.0391 \textit{14} (\href{https://www.nndc.bnl.gov/nsr/nsrlink.jsp?1975Ca28,B}{1975Ca28}: \ensuremath{^{\textnormal{19}}}F(p,n)).\vspace{0.1cm}}&\\
\end{longtable}
\parbox[b][0.3cm]{17.7cm}{\makebox[1ex]{\ensuremath{^{\hypertarget{NE40LEVEL0}{a}}}} From the \ensuremath{^{\textnormal{19}}}Ne Adopted Levels.}\\
\vspace{0.5cm}
\clearpage
\subsection[\hspace{-0.2cm}Ca(p,\ensuremath{^{\textnormal{19}}}Ne)]{ }
\vspace{-27pt}
\vspace{0.3cm}
\hypertarget{NE41}{{\bf \small \underline{Ca(p,\ensuremath{^{\textnormal{19}}}Ne)\hspace{0.2in}\href{https://www.nndc.bnl.gov/nsr/nsrlink.jsp?2008Ge07,B}{2008Ge07},\href{https://www.nndc.bnl.gov/nsr/nsrlink.jsp?2011Ma48,B}{2011Ma48}}}}\\
\vspace{4pt}
\vspace{8pt}
\parbox[b][0.3cm]{17.7cm}{\addtolength{\parindent}{-0.2in}Spallation reaction.}\\
\parbox[b][0.3cm]{17.7cm}{\addtolength{\parindent}{-0.2in}J\ensuremath{^{\ensuremath{\pi}}}(\ensuremath{^{\textnormal{40}}}Ca\ensuremath{_{\textnormal{g.s.}}})=0\ensuremath{^{\textnormal{+}}} and J\ensuremath{^{\ensuremath{\pi}}}(p)=1/2\ensuremath{^{\textnormal{+}}}.}\\
\parbox[b][0.3cm]{17.7cm}{\addtolength{\parindent}{-0.2in}\href{https://www.nndc.bnl.gov/nsr/nsrlink.jsp?2005Ge06,B}{2005Ge06}, \href{https://www.nndc.bnl.gov/nsr/nsrlink.jsp?2008Ge07,B}{2008Ge07}, \href{https://www.nndc.bnl.gov/nsr/nsrlink.jsp?2011Ma48,B}{2011Ma48}: CaO(p,\ensuremath{^{\textnormal{19}}}Ne) E=1.4 GeV; \ensuremath{^{\textnormal{19}}}Ne was produced from spallation reaction in a CaO target using the}\\
\parbox[b][0.3cm]{17.7cm}{ISOLDE on-line isotope separator; ionized and accelerated \ensuremath{^{\textnormal{19}}}Ne to 60 keV; and mass analyzed it by the ISOLDE GPS separator.}\\
\parbox[b][0.3cm]{17.7cm}{(\href{https://www.nndc.bnl.gov/nsr/nsrlink.jsp?2008Ge07,B}{2008Ge07}) measured mass of \ensuremath{^{\textnormal{19}}}Ne using a Penning trap. (\href{https://www.nndc.bnl.gov/nsr/nsrlink.jsp?2005Ge06,B}{2005Ge06}), (\href{https://www.nndc.bnl.gov/nsr/nsrlink.jsp?2008Ge07,B}{2008Ge07}), and (\href{https://www.nndc.bnl.gov/nsr/nsrlink.jsp?2011Ma48,B}{2011Ma48}) measured isotope shift by}\\
\parbox[b][0.3cm]{17.7cm}{exciting the atomic meta-stable 2\textit{p}\ensuremath{^{\textnormal{5}}}3\textit{s}[3/2]\ensuremath{_{\textnormal{2}}} state to the excited 2\textit{p}\ensuremath{^{\textnormal{5}}}3\textit{p}[3/2]\ensuremath{_{\textnormal{2}}} atomic state using the collinear fast-beam laser}\\
\parbox[b][0.3cm]{17.7cm}{spectroscopy combined with ion detection of optical resonance. (\href{https://www.nndc.bnl.gov/nsr/nsrlink.jsp?2005Ge06,B}{2005Ge06}) deduced magnetic dipole moment of \ensuremath{^{\textnormal{19}}}Ne. (\href{https://www.nndc.bnl.gov/nsr/nsrlink.jsp?2008Ge07,B}{2008Ge07},}\\
\parbox[b][0.3cm]{17.7cm}{\href{https://www.nndc.bnl.gov/nsr/nsrlink.jsp?2011Ma48,B}{2011Ma48}) deduced isotope shift, and the change in mean square charge radius of \ensuremath{^{\textnormal{19}}}Ne relative to that in \ensuremath{^{\textnormal{20}}}Ne. (\href{https://www.nndc.bnl.gov/nsr/nsrlink.jsp?2008Ge07,B}{2008Ge07})}\\
\parbox[b][0.3cm]{17.7cm}{deduced the charge radius of \ensuremath{^{\textnormal{19}}}Ne. Comparison with several theoretical model calculations is discussed in (\href{https://www.nndc.bnl.gov/nsr/nsrlink.jsp?2011Ma48,B}{2011Ma48}).}\\
\vspace{12pt}
\underline{$^{19}$Ne Levels}\\

\parbox[b][0.3cm]{17.7cm}{\makebox[1ex]{\ensuremath{^{\hypertarget{NE41LEVEL0}{a}}}} From the \ensuremath{^{\textnormal{19}}}Ne Adopted Levels.}\\
\vspace{0.5cm}
\end{center}
\clearpage
\newpage
\pagestyle{plain}
\section[References]{ }
\vspace{-30pt}
%
\end{document}